\documentclass[12pt]{book}
\usepackage[letterpaper]{meta-donnees}
\usepackage[margin=1.0in]{geometry}
\usepackage{hyperref}
\usepackage{amsmath,amssymb}
\usepackage{graphicx}
\usepackage{cite} 
\usepackage{rotating} 
\usepackage{placeins} 
\usepackage{bm}
\usepackage[nottoc,numbib]{tocbibind} 
\usepackage{mathrsfs}
\usepackage{slashed} 
\usepackage{epigraph} 
\usepackage[T1]{fontenc} 

\usepackage[normalem]{ulem} 

\newcommand{\beq}{\begin{equation}}
\newcommand{\eeq}{\end{equation}}

\def\lsim{\mathrel{\raise.3ex\hbox{$<$\kern-.75em\lower1ex\hbox{$\sim$}}}}
\def\gsim{\mathrel{\raise.3ex\hbox{$>$\kern-.75em\lower1ex\hbox{$\sim$}}}}

\def\stilde{\widetilde}

\def\CU{C_U}
\def\CD{C_D}
\def\CL{C_L}
\def\CV{C_V}
\def\CG{C_g}
\def\CP{C_\gamma}
\def\cu{\CU}
\def\cd{\CD}
\def\cv{\CV}
\def\cg{\CG}
\def\cp{\CP}
\def\cl{\CL}
\def\sina{\sin\alpha}
\def\cosa{\cos\alpha}
\def\tanb{\tan\beta}
\def\sinb{\sin\beta}
\def\cosb{\cos\beta}
\def\cotb{\cot\beta}
\def\mw{m_W}
\def\mz{m_Z}
\def\gam{\gamma}
\def\anti{\overline}
\def\ie{{\it i.e.}}
\def\eg{{\it e.g.}}
\def\etc{{\it etc.}}
\def\dcg{\Delta \CG}
\def\dcp{\Delta \CP}
\def\gev{~{\rm GeV}}
\def\Eq#1{Eq.~(\ref{#1})}
\def\what{\widehat}
\def\fbi{~{\rm fb}^{-1}}
\def\chisq{\chi^2}
\def\chimin{\chi^2_{\rm min}}
\def\ggf{{\rm ggF}}
\def\dof{{\rm d.o.f.}}
\def\cpb{\anti \CP}
\def\cgb{\anti \CG}
\def\tth{{\rm ttH}}
\def\vbf{{\rm VBF}}
\def\vh{{\rm VH}}
\def\dchisq{\Delta\chisq}
\def\lam{\lambda}

\def\mneut{m_{\tilde{\chi}^0_1}}
\def\tev{~{\rm TeV}}
\def\ma{M_{A^0}}
\def\mcha{m_{\tilde{\chi}_1^\pm}}
\def\mstaua{m_{\stau_1}}
\def\stau{\tilde{\tau}}
\def\neut{{\tilde{\chi}^0_1}}
\def\mev{\text{MeV}}
\def\mchar{m_{\tilde{\chi}^\pm_1}}
\def\charg{{\tilde{\chi}^+_1}}
\def\lsp{{\tilde{\chi}^0_1}}

\def\sigmabar{\overline\sigma}

\def\chitz{\widetilde{\chi}^0_2}
\def\chiz{\widetilde{\chi}^0_1}
\def\chipm{\widetilde{\chi}^\pm_1}
\def\omhsq{\Omega_{\tilde \chi_1^0}h^2}
\def\sigsi{\sigma^{\rm SI}(\chiz p)}

\newcommand{\madanalysis}{{\tt MadAnalysis~5}}
\newcommand{\sampleanalyzer}{{\tt SampleAnalyzer}}
\newcommand{\python}{{\tt Python}}
\newcommand{\cpp}{{\tt C++}}

\newcommand{\be}{\begin{equation}}
\newcommand{\ee}{\end{equation}}
\def\bsp#1\esp{\begin{split}#1\end{split}}

\newcommand{\nn}{\nonumber}

\newcommand{\WW}{W\hspace{-.5pt}W}
\newcommand{\BB}{B\hspace{-.5pt}B}
\newcommand{\WB}{W\hspace{-1.5pt}B}

\renewcommand{\O}{\mathcal O}

\newcommand\invisiblesection[1]{
  \refstepcounter{chapter}
  \chaptermark{#1}}

\def\br{{\mathcal B}}
\def\brinv{\br_{\rm inv}}
\def\st{\tilde{t}}
\def\cnone{\tilde \chi_1^0}

\begin{document}

\Specialite{Physique th\'eorique}
\Arrete{7 ao\^ut 2006}
\Auteur{B\'eranger Dumont}
\Directeur{Sabine Kraml}
\Laboratoire{LPSC Grenoble}
\EcoleDoctorale{l'\'ecole doctorale de physique}         
\Titre{Higgs, supersymmetry and dark matter after Run~I of the LHC}
\Depot{24/09/2014}       

\Jury{
\UGTPresident{Pr.~Johann Collot}{Professeur, LPSC Grenoble, Universit\'e de Grenoble}

\UGTRapporteur{Dr.~Abdelhak Djouadi}{Directeur de recherche CNRS, LPT Orsay, Universit\'e Paris-Sud XI}
\UGTRapporteur{Pr.~Manuel Drees}{Professor, Bethe Center for Theoretical Physics, Universit\"at Bonn}

\UGTExaminatrice{Dr.~Genevi\`eve B\'elanger}{Directeur de recherche CNRS, LAPTh, Universit\'e de Savoie} 
\UGTExaminatrice{Dr.~Veronica Sanz}{Lecturer, University of Sussex}
\UGTExaminateur{Dr.~Cyril Hugonie}{Ma\^itre de conf\'erences, LUPM, Universit\'e Montpellier 2}

\UGTDirecteur{Dr.~Sabine Kraml}{Directeur de recherche CNRS, LPSC Grenoble, Universit\'e de Grenoble}
}

\MakeUGthesePDG

\pagenumbering{roman}
\setcounter{page}{2}

\ 
\clearpage

\setlength{\epigraphwidth}{.75\textwidth}
\renewcommand{\textflush}{flushepinormal}

\vspace*{\fill} 
\epigraph{And here, poor fool! with all my lore \\ I stand, no wiser than before}{Johann Wolfgang von Goethe, {\it Faust}, 1808}

\setlength{\epigraphwidth}{.75\textwidth}

\epigraph{We shed as we pick up, like travellers who must carry everything in their arms, and what we let fall will be picked up by those behind. The procession is very long and life is very short. We die on the march. But there is nothing outside the march so nothing can be lost to it.}{Tom Stoppard, {\it Arcadia}, Act I, Sc.\ 3, 1993}

\vspace*{\fill}

\chapter*{Remerciements}

La thèse est une expérience dont on ne peut désapprendre les principaux enseignements. Elle laisse le temps de se développer et de réaliser une diversité de projets, mais passe également en un clin d'\oe{}il. Ainsi, à l'heure du bilan, il m'est difficile de trouver le juste équilibre pour remercier et rendre hommage à tous ceux qui m'ont aidé et ont accompagné mon développement de chercheur.

Je ne peux commencer, bien sûr, que par Sabine. Le mérite de ne pas m'avoir fait regretter mon exil dans les montagnes grenobloises pendant trois ans et demi, pour le stage de Master 2 puis pour la thèse, lui revient très largement. Je pense pouvoir dire que son rôle de \guillemotleft\ conseillère de thèse \guillemotright\ a été rempli à la perfection, grâce à son dynamisme et à sa détermination à donner les meilleures chances à ses étudiants (ainsi que sa compréhension pour mes horaires légèrement décalés).
Les nombreux workshops et conférences auxquels j'ai pu assister, et pour lesquels j'ai été invité par Sabine à participer dès les débuts de la thèse, ont été un grand atout. Au-delà de cela, j'ai pu apprécier le juste équilibre entre le souci apporté à la liberté et l'autonomie du jeune chercheur d'un côté, et le souci de la production scientifique de qualité de l'autre. C'est un encadrement idéal à l'émancipation et permettant de devenir un chercheur productif et visible, bien adapté à la compétition internationale. Mon principal regret est de ne pas avoir assez saisi les opportunités de faire de l'escalade, et de ne pas avoir fait la dégustation de vin prévue, mais cela pourrait se régler lors de mon retour à Grenoble !

Je tiens également à remercier tous mes collaborateurs pour les échanges fructueux que nous avons eus, ainsi que les membres de mon jury de thèse pour m'avoir fait l'honneur de leur présence et pour leurs remarques pertinentes sur le manuscrit de thèse, qui ont permis de l'améliorer. J'ai une pensée particulière pour Geneviève, présente dans le jury en tant qu'examinatrice mais qui aurait tout aussi bien pu apparaître en tant que co-directrice de thèse ! Nos collaborations régulières, dès mon arrivée à Grenoble, ont toujours été très agréables. Je voudrais par ailleurs saluer le professionnalisme et le plaisir que j'ai eu à travailler avec Jérémy Bernon, actuellement étudiant en thèse au LPSC. Je lui souhaite une expérience de recherche pleinement épanouissante. Enfin, je remercie John F.~Gunion pour son invitation à Davis ainsi que pour sa lettre de recommandation.

Le travail effectué lors de ces trois ans et demi au sein du groupe théorie du LPSC aurait cependant été bien fade sans les pauses cafés et les fameux birthday/name day/paper cakes. Sans prétention d'exhaustivité, je tiens à saluer Ak\i n, Dipan, Florian, Guillaume, Ingo (qui a également été mon tuteur de monitorat), Jérémy, Josselin, Quentin, Suchita, Tom\'a\v{s} et Zhao-ting, tous (ex-)membres du groupe, pour les discussions amicales. Il me faut également mentioner le très bon accueil du groupe lors de mon arrivée, et l'aide reçue à mes débuts, notamment de la part de Tom\'a\v{s}.

Je n'oublie pas que la thèse est aussi l'achèvement du cursus universitaire, commencé il y a maintenant huit ans à l'université Montpellier 2. Je garde un souvenir très riche du Master \guillemotleft\ Cosmos, Champs et Particules \guillemotright,\ et tout particulièrement des cours de qualité de Cyril Hugonie et Gilbert Moultaka en physique des particules, et de Henri Reboul et David Polarski en cosmologie. J'en profite pour faire un salut amical à tous les membres de ma promotion, en particulier Pablo, Michaël et Léo.

Enfin, l'achèvement de cette thèse est l'une des trop rares occasions qui me permettent de vraiment remercier ceux qui me sont chers et m'accompagnent depuis toujours. Le deuxième docteur de la famille vous doit beaucoup. Merci enfin, et par dessus tout, à Flora pour sa patience et son soutien sans faille, pour ses aides en anglais (britannique), et pour partager ma vie.

\tableofcontents

\mainmatter

\chapter{Introduction}

Since the confirmation of the existence of atoms in the 19th century, we know that ordinary matter is associated with a scale that determines its chemical properties. Below this scale, matter can no longer be divided without a dramatic modification of its properties, and has been shown not to obey the laws of classical mechanics.
The birth of modern particle physics, in the 20th century, is associated with the exploration of the subatomic scale and its description in terms of a relativistic quantum field theory. The development of particle physics is intimately related to high-energy physics because energetic collisions are required to probe the structure of matter ({\it i.e.}, smaller scales) in a direct way. To this aim, facilities where particles are accelerated up to the relativistic regime before colliding have been built. The outcome of the collisions is recorded by a detector located around the interaction point, and can be used to discriminate between models of particle physics.

As the center-of-mass energy of the collisions was pushed higher and higher, more and more short-lived particles that did not seem to constitute the ordinary matter were observed, with various masses, lifetimes, electric charges and decay patterns.
It became evident that this wealth of particles was hiding a more fundamental structure to be revealed after classification, as was done for the chemical elements by Mendeleev.
Detailed study of these new states lead to the identification of relevant quantum numbers, later identified with the number and flavor of quarks constituting the observed particles called hadrons (because bound by the strong interaction). Moreover, it was shown that quarks should possess an additional quantum number, the ``color'', and lie in the fundamental representation of a non-abelian gauge group associated to color, ${\rm SU}(3)_C$. The gauge theory based on ${\rm SU}(3)_C$, quantum chromodynamics (QCD), predicts massless vector gauge bosons called gluons and determines the dynamics of strong interactions. QCD was also shown to predict the confinement of quarks inside the hadrons as well as asymptotic freedom~\cite{Gross:1973id,Politzer:1973fx,Gross:1973ju,Gross:1974cs}, {\it i.e.}\ the decrease of the coupling constant at high energies, making it possible to test the theory via perturbative calculations.

In parallel to these findings, not only electromagnetic and strong interactions were observed but also weak interactions, in particular in $\beta$ decays of nuclei and in muon decays. Initially explained with four-fermion contact interactions with strength $G_F$ by Fermi in the 1930's~\cite{Fermi:1934hr}, it was later identified with a vector minus axial vector tensor structure, $V-A$~\cite{Feynman:1958ty}. However, since the four-fermion operator has mass dimension 6 it is an effective operator, and must be suppressed by a factor of $\Lambda^2$, where $\Lambda$ is the scale where new interactions should appear. Assuming coefficients of order unity, this leads to
\begin{equation}
G_F = \frac{{\cal O}(1)}{\Lambda^2} \approx 10^{-5}~{\rm GeV}^{-2} \Rightarrow  \Lambda \approx 300~{\rm GeV} \,.
\end{equation}
This simple dimensional analysis hints the true scale of weak interactions. A complete description of the weak interactions only came in the 1960--1970's~\cite{Glashow:1961tr,Weinberg:1967tq,Salam:1968rm,Glashow:1970gm}, after unification with the electromagnetic interaction (collectively called electroweak interaction), and was the last major milestone in the formulation of the Standard Model (SM) of particle physics. It predicted neutral weak currents in addition to the charged weak currents that were observed so far, and had a spectacular confirmation in 1974 in Gargamelle~\cite{Hasert:1973ff}, and finally with the discovery of the $W$ and $Z$ bosons in the UA1 and UA2 experiments at CERN in 1983~\cite{Arnison:1983rp,Arnison:1983mk,Banner:1983jy,Bagnaia:1983zx}. The SM was then completed at LEP, Tevatron and the LHC with precision measurements and the discovery of the top quark and Higgs boson.

The Lagrangian formulation of the SM and a brief summary of the (consequences of the) breaking of the electroweak symmetry will be given in Section~\ref{sec:intro-sm}. As we will see in Section~\ref{sec:intro-bsm}, the SM has limitations and in particular one can expect new physics to arise in the Higgs sector, motivating beyond the SM (BSM) theories. In Section~\ref{sec:susy}, a possible solution to these problems will then be discussed in the context of supersymmetric extensions to the SM. Finally, a short review on a class of dark matter (DM) candidates, the WIMPs, will be given in Section~\ref{sec:intro-dm}.

\section{A brief overview of the Standard Model} \label{sec:intro-sm}

The Standard Model is a quantum field theory based on the Poincar\'e symmetry and the ${\rm SU}(3)_C \times {\rm SU}(2)_L \times {\rm U}(1)_Y$ local symmetry. Its field content is given in Table~\ref{tab:smfields}.
The fermions of the SM, given in terms of left-handed Weyl spinors, can be divided into two categories: up- and down-type quarks, which are colored particles (they transform under the irreducible representation of ${\rm SU}(3)_C$), and charged leptons and neutrinos which are uncolored particles. Out of the three generations of fermions, that transform under the same representations but will later be given different masses, only one generation is shown in Table~\ref{tab:smfields}.
The gauge bosons associated with each of the three gauge groups of the SM (and that transform under the adjoint representation) are also given in Table~\ref{tab:smfields}. Finally, there is a complex scalar Higgs field, which transforms as a doublet under ${\rm SU}(2)_L$.

\renewcommand{\arraystretch}{1.5}
\begin{table}[ht]
\center
\begin{tabular}{c|c|c}
\hline
\multicolumn{2}{c|}{Field} & ${\rm SU}(3)_C$, ${\rm SU}(2)_L$, ${\rm U}(1)_Y$ \\
\hline
quarks & $q_L = (u_L\ d_L)$ & $({\bm 3}, {\bm 2}, \frac{1}{6})$ \\
($\times3$ generations) & $u_R^\dagger$ & $(\bar{{\bm 3}}, {\bm 1}, -\frac{2}{3})$ \\
& $d_R^\dagger$ & $(\bar{{\bm 3}}, {\bm 1}, \frac{1}{3})$ \\
\hline
leptons & $l = (\nu_e\ e_L)$ & $({\bm 1}, {\bm 2}, -\frac{1}{2})$ \\
($\times3$ generations) & $e_R^\dagger$ & $({\bm 1}, {\bm 1}, 1)$ \\
\hline \hline
gluon & $g$ & $({\bm 8}, {\bm 1}, 0)$ \\
$W$ & $(W^+\ W^0\ W^-)$ & $({\bm 1}, {\bm 3}, 0)$ \\
$B$ & $B^0$ & $({\bm 1}, {\bm 1}, 0)$ \\
\hline \hline
Higgs & $\Phi = (\phi^+\ \phi^0)$ & $({\bm 1}, {\bm 2}, \frac{1}{2})$ \\
\hline
\end{tabular}
\caption{Field content of the Standard Model. The fermionic (quark and lepton) content of the theory is given in terms of left-handed Weyl spinors. The normalization of the weak hypercharge $Y$ is chosen so that the electric charge is $Q = T_3 + Y$, where $T_3$ is the third component of the weak isospin.} \label{tab:smfields}
\end{table}
\renewcommand{\arraystretch}{1.0}

The Lagrangian of the SM can be written in a compact form as follows:
\begin{equation}
  {\cal L} = -\frac{1}{4}F_{\mu\nu}^aF^{a\,\mu\nu} + i \chi^\dagger \slashed{D} \chi + |D_\mu \Phi|^2 + (y_{ij} \chi_i \chi_j \Phi + {\rm h.c.}) - V(\Phi) \,, \label{eq:smlagr}
\end{equation}
where $\chi$ is a left-handed Weyl fermion from Table~\ref{tab:smfields}. It is the most general renormalizable Lagrangian based on the considered global and local symmetries and on the fields present in Table~\ref{tab:smfields}.
The first three terms of the Lagrangian contain the kinetic terms for the gauge, fermionic, and Higgs fields, respectively. In the second and third terms, the covariant derivative, $D_\mu$, comes from the requirement of gauge invariance, and encapsulates the gauge interactions of the fermions and of the Higgs field. The next-to-last term of the Lagrangian controls the Yukawa couplings of the Higgs to fermions. Writing down the ${\rm SU}(2)_L$ invariants explicitly, we get
\begin{equation}
  {\cal L}_{\rm Yukawa} = y_{ij} \chi_i \chi_j \Phi + {\rm h.c.} = 
  y^u_{ij} q_{L,i} \varepsilon \Phi^T u^\dagger_{R,j}  + y^d_{ij} q_{L,i} \Phi^\dagger d^\dagger_{R,j} + y^l_{ij} l_i \Phi^\dagger e^\dagger_{R,j} + {\rm h.c.} \,, \label{eq:yukawalagr}
\end{equation}
where $y^u_{ij}$, $y^d_{ij}$ and $y^l_{ij}$ are $3 \times 3$ complex matrices in flavor space.
Finally, the last term in Eq.~\eqref{eq:smlagr} is the Higgs potential, defined as
\begin{equation}
  V(\Phi) = \mu^2 \Phi^\dagger \Phi + \lambda (\Phi^\dagger \Phi)^2 \,, \label{eq:higgspot}
\end{equation}
where it can be noted that $\mu$ is the only dimensionful parameter of the SM.

The Lagrangian given in Eq.~\eqref{eq:smlagr} seems to be in gross contradiction with basic experimental results. Indeed, we do not observe two independent massless electrons ($e_L$ and $e_R$), and the photon is simply absent from the list of fields given in Table~\ref{tab:smfields}.
A simple solution would be to add explicit mass terms for the fermions and $W$ and $Z$ bosons in the Lagrangian, but this is not invariant under ${\rm SU}(2)_L$, hence would break the electroweak symmetry. Therefore, these fields need to be given mass in an indirect way. The simplest solution, realized in the SM, is the Higgs (or Brout--Englert--Higgs) mechanism~\cite{Englert:1964et,Higgs:1964ia,Higgs:1964pj,Guralnik:1964eu,Higgs:1966ev,Kibble:1967sv}, and involve the scalar field $\Phi$, whose potential is given in Eq.~\eqref{eq:higgspot}.

Taking $\mu^2 < 0$ and $\lambda > 0$ in Eq.~\eqref{eq:higgspot}, the Higgs potential takes on a ``mexican hat'' shape, with a local maximum at zero field value and degenerate global minima for $\Phi^\dagger \Phi = - \mu^2 / (2\lambda) \equiv v^2$, where $v$ is the vacuum expectation value (vev) of the Higgs field.
This results in a spontaneous breaking of the electroweak symmetry, while preserving the conservation of the electric charge related to the ${\rm U}(1)_{\rm e.m.}$ gauge group: ${\rm SU}(2)_L \times {\rm U}(1)_Y \to {\rm U}(1)_{\rm e.m.}$. Unlike the spontaneous breaking of a global symmetry that gives rise to massless Nambu--Goldstone modes~\cite{Nambu:1960xd,Goldstone:1961eq,Goldstone:1962es}, in the breaking of the electroweak symmetry three of the four would-be Goldstone bosons are ``eaten'' by the gauge bosons, and give longitudinal degrees of freedom ({\it i.e.}, masses) to the $W^+$, $W^-$ and $Z^0$ bosons (the $W^0$ and $B^0$ bosons mix after electroweak symmetry breaking to give a massive $Z^0$ boson and a massless photon),
\begin{equation}
M_W = \frac{g v}{2}\,, \qquad M_Z = \frac{\sqrt{g^2+g'^2}\,v}{2} \,.
\end{equation}
The remaining degree of freedom can be written as 
\begin{equation}
  \Phi = \frac{1}{\sqrt{2}} \begin{pmatrix} 0 \\ v + H(x) \end{pmatrix} \,,
\end{equation}
where $H$ is a real neutral scalar, the massive physical Higgs field.

In addition to giving masses to the weak gauge bosons, the Higgs mechanism also gives masses to the fermions. Indeed, after electroweak symmetry breaking the Yukawa terms in the Lagrangian (see Eq.~\eqref{eq:yukawalagr}) generate Dirac mass terms for the fermions with the replacement $\phi^0 \rightarrow (v + H)/\sqrt{2}$. A connection is thus made between two Weyl fermions, which form Dirac fermions, for the up- and down-type quarks and for the charged leptons: $(u_L\ u_R)$, $(d_L\ d_R)$, and $(e_L\ e_R)$, respectively. All fermion masses can be obtained after diagonalization of the corresponding $y_{ij}$ matrix, leading to $m_{f,i} = y_i \times v / \sqrt{2}$ (with $f = \{u,d,e\}$ and $i=1,2,3$). However, in general the up- and down-type Yukawa matrices cannot be diagonalized simultaneously, leading to a mismatch between the mass eigenstates (that propagate freely) and the weak eigenstate (where charged currents couple to the $u_L$ and $d_L$ fields).
This mismatch is specified with the Cabibbo--Kobayashi--Maskawa (CKM) matrix~\cite{Cabibbo:1963yz,Kobayashi:1973fv}, a $3 \times 3$ unitary matrix parameterized by three mixing angles and a CP-violating phase.

In total, the SM has 19 free parameters: nine fermion masses (six for the quarks, three for the charged leptons), four parameters for the CKM matrix, three gauge couplings ($g_s$, $g$ and $g'$ for ${\rm SU}(3)_C$, ${\rm SU}(2)_L$ and ${\rm U}(1)_Y$, respectively), $\mu^2$ and $\lambda$ (or, equivalently, $v$ and $m_H$), and the QCD vacuum angle $\theta_{\rm QCD}$ controlling CP violation in the strong sector.
However, since the discovery of neutrino oscillations~\cite{Fukuda:1999pp} we know that at least two neutrinos must have a mass. Therefore an extension of the SM where Majorana and/or Dirac mass terms are given to the neutrinos is required, the latter solution involving an additional neutrino field, $\nu_{eR}$, that is sterile (not charged under any of the SM gauge groups). Neutrino masses also lead to the equivalent of the CKM matrix in the lepton sector, the Pontecorvo--Maki--Nakagawa--Sakata (PMNS) matrix~\cite{Pontecorvo:1957cp,Maki:1962mu,Pontecorvo:1967fh}.

\begin{figure}[ht]
\begin{center}
\includegraphics[width=0.5\textwidth]{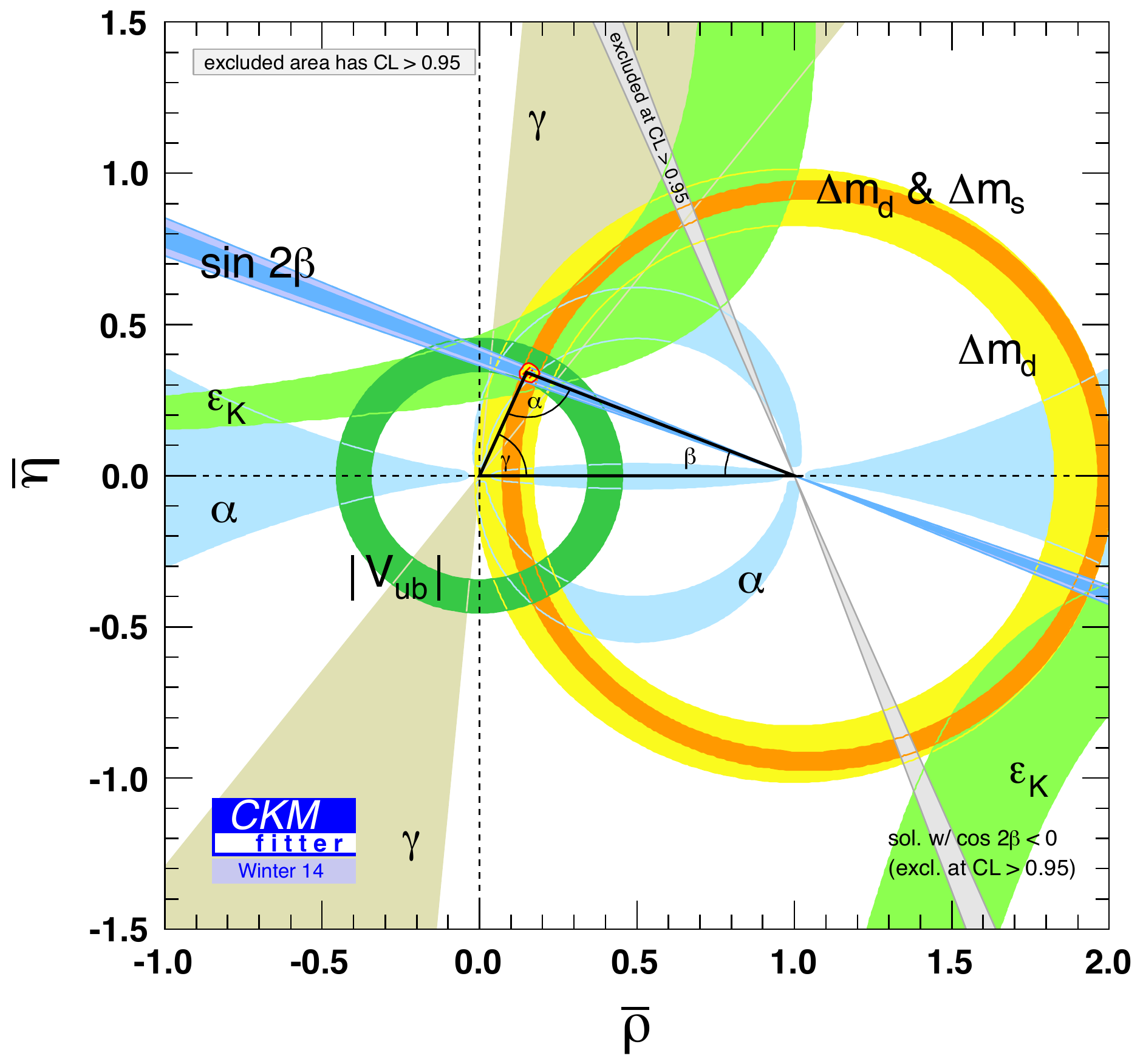}
\end{center}
\caption{Constraints on the CKM matrix in the $(\bar\rho, \bar\eta)$ plane as performed by the CKMfitter group and taken from Ref.~\cite{ckmfitter}.
The individual constraints from charmless semileptonic $B$ decays and $B \to \tau\nu$ ($|V_{ub}|$), from mass differences in the $B$ ($\Delta m_d$) and $B_s$ ($\Delta m_s$) neutral meson systems, and from CP violation in kaon systems ($\varepsilon_K$), in $B \to \psi K$ ($\sin 2\beta$), $B \to \pi\pi,\rho\pi,\rho\rho$ ($\alpha$) and $B^\pm \to DK^\pm$ ($\gamma$) are superimposed.}
\label{fig:CKMfit}
\end{figure}

The predictions of the SM can be tested in many ways. First, within the SM all flavor-changing processes in the quark sector must be explained with the CKM matrix, where moreover CP violation is parameterized with only one phase. BSM physics generically induces or contributes to flavor-changing and CP-violating processes, and might spoil the unitarity of the CKM matrix. For example, some of the flavor physics measurements can be shown in the $(\bar\rho, \bar\eta)$ plane (two parameters of the CKM matrix derived from the Wolfenstein parameterization~\cite{Wolfenstein:1983yz}) and can be checked for consistency. This was done by the CKMfitter group~\cite{Hocker:2001xe,ckmfitter} using the most recent experimental data and is shown in Fig.~\ref{fig:CKMfit} (see also global fits by the UTfit collaboration~\cite{Ciuchini:2000de,utfit}).
Excellent agreement is observed for $(\bar\rho,\bar\eta) \approx (0.15,0.34)$ from all considered measurements, as well as in the most general flavor fits.
Note that several flavor measurements constrain new physics much above the TeV scale, depending on the flavor structure of the theory~\cite{Bona:2007vi,Isidori:2010kg}. This is a strong indication that new physics must be either beyond the LHC reach, or with a non-generic flavor structure.
(Strong experimental bounds also exist in the lepton sector, in particular on the flavor-changing $\mu \to e\gamma$ process~\cite{Adam:2013mnn} and on the electric dipole moment of the electron~\cite{Baron:2013eja}.)
Finally, all searches for quark and lepton substructure (see, {\it e.g.}, Refs.~\cite{Aad:2013jja,Khachatryan:2014aka}), as well as searches for new chiral and vector-like fermions turned out negative.

The SM was also tested with high precision in the electroweak sector, in particular using observables at the $Z^0$ pole measured at LEP~\cite{ALEPH:2005ab}. This will be briefly presented in Section~\ref{sec:higgs-prelhc} (see in particular Fig.~\ref{fig:LEPew}; also Ref.~\cite{Baak:2014ora}). Moreover, the recent measurements of the Higgs boson performed at CERN's Large Hadron Collider (LHC) start to probe the Higgs sector of the SM with a good precision, see Chapter~2 from Section~\ref{sec:higgs-measlhc} onwards. In all cases, no significant deviation from the SM predictions was observed, and all direct searches for new gauge bosons and extra Higgs bosons also turned out negative so far. To date, the most serious discrepancy from the SM predictions in beam-based experiments may come from the the anomalous magnetic moment of the muon, $(g-2)_\mu$, where a $\sim 3\sigma$ discrepancy is found between the experimental measurement~\cite{Bennett:2006fi} and the predicted SM value~\cite{Hagiwara:2011af}. However, the treatment of the hadronic contributions to the magnetic moment, and their associated uncertainties, is still debated; new measurements and improved SM calculations should clarify the situation in the future.

In summary, significant efforts were made in testing predictions of the SM with high precision in the past twenty years in the different sectors of the theory. So far, all measurements show a good agreement with the SM predictions. This spectacular success is even more remarkable given the construction of the SM, directly obtained from symmetry principles and the fields content of Table~\ref{tab:smfields}. To say the least, the SM seems to be an excellent effective field theory at the electroweak scale, and possibly beyond.
	
\section{The need for BSM physics} \label{sec:intro-bsm}

Although the Standard Model successfully describes all phenomena observed in collider-based experiments so far, it cannot be thought of as a theory of everything. Obviously, a first limitation of the SM is that it does not incorporate gravity. Gravity is associated with the Planck scale, $M_{\rm Pl} \approx 10^{19}$~GeV, which sets an upper bound on the validity scale of the SM. Because of the extremely large difference between the Planck scale and the TeV scale that is directly accessible at the LHC---sixteen orders of magnitude---quantum effects of gravity, suppressed by powers of the Planck scale, should be out of reach of the LHC and of any foreseeable human-based experiment.  

Should we also expect beyond the SM physics below the Planck scale? As we will see, there are very good reasons to believe so. Although being a successful and consistent theory, the SM leaves many questions unanswered and exhibits intriguing features that call for an extension to the SM. This will be discussed in Section~\ref{sec:intro-bsm-hierarchy}. Certainly, the most pressing issue is the hierarchy problem in the Higgs sector, that gives a strong motivation for BSM physics around the electroweak scale. In addition to these theoretical problems, the SM is known to fail at explaining a number of astrophysical and cosmological observations. Of particular importance is the necessity of a BSM particle candidate to account for the density of dark matter that is observed. This will be presented in Section~\ref{sec:intro-bsm-obs}, together with a brief discussion on other observational problems.

\subsection{Theoretical problems: hierarchy and aesthetics} \label{sec:intro-bsm-hierarchy}

\begin{figure}[ht]
\begin{center}
\includegraphics[width=0.3\textwidth]{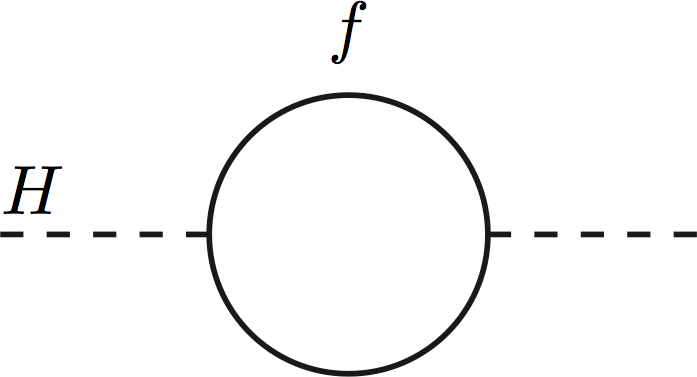}
\end{center}
\caption{One-loop self-energy correction to the Higgs mass due to a Dirac fermion $f$.}
\label{fig:higgscorr}
\end{figure}

What is commonly believed to be the most severe problem of the SM is the hierarchy problem in the Higgs sector, {\it i.e.}~the understanding of the very large difference between the electroweak and the Planck scale.
As was shown in Section~\ref{sec:intro-sm}, the Higgs mass is a free parameter of the model. However, it was known long before its discovery that it should be of the order of 100~GeV, from consistency reasons and from precision measurements in the electroweak sector (see Section~\ref{sec:higgs-prelhc}).
Therefore, a natural value for the $\mu^2$ parameter appearing in the Higgs potential would be of order $-(100~{\rm GeV})^2$. At the quantum level, however, one needs to take into account self-energy corrections to the Higgs mass, as shown in Fig.~\ref{fig:higgscorr} for a Dirac fermion $f$.

Following Ref.~\cite{sparticles}, the one-loop self-energy correction shown in Fig.~\ref{fig:higgscorr} can be written
\begin{equation}
\Pi^f_{hh}(0) = -2y_f^2 \int \frac{d^4k}{(2\pi)^4}\left[\frac{1}{k^2 - m_f^2} + \frac{2m_f^2}{(k^2 - m_f^2)^2}\right] \,,
\end{equation}
where it is manifest that the first term of the loop integral is quadratically divergent. Using momentum cutoff regularization, one finds $\delta m_H^2 \propto \Lambda^2$, where $\Lambda$ is the scale of new physics. Assuming that the SM remains valid up to the Planck scale ({\it i.e.}, taking $\Lambda = M_{\rm Pl}$) would result in a correction to the Higgs mass of the order of the Planck mass, in clear contradiction with the desired value of 100~GeV.
At this point, a remark is in order. While momentum cutoff regularization makes it directly explicit that the Higgs mass is sensitive to high scales, using dimensional regularization instead one would obtain a correction $\delta m_H^2 \propto m_f^2$.
Since the heaviest fermion in the SM is the top quark, with mass of about 173~GeV, this would only amount to a reasonable correction to the Higgs mass.

In both cases, however, the extreme sensitivity to high scales remains since the Higgs mass is driven by the mass of the heaviest particle 
that couples to the Higgs. This remains true even if the coupling is not direct, for example in the case of a heavy BSM particle that acquires mass in a different way and only couples to the Higgs via gauge bosons. Unlike the other particles of the SM, the mass of a scalar is not protected by any symmetry, explaining these disastrously large corrections. All known solutions to the hierarchy problem invoke new physics at a rather low scale (not too far from the electroweak scale) in order to avoid large fine-tuning cancellations. 
This is the strongest argument in favor of new physics at the TeV scale, accessible at colliders such as the LHC.

Possible alternatives would be {\it (i)}~there is no new physics up to the Planck scale, at which a yet unknown mechanism would protect the Higgs mass from receiving any large correction from Planck-scale physics, {\it (ii)}~to simply accept the enormous fine-tuning it implies, or {\it (iii)}~to explain the fine-tuning with the anthropic principle ({\it e.g.}, in the context of a multiverse, possibly motivated by the string theory landscape).
The importance to be given to the hierarchy problem is of course a personal choice, as long as the ``fine-tuning price'' to pay in order to not have physics around the electroweak scale is kept in mind. My personal opinion is that {\it (i)}~is highly unlikely already in light of the other limitations of the SM, given in the rest of the section, that also motivate new physics. The views {\it (ii)} and {\it (iii)} are acceptable and can be compared to the tuning of the cosmological constant, but are much less satisfactory and less appealing than solutions to the hierarchy problem. Incidentally, all three alternatives would greatly reduce the expectations for new physics at the TeV scale.

\begin{figure}[ht]
\begin{center}
\includegraphics[width=0.5\textwidth]{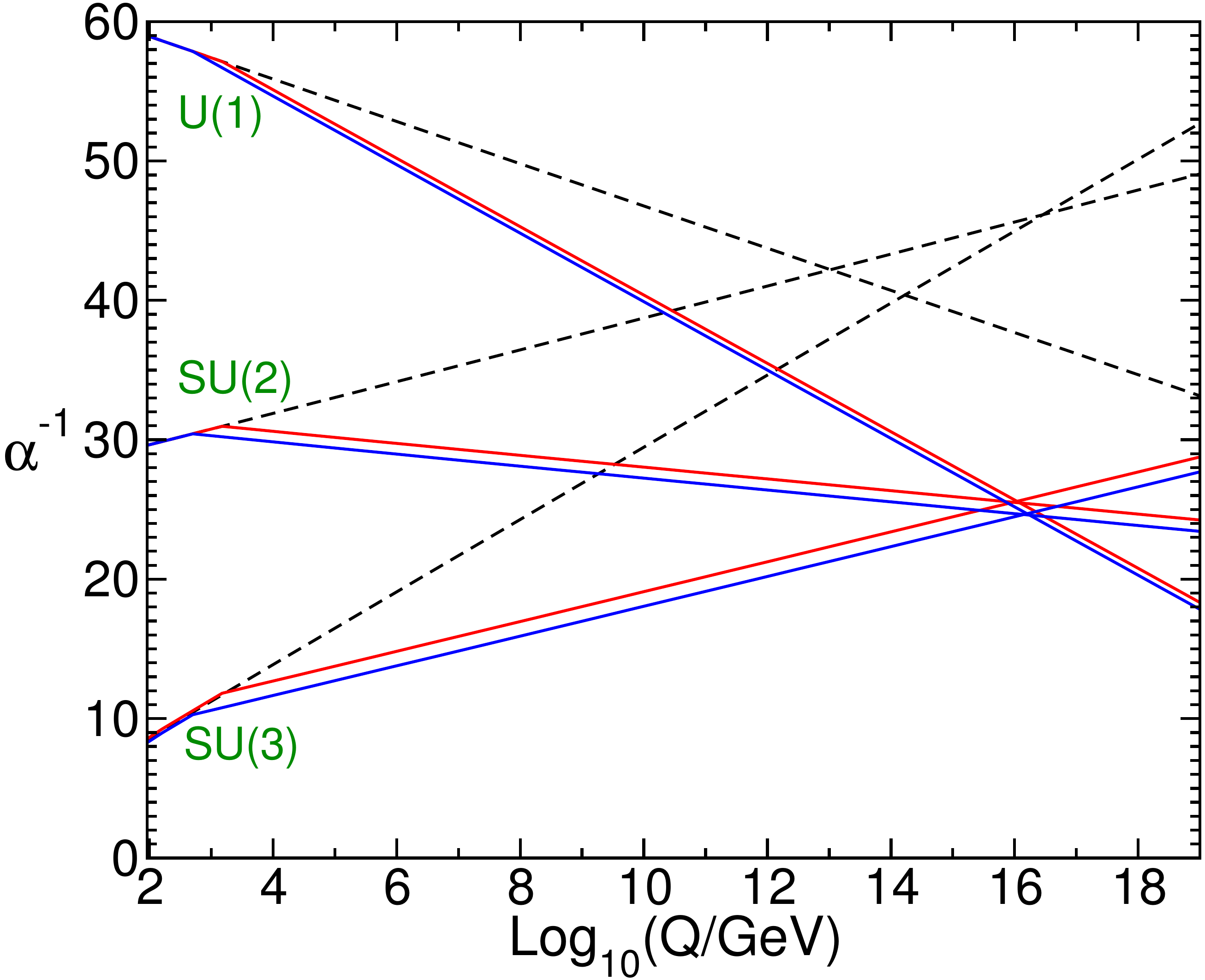}
\end{center}
\caption{Two-loop renormalization group evolution of the three gauge couplings, in terms of $\alpha_i^{-1} = (g_i^2/4\pi)^{-1}$ (where $g_1 = \sqrt{\frac{5}{3}}g'$, $g_2 = g$ and $g_3 = g_s$), taken from Ref.~\cite{Martin:1997ns}. The case of the Standard Model (MSSM) is shown in dashed (solid) lines. The blue and red lines correspond to a common threshold for the sparticle masses of 500~GeV and 1.5~TeV, respectively.}
\label{fig:rgerunning}
\end{figure}

Turning to other theoretical problems in the SM, several coincidences and intriguing features can be noted and seen as a motivation to go beyond the SM. First, after the unification of electricity and magnetism into electromagnetism, and the (incomplete) unification of the electromagnetic and weak interactions, it is natural to go a step further and try to unify the electroweak and strong interactions. In that case, the SM, based on the direct product of three gauge groups, would be embedded into a so-called Grand Unified Theory (GUT) based on a single gauge group. It was shown that the ${\rm SU}(5)$ and ${\rm SO}(10)$ groups contain the Standard Model as a subgroup and fit directly the fermionic content of Table~\ref{tab:smfields} into simple representations. This is a remarkable feature given the complicated fermionic structure of the theory, and can be seen as a clear hint of the existence of a GUT. However, in order to have a GUT the three gauge couplings need to unify at some energy scale. Using the renormalization group equations (RGEs) of the SM, the three couplings can be extrapolated to high energies.\footnote{In the SM, the normalization of the hypercharge is free since there is no requirement from anomaly cancellation and only the product $g' Y$ appears. On the other hand, assuming that the SM gauge groups are obtained from ${\rm SU}(5)$ and/or ${\rm SO}(10)$ fixes the normalization of the hypercharge, which is necessary for testing gauge coupling unification.} The evolution of the gauge couplings is shown in Fig.~\ref{fig:rgerunning} (taken from Ref.~\cite{Martin:1997ns}), where the dashed lines correspond to the SM. Clearly, the gauge couplings do not unify in the SM. This can be interpreted 
either as an objection against GUTs, or, more optimistically, as the indication that new physics is needed below $10^{13}$~GeV to deflect the running of the gauge couplings and make them unify. An example of such a model of new physics is given in the blue and red lines of Fig.~\ref{fig:rgerunning}. This corresponds to a supersymmetric model that will be presented in Section~\ref{sec:susy}.

The presence of tiny neutrino masses can also be seen as an argument in favor of new physics. While Dirac masses can be given to the neutrinos (with the addition of sterile neutrinos to the theory) in the exact same way as for the other fermions, this is often found to be unnatural because the upper bound of $m_{\nu} \lesssim 1$~eV implies $y_{\nu} \lesssim 10^{-11}$,  which is an uncomfortably small number. The alternative is to give Majorana masses to neutrinos, but they do not appear in the Lagrangian given in Eq.~\eqref{eq:smlagr} because they are not gauge invariant. A more natural explanation requires new physics, and it was found in particular that tiny neutrino masses can be explained from the large difference between the electroweak and the GUT scale (known as see-saw mechanisms), providing a supporting evidence for the existence of a GUT. Note that the Dirac or Majorana nature of the neutrinos can be tested experimentally from the (non-)detection of neutrinoless double beta decay, which only occurs if neutrinos are Majorana fermions (for recent experimental results, see Ref.~\cite{Albert:2014awa}).

Furthermore, in the SM not only the generation of neutrino masses is explained by an unnaturally small value or by new physics. In the QCD Lagrangian, a CP-violating term of the form $\theta_{\rm QCD} F_{\mu\nu}^a\tilde F^{\mu\nu\,a}$ should be present, where $\theta_{\rm QCD}$ is the QCD vacuum angle. While this parameter is expected to be of order 1, upper bounds on the electric dipole moment of the neutron~\cite{Baker:2006ts} translate into an upper bound on $\theta_{\rm QCD}$~\cite{Pospelov:1999ha} of about $10^{-10}$. This is called the strong CP problem. A solution to this problem, the Peccei--Quinn theory~\cite{Peccei:1977hh,Peccei:1977ur}, involves a new field, the axion, that may account for the dark matter in the Universe (see Section~\ref{sec:intro-bsm-obs} below).

Finally, the SM can be seen as an incomplete theory of electroweak and strong interactions because of the relatively large number of free parameters (19 without giving masses to the neutrinos), among which 13 are related to fermion masses and mixings. The flavor structure given by the CKM and PMNS matrices is non-trivial, but is not explained by any widely accepted flavor symmetry. A number of desirable features for describing our Universe also turn out to be ``accidentally'' fulfilled in the SM. This is the case of the conservation of the baryon number and of the lepton numbers. Also, the field content of the SM miraculously cancels all gauge anomalies. These are extra arguments in favor of the SM being a residual symmetry of a more fundamental theory.

Needless to say, a number of arguments mentioned above rely on the idea of having, ultimately, a ``theory of everything'' that would explain everything from symmetry principles and involve no more than a handful of free parameters. This is a normal attitude after the spectacular success of the SM, but gives no guarantee for BSM physics. When asking ourselves why things are the way they are, at a certain point there might no longer be an answer to be found---and at this point we should simply be satisfied with an accurate description of the microscopic world.

\subsection{Observational problems: dark matter and others} \label{sec:intro-bsm-obs}

While no significant deviation from the SM was found in any collider-based experiment, pushing the scale of new physics higher and higher, a number of astrophysical and cosmological observations clearly require new physics beyond the SM---without, however, hinting for any particular energy scale. A very important observation is the existence of dark matter, a weakly interacting form of matter that is present in our galaxy and in the Universe at large scales (for a review, see, {\it e.g.}, Refs.~\cite{Bertone:2004pz,Bertone:2010zz}). Some observational evidence will be presented in this section, while in Section~\ref{sec:intro-dm} a particular class of DM candidates, the Weakly Interacting Massive Particles (WIMPs), will be discussed.

\begin{figure}[ht]
\begin{center}
\includegraphics[width=0.50\textwidth]{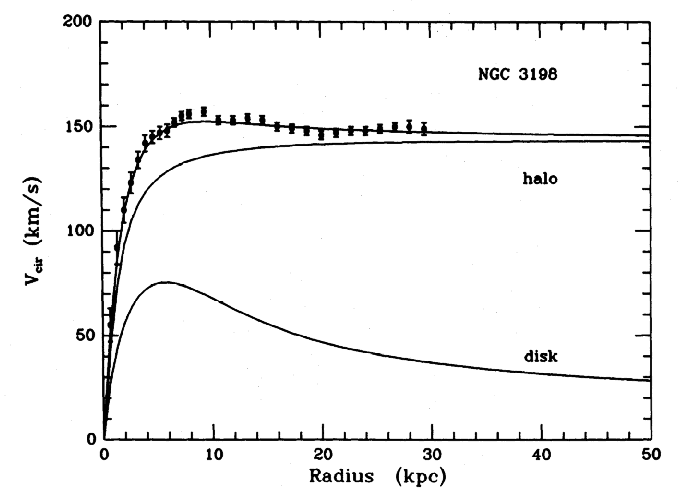}
\includegraphics[width=0.48\textwidth]{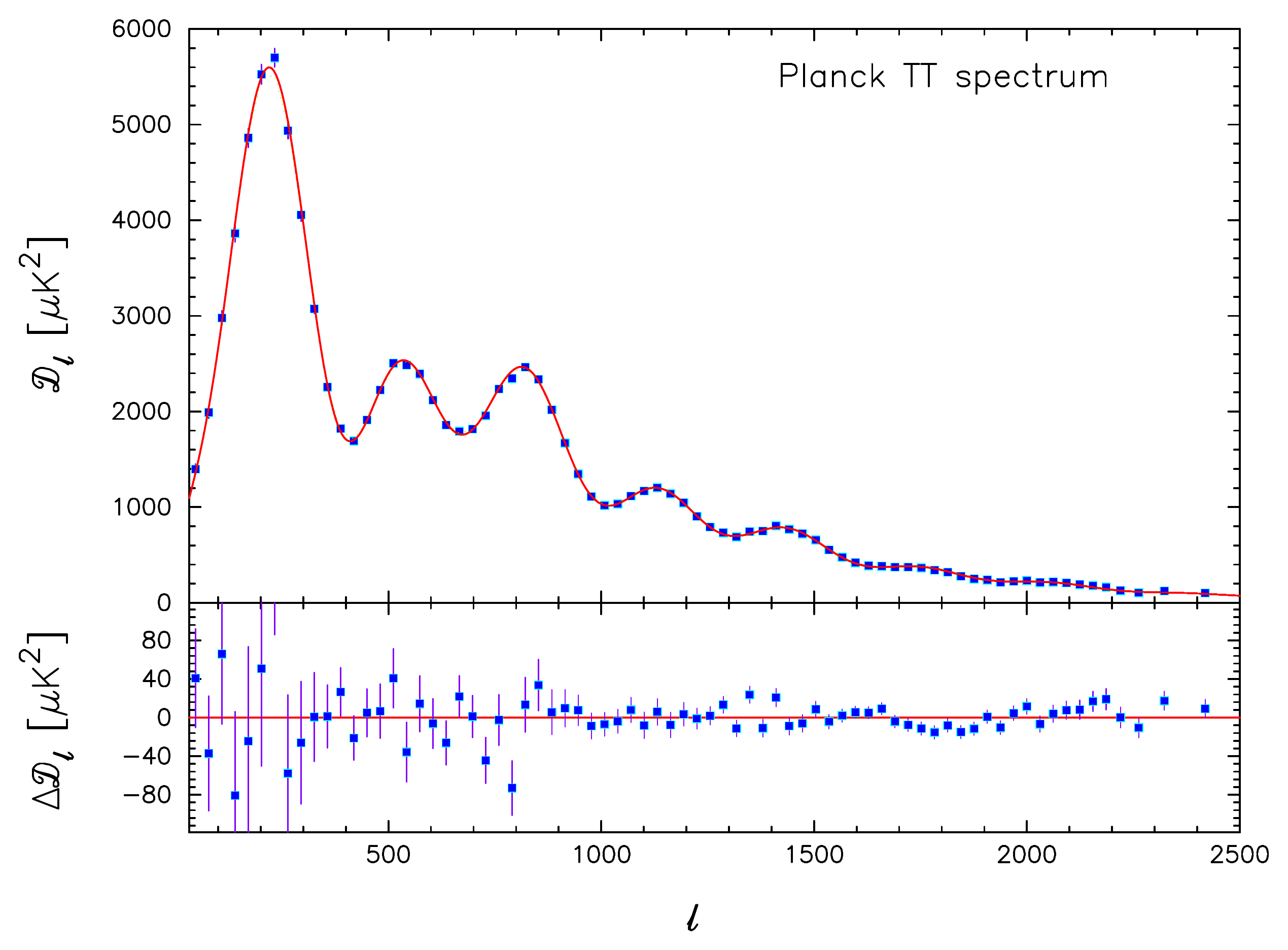}
\end{center}
\caption{Left: rotation curve of the galaxy NGC-3198, as was reported in Ref.~\cite{vanAlbada:1984js}. The expectations from visible matter only (``disk'') and dark matter only (``halo'') are given in addition to the velocity measurements. 
Right: Power spectrum of the CMB, taken from Ref.~\cite{Ade:2013zuv}, derived from the observations of the Planck satellite.}
\label{fig:cmbrot}
\end{figure}

The first evidence for DM was found in the 1930's in the observation of galaxies and clusters of galaxies. Gravitational effects inferred from the visible matter were found to be insufficient to explain the dynamics of these objects, leading to the assumption that most of the mass is not coming from the visible matter, but instead from dark matter~\cite{Zwicky:1933gu}.
In this assumption, each galaxy would be associated with a DM halo (possibly spherical) that extends beyond the visible matter, and that accounts for most of the mass of the galaxy~\cite{Rubin:1980zd}. The classic example of the rotation curve of the galaxy NGC-3198~\cite{vanAlbada:1984js} is shown in the left panel of Fig.~\ref{fig:cmbrot}. The ``disk'' line corresponds to the expectation from visible matter only (concentrated near the center of the galaxy), and neither fits the shape nor the normalization of the observed galaxy rotation curve. Another component is thus needed to explain the presence of a plateau away from the galactic center: the DM halo.

The presence of DM can also be probed at the cosmological scale from the study of the thermal anisotropies in the Cosmic Microwave Background (CMB) that were observed with high precision by the WMAP and Planck satellites~\cite{Komatsu:2010fb,Ade:2013zuv}. These 
temperature fluctuations provide information on the early Universe, including the density of DM. The angular power spectrum of the CMB anisotropies as obtained by Planck~\cite{Ade:2013zuv} is shown in the right panel of Fig.~\ref{fig:cmbrot}. The location and amplitude of the acoustic peaks, and their relative ratios, can be used to infer the value of parameters of a given cosmological model. Within the $\Lambda$CDM model, the standard model of cosmology, a very precise determination of the dark matter relic density is obtained from the Planck measurements (in combination with other cosmological observations): $\Omega_{\rm DM}h^2 = 0.1187 \pm 0.0017$~\cite{Ade:2013zuv}, where $h$ is the reduced Hubble constant and $\Omega_{\rm DM} \equiv \rho_{\rm DM} / \rho_c$ (the ratio of the observed DM density to the critical density of the Universe). This corresponds to about five times the baryonic density and is a very strong evidence in favor of dark matter.

From the first pieces of evidence in the 1930's to the present day, dramatic progress have been made in the observations. There is strong evidence for dark matter from the galactic to the cosmological scale. However, so far all direct observations rely on the gravitational interactions of dark matter. A sensible alternative is thus to modify gravitational interactions in order to account for the modified dynamics at and beyond the galactic scale (as well as observations using gravitational lensing), and explain the CMB anisotropies without dark matter. It is very challenging to explain the observations made at different scales with modified gravity instead of dark matter, but does not constitute a no-go theorem.
However, in models of modified gravity the presence of mass should necessarily be correlated with baryonic matter, while dark matter may decouple from baryonic matter. In the observation of a collision between two clusters, this difference can be used to discriminate between the two hypotheses. Indeed, when two such objects collide the average velocity of baryonic matter should considerably reduce, while the dark matter halo, if not made of strongly self-interacting particles, is expected to be mostly unaffected by the collision. The most famous example is the observation of the bullet cluster, from which it was shown that the spatial distribution of the baryonic matter (inferred from X-rays) is well separated from the mass (inferred from weak gravitational lensing)~\cite{Clowe:2003tk,Clowe:2006eq}. This is a clear argument in favor of the particle nature of dark matter.

So there should be a dark matter particle, but can such a particle be found? In the SM, the only stable and electrically neutral particles are the neutrinos. However, both CMB anisotropies and structure formation are not only sensitive to the density of dark matter, but also to its temperature. Cold (non-relativistic) dark matter is favored by the data, while neutrinos are hot dark matter that cannot constitute more than a small fraction of the dark matter.
New physics beyond the SM is thus needed to accommodate the observations. The possibility of axion dark matter has already been mentioned in Section~\ref{sec:intro-bsm-hierarchy},  but many other possibilities have been proposed over the past decades. In Section~\ref{sec:intro-dm}, an important class of dark matter candidates, the WIMPs, will be presented.

The dark matter problem is not the only observational issue of the SM. It is clear that all the matter we are made of and surrounded with is made of particles, and not of antiparticles. Moreover, so far the observations indicate that all objects in the observable Universe are made of matter and not of antimatter. This imbalance can be explained by an asymmetry between matter and antimatter, which is necessary for a successful baryogenesis ({\it i.e.}, the domination of matter over antimatter from processes in the early Universe).
This can be achieved if all three Sakharov conditions~\cite{Sakharov:1967dj} are satisfied: violation of the baryon number $B$, C and CP violation, and interactions outside of thermal equilibrium. In the SM, the first condition does not seem to be satisfied at first sight as there are accidental symmetries associated with the conservation of the baryon and lepton number ($B$ and $L$). However, the violation of the baryon number (while conserving $B-L$) is achieved through non-perturbative processes, for a saddle-point solution to the electroweak field equations called sphaleron. The second condition is satisfied from CP violation in the weak sector, while the third one requires that the electroweak phase transition is of strong first order, which is only the case if the Higgs is very light~\cite{Csikor:1998eu}, which was ruled out by LEP~\cite{Barate:2003sz}. BSM physics is thus needed for baryogenesis, which could be explained from an asymmetry in the leptonic sector (known as leptogenesis) that is converted into an asymmetry in the baryonic sector via, {\it e.g.}, sphalerons.

Finally, the last two problems do not really come from direct observation of phenomena that cannot be explained within the SM, but are well-established paradigms in cosmology that generally require new fields: dark energy and inflation. They have been invoked to explain, respectively, the accelerated expansion of the Universe at present times, and the flatness, homogeneity, and isotropy of the Universe. Nonetheless, these phenomena can simply be explained by the Friedmann--Lema\^itre equations which are solution to Einstein's equation in the case of a homogeneous and isotropic universe that is expanding or contracting, described by the Friedmann--Lema\^itre--Robertson--Walker metric. The acceleration of the Universe, motivating dark energy, can be explained with a non-vanishing $\Lambda g_{\mu\nu}$ term in Einstein's equation, where $\Lambda$ is known as the cosmological constant. Besides, the flatness of the Universe can be set by hand, and the homogeneity and isotropy at large scales might come from initial conditions at the Planck scale.

However, there are good reasons to believe that these direct explanations to the accelerated expansion, and to the flatness, homogeneity, and isotropy of the Universe are not really satisfactory. In order to match the observed expansion of the Universe, the cosmological constant $\Lambda$---that can either be seen as a geometrical factor or as a contribution to the stress--energy tensor---needs to be fine-tuned to an extremely small value, of the order of $10^{-52}$~m$^2$. Several proposals for a dynamical origin of the accelerated expansion of the Universe, involving new fields, have been made.
Moreover, the observed flatness of the Universe (corresponding to $\rho_{\rm tot} \approx \rho_c$, the total energy density of the Universe being close to its critical density) imply that in the early Universe the total energy density was extremely close to the critical density, requiring a fine-tuning of the cosmological parameters.
Finally,  the homogeneity and uniformity of the Universe at large scales is difficult to understand as regions of the sky separated by more than about one degree could not have been in causal contact.
A solution of the flatness, homogeneity and uniformity problems is inflation, a period of exponential expansion in the very early universe that involved an inflaton field. Recently, a signal was reported by the BICEP2 experiment in the search for inflationary gravitational waves in the polarization of the CMB~\cite{Ade:2014xna}, which would hint at an inflation scale of about $10^{16}$~GeV (remarkably close to the GUT scale in the MSSM, see next section). This claim should be clarified in light of a precise estimate of the foreground dust contamination in the coming months and years.

\section{Supersymmetry} \label{sec:susy}

The SM remains unchallenged in collider-based experiments, but in the previous section we have seen that there are a number of motivations for going beyond the SM, from theoretical arguments as well as from cosmological observations.
In light of the TeV energy scale being explored by the LHC, solutions to the hierarchy problem deserve special care as they should involve new particles and phenomena near the electroweak scale. In my opinion, this represents the greatest hope to shed light on BSM physics in the coming years, while all other discussed problems may or may not have anything to do with the TeV scale.

In Section~\ref{sec:intro-bsm-hierarchy}, we have seen that quadratic divergences appear in the contribution from fermions (and analogously from the gauge and Higgs bosons) to the self-energy correction to the Higgs mass. This extreme sensitivity to high scale physics can be canceled to a great extent by BSM particles with the appropriate couplings, but these cancellations generally hold only at a given order in perturbation theory, and reintroduce the hierarchy problem at the next order. In order to cancel the large loop corrections in a systematic way at all orders in perturbation theory, a new symmetry is needed, and is called supersymmetry.

Supersymmetry (SUSY) is an extension of the Poincar\'e symmetry that groups fermions and scalars into chiral supermultiplets, and vector bosons and fermions into gauge supermultiplets. The Coleman--Mandula no-go theorem~\cite{Coleman:1967ad}, that prevents from mixing the Poincar\'e group and gauge groups, is circumvented with the introduction of anticommuting spinor generators.  
The extension of the Coleman--Mandula theorem to the case of supersymmetry is the Haag--Lopuszanski--Sohnius theorem~\cite{Haag:1974qh}.
In $N=1$ supersymmetry (which will be considered in this thesis), two spinor generators, $Q$ and $Q^\dagger$, are introduced and transform a boson into a fermion and vice versa.

Supersymmetric extensions of the SM have been studied intensively for more than thirty years. The main reason 
is that SUSY elegantly solves the hierarchy problem by canceling all quadratic divergences, providing the symmetry that was missing to protect the Higgs mass from receiving large corrections. In a supersymmetric extension of the SM, the right amount of scalar degrees of freedom is introduced and cancels the quadratic divergences coming from fermions at all orders in perturbation theory.
A number of other problems can easily be solved in a supersymmetric context. In Section~\ref{sec:intro-bsm-hierarchy}, we have seen that there are good motivations for the existence of a GUT. However, in the SM the gauge couplings do not unify. In the minimal supersymmetric extension of the SM, the Minimal Supersymmetric Standard Model (MSSM), if supersymmetric partners of the SM particles are around the TeV scale, the unification of the gauge couplings is successfully achieved at a scale of about $2 \times 10^{16}$~GeV~\cite{Ellis:1990wk,Amaldi:1991cn,Langacker:1991an,Giunti:1991ta}, as can be seen in Fig.~\ref{fig:rgerunning} (solid lines).
Moreover, a dark matter candidate emerges naturally in supersymmetric extensions~\cite{Goldberg:1983nd,Ellis:1983ew}, as will be seen in Section~\ref{sec:susy-mssm}.

In Section~\ref{sec:susy-lagrangian}, the construction of a supersymmetric Lagrangian will be briefly reviewed together with the issue of SUSY breaking. The MSSM Lagrangian and some direct phenomenological consequences will then be seen in Section~\ref{sec:susy-mssm}. This section aims at introducing notations that will be used in the rest of the thesis when testing the MSSM against the results from the LHC and dark matter experiments. Reviews on supersymmetric extensions of the SM can be found in, {\it e.g.}, Refs.~\cite{Martin:1997ns,sparticles}, on which part of this section was based. In this section, most notations correspond to the ones of Ref.~\cite{Martin:1997ns}.

\subsection{Supersymmetric Lagrangians} \label{sec:susy-lagrangian}

Since supersymmetry is an extension of the space-time symmetry, a supersymmetric model is most naturally defined in superspace, where in addition to the usual ``bosonic'' space-time coordinates, $x^\mu$, four fermionic coordinates are added and expressed as two complex anticommuting two-component spinors: $\theta^\alpha$ and $\theta^\dagger_{\dot\alpha}$ (where $\alpha$ and $\dot\alpha = 1,2$). In this formalism, a chiral superfield can be conveniently expressed as
\begin{equation}
\Phi(y,\theta,\theta^\dagger) =
\phi(y) + \sqrt{2}\theta \psi(y) + \theta\theta F(y) \,,
\label{eq:chiralsuperf}
\end{equation}
where $y^\mu \equiv x^\mu + i\theta^\dagger \sigmabar^\mu \theta$ (with $\sigmabar = ({\bf 1}, -\sigma^i)$). $\phi$ and $\psi$ are a complex scalar and a left-handed Weyl fermion, respectively, and $F$ is a complex scalar auxiliary field (that will not be given a kinetic term). In turn, a gauge superfield is given by
\begin{equation}
V(x,\theta,\theta^\dagger) =
\theta^\dagger \sigmabar^\mu \theta  A_\mu(x) 
+ \theta^\dagger \theta^\dagger \theta \lambda(x) 
+ \theta \theta  \theta^\dagger \lambda^\dagger(x)  
+ \frac{1}{2} \theta\theta \theta^\dagger \theta^\dagger D(x) \,,
\label{eq:gaugesuperf}
\end{equation}
in the Wess--Zumino supergauge. $A_\mu$ and $\lambda$ are a gauge boson and a Weyl fermion, respectively, and $D$ is a bosonic auxiliary field.

One can write a supersymmetric Lagrangian based on superfields after integration over the fermionic coordinates,
\begin{equation}
{\cal L} = \int d^2\theta d^2\theta^\dagger S(x,\theta,\theta^\dagger) \,,
\end{equation}
for a given superfield $S$. For chiral and gauge superfields, $\Phi$ and $V$, one usually defines
\begin{align}
[\Phi]_F + {\rm c.c.} &=  \int d^2\theta d^2\theta^\dagger
\left[
\delta^{(2)}(\theta^\dagger)\Phi(x,\theta,\theta^\dagger) +
\delta^{(2)}(\theta)\Phi^*(x,\theta,\theta^\dagger)
\right] = F + F^* \,,
\label{eq:ftermlagr}
\\
[V]_D &= \int d^2\theta d^2\theta^\dagger V(x,\theta,\theta^\dagger) = \frac{1}{2}D \,.
\label{eq:dtermlagr}
\end{align}
These are called $F$- and $D$-terms contributions to the Lagrangian. In building a supersymmetric Lagrangian, one should usually consider that the superfields $\Phi$ and $V$ appearing in Eqs.~\eqref{eq:ftermlagr}--\eqref{eq:dtermlagr} are not fundamental superfields (as in in Eqs.~\eqref{eq:chiralsuperf}--\eqref{eq:gaugesuperf}), but instead combinations of the fundamental fields.

A general Lagrangian for a gauge theory can finally be written as
\begin{equation}
{\cal L} =  \frac{1}{4} 
\left [
{\cal W}^{a\,\alpha} {\cal W}^a_\alpha 
\right ]_F
+ {\rm c.c.}
+
\left [
\Phi^{*i} (e^{2g_a T^a V^a})_i{}^j \Phi_j \right]_D
+ 
\left ( [W(\Phi_i)]_F + {\rm c.c.} \right ) \,.
\label{eq:lagrsuperf}
\end{equation}
The first term contains ${\cal W}$, a supersymmetric generalization of the field strength tensor ($a$ labels elements of the adjoint representation of the gauge group and $\alpha$ is a spinor index).
Kinetic terms and gauge interactions for the chiral superfields are given with the second term, from the expansion of the exponential. Finally, $W$ is the superpotential, an important object that sets all interactions between scalars and fermions of the chiral superfields (including the scalar potential of the theory). It is a holomorphic function of the chiral superfields ({\it i.e.}, a function of the $\Phi_i$ but not $\Phi^*_i$). Considering only renormalizable terms, its general form is $W = L^i \Phi_i + \frac{1}{2} M^{ij}\Phi_i\Phi_j + \frac{1}{6}y^{ijk} \Phi_i\Phi_j\Phi_k$.

An immediate prediction of supersymmetry is that equal masses are given to the scalar and fermion fields within a chiral superfield. If that were true, it would be extraordinarily easy to observe SUSY partners of the SM fermions, and for instance the scalar partners of the electron would have a mass of 511~keV. This is why SUSY cannot be an exact symmetry and needs to be broken in any realistic supersymmetric extension of the SM.
The breaking of supersymmetry and the mediation of SUSY breaking to the superfields associated with the SM particles is an important topic that generated a lot of activity. There are different possibilities for obtaining a viable particle spectrum, that may induce a very different phenomenology at colliders and different dark matter properties. Clearly, if SUSY partners are discovered at the LHC a prime goal will be to understand supersymmetry breaking.

In order to be viable, SUSY breaking must be realized in a hidden sector, {\it i.e.}\ from fields that do not share direct interactions with the superfields that play a phenomenological role (and which make the visible sector). 
The scale of SUSY breaking cannot be obtained from the masses and mixings of the supersymmetric partners of the SM particles in a general way, as it depends on how SUSY breaking is mediated to the visible sector. Possible solutions include Planck-scale and gauge interactions, and have important phenomenological implications. Taken as a local symmetry (known as supergravity), supersymmetry predicts a spin-$3/2$ massless fermion associated with the graviton: the gravitino. Supersymmetry breaking involves (at least) one scalar field, to which is associated a fermionic partner, the goldstino. Analogously to the generation of $W$ and $Z$ boson masses after electroweak symmetry breaking in the SM, the gravitino absorbs the would-be goldstino and acquires mass after SUSY breaking. This generalization of the Higgs mechanism is known as the super-Higgs mechanism.
If the gravitino is the lightest supersymmetric particle and if $R$-parity is conserved (see Section~\ref{sec:susy-mssm}), it plays a significant phenomenological role. This is expected, in particular, in gauge mediated SUSY breaking. If, on the other hand, the gravitino is not the lightest particle of the SUSY spectrum it is usually not taken into account at all. Indeed, its interactions with superfields of the visible sector are suppressed by a power of the SUSY breaking scale, and can generally be neglected for the phenomenology at colliders. The latter assumption will be made in the rest of this thesis.

It is possible to encompass all possibilities for SUSY breaking and mediation to the visible sector using an effective explicit parameterization of SUSY breaking, denoted as {\it soft} because it does not reintroduce quadratic divergences, whose cancellation was the main motivation for a supersymmetric extension of the SM. The soft SUSY breaking terms can be written
\begin{equation}
{\cal L}_{\rm soft} =
-\left (
\frac{1}{2} M_a\, \lambda^a\lambda^a 
+ \frac{1}{6}a^{ijk} \phi_i\phi_j\phi_k 
+ \frac{1}{2} b^{ij} \phi_i\phi_j 
\right )
+ {\rm c.c.}
- (m^2)_j^i \phi^{j*} \phi_i \,.
\label{eq:lagrsoft}
\end{equation}
This introduces mass splittings between the fields within a superfield, as well as trilinear couplings for the scalars.

\subsection{The MSSM} \label{sec:susy-mssm}

Let us now turn to the case of the minimal phenomenologically viable supersymmetric extension of the SM, the Minimal Supersymmetric Standard Model or MSSM. The superfield content of the MSSM is very simple: all fermion fields shown in Table~\ref{tab:smfields} are promoted to chiral superfields, and all gauge bosons are promoted to gauge superfields. In the Higgs sector, however, a non-minimal extension is needed and two chiral superfields are introduced, $H_u$ and $H_d$, with hypercharge of $+\frac{1}{2}$ and $-\frac{1}{2}$, respectively.
This can be understood in two ways. First, by construction the superpotential $W(\Phi_i)$ appearing in Eq.~\eqref{eq:lagrsuperf} is a function of the chiral superfields that is holomorphic. This means that the complex conjugation of a Higgs field, as would be needed to give masses to all fermions as in the SM (see Eq.~\eqref{eq:yukawalagr}, where $\Phi$ is the ${\rm SU}(2)_L$ Higgs doublet), cannot be present in the superpotential. Therefore masses cannot be given to up- and down-type fermions at the same time with a single Higgs superfield. In addition, having only one Higgs superfield would lead to a gauge anomaly because the fermion associated with the scalar Higgs field would have no counterpart. Both problems are solved in the presence of two chiral Higgs superfields, $H_u$ and $H_d$ (coupling to up- and down-type fermions, respectively), with opposite hypercharge.

The naming scheme of the particles of the MSSM is as follows. The name of the scalar partners of the SM fermions is obtained with the addition of a ``s-'' prefix ({\it e.g.}, selectron and sbottom) and are collectively called sfermions (squarks and sleptons). The name of the fermionic partners of known bosons (gauge bosons, Higgs boson and graviton) is obtained with the addition of the ``-ino'' suffix ({\it e.g.}, bino, higgsino, gravitino); the partners of gauge bosons are collectively called gauginos, and wino, higgsino and bino are usually referred to as electroweak-inos. 
SUSY partners of the SM particles are given the same symbol with the addition of a tilde.
For instance, $\tilde e_R$ is the right-handed selectron,\footnote{Scalar fields do not have chirality, and in this case ``right-handed'' is just a reminder for the quantum numbers of the fermionic partner.} and $\tilde g$ is the gluino.
The superfield content of the MSSM is summarized in Tables~\ref{tab:chiralsuper} and~\ref{tab:gaugesuper}.

\renewcommand{\arraystretch}{1.5}
\begin{table}[ht]
\center
\begin{tabular}{c|c|c|c|c}
\hline
\multicolumn{2}{c|}{Chiral superfield} & spin-0 & spin-$1/2$ & ${\rm SU}(3)_C$, ${\rm SU}(2)_L$, ${\rm U}(1)_Y$ \\
\hline
(s)quarks & $Q$ & $\tilde q_L = (\tilde u_L\ \tilde d_L)$ & $q_L = (u_L\ d_L)$ & $({\bm 3}, {\bm 2}, \frac{1}{6})$ \\
($\times3$ generations) & $\overline U$ & $\tilde u_R^*$ & $u_R^\dagger$ & $(\bar{{\bm 3}}, {\bm 1}, -\frac{2}{3})$ \\
& $\overline D$ & $\tilde d_R^*$ & $d_R^\dagger$ & $(\bar{{\bm 3}}, {\bm 1}, \frac{1}{3})$ \\
\hline
(s)leptons & $L$ & $\tilde l = (\tilde\nu_e\ \tilde e_L)$ & $l = (\nu_e\ e_L)$ & $({\bm 1}, {\bm 2}, -\frac{1}{2})$ \\
($\times3$ generations) & $\overline E$ & $\tilde e_R^*$ & $e_R^\dagger$ & $({\bm 1}, {\bm 1}, 1)$ \\
\hline
Up-type Higgs(ino) & $H_U$ & $H_u = (H_u^+\ H_u^0)$ & $\stilde H_u = (\stilde H_u^+\ \stilde H_u^0)$ & $({\bm 1}, {\bm 2}, \frac{1}{2})$ \\
Down-type Higgs(ino) & $H_D$ & $H_d = (H_d^0\ H_d^-)$ & $\stilde H_d = (\stilde H_d^0\ \stilde H_d^-)$ & $({\bm 1}, {\bm 2}, -\frac{1}{2})$ \\
\hline
\end{tabular}
\caption{Chiral superfield content of the MSSM, given in terms of left-handed superfields. The name of the SUSY partner of the SM particle is given in brackets. Same conventions as in Table~\ref{tab:smfields}.} \label{tab:chiralsuper}
\end{table}
\renewcommand{\arraystretch}{1.0}

\renewcommand{\arraystretch}{1.5}
\begin{table}[ht]
\center
\begin{tabular}{c|c|c|c}
\hline
Gauge superfield & spin-1/2 & spin-1 & ${\rm SU}(3)_C$, ${\rm SU}(2)_L$, ${\rm U}(1)_Y$ \\
\hline
gluino, gluon & $\tilde g$ & $g$ & $({\bm 8}, {\bm 1}, 0)$ \\
wino, $W$ & $(\stilde W^+\ \stilde W^0\ \stilde W^-)$ & $(W^+\ W^0\ W^-)$ &  $({\bm 1}, {\bm 3}, 0)$ \\
bino, $B$ & $\stilde B^0$ & $B^0$ & $({\bm 1}, {\bm 1}, 0)$ \\
\hline
\end{tabular}
\caption{Gauge superfield content of the MSSM. Same conventions as in Table~\ref{tab:smfields}.} \label{tab:gaugesuper}
\end{table}
\renewcommand{\arraystretch}{1.0}

The superpotential of the MSSM determines all SUSY-conserving couplings between the chiral superfields. It includes the Yukawa terms that give masses to the fermions.
Writing by convention the superfields with capital letters, and denoting conjugation of the components of a superfield with a bar (for instance, $\overline U$ contains $u_R^\dagger$ and $\tilde u_R^*$, see Table~\ref{tab:chiralsuper}), it reads
\begin{equation}
W_{\rm MSSM} =
\overline U {\bf y_u} Q H_U -
\overline D {\bf y_d} Q H_D -
\overline E {\bf y_e} L H_D +
\mu H_U H_D \,,
\label{eq:MSSMsuperpot}
\end{equation}
where the ${\bf y_i}$ are $3 \times 3$ matrices in flavor space as in the SM (see Eq.~\eqref{eq:yukawalagr}), and $\mu$ is a dimensionful parameter, the higgsino mass parameter. Note that masses are not given to the neutrinos, but can be trivially added.
The soft supersymmetry breaking terms of the MSSM are
\begin{align}
{\cal L}_{\rm soft}^{\rm MSSM} &= -\frac{1}{2}\left (
M_1 \stilde B\stilde B 
+ M_2 \stilde W \stilde W
+ M_3 \stilde g  \stilde g
+ {\rm c.c.} \right )
\nonumber
\\
& -\left ( \stilde {\overline U} \,{\bf a_u}\, \stilde Q H_u
- \stilde {\overline D} \,{\bf a_d}\, \stilde Q H_d
- \stilde {\overline E} \,{\bf a_e}\, \stilde L H_d
+ {\rm c.c.} \right ) 
\nonumber
\\
& -\stilde Q^\dagger \, {\bf m^2_{Q}}\, \stilde Q
-\stilde L^\dagger \,{\bf m^2_{L}}\,\stilde L
-\stilde {\overline U} \,{\bf m^2_{{\overline U}}}\, {\stilde {\overline U}}^\dagger
-\stilde {\overline D} \,{\bf m^2_{{\overline D}}} \, {\stilde {\overline D}}^\dagger
-\stilde {\overline E} \,{\bf m^2_{{\overline E}}}\, {\stilde {\overline E}}^\dagger
\nonumber
\\
& - \, m_{H_u}^2 H_u^* H_u - m_{H_d}^2 H_d^* H_d
\nonumber
\\
& - \left ( b H_u H_d + {\rm c.c.} \right ) \,,
\label{eq:MSSMsoft}
\end{align}
where $M_1$, $M_2$ and $M_3$ are mass terms for the bino, wino and gluino, respectively, and all ${\bf a_i}$ and ${\bf m_i^2}$ matrices are possibly complex and $3 \times 3$ in flavor space.
The trilinear terms are often rewritten $(a_{u,d,e})_{ij} = (A_{u,d,e})_{ij} \times y_{ij}$ and the $b$ parameter can be seen elsewhere as $B\mu$ or $m_3^2$.
Matching the soft terms of the MSSM with the generic expression given in Eq.~\eqref{eq:lagrsoft} can also be instructive. The first line contains $M_a\, \lambda^a\lambda^a$ terms, the second one $a^{ijk} \phi_i\phi_j\phi_k $ terms, the third and fourth lines contain terms of the $(m^2)_j^i \phi^{j*} \phi_i$ type, and finally the last line contains a $b^{ij} \phi_i\phi_j $ term (the only one allowed by the gauge symmetries).

A comparison between Eq.~\eqref{eq:MSSMsuperpot} and~\eqref{eq:MSSMsoft} shows clearly the problem of having a generic parameterization of SUSY breaking. While the SUSY-conserving part is extremely simple and predictive, more than one hundred free parameters are introduced with the soft breaking of SUSY~\cite{Dimopoulos:1995ju}.
However, if sparticles are around the TeV scale we know that the flavor structure of the soft terms is strongly constrained by the flavor physics measurements on flavor-changing and CP-violating processes (see Section~\ref{sec:intro-sm}). Therefore, most of the parameters present in Eq.~\eqref{eq:MSSMsoft} cannot be completely arbitrary, and it is commonly assumed that SUSY breaking is mediated by flavor-blind interactions and that the only source of CP violation comes from the CKM matrix (as in the SM). This drastically reduces the number of free parameters, which is needed for carrying out a global study of the model. A small number of extra assumptions lead to the 19-parameter phenomenological MSSM (pMSSM)~\cite{Djouadi:1998di}, an agnostic parameterization of the MSSM that will be defined and studied in light of experimental constraints in Section~\ref{sec:pmssm}.

In the MSSM, the breaking of the electroweak symmetry is slightly more complicated due to the presence of two Higgs doublets. The tree-level potential for the Higgs scalars involve the parameters $|\mu|^2$, $m_{H_u}^2$, $m_{H_d}^2$, and $b$ (while only three (combination of) parameters will be relevant). For reviews, see, {\it e.g.}, Refs.~\cite{Martin:1997ns,Djouadi:2005gj,sparticles}.
Under certain requirements on these parameters (in particular, the potential must be bounded from below and a stable minimum should be away from $H_u^0 = H_d^0 = 0$) spontaneous breaking of the electroweak symmetry occurs.
Interestingly, starting from positive parameters at the GUT scale electroweak symmetry breaking can be radiatively induced from the large contributions of stops to $m_{H_u}^2$. 
As in the case of the SM, three would-be Goldstone bosons are ``eaten'' and give masses to the $W^+$, $W^-$ and $Z^0$ bosons. However, with two Higgs doublets there are five remaining degrees of freedom left instead of just one. This leads to five massive physical Higgs states after electroweak symmetry breaking: two neutral CP-even, $h^0$ and $H^0$, one neutral CP-odd, $A^0$, and two charged Higgses, $H^+$ and $H^-$.
Two parameters of interest moreover need to be defined: $\tanb \equiv v_u/v_d$, the ratio of the vacuum expectation values of $H_u$ and $H_d$ (with $v^2 = v_u^2 + v_d^2 \approx (246~{\rm GeV})^2$), and $\alpha$, the CP-even Higgs mixing angle. Note that, for convenience, the four parameters appearing in the Higgs potential $(\mu, m_{H_u}^2, m_{H_d}^2, b)$ can be traded for $(\mu, m_A, \tanb, v)$, and that the tree-level couplings of the Higgs bosons to SM particles are completely determined by the new parameters $\alpha$ and $\tan\beta$.

An interesting consequence of electroweak symmetry breaking in the supersymmetric case is that the mass of the lightest Higgs boson, $m_{h^0}$, is bounded from above (contrary to the other Higgs boson masses). At tree-level, one finds a severe constraint:
\begin{equation}
m_{h^0} < m_{Z^0} |\! \cos(2\beta)| \,,
\end{equation}
which is in contradiction with the observed Higgs boson mass of about 125~GeV at the LHC (see Section~\ref{sec:higgs-measlhc}). Fortunately, loop corrections, in particular involving the stops (superpartners of the top quark), can induce sizable corrections to the Higgs mass and explain a 125~GeV Higgs boson (depending on the masses and mixings of the stops---see, {\it e.g.}, Refs.~\cite{Djouadi:2013lra,Brummer:2012ns}).

After supersymmetry and electroweak symmetry breaking, not only the Higgs fields but also sfermions and electroweak-inos of the theory mix. First, the mixing between ``left-'' and ``right-handed'' sfermions is controlled by the ${\bf a_i}$ soft terms, which are often taken to be proportional to the Yukawa couplings. In the case of stops, for instance, a significant mixing can be expected and leads to two physical states, $\tilde t_1$ and $\tilde t_2$ (with $m_{\tilde t_1} < m_{\tilde t_2}$), assuming no mixing with other sfermions. Second, the four neutral electroweak-inos ($\stilde B^0$, $\stilde W^0$, $\stilde H^0_u$, and $\stilde H^0_d$) mix into so-called neutralinos ($\tilde \chi^0_i$, $i=1,2,3,4$) and the four charged electroweak-inos ($\stilde W^+$, $\stilde H^+_u$, $\stilde W^-$, and $\stilde H^-_d$) mix into charginos ($\tilde \chi^\pm_i$, $i=1,2$).

An undesirable feature of the MSSM has been ignored so far. As was discussed in Section~\ref{sec:intro-bsm}, in the SM there are accidental symmetries that conserve the baryon number and the lepton numbers. This is important as it ensures stability of the proton (which is experimentally very well measured~\cite{Nishino:2012ipa}). However, this accidental symmetry does not exist in the MSSM, and terms that violate $B$ or $L$ by one unit could be added to the superpotential. Assuming extra ${\rm U}(1)_B$ and ${\rm U}(1)_L$ global symmetries is not really a satisfactory solution: it is {\it ad hoc}, and moreover we know that $B$ and $L$ symmetries (but not $B-L$) are violated by non-perturbative processes in the electroweak sector of the SM.
In the MSSM, the most popular solution to this problem is to introduce a discrete symmetry, the $R$-parity~\cite{Farrar:1978xj}. A multiplicatively conserved quantum number, $P_R$, is given to all particles of the MSSM, and is defined as
\begin{equation}
P_R = (-1)^{3(B-L)+2s} \,,
\end{equation}
where $s$ is the spin of the particle. Thus, all SM particles and Higgs bosons have $P_R = +1$, while all superpartners have $P_R = -1$. An exactly conserved $R$-parity can be obtained from a ${\rm U}(1)_{B-L}$ gauge symmetry broken at a high scale. In addition to forbidding the dangerous terms that lead to proton decay, this discrete symmetry has very important phenomenological consequences.
First, it means that the lightest supersymmetric particle (LSP) is stable and can be a dark matter candidate (see next section). In the MSSM, there are three kind of electrically and color neutral particles that could be candidate for dark matter: the lightest neutralino, $\tilde\chi^0_1$, the sneutrinos, $\tilde \nu$, and the gravitino, $\stilde G$.
Second, $R$-parity conservation implies that SUSY partners are pair-produced at colliders, and then decay into one SUSY ($P_R$-odd) particle and one or several SM ($P_R$-even) particles until the LSP is produced. This offers distinctive signatures: there must be at least two invisible particles (one at the end of each decay chain), which correspond to missing energy that can be used to discriminate SUSY from SM events in the search for supersymmetry at colliders.

In order to be detectable at colliders, and in particular at the LHC, at least some of the SUSY partners must have a mass of the order of a TeV. However, the hierarchy problem in the Higgs sector is solved (the quadratic divergences are removed) when introducing supersymmetry with a soft breaking. Why should SUSY partners moreover be light enough to be detected and not anywhere between the electroweak and the Planck scale? A first reason is that gauge coupling unification (see Section~\ref{sec:intro-bsm-hierarchy}) in the MSSM requires SUSY partners not too far away from the TeV scale. This is, however, not a very strong argument: SUSY could certainly be out of reach of the LHC while yielding a convincing unification, and there could be threshold corrections anywhere between the SUSY and the GUT scale that would help the three gauge couplings to meet.
Instead, the strongest argument for having light SUSY partners comes from naturalness of the breaking of the electroweak symmetry. In the MSSM, the $Z^0$ boson mass is given by
\begin{equation}
\frac{m_{Z}^2}{2} = m^2_{H_u} + |\mu|^2 \,,
\end{equation}
in the large $\tan \beta$ limit and at tree-level. In order to escape experimental constraints, the most simple solutions are typically found for $m^2_{H_u}, |\mu|^2 \gg m_{Z}^2$, requiring a fine-tuning cancellation to obtain the $Z$ mass of about 91~GeV.
This argument extends beyond tree-level, since there are important corrections to $m^2_{H_u}$, at one-loop from stops and at two-loop from gluinos (see Ref.~\cite{Papucci:2011wy} and references therein). In order to avoid fine-tuned cancellations, higgsinos certainly need to be light ($\mu$ to be small), as well as stops and to a lesser extent gluinos. This scenario is known as ``natural SUSY'', and is the best reason why SUSY is expected to appear at the TeV scale which will continue to be explored at Run~II of the LHC.

\section{Dark matter: the last gasp of WIMPs?} \label{sec:intro-dm}

That the Higgs boson was discovered nearly fifty years after its theoretical prediction may seem excessively long.
Unfortunately, this situation is not an exception in contemporary particle physics and might not improve in the future as the argument of the necessity of new physics around the electroweak scale is more and more put into question by the negative BSM search results at the LHC. A particularly remarkable example of a long-sought particle is dark matter. 
We have seen in Section~\ref{sec:intro-bsm-obs} that there are many different observations of dark matter. However, they 
do not give any hint on the nature of the dark matter {\it particle} itself apart from 
stability on the cosmological time scale and 
modest interactions with the ordinary matter. More concretely, the dark matter particle(s) needs to be stable, electrically and color neutral, and may or may not interact with the massive vector bosons, with the Higgs boson, or with itself (or the other components of dark matter).
This leaves us with a plethora of viable dark matter candidates (for which the relic density and the temperature matches the observations) spanning mass ranges from the sub-eV to the GUT scale.

In the past decades, one particular class of dark matter candidates attracted a lot of attention: the WIMPs, weakly interacting thermal relics with mass of the order of 100~GeV. The most popular WIMP candidate is the lightest neutralino of the $R$-parity conserving MSSM (see Section~\ref{sec:susy-mssm}), $\tilde{\chi}^0_1$, which is a prime candidate for the LSP and, as we will see, can be a very good dark matter candidate. Reviews on (supersymmetric) WIMP dark matter can be found in Refs.~\cite{Jungman:1995df,Bertone:2004pz,Bertone:2010zz,Kolb:1990vq}.
The general picture is the following: the WIMP candidate is in thermal equilibrium in the early Universe, and eventually decouples from the thermal bath when the interaction rate cannot compensate for the expansion of the Universe. This is called freeze-out, and is necessary to account for the observed dark matter density. Indeed, as temperature drops the WIMP candidate becomes non-relativistic (for $T \lesssim m_\chi$, where $T$ is the temperature and $\chi$ is a generic WIMP) and the number density of dark matter particles becomes Boltzmann suppressed and quickly vanishes. When freeze-out occurs, the WIMP becomes decoupled. The remaining dark matter density is called relic density, and should match (or at least not exceed) the observed cold dark matter density obtained from cosmological observations, see Section~\ref{sec:intro-bsm-obs}.

\begin{figure}[ht]
\begin{center}
\includegraphics[width=0.47\textwidth]{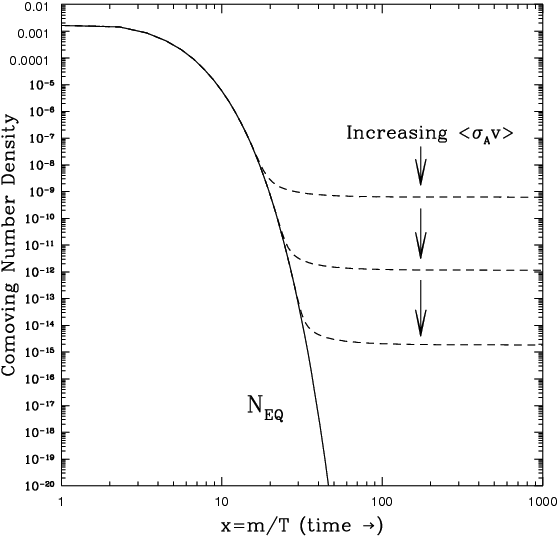}
\includegraphics[width=0.52\textwidth]{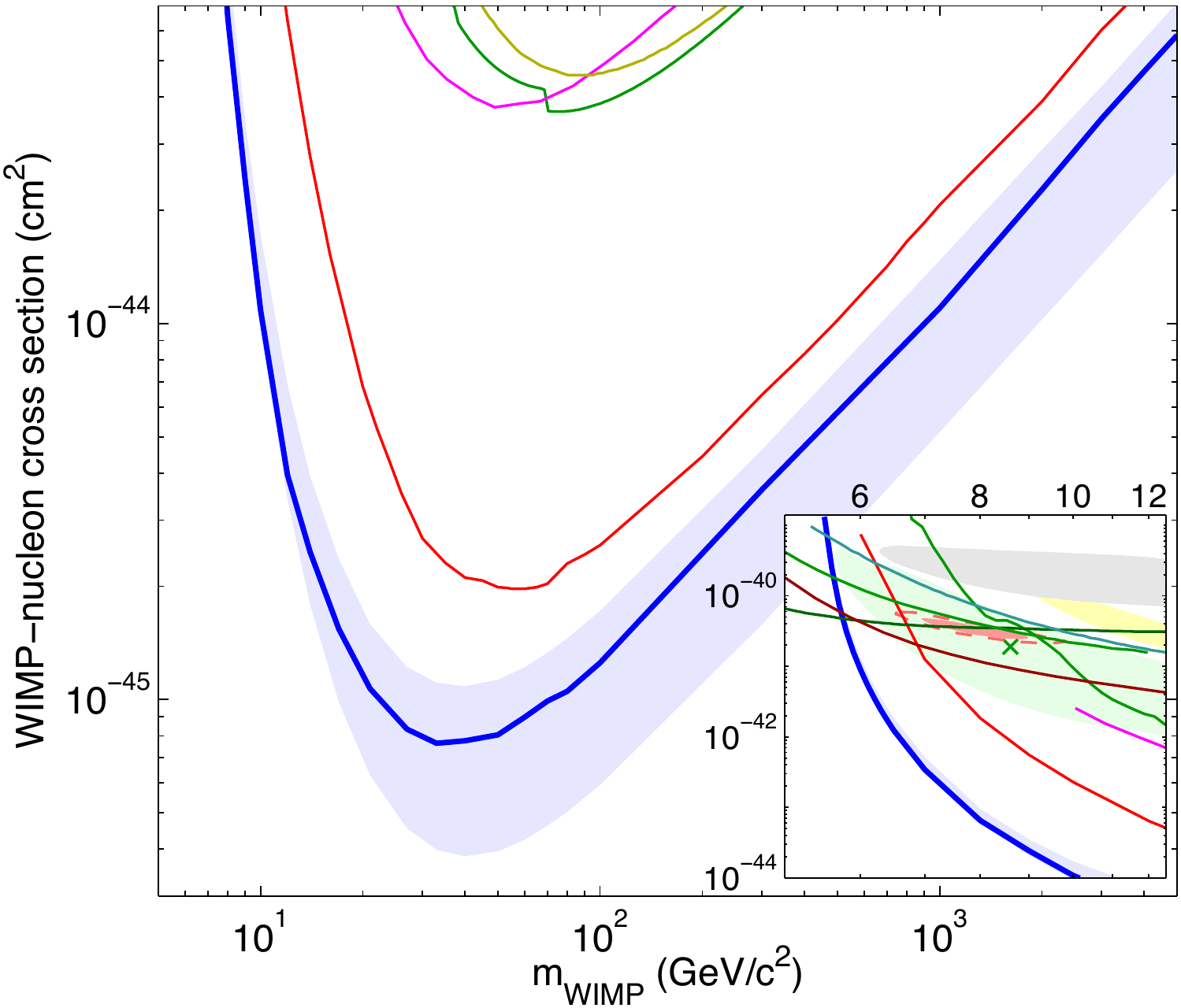}
\end{center}
\caption{Left: comoving number density of a WIMP in the early Universe as a function of $m_\chi / T$, taken from Ref.~\cite{Kolb:1990vq}. Right: upper limit at 90\%~CL on the DM-nucleon spin-independent scattering cross section as a function of the WIMP mass, taken from Ref.~\cite{Akerib:2013tjd}. The blue line is the 2013 limit from the LUX experiment.}
\label{fig:dmfigs}
\end{figure}

The evolution of the number density of dark matter ($n_\chi$) is obtained by the Boltzmann equation
\begin{equation}
\frac{{\rm d}n_\chi}{{\rm d}t} + 3Hn_\chi = - \langle \sigma_A v \rangle \left[ (n_\chi)^2 - (n_{\chi}^{\rm eq})^2 \right] \,,
\end{equation}
where $H$ is the Hubble expansion rate, $\langle \sigma_A v \rangle$ is the thermally averaged annihilation cross section, and $n_{\chi}^{\rm eq}$ is the number density at thermal equilibrium. 
While there is no closed-form analytical solution, the relic density can be obtained from approximate solutions that are standard~\cite{Jungman:1995df,Bertone:2004pz,Bertone:2010zz,Kolb:1990vq}, and numerical solutions that are fast and accurate have been implemented in public tools such as {\tt DarkSUSY}~\cite{Gondolo:2004sc} and {\tt SuperIso Relic}~\cite{Arbey:2009gu,Arbey:2011zz} for SUSY models, and {\tt micrOMEGAs}~\cite{Belanger:2008sj,Belanger:2013oya} for any model.
Numerical solutions to the Boltzmann equation are shown in the left panel of Fig.~\ref{fig:dmfigs}, giving the evolution of the comoving number density (constant if the number of particles is conserved) of a WIMP in the early Universe. The solid line corresponds to thermal equilibrium, and quickly drops to very small values. When freeze-out occurs (dashed lines), the comoving number density becomes constant and corresponds to the dark matter relic. It can be seen that larger $\langle \sigma_A v \rangle$ imply a later decoupling from the thermal bath, and therefore a reduced relic density. A useful order of magnitude estimate is
\begin{equation}
\Omega_{\rm DM}h^2 \approx \frac{3 \times 10^{-27}~{\rm cm}^3{\rm s}^{-1}}{\langle \sigma_A v \rangle} \,,
\end{equation}
and a typical freeze-out temperature is $T_f \approx m_\chi / 20$, meaning that WIMPs are already non-relativistic when they decouple.

WIMP scenarios are particularly interesting under the assumption of new physics close to the electroweak scale. Indeed, with a mass of the order of 100~GeV and typical couplings of the weak interaction, one obtains the correct order of magnitude for the relic density. This is called the ``WIMP miracle''. Therefore, it is tempting to link the dark matter problem with electroweak symmetry breaking, as can be the case in the MSSM (however, the ``WIMP miracle'' argument is certainly not strong enough to overlook alternative DM scenarios). 
The lightest neutralino of the MSSM, $\tilde \chi^0_1$, can be a good dark matter candidate depending on the bino, wino and higgsino admixture. This will be developed in this thesis in light of the experimental constraints in Sections~\ref{sec:pmssm} and~\ref{sec:simpmod-lightneutralino}, while another supersymmetric WIMP candidate, the mixed sneutrino, will be studied in Section~\ref{sec:simpmod-sneutrino}.

Apart from relic density, a number of other constraints apply on WIMP scenarios.
First, several direct detection experiments attempt to observe the nuclear recoils produced by WIMP scattering off nucleons. In order to reduce the background from atmospheric muons, such experiments are usually operating in underground facilities. Depending on the nuclei that constitute the target, direct detection experiments primarily target either spin-independent (scalar) or spin-dependent (axial-vector) interactions. The strongest constraints on WIMP models usually come from the results on spin-independent scattering, in experiments based on xenon or germanium. No signal was observed in the direct detection of dark matter (excesses were claimed in the low-mass region, around 10~GeV, but partially disagree with each other and have been ruled out by more precise measurements). The best limits on the DM-nucleon spin-independent scattering cross section, obtained by the LUX experiment~\cite{Akerib:2013tjd}, are shown in the right panel of Fig.~\ref{fig:dmfigs}. These negative results have a significant impact on various WIMP scenarios (see examples in Sections~\ref{2013c-sec:DMinterplay},~\ref{sec:pmssm},~\ref{sec:simpmod-lightneutralino}, and~\ref{sec:simpmod-sneutrino}). Future results in direct detection experiments should significantly improve sensitivity to low cross sections, in particular with the next generation of experiments at the ton scale. However, note that direct detection experiments will ultimately be limited by an irreducible neutrino background. This will make further improvements in direct detection experiments very difficult (see Ref.~\cite{Billard:2013qya} for a recent discussion).

Second, dark matter could be seen in an indirect way, from the observation of the stable annihilation products of dark matter (photons, neutrinos, positrons, and antiprotons). The first two messengers, being neutral particles, do not necessarily require to model the complex propagation of cosmic rays in the galaxy, thus constitute {\it a priori} cleaner probes. Annihilation products of dark matter can be looked for in very different places. While the largest local signal is expected in the direction of the galactic center, the limited astrophysical understanding of the center of our galaxy makes it very difficult to interpret the observations in terms of SM or new physics. Notable recent results include the $Fermi$-LAT limit on dark matter annihilation using gamma-rays from Milky Way satellite galaxies~\cite{Ackermann:2013yva}. Moreover, recent neutrino results from the IceCube experiment show promising prospects~\cite{Aartsen:2013mla}.

Third, WIMP dark matter (if light enough) could be pair-produced  at colliders such as the LHC. However, it is impossible to trigger on purely invisible final states, requiring the presence of an additional particle that is visible to constrain WIMP pair production. Possibilities include, in particular, a jet or photon initial state radiation recoiling against the pair of WIMPs. At the LHC, limits on dark matter production have recently been set from monojet, monophoton and mono-lepton events~\cite{ATLAS-CONF-2012-147,CMS-PAS-EXO-12-048,CMS-PAS-EXO-12-047,CMS-PAS-EXO-13-004}.

Finally, one should stress again that although WIMP dark matter fits very well in the context of electroweak symmetry breaking, the dark matter particle(s) may not have any direct connection with the electroweak scale. For many other viable candidates, the dark matter is not a thermal relic; for instance it could be produced from the decay of heavier particles which are in thermal equilibrium in the early Universe. For more information, see Refs.~\cite{Bertone:2004pz,Bertone:2010zz,Baer:2008uu,Baer:2014eja} and references therein.

\newpage

\chapter{At least a Higgs boson} \label{sec:higgs}

On July 4th, 2012 the ATLAS and CMS collaborations at CERN's Large Hadronic Collider reported the observation of a new particle with properties consistent with those expected from the SM Higgs~\cite{Aad:2012tfa,Chatrchyan:2012ufa}. This discovery is truly remarkable, first because the Higgs boson was the last elementary particle predicted by the SM remaining to be observed after many years of unsuccessful searches. But most of all, it is the key role of the Higgs field in the SM---triggering the breaking of the electroweak symmetry and giving masses to the elementary particles, as explained in Section~\ref{sec:intro-sm}---that makes the discovery of the Higgs boson so special. This has been rewarded in 2013 with the Nobel Prize in physics for Fran\c{c}ois Englert and Peter Higgs for the ``theoretical discovery'' of what is now called the Higgs (or Brout--Englert--Higgs) mechanism~\cite{Englert:1964et,Higgs:1964ia,Higgs:1964pj,Guralnik:1964eu,Higgs:1966ev,Kibble:1967sv}.

This new particle certainly represents the ultimate triumph of the SM. In the difficult quest for new physics beyond the SM, motivated by several problems for which I give a short and personal review in Section~\ref{sec:intro-bsm}, we, particle physicists, sometimes forget to look back and admire the fact that such a simple theory, based on symmetry considerations, suffices to describe all observed phenomena in the microscopic world with an excellent precision, to say the least. This being said, the LHC would not have been built only to fix the last free parameter of the SM. Its raison d'\^etre is the search for new physics, hence the discovery of this Higgs boson should also be viewed as an opportunity to discover---or at least corner---new physics. Hopefully, the first observation of a Higgs boson will later be seen as the opening of a new chapter in particle physics and not only as the closing of the SM one.

Indeed, the measurements of the properties of the Higgs boson (starting from its mass) are of immediate relevance for models aiming at solving the hierarchy problem, and in general for any model predicting a modified or extended Higgs sector and/or new particles coupling to the Higgs field. As we will see in several examples, the information that is obtained is complementary to the direct searches for new particles at the LHC, or even to direct searches for dark matter. For that reason, the discovery of a Higgs-like boson and the subsequent measurement of its properties at the LHC have generated an intense activity in the theory community---to which I started to participate in mid-2012, shortly after the discovery.

This chapter will be divided as follows. In Section~\ref{sec:higgs-prelhc}, I will start by the pre-LHC constraints on an SM(-like) Higgs boson, before turning in Section~\ref{sec:higgs-smprop} to a brief review of the properties of the Higgs boson, as predicted in the SM. I will then list and discuss the various measurements performed at the LHC in Section~\ref{sec:higgs-measlhc}. The way this experimental information is used to constrain new physics scenarios is described in Section~\ref{sec:higgs-npconstlhc}. I will then present the various works I have been involved in. The physics studies are separated into two categories: the model-independent studies in which new physics is constrained from an effective approach are presented in Sections~\ref{sec:higgs2012}, \ref{sec:higgs2013}, and \ref{sec:higgsdim6}, while the complete study of the constraints on a specific new physics scenario, the phenomenological MSSM (pMSSM), will be presented in Section~\ref{sec:pmssm}. In Section~\ref{sec:lilith}, I will present {\tt Lilith}, a new public tool that provides an approximation to the Higgs likelihood in order to constrain generic BSM scenarios. Finally, the future of Higgs constraints on new physics will be discussed in Section~\ref{sec:higgsfuture}.

For all of this, I would like to acknowledge collaboration with Genevi\`eve B\'elanger, J\'er\'emy Bernon, Ulrich Ellwanger, Sylvain Fichet, Gero von Gersdorff, John F.~Gunion, Sabine Kraml, and Sezen Sekmen as well as useful discussions with, in particular, Guillaume Drieu La Rochelle.

\section{Pre-LHC constraints on the Higgs boson} \label{sec:higgs-prelhc}
While being crucial, it should be noted that the discovery of a Higgs boson at the LHC did not come as a surprise. Mass terms for the $W^\pm$ and $Z^0$ bosons---as well as mass terms for the fermions---break the $SU(2)_L \times U(1)_Y$ symmetry that successfully describes the electroweak interactions. Ignoring aesthetic considerations and breaking explicitly this symmetry, {\it i.e.}~putting directly a mass term $M_W^2 W^\mu W_\mu$ in the Lagrangian, leads to the violation of unitarity in the $W_L W_L$ scattering process ($W_L$ being the longitudinal component of the $W^{\pm}$ bosons) at center-of-mass energies $\sqrt{s} \gtrsim 1$~TeV, which would be a clear indication of the presence of a new particle at or below the TeV scale, see {\it e.g.}~\cite{Chanowitz:1998wi} and references therein.
Fortunately, all these problems can be solved at once if the electroweak symmetry is {\it spontaneously broken} via the Higgs mechanism, as explained in Section~\ref{sec:intro-sm}. The simplest solution involves only one elementary scalar field, the Higgs field. This is what is done in the SM.

\begin{figure}[ht]
\begin{center}
\includegraphics[width=0.49\textwidth]{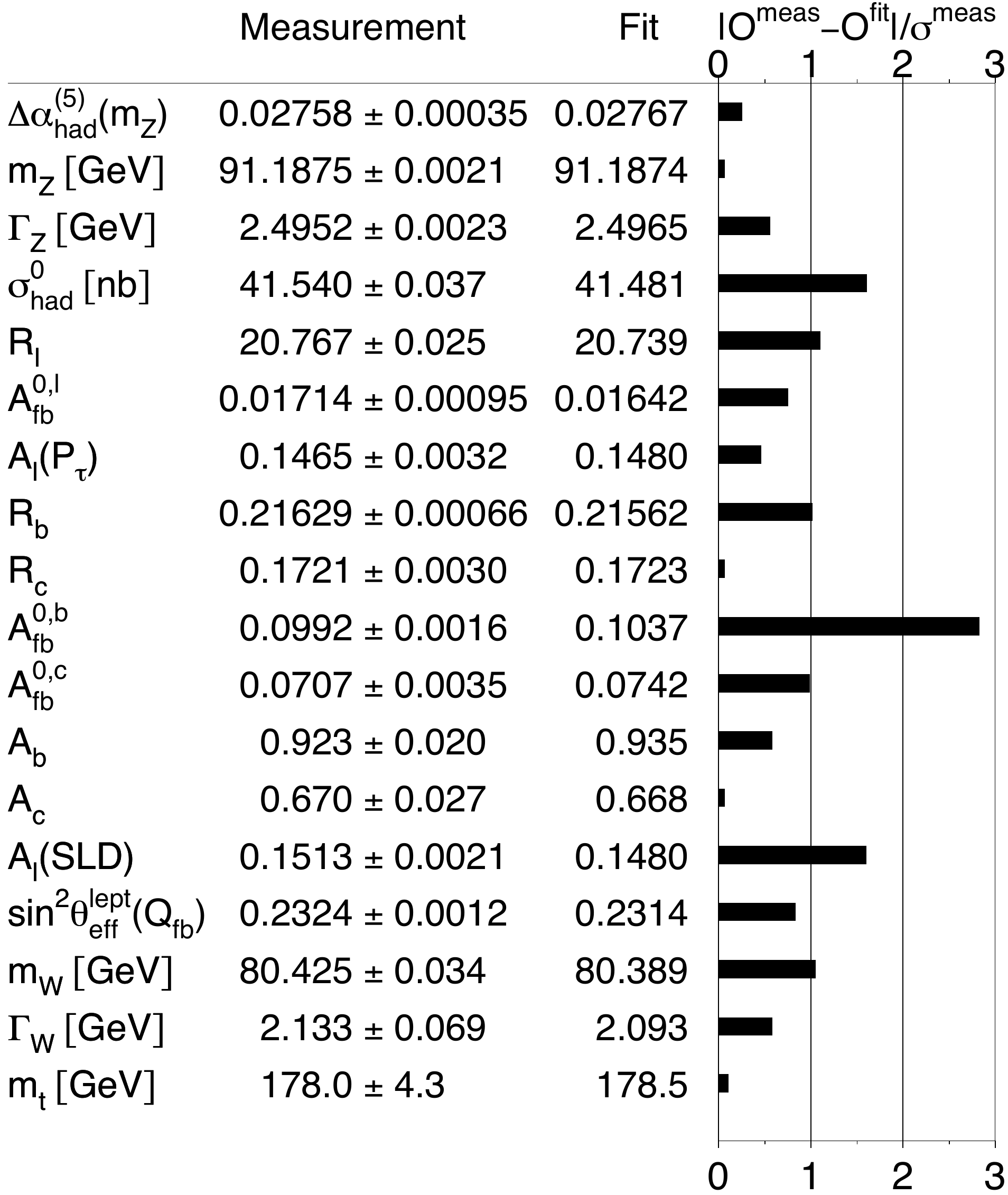}
\includegraphics[width=0.49\textwidth]{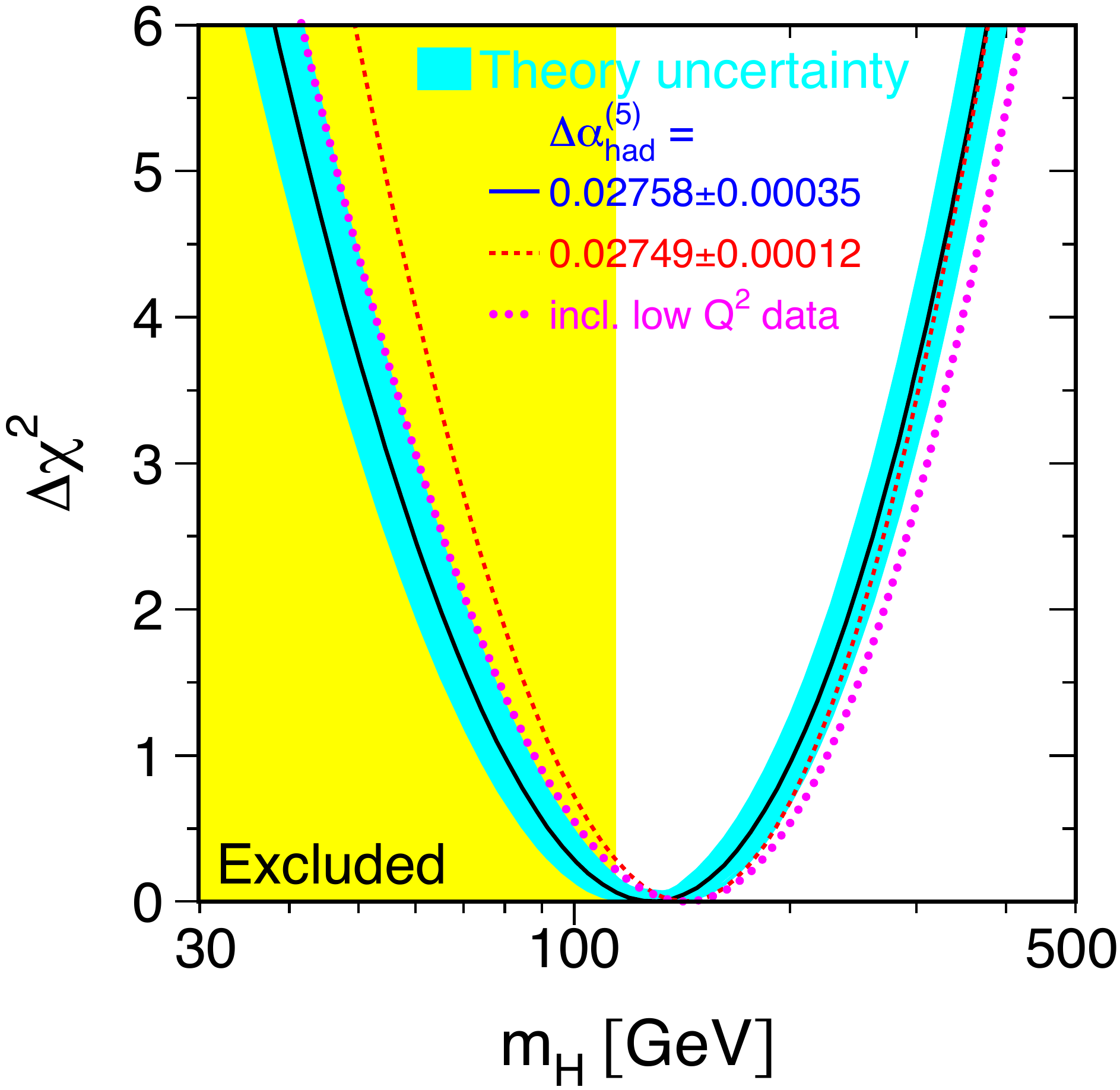}
\end{center}
\caption{Electroweak fit using mostly observables at the $Z^0$ pole, as was performed by the LEP collaborations in 2005~\cite{ALEPH:2005ab}. Left: pull comparison of the fit results with the measurements of the observables. Right: fit of the Higgs boson mass from the observables shown in the left panel.}
\label{fig:LEPew}
\end{figure}

This picture is strengthened by the precision tests of the electroweak sector. In particular, the precise measurements performed on the $Z^0$ resonance at the LEP experiment~\cite{ALEPH:2005ab} constitute a stringent test of the SM. From these data, a fit to a relevant subset of the SM parameters that enters the loop corrections to the observables (listed in the left panel of Fig.~\ref{fig:LEPew}) was performed by the collaborations at LEP, thus making it possible to check the consistency of the SM with high accuracy.\footnote{The five relevant SM parameters for the calculation of $Z^0$-pole observables were identified to be the QED and QCD coupling constants $\alpha_s(m_Z^2)$ and $\alpha(m_Z^2)$, the masses of the $Z^0$ and Higgs bosons $m_Z$ and $m_H$, and the top mass $m_t$.} An overall excellent agreement is found, as can be seen in the left panel of Fig.~\ref{fig:LEPew}. The largest deviation with respect to SM expectations comes from the forward-backward asymmetry in $b$-quark production, $A_{\rm FB}^{0,b}$, and is at the level of 2.8 standard deviations. From this SM fit, a prediction on the Higgs boson mass can be made since it enters (however, only logarithmically) in the loop corrections. The result is shown in the right panel of Fig.~\ref{fig:LEPew}. This corresponds to an imprecise yet relevant information on the Higgs mass in the SM, being $m_H = 129^{+74}_{-49}$~GeV at 68\% confidence level (CL). Note that, in addition to the LEP I results at the $Z^0$ resonance, some results from LEP II and from Tevatron Run~I were used in this fit. This electroweak fit has been regularly updated by the LEP electroweak working group~\cite{lep-ew-wg}, and the Gfitter group is also performing electroweak fits using all publicly available data~\cite{Flacher:2008zq}, also including the Higgs mass since its discovery~\cite{Baak:2012kk,Baak:2014ora}. The latest update still shows a very good agreement with the SM predictions, while the determination of the Higgs boson mass (not including LHC data) is significantly improved: $m_H = 94^{+25}_{-22}$~GeV. This is mostly coming from the precise measurement of the top mass and of the $W$ mass using the full statistics collected in the $p \bar p$ collisions at Tevatron.

Electroweak fits show the perfect consistency of the SM and give the expected range for the Higgs boson mass. However, and even if all other searches for BSM physics only produced negative results so far, there were very good reasons to search for the Higgs boson itself, beyond the simple determination of its mass. The hierarchy problem, which is one of the most pressing issues of the SM as explained in Section~\ref{sec:intro-bsm-hierarchy}, is believed to be most naturally solved by TeV-scale new physics. First of all, the value of the Higgs mass can have substantial implications on these models, and in particular supersymmetry, since it is no longer a free parameter but rather comes as a prediction. Moreover, the new, BSM particles couple to or mix with the Higgs boson, and can dramatically change its properties compared to the SM predictions. Therefore the study of the properties of the Higgs boson gives an additional insight on the electroweak symmetry breaking, complementary to the direct searches for these new TeV-scale particles at colliders. The Higgs boson can also have profound implications on cosmology since if dark matter is made of WIMPs, the observed relic density of dark matter is likely to be explained from the interactions with the Higgs.

\begin{figure}[ht]
\begin{center}
\includegraphics[width=0.49\textwidth]{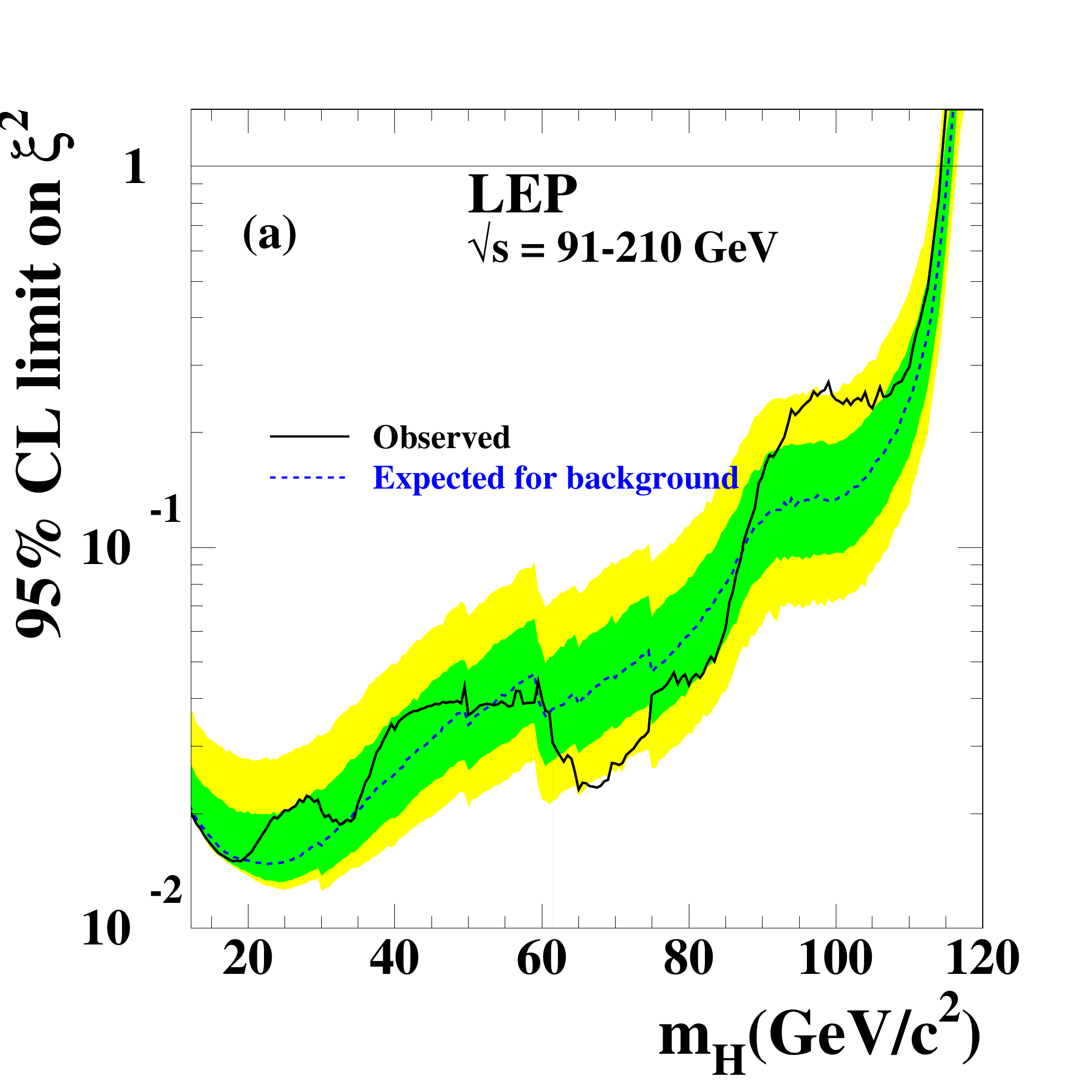}
\includegraphics[width=0.49\textwidth]{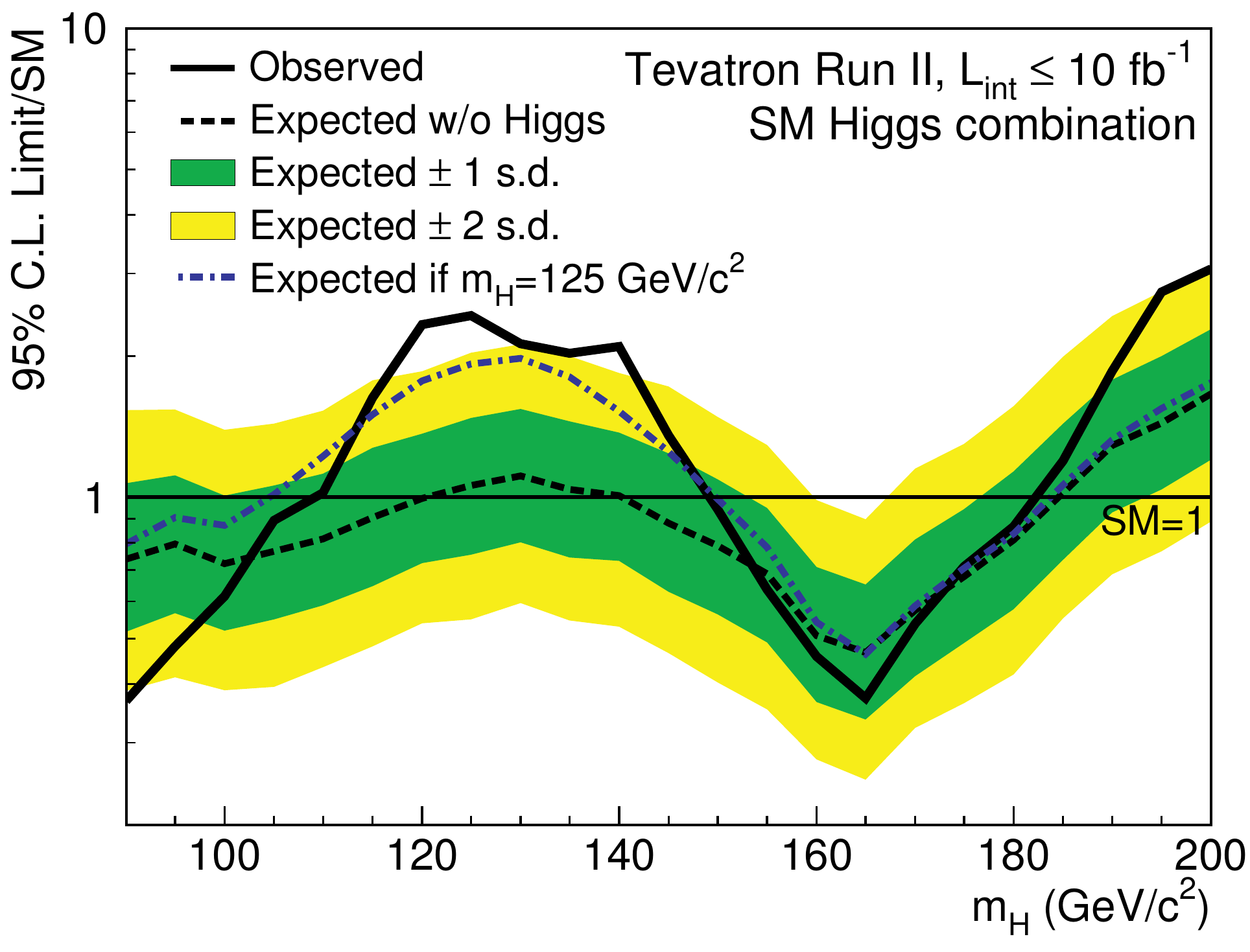}
\end{center}
\caption{Final results in the search for the Higgs boson at LEP~\cite{Barate:2003sz} (left) and at Tevatron~\cite{Aaltonen:2013kxa} (right). In the left panel, $\xi$ is defined as a reduced $HZZ$ coupling, $g_{HZZ} / g^{\rm SM}_{HZZ}$.}
\label{fig:leptevdirecthiggs}
\end{figure}

Before the LHC, direct searches for the Higgs boson were performed at LEP~\cite{Barate:2003sz} and at Tevatron~\cite{Aaltonen:2013kxa}. Searches performed at LEP, being an $e^+e^-$ collider, are sensitive to the production of a Higgs boson in association with a $Z^0$, with the decay $H \to b\bar{b}$ or $H \to \tau^+\tau^-$. Tevatron searches involve many different categories sensitive to different final states and production modes, but in the light region ($m_H \lesssim 140$~GeV) the results are mostly driven by the production of a Higgs boson in association with a vector boson ($W^\pm$ or $Z^0$), with the decay $H \to b\bar{b}$, subsequently denoted as $VH \to b\bar{b}$. The final results from the two experiments are shown in Fig.~\ref{fig:leptevdirecthiggs}. In both cases the 95\%~CL upper bounds on the ratio of production cross sections $\sigma/\sigma^{\rm SM}$ are given as a function of the Higgs mass. From LEP results, a lower bound of 114.4~GeV is set on the SM Higgs mass, while 2012 Tevatron results moreover excludes the $[149,182]$~GeV range and observes an excess of events between 115 and 140~GeV, with a local significance of $3\sigma$ at $m_H = 125$~GeV. As can be seen, the situation before the LHC was quite undecided and a large range of the possible Higgs masses remained unexplored.

\begin{figure}[ht]
\begin{center}
\includegraphics[width=0.49\textwidth]{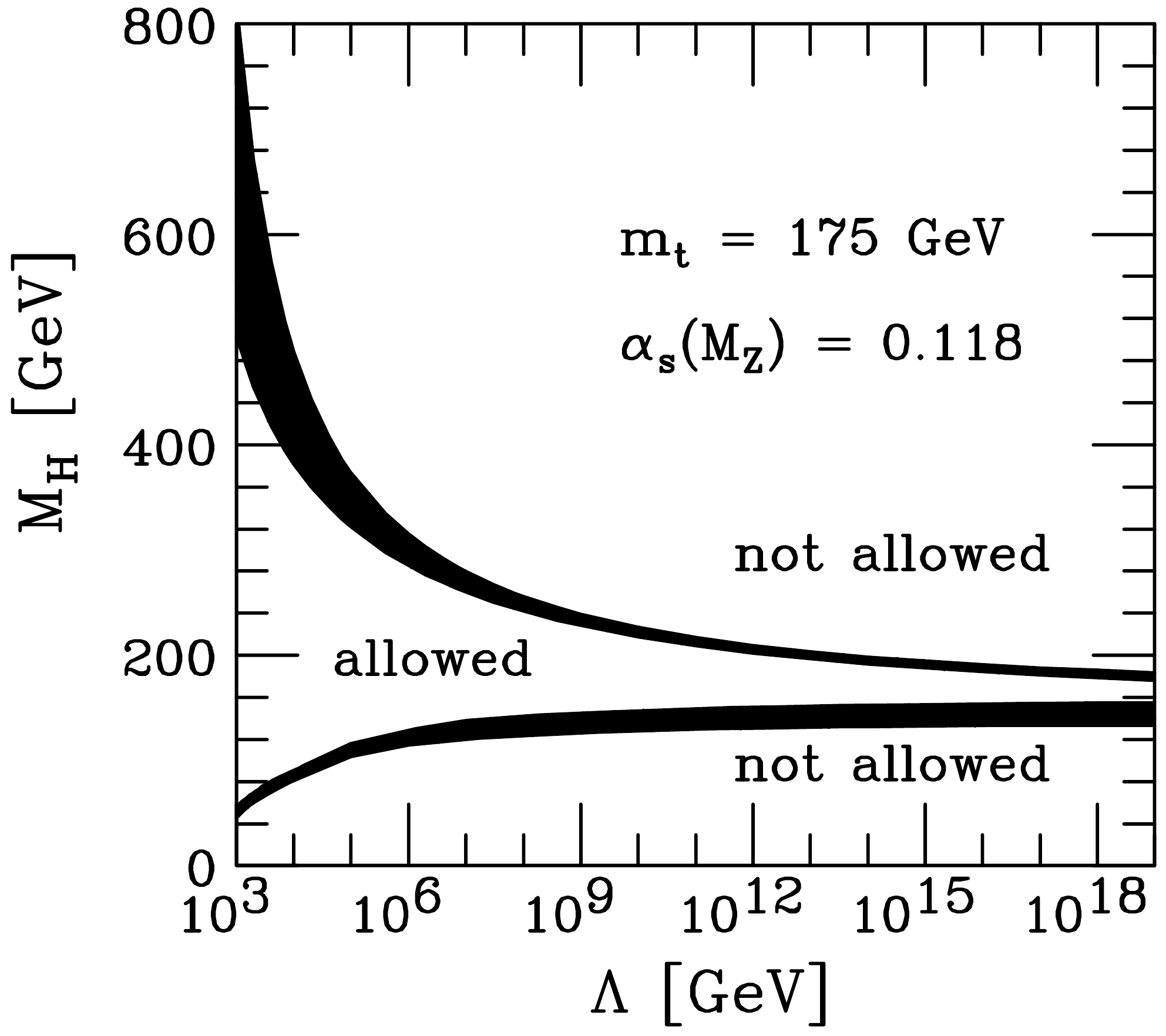}
\includegraphics[width=0.49\textwidth]{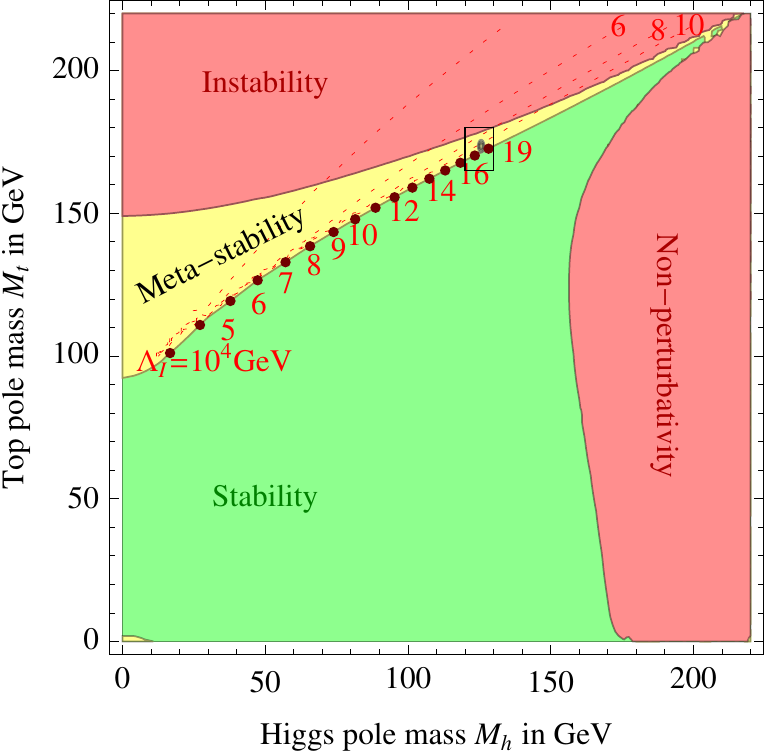}
\end{center}
\caption{Left: the maximum validity of the SM as a function of the Higgs mass, from Ref.~\cite{Hambye:1996wb}. The lower curve is the stability bound, while the upper one is the triviality bound. Right: regions of absolute stability, meta-stability and instability of the SM vacuum in the $m_t$ versus $m_H$ plane, from Ref.~\cite{Buttazzo:2013uya}. The dashed red lines show the instability scale, denoted as $\Lambda_I$.}
\label{fig:stabilityhiggs}
\end{figure}

Before turning to the properties of the SM Higgs that will be necessary to understand and interpret the results from the LHC, it is worth mentioning that there are other, theoretical constraints on the SM Higgs boson mass coming from the requirement of validity up to a certain physical scale (that can be pushed up to the Planck scale, $M_{\rm Pl} \approx 10^{19}$~GeV). Indeed, 2-loop calculations of the RG running of the $\lambda$ parameter in the Higgs potential show that $\lambda$ can become negative (leading to instability of the potential) or non-perturbative at a certain scale. 
The maximum validity scale $\Lambda$ as a function of the Higgs mass, as derived in Ref.~\cite{Hambye:1996wb}, is shown in the left panel of Fig.~\ref{fig:stabilityhiggs}. It shows that a heavy Higgs boson, which is already disfavored from the electroweak fit of the SM, must come with new physics at a relatively low scale, and that light Higgses, just above the LEP bound, may suffer from the stability bound. These has been updated with 3-loop results and a better precision on the SM input parameters in Refs.~\cite{Degrassi:2012ry,Buttazzo:2013uya}. The result is given in the right panel of Fig.~\ref{fig:stabilityhiggs}, showing in addition a meta-stability region, in which the potential is unstable but has a very small probability of quantum tunneling such that the lifetime of the electroweak vacuum exceeds the one of the Universe. The currently favored region after the discovery at the LHC is inside the black rectangle, and is close to the stable region but seems to lie in the meta-stable region (depending however on the top quark mass, see also Ref.~\cite{Alekhin:2012py}). 

\section{Production and decay of the SM Higgs boson at the LHC} \label{sec:higgs-smprop}

\begin{figure}[ht]
\begin{center}
\includegraphics[width=0.56\textwidth]{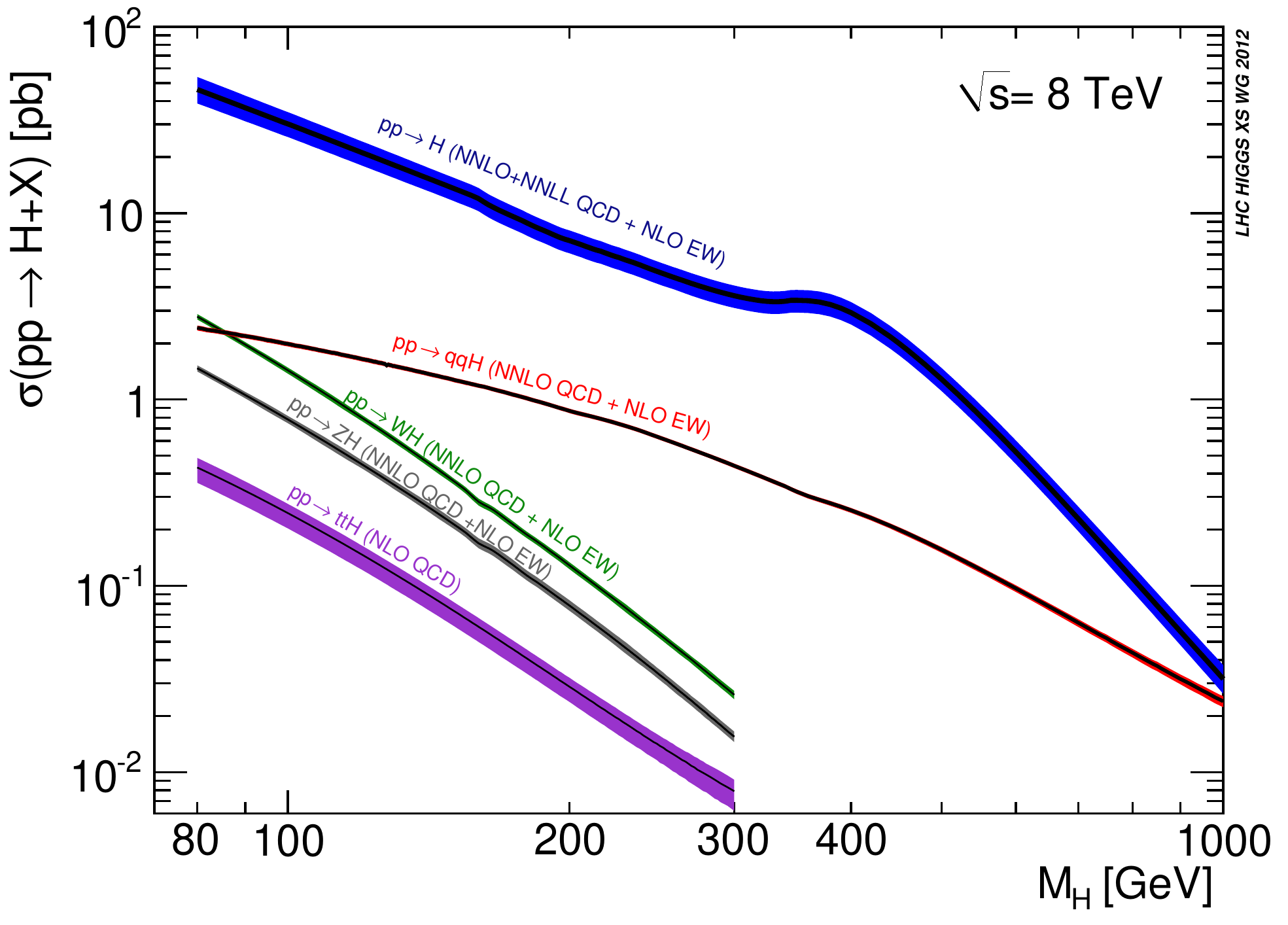}
\includegraphics[width=0.42\textwidth]{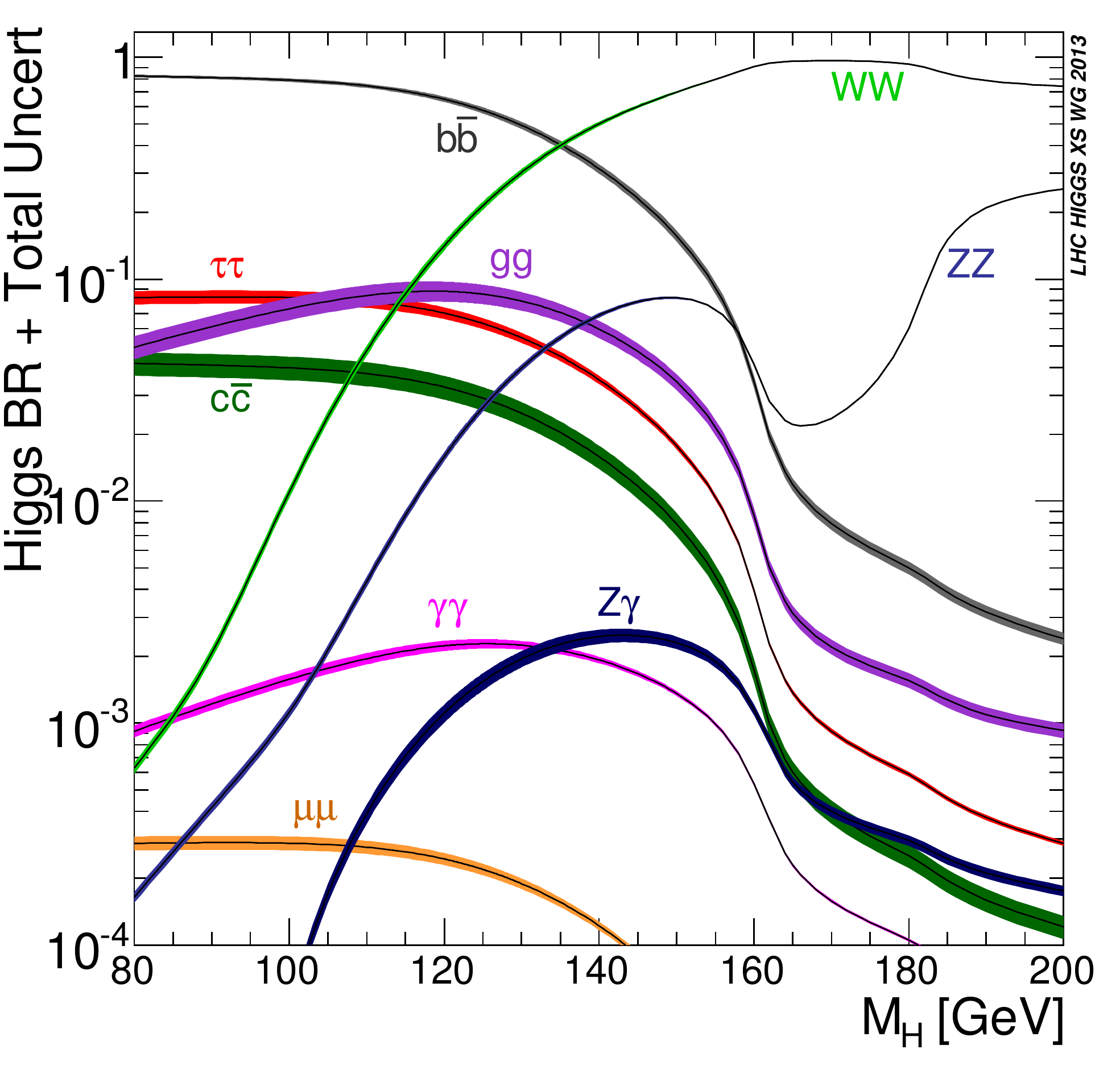}
\end{center}
\caption{Left: Production modes and associated cross sections for the SM Higgs boson at the LHC for $\sqrt{s} = 8$~TeV. Right: decay branching fractions of the SM Higgs boson. These results are given as a function of the Higgs mass, and are provided by the LHC Higgs Cross Section Working Group~\cite{Heinemeyer:2013tqa}. In both cases, the width of the band around each line gives an estimate of the theoretical uncertainty.}
\label{fig:proddecayhiggs}
\end{figure}

Searches for the Higgs boson at LEP turned out unsuccessful, while the Tevatron final results only exhibit a $3\sigma$ evidence around $m_H = 125$~GeV but no discovery. Since almost all results that will be used to constrain new physics from the properties of the observed Higgs boson come from LHC searches, it is interesting to first look in detail at the various production mechanisms at a 8~TeV $pp$ collider (corresponding to most of the data collected during Run~I of the LHC) and at the accessible decay modes. A comprehensive introduction to the production and decay modes of the SM Higgs boson can be found in Ref.~\cite{Djouadi:2005gi}, while a summary of the latest SM predictions is provided by the LHC Higgs Cross Section Working Group~\cite{Heinemeyer:2013tqa} (see also~\cite{Baglio:2010ae}) and shown here in Fig.~\ref{fig:proddecayhiggs}.

First of all, it should be noted that searches for the Higgs boson at a hadron collider such as the LHC are notoriously difficult due to the small cross sections, which are at most several dozens of pb in the low-mass region. This is in strong contrast with $W$ and $Z$ production, where the cross sections are of several dozens of nb, {\it i.e.}\ three orders of magnitude larger.
This can be understood because the couplings of the SM Higgs boson to particles inside the proton are either tiny or simply absent at tree-level. Indeed, the Higgs boson couples to fermions as $g_{Hff} = m_f/v$, which is very small for the light quarks constituting the proton, while gluons only couple to the Higgs boson at loop level because it is a color-neutral particle.
The two remaining possibilities for a sizable production of Higgs bosons at the LHC are {\it i)} via the coupling of the Higgs to $W$ and $Z$ vector bosons, or {\it ii)} via indirect or loop-induced processes originating from gluons. The latter possibility turns out to be the dominant production mechanism at the LHC, because of the strong, ${\cal O}(1)$ coupling of the Higgs to top quarks running in a one-loop (``triangle'') diagram originating from two gluons. This process will be subsequently denoted as gluon fusion (ggF), and constitute the dominant contribution to the blue, $pp \to H$ curve in the left panel on Fig.~\ref{fig:proddecayhiggs}.

Production of the Higgs boson via $W$ and $Z$ can be separated into two categories, leading to experimentally distinct signatures. First of all two vector bosons emitted from two distinct quarks can fuse into a Higgs boson, via its relatively large couplings to vector bosons, being $g_{HVV} = 2m_V^2/v$. This is known as vector boson fusion (VBF), and corresponds to a $pp \to qqH$ final state, where two quarks are emitted in the forward directions of the detector, allowing for a discrimination against the QCD background. The second possibility is to produce a Higgs boson in association with a $W$ or $Z$ vector boson, via an off-shell vector boson produced from quarks. The (possibly leptonic) decay of the vector boson can then be used to trigger the event and/or discriminate it against the background or other Higgs production modes. This will be further referred to as WH and ZH, commonly denoted as VH.
Finally, the production mechanism with the lowest cross section at the LHC, while still being accessible, involves the fusion of two top quarks into a Higgs boson, with the top quarks coming from the splitting of two gluons into a pair of top-antitop. This is denoted as ttH, production of the Higgs boson in association with a top quark pair. This rare production mechanism is however crucial to access directly the coupling of the Higgs to top quarks, as relatively light Higgs bosons can not decay into top quarks.

Turning to the possible decays of the Higgs, as shown in the right panel of Fig.~\ref{fig:proddecayhiggs}, two distinct regions can be seen. For $m_H \lesssim 160$~GeV, decays into fermions and gluons constitute a sizable or dominant part of the decay width, while above this threshold the decays into $WW$ and $ZZ$ completely dominate. This comes from the transition to the on-shell decay into vector bosons, with $\Gamma(H \to VV) \propto m_H^3$ at high mass while the decay width into fermions always scales as $m_H$. In order to constrain the properties of the Higgs boson, the most interesting region is at low-mass, $m_H \lesssim 160$~GeV, where complementary measurements can be made in many different final states. It is thus fortunate that an SM-like Higgs with a mass of about 125~GeV was found at the LHC, as we will see in Section~\ref{sec:higgs-measlhc}.

In spite of a relatively large $\sigma \times {\rm BR}$, some of the production $\times$ decay modes are difficult or impossible to access at the LHC because of the very large QCD background. In contrast, some rare decays were already observed with a good accuracy from the data collected during Run~I of the LHC since they correspond to ``clean'' final state, {\it i.e.}\ easy to distinguish from the SM background. For a 125~GeV SM-like Higgs boson, we expect $N = \sigma_{\rm tot} \times \mathscr{L} \approx 20~{\rm pb} \times 20~{\rm fb}^{-1} = 400000$ Higgs bosons to be produced  with the 20 fb$^{-1}$ of data collected at $\sqrt{s} = 8$~TeV at the LHC. It is interesting to note that about 50\% of these events correspond to $gg \to H \to b\bar b$, which cannot be distinguished from the overwhelming QCD background. At the LHC, the decay of Higgs bosons into a pair of $b$-quarks can only be probed when the Higgs boson is produced in association with other particles which can be triggered. A prime candidate is VH with leptonic decays of the massive vector bosons, {\it i.e.}\ $W^{\pm} \to \ell^{\pm}\nu$ or $Z^0 \to \ell^+\ell^-$, with $\ell \equiv e,\mu$.

The observation of $H \to gg$, loop-induced decay of the Higgs into gluons, and $H \to c\bar c$ processes at LHC is much more difficult. The large QCD background already mentioned for $H \to b\bar b$ remain as a problem, but in addition {\it i)} branching fractions, hence signal over background ratios, are much smaller, and {\it ii)} from the tagging of $b$-jets a discrimination can be made between $H \to b\bar b$ and most of the QCD background, while the tagging of $c$-jets and, even worse, gluon jets is extremely challenging. For these reasons, the expected sensitivity to these channels is very low at the LHC, even in association with a vector boson.\footnote{Modifications of the Higgs coupling to charm quarks were discussed recently in Ref.~\cite{Delaunay:2013pja}.} Finally, $H \to \mu\mu$ has a small, ${\cal O}(10^{-4})$ branching fraction hence its observation will require a lot of statistics at the LHC~\cite{ATL-PHYS-PUB-2012-004,CMS-NOTE-2012-006}. As a consequence, for an SM-like Higgs only the couplings to the $3^{\rm rd}$ generation of quarks and leptons are accessible on the middle term.

\section{Discovery and measurements at the LHC} \label{sec:higgs-measlhc}

ATLAS and CMS are the two main, multi-purpose detectors at the LHC. The physics operations at the LHC started in 2010, but most of the data was accumulated in 2011 ($\sim$4.7~fb$^{-1}$ at $\sqrt{s} = 7$~TeV) and in 2012 ($\sim$20.3~fb$^{-1}$ at $\sqrt{s} = 8$~TeV).
The observation of a new particle in the search for the Higgs boson was announced jointly by the ATLAS and CMS collaborations on July 4th, 2012, and published shortly after~\cite{Aad:2012tfa,Chatrchyan:2012ufa,Chatrchyan:2013lba}. This was based on the full statistics collected at 7~TeV plus about 5.5~fb$^{-1}$ of data at 8~TeV, and resulted for each experiment from the combination of searches for the following five final states: $H \to \gamma\gamma$, $H \to Z Z^{(*)} \to 4\ell$, $H \to W W^{(*)} \to 2\ell2\nu$, $H \to \tau\tau$ and $VH \to b\bar b$.

\begin{figure}[ht]
\begin{center}
\includegraphics[width=0.49\textwidth]{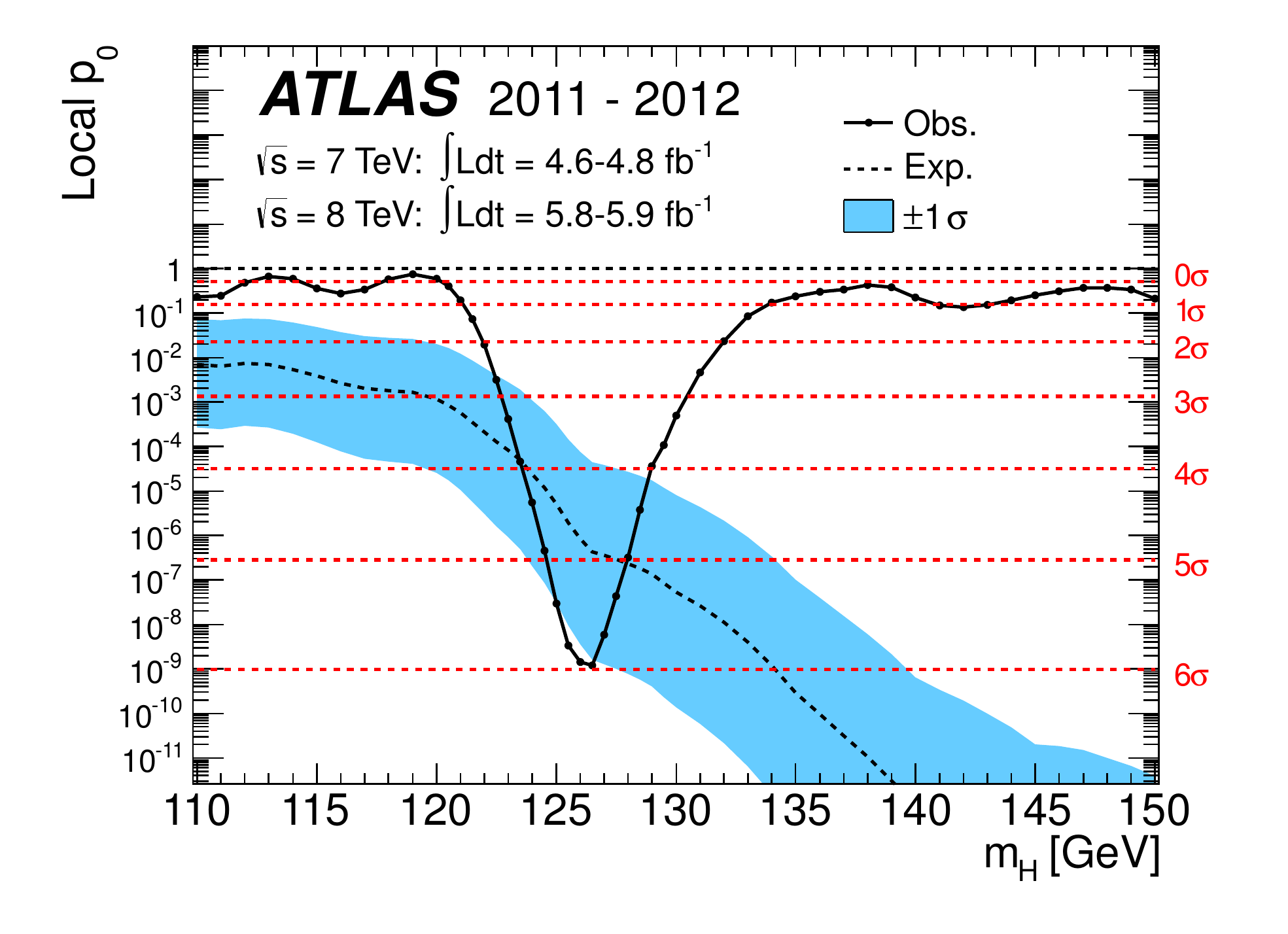}
\includegraphics[width=0.49\textwidth]{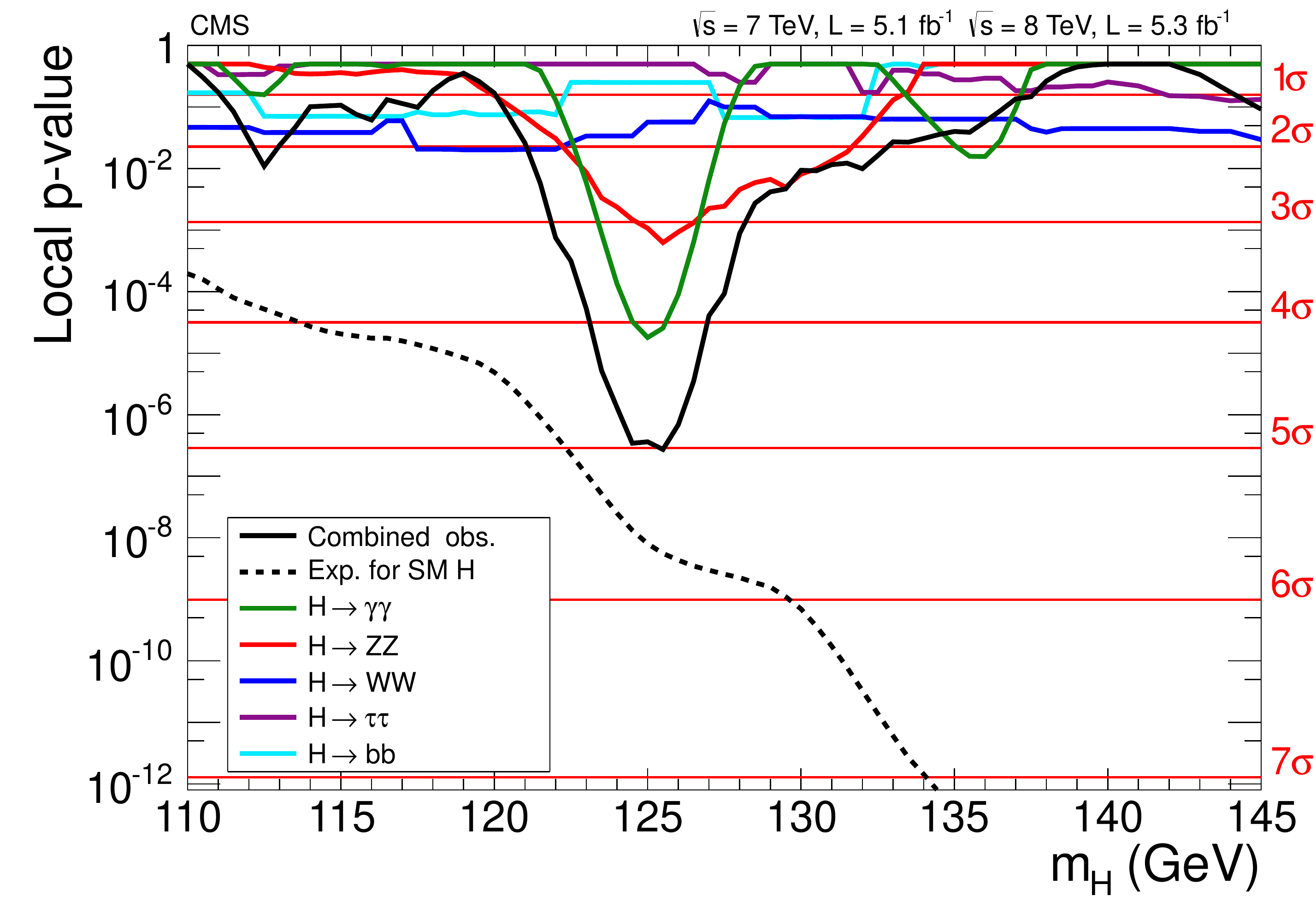}
\end{center}
\caption{Observed and expected $p$-value as a function of the Higgs mass in the search for the Higgs boson with the ATLAS (left) and CMS (right) detectors in July 2012. In both cases the black line corresponds to the combination of the search for the Higgs boson in different final states, which is shown explicitly for CMS.}
\label{fig:lhcdiscoverypvalue}
\end{figure}

In high energy physics, the statistical significance of a new phenomenon is expressed as a $p$-value, corresponding to the degree to which a given null hypothesis (in this case, SM with no Higgs boson) is incompatible with the data. More precisely, it quantifies the probability of obtaining a result at least as extreme as the one that was actually observed, assuming that the null hypothesis is true~\cite{Beringer:1900zz}. The $p$-value is commonly expressed in units of standard deviation of a normal distribution, or ``number of sigmas''. From the $p$-value, this equivalent significance $Z$ is given by $Z = \Phi^{-1}(1-p)$, where $\Phi^{-1}$ is the inverse of the cumulative distribution function of the normal distribution. The commonly accepted criteria to declare discovery is five-sigma, $Z=5$, which corresponds to an extremely small $p$-value of $3 \times 10^{-7}$. As can be seen in Fig.~\ref{fig:lhcdiscoverypvalue}, this local significance of $5\sigma$ was observed by the ATLAS and CMS collaborations, at around the same mass of $\sim 125$~GeV. One can directly see that the CMS discovery was mostly driven by the $H \to \gamma\gamma$ and $H \to Z Z^* \to 4\ell$ channels, which is also true for ATLAS but not explicitly shown here. This is an interesting observation because, at 125~GeV, ${\rm BR}(H \to \gamma\gamma) = 2.3 \times 10^{-3}$ and ${\rm BR}(H \to Z Z^* \to 4\ell) = 1.3 \times 10^{-4}$~\cite{Heinemeyer:2013tqa,Spira:1996if,Djouadi:1997yw}, meaning that the Higgs boson was discovered from rare decays. This latter decay mode was often called the ``gold plated mode'' for the discovery of the Higgs boson as it gives a very clear signature with a low background.

\begin{figure}[ht]
\begin{center}
\includegraphics[width=0.55\textwidth]{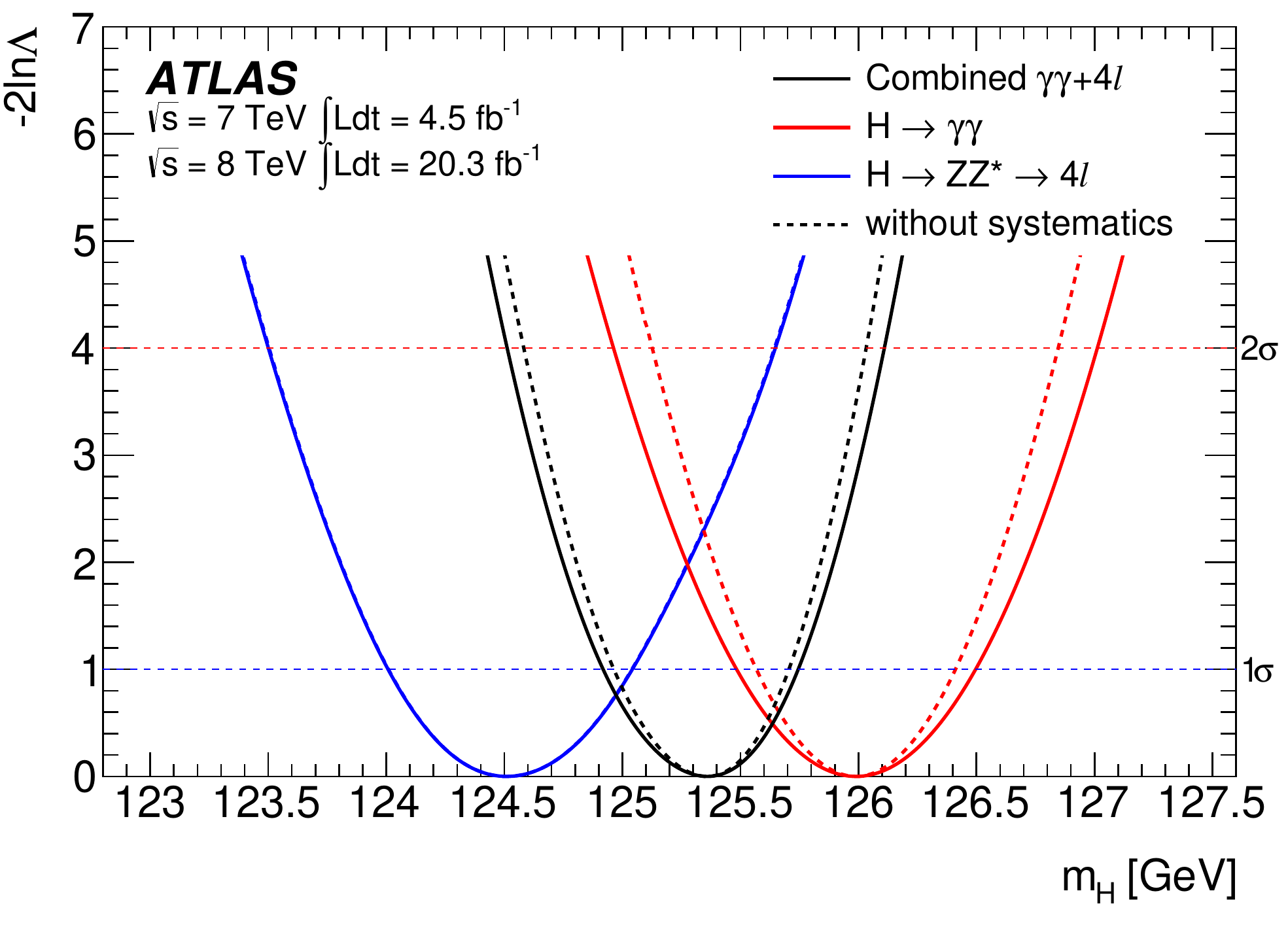}
\includegraphics[width=0.43\textwidth]{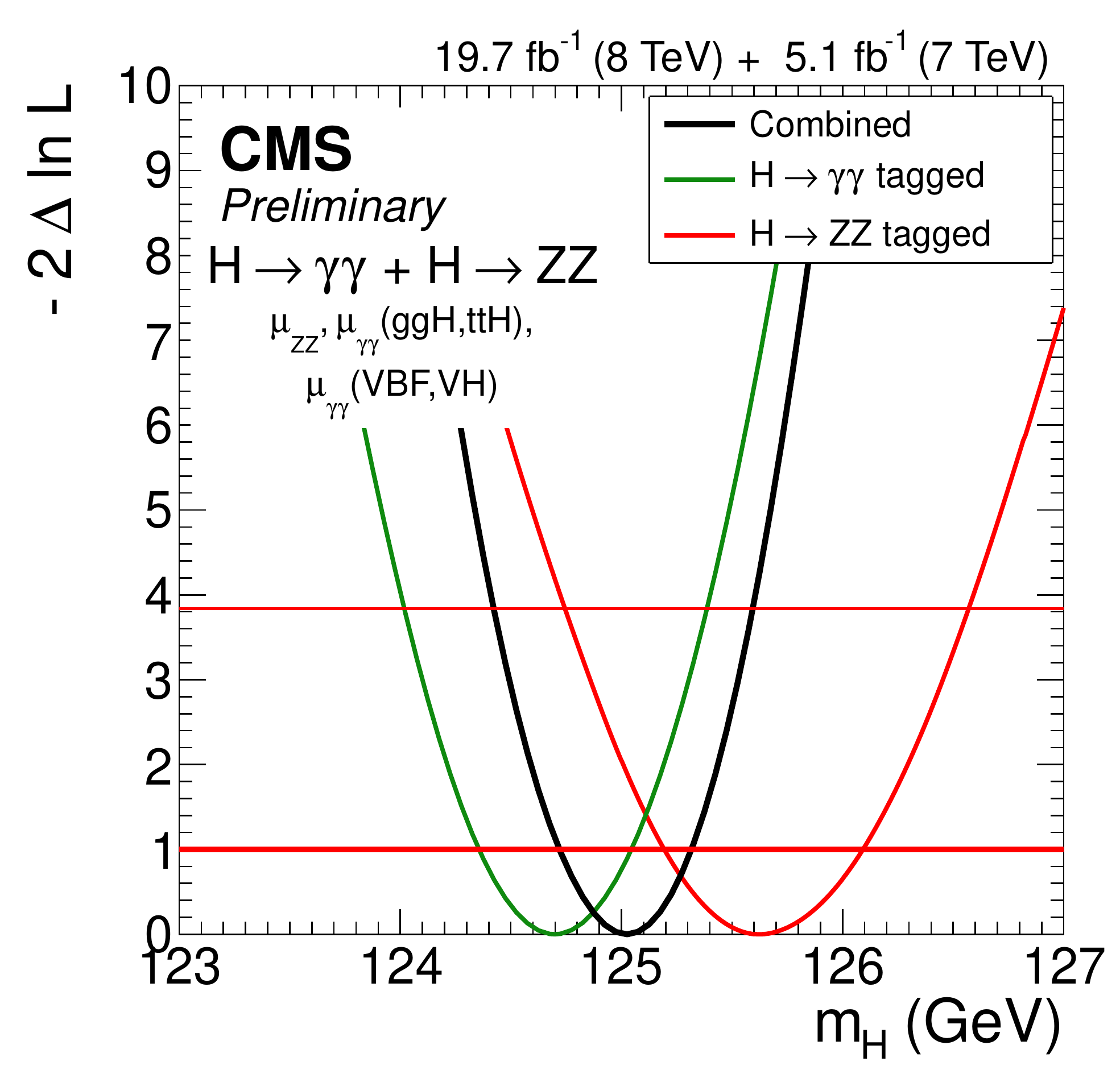}
\end{center}
\caption{Contraints on the Higgs mass from the measurements in the $H \to \gamma\gamma$ and $H \to ZZ^* \to 4\ell$ channels, as was performed by the ATLAS collaboration (left) and by the CMS collaboration (right). The y-axis is $-2 \, \Delta \log L$, where $L$ is the likelihood profiled over all parameters except $m_H$.}
\label{fig:lhchmass}
\end{figure}

Turning to results using the full luminosity collected during Run~I of the LHC, let us first look at the mass measurement. This is done using the two ``high-resolution'' channels,  $H \to \gamma\gamma$ and $H \to Z Z^* \to 4\ell$. Indeed, the estimate from $H\to WW^* \to 2\ell2\nu$  is very imprecise due to the presence of neutrinos in the final state, while mass measurements from $H \to b \bar b$ are affected by larger uncertainties due to the showering and hadronization of $b$ quarks. Finally, $H \to \tau\tau$ suffers from one or both problems, depending on the leptonic or hadronic nature of the $\tau$ decays. The latest results on the measurement of the Higgs mass from $H \to \gamma\gamma$ and $H \to Z Z^* \to 4\ell$ in ATLAS and CMS are shown in Fig.~\ref{fig:lhchmass}. ATLAS results are taken from the final Run~I mass measurement~\cite{Aad:2014aba}, while CMS results are a preliminary combination of the $H \to \gamma\gamma$ and $H \to ZZ^*$ results~\cite{CMS-PAS-HIG-14-009}.
The combination of these two channels in ATLAS gives a mass of $125.36 \pm 0.37 {\rm ~(stat.)} \, \pm 0.18 {\rm ~(sys.)}$~GeV, in perfect agreement with the CMS result of $125.03 ^{+0.26}_{-0.27} {\rm ~(stat.)} \, ^{+0.13}_{-0.15} {\rm ~(sys.)}$~GeV~\cite{CMS-PAS-HIG-13-005}. The discrepancy between the two channels is at the level of $2.0\sigma$ in ATLAS, which is a reduced tension compared to the $2.4\sigma$ discrepancy found in a previous mass estimate~\cite{ATLAS-CONF-2013-014}. Both ATLAS and CMS measurements are dominated by statistical uncertainties and should improve significantly during Run~II of the LHC. Finally, note that $H \to Z\gamma$ (with $Z \to \ell^+\ell^-$) is also a high-resolution channel, 
but it is not yet accessible at the LHC (the current limit on $H \to Z\gamma$ at $m_H = 125.5$~GeV is of about 10 times the SM expectation~\cite{Aad:2014fia,Chatrchyan:2013vaa}).

Beyond the mass of the observed state, the various searches at the LHC---thanks to the many decay modes accessible for a 125~GeV Higgs boson---open up the possibility to test in detail the properties of the Higgs boson and its couplings to other particles. This is crucial information to constrain new physics, as we will see in concrete examples in Sections~\ref{sec:higgs2012} to \ref{sec:pmssm}. The results of the Higgs searches at the LHC are given in terms of signal strengths, $\mu$, which scale the number of signal events expected for the SM Higgs, $n_s$. For a given set of selection criteria (or ``cuts''), the expected number of events is therefore $\mu \cdot n_s + n_b$, where $n_b$ is the expected number of background events, so that $\mu = 0$ corresponds to the no-Higgs scenario and $\mu = 1$ to an SM-like Higgs. Equivalently, signal strengths can be expressed as
\beq
\mu = \frac{\sigma \times A \times \varepsilon}{[\sigma \times A \times \varepsilon]_{\rm SM}} \,,
\label{eq:signalstr1}
\eeq
{\it i.e.}\ the ``visible cross section'' $\sigma_{\rm vis} = \sigma \times A \times \varepsilon$ divided by its SM expectation. Here, $A$ is the (geometrical) acceptance factor, {\it i.e.}\ the fraction of produced events that will cross the detector, while $\varepsilon$ is the efficiency of the cuts.

The presentation of the results in terms of signal strengths has non-trivial implications when using these results to constrain new physics affecting (or possibly faking) the Higgs boson. We will discuss that shortly after, in Section~\ref{sec:higgs-npconstlhc}; here we first focus on the status of the Higgs measurements with the full statistics collected during Run~I of the LHC. Experimental results on the Higgs boson are usually divided into publications presenting searches targeting one decay mode. This is well justified because in most cases the experimental signatures are clearly distinct, {\it i.e.}\ no or little contamination is expected between searches. For a given decay mode, there are usually several ``categories'' (or ``subchannels'' or ``signal regions''), corresponding to a set of cuts and for which results are given in terms of observed signal strengths as shown in Eq.~\eqref{eq:signalstr1}. These categories can be defined according to the nature of the final state particles produced from the decay of the Higgs ({\it e.g.}\ $H \to ZZ^* \to 4\ell$ gives $4e$, $4\mu$ or $2e2\mu$), or categories can be defined in order to improve sensitivity to given production modes (among ggF, VBF, WH, ZH and ttH) or from other properties of the final state objets ({\it e.g.} which part of the detector is involved).

\begin{figure}[ht]
\begin{center}
\includegraphics[width=0.45\textwidth]{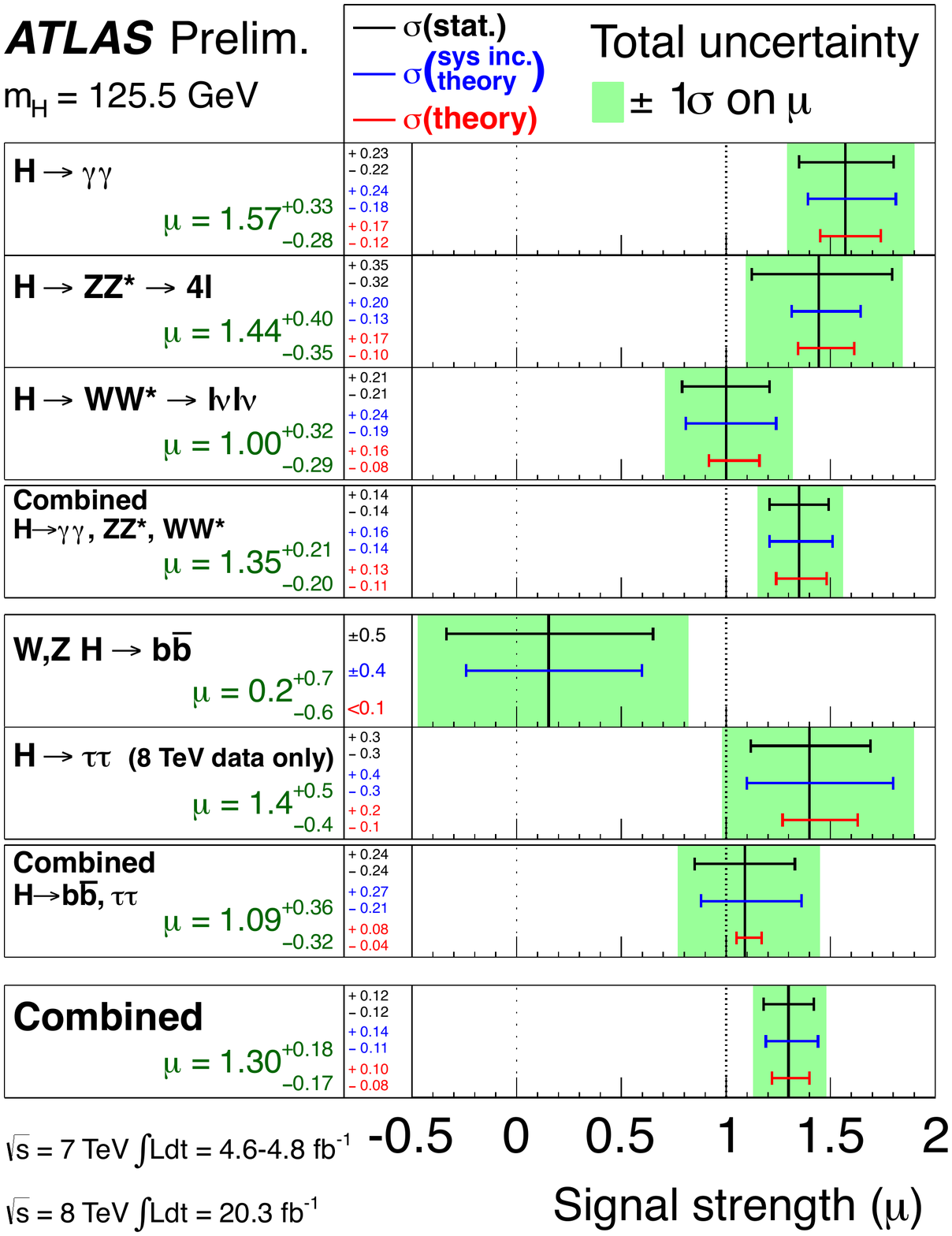}
\includegraphics[width=0.53\textwidth]{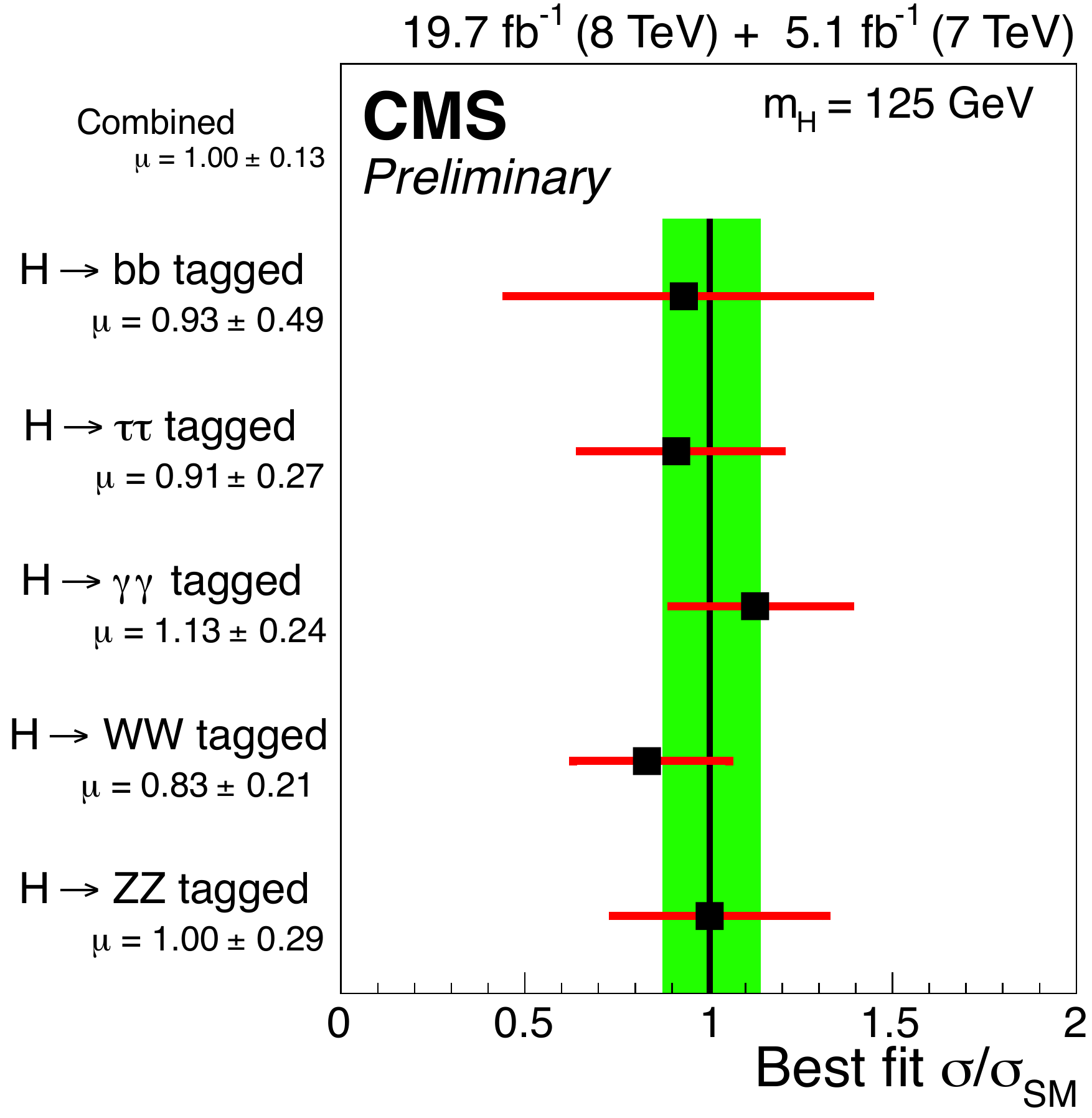}
\end{center}
\caption{Results in the search for the Higgs boson from ATLAS~\cite{ATLAS-CONF-2014-009} (left) and CMS~\cite{CMS-PAS-HIG-14-009} (right), given in terms of combined signal strengths (denoted as $\mu$) for the main decay modes of the Higgs boson: $\gamma\gamma$, $ZZ^*$, $WW^*$, $b\bar b$, $\tau\tau$. Also shown is the combination of several or all decay modes into one signal strength.}
\label{fig:lhchiggsresults}
\end{figure}

For each individual decay mode a combination of the categories can be made, but only {\it under the assumption of a universal rescaling of the production cross sections}. The results then corresponds to ``combined'' signal strengths. Due to this underlying assumption, it should be stressed that while a significant deviation of these combined $\mu$ from 1 must indicate the presence of new physics, a value of $\mu = 1$ could result from new physics enhancing the signal in some channels while reducing it in others. As we will see in Section~\ref{sec:higgs-npconstlhc}, this is the reason why we do not use this information to constrain new physics. However, these combined signal strengths give useful information on the current precision with which Higgs properties are measured. This information is shown in Fig.~\ref{fig:lhchiggsresults} for the latest ATLAS and CMS Higgs results~\cite{ATLAS-CONF-2014-009,CMS-PAS-HIG-14-009}. All the combined signal strengths show a very good consistency with the SM prediction. This leaves very little doubt that the observed state really is the Higgs boson from the scalar field\footnote{The spin-parity of the observed state induces significant changes in the kinematic distributions and has been constrained at the LHC in the diboson final states~\cite{Aad:2013xqa,Chatrchyan:2012jja,Chatrchyan:2013mxa,Chatrchyan:2013iaa,CMS-PAS-HIG-13-005}. The SM value of $J^P = 0^+$ is favored against all other tested hypotheses.} mostly responsible for the breaking of the electroweak symmetry. As can be seen, the diboson final states ($\gamma\gamma$, $ZZ^*$ and $WW^*$) are measured with a much better precision than the fermionic ($b\bar b$, $\tau\tau$) ones. Each of these three diboson final states are observed with a significance of more than $3\sigma$ by ATLAS or CMS alone, and uncertainties in the $\gamma\gamma$ and $WW^*$ channels start to be dominated by systematic uncertainties (in which theory uncertainties have a large part) while the $ZZ^*$ channel is still dominated by statistical uncertainties~\cite{Aad:2013wqa,Chatrchyan:2013mxa,Chatrchyan:2013iaa,Khachatryan:2014ira}. Turning to the fermionic channels, while the decay of the Higgs into $\tau\tau$ has been observed with more than $3\sigma$ significance in both experiments~\cite{ATLAS-CONF-2013-108,Chatrchyan:2014nva}, measurements on $H \to b\bar b$ remain very imprecise~\cite{ATLAS-CONF-2013-079,Chatrchyan:2013zna}. Fortunately, Tevatron final result exhibit a $3\sigma$ excess around 125~GeV in the $VH \to b\bar b$ channel~\cite{Aaltonen:2012qt}, leading to $\mu = 1.59^{+0.69}_{-0.72}$ at $m_H = 125$~GeV~\cite{Aaltonen:2013kxa} and to a constraint complementary to the LHC ones in this channel. Finally, as can be seen in Fig.~\ref{fig:lhchiggsresults}, combination of several (or all) decay modes is also performed by the collaborations. This however carries little information and is not expected to be useful in constraining new physics because a completely universal rescaling of all number of signal events has to be assumed.

As mentioned, 
the information given in Fig.~\ref{fig:lhchiggsresults} is not sufficient to constrain new physics. Indeed, as we will see even simple extensions of the SM make it possible, for instance, to reduce VBF and VH production modes while having enhanced or SM-like gluon fusion. Given the various categories defined for each decay mode, experiments have sensitivity to various production channels. In order to capture these effects with a few numbers, one can define new scale factors with respect to the SM Higgs as
\beq
\mu(X,Y)\equiv \frac{\sigma(X){\rm BR}(H\to Y)}{\sigma(X_{\rm SM} ){\rm BR}(H_{\rm SM}\to Y)}\,, \label{eq:theomu}
\eeq
where $X$ are the production modes (ggF, VBF, WH, ZH, ttH) and $Y$ are the decay modes (mainly $\gamma\gamma$, $ZZ^*$, $WW^*$, $b\bar b$, $\tau\tau$) of the SM Higgs boson. As discussed in the next Section, the $\sigma(X)$ and ${\rm BR}(H\to Y)$ are assumed to be simple rescalings of the SM value.
For all accessible decay modes, the ATLAS and CMS collaborations are showing results in terms of these ``signal strengths in the theory plane''. The five production modes of the SM are usually combined to form just two effective $X$ modes, VBF + VH (both of which depend on the $HVV$ coupling at tree-level) and ggF + ttH. The relevance of this combination will be discussed in the next Section. The likelihood can then be shown in the $(\mu({\rm ggF+ttH}, Y), \mu({\rm VBF+VH}, Y))$ plane for each decay mode $Y$.

\begin{figure}[ht]
\begin{center}
\includegraphics[width=0.54\textwidth]{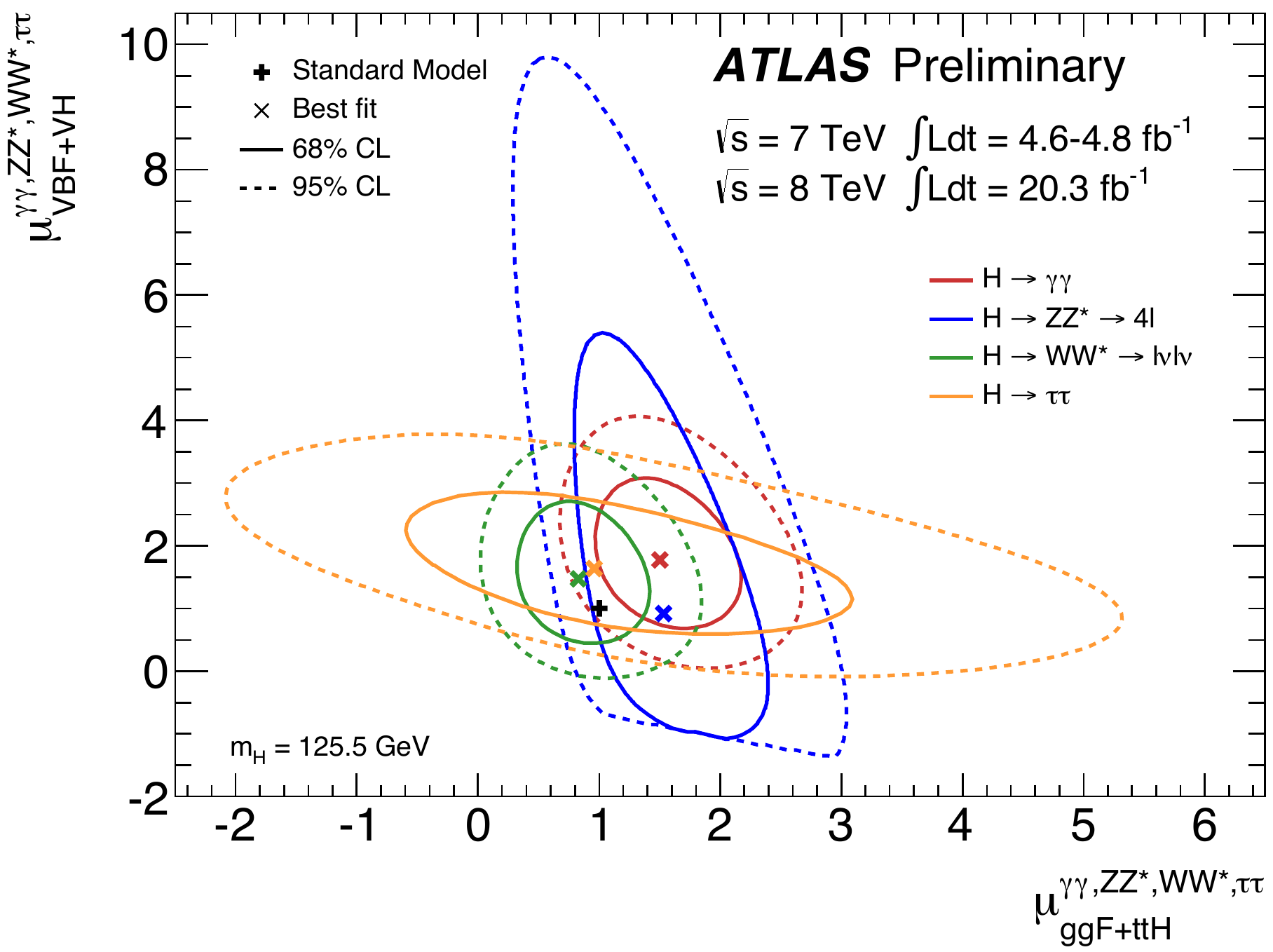}
\includegraphics[width=0.44\textwidth]{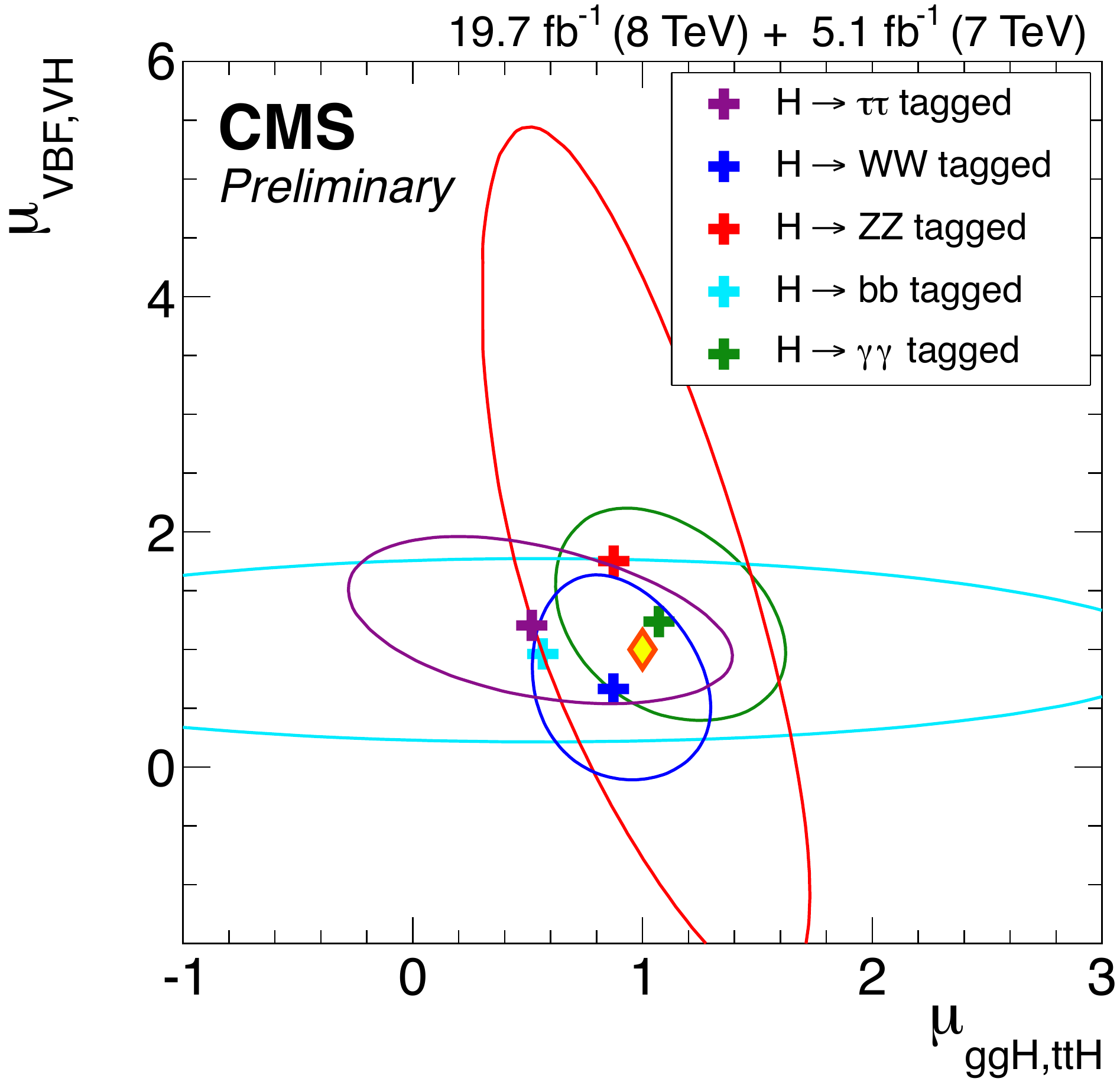}
\end{center}
\caption{Summary of the searches for the Higgs boson using the full luminosity collected during Run~I of the LHC with the ATLAS~\cite{ATLAS-CONF-2014-009} (left) and CMS~\cite{CMS-PAS-HIG-14-009} (right) detectors, in the plane $(\mu({\rm ggF+ttH}, Y), \mu({\rm VBF+VH}, Y))$ where $Y = \gamma\gamma, ZZ^*, WW^*, \tau\tau$, and (for CMS only) $b\bar b$. Contours of constant likelihood are shown, with the solid line corresponding to 68\%~CL and the dashed line (for ATLAS only) to 95\%~CL.}
\label{fig:lhchiggsresults2}
\end{figure}

The latest results from the LHC in this 2D plane~\cite{ATLAS-CONF-2014-009,CMS-PAS-HIG-14-009} are shown in Fig.~\ref{fig:lhchiggsresults2}. The excellent agreement with the SM predictions, already discussed from Fig.~\ref{fig:lhchiggsresults}, is even more manifest here. It is important to keep in mind that in these results, $\mu({\rm ggF+ttH}, Y)$ always reduces to a very good approximation to $\mu({\rm ggF}, Y)$, except for $Y = b\bar b$ (only shown for CMS), where the (very weak) constraint comes from ttH alone because ggF is not accessible. The current experimental results on ttH are imprecise (see, in particular, \cite{ATLAS-CONF-2014-011,ATLAS-CONF-2014-043, cms-tth-comb}) and never compete with the constraints from ggF, when available. Production of the Higgs boson in association with a pair of top quarks is experimentally challenging, as was discussed in Section~\ref{sec:higgs-smprop}, because of its small cross section, being 150 times lower than ggF for $m_H = 125$~GeV at $\sqrt{s} = 8$~TeV. Besides, note that there is a clear correlation between $\mu({\rm ggF+ttH}, Y)$ and $\mu({\rm VBF+VH}, Y)$ in almost all contours. This is because the categories designed to improve sensitivity to a given production mode within a given search will always contain some (often large) contamination from the other production modes. For example, requiring two additional forward jets in the $H \to \gamma\gamma$ analysis clearly improves the sensitivity to VBF production, but also has contamination from ggF (through $gg \to H + 2$~jets) and from VH (when the vector boson decays hadronically).

A comment on the $H \to ZZ^*$ decay mode can also be made. This decay is clearly mostly constrained through ggF and not via VBF and VH. This is a direct consequence of the lack of statistics in this channel (only 32 events in the $[120,130]$~GeV window in ATLAS~\cite{Aad:2013wqa}, 25 events in the $[121.5, 130.5]$~GeV window in CMS~\cite{Chatrchyan:2013mxa}), thus efficiently constraining only the dominant production mode, gluon fusion. (The sharp cut seen at $\mu({\rm VBF+VH}, ZZ^*) \approx -1$ in the ATLAS results comes from the impossibility to test a total number of events, $n_{\rm tot} = \mu \cdot n_s + n_b$, that is negative.)

\section{Constraining new physics with the LHC Higgs results} \label{sec:higgs-npconstlhc}

Since all results are expressed in terms of signal strengths, it is interesting to go back to the general expression given in Eq.~\eqref{eq:signalstr1} and ask ourselves how we can constrain new physics from it. As reminder, we have
\beq
\mu = \frac{\sigma \times A \times \varepsilon}{[\sigma \times A \times \varepsilon]_{\rm SM}} \,,
\label{eq:signalstr1bis}
\eeq
where, in general, $A \times \varepsilon$ and $[A \times \varepsilon]_{\rm SM}$ will not be the same. Indeed, in most experimental categories several production modes contribute to the expected Higgs signal, and new physics typically affect these production modes in a different way. For instance, suppressing gluon fusion while enhancing vector boson fusion may lead to the SM cross section while having $\mu \neq 1$ because of the change in the total fraction of events passing the cuts.
Moreover, as we will see in Section~\ref{sec:higgsdim6}, new physics could lead to modifications in the structure of the Higgs couplings to SM particles that lead to a change in the kinematic distributions, hence to a modification in $A \times \varepsilon$. However, an SM-like structure for the Higgs couplings happens to be an excellent approximation in most of the new physics models. In this case, {\it i.e.}\ modification of the strengths of the couplings only, the general expression for the signal strengths is
\beq
\mu = \frac{\sum_{X,Y} (A \times \varepsilon)_{XY} \sigma_X {\rm BR}_Y}{\sum_{X,Y} (A \times \varepsilon)_{XY} \sigma_X^{\rm SM} {\rm BR}^{\rm SM}_Y}
\,.  \label{eq:signalstr2}
\eeq
We recall that $X$ are the production modes (ggF, VBF, WH, ZH, ttH) and $Y$ are the decay modes (mainly $\gamma\gamma$, $ZZ^*$, $WW^*$, $b\bar b$, $\tau\tau$) of the SM Higgs boson.
In most cases, experimental categories are only sensitive to one decay mode of the Higgs, and Eq.~\eqref{eq:signalstr2} reduces to
\begin{align}
\mu = \frac{\sum_X (A \times \varepsilon)_X \sigma_X}{\sum_X (A \times \varepsilon)_X \sigma_X^{\rm SM}} \times \frac{{\rm BR}_Y}{{\rm BR}^{\rm SM}_Y}
&= \frac{\sum_X (A \times \varepsilon)_X \sigma_X^{\rm SM} C_X^2}{\sum_X (A \times \varepsilon)_X \sigma_X^{\rm SM}} \times \frac{{\rm BR}_Y}{{\rm BR}^{\rm SM}_Y} \nonumber \\
&= \sum_X {\rm eff}_X C_X^2  \times \frac{{\rm BR}_Y}{{\rm BR}^{\rm SM}_Y}
\,, \label{eq:signalstr3}
\end{align}
where the $C_X^2$ are factors scaling the cross sections compared to the SM expectation for each process $X$, and ${\rm eff}_X$ are ``reduced efficiencies'' that add up to 1. In the case of an inclusive search ({\it i.e.}\ $\forall X, (A \times \varepsilon)_X = (A \times \varepsilon)$), the ${\rm eff}_X$ are equal to the ratio of SM cross sections, $\sigma^{\rm SM}_X / (\sum_X \sigma^{\rm SM}_X)$.

While the LHC is constraining $\sigma \times {\rm BR}$, new physics affecting the decays of the Higgs correspond to a modification of the partial decays widths. In the approximation of an SM-like structure, the partial widths simply are simply scaled as $\Gamma_Y = \Gamma^{\rm SM}_Y C_Y^2$. Defining $\Gamma_H$ as the total decay width of the Higgs, signal strengths can then be expressed as
\begin{align}
\mu = \sum_X {\rm eff}_X C_X^2  \times \frac{{\rm BR}_Y}{{\rm BR}^{\rm SM}_Y} &= \sum_X {\rm eff}_X C_X^2  \times \frac{\Gamma_Y}{\Gamma^{\rm SM}_Y} \times \frac{\Gamma^{\rm SM}_H}{\Gamma_H} \nonumber \\
&= \sum_X {\rm eff}_X C_X^2  \times \frac{\Gamma^{\rm SM}_Y C_Y^2}{\Gamma^{\rm SM}_Y} \times \frac{\Gamma^{\rm SM}_H}{\sum_Y \Gamma^{\rm SM}_Y C_Y^2} \nonumber \\
&= \frac{1}{\sum_Y {\rm BR}^{\rm SM}_Y C^2_Y} \sum_X {\rm eff}_X C_X^2 C_Y^2
\,. \label{eq:signalstr4}
\end{align}
This means that a modification of a single decay width will impact all channels. This is particularly significant for $H \to b\bar b$, as we will see in several examples, because it has the largest branching fraction (57\% at $m_H = 125.5$~GeV).

We now have the general procedure for matching new physics modifications to the Higgs couplings---assuming SM-like coupling structure---with the measurement of a signal strength in a given channel.
In order to assess the compatibility of a given set of $(C_X,C_Y)$ with a single experimental result, we need to define a likelihood function $L(\mu, \boldsymbol{\nu}$), where $\boldsymbol{\nu}$ are the nuisance parameters, whose values are known with a limited accuracy from auxiliary or control measurements. The nuisance parameters model detector effects (affecting the identification and reconstruction of the particles) but also theoretical uncertainties, coming from uncertainties in the parton distribution functions (PDF), from the imperfect knowledge of the value of the SM parameters and from the missing higher-order corrections in the calculation of the SM cross sections and branching fractions. In the latter case, the auxiliary measurements do not truly exist but are introduced in the likelihood for convenience. The full likelihood function depends on the internal modeling of all these effects and is almost never provided by the experimental collaborations. It is however possible to reconstruct a simple likelihood, $L(\mu)$, from two information given in the experimental publications: the best fit to the data, denoted as $\hat\mu$, and the uncertainty at 68\%~CL (or $1\sigma$, also called the standard error), $\Delta \mu$. Assuming that the measurements are Gaussian, $- 2 \log L(\mu)$ follows a $\chi^2$ law, which is expressed as
\beq
- 2 \log L(\mu) = \chi^2(\mu) = \left(\frac{\mu - \hat\mu}{\Delta \mu}\right)^2 \,. \label{eq:Lchi2}
\eeq

While this is often a valid approximation, it needs to be pointed out that measurements are not necessarily Gaussian, depending on the size of the sample (which is currently a problem for $H \to ZZ^*$) and on the modeling of the systematic uncertainties. For example, it can be seen from the category results of $H \to \gamma\gamma$ from both experiments~\cite{ATLAS-CONF-2013-012,Khachatryan:2014ira} that the error bars are not necessarily symmetric around the best fit point, which indicates non-Gaussianities. Furthermore, even if the Gaussian approximation holds around the best fit point it may be inaccurate when testing signal strengths well beyond the standard error. It is also worth noting that this likelihood is only a function of $\mu$ and not of the nuisance parameters. It comes from the presentation of the experimental result, in which nuisance parameters have been removed from the full likelihood by constructing a profile likelihood,
\beq
L(\mu) = L(\mu, \widehat{\widehat{\boldsymbol{\nu}}}(\mu)) \,,
\eeq
where $\widehat{\widehat{\boldsymbol{\nu}}}(\mu)$ is given by the $\boldsymbol{\nu}$ that maximizes the likelihood for fixed $\mu$~\cite{Beringer:1900zz}.
Working with an approximation of the likelihood profiled over the nuisance parameters is not an issue, but it removes some freedom a theorist would like to have when using the Higgs data. In particular, there is no universal agreement on the treatment of theoretical uncertainties at the LHC and one might want to change it. Moreover, theoretical uncertainties will reduce in the future with more precise calculations of the Higgs production and decay processes and with the inclusion of new data into the PDF sets. Having only the likelihood profiled over all parameters except $\mu$ makes it very difficult to take into account these future improvements or simply to test an alternative treatment of the theoretical uncertainties. Finally, there should also be a dependence on the Higgs mass, while these individual results are usually only given for a single choice of Higgs mass.

Barring these limitations, Eq.~\eqref{eq:Lchi2} can be used to constrain new physics. However, it requires that at least $\hat\mu$, $\Delta \mu$, and also the reduced efficiencies ${\rm eff}_X$ (see Eq.~\eqref{eq:signalstr3}) be provided by the experimental collaborations for every individual category. This is unfortunately not always the case. Categories are sometimes defined without giving the corresponding signal efficiencies ({\it e.g.}~the ATLAS $H \to WW^*$ analysis~\cite{ATLAS-CONF-2013-030}), and/or the result is given as a (set of) ``combined'' signal strength(s) but not in terms of signal strengths category per category ({\it e.g.}~the ATLAS $ZZ^*$ analysis~\cite{ATLAS-CONF-2013-013} and the CMS $\tau\tau$ analysis~\cite{Chatrchyan:2014nva}). Such combined $\mu$ should in general not be used because they have been obtained under the assumption of SM-like production of the Higgs boson.
 Whenever the ${\rm eff}_X$ are not given in the experimental publications it is in principe possible to obtain estimates from a reproduction of the analysis cuts applied on signal samples generated by Monte Carlo (MC) simulation. However, this turns out to be a very difficult or impossible task. Indeed, the discovery of the Higgs boson and the measurement of its properties were a top priority of the LHC physics program, hence experimentalists prepared complex search strategies to optimize the sensitivity. They often rely on multivariate analyses (MVA) techniques that are impossible to reproduce in practice. Whenever the information on reduced efficiencies is not available we are left to guesswork, with a natural default choice being that ${\rm eff}_X = \sigma^{\rm SM}_X / (\sum_X \sigma^{\rm SM}_X)$, corresponding to an inclusive analysis.

We have just discussed the constraints on new physics from one LHC Higgs channel. While this is already a non-trivial task, complications arise when using several categories/searches at the same time, which is our goal ultimately. The simplest solution is to define the full likelihood as the product of the individual likelihoods,
\beq
L(\boldsymbol{\mu}) = \prod_{i=1}^{n} L(\mu_i) \quad \Rightarrow \quad \chi^2(\boldsymbol{\mu}) = \sum_{i=1}^{n} \chi^2(\mu_i) = \sum_{i=1}^{n} \left(\frac{\mu_i - \hat\mu_i}{\Delta \mu_i}\right)^2 \,. \label{eq:likesimpleprod}
\eeq
However, this assumes that all the measurements are completely independent. We know that this is not the case and that the various individual measurement share common systematic uncertainties. They are divided into two categories: the shared experimental uncertainties, coming from the presence of the same final state objects and from the estimation of the luminosity, and the shared theoretical uncertainties, dominated by the contributions from identical production and/or decay modes to the expected Higgs signal in different categories. The estimation of the experimental uncertainties in ATLAS should be largely independent from the one in CMS, hence these correlations can be treated separately for measurements performed by one collaboration or the other. Conversely, the estimation of the theoretical uncertainties are the same in ATLAS and CMS and should be correlated between all measurements.

In the case where all measurements are Gaussian, it is possible to take these correlations into account in a simple way, defining our likelihood as
\beq
-2 \log L(\boldsymbol{\mu}) = \chi^2(\boldsymbol{\mu}) = (\boldsymbol{\mu} - \hat{\boldsymbol{\mu}})^T V^{-1} (\boldsymbol{\mu} - \hat{\boldsymbol{\mu}}) \,, \label{eq:gausscovmatrix}
\eeq
where $V^{-1}$ is the inverse of the $n \times n$ covariance matrix~\cite{Beringer:1900zz}, with $V_{ij} = {\rm cov}[\hat\mu_i,\hat\mu_j]$ (leading to $V_{ii} = \sigma_i^2$). Unfortunately, the off-diagonal elements of this matrix are not given by the experimental collaborations and are very difficult to estimate from outside the collaboration. This remarkably simple and compact expression for the likelihood (a $n \times n$ matrix) is only valid under the Gaussian approximation; beyond that the expression and the communication of the likelihood become more complicated.

In spite of the various caveats and difficulties due to missing information from the experiments, this simple procedure typically gives an acceptable approximation to the experimental likelihood with the current data (as can be seen, {\it e.g.}, in Fig.~\ref{fig:reconstruction} below in cases where all the information on reduced efficiencies is available) and has been used in countless papers to fit the couplings of the Higgs and apply constraints on new physics models. There is however an alternative experimental input one can use to constrain new physics: the signal strengths in the theory plane, previously defined as
\beq
\mu(X,Y)\equiv \frac{\sigma(X){\rm BR}(H\to Y)}{\sigma(X_{\rm SM} ){\rm BR}(H_{\rm SM}\to Y)}\,.
\eeq
Results in terms of theoretical signal strengths are (most of the time) shown in the plane $(\mu({\rm ggF+ttH}, Y), \mu({\rm VBF+VH}, Y))$, as can be seen in Fig.~\ref{fig:lhchiggsresults2}. The combination of the VBF, WH and ZH production modes is well justified from the theory point of view. Indeed, one can generate different scaling factors ($C_{\rm VBF} \neq C_{\rm WH} \neq C_{\rm ZH}$) from
\beq
\left[ C_Z \frac{m_Z^2}{v} (Z_\mu)^2 + C_W \frac{2 m_W^2}{v} W_\mu^+W_\mu^- \right] H
\eeq
terms in the Lagrangian, with $C_W \neq C_Z$, which can be induced from the dimension-6 operator ${\cal O}'_{D^2} = |H^\dag D_\mu H|^2$, as we will see in Section~\ref{sec:higgsdim6}. This corresponds to a violation of the custodial symmetry which leads to large corrections to the Peskin-Takeuchi $T$ parameter~\cite{Peskin:1990zt,Peskin:1991sw}, which is constrained to be very small from the electroweak precision measurements at LEP ~\cite{ALEPH:2005ab} (see Ref.~\cite{Baak:2014ora} for the latest results from a global electroweak fit).
While there are ways to generate $C_{\rm VBF} \neq C_{\rm WH} \neq C_{\rm ZH}$ without violating the custodial symmetry in an effective approach (see Section~\ref{sec:higgsdim6}), these effects are usually small and therefore grouping together VBF, WH and ZH is not a problem for testing the vast majority of the new physics models.

The combination of the ggF and ttH production modes might be more problematic. In the SM, gluon fusion is dominated by the top quark contribution~\cite{Djouadi:2005gi}. We have $C_{\rm ggF} \approx C_{\rm ttH}$ in models of new physics where it is still the case, but {\it i)} this is only an approximation where the contributions from the bottom quark are neglected, and {\it ii)} this can be drastically modified with new physics affecting the Higgs boson. For instance, in the Two-Higgs-doublet-model (2HDM) of Type~II, tree-level couplings of the Higgs to top quarks and to bottom quarks are rescaled independently. Moreover, new particles could enter the gluon fusion loop (such as the stops in SUSY, the superpartners of the top quark) and change the ggF scaling factor independently from the ttH one.
``Fortunately'', as already discussed in Section~\ref{sec:higgs-measlhc}, for all decay modes except $H \to b\bar b$ (where gluon fusion initiated production of the Higgs is not accessible) the ttH production mode is currently constrained with much poorer precision than ggF because of its small cross section. Therefore, with the current data it is justified to take $\mu({\rm ggF+ttH}, Y) = \mu({\rm ggF}, Y)$ for all channels except $H \to b\bar b$, and $\mu({\rm ggF+ttH}, Y) = \mu({\rm ttH}, Y)$ for $Y = b\bar b$.

The information presented in Fig.~\ref{fig:lhchiggsresults2} consists in 68\%~CL contours in the 2D plane $(\mu({\rm ggF+ttH}, Y), \mu({\rm VBF+VH}, Y))$ (supplemented by the corresponding 95\%~CL contours in the case of ATLAS).
At this point, a comment is in order. In order to obtain this result for each decay mode $Y$, the experiments have defined as test statistic the profile likelihood ratio, defined as
\beq
\Lambda(\boldsymbol{\mu})
= \frac{L(\boldsymbol{\mu}, \widehat{\widehat{\boldsymbol{\nu}}}(\boldsymbol{\mu}))}{L(\widehat{\boldsymbol{\mu}}, \widehat{\boldsymbol{\nu}})} \,, \label{eq:profilelikeratio}
\eeq
where $\boldsymbol{\mu} = \begin{pmatrix} \mu({\rm ggF+ttH}, Y) \\ \mu({\rm VBF+VH}, Y) \end{pmatrix}$ and $L(\widehat{\boldsymbol{\mu}}, \widehat{\boldsymbol{\nu}})$ corresponds to the (global) maximum of the full likelihood~\cite{Beringer:1900zz,ATLAS-CONF-2014-009,CMS-PAS-HIG-14-009}. All other parameters are treated as nuisance parameters. According to asymptotic properties of the profile likelihood ratio, $- 2 \log \Lambda(\boldsymbol{\mu})$ is supposed to be distributed as a $\chi^2$ distribution with $n$ degrees of freedom, with $n = {\rm dim}(\boldsymbol{\mu}) = 2$ in our case. It is thus possible to directly match a value of $- 2 \log \Lambda(\boldsymbol{\mu})$ with a confidence level from the cumulative distribution function of the $\chi^2$ distribution. The most common values are tabulated, and for $n=2$, the 68\%~CL and 95\%~CL contours correspond to $- 2 \log \Lambda(\boldsymbol{\mu}) = 2.3$ and 6.0, respectively.

\begin{figure}[t]
	\centering
		\includegraphics[width=5cm]{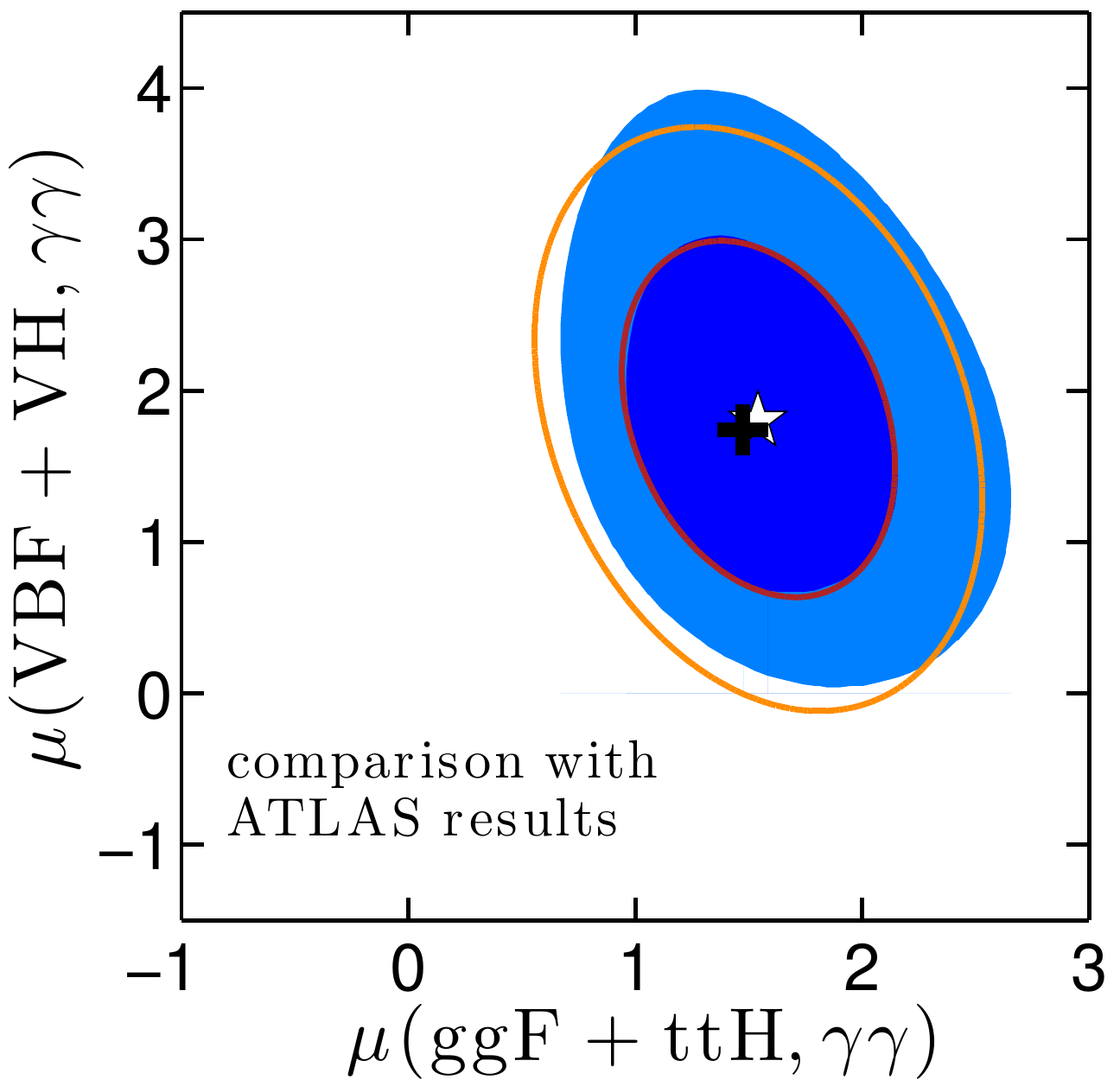}\quad
		\includegraphics[width=5cm]{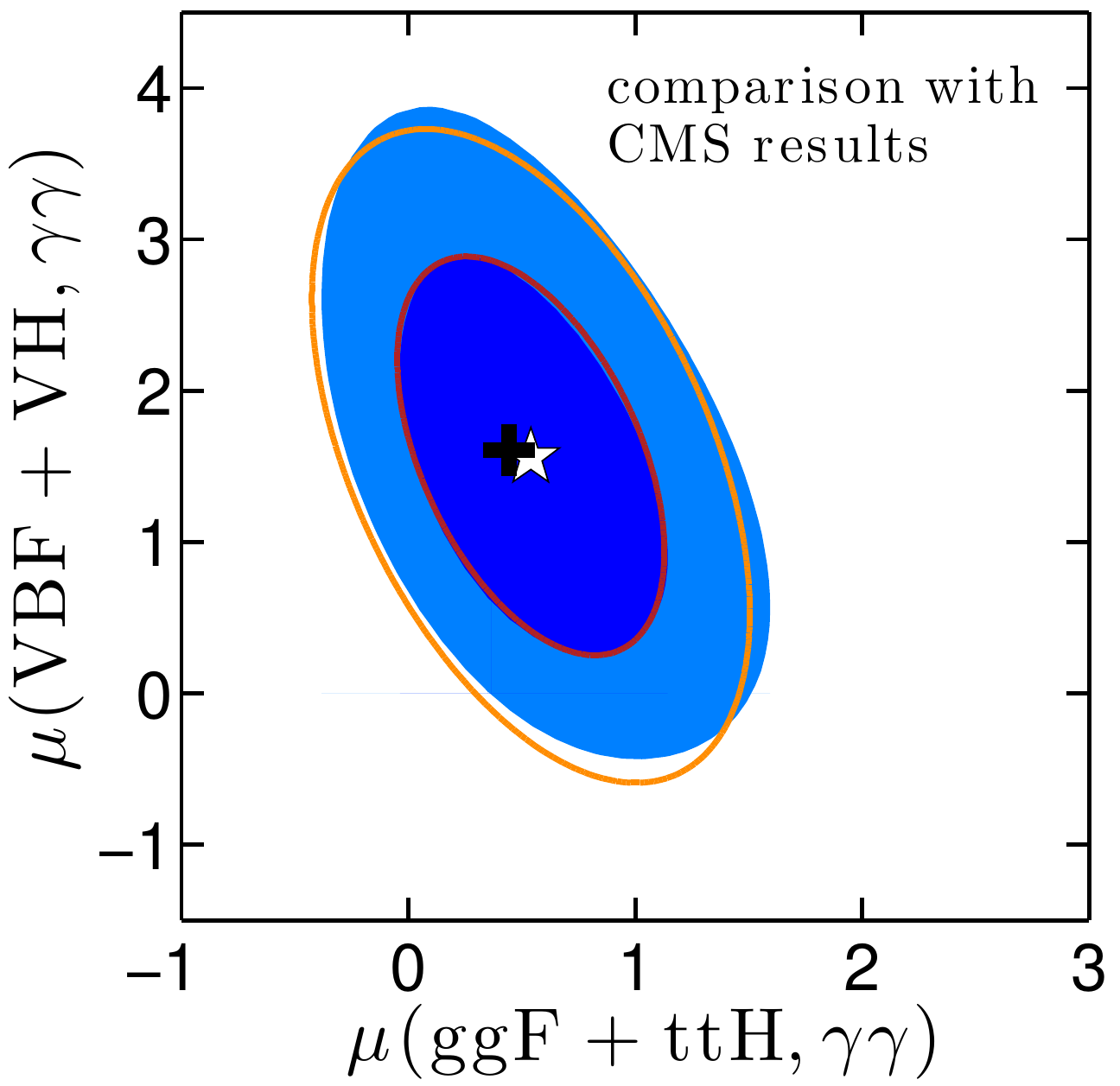}\quad
		\includegraphics[width=4.9cm]{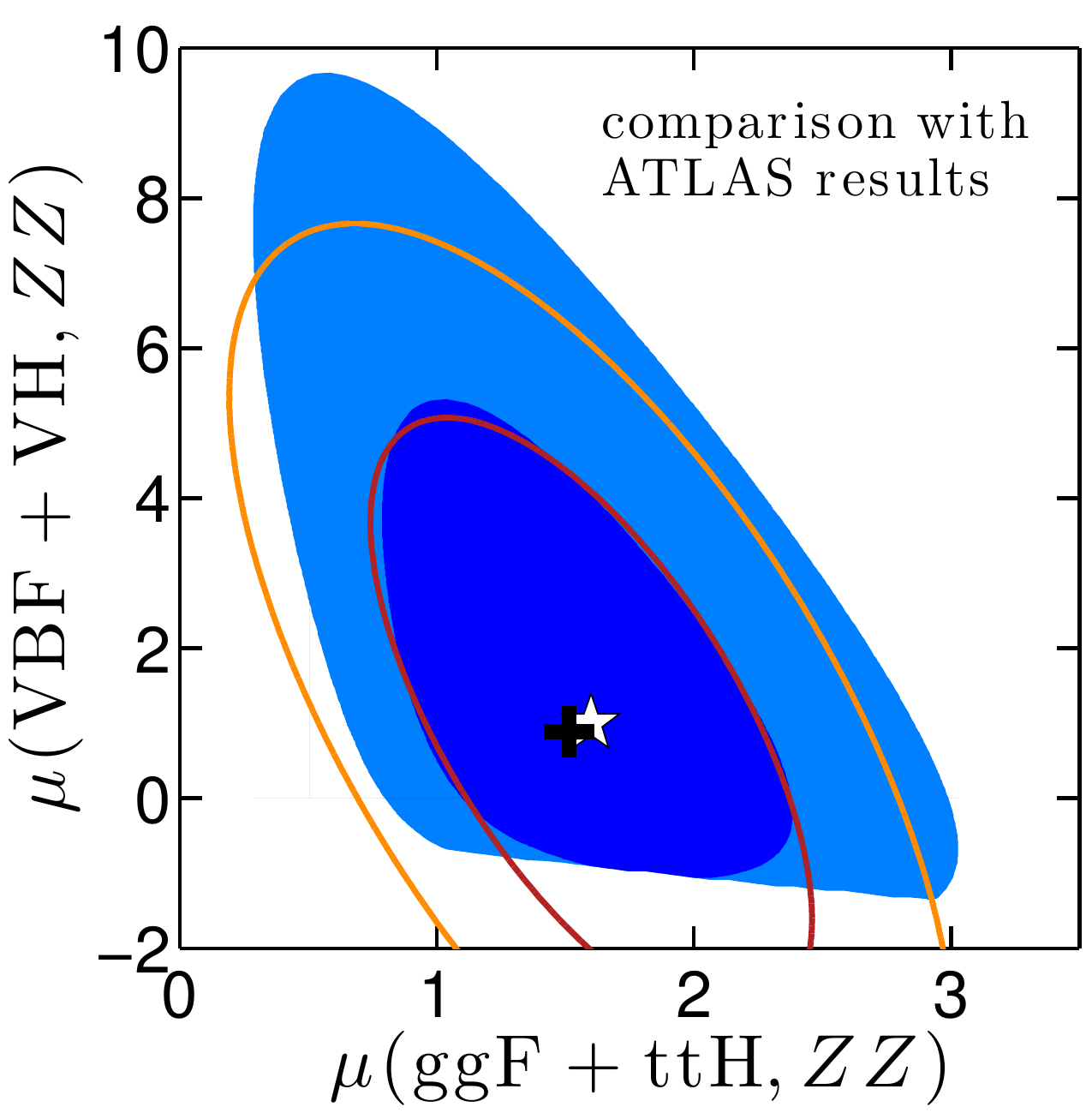}
	\caption{Gaussian fit to signal strenghts in the $(\mu({\rm ggF+ttH}, Y), \mu({\rm VBF+VH}, Y))$ plane, from left to right for the ATLAS $H \rightarrow \gamma\gamma$~\cite{Aad:2013wqa}, CMS $H \rightarrow \gamma\gamma$ (preliminary)~\cite{CMS-PAS-HIG-13-001}, and ATLAS $H \rightarrow ZZ^*$~\cite{Aad:2013wqa} channels. The dark and light blue filled areas are the 68\% and 95\%~CL regions given by the experiments, the red and orange lines show the fitted ones. In all three cases, we approximately reconstruct the likelihood by fitting a bivariate normal distribution to the 68\%~CL contour given by the collaboration. The black crosses are the experimental best fit points, while the white stars are the mean values from the fit. 
\label{fig:GaussianFit}}
\end{figure}

\begin{figure}[t]
		\includegraphics[width=4.95cm]{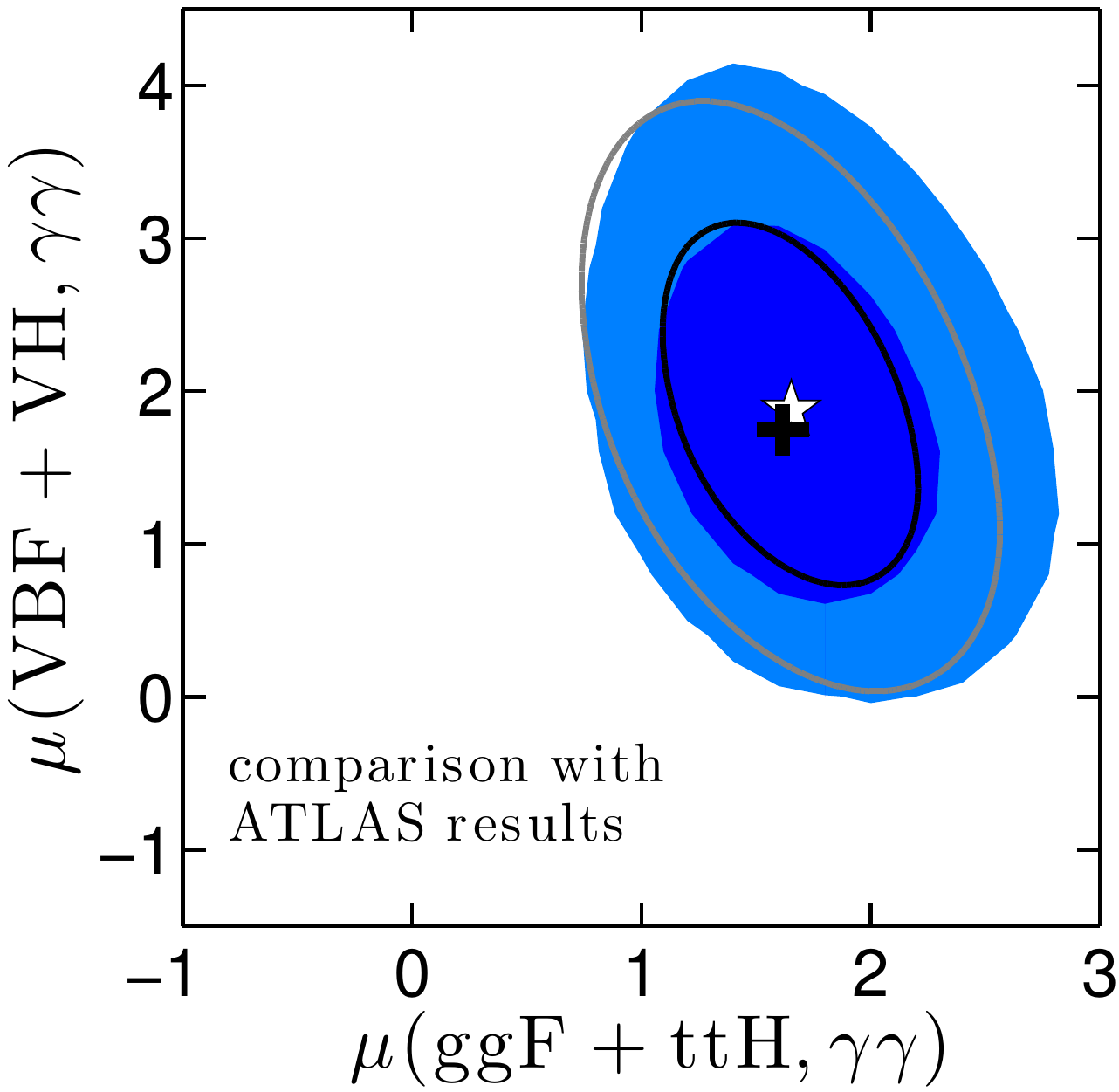}\quad
		\includegraphics[width=5.0cm]{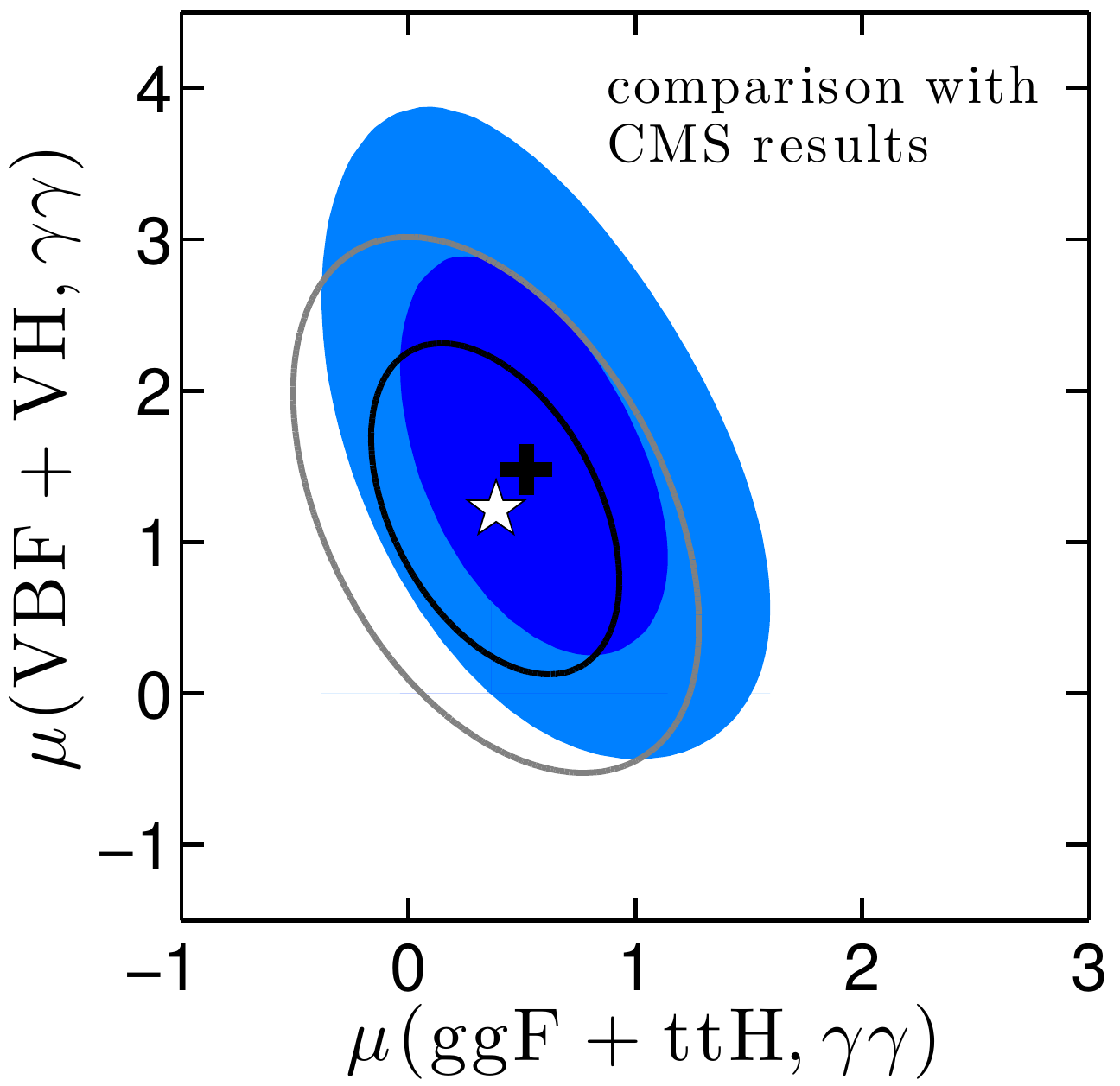}\quad
		\includegraphics[width=5.0cm]{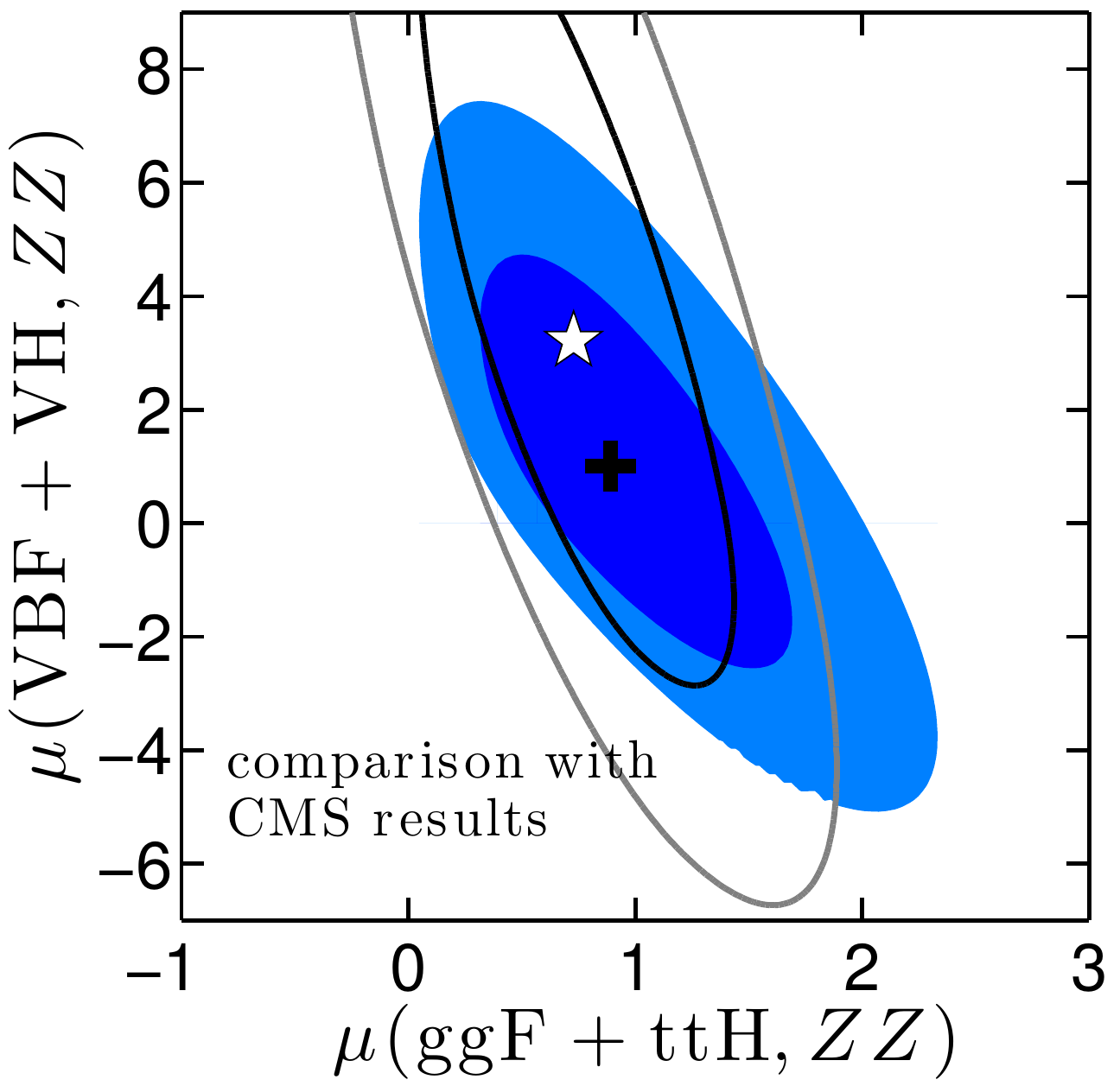}
	\caption{Reconstructing the likelihood from subchannel information. The black and gray lines show the 68\% and 95\%~CL contours in the $(\mu({\rm ggF+ttH}, Y), \mu({\rm VBF+VH}, Y))$ plane, reconstructed from signal strengths and efficiencies for the experimental categories $I$ in each final state; from left to right for the ATLAS $H \rightarrow \gamma\gamma$~\cite{ATLAS-CONF-2013-012}, CMS $H \rightarrow \gamma\gamma$~\cite{CMS-PAS-HIG-13-001}, and CMS $H \rightarrow ZZ^*$ (preliminary)~\cite{CMS-PAS-HIG-13-002} channels.  For comparison, the dark and light blue filled areas show the 68\% and 95\%~CL regions directly given by the collaborations. The black crosses are the experimental best fit points,  the white stars are the reconstructed ones. 
\label{fig:reconstruction}}
\end{figure}

In order to constrain new physics, we need for each experiment and for each decay mode the full information in this 2D plane and not only one (or two) contours. Hopefully, this information will be released systematically by the experimental collaborations. At the moment it is available for the $H \to \gamma\gamma$, $H \to ZZ^*$ and $H \to WW^*$ final states in ATLAS under a convenient format available on HepData~\cite{atlasgamgamgrid,atlasZZgrid,atlasWWgrid}, and a ``temperature plot'' in this 2D plane (with the color indicating the value of the likelihood) is given for the CMS $H \to \gamma\gamma$ results~\cite{Khachatryan:2014ira}. 
Whenever this information is not available the way out is to fit the 68\%~CL contour, corresponding to $- 2 \log \Lambda(\boldsymbol{\mu}) = 2.3$, with a bivariate normal distribution. Using the shorthand ggF for ggF+ttH, and VBF for VBF+VH, the likelihood for a given decay mode $Y$ can be expressed as 
\beq
- 2 \log \Lambda(\boldsymbol{\mu}) =
(\boldsymbol{\mu} - \hat{\boldsymbol{\mu}})^T
\begin{pmatrix} \sigma_{\rm ggF}^2 & \rho \sigma_{\rm ggF} \sigma_{\rm VBF} \\ \rho \sigma_{\rm ggF} \sigma_{\rm VBF} & \sigma_{\rm VBF}^2 \end{pmatrix}^{-1}
(\boldsymbol{\mu} - \hat{\boldsymbol{\mu}}) \,, \label{eq:mu2d}
\eeq
which is equal to 2.3 for the points $\boldsymbol{\mu} = \begin{pmatrix} \mu_{\rm ggF} \\ \mu_{\rm VBF} \end{pmatrix}$ sitting on the 68\%~CL contour. From this expression, and using a digitized version of the contour, one can fit the five parameters $\hat{\mu}_{\rm ggF}$, $\hat{\mu}_{\rm VBF}$, $\sigma_{\rm ggF}$, $\sigma_{\rm VBF}$, and $\rho$ (the correlation between the measurements). Several checks can then be made: ($\hat{\mu}_{\rm ggF}$, $\hat{\mu}_{\rm VBF}$) has to be close to the position of the true best fit point, agreement between the fitted 68\%~CL contour and the one from ATLAS or CMS should be good, and if a 95\%~CL contour is available one can evaluate the importance of non-Gaussianities further away from the best fit region. This is shown for the ATLAS $H \rightarrow \gamma\gamma$~\cite{Aad:2013wqa}, CMS $H \rightarrow \gamma\gamma$ (preliminary)~\cite{CMS-PAS-HIG-13-001}, and ATLAS $H \rightarrow ZZ^*$~\cite{Aad:2013wqa} results in Fig.~\ref{fig:GaussianFit}. The agreement is excellent except for ATLAS $H \to ZZ^*$, as could be expected given the small number of events. However, the discrepancy is significant only in extreme regions ($\mu \gg 1$) that are likely to be excluded from other measurements (and first of all from the results in $H \to WW^*$); also for this channel the full likelihood is now available~\cite{atlasZZgrid}.

This procedure can be applied to all accessible final state in each experiment, and the final likelihood can be expressed as
\begin{align}
- 2 \log L &= \sum_{i=1}^{n} (- 2 \log \Lambda(\boldsymbol{\mu}_i))
+ \sum_{j=1}^{m} (- 2 \log L_{\rm full}(\boldsymbol{\mu}_j)) \label{eq:ourbestlike} \\
&= \sum_{i=1}^{n} (\boldsymbol{\mu}_i - \hat{\boldsymbol{\mu}}_i)^T
\begin{pmatrix} \sigma_{{\rm ggF},i}^2 & \rho_i \sigma_{{\rm ggF},i} \sigma_{{\rm VBF},i} \\ \rho_i \sigma_{{\rm ggF},i}\sigma_{{\rm VBF},i} & \sigma_{{\rm VBF},i}^2 \end{pmatrix}^{-1}
(\boldsymbol{\mu}_i - \hat{\boldsymbol{\mu}}_i)
+ \sum_{j=1}^{m} (- 2 \log L_{\rm full}(\boldsymbol{\mu}_j)) \,, \nonumber
\end{align}
where the index $i$ runs over the $n$ 2D measurements taken into account in the Gaussian approximation, while the index $j$ runs over the $m$ 2D measurements for which the full likelihood is available and taken into account.
In order to constrain new physics, taking the results in the $(\mu({\rm ggF+ttH}, Y), \mu({\rm VBF+VH}, Y))$ plane has several advantages compared to using the signal strength information from each category. First of all, the problems related to missing information from the experimental side are largely solved: in this approach, the reduced efficiencies ${\rm eff}_X$ are not needed and there is no concern related to the use of signal strengths before any combination. But the main reason for the likelihood defined in Eq.~\eqref{eq:ourbestlike} to be a better approximation to the full likelihood than the one in Eq.~\eqref{eq:likesimpleprod} is the complete treatment of the correlations between all systematic uncertainties---for a given decay mode $Y$ in a given experiment. Last but not least, this approach depends less on the Gaussian approximation. This is obvious when the full likelihood in this 2D plane is available, but it generally remains true even when the 2D likelihood is reconstructed from the 68\%~CL contour under the (bivariate) Gaussian approximation. Typically, we are losing less information on non-Gaussianities from a single 2D measurement than when using several measurements (if more than two) under the Gaussian approximation.

The results in the $(\mu({\rm ggF+ttH}, Y), \mu({\rm VBF+VH}, Y))$ plane can be reconstructed from the information in the individual categories, thus allowing us to check against the results from ATLAS or CMS in the same plane. This is shown in Fig.~\ref{fig:reconstruction}, in three example cases (ATLAS $H \rightarrow \gamma\gamma$~\cite{ATLAS-CONF-2013-012}, CMS $H \rightarrow \gamma\gamma$ (preliminary)~\cite{CMS-PAS-HIG-13-001}, and CMS $H \rightarrow ZZ^*$ (preliminary)~\cite{CMS-PAS-HIG-13-002}) where the information on the efficiencies is clearly given.
While the ATLAS $H \to \gamma\gamma$ results are reproduced reasonably well, large discrepancies appear in the two other examples.
It should be said that, being closer to the experimental selections, the use of signal strengths from categories should offer more flexibility in testing models, beyond a simple scaling of the SM production and decay modes. However, it would require to reproduce the cuts of the analysis in order to estimate the acceptance$\times$efficiency factor, $A \times \varepsilon$, that appears in Eq.~\eqref{eq:signalstr1}. In most cases this is impossible given the complexity of the current analyses.

In Sections~\ref{sec:higgs2012}, \ref{sec:higgs2013}, \ref{sec:higgsdim6} and \ref{sec:pmssm}, the impact of the LHC Higgs results on new physics will be studied in an effective approach and on explicit new physics scenario. In all cases, the experimental input will be taken from the $(\mu({\rm ggF+ttH}, Y), \mu({\rm VBF+VH}, Y))$ plane whenever available. However, this approximation to the full Higgs likelihood can and should be improved in the future. This is in particular crucial for a complete treatment of the theoretical uncertainties and of their correlations. This will be discussed when presenting the public tool {\tt Lilith} in Section~\ref{sec:lilith}. Possible future improvements will be discussed in Section~\ref{sec:higgsfuture}.


\section[The excitement about an excess in the diphoton channel in 2012]{The excitement about an excess in the diphoton channel in 2012%
\sectionmark{The excess in the diphoton channel at the end of 2012}}
\sectionmark{The excess in the diphoton channel at the end of 2012} \label{sec:higgs2012}

\begin{figure}[ht]
	\centering
		\includegraphics[width=0.49\textwidth]{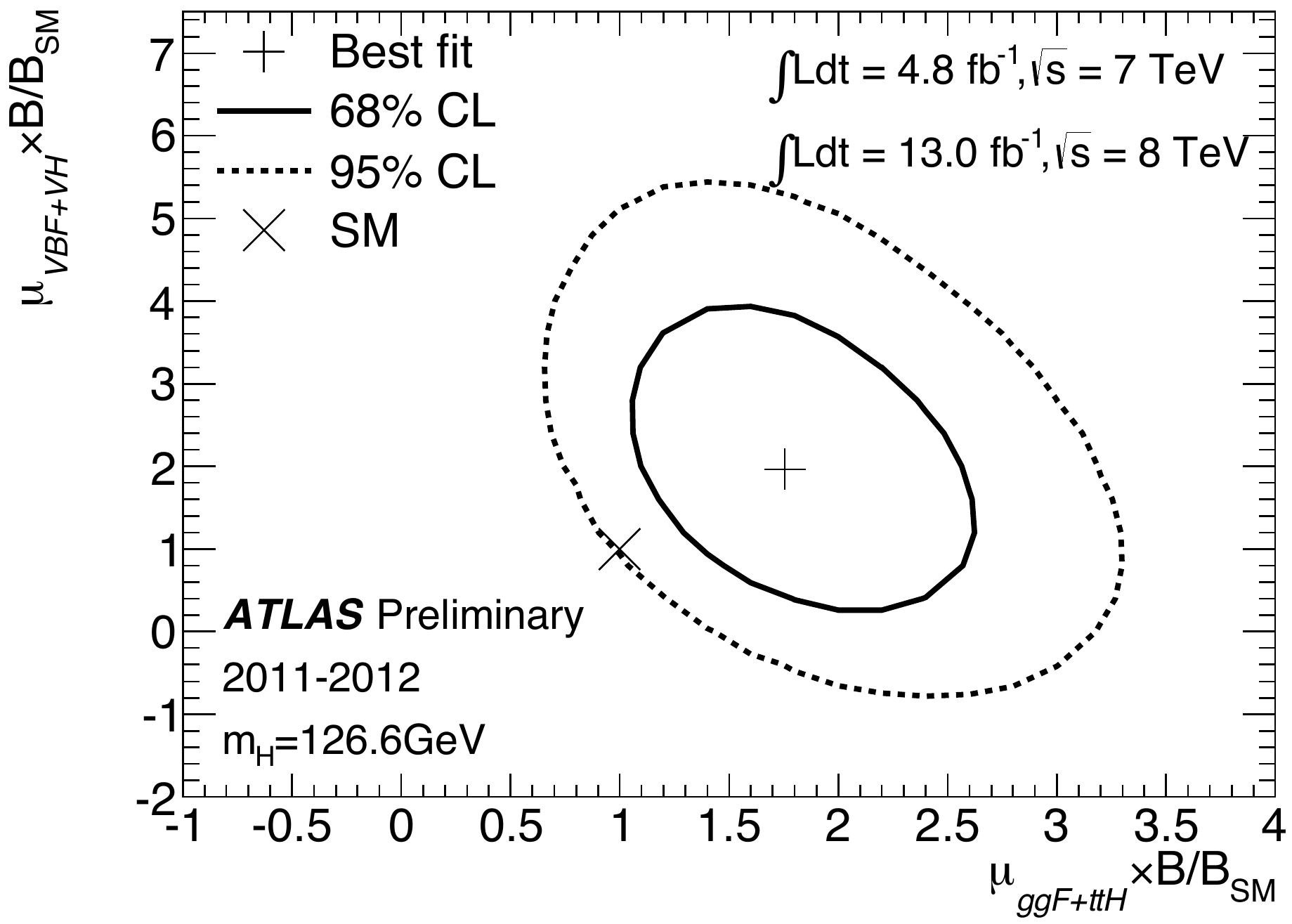}\quad
		\includegraphics[width=0.42\textwidth]{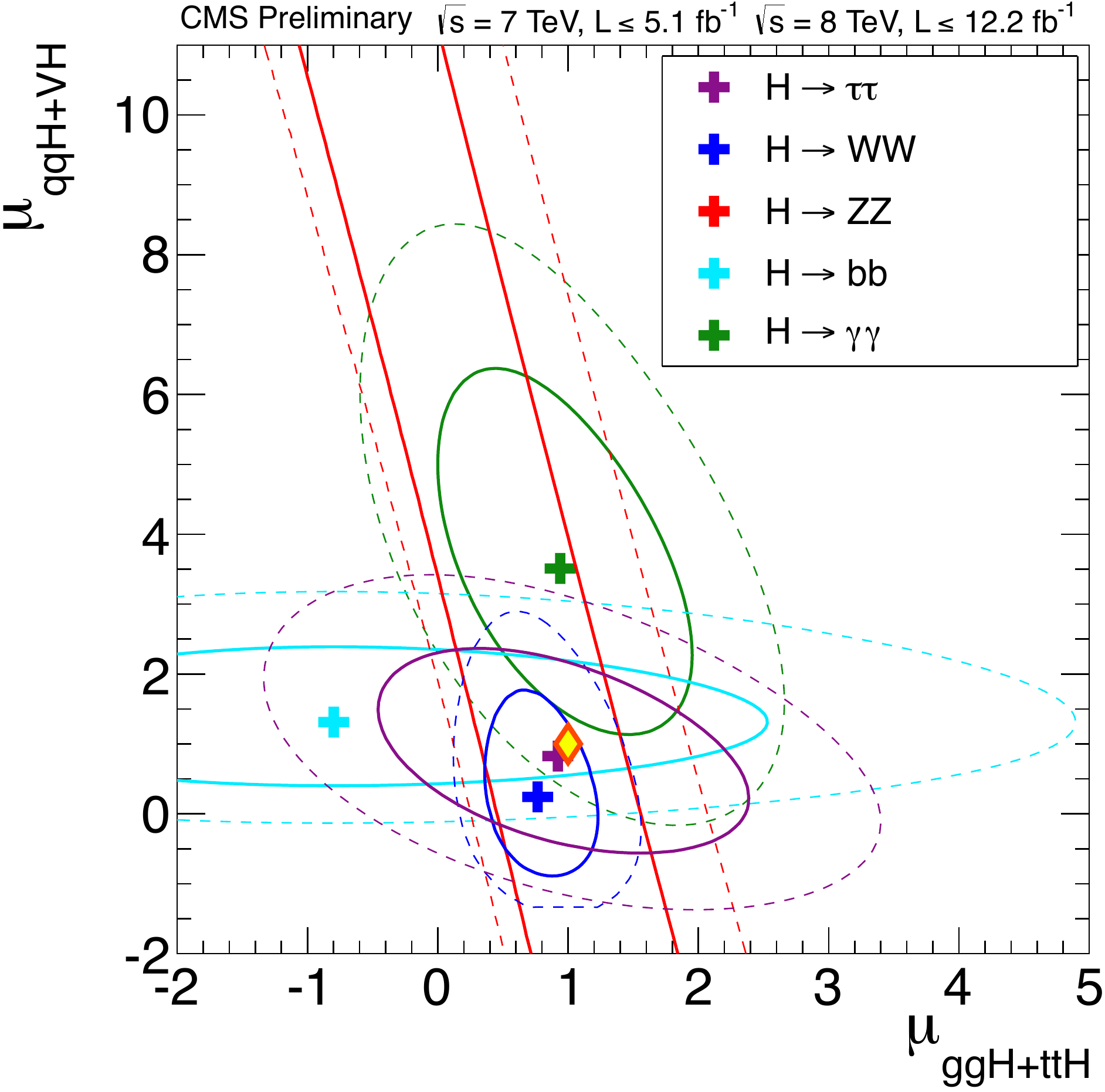}
	\caption{Results from the searches of the Higgs boson decaying into two photons in 2012, from the ATLAS~\cite{ATLAS-CONF-2012-168} (left) and CMS~\cite{CMS-PAS-HIG-12-015,Chatrchyan:2012ufa,CMS-PAS-HIG-12-045} (right) collaborations, given in the plane $(\mu({\rm ggF+ttH}, \gamma\gamma), \mu({\rm VBF+VH}, \gamma\gamma))$. In addition to the full luminosity at 7~TeV, 13.0~fb$^{-1}$ (ATLAS) and 5.3~fb$^{-1}$ (CMS) of data at 8~TeV are used. Shown are contours of constant likelihood, with the solid line corresponding to 68\%~CL and the dashed line to 95\%~CL.
\label{fig:2012gamgamstatus}}
\end{figure}

In the latest LHC Higgs results presented in Section~\ref{sec:higgs-measlhc}, no significant deviation from the SM value $\mu = 1$ can be seen. The situation was certainly different at the end of 2012 and at the beginning of 2013. Indeed, the preliminary results from ATLAS and CMS, using the full statistics collected at $\sqrt{s} = 7$~TeV and 5 to 13~fb$^{-1}$ of data at $\sqrt{s} = 8$~TeV (over the 20~fb$^{-1}$ collected in total in 2012) suggested the presence of an excess in the diphoton channel compared to SM expectations. The experimental situation at the end of 2012 in the $(\mu({\rm ggF+ttH}, \gamma\gamma), \mu({\rm VBF+VH}, \gamma\gamma))$ plane is shown in Fig.~\ref{fig:2012gamgamstatus}. The ATLAS results were updated with 13~fb$^{-1}$ of data at 8~TeV~\cite{ATLAS-CONF-2012-168} at the Open Session of the CERN Council in December 2012~\cite{cerncouncil}, while CMS results~\cite{CMS-PAS-HIG-12-045} were presented in this plane at the Hadron Collider Physics Symposium in Nov.\ 2012 (HCP2012)~\cite{HCP}, but correspond to the analysis presented in Refs.~\cite{CMS-PAS-HIG-12-015,Chatrchyan:2012ufa} for 7~TeV data and 5.3~fb$^{-1}$ at 8~TeV.
A more than $2\sigma$ excess can be seen in Fig.~\ref{fig:2012gamgamstatus}, and is mostly driven by the ATLAS results. (The excess in the ATLAS data was already present at the time of the discovery~\cite{Aad:2012tfa} and was slightly strengthen with the update presented in December.)

At the end of the summer 2012, Genevi\`eve B\'elanger, Ulrich Ellwanger, John F.~Gunion, Sabine Kraml and myself started working on possible implications of the LHC Higgs results on new physics---including the $H \to \gamma\gamma$ results in addition to all other available results---in an effective approach and in the 2HDM. This lead to the paper ``Higgs Couplings at the End of 2012'', Ref.~\cite{Belanger:2012gc}, that was submitted to arXiv on December 20, 2012 and published in JHEP in February 2013. In the rest of the Section the methodology and the main results presented in Ref.~\cite{Belanger:2012gc} will be given. In 2012 and 2013, there has been a lot of activity from many different groups in fitting the Higgs couplings to the LHC data, using various parametrizations. An extensive (yet probably incomplete) list of work on this topic up to now can be found in~\cite{Carmi:2012yp,Azatov:2012bz,Espinosa:2012ir,Klute:2012pu,Azatov:2012wq,
Carmi:2012zd,Low:2012rj,Corbett:2012dm,Giardino:2012dp,Ellis:2012hz,Montull:2012ik,
Espinosa:2012im,Carmi:2012in,Banerjee:2012xc,Bonnet:2012nm,Plehn:2012iz,
Espinosa:2012in,Elander:2012fk,Djouadi:2012rh,Altmannshofer:2012ar,Dobrescu:2012td,Chang:2012ve,
Moreau:2012da,Cacciapaglia:2012wb,Bechtle:2012jw,Corbett:2012ja,Masso:2012eq,Azatov:2012qz,
Giardino:2013bma,Falkowski:2013dza,Alanne:2013dra,Djouadi:2013qya,Corbett:2013pja,Cacciapaglia:2013ora,Bechtle:2014ewa}; theoretical uncertainties were also discussed in~\cite{Baglio:2012et}.

The section is organized as follows. The framework that is used is presented in Section~\ref{sec:2012framework}, while the experimental inputs and the fitting procedure are described in Section~\ref{sec:2012expinput}. The results of three generic fits are presented in Section~\ref{sec:2012coupfit} together with the results of a fit in Two-Higgs-Doublet models in Section~\ref{sec:20122HDMfit}. Section~\ref{sec:2012conclusions} contains our conclusions.

\subsection{Framework} \label{sec:2012framework}

With the measurements in various channels, a comprehensive study of the properties of the Higgs-like state becomes possible and has the potential for revealing whether or not the Higgs sector is as simple as envisioned in the SM.  In particular it is crucial to determine the Higgs couplings to gauge bosons and to fermions as defined by the Lagrangian
\begin{equation}
  {\cal L} =  g\left[  C_W\,  \mw W_\mu W^\mu + C_Z \frac{\mz}{\cos\theta_W} Z_\mu Z^\mu - \sum_F C_F \frac{m_F}{2\mw} \bar F F\, \right] H\,, 
\label{eq:1212.5244ldef}
\end{equation}
where the $C_I$ are scaling factors for the couplings relative to their SM values, introduced to test possible deviations in the data from SM expectations. In principle all the $C_I$ are independent, in particular the $C_F$ can be different for up- and down-type quarks and/or leptons. A significant deviation of any $C_I$ from unity would imply new physics beyond the SM. 

While fits to various combinations of $C_I$'s are performed by the experimental collaborations 
themselves~\cite{ATLAS-CONF-2012-127, CMS-PAS-HIG-12-045}, 
we find it important to develop our own scheme in order to bring all results from ATLAS, CMS 
and the Tevatron experiments together and test not only the SM but also specific models beyond.  
In the present study, we include all publicly available data as of the end of 2012. In particular we take into account the updates presented at HCP2012 in Nov.~2012~\cite{HCP} and at the Open Session of the CERN Council in Dec.~2012~\cite{cerncouncil}.

Our parametrization is as follows. We treat the couplings to up-type and down-type fermions, $\cu$ and $\cd$, as independent parameters (but we  only consider the case $\cl=\cd$, and we assume that the $C_F$ are family universal). 
Moreover, we assume a custodial symmetry in employing a single $C_W=C_Z\equiv\CV$ in Eq.~(\ref{eq:1212.5244ldef}). 
The structure we are testing thus becomes
\begin{equation}
   {\cal L} =  g\left[  \CV \left(\mw W_\mu W^\mu+\frac{\mz}{\cos\theta_W} Z_\mu Z^\mu \right)  
   - \CU \frac{m_t}{2\mw} \bar tt  - \CD    \frac{m_b}{2\mw} \bar bb - \CD \frac{m_\tau}{2 \mw}\bar\tau\tau \right ]H\,.
   \label{eq:1212.5244ourldef}
\end{equation}
In general, the $C_I$ can take on negative as well as positive values; there is one overall sign ambiguity which we fix by taking $\CV>0$.
Even in this restricted context, various types of deviations of these three  $C_I$ from unity are possible in extended theories such as 2HDMs, models with singlet-doublet mixing, and supersymmetric models such as the MSSM and the Next-to-MSSM (NMSSM). 

In addition to the tree-level couplings given above, the $H$ has couplings to $gg$ and $\gam\gam$ that are first induced at one loop and are completely computable in terms of $\cu$, $\cd$ and $\cv$ if only loops containing SM particles are present. We define $\anti \cg$ and $\anti \cp$ to be the ratio of these couplings so computed to the  SM (\ie\ $\cu=\cd=\cv=1$) values.
However, in some of our fits we will also allow for additional loop contributions $\dcg$ and $\dcp$ from new particles; in this case $\cg=\anti \cg+\dcg$ and $\cp=\anti \cp +\dcp$. The largest set of independent parameters in our fits is thus
\begin{equation}
   \cu,~\cd,~\cv,~\dcg,~\dcp\,.
\end{equation}

In this study, we focus on models in which the Higgs decays only to SM particles, in particular not allowing for invisible ({\it e.g.}\ $H\to \tilde \chi^0_1 \tilde \chi^0_1$, where $\tilde \chi^0_1$ is the lightest SUSY particle) or undetected decays (such as $H\to aa$, where $a$ is a light CP-odd, perhaps singlet scalar). This approach, when we allow in the most general case for the $\cu$, $\cd$, $\cv$, $\cp$ and $\cg$ couplings to be fully independent, encompasses a very broad range of models, including in particular those  in which the Higgs sector consists of any number of doublets  + singlets, the only proviso being the absence of decays of the observed $\sim 125\gev$ state to non-SM final states. 
(A fit for invisible Higgs decays was performed early on in \cite{Espinosa:2012vu}.) 
This approach however does not cover models  such as composite models and Higgs-radion mixing models
for which the $VVH$ coupling has a more complicated tensor structure than that given in \Eq{eq:1212.5244ourldef}.
Our procedure will also be inadequate should the observed signal at $\sim 125\gev$ actually arise from two 
or more degenerate Higgs bosons (see {\it e.g.}~\cite{Gunion:2012gc,Ferreira:2012nv}).   
Although the success of our fits implies that there is no need for such extra states, the explicit tests for degenerate states 
developed in \cite{Gunion:2012he} should be kept in mind as a means to test directly for two or more Higgs bosons
contributing to the signal at 125~GeV. Note that the presence of two near mass-degenerate states has already been tested by the CMS collaboration in the $H \to \gamma\gamma$ channel~\cite{CMS-PAS-HIG-13-016}.

\subsection{Experimental inputs and fitting procedure} \label{sec:2012expinput}

We perform fits employing all production/decay channels for which results are available from the ATLAS and CMS collaborations at the LHC, as well as the Tevatron CDF+D0 Higgs results. 
The values for the signal strengths in the various (sub)channels as reported by the experiments and used in this analysis, together with the estimated decompositions into production channels are given in Tables~\ref{1212.5244ATLASresults}--\ref{1212.5244Tevatronresults}. Note that all measurements are only given in the Gaussian approximation; a $\chi^2$ is computed using the method explained in Section~\ref{sec:higgs-npconstlhc}. Measurements in the $(\mu({\rm ggF+ttH}, Y), \mu({\rm VBF+VH}, Y))$ plane were not yet systematically available and have not always been favored over the category results. 
For the computation of the various $\mu(X, Y)$ from the reduced couplings including next-to-leading order (NLO) corrections we follow the procedure recommended by the 
LHC Higgs Cross Section Working Group in~\cite{LHCHiggsCrossSectionWorkingGroup:2012nn}. In particular 
we include all the available QCD corrections for $C_g$ using \texttt{HIGLU}~\cite{Spira:1995rr,Spira:1995mt,Spira:1996if} 
and for $C_\gamma$ using \texttt{HDECAY}~\cite{Spira:1996if,Djouadi:1997yw}, and we switch off the
electroweak corrections. 
The reduced efficiencies ${\rm eff}_X$ are specific to every analysis and hence differ from experiment to experiment. Whenever these are not given, we assume that the search is inclusive.


\begin{table}\centering
\begin{tabular}{|c|c|c|cccc|}
\hline
Channel & Signal strength $\mu$ & $m_H$ (GeV) & \multicolumn{4}{c|}{Reduced efficiencies} \\
& & & ggF & VBF & VH & ttH \\
\hline
\multicolumn{7}{|c|}{$H \rightarrow \gamma\gamma$ ($4.8\fbi$ at 7~TeV + $13.0\fbi$ at 8~TeV)~\cite{ATLAS-CONF-2012-168}} \\
\hline
$\mu({\rm ggF+ttH},\gamma\gamma)$ & $1.85 \pm 0.52$ & 126.6 & 100\% & -- & -- & -- \\
$\mu({\rm VBF+VH} ,\gamma\gamma)$ & $2.01 \pm 1.23$ & 126.6 & -- & 60\% & 40\% & -- \\
\hline
\multicolumn{7}{|c|}{$H \rightarrow ZZ$ ($4.6\fbi$ at 7~TeV + $13.0\fbi$ at 8~TeV)~\cite{ATLAS-CONF-2012-169,ATLAS-CONF-2012-170}} \\
\hline
Inclusive & $1.01^{+0.45}_{-0.40}$  & 125 & 87\% & 7\% & 5\% & 1\% \\
\hline
\multicolumn{7}{|c|}{$H \rightarrow WW$ $(13.0\fbi$ at 8~TeV)~\cite{ATLAS-CONF-2012-158,ATLAS-CONF-2012-170}} \\
\hline
$e\nu\mu\nu$ & $1.42^{+0.58}_{-0.54}$ & 125.5 & 95\% & 3\% & 2\% & -- \\
\hline
\multicolumn{7}{|c|}{$H \rightarrow b\bar{b}$ ($4.7\fbi$ at 7~TeV + $13.0\fbi$ at 8~TeV)~\cite{ATLAS-CONF-2012-161,ATLAS-CONF-2012-170}} \\
\hline
VH tag & $-0.39 \pm 1.02$ & 125.5 & -- & -- & 100\% & -- \\
\hline
\multicolumn{7}{|c|}{$H \rightarrow \tau\tau$ ($4.6\fbi$ at 7~TeV + $13.0\fbi$ at 8~TeV)~\cite{ATLAS-CONF-2012-160}} \\
\hline
$\mu({\rm ggF},\tau\tau)$ & $\phantom{-}2.41 \pm 1.57$ & 125 & 100\% & -- & -- &-- \\
$\mu(\mathrm{VBF}+\mathrm{VH},\tau\tau)$ & $-0.26 \pm 1.02$ & 125 & -- & 60\% & 40\% & -- \\
\hline
\end{tabular}
\caption{ATLAS results as employed in this analysis. The correlations included in the fits are $\rho = -0.37$ for the $\gamma\gamma$ and $\rho = -0.50$ for the $\tau\tau$ channels.}
\label{1212.5244ATLASresults}
\end{table}

\begin{table}\centering
\begin{tabular}{|c|c|c|cccc|}
\hline
Channel & Signal strength $\mu$ & $m_H$ (GeV) & \multicolumn{4}{c|}{Reduced efficiencies} \\
& & & ggF & VBF & VH & ttH \\
\hline
\multicolumn{7}{|c|}{$H \rightarrow \gamma\gamma$ ($5.1\fbi$ at 7~TeV + $5.3\fbi$ at 8~TeV)~\cite{Chatrchyan:2012ufa,CMS-PAS-HIG-12-015,CMS-PAS-HIG-12-045}} \\
\hline
$\mu({\rm ggF+ttH},\gamma\gamma)$ & $0.95 \pm 0.65$ & 125.8 & 100\% & -- & -- & -- \\
$\mu({\rm VBF+VH},\gamma\gamma)$ & $3.77 \pm 1.75$ & 125.8 & -- & 60\% & 40\% & --\\
\hline
\multicolumn{7}{|c|}{$H \rightarrow ZZ$ ($5.1\fbi$ at 7~TeV + $12.2\fbi$ at 8~TeV)~\cite{CMS-PAS-HIG-12-045,CMS-PAS-HIG-12-041}} \\
\hline
Inclusive & $\phantom{-}0.81^{+0.35}_{-0.28}$ & 125.8 & 87\% & 7\% & 5\% & 1\% \\
\hline
\multicolumn{7}{|c|}{$H \rightarrow WW$ (up to $4.9\fbi$ at 7~TeV + $12.1\fbi$ at 8~TeV)~\cite{CMS-PAS-HIG-12-039,CMS-PAS-HIG-12-042,CMS-PAS-HIG-12-045}} \\
\hline
0/1 jet & $\phantom{-}0.77^{+0.27}_{-0.25}$ & 125.8 & 97\% & 3\% & -- & -- \\
VBF tag & $-0.05^{+0.74}_{-0.55}$ & 125.8 & 17\% & 83\% & -- & -- \\
VH tag & $-0.31^{+2.22}_{-1.94}$ & 125.8 & -- & -- & 100\% & -- \\
\hline
\multicolumn{7}{|c|}{$H \rightarrow b\bar{b}$ (up to $5.0\fbi$ at 7~TeV + $12.1\fbi$ at 8~TeV)~\cite{CMS-PAS-HIG-12-044,CMS-PAS-HIG-12-025,CMS-PAS-HIG-12-045}} \\
\hline
VH tag & $\phantom{-}1.31^{+0.65}_{-0.60}$ & 125.8 & -- & -- & 100\% & -- \\
ttH tag & $-0.80^{+2.10}_{-1.84}$ & 125.8 & -- & -- & -- & 100\% \\
\hline
\multicolumn{7}{|c|}{$H \rightarrow \tau\tau$ (up to $5.0\fbi$ at 7~TeV + $12.1\fbi$ at 8~TeV)~\cite{CMS-PAS-HIG-12-043,CMS-PAS-HIG-12-051,CMS-PAS-HIG-12-045}} \\
\hline
0/1 jet & $\phantom{-}0.85^{+0.68}_{-0.66}$ & 125.8 & 76\% & 16\% & 7\% & 1\% \\
VBF tag & $\phantom{-}0.82^{+0.82}_{-0.75}$ & 125.8 & 19\% & 81\% & -- & -- \\
VH tag & $\phantom{-}0.86^{+1.92}_{-1.68}$ & 125.8 & -- & -- & 100\% & -- \\
\hline
\end{tabular}
\caption{CMS results as employed in this analysis. The correlation included for the $\gamma\gamma$ channel is $\rho = -0.54$.}
\label{1212.5244CMSresults}
\end{table}

\begin{table}\centering
\begin{tabular}{|c|c|c|cccc|}
\hline
Channel & Signal strength $\mu$ & $m_H$ (GeV) & \multicolumn{4}{c|}{Reduced efficiencies} \\
& & & ggF & VBF & VH & ttH \\
\hline
\multicolumn{7}{|c|}{$H \rightarrow \gamma\gamma$~\cite{HCPHiggsTevatron}} \\
\hline
Combined & $6.14^{+3.25}_{-3.19}$ & 125 & 78\% & 5\% & 17\% & -- \\
\hline
\multicolumn{7}{|c|}{$H \rightarrow WW$~\cite{HCPHiggsTevatron}} \\
\hline
Combined & $0.85^{+0.88}_{-0.81}$ & 125 & 78\% & 5\% & 17\% & -- \\
\hline
\multicolumn{7}{|c|}{$H \rightarrow b\bar{b}$~\cite{HCPtevBBtalk}} \\
\hline
VH tag & $1.56^{+0.72}_{-0.73}$ & 125 & -- & -- & 100\% & -- \\
\hline
\end{tabular}
\caption{Tevatron results for up to $10\fbi$ at $\sqrt{s} = 1.96$~TeV, as employed in this analysis.}
\label{1212.5244Tevatronresults}
\end{table}


With this framework programmed,  our fitting procedure is as follows.  We first scan over a fine grid of the free 
parameters of the scenario considered, for example, $\CU$, $\CD$, $\CV$ with $\CG,\CP=\anti\CG,\anti\CP$ 
as computed from the SM-particle loops (this will be Fit {\bf II} below). 
We obtain the value of $\chisq$ associated with each point in the grid 
and thus determine the values of the parameters associated with the approximate minimum (or minima).  
To get the true minimum $\chisq$, $\chimin$, and the associated ``best-fit" values and the one-standard 
deviation ($1\sigma$) errors on them we employ MINUIT~\cite{James:1975dr}.  
(The errors on parameters which are not input, \ie\ $\CG$ and $\CP$, are determined from the grid data.)
For plotting distributions of $\chisq$ as a function of any one variable, we use the above grid data 
together with the best fit value, to profile the minimal $\chisq$ value with respect to the remaining 
unconstrained parameters. The 68\%, 95\% and 99.7\% CL intervals are then given 
by $\chisq=\chimin+1$, $+4$ and  $+9$, respectively. 
Two-dimensional $\chisq$ distributions are obtained analogously from a grid in the two parameters of interest, profiling over the other, unseen parameters;
in this case, we show contours of $\chisq$ corresponding to the 68\% ($\chisq=\chimin+2.30$), 
95\% ($\chisq=\chimin+6.18$) and 99.7\% ($\chisq=\chimin+11.83$) confidence levels for 2 parameters treated jointly. Note that it corresponds to a profile likelihood ratio and that the same procedure is used by the experimental collaborations to perform coupling fits and also derive the results in the $(\mu({\rm ggF+ttH}, Y), \mu({\rm VBF+VH}, Y))$ plane, as was explained in Eq.~\eqref{eq:profilelikeratio} in Section~\ref{sec:higgs-npconstlhc}.

Before presenting our results, a couple of comments are in order. First of all, we stress that in models of new physics beyond-the-SM (BSM), both the branching fractions and the production cross sections  and distributions (and indeed the number of Higgs particles) may differ from SM expectations.
For any BSM interpretation of the Higgs search results  it is absolutely crucial to have as precise and complete channel-by-channel information as possible~\cite{Kraml:2012sg}. 
Unfortunately, not all the experimental analyses give all the necessary details. 
Below we comment on how we use the currently available information from the experiments.

\subsubsection*{ATLAS}

\begin{itemize}
\item $H \to \gamma\gamma$: we fit the 68\%~CL contour in the $(\mu({\rm ggF+ttH}, Y), \mu({\rm VBF+VH}, Y))$ plane from Fig.~4 of~\cite{ATLAS-CONF-2012-168} as explained around Eq.~\eqref{eq:mu2d} in Section~\ref{sec:higgs-npconstlhc}.
We note that while Fig.~4 of~\cite{ATLAS-CONF-2012-168} is for 126.6~GeV, Fig.~12 (right) in the same paper shows that there is a broad ``plateau''  as a function of the mass when the energy scale uncertainty is taken into account, implying that  the results should not depend too much on the mass. 
\item $H \to ZZ$: the signal strength in this channel reported by ATLAS \cite{ATLAS-CONF-2012-169,ATLAS-CONF-2012-170}
is $\mu=1.3^{+0.53}_{-0.48}$ with a best fit mass of $m_H = 123.5\pm 0.9\ {\rm (stat.)} \pm 0.3\ {\rm (sys.)}$~GeV. 
At $m_H=125$~GeV, the signal strength is $\mu=1.01^{+0.45}_{-0.40}$, see Fig.~10 in \cite{ATLAS-CONF-2012-170}.  
Assuming that the discrepancy in the Higgs mass determined from the $\gam\gam$ and the 4~lepton final states is 
due to a statistical fluctuation (rather than unknown systematics) we use the inclusive $\mu(H \to ZZ)$ results at $m_H=125$~GeV, \ie\  
close to the combined best fit mass from ATLAS, in our fits. 
Alternatively, one could rescale the value of $\mu=1.3^{+0.53}_{-0.48}$ at $m_H=123.5$~GeV for a Higgs mass of 125~GeV. This would give $\mu(H \to ZZ)=1.15^{+0.53}_{-0.48}$ at $m_H=125$~GeV (or $\mu(H \to ZZ)=1.11^{+0.53}_{-0.48}$ at $m_H=125.5$~GeV). We checked that taking this alternative approach has only marginal influence on our results.  
Regarding the decomposition in production modes, no statement is made in the conference note or paper. 
However, as it is an inclusive analysis, 
we take the relative ratios of production cross sections for an SM Higgs as a reasonable approximation. 
To this end, we use the ratios given by the LHC Higgs Cross Section Working Group~\cite{HXSWG}.  
\item $H \to WW$: we adopt relative contributions of 95\% \ggf\ and 5\% VBF~\cite{ATLAS-CONF-2012-158}. We do not include any result for 7~TeV because the update presented at HCP is a combination of 7 and 8~TeV.
\end{itemize}

\subsubsection*{CMS}

\begin{itemize}
\item $H \to \gamma\gamma$: we follow the same procedure as for ATLAS $H \to \gamma \gamma$, using Fig.~11 from~\cite{CMS-PAS-HIG-12-045}. The correlation is $\rho = -0.54$.
\item $H \to ZZ$: no decomposition with respect to production modes is given in the conference note or paper. 
As it is a fully inclusive analysis, we use the relative ratios of production cross sections given by the LHC 
Higgs Cross Section Working Group~\cite{HXSWG} as a good approximation~\cite{albert}. 
\item $H \rightarrow WW$: the information provided in the conference note and papers is incomplete; our decomposition into 
production modes is based on \cite{albert}. 
Our combination (weighted mean) agrees within 9\% with that given by CMS ($\mu_{\rm comb} = 0.64 \pm 0.24$ instead of $0.70^{+0.24}_{-0.23}$).
\item $H \rightarrow b\bar{b}$: 
as there is no information on possible contaminations by other production modes, we assume 100\% VH or 100\% ttH production for the respective categories.
\item $H \rightarrow \tau\tau$: 
for the 0/1 jet and VBF tag categories we extract the decomposition into production modes from~\cite{CMS-PAS-HIG-12-043}, assuming that there is no significant change in the efficiencies between $m_H = 125$~GeV and $m_H = 125.8$~GeV. We use the efficiencies from the first three categories ($\mu\tau_{h}+X$, $e\tau_{h}+X$ and $e\mu+X$) because they are the most sensitive ones; they lead to very similar decompositions which we use in our analysis.  
Our combination (weighted mean) agrees within 6\% with that given by CMS ($\mu_{\rm comb} = 0.83 \pm 0.49$ instead of $0.88^{+0.51}_{-0.48}$). 
\end{itemize}

\subsubsection*{Tevatron}

\begin{itemize}
\item $H \to \gamma\gamma$ and $H\to WW$: no decomposition into production modes is given by the experiments. 
We assume that the analyses are inclusive and we thus employ the ratios of the  theoretical predictions for the (SM) Higgs production cross sections.  
\item $H \to b\bar{b}$: we use the results presented at HCP2012~\cite{HCPtevBBtalk} assuming 100\% VH.
\end{itemize}

\subsection{Fits to reduced Higgs couplings} \label{sec:2012coupfit}
%

\subsubsection*{\boldmath Fit I: $\CU=\CD=\CV=1$, $\dcg$ and $\dcp$ free}

For a first test of the SM nature of the observed Higgs boson, we take $\cu=\cd=\cv=1$ (\ie\ quark, lepton and $W,Z$ vector boson couplings to the Higgs 
are required to be SM-like) but we allow for additional new physics contributions  to the $\gam\gam$ and $gg$ couplings, parameterized by $\dcg$ and $\dcp$, coming from loops involving non-SM particles. 
This fit, which we refer to as Fit~{\bf I}, is designed to determine if the case where all tree-level Higgs couplings 
are equal to their SM values can be consistent with the data.    For example, such a fit is relevant in the context of UED models where the tree-level couplings of the Higgs are SM-like~\cite{Petriello:2002uu,Belanger:2012mc}.

\begin{figure}[ht]\centering
\includegraphics[scale=0.40]{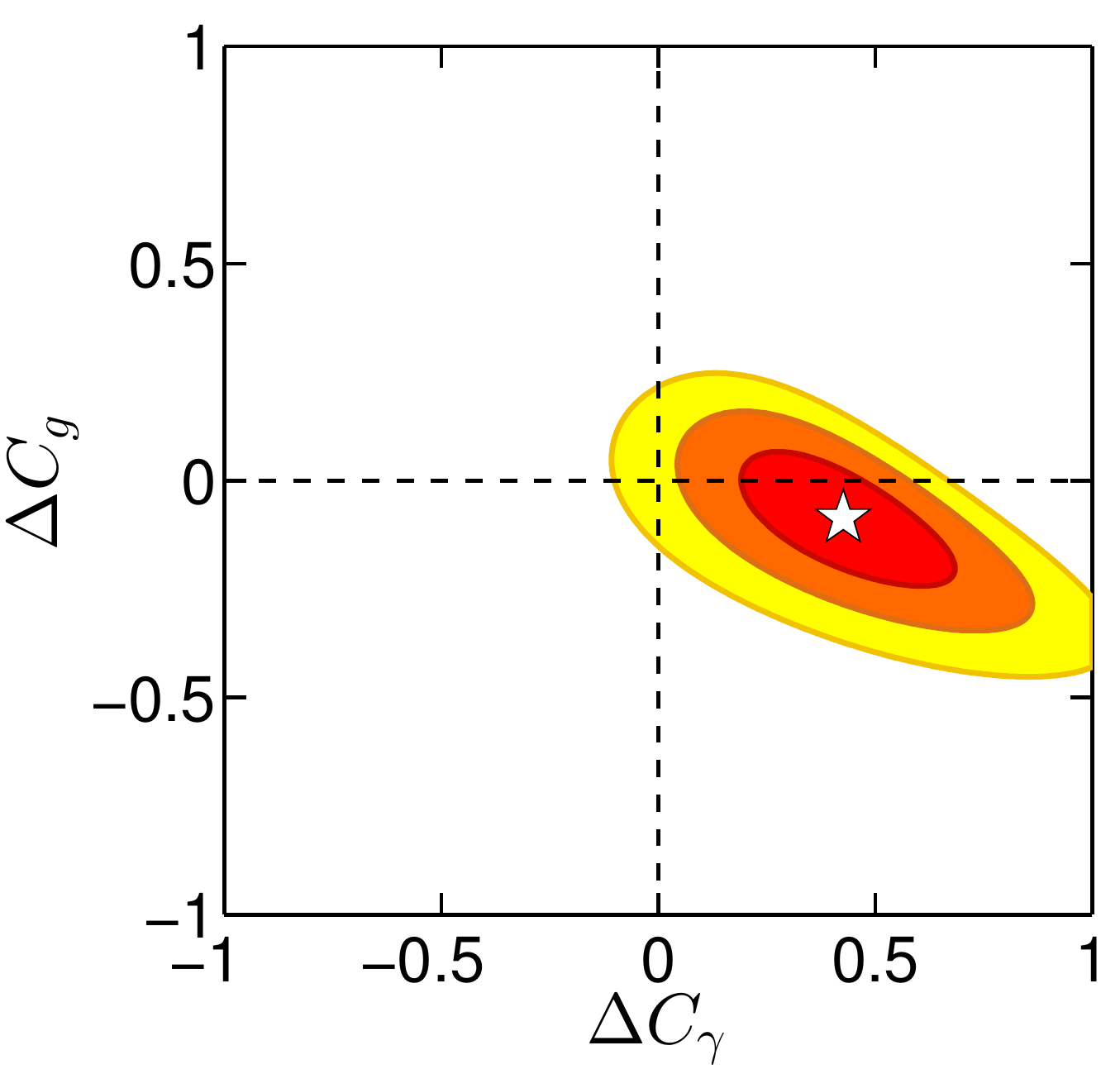}
\caption{Two parameter fit of $\Delta C_\gamma$ and $\Delta C_g$, assuming $\CU=\CD=\CV=1$ (Fit~{\bf I}). 
The red, orange and yellow ellipses show the 68\%, 95\% and 99.7\%  CL regions, respectively. 
The white star marks the best-fit point $\dcp=0.426$, $\dcg=-0.086$.
\label{1212.5244fit1} }
\end{figure}

Fig.~\ref{1212.5244fit1} displays the results of this fit in the  $\dcg$ versus $\dcp$ plane.  
The best fit is obtained for $\dcp\simeq0.43$, $\dcg\simeq-0.09$,  and has $\chimin=12.31$ for 19 degrees of freedom (\dof), 
giving a $p$-value of $0.87$. 
The results of this fit are summarized in Table~\ref{1212.5244chisqmintable}, together with the results of the other fits of this section.

We note that the SM (\ie\ $\CU=\CD=\CV=1$, $\dcg=\dcp=0$) has $\chi^2= 20.2$ and is hence more 
than $2\sigma$ away from the best fit in Fig.~\ref{1212.5244fit1}.
The number of degrees of freedom for the SM fit is 21,  
implying a $p$-value of $0.51$. The largest  $\chi^2$ contributions come from the $H\to\gamma\gamma$ channels 
from ATLAS ($\Delta\chi^2=5.06$), CMS ($\Delta\chi^2=3.36$) and Tevatron ($\Delta\chi^2=2.60$), followed 
by the VBF result for $H\to WW$ from CMS with $\Delta\chi^2=2.01$. 

\subsubsection*{\boldmath Fit~II: varying $\cu$, $\cd$ and $\cv$ ($\dcp=\dcg=0$)}

\begin{figure}[ht] 
\includegraphics[scale=0.32]{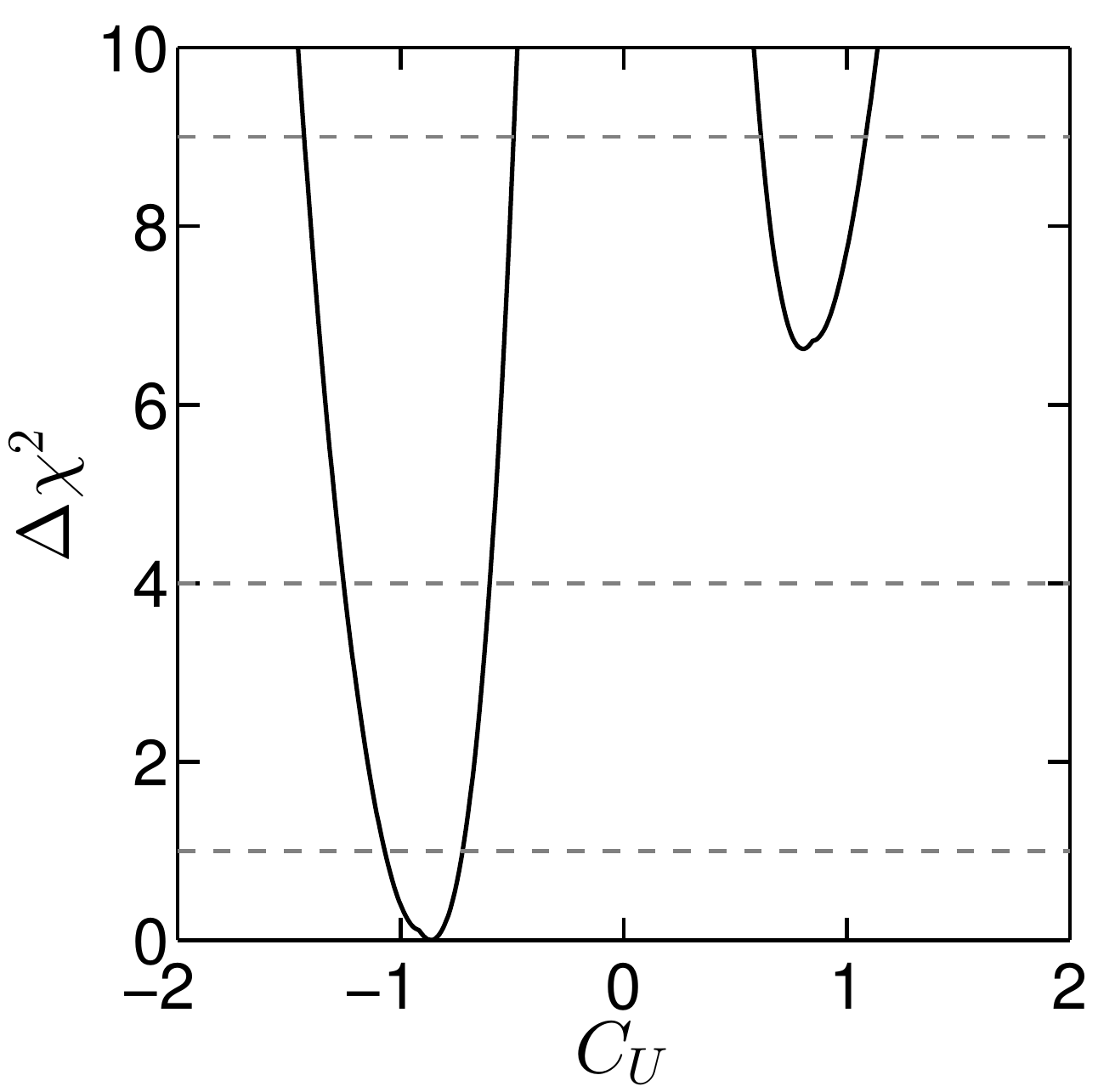}\includegraphics[scale=0.32]{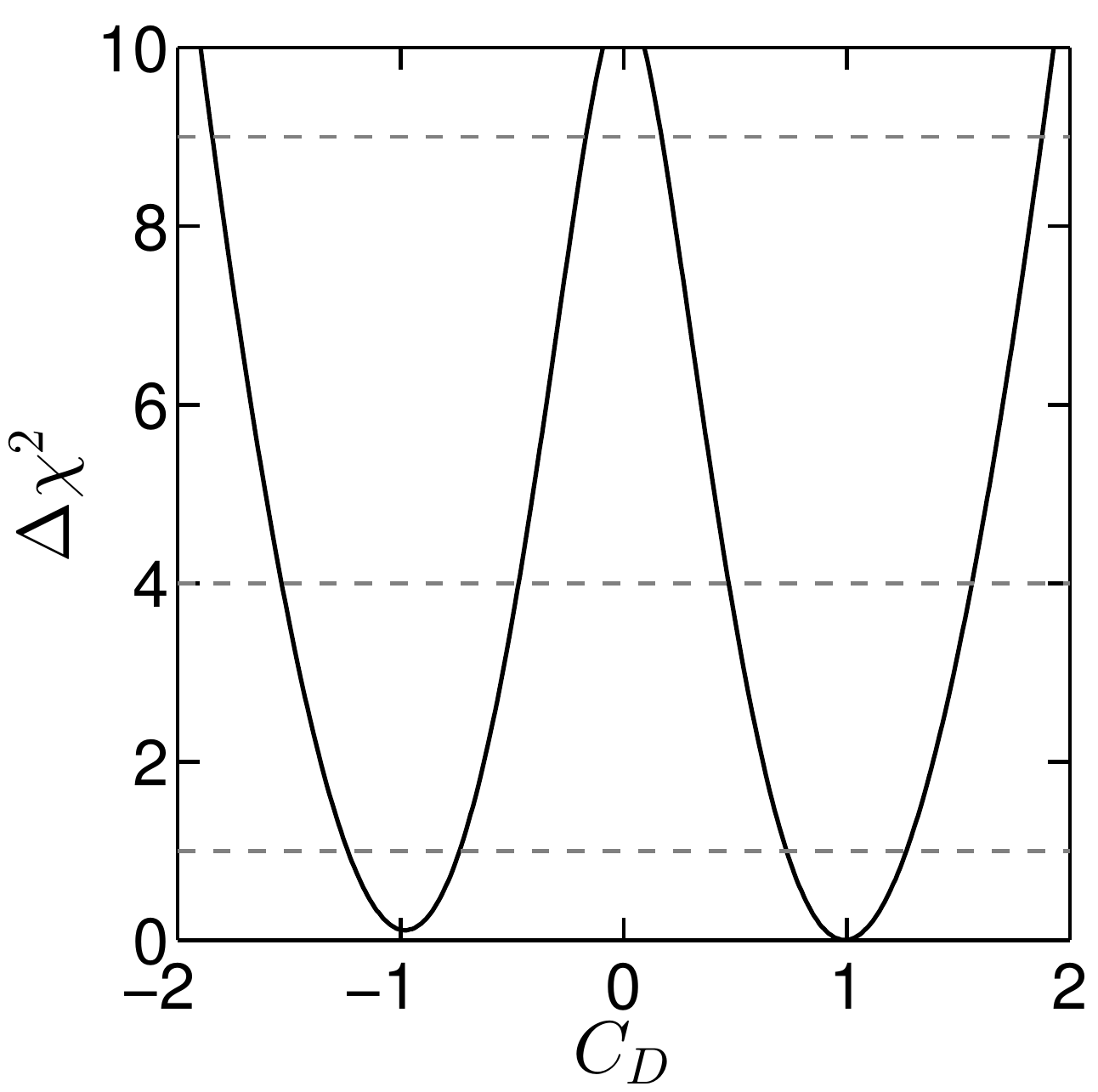}\includegraphics[scale=0.32]{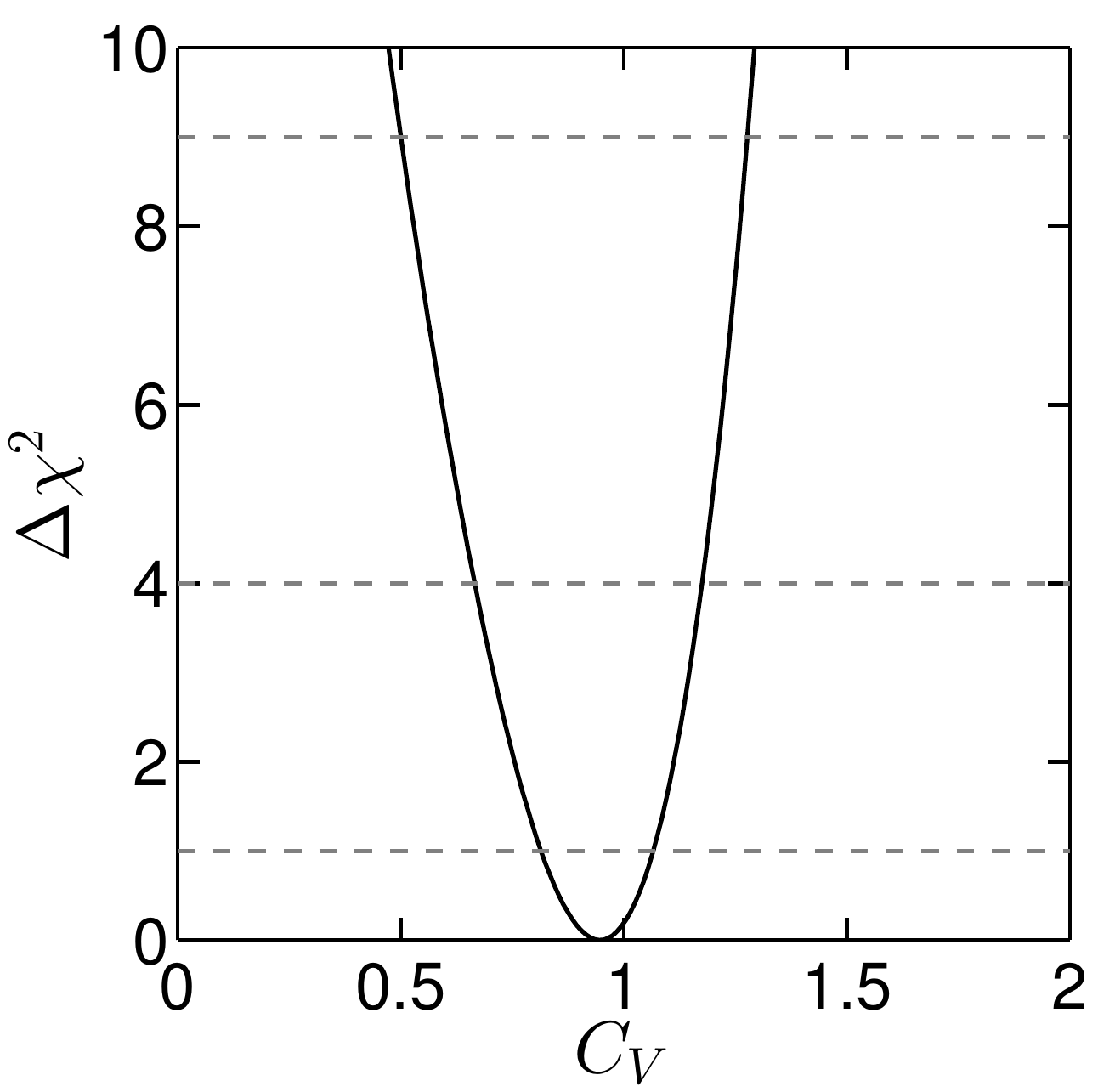}\includegraphics[scale=0.32]{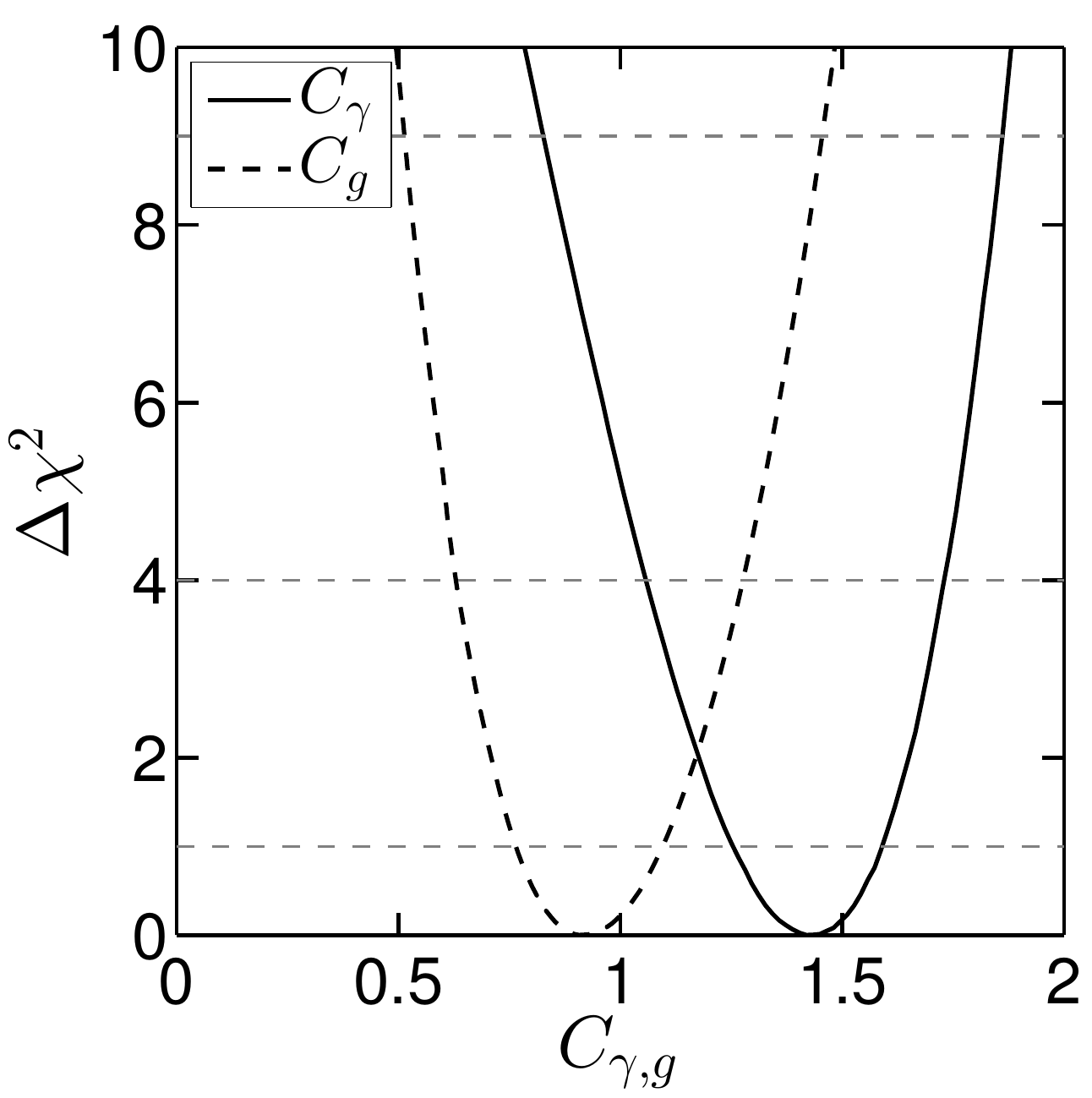}
\caption{One-dimensional $\chisq$ distributions for the three parameter fit, Fit~{\bf II},  of $\CU$, $\CD$, $\CV$ with $\cp=\cpb$ and $\cg=\cgb$ as computed in terms of $\cu,\cd,\cv$. 
\label{1212.5244fit2-1d} }
\end{figure}

\begin{figure}[ht]\centering
\includegraphics[scale=0.4]{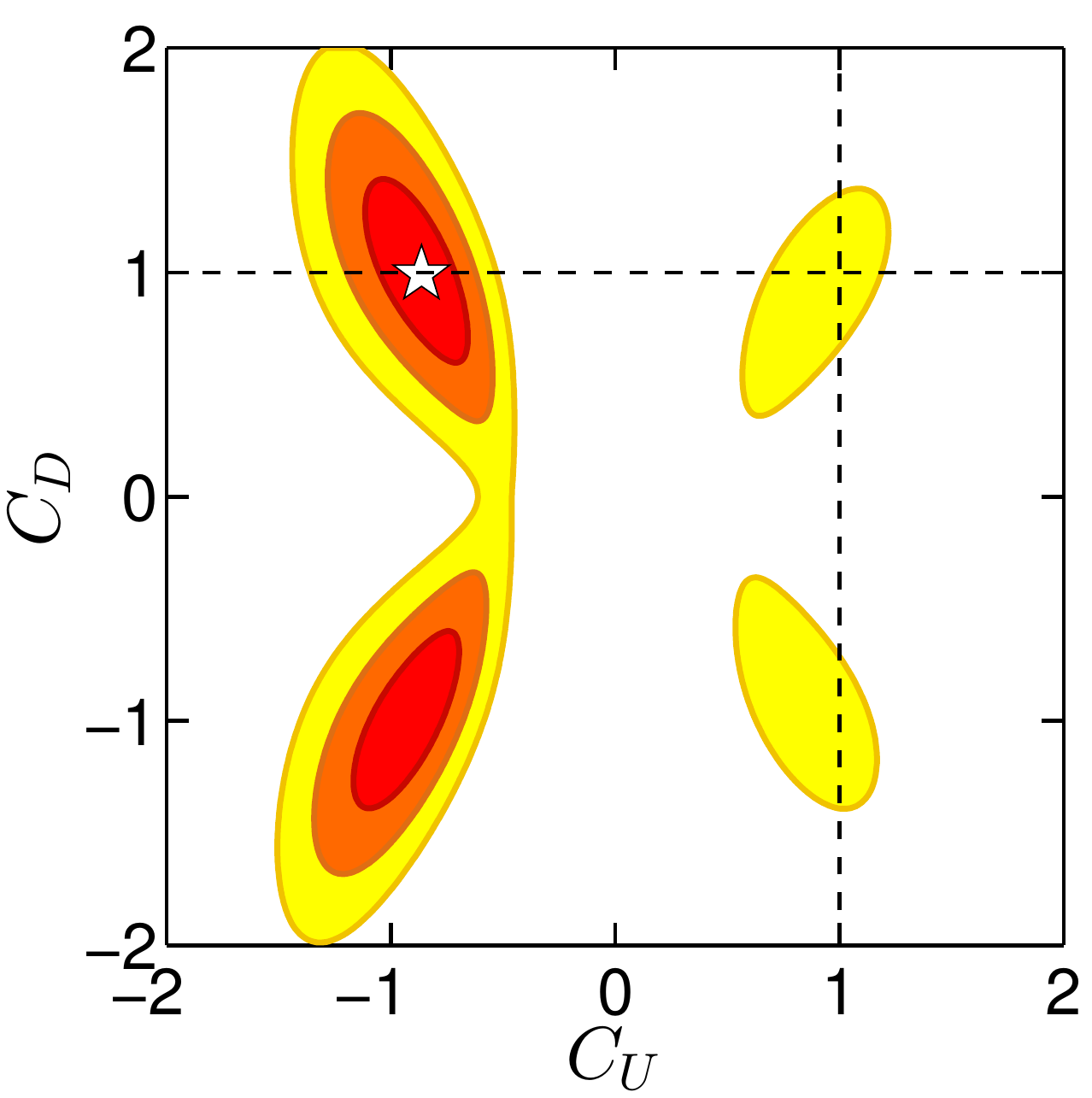}\includegraphics[scale=0.4]{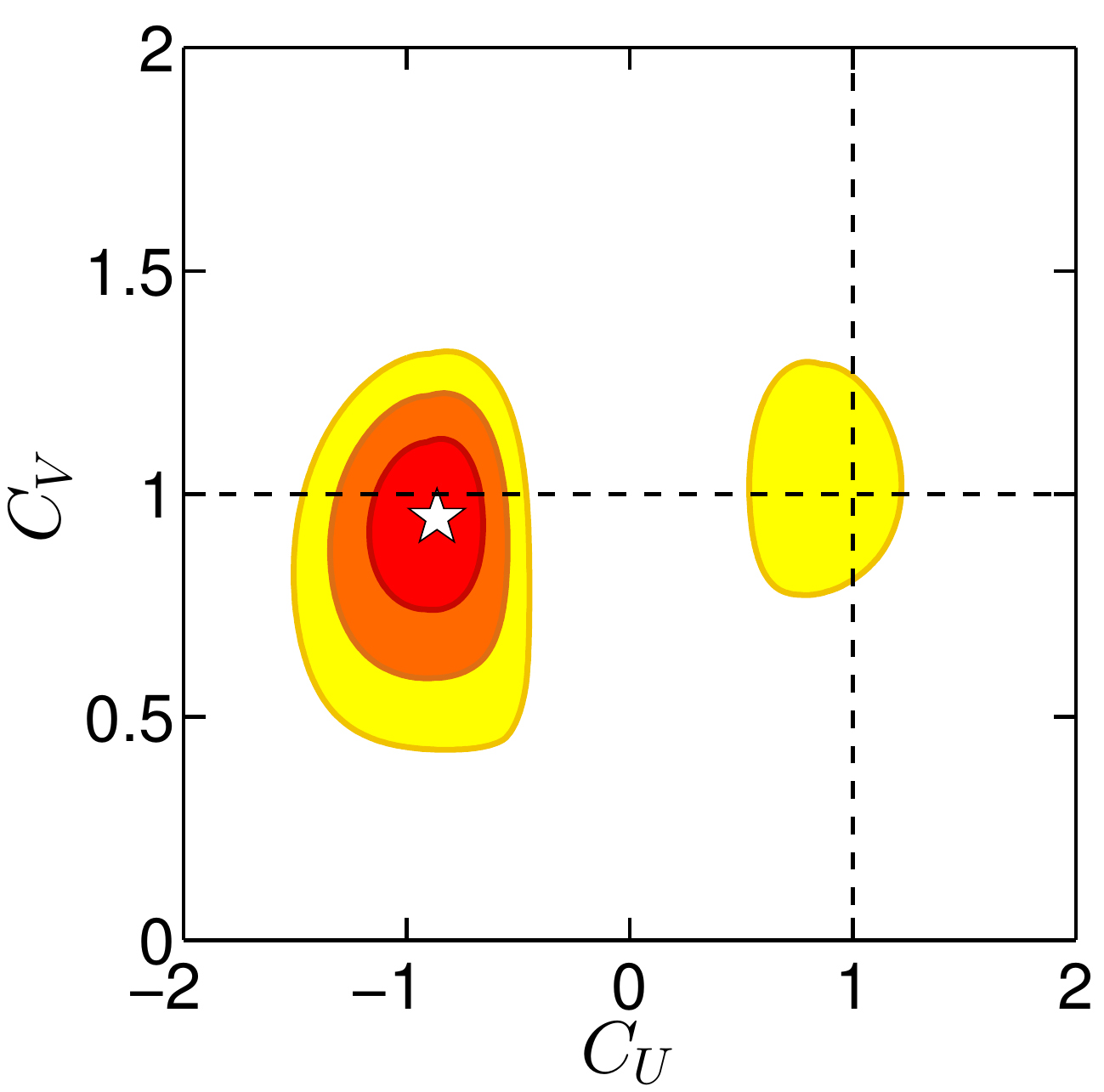}\includegraphics[scale=0.4]{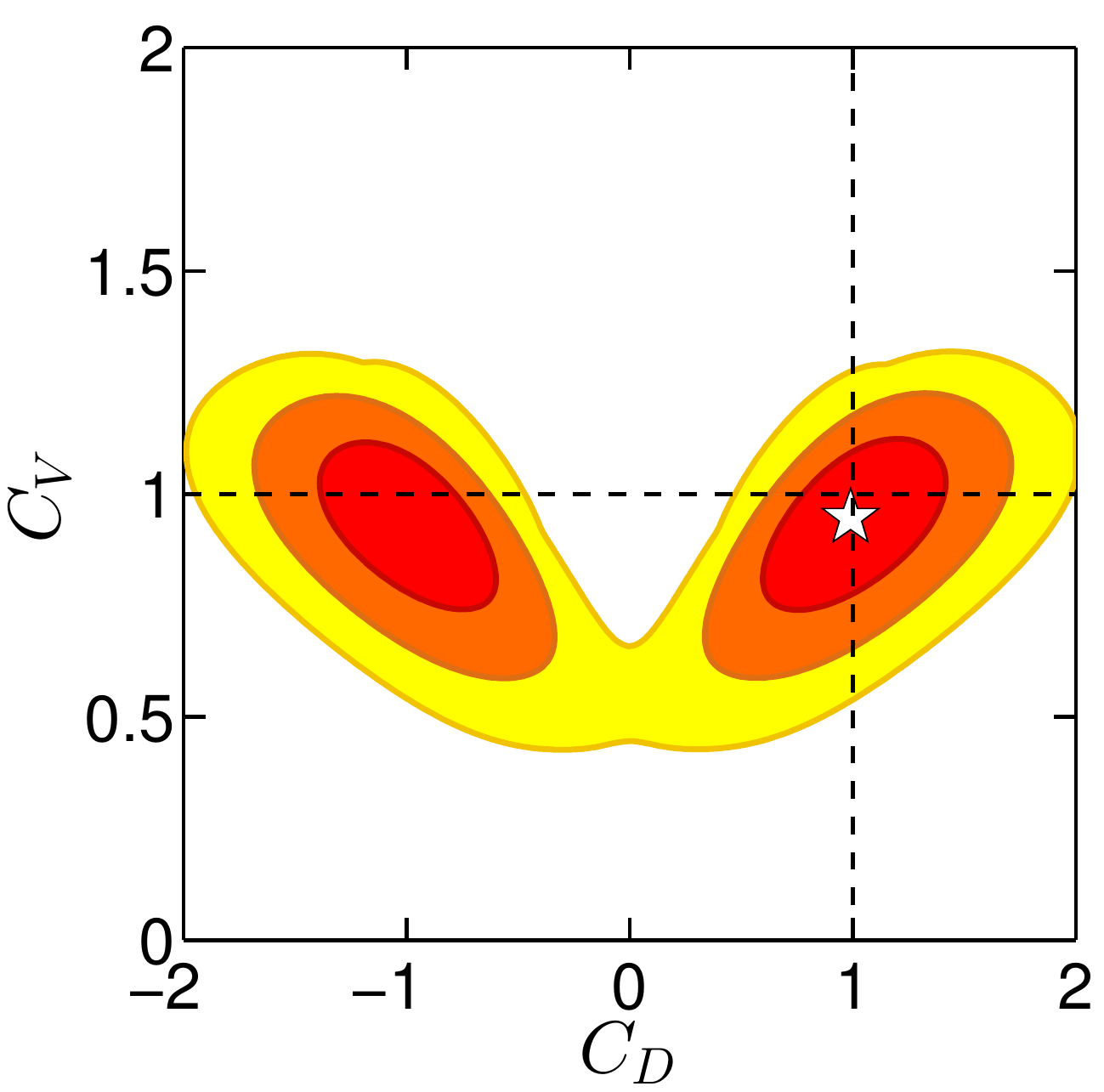}\\
\includegraphics[scale=0.4]{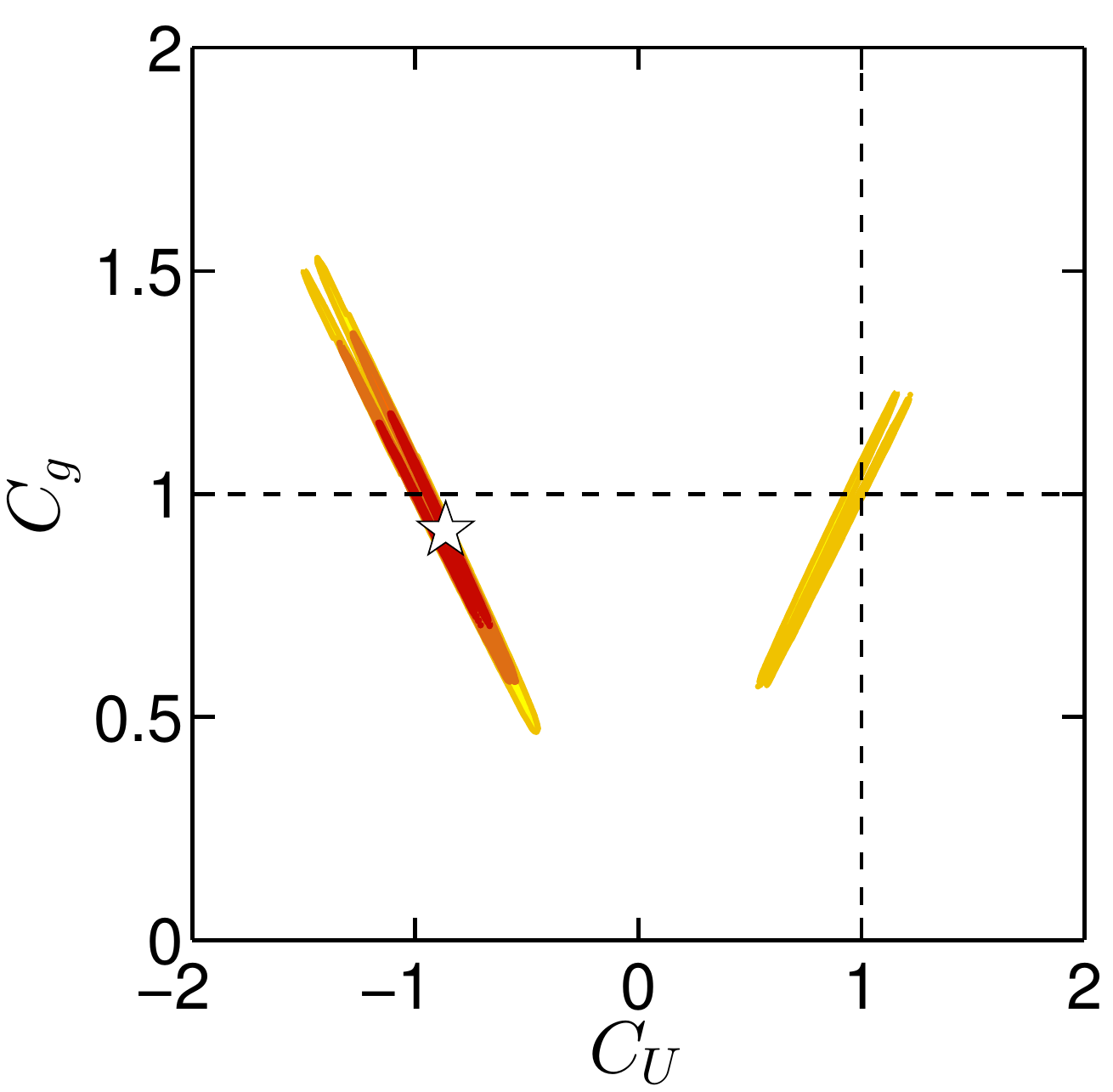}\includegraphics[scale=0.4]{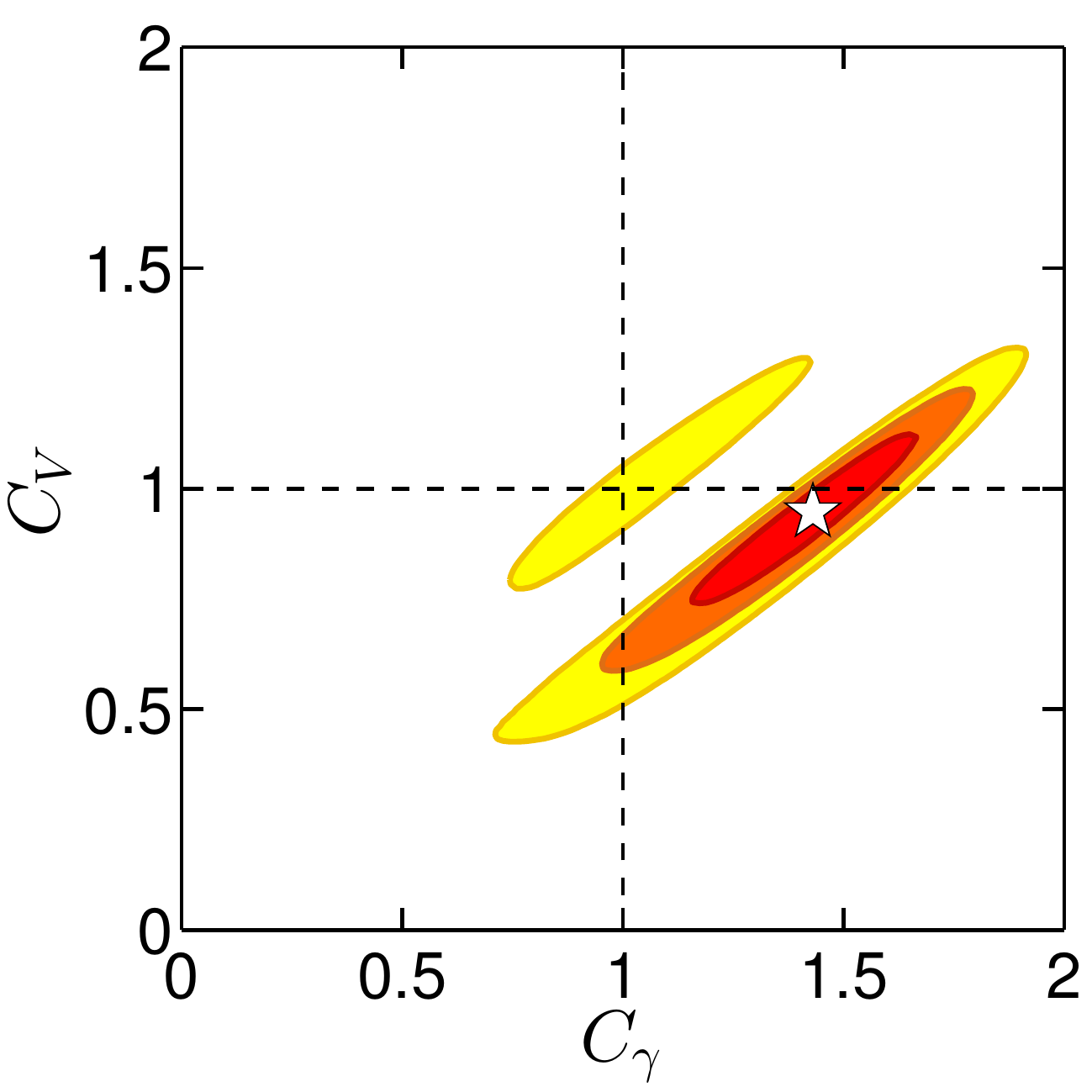}\includegraphics[scale=0.4]{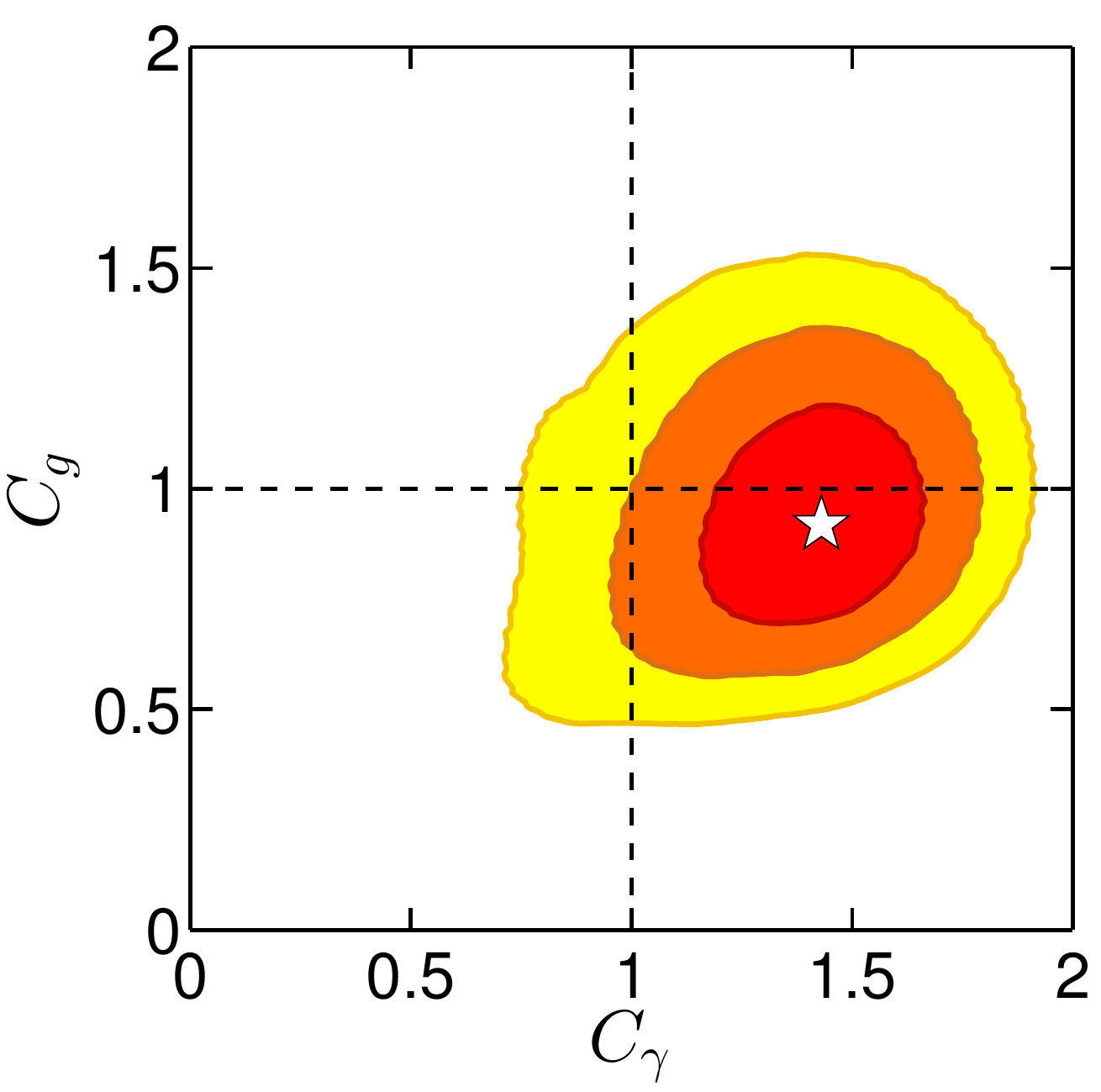}
\caption{Two-dimensional $\chisq$ distributions for the three parameter fit, Fit~{\bf II}, of $\CU$, $\CD$, $\CV$ with $\cp=\cpb$ and $\cg=\cgb$ as computed in terms of $\cu,\cd,\cv$. The red, orange and yellow ellipses show the 68\%, 95\% and 99.7\%  CL regions, respectively. The white star marks the best-fit point. Details on the minima in different sectors of the ($\cu$,\,$\cd$) plane can be found in Table~\ref{tab:1212.5244fit2}. 
\label{1212.5244fit2-2d} }
\end{figure}

Next, we let  $\CU$, $\CD$, $\CV$ vary,  assuming there are no new particles contributing to the effective 
Higgs couplings to gluons and photons, \ie\  we take $\dcp=\dcg=0$ implying $\CG=\anti\CG$, $\CP=\anti\CP$ 
as computed from the SM-particle loops.  
The results for the one-dimensional and two-dimensional $\chi^2$ distributions are shown in Figs.~\ref{1212.5244fit2-1d} and \ref{1212.5244fit2-2d}. 
The value of $\cv$ is rather well determined to be close to unity. It is intriguing that the best fit of $\cv$ is 
indeed just slightly below 1, as any model with only Higgs doublets or singlets requires $\cv\le1$. The best fit values for $\cd$ and $\cu$ are SM-like in that they have magnitudes that are close to one. However, the best fit $\cu$ value is opposite in sign to the SM Higgs case. 
The preference for $\cu<0$ is at the level of  
$2.6\sigma$ --- see the first plot in Fig.~\ref{1212.5244fit2-1d}. This results from the fact that an enhanced $\gam\gam$ rate (as observed in the experimental data) is obtained by changing the sign of the top-loop contribution so that it adds, rather than subtracts, from the $W$ loop.  
In contrast, in the case of $\cd$ almost equally good minima are found with $\cd<0$ and $\cd>0$. 
Details on the minima in different sectors of the ($\cu$,\,$\cd$) plane are given in Table~\ref{tab:1212.5244fit2}.
Note that, for the best fit point, the resulting $\CP$ and $\CG$ are in good agreement with the result of Fit~{\bf I} above, for which $\cp=1.43$ and $\cg=0.91$. Here, however, the enhanced $\cp$ value derives from $\cu<0$ rather than from $\dcp\neq 0$. The best fit results are again tabulated in Table~\ref{1212.5244chisqmintable}.

\begin{figure}[ht] 
\includegraphics[scale=0.32]{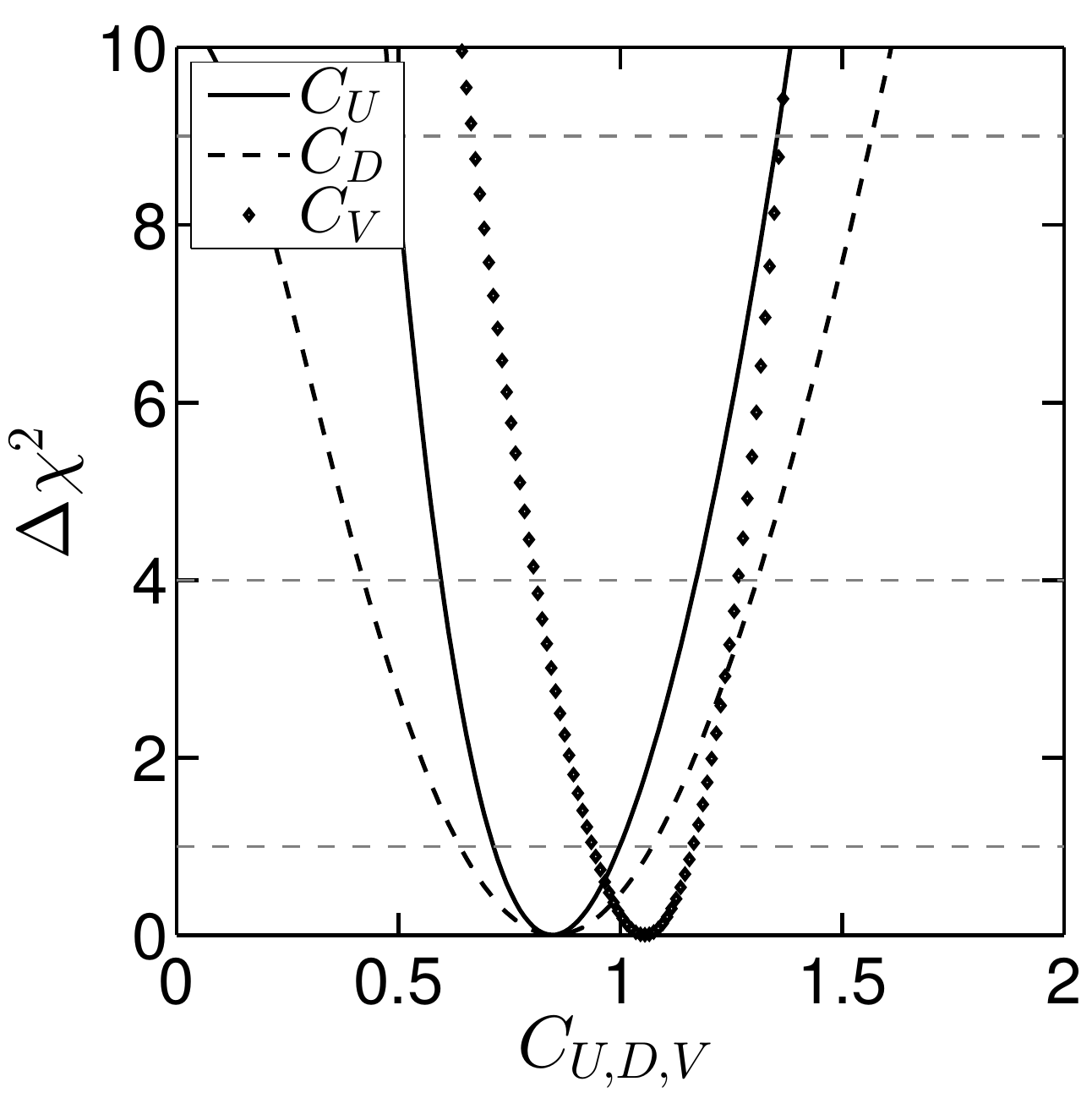}\includegraphics[scale=0.32]{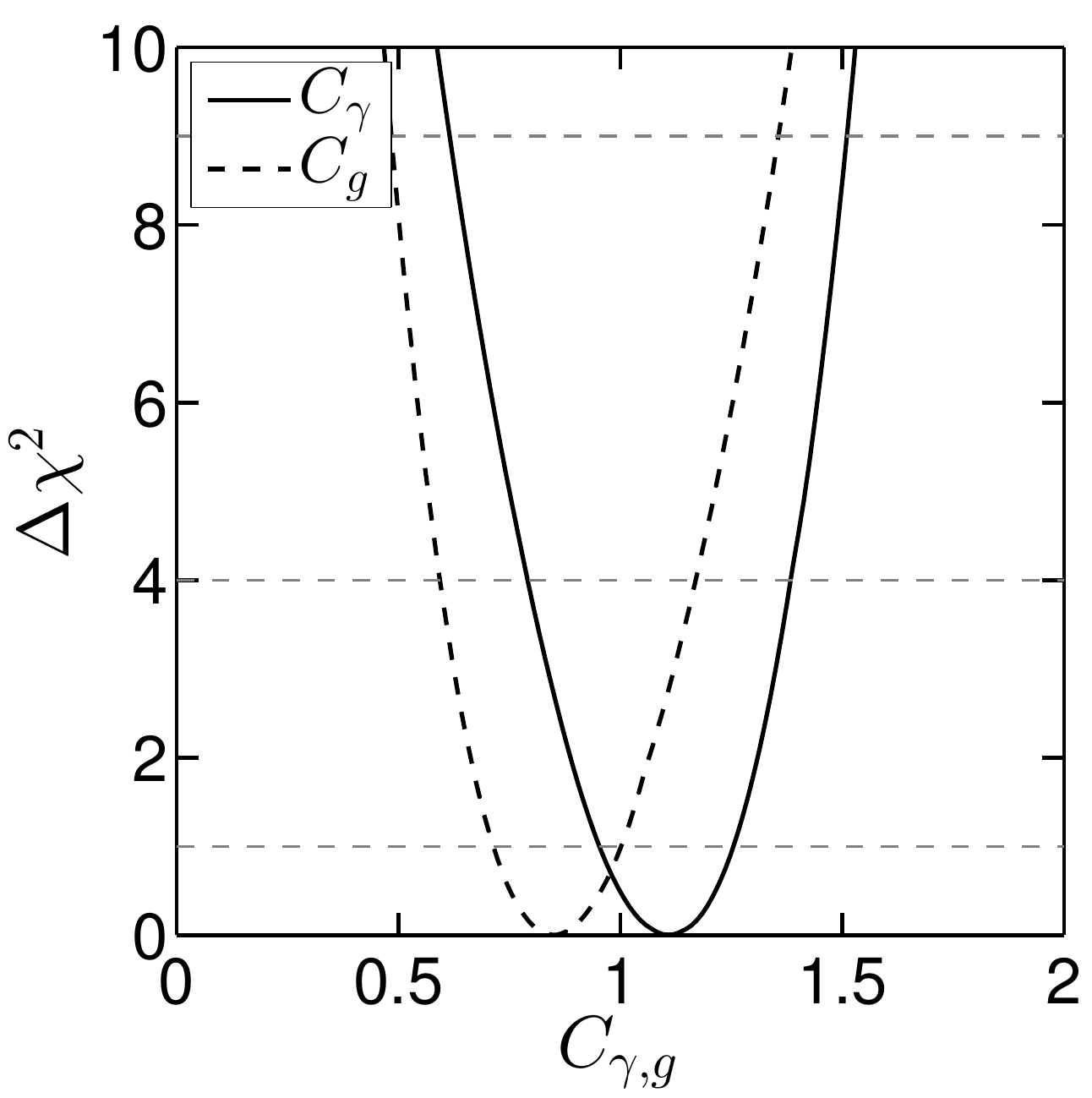}\includegraphics[scale=0.32]{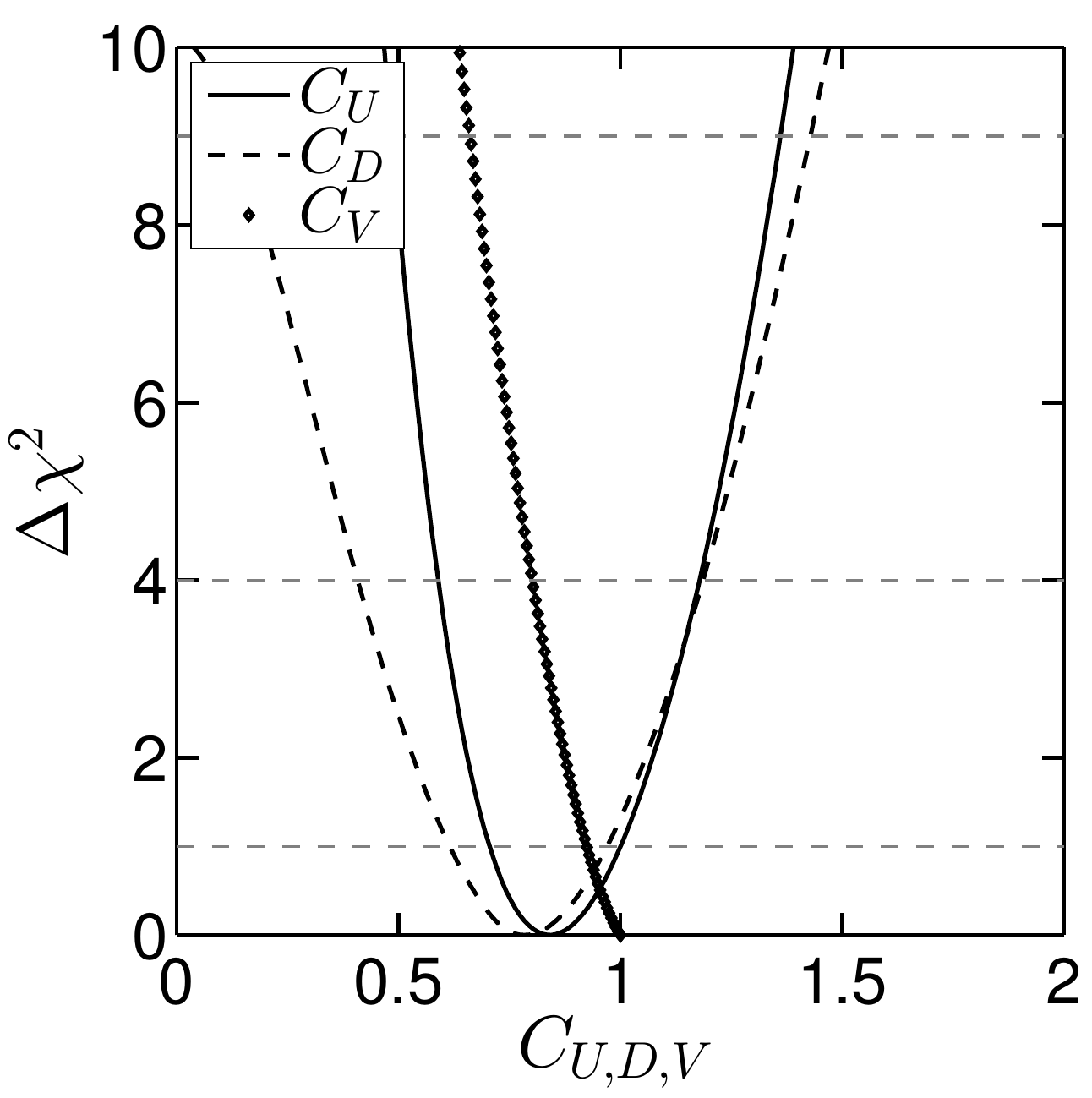}\includegraphics[scale=0.32]{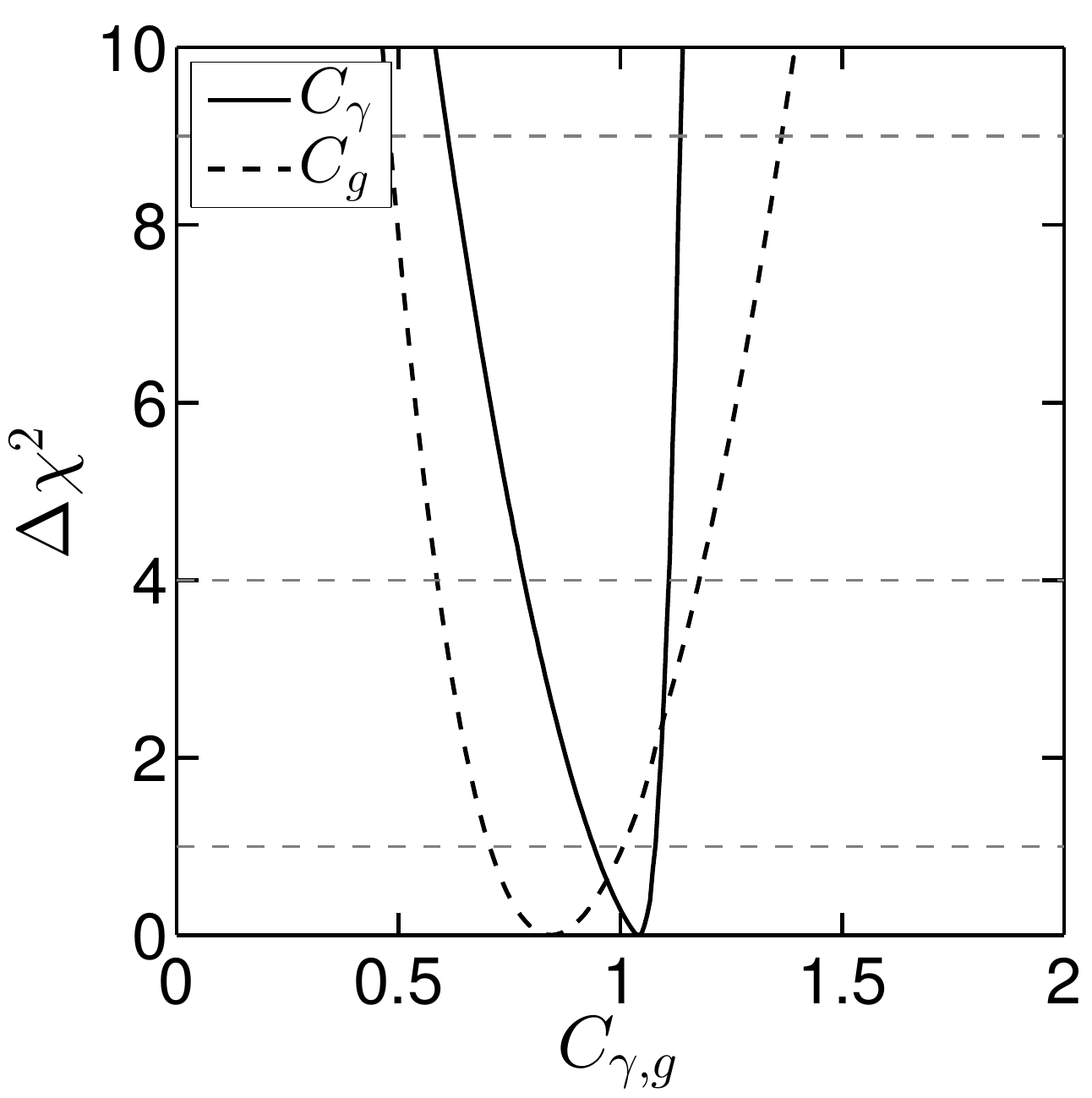}
\caption{One-dimensional $\chisq$ distributions for the three parameter fit, Fit~{\bf II},  
but imposing $\CU>0$, $\CD>0$; the left two plots allow for $\CV>1$ ($\chisq_{\rm min}=18.66$), while in the right two plots $\CV\le1$ ($\chisq_{\rm min}=18.89$). 
\label{1212.5244fit2-1dpos} }
\end{figure}

\begin{figure}[ht]\centering
\includegraphics[scale=0.4]{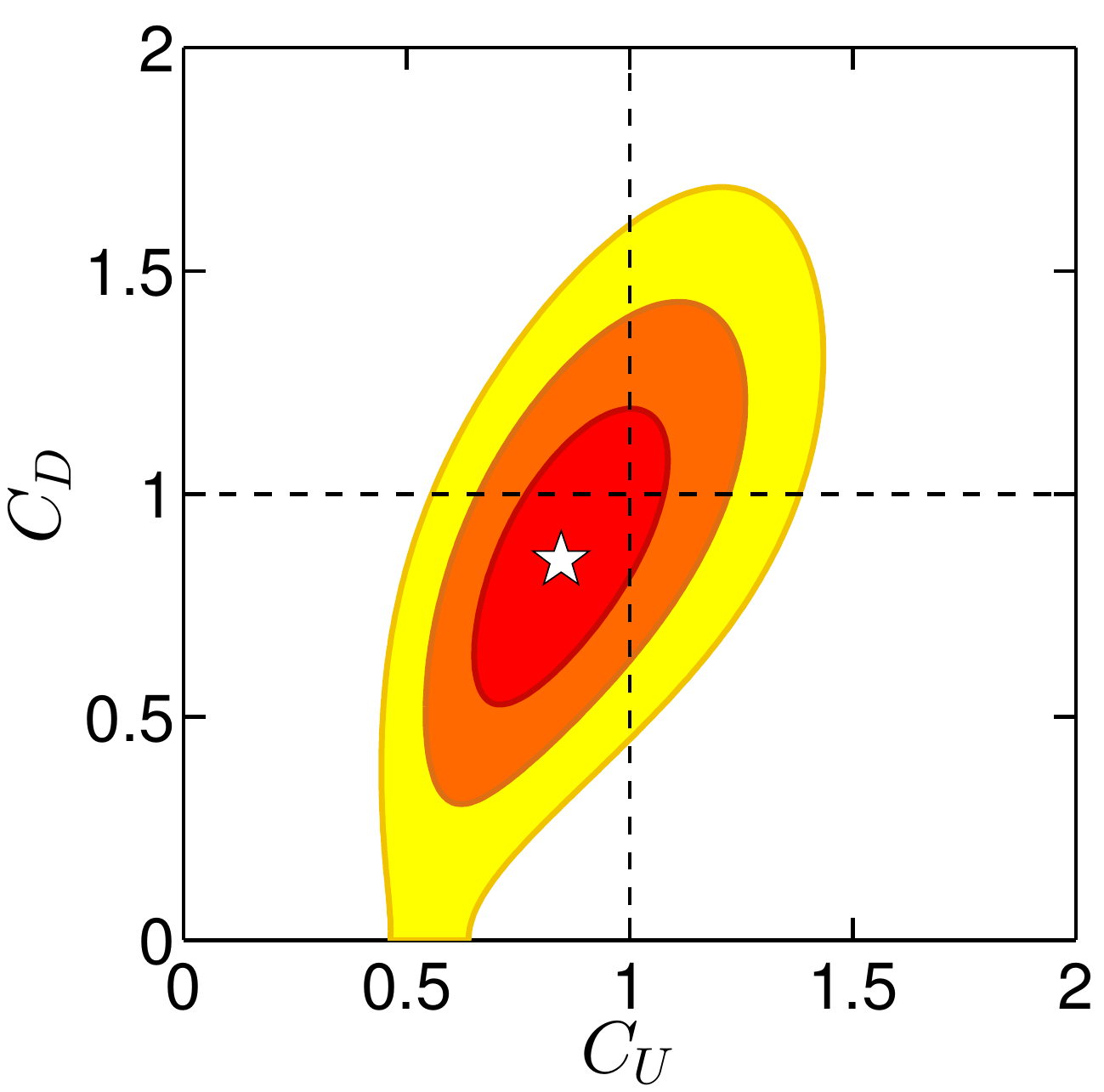}\includegraphics[scale=0.4]{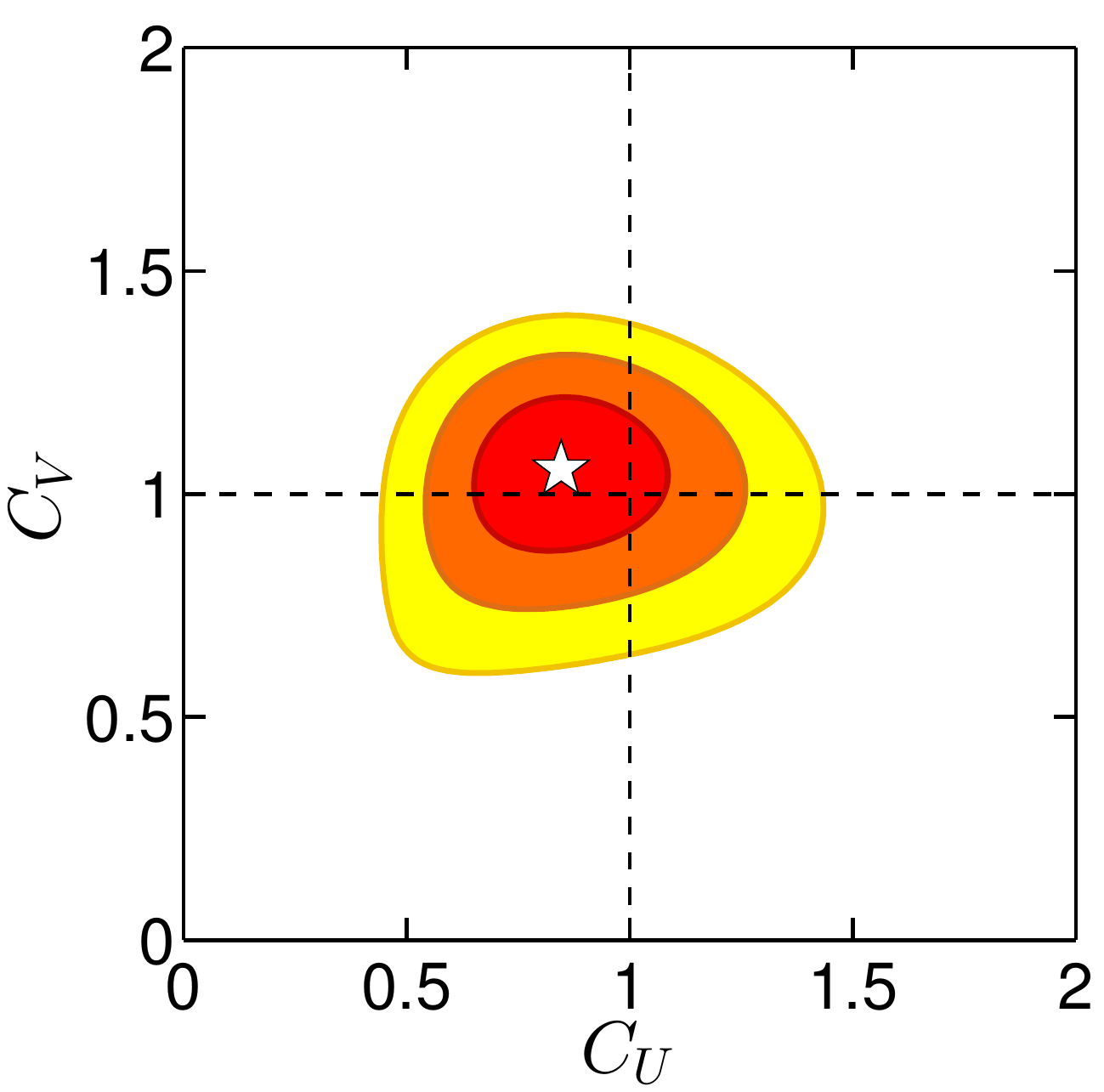}\includegraphics[scale=0.4]{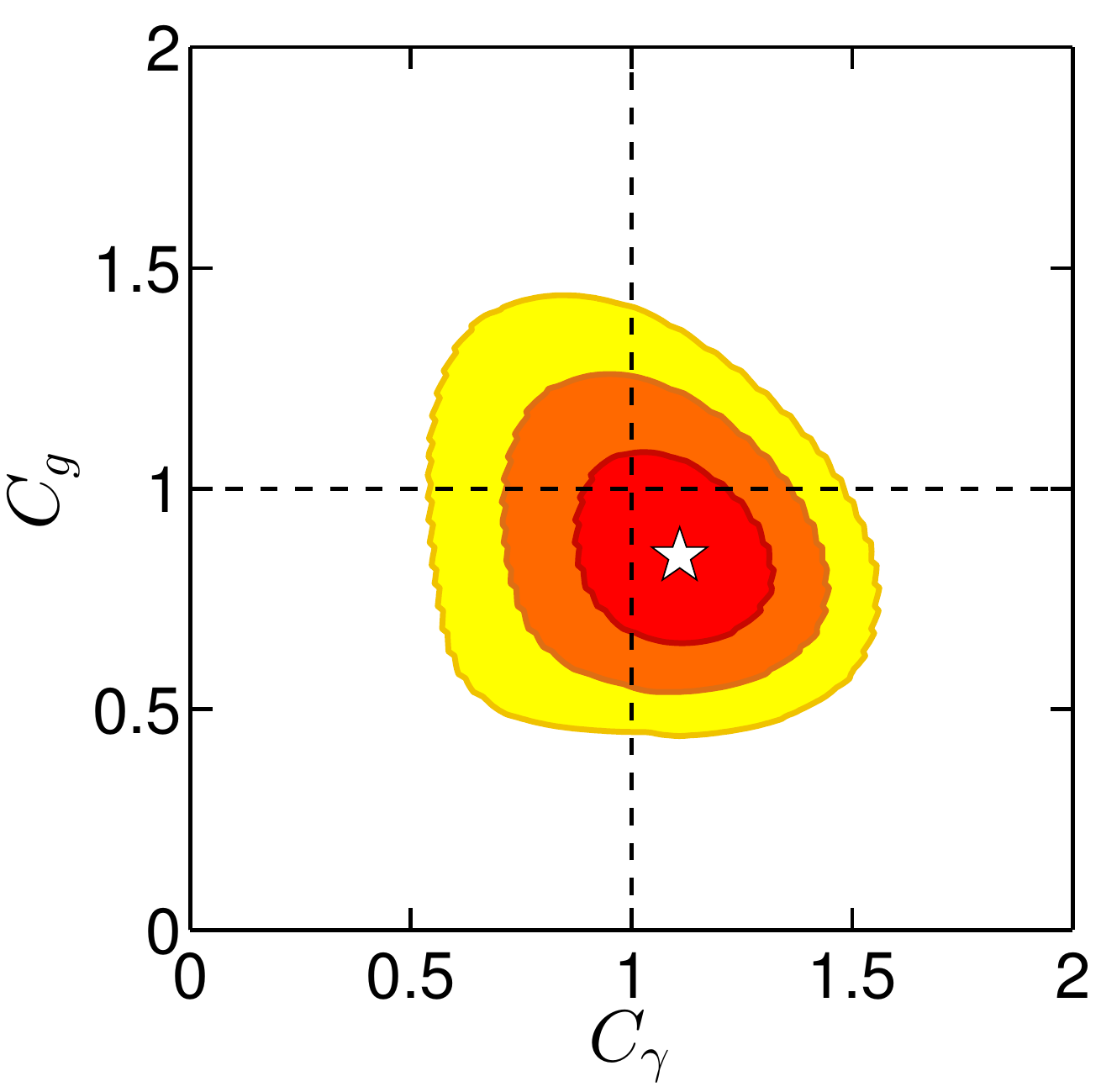}\\\includegraphics[scale=0.4]{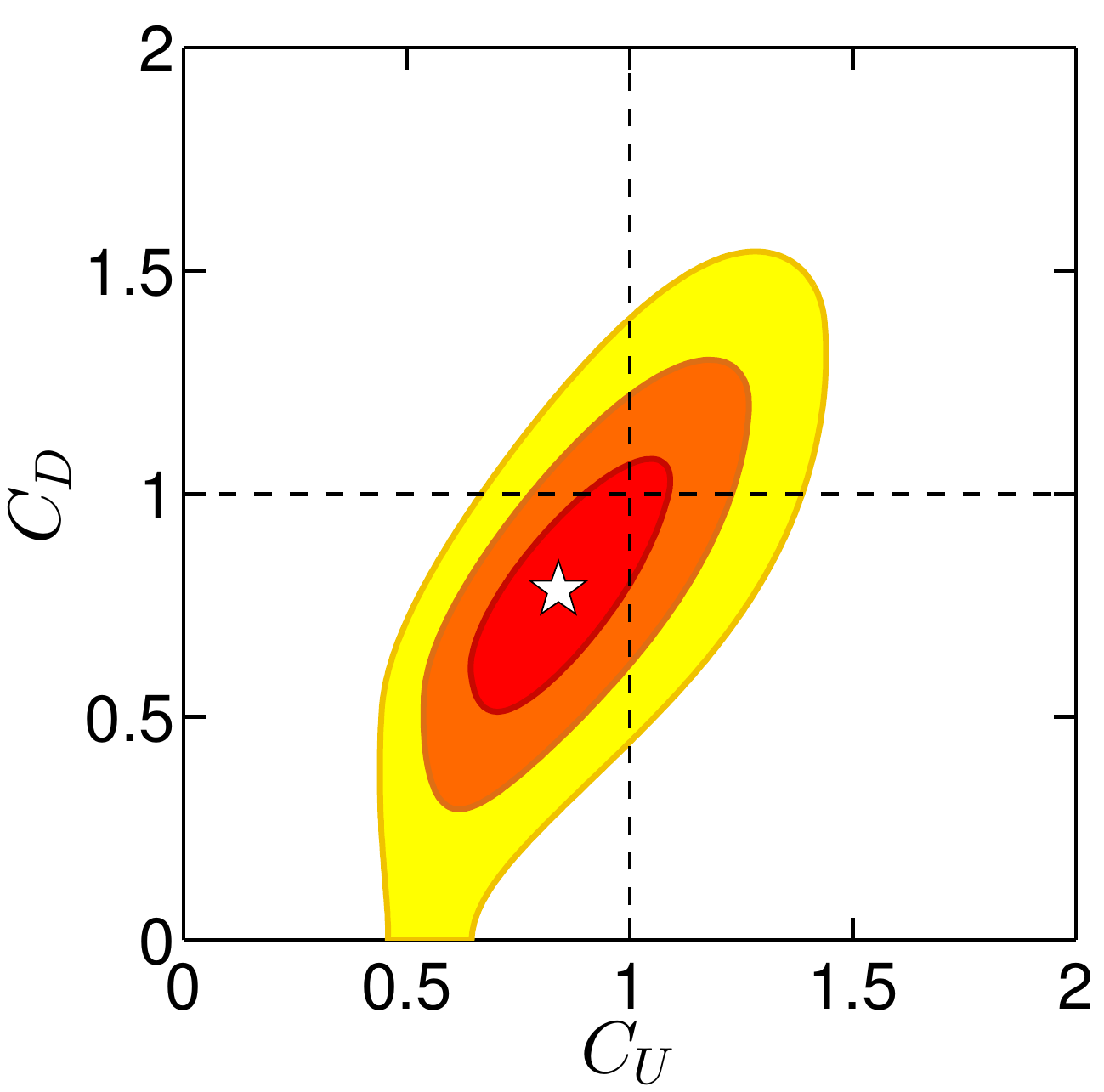}\includegraphics[scale=0.4]{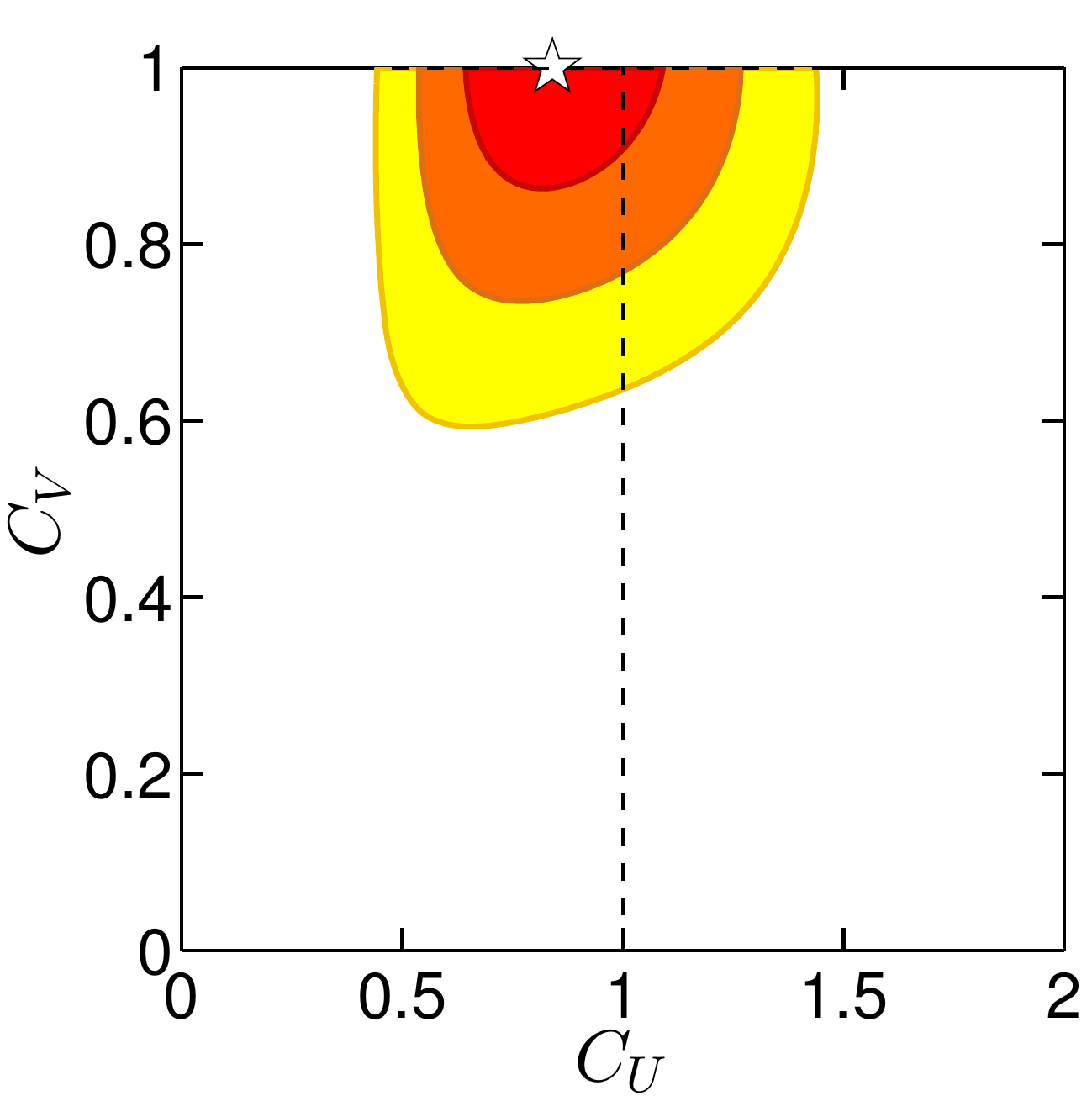}\includegraphics[scale=0.4]{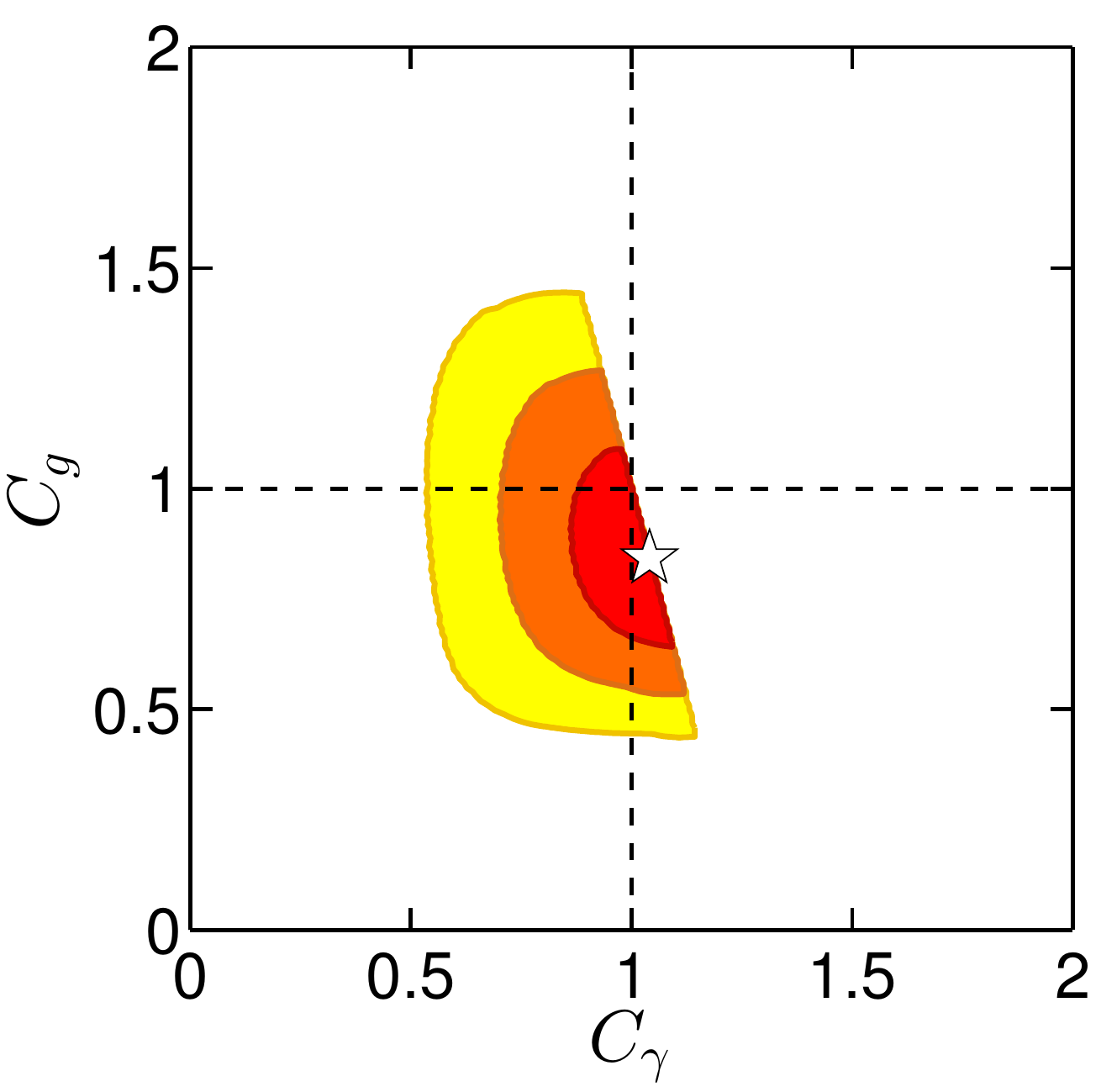}
\caption{Two-dimensional $\chisq$ distributions for the three parameter fit, Fit~{\bf II}, as in Fig.~\ref{1212.5244fit2-2d} 
but with $\CU>0$, $\CD>0$, $\CV>0$. 
The upper row of plots allows for $\cv>1$, while in the lower row of plots $\cv\le1$ is imposed. 
\label{1212.5244fit2-2dpos} }
\end{figure}

A negative sign of $\CU$---while maintaining a positive sign of
$m_t$---is actually not easy to achieve. (A sign change of
both $\CU$ and $m_t$ would have no impact on the top quark
induced loop amplitudes.) It would require that $m_t$ is
induced dominantly by the vev of a Higgs boson which is \emph{not} the
Higgs boson considered here. Hence, we have $\CU > 0$ in most models, implying that it
 is important to study the impact of this constraint on our fits.
The fit results when requiring $\CU,\CD > 0$ are shown in the left two plots of Fig.~\ref{1212.5244fit2-1dpos} 
and the top row of Fig.~\ref{1212.5244fit2-2dpos}; see also Table~\ref{tab:1212.5244fit2}.  
We observe that  for this quadrant the results are consistent with SM expectations (\ie\ within $\sim1\sigma$).
Interestingly the fit is not better than the SM itself: $\chimin=18.66$ for $21-3=18$~\dof, corresponding to $p=0.41$.

Another possible model constraint is to require $\cv\leq 1$ (recall that $\cv>0$ by  convention).  
This constraint applies to any model containing only Higgs doublets and singlets. 
The 1d results for the combined requirement of $\cu,\cd>0$ and $\cv\leq 1$ are shown 
in the right two plots of Fig.~\ref{1212.5244fit2-1dpos}, and  in the bottom-row plots of Fig.~\ref{1212.5244fit2-2dpos}.
We observe that the best fit values for $\cu$ and $\cd$ are only slightly shifted relative those found 
without constraining $\cv$, and that accordingly the $\cp=\cpb$ and $\cg=\cgb$ at 
the best fit point are only slightly shifted.  However, the $\cv\leq 1$ constraint does 
severely change the upper bound on $\cp$, which for $\cu>0$ and $\dcp=0$ mostly depends 
on the $W$-boson loop contribution. 
The apparent sharpness of the boundary in the $\cg$ vs.\ $\cp$ plane is a result of the fact that these 
two quantities really only depend on $\cu$ for $\cv=1$. 

Finally note that it has been shown in~\cite{Biswas:2012bd,Farina:2012xp} that single top production in
association with a Higgs is greatly enhanced when $\cu$,$\cv$ have
opposite signs. Thus, the possibility of $\cu<0$ should be further
scrutinized by precision measurements of the single top production cross
section at the LHC.

\renewcommand{\arraystretch}{1.3}
\begin{table}\centering
\begin{tabular}{|c|c|c|c|c|}
\hline
Fit  & {\bf I} & {\bf II} & {\bf III}, 1st min. & {\bf III}, 2nd min. \\
\hline 
   $\CU$  &  $1$  & $-0.86_{-0.16}^{+0.14}$ & $-0.06\pm1.30$\; & $\phantom{-}0.06\pm1.30$\; \\
   $\CD$  &  $1$  & $\phantom{-}0.99_{-0.26}^{+0.28}$ & $\phantom{-}1.00_{-0.26}^{+0.28}$ &  $-1.00_{-0.28}^{+0.26}$  \\ 
   $\CV$  &  $1$  & $\phantom{-}0.95_{-0.13}^{+0.12}$ & $\phantom{-}0.93_{-0.14}^{+0.12}$ &  $\phantom{-}0.93_{-0.14}^{+0.12}$  \\
   $\Delta\CP$   &  $\phantom{-}0.43_{-0.16}^{+0.17}$ & -- & $\phantom{-}0.16_{-0.36}^{+0.38}$ & $\phantom{-}0.21_{-0.39}^{+0.37}$ \\ 
   $\Delta\CG$   &  $-0.09 \pm 0.10$ & -- & $\phantom{-}0.83_{-1.17}^{+0.24}$& $\phantom{-}0.83_{-1.17}^{+0.24}$ \\
\hline
  $\CP$ & $\phantom{-}1.43_{-0.16}^{+0.17}$ & $\phantom{-}1.43\pm 0.17$ & $\phantom{-}1.36_{-0.23}^{+0.26}$ & $\phantom{-}1.36_{-0.23}^{+0.26}$ \\ 
  $\CG$ & $\phantom{-}0.91 \pm 0.10$ & $\phantom{-}0.92_{-0.15}^{+0.17}$ & $\phantom{-}0.95_{-0.23}^{+0.26}$ &  $\phantom{-}0.95_{-0.23}^{+0.26}$\\
\hline
 $\chisq_{\rm min}$ &  $12.31$ & $11.95$  & $11.46$ & $11.46$ \\
 $\chisq_{\rm min}/\dof$ &  $0.65$ & $0.66$ & $0.72$ & $0.72$ \\
\hline
\end{tabular}
\caption{Summary of results for Fits {\bf I}--{\bf III}.  For Fit {\bf II}, the tabulated results are from the best fit, cf.\ column 1 of Table~\ref{tab:1212.5244fit2}.}
\label{1212.5244chisqmintable}
\end{table}

\renewcommand{\arraystretch}{1.3}
\begin{table}[t]\centering
\begin{tabular}{|c|c|c|c|}
\hline
Sector  & $\cu<0,\,\cd>0$ & $\cu,\,\cd<0$ & $\cu,\,\cd>0$ \\
\hline 
   $\CU$  & $-0.86_{-0.16}^{+0.14}$ & $-0.91_{-0.17}^{+0.15}$ & $0.85_{-0.13}^{+0.15}$ \\
   $\CD$  & $\phantom{-}0.99_{-0.26}^{+0.28}$ & $-0.98_{-0.27}^{+0.26}$ & $0.85_{-0.21}^{+0.22}$ \\ 
   $\CV$  & $\phantom{-}0.95_{-0.13}^{+0.12}$ & $\phantom{-}0.94_{-0.13}^{+0.12}$ & $1.06_{-0.12}^{+0.11}$ \\
\hline
  $\CP$ & $\phantom{-}1.43 \pm 0.17$ & $\phantom{-}1.43_{-0.17}^{+0.16}$ & $1.11_{-0.16}^{+0.15}$ \\ 
  $\CG$ & $\phantom{-}0.92_{-0.15}^{+0.17}$ & $\phantom{-}0.91_{-0.15}^{+0.17}$ & $0.85_{-0.13}^{+0.16}$ \\
\hline
 $\chisq_{\rm min}$ & 11.95 & 12.06 &  18.66 \\
 $\chisq_{\rm min}/\dof$ & 0.66 & 0.67 &  1.04 \\
\hline
\end{tabular}
\caption{Results for Fit {\bf II} in different sectors of the ($\cu$,\,$\cd$) plane.}
\label{tab:1212.5244fit2}
\end{table}
\renewcommand{\arraystretch}{1.0}

\subsubsection*{\boldmath Fit III: varying $\cu$, $\cd$, $\cv$, $\dcp$ and $\dcg$}

\begin{figure}[ht]
\includegraphics[scale=0.4]{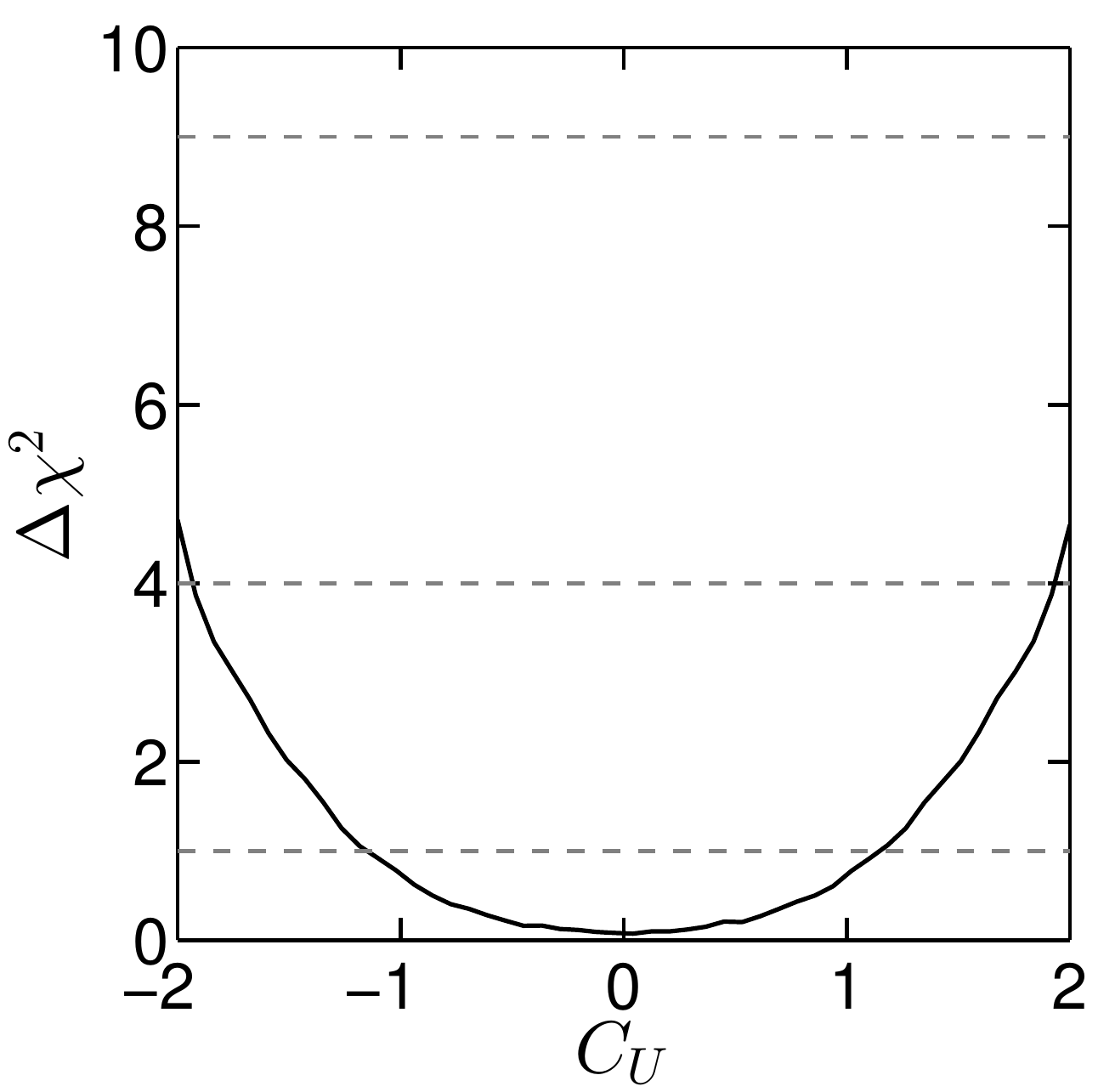}\includegraphics[scale=0.4]{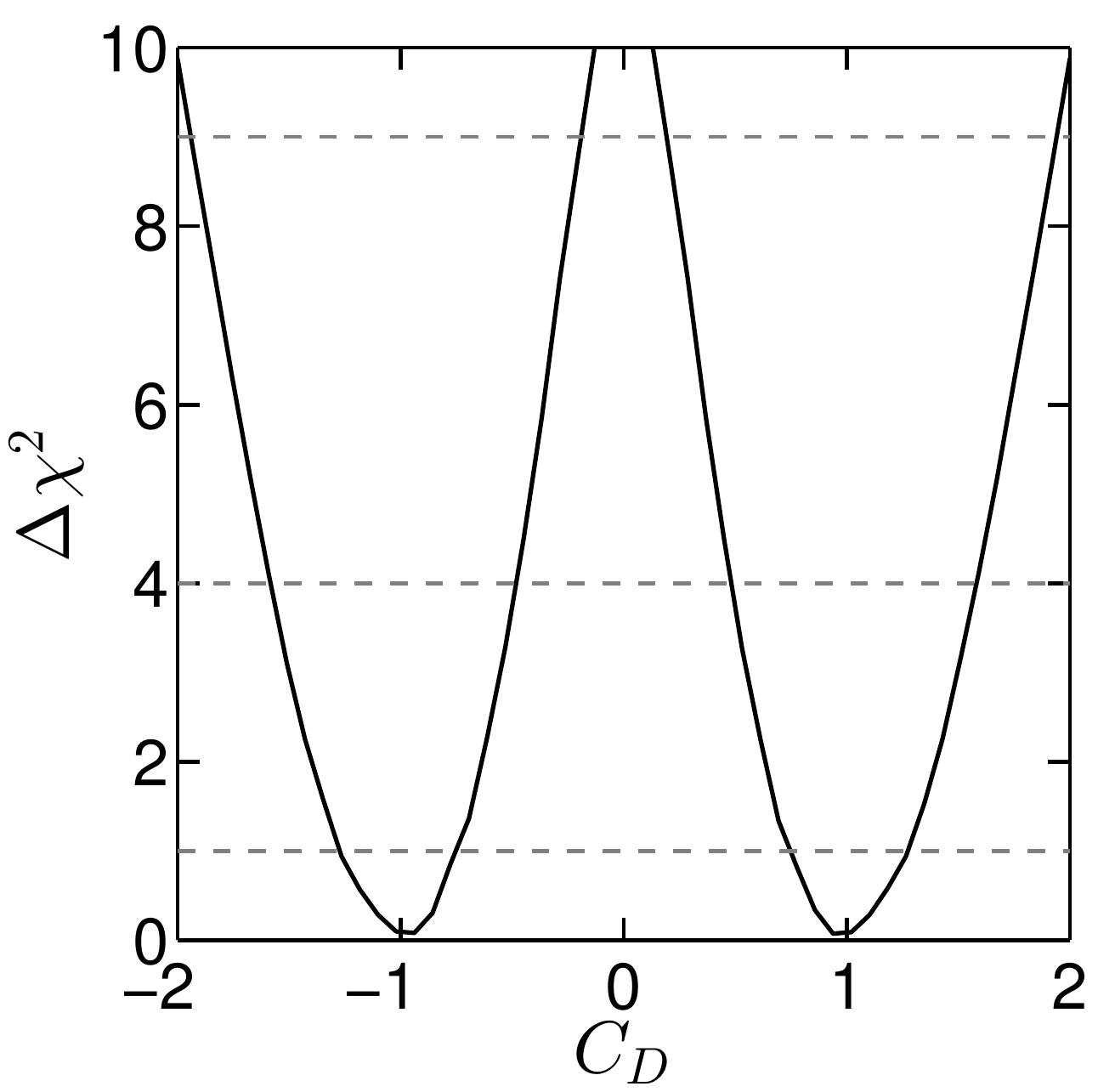}\includegraphics[scale=0.4]{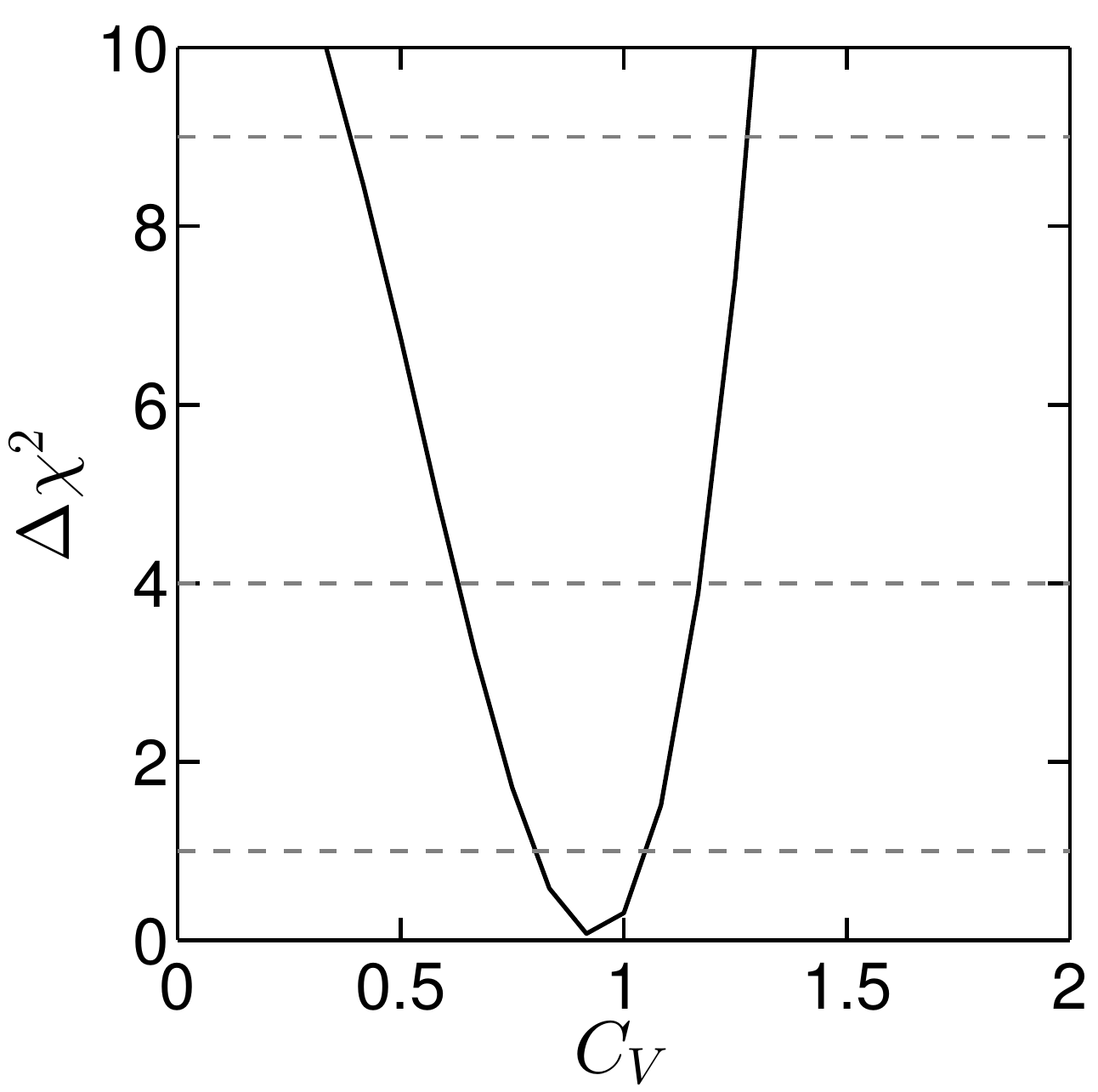}
\includegraphics[scale=0.4]{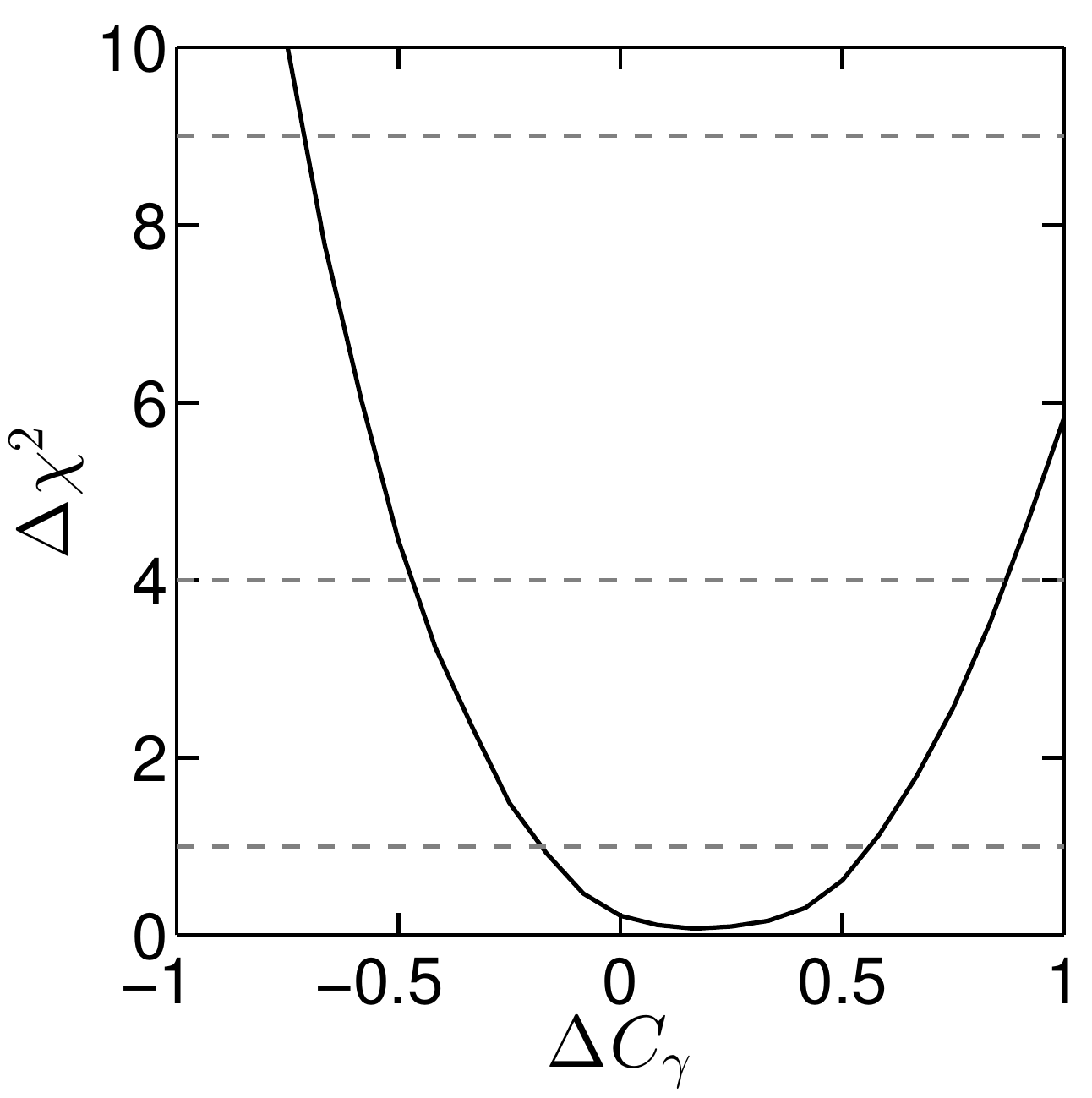}\includegraphics[scale=0.4]{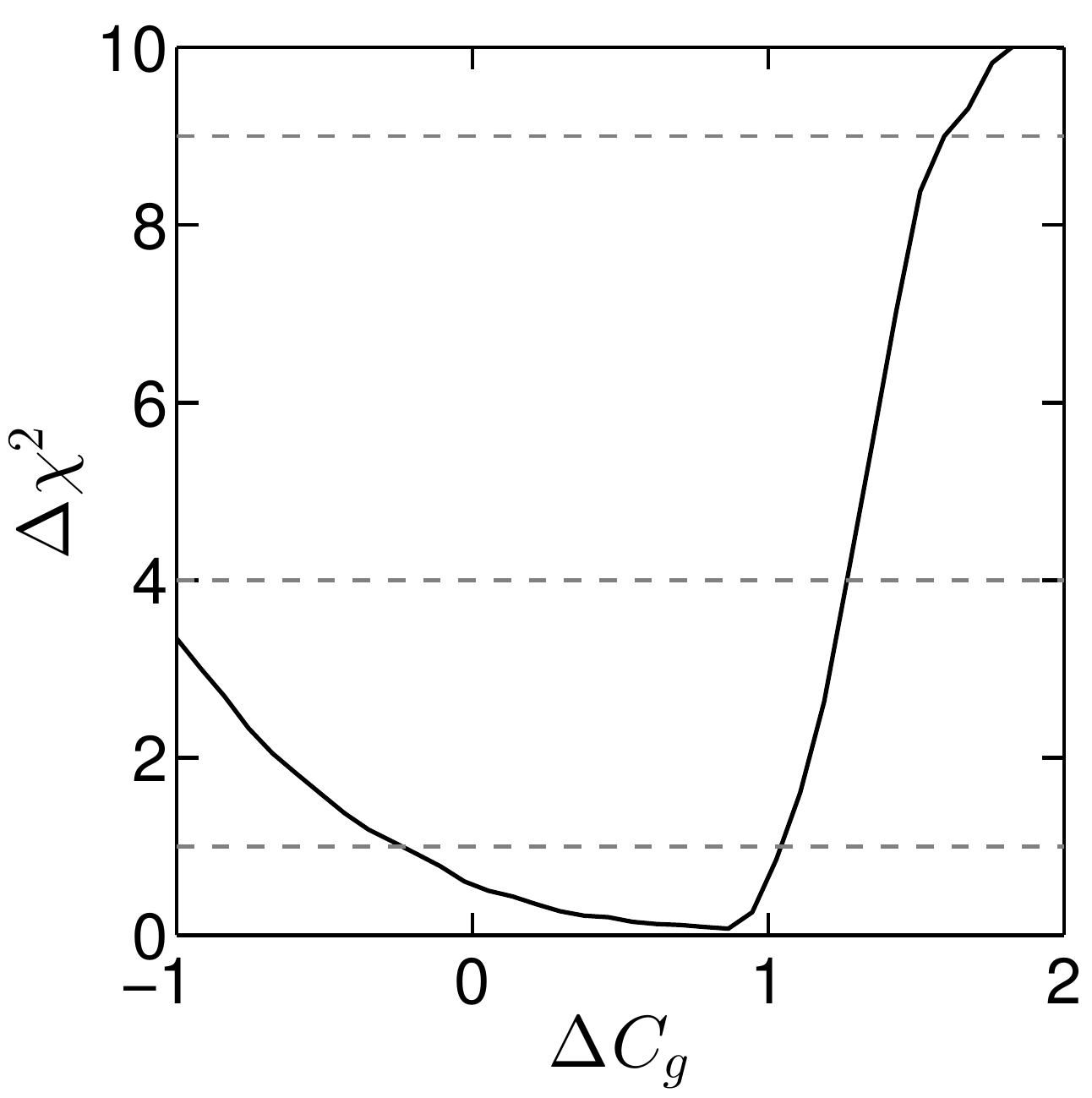}\includegraphics[scale=0.4]{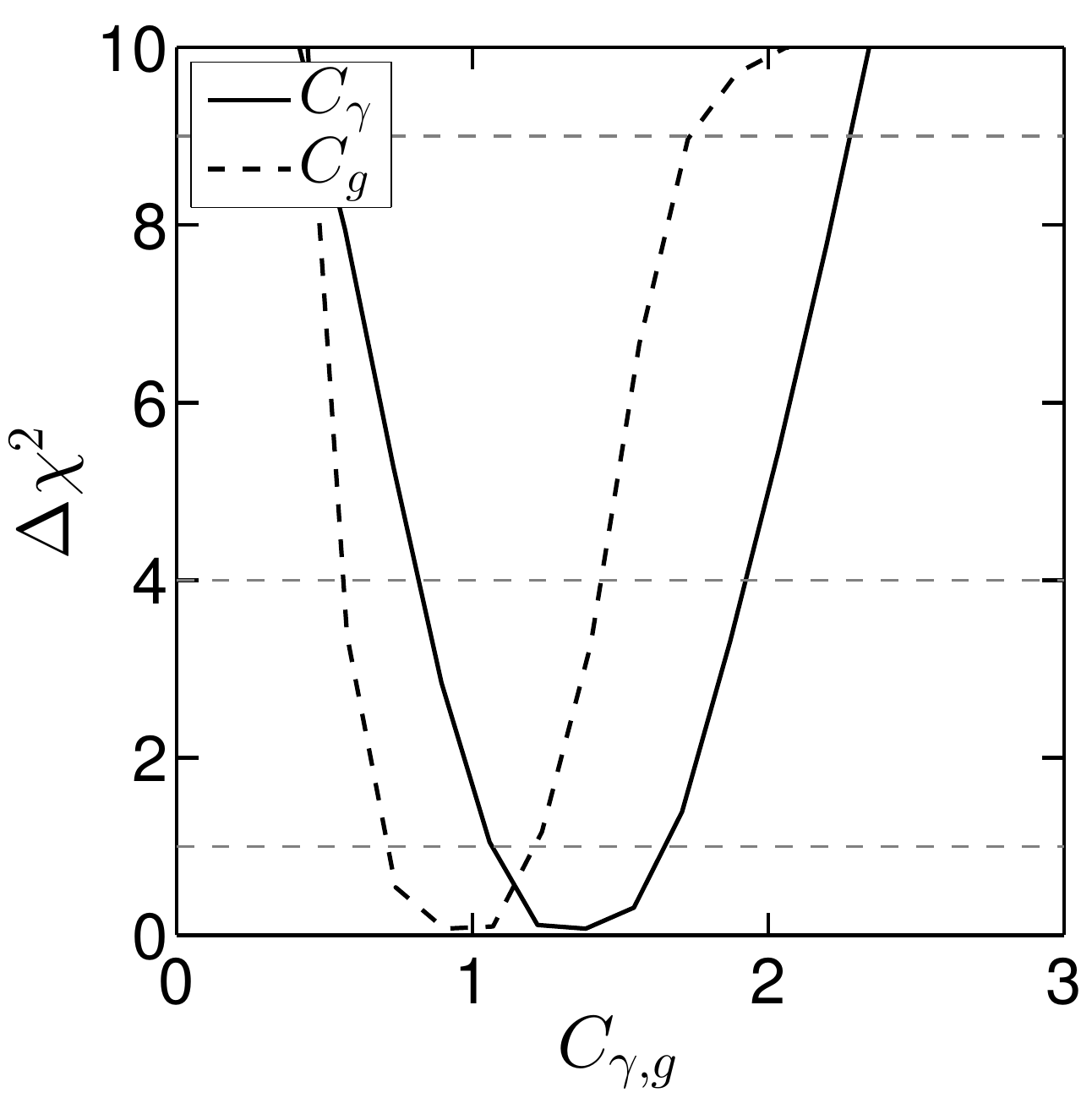}
\caption{One-dimensional $\chisq$ distributions for the five parameter fit of $\CU$, $\CD$, $\CV$, $\Delta\CP$ and $\Delta\CG$ (Fit~{\bf III}). Details regarding the best fit point are given in Table~\ref{1212.5244chisqmintable}.
\label{1212.5244fit3-1d} }
\end{figure}

\begin{figure}[ht]\centering
\includegraphics[scale=0.4]{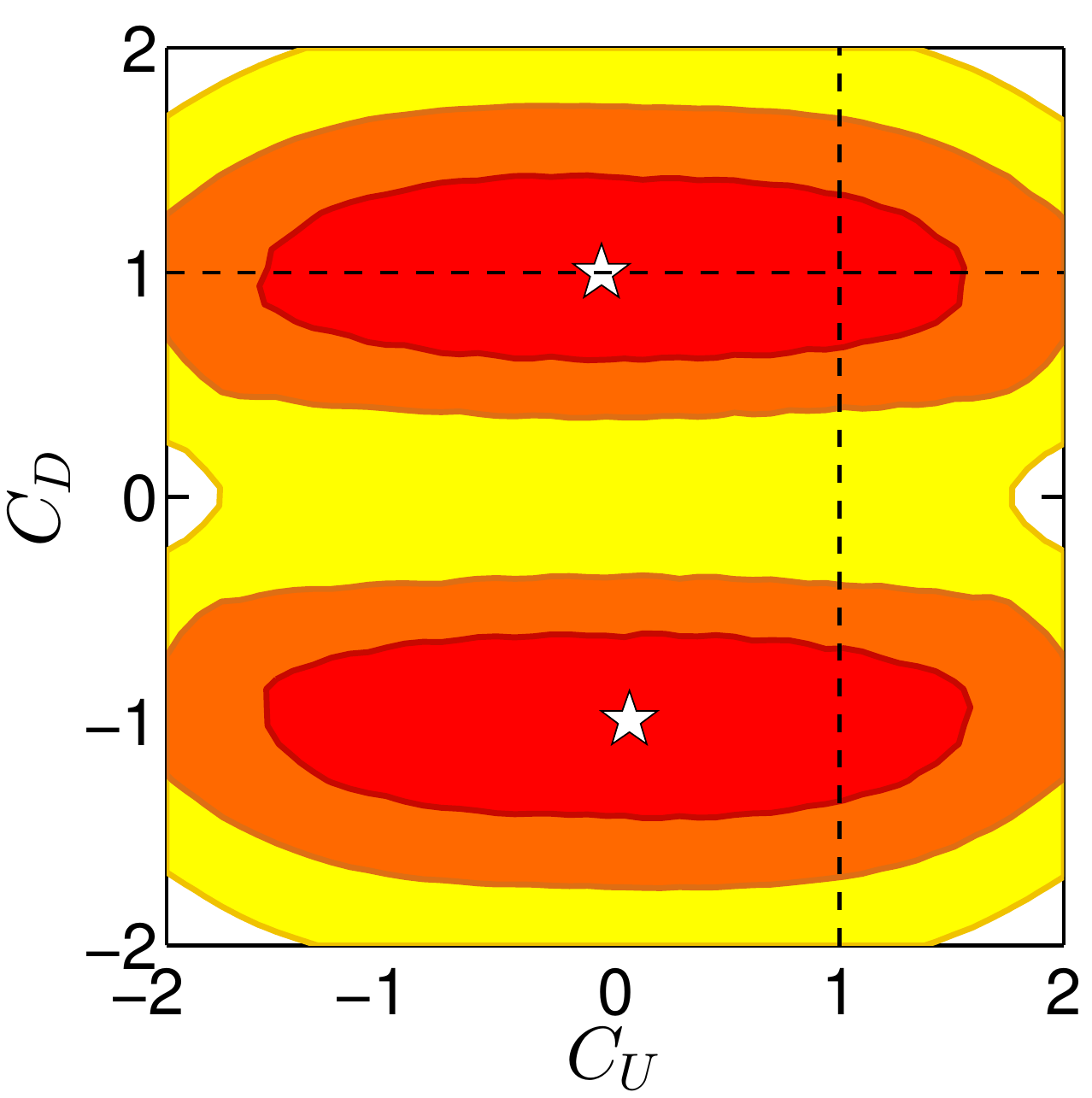}\includegraphics[scale=0.4]{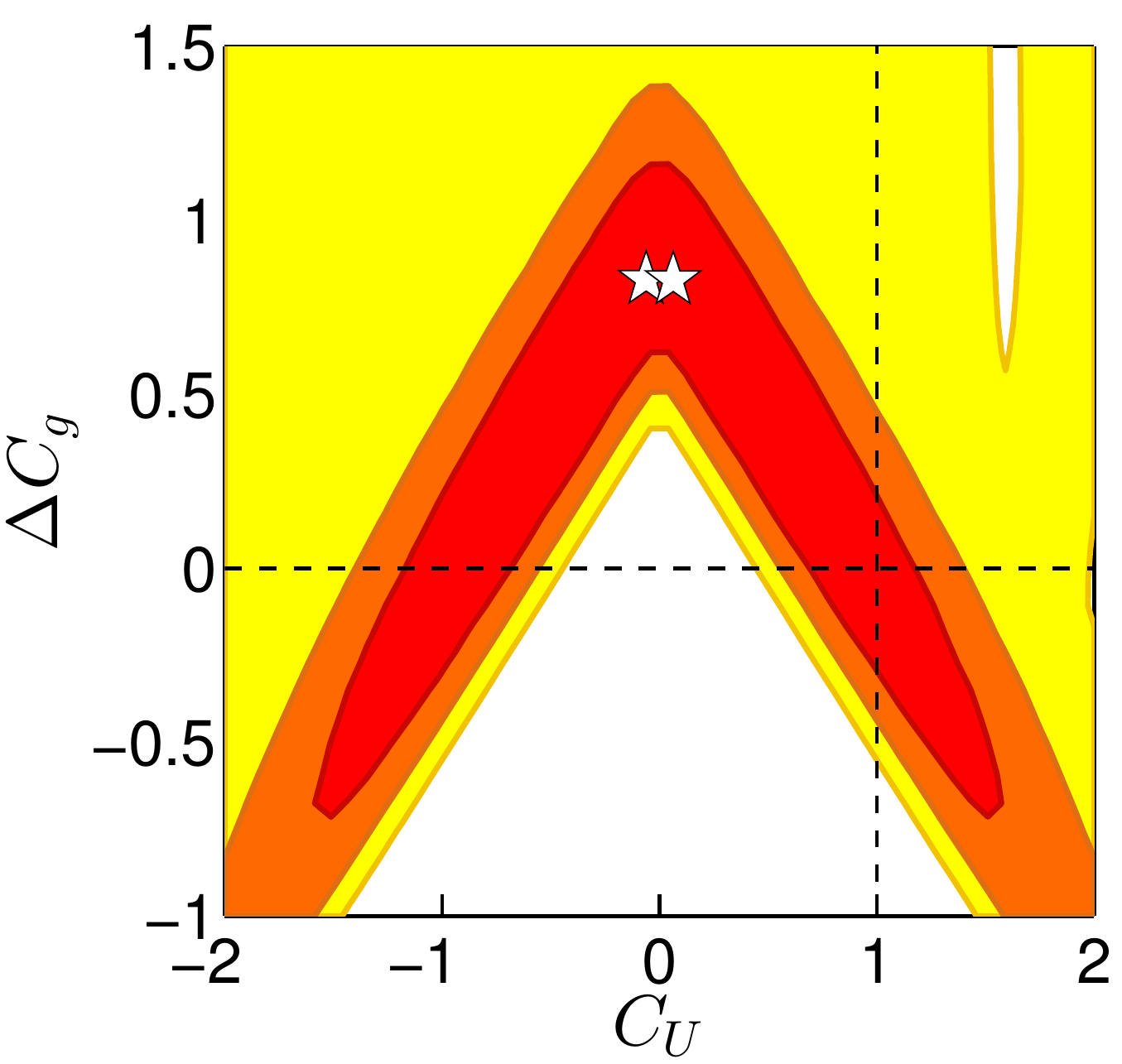}\includegraphics[scale=0.4]{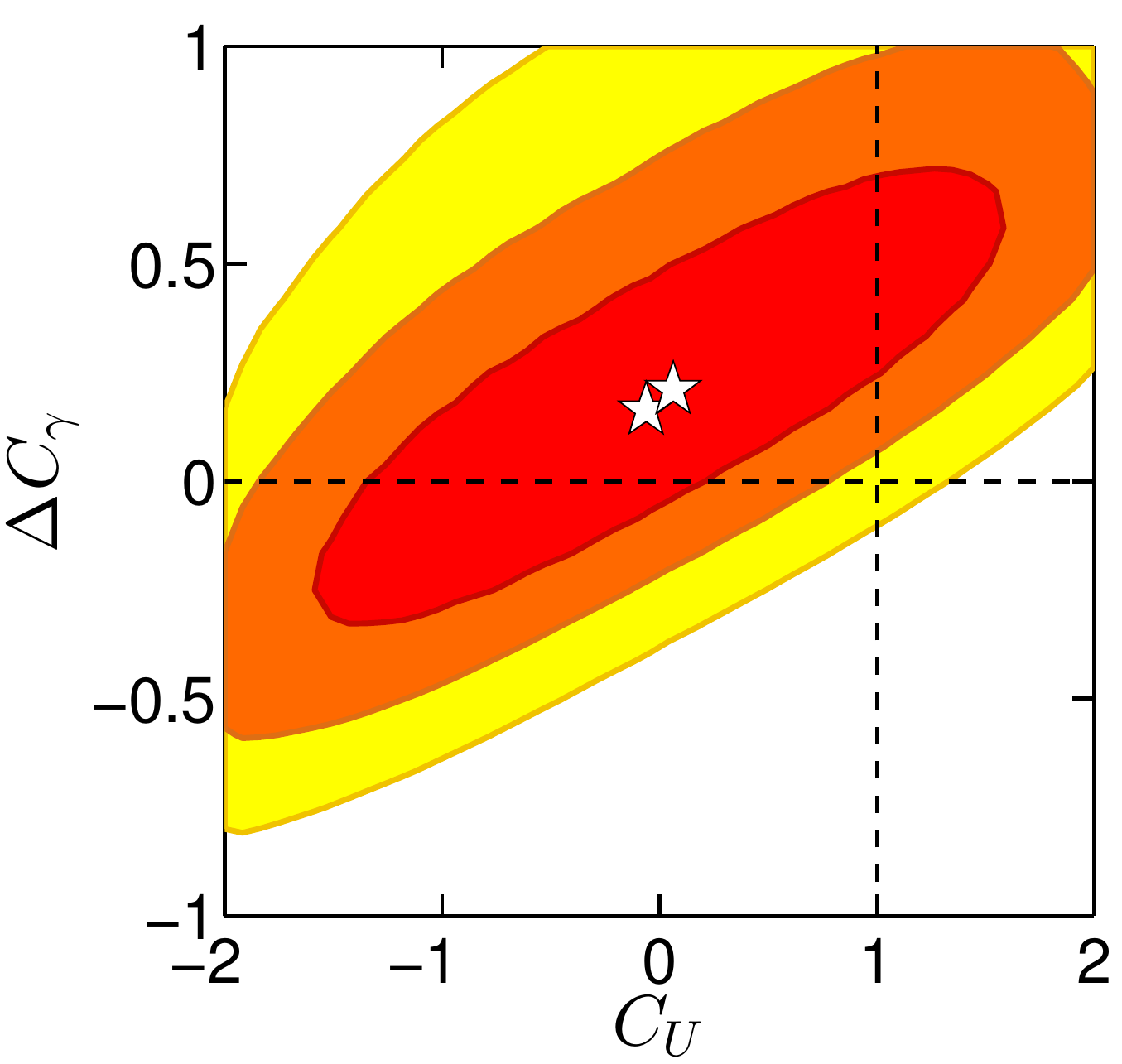}
\includegraphics[scale=0.4]{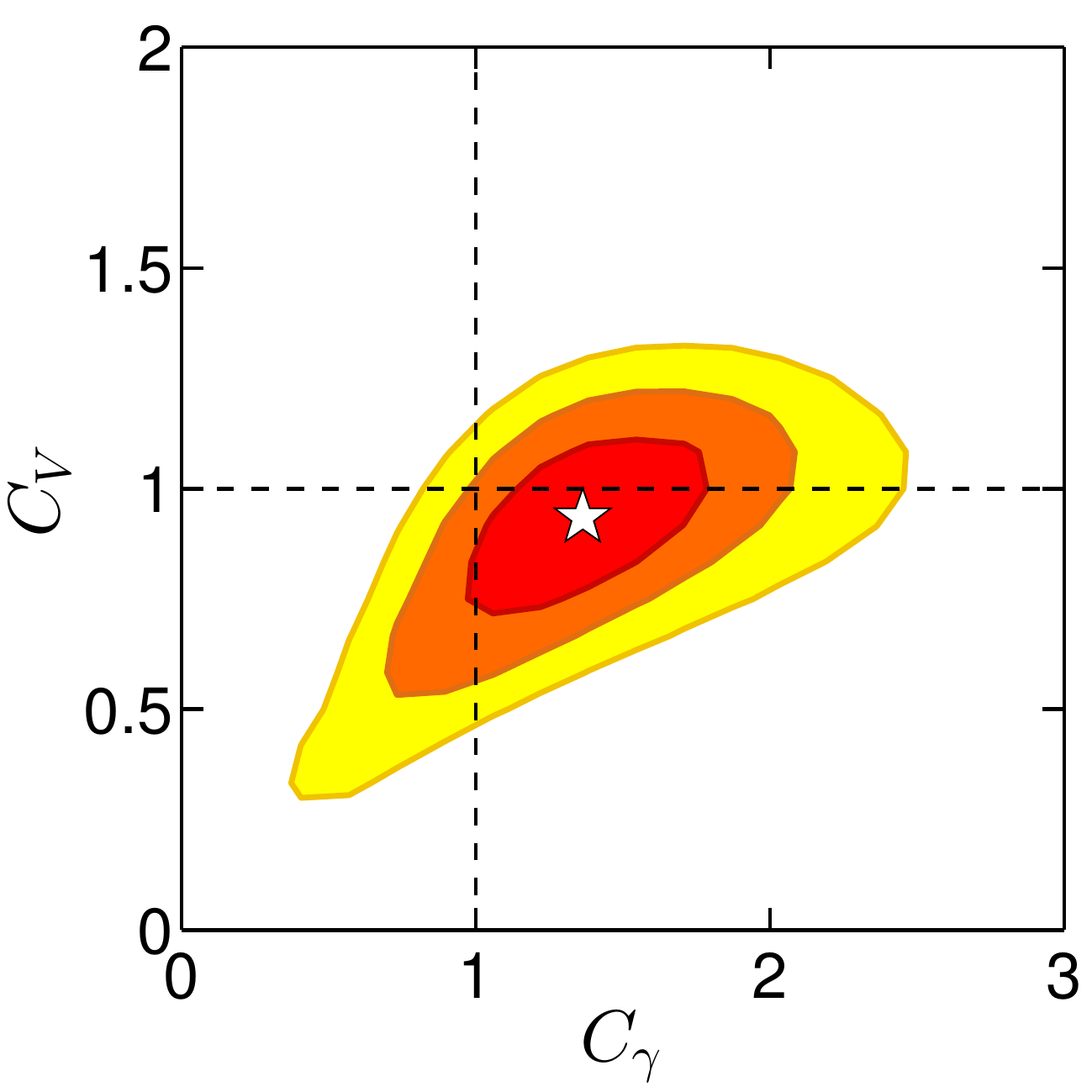}\includegraphics[scale=0.4]{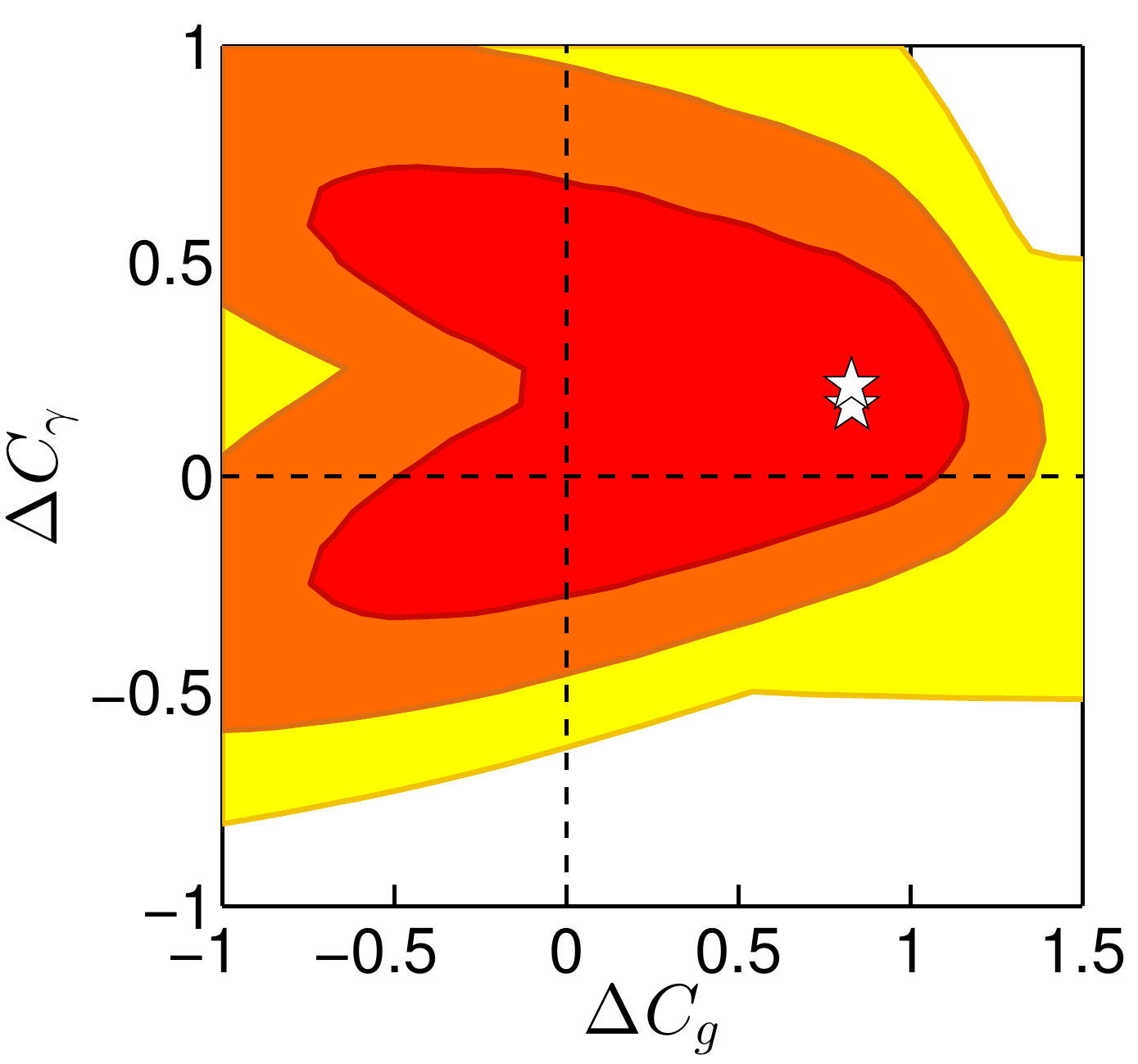}\includegraphics[scale=0.4]{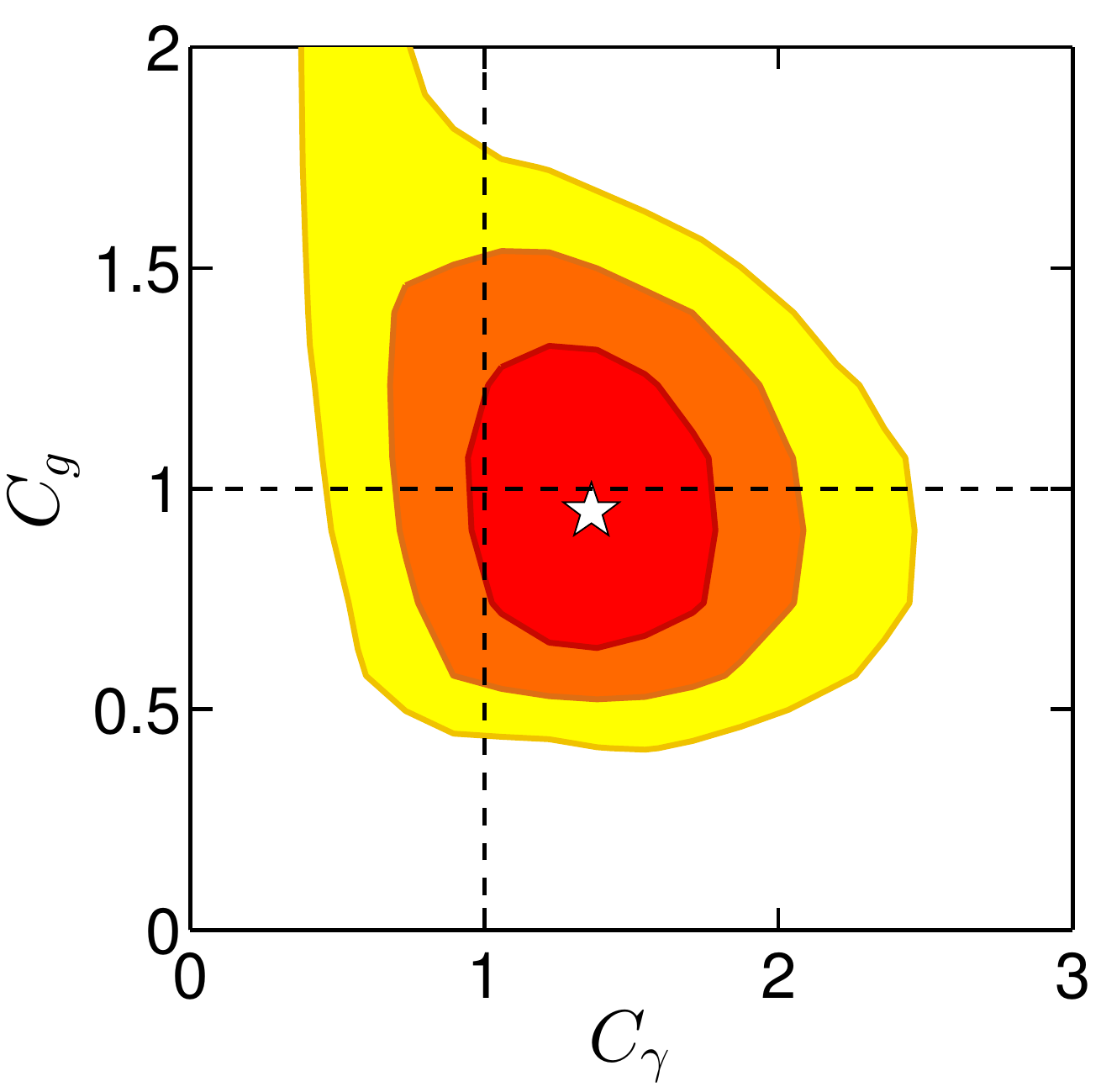}
\caption{Two-dimensional distributions for the five parameter fit of $\CU$, $\CD$, $\CV$, $\Delta\CP$ and $\Delta\CG$ (Fit~{\bf III}). Details regarding the best fit point are given in Table~\ref{1212.5244chisqmintable}.
\label{1212.5244fit3-2d} }
\end{figure}

Finally, in Fit~{\bf III}, we allow the $\dcg$ and $\dcp$ additions to $\cgb$ and $\cpb$, fitting 
therefore to five free parameters: $\CU$, $\CD$, $\CV$, $\dcg$, and $\dcp$. 
The associated 1d and 2d plots are given in Figs.~\ref{1212.5244fit3-1d} and \ref{1212.5244fit3-2d}.  
There are two main differences as compared to Fit~{\bf II}.  
On the one hand, the preference for $\cp>1$ does does not necessarily imply a negative value for $\cu$, 
since a  positive value for $\dcp$ can contribute to an increase in $\cp$ even when the top-quark loop 
interferes destructively with the $W$ loop. (This is obviously already expected from  Fit~{\bf I}.) 
On the other hand, both $\cu$ and $\dcg$ feed into the effective $\cg$, and if one of them is large the other 
one has to be small to result in a near SM-like $gg\to H$ cross section. 
This anti-correlation between $|\cu|$ and $\dcg$ can be seen in the center-top  plot in Fig.~\ref{1212.5244fit3-2d}.  
The best fit is actually obtained for $\cu\approx 0$, with $\dcg\approx 1$ in order to compensate for the very 
suppressed top-loop contribution to ggF.  However, it is also apparent that the minimum at $\cu=0$ is quite 
shallow (cf.\ the top left plot in Fig.~\ref{1212.5244fit3-1d}) and that a fit with $\cu\approx 1$ with small $\dcg$ is well 
within the $68\%$ contour (as should indeed be the case for consistency with Fits {\bf I} and {\bf II}).  

We also note that at the best fit, \ie\ that with $\cu\approx 0$, one finds $\cp\sim \cpb>1$ by virtue of the fact that the 
$W$ loop is not partially 
canceled by the top loop and only a small $\dcp\sim 0.16$--$0.21$ is needed to further enhance the $\gam\gam$ final state  
and bring $\mu(\gam\gam)$ into agreement with observations; see top-right and bottom-right plots of Fig.~\ref{1212.5244fit3-2d}. 
If we move to the SM value of $\cu=1$ then 
$\dcp\sim 0.45$ is needed to fit the $\gam\gam$ rate. The best fit results are tabulated in Table~\ref{1212.5244chisqmintable}.

\begin{figure}[ht]\centering
\includegraphics[scale=0.4]{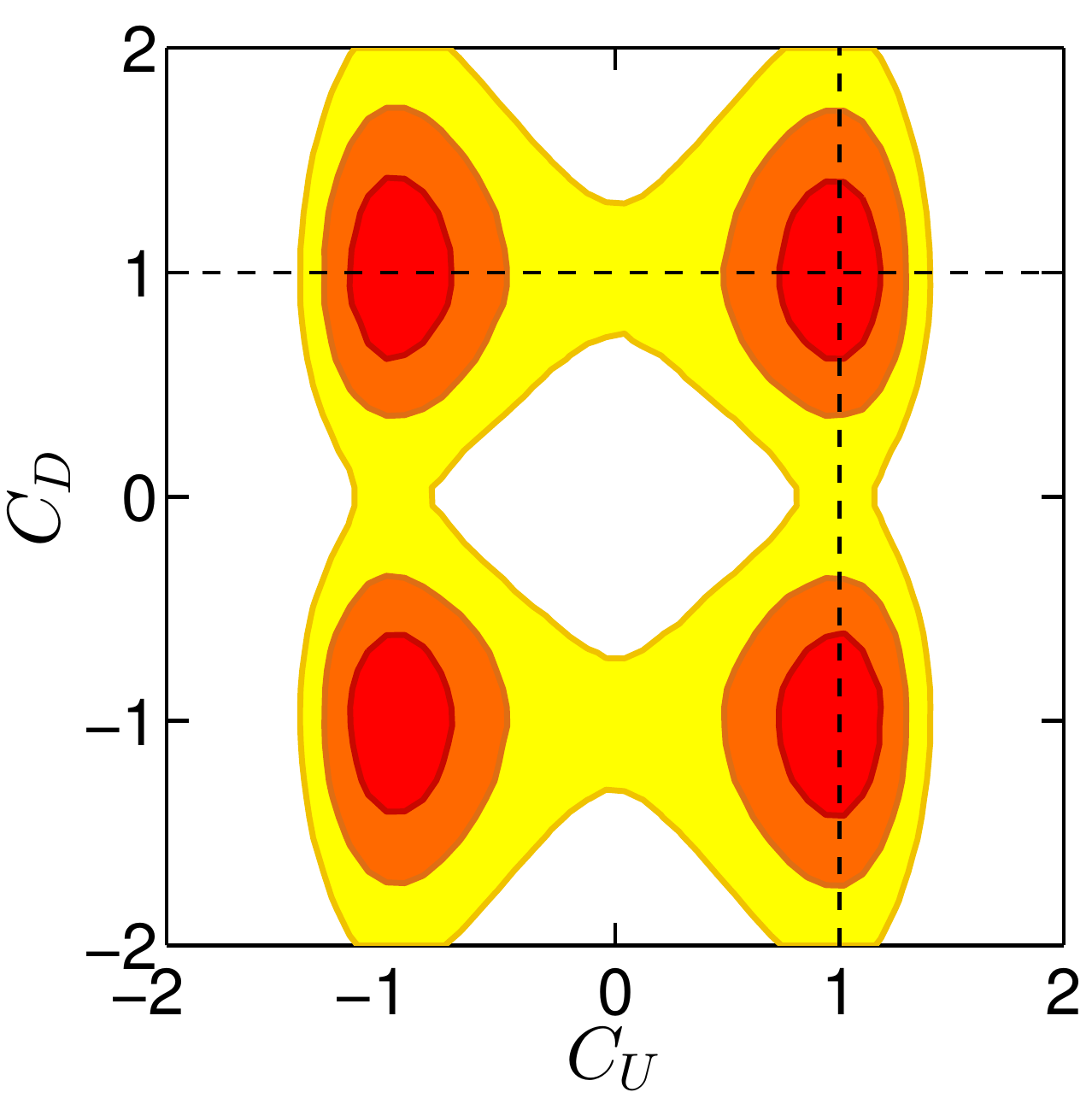}\includegraphics[scale=0.4]{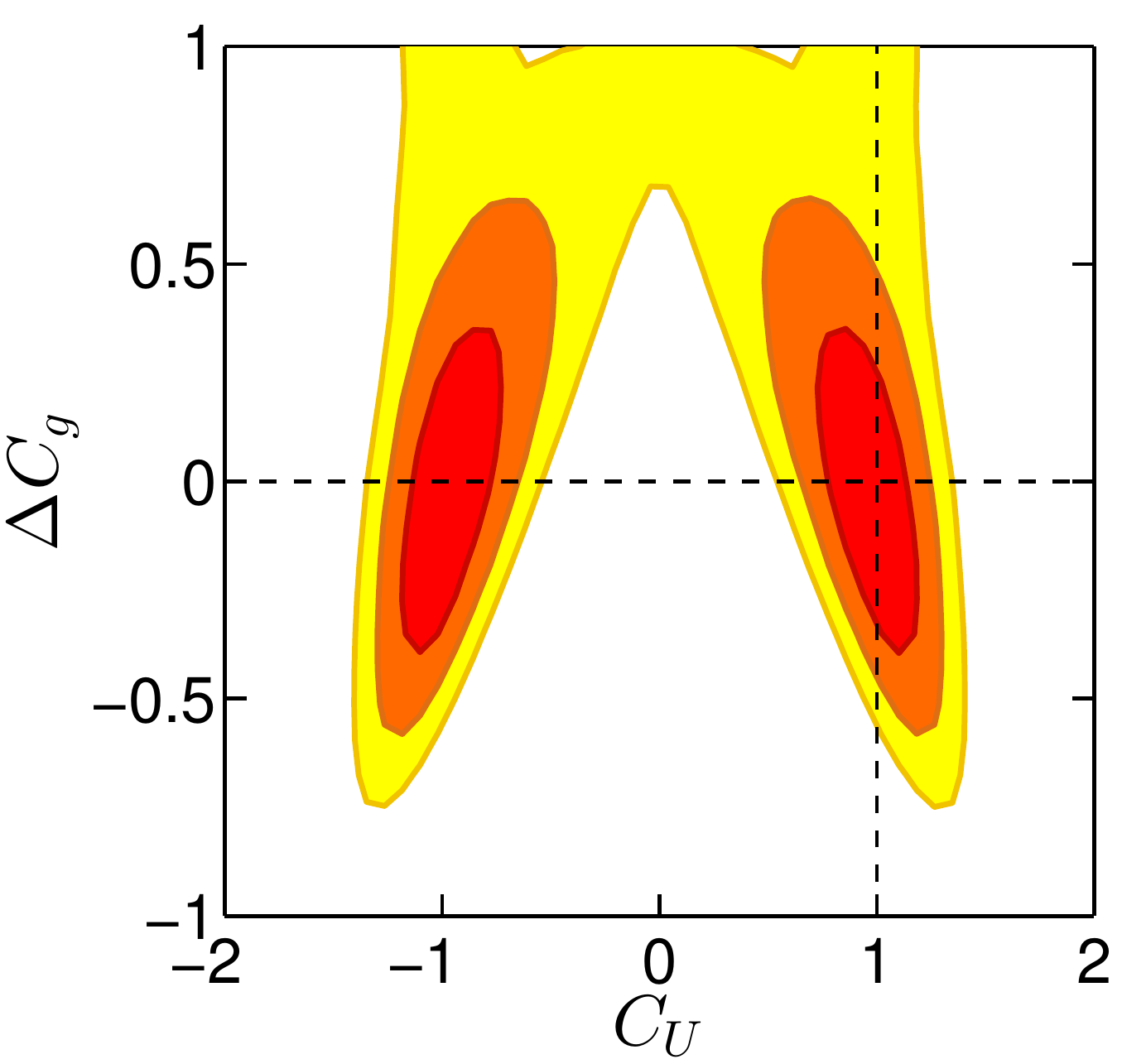}
\caption{Lifting of the degeneracy in $\cu$ and $\dcg$ in Fit~{\bf III} when $t\bar tH$ is measured to 30\% accuracy ($\mu(\tth)=1\pm 0.3$). These two plots should be compared to the top left and top middle plots of Fig.~\ref{1212.5244fit3-2d}. See text for details. 
\label{1212.5244fit3-2d-with-ttH} }
\end{figure}

A way to lift the degeneracy in $\cu$ and $\dcg$ would be to have an independent determination of $\cu$. 
This might be achieved by an accurate measurement of the $\tth$ channel, as illustrated in Fig.~\ref{1212.5244fit3-2d-with-ttH}. 
This figure assumes that $\mu(\tth)$ will eventually be measured with 30\% accuracy --- more concretely, the figure assumes 
$\mu(\tth)=1\pm 0.3$. This is certainly a very challenging task. 
For comparison, CMS currently gives $\mu(\tth)\approx -0.8^{+2.2}_{-1.8}$~\cite{CMS-PAS-HIG-12-045}. 
Finally, as mentioned above, $\cu$ may also be constrained by the associated production of a single top and a Higgs~\cite{Biswas:2012bd,Farina:2012xp}.

\subsection{Application to two-Higgs-Doublet Models} \label{sec:20122HDMfit}
 
So far our fits have been model-independent, relying only on the Lagrangian structure of \Eq{eq:1212.5244ourldef}. 
Let us now turn to the concrete examples of Two-Higgs-Doublet Models (2HDMs) of Type~I and Type~II. 
In both cases, the basic parameters describing the coupling of either the light $h$ or heavy $H$ CP-even 
Higgs boson are only two: $\alpha$ (the CP-even Higgs mixing angle) and $\tanb=v_u/v_d$, where $v_u$ 
and $v_d$ are the vacuum expectation values of the Higgs field that couples to up-type quarks and down-type 
quarks, respectively.  The Type~I and Type~II models are distinguished by the pattern of their fermionic couplings 
as given in Table~\ref{1212.5244fermcoups}.  The SM limit for the $h$ ($H$) in the case of both Type~I and Type~II models 
corresponds to $\alpha=\beta-\pi/2$ ($\alpha=\beta$).  
In our discussion below, we implicitly assume that there are no contributions from non-SM particles to the loop 
diagrams for $\cp$ and $\cg$.  In particular, this means our results correspond to the case where the charged 
Higgs boson, whose loop might contribute to $\cp$, is heavy.  

\begin{table}[t]
\begin{center}
\begin{tabular}{|c|c|c|c|c|c|}
\hline
\ & Type~I and II  & \multicolumn{2}{c|}  {Type~I} & \multicolumn{2}{c|}{Type~II} \cr
\hline
Higgs & VV & up quarks & down quarks \& & up quarks & down quarks \&  \cr
&  &  &  leptons &  &  leptons \cr
\hline
 $h$ & $\sin(\beta-\alpha)$ & $\cosa/ \sinb$ & $\cosa/ \sinb$  &  $\cosa/\sinb$ & $-{\sina/\cosb}$   \cr
\hline
 $H$ & $\cos(\beta-\alpha)$ & $\sina/ \sinb$ &  $\sina/ \sinb$ &  $\sina/ \sinb$ & $\cosa/\cosb$ \cr
\hline
 $A$ & 0 & $\cotb$ & $-\cotb$ & $\cotb$  & $\tanb$ \cr
\hline 
\end{tabular}
\end{center}
\vspace{-.15in}
\caption{Tree-level vector boson couplings $C_V^{h_i}$ ($V=W,Z$) and fermionic couplings $C^{h_i}_{F}$
normalized to their SM values for the Type~I and Type~II two-Higgs-doublet models. }
\label{1212.5244fermcoups}
\end{table}

\begin{figure}[ht]\centering
\includegraphics[scale=0.44]{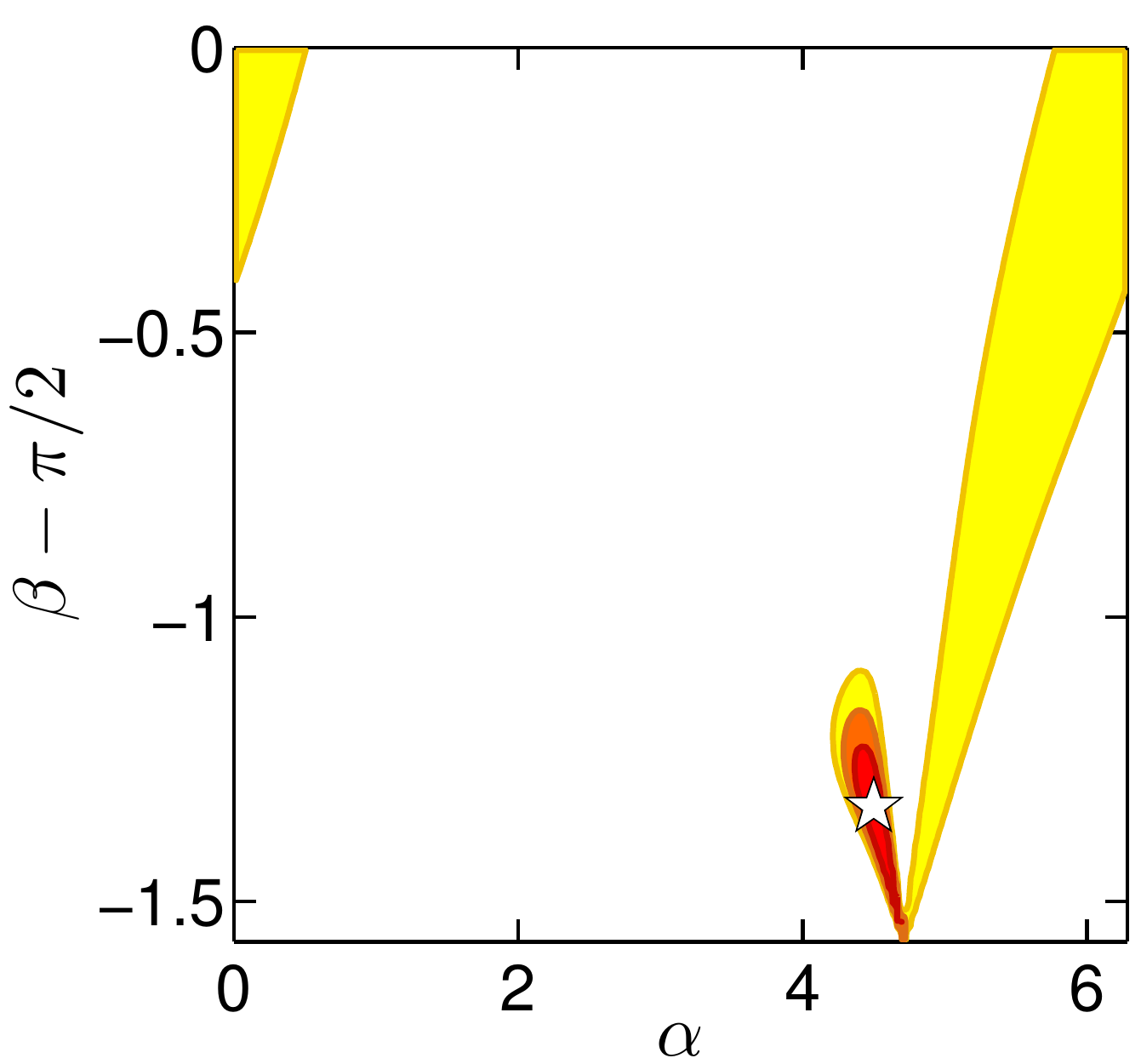}\qquad\includegraphics[scale=0.44]{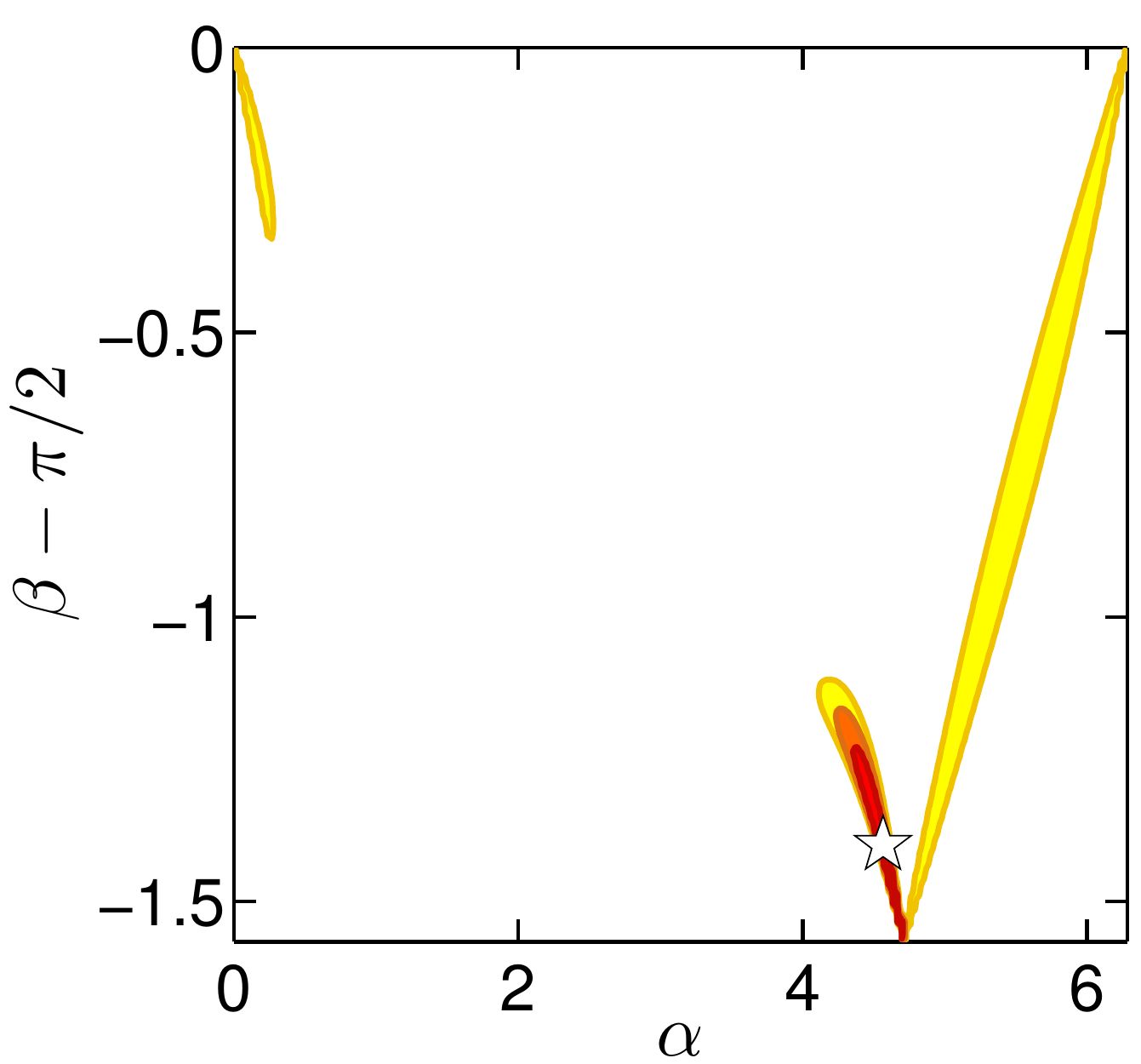}
\includegraphics[scale=0.44]{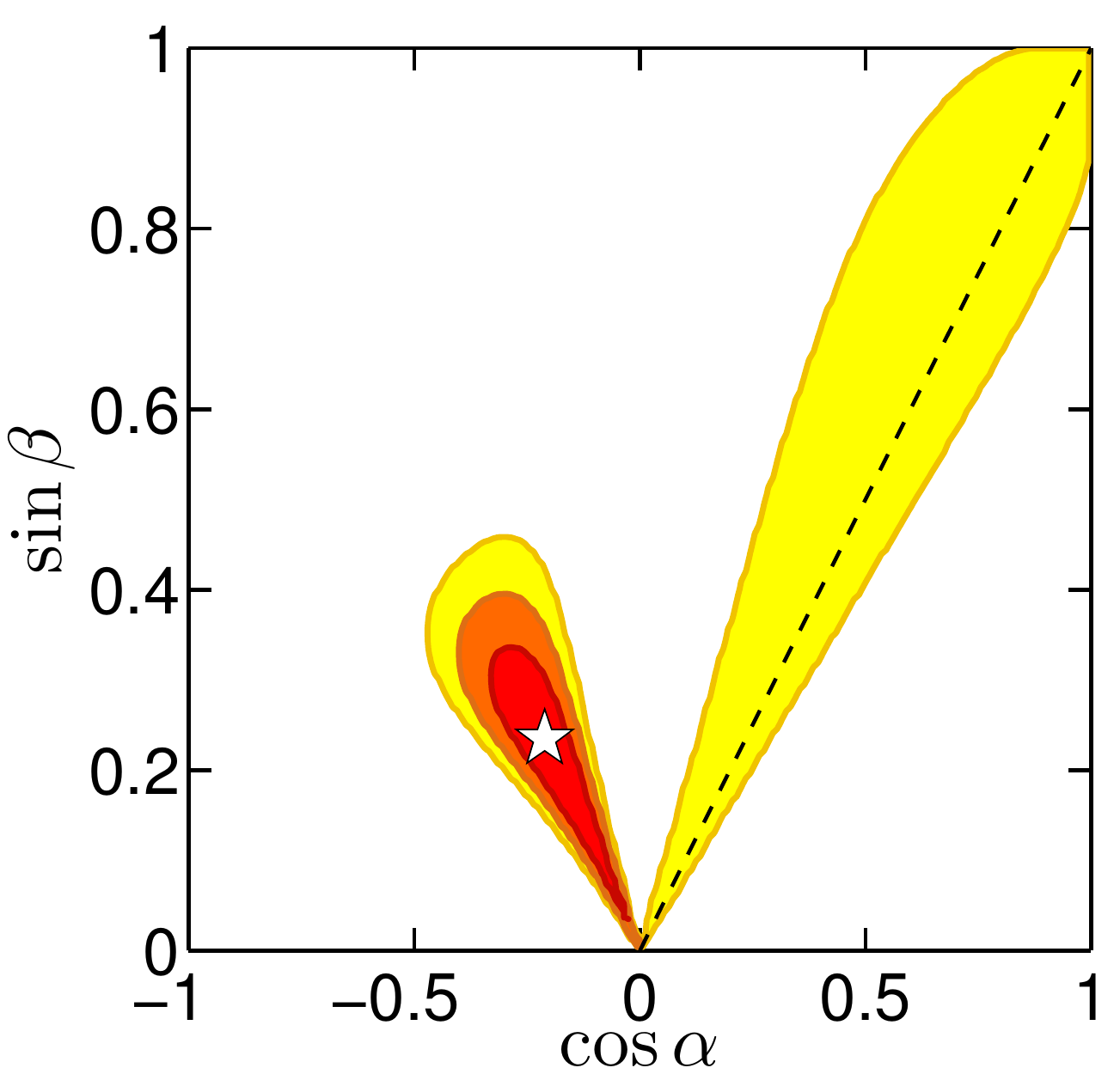}\qquad\includegraphics[scale=0.44]{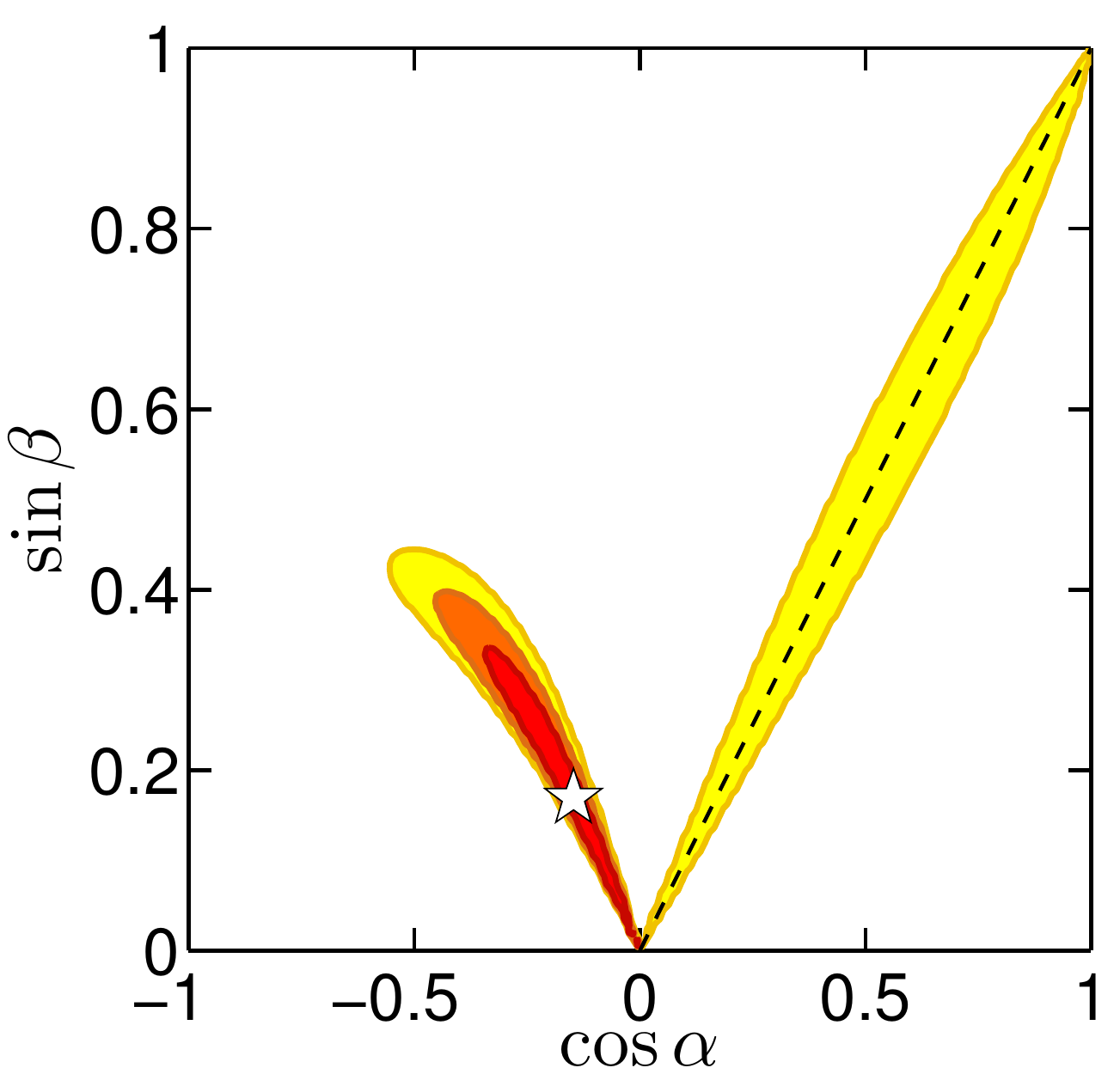}
\caption{2HDM fits for the $h$ in the Type~I (left) and Type~II (right) models.  
The upper row shows the fit results in the $\beta-\pi/2$ vs.\ $\alpha$ plane, while the 
lower row shows the $\sin\beta$ vs.\ $\cos\alpha$ plane. 
The dashed lines indicate the SM limit. 
The same results are obtained for the heavier $H$ with the replacements 
$\beta-\pi/2\to \beta$ and $\alpha\to \alpha+\pi/2$ ($\sin\beta\to-\cos\beta$, $\cos\alpha\to\sin\alpha$).
\label{1212.5244fit-2d-2hdmh} }
\end{figure}

The results of the 2HDM fits are shown in Fig.~\ref{1212.5244fit-2d-2hdmh} for the case that the state near 125~GeV
is the lighter CP-even $h$. The figure also applies for the case of the heavier $H$ being identified with the 
$\sim 125\gev$ state with the replacement rules given in the figure caption.\footnote{Since the $\sim 125\gev$ state clearly couples to $WW,ZZ$ we do not consider the case where the $A$ 
is the only state at $\sim 125\gev$.  We also do not consider the cases where the $\sim125\gev$ peak comprises 
degenerate $(h,H)$, $(h,A)$ or $(H,A)$ pairs.} 
Note that the convention $\cv>0$ implies $\sin(\beta-\alpha)>0$ for the $h$ 
and $\cos(\beta-\alpha)>0$ for the $H$.  
Moreover, the requirement $\tan\beta>0$ restricts $\beta\in [0,\pi/2]$.  
The best fit values and $1\sigma$ ranges for $\alpha$ and $\beta$, together with the corresponding 
values for $\cu$, $\cd$, $\cv$, $\cg$ and $\cg$, are listed in Table~\ref{table:1212.5244THDMfits}. These numbers are  
again for the case of $h$ being the state near 125~GeV. Replacing $h$ by $H$ amounts to a shift in 
$\alpha\to \alpha+\pi/2$; thus we find $\alpha=6.07_{-0.08}^{+0.09}$ 
($\cos\alpha = 0.98 \pm 0.02$) for the 2HDM-I and $\alpha=6.14_{-0.14}^{+0.15}$ 
($\cos\alpha = 0.99_{-0.03}^{+0.01}$) for the 2HDM-II, 
while the values for $\tan\beta$, $\cu$, $\cd$, $\cv$, {\it etc.}\ do not change.

\renewcommand{\arraystretch}{1.3}
\begin{table}[ht]
\centering
\begin{tabular}{|c||c|c||c|c|}
\hline
Fit  & 2HDM-I & 2HDM-II & 2HDM-I, $\tan\beta>1$  & 2HDM-II, $\tan\beta>1$  \\
 \hline 
 $\alpha$ [rad] & $\phantom{-}4.50_{-0.08}^{+0.09}$ & $\phantom{-}4.56_{-0.14}^{+0.15}$ & $5.37_{-0.13}^{+1.11}$ & $6.28_{-0.83}^{+0.17}$ \\		
 $\beta$ [rad]   & $\phantom{-}0.24_{-0.10}^{+0.07}$ & $\phantom{-}0.17_{-0.17}^{+0.12}$ & $[\pi/4,\,\pi/2]$ & $1.56_{-0.78}^{+0.01}$ \\
 \hline
 $\cos\alpha$ & $-0.21_{-0.08}^{+0.09}$ & $-0.15_{-0.13}^{+0.15}$ & $0.61_{-0.11}^{+0.39}$ & $1.00_{-0.67}$ \\		
 $\tan\beta$    & $\phantom{-}0.24_{-0.10}^{+0.08}$ & $\phantom{-}0.17_{-0.17}^{+0.13}$ & $[1,\,+\infty [$ & $[1,\,+\infty [$ \\
 \hline
 $\cu$ & $-0.90_{-0.19}^{+0.17}$ & $-0.87_{-0.13}^{+0.12}$ & $0.87_{-0.15}^{+0.17}$ & $1.02_{-0.07}^{+0.05}$ \\
 $\cd$ & $-0.90_{-0.19}^{+0.17}$ & $\phantom{-}1.00_{-0.01}$ & $0.87_{-0.15}^{+0.17}$ & $0.94_{-0.11}^{+0.13}$ \\
 $\cv$ & $\phantom{-}0.90 \pm 0.07$ & $\phantom{-}0.95_{-0.12}^{+0.05}$ & $0.99_{-0.04}^{+0.01}$ & $1.00_{-0.05}$ \\
 $\cp$ & $\phantom{-}1.37_{-0.10}^{+0.09}$ & $\phantom{-}1.44_{-0.13}^{+0.08}$ & $1.03_{-0.06}$ & $1.01_{-0.09}^{+0.01}$ \\
 $\cg$ & $\phantom{-}0.90_{-0.16}^{+0.19}$ & $\phantom{-}0.92_{-0.11}^{+0.13}$ & $0.87_{-0.15}^{+0.16}$ & $0.99_{-0.04}^{+0.08}$ \\
\hline
 $\chimin$ & 12.20 & 11.95 & 19.43 & 19.88 \\
 \hline
 \end{tabular}
\caption{Summary of fit results for the $h$ in 2HDMs of Type~I and Type~II.}
\label{table:1212.5244THDMfits}
\end{table}
\renewcommand{\arraystretch}{1.0}

Note that for both the Type~I and the Type~II model, the best fits are quite far from the SM limit in parameter space.  
In particular, since we do not include any extra loop contributions to $\cp$, we end up with negative $\cu$ 
close to $-1$ as in Fit~{\bf II}.  
Demanding $\cu>0$ (\ie\ $\cos\alpha>0$ for $h$, $\sin\alpha>0$ for $H$), 
one ends up in a long `valley' along the decoupling limit 
where the Higgs couplings are SM like, see Fig.~\ref{1212.5244fit-2d-2hdmh};  
this is however always more than $2\sigma$ away from the best fit.
Furthermore, solutions with very small $\tan\beta<1$ are preferred at more than $2\sigma$.  
Since such small values of $\tanb$ are rather problematic 
(in particular $\tanb<0.5$ is problematical for maintaining a perturbative magnitude for the 
top-quark Yukawa coupling) 
we also give in Table~\ref{table:1212.5244THDMfits} the corresponding fit results requiring  $\tan\beta>1$. 
These results come quite close to the SM limit, and accordingly 
have a $\chimin$ of about 19--20 (recall that for the SM we find $\chi^2\simeq20.2$). 
2HDMs with $\tan\beta>1$ hence do not provide a better fit than the SM itself. 

A couple of more comments are in order. 
First, an important question that we 
leave for future work 
is whether other --- \eg\ stability, unitarity, perturbativity (SUP) and precision electroweak (PEW) --- 
constraints are obeyed at the best-fit points, or the 68\%~CL regions.  
Here we just note that according to Fig.~1 of \cite{Drozd:2012vf}, the SUP and PEW constraints 
do not seem problematic for Type~II, but may play a role for Type~I models at low $\tanb$.   

Second, the best fits correspond to very small $\tanb$ (small $\beta$) values that are potentially constrained 
by limits from B-physics, in particular from $\Delta M_{Bs}$ and $Z\to b\bar b$ . 
The B-physics constraints are summarized in Figs.~15 and 18 of \cite{Branco:2011iw} 
for Type~II and Type~I, respectively.   
Fig.~18 for Type~I places a lower bound on $\tanb$ as a function of the charged Higgs mass which excludes 
small $\tanb<1$ unless the charged Higgs is {\it very} heavy, something that is possible but somewhat unnatural.  
Fig.~15 for Type~II places a substantial lower bound on the charged Higgs 
mass for all $\tanb$, but such a constraint does not exclude the 68\%~CL region.  

\begin{figure}[ht]\centering
\includegraphics[scale=0.4]{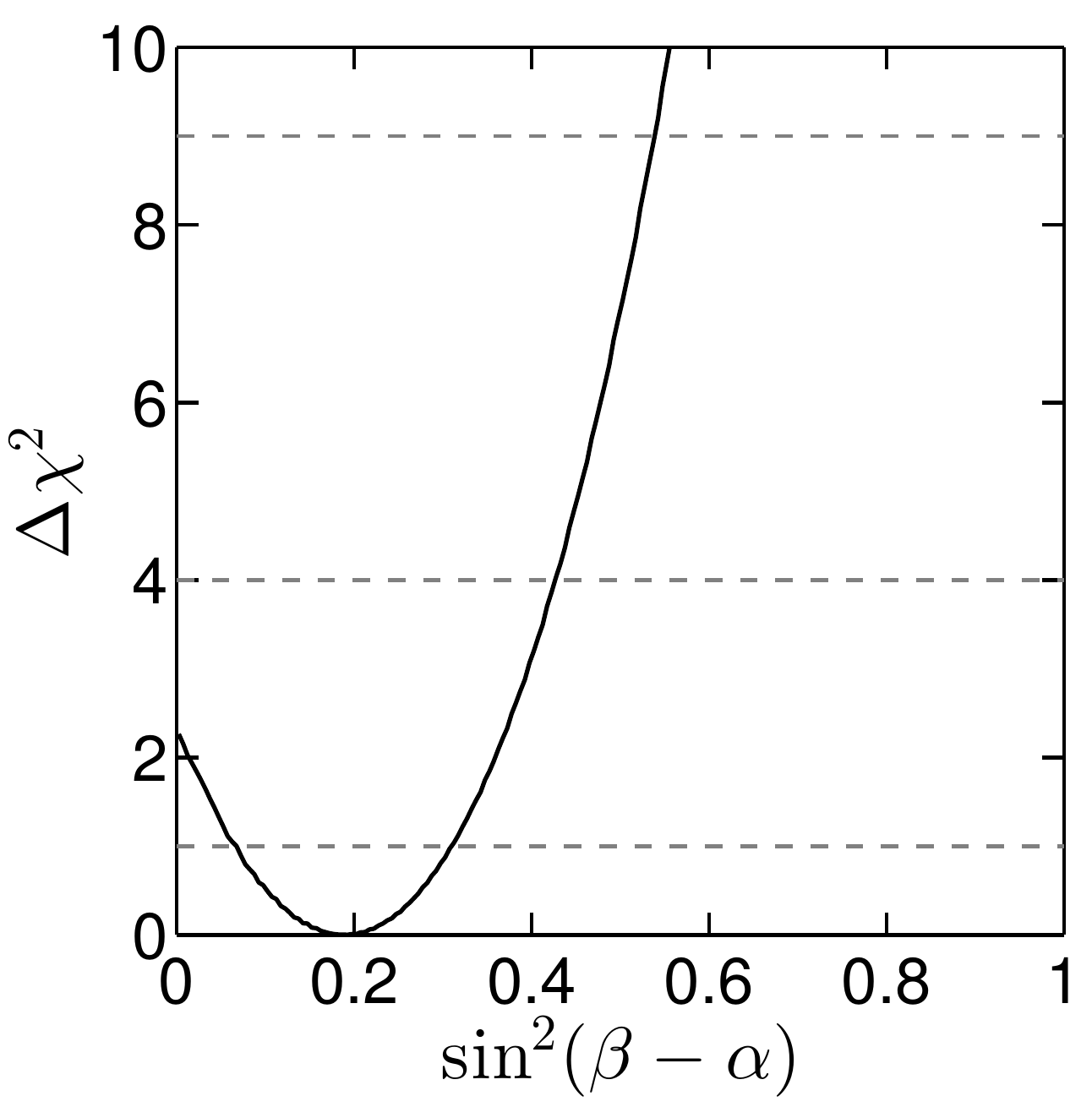}\qquad\includegraphics[scale=0.4]{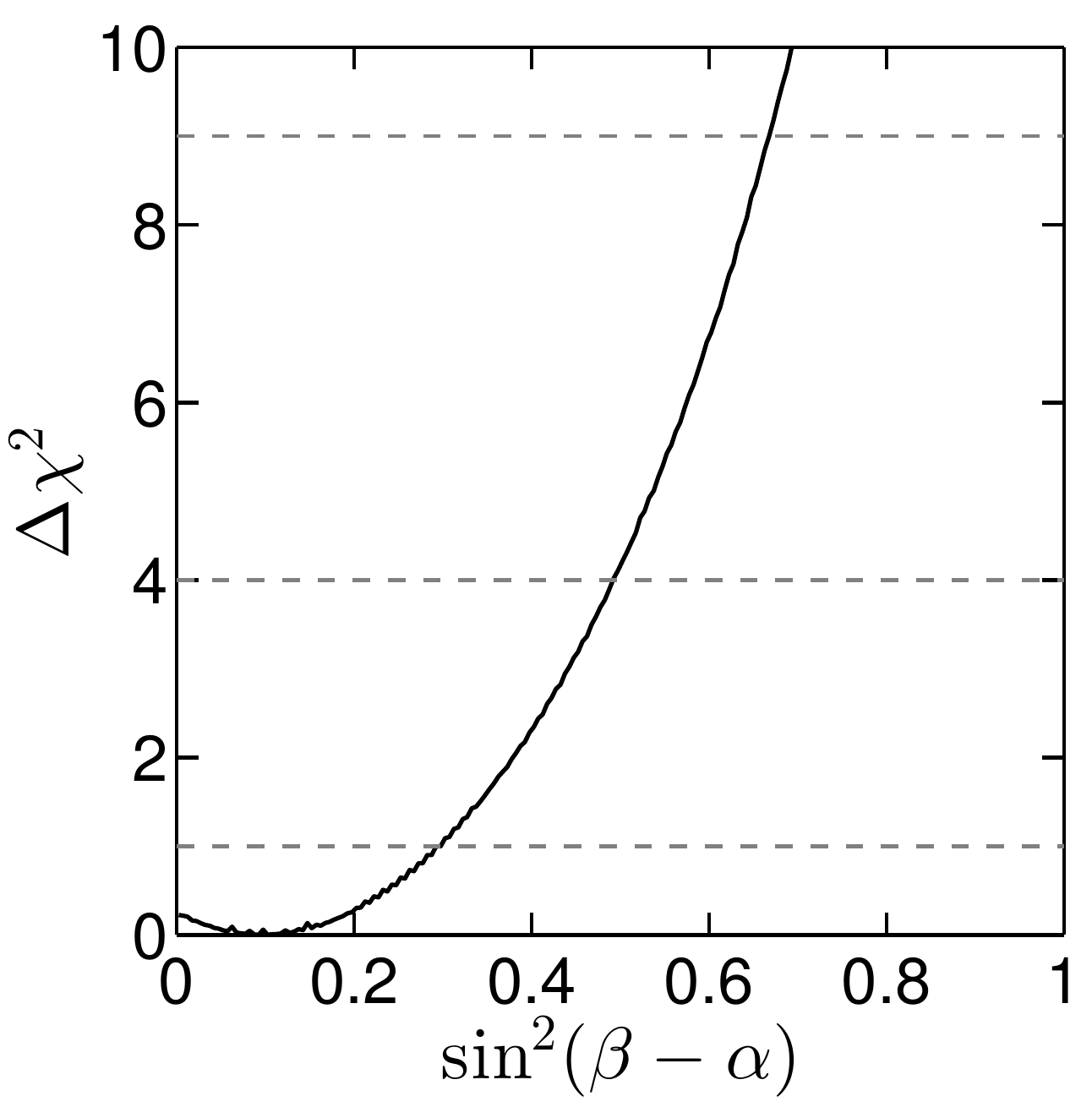}
\caption{$\Delta\chi^2$ distribution of $\sin^2(\beta-\alpha)$ in the Type~I (left) and Type~II (right) models 
for the case that $H$ is the observed state near 125~GeV.   
\label{1212.5244sinbma} }
\end{figure}

Third, we remind the reader that in the 2HDMs, the soft $Z_2$-symmetry-breaking $m_{12}^2$ and 
the other Higgs masses ($m_h$, $m_H$ and $M_A$) are independent parameters.  
It is thus possible to have either $m_h$ or $m_H\sim 125\gev$ without violating constraints from direct 
searches for the charged Higgs whose mass is related to $m_A$. However, in the case of  $m_H\sim 125\gev$, 
one has to avoid the LEP limits for the lighter $h$, which severely constrain the $h$ coupling to $ZZ$ in case   
of $m_h<114\gev$~\cite{TeixeiraDias:2008fc}. So either $m_h\gtrsim 114 \gev$ for $m_H\approx 125\gev$, or 
$\sin^2(\beta-\alpha)$ needs to be small (\eg\ $\sin^2(\beta-\alpha)\lesssim 0.3$ for $m_h\approx 100\gev$, or 
$\sin^2(\beta-\alpha)\lesssim 0.1$ for $m_h<90\gev$). 
The $\Delta\chi^2$ distributions of $\sin^2(\beta-\alpha)$ for Type~I and Type~II with $m_H\sim 125\gev$ 
are shown in Fig.~\ref{1212.5244sinbma}. Interestingly,  around the best fit the $h$ coupling to $ZZ$ is sufficiently suppressed 
to allow for $m_h$ of the order of $100\gev$ (or lower in Type~II).

\subsection{Conclusions} \label{sec:2012conclusions}

We assessed to which extend the available data on the Higgs-like scalar at the end of 2012 constrain the Higgs couplings. To this end we performed fits to all public data from the LHC and the Tevatron experiments. 

First, we employed a general 
parametrization of the Higgs couplings based on an SM-like Lagrangian, but allowing for extra contributions to the loop-induced couplings of the Higgs-like scalar to gluons and photons. While the SM does not provide a bad fit ($\chi^2/\dof=0.96$), 
it is more than $2\sigma$ away from our best fit solutions. The main pull comes from the enhanced $H\to\gamma\gamma$ 
rates observed by ATLAS and CMS, as well as from the Tevatron experiments. 
The best fits are thus obtained when either $\cu\sim -1$ (\ie\ opposite in sign to the SM expectation) or there is a large BSM contribution to the $\gam\gam$ coupling of the Higgs. In short, significant deviations from the SM values are preferred by the currently available data and should certainly be considered viable. Since having $\cu\sim -1$ (in the $\cv>0$ convention)  is not easy to achieve in a realistic model context, and leads to unitarity violation in $WW\rightarrow t\bar{t}$ scattering at scales that can be as low as 5~TeV~\cite{Choudhury:2012tk,Bhattacharyya:2012tj},
it would seem that new physics contributions to the effective couplings of the Higgs to gluons and photons are the preferred option.  (The possibility of a second, degenerate Higgs boson contributing to the observed signal remains another interesting option, not considered here.) 

Second, we examined how well 2HDM models of Type~I and Type~II fit the data. 
We found that it is possible to obtain a good fit in these models with $\sin(\beta-\alpha)$ ($\cos(\beta-\alpha)$), in the $h$ ($H$) cases, respectively, not far from $1$.  However, the best fit values for the individual $\cu$, $\cd$, $\cp$ and $\cg$ parameters lie far from their SM values. Further, the best fits give $\tan\beta<1$, which is disfavored from the theoretical point of view if we want perturbativity up to the GUT scale. 
Requiring $\tan\beta>1$ (or simply $\cu>0$) pushes the fit into the SM
`valley' and no improvement over the pure SM solution is obtained. In
particular the $\chi^2$ obtained in this region is substantially
larger than that for the best fit, and not far from the $\chi^2$ found
for the SM.

\begin{figure}[ht]\centering
\includegraphics[scale=0.56]{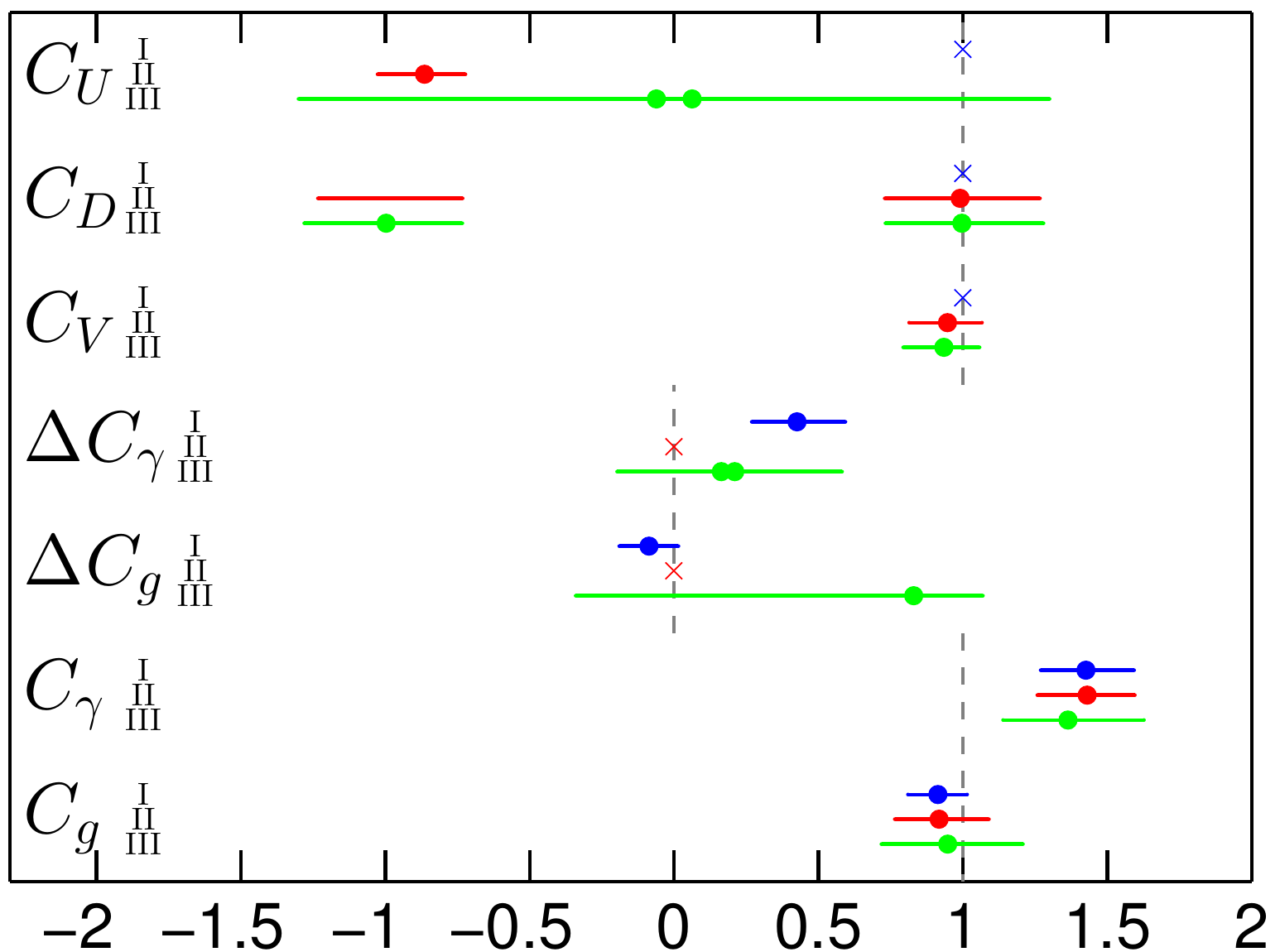}
\caption{Graphical representation of the best fit values for $\cu$, $\cd$, $\cv$, $\dcp$ and $\dcg$ of Table~\ref{1212.5244chisqmintable}.   The labels refer to the fits discussed in the text.  The dashed lines indicate the SM value for the given quantity. 
The $\times$'s indicate cases where the parameter in question was fixed to its SM value. 
}
\label{1212.5244bestplot}
\end{figure}

We once again refer the reader to Tables~\ref{1212.5244chisqmintable}, \ref{tab:1212.5244fit2} and \ref{table:1212.5244THDMfits} which 
summarize the best fit values and $1\sigma$ errors for the parameters for the various cases considered. 
In Fig.~\ref{1212.5244bestplot}  we show some of these results graphically.  Moreover, in order to assess the physics associated 
with our best fit points, we give in Tables~\ref{1212.5244chisqminmutable}
the values of the derived (theory level) signal strengths $\mu(\ggf,\gam\gam)$, $\mu(\ggf,ZZ)$, $\mu(\ggf,b\anti b)$, $\mu(\vbf,\gam\gam)$, $\mu(\vbf,ZZ)$, and $\mu(\vbf,b\anti b)$  for the best fit point in the various coupling fits we have considered.  (These are a complete set since for the models we consider $\mu(X,\tau\tau)=\mu(X,b\anti b)$, $\mu (X,WW)=\mu(X,ZZ) $ and $\mu(\vbf,Y)=\mu(\vh,Y)$.) 
We see that in the general case both $\mu(\ggf,\gam\gam)$ and $\mu(\vbf,\gam\gam)$ are enhanced by factors 1.7--2.1, while the other signal strengths tend to be $\lesssim 1$. When demanding $\cu>0$ without allowing for extra contributions from new particles, then only very small enhancements of 
$\mu(\vbf,\gam\gam)$ and $\mu(\vbf,ZZ)$ of the order of 1.2--1.3 are found.

Using the same framework, our group also investigated the extent to which 2012 data constrain invisible (or undetected) decays of the Higgs boson. This results were presented in the letter ``Status of invisible Higgs decays'', Ref.~\cite{Belanger:2013kya} that was submitted to arXiv on February 22, 2013 and published in PLB in June 2013. We found in particular the limit ${\rm BR}(H \to {\rm invisible}) < 0.23$ at 95\%~CL for SM-like Higgs couplings, but up to 60\% invisible/undetected decays of the Higgs were allowed at 95\%~CL in Fit {\bf I}. These results have been obtained without taking account any direct search for $H \to {\rm invisible}$, for which no result was available in 2012.

%
%

\renewcommand{\arraystretch}{1.3}
\begin{table}[ht]
\centering
\begin{tabular}{|l|c|c|c|c|}
\hline
Fit  & {\bf I} & {\bf II}, $\cu<0$ & {\bf II}, $\cu>0$ & {\bf III} \\
\hline 
 $\mu(\ggf,\gam\gam)$ & $1.71_{-0.32}^{+0.33}$ & $1.81_{-0.41}^{+0.43}$ & $1.07 \pm 0.18$ & $1.79_{-0.34}^{+0.36}$ \\		
 $\mu(\ggf,ZZ)$ & $0.84_{-0.17}^{+0.18}$ & $0.79 \pm 0.15$ & $0.97 \pm 0.20$ & $0.84_{-0.18}^{+0.21}$ \\
 $\mu(\ggf,b\anti b)$ & $0.84_{-0.17}^{+0.18}$ & $0.87_{-0.40}^{+0.57}$ & $0.63_{-0.26}^{+0.36}$ & $0.96_{-0.43}^{+0.59}$ \\
 $\mu(\vbf,\gam\gam)$ & $2.05_{-0.44}^{+0.54}$ & $1.92_{-0.68}^{+0.78}$ & $1.66_{-0.63}^{+0.70}$ & $1.74_{-0.73}^{+0.84}$ \\
 $\mu(\vbf,ZZ)$ & $1.00 \pm 0.02$ & $0.84_{-0.36}^{+0.42}$ & $1.50_{-0.46}^{+0.50}$ & $0.82_{-0.35}^{+0.38}$ \\
 $\mu(\vbf,b\anti b)$ & $1.00 \pm 0.02$ & $0.92 \pm 0.30$ & $0.98 \pm 0.32$ & $0.93_{-0.29}^{+0.25}$ \\
 \hline
 \end{tabular}
\caption{Summary of $\mu$ results for Fits {\bf I}--{\bf III}.  
For Fit {\bf II}, the tabulated results are for the best fit with $\cu<0$, column 1 of Table~\ref{tab:1212.5244fit2}, and for the 
case $\cu,\cd>0$, column 3 of Table~\ref{tab:1212.5244fit2}. }
\label{1212.5244chisqminmutable}
\end{table}

\clearpage

\section{The status of Higgs couplings after Moriond 2013} \label{sec:higgs2013}

\begin{figure}[ht]
	\centering
		\includegraphics[scale=0.4]{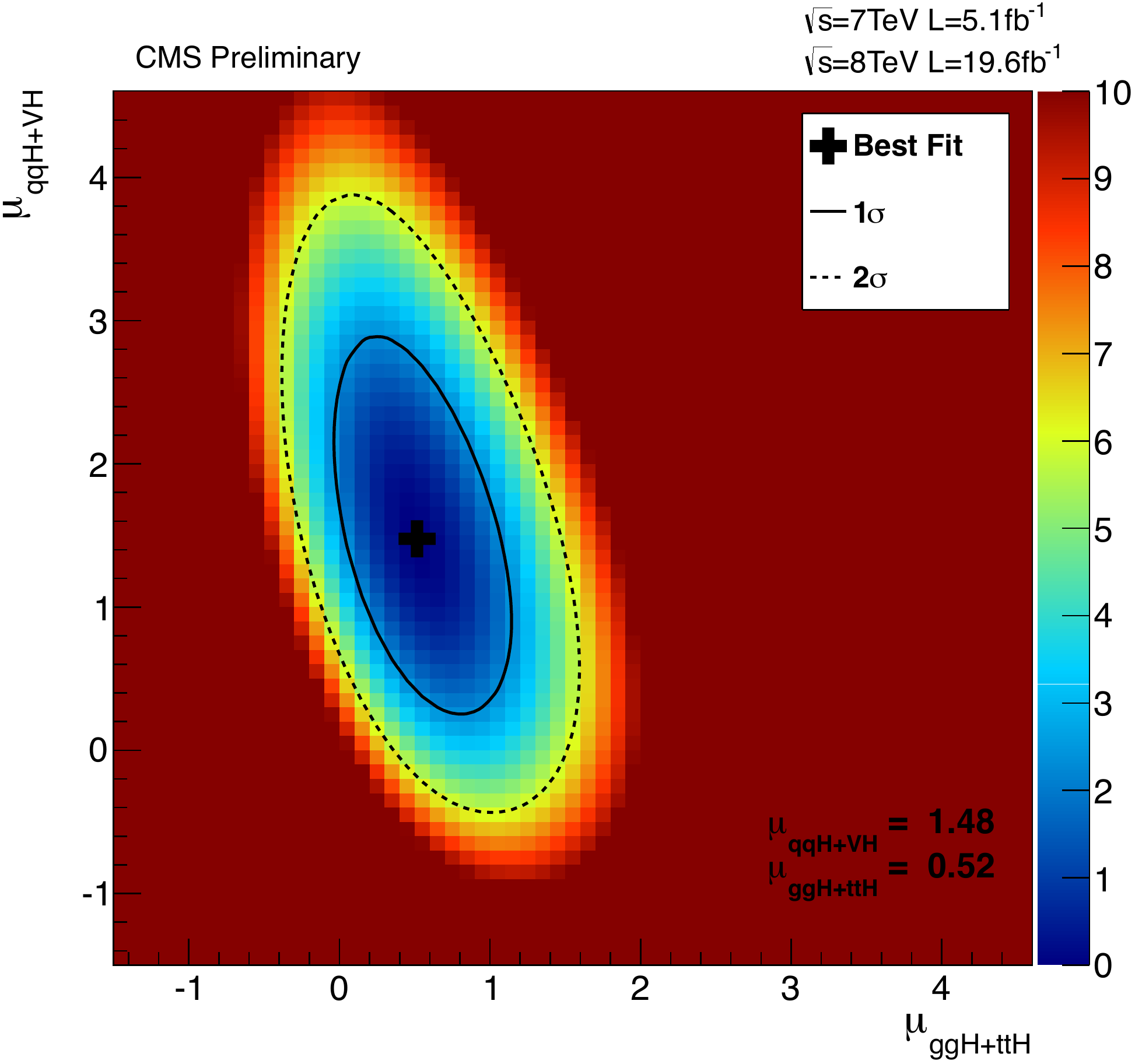}
	\caption{Preliminary CMS results for $H \to \gamma\gamma$~\cite{CMS-PAS-HIG-13-001} (MVA), using the full statistics collected during Run~I of the LHC and given in the plane $(\mu({\rm ggF+ttH}, \gamma\gamma), \mu({\rm VBF+VH}, \gamma\gamma))$. The color code gives the value of $- 2 \log L$. Contours of constant likelihood are shown, with the solid line corresponding to 68\%~CL and the dashed line to 95\%~CL.
\label{fig:2013CMSgamgam}}
\end{figure}

At the 48th Rencontres de Moriond in March 2013~\cite{Moriond2013EW,Moriond2013QCD}, preliminary results using the full statistics collected at $\sqrt{s} = 7$ and 8~TeV were given for various channels. This includes the search for $H \to \gamma\gamma$ in CMS~\cite{CMS-PAS-HIG-13-001}, for which in addition to the main analysis, using MVA techniques, a cut-based version of the analysis was given. The result of the MVA analysis in the $(\mu({\rm ggF+ttH}, \gamma\gamma), \mu({\rm VBF+VH}, \gamma\gamma))$ plane is shown in Fig.~\ref{fig:2013CMSgamgam}. The best fit point in this 2D plane is at $(0.52, 1.48)$, which is in strong contrast with the 2012 results shown in Fig.~\ref{fig:2012gamgamstatus}, where the best fit point was located at $(0.95, 3.77)$. The new physics interpretations using 2012 data, presented in the previous section, were mostly driven by the excess in $H \to \gamma\gamma$. This new CMS result is thus expected to have a large impact on the determination of the couplings of the Higgs boson and on the favored regions in new physics scenarios.

After Moriond 2013, the same team involved in Ref.~\cite{Belanger:2012gc,Belanger:2013kya} (Genevi\`eve B\'elanger, Ulrich Ellwanger, John F.~Gunion, Sabine Kraml and myself) worked on updating the fits previously presented with the latest results using the full luminosity at 8~TeV. We went beyond our previous works in several ways. First, we found it interesting to show results in terms of combined signal strengths in the $(\mu({\rm ggF+ttH}, Y), \mu({\rm VBF+VH}, Y))$ plane, for $Y = \gamma\gamma, VV, b\bar b$, and $\tau\tau$. From these results we obtained a simple $\chi^2$ formula that can easily be used to constrain a large class of new physics models.
Second, we studied the implications for various new physics scenarios with extended Higgs sectors: in addition to the 2HDM of Type~I and Type~II, we studied the Inert Doublet model (IDM), and the Georgi--Machacek triplet Higgs model.
The paper ``Global fit to Higgs signal strengths and couplings and implications for extended Higgs sectors'', Ref.~\cite{Belanger:2013xza}, was submitted to arXiv on June 12, 2013 and published in PRD in October 2013. In addition, I made a contribution to the proceedings of the 2013 Moriond conference based on this work~\cite{Dumont:2013mba}. 

The rest of this section is largely based on~\cite{Belanger:2013xza}. Section~\ref{2013c-sec:combinedss} explains our procedure for deriving the combined signal strengths in the $(\mu({\rm ggF+ttH}, Y), \mu({\rm VBF+VH}, Y))$ plane and give the updated experimental results we use compared to the end of 2012. We parametrize the signal strengths $\mu_i$ in terms of
various sets of Higgs couplings and show results in Section~\ref{2013c-sec:coupfit}. The implications for dark matter of the limits on invisible decays of the Higgs (originally presented in Ref.~\cite{Belanger:2013kya}) will be discussed in Section~\ref{2013c-sec:DMinterplay}. The impact of Higgs searches will then be discussed in the context of the 2HDM in Section~\ref{2013c-sec:2HDM} and of the IDM in Section~\ref{2013c-sec:IDM} (constraints on triplet Higgs models, presented in~\cite{Belanger:2013xza}, will not be reproduced here). Finally, our conclusions are given in Section~\ref{2013c-sec:conclusions}.

\subsection{Methodology and combined signal strengths ellipses} \label{2013c-sec:combinedss}

Our first purpose is to combine the information
provided by ATLAS, CMS and the Tevatron experiments on the 
$\gam\gam$, $ZZ^{(*)}$, $WW^{(*)}$, $b\bar{b}$ and $\tau\tau$  final states 
including the error correlations
among the (VBF+VH) and (ggF+ttH) production modes.
Using a Gaussian approximation, we derive 
for each final state a combined likelihood in the  
$\mu({\rm ggF + ttH})$ versus $\mu({\rm VBF + VH})$ plane, 
which can then simply be expressed as a $\chi^2$. From the general expression of the likelihood given in Eq.~\eqref{eq:ourbestlike}, we obtain
\begin{align}
\chi^2_i &= (\boldsymbol{\mu}_i - \hat{\boldsymbol{\mu}}_i)^T
\begin{pmatrix} \sigma_{{\rm ggF},i}^2 & \rho_i \sigma_{{\rm ggF},i} \sigma_{{\rm VBF},i} \\ \rho_i \sigma_{{\rm ggF},i}\sigma_{{\rm VBF},i} & \sigma_{{\rm VBF},i}^2 \end{pmatrix}^{-1}
(\boldsymbol{\mu}_i - \hat{\boldsymbol{\mu}}_i) \label{eq:bestlikegaussian} \\
&= (\boldsymbol{\mu}_i - \hat{\boldsymbol{\mu}}_i)^T
\begin{pmatrix} a_i & b_i \\ b_i & c_i \end{pmatrix}
(\boldsymbol{\mu}_i - \hat{\boldsymbol{\mu}}_i) \nonumber \\
& = a_i(\mu_{{\rm ggF},i}-\hat{\mu}_{{\rm ggF},i})^2
+2b_i(\mu_{{\rm ggF},i}-\hat{\mu}_{{\rm ggF},i})
(\mu_{{\rm VBF},i}-\hat{\mu}_{{\rm VBF},i})
+c_i(\mu_{{\rm VBF},i}-\hat{\mu}_{{\rm VBF},i})^2 \,, \nonumber 
\end{align}
where the indices ggF and VBF stand for (ggF+ttH) and (VBF+VH), respectively, 
and the index $i$ stands for $\gam\gam$, $VV^{(*)}$, $b\bar{b}$ and $\tau\tau$ (or $b\bar{b}=\tau\tau$),  
and $\hat{\mu}_{{\rm ggF},i}$ and $\hat{\mu}_{{\rm VBF},i}$ denote the
best-fit points obtained from the measurements.
We thus obtain ``combined likelihood ellipses'', which can be used in a simple, generic way to 
constrain non-standard Higgs sectors and new contributions to the loop-induced processes, provided they 
have the same Lagrangian structure as the SM.
In particular, these likelihoods can be used to derive constraints on a
model-dependent choice of generalized Higgs couplings, the implications of which we study
subsequently for several well-motivated models. The choice of models is
far from exhaustive, but we present our results for the likelihoods as a
function of the independent signal strengths $\mu_i$ in such a manner that these
can easily be applied to other models.

\begin{figure}[ht]\centering
\includegraphics[width=6cm]{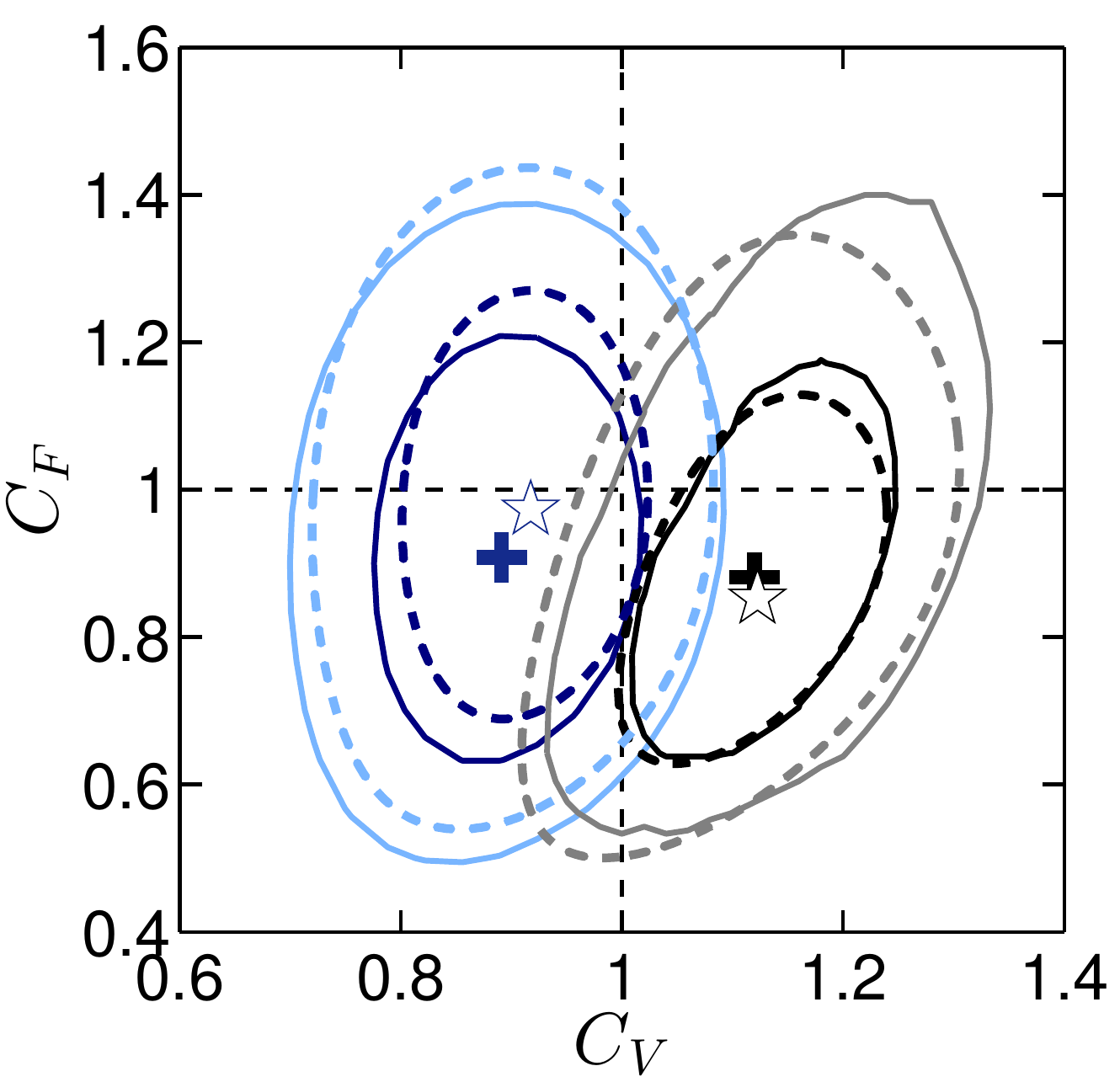}\quad
\includegraphics[width=5.85cm]{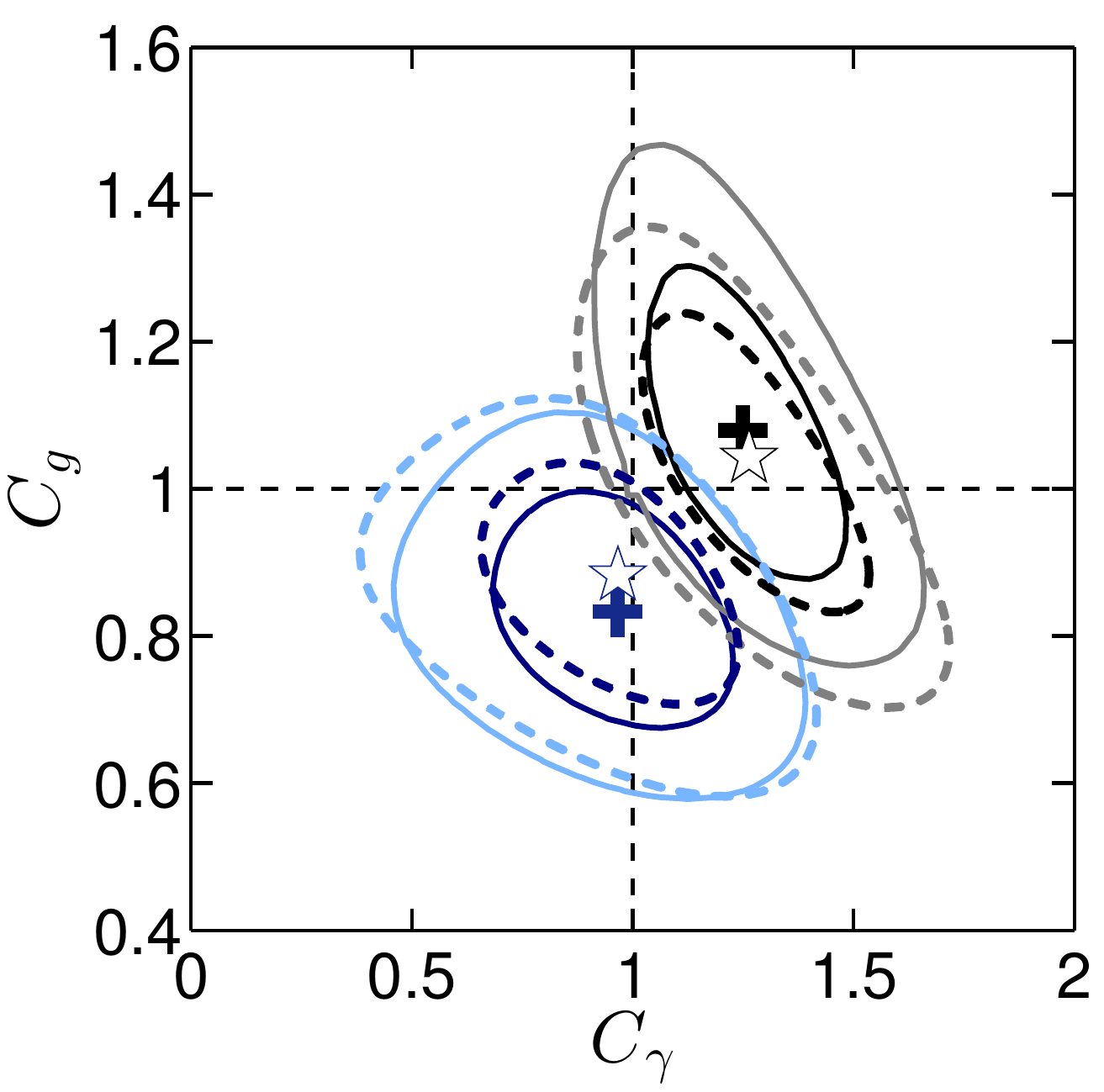}
\caption{Fit to the couplings $(C_F,C_V)$ (left) and $(C_g,C_\gamma)$
(right) using separately results from ATLAS and CMS up to the Moriond
2013 conference. The black and gray (dark blue and light blue) contours
show the 68\% and 95\%~CL regions for ATLAS (CMS), respectively. The
solid contours correspond to the results published by the experimental
collaborations, while dashed contours have been obtained using the
fitted signal strength ellipses as determined using the separate data
for ATLAS (CMS) in the manner described in
this section.\label{2013c-fig:lhc-check} }
\end{figure}

As was mentioned in Section~\ref{sec:higgs-npconstlhc}, in this approach we do not account for correlations between different final states (but identical production modes) which originate from common theoretical errors on the production cross 
sections~\cite{Giardino:2013bma,Bechtle:2013xfa}, nor for correlations between systematic errors due to common detector components (like EM calorimeters) sensitive to different final states (such as
$\gam\gam$ and $e^-$ from $ZZ^{(*)}$ and $WW^{(*)}$). A discussion on the precise treatment of these `2nd order' corrections to our likelihood will be made in Section~\ref{sec:higgsfuture}.
It is however possible to estimate their importance,
\eg, by reproducing the results of coupling fits performed by ATLAS
and CMS using all available results up to the Moriond 2013 conference~\cite{ATLAS-CONF-2013-034,CMS-PAS-HIG-13-005}.
For the aim of comparison, 
we have performed fits to the $(C_F,\,C_V)$ and $(C_g,\,C_\gamma)$ couplings, using separately only ATLAS or CMS data up to the Moriond 2013 conference. Fig.~\ref{2013c-fig:lhc-check} compares our results to those published by ATLAS~\cite{ATLAS-CONF-2013-034} and CMS~\cite{CMS-PAS-HIG-13-005}. 
We obtain good agreement  in all four cases. The ATLAS (CMS) best fit points are at distances of $\sqrt{(\Delta C_V)^2 + (\Delta C_F)^2} = 0.03$ (0.07) and $\sqrt{(\Delta C_\gamma)^2 + (\Delta C_g)^2} = 0.04$ (0.05) from the reconstructed best fit points, and good coverage of the 68\% and 95\%~CL regions is observed.

Once the expressions for the various $\chi_i^2$ are given in the form of
Eq.~(\ref{eq:bestlikegaussian}), it becomes straightforward to evaluate the numerical
value of $\chi^2=\sum_i \chi_i^2$ in any theoretical model with SM-like Lagrangian structure, 
in which predictions for the Higgs branching fractions and the (VBF+VH) and (ggF+ttH) production 
modes (relative to the SM) can be made.
From the corresponding information provided by the experimental collaborations one
finds that the Gaussian approximation is justified in the neighborhood (68\%~CL
contours) of the best fit points. Hence we parametrize these 68\%~CL contours, 
separately for each experiment, as in Eq.~(\ref{eq:bestlikegaussian}), using the procedure explained in Section~\ref{sec:higgs-npconstlhc} around Eq.~\eqref{eq:mu2d}. Occasionally, only a single
signal rate including error bars for a specific final state is given. Using the relative contributions from the various production modes, this kind of information can still be represented in the form of Eq.~(\ref{eq:bestlikegaussian}), leading to an ``ellipse'' which reduces to a strip in the plane of the (VBF+VH) and (ggF+ttH) production modes.
Subsequently these expressions can easily be combined and be represented again in the form of Eq.~(\ref{eq:bestlikegaussian}). 
We expect that the result is reliable up to $\chi_i^2 \lsim 6$ (making it possible to derive
95\%~CL contours), but its extrapolation to (much) larger values of
$\chi_i^2$ should be handled with care.

Starting with the $H\to \gamma\gamma$ final state, we treat in this way
the 68\%~CL contours given by ATLAS in
\cite{ATLAS-CONF-2013-012,ATLAS-CONF-2013-014, ATLAS-CONF-2013-034}, by
CMS in \cite{CMS-PAS-HIG-13-001,CMS-PAS-HIG-13-005,
CMS-PAS-HIG-13-015}\footnote{Note that we are using the MVA analysis for CMS $H \to \gamma\gamma$. The cut-based analysis (CiC) also presented by CMS~\cite{CMS-PAS-HIG-13-001}---that leads to higher but compatible signal strengths---is unfortunately not available in the form of contours in the plane of the (VBF+VH) and (ggF+ttH) production modes. Moreover, no information is given on the sub-channel decomposition, so in fact the CMS CiC analysis cannot be used for our purpose.}
and the Tevatron in \cite{Aaltonen:2013kxa}. 
(In the case of the Tevatron, for all final states only a strip in the plane of the
(VBF+VH) and (ggF+ttH) production modes is defined.) 
For the combination of the $ZZ$ and $WW$ final states, we use the 68\%~CL 
contours given by ATLAS for $ZZ$ in \cite{ATLAS-CONF-2013-013,ATLAS-CONF-2013-014,ATLAS-CONF-2013-034}, by CMS for $ZZ$ in \cite{CMS-PAS-HIG-13-002,CMS-PAS-HIG-13-005},
by ATLAS for $WW$ in \cite{ATLAS-CONF-2013-030,ATLAS-CONF-2013-034}, by CMS for $WW$
in \cite{CMS-PAS-HIG-13-005,CMS-PAS-HIG-13-003,CMS-PAS-HIG-13-009} and by the Tevatron for $WW$ in
\cite{Aaltonen:2013kxa}. 
For the combination of the $b\bar{b}$ and $\tau\tau$ final states, we use the
``strip'' defined by the ATLAS result for $b\bar{b}$ in associated VH
production from \cite{ATLAS-CONF-2012-161}, the 68\%~CL contour
given by CMS for $b\bar{b}$ in \cite{CMS-PAS-HIG-13-012}, the Tevatron result
for $b\bar{b}$ from \cite{Aaltonen:2013kxa} and combine
them with the ATLAS 68\%~CL contour for $\tau\tau$ from
\cite{ATLAS-CONF-2012-160,ATLAS-CONF-2013-034} and the CMS 68\%~CL contours for
$\tau\tau$ from \cite{CMS-PAS-HIG-13-005,CMS-PAS-HIG-13-004}.
We also use the ATLAS search for $ZH\to \ell^+\ell^-\!+{\rm invisible}$, extracting the likelihood from Fig.~10b of \cite{ATLAS-CONF-2013-011}. 
All the above 68\%~CL likelihood contours are parametrized by
ellipses (or strips) in
$\chi^2$ as in Eq.~(\ref{eq:bestlikegaussian}), which can subsequently be
combined. (The analytical expressions are given in Appendix~A of Ref.~\cite{Belanger:2013xza}.)
While ATLAS searches for $H \to b\bar b$ and $H \to \tau\tau$ still correspond to only 13 fb$^{-1}$ at $\sqrt{s} = 8$~TeV, almost all of other results correspond to the full luminosity collected during Run~I of the LHC. Therefore, and while notable changes can be seen between the preliminary and the published results, the results that will be shown in the rest of the Section are expected to remain largely valid until the first results from Run~II of the LHC.

\begin{figure}[ht]\centering
\includegraphics[scale=0.4]{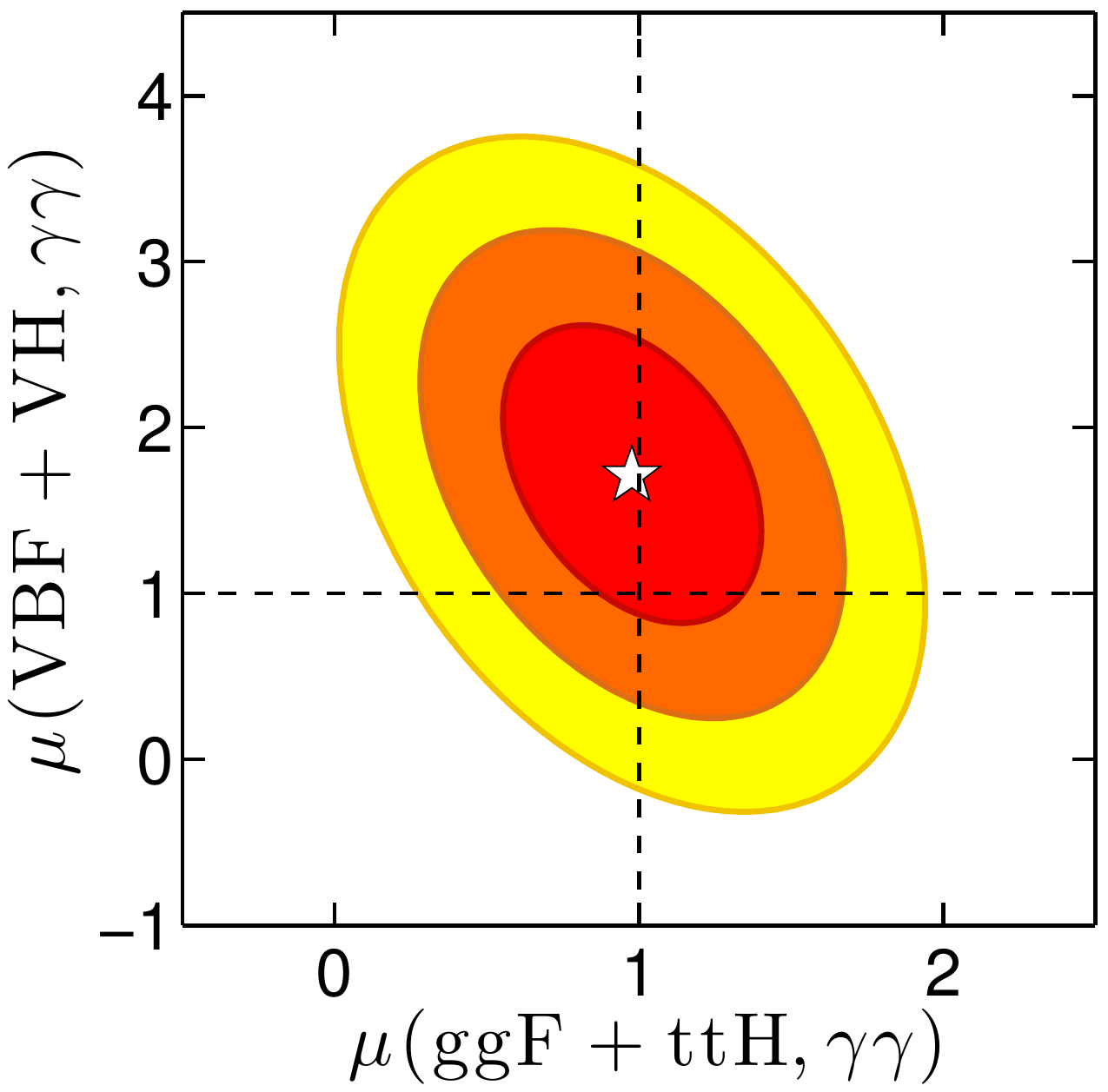}
\includegraphics[scale=0.4]{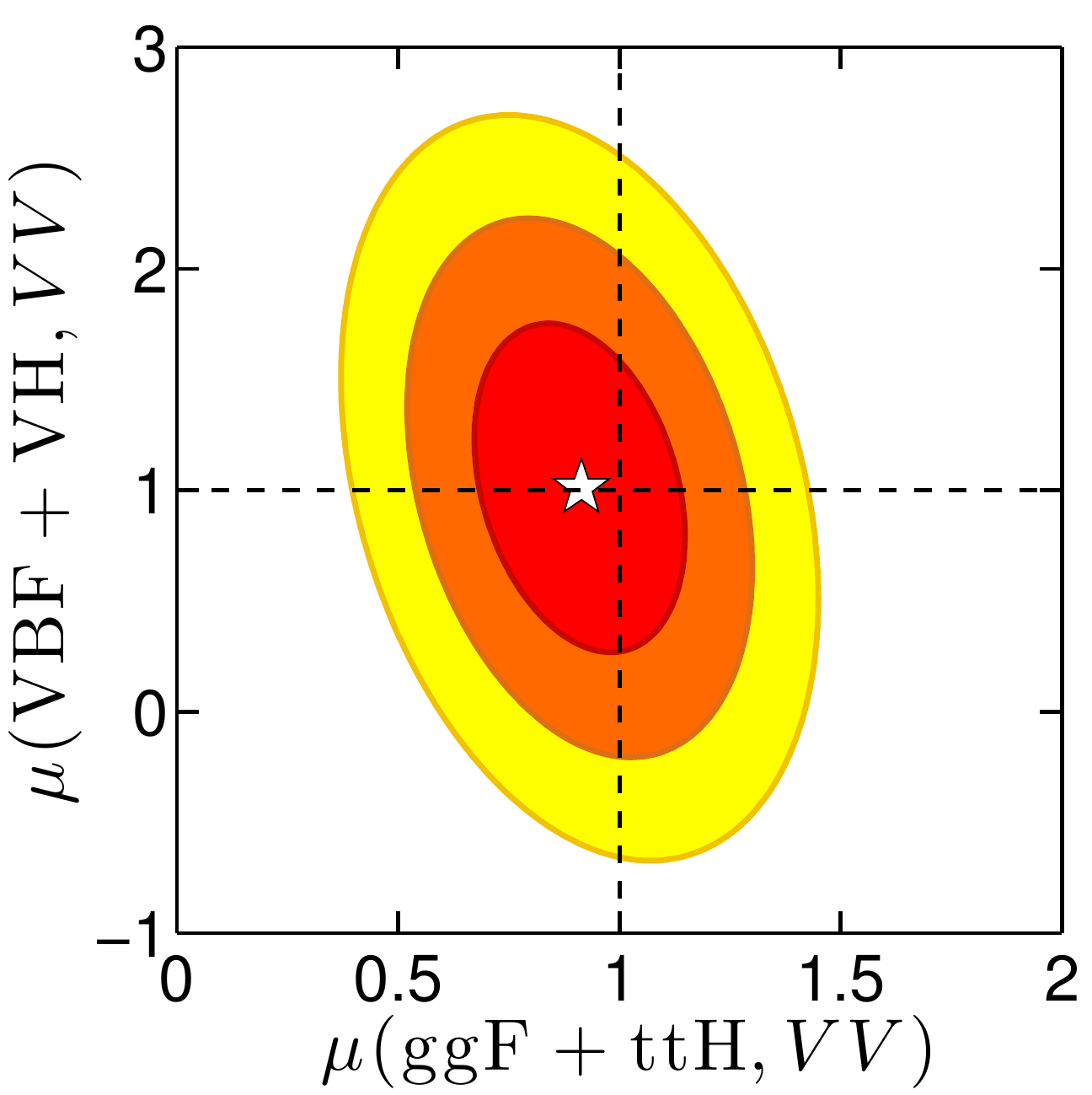}
\includegraphics[scale=0.4]{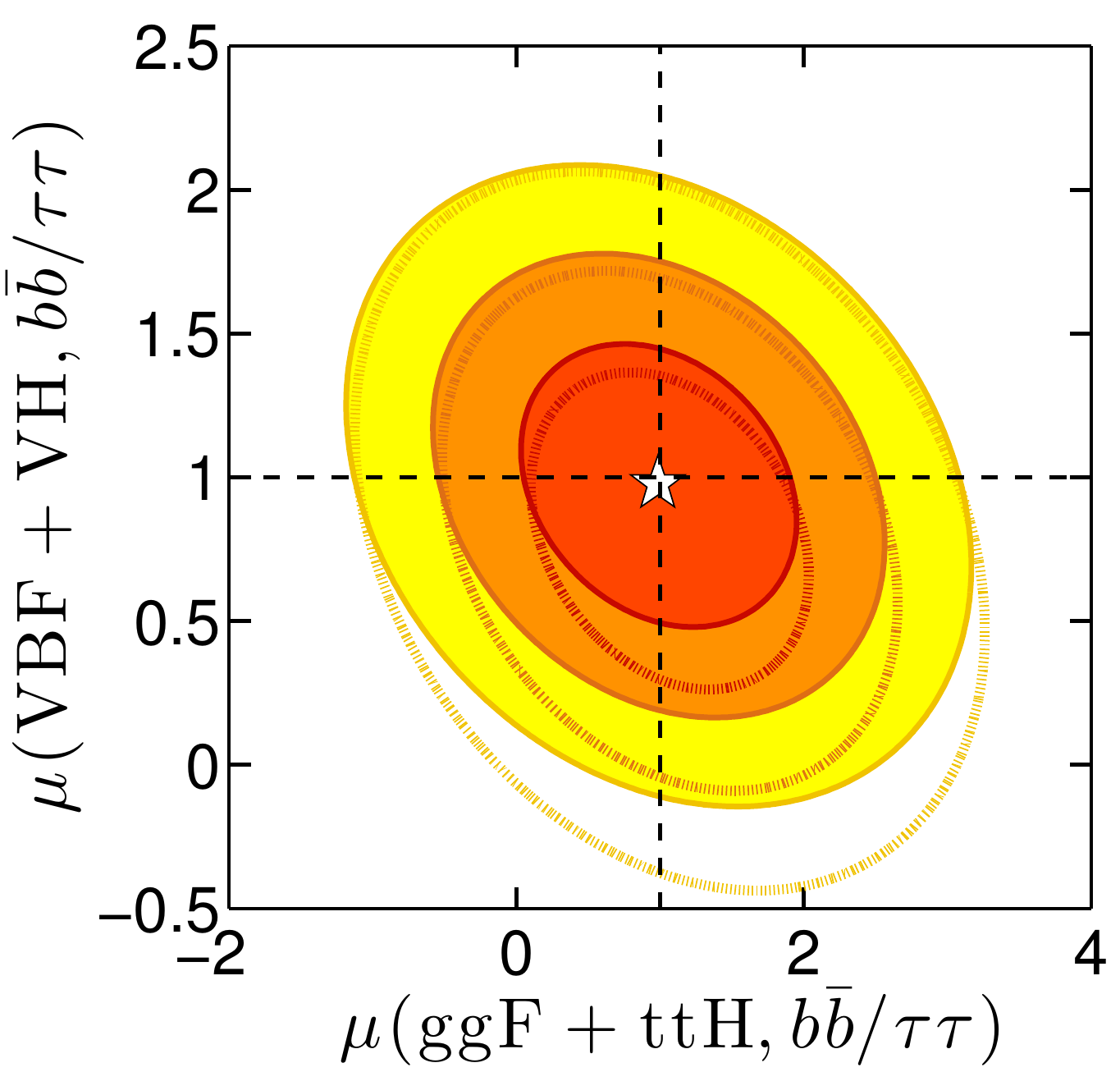}
\caption{Combined signal strength ellipses for the $\gamma\gamma$, $VV=ZZ,WW$ and $b\bar b=\tau\tau$ channels. 
The filled red, orange and yellow ellipses show the 68\%, 95\% and 99.7\%  CL regions, respectively, derived by combining the ATLAS, CMS and Tevatron results. The red, orange and yellow line contours in the right-most plot show how these ellipses change  when neglecting the Tevatron results. 
The white stars mark the best-fit points.
\label{2013c-fig:ellipses1} }
\end{figure}

The resulting parameters $\hat{\mu}_{\rm{ggF}}$, $\hat{\mu}_{\rm{VBF}}$, $a$, $b$ and
$c$ for Eq.~(\ref{eq:bestlikegaussian}) (and, for completeness, the correlation coefficient $\rho$) for the different 
final states are listed in Table~\ref{2013c-tab:1}. The corresponding 68\%, 95\% and 99.7\%  CL ellipses are represented graphically in Fig.~\ref{2013c-fig:ellipses1}.
We see that, after combining different experiments, the best fit signal
strengths are astonishingly close to their SM values, the only 
exception being the $\gamma\gamma$ final state produced via (VBF+VH) for which
 the SM is, nonetheless, still within the 68\%~CL contour. Therefore, these
results serve mainly to constrain BSM contributions to the properties of the Higgs boson.

\begin{table}[ht]
\center
\renewcommand{\arraystretch}{1.1}
\begin{tabular}{|c|c|c|c||c|c|c|}
\hline
& $\hat{\mu}_{\rm{ggF}}$ & $\hat{\mu}_{\rm{VBF}}$ & $\rho$ & $a$ & $b$ & $c$ \\
\hline 
$\gamma\gamma$ & $\phantom{-}0.98 \pm 0.28$ & $1.72 \pm 0.59$ & $-0.38$ & 14.94 & 2.69 & 3.34 \\
\hline
$VV$ & $\phantom{-}0.91 \pm 0.16$ & $1.01 \pm 0.49$ & $-0.30$ & 44.59 & 4.24 & 4.58 \\
\hline
$b\bar{b}/\tau\tau$ & $\phantom{-}0.98 \pm 0.63$ & $0.97 \pm 0.32$ & $-0.25$ & \phantom{0}2.67 & 1.31 &
10.12 \\
\hline
$b\bar{b}$ & $-0.23 \pm 2.86$ & $0.97 \pm 0.38$ & $0$ & \phantom{0}0.12 & 0 & 7.06 \\
\hline
$\tau\tau$ & $\phantom{-}1.07 \pm 0.71$ & $0.94 \pm 0.65$ & $-0.47$ & \phantom{0}2.55 & 1.31 & 3.07 \\
\hline
\end{tabular}
\caption{Combined best-fit signal strengths $\hat{\mu}_{\rm{ggF}}$, $\hat{\mu}_{\rm{VBF}}$ 
and correlation coefficient $\rho$ for various final states, as well as the coefficients 
$a$, $b$ and $c$ for the $\chi^2$ in Eq.~(\ref{eq:bestlikegaussian}).}
\label{2013c-tab:1}
\end{table}

\begin{figure}[t]\centering
\includegraphics[width=5cm]{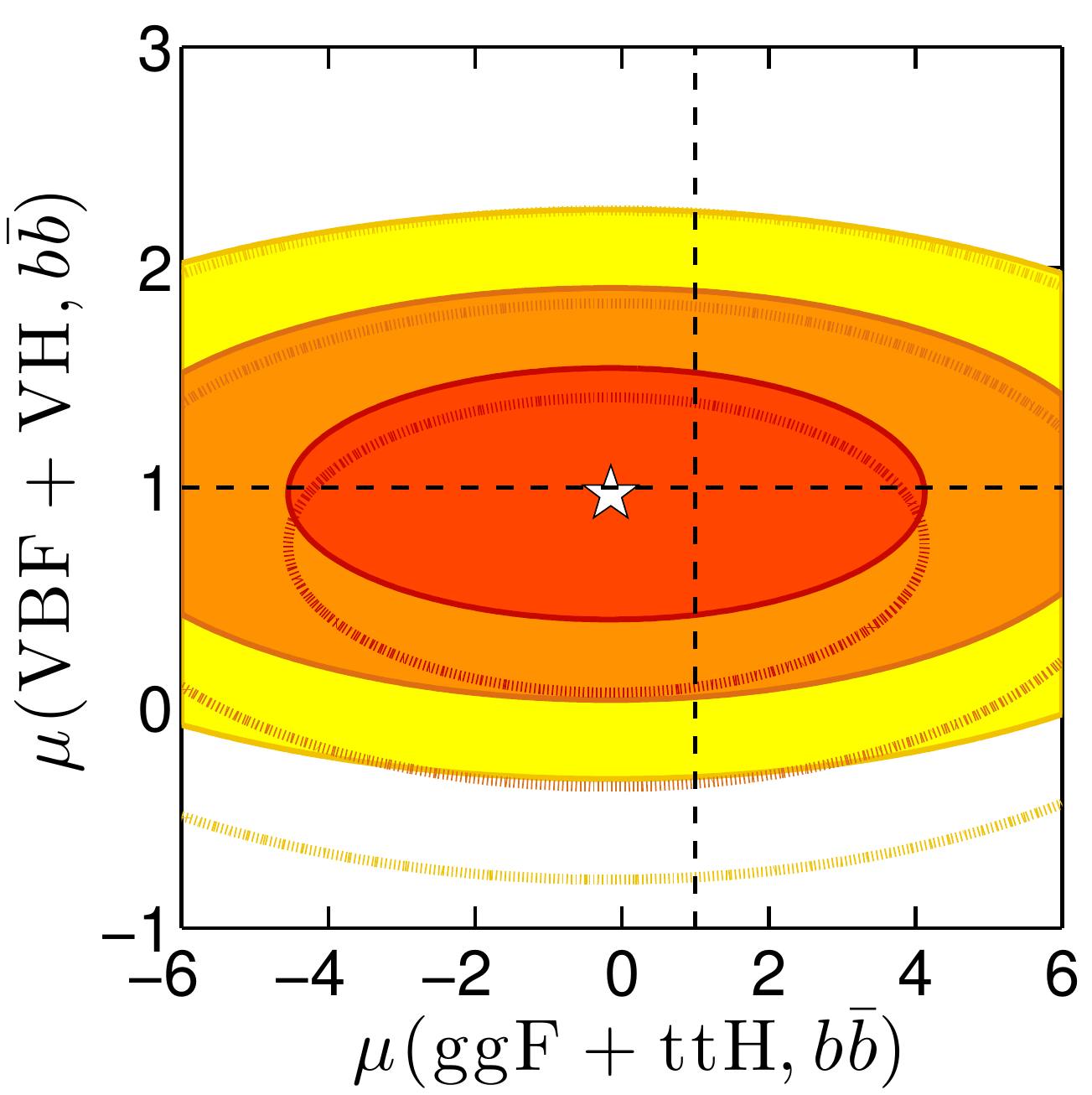}\quad
\includegraphics[width=5cm]{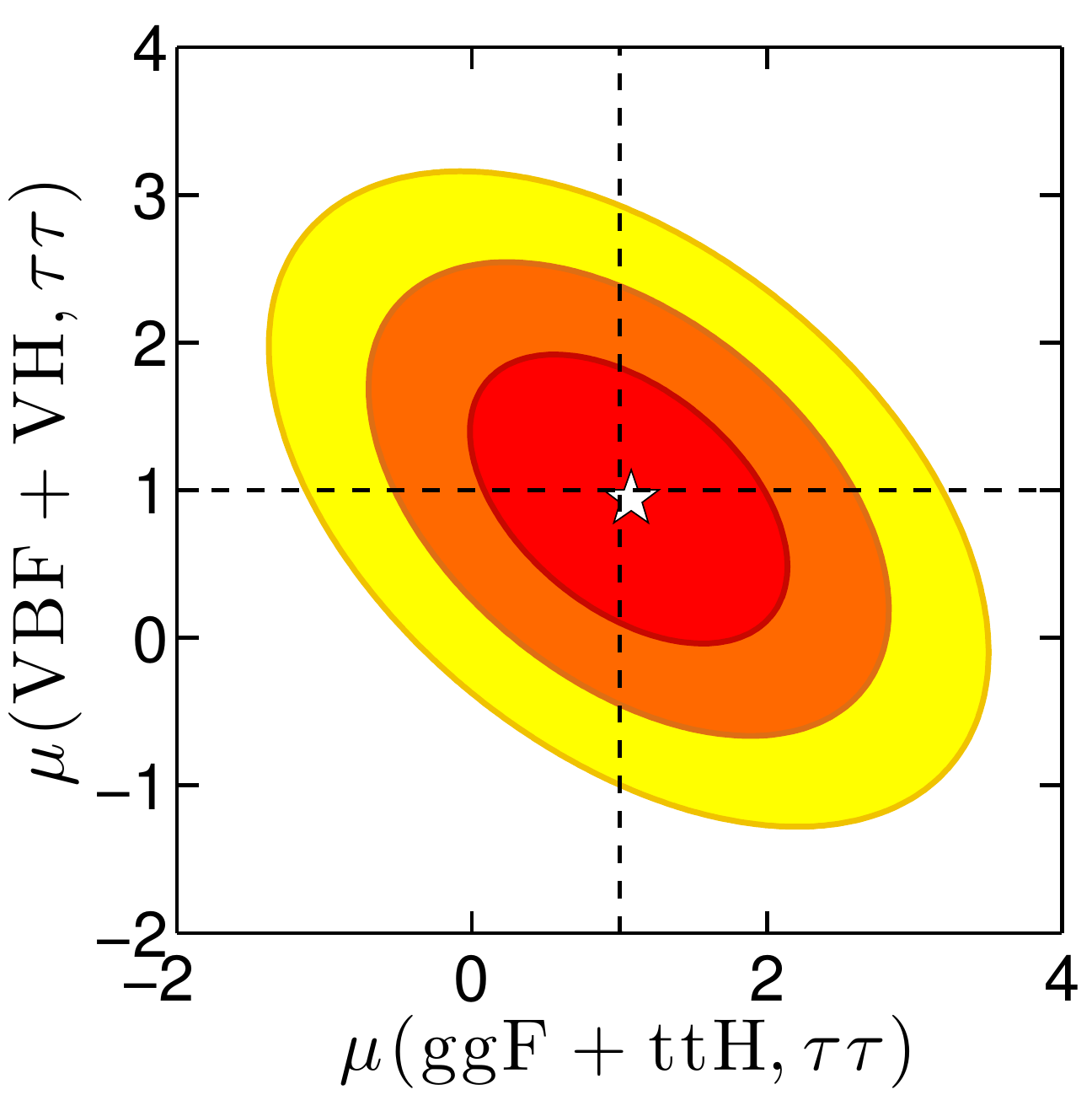}
\caption{Combined signal strength ellipses as in Fig.~\ref{2013c-fig:ellipses1} but treating the couplings 
to $b\bar b$ and $\tau\tau$ separately.
\label{2013c-fig:ellipses2} }
\end{figure}

The combination of the $b\bar{b}$ and $\tau\tau$ final states is justified, in principle,
in models where one specific Higgs doublet has the same reduced couplings (with respect
to the SM) to down-type quarks and leptons. However, even in this case QCD corrections
and so-called $\Delta_b$ corrections (from radiative corrections, notably at large $\tan\beta$, 
inducing couplings of another Higgs doublet to $b$~quarks, see {\it e.g.} \cite{Carena:1999py,Eberl:1999he}) 
can lead to deviations of the reduced $Hbb$ and $H\tau\tau$ couplings from a common value. Therefore, for completeness we show
the result for the $b\bar{b}$ final state only (combining ATLAS, CMS and
Tevatron results as given in the previous paragraph) in the fourth line of
Table~\ref{2013c-tab:1}, and the resulting 68\%, 95\% and 99.7\%  CL contours in the left plot in  
Fig.~\ref{2013c-fig:ellipses2}. The result for the $\tau\tau$ final state only (combining ATLAS and CMS results as given in the previous paragraph) is shown in the fifth line of
Table~\ref{2013c-tab:1}, and the resulting 68\%, 95\% and 99.7\%  CL contours in the right plot in Fig.~\ref{2013c-fig:ellipses2}.
Before proceeding, a comment is in order regarding the impact of the Tevatron results. 
While for the $\gamma\gamma$ and $VV$ final states, our combined likelihoods are completely 
dominated by the LHC measurements, to the extent that they are the same with or without including 
the Tevatron results, this is not the case for the $b\bar{b}$ final state. For illustration, in the plots for the 
$b\bar{b}$ final state in Figs.~\ref{2013c-fig:ellipses1} and \ref{2013c-fig:ellipses2} we also show 
what would be the result neglecting the Tevatron measurements.

\subsection{Fits to reduced Higgs couplings} \label{2013c-sec:coupfit}

Using the results of the previous section, it is straightforward to determine constraints 
on the couplings of the observed Higgs boson to various particle pairs, 
assuming only an SM-like Lagrangian structure. 
As in Section~\ref{sec:higgs2012}, we define $\cu$, $\cd$ and $\cv$ to be
ratios of the $H$ coupling to up-type quarks, down-type quarks and
leptons, and vector boson pairs, respectively, relative  to that
predicted in the case of the SM Higgs boson (with $\CV>0$ by convention).
In addition to these tree-level couplings there are also the one-loop
induced couplings of the $H$ to $gg$ and $\gam\gam$.  Given values for
$\cu$, $\cd$ and $\cv$ the contributions of SM particles to the $gg$ and
$\gam\gam$ couplings, denoted $\anti\cg$ and $\anti \cp$ respectively,
can be computed. We take into account NLO corrections to $\anti\cg$ and $\anti \cp$ as recommended by the  LHC Higgs Cross Section Working Group~\cite{LHCHiggsCrossSectionWorkingGroup:2012nn}. 
In particular we include all the available QCD corrections for $C_g$ using \texttt{HIGLU}~\cite{Spira:1995rr,Spira:1995mt,Spira:1996if} 
and for $C_\gamma$ using \texttt{HDECAY}~\cite{Spira:1996if,Djouadi:1997yw}, and we switch off the
electroweak corrections. 
In some of the fits below, we will also allow for
additional new physics contributions to $\cg$ and $\cp$ by writing
$\cg=\anti\cg+\dcg$ and $\cp=\anti\cp+\dcp$. 

We note that in presenting one- (1D) and two-dimensional (2D) distributions of $\dchisq$, those quantities among $\cu$, $\cd$, $\cv$, $\dcg$ and $\dcp$ not plotted, but that are treated as variables, are being profiled over.
The fits presented below will be performed with and without allowing for invisible decays of the Higgs boson. In the latter case, only SM decay modes are present. In the former case, the new decay modes are assumed to produce invisible or undetected particles that would be detected as missing transverse energy at the LHC. A direct search for invisible decays of the Higgs boson has been performed by ATLAS in the $ZH \to \ell^+\ell^- + E_T^{\rm miss}$ channel~\cite{ATLAS-CONF-2013-011} and is implemented in the analysis. Thus, the total width is fully calculable from the set of $C_i$ and ${\cal B}(H \to {\rm invisible})$ in all the cases we consider. (We will come back to this at the end of this section.)

\begin{figure}[t]\centering
\includegraphics[width=4.75cm]{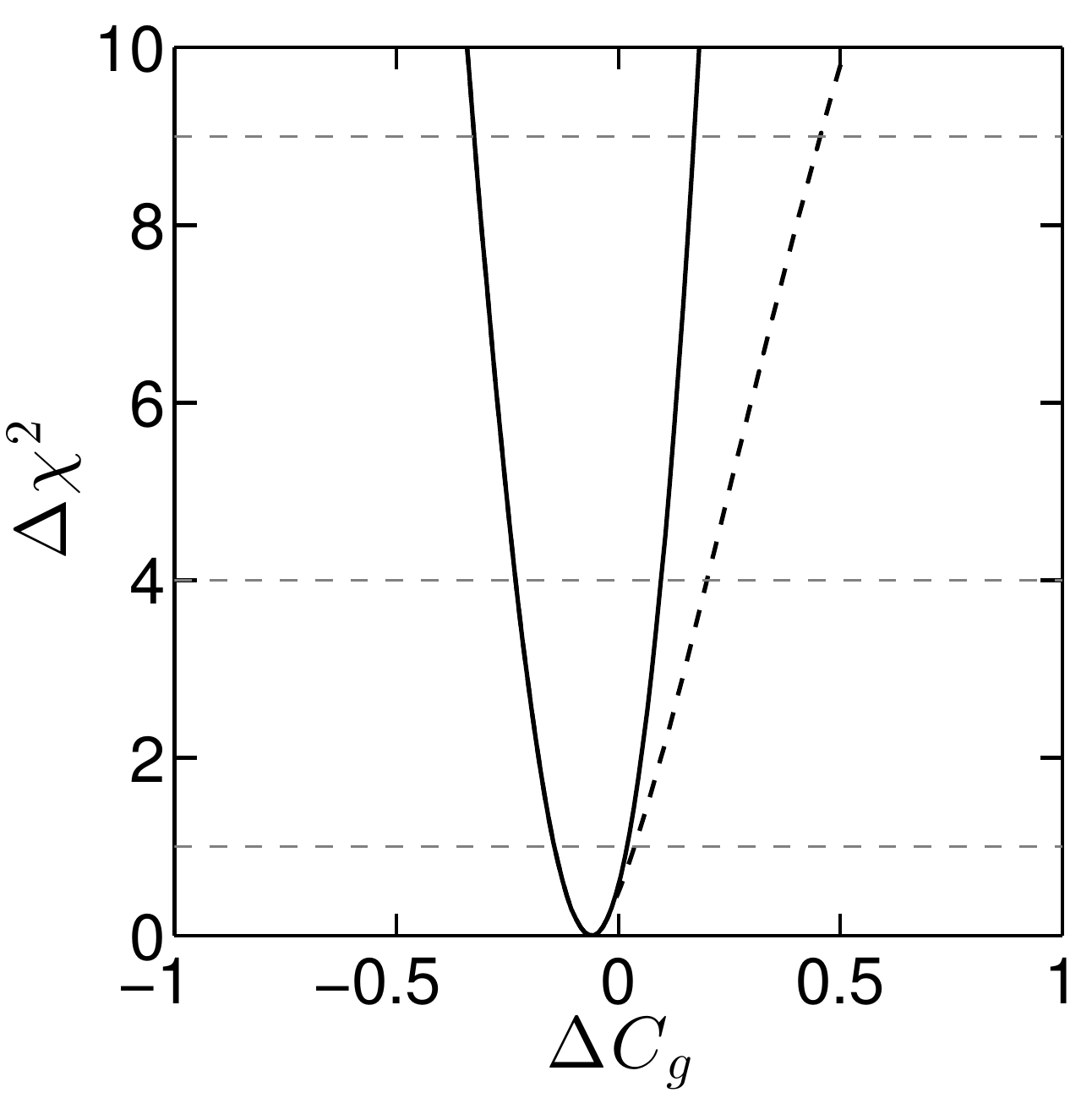}\quad
\includegraphics[width=5cm]{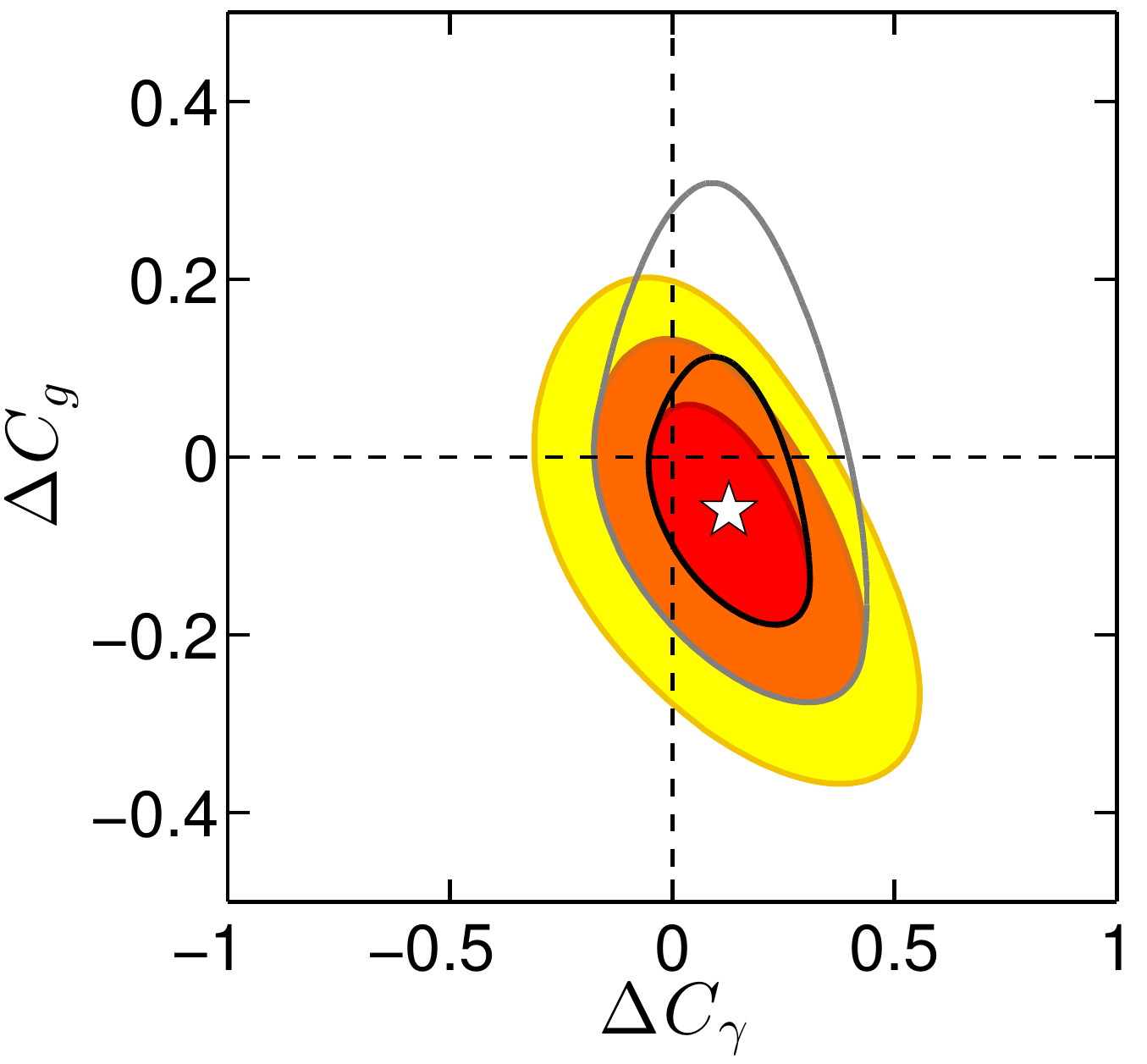}\quad
\includegraphics[width=4.75cm]{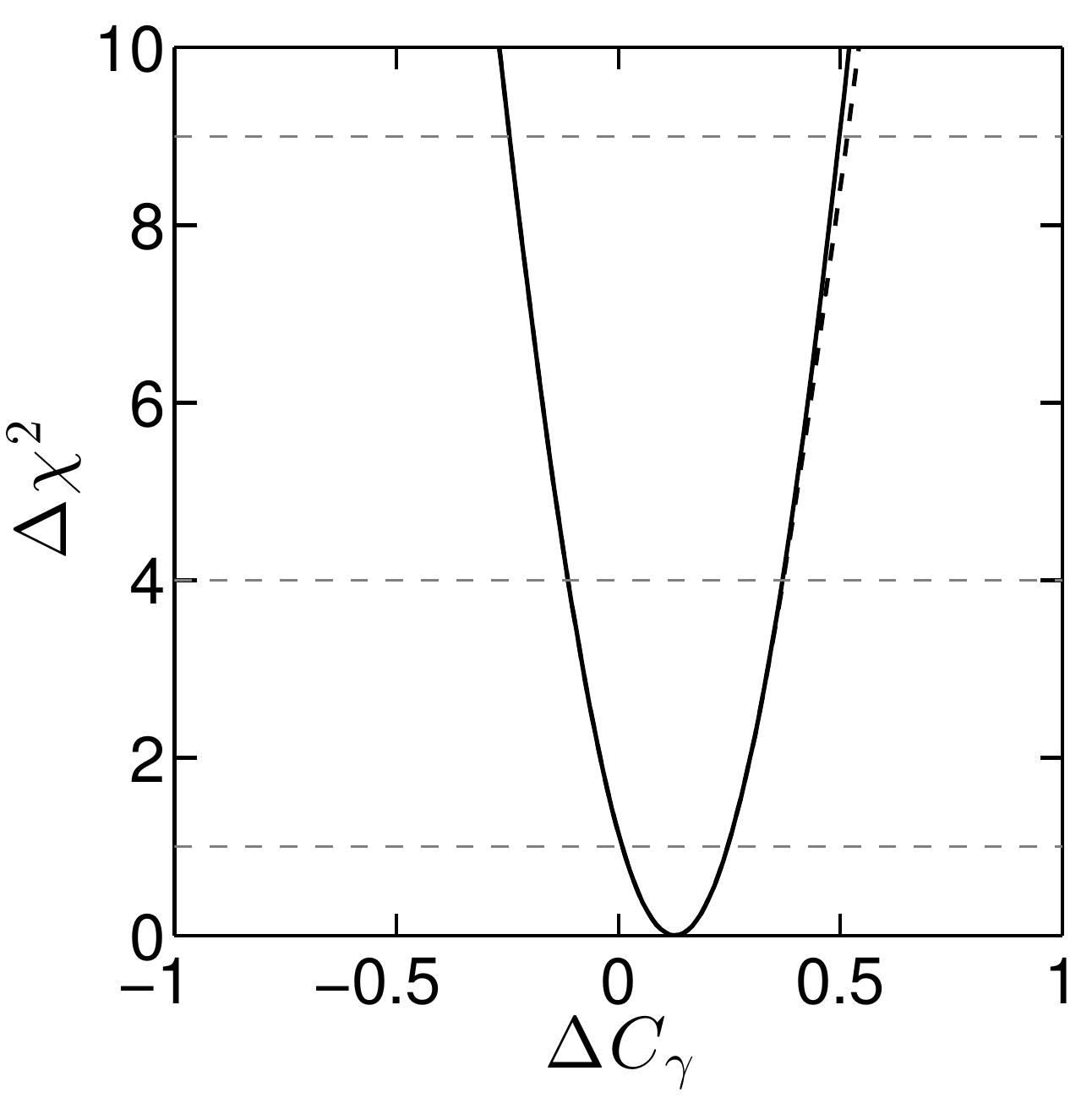}\quad
\caption{$\Delta \chi^2$ distributions in 1D and 2D for the fit of $\dcg$ and $\dcp$ for $\CU=\CD=\CV=1$. 
In the 1D plots, the solid (dashed) lines are for the case that invisible/unseen decays are absent (allowed).
In the 2D plot, the red, orange and yellow areas are the 68\%, 95\% and 99.7\% CL regions, respectively, assuming invisible decays are absent. The white star marks the best-fit point. The black and gray lines show the 68\% and 95\%~CL contours when allowing for invisible decays.
\label{2013c-fig:CPadd-CGadd} }
\end{figure}

We begin by taking SM values for the tree-level couplings to fermions and vector bosons, \ie\  $\cu=\cd=\cv=1$, 
but allow for new physics contributions to the couplings to $gg$ and $\gam\gam$.  The fit results with and without 
allowing for invisible/unseen Higgs decays are shown in  Fig.~\ref{2013c-fig:CPadd-CGadd}. We observe that the SM point of $\dcg=\dcp=0$ is well within the 68\% contour with the best fit points favoring a slightly positive (negative) value for $\dcp$ ($\dcg$). 
Allowing for invisible/unseen decays expands the 68\%, 95\% and 99.7\% CL regions by only a modest amount. 
This is in contrast to the situation at the end of 2012 (see Section~\ref{sec:higgs2012} and Ref.~\cite{Belanger:2013kya}), where some new physics contribution to both $\dcg$ and $\dcp$ was preferred, and allowing for invisible decays had a large effect; 
with the higher statistics and with the reduced $\gamma\gamma$ 
signal strength from CMS~\cite{CMS-PAS-HIG-13-001}, $\dcg$ and $\dcp$ are now much more constrained. 
The best fit is obtained for $\dcg=-0.06$, $\dcp=0.13$, $\brinv\equiv \br({H\to\rm invisible})=0$ and has $\chimin=17.71$ for 21 d.o.f.\ (degrees of freedom)\footnote{There are in total 23 measurements entering our fit, and we adopt the simple definition of the number of d.o.f.\ as number of measurements minus number of parameters.}, 
as compared to $\chisq=18.95$ with 23 d.o.f.\ for the SM, so allowing for additional loop contributions  
does not 
improve the fit.

\begin{figure}[t!]\centering
\includegraphics[width=5cm]{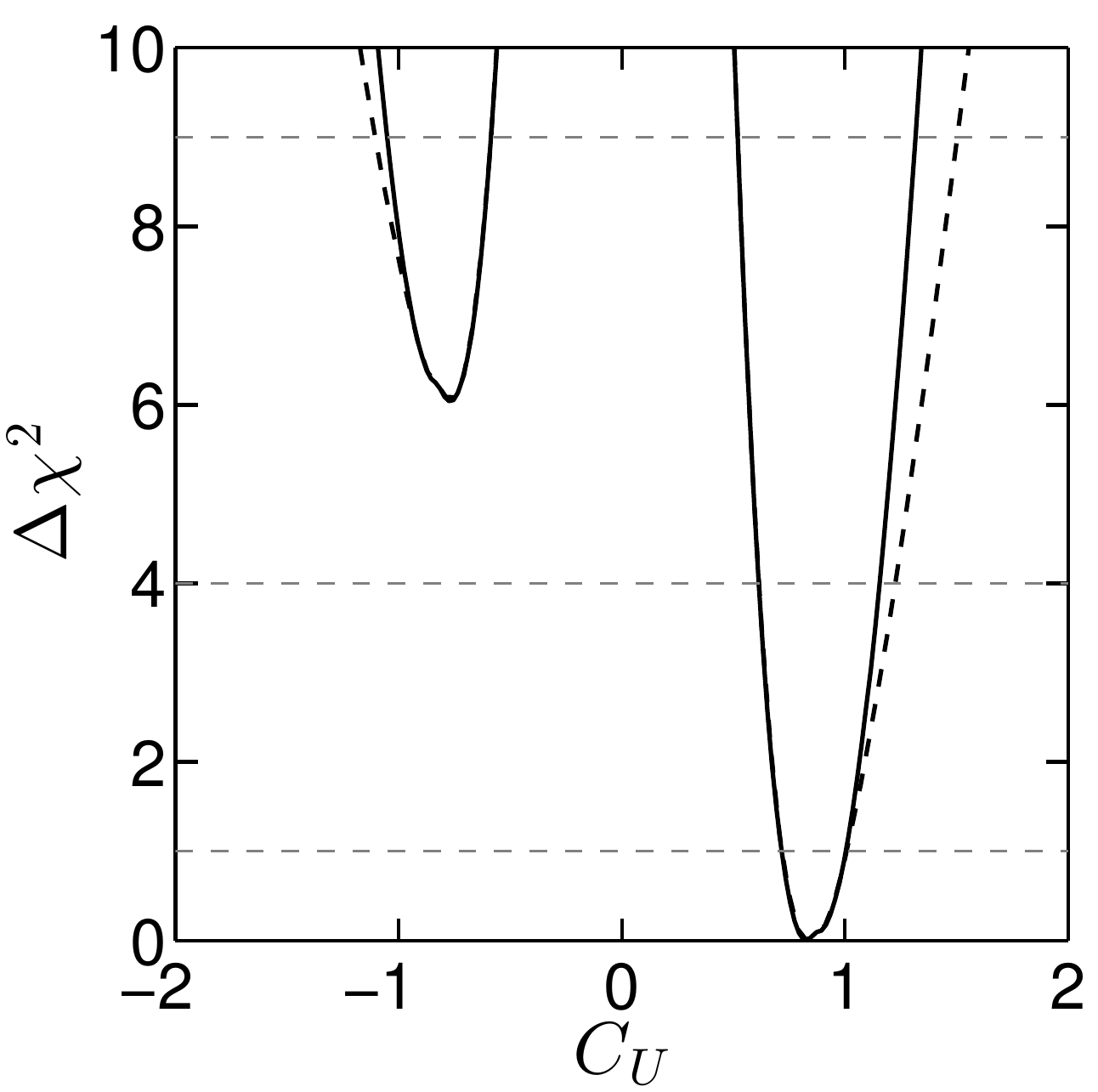}\quad
\includegraphics[width=5cm]{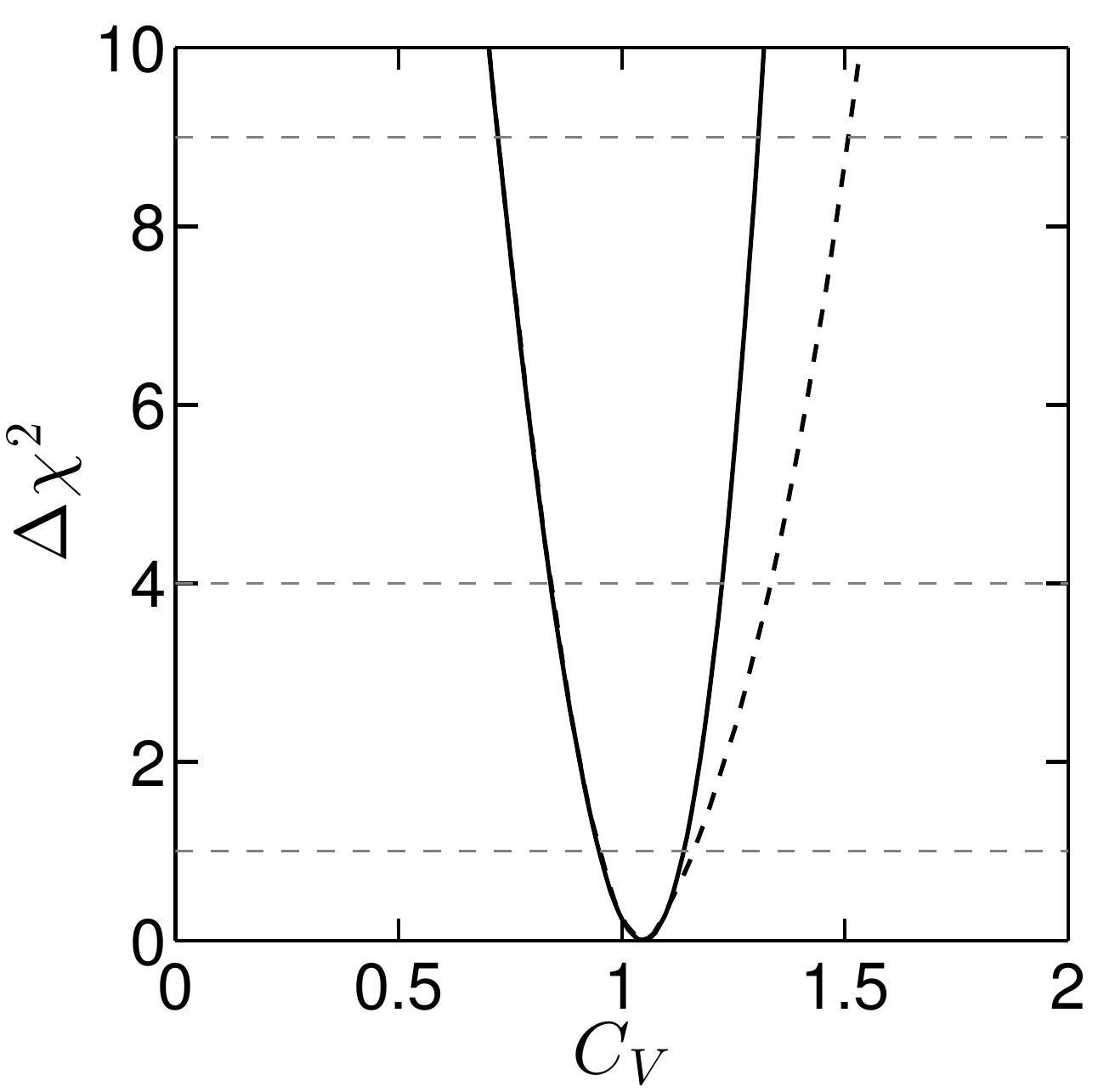}
\caption{Fit of $\CU$, $\CD$, $\CV$ for $\dcg=\dcp=0$. The plots show the 1D $\dchisq$ distribution as a function of $\cu$ (left) and $\cv$ (right). The solid (dashed) lines are for the case that invisible/unseen decays are absent (allowed).
\label{2013c-fig:CU-CD-CV} }
\end{figure}

Next, we allow $\cu$, $\cd$ and $\cv$ to vary but assume that there is no new physics in the $gg$ and $\gam\gam$ loops, \ie\ we take $\dcg=\dcp=0$. Results for this case are shown in Fig.~\ref{2013c-fig:CU-CD-CV}.  We observe that, contrary to the situation at the end of 2012, the latest data prefer a positive value of $\cu$ close to 1. 
This is good news, as a negative sign of $\cu$---in the convention where $m_t$ is positive---is quite problematic in the context of most theoretical models.\footnote{If the top quark and Higgs
bosons are considered as fundamental fields, it would require that the
top quark mass is induced dominantly by the vev of at least one
additional Higgs boson which is not the Higgs boson considered here, and 
typically leads to various consistency problems as discussed, \eg, in
\cite{Choudhury:2012tk}.}
(We do not show the distribution for $\cd$ here but just remark that
$|\CD|\simeq 1\pm0.2$ with a sign ambiguity following from
the weak dependence of the $gg$ and $\gam\gam$ loops on the bottom-quark
coupling.)  For $\CV$, we find a best-fit value slightly above 1, at
$\CV=1.07$, but  with  the SM-like value of $\cv=1$ lying well within one
standard deviation.

\begin{figure}[t!]\centering
\includegraphics[width=5cm]{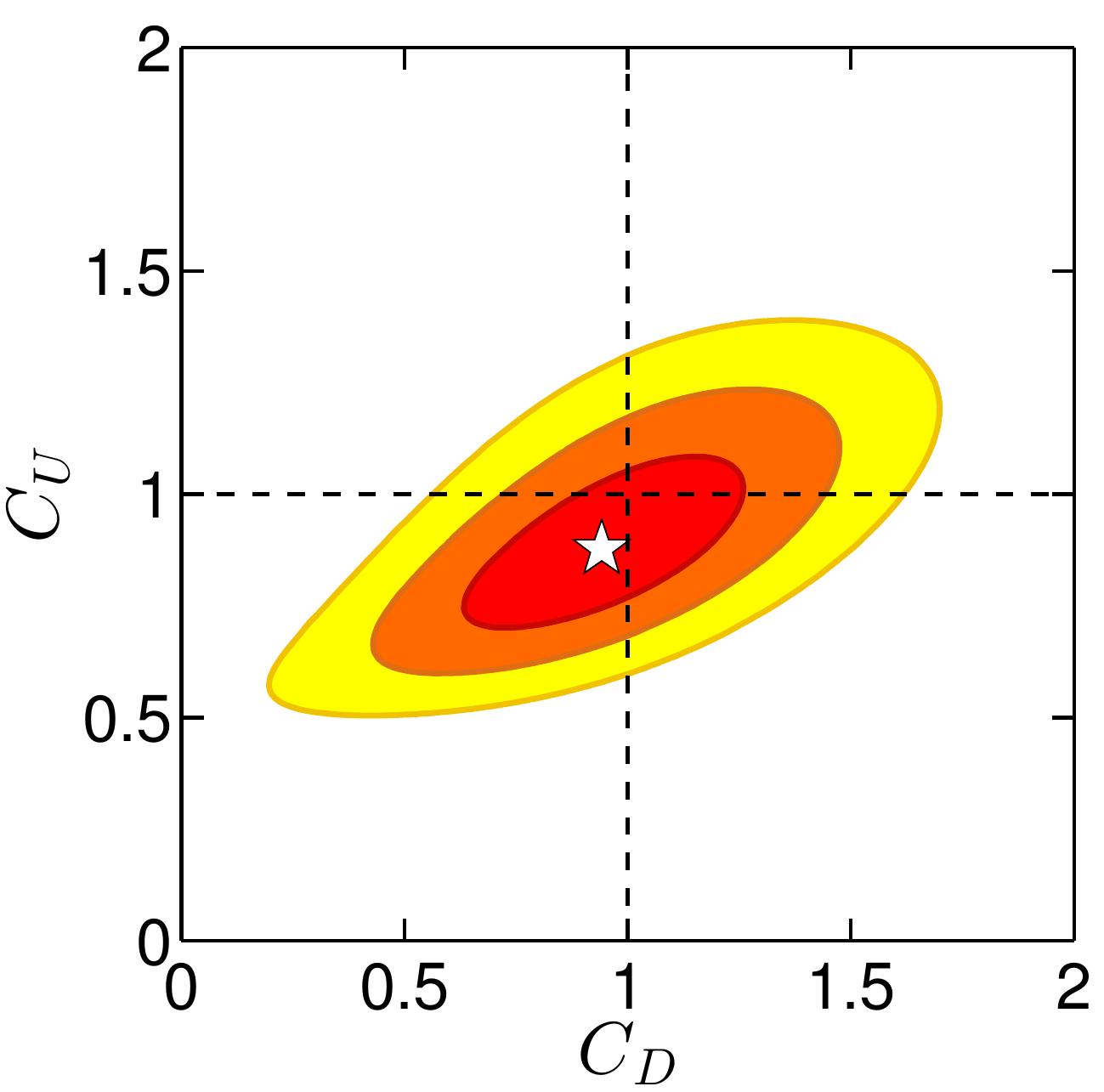}\quad
\includegraphics[width=5cm]{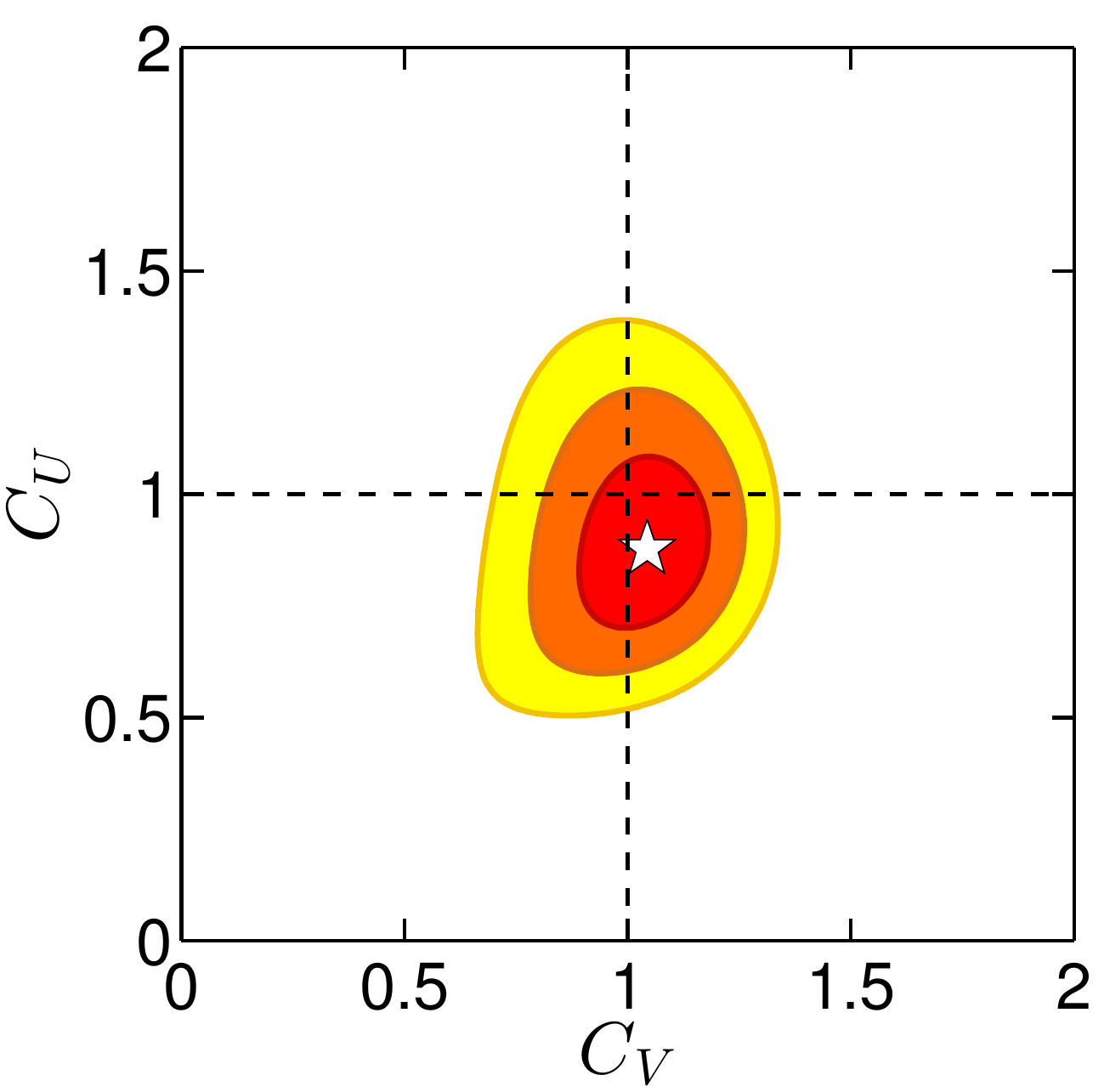}\quad
\includegraphics[width=5cm]{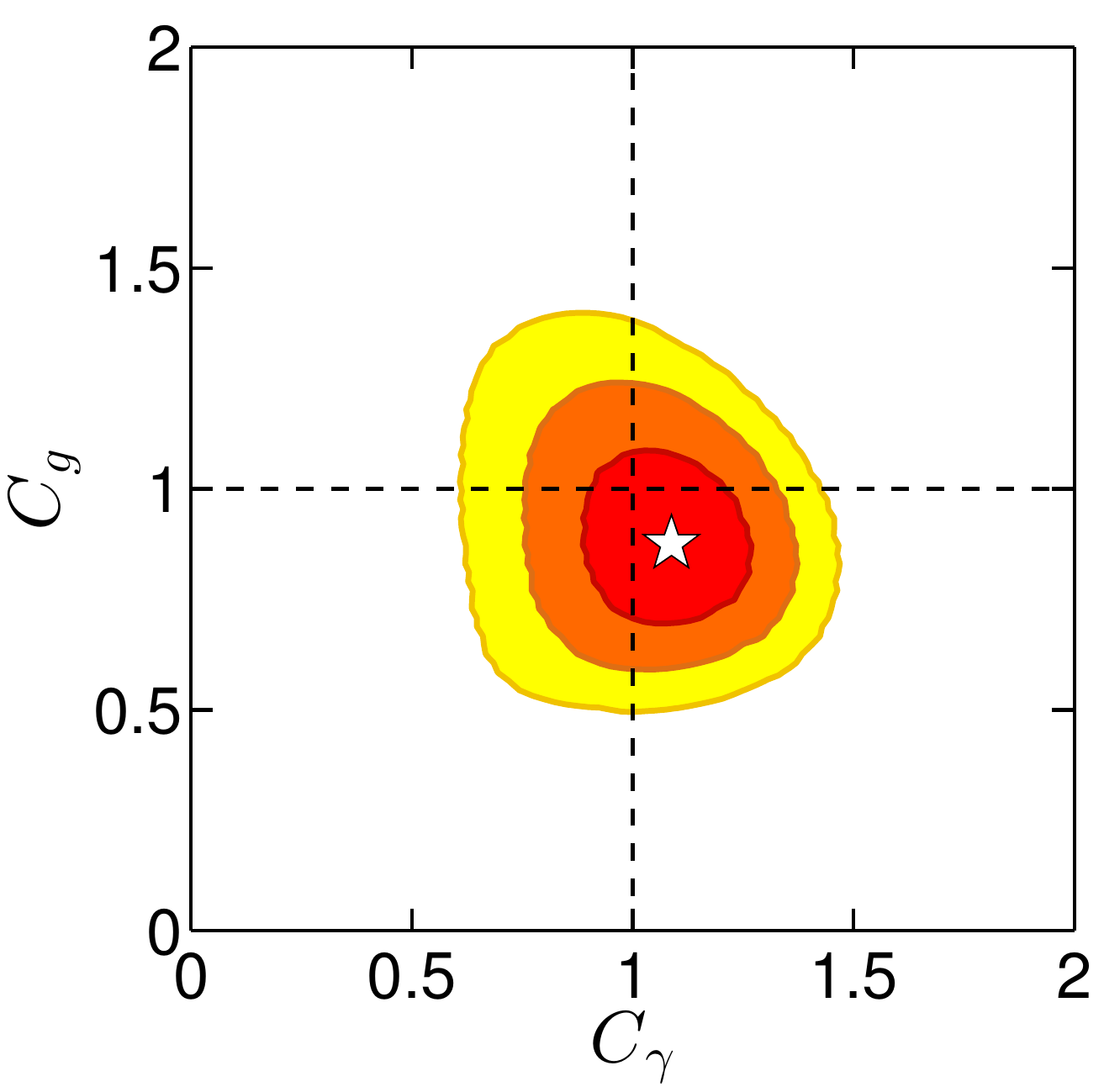}\quad
\caption{Fit of $\CU>0$, $\CD>0$ and $\CV$ for $\dcg=\dcp=0$. 
The red, orange and yellow areas are the 68\%, 95\% and 99.7\% CL regions, respectively, assuming 
invisible decays are absent. The white star marks the best-fit point. 
\label{2013c-fig:CUpos-CDpos-CV} }
\end{figure}

Since $\cu<0$ is now disfavored and the sign of $\CD$ is irrelevant, we
confine ourselves subsequently to $\cu,\cd>0$. In
Fig.~\ref{2013c-fig:CUpos-CDpos-CV}  we show $\dchisq$ distributions in 2D
planes confined to this range, still assuming $\dcg=\dcp=0$.
The mild correlation between $\CU$ and $\CD$ in the leftmost plot of
Fig.~\ref{2013c-fig:CUpos-CDpos-CV} follows from the very SM-like signal
rates in the $VV$ and $\gamma\gamma$ final states in ggF: varying $\CD$
implies a variation of the partial width $\Gamma(H\to bb)$ which
dominates the total width. Hence, the branching fractions $\br(H\to VV)$
and $\br(H\to\gamma\gamma)$ change in the opposite direction, decreasing
with increasing total width (\ie\ with increasing $\CD$) and vice versa. In
order to keep the signal rates close to~1, the ggF production cross
section, which is  roughly proportional to $ \CU^2$,  has to vary in the same direction as $\CD$.
The best fit is obtained for $\CU=0.88$, $\CD=0.94$, $\CV=1.04$, $\cp=1.09$, $\cg=0.88$
(and, in fact, $\brinv=0$).  
Note that if $\cv>1$ were confirmed, this would imply that the observed Higgs boson must have a significant  triplet (or higher representation) component~\cite{Logan:2010en,Falkowski:2012vh}.
Currently the coupling fits are, however, perfectly consistent with SM values. 
Again, with a $\chimin=17.79$ (for 20 d.o.f.) as compared to $\chisq=18.95$ for the SM, allowing for deviations from the SM does not significantly improve the fit.

\begin{figure}[t!]\centering
\includegraphics[width=5cm]{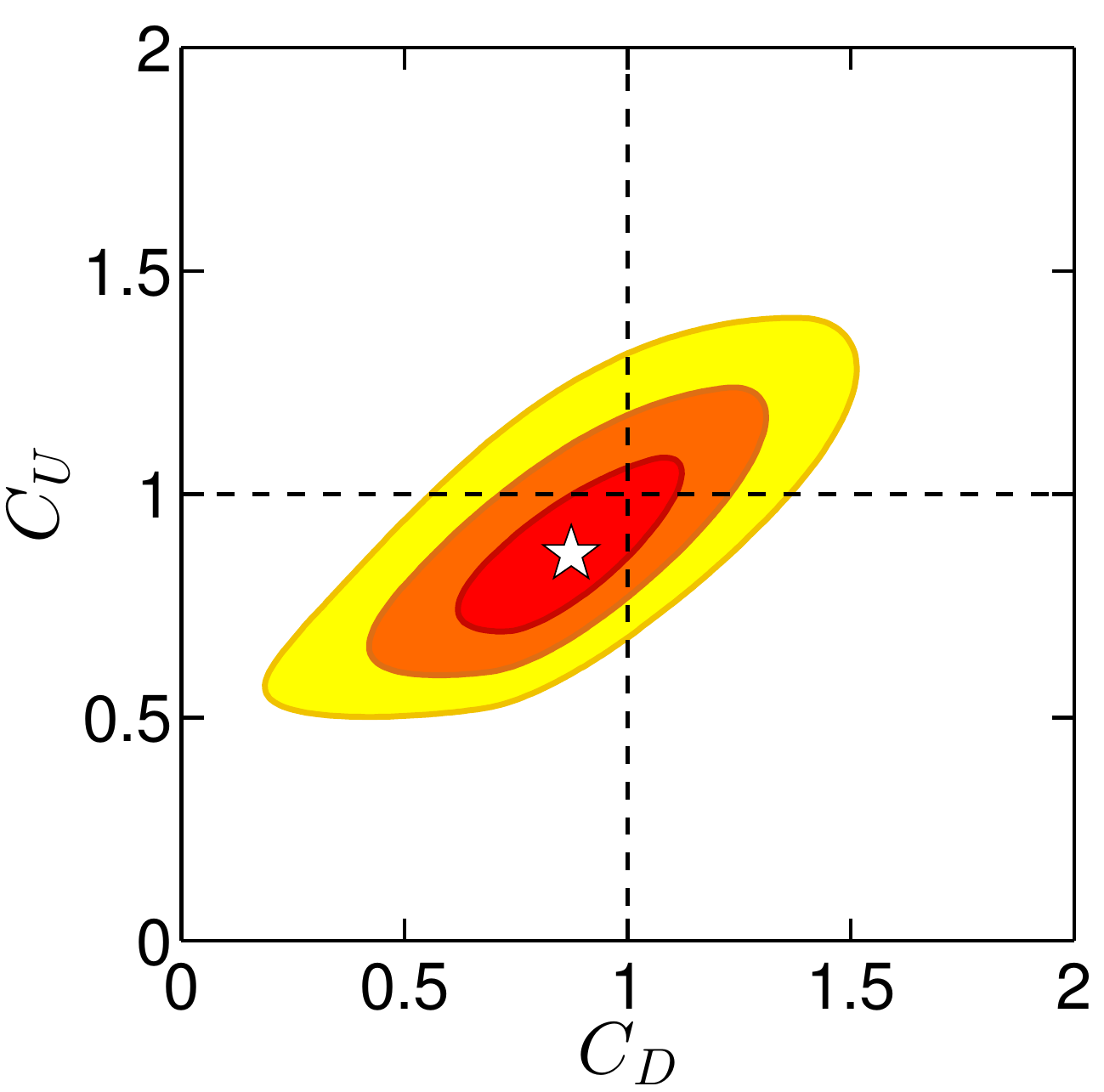}\quad
\includegraphics[width=5cm]{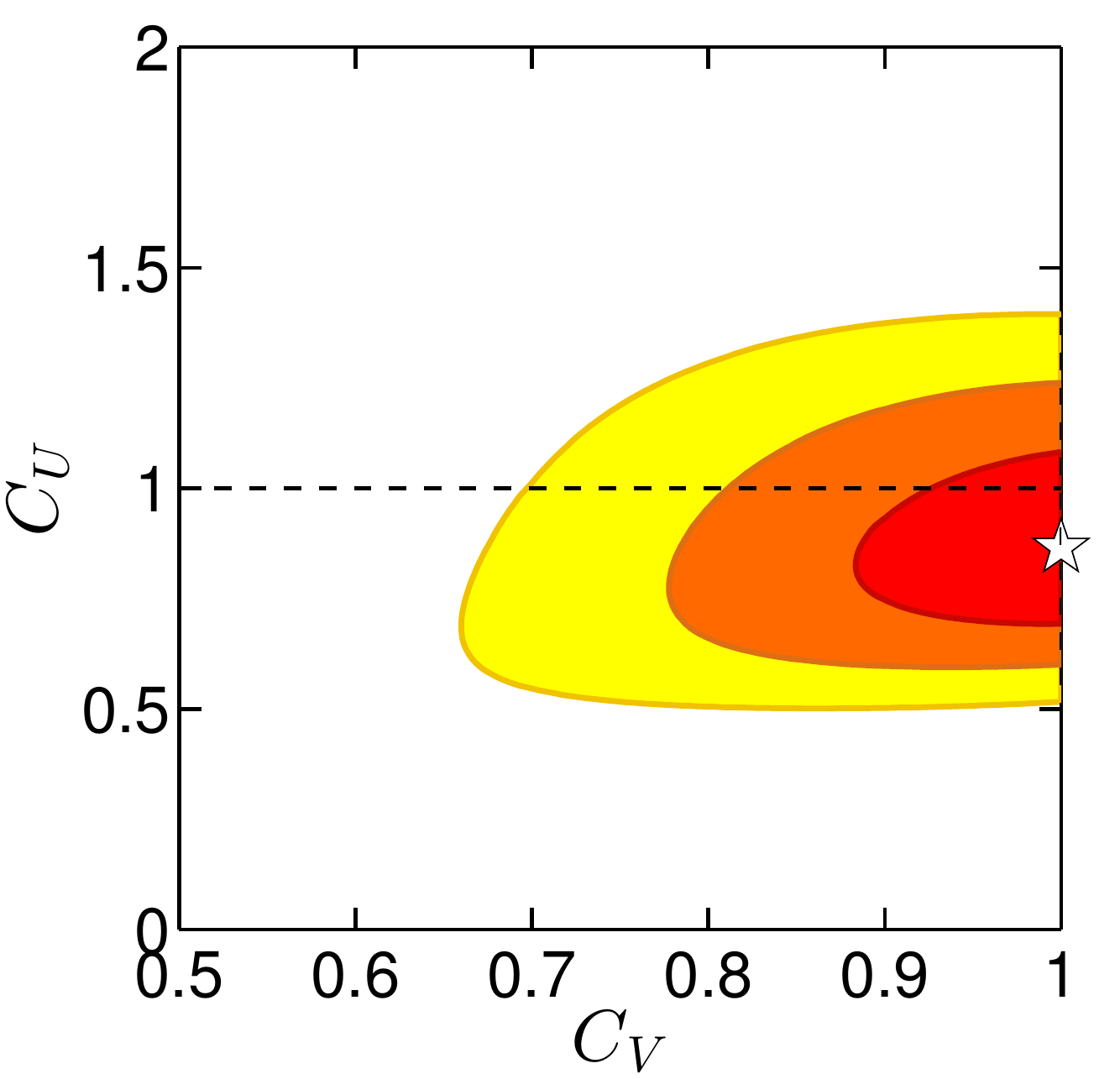}\quad
\includegraphics[width=5cm]{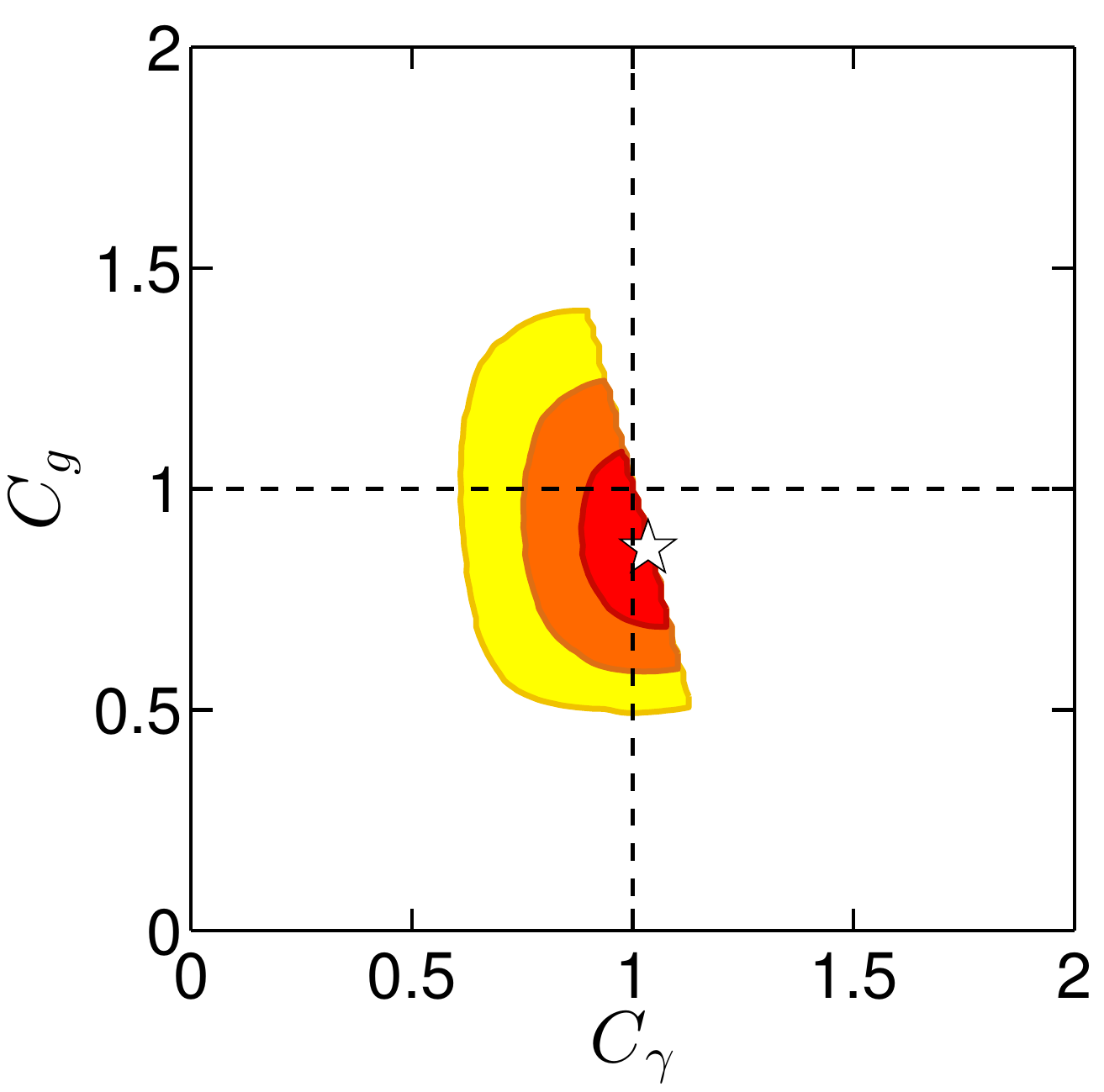}\quad
\caption{As in Fig.~\ref{2013c-fig:CUpos-CDpos-CV} but for $\CV\leq 1$.
\label{2013c-fig:CUpos-CDpos-CVle1} }
\end{figure}

\begin{figure}[t!]\centering
\includegraphics[width=5cm]{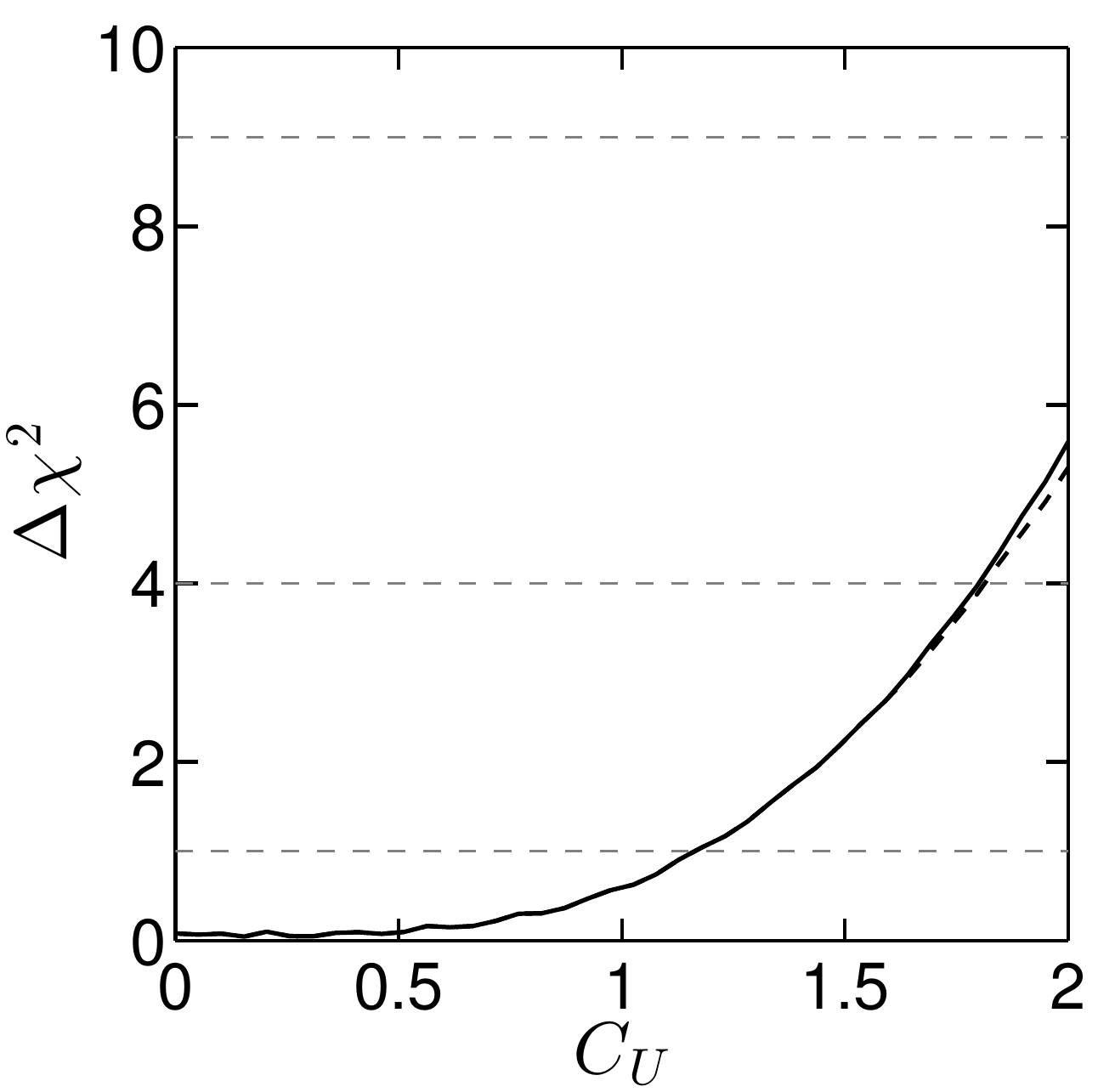}\quad
\includegraphics[width=5cm]{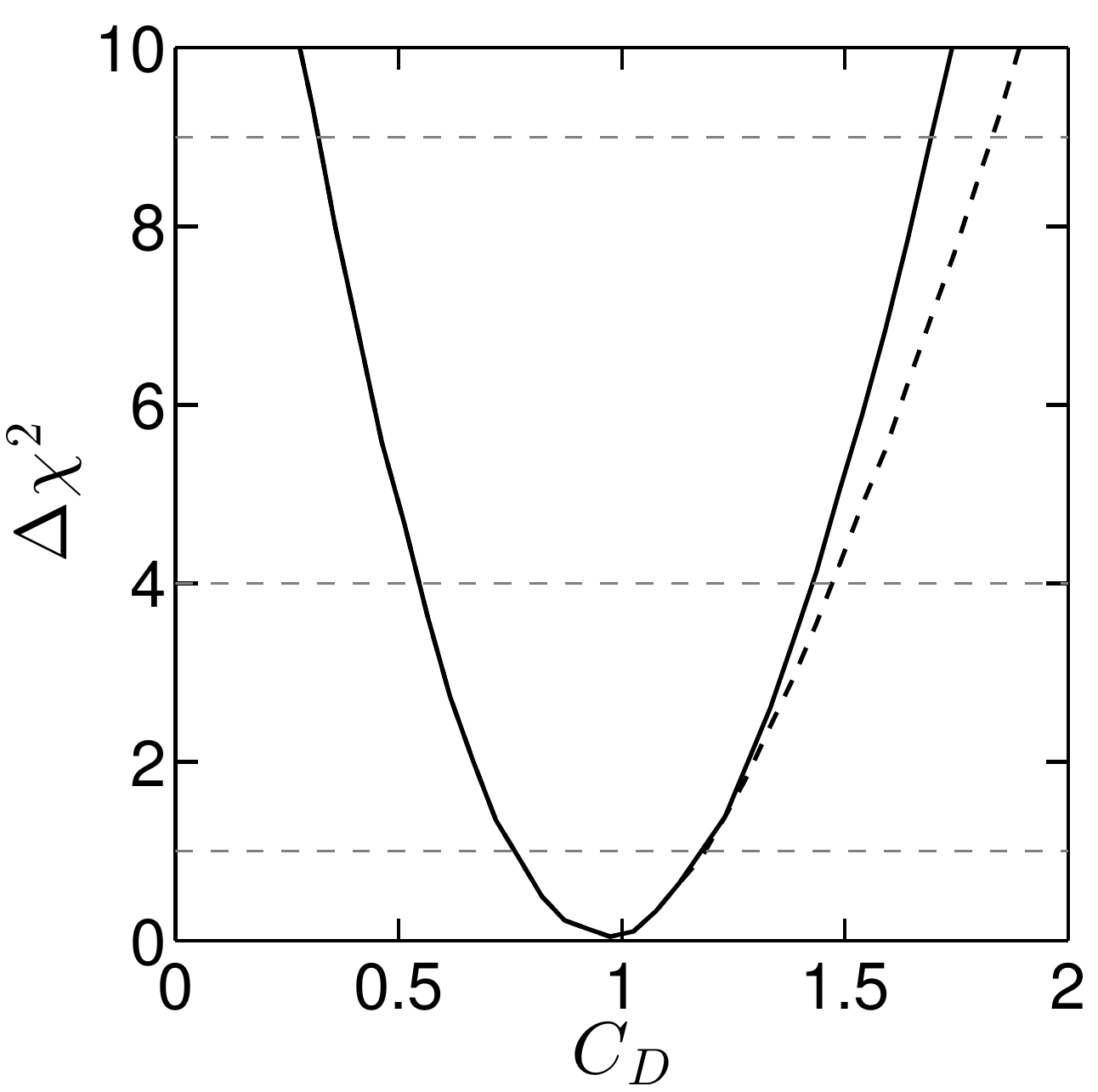}\quad
\includegraphics[width=5cm]{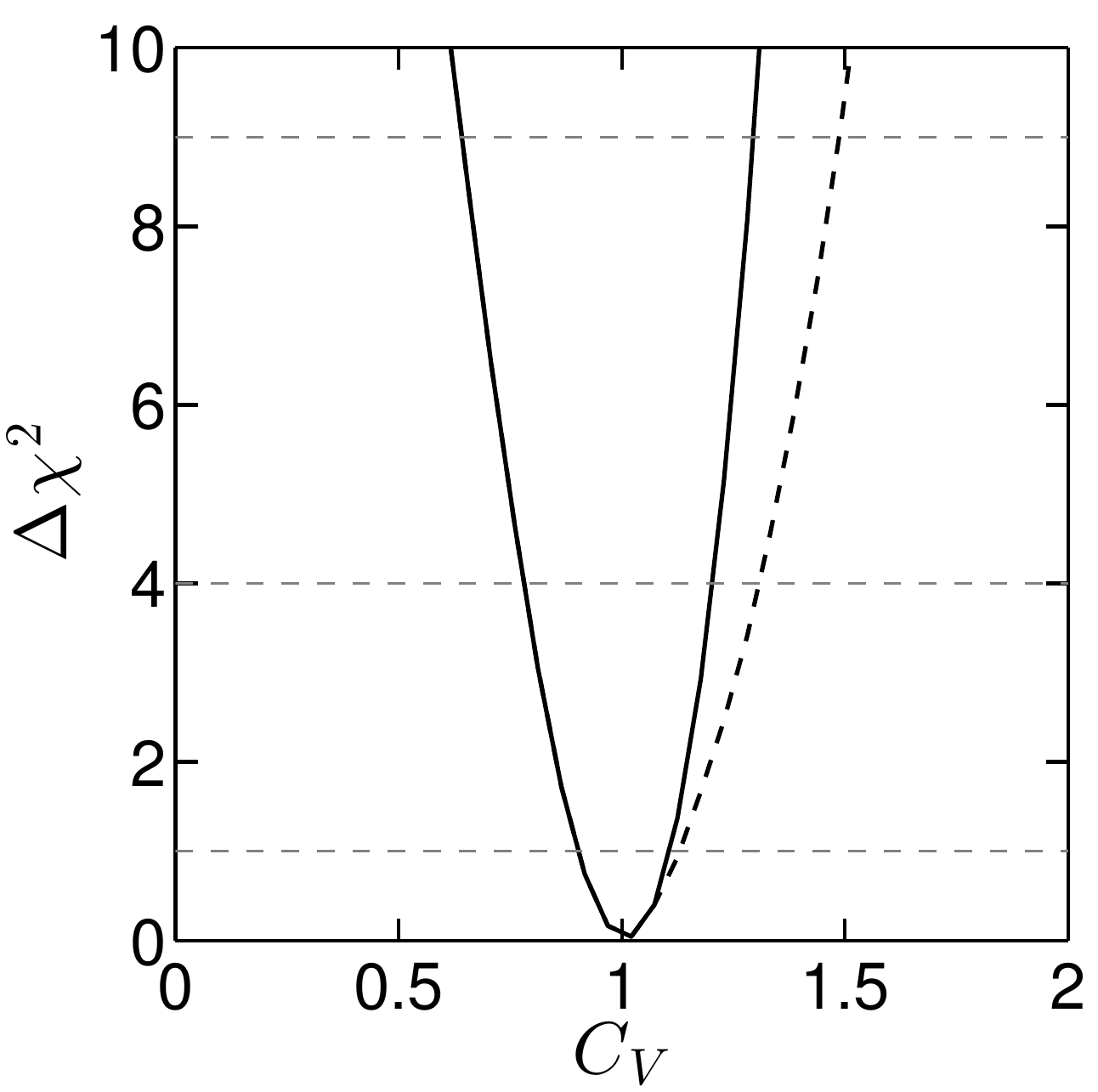}\quad
\includegraphics[width=5cm]{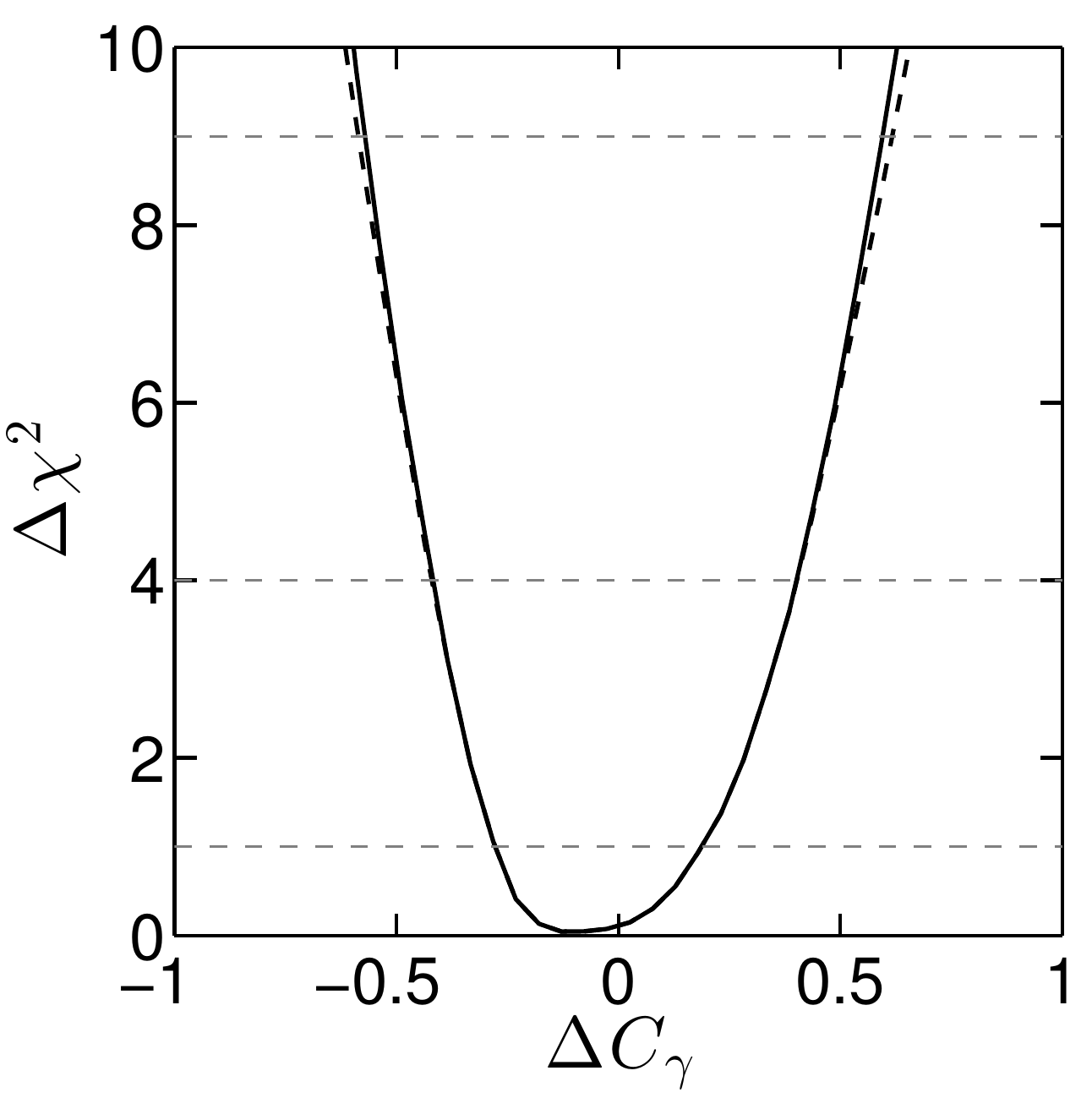}\quad
\includegraphics[width=5cm]{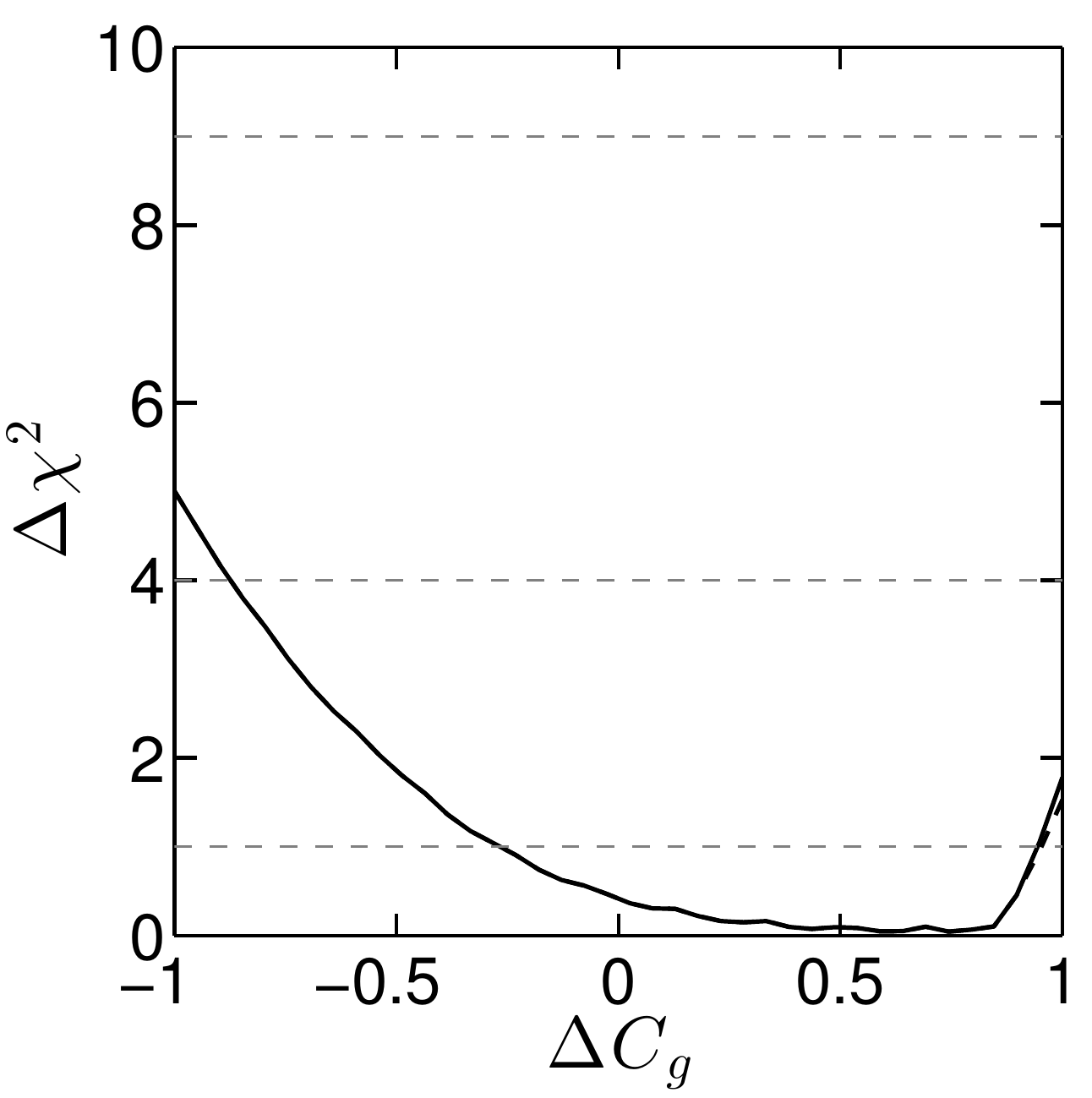}\\
\caption{Five (six) parameter fit of $\CU$, $\CD$, $\CV$,  $\dcg$ and $\dcp$; the solid (dashed) curves 
are those obtained when invisible/unseen decay modes are not allowed (allowed) for.
\label{2013c-fig:5param} }
\end{figure}

In models where the Higgs sector consists of doublets+singlets only one always
obtains  $\cv\le1$.   Results for this
case are shown in Fig.~\ref{2013c-fig:CUpos-CDpos-CVle1}. Given the slight
preference for $\cv>1$ in the previous free-$\cv$ plots, it is no
surprise the $\cv=1$ provides the best fit along with $\CU=\cg=0.87$,
$\CD=0.88$ and $\cp=1.03$. Of course, the SM is again well within the
$68\%$ CL zone.
The general case of free parameters $\cu$, $\cd$, $\cv$, $\dcg$ and $\dcp$
is illustrated in Fig.~\ref{2013c-fig:5param}, where we show the 1D $\dchisq$
distributions for these five parameters (each time profiling over the
other four parameters).  As before, the solid (dashed) lines indicate
results not allowing for (allowing for) invisible/unseen decay modes of
the Higgs.  Allowing for invisible/unseen decay modes again relaxes the
$\dchisq$ behavior only modestly.  The best fit point always corresponds
to $\brinv=0$.

\begin{figure}[t!]\centering
\includegraphics[width=6cm]{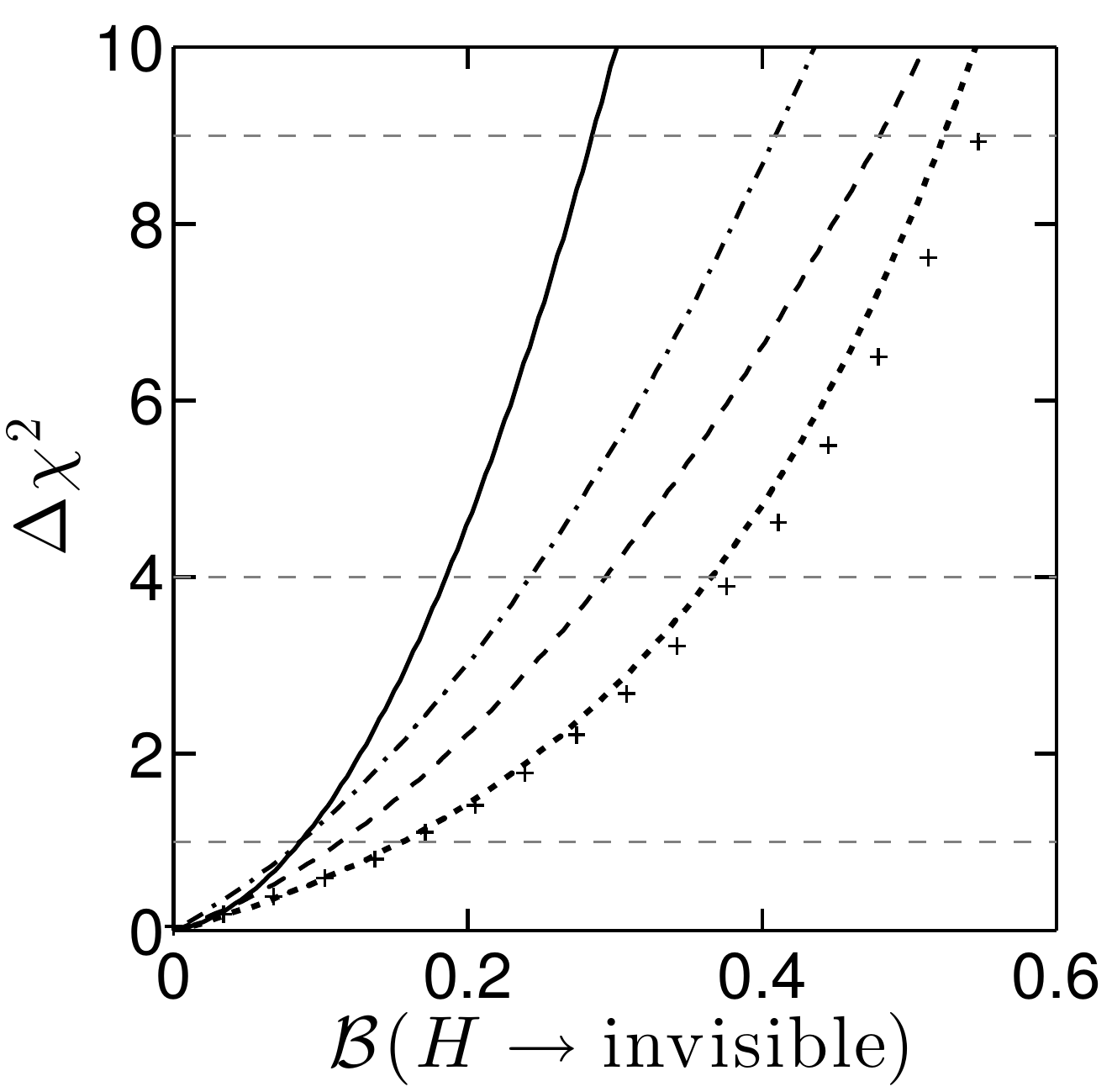}\\
\caption{$\Delta\chi^2$ distributions for the branching fraction of
invisible Higgs decays for various cases.  
Solid: SM+invisible. 
Dashed: varying $\dcg$ and $\dcp$ for $\CU=\CD=\CV=1$.
Dotted: varying $\CU$, $\CD$ and $\CV$ for $\dcg=\dcp=0$. 
Dot-dashed: varying $\CU$, $\CD$ and $\CV\leq 1$ for $\dcg=\dcp=0$.
Crosses: varying $\CU$, $\CD$, $\CV$,  $\dcg$ and $\dcp$.
\label{2013c-fig:BRinv} }
\end{figure}

An overview of the current status of invisible decays is given in Fig.~\ref{2013c-fig:BRinv}, which shows 
the behavior of $\dchisq$ as a function of $\brinv$ for various different cases of interest: \\
\indent a) SM Higgs with allowance for invisible decays --- one finds $\brinv<0.09$ (0.19); \\
\indent b) $\cu=\cd=\cv=1$ but $\dcp,\dcg$ allowed for --- $\brinv< 0.11$ (0.29); \\
\indent c) $\cu,\cd,\cv$ free, $\dcp=\dcg=0$, --- $\brinv<0.15$ (0.36); \\
\indent d) $\cu,\cd$ free, $\cv\leq 1$, $\dcp=\dcg=0$ --- $\brinv<0.09$ (0.24); \\
\indent e) $\cu,\cd,\cv,\dcg,\dcp$ free --- $\brinv<0.16$ (0.38).  \\
(All $\brinv$ limits are given at 68\% (95\%) CL.) 
Thus, while $\brinv$ is certainly significantly limited by the current data set, there remains ample room for invisible/unseen decays.  At 95\%~CL, $\brinv$ as large as $\sim 0.38$ is possible.
Here, we remind the reader that the above results are obtained after fitting the $125.5\gev$ data {\em and} inputting the experimental results for the $(Z \to \ell^+\ell^-) \;+$ invisible direct searches. When $C_V \leq 1$, $H\to\,$invisible is much more constrained by the global fits to the $H$ properties than by the direct searches for invisible decays, {\it cf.}\ the solid, dashed and dash-dotted lines in Fig.~\ref{2013c-fig:BRinv}. 
For unconstrained  $C_U$, $C_D$ and $C_V$, on the other hand, {\it cf.}\ dotted line and crosses in Fig.~\ref{2013c-fig:BRinv}, the limit comes from the direct search for invisible decays in the $ZH$ channel.  
 
A comment is in order here. In principle there is a flat direction in the unconstrained LHC Higgs coupling fit when unobserved decay modes are present: setting $\cu = \cd = \cv \equiv C$, so that ratios of rates remain fixed, all the Higgs production$\times$decay rates can be kept fixed to the SM ones by scaling up $C$ while adding a new, unseen decay mode with branching fraction ${\cal B}_{\rm new}$ according to $C^2 = 1/(1 - {\cal B}_{\rm new}$)~\cite{Zeppenfeld:2000td,Djouadi:2000gu}, see also \cite{Duhrssen:2004cv}.
In~\cite{Belanger:2013kya} we found that it is mainly $\cv$ which is critical here, because of the rather well measured ${\rm VBF}\to H\to VV$ channel.  Therefore limiting $\cv\le 1$ gives a strong constraint on ${\cal B}_{\rm new}$, 
similar to the case of truly invisible decays. Concretely we find at 95\%~CL:  
{\it i)}  ${\cal B}_{\rm new}<0.21$ for an SM Higgs with allowance for unseen decays; 
{\it ii)}  ${\cal B}_{\rm new}<0.39$ for $\cu=\cd=\cv=1$ but $\dcp,\dcg$ allowed for; and 
{\it iii)}  ${\cal B}_{\rm new}<0.31$ for $\cu,\cd$ free, $\cv\leq 1$ and $\dcp=\dcg=0$. 
For unconstrained  $C_U$, $C_D$ and $C_V$, however, there is no limit on ${\cal B}_{\rm new}$.

\begin{figure}[t!]\centering
\includegraphics[width=6cm]{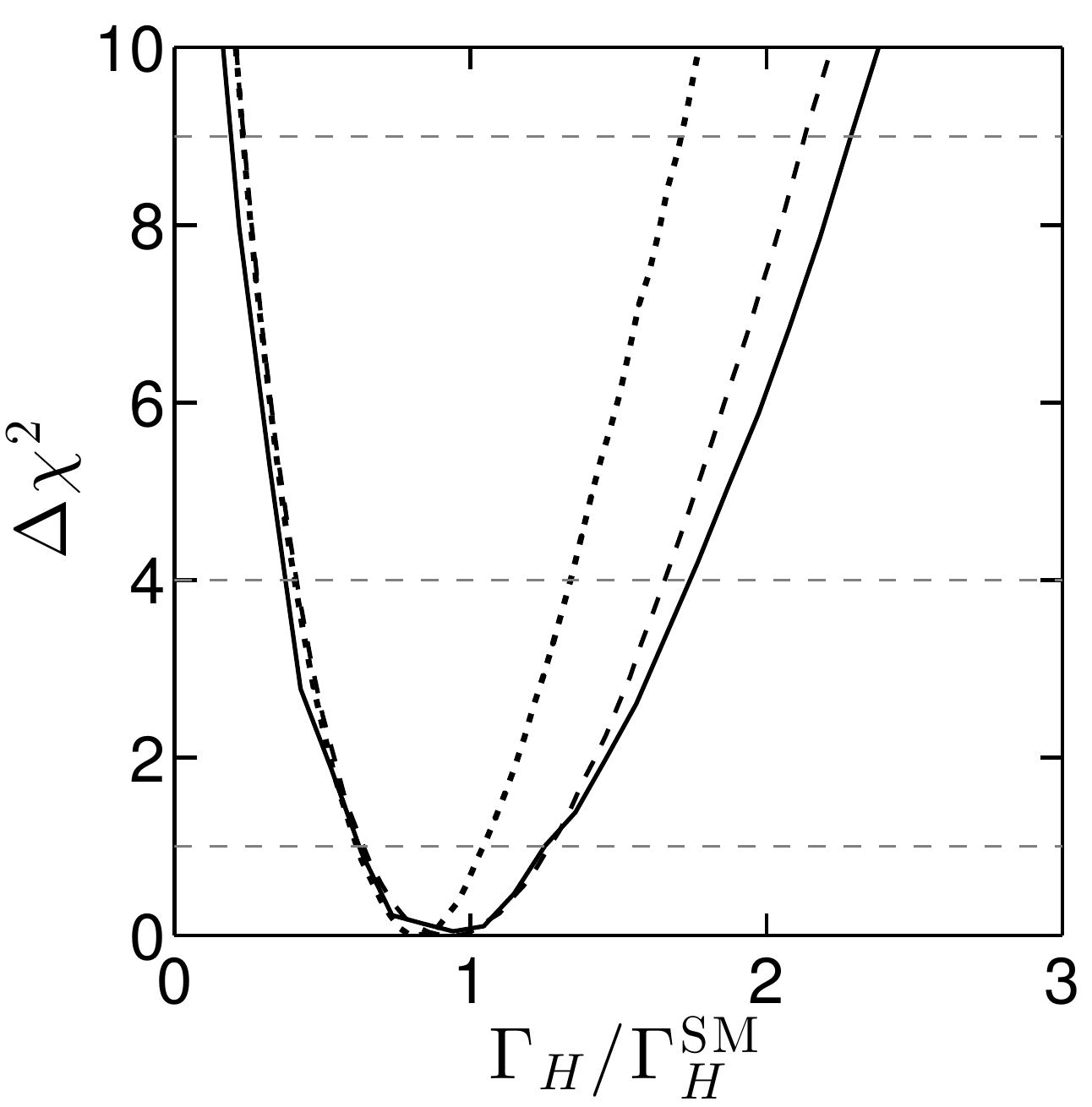}\quad
\includegraphics[width=6cm]{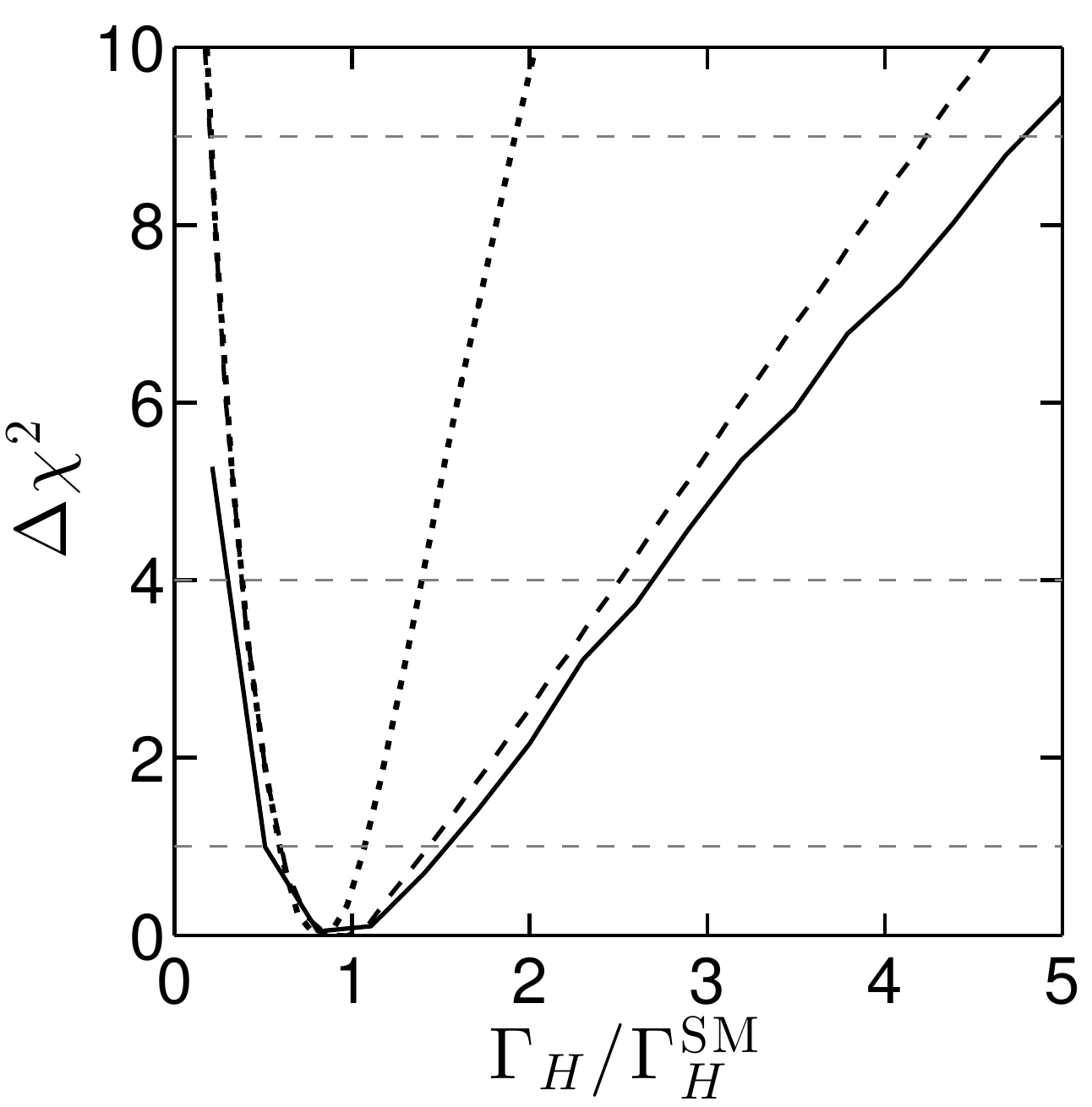}
\caption{$\Delta\chi^2$ distributions for the total Higgs decay width relative to SM, $\Gamma_{\rm tot}/\Gamma_{\rm tot}^{\rm SM}$, in the left panel without invisible decays, in the right panel including $\brinv$ as a free parameter in the fit. The lines are for:   
$\CU$, $\CD$ and $\CV\le1$ (dotted);
$\CU$, $\CD$ and free $\CV$ (dashed); and 
$\CU$, $\CD$, free $\CV$, $\dcg$, $\dcp$ (solid).
\label{2013c-fig:Rwidth} }
\end{figure}

With this in mind, 
the global fit we perform here also makes it possible to constrain the Higgs boson's total decay width, $\Gamma_{\rm tot}$, 
a quantity which is not directly measurable at the LHC. 
For SM + invisible decays, we find 
$\Gamma_{\rm tot}/\Gamma_{\rm tot}^{\rm SM}<1.11$ (1.25) at 68\% (95\%) CL. 
Fig.~\ref{2013c-fig:Rwidth} shows the $\Delta\chi^2$ as function of 
$\Gamma_{\rm tot}/\Gamma_{\rm tot}^{\rm SM}$ for the fits of:  $\CU$, $\CD$, and $\CV\le1$; 
$\CU$, $\CD$, and $\CV$ free; and $\CU$, $\CD$, $\CV$, $\dcg$, $\dcp$.  
The case of $\dcg$, $\dcp$ with $\cu=\cd=\cv=1$ is not shown;  without invisible decays we find 
$\Gamma_{\rm tot}/\Gamma_{\rm tot}^{\rm SM}=[0.98,1.0]$ ($[0.97,1.02]$) at 68\% (95\%) CL 
in this case. Allowing for invisible decays this changes to 
$\Gamma_{\rm tot}/\Gamma_{\rm tot}^{\rm SM}=[0.97,1.14]$, ($[0.96,1.46]$), \ie\ it is very close 
to the line for $\CU$, $\CD$, $\CV\le1$ in the right plot of Fig.~\ref{2013c-fig:Rwidth}.

\subsection{Interplay with direct dark matter searches} \label{2013c-sec:DMinterplay}

Assuming that the invisible particle which the Higgs potentially decays into is the dark matter of the Universe, 
the LHC bounds on $\brinv$ can be turned into bounds on the DM scattering off nucleons, mediated by Higgs exchange, {\it cf.}~\cite{Burgess:2000yq,Kanemura:2010sh,He:2011de,Mambrini:2011ik,Fox:2011pm,Djouadi:2011aa,Djouadi:2012zc}. These bounds are often much stronger than the current limits from XENON100~\cite{Aprile:2012nq} and LUX~\cite{Akerib:2013tjd} for $m_{\rm DM}<62$~GeV ({\it i.e.}, $m_H/2$).
Both the invisible width of the Higgs and the spin-independent cross section for scattering on protons depend on the 
square of the Higgs--DM--DM coupling $C_{\rm DM}$.  
If the DM is a Majorana fermion, $\chi$, the invisible width arising from $H\to \chi\chi$ decays is given by
\begin{equation}
   \Gamma_{\rm inv}=\Gamma(H \rightarrow \chi\chi) =\frac{g^2}{16\pi} m_H C_\chi^2 \beta^3 \, ,
\end{equation}
where $\beta=(1-4m_\chi^2/m_H^2)^{1/2}$ and $C_\chi$ is defined by ${\cal L}=gC_\chi \bar{\chi}\chi H$. 
In case of the DM being a real scalar, $\phi$, we have ${\cal L}=g m_\phi C_\phi \phi\phi H$ and 
\begin{equation}
   \Gamma_{\rm inv}=\Gamma(H \rightarrow \phi\phi) =\frac{g^2}{32\pi} \frac{m_\phi^2C_\phi^2}{m_H} \beta \, .
\end{equation}

The spin-independent cross section for scattering on a nucleon, considering only the Higgs exchange diagram, 
can then be  directly related to the invisible width of the Higgs: 
\begin{equation}
   \sigma_{\rm SI}= \eta \mu_r^2 m_p^2  \frac{g^2}{M_W^2}\, \Gamma_{\rm inv} 
   \left[  C_U (f_u^N+f_c^N+f_t^N) +C_D (f_d^N+f_s^N+f_b^N)+ \frac{\Delta C_g}{\what C_g}f_g^N\right]^2 \,,
   \label{eq:sigSI}
\end{equation}
with $\eta=4/(m_H^5 \beta^3)$ for a Majorana fermion and 
$\eta=2/(m_H^3 m_\phi^2\beta)$ for a real scalar;  
$\mu_r$ is the reduced mass and $f_q^N\,(f_g^N)$ are the quark (gluon) coefficients in the nucleon. 
We take the values  $f_s^p=0.0447$, $f_u^p=0.0135$, and $f_d^p=0.0203$ 
from an average of recent lattice results~\cite{Junnarkar:2013ac,Belanger:2013oya}. The gluon and heavy quark ($Q=c,b,t$) coefficients are related to those of light quarks, and $f_Q^p= 2/27 f_g^p = 2/27(1-\sum_{q=u,d,s} f_q^p)$ at leading order. 
Since the contribution of heavy quarks to the scattering amplitude originates from their contribution to the $Hgg$  coupling, 
we write the effect of $\dcg$, the last term in Eq.~(\ref{eq:sigSI}), in terms of an additional top quark contributing to the $Hgg$ coupling; numerically  $\what C_g=\overline C_g=1.052$ with only the SM top-quark contribution taken into account for computing $\overline C_g$.

For the numerical evaluation of $\sigma_{\rm SI}$, we use {\tt micrOMEGAs}~\cite{Belanger:2008sj,Belanger:2013oya} in which the relation between the heavy quark coefficients and the light ones are modified by QCD corrections.  This amounts to taking
\begin{equation}
C_Q f_Q^p\rightarrow  C_Q \left (1+\frac{35 \alpha_s(m_Q)}{36 \pi} \right) f_Q^p \,, \qquad \Delta C_g f_g^p \rightarrow \Delta C_g \left(1-\frac{16 \alpha_s(m_t)}{9 \pi} \right) f_g^p \,.
\end{equation}

\begin{figure}[t]\centering
\includegraphics[scale=0.5]{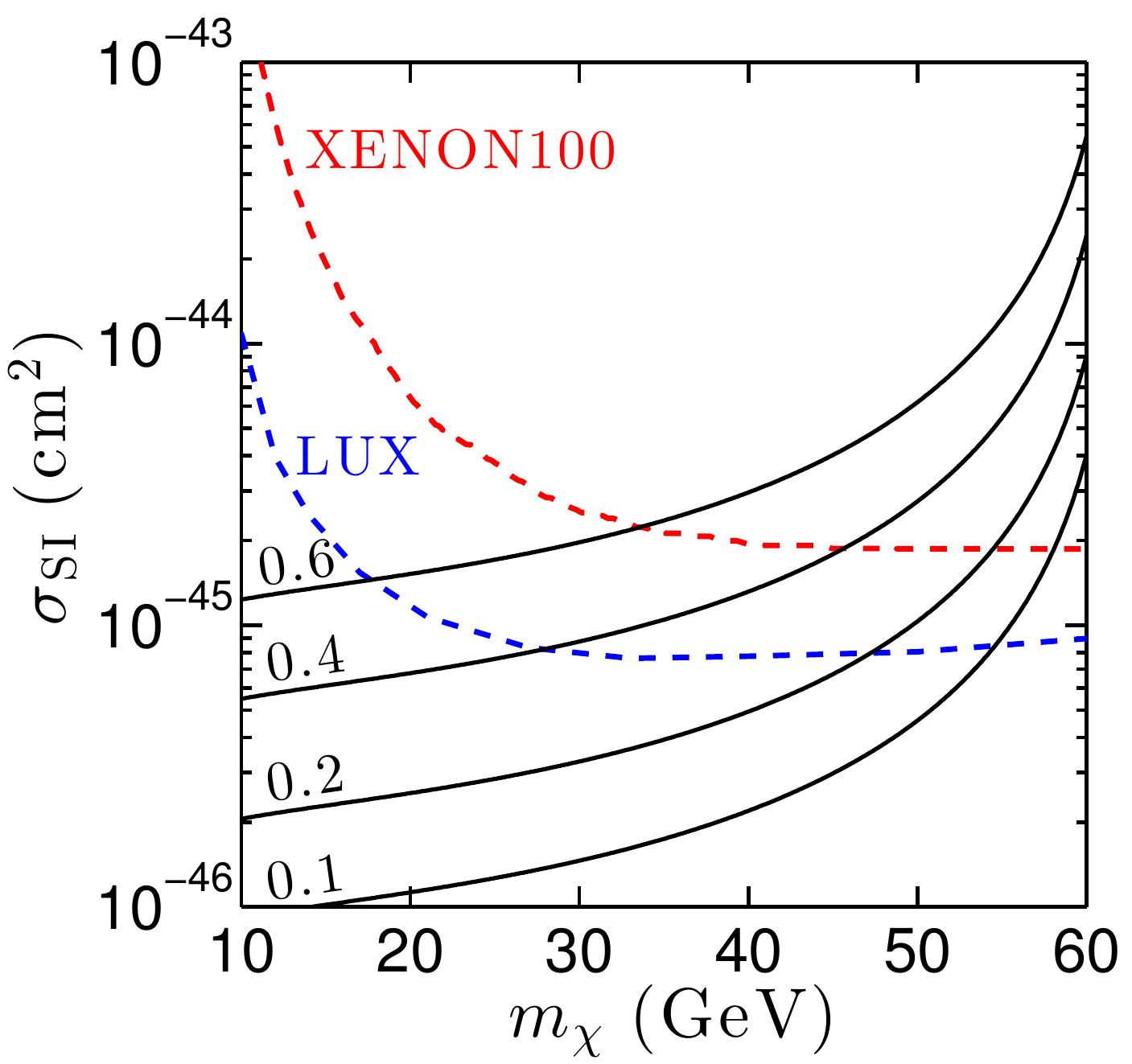}\quad\includegraphics[scale=0.5]{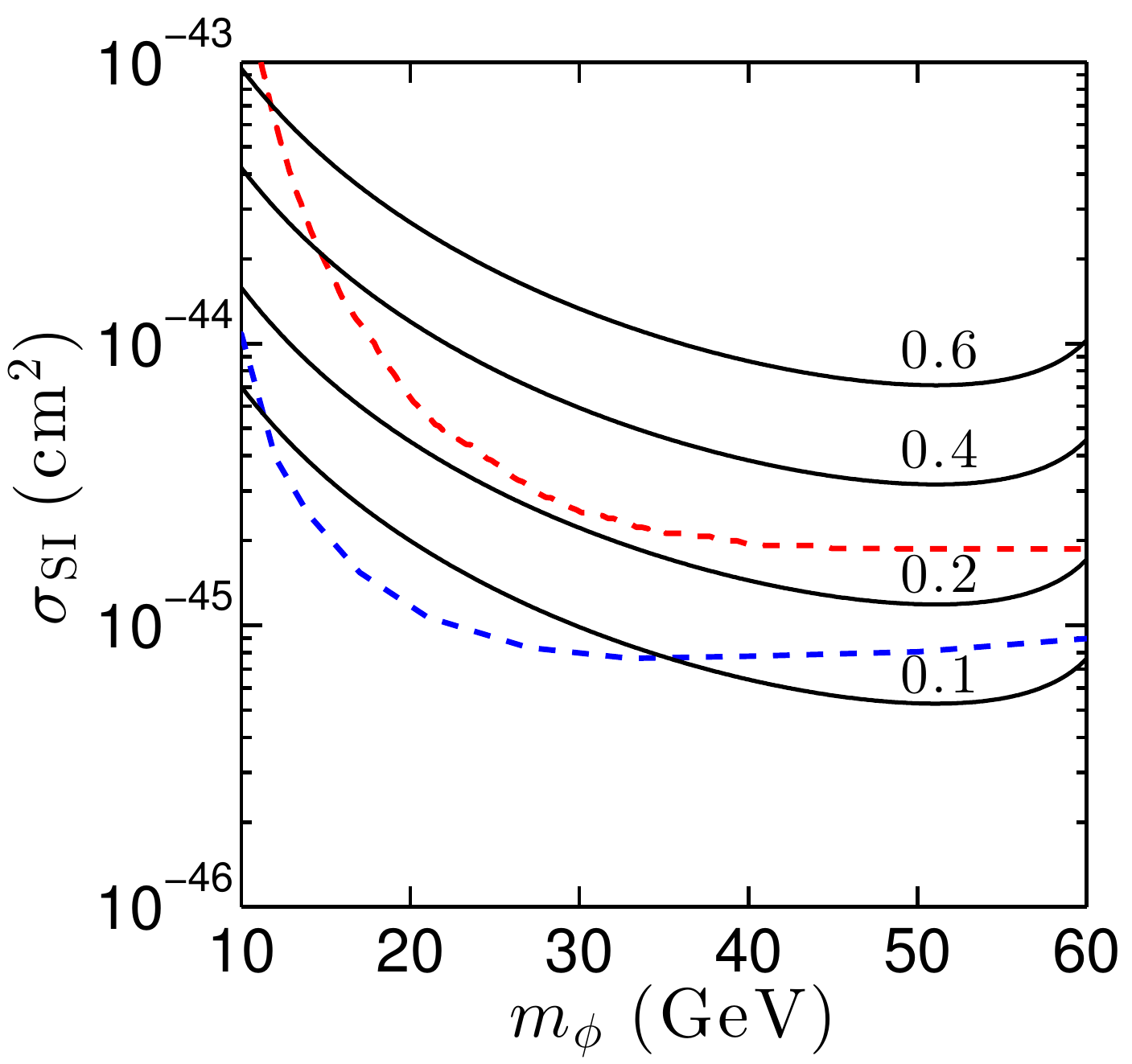}
\caption{$\sigma_{\rm SI}$ as a function of the mass of the DM particle, for  $\brinv=0.1,0.2,0.4,0.6$ (from bottom to top) for the case of a Majorana $\chi$ (left panel) or a real scalar $\phi$ (right panel) when $C_U=C_D=C_V=1$ and $\Delta\cg=\Delta\cp=0$, \ie\ an SM Higgs plus invisible decays. 
The red (blue) dashed curves show the XENON100 (LUX) exclusion limit at 90\%~CL.
\label{BRinv-sigmaSI} }
\end{figure}

The results for $\sigma_{\rm SI}$ versus the DM mass and for different $\brinv$ are displayed in Fig.~\ref{BRinv-sigmaSI} for a Majorana fermion (left panel)\footnote{For a Dirac fermion, the cross sections are a factor 1/2 smaller.}
and a real scalar (right panel) assuming SM-like couplings of the Higgs boson. 
As can be seen, for a Majorana fermion the current LUX limits~\cite{Akerib:2013tjd} exclude, for example, 
$\brinv>0.4$ when $28~{\rm GeV} < m_\chi < m_H/2$. For scalar DM, the cross sections are larger, and 
LUX excludes $\brinv>0.2$ for any $m_\phi$ in the $[10~{\rm GeV},m_H/2]$ range. 
These limits become much stronger when $C_U$ and/or $C_D$ are large provided they have the same sign. 
Further, these limits become stronger  when we include a non-zero value of $\Delta C_g$. For example, for  $\Delta C_g=1$  we find that  $\sigma_{\rm SI}$ increases by a factor 1.8 as compared to the case $\Delta C_g=0$  for any given value of $\brinv$. This increase is due in part to the new contribution in Eq.~(\ref{eq:sigSI}) and in part because  a larger coupling of the DM to the Higgs is necessary to maintain a constant value of $\brinv$.
Note that  imposing universality of quark couplings to the Higgs  has an impact on our  predictions for 
$\sigma_{\rm SI}$ since all quark flavors contribute to this observable,  whereas  universality plays basically no role for Higgs decays as only the third generation is important.  
 
When $C_U<0$ and $C_D>0$, 
there is a destructive interference between the $u$-type and $d$-type quark  contributions such that $\sigma_{\rm SI}$ is much below the current limit. Note however that this is clearly disfavored by the latest data. 
When the DM candidate is a Dirac fermion and one assumes the same amount of matter and anti-matter in the early Universe, the results for $\sigma_{\rm SI}$ are  simply a factor $1/2$ lower then those obtained in  the Majorana case. However if this fermion also couples to the $Z$, this gives an additional positive contribution to  $\sigma_{\rm SI}$, thus leading to stronger constraints from direct detection experiments.  Similar arguments hold for the case of a complex scalar, as compared to a real scalar.

\subsection{Application to two-Higgs-Doublet Models} \label{2013c-sec:2HDM}

So far our fits have been largely model-independent, relying only on assuming the 
Lagrangian structure of the SM. Let us now apply our fits to some concrete examples 
of specific models in which there are relations between some of the coupling factors $C_I$.
As a first example, we consider Two-Higgs-Doublet Models (2HDMs) of Type~I and Type~II 
(see also \cite{Altmannshofer:2012ar,Chang:2012ve,Chen:2013kt,Celis:2013rcs,Grinstein:2013npa,Coleppa:2013dya,Chen:2013rba,Eberhardt:2013uba,Craig:2013hca,Maiani:2013nga} %
for other 2HDM analyses in the light of recent LHC data). 
In both cases, the basic parameters describing the coupling of either the light $h$ or heavy $H$ CP-even 
Higgs boson are only two: $\alpha$ (the CP-even Higgs mixing angle) and $\tanb=v_2/v_1$, where $v_2$ and $v_1$ are the two vacuum expectation values. 
 The Type~I and Type~II models are distinguished by the pattern of their fermionic couplings 
as given in Table~\ref{1212.5244fermcoups}.  The SM limit for the $h$ ($H$) in the case of both Type~I and Type~II models corresponds to $\alpha=\beta-\pi/2$ ($\alpha=\beta$).  
We implicitly assume that there are no contributions from non-SM particles to the loop 
diagrams for $\cp$ and $\cg$.  In particular, this means our results correspond to the case where the charged 
Higgs boson, whose loop might contribute to $\cp$, is heavy.

\begin{figure}[t!]\centering
\includegraphics[width=6.1cm]{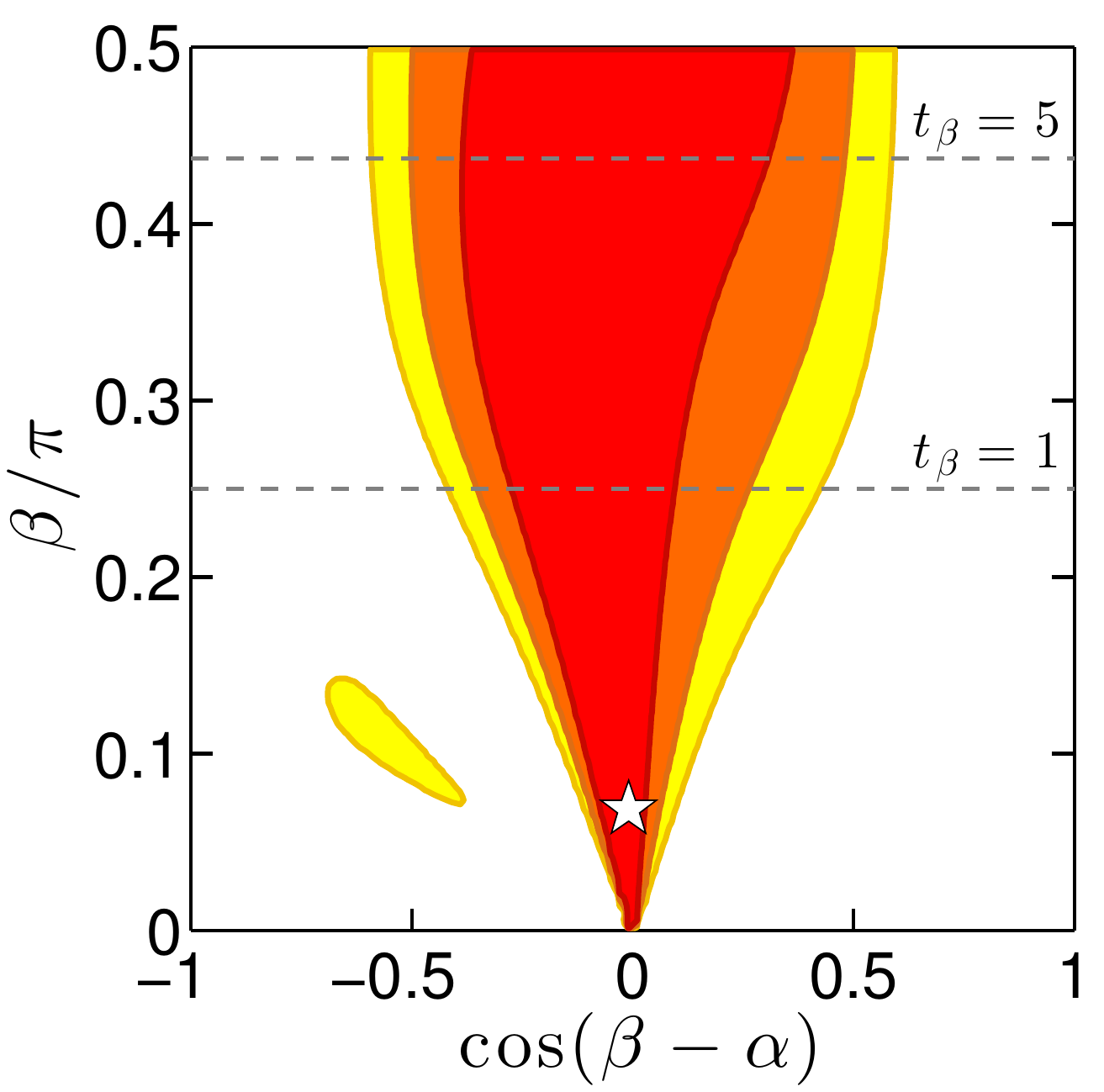}\quad
\includegraphics[width=6.1cm]{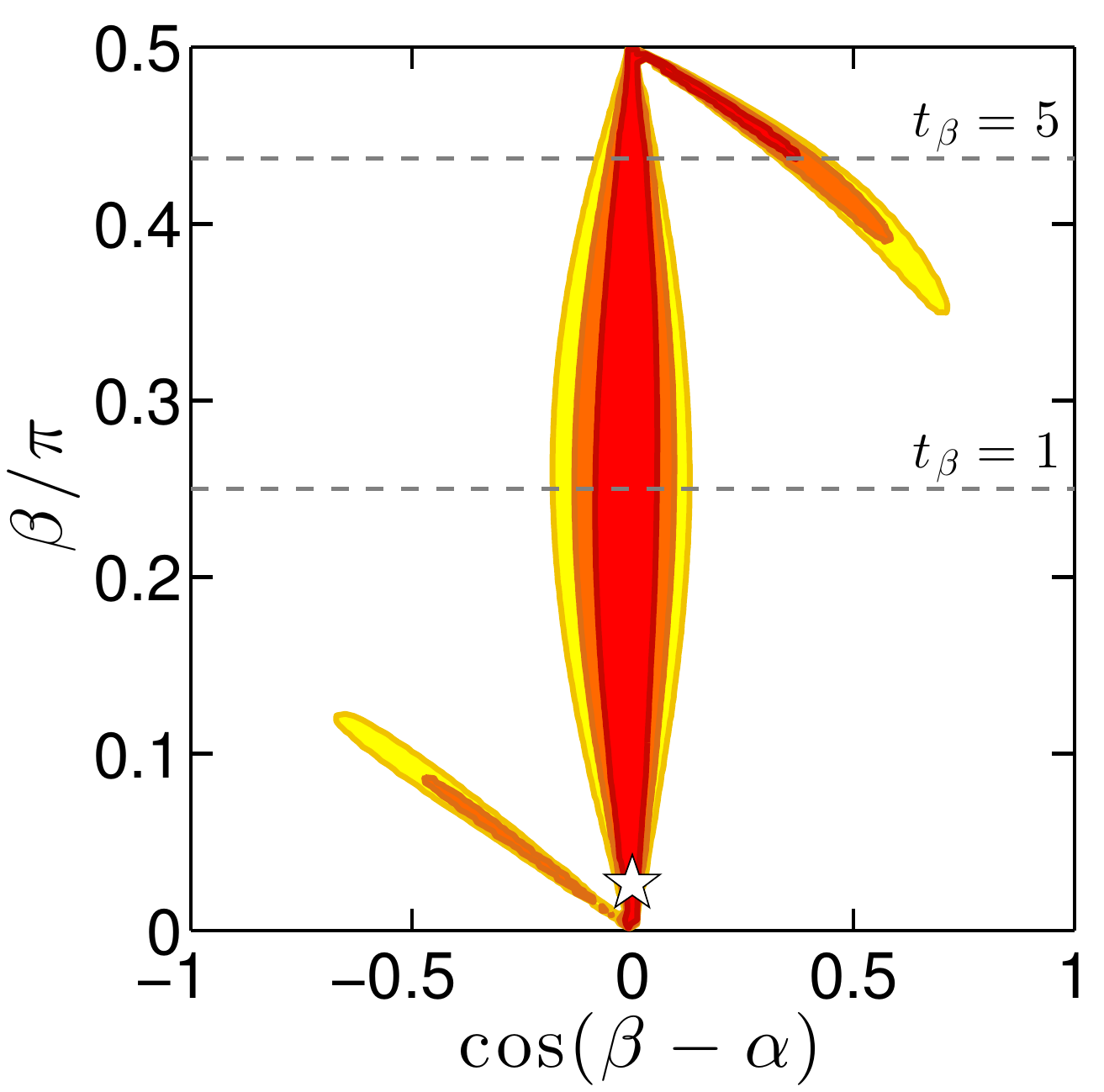}\\
\includegraphics[width=6cm]{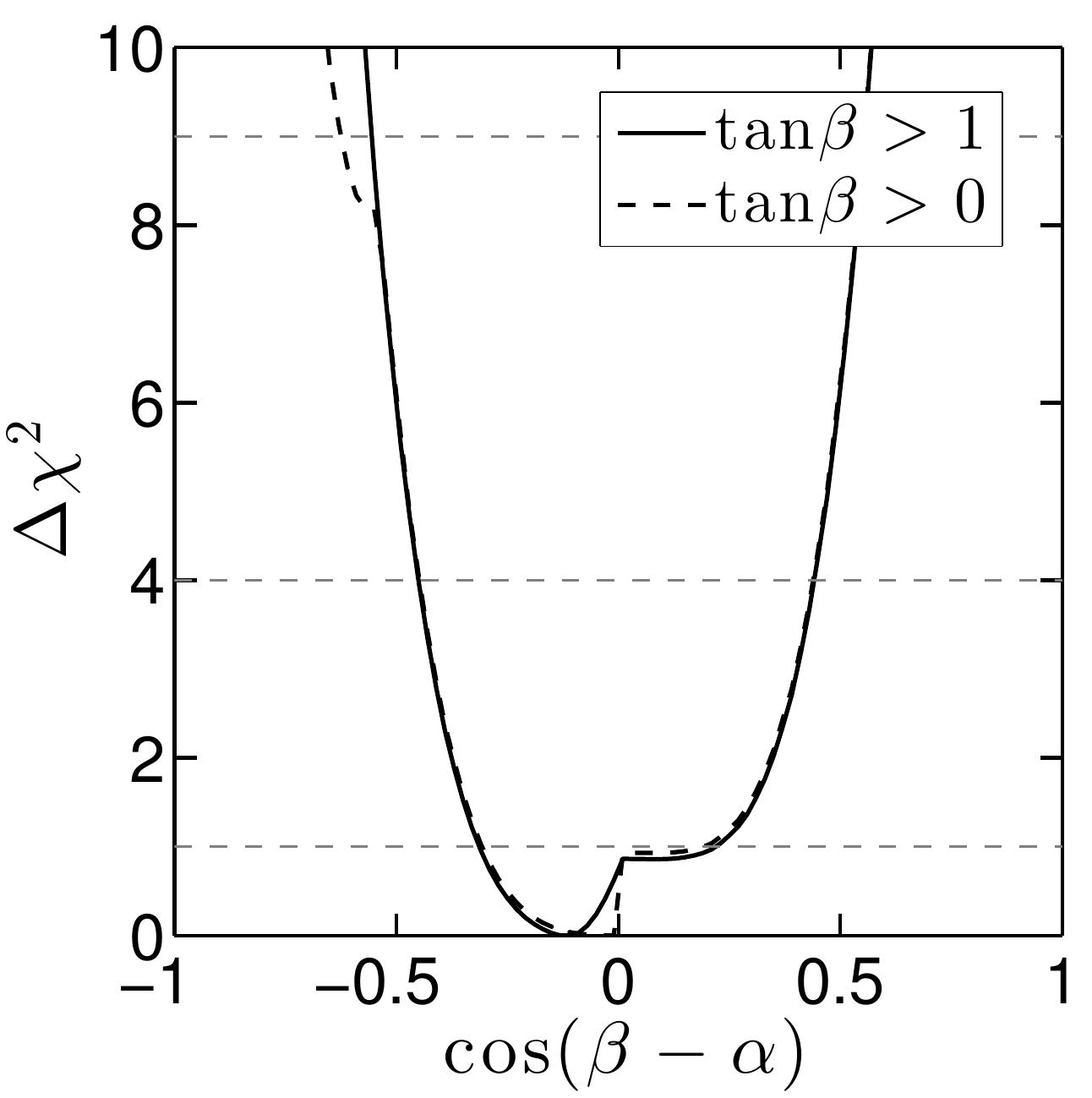}\quad
\includegraphics[width=6cm]{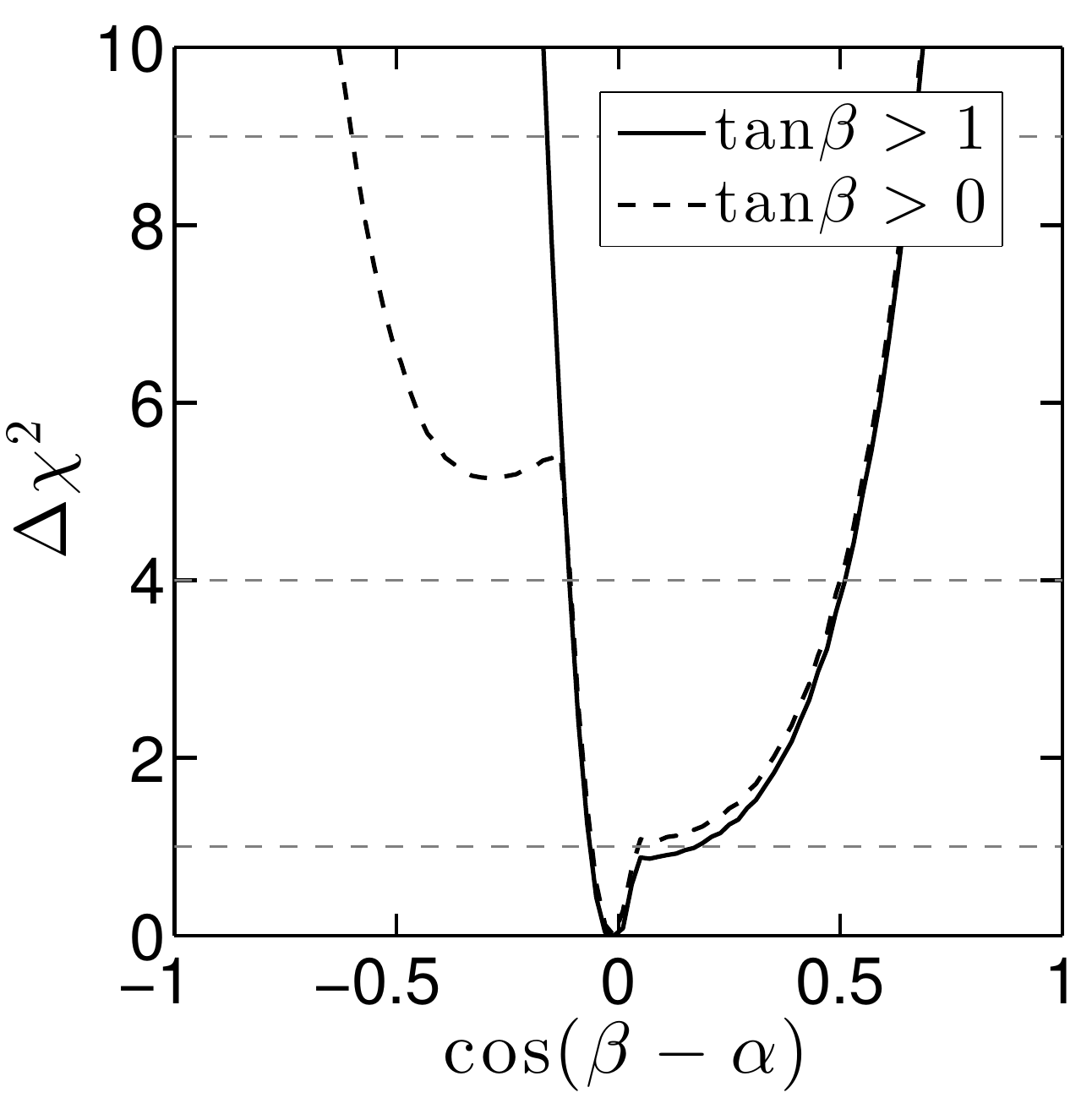}
\caption{Fits for the  2HDM Type~I (left) and type~II (right) models for
$m_h=125.5$~GeV.   See text for details. \label{2013c-fig:2hdm} 
}
\end{figure}

The results of the 2HDM fits are shown in Fig.~\ref{2013c-fig:2hdm} for the case that the state near 125~GeV
is the lighter CP-even $h$. To be precise, the top row shows $\Delta\chi^2$ contours in the 
$\beta$ versus $\cos(\beta-\alpha)$ plane while the bottom row shows the 1D projection of   
$\Delta\chi^2$ onto  $\cos(\beta-\alpha)$ with $\beta$ profiled over. 
For identifying the heavier $H$ with the state near 125~GeV, replace $\cos(\beta-\alpha)$ by $\sin(\beta-\alpha)$ in the 1D plots. (Since the $\sim 125\gev$ state clearly couples to $WW,ZZ$ we do not consider the case where the $A$ is the only state at $\sim 125\gev$.)

In the case of the Type~I model, we note a rather broad valley along the SM limit of $\cos(\beta-\alpha)=0$, 
which is rather flat in $\tan\beta$; the 68\% (95\%) CL region extends to 
$\cos(\beta-\alpha) = [-0.31,\, 0.19]$ ($[-0.45,\,0.44]$). 
The best fit point lies at $\beta\simeq0.02\pi$ and $\alpha\simeq 1.52\pi$ 
with $\chimin=18.01$ for 21 d.o.f.\ (to be compared to the SM $\chimin=18.95$). 
Requiring $\tan\beta>1$, this moves to $\beta\simeq0.25\pi$, 
\ie\ $\tan\beta$ just above 1, with  $\alpha\simeq 1.71\pi$ and $\chimin=18.08$.
At 99.7\% CL, there is also a small island at $\cos(\beta-\alpha)\approx -0.5$ and $\tan\beta<1$, 
which corresponds to the $\CU<0$ solution. (This is responsible for the splitting of the two lines at $\cos(\beta-\alpha)\lesssim-0.5$ in the 1D plot.)

In contrast, for the Type~II  model, we observe two narrow 68\%~CL valleys in the $\beta$ versus 
$\cos(\beta-\alpha)$ plane, one along the SM solution with the minimum again very close 
to $\beta\approx 0$ and a second banana-shaped one with $\tan\beta\gtrsim 5$ (3) and 
$\cos(\beta-\alpha)\lesssim 0.4$ (0.6) at 68\% (95\%) CL. 
This second valley is the degenerate solution with $\CD\approx-1$; it does not appear in Fig.~3 
of \cite{Craig:2013hca} because there $\CU,\CD>0$ was implicitly assumed.
The best fit point is very similar to that for Type~I: $\beta\simeq0.01\pi$ ($0.25\pi$) and $\alpha\simeq 1.5\pi$ ($1.75\pi$) with $\chimin=18.68$ ($18.86$) for 21 d.o.f.\ for arbitrary $\tan\beta$ ($\tan\beta>1$).  
Again, there is an additional valley very close to $\beta\sim 0$, extending into the negative $\cos(\beta-\alpha)$ direction, which however does not have a 68\%~CL region. 
In 1D, we find $\cos(\beta-\alpha) = [-0.11,\,0.50]$ at 95\%~CL.

Let us end the 2HDM discussion with some comments regarding the ``other''
scalar and/or the pseudoscalar $A$. To simplify the discussion, we will focus on the $m_h=125.5\gev$ case.
First, we note that if the $H$ and $A$ are heavy enough (having masses greater than roughly $600\gev$) then their properties are unconstrained by LHC data and the global fits for the $h$ will be unaffected. If they are lighter then it becomes interesting to consider constraints that might arise from not having observed them. Such constraints will, of course, depend upon their postulated masses, both of which are independent parameters in the general 2HDM. For purposes of discussion, let us neglect the possibly very important decays into the 125.5~GeV Higgs boson (such as $H\to hh$).  The most relevant final states are then $H\to VV$ and $H,A\to \tau\tau$.

With regard to observing the heavy Higgs in the $H \to VV$ channels, we note that for the $H$ our fits predict the $VV$ coupling to be very much suppressed in a large part (but not all) of the 95\%~CL allowed region. While this implies suppression of the VBF production mode for the $H$ it does not affect the ggF production mode and except for very small $VV$ coupling the branching fraction of the $H$ to $VV$ final states declines only modestly. As a result, the limits in the $ZZ\to4\ell$ channel \cite{ATLAS-CONF-2013-013}, 
which already extend down to about $0.1\times$SM 
in the mass range $m_H\approx 180-400$~GeV, and to about  $0.8\times$SM 
at $m_H\approx 600$~GeV, can be quite relevant. 
For instance, for a heavy scalar $H$ of mass $m_H=300$~GeV, in the 95\%~CL region of our fits 
the signal strength in the $gg\to H\to ZZ$
channel ranges from 0 to 5.4 in Type~I and from 0 to 33 in Type~II. For $m_H=600$~GeV, 
we find $\mu(gg\to H\to ZZ)\lesssim 1.1$ (0.6) in Type~I (II).  Further, at the best-fit point for $\tanb>1$, $\mu(gg\to H\to ZZ)=1.10~(0.08)$ at $m_H=300~(600)\gev$ in Type~I and  $\mu(gg\to H\to ZZ)=0.12~(0.001)$ at $m_H=300~(600)\gev$ in Type~II, which violate the nominal limits at $m_H=300\gev$ in both models.  
Note, however, that it is possible to completely evade the $4\ell$ bounds
if $H\rightarrow hh$ decays are dominant.

Moreover, both the $H$ and the $A$, which has no tree-level couplings to $VV$, may 
show up in the $\tau\tau$ final state through ggF. Limits from ATLAS \cite{Aad:2012cfr} 
range (roughly) from $\mu(gg\to H,A \to \tau\tau)$   $< 2500$ at $m_{H,A}=300\gev$ to $<21000$ at $m_{H,A}=500\gev$. 
These may seem rather weak limits, but in fact the signal strengths for $H\to\tau\tau$ and $A\to\tau\tau$ 
(relative to $H_{\rm SM}$) can be extremely large. In the case of the $A$, this is because the $A\to\tau\tau$ branching fraction is generically much larger than the $H_{\rm SM}\to\tau\tau$ branching fraction, the latter being dominated by $VV$ final states at high mass.  In the case of the $H$, the same statement applies whenever its $VV$ coupling  is greatly suppressed.
We find that only the Type~I model with $\tan\beta>1$ completely evades the $\tau\tau$ 
bounds throughout the 95\%~CL region of the $h$ fit since both the fermionic couplings of $H$ and $A$ 
are suppressed by large $\tan\beta$.
In the Type~II model, $gg\to A\to\tau\tau$ satisfies the $\tau\tau$ bounds at 95\%~CL, but 
$gg\to H\to\tau\tau$  can give a very large signal. However, the best fit $h$ point for $\tanb>1$ in Type~II 
predicts  $\mu(gg\to H\to\tau\tau)$ values of $674$ and $6.4$ at $300$ and $500\gev$, both of which  
satisfy the earlier-stated bounds. We also stress that no bounds are available in the $\tau\tau$ 
channel above 500~GeV. 

Clearly, a full study is needed to ascertain the extent to which limits in the $H\to ZZ$ and $H,A\to \tau\tau$ 
channels will impact the portion of the $\alpha$ --- $\beta$ plane allowed at 95\%~CL after taking into account  
Higgs-to-Higgs decays, which are typically substantial.
This is beyond the scope of this section and will be presented elsewhere~\cite{Dumont:2014wha}.

\subsection{Application to the Inert Doublet Model} \label{2013c-sec:IDM}

In the Inert Doublet Model (IDM)~\cite{Deshpande:1977rw}, 
a Higgs doublet $\tilde H_2$ which is odd 
under a $Z_2$ symmetry is added to the SM leading to four new particles: a scalar $\tilde{H}$, 
a pseudoscalar $\tilde{A}$,  and two charged states $\tilde{H}^\pm$ in addition to the SM-like Higgs $h$.\footnote{For distinction with the 2HDM, 
we denote all IDM particles odd under $Z_2$ with a tilde.}
All other fields being even, this discrete symmetry not only guarantees that the lightest inert Higgs particle is stable, and thus a suitable dark matter candidate~\cite{Ma:2006km,Barbieri:2006dq,LopezHonorez:2006gr,Krawczyk:2013jta}, but also  prevents the coupling of any of the inert doublet particles to pairs of SM particles.
Therefore, the only modification to the SM-like Higgs couplings is through the charged Higgs contribution 
to $\Delta C_\gamma$. The scalar potential of the IDM  is given by
\begin{align}
V &= \mu_1^2 |H_1|^2 + \mu_2^2 |\tilde{H}_2|^2 +\lambda_1 |H_1|^4+\lambda_2|\tilde{H}_2|^4+\lambda_3|H_1|^2 |\tilde{H}_2|^2\nonumber\\
&  +\, \lambda_4 |H_1^\dagger \tilde{H}_2|^2 
         +\frac{\lambda_5}{2} \left[  \left( H_1^\dagger \tilde{H}_2 \right)^2 + {\rm h.c.}\right] \,,
\end{align}
where $\mu_2^2>-\lambda_3 v^2/2$ is required in order that $\tilde H_2^0$ not acquire a non-zero vev (which would violate the symmetry needed for $\tilde H$ to be a dark matter particle).
The crucial interactions implied by this potential are those coupling the light Higgs $h$ associated with the $H_1$ field to pairs of Higgs bosons coming from the $\tilde H_2$ field.
These are given by:  $-(2 m_W/g) \lambda_3 h \tilde{H}^+ \tilde{H}^-$,
 $-(2 m_W/g) \lambda_L h \tilde{H} \tilde{H}$ and  $-(2 m_W/g) \lambda_S h \tilde{A} \tilde{A}$ for the charged, scalar and pseudo scalar, respectively, where
\begin{equation}
\lambda_{L,S}= \frac{1}{2}(\lambda_3+\lambda_4\pm\lambda_5)\,.
\end{equation}
With these abbreviations, the Higgs masses at tree-level can be written as 
\beq
   m^2_h = \mu_1^2 + 3\lambda_1v^2,\quad
   m^2_{\tilde H,(\tilde A)} = \mu_2^2 + \lambda_{L(S)}\,v^2,\quad
   m^2_{\tilde H^\pm} = \mu_2^2 + \frac{1}{2}\lambda_3v^2\,.
   \label{2013c-eq:idm-mh}
\eeq
Moreover, the couplings to the inert charged and neutral Higgses are related by
\begin{equation}
\frac{\lambda_3}{2}=\frac{1}{v^2}\left( m_{\tilde{H}^+}^2-m_{\tilde{H}}^2 \right) + \lambda_L \,.
\label{2013c-eq:lambda3}
\end{equation}
It is important to note that a priori  $m^2_{\tilde H,\tilde A,\tilde H^+}$ are each free parameters and could be small enough that $h$ decays to a pair of the dark sector states would be present and possibly very important.  The $h\to \tilde H\tilde H$ and $h\to \tilde A\tilde A$ decays would be invisible and contribute to $\brinv$ for the $h$; $h\to \tilde H^+\tilde H^-$ decays would generally be visible so long as the $\tilde H^+$ was not closely degenerate with the $\tilde H$.

Theoretical constraints  impose some conditions on the couplings. Concretely, we assume a generic 
perturbativity upper bound $|\lambda_i|<4\pi$, which, when coupled with the vacuum stability 
and perturbative unitarity conditions on the potential, leads to $\lambda_3>-1.5$ and 
$\mu_2^2\gtrsim-4.5\times 10^4$~GeV$^2$~\cite{Krawczyk:2013jta,Swiezewska:2012ej}. We also adopt
a lower bound of $m_{\tilde{H}^\pm}>70$~GeV, as  derived from chargino limits at LEP~\cite{Pierce:2007ut,Lundstrom:2008ai}.
Note however that LHC exclusions for the SM Higgs do not apply to members of the inert doublet because 
{\em i)} they do not couple to fermions and {\em ii)} trilinear and quartic couplings to gauge bosons involve two inert Higgses. 

\begin{figure}[t]\centering
\includegraphics[width=6cm]{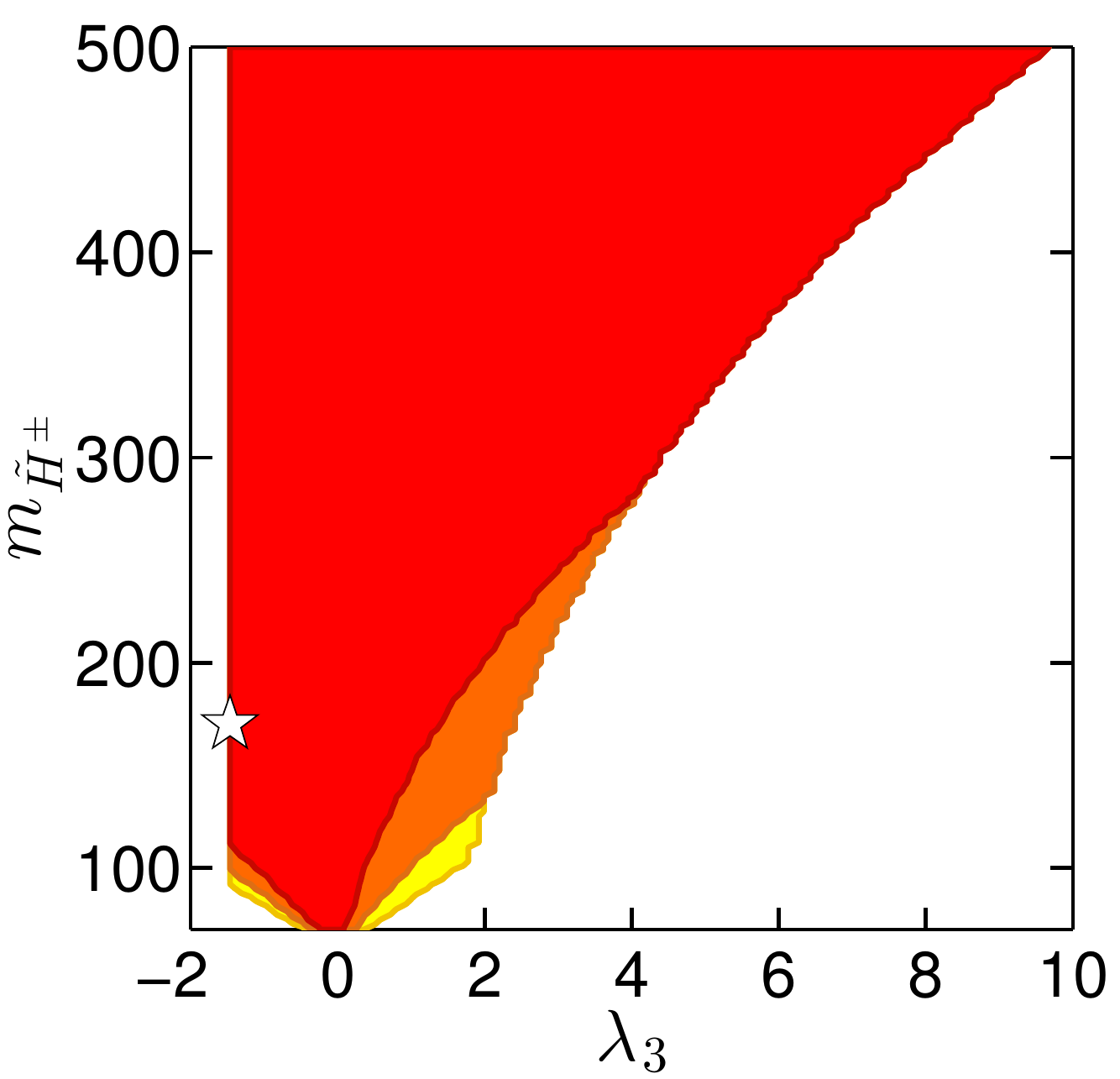}
\caption{Contours of 68\%, 95\%, 99.7\% CL in the $m_{\tilde{H}^\pm}$ versus $\lambda_3$ plane for the 
IDM assuming that there are no invisible decays of the SM-like Higgs $h$. 
\label{2013c-fig:idm1} }
\end{figure}

Let us now turn to the fit results.\footnote{In our IDM fits, the $h\gamma\gamma$ coupling is computed with {\tt micrOMEGAs~3}~\cite{Belanger:2013oya}.} 
First, we consider the case where $m_{\tilde{H}},m_{\tilde{A}}>m_h/2$---the only deviation from the SM then 
arises from the charged Higgs contribution to  $\Delta C_\gamma$ parametrized by $\lambda_3$ and $m_{\tilde{H}^\pm}$. 
The general one-parameter  fit to the Higgs couplings leads to the bounds  
$-0.02~(-0.13) <\Delta C_\gamma < 0.17~(0.26)$ at $1\sigma$ $(2\sigma)$. 
The corresponding contours in the $m_{\tilde{H}^\pm}$ versus $\lambda_3$ plane are shown in Fig.~\ref{2013c-fig:idm1}.  
Note that the 3rd equality of Eq.~(\ref{2013c-eq:idm-mh}) and the lower bound of $\mu_2^2\gtrsim-4.5\times 10^4$~GeV$^2 $ imply an upper bound on $\lam_3$ for any  given $m_{\tilde{H}^\pm}$.  This excludes the large-$\lam_3$ region when $m_{\tilde H^+}\gsim 130\gev$.  
The impact of the global fit is confined to the region $m_{\tilde H^+}\lsim 130\gev$ and $|\lam_3|\lsim 2$ (at 95\%~CL). The best fit point lies at $m_{\tilde H^+}=170$~GeV and $\lam_3=-1.47$.

\begin{figure}[t]\centering
\includegraphics[width=5.2cm]{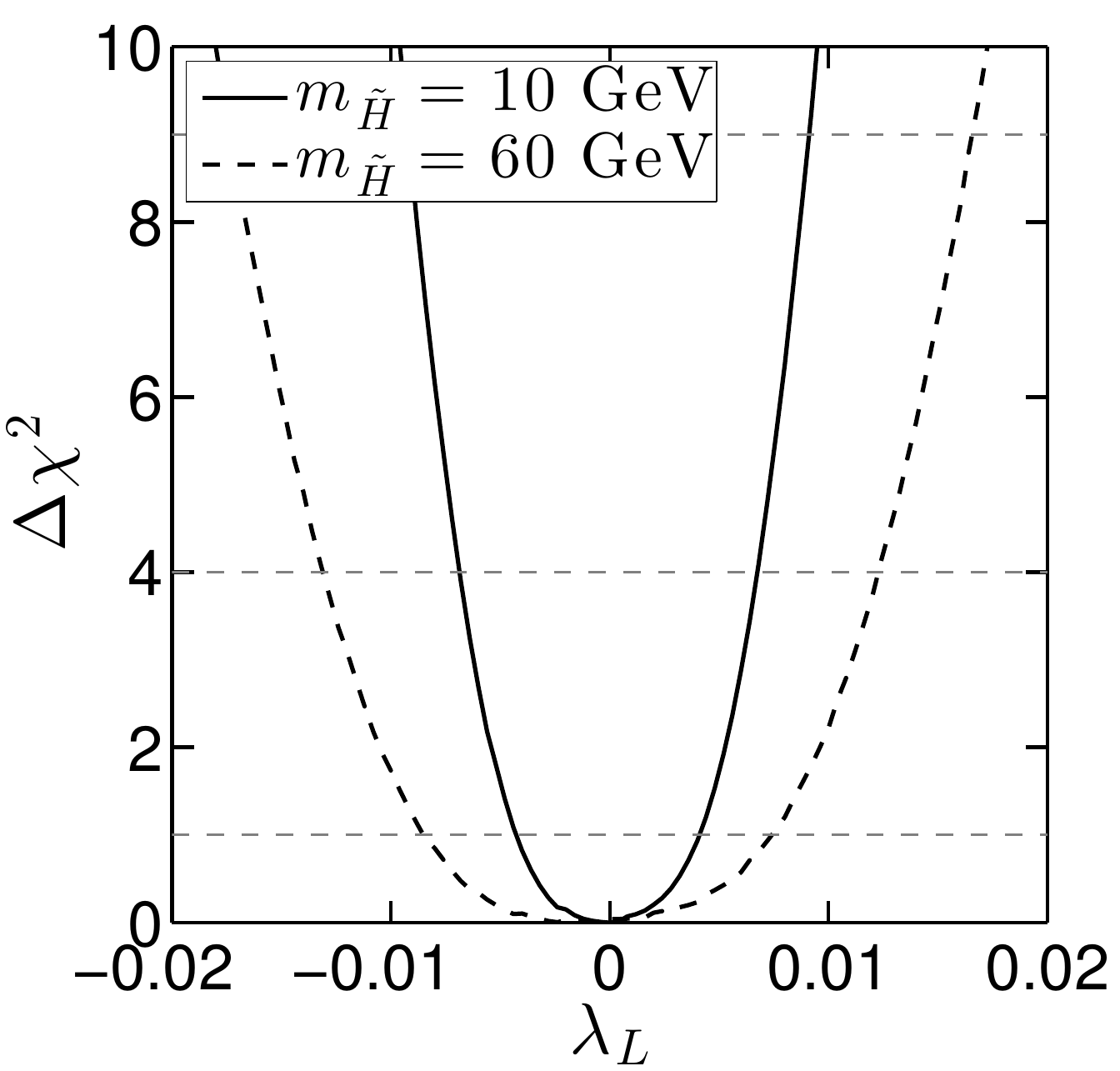}
\includegraphics[width=5cm]{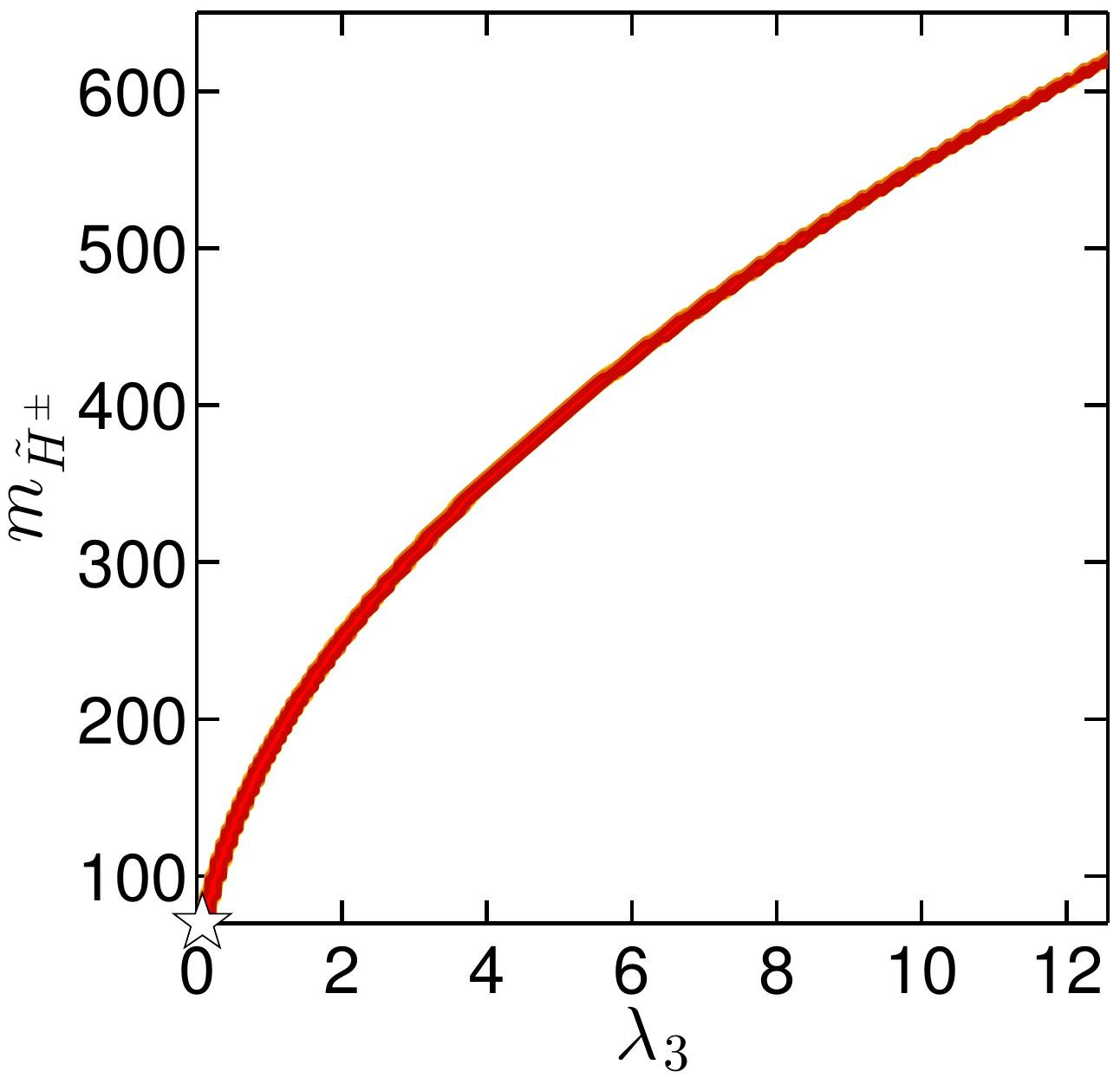}
\includegraphics[width=5cm]{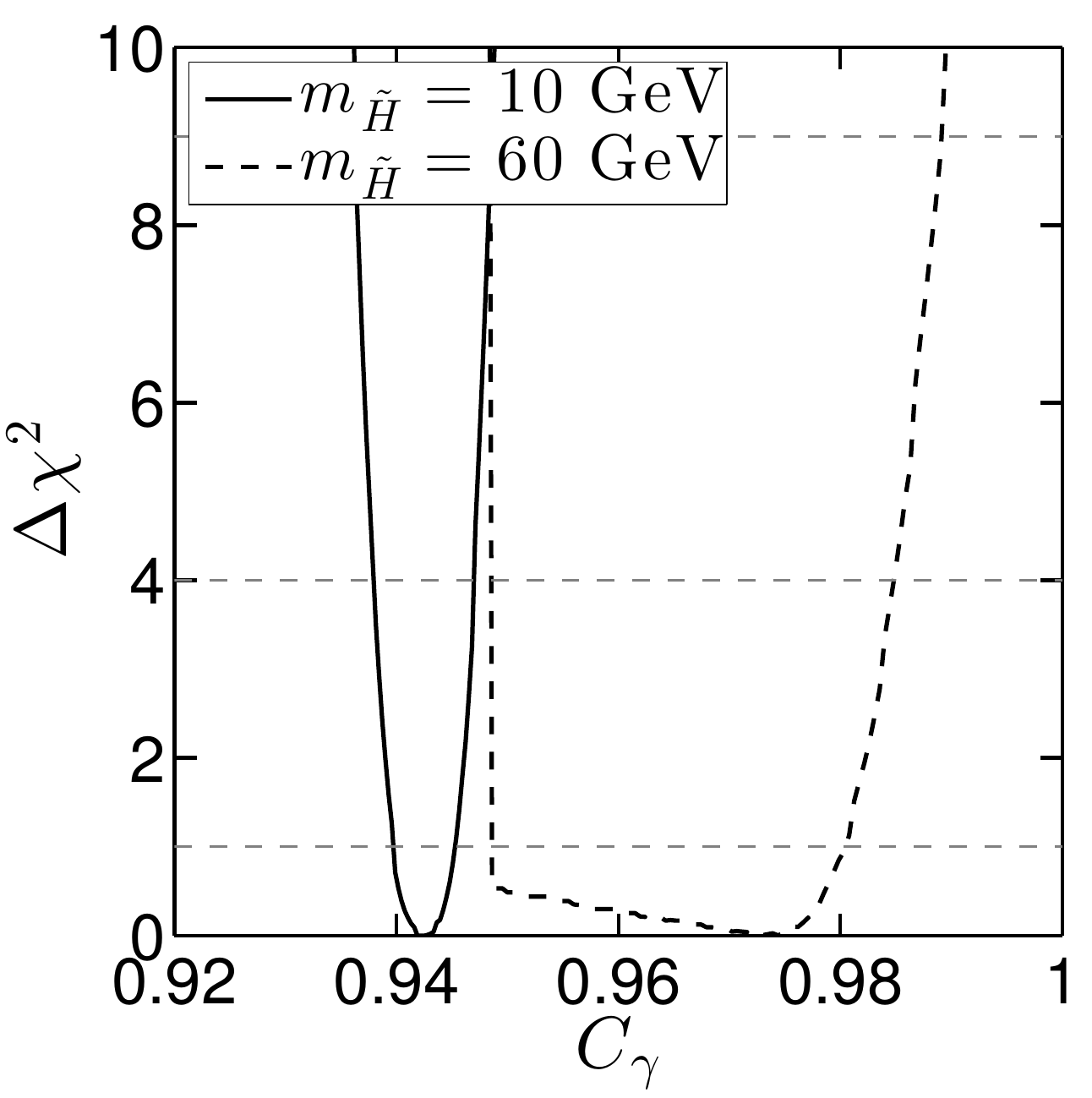}
\caption{Left panel: $\Delta\chi^2$ distribution of $\lambda_L$ for $m_{\tilde H}=10$~GeV  (full line)  and 60~GeV (dashed line) with $m_{\tilde{H^+}}$ profiled over its whole allowed range. 
Middle panel: relation between $m_{\tilde{H}^\pm}$ and $\lambda_3$ with $m_{\tilde H}$ profiled over from 1 to 60~GeV. 
Right panel: $\Delta\chi^2$ as function of $C_\gamma$ for $m_{\tilde H}=10$~GeV  (full line)  and 60~GeV (dashed line) with $m_{\tilde{H}^\pm}$ profiled over.  
\label{2013c-fig:idm2} }
\end{figure}

Second, we consider the case where the inert scalar is light and examine how invisible 
$h\to \tilde H\tilde H$ decays further constrain the parameters.
The bounds on the invisible width actually lead to a strong constraint on the coupling $\lambda_L$.  
The $1\sigma$ ($2\sigma$) allowed range is roughly $\lambda_L \times 10^3=\pm 4$  ($\pm 7$) 
for $m_{\tilde{H}}=10~{\rm GeV}$. This bound weakens only when the invisible decay is suppressed by 
kinematics; for  $m_{\tilde{H}}=60~{\rm GeV}$, we find $\lambda_L\times 10^{3}=[-9,7]$ ($[-13,12]$) at $1\sigma$ ($2\sigma$).  
The $\Delta\chi^2$ distributions of $\lambda_L$ for $m_{\tilde{H}}=10$ and 60~GeV are shown in the left panel 
in Fig.~\ref{2013c-fig:idm2}, with $m_{\tilde H^\pm}$ profiled over from 70~GeV to about 650~GeV (the concrete upper 
limit being determined by the perturbativity constraint). 
This strong constraint on $\lambda_L$ implies that it can be neglected in Eq.~(\ref{2013c-eq:lambda3}) and 
that the charged Higgs coupling $\lambda_3$ is directly related to $m_{\tilde{H}^\pm}$ for a given 
$m_{\tilde H}$,  as illustrated in the middle panel of Fig.~\ref{2013c-fig:idm2} 
(here, the mass of the inert scalar is profiled over in the range $m_{\tilde H} \in [1,\,60]$~GeV). 
As a result the value of $C_\gamma$ is also strongly constrained from the upper bound on the invisible width. 
For example for $m_{\tilde{H}}=10$~GeV, we find that $C_\gamma =[0.940,0.945]$ at 68\%~CL. 
Note that because $m_{\tilde{H}^\pm}> m_{\tilde{H}}$ is needed in order to have a neutral dark matter candidate, $\lambda_3$ is 
always positive  and therefore $C_\gamma<1$. 
To approach $C_\gamma\simeq 1$, the inert Higgs mass has to be close to the kinematic threshold,  $m_{\tilde H}\to m_h/2$ so that the constraint on $\lambda_L$ is relaxed. For illustration, see the right panel in Fig.~\ref{2013c-fig:idm2}. 
These results imply that with an improved accuracy on the measurements of the Higgs coupling, for example showing that $C_\gamma> 0.95$, it would be possible to exclude light dark matter  ($m_{\tilde H}<10$~GeV) in the IDM.
Another consequence is that for a given $m_{\tilde H}$ the perturbativity limit $\lambda_3<4\pi$ implies an upper bound on the charged Higgs mass. For $m_{\tilde H} \in [1,\,60]$~GeV we obtain $m_{\tilde{H}^\pm}<620$~GeV.

Finally note that the  case where $\tilde{A}$ is the lightest neutral state and $m_{\tilde{A}}< m_h/2$ 
is analogous to the $\tilde H$ case just discussed, with $m_{\tilde{H}}\rightarrow
m_{\tilde{A}}$ and $\lambda_L \rightarrow \lambda_S$ and leads to analogous conclusions. 
Analyses of the Higgs sector of the Inert Doublet Model were also performed recently  in~\cite{Krawczyk:2013jta,Goudelis:2013uca,Arhrib:2012ia,Gustafsson:2012aj,Swiezewska:2012eh}.

\subsection{Conclusions} \label{2013c-sec:conclusions}

The most general fits considered were those in which $\cu,\cd,\cv,\dcg,\dcp$ were all allowed to vary freely. 
If there are no unseen (as opposed to truly invisible) decay modes of the Higgs,
one finds that the observed $125.5\gev$ state prefers to have quite SM-like couplings 
whether or not  $\brinv=0$ is imposed --- 
more constrained fits, for example taking $\dcg=\dcp=0$ while allowing $\cu,\cd,\cv$ to vary, inevitably imply that the other parameters must lie even closer to their SM values.   

Allowing for invisible decays of the $125.5\gev$ state through $\brinv>0$ does not change the best-fit parameter values but does widen the $\dchisq$ distributions somewhat leading to important implications, \eg, 
for decays into dark matter particles.  In particular, we found that at 95\%~CL there is  still considerable room for such Higgs decays, up to $\brinv\sim 0.38$ when $\cu,\cd,\cv,\dcg,\dcp$ are all allowed to vary independently of one another. In comparison, a fit for which $\cu,\cd$ are allowed to vary freely, but $\cv\leq 1$ is required (as appropriate for any doublets+singlets model) and $\dcg=\dcp=0$ is imposed, yields $\brinv\lsim 0.24$ at 95\%~CL. Even requiring completely SM couplings for the Higgs ($\cu=\cd=\cv=1$, $\dcg=\dcp=0$) still allows $\brinv\leq 0.19$ at 95\%~CL.
It is worthwhile noting that for $\cv \leq 1$, the limits on $\brinv$ from global coupling fits are currently more constraining than those from direct searches for invisible decays, \eg, in the $ZH\to\ell^+\ell^- + E_T^{\rm miss}$ mode;  
thus for $\cv \leq 1$ the limits on  merely unseen (\ie\ not strictly invisible) decays are similar to the ones on $\brinv$.

As part of the fitting procedure, the total width of the Higgs relative to the SM prediction is computed as a function of the parameters and a $\dchisq$ distribution for $\Gamma_{\rm tot}/ \Gamma_{\rm tot}^{\rm SM}$ is obtained.  Assuming no unseen, but potentially visible, decays, we found $\Gamma_{\rm tot}/ \Gamma_{\rm tot}^{\rm SM}\in[0.5,2]$ at 95\%~CL for the case where $\cu,\cd,\cv,\dcg,\dcp$ and $\brinv$ are all allowed to vary freely, while $\Gamma_{\rm tot}/ \Gamma_{\rm tot}^{\rm SM}\in[1,1.25]$ at 95\%~CL if $\cu=\cd=\cv=1$, $\dcg=\dcp=0$ are imposed and only $\brinv\geq 0$ is allowed for.  These are useful limits given the inability to directly measure $\Gamma_{\rm tot}$ at the LHC.
Of course, if there are unseen (but not invisible) decays, there is a flat direction that would prevent setting limits on the total width.

Finally, we have also shown that if $\brinv\neq0$ is due to $H$ decays to a pair of DM particles, there are significant constraints on the size of $\brinv$ from the non-observation of spin-independent DM scattering on nucleons, the most important such limits currently being those from the LUX experiment.  
These constraints are much stronger for scalar DM than for Majorana or Dirac fermions. Overall, 
our results suggest a continued competition between limits on $\sigma_{\rm SI}$ and those on $\brinv$ as direct detection experiments achieve improved sensitivity and increasingly accurate measurements of the properties of the $H$ become available with future LHC running.

In the second part of the Section, we then examined implications of these results in the context of some simple concrete models with an extended Higgs sector: the Type~I and Type~II Two-Higgs-doublet models, and the Inert Doublet Model, using the combined likelihood ellipses to constrain the parameter spaces.
In the 2HDM, enhancement of the signal strength for a 2nd neutral (scalar or pseudoscalar) Higgs boson with mass above $125.5\gev$ can occur in both the $4\ell$ and $\tau\tau$ channels. Therefore additional constraints on $\alpha$ and $\beta$ can be set unless the decay of the heavier Higgs to a pair of the $125.5\gev$ states dominates.  Generally the signals in both channels can be  at a level of interest for future LHC runs.
In the Inert Doublet Model, the inert Higgs states can only be pair-produced and therefore are not currently constrained. However, we showed that the bound on the invisible decay of the $125.5$~GeV SM-like Higgs, relevant when one inert Higgs is lighter than $\approx 60$~GeV,  constrains the allowed range for the two-photon width.  Thus, a precise determination of $C_\gamma$  could rule out light inert Higgs dark matter.


\section[A Bayesian view of the Higgs sector with higher dimensional operators]{A Bayesian view of the Higgs sector with higher dimensional operators%
\sectionmark{A Bayesian view of the Higgs sector with HDO}}
\sectionmark{A Bayesian view of the Higgs sector with HDO} \label{sec:higgsdim6}


\sectionmark{A Bayesian view of the Higgs sector with HDO}

So far, constraints on new physics have been parametrized in the context of the Lagrangian shown in Eq.~\eqref{eq:1212.5244ldef}, which introduced reduced couplings to vector bosons ($C_W$, $C_Z$) and to fermions ($C_F$). In addition, there are {\it i)}\ effective contributions to the $\gamma\gamma$ and $gg$ loop-induced couplings, entering the Lagrangian as $H(F_{\mu\nu})^2$ and $H(G_{\mu\nu})^2$, respectively, and {\it ii)}\ generic invisible/undetected decays of the Higgs boson. This framework is fully justified and applicable to a very wide class of new physics models. It is however possible to take a different perspective in light of the negative LEP, Tevatron and LHC results in the search for new physics at or close to the electroweak scale.
If new physics is indeed present and is somehow separated from the electroweak scale, the couplings of the Higgs boson will be close to those of the SM and will only be modified by the effect of a few higher dimensional operators (HDOs).
In this section, we will explore an effective field theory (EFT) with only relatively few new parameters. As we will see, most higher dimensional operators will not only have an effect on the Higgs phenomenology but also on measured quantities related to the electroweak sector, such as the Peskin--Takeuchi $S$ and $T$ parameters~\cite{Peskin:1990zt,Peskin:1991sw}. This will be taken into account and the interplay between the different observables will be discussed.

The statistical procedure is another difference with the work presented in Sections~\ref{sec:higgs2012} and \ref{sec:higgs2013}, where we considered a fully frequentist approach and derived confidence intervals for the parameters of interest after profiling over the other, unseen parameters. 
In this section, we will adopt the statistical framework of Bayesian inference, which allows us to assign probabilies to our parameters and to deal with partially constrained problems.
Another interesting property is that the unnatural (\ie~fine-tuned) character of precise cancellations which may occur between HDO contributions is built-in in this framework. Indeed, regions of the parameter space in which precise cancellations occur have by construction a weak statistical weight.

The work presented in this section has been conducted in collaboration with Sylvain Fichet and Gero von Gersdorff at the end of 2012 and at the beginning of 2013, in parallel to work presented in Sections~\ref{sec:higgs2012} and \ref{sec:higgs2013}. This lead to the paper ``A Bayesian view of the Higgs sector with higher dimensional operators'', Ref.~\cite{Dumont:2013wma}, that was submitted to arXiv on April 11, 2013 and published in JHEP in July 2013. 

The outline of the Section is as follows. In Section \ref{2013eft-se_HDOs}, we lay out the formalism for higher-dimension operators in the  electroweak sector. In Section \ref{2013eft-se_data}, we present the dataset used for the analysis and the measurements entering the likelihood functions. The peculiar case of observables sensitive to tensorial couplings relating Higgs and weak bosons is investigated in Section \ref{2013eft-se_tensorials}. In Section \ref{2013eft-se_deviations}, we derive the observable deviations from the SM induced by the higher dimensional operators, taking into account leading NLO QCD effects. Section \ref{2013eft-se_bayesian_setup} presents the setup of our Bayesian analysis. Section \ref{2013eft-se_results} is devoted to our results. Our conclusions are given in Section \ref{2013eft-se_conclusions}.

\subsection{Electroweak higher-dimension operators}  \label{2013eft-se_HDOs}

We consider that new states appear at a typical scale $\Lambda$ substantially larger than the electroweak scale. For physical processes involving an energy scale smaller than $\Lambda$, new physics can be integrated out. As a consequence of this hypothesis, the resulting low-energy effective theory consists in the Standard Model, supplemented by infinite series of local operators with higher dimension, which involves negative powers of $\Lambda$, 
\beq
\mathcal{L}_{\rm eff}=\mathcal{L}_{\rm SM}+\sum_{i} \frac{\alpha_{i}}{\Lambda^{n_i}} \mathcal{O}_{i}\,. \label{2013eft-L_eff}
\eeq
The effects of such higher dimensional operators have been investigated in many contexts such as flavor physics, or the study of the properties of the electroweak gauge bosons through LEP precision measurements. The purpose of this work is to study the electroweak sector again, which now includes new Higgs observables. For our analysis, we only have to consider the leading HDOs. The only operator with $n_i=1$ is the one giving a Majorana mass to the neutrino, and is not relevant for our study. We will thus be exclusively interested  in the $n_i=2$ terms, \ie~dimension-$6$ operators.

In this section, we define the basis of dimension-6 operators  supplementing the renormalizable electroweak sector of  the SM Lagrangian.
We refer to \cite{Buchmuller:1985jz, Grzadkowski:2010es} for further details on the Standard Model HDOs. 
A basis of CP-even operators not involving fermions can be chosen as\footnote{The operator $\mathcal O_6$ plays no role in what follows and is listed here only for completeness.}
\begin{align}
\O_6=|H|^6\,,\qquad \O_{D^2}&=|H|^2|D_\mu H|^2\,,\qquad \O'_{D^2}=|H^\dagger D_\mu H|^2\,, \label{2013eft-HDO_O_6} \\
\O_{\WW}=H^\dagger H\,(W_{\mu\nu}^a)^2\,,\qquad 
\O_{\BB}&=H^\dagger H\,(B_{\mu\nu})^2\,,\qquad
\O_{\WB}=H^\dagger\,W_{\mu\nu} H\,B_{\mu\nu}\,, \label{2013eft-HDO_O_WW} \\
\O_{GG}&=H^\dagger H\,(G_{\mu\nu}^a)^2\,. \label{2013eft-HDO_O_GG}
\end{align}
Any other operator can be reduced to these via integration by parts and the use of the SM equations of motion for the Higgs and gauge fields, possibly generating operators involving fermions. 
Amongst the latter, only a limited set will be relevant for our purpose. 
Operators of the form $J_H\cdot J_{f}$, where $J_H$ and $J_{f}$ are $SU(2)$ or $U(1)_Y$ currents involving Higgs field and fermion $f$  respectively, will in general contribute to FCNC as well as electroweak non-oblique corrections (\eg, non-universal couplings of fermions to gauge bosons).\footnote{The non-universal corrections to the weak bosons couplings of the top quark are only very mildly constrained by EW data, and it is a priori not justified to set them to zero. 
However the only effect to Higgs observables at leading order is a modification of the top loop contribution to the $h\to Z\gamma$ decay due to the anomalous $Ztt$ vertex. The top contribution is however about one order of magnitude  smaller than the leading contribution from the $W$ loop \cite{Djouadi:2005gi}.  
We will therefore only consider universal (oblique) corrections to EW data.}
 However, the operators
\beq
\O_D=J^a_{H\,\mu}\,J^a_{\mu}\,,\qquad
\O_{D}'=J^Y_{H\,\mu}\,J^Y_{\mu}\,,
\eeq
where $J=\sum_f J_{f}$ are the SM fermion currents coupling to $B_\mu$ and $W_\mu$,
are flavor diagonal and only result in universal corrections to gauge couplings and should hence be viewed as contributing to $S$ and $T$.\footnote{In fact this is the way how contributions to $S$ and $T$ can arise in theories with new spin-1 states, such as in warped extra dimensions \cite{Davoudiasl:2009cd,Cabrer:2011fb}.}
We will also need to consider Yukawa corrections of the form
\beq
\O_f= 2 y_f\,|H|^2 \, H\bar f_L f_R \,, \label{2013eft-HDO_O_f}
\eeq
where $f_R=t_R,\ b_R,\ \tau_R$ and $f_L$ the corresponding doublet ($\bar f^a_L=\epsilon^{ab} \bar q^{b,3}_L,\ \bar q_L^{a,3},\bar\ell_L^{a,3}$) and $y_f$ the Yukawa coupling.

Note that the operators $\O_D$ and $\O_D'$ could be traded for the operators 
\beq
\O_{W}=(D_\mu H)^\dagger W_{\mu\nu}D_\nu H\,,\qquad \O_B=(D_\mu H)^\dagger D_\nu H\,B_{\mu\nu}\,,
\eeq
by use of the SM equations of motion for $B$ and $W$. While $\O_D$ and $\O_D'$ contribute to $S$ and $T$ but not to the modified Higgs couplings, for $\O_B$ and $\O_W$ it is the other way around. Both choices of basis are physically equivalent.  
Before passing from a general redundant set of operators to a convenient irreducible basis via the equations of motion, it is useful to first identify the operators that cannot be generated at tree-level.\footnote{A detailed study about perturbative generation of HDOs can be found in \cite{Arzt:1994gp}. } This is valuable information and we would like to avoid it to be lost in the course of the reduction. However this is what would happen if we eliminated $\O_D$ and $\O_D'$ in favor of $\O_W$ and $\O_B$. Indeed, this would cause the coefficient of \eg~$\O_{\WB}$ (which cannot be generated at tree-level) to be shifted by the coefficient of $\O_D$  (which can be generated at the tree-level via exchange of spin-one states).  This is why we choose this basis.

The only remaining two-fermion operators are of the dipole type. These operators are tightly constrained by FCNC as well as by their contributions to electric and magnetic dipole moments. Moreover, they are necessarily generated at the loop-level, and only affect Higgs couplings to gauge bosons by modifying existing SM loops. They will not have any impact on our results, therefore we can neglect them entirely.

We do not take into account CP-violating HDOs. These operators are constrained  by observables such as electric dipole moments. If we choosed to include these CP-odd HDOs in our analysis, we would also need to consider the whole set of data sensitive to CP violation. Although there is no fundamental problem with such extended analysis, that is beyond the scope of the present work. Moreover, the effects induced by CP-violating HDOs are often subleading with respect to the effects of CP-even operators, unless the latter are sufficiently suppressed. This is the case for Higgs decays, because CP-violating amplitudes do not interfere with SM amplitudes, whereas CP-conserving amplitudes do interfere with SM amplitudes \cite{Manohar:2006gz}. In the following we will derive observable deviations from the Standard Model using the full set of HDOs, and perform the analysis presented in Section~\ref{2013eft-se_bayesian_setup} taking into account only operators that respect custodial symmetry.

We could also rigorously take into account the running of the HDO coefficients $\alpha_i$ from the scale $\Lambda$ to the low scale ($m_h$ or $\sqrt{s}$, depending on the process considered), see for example \cite{Grojean:2013kd}.  However, the consequences of this running are rather mild so we will neglect them in this study. Notice that the strong effect of the operator $O_{\WB}$ on the $h\gamma\gamma$ vertex found in \cite{Grojean:2013kd} requires large enhancement of $\alpha_{\WB}$ with respect to $ s_w^2\,\alpha_{\WW}+ c_w^2\, \alpha_{\BB}-\frac{1}{2} s_w c_w\,\alpha_{\WB}$ (the coefficient of $h\,F_{\mu\nu}F^{\mu\nu}$, see Eq.~(\ref{2013eft-lambda_gamma_tens})). As is evident in our basis, this cannot be explained by a relative loop factor as none of these operators receive contributions at tree-level. Moreover, it has been shown in Ref.~\cite{Elias-Miro:2013gya} that operators that can be generated at tree-level (such as $\mathcal O_{D}$) do not mix with the loop-suppressed operators such as $\mathcal O_{VV}$ in the renormalization group flow.
In the absence of large hierarchies in the couplings of new physics states, we conclude that operator mixing does not lead to a large enhancement of the $h \rightarrow \gamma\gamma$ rate. 

\subsubsection{Effective Lagrangian}

In this section we will present the  effect of the HDOs on the SM tree-level couplings. Loops involving SM particles are considered in Section~\ref{2013eft-loops}.  We define the physical Higgs field $h$ as
\beq
H=\left(\begin{array}{c}0\\ \frac{1}{\sqrt{2}}\,(\tilde v+h)\end{array}\right)\,,
\eeq
and parametrize the couplings of $h$ to gauge bosons and fermions as\footnote{For the $hVV$ couplings with different tensor structure see below.}  
\beq
\mathcal L^{\rm tree}_{v,f}=\lambda_{Z}\,h\,(Z_\mu)^2+\lambda_{W}\, h\,W_{\mu}^+W_{\mu}^-  +\sum_f\lambda_f\, h\,\bar f_Lf_R \,. \label{2013eft-L_vect}
\eeq
The SM tree-level predictions for these quantities are given in terms of the SM input parameters $\tilde g$, $\tilde v$ and $\tilde s_w^2\equiv \tilde g'^2/(\tilde g^2+\tilde g'^2)$:
\beq
\lambda_{Z}=\frac{\tilde g^2\,\tilde v}{4\,\tilde c_w^2}\equiv\frac{\tilde m_{Z}^2}{\tilde v}\,,\qquad\lambda_{W}=\frac{\tilde g^2\,\tilde v}{2}\equiv\frac{2\,\tilde m_{W}^2}{\tilde v}\,,\qquad 
\lambda_{f}=-\frac{\tilde y_{f}}{\sqrt{2}}\equiv-\frac{\tilde m_{f}}{\tilde v} \,, \label{2013eft-lambda_vect}
\eeq
where the quantities with a tilde are the ones that appear in the SM part of the Lagrangian. For instance, $\tilde g$ and $\tilde g'$ are the couplings appearing in the covariant derivatives. However, these couplings do not take the same values as in the SM, since there are corrections from HDOs. 
There are distinct effects, as follows (see Ref.~\cite{Burgess:1993vc} for an analogous discussion on fermion couplings).
\begin{itemize}
\item
Operators such as $\O_{D^2}$ correct directly the tree-level SM vertices.
\item
Some operators (\eg~$\O_{D^2}$, $\O_{\WW}$) modify the kinetic terms of Higgs and gauge fields and thus indirectly lead to the rescaling of some couplings.
\item
Finally, there can be indirect effects from input parameters. They are taken to be the fine-structure constant $\alpha$, the $Z$ boson mass $m_Z$ and the Fermi constant $G_F$, as well as the physical fermion masses $m_f$ and the strong coupling constant $\alpha_s$. These quantities receive corrections from HDOs but must be held fixed in the analysis. Yet, this causes the SM parameters $\tilde g$, $\tilde v$ and $\tilde s_w$ to become functions of the HDO coefficients.
\end{itemize}

The last point is sometimes not taken into account in the literature. Let us focus on it and {\em define} the quantities $v$, $g$ and $s_w$ via
\beq
4\pi\alpha \equiv s_w^2 g^2\,,\qquad
m_Z^2\equiv\frac{ v^2\, g^2}{4\,  c_w^2}\,,\qquad 
G_F\equiv\frac{1}{\sqrt{2}\, v^2} \,.
\label{2013eft-hat}
\eeq
These quantities can be viewed as the ``familiar'' numbers from the SM (\eg~$v=246$~GeV). Like the input parameters they stay fixed in our analysis. On the other hand, the parameters $\tilde g$, $\tilde s_w$ and $\tilde v$ are the gauge couplings appearing in the covariant derivatives and the vacuum expectation value (vev) of the Higgs field, and must be expressed in terms of the HDO coefficients. The details of this procedure are presented in Appendix~A of~\cite{Dumont:2013wma}. Taking into account all the above effects, we obtain
\beq
\lambda_{Z}=a_Z\,\frac{m_Z^2}{ v}\,,\qquad
\lambda_{W}=a_W\,\frac{2\,m_W^2}{ v}\,,\qquad
\lambda_{f}=-c_f\,\frac{m_f}{ v} \,,
\eeq
where $m_f$ and $m_W$ are the {\em physical} masses. In particular, $m_W$ is given by\footnote{Unlike $m_Z$ and $m_f$, which are input parameters, the $W$ mass is a prediction in terms of input parameters and HDO coefficients.}
\begin{align}
m_W^2&=\frac{ g^2\, v^2}{4}\left(1+\left(\frac{1}{2}\alpha_D-\frac{ c_w^2}{2( c_w^2- s_w^2)}[\alpha_{D^2}'+\alpha_D]-
\frac{ c_w s_w}{ c_w^2- s_w^2}\alpha_{\WB}
\right) \frac{ v^2}{\Lambda^2} \right)\nn\\
&=\frac{ g^2\, v^2}{4}\left(1-\frac{\alpha S}{2( c_w^2- s_w^2)}
+\frac{ c_w^2\,\alpha T}{ c_w^2- s_w^2}
\right) \, . \label{2013eft-m_W_ST}
\end{align}
In the last row we have used Eqs.~(\ref{2013eft-S_tree}) and (\ref{2013eft-T_tree}) in order to compare our derivation of $m_W$ with the one in \cite{Burgess:1993vc}. The SM prediction of $m_W$ is thus only corrected by the oblique parameters. 
In this parametrization, the rescaling factors $a_Z$, $a_W$ and $c_f$ are given by
\begin{align}
a_Z&=1+\left(\frac{1}{2}\alpha_{D^2}\, -\frac{1}{4}\alpha_D\,  +\frac{1}{4}\alpha_{D^2}'\, \right) \frac{ v^2}{\Lambda^2} \, , \nn\\
a_W&=1+\left( \frac{1}{2}\alpha_{D^2}\,-\frac{1}{4}\alpha_D\,-\frac{1}{4}\alpha'_{D^2}\, 
\right) \frac{ v^2}{\Lambda^2} \, , \nn\\
c_f&=1-\left(\frac{1}{4}\alpha'_{D^2}\,-\frac{1}{4}\alpha_D\, -\alpha_f\right) \frac{ v^2}{\Lambda^2} \, .
\label{2013eft-anomalous1}
\end{align}
As a non-trivial consistency check, note that the vector anomalous couplings are rescaled in a custodially symmetric way $(a_Z=a_W)$ once the custodial-symmetry violating operator $\O_{D^2}'$ is turned off.

To conclude this subsection we compute the direct tree-level HDO contribution to the tensor couplings,
\beq
\mathcal L^{\rm tree}_{t}=
\zeta_\gamma\,h\,(F_{\mu\nu})^2+\zeta_g\,h\,(G_{\mu\nu})^2
+\zeta_{Z \gamma}\,h\,F_{\mu\nu}Z_{\mu\nu}\,+\zeta_{Z}\,h\,(Z_{\mu\nu})^2+\zeta_{W}\, h\,W_{\mu\nu}^+W_{\mu\nu}^-\,, \label{2013eft-L_tens}
\eeq
which are all zero in the SM at tree-level. One finds
\beq
\zeta_\gamma=\left( s_w^2\,\alpha_{\WW}+ c_w^2\, \alpha_{\BB}-\frac{1}{2} s_w c_w\,\alpha_{\WB}\right)\,\frac{ v}{\Lambda^2}\,,\qquad
\zeta_g=\alpha_{GG}\, \frac{ v}{\Lambda^2} \, , \label{2013eft-lambda_gamma_tens}
\eeq
\beq
\zeta_{Z \gamma}=\left(2 c_w s_w\,\alpha_{\WW}-2 c_w s_w\,\alpha_{\BB}-\frac{1}{2}( c_w^2- s_w^2)\,\alpha_{\WB}\right) \,\frac{ v}{\Lambda^2} \, , \label{2013eft-lambda_Zgamma_tens}
\eeq
\beq
\zeta_{Z}=  \left( c_w^2\,\alpha_{\WW} +  s_w^2\,\alpha_{\BB} + \frac{1}{2} c_w s_w\, \alpha_{\WB}\right)\,\frac{ v}{\Lambda^2}\,, \label{2013eft-lambda_WZ_tens}
\qquad
\zeta_{W}=  2\,\alpha_{\WW}\,\frac{ v}{\Lambda^2}\,. 
\eeq
The first two quantities constitute important corrections to the production and decay of the Higgs boson.
The last two corrections modify the tensorial structure of the SM Higgs--weak bosons coupling in a non-trivial way, which is discussed in detail in Section \ref{2013eft-se_tensorials}.

\subsubsection{Standard Model loop-induced HDOs}
\label{2013eft-loops}

In this section we compute the  Standard Model loop-induced operators relevant for Higgs physics. These operators contain indirect modifications due to couplings modified by the HDOs considered in the previous subsection. We want to make sure that we do not double-count possible new physics contribution to the Higgs couplings. 
In order to have a well-defined HDO framework at loop-level, we should consider that the HDOs we present in Eqs.~\eqref{2013eft-HDO_O_6}--\eqref{2013eft-HDO_O_f} are generated exclusively through new physics states at leading order, and enclose higher-order SM corrections only from irreducible loops.\footnote{This last point is important for NLO QCD corrections, see Section \ref{2013eft-se_deviations}.} Hence, the modified SM loops are not included in the tree-level contributions computed in the previous subsection. Our strategy is thus to compute the one-loop corrections to $\mathcal L^{\rm tree}$ using the couplings shown in Eq.~(\ref{2013eft-anomalous1}). 

The one-loop Lagrangian is parametrized as
\beq
\mathcal L^{1-{\rm loop}}=\lambda_\gamma\,h\,(F_{\mu\nu})^2+\lambda_g\,h\,(G_{\mu\nu})^2
+\lambda_{Z \gamma}\,h\,F_{\mu\nu}Z_{\mu\nu} \,.
\eeq
Let us decompose these couplings according to the particle in the loop, $\lambda_i=\sum_X\lambda_i^X$. We find\footnote{Note that in Eqs.~(\ref{2013eft-lambda_loop_gammaW})--(\ref{2013eft-lambda_loop_Zgammaf}) only the quantites with a tilde appear. Besides the modified Higgs couplings, the HDOs we consider only affect the couplings of the fermions to the $W$ and $Z$ bosons, precisely via the oblique parameters $S$ and $T$. The latter would in fact only show up in $\lambda_{Z \gamma}^f$. However these corrections are subleading and rather small (few percents at most), so that it is safe to neglect them.}
\beq
\lambda_\gamma^W=a_W\,\lambda_\gamma^{W, {\rm SM}}= \frac{7}{2}\, \frac{ g^2\, s_w^2}{16\pi^2} \, \frac{a_W}{ v} A_v(\tau_W) \,, \label{2013eft-lambda_loop_gammaW}
\eeq
\beq
\lambda_\gamma^f=c_f\,\lambda_\gamma^{f, {\rm SM}}= -\frac{2}{3}\, N^c_{f}\, e_f^2 \,\frac{ g^2  s_w^2}{16\pi^2} \, \frac{c_f}{ v}\, A_f(\tau_f)\,,\qquad
\lambda_g^f=c_f\,\lambda_g^{f,{\rm SM}}=-\frac{1}{3}\,\frac{ g^2_s}{16\pi^2}\, \frac{c_f}{ v} \,A_f(\tau_f) \,,\label{2013eft-lambda_loop_gf}
\eeq
\begin{multline}
\lambda_{Z \gamma}^W=a_W\,\lambda_{Z \gamma}^{W, {\rm SM}}= \frac{ e^2}{16\pi^2} \, \frac{a_W}{ v}\,
 t_w^{-1} \left( 2\left[ t^2_w -3\right] A_{Z \gamma}(\tau_W,\kappa_W) \phantom{\int} \right. \\ \left. + \left[
\frac{5- t_w^2}{2}+\frac{1- t_w^2}{\tau_W}
\right]B_{Z \gamma}(\tau_W,\kappa_W) \right)\,,
\end{multline}
\beq
\lambda_{Z \gamma}^f=-c_f\,\lambda_{Z \gamma}^{f, {\rm SM}}= \frac{ e^2}{16\pi^2} \, \frac{c_f}{ v}\,
N^c_{f}\,\frac{e_f^2\, (T^{3L}_f-2e_f  s^2_w)}{ s_w  c_w} \bigl(B_{Z \gamma}(\tau_f,\kappa_f) - A_{Z \gamma}(\tau_f,\kappa_f)\bigr)\,.
\label{2013eft-lambda_loop_Zgammaf}
\eeq
where $N^c_f$ and $e_f$ are the number of colors and the fraction of electric charge of the fermion running in the loop, respectively.
We define $\tau_i=4m^2_i/m_h^2$, $\kappa_i=4m^2_i/m_Z^2$ . The form factors $A_i, B$ are given in Appendix~B of~\cite{Dumont:2013wma}. They are defined so that in the decoupling limit, $A_{f,v}\rightarrow 1$ when $\tau\rightarrow \infty$, and $A_{Z \gamma}\rightarrow 1$, $B_{Z \gamma}\rightarrow 0$  when $\tau, \kappa\rightarrow \infty$.

\subsubsection{Trilinear gauge boson vertices}

The higher dimensional operators that we are considering also affect charged triple gauge boson vertices (TGV).
In the parametrization of Ref.~\cite{Hagiwara:1986vm},
\beq
\mathcal L_{\rm TGV}=
-i\, e\, \kappa_\gamma F_{\mu\nu}\,W_\mu^-W_\nu^+
-i\, g\, c_w\,\kappa_Z Z_{\mu\nu}\,W_\mu^-W_\nu^+
-i\, g\, c_w\,g_1^Z\left[ W_{\mu\nu}^+W_\nu^-- W_{\mu\nu}^-W_\nu^+\right]
 Z_{\mu} \,,
\eeq
the deviations from the Standard Model can be expressed in terms of the HDO coefficients as follows:
\begin{align}
\kappa_\gamma&=1+\frac{\alpha_{\WB} }{2 t_w} \frac{v^2}{\Lambda^2} \,, \nn\\
\kappa_Z&=1-\left(\frac{ s_w c_w}{( c_w^2- s_w^2)}\alpha_{\WB} +\frac{1}{4( c_w^2- s_w^2)}[\alpha'_{D^2}+\alpha_D] \right) \frac{v^2}{\Lambda^2} \,, \nn\\
g_1^Z&= 1-\left(\frac{ s_w}{2 c_w( c_w^2- s_w^2)}\alpha_{\WB} +\frac{1}{4( c_w^2- s_w^2)}[\alpha'_{D^2}+\alpha_D] \right) \frac{v^2}{\Lambda^2} \,,
\label{2013eft-eq:tgv_hdo}
\end{align}
where again some indirect effects from fixing input parameters were taken into account.
Gauge invariance implies the relation $\kappa_Z=g_1^Z-(\kappa_\gamma-1) t_w^2$  and one can check that it is indeed fulfilled. We then choose $\kappa_\gamma$ and $g_1^Z$ as independent couplings.

\subsection{Data treatment} \label{2013eft-se_data}

We exploit the results from Higgs searches at the LHC and at Tevatron as well as electroweak precision observables and trilinear gauge couplings. Starting from the Higgs searches, as in Section~\ref{sec:higgs2013} we use the results given in the $(\mu({\rm ggF+ttH}, Y), \mu({\rm VBF+VH}, Y))$ plane when available. The values for the signal strengths in the various (sub)channels as reported by the experiments and used in this analysis, together with the estimated decompositions into production channels are given in Tables~\ref{2013eft-ATLASresults}--\ref{2013eft-Tevatronresults}. Some of the decompositions into production channels are taken from~\cite{Belanger:2012gc}. In case of missing information, we take the relative ratios of production cross sections for an SM Higgs as a reasonable approximation, \ie~we assume that the experimental search is fully inclusive and compute the signal strength modified by HDOs accordingly. To this end, we use the latest predictions of the cross sections at the LHC~\cite{HXSWG} and at Tevatron~\cite{Group:2012zca}.
In our analysis, the Higgs mass is set to $m_h = 125.5$~GeV (close to the combined mass measurement from the two experiments) since it is not yet possible to take it as a nuisance parameter without losing the correlations between production channels. We consider experimental measurements of the signal strengths as close as possibe to this value.

\begin{table}\centering
\begin{tabular}{|c|c|c|ccccc|}
\hline
Channel & Signal strength $\mu$ & $m_h$ (GeV) & \multicolumn{5}{c|}{Reduced efficiencies} \\
& & & ggF & VBF & WH & ZH & ttH \\
\hline
\multicolumn{8}{|c|}{$h \rightarrow \gamma\gamma$ (4.8 fb$^{-1}$ at 7~TeV + 20.7 fb$^{-1}$ at 8~TeV)~\cite{ATLAS-CONF-2013-012,ATLAS-CONF-2013-014}} \\
\hline
$\mu({\rm ggF+ttH},\gamma\gamma)$ & $1.60 \pm 0.41$ & 125.5 & 100\% & -- & -- & -- & -- \\
$\mu({\rm VBF+VH} ,\gamma\gamma)$ & $1.94 \pm 0.82$ & 125.5 & -- & 60\% & 26\% & 14\% & -- \\
\hline
\multicolumn{8}{|c|}{$h \rightarrow ZZ$ (4.6 fb$^{-1}$ at 7~TeV + 20.7 fb$^{-1}$ at 8~TeV)~\cite{ATLAS-CONF-2013-013,ATLAS-CONF-2013-014}} \\
\hline
$\mu({\rm ggF+ttH},ZZ)$ & $1.51 \pm 0.52$ & 125.5 & 100\% & -- & -- & -- & -- \\
$\mu({\rm VBF+VH} ,ZZ)$ & $1.99 \pm 2.12$ & 125.5 & -- & 60\% & 26\% & 14\% & -- \\
\hline
\multicolumn{8}{|c|}{$h \rightarrow WW$ (4.6 fb$^{-1}$ at 7~TeV + 20.7 fb$^{-1}$ at 8~TeV)~\cite{ATLAS-CONF-2013-030,ATLAS-CONF-2013-034}} \\
\hline
$\mu({\rm ggF+ttH},WW)$ & $0.79 \pm 0.35$ & 125.5 & 100\% & -- & -- & -- & -- \\
$\mu({\rm VBF+VH} ,WW)$ & $1.71 \pm 0.76$ & 125.5 & -- & 60\% & 26\% & 14\% & -- \\
\hline
\multicolumn{8}{|c|}{$h \rightarrow b\bar{b}$ (4.7 fb$^{-1}$ at 7~TeV + 13.0 fb$^{-1}$ at 8~TeV)~\cite{ATLAS-CONF-2012-161,ATLAS-CONF-2013-014}} \\
\hline
VH tag & $-0.39 \pm 1.02$ & 125.5 & -- & -- & 64\% & 36\% & -- \\
\hline
\multicolumn{8}{|c|}{$h \rightarrow \tau\tau$ (4.6 fb$^{-1}$ at 7~TeV + 13.0 fb$^{-1}$ at 8~TeV)~\cite{ATLAS-CONF-2013-014}} \\
\hline
$\mu({\rm ggF+ttH},\tau\tau)$ & $2.31 \pm 1.61$ & 125.5 & 100\% & -- & -- & -- & -- \\
$\mu(\mathrm{VBF}+\mathrm{VH},\tau\tau)$ & $-0.20 \pm 1.06$ & 125.5 & -- & 60\% & 26\% & 14\% & -- \\
\hline
\end{tabular}
\caption{ATLAS results, as employed in this analysis. The following correlations are included in the fit: $\rho_{\gamma\gamma} = -0.27$, $\rho_{ZZ} = -0.50$, $\rho_{\WW} = -0.18$, $\rho_{\tau\tau} = -0.49$.}
\label{2013eft-ATLASresults}
\end{table}

\begin{table}\centering
\begin{tabular}{|c|c|c|ccccc|}
\hline
Channel & Signal strength $\mu$ & $m_h$ (GeV) & \multicolumn{5}{c|}{Reduced efficiencies} \\
& & & ggF & VBF & WH & ZH & ttH \\
\hline
\multicolumn{8}{|c|}{$h \rightarrow \gamma\gamma$ (5.1 fb$^{-1}$ at 7~TeV + 19.6 fb$^{-1}$ at 8~TeV)~\cite{CMS-PAS-HIG-13-001}} \\
\hline
$\mu({\rm ggF+ttH},\gamma\gamma)$ & $0.49 \pm 0.39$ & 125 & 100\% & -- & -- & -- & -- \\
$\mu({\rm VBF+VH},\gamma\gamma)$ & $1.65 \pm 0.87$ & 125 & -- & 60\% & 26\% & 14\% & --\\
\hline
\multicolumn{8}{|c|}{$h \rightarrow ZZ$ (5.1 fb$^{-1}$ at 7~TeV + 19.6 fb$^{-1}$ at 8~TeV)~\cite{CMS-PAS-HIG-13-002}} \\
\hline
$\mu({\rm ggF+ttH},ZZ)$ & $0.99 \pm 0.46$ & 125.8 & 100\% & -- & -- & -- & -- \\
$\mu({\rm VBF+VH} ,ZZ)$ & $1.05 \pm 2.38$ & 125.8 & -- & 60\% & 26\% & 14\% & -- \\
\hline
\multicolumn{8}{|c|}{$h \rightarrow WW$ (up to 4.9 fb$^{-1}$ at 7~TeV + 19.5 fb$^{-1}$ at 8~TeV)~\cite{CMS-PAS-HIG-13-003, CMS-PAS-HIG-12-039,CMS-PAS-HIG-12-042,CMS-PAS-HIG-12-045}} \\
\hline
0/1 jet & $0.76 \pm 0.21$ & 125 & 97\% & 3\% & -- & -- & -- \\
VBF tag & $-0.05^{+0.74}_{-0.55}$ & 125.8 & 17\% & 83\% & -- & -- & -- \\
VH tag & $-0.31^{+2.22}_{-1.94}$ & 125.8 & -- & -- & 64\% & 36\% & -- \\
\hline
\multicolumn{8}{|c|}{$h \rightarrow b\bar{b}$ (up to 5.0 fb$^{-1}$ at 7~TeV + 12.1 fb$^{-1}$ at 8~TeV)~\cite{CMS-PAS-HIG-12-044,CMS-PAS-HIG-12-025,CMS-PAS-HIG-12-045}} \\
\hline
${\rm Z}(\ell^-\ell^+){\rm h}$ & $1.55_{-1.07}^{+1.20}$ & 125 & -- & -- & -- & 100\% & -- \\
${\rm Z}(\nu\bar{\nu}){\rm h}$ & $1.79_{-1.02}^{+1.11}$ & 125 & -- & -- & -- & 100\% & -- \\
${\rm W}(\ell\nu){\rm h}$ & $0.69_{-0.88}^{+0.91}$ & 125 & -- & -- & 100\% & -- & -- \\
ttH tag & $-0.80^{+2.10}_{-1.84}$ & 125.8 & -- & -- & -- & -- & 100\% \\
\hline
\multicolumn{8}{|c|}{$h \rightarrow \tau\tau$ (4.9 fb$^{-1}$ at 7~TeV + 19.4 fb$^{-1}$ at 8~TeV)~\cite{CMS-PAS-HIG-13-004}} \\
\hline
0/1 jet & $0.76_{-0.52}^{+0.49}$ & 125 & 76\% & 16\% & 4\% & 3\% & 1\% \\
VBF tag & $1.40_{-0.57}^{+0.60}$ & 125 & 19\% & 81\% & -- & -- & -- \\
VH tag & $0.77_{-1.43}^{+1.48}$ & 125 & -- & -- & 64\% & 36\% & -- \\
\hline
\multicolumn{8}{|c|}{$h \rightarrow Z\gamma$ (5.0 fb$^{-1}$ at 7~TeV + 19.6 fb$^{-1}$ at 8~TeV)~\cite{CMS-PAS-HIG-13-006}} \\
\hline
Inclusive  &  $<9.3$ at 95\%~CL   &  125.5   & 87\%  &7\%  & 3\% & 2\% & 1\% \\
\hline
\end{tabular}
\caption{CMS results, as employed in this analysis. The following correlations are included in the fit: $\rho_{\gamma\gamma} = -0.50$, $\rho_{ZZ} = -0.73$.}
\label{2013eft-CMSresults}
\end{table}

\begin{table}\centering
\begin{tabular}{|c|c|c|ccccc|}
\hline
Channel & Signal strength $\mu$ & $m_h$ (GeV) & \multicolumn{5}{c|}{Reduced efficiencies} \\
& & & ggF & VBF & WH & ZH & ttH \\
\hline
\multicolumn{8}{|c|}{$h \rightarrow \gamma\gamma$~\cite{HCPHiggsTevatron}} \\
\hline
Combined & $6.14^{+3.25}_{-3.19}$ & 125 & 78\% & 5\% & 11\% & 6\% & -- \\
\hline
\multicolumn{8}{|c|}{$h \rightarrow WW$~\cite{HCPHiggsTevatron}} \\
\hline
Combined & $0.85^{+0.88}_{-0.81}$ & 125 & 78\% & 5\% & 11\% & 6\% & -- \\
\hline
\multicolumn{8}{|c|}{$h \rightarrow b\bar{b}$~\cite{HCPtevBBtalk}} \\
\hline
VH tag & $1.56^{+0.72}_{-0.73}$ & 125 & -- & -- & 62\% & 38\% & -- \\
\hline
\end{tabular}
\caption{Tevatron results for up to $10\fbi$ at $\sqrt{s} = 1.96$~TeV, as employed in this analysis.}
\label{2013eft-Tevatronresults}
\end{table}

We take into account the electroweak precision observables using the Peskin--Takeuchi $S$ and $T$ parameters~\cite{Peskin:1990zt,Peskin:1991sw}. 
Beyond $S$ and $T$, the $W$ and $Y$ parameters~\cite{Barbieri:2004qk} should be used in the HDO framework. However we find that constraints arising from these parameters are by far subleading with respect to our other constraints. 
 Experimental values of $S$ and $T$ are taken from the latest electroweak fit of the SM done by the Gfitter Group~\cite{Baak:2012kk}: $S=0.05 \pm 0.09$ and $T= 0.08 \pm 0.07$ with a correlation coefficient of 0.91.
Regarding constraints on TGV, we take into account the LEP measurements~\cite{TGV}:
\begin{align}
\kappa_\gamma &= 0.973^{+0.044}_{-0.045} \,, \nn \\
g_1^Z &= 0.984^{+0.022}_{-0.019} \,.
\label{2013eft-eq:lep_tgv}
\end{align}

The global likelihood function is defined as the product of the likelihoods associated to the various observables,
\beq
L = L_{\rm Higgs} \times L_{S,T} \times L_{\rm TGV} \,,
\label{2013eft-eq:likelihood}
\eeq
where $L_{\rm Higgs}$ is the likelihood as given in Eq.~\eqref{eq:ourbestlike}.
The likelihood associated to the measurement of an observable $\hat O$, given as a central value $O$ and a symmetric uncertainty $\sigma$, is modeled by a normal law,
\beq
L_{O}\propto e^{-(O-\hat O)^2/2\sigma^2}\,.
\eeq
When uncertainties are asymmetric, we use the positive error bar if $(\hat{O} - O) > 0$, whereas we use the negative error bar if $(\hat{O} - O) < 0$.
Finally, the CMS bound on the decay channel $h \rightarrow Z\gamma$ is implemented as a step function, 
\beq
L_{\mu_{Z\gamma}}\propto
\begin{cases}
  1~  &\textrm{if} ~~ \hat\mu_{Z\gamma}<9.3 \,,\\
  0 ~  &\textrm{otherwise}\,.
\end{cases} 
\eeq

We will now derive the deviations induced by the HDOs to the observables presented in Section \ref{2013eft-se_data}. We first discuss the particular treatment of tensorial couplings. All formulas are given in the following section.

\subsection{On weak bosons tensorial couplings} \label{2013eft-se_tensorials}

Because of electroweak symmetry breaking, the $W,Z\equiv V$ bosons generally couple to the Higgs through two  different Lorentz structures. The coupling can be vectorial, $\propto g^{\mu\nu}$, or it can be tensorial with a vertex $\propto (g^{\mu\nu}-\frac{q_1^\mu q_2^\nu}{q_1.q_2})$, where $q_1$, $q_2$ are the momenta of the two gauge bosons. The leading SM couplings $\lambda_{W},\lambda_{Z}$ given in Eqs.~\eqref{2013eft-L_vect} and \eqref{2013eft-lambda_vect} are vectorial. Tensorial couplings are generated only at one-loop and are $\mathcal{O}(\alpha) \sim 10^{-2}$. 

Once HDOs are taken into account, the relative importance of the vectorial and tensorial terms is modified. On one hand vectorial couplings are rescaled by the coefficients $a_{W,Z}$. On the other hand new tensorial contributions $\zeta_{W}$, $\zeta_{Z}$ are generated following Eq.~\eqref{2013eft-lambda_WZ_tens}. The amplitude associated to a $hVV$ vertex (with the $V$'s possibly off-shell) is in general
\beq
\mathcal{M}(h VV)^{\lambda_1,\lambda_2}=e^{\mu (*)}_{\lambda_1} e^{\nu (*)}_{\lambda_2} \left( ia_V \lambda_V^{\rm SM}g^{\mu\nu} -i2\zeta_{V}q_1.q_2\left[g^{\mu\nu}-\frac{q_1^\mu q_2^\nu}{q_1.q_2}\right]  \right)\,,
\label{2013eft-M_hvv}
\eeq
where $\mathcal{M}^{0,0}$ and $\mathcal{M}^{\pm,\pm}$ are the longitudinal and transverse helicities amplitudes, respectively. Interferences among helicity amplitudes then determine angular distributions (see \eg~\cite{Hagiwara:2009wt}). 
In this work, we consider that the SM contribution to the tensorial coupling is small with respect to the one induced by new physics.
The relative magnitude of the longitudinal and transverse amplitudes in case of a vectorial coupling is given by
\beq
r_v=\left|\frac{\mathcal{M}_v^{0,0}}{\mathcal{M}_v^{\pm,\pm}}\right|=\frac{\left|m_h^2 - q_1^2-q_2^2\right|}{2|q_1||q_2|}\,,
\eeq
while it is the inverse in case of a tensorial coupling,
\beq
r_t=\left|\frac{\mathcal{M}_t^{0,0}}{\mathcal{M}_t^{\pm,\pm}}\right|=\frac{2|q_1||q_2|}{\left|m_h^2 - q_1^2-q_2^2\right|}\,.
\eeq
The two vector bosons can be off-shell in the above expression, while the Higgs is on-shell.

As $r_v\neq r_t$, the two Lorentz structures imply generally different angular distributions. Moreover, even for unpolarized processes, the energy dependence in  Eq.~\eqref{2013eft-M_hvv} is different for both contributions, such that also energy distributions are modified.
Because of this different energy dependence, kinematic cuts prepared for the SM are generally unadapted to such a non-trivial modification.
That is, in Eq.~\eqref{eq:signalstr1}, $A \times \varepsilon \neq [A \times \varepsilon]_{\rm SM}$.  The consequences may be an incorrect estimation of the signal strength and of the Higgs mass.  To perform an exact analysis, one should redo the fits to LHC data taking into account the modified Lorentz structure in the expected signal. Such work is clearly beyond the scope of our present study.
Instead we will show that under reasonable  approximations we can use  $A \times \varepsilon = [A \times \varepsilon]_{\rm SM}$ in the present analysis.

There are three processes sensitive to the $\zeta_{V}$ tensorial couplings in the context of the searches for the Higgs boson at around 125~GeV: the leading decay to weak bosons $h\rightarrow VV^*$, and the VBF and VH production modes.  We now discuss how we treat these three tensorial contributions.

\subsubsection{Decay into vector bosons}

In the case of a light Higgs boson, the leading decay occurs with one of the $V$ off the mass shell. The weak bosons then decay into fermions. For massless fermions, the kinematic bounds on the on-shell boson energy  $E_V$ are $m_V<E_V<(m_h^2+m_V^2 ) /2 m_h$ in the rest frame of the Higgs. Because of the $V^*$ propagator, the lower bound $E_V=m_V$ is favored, implying that both weak bosons are preferentially produced at rest. Longitudinal and transverse amplitudes are then equally populated, $r_v=1$. Therefore, one has $r_t=1$ as well, such that one can see qualitatively that a tensorial contribution  cannot radically modify  angular distributions. This is confirmed with the exact  angular and invariant mass distributions among leptons induced  by pure vectorial and pure tensorial couplings \cite{Gao:2010qx, Stolarski:2012ps}.\footnote{
Overall, the situation is much less striking than for a CP-violating contribution, which forbids the decay to the longitudinal polarization state. } 
In our study, the tensorial contributions are constrained to be subleading with respect to the vectorial contributions, such that the deviations induced on angular and invariant mass distributions can easily be smaller than the current statistical uncertainty. In addition, they could also be misidentified with the background. For example, in $h\rightarrow VV^*$, the distribution of the most discriminant observable, ``lepton-opposite $Z$ momentum angle'', is very similar to the distribution of the irreducible background  $q\bar q\rightarrow ZZ^*$ (see Fig.~3 in \cite{Gainer:2011xz}).

Following what discussed above, we can reasonably assume that angular and invariant mass distributions are not affected by the presence of tensorial couplings given the current level of precision.
Polarization of the on-shell $V$ can thus be averaged, and we are left with  a matrix element scaling as 
\beq
|\mathcal{M}|^2=|\mathcal{M}_v+\mathcal{M}_t|^2\propto\left| a_V\lambda_{V}^{\rm SM}-2\zeta_{V}q_1.q_2 \right|^2  \,, \label{2013eft-amplitude_vect_tens}
\eeq
where $q_1$, $q_2$ are the momenta of the two vector bosons.  In the Higgs rest frame, one has $q_1.q_2=m_h E_V-m_V^2$, which is bounded as
\beq
m_V(m_h-m_V)< q_1.q_2< \frac{m_h^2-m^2_V}{2} \,. \label{2013eft-eq_bound_EV}
\eeq 
 The exact tensorial contributions to the total decay widths are given in Appendix~C of~\cite{Dumont:2013wma}. We introduce the dimensionless positive quantity
\beq \nu_{VV}=q_1.q_2/m_h^2\,, \eeq
with $V\equiv W,Z$.
Defining
\beq
\langle \nu_{VV} \rangle= \frac{ \int \nu_{VV} \mathcal{M}_v \mathcal{M}_t^* dPS}{\int \mathcal{M}_v \mathcal{M}_t^* dPS}\,,\,\,\,\,\,\,
\langle \nu_{VV}^2\rangle = \frac{ \int \nu^2_{VV} |\mathcal{M}_t|^2dPS}{\int  |\mathcal{M}_t|^2 dPS} \,, \label{2013eft-nu_VV}
\eeq
the vector-tensor interference term will be $\propto \zeta_{V} \langle \nu_{VV}\rangle$ and the pure tensor contribution will be $\propto|\zeta_{V}|^2 \langle \nu^2_{VV}\rangle $. 
For $m_h=125.5\,\textrm{GeV}$, $m_Z=91\,\textrm{GeV}$, $m_W=80\,\textrm{GeV}$,  one gets $\langle \nu_{ZZ}\rangle=0.2209$, $\langle \nu^2_{ZZ}\rangle^\frac{1}{2}=0.2211$, $\langle \nu_{\WW}\rangle=0.2653$, $\langle\nu^2_{\WW}\rangle^\frac{1}{2}=0.2659$. In the following we will make the approximation $\langle \nu_{VV}^2\rangle\approx \langle \nu_{VV}\rangle^2$.

\subsubsection{VBF production mode} \label{2013eft-se_vbf}

For the VBF process, both ATLAS and CMS apply hard cuts on the outgoing jets rapidities and their difference. The rapidity distributions of the two jets are similar in presence of a tensorial coupling, just like in the decay into two photons or in the production via gluon-gluon fusion, such that one can assume that cut efficiency is the same. The crucial change lies in the azimuthal angle $\phi_{jj}$ between the two tagging jets  (see \eg~\cite{Hagiwara:2009wt} and references therein).
Indeed, both weak bosons are space-like, with virtualities considerably smaller than $m_h^2$. Such values are favored to balance the space-like $V$ and the outgoing jets virtualities. As a result one has typically $r_v\gg 1$, $r_t\ll 1$ \ie~vectorial and tensorial amplitudes are mostly longitudinal and transverse, respectively.  Consequently, the $\phi_{jj}$ distribution is almost flat for a pure vectorial coupling, and strongly peaked at $\pi/2$ for a pure tensorial coupling. For a large enough HDO contribution to the tensorial coupling, an anomalous $\phi_{jj}$ distribution could thus be observed.
 However, this variable is not used for the selection of the events in the experimental analyses we consider. Therefore, the selection efficiencies are also suitable in the case of large tensorial contributions, and one has $\varepsilon_{\rm SM}=\varepsilon_{\rm SM+HDO}$. One can average over the polarizations, and the squared amplitude is then simply rescaled by a factor 
$\left| a_V\lambda_{VBF}^{\rm SM}-2\zeta_{V}q_1.q_2 \right|^2 $.

We still have to determine the magnitude of the tensorial contribution.
In this process, the scalar product of the weak boson momenta $q_1.q_2$ is related to the incoming and outgoing quarks as $q_1.q_2=m_h^2/2+p_1.p_3+p_2.p_4$. The outgoing quarks are highly energetic with respect to the amount of $p_T$ they receive from the $V$ fusion, such that one has $\bf{\left|p_1\right|\simeq\left|p_3\right|}$ and $\bf{\left|p_2\right|\simeq\left|p_4\right|}$. In terms of the $p_T$ and rapidities of the outgoing quarks we have then 
\beq
q_1.q_2=\frac{m_h^2}{2}+ |p_{T,3}|^2\frac{1+e^{-\eta_{3}}}{2}+ |p_{T,4}|^2\frac{1+e^{-\eta_{4}}}{2}\,.
\eeq 
Without the tensorial contribution, the $p_T$ distribution peaks typically at values smaller than $m_V$.  The tails of the $p_T$ distributions drop quickly for higher energies \cite{Aad:2009wy}, with typically one jet at a time getting a large $p_T$ \cite{Djouadi:2005gi}. One can thus assume $q_1.q_2\approx m_h^2/2$  to a good approximation.
Once the tensorial coupling is taken into account, a deviation from the expected SM distributions might be present in the high-$p_T$ tails, as $q_1.q_2$ is enhanced at large $p_T$.
However, as long as one counts the total number of events, \ie~the integral of the distribution, this enhancement of $q_1.q_2$ has a small weight and can be safely neglected.
Finally, defining the dimensionless positive quantity \beq\nu_{\rm VBF}=q_1.q_2/m_h^2\,,\eeq
with $V=W,Z$, we have thus $\nu_{\rm VBF}\approx 1/2$ after phase space integration.

\subsubsection{VH production mode}

In the case of the associated production with an electroweak gauge boson, the scalar product of the momenta of the weak bosons is given by
\beq
q_1.q_2=\frac{s+m_V^2-m_h^2}{2}\,,
\eeq
where $\sqrt{s}$ is the partonic center-of-mass energy, which can much larger than $m_V + m_h$ at the LHC.  
Therefore, contrary to the two other processes, the product $q_1.q_2$ can be large. The tensorial contribution can then be substantially enhanced in this process, and lead to modifications of the angular distributions. 

However, it turns out that for both polar and angular distributions, the angular effects can be neglected. We refer to \cite{Kilian:1996wu} and references therein for the expressions. Although results are given for $e^+ e^-$ collisions, they can be trivially generalized in the case of the LHC. For the distribution of the polar angle of the vector boson in the laboratory frame, it is the longitudinal component of $V$ which enters mainly, such that the tensorial contribution to the distribution is suppressed by an additional factor $\mathcal{O}(m_V^2/s)$. 
For the azimuthal distributions, the tensorial contributions can be sizable, but the whole distribution tends to be flat for $s\gg m_V^2$, with non-flat terms suppressed by powers of $m_V/\sqrt{s}$. 
As a result, although various pieces of angular information are used in event selection for this  mode of production, we can safely neglect the angular effects of the tensorial coupling. 

Concerning the magnitude of the tensorial contribution, it appears that it reduces to a simple rescaling $\propto \lambda_V^{\rm SM}+ 12\zeta_V m_V^2$  in the limit $s\gg m_V^2$. The rescaling is exact up to a subleading term $\mathcal{O}(12 m_V^2/s)\approx 0.1$. To include the subleading $s$-dependent terms, an integration over the partonic density functions would be necessary.

\subsection{Deviations caused by new physics} \label{2013eft-se_deviations}

\subsubsection{Higgs signal strengths}

In all generality, efficiencies in the SM with and without HDOs are not necessarily the same, \ie~$A \times \varepsilon \neq [A \times \varepsilon]_{\rm SM}$, because kinematic distributions can be modified in a non-trivial way by HDOs.  The selection criteria calibrated on the SM expectations are then unadapted in such situation, which complicates the interpretation of the signal strengths. 
However, we have seen in  Section~\ref{2013eft-se_tensorials} that one can safely ignore these possibilities of HDOs affecting the kinematic distributions, given the current precision of the experimental searches. It is therefore a good approximation to set $\varepsilon_{\textrm{SM+HDO}} =\varepsilon_{\textrm{SM}}$. Thus, for each signal strength, one can simply incorporate the contributions coming from the tensorial couplings in the rescaling of the Standard Model signal strength.

The gluon-gluon fusion process is modified both by the tree-level HDO contribution $\zeta_g$ and the anomalous Higgs--fermion couplings $c_f$. Keeping only the third generation, we get
\beq
\sigma_{\rm ggF}=\sigma^{\rm SM}_{\rm ggF} \left|\frac{c_t\lambda_g^{t, {\rm SM}}+c_b\lambda_g^{b, {\rm SM}}+\zeta_g}{\lambda_g^{t, {\rm SM}}+\lambda_g^{b, {\rm SM}}} \right|^2\,.
\eeq
Vector boson fusion is modified by the anomalous vectorial couplings $a_{W,Z}$ and by $\zeta_{W,Z}$. Denoting by $\lambda_{\rm VBF}^{\rm SM}$ the effective SM couplings, one has
\beq
\sigma_{\rm VBF}=\sigma_{\rm VBF}^{\rm SM} \left| \frac{a_W\lambda_{W}^{\rm SM} + a_Z\lambda_{Z}^{\rm SM} - 2 \, \nu_{\rm VBF} \, m_h^2 \, (\zeta_{W}+\zeta_{Z})}{\lambda_{W}^{\rm SM}+\lambda_{Z}^{\rm SM}} \right|^2\,.
\eeq
The parameter $\nu_{\rm VBF}$ is defined in Section~\ref{2013eft-se_vbf}. We take $\nu_{\rm VBF} = 1/2$.
The associated production with an electroweak gauge boson is modified as 
\beq
\sigma_{\rm VH}=\sigma_{\rm VH}^{\rm SM}\left|\frac{a_V\lambda_V^{\rm SM}+12\zeta_V m_V^2}{\lambda_V^{\rm SM}}   \right|^2\,,
\label{2013eft-eq:VH}
\eeq
where $V=W,Z$. Finally, the associated production with a $t\bar{t}$ pair is rescaled as  
\beq
\sigma_{\rm ttH} = |c_t|^2 \sigma_{\rm ttH}^{\rm SM}\,.
\eeq
The decays of the Higgs boson into fermions are modified as
\beq
\Gamma_{ff}=|c_f|^2\Gamma^{\rm SM}_{ff}\,.
\eeq
The tree-level decays to vector bosons are modified as
\beq
 \Gamma_{VV}=\left|\frac{a_V\,\lambda_{V}^{\rm SM}-2\,\zeta_{V} m_h^2 \,\langle\nu_{VV}\rangle }{\lambda_{V}^{\rm SM}
 }\right|^2 \Gamma^{\rm SM}_{VV}\,,
\eeq
where the  parameter $\langle \nu_{VV}\rangle$,
 defined in Eq.~\eqref{2013eft-nu_VV}, encodes the modification of phase space integrals.
Loop-induced decays are sensitive to more deviations, 
\beq
 \Gamma_{\gamma\gamma}= \Gamma_{\gamma\gamma}^{\rm SM} \left|\frac{a_W\lambda_\gamma^{W, {\rm SM}}+c_t\lambda_\gamma^{t, {\rm SM}}+c_b\lambda_\gamma^{b, {\rm SM}}+c_\tau\lambda_\gamma^{\tau, {\rm SM}}+\zeta_\gamma}{\lambda_\gamma^{W, {\rm SM}}+\lambda_\gamma^{t, {\rm SM}}+\lambda_\gamma^{b, {\rm SM}}+\lambda_\gamma^{\tau, {\rm SM}}} \right|^2\,,
\eeq
\beq
 \Gamma_{Z \gamma}= \Gamma_{Z \gamma}^{\rm SM} \left|\frac{a_W\lambda_{Z \gamma}^{W, {\rm SM}}+c_t\lambda_{Z \gamma}^{t, {\rm SM}}+c_b\lambda_{Z \gamma}^{b, {\rm SM}}+c_\tau\lambda_{Z \gamma}^{\tau, {\rm SM}}+\zeta_{Z \gamma}}{\lambda_{Z \gamma}^{W, {\rm SM}}+\lambda_{Z \gamma}^{t, {\rm SM}}+\lambda_{Z \gamma}^{b, {\rm SM}}+\lambda_{Z \gamma}^{\tau, {\rm SM}}} \right|^2\,.
\eeq
In such cases, the tensorial couplings can compete with the SM effective couplings.

\subsubsection{QCD radiative corrections}
 
Many of the above described processes receive leading radiative corrections from QCD loops. For all the tree-level processes, the structure of loop diagrams is not modified by the insertion of HDOs, including the tensorial couplings, such that radiative corrections factorize up to higher order corrections. It is thus straightforward to take them into account, simply using the NLO predictions of $\sigma^{\rm SM}$ and $\Gamma^{\rm SM}$.

The situation is more involved in the case of the loop-induced processes ($h \rightarrow \gamma \gamma$, $h \rightarrow Z \gamma$, and $gg \rightarrow h$) because this time the tensorial coupling is competing with the SM loops. Hence the effects of the $\zeta$'s may be very large in these processes, such that it is important to properly take into account the radiative corrections.
As stated in Section~\ref{2013eft-se_HDOs}, the HDOs implicitely contain higher-order corrections from irreducible SM loops. These contributions therefore have to be taken into account for the SM effective couplings and not for the $\zeta$ couplings.\footnote{We are grateful to M. Spira for enlightening discussion on this subject.} 
 
The processes $h \rightarrow \gamma \gamma$ and $h\rightarrow Z \gamma$ only receive virtual NLO QCD corrections. For $h \rightarrow \gamma \gamma$, we take into account the exact values of the correction factor to the quark effective couplings 
 \beq
 \lambda_\gamma^{q, {\rm SM}}=\lambda_\gamma^{q, {\rm SM}}|_{\rm LO} \left( 1+\frac{\alpha_s}{\pi}C_H(\tau_q) \right)\,,
 \eeq   
 where the $C_H$ function can be found in \cite{Djouadi:2005gi}. For $h\rightarrow Z \gamma$, one can take the correction in the heavy top limit as a good approximation~\cite{Djouadi:2005gi},
 \beq
 \lambda_\gamma^{t, {\rm SM}}=\lambda_\gamma^{t, {\rm SM}}|_{\rm LO} \left( 1-\frac{\alpha_s}{\pi} \right)\,.
 \eeq   

The situation  is more subtle for the ggF process, because of the presence of important NLO real corrections. Introducing the tensorial coupling leads generally to non-trivial modifications of the integrals over parton densities for real emissions. However, in the heavy-top limit and neglecting the small bottom quark contribution, the QCD corrections to the SM loop and to the tensorial coupling $\zeta_g$ become similar and factorize. Adopting this fairly good approximation, the SM effective coupling are rescaled as 
 \beq
 \lambda_g^{t, {\rm SM}}=\lambda_g^{t, {\rm SM}}|_{\rm LO} \left( 1+\frac{11}{4}\frac{\alpha_s}{\pi} \right)\,.
 \eeq

\subsubsection{$S$ and $T$ parameters}

The electroweak precision observables are affected in the presence of the HDOs. At tree-level the $S$ and $T$ parameters are related to the HDO coefficients as follows:
\begin{align}
\alpha\,S&=\biggr(2\, s_w c_w\,\alpha_{\WB}+ s_w^2\,\alpha_D+ c_w^2\,\alpha_{D}' \biggr)\frac{ v^2}{\Lambda^2}\,, \label{2013eft-S_tree}\\
\alpha\,T&=\left(-\frac{1}{2}\,\alpha_{D^2}'+\frac{1}{2}\,\alpha_{D}' \right)\frac{ v^2}{\Lambda^2}\,. \label{2013eft-T_tree}
\end{align}
Moreover, the SM loops are modified by the HDOs. The $T$ parameter receives  new divergent contributions from the modified SM couplings $a_Z$ and $a_W$ in Eq.~(\ref{2013eft-anomalous1}).  A quadratic divergence,
\beq
\alpha \Delta T=-\frac{\Lambda^2}{16 \,\pi^2\, v^2} \frac{\alpha_{D^2}' v^2}{\Lambda^2} \,,
\eeq
arises from custodial breaking \cite{Farina:2012ea}. Dropping other terms that are proportional to $\alpha_{D}'$ and $\alpha_{D^2}'$ (that already appear at tree-level) 
we can take the result from Ref.~\cite{Espinosa:2012im},
\beq
\alpha \Delta T=-\frac{3\, e^2}{32\, \pi^2\, c_w^2} \left(\alpha_{D^2}-\frac{1}{2}\alpha_D\right) \frac{ v^2}{\Lambda^2}
\log\left(\frac{m_h}{\Lambda}\right) \,.
\eeq 
Similarly, the $S$ parameter receives corrections due to the modified Higgs coupling $\alpha_Z$ \cite{Espinosa:2012im}, 
hence it is expected to get new contributions proportional to $\alpha_{D^2}$ and $\alpha_{D^2}'$. 
Finally, the tensor couplings $\zeta_{V}$ can also generate new SM loop contributions which have been given in Ref.~\cite{Alam:1997nk}, 
\beq
\alpha \Delta S=
\frac{ e^2}{24 \pi^2}\left(\alpha_{D^2}+\frac{1}{2}\alpha_{D^2}'\right)\frac{ v^2}{\Lambda^2}\log\left(\frac{m_h}{\Lambda}\right)
+\frac{ e^2}{2\pi^2}(\alpha_{\BB}+\alpha_{\WW})\frac{ v^2}{\Lambda^2}
\log\left(\frac{m_h}{\Lambda}\right)
\,.
\label{2013eft-eq:DeltaS}
\eeq
Finally we neglect the contraints coming from the $W$ and $Y$ parameters \cite{Barbieri:2004qk} as they are expected to have a small impact on our results.

\subsection{Bayesian setup and low-$\Lambda$ scenario} \label{2013eft-se_bayesian_setup}

\subsubsection{Bayesian inference}

We are working in the framework of Bayesian statistics (see \cite{Trotta:2008qt} for an introduction). In this approach, a probability is interpreted as a measure of the degree of belief about a proposition.
Our study lies in the domain of Bayesian inference, which is based on the relation
\beq
p(\theta |d,\mathcal{M})\propto p(d|\theta,\mathcal{M})p(\theta|\mathcal{M})\,,
\eeq
where $\theta\equiv\{\theta_{1\ldots n}\}$ are the parameters of the model $\mathcal{M}$, and $d$ denotes the experimental data. The distribution $p(\theta |d,\mathcal{M})$ is the so-called posterior probability density function (PDF), $p(d|\theta,\mathcal{M})\equiv L(\theta)$ is the likelihood function enclosing experimental data, and  $p(\theta|\mathcal{M})$ is the prior PDF, which represents our a priori degree of belief on the parameters. The model $\mathcal{M}$ is in our case the Standard Model extended with higher dimensional operators. The likelihood is defined in Section~\ref{2013eft-se_data} (see Eq.~(\ref{2013eft-eq:likelihood})) and the theoretical expressions for the HDO modified signal strengths are given in Section~\ref{2013eft-se_deviations}.
The prior PDF is discussed in the next subsection.

The posterior PDF is the core of our results. Integrating the posterior over a subset $\lambda$ of the parameter set $\theta\equiv \{\psi,\lambda\}$,
\beq p(\psi|d,\mathcal{M})\propto\int d\lambda\, p(\psi,\lambda|\mathcal{M})L(\psi,\lambda)\,,\eeq
leads to inference on the parameters $\psi$.

Also the notion of naturalness and fine-tuning are built-in \cite{Fichet:2012sn} in the Bayesian approach. This is relevant for our study, in which precise (``fine-tuned'') cancellations between various HDO contributions can happen. Intrinsically, the regions of parameter space in which precise cancellations occur have a weak statistical weight, such that they are flushed away after integration. The results we will present can thus be considered as generic, \ie~free of improbable cancellations. 

We will consider uniform (flat) priors for the quantities
\beq
\beta_i\equiv \alpha_i\frac{v^2}{\Lambda^2}
\eeq
and demand $|\beta_i|<1$.
Moreover, we will fix the cutoff scale to be $\Lambda=4\pi v$. In the following we will justify these choices and argue that it ensures in particular convergence of the HDO expansion as well as perturbativity of the UV theory, and minimizes the dependence on the choice of the HDO basis.

\subsubsection{Priors and low-$\Lambda$ scenarios}

The prior distributions associated to our parameters is a key feature of Bayesian inference. We follow the ``principle of indifference'' \cite{press,jaynes} that maximizes the objectiveness of the priors. Once a transformation law $\gamma =f(\theta)$ irrelevant  for a given problem is identified, this principle let us find the most objective prior by identifying $p_\Theta\equiv p_\Gamma$ in the relation $p_\Theta(\theta)d\theta=p_\Gamma(\gamma)d\gamma$.

The cutoff scale $\Lambda$ is given a logarithmically uniform PDF,
\beq p(\Lambda)\propto \frac{1}{\Lambda}\,. \label{2013eft-prior_lambda}\eeq
By doing so, all order of magnitudes are given the same probability density. 
Regarding the dimensionless coefficients $\alpha$, note that the choice of the HDO basis should be irrelevant for the conclusions of our study. Given that coefficients in different basis are related through linear transformations, the most objective prior to associate to each $\alpha_i$ is the uniform PDF,\footnote{Here the principle of indifference sets the shape of the PDFs but does not set the bounds. One can see that ranges on $\alpha$'s are not conserved from one basis to another. In the scenario of democratic HDOs, this issue will be automatically solved, as one relies only on  perturbativity of the HDO expansion to set the bounds on $\alpha$'s. In the scenario of loop suppressed $\mathcal{O}_{FF}$'s, one takes advantage of a particular choice of basis, so the same argument does not apply in that case.  
 } 
\beq
p(\alpha_i)\propto 1\,.\label{2013eft-prior_ci}
\eeq 
This choice of prior is well justified, however, one should keep in mind that other possibilities still exist.

Let us emphasize that in our general framework, the following hypotheses need to be scrutinized.
\begin{itemize}
\item Perturbativity of the HDO expansion, $|\alpha_i|/\Lambda^2 < \mathcal{O}( 1/v^2)$,
\item Perturbativity of the couplings expansions in the UV theory, $|\alpha_i| < \mathcal{O}(16\pi^2)$,
\item HDO generation by loops, 
\item Custodial symmetry.
\end{itemize}

In the present work, we investigate scenarios of low-scale new physics, with values of $\Lambda$ going up to  ${\cal O}(4\pi v)$. We take  custodial symmetry  to be an exact symmetry of the theory.
This forbids the presence of the operators $\mathcal O'_{D^2}$ and $\mathcal O'_{D}$.  As a consequence, one has $a_W=a_Z\equiv a_V$ and some contributions to the EW precision observables are suppressed including the potentially large quadratic divergence in $T$. Recall that $\mathcal O_{\WW}$, $\mathcal O_{\WB}$, and $\mathcal O_{\BB}$ are all independently custodially symmetric. This generally implies that processes involving the $W$ and $Z$ are not identically rescaled, for instance
\beq
\frac{\sigma_{\rm WH}}{\sigma_{\rm WH}^{\rm SM}} \neq \frac{\sigma_{\rm ZH}}{\sigma_{\rm ZH}^{\rm SM}} \,.
\eeq
Our approach goes therefore beyond the fits involving pure rescalings induced by anomalous couplings.

Over this range of $\Lambda$, perturbativity of the HDO expansion is the dominant constraint as it requires $|\alpha_i|<\Lambda^2/v^2$ which automatically implies $|\alpha_i|<16 \pi^2$ and hence perturbativity of the couplings expansions in the UV theory.

When the HDOs are generated within a perturbative UV theory, none of the field strength--Higgs operators ${\cal O}_{FF}\equiv {\cal O}_{\WW,\,\WB,\,\BB,\,GG}$ (see Eqs. \eqref{2013eft-HDO_O_WW} and \eqref{2013eft-HDO_O_GG}) can be generated at tree-level. Because of our appropriate choice of basis, these loop-generated HDOs are exactly the ones associated with the tensorial couplings $\zeta_{g,\gamma,Z\gamma}$. We will therefore distinguish between two scenarios, depending on whether or not the $\mathcal O_{FF}$'s are loop suppressed with respect to the other HDOs. Given that tensorial couplings can play an important role, this distinction is particularly crucial. The two scenarios, denoted by I and II, are respectively dubbed ``democratic HDOs'' and ``loop-suppressed $\mathcal O_{FF}$'s''. The main features are summarized in Table \ref{2013eft-tab_scenarios}.
These two scenarios are generic, in the sense that they encompass all known UV models in addition to the ones not yet thought of. This implies that features predicted only by specific UV models--\textit{e.g.} suppression of HDOs or precise cancellations between HDOs--will get a small statistical weight, as we consider the whole set of UV realizations. Finally, we emphasize that the interpretation of $\Lambda$ as a true new physics scale also depends at which order  the whole set of HDOs is generated. For instance, in the $R$-parity conserving MSSM, the whole set of HDOs is generated only at one-loop order, such that the actual NP scale should be $\mathcal{O}(4\pi \Lambda)$.

A parameterization particularly adapted to low-$\Lambda$ scenarios is as follows. Defining the parameters
\beq
\beta_i=\alpha_i \frac{v^2}{\Lambda^2}\,,
\eeq
it follows that the $\beta$'s and $\Lambda$ are independent, \ie~$p(\alpha_i,\Lambda)=p(\Lambda)p(\beta_i)$. 
The $\beta$'s prior is the uniform PDF over $[-1;1]$, noted $U(\beta_i)$. The prior of $\Lambda$ is $p(\Lambda)\propto\Lambda^{2n-1}$, where $n$ is the number of  $\beta$'s. In our case, $n=9$ is large enough such that this prior is essentially peaked at $\Lambda_{\rm max}$, $p(\Lambda)\approx\delta(\Lambda-\Lambda_{\rm max})$. We have therefore 
\beq
p(\alpha_i,\Lambda)=\delta(\Lambda-\Lambda_{\rm max})U(\beta_1)\ldots U(\beta_n)\,.
\eeq

This factorization allows us to marginalize over $\Lambda$, and to present our results in terms of $\beta$'s, which contain all the relevant information. A mild dependence on $\Lambda$ will remain through loop-level $\mathcal{O}(\log \Lambda)$ terms in the $S$ and $T$ parameters, that will be discussed below. The fact that $\beta$'s prior is  uniform and spans a constant range is essential  to facilitate interpretation of the posterior PDFs. The fact that $\Lambda\approx \Lambda_{\rm max}$ is also useful, as it renders straightforward the evaluation of the few $\Lambda$-dependent terms.

\begin{table}
\center
\begin{tabular}{|c|c|c|}
\hline
& I) Democratic HDOs & II) Loop-suppressed $\mathcal O_{FF}$'s \\
\hline
$\Lambda$   &  $4\pi v$ &  $4\pi v$ \\
\hline
$\beta_{FF}$   &  $[-1,1]$ &  $[-1/16\pi^2,1/16\pi^2]$ \\
\hline
Other $\beta$   &  $[-1,1]$ & $[-1,1]$ \\
\hline
\end{tabular}
\caption{Summary of the setup of the scan in the two scenarios we consider. The $\beta_{FF} \equiv \alpha_{FF} \, v^2 / \Lambda^2$ coefficients (where $FF = \WW,\,WB,\,BB,\,GG$) correspond to the field-strength--Higgs operators. In both cases we take custodial symmetry to be an unbroken symmetry.} \label{2013eft-tab_scenarios}
\end{table}

This parameterization turns out to be convenient in order to extract information about HDOs  in a scale independent way, up to a mild ${\cal O}(\log\Lambda_{\rm max})$ dependence. For example, for a given $\Lambda$, one can directly read the values of $\alpha$'s on the $\beta$'s plot. Similarly, for given $\alpha$'s, one can deduce the allowed $\Lambda$ values from the plots. 
This parameterization is appropriate at low $\Lambda$, up to $\Lambda={\cal O}(4\pi v)$. Beyond this scale, the bound from HDO perturbative expansion competes with the bound from the perturbative expansion of the couplings. Once the latter dominates, the features of factorization no longer hold.

\subsubsection{The MCMC setup}

We evaluate posterior PDFs by means of a Markov Chain Monte Carlo (MCMC) method. The basic idea of a MCMC is setting a random walk in the parameter space such that the density of points asymptotically reproduces the posterior PDF. Any marginalisation is then reduced to a summation over the points of the Markov chain. We refer to~\cite{Allanach:2005kz,Trotta:2008qt} for details on MCMCs and Bayesian inference. Our MCMC method uses the Metropolis-Hastings algorithm with a symmetric, Gaussian proposal function.
We run respectively 50 and 15 chains with $\mathcal{O}(10^8)$ iterations each for the democratic HDOs case and the loop-suppressed $\mathcal{O}_{FF}$'s case.
Finally, we check the convergence of our chains using an improved Gelman and Rubin test with multiple chains~\cite{Gelman:1992zz}. The first $10^4$ iterations are discarded (burn-in).

\subsection{Inference on HDOs} \label{2013eft-se_results}

In this section we present and analyze the posterior PDFs arising in our scenarios of democratic HDOs and loop-suppressed $\mathcal{O}_{FF}$'s, denoted by I and II, respectively. Our results will be shown in terms of the $\beta_i \equiv \alpha_i v^2 / \Lambda^2$ parameters, which encode information about the fundamental parameters. Recall that the $\beta$'s prior PDF is uniform, and that the $\beta$ PDFs we show are valid for any value of the cutoff scale $\Lambda<\mathcal{O}(4\pi v)$, up to a mild $\log\Lambda$ dependence.
Moreover, this parameterization sets $\Lambda\approx \Lambda_{\rm max}$.
The posterior PDF we present is computed for $\Lambda=4\pi v\approx 3\,$TeV. For smaller $\Lambda$, we expect the $\propto \log\Lambda$ constraints from $\Delta S$ and $\Delta T$ to mildly relax. 
We will comment below on this effect.

We will also discuss deviations from the SM cross sections and decay widths, defining
\beq
R_X = \frac{\sigma_X}{\sigma^{\rm SM}_X}\,, \quad R_Y=\frac{\Gamma_{Y}}{\Gamma^{\rm SM}_{Y}}\,, \quad R_{\rm width}=\frac{\Gamma_h}{\Gamma^{\rm SM}_h}\,,
\label{2013eft-eq:Rs}
\eeq
where $X = {\rm ggF}, {\rm VBF}, {\rm WH}, {\rm ZH}, {\rm ttH}$, and $Y = \gamma\gamma$, $ZZ$, $Z\gamma$, $WW$, $b\bar{b}$, $\tau\tau$. Note that the observables are the signal strengths $\mu(X,Y)$ rather than the individual $R_X$ and $R_Y$.

\begin{figure}
	\centering
	\includegraphics[width=4.5cm]{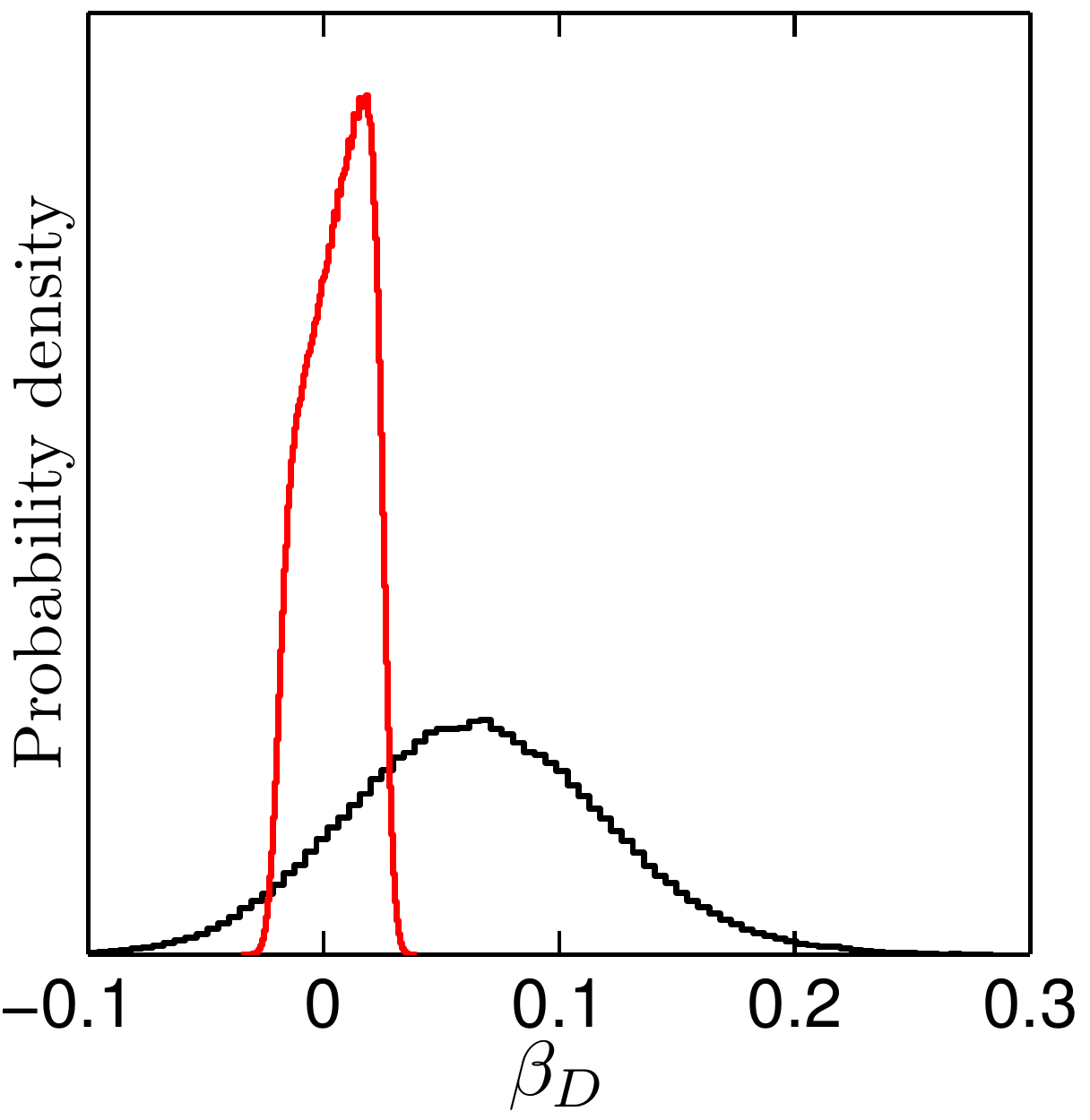}
    \includegraphics[width=4.5cm]{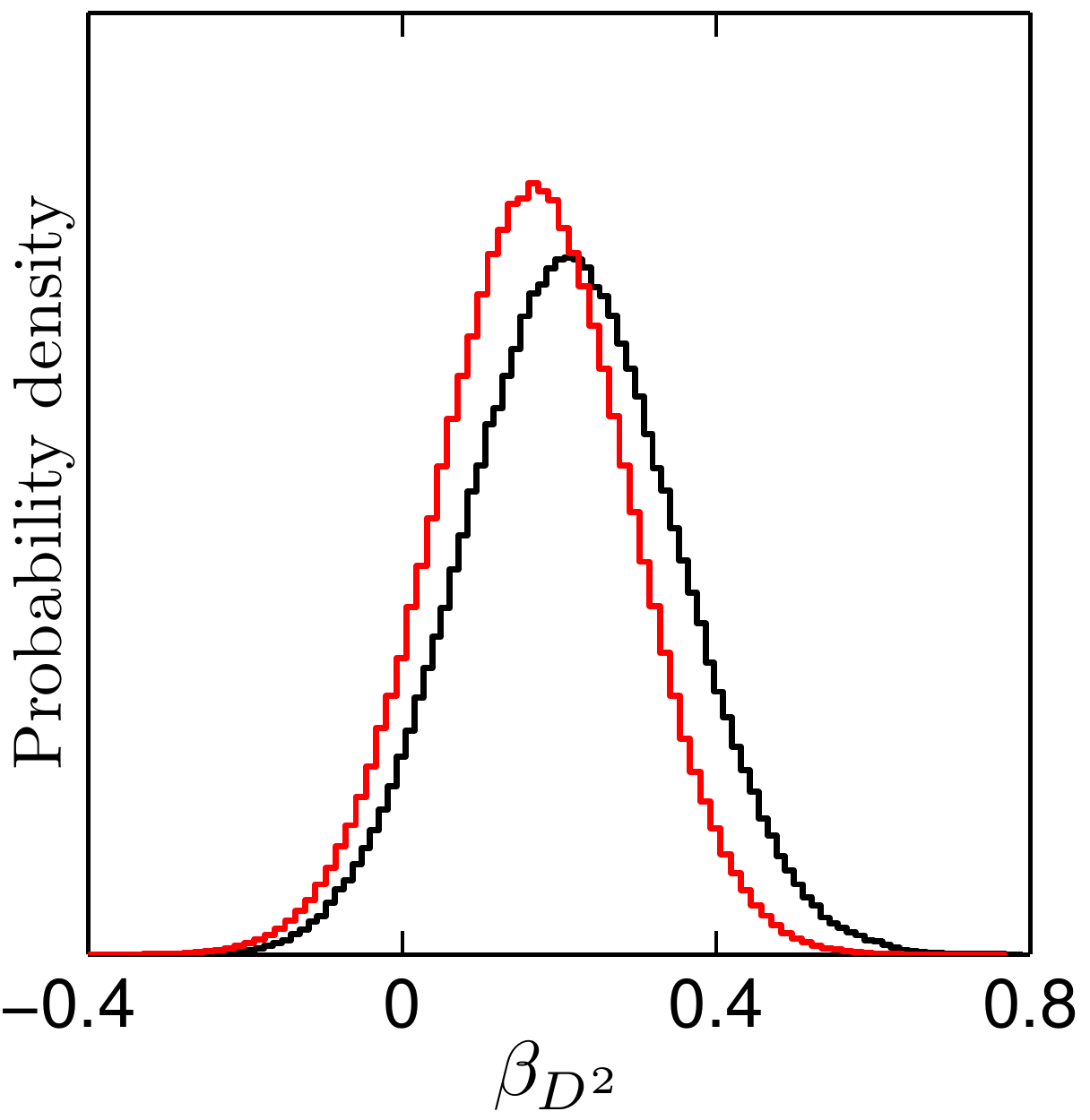}						
	\includegraphics[width=4.4cm]{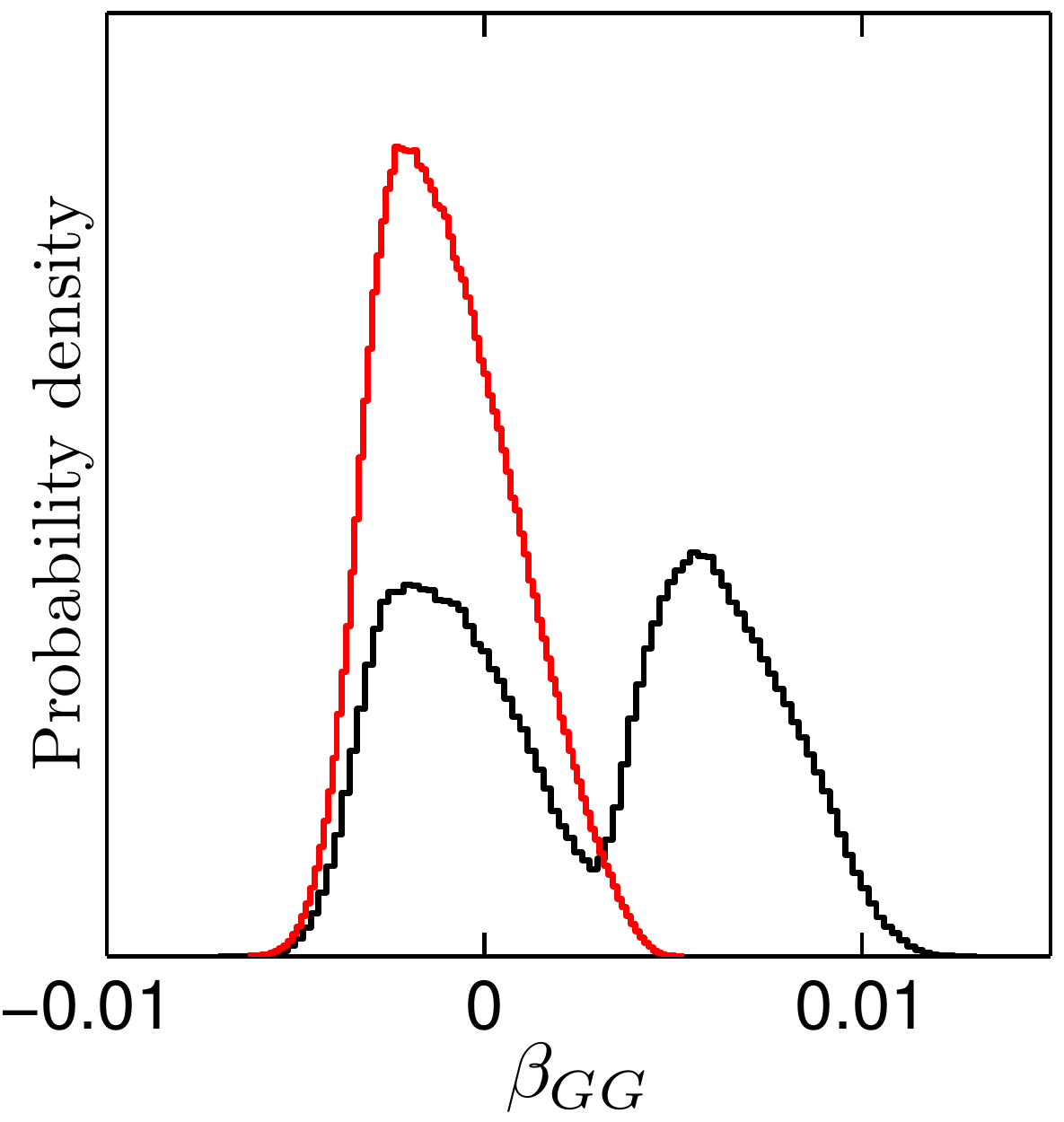}
	\includegraphics[width=4.5cm]{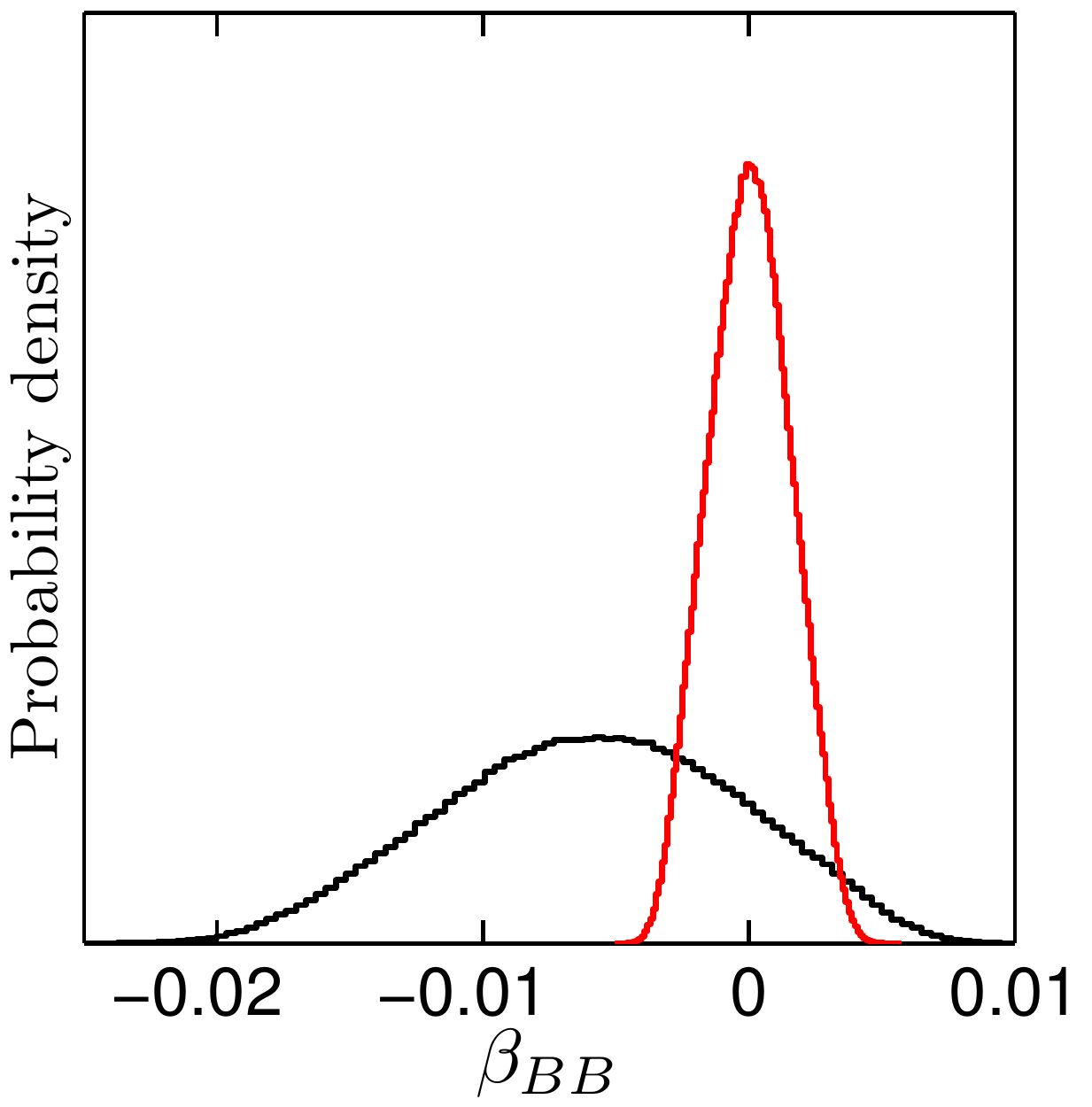}
	\includegraphics[width=4.6cm]{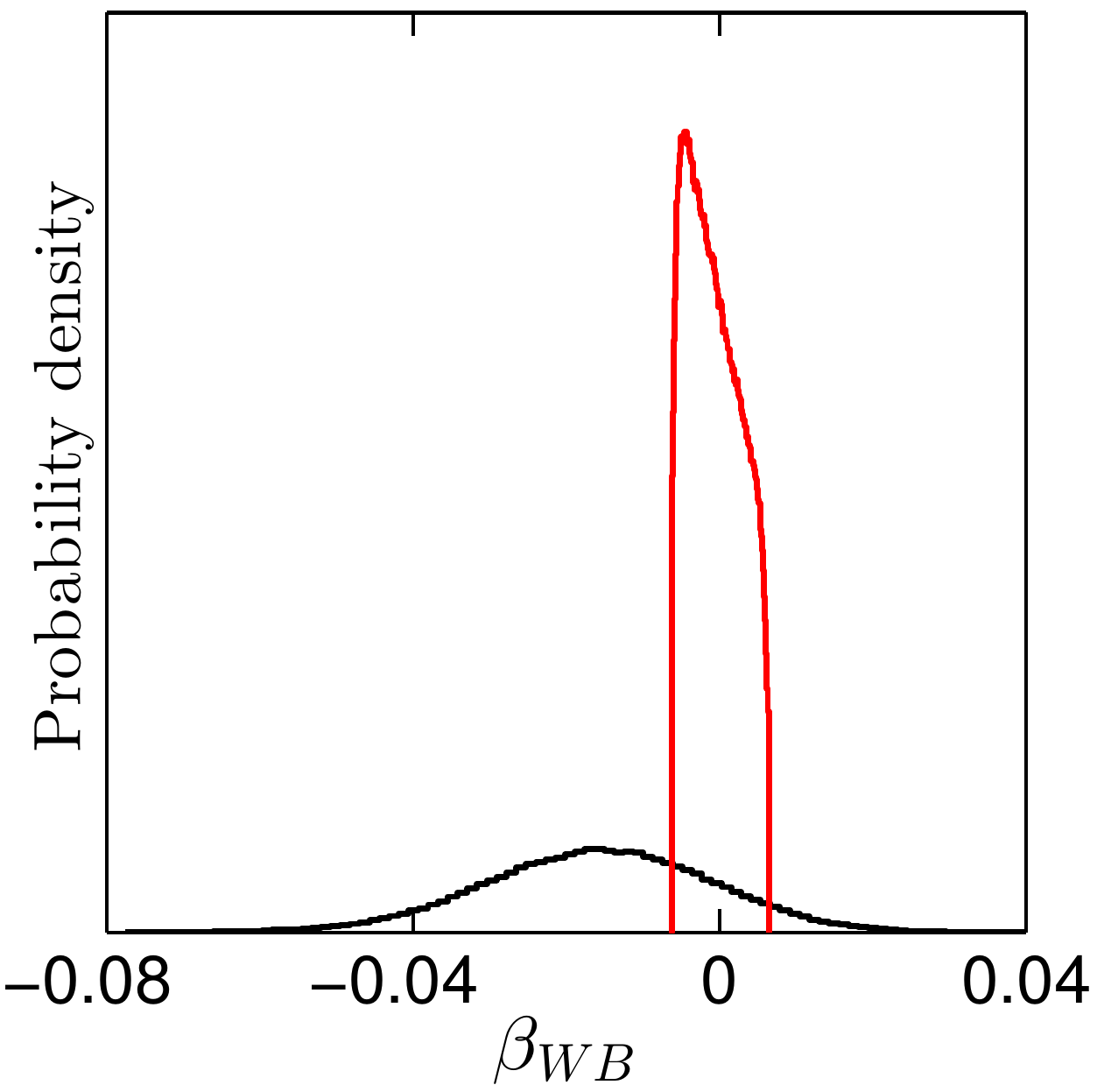}
	\includegraphics[width=4.5cm]{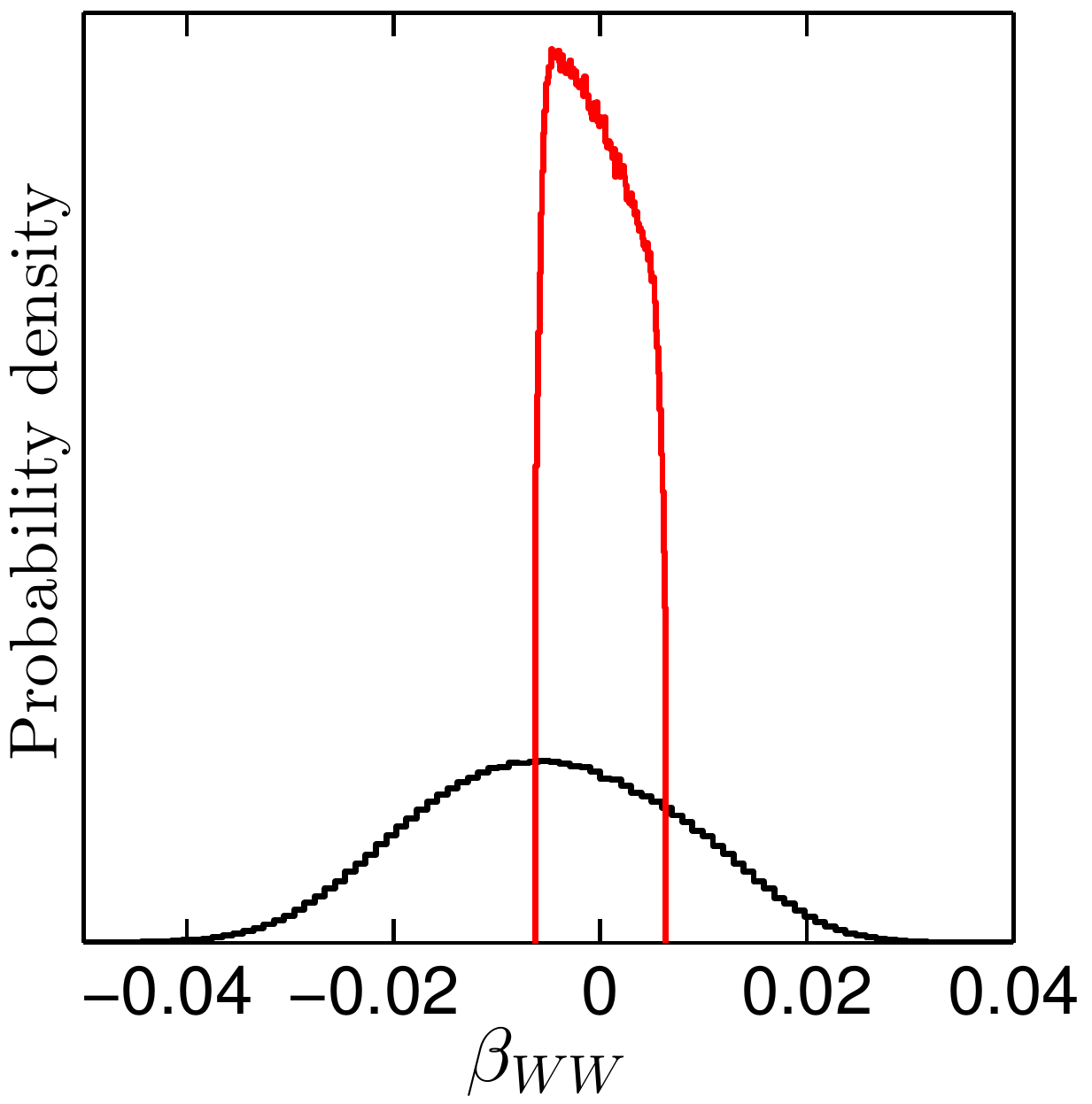}
	\includegraphics[width=4.5cm]{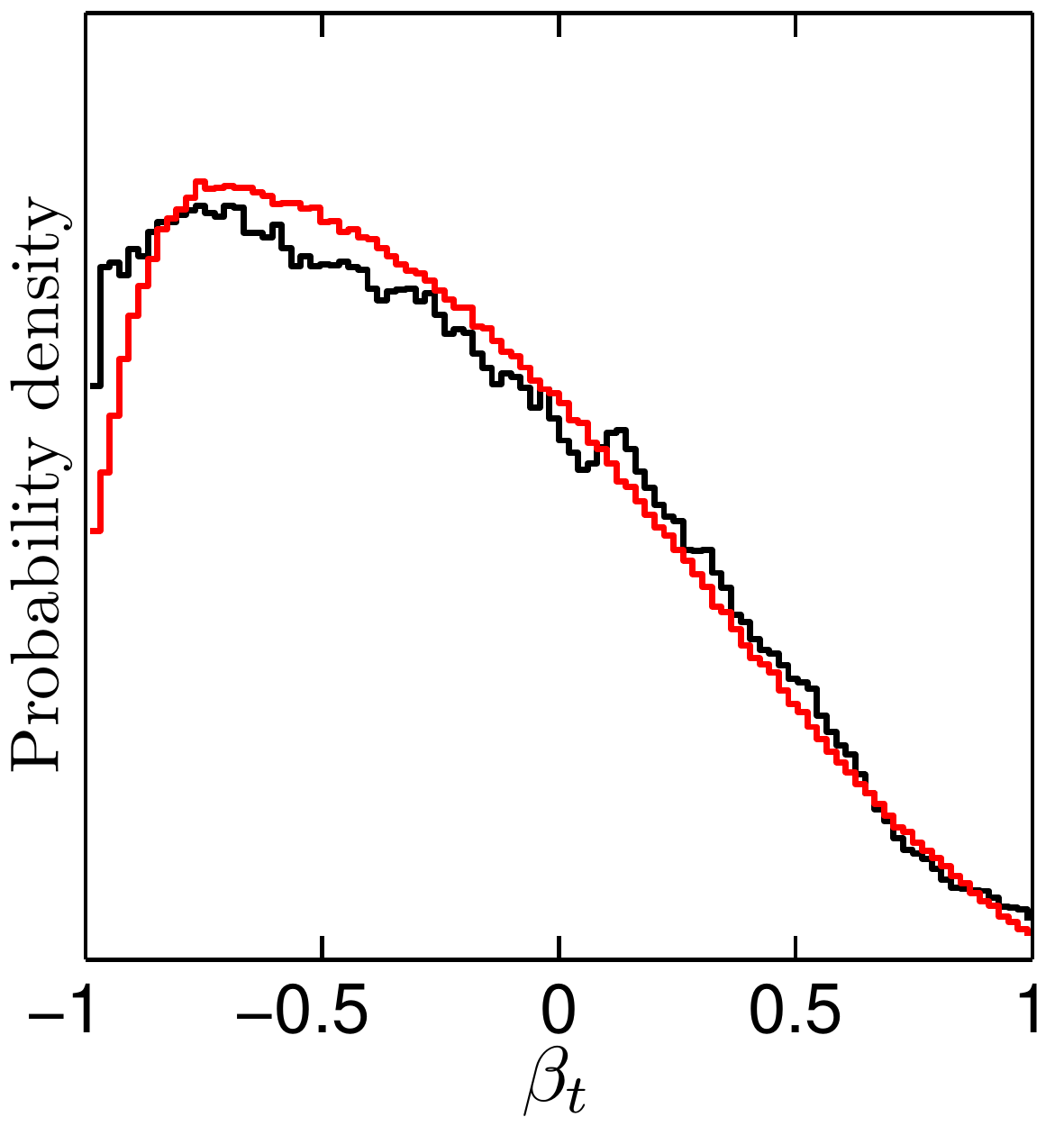}
	\includegraphics[width=4.5cm]{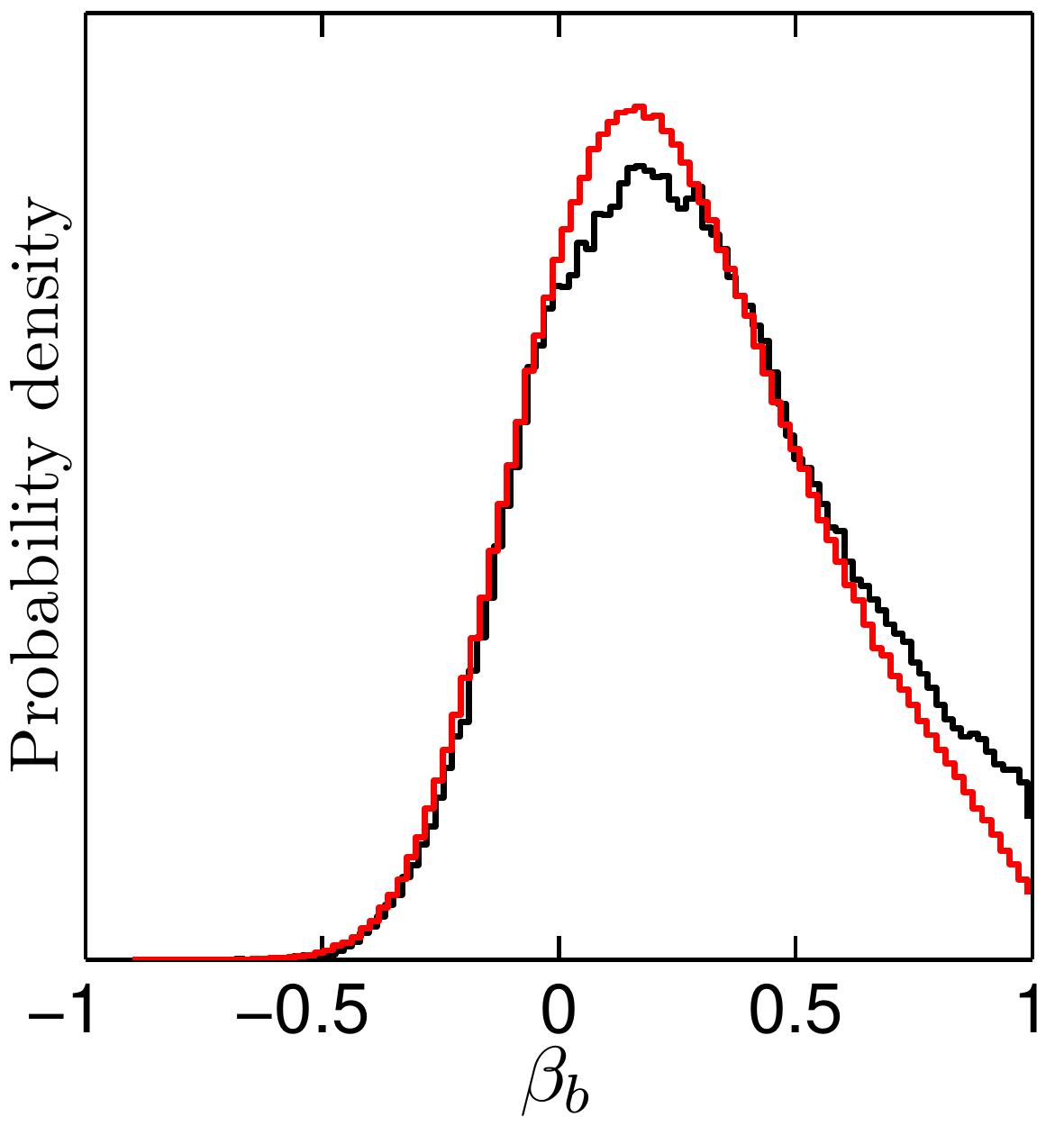}
	\includegraphics[width=4.5cm]{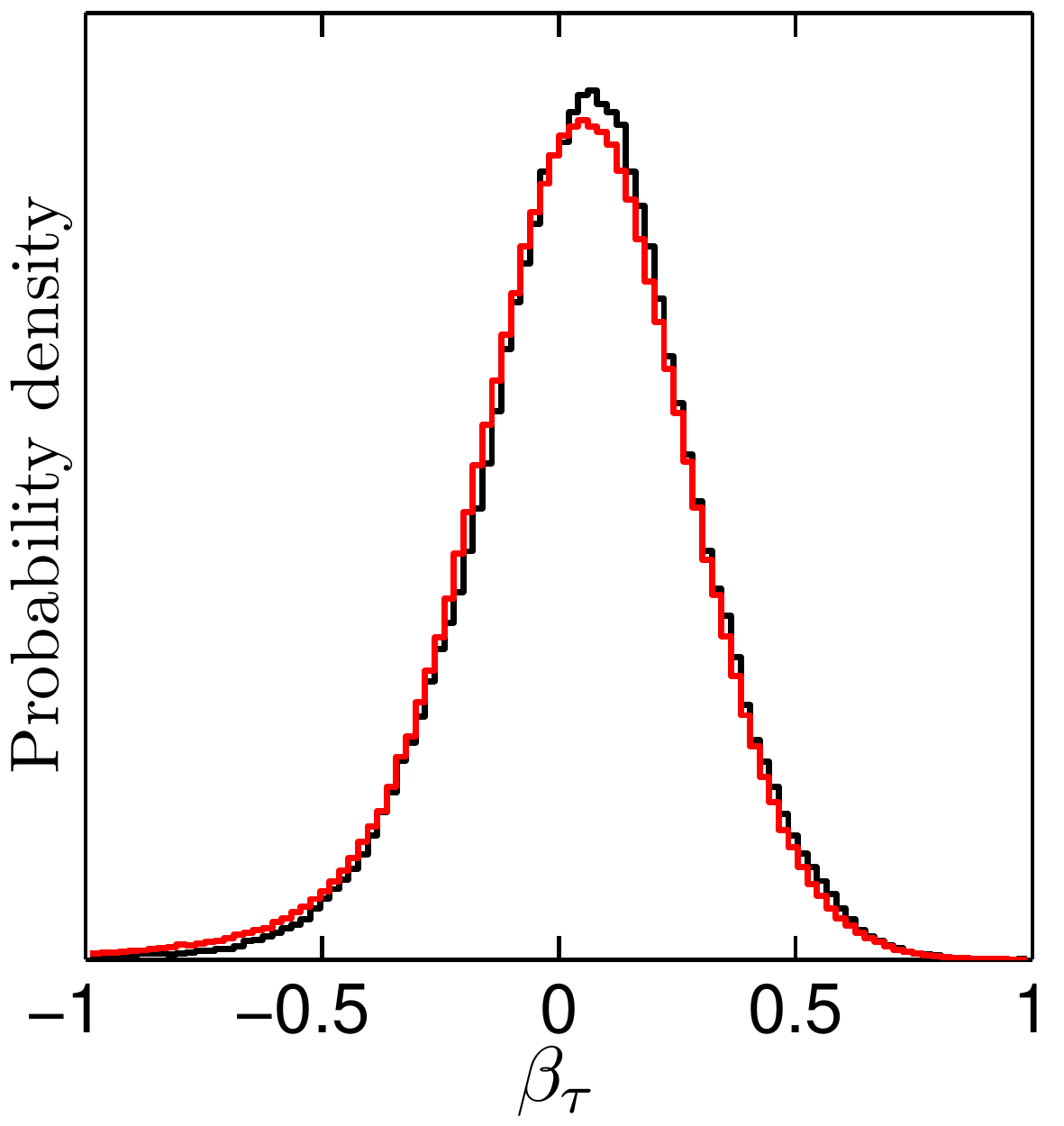}	
	\caption{Posterior PDFs of the 9 fundamental parameters, $\beta_i \equiv \alpha_i v^2/\Lambda^2$, in scenario I (black) and scenario II (red). \label{2013eft-fig:betas}}
\end{figure}

\begin{figure}
	\centering
     	\includegraphics[width=5cm]{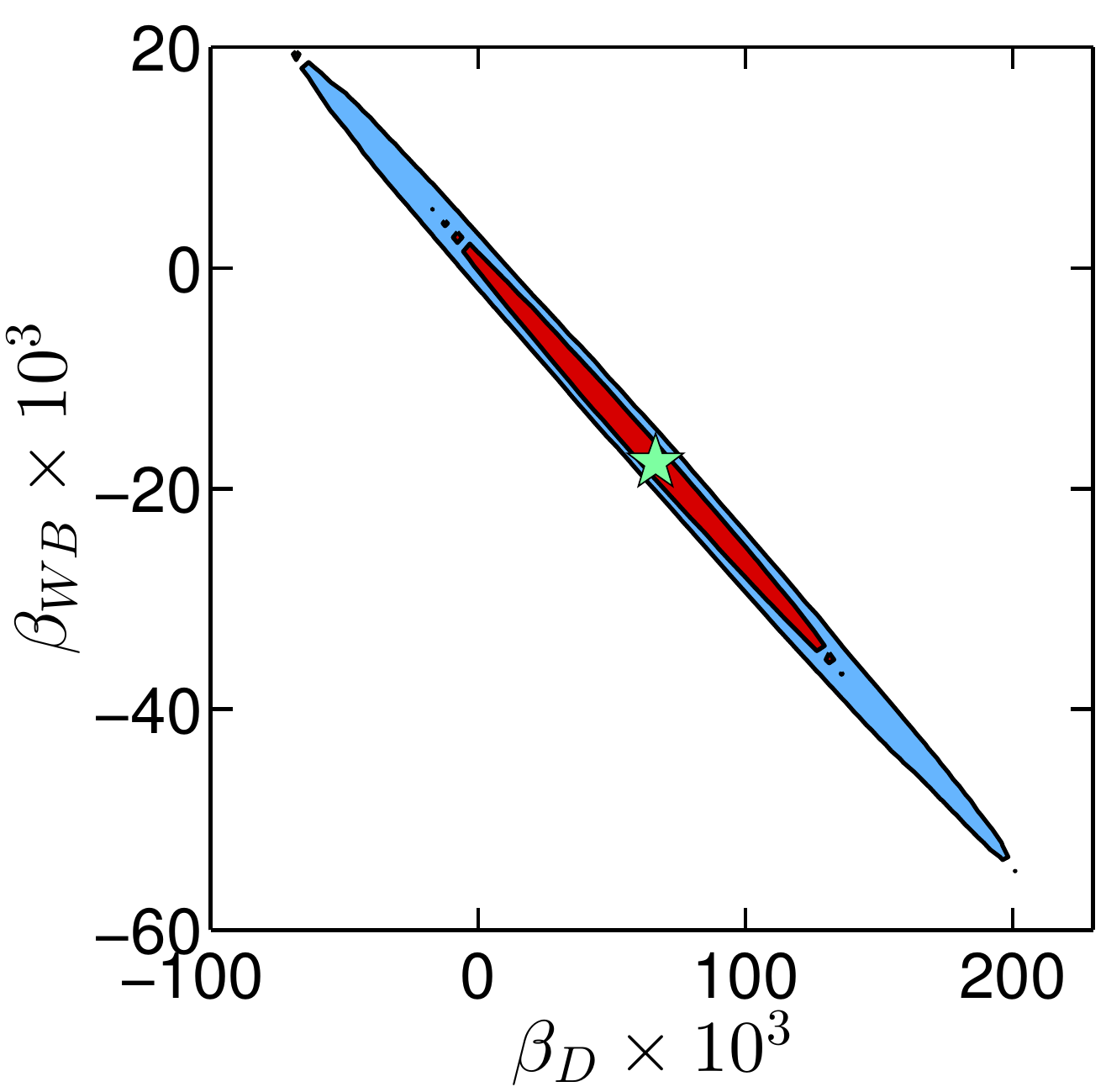}	
		\includegraphics[width=5cm]{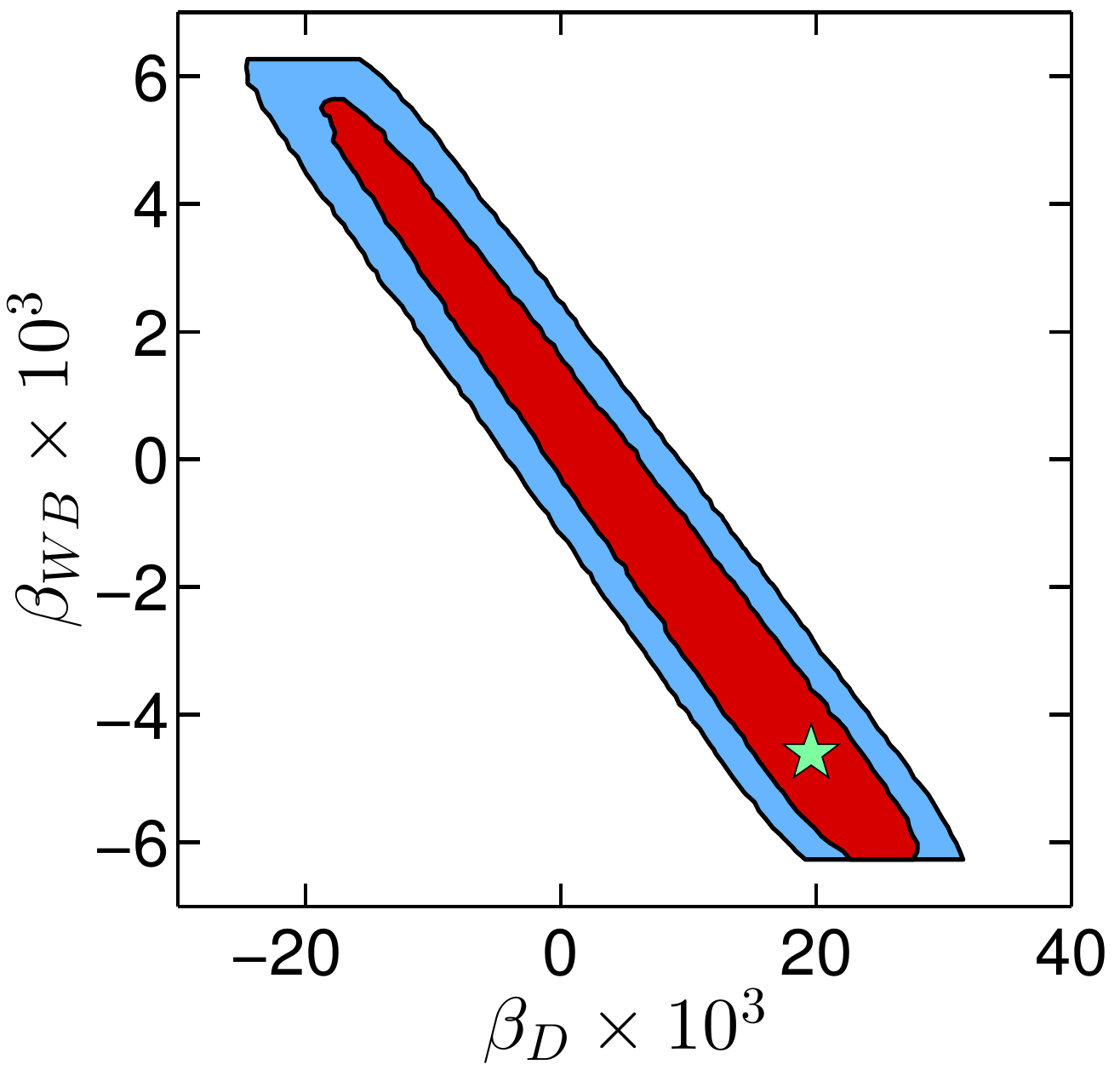}
	\caption{Posterior PDFs of $\beta_{\WB}$ versus $\beta_{D}$ in scenario I (left) and scenario II (right). The red and blue regions correspond to the 68\% and 95\% Bayesian credible regions (BCRs). The green star indicates the maximum of our posterior PDF. \label{2013eft-fig:corr_D_WB}}
\end{figure}

We present one-dimensional PDFs of the fundamental parameters $\beta_i$ for both scenarios in Fig.~\ref{2013eft-fig:betas}.
Moreover, in Table~\ref{2013eft-tab:BCRs_7} we report the $68\%$ and $95\%$ Bayesian credible intervals (BCIs) for these quantities.  We also present the BCIs for the other, dependent quantities, \ie~the anomalous couplings $a_V$ and $c_f$, the tensorial couplings $\zeta_i$, and the various $R$'s. 

\begin{table}[ht]
\center
\begin{tabular}{|c|c|c|c|c|}
\hline
& \multicolumn{2}{|c|}{scenario I} & \multicolumn{2}{|c|}{scenario II} \\
& 68\% BCI & 95\% BCI & 68\% BCI & 95\% BCI \\
\hline
$\beta_D \times 10^3$ & $[10,120]$ & $[-50,180]$ & $[-6, 23]$ & $[-19,26]$ \\
$\beta_{D^2} \times 10^3$ & $[70,350]$ & $[-50,480]$ & $[40,290]$ & $[-90,400]$ \\
$\beta_t \times 10^3$ & $[-1000,110]$ & $[-1000,610]$ & $[-930,10]$ & $[-1000,590]$ \\
$\beta_b \times 10^3$ & $[-10,530]$ & $[-220,930]$ & $[-110,500]$ & $[-280,860]$ \\
$\beta_\tau \times 10^3$ & $[-170,300]$ & $[-420,510]$ & $[-190,270]$ & $[-450,510]$ \\
$\beta_{GG} \times 10^3$ & $[-3.2,8.0]$ & $[-4.0,9.6]$ & $[-3.3,0.6]$ & $[-4.2,2.7]$ \\
$\beta_{\WW} \times 10^3$ & $[-19,7]$ & $[-30,18]$ & $[-5.6,2.3]$ & $[-6.0,5.6]$ \\
$\beta_{\WB} \times 10^3$ & $[-32,1]$ & $[-49,13]$ & $[-6.0,1.6]$ & $[-6.3,5.3]$ \\
$\beta_{\BB} \times 10^3$ & $[-12,0]$ & $[-17,4]$ & $[-1.7,1.6]$ & $[-2.9,3.0]$ \\
\hline
$a_{V}$ & $[1.02, 1.15]$ & $[0.96,1.21]$ & $[1.02,1.14]$ & $[0.96,1.20]$ \\
$c_t$ & $[0.05,1.14]$ & $[0.03, 1.63]$ & $[0.06,1.01]$ & $[0.04,1.60]$ \\
$c_b$ & $[0.90,1.54]$ & $[0.79,1.96]$ & $[0.89,1.50]$ & $[0.72,1.86]$ \\
$c_\tau$ & $[0.84,1.31]$ & $[0.58, 1.53]$ & $[0.81,1.27]$ & $[0.55,1.51]$ \\
\hline

$\zeta_g \, v \times 10^3$ & $[-3.2,8.0]$ & $[-4.0,9.6]$ & $[-3.3,0.6]$ & $[-4.2,2.7]$ \\
$\zeta_\gamma\, v \times 10^3$ & $[-5.5,0.5]$ & $[-6.1,0.9]$ & $[-0.33,0.46]$ & $[-0.69,0.86]$ \\
$\zeta_{Z\gamma}\, v \times 10^3$ & $[-13,18]$ & $[-18,30]$ & $[-4.9,4.4]$ & $[-7.6,7.9]$ \\
$\zeta_Z\, v \times 10^3$ & $[-20,2]$ & $[-31,11]$ & $[-3.4,2.3]$ & $[-5.1,4.4]$ \\
$\zeta_W\, v \times 10^3$ & $[-39,15]$ & $[-59,37]$ & $[-11,5]$ & $[-12,11]$ \\
\hline
$R_{\rm ggF}$ & $[0.6,1.3]$ & $[0.5,2.0]$ & $[0.6,1.3]$ & $[0.4,2.0]$ \\
$R_{\rm VBF}$ & $[1.0,1.4]$ & $[0.9,1.6]$ & $[1.0,1.3]$ & $[0.9,1.4]$ \\
$R_{\rm WH}$ & $[0.7,1.3]$ & $[0.5,1.7]$ & $[1.0,1.3]$ & $[0.9,1.4]$ \\
$R_{\rm ZH}$ & $[0.7,1.2]$ & $[0.5,1.5]$ & $[1.0,1.3]$ & $[0.9,1.4]$ \\
$R_{\rm ttH}$ & $[0.02,1.0]$ & $[0.02,2.6]$ & $[0,0.9]$ & $[0,2.5]$ \\
\hline
$R_{\gamma\gamma}$ & $[1.1,1.9]$ & $[0.8,2.5]$ & $[1.1,1.8]$ & $[0.8,2.3]$ \\
$R_{Z\gamma}$ & $[0,5.2]$ & $[0,12.0]$ & $[0,2.2]$ & $[0,4.3]$ \\
$R_{ZZ}$ & $[1.0,1.3]$ & $[0.9,1.5]$ & $[1.0,1.3]$ & $[0.9,1.4]$ \\
$R_{\WW}$ & $[1.0,1.3]$ & $[0.9,1.5]$ & $[1.0,1.3]$ & $[0.9,1.4]$ \\
$R_{b\bar{b}}$ & $[0.7,2.2]$ & $[0.5,3.6]$ & $[0.7,2.1]$ & $[0.4,3.3]$ \\
$R_{\tau\tau}$ & $[0.6,1.6]$ & $[0.3,2.2]$ & $[0.6,1.5]$ & $[0.2,2.1]$ \\
\hline
$R_{\rm width}$ & $[0.8, 1.9]$ & $[0.7,2.7]$ & $[0.8,1.8]$ & $[0.6,2.5]$ \\
\hline
\end{tabular}
\caption{$68\%$ and $95\%$ Bayesian credible intervals (BCIs) for the democratic HDOs case (scenario I) and for the loop-suppressed $\mathcal{O}_{FF}$'s case (scenario II).}
\label{2013eft-tab:BCRs_7}
\end{table}

One can first remark that all of our HDO coefficients except $\beta_t$ and $\beta_b$ are constrained enough to stay within the bound $|\beta_i| <1$, as required for the convergence of the HDO expansion. Furthermore, the $\beta_{FF} \equiv \beta_{\WW,\,WB,\,BB,\,GG}$ coefficients are ${\cal O}(0.01)$ in both scenarios.
$\beta_D$ and $\beta_{\WB}$ are strongly correlated in both scenarios as they appear in the $S$ parameter at tree-level, see Eq.~\eqref{2013eft-S_tree} (we recall that we fix $\alpha'_{D} = \alpha'_{D^2} = 0$ in order to preserve custodial symmetry). We thus have $2\,c_w \, \beta_{\WB}\approx-s_w\, \beta_{D}$ as can be seen in Fig.~\ref{2013eft-fig:corr_D_WB}.
The TGV observables also involve $\beta_{D}$ and $\beta_{\WB}$ (see Eq.~\eqref{2013eft-eq:tgv_hdo}), and thus provide an independent constraint on $\beta_{D}$ (or equivalently $\beta_{\WB}$). The slight deficit in $\kappa_{\gamma}$ and $g_1^Z$ as measured by LEP, see Eq.~\eqref{2013eft-eq:lep_tgv}, tend to favors positive (negative) $\beta_{\WB}$ ($\beta_D$). Finally, note that in scenario II the PDF of $\beta_{\WB}$ is limited to the $[-1/16\pi^2,1/16\pi^2]$ range since we consider that the operator ${\cal O}_{\WB}$ is loop-suppressed. This in turn fixes the allowed range for $\beta_{D}$.

The $\beta_{D^2}$ coefficient is allowed to deviate significantly from 0 as it only appears in loop contributions to $S$ and $T$ and in $a_V$. The probability of having $\beta_{D^2} > 0$ is 94\% (90\%)  in scenario I (II) and comes from $T$, as well as VBF and VH production modes and $h\to VV$ decays. A value for $a_V > 1$ leads to a positive contribution to $T$, as well as an enhancement of the VBF and VH production processes, the $h \rightarrow VV^*$ decays, and also to the loop-induced decay rates, $h \rightarrow \gamma\gamma$ and $h \rightarrow Z\gamma$.

$\beta_{\WW}$ and $\beta_{\BB}$ are mainly constrained by the searches for $h \rightarrow \gamma\gamma$ and $h \rightarrow Z\gamma$  as they contribute to the tensorial couplings $\zeta_\gamma$ and $\zeta_{Z\gamma}$, see Eq.~\eqref{2013eft-lambda_gamma_tens} and \eqref{2013eft-lambda_Zgamma_tens}. Given the large allowed range for $\beta_{\WB}$, a cancellation has to occur with $\beta_{\WW}$ and $\beta_{\BB}$ in order to achieve a $h \rightarrow \gamma\gamma$ rate compatible with experiment.
As a result, negative values of $\beta_{\WW}$ and $\beta_{\BB}$ are favored. 
Moreover, $\beta_{\WW}$ is also constrained from VH production processes via the quantities $\zeta_V$. In contrast $\beta_{\BB}$ and $\beta_{\WB}$ play no role for these measurements, as they are more strongly constrained by the other effects mentioned above. 
The contributions of $\beta_{\WW}$ and $\beta_{\BB}$ to $S$ are up to $\mathcal O(0.03)$ and do not impact the PDFs. This effect is even smaller in scenario II, and is also smaller if we take $\Lambda < 4 \pi v$ due to the $\log(m_h/\Lambda)$ factor in Eq.~(\ref{2013eft-eq:DeltaS}). 
We note that in scenario II the PDFs for $\beta_{\WW}$ and $\beta_{\WB}$ can easily reach the bounds set by the priors, while $\beta_{\BB}$ is more strongly constrained by the data. This is due to the fact that $\beta_{\BB}$ enters in $\zeta_\gamma$ with a coefficient roughly four times larger than the other two.

\begin{figure}
	\centering
		\includegraphics[width=4.5cm]{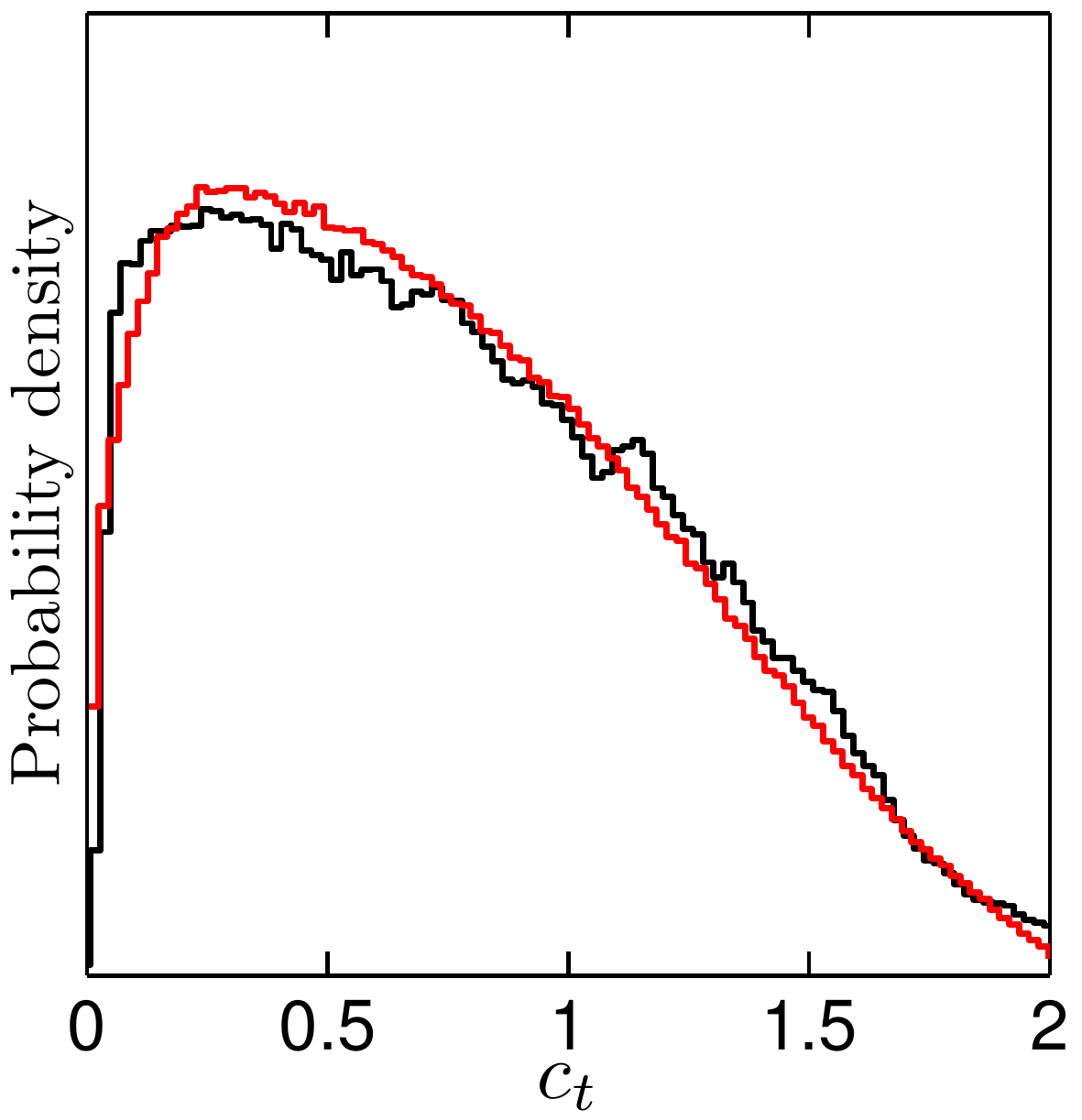}		
		\includegraphics[width=4.4cm]{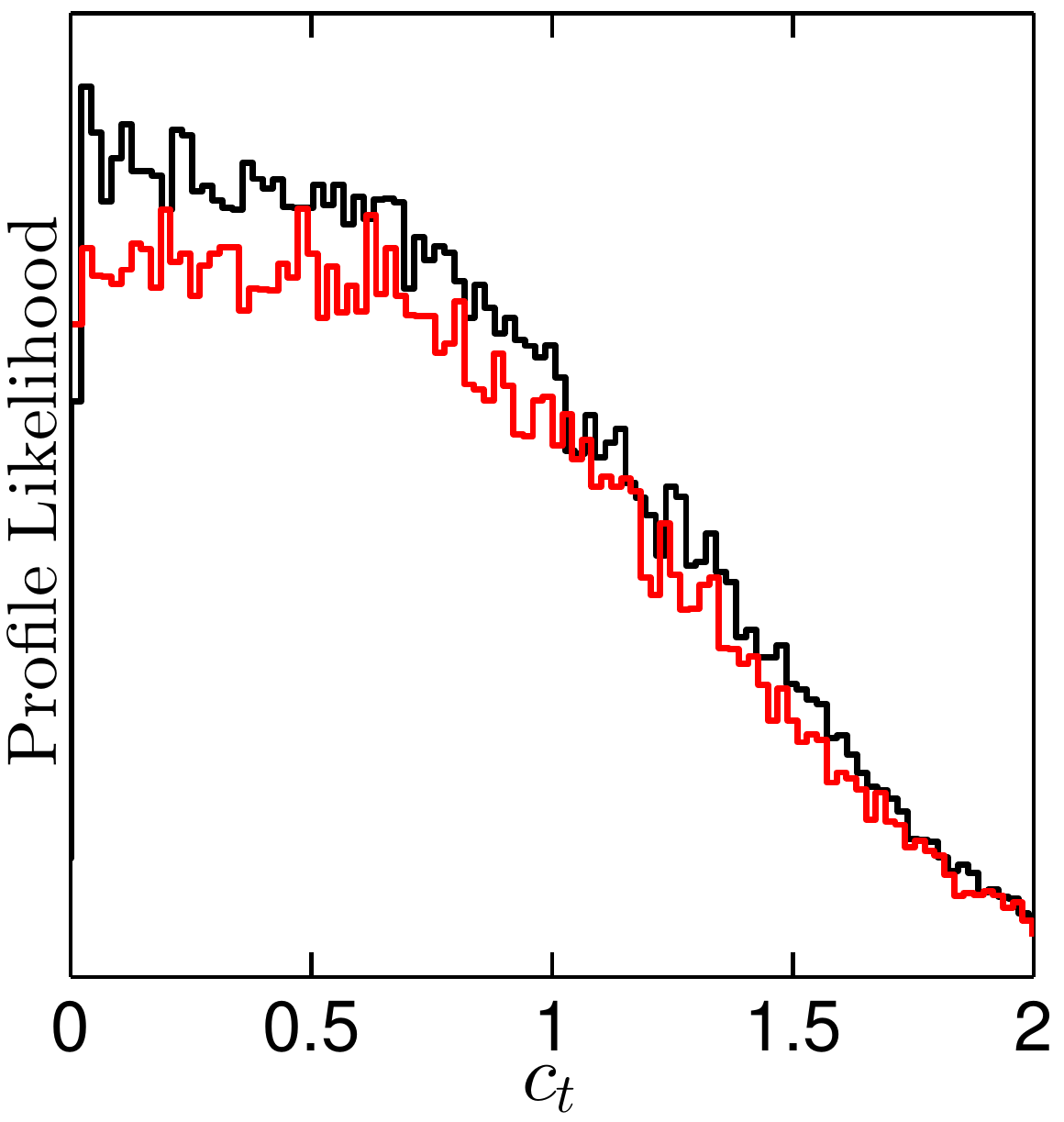}
	\caption{In the left panel, posterior PDF of $c_t$ in scenario I (black) and scenario II (red). In the right panel, profile likelihood along the $c_t$ axis in scenario I and scenario II (same color code). \label{2013eft-ct_profile}}
\end{figure}

Finally, the Yukawa corrections parametrized by $\beta_f$ ($f = t,b,\tau$) are much less constrained as they only contribute to the rescaling factors $c_f$, but account for most of the deviations of $c_f$ from 1, such that we often have $|\beta_f| \gg |\beta_D/4|$ and thus $c_f \approx 1 + \beta_f$.
It is worth noting that $\beta_t$ has a fairly large probability of being close to $-1$, which leads to small or vanishing $c_t$. The posterior PDF of $c_t$ is shown in the left panel of Fig.~\ref{2013eft-ct_profile}.

In such case, one may wonder whether or not the preference for small $c_t$ is due to a volume effect caused by the process of marginalization. To this end, we display in the right panel of Fig.~\ref{2013eft-ct_profile} the profile likelihood for the parameter $c_t$, \ie~the likelihood for given $c_t$, maximized over all the other parameters. We conclude that in both scenarios, the preference for small $c_t$ originates from the likelihood and not from a volume effect.\footnote{We have also checked that no volume effects appear for any of the other posterior PDFs either.}

\begin{figure}
	\centering
	\includegraphics[width=5.1cm]{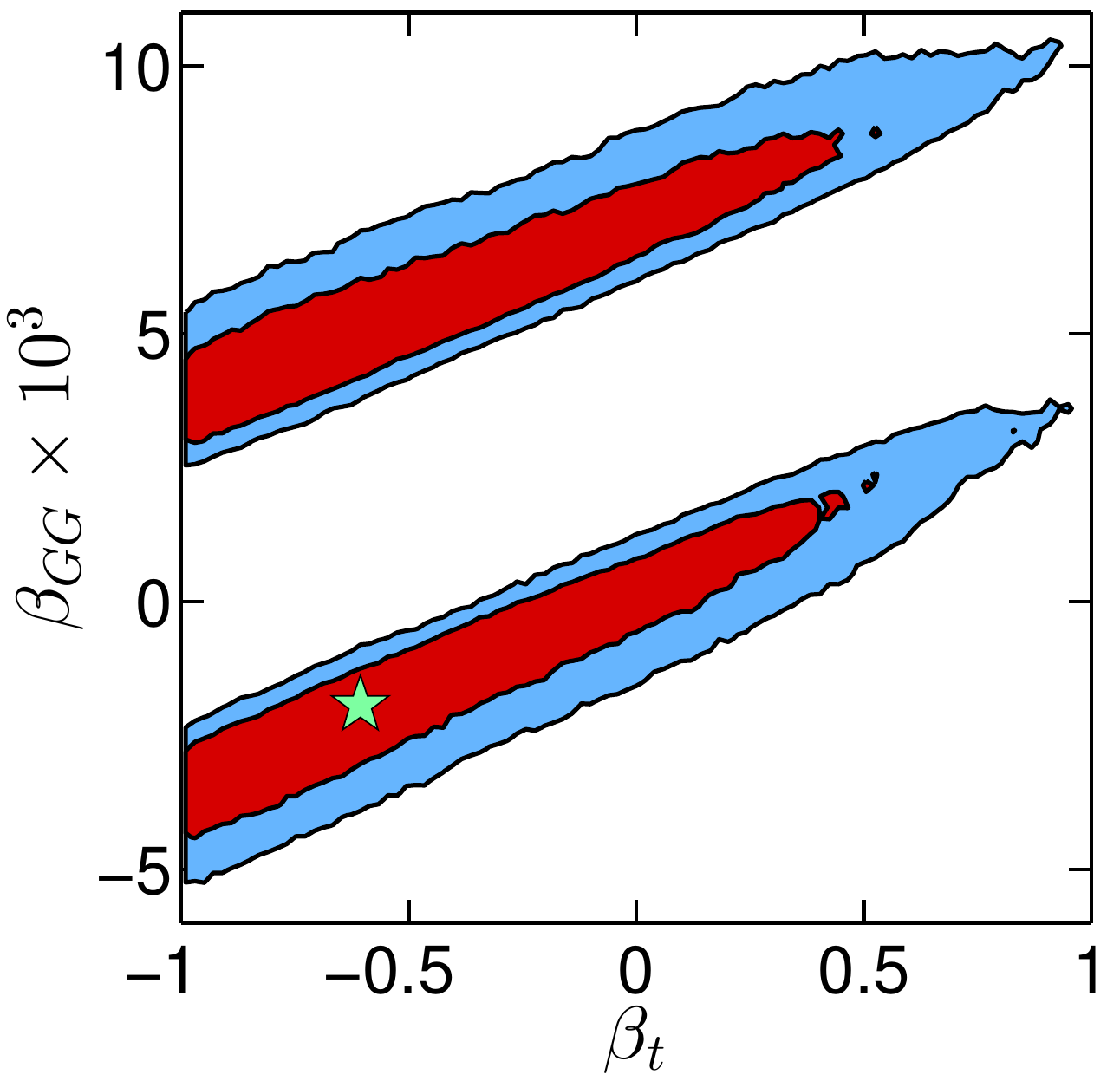}	
	\includegraphics[width=5.1cm]{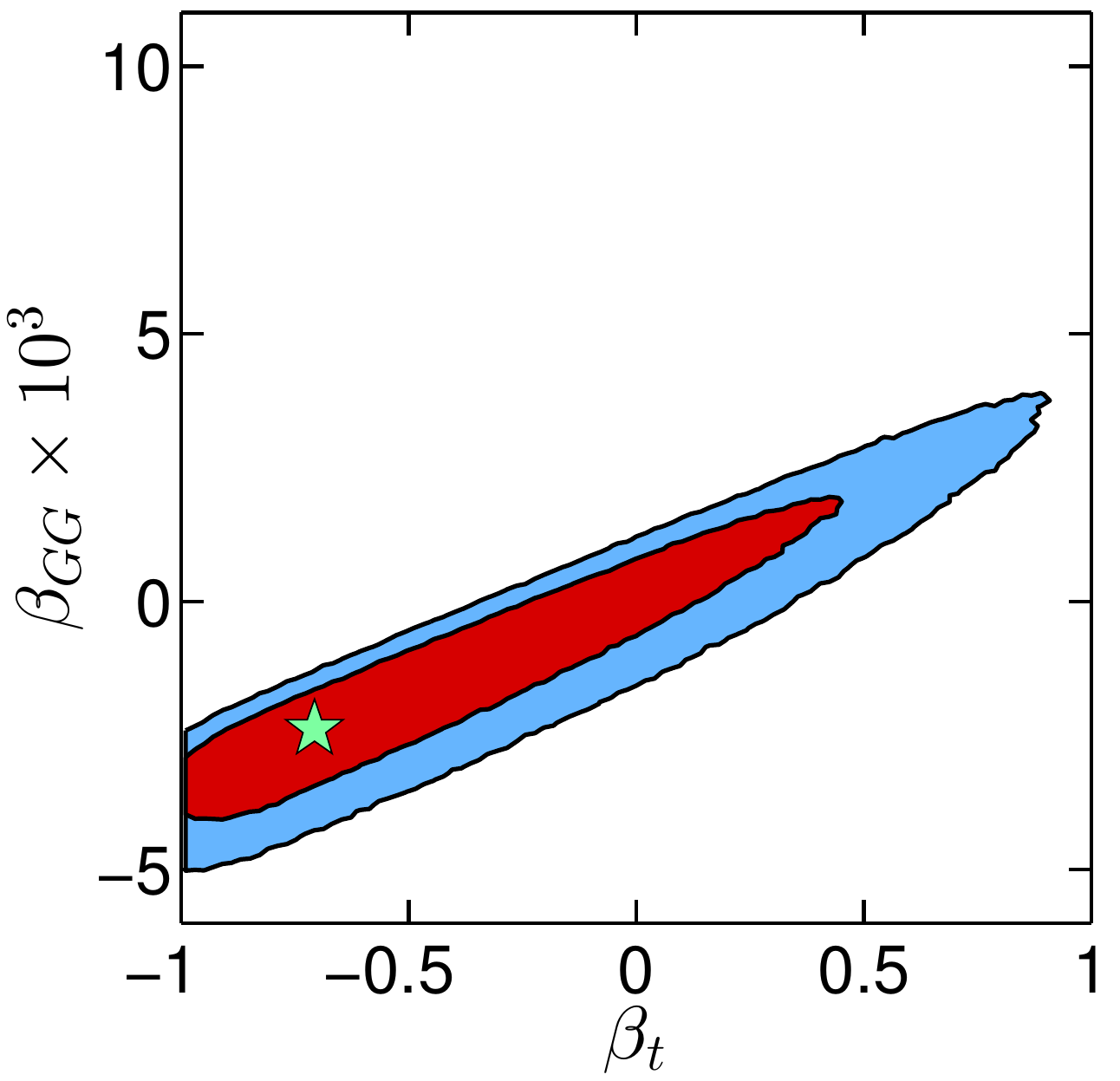} 							
	\caption{Posterior PDF of $\beta_{GG}$ versus $\beta_t$ in scenario I (left) and scenario II (right). Color code as in
Fig.~\ref{2013eft-fig:corr_D_WB}.
  \label{2013eft-fig:beta_t_beta_GG}}
\end{figure}

\begin{figure}
	\centering
     	\includegraphics[width=4.45cm]{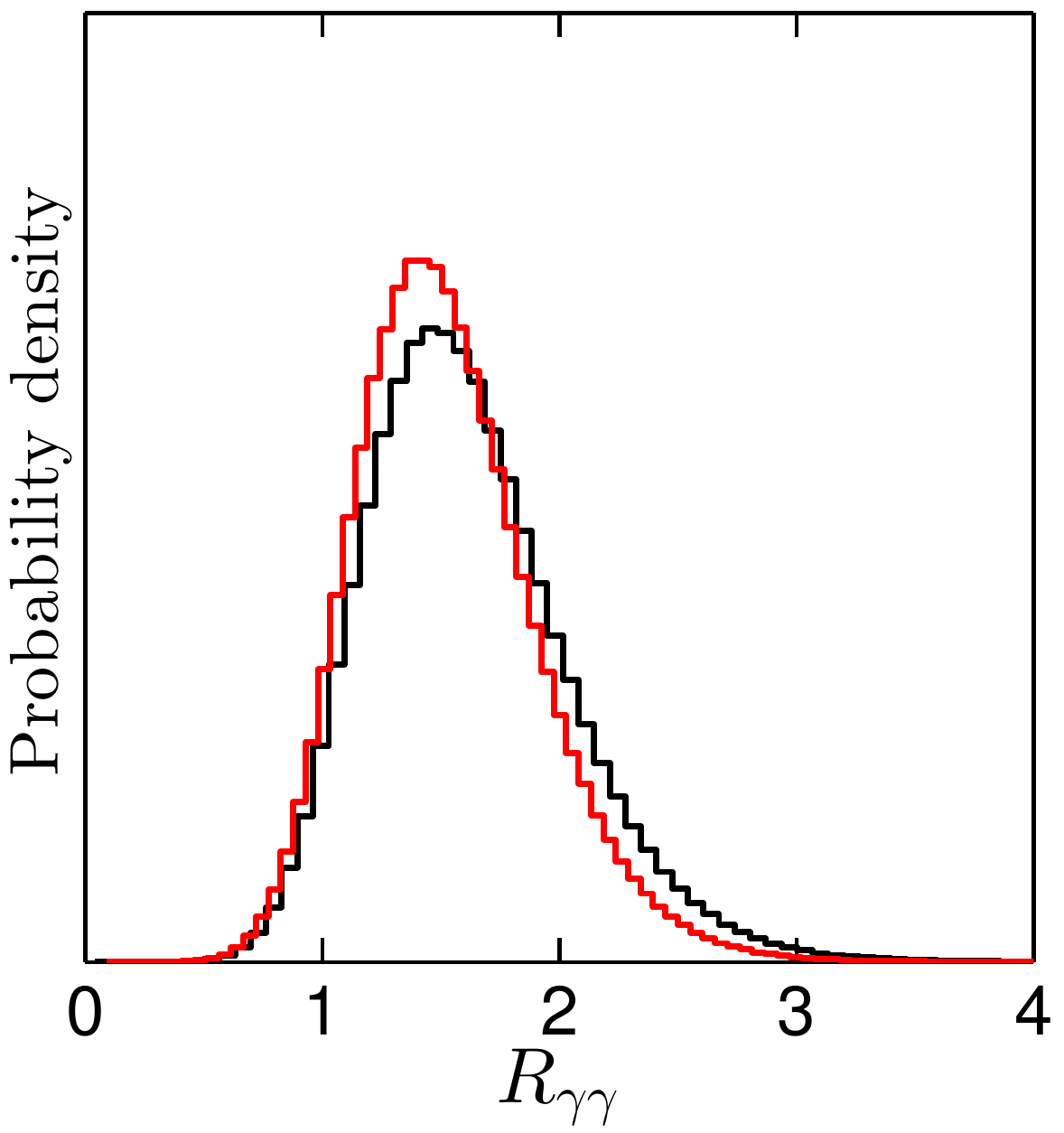}
     	\includegraphics[width=4.9cm]{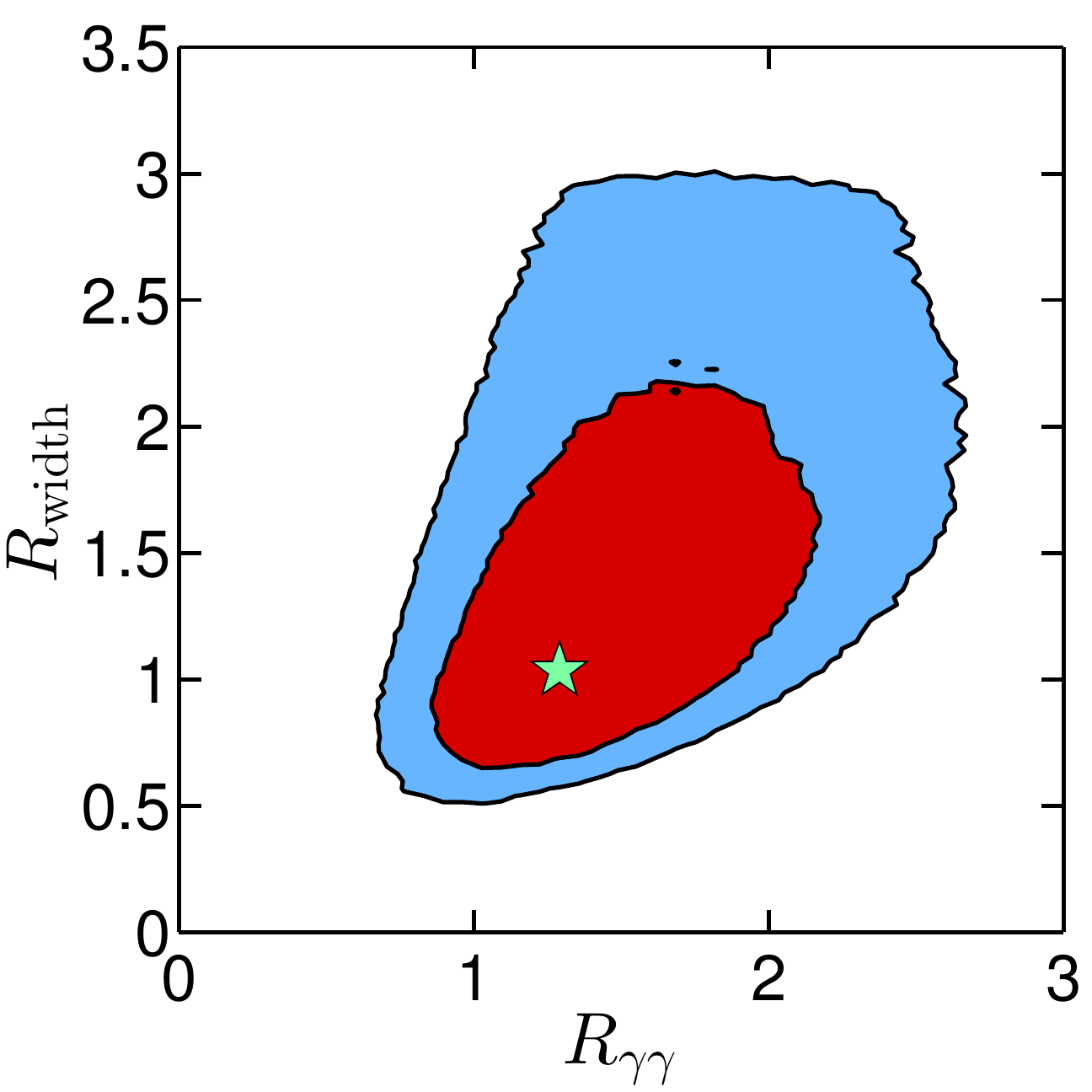}
     	\includegraphics[width=4.9cm]{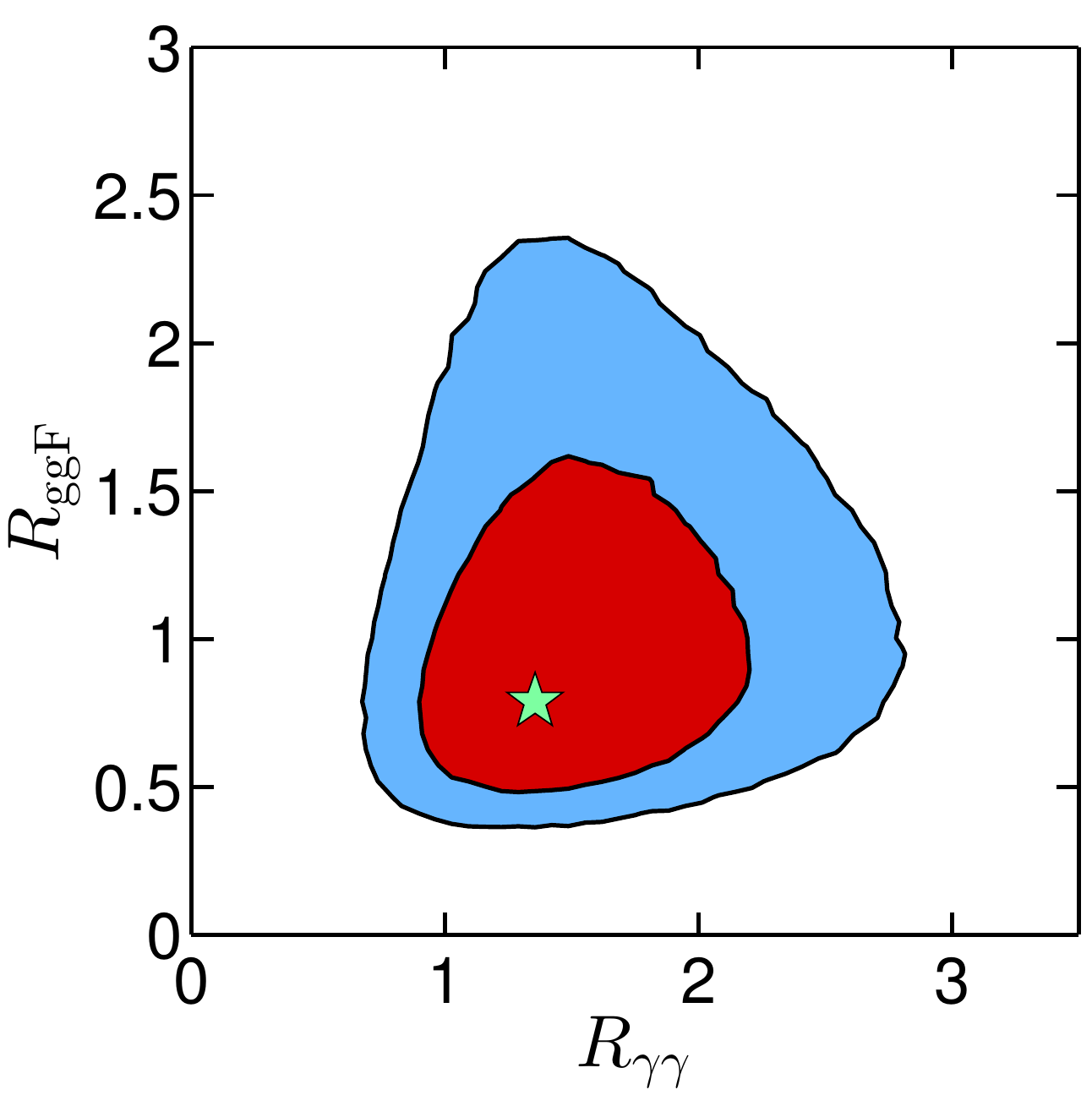}
	\caption{On the left, posterior PDF of $R_{\gamma\gamma}$ in scenario I (black) and scenario II (red). Also shown are the 2D posterior PDFs of $R_{\rm width}$ versus $R_{\gamma\gamma}$ (middle) and $R_{\rm ggF}$ versus $R_{\gamma\gamma}$ (right) in scenario I. Color code as in the previous figure. \label{2013eft-fig:R_2gamma}}
\end{figure}

The shapes of the PDF and profile likelihood for $c_t$ in Fig.~\ref{2013eft-ct_profile} are in fact a direct consequence of the signal strength measurement $\mu({\rm ttH},b\bar b)$ by CMS \cite{CMS-PAS-HIG-12-025}, see Table~\ref{2013eft-CMSresults}. Notice that the latter is so far the only analysis sensitive to the ttH production mode.
In spite of its large error, the low central value drives $c_t$ efficiently to small values because of the relation $R_{\rm ttH}=c_t^2$. Although small $c_t$ decreases (increases) the value of $R_{\rm ggF}$  ($R_{\rm \gamma\gamma}$), these changes can be compensated for  without decreasing the likelihood.
In the case $c_t \approx 0$, the gluon-gluon fusion (ggF) process is mainly driven by the tensorial coupling $\zeta_g \equiv \beta_{GG}/v$. We show in Fig.~\ref{2013eft-fig:beta_t_beta_GG} the correlation between $\beta_{GG}$ and $\beta_t$, which is needed to reproduce the observed ggF rate.
For the decay $h\to\gamma\gamma$, we observe an increased rate $R_{\gamma\gamma} > 1$, which can be seen in Fig.~\ref{2013eft-fig:R_2gamma}. Indeed, in the SM the $h \rightarrow \gamma\gamma$ process is dominated by the $W$ loops, and there is a destructive interference between the $t$ and $W$ contributions. Therefore, the suppression of $c_t$ helps increasing $R_{\gamma\gamma}$. To better understand this enhanced rate, notice that naively combining the data in Table~\ref{2013eft-ATLASresults} -- \ref{2013eft-Tevatronresults} one obtains $\mu({\rm ggF+ttH},\gamma \gamma)=1.05\pm 0.28$ and $\mu({\rm VBF+VH},\gamma\gamma)=1.8\pm 0.6$. It turns out that these different values are then realized with a slighly reduced $R_{\rm ggF}$ and an increased $R_{\gamma\gamma}$.


\begin{figure}
	\centering
	\includegraphics[width=4.5cm]{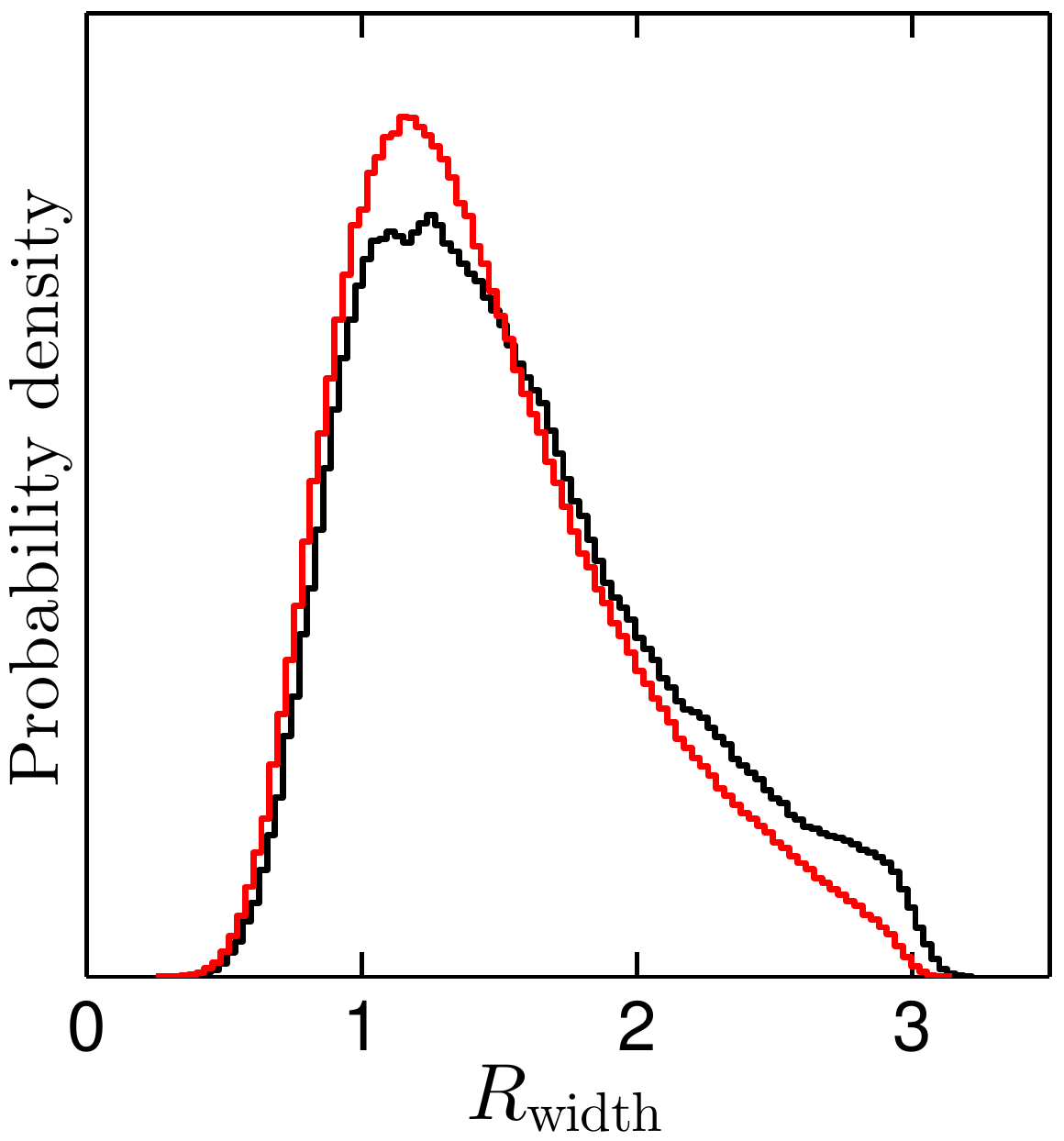}	
	\includegraphics[width=5.1cm]{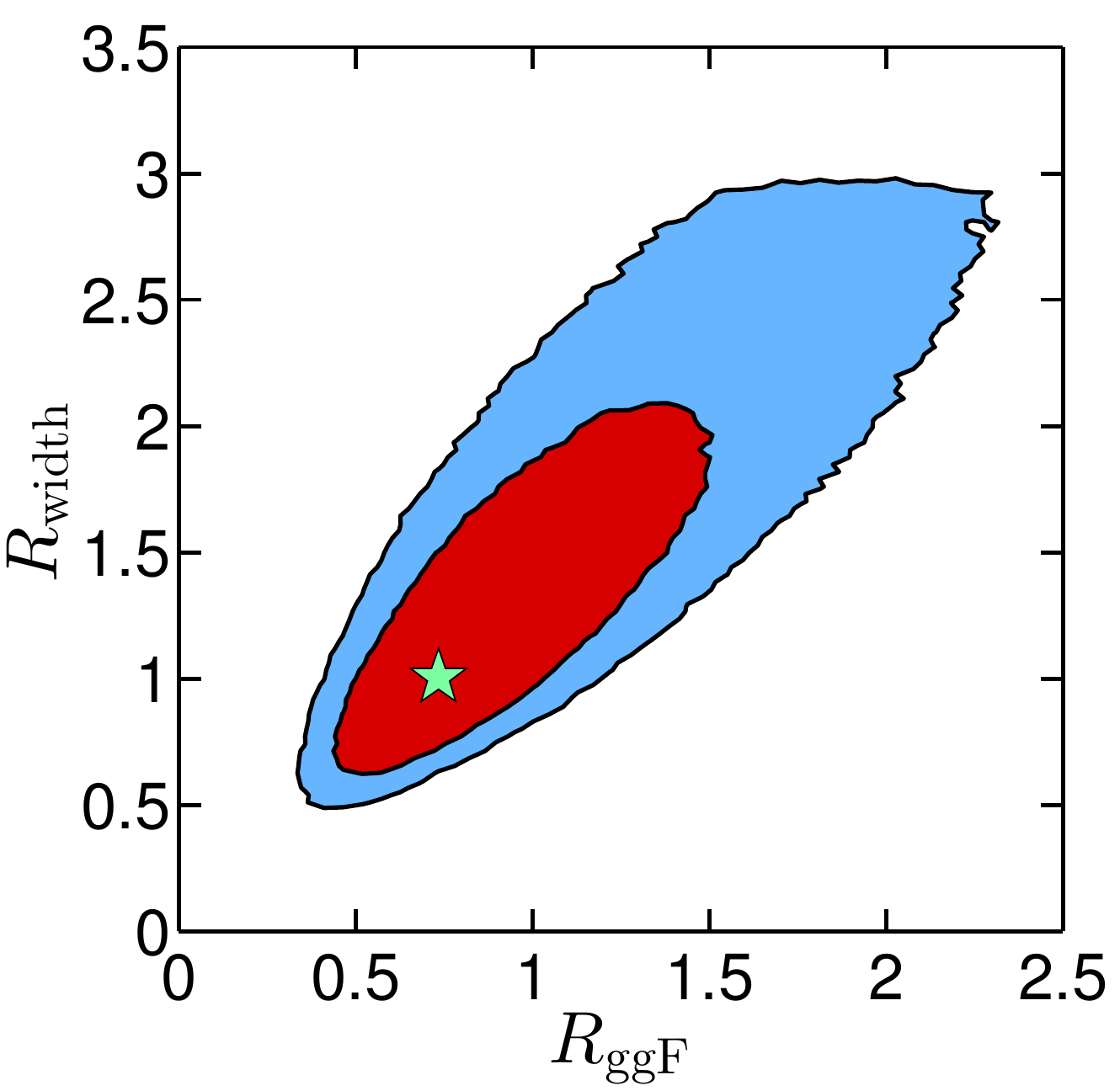}
	\includegraphics[width=5.1cm]{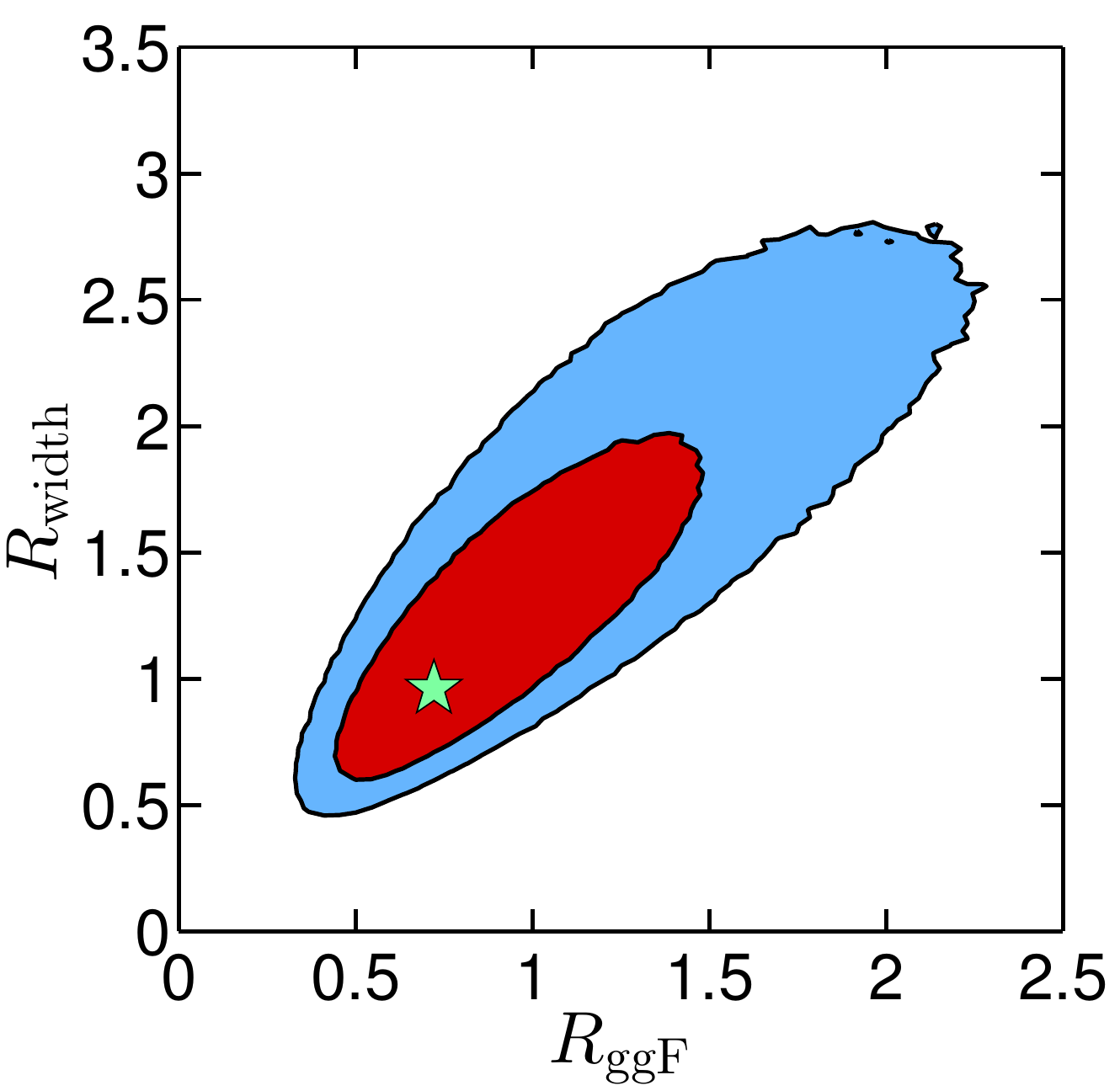}
	\caption{In the left panel, posterior PDF of $R_{\rm width} = \Gamma_h / \Gamma^{\rm SM}_h$ in scenario I (black) and scenario II (red). Also shown is the 2D posterior PDF of $R_{\rm width}$ versus $R_{\rm ggF}$ in scenario I (middle) and scenario II (right). Color code as in the previous figures. \label{2013eft-fig:Rwidth}}
\end{figure}

The PDF of $\beta_b$ is asymmetric with a longer right tail. $\beta_b$ appears mainly in the $h \rightarrow b\bar{b}$ decay rate, \ie~in $R_{b\bar{b}}$.\footnote{$b$ quark contributions in the loop-induced processes (ggF, $\gamma\gamma$, $Z\gamma$) are small and can be disregarded.} The reason of the asymmetry is the following: as the branching fraction ${\cal B}(h \rightarrow b\bar{b}) = 57\%$ in the SM, a deviation of $\beta_b$ from 0 (hence $c_b$ from 1) results in a sizable modification of the total width of the Higgs.
Our signal strengths are expressed as $\hat\mu(X, b\bar{b}) = R_X R_{b\bar{b}} / R_{\rm width}$ and contain a strong correlation between $R_{b\bar{b}}$ and $R_{\rm width}$. As $R_{\rm width}$ significantly increases with $R_{b\bar{b}}$, the deviations from $\mu_{b\bar{b}} = 1$ are smaller than what we could naively expect, allowing large values of $R_{b\bar{b}}$, hence $\beta_b$. This explains the tails of the PDF of $\beta_b$.

The 1D and 2D PDFs of $R_{\rm width}$ are shown in the left panel of Fig.~\ref{2013eft-fig:Rwidth}. It turns out that a large increase of $R_{\rm width}$ is not forbidden by the measurements of other channels, in which this effect is compensated by an increase of the decay or production rates, in particular ggF. The upper bound on $R_{\rm width}$, $R_{\rm width} \lesssim 3$, comes from the requirement $\beta_b<1$.

\begin{figure}[ht]
	\centering
		\includegraphics[width=4.5cm]{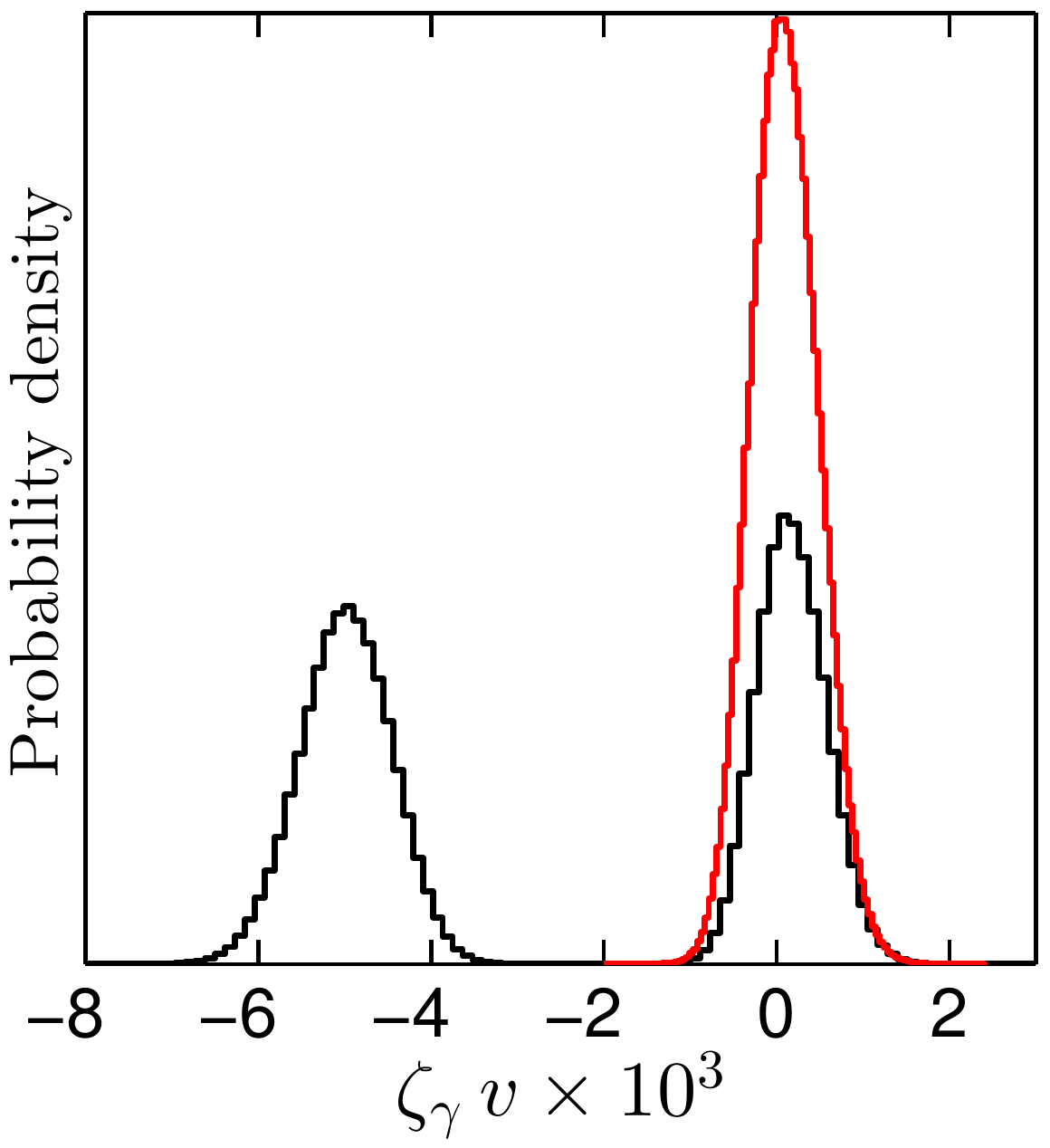}
        \includegraphics[width=4.5cm]{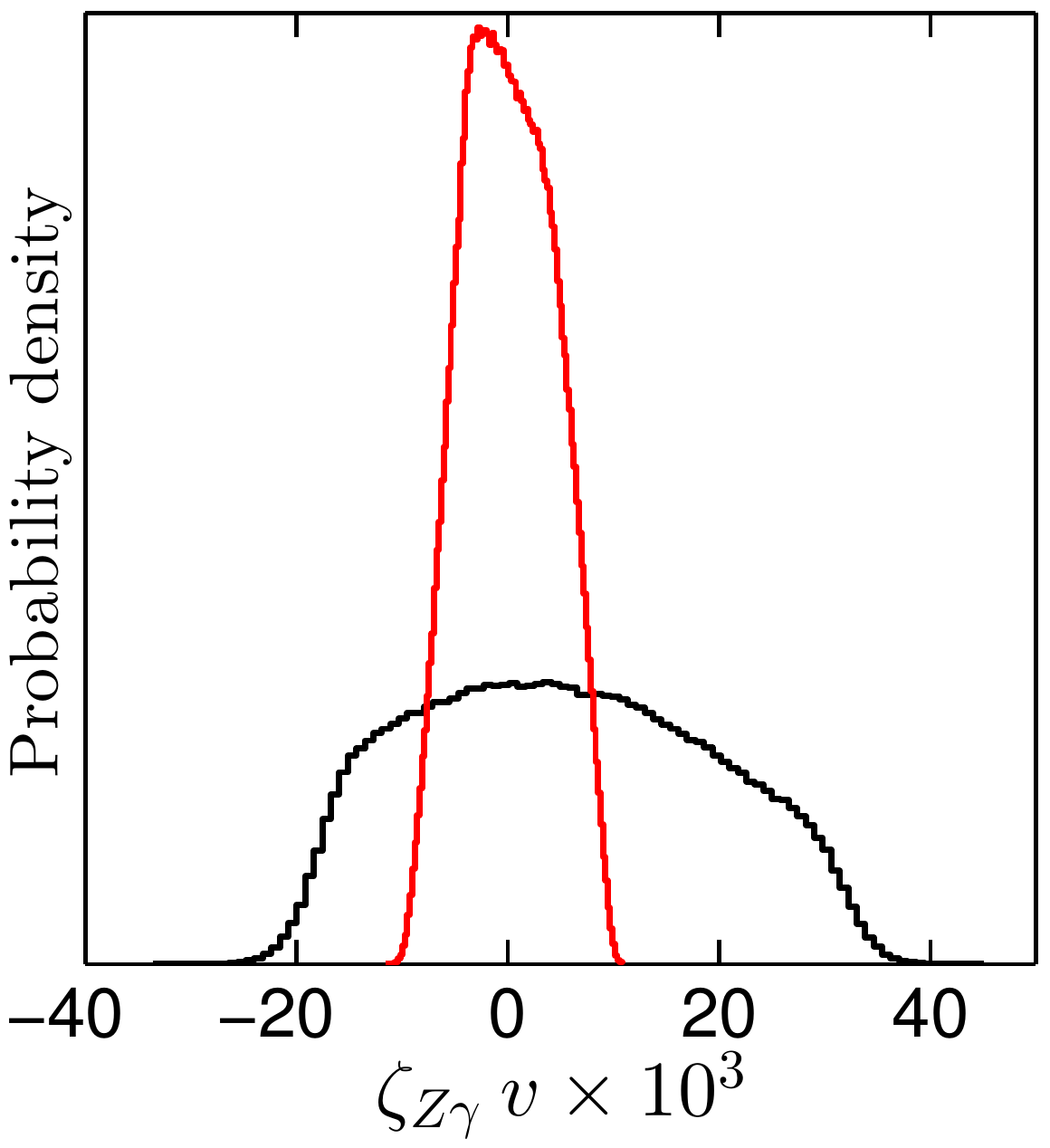}			
		\includegraphics[width=4.7cm]{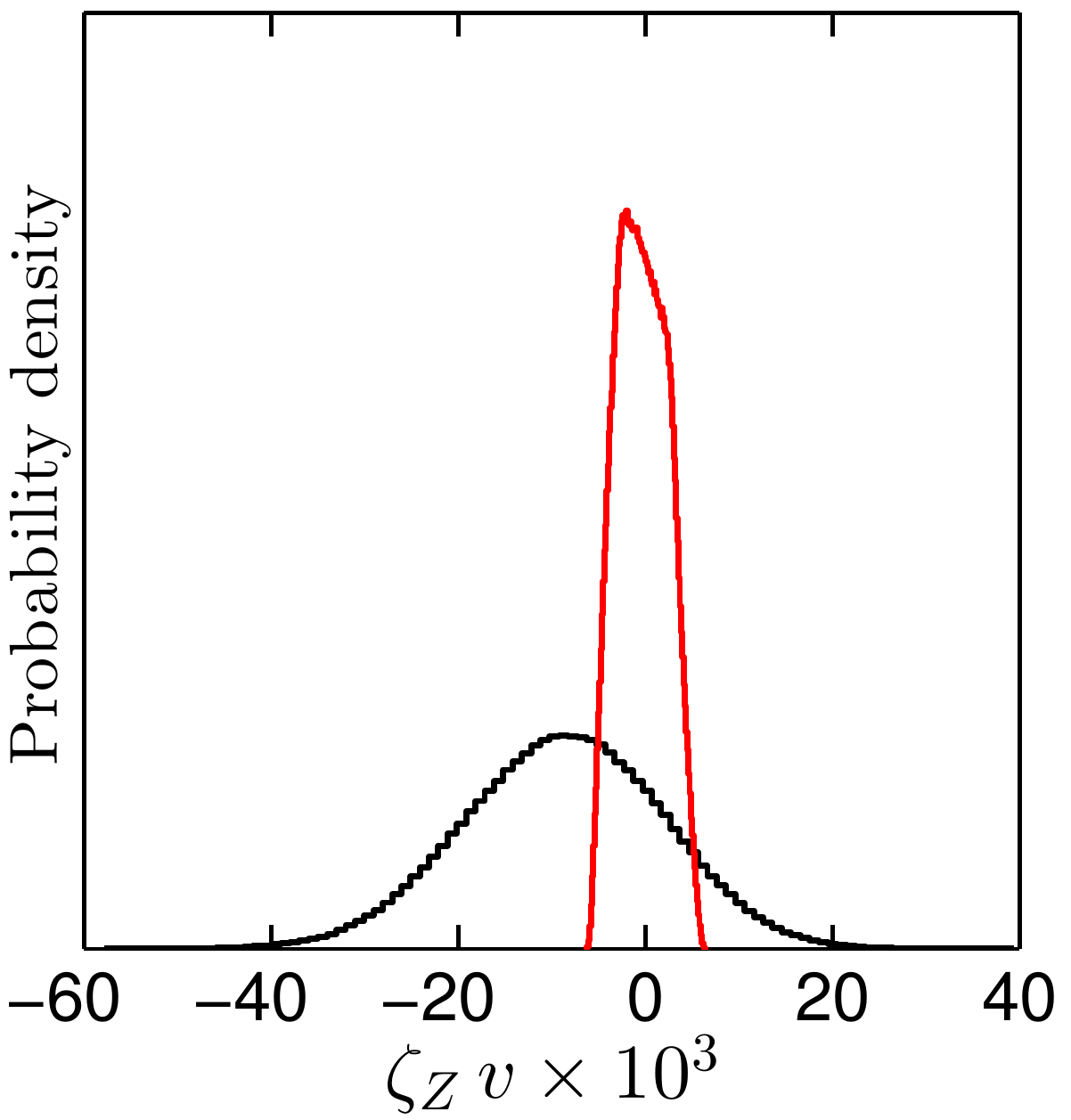}					
	\caption{Posterior PDFs of $\zeta_\gamma$, $\zeta_{Z\gamma}$, and $\zeta_Z$ in scenario I (black) and scenario II (red). \label{2013eft-fig:zeta_g_Zg_Z}}
\end{figure}

\begin{figure}[ht]
	\centering
	\includegraphics[width=5cm]{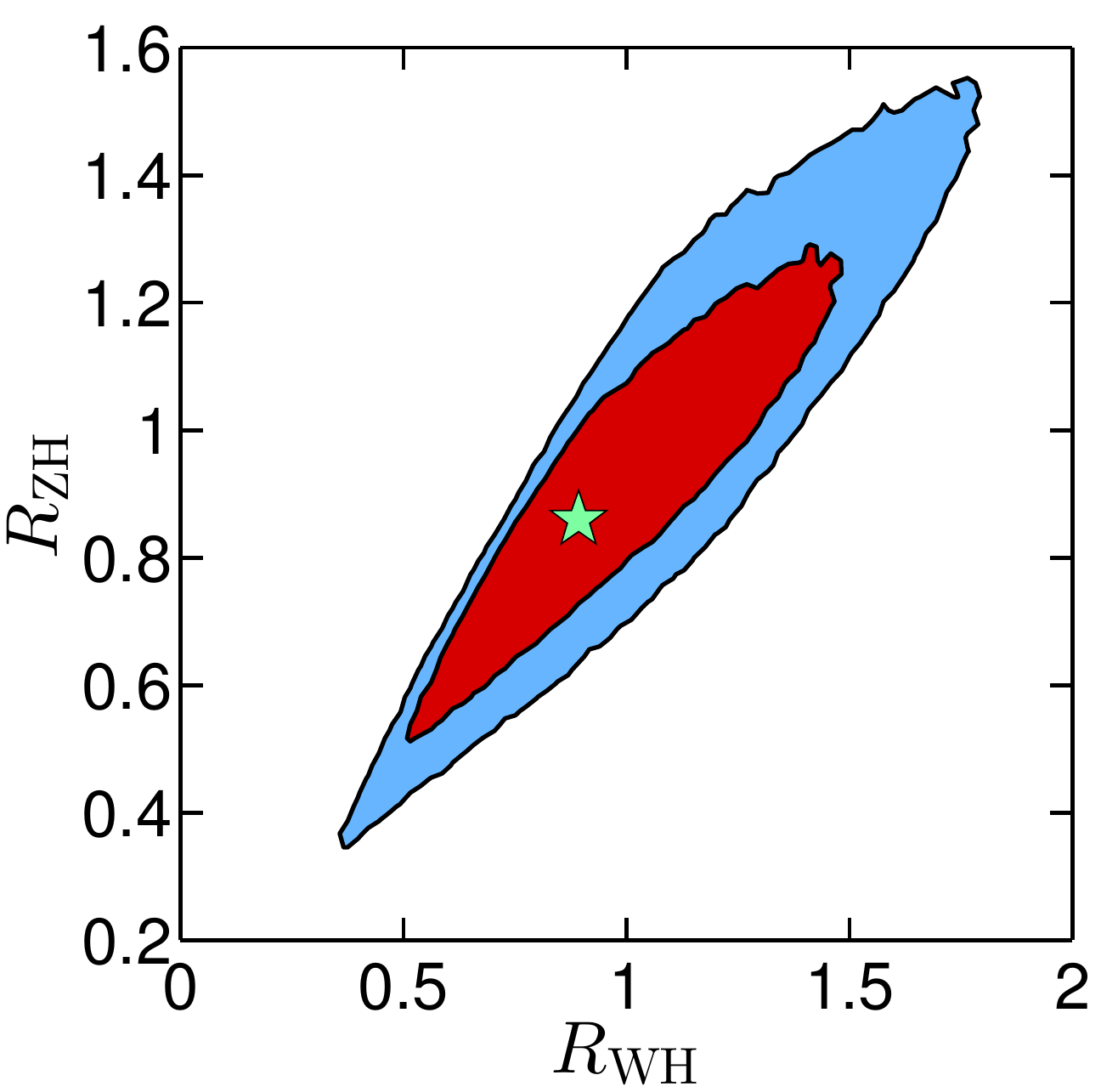}
	\includegraphics[width=5.1cm]{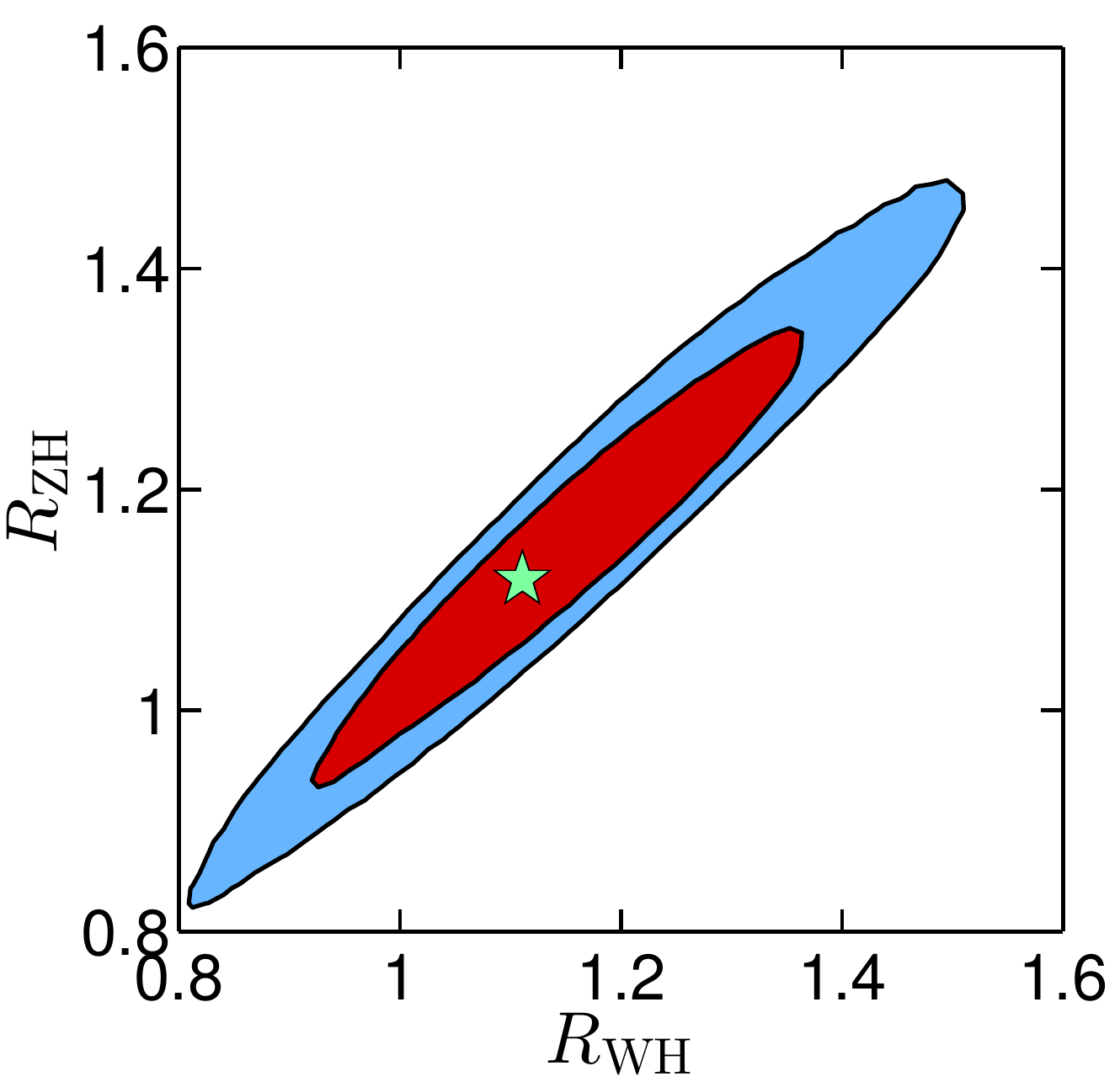}
	\caption{Posterior PDF of $R_{\rm ZH}$ versus $R_{\rm WH}$ in scenario I (left) and scenario II (right). Color code as in
the previous figures. \label{2013eft-fig:RWH_RZH}}
\end{figure}

In Fig.~\ref{2013eft-fig:zeta_g_Zg_Z}, we show  the PDFs of the tensorial couplings $\zeta_\gamma$, $\zeta_{Z\gamma}$, $\zeta_Z$ for scenario I.
The PDF of $\zeta_\gamma$ is constrained to small values in order to have the correct $H \rightarrow \gamma\gamma$ rate.
The PDF of $Z\gamma$ is much broader because of the weak experimental sensitivity to the $Z\gamma$ rate. The distribution for $\zeta_Z$ (and similarly $\zeta_W$) is mainly due to indirect effects on the fundamental parameters $\beta_{VV}$ ($\gamma\gamma$ and $Z\gamma$ rates, as well as TGVs) rather than because of direct experimental constraints. 
Notice that even with the assumption of custodial symmetry (which enforces $a_W=a_Z$), Eq.~(\ref{2013eft-eq:VH}) allows the rates for associated production to be different for $Z$ and $W$ because of the contribution of the tensorial couplings. 
It turns out that $\zeta_W$ and $\zeta_Z$ can be large enough in scenario I to induce a substantial deviation from $R_{\rm WH}=R_{\rm ZH}$. This is shown in Fig.~\ref{2013eft-fig:RWH_RZH}. 
This effect is also present in scenario II to a lesser extent. 

In scenario I, we observe two peaks of opposite signs for the tensorial couplings $\zeta_g$ and $\zeta_\gamma$. These features appear because of the competition between the tree-level $\zeta_{g,\gamma}$ and the loop-level SM couplings in the ggF and diphoton amplitudes. In addition to the classical region  where $\zeta$ adds up to the SM coupling and cannot be very large, regions with $\zeta=\mathcal{O}(-2\lambda^{\rm SM})$ are also allowed. 
Note that  $\zeta_\gamma$ is a linear combination of $\beta_{\WW}$, $ \beta_{\WB}$, $ \beta_{\BB}$, but that the two $\zeta_\gamma$ peaks cannot be seen in the PDFs of these parameters.
These four regions do not show up in scenario II, because the $\zeta_i$ are loop-suppressed and thus cannot be large enough to cancel the SM couplings.

\begin{figure}[ht]
	\centering
	\includegraphics[width=4.7cm]{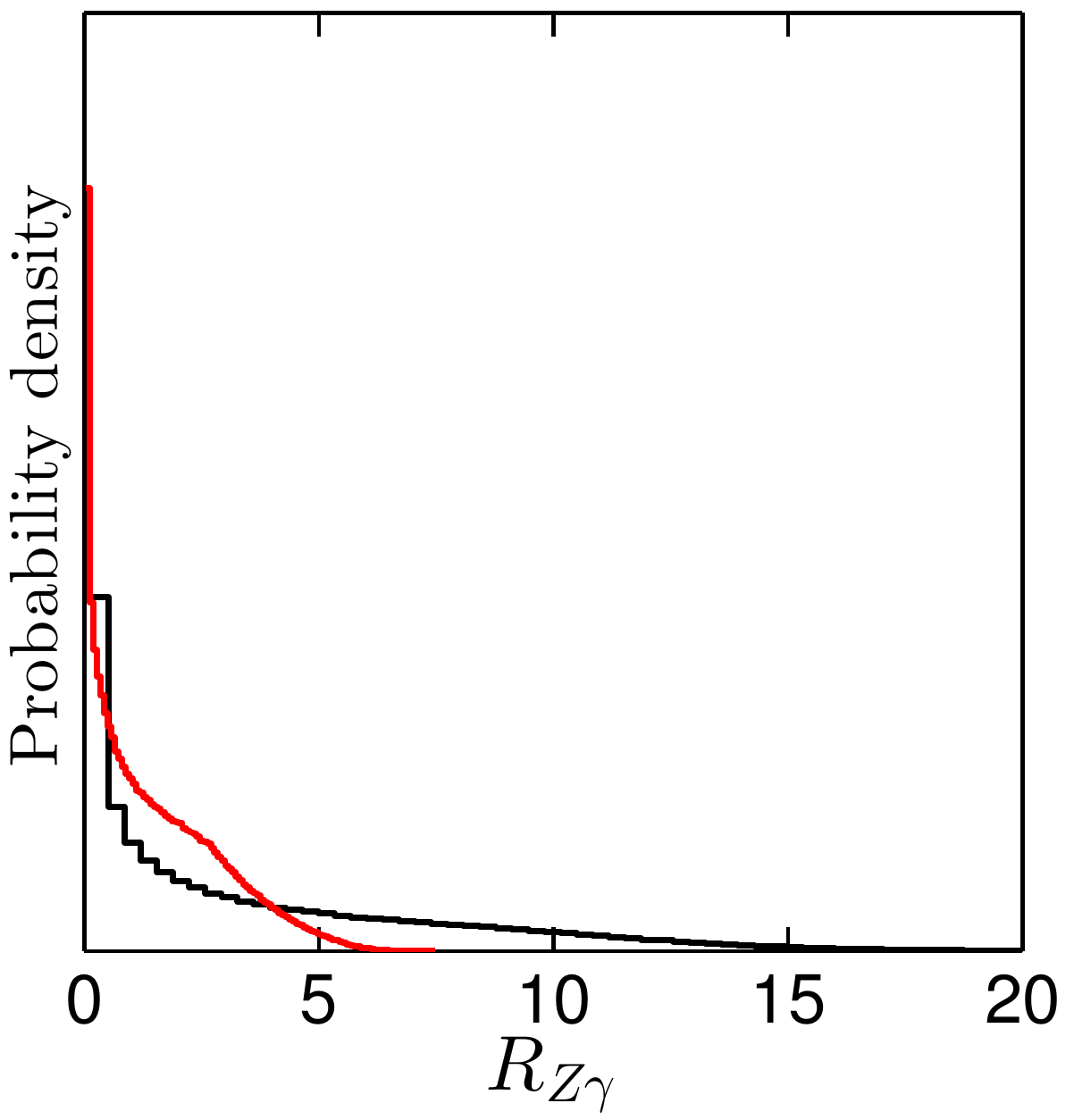}							
	\includegraphics[width=4.7cm]{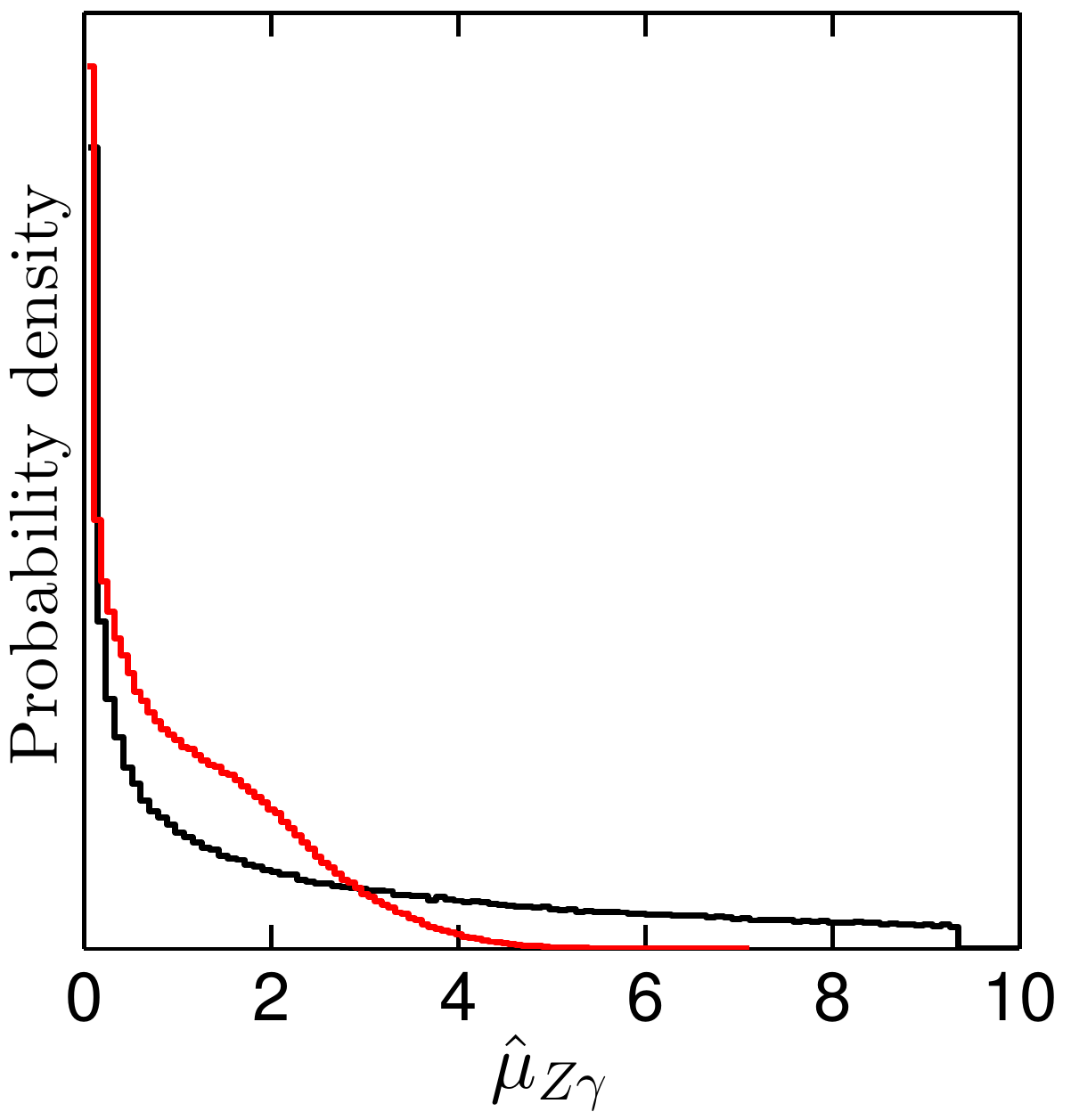}								
	\caption{Posterior PDFs of $R_{Z\gamma} \equiv \Gamma(h\rightarrow Z\gamma)/\Gamma^{\rm SM}(h\rightarrow Z\gamma)$ (left) and $\hat \mu_{Z\gamma}$ (right) in scenario I (black) and scenario II (red). \label{2013eft-fig:RZgam_1D}}
\end{figure}

A feature of the PDF of $\zeta_{Z\gamma}$ is that, in spite of the various constraints on $\beta_{\WW}$, $\beta_{\WB}$, and $\beta_{\BB}$, the $Z\gamma$ rate can still be considerably enhanced.
The shape of the $\zeta_{Z\gamma}$ PDF is mainly constrained by the CMS bound $\hat \mu_{Z\gamma}<9.3$ in scenario I, while indirect constraints from the $S$ parameter and trilinear gauge vertices dominate in scenario II. 
In the enhanced rate, the HDO contribution dominates, such that $R_{Z\gamma}$ is mostly proportional to $(\zeta_{Z\gamma})^2$.
This happens in scenario I, but also in scenario II although the $\zeta_i$ are smaller.
The PDF of the ratio $R_{Z\gamma}$ can be seen in Fig.~\ref{2013eft-fig:RZgam_1D}. The $95\%$ Bayesian credible intervals are $[0, 12.0]$ for scenario I and $[ 0, 4.3]$ for scenario II. As large deviations are allowed in this channel within this framework, it is therefore particularly promising for the discovery of a NP signal.

In scenario I, $\zeta_{Z\gamma}$ is sufficiently large to cancel the SM coupling, such that enhancement with both signs of $\zeta_{Z\gamma}$ is realized. In contrast, for scenario II, only the branch with constructive interference $\zeta_{Z\gamma}<0$ can  enhance $R_{Z\gamma}$. In both scenarios, $\zeta_{Z\gamma}$ can cancel the SM coupling, such that having a small or vanishing $R_{Z\gamma}$ is also possible.\footnote{We do not focus on this aspect as the direct searches at the LHC are still far from this level of precision.
The shape of the $R_{Z\gamma}$ PDF follows the distribution of $|\zeta_{Z\gamma}+\lambda_{Z\gamma}|^2$, which presents a peak in 0. Schematically, for a uniform distribution of $\zeta_{Z\gamma}$, the peak behaves as $\zeta_{Z\gamma}^{-1/2}$. One can observe a similar behaviour for $R_{\rm ttH}$.}

Finally, we compute the signal strength of $h \rightarrow Z\gamma$ in case of a fully inclusive analysis at the LHC. The PDFs are shown in the right panel of Fig.~\ref{2013eft-fig:RZgam_1D} for both scenarios. In scenario I, the distribution reaches the CMS 95\% C.L.~bound $\hat\mu_{Z\gamma}<9.3$, while it vanishes before in scenario II. The $68\%$ and $95\%$ BCIs are $[0,3.6]$, $[0,8.1]$ in scenario I and $[0,1.6]$, $[0,3.2]$ in scenario II. 

\begin{figure}[ht]
	\centering
		\includegraphics[width=5.1cm]{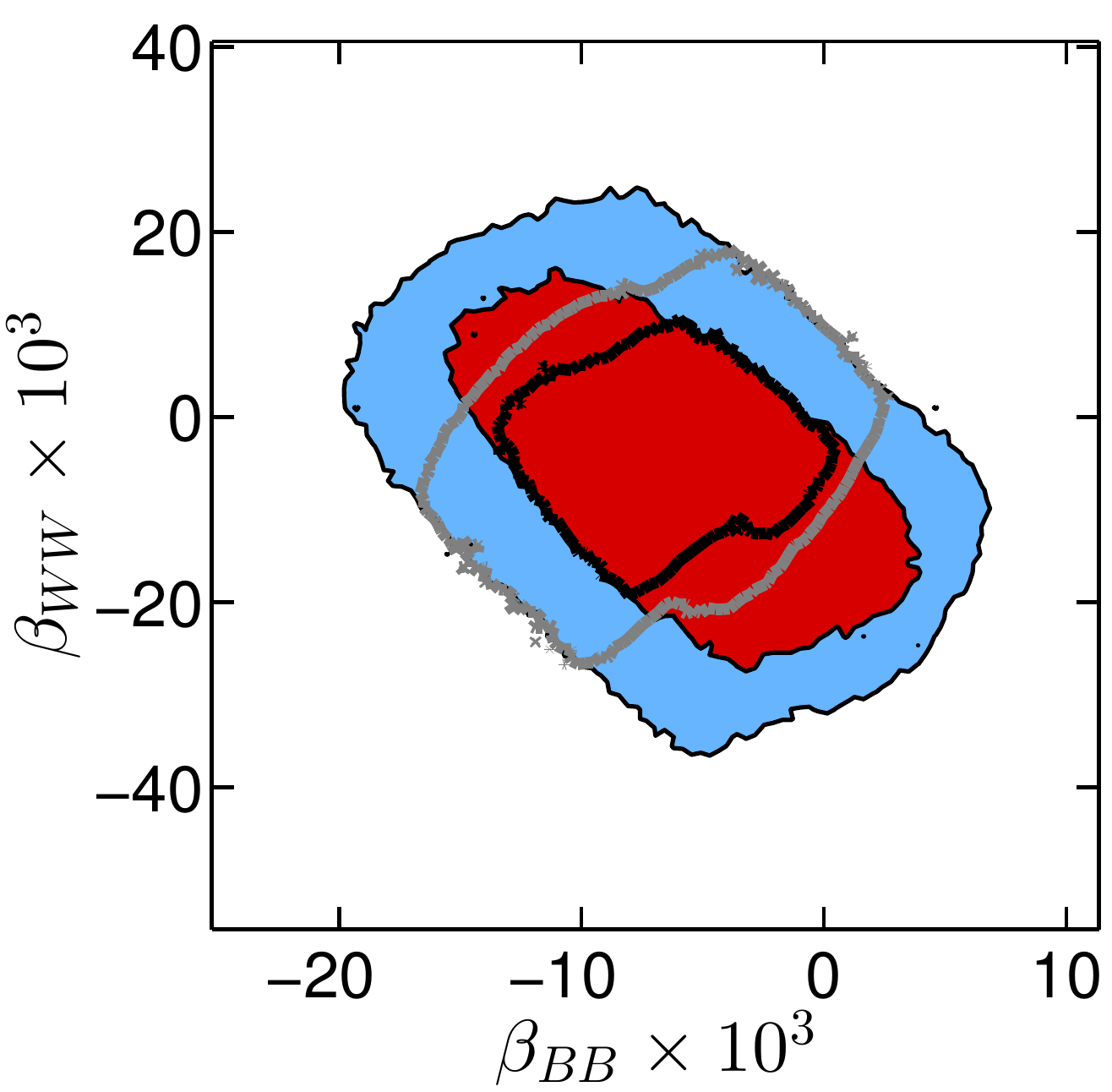}
		\includegraphics[width=4.95cm]{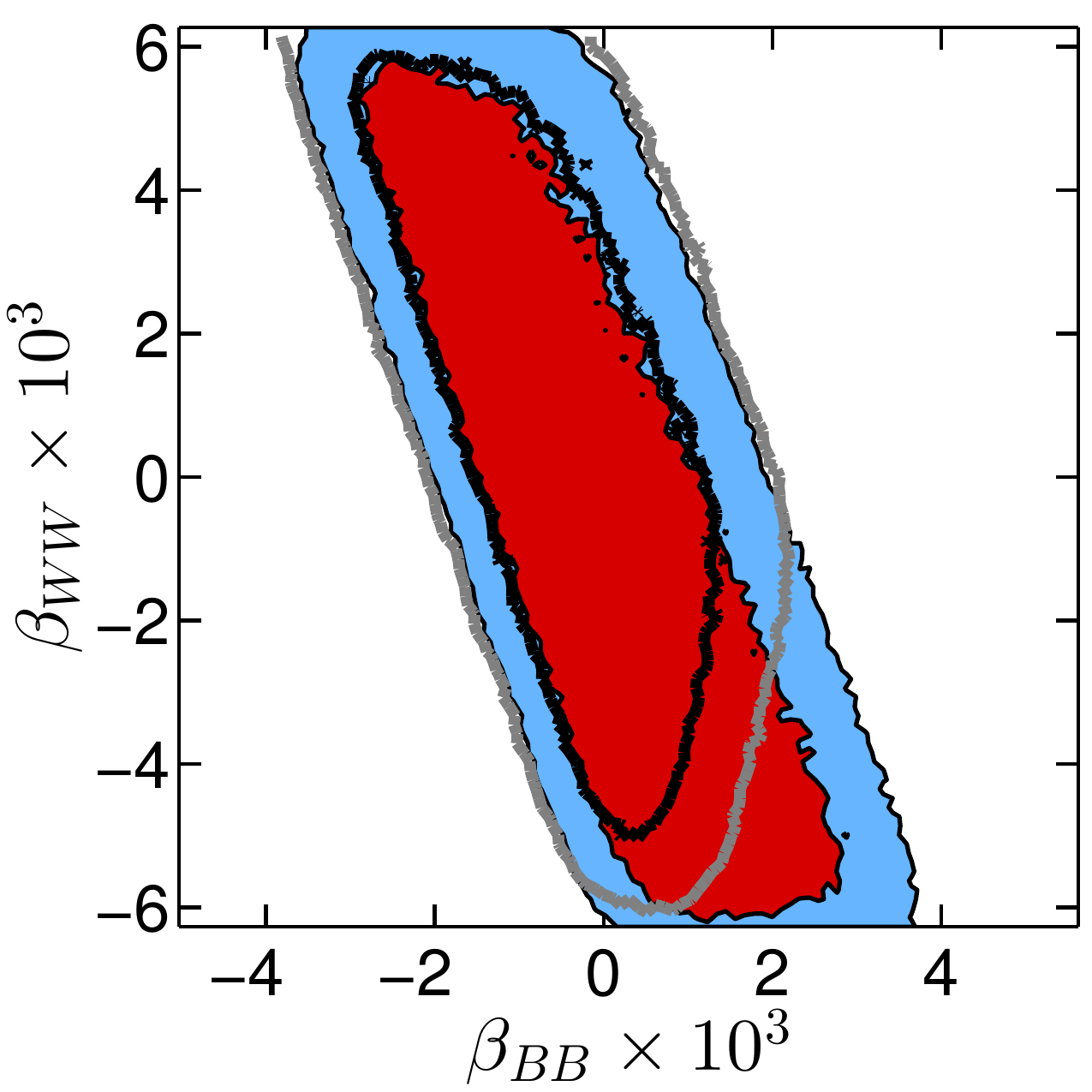}		
	\caption{Posterior PDF of $\beta_{\WW}$ versus $\beta_{\BB}$ in scenario I (left) and scenario II (right). The red and blue regions correspond to the 68\% and 95\% BCRs from the current measurements, while the black and gray contours correspond to the 68\% and 95\% BCRs assuming in addition that $\hat\mu_{Z\gamma}<2$.
\label{2013eft-fig:Zgam}}
\end{figure}

Given that the 13/14~TeV LHC has a good potential to constrain the $h \rightarrow Z\gamma$ rate, one may wonder about the impact of a more precise measurement on our results. Therefore, we investigate the possibility of having $\hat\mu_{Z\gamma}<2$ at 95\%~CL, and we implement this bound as a step function.\footnote{Note that the relative SM production rates $\sigma^{\rm SM}_X / \sum_X \sigma^{\rm SM}_X$ do not change significantly between 8~TeV and 14~TeV. Thus, assuming a fully inclusive analysis, we can take the decomposition into production channels as given in Section~\ref{2013eft-se_data}, Table~\ref{2013eft-CMSresults}.} 
It mainly results in a better determination of $\beta_{\BB}$ and $\beta_{\WW}$ in both scenarios, as can be seen in Fig.~\ref{2013eft-fig:Zgam}.
This new limit has an effect on the Higgs phenomenology in scenario I only. It leads to a slightly better prediction of $R_{\rm WH}$: the 95\% BCI is $[0.7,1.5]$, instead of $[0.5,1.7]$ when we take into account the current limit on $h \rightarrow Z\gamma$.

\subsection{Conclusions} \label{2013eft-se_conclusions}

In this section, we used a complete basis of dimension-six operators encoding NP effects in an effective Lagrangian in which all tensorial couplings are taken into account. The basis was chosen such that field-strength--Higgs operators ($\mathcal{O}_{FF}$) are exactly mapped into tensorial couplings. In this basis it is straightforward to study the well-motivated hypothesis of loop-suppression of these operators. 

The data taken into account in our analysis are the whole set of results from ATLAS and CMS, including all available correlations, as well as Tevatron data. Trilinear gauge vertices measurements and constraints on electroweak precision observables are also included in our study.

It turns out that weak bosons tensorial couplings can in principle modify non-trivially the kinematic structure of the  VBF, VH and $h\rightarrow VV$ processes, and thus the efficiency of the kinematic selections. Scrutinizing the experimental analyses, we find that one can consider unchanged kinematic cuts efficiencies for the rates of these processes to a good approximation.
The effect of tensorial couplings is crucial in the ggF, $h \rightarrow \gamma\gamma$, $h \rightarrow Z\gamma$ processes, where they compete with the one-loop SM couplings. In our predicted rates, leading loop corrections are taken into account, consistently  with respect to the loop-level HDO framework. 

In order to put constraints on the higher dimensional operators we consider, we carry out a global analysis in the framework of Bayesian inference. We find this approach particularly appropriate as it deals with weakly constrained problems and naturally takes into account fine-tuning, such that the results we present are free of improbable, \ie~fine-tuned, cancellations. Markov Chain Monte Carlo techniques are used to perform the numerical integrations.

Our analysis is centered on new physics arising at $\lesssim$~3~TeV. 
It allows us to express our results in terms of $\beta_i \equiv \alpha_i\, v^2 / \Lambda^2$, {\it i.e.}\ to factorize the effect of the coefficients $\alpha_i$ and of the scale of new physics $\Lambda$.
Among other things, this parameterization is $\Lambda$-independent up to a small $\log\Lambda$ dependence. Distributions of HDO coefficients at any low $\Lambda$ can be easily mapped from our results.
Two general scenarios are considered: I) democratic HDOs, where the coefficients of all the dimension-six operators are essentially unconstrained, and II) loop-suppressed $\mathcal{O}_{FF}$'s, where the field-strength--Higgs operators are loop suppressed with respect to the other HDOs.

We find overall a substantial amount of freedom in both of these scenarios. For instance, the coupling to the top quark still allows for $\mathcal O(1)$ deviations, while the couplings to bottom and tau are slightly more constrained. We report in both scenarios a large and natural region at small or vanishing $c_t$, favored by goodness of fit, which we trace back to the slight deficit observed in the $b\bar b$ channel with the ttH production mode, as reported by CMS.
Regarding the couplings of the vector bosons, we find that only small deviations are allowed, of the order of $10\%-20\%$. 
Despite the fact that the inclusive rate of the $h\to\gamma\gamma$ channel is close to one, we observe a slightly enhanced $R_{\gamma\gamma}=\Gamma_{\gamma\gamma}/\Gamma^{SM}_{\gamma\gamma}$, with the 95\% Bayesian credible interval (BCI) for $R_{\gamma\gamma}$ given by $[0.8,2.5]$ ($[0.8,2.3]$) in scenario I (II). Correlations with $R_{\rm width} = \Gamma_h / \Gamma^{\rm SM}_h$ as well as the deviations of the various production modes from the SM are then identified to be responsible for fitting the experimental signal strengths.
We also find that an enhanced coupling to the $b$ quark is likely, leading to a larger total width, with the 95\% BCI for $R_{\rm width}$ given by $[0.7,2.7]$ ($[0.6,2.5]$) in scenario I (II). A strong correlation with $R_{\rm ggF}$ exists and makes the predicted signal strengths compatible with the data.

Overall, it appears that the tensorial couplings play a crucial role in the SM loop-induced processes. In particular, after taking into account all the constraints we find that the Higgs boson decay width into $Z\gamma$ can be enhanced by up to a factor 12 (4) in scenario I (II), within 95\% BCI. Conversely, future measurements in the $Z\gamma$ channel will provide important bounds on the coefficients of the higher dimensional operators we consider.

\clearpage

\section[The phenomenological MSSM in view of the 125~GeV Higgs data]{The phenomenological MSSM in view of the \\ 125~GeV Higgs data} \label{sec:pmssm}

\sectionmark{The pMSSM in view of the 125~GeV Higgs data}

So far, in this chapter we have seen the impact of the LHC Higgs measurements on effective parameterizations of new physics. While the impact on models with extended Higgs sectors was examined in Sections~\ref{sec:20122HDMfit} and \ref{2013c-sec:2HDM} for the 2HDM, and in Section~\ref{2013c-sec:IDM} for the IDM, it remained largely restricted to the properties of the Higgs boson with mass around 125~GeV. In particular, the other physical Higgs states were ignored and other phenomenological aspects, such as the viability of possible dark matter candidates, were not explored. In this section, a complete study of the phenomenological MSSM (pMSSM), a 19-dimensional parametrization of the weak-scale Lagrangian of the MSSM (see Section~\ref{sec:susy-mssm}), will be presented. 

In the MSSM, a light Higgs mass of the order of 125~GeV requires that stops be either very heavy or near-maximally mixed (see, {\it e.g.}, Refs.~\cite{Djouadi:2013lra,Brummer:2012ns}).
In addition to a modification of the electroweak symmetry breaking sector by the presence of a second Higgs doublet, the MSSM predicts a wealth of new particles that couple to the light Higgs boson. These can, depending on their masses and mixings, modify the Higgs couplings and consequently the production and decay rates of the Higgs boson in various channels. It is thus interesting to ask whether, besides the measured Higgs mass, the Higgs signal strengths provide constraints on the MSSM and may thus be used as a guide for where to look for SUSY. 

Indeed, the apparent excess in the diphoton channel reported by both ATLAS and CMS 
in 2012~\cite{Aad:2012tfa,Chatrchyan:2012ufa} 
motivated scenarios with light staus in the MSSM \cite{Carena:2012gp} or 
small $\tan\beta$/large $\lambda$ in the next-to-MSSM \cite{Ellwanger:2011aa,Gunion:2012zd} 
(see also \cite{Kraml:2013wna,Arbey:2012bp}). 
As we saw in Section~\ref{sec:higgs2013}, this drastically changed with the updated results presented at the Moriond 2013 conference and thereafter, which point towards a very SM-like Higgs boson, without the need of any modifications of the couplings due to new, beyond-the-SM particles.  

The implications of the latest Higgs data for the MSSM were discussed recently in \cite{Djouadi:2013uqa,Cahill-Rowley:2013vfa}. Ref.~\cite{Djouadi:2013uqa} concentrated on describing (the consequences for) the heavy Higgs 
states in the limit of heavy SUSY particles; the best coupling fit was found at low $\tan\beta$, $\tan\beta\approx 1$,  
with a not too high CP-odd Higgs mass of $m_A\approx 560$~GeV. 
Ref.~\cite{Cahill-Rowley:2013vfa} analyzed the consequences of the SUSY null-searches on the one hand 
and of the measurements of the Higgs properties on the other hand based on flat random scans of the so-called phenomenological MSSM (pMSSM) with the conclusion that SUSY searches and Higgs boson properties are to a very good approximation orthogonal.  More concretely, Ref.~\cite{Cahill-Rowley:2013vfa}
concluded that Higgs coupling measurements at the 14~TeV LHC, and particularly at a 500~GeV International Linear Collider (ILC), will be sensitive to regions of the pMSSM space that are not accessible to direct SUSY searches.  

In this section, we follow a different approach. Performing a Bayesian analysis of  the pMSSM parameter space by means of a Markov Chain Monte Carlo analysis (as in Section~\ref{sec:higgsdim6}), we investigate how the latest LHC results on the properties of the 125~GeV Higgs state impact the probability distributions of the pMSSM parameters, masses
and other observables.
In doing so, we take into account Higgs measurements as in Section~\ref{sec:higgs2013}, on top of constraints from LEP searches and low-energy observables.
In addition, we explore consequences for our probability distributions from the latest dark matter constraints 
and discuss prospects for measurements of the Higgs signal at the next run of the LHC at 13--14~TeV.
Our results are orthogonal and directly comparable to the pMSSM interpretation of the CMS SUSY searches~\cite{CMS-PAS-SUS-12-030,CMS-PAS-SUS-13-020}.

This work was done in collaboration with John F.~Gunion, Sabine Kraml and Sezen Sekmen. The rest of the section will reproduce results from the paper ``The phenomenological MSSM in view of the 125~GeV Higgs data'', Ref.~\cite{Dumont:2013npa}, that was submitted to arXiv on December 25, 2013 and published in PRD in March 2014. It was also summarized in a contribution to proceedings of the DIS~2014 conference~\cite{Dumont:2014ura}.

\subsection{Analysis}
%

\subsubsection{Definition of the phenomenological MSSM (pMSSM)}
\label{pmssm-sec:pmssm}

The purpose of this study is to assess what current Higgs data tell us, and do not tell us, about the 
MSSM at the weak scale, without any assumption as to the SUSY-breaking scheme.  
A priori, the weak-scale MSSM has 120 free parameters, assuming that $R$-parity is conserved (to avoid proton decay and to ensure that the lightest SUSY particle, the lightest supersymmetric particle (LSP), is stable) and assuming that the gravitino is heavy. 
This is clearly too much for any phenomenological study. 
However, most of these parameters are associated with CP-violating phases and/or flavor changing neutral currents (FCNC), which are severely constrained by experiment.  A few reasonable assumptions about the flavor and CP structure therefore allow us to reduce the number of free parameters by a factor 6, without imposing any SUSY-breaking scheme. Working with parameters defined at the weak scale is indeed of great advantage for our purpose, because models of SUSY breaking always introduce relations between the soft terms that need not  hold in general. 

Concretely, the only generic way to satisfy very strong constraints on CP violation is to take all parameters to be real. 
FCNC constraints are satisfied in a generic way by taking all sfermion mass matrices and trilinear couplings to be 
flavor-diagonal. As a further simplification, the various independent  sfermion masses for the 2nd generation  are taken to be equal to their counterparts for the 1st generation. 
Regarding the trilinear $A$-terms of the first two generations, these only enter phenomenology multiplied by the associated very small Yukawa couplings and are thus not experimentally relevant unless unreasonably large. Only the 3rd generation
parameters $A_t$, $A_b$ and $A_\tau$ have observational impact.

This leaves us with 19 real, weak-scale SUSY Lagrangian parameters---the
so-called p(henomenological) MSSM~\cite{Djouadi:1998di}. 
As mentioned, the pMSSM captures most of the phenomenological 
features of the $R$-parity conserving MSSM and, most importantly, encompasses and goes beyond a broad 
range of more constrained SUSY models.  
The free parameters of the pMSSM are the following: 
\begin{itemize}
   \item the gaugino mass parameters $M_1$, $M_2$, and $M_3$; 
   \item the ratio of the Higgs vacuum expectation values (vevs), $\tan\beta=v_2/v_1$;
   \item the higgsino mass parameter $\mu$ and 
            the pseudo-scalar Higgs mass $m_A$;
    \item 10 sfermion mass parameters $m_{\tilde{F}}$, where 
         $\tilde{F} = \tilde{Q}_1, \tilde{U}_1, \tilde{D}_1, 
                      \tilde{L}_1, \tilde{E}_1, 
                      \tilde{Q}_3, \tilde{U}_3, \tilde{D}_3, 
                      \tilde{L}_3, \tilde{E}_3$\\ 
(with 2nd generation sfermion masses equal to their 1st generation counterparts, {\it i.e.}\ $m_{\tilde{Q}_1}\equiv m_{\tilde{Q}_2}$, 
           $m_{\tilde{L}_1}\equiv m_{\tilde{L}_2}$, {\it etc.}), and          
   \item the trilinear couplings $A_t$, $A_b$ and $A_\tau$\,,               
\end{itemize}
in addition to the SM parameters.  
To minimize theoretical uncertainties in the Higgs sector, these parameters are conveniently defined 
at the scale $M_{\rm SUSY} \equiv \sqrt{m_{\tilde t_1}m_{\tilde t_2}}$, often also referred to as the 
EWSB scale.

The pMSSM parameter space is constrained by a number of theoretical requirements.  
In particular, the Higgs potential must be bounded from below and lead to consistent EWSB, and 
the sparticle spectrum must be free of tachyons.
Moreover, in this study, we require that the LSP is the lightest neutralino, $\tilde\chi^0_1$. 
These requirements we refer to as theoretical constraints. 
Note that we do not check for charge and/or color breaking minima beyond warnings from the spectrum generator; this could be done, \eg, using {\tt Vevacious}~\cite{Camargo-Molina:2013qva}, but would require too much CPU time for this  study.

\subsubsection{Construction of the pMSSM prior} 
\label{pmssm-sec:prior}

We perform a global Bayesian analysis that yields posterior probability densities of model parameters, masses and observables.  
We allow the pMSSM parameters to vary within the following ranges:
\begin{eqnarray}
\nonumber &-3\,{\rm TeV} \le M_1,\, M_2,\, \mu \le 3\,{\rm TeV}\,;& \\
\nonumber &0 \le M_3,  m_{\tilde{F}}, m_A \le 3\,{\rm TeV}\,;& \\
\nonumber &-7\,{\rm TeV} \le A_t, A_b, A_\tau \le 7\,{\rm TeV}\,;& \\
 &2 \le \tan\beta \le 60\,.& 
\label{pmssm-eq:subspace}
\end{eqnarray}
A point in this space will be denoted by $\theta$. In addition, we treat the 
SM parameters $m_t$, $m_b(m_b)$ and $\alpha_s(M_Z)$ as nuisance parameters, 
constrained with a likelihood.
For each pMSSM point, we use  
{\tt SoftSUSY\_3.3.1}~\cite{Allanach:2001kg} to compute the SUSY spectrum,
{\tt SuperIso\_v3.3}~\cite{Mahmoudi:2008tp} to compute the low-energy constraints, and 
{\tt micrOMEGAs\_2.4.5}~\cite{Belanger:2001fz} to compute the neutralino relic density $\omhsq$, direct detection cross sections and to check compatibility with various pre-LHC sparticle mass limits.
Moreover, we use 
{\tt SDECAY\_1.3b}~\cite{Muhlleitner:2003vg} and {\tt HDECAY\_5.11}~\cite{Spira:1996if,Djouadi:1997yw} to produce SUSY and Higgs decay tables.
The various codes are interfaced using the SUSY Les Houches Accord (SLHA)~\cite{Skands:2003cj,Allanach:2008qq}. 

The posterior density of $\theta$ given data $D$ is given by
\begin{equation}
p(\theta | D) \sim L(D | \theta)\, p_0(\theta)\,,
\label{pmssm-pchain}
\end{equation}
where $L(D | \theta)$ is the likelihood and $p_0(\theta)$ is the prior probability density, or
prior for short. Beginning with a flat distribution in the parameters within the ranges defined by Eq.~(\ref{pmssm-eq:subspace}),  $p_0(\theta)$ is obtained by incorporating  the theoretical constraints noted above.  In other words, $p_0(\theta)$ is the result of sculpting the flat parameter distributions by the requirements related to theoretical consistency and $\tilde{\chi}^0_1$ being the LSP. 
This $p_0(\theta)$ defines the starting prior, which will be modified by actual data using  Eq.~(\ref{pmssm-pchain}).    
Since we consider multiple independent 
measurements $D_i$, the combined likelihood is given by $L(D | \theta) = \prod_{i}L(D_i | \theta)$.  

We partition the data into two parts: 
\begin{enumerate}
\item a set of constraints, listed in Table~\ref{pmssm-tab:preHiggs}, which are independent of the Higgs measurements; these constraints are used for the MCMC sampling and are collectively referred to by the label ``preHiggs'', and
\item the Higgs  measurements, which include the Higgs mass 
window, $m_h=123-128$~GeV, and the signal strength likelihood as derived in Section~\ref{sec:higgs2013}.
\end{enumerate}
With this partitioning, the posterior density becomes
\begin{equation}
p(\theta | D) \sim L(D^{\rm Higgs} | \theta) \, L(D^{\rm preHiggs} | \theta) \, p_0(\theta) = L(D^{\rm Higgs} | \theta) \, p^{\rm preHiggs}(\theta)\,,
\end{equation}
where $p_0(\theta)$ is the prior (as defined earlier) at the start of the inference chain and $p^{\rm preHiggs}(\theta) \sim L(D^{\rm preHiggs} | \theta) \, p_0(\theta)$ can be viewed as a prior that encodes the information from the preHiggs-measurements as well as the theoretical consistency requirements.  This partitioning allows us to assess the impact of the Higgs  results on the pMSSM parameter space while being consistent with constraints from the previous measurements. Note that at this stage we do not consider the direct limits from SUSY 
searches from ATLAS or CMS.

\begin{sidewaystable}[tbp]
\caption{The measurements that are the basis of our pMSSM prior $p^{\rm preHiggs}(\theta)$.  
All measurements were used to sample points from the pMSSM parameter space via MCMC methods. 
The likelihood for each point was reweighted post-MCMC based on better determinations of 
$\br(b \rightarrow s\gamma)$, $\br(B_s \rightarrow \mu \mu)$, $R(B_u \rightarrow \tau \nu)$, and $m_t$. }
\begin{center}
\begin{tabular}{|c|c|c|c|c|}
\hline
$i$     & Observable    & Constraint   & Likelihood function & MCMC / \\
        & $\mu_j(\theta)$               & $D^{\rm preHiggs}_j$             &  $L(D^{\rm preHiggs}_j|\mu_j(\theta))$  & post-MCMC \\
\hline\hline
1a & $\br(b \rightarrow s\gamma)$~\cite{Amhis:2012bh,Misiak:2006zs} & $(3.55 \pm 0.24^{\rm stat} \pm 0.23^{\rm th} \pm 0.09^{\rm sys})\times 10^{-4}$ & Gaussian & MCMC \\
1b & $\br(b \rightarrow s\gamma)$~\cite{HFAG2013} & $(3.43 \pm 0.21^{\rm stat} \pm 0.23^{\rm th} \pm 0.07^{\rm sys})\times 10^{-4}$ & Gaussian & reweight \\
\hline
2a & $\br(B_s \rightarrow \mu \mu)$~\cite{oldbsmm,Akeroyd:2011kd} & observed ${\rm CL}_s$ curve from \cite{oldbsmm} & $d(1 - {\rm CL}_s)/d(BR(B_s \rightarrow \mu\mu))$ & MCMC \\
2b & $\br(B_s \rightarrow \mu \mu)$~\cite{CMSandLHCbCollaborations:2013pla,Akeroyd:2011kd} & $(2.9 \pm 0.7 \pm 0.29^{\rm th})\times 10^{-9}$ & Gaussian & reweight \\
\hline
3a & $R(B_u \rightarrow \tau \nu)$\cite{Beringer:1900zz} & $1.63\pm 0.54$ & Gaussian & MCMC \\
3b & $R(B_u \rightarrow \tau \nu)$\cite{HFAG2013} & $1.04\pm 0.34$ & Gaussian & reweight \\
\hline
4 & $\Delta a_\mu$~\cite{Bennett:2006fi,Hagiwara:2011af,Stockinger:2006zn} & $(26.1 \pm 8.0^{\rm exp}\pm 10.0^{\rm th})\times 10^{-10}$ & Gaussian & MCMC \\
\hline
5a & $m_t$~\cite{topmasslhc} & $173.3\pm0.5^{\rm stat}\pm1.3^{\rm sys}$~GeV & Gaussian & MCMC \\
5b & $m_t$~\cite{CDF:2013jga} & $173.20\pm0.87$~GeV & Gaussian & reweight \\
\hline
6 & $m_b(m_b)$~\cite{Beringer:1900zz} & $4.19^{+0.18}_{-0.06}$~GeV & Two-sided Gaussian & MCMC \\
\hline
7 & $\alpha_s(M_Z)$~\cite{Beringer:1900zz} & $0.1184 \pm 0.0007$ & Gaussian & MCMC \\
\hline
8 & sparticle & LEP \cite{lepsusy} & 1 if allowed & MCMC \\
   & masses & (via {\tt micrOMEGAs}~\cite{Belanger:2001fz}) & 0 if excluded & \\
\hline
\end{tabular}
\label{pmssm-tab:preHiggs}
\end{center}
\end{sidewaystable}

In addition to the experimental results included in our calculation of the prior $p^{\rm preHiggs}
(\theta)$, Table~\ref{pmssm-tab:preHiggs}  lists the corresponding likelihood $L(D^{\rm preHiggs}_{j} | \mu_j(\theta))$ for each observable $j$, where $\mu_j(\theta)$ denotes the model prediction for the observable $j$,  such as $\br(b \rightarrow s\gamma)$ for a given $\theta$.    
We obtained a discrete representation of the prior $p^{\rm preHiggs}(\theta)$ within the
sub-space defined in Eq.~(\ref{pmssm-eq:subspace}) by sampling points from $p^{\rm preHiggs}(\theta)$ using a MCMC method (for an introduction see, \eg, \cite{Trotta:2008qt}).  
By construction, this method produces a sample of points whose density in the neighborhood  of 
$\theta$ is $\propto p^{\rm preHiggs}(\theta)$, \ie\ the sampled points will constitute a
discrete representation of the preHiggs likelihood as a function of the pMSSM parameters
$\theta$. 

Our study is based on approximately $2\times 10^6$ MCMC points, which were originally sampled  
for the CMS study \cite{CMS-PAS-SUS-12-030} in which some of us participated. 
(The CMS study then used a random sub-sample of 7205 points from this data.) 
In the meanwhile, several experimental constraints that enter the preHiggs likelihood function have been updated. 
For example, first evidence for the decay $B_s \rightarrow \mu \mu$ was reported by the LHCb collaboration in~\cite{Aaij:2012nna} and recently new improved measurements have become available by CMS and LHCb~\cite{CMSandLHCbCollaborations:2013pla}.  We have taken the up-to-date value into account by reweighting each sampled point by the ratio of 
the new $\br(B_s \rightarrow \mu \mu)$ likelihood, 2b,  to the old likelihood, 2a, in Table~\ref{pmssm-tab:preHiggs}. 
Analogous reweighting was performed to take into account the updated values of $\br(b \rightarrow s\gamma)$,
$R(B_u \rightarrow \tau \nu)$, and $m_t$.

\subsubsection{Higgs likelihood} 
\label{pmssm-sec:higgslikeli}

For fitting the properties of the observed Higgs boson, we use the information presented in terms of combined ellipses in Section~\ref{2013c-sec:combinedss}, as well as the preliminary ATLAS results on invisible decays from $ZH$ associated production with $Z\to \ell^+\ell^-$ and $H \to {\rm invisible}$, extracting the likelihood from Fig.~10b of \cite{ATLAS-CONF-2013-011}. All these results are combined into the ``Higgs signal likelihood'' $L(D^{\rm Higgs} | \theta)$.

For the concrete calculation, we use {\tt HDECAY\_5.11} and approximate $\sigma(gg \to h)/\sigma(gg \to H_{\rm SM}) \simeq \Gamma(h \to gg)/\Gamma(H_{\rm SM} \to gg)$. Moreover,  for computing the SM results entering the calculation of signal strengths, we use the MSSM decoupling limit with $m_A$ and the relevant SUSY masses set to 4~TeV. This ensures completely SM-like Higgs boson couplings at tree-level, as well as vanishing  
radiative contributions from the SUSY particles (including non-decoupling effects). We choose this procedure in order to guarantee that the radiative corrections being included are precisely the same for the numerator and denominator in Eq.~\eqref{eq:signalstr2}.

For completeness, we also take into account the limits  
from the $H,A\to\tau\tau$ searches in the MSSM~\cite{CMS-PAS-HIG-13-021}. These limits are implemented in a binary fashion: we set the likelihood from each of these constraints to 1 when the 95\%~CL limit is obeyed and to 0 when it is violated. (Including or not including this limit however has hardly any visible effect on the posterior distributions.)

\subsubsection{Dark matter constraints} 
\label{pmssm-sec:dmconst}

The calculation of the properties of the neutralino LSP as a thermal cold dark matter (DM) candidate 
(or one of the cold DM components) depends on a number of cosmological assumptions, like 
complete thermalization, no non-thermal production, no late entropy production, \etc\ 
In order to be independent of these assumptions, 
we will show results with and without requiring consistence with DM constraints. 
When we do apply DM constraints, we adopt the following procedure.
For the relic density, we apply an upper bound as a smoothed step function at the Planck 
value of $\Omega h^2=0.1189$~\cite{Ade:2013zuv}, accounting for a 10\% theory-dominated uncertainty. 
Concretely, we take
\begin{equation}
L=
\left\lbrace
\begin{array}{ccc}
1 & \mbox{if} & \Omega h^2 < 0.119\,, \\
\exp[(0.119-\Omega h^2)/0.012)^2 / 2] & \mbox{if} & \Omega h^2 > 0.119\,.\end{array}\right.
\end{equation}
For  the spin-independent scattering cross section off protons, we use the 90\%~CL limit from LUX~\cite{Akerib:2013tjd}, 
rescaling the computed $\sigsi$ by a factor $\xi=\omhsq/0.119$ to account for the lower local density when the neutralino is only part of the DM. (The alternative would be to assume that the missing amount of $\omhsq$ is substituted by non-thermal production, which would make the direct detection constraints more severe. 
Our approach is more conservative in the sense of not being overly restrictive.)

\subsubsection{Prompt chargino requirement} 
\label{pmssm-sec:promptcharg}

Before presenting the sampled distributions, another comment is in order. 
Letting $M_1$, $M_2$ and $\mu$, vary freely over the same range implies that about $2/3$ of the time 
$M_2$ or $\mu$ will be the smallest mass parameter in the neutralino mass matrix. 
This implies that  in a considerable portion of the pMSSM parameter space the 
$\tilde{\chi}_1^{\pm}$ and $\tilde{\chi}_2^0$ are close in mass or almost degenerate 
with the LSP, $\tilde{\chi}_1^0$~\cite{Gunion:1987yh}.
When the $\tilde{\chi}_1^{\pm}$--$\tilde{\chi}_1^0$ mass difference becomes very small, 
below about 300~MeV,  
the charginos are long-lived and can traverse the detector before they decay.  
This typically occurs for wino-LSP scenarios with $|M_2|\ll |M_1|,\, |\mu|$. 
Since long-lived heavy charged particles were not considered in the SUSY searches used in \cite{CMS-PAS-SUS-12-030}, 
charginos were required to decay promptly; in practice this means a cut on the average proper lifetime of $c\tau < 10$~mm. 
In order to be able to directly compare our results (based on the Higgs measurements) 
with the CMS study (based on SUSY search results) \cite{CMS-PAS-SUS-12-030} and its up-coming update \cite{CMS-PAS-SUS-13-020}, 
we also require ``prompt'' chargino decays, \ie\  $c\tau < 10$~mm. 
Most of our conclusions are insensitive to this requirement.
Wherever it matters, we will however also show the results obtained without imposing the 
$c\tau$ cut.

\subsection{Results}
%

\subsubsection{Pre-Higgs distributions and impact of the Higgs mass}\label{pmssm-sec:YellowPlots}

\begin{figure}[htbp]
\begin{center}
\includegraphics[width=0.24\linewidth]{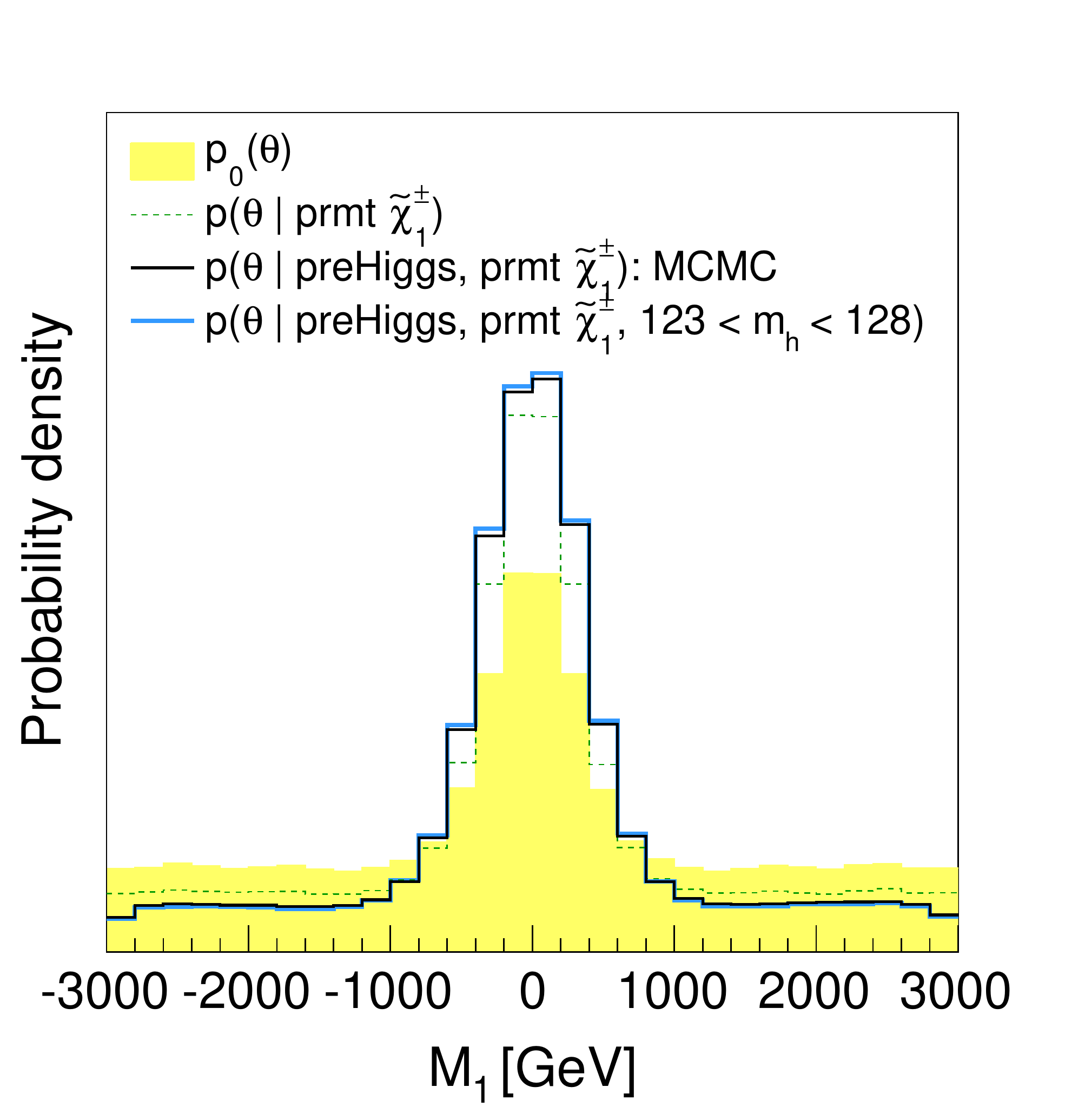}
\includegraphics[width=0.24\linewidth]{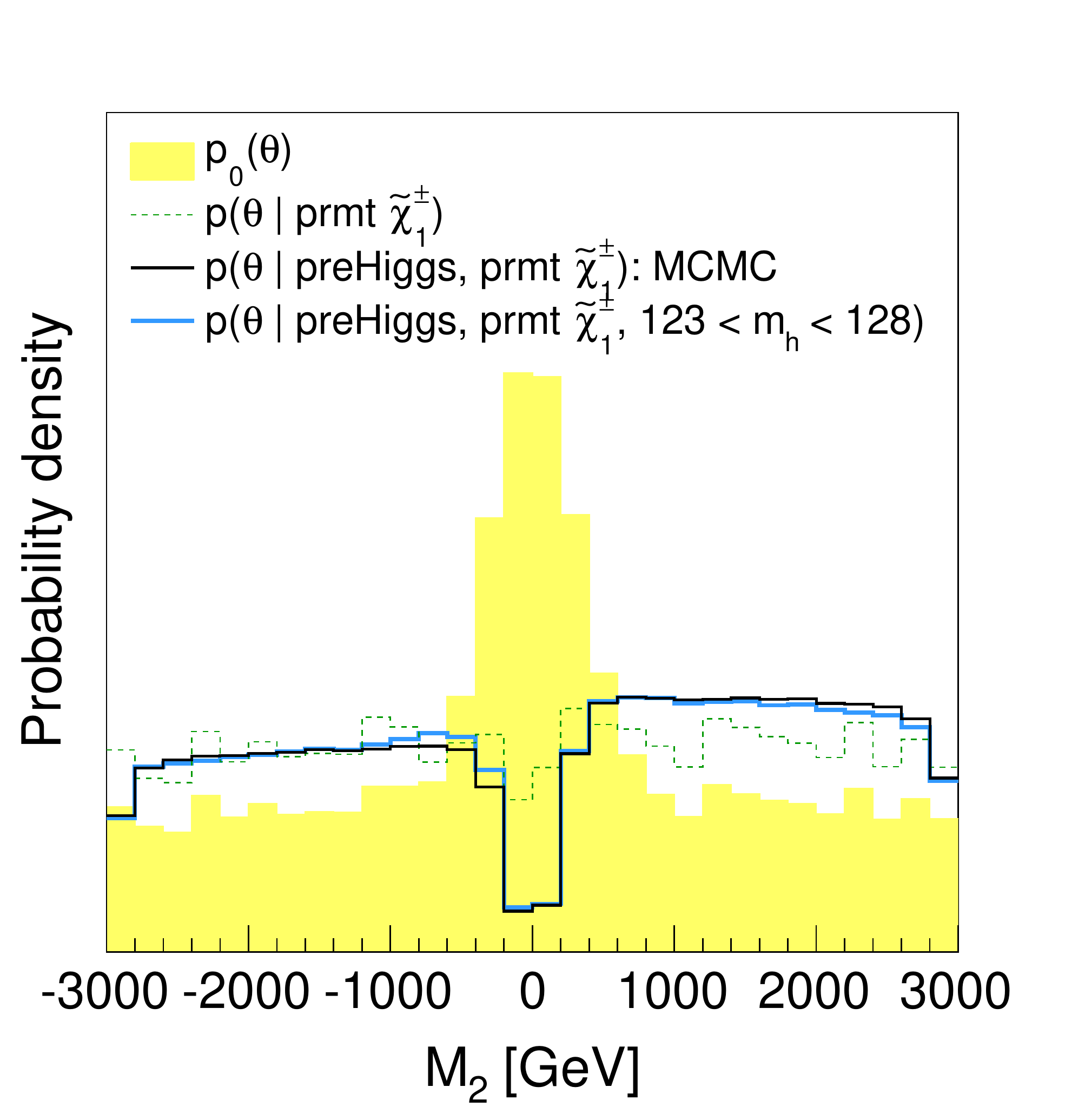}
\includegraphics[width=0.24\linewidth]{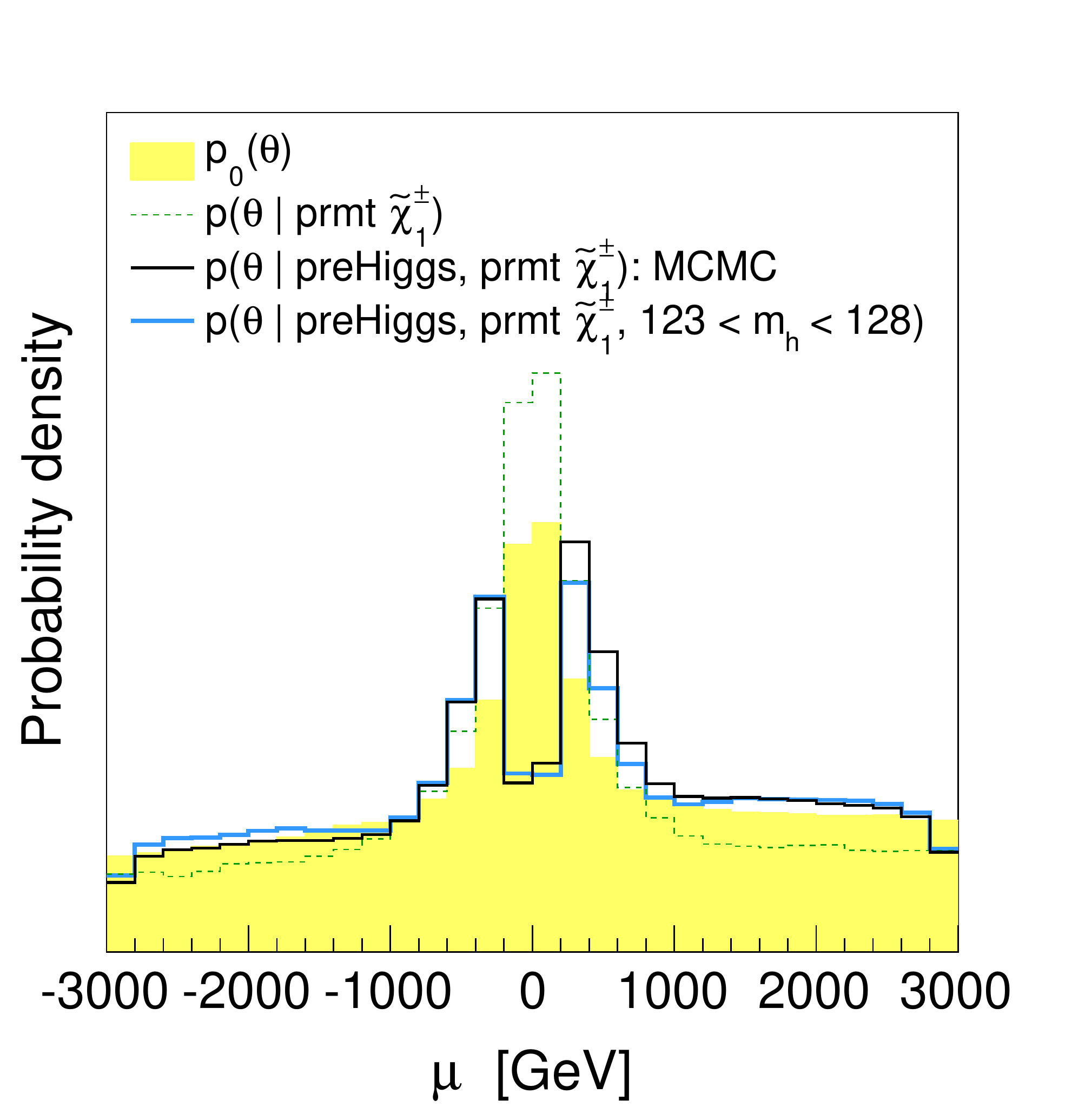} 
\includegraphics[width=0.24\linewidth]{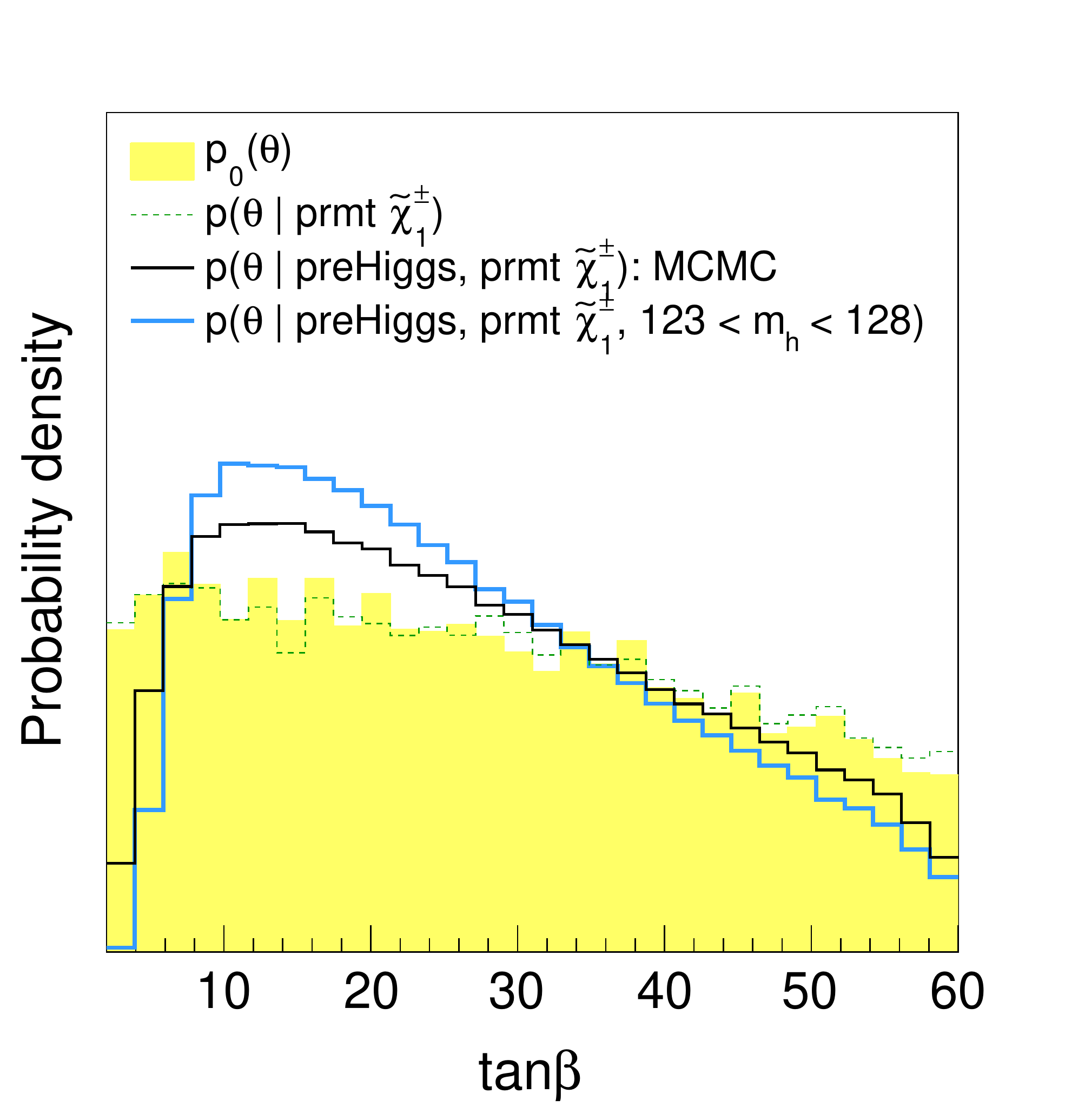}
\includegraphics[width=0.24\linewidth]{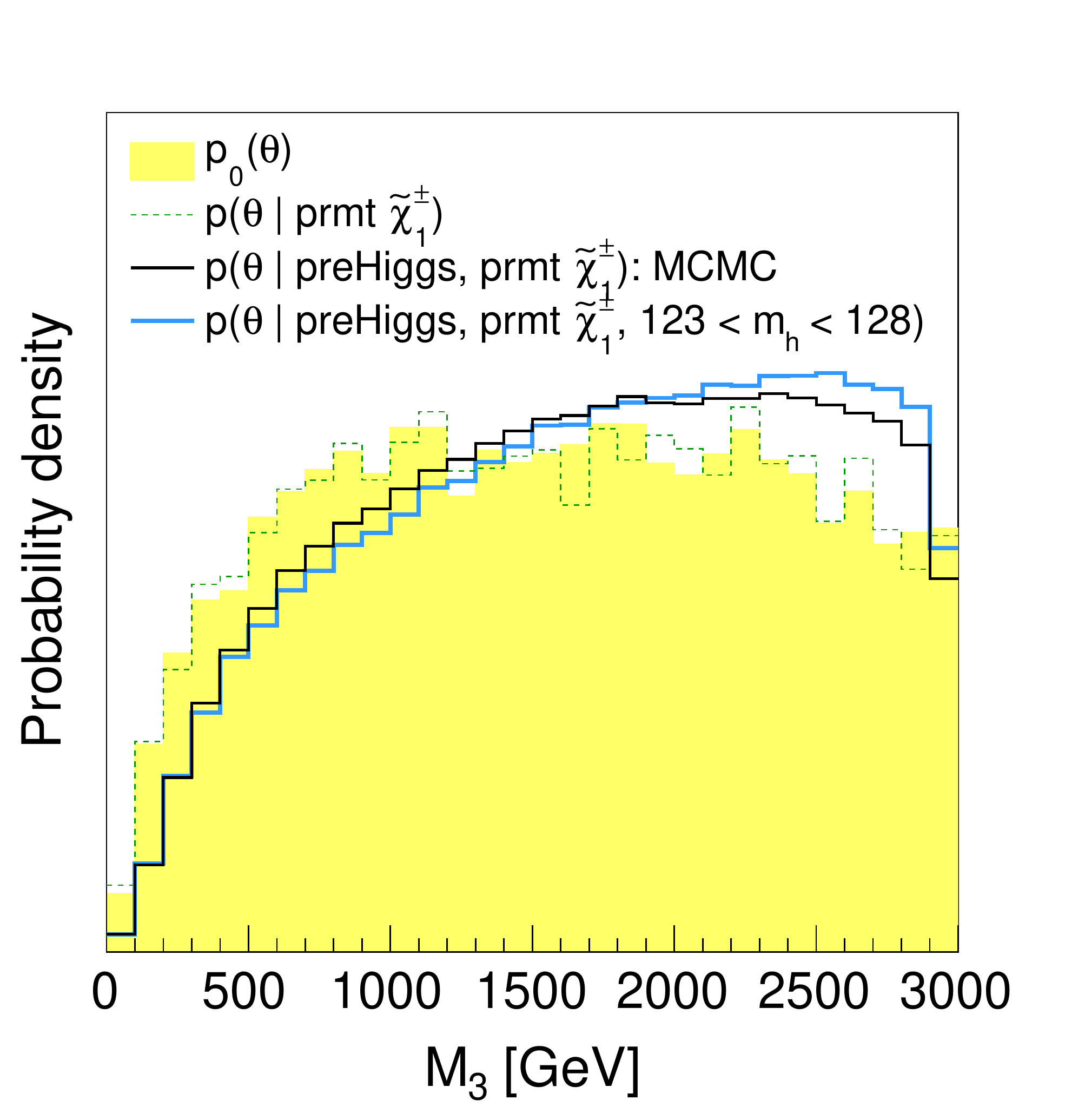}
\includegraphics[width=0.24\linewidth]{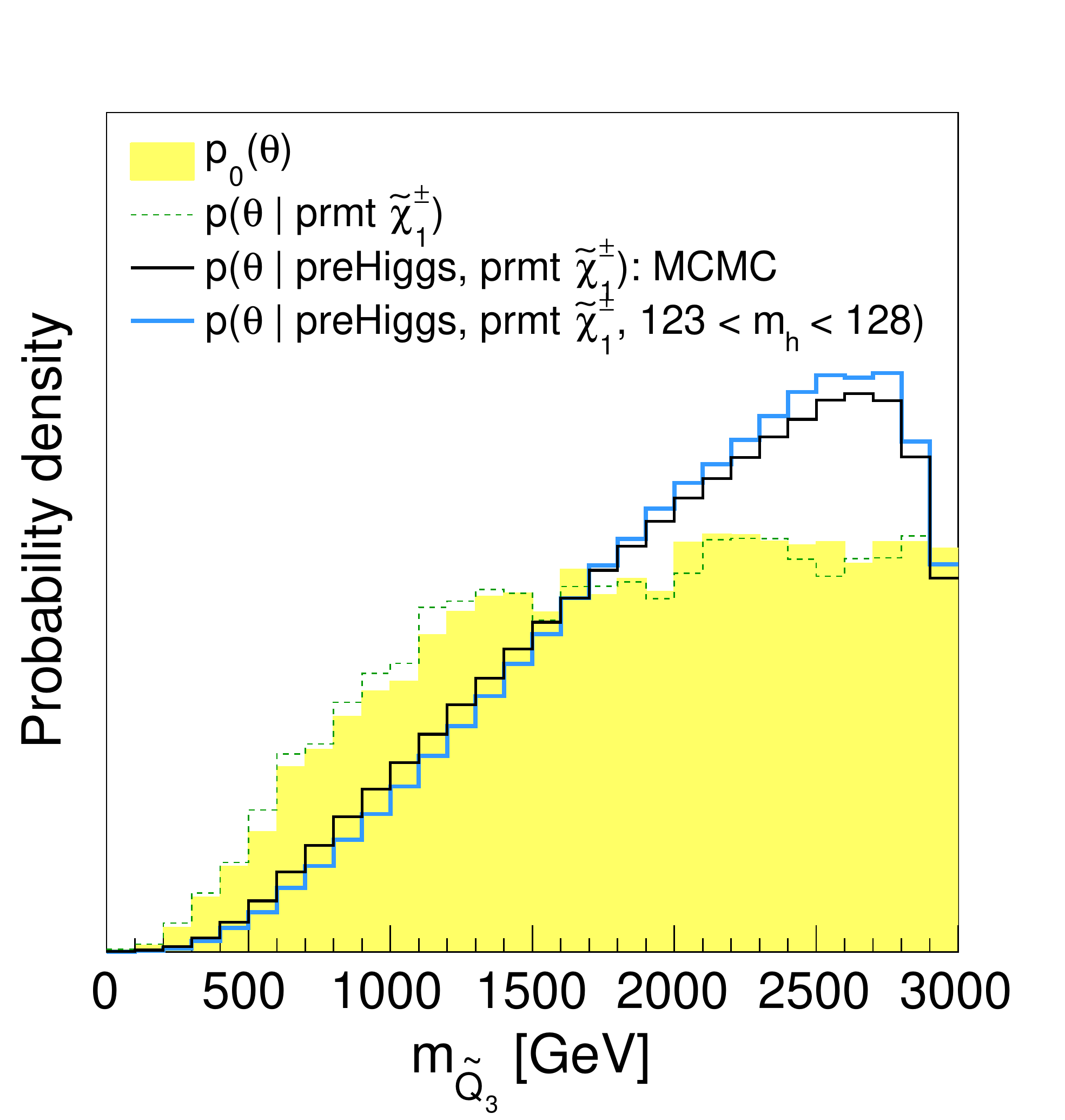} 
\includegraphics[width=0.24\linewidth]{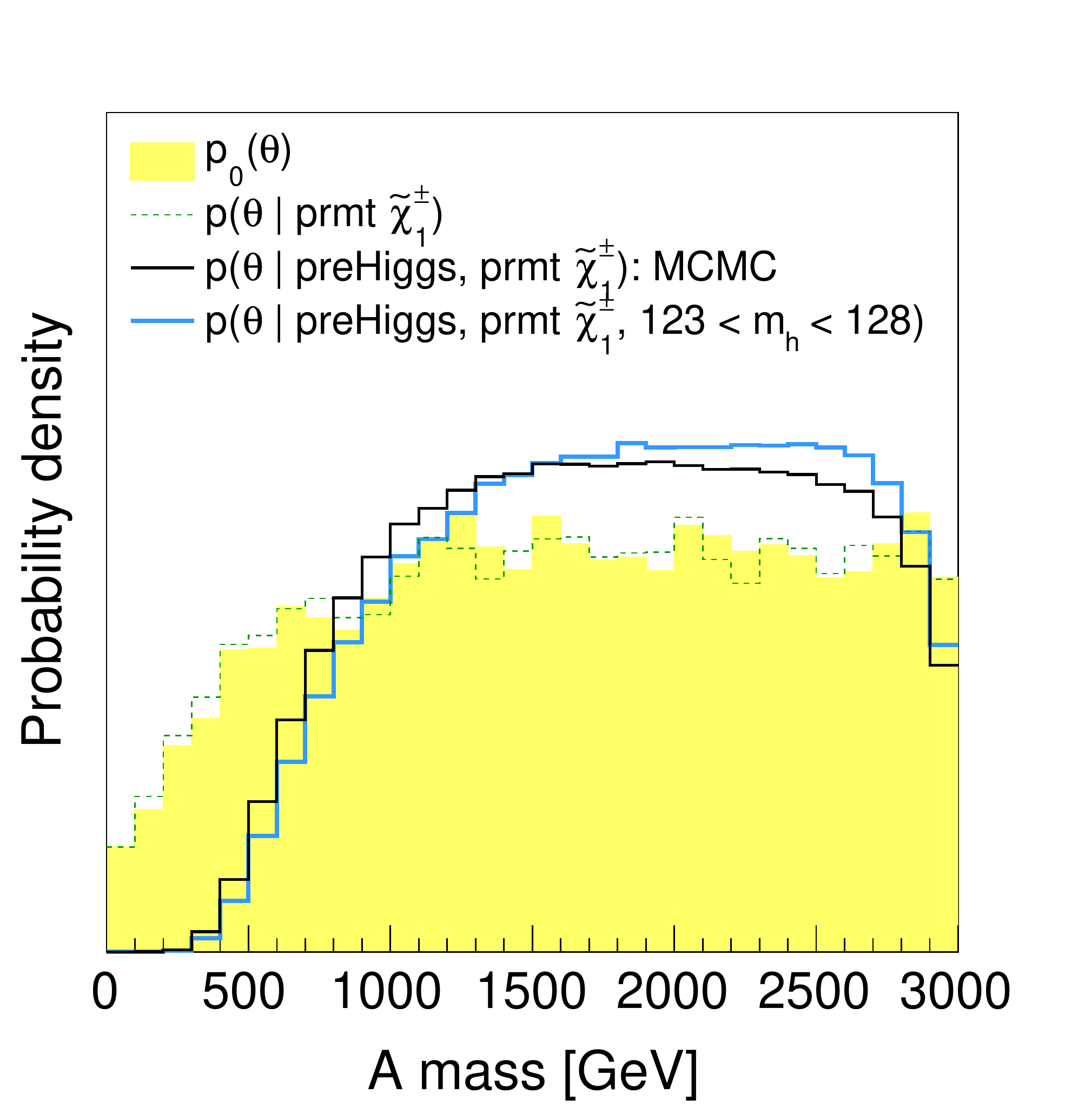} 
\includegraphics[width=0.24\linewidth]{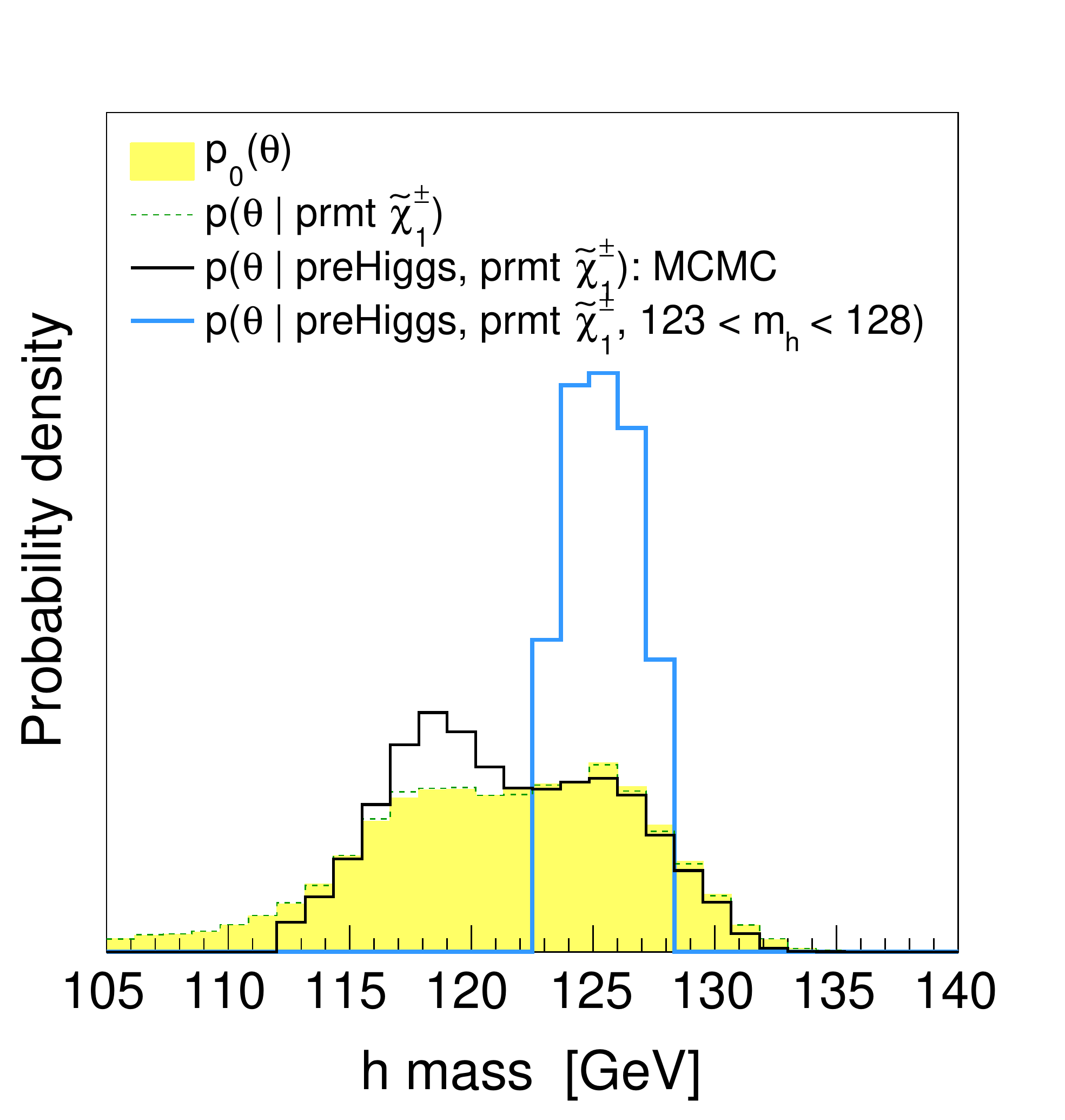}
\includegraphics[width=0.24\linewidth]{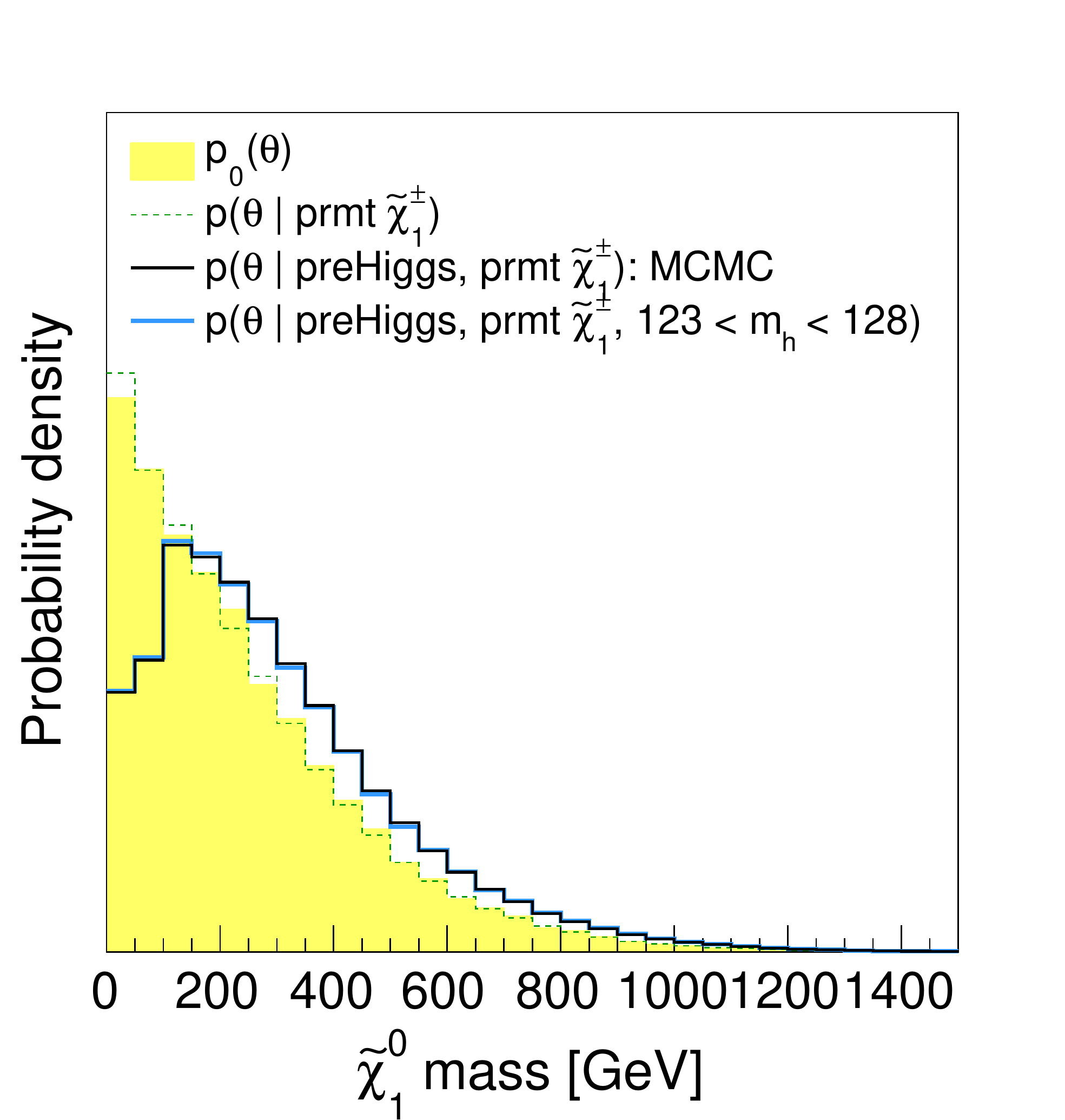}
\includegraphics[width=0.24\linewidth]{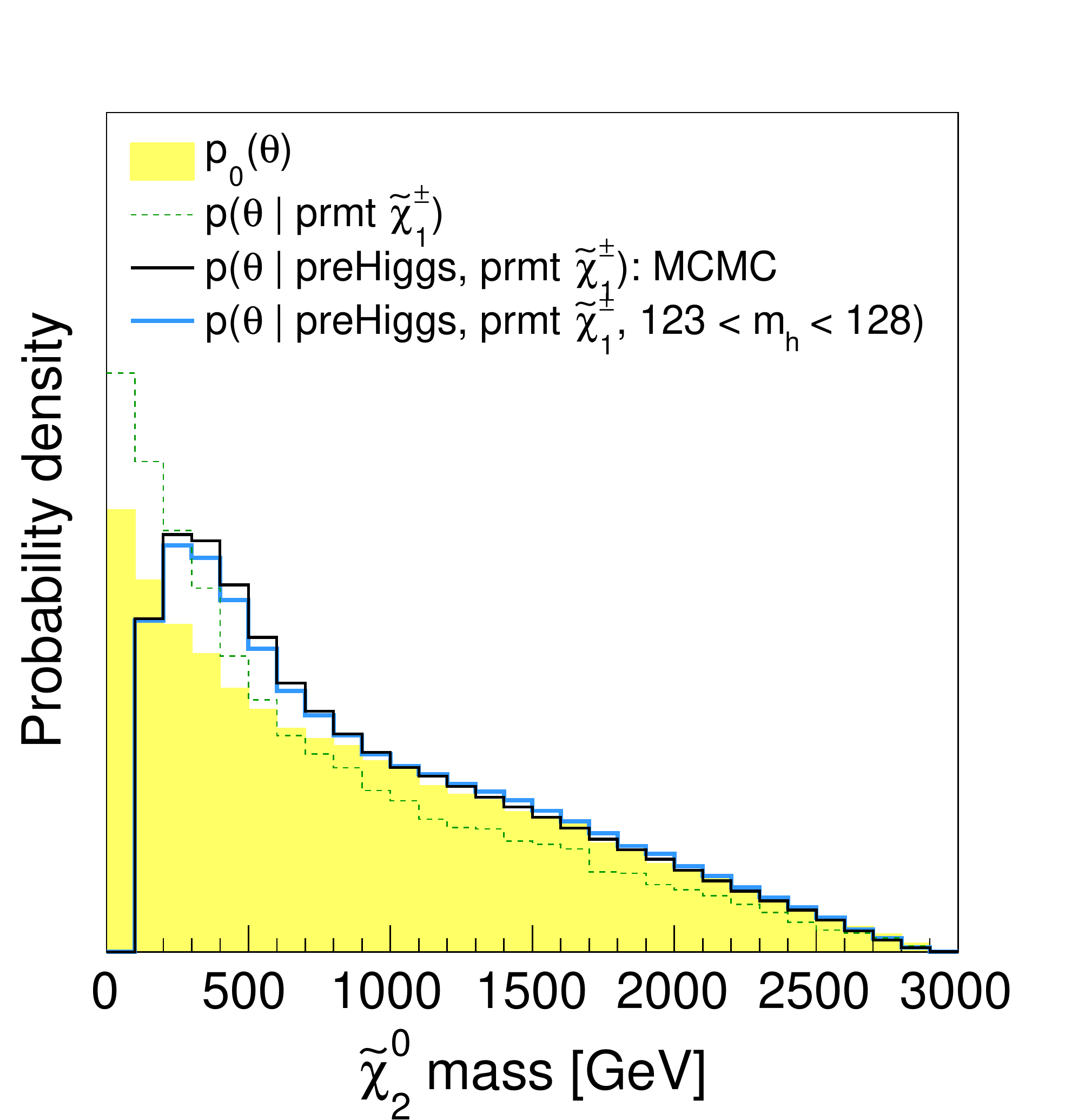} 
\includegraphics[width=0.24\linewidth]{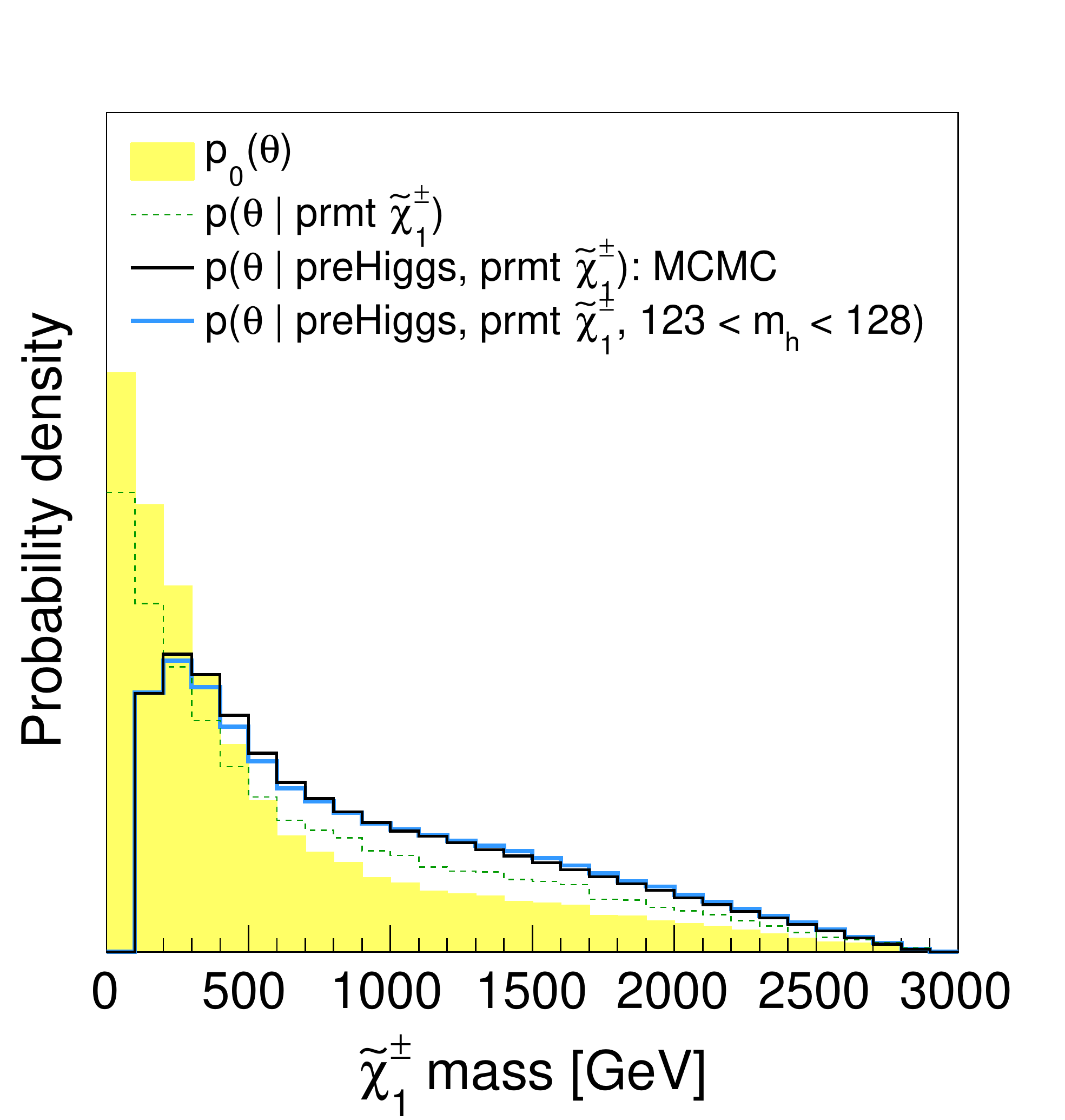}
\includegraphics[width=0.24\linewidth]{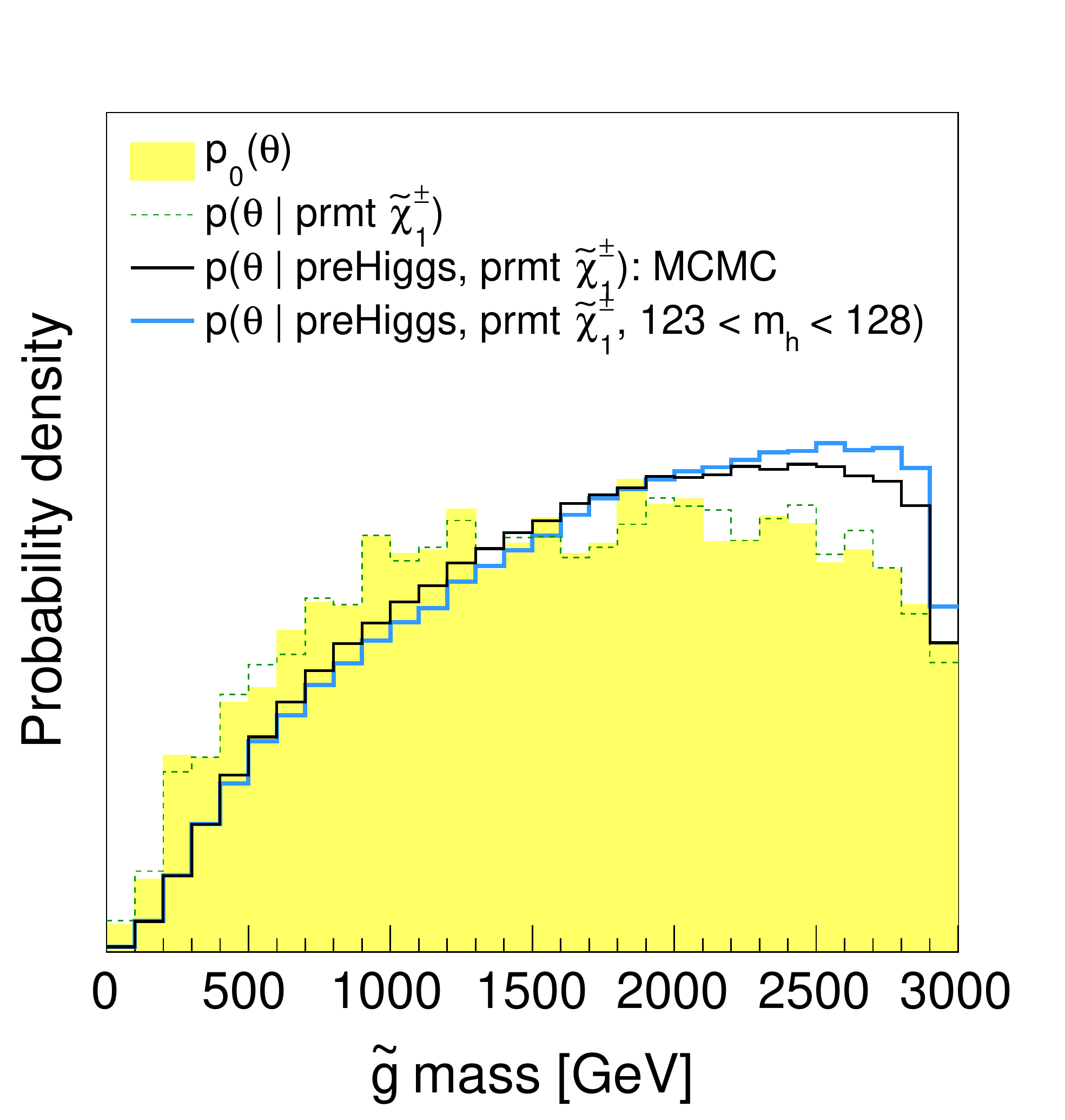}
\includegraphics[width=0.24\linewidth]{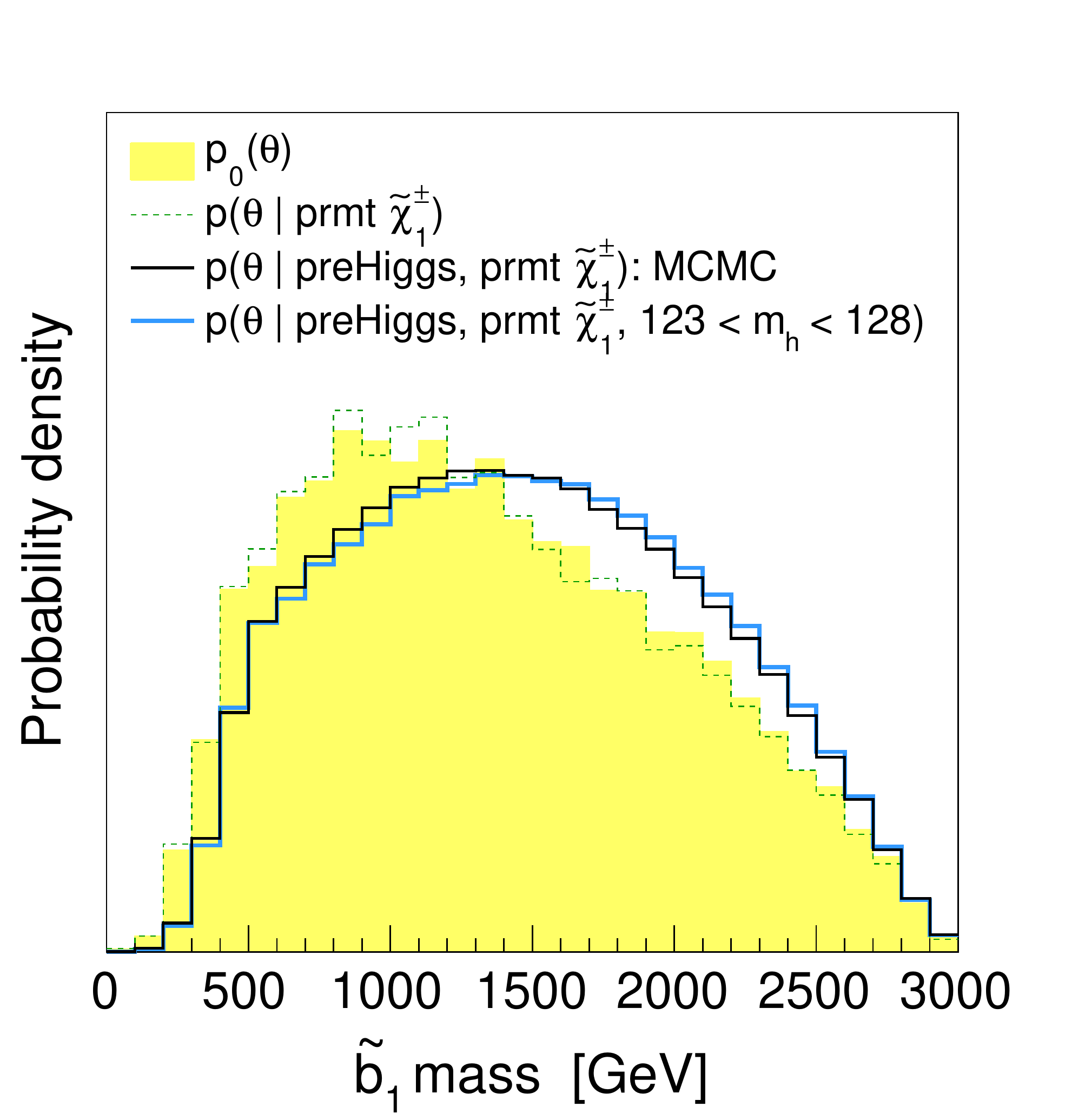} 
\includegraphics[width=0.24\linewidth]{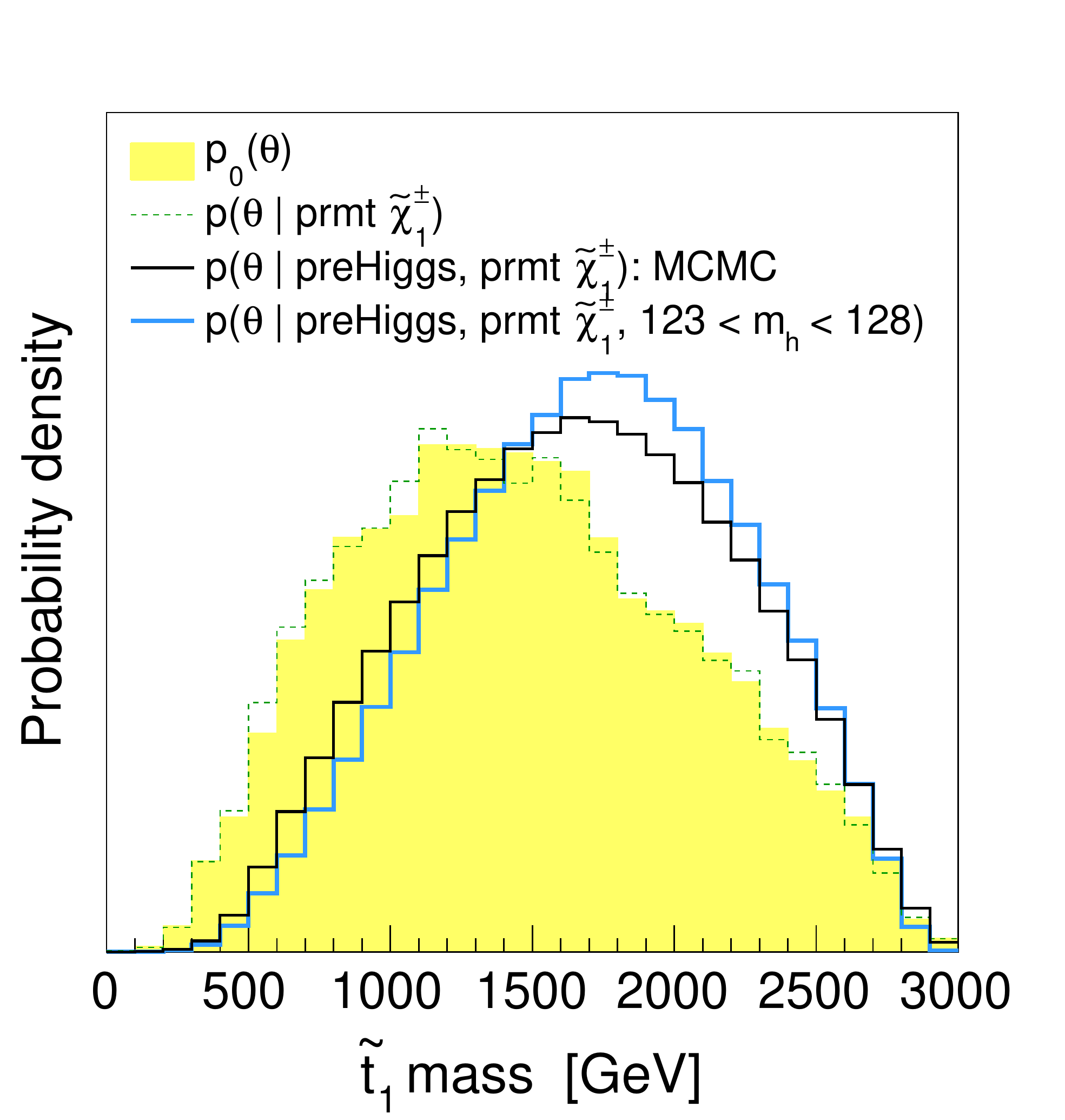}
\includegraphics[width=0.24\linewidth]{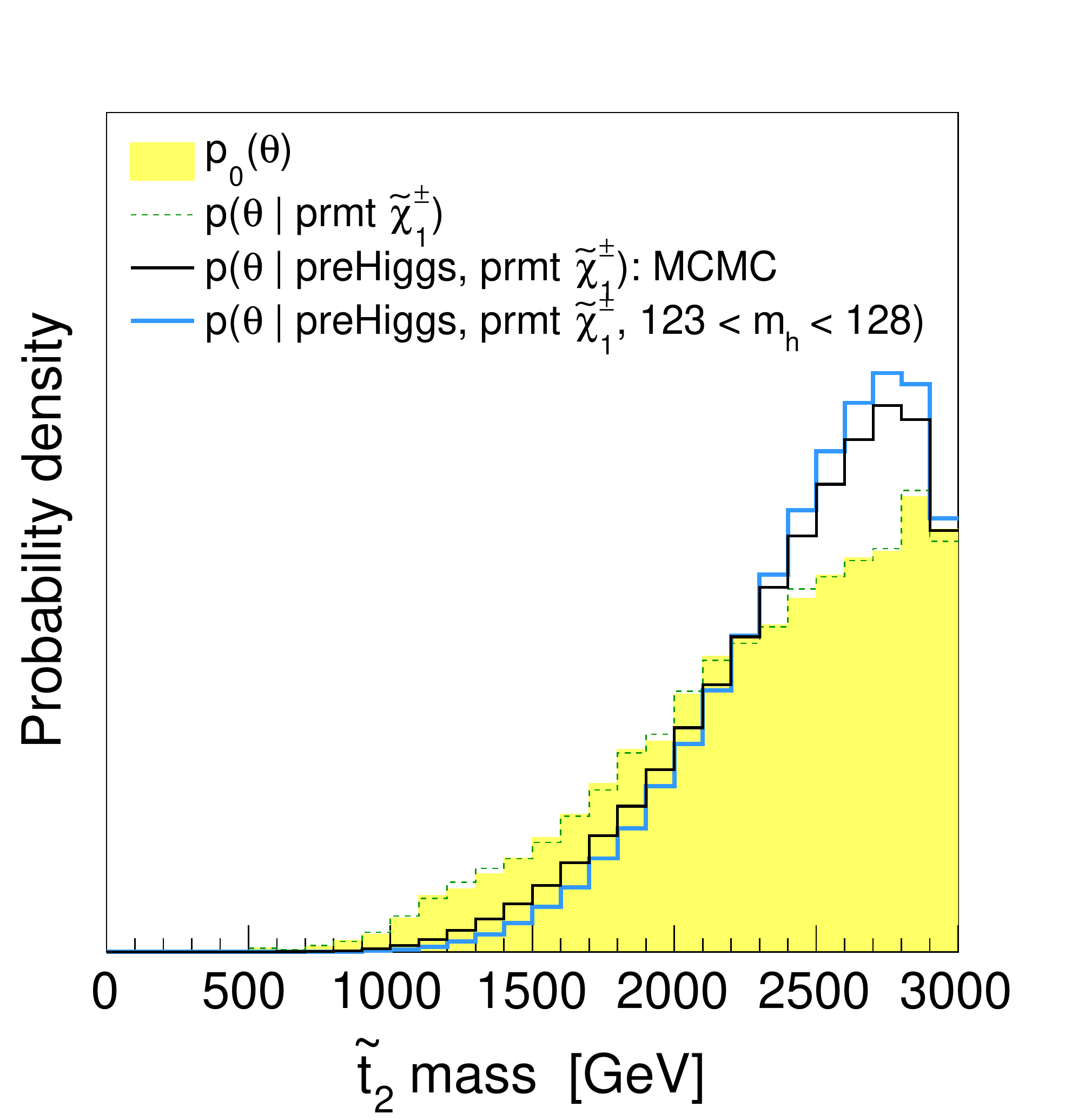}
\includegraphics[width=0.24\linewidth]{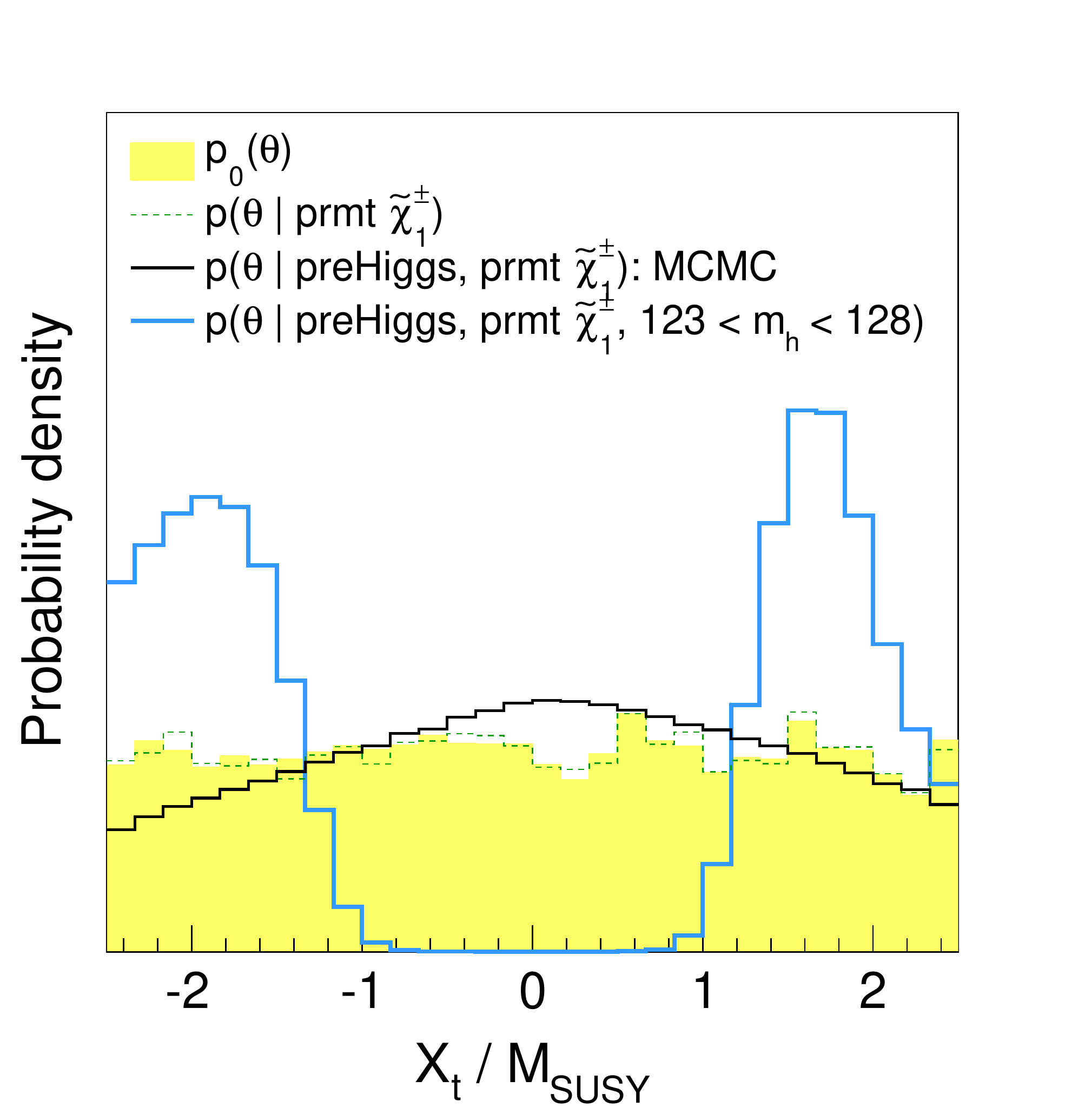} 
\caption{Marginalized 1D posterior densities for selected parameters and masses.
The yellow histograms show the sampled distributions, $p_0(\theta)$, as obtained after imposing theoretical constraints starting from a flat scan in the parameter ranges specified by Eq.~(\ref{pmssm-eq:subspace}). The dashed 
green lines are the distributions  after requiring prompt charginos (prmt), the full black lines show  
the distributions  based on the ``preHiggs'' measurements of Table~\ref{pmssm-tab:preHiggs}, 
and the full blue lines the ones when requiring $m_h=[123,\,128]$~GeV in addition 
to ``prmt'' and ``preHiggs''  constraints.
The  bottom right plot of $X_t/M_{\rm SUSY}$ shows that large (but not maximal) 
stop mixing is favored by the $m_h=123-128\gev$ requirement. }
\label{pmssm-fig:sampling1}
\end{center}
\end{figure}

We begin our discussion by showing in Fig.~\ref{pmssm-fig:sampling1} the sampled distributions of selected parameters and masses and the effect of the model prior.  All distributions except that of the pMSSM prior $p_0(\theta)$ include the prompt chargino requirement; as can be seen, this requirement substantially alters the probability distributions for 
the parameters $M_1$, $M_2$, and $\mu$ and the chargino and neutralino masses  relative to the $p_0(\theta)$ distributions, but has very little impact on the other parameters or masses. Further, in all the plots we observe that the preHiggs measurements incorporated in the MCMC influence the probability distributions relative to the simple prompt-chargino-decay distributions quite significantly, in particular shifting the neutralino, chargino, gluino, and also the stop/sbottom masses to higher values.

Also shown is the impact of requiring,  in addition,  that the mass of the light $h$ fall in the window $123~{\rm GeV} \leq m_h \leq 128~{\rm GeV}$. This Higgs mass constraint strongly affects the stop mixing parameter $X_t /M_{\rm SUSY} \equiv (A_t - \mu / \tan \beta)/\sqrt{m_{\st_1}m_{\st_2}}$, whose distribution takes on a two-peak structure emphasizing larger absolute values. More precisely, values around $|X_t /M_{\rm SUSY}|\approx 2$, \ie\ large but not maximal stop mixing is preferred. (Maximal stop mixing would mean $|X_t /M_{\rm SUSY}|=\sqrt{6}$; for a detailed discussion of the relation between $|X_t /M_{\rm SUSY}|$ and $m_h$ see, \eg,  \cite{Djouadi:2005gj,Brummer:2012ns}). 
It is interesting to note here that, in view of naturalness, the optimal stop mixing is indeed somewhat shy of maximal~\cite{Wymant:2012zp}. The optimal value is actually quite close to that which has the highest probability in the pMSSM context, despite the fact that no measure of naturalness is input into the pMSSM likelihood analyses.    
The Higgs mass window requirement also results in a shift of the $\tilde t_1$ mass distribution to slightly larger values; however, compared to the impact of the preHiggs constraints the effect is quite small.  Aside from an increased preference for values of $\tan\beta\approx 10-20$, 
the other parameters and masses are hardly affected by the Higgs mass window. 

\begin{figure}[t!]
\begin{center}
\includegraphics[width=0.3\linewidth]{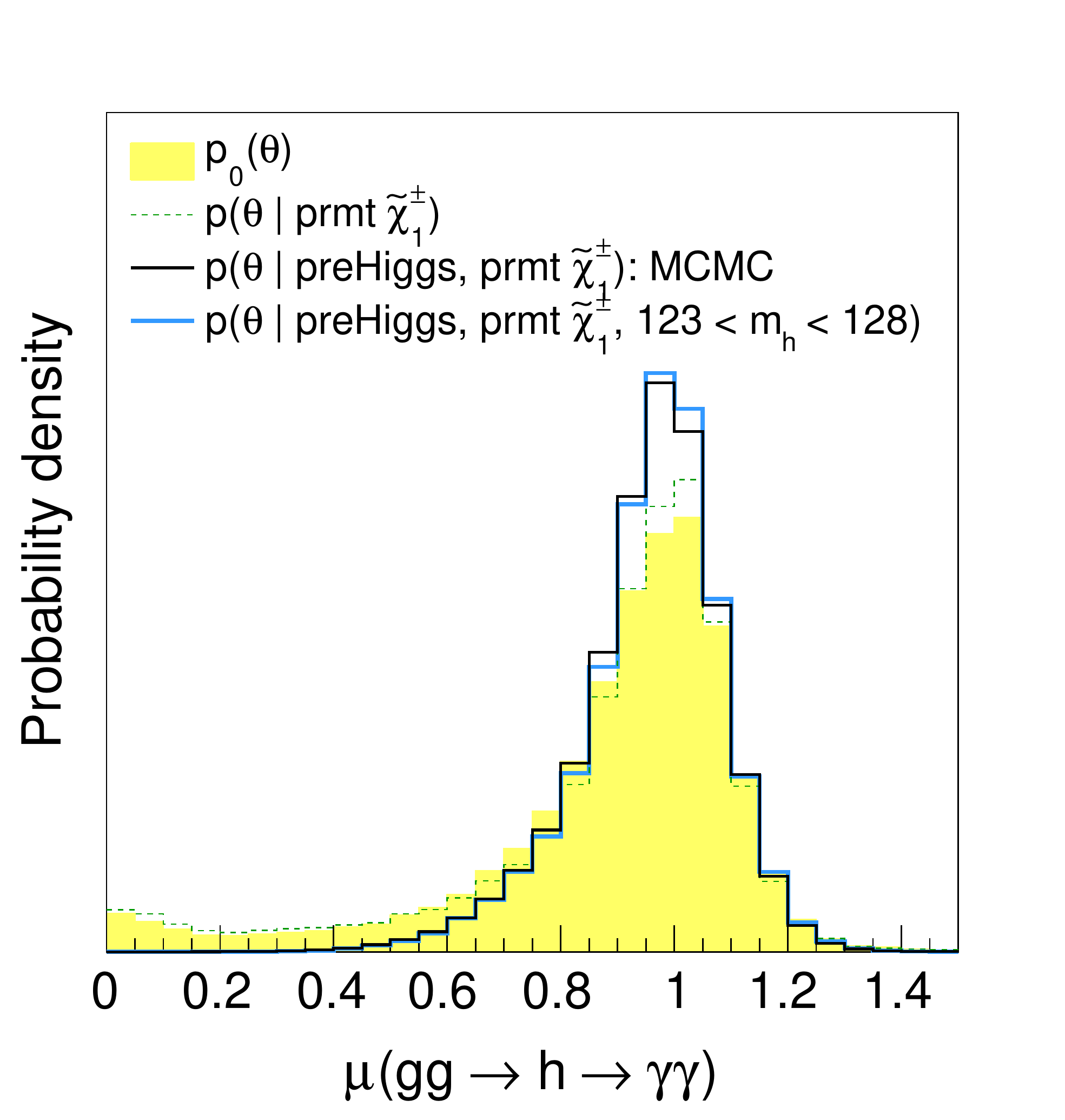}
\includegraphics[width=0.3\linewidth]{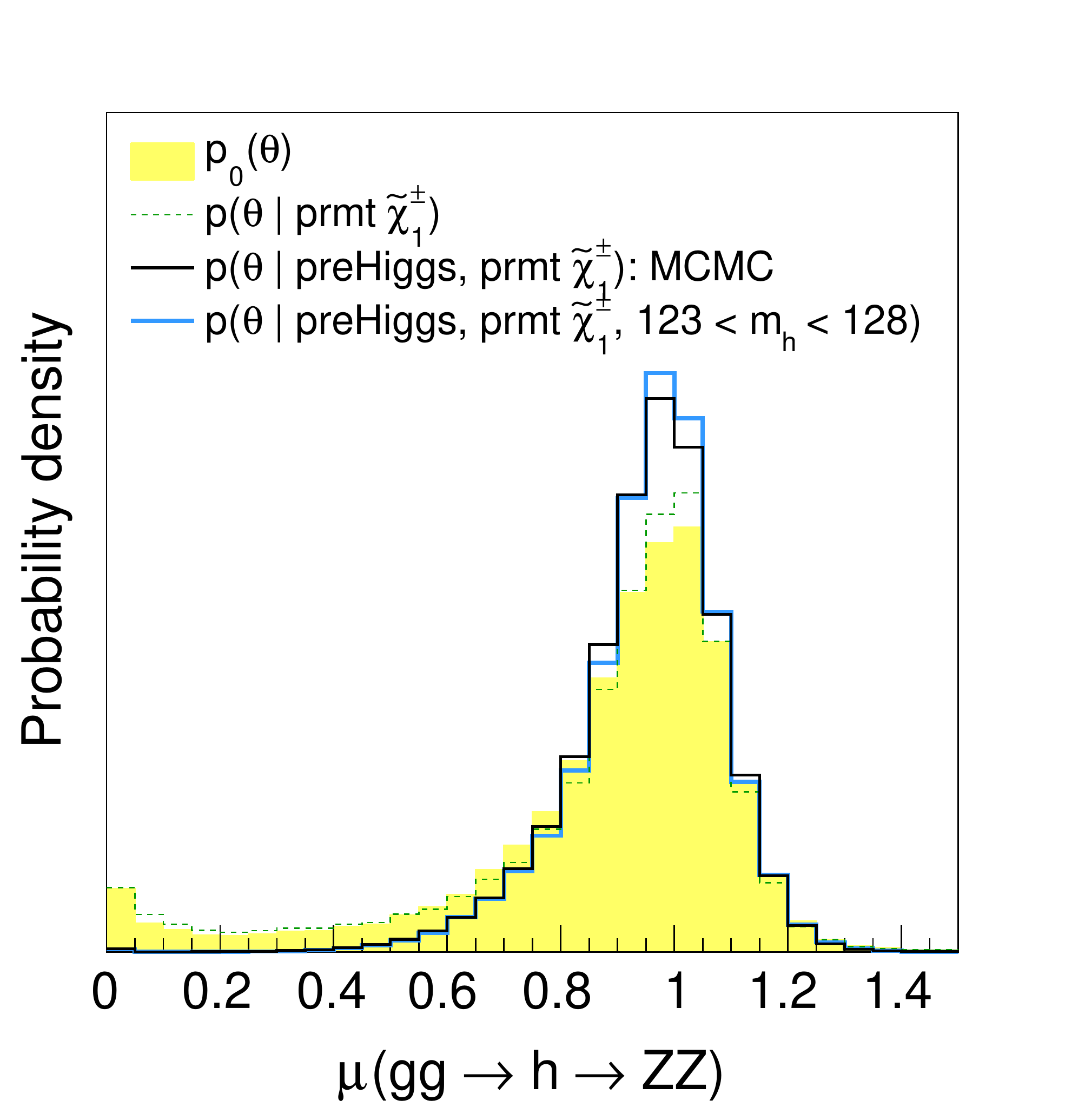}
\includegraphics[width=0.3\linewidth]{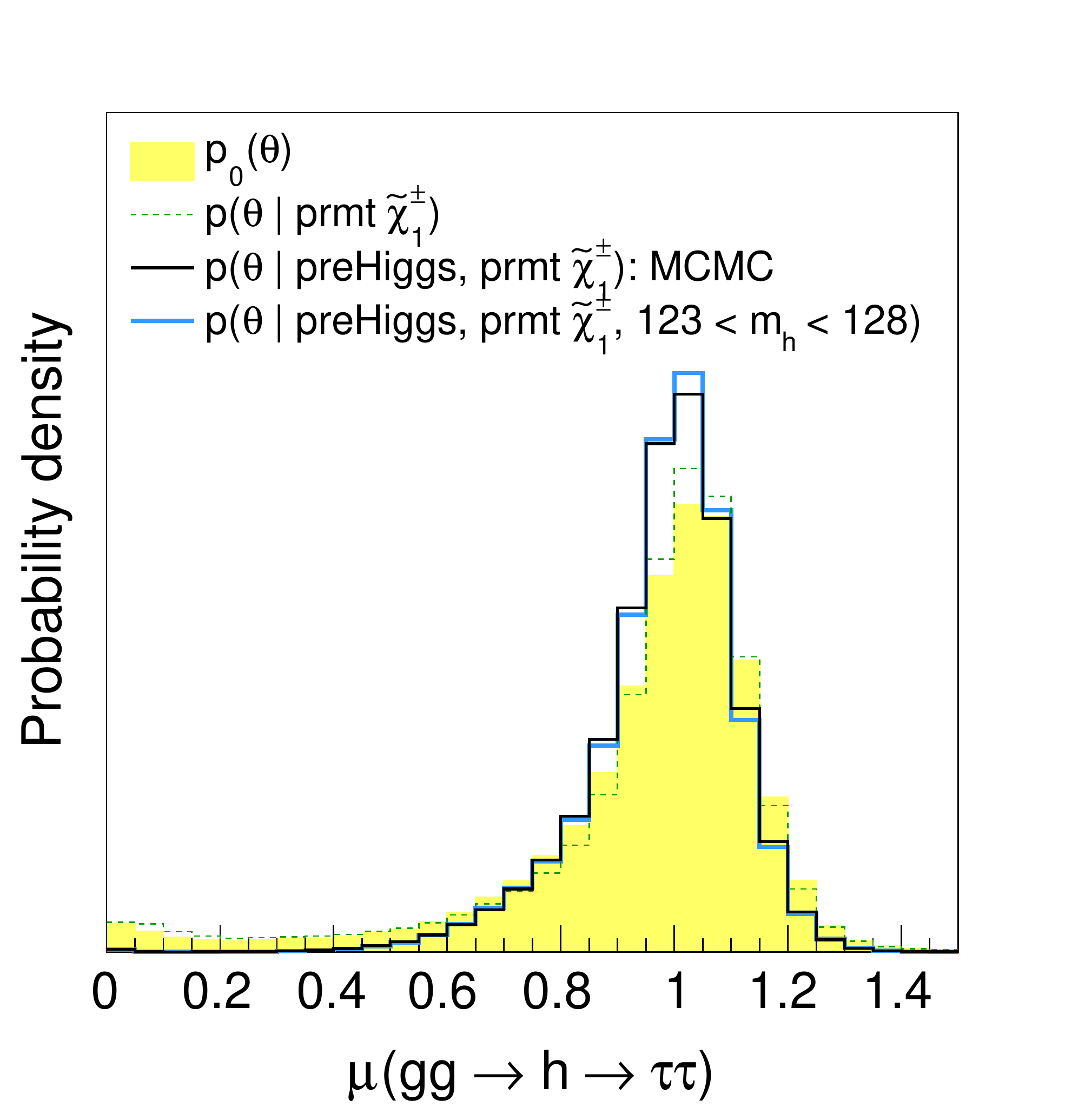}
\includegraphics[width=0.3\linewidth]{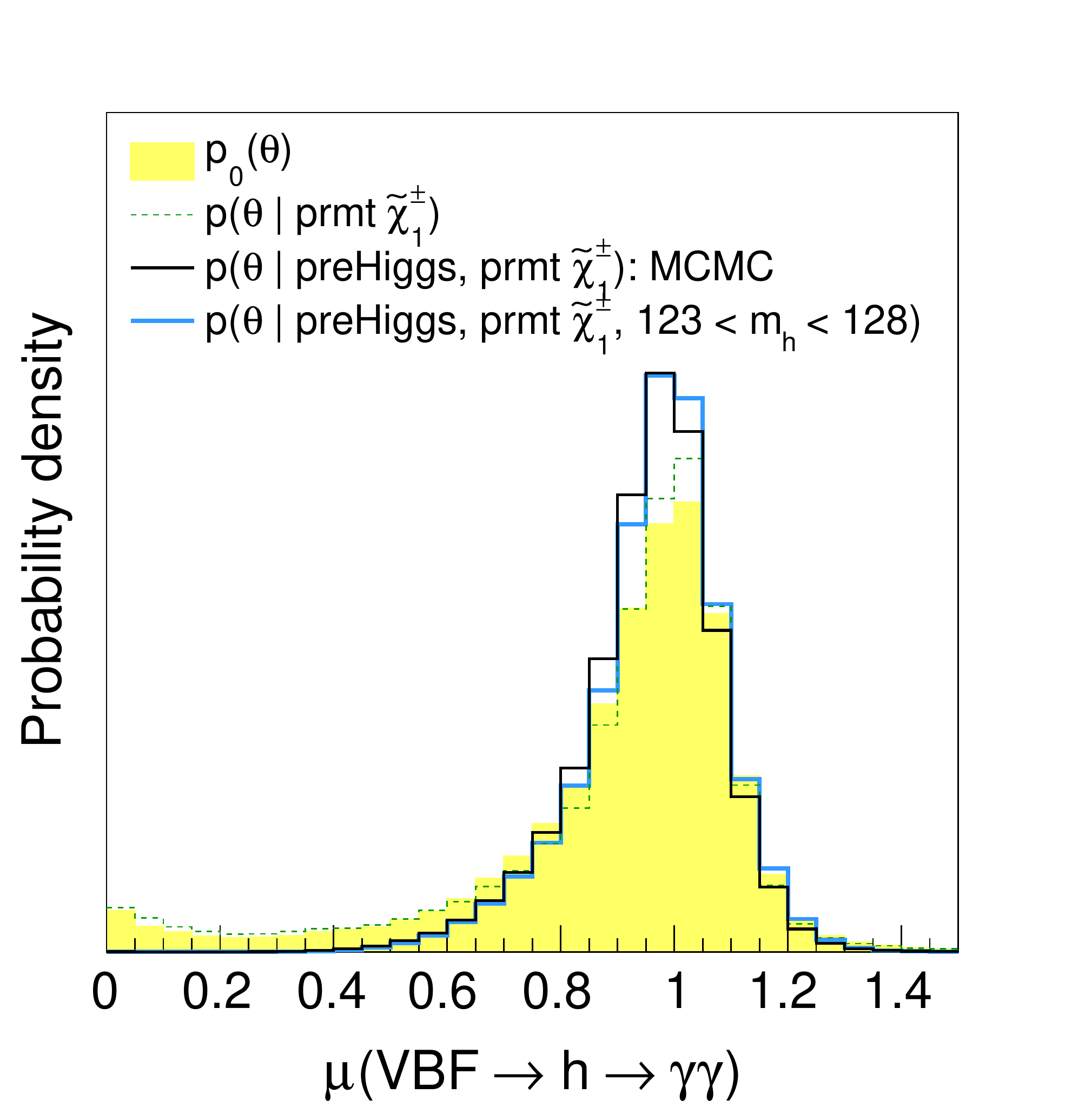}
\includegraphics[width=0.3\linewidth]{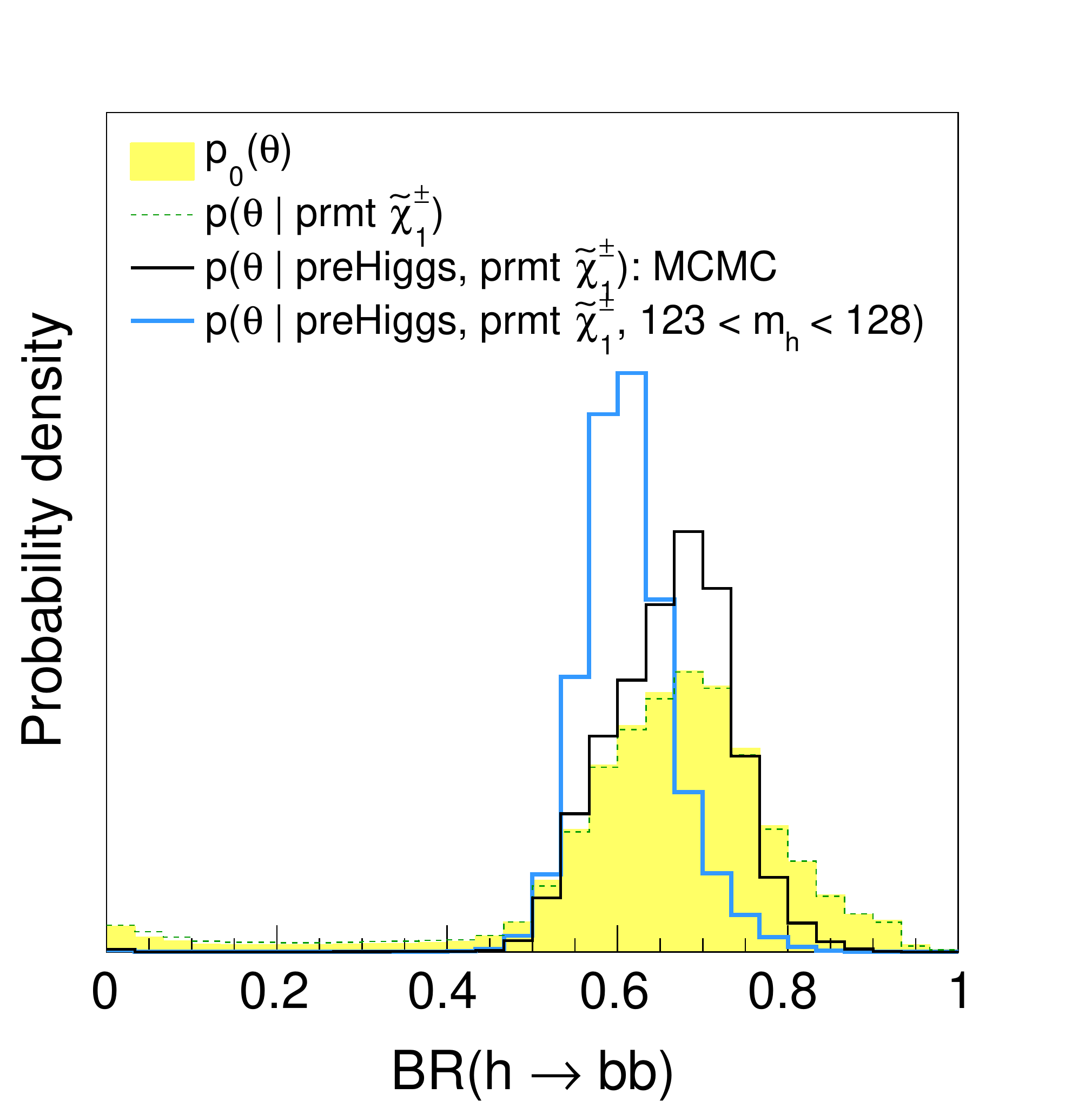}
\includegraphics[width=0.3\linewidth]{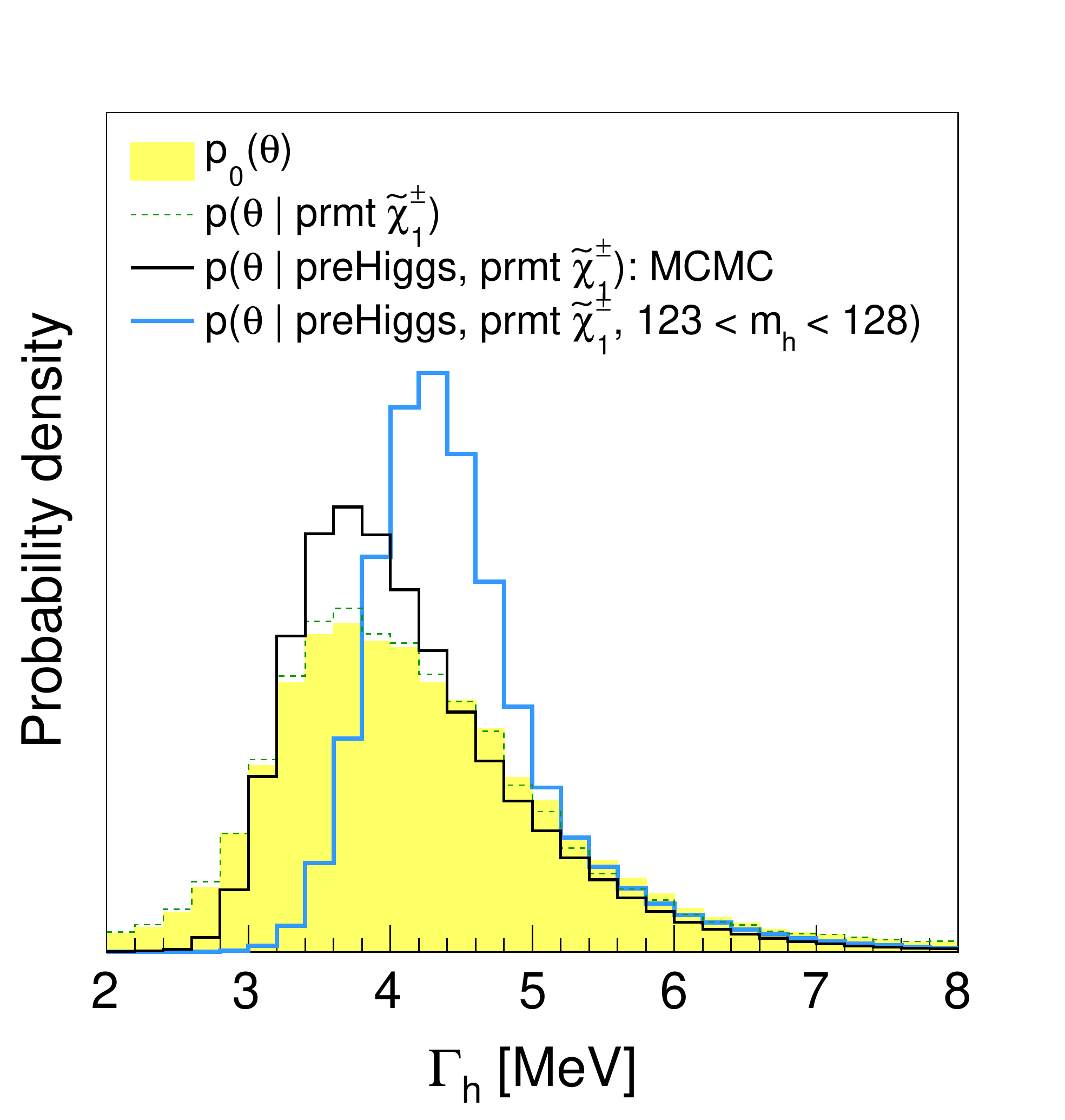}
\caption{Same as Fig.~\ref{pmssm-fig:sampling1} but for selected $h$ signal strengths, $\br(h\to b\bar b)$ and the total decay width $\Gamma_h$. The VBF distributions look practically the same as the ggF distributions, as exemplified for the 
${\rm VBF}\to h\to \gamma\gamma$ case, though they show a slightly larger effect from requiring $m_h=123-128$~GeV than the ggF distributions. 
}
\label{pmssm-fig:sampling3}
\end{center}
\end{figure}

It is also interesting to consider the $h$ signal at this level. Some relevant distributions are shown in Fig.~\ref{pmssm-fig:sampling3}.  While generically the $h$ signal strength can go down to zero in the MSSM, already the ``preHiggs'' constraints eliminate very small values below $\mu\approx 0.6$ and narrow the signal strength distributions to a range of $\mu\approx 1\pm0.4$. This is coming from two different effects. First, in the low-$m_A$ region the heavier scalar $H$ can be more SM-like than $h$. Second, in the region where the LSP is light ($m_{\tilde \chi^0_1} \lesssim 65$~GeV) a large increase of the total width, resulting in  reduced signal strengths, is possible through $h \to \tilde \chi^0_1 \tilde \chi^0_1$.
The low-$m_A$ region is mostly disfavored from flavor constraints, while a light neutralino---if mainly wino or higgsino---is excluded by the LEP bound on charginos. In both cases, requiring $m_h=123-128$~GeV only has a very small additional effect.

One might expect that the influence of the Higgs mass is larger in the ggF channels than in the VBF channels (because of the negative loop contribution from maximally mixed stops affecting the former) but, in fact, the effect is very small and goes in the opposite direction,  as can be seen by comparing the top-left and the bottom-left plots in Fig.~\ref{pmssm-fig:sampling3}. 
The observables which are really influenced by the Higgs mass are  the branching fraction into $b\bar b$, which becomes  centered around $\br(h\to b\bar b)\approx 0.6$, and the $h$ total width, for which the most likely value is shifted a bit upwards to $\Gamma_h\approx4$--5~MeV. However, this is not really a SUSY effect: the same happens for the SM Higgs when going from $m_H\lesssim 120$~GeV to $m_H\approx 125$~GeV.

\subsubsection{Impact of Higgs signal strengths}\label{pmssm-sec:BluePlots}

\begin{figure}[htbp]
\begin{center}
\includegraphics[width=0.24\linewidth]{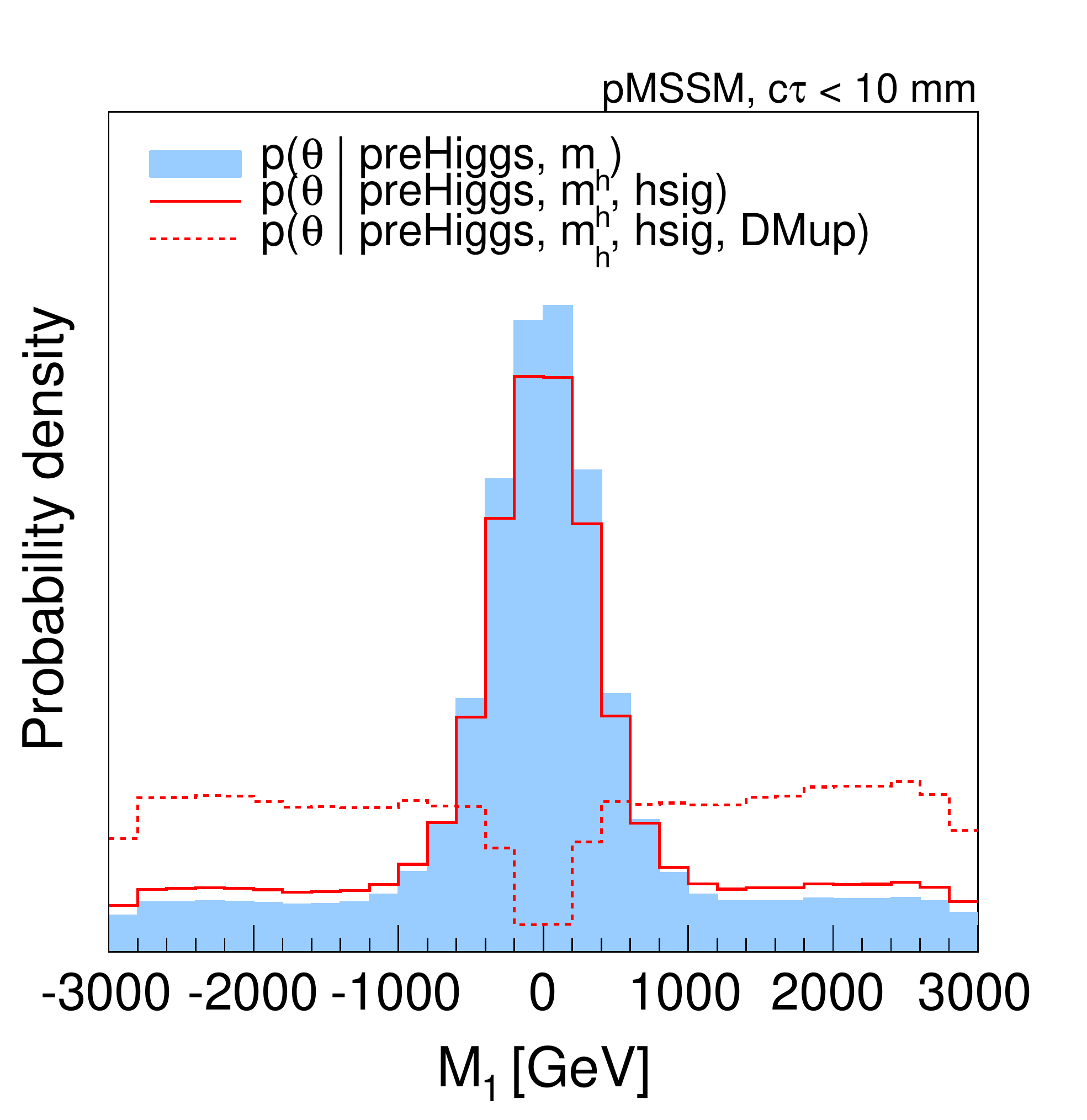}
\includegraphics[width=0.24\linewidth]{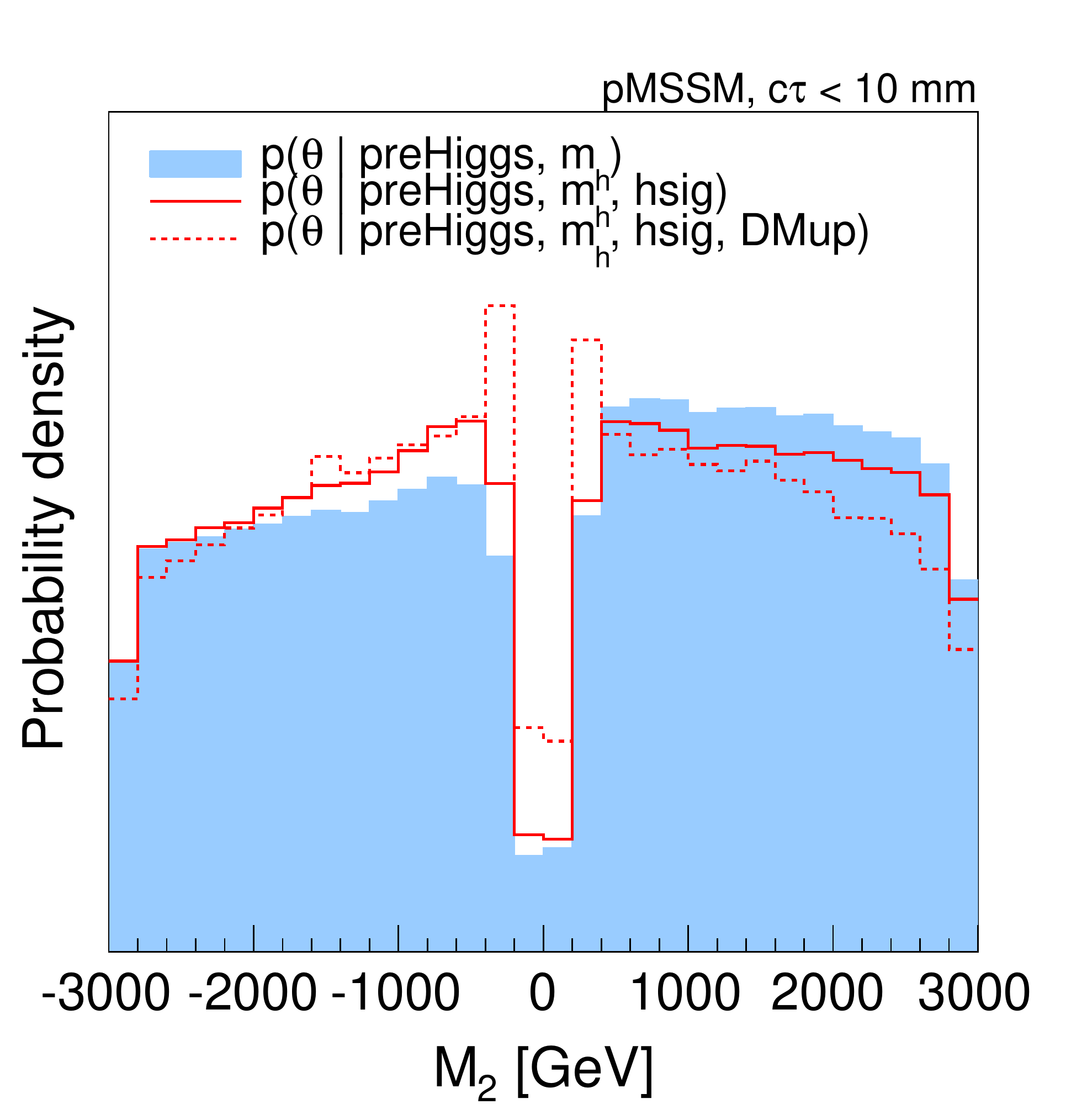}
\includegraphics[width=0.24\linewidth]{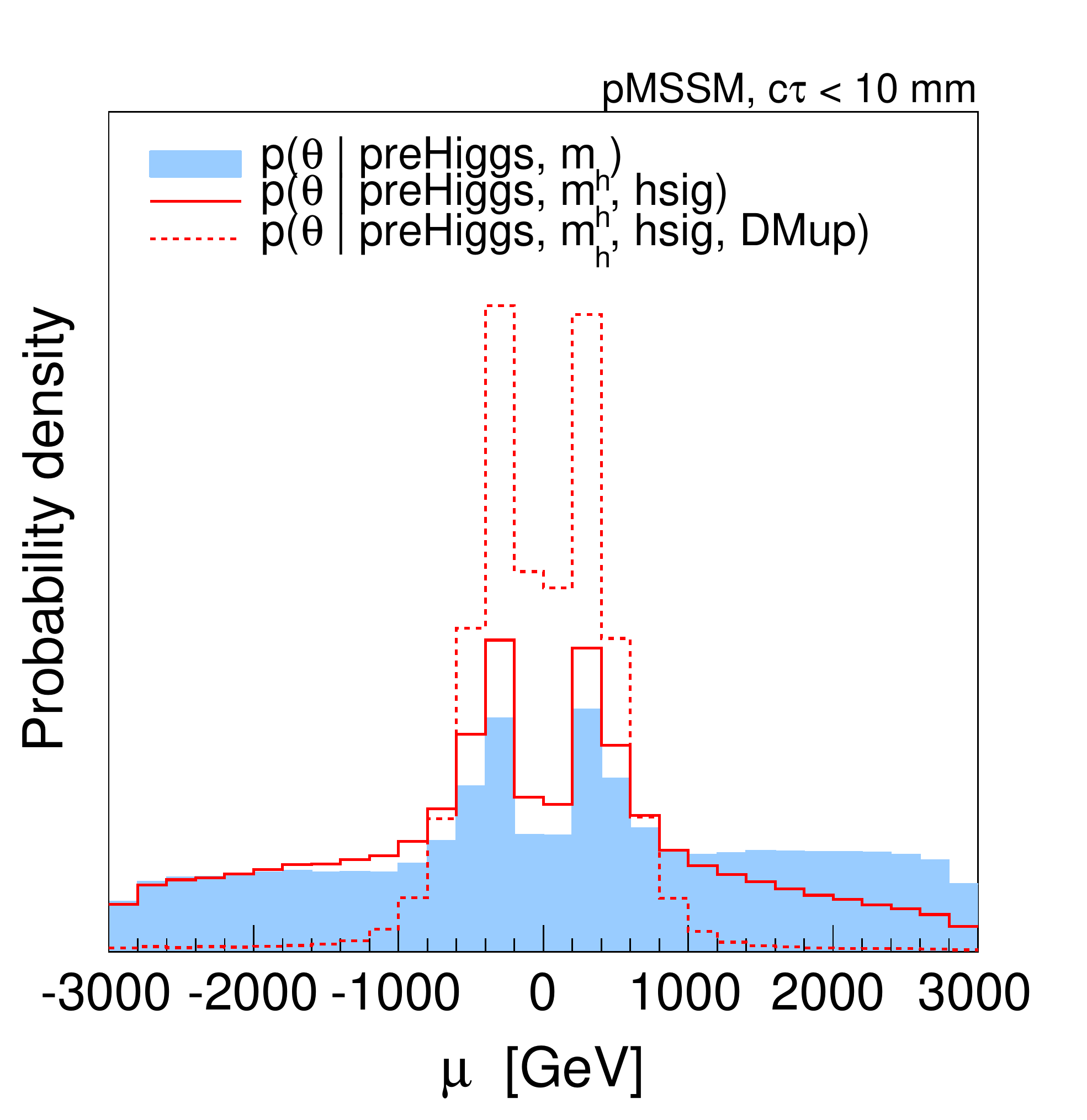} 
\includegraphics[width=0.24\linewidth]{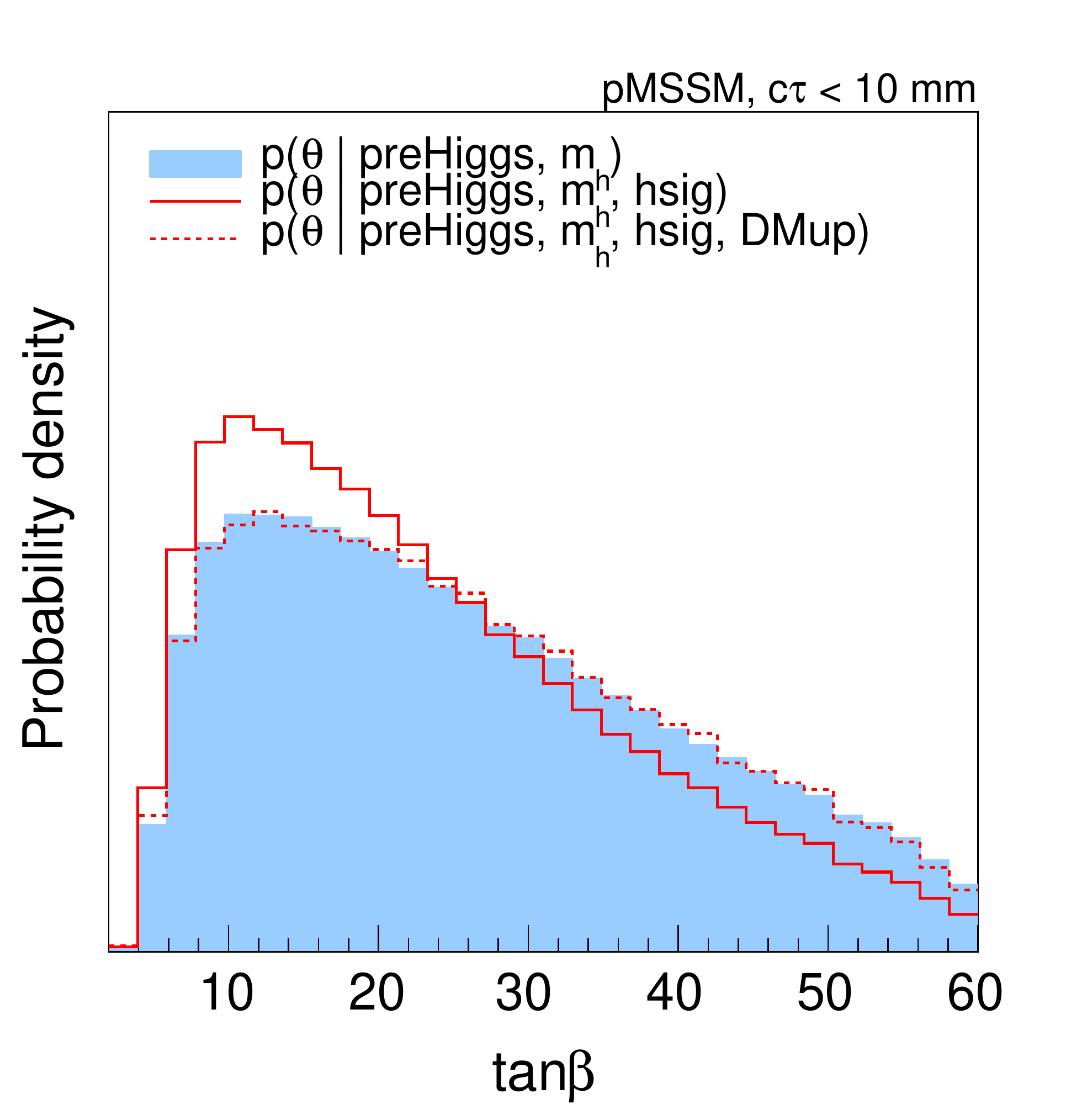}
\includegraphics[width=0.24\linewidth]{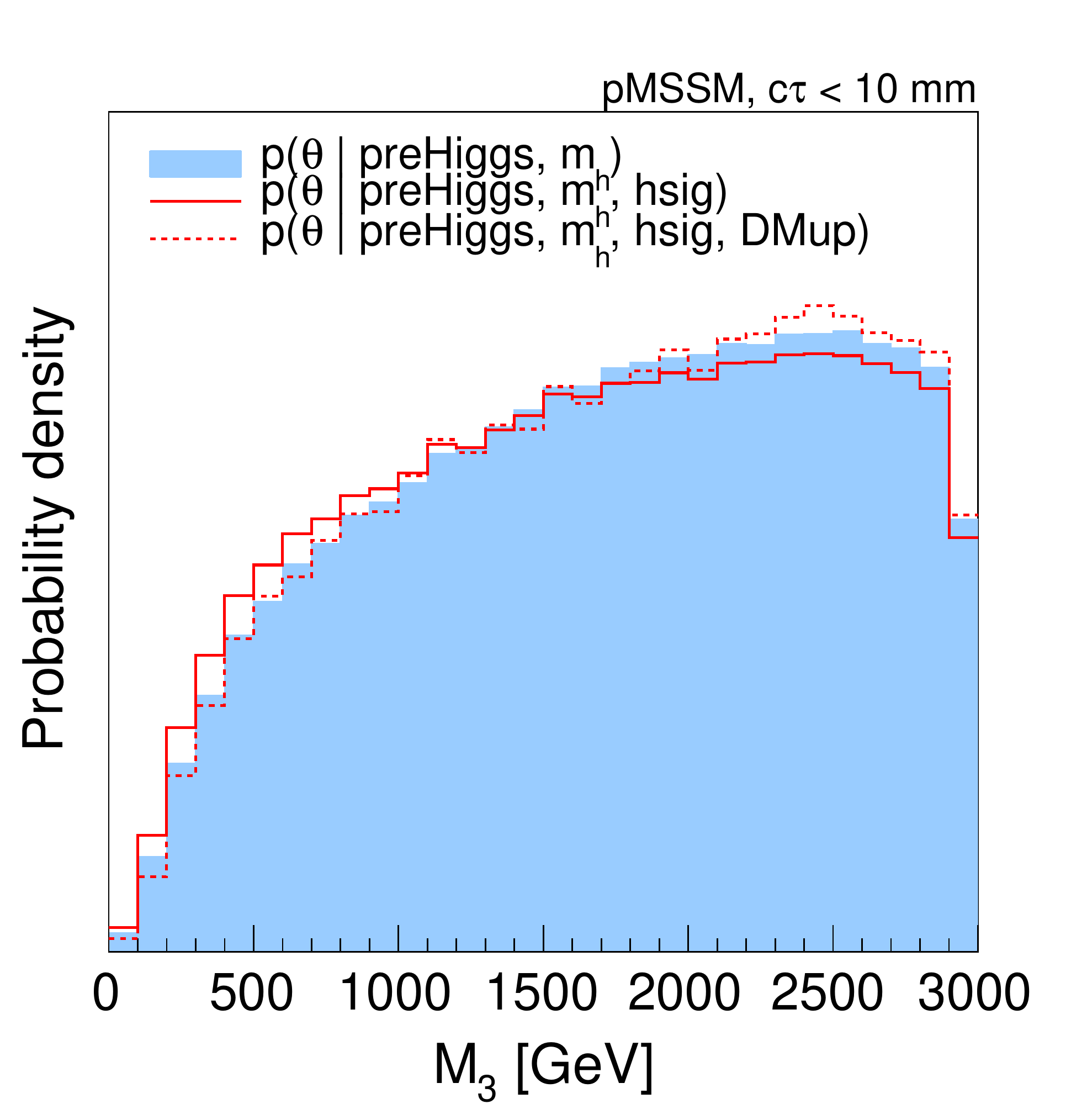}
\includegraphics[width=0.24\linewidth]{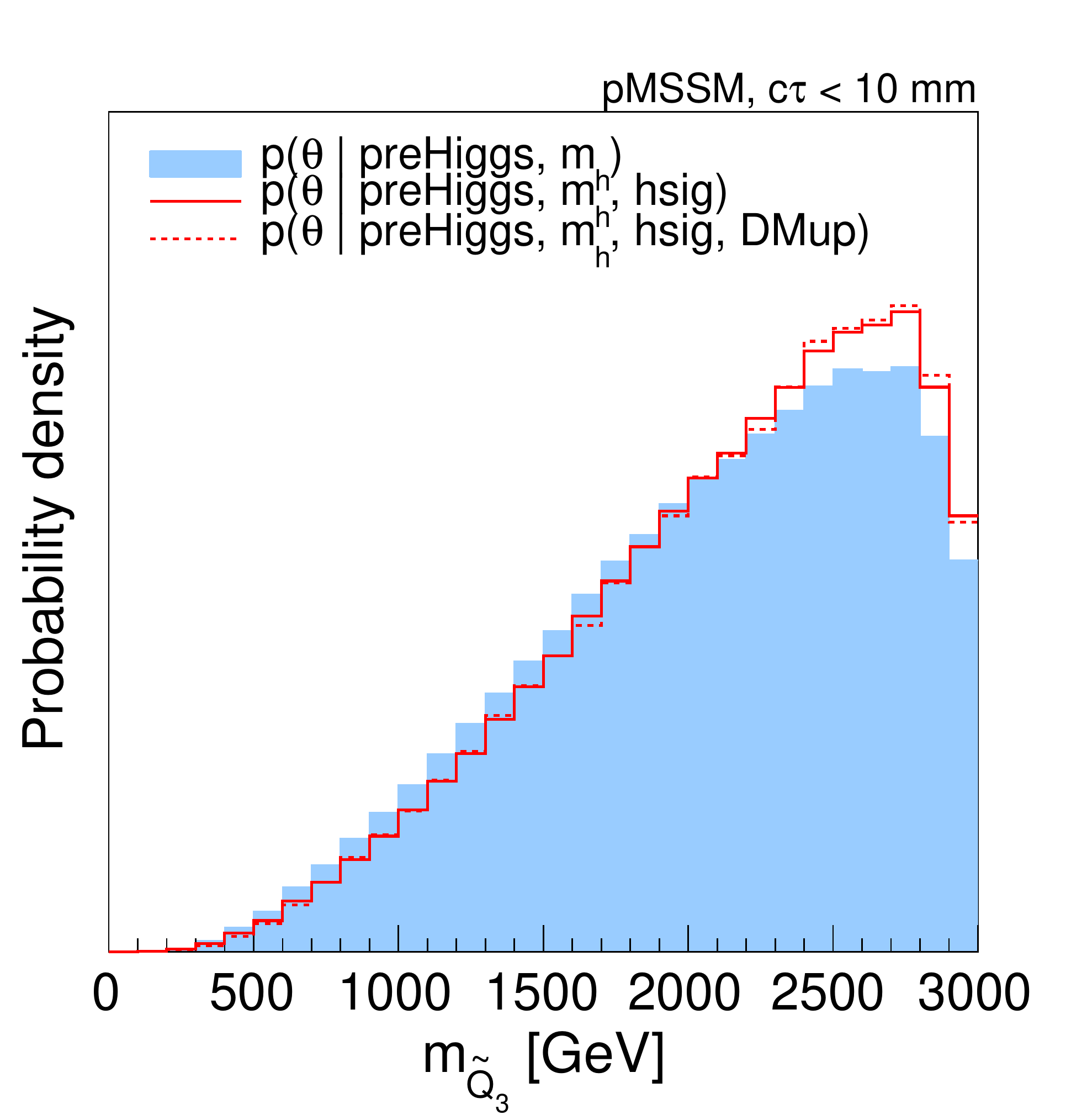}
\includegraphics[width=0.24\linewidth]{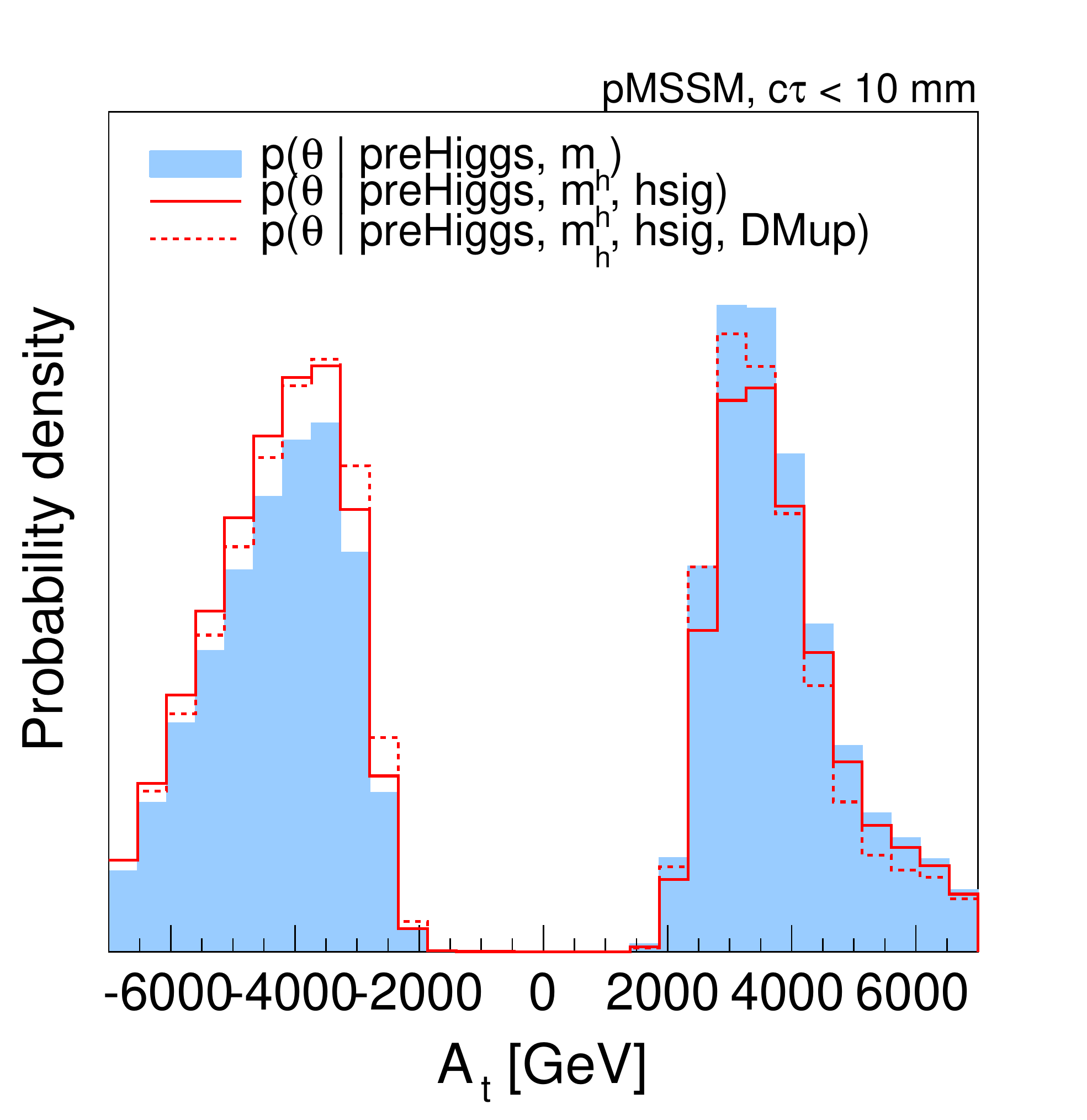}
\includegraphics[width=0.25\linewidth]{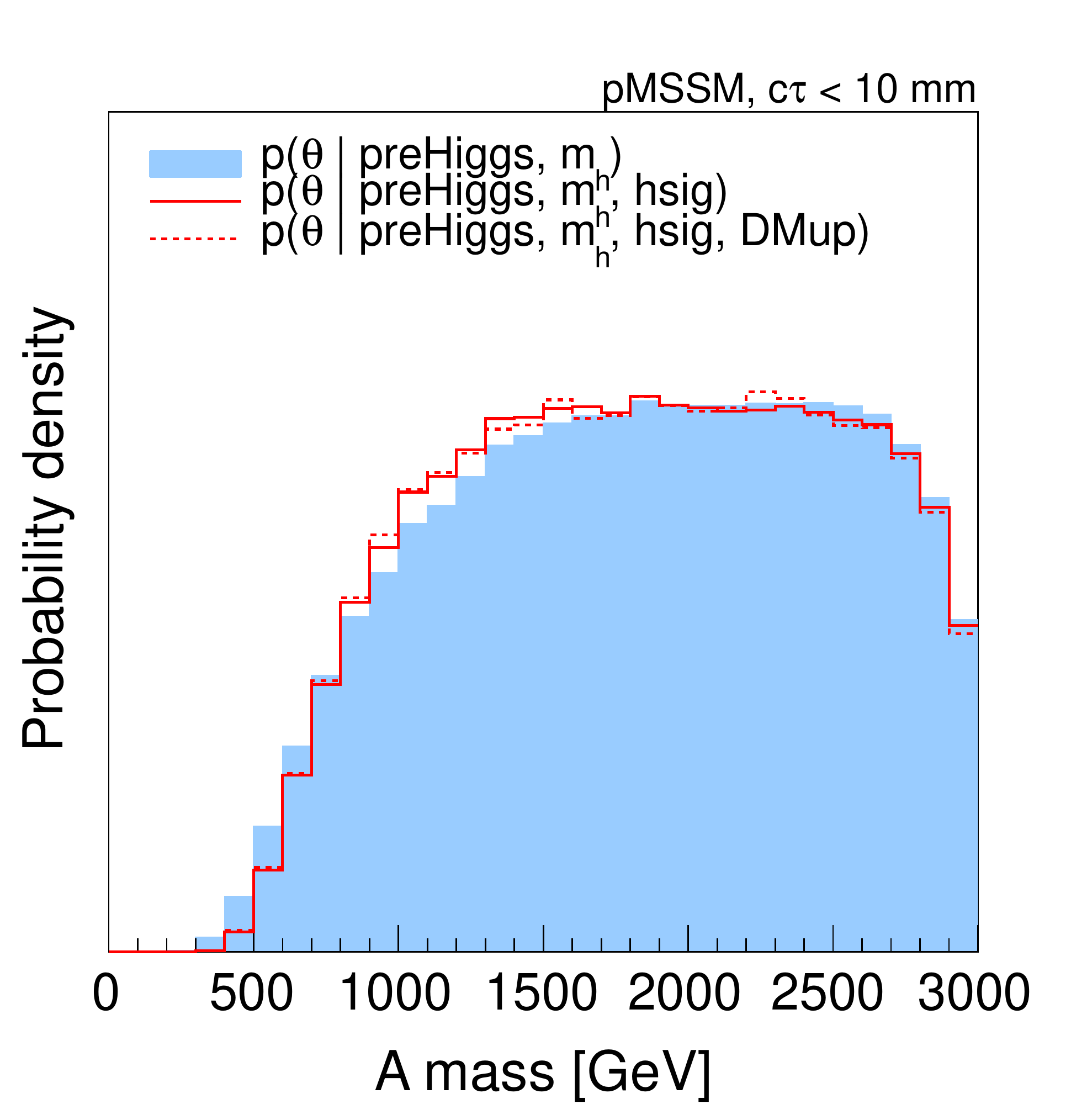}
\includegraphics[width=0.24\linewidth]{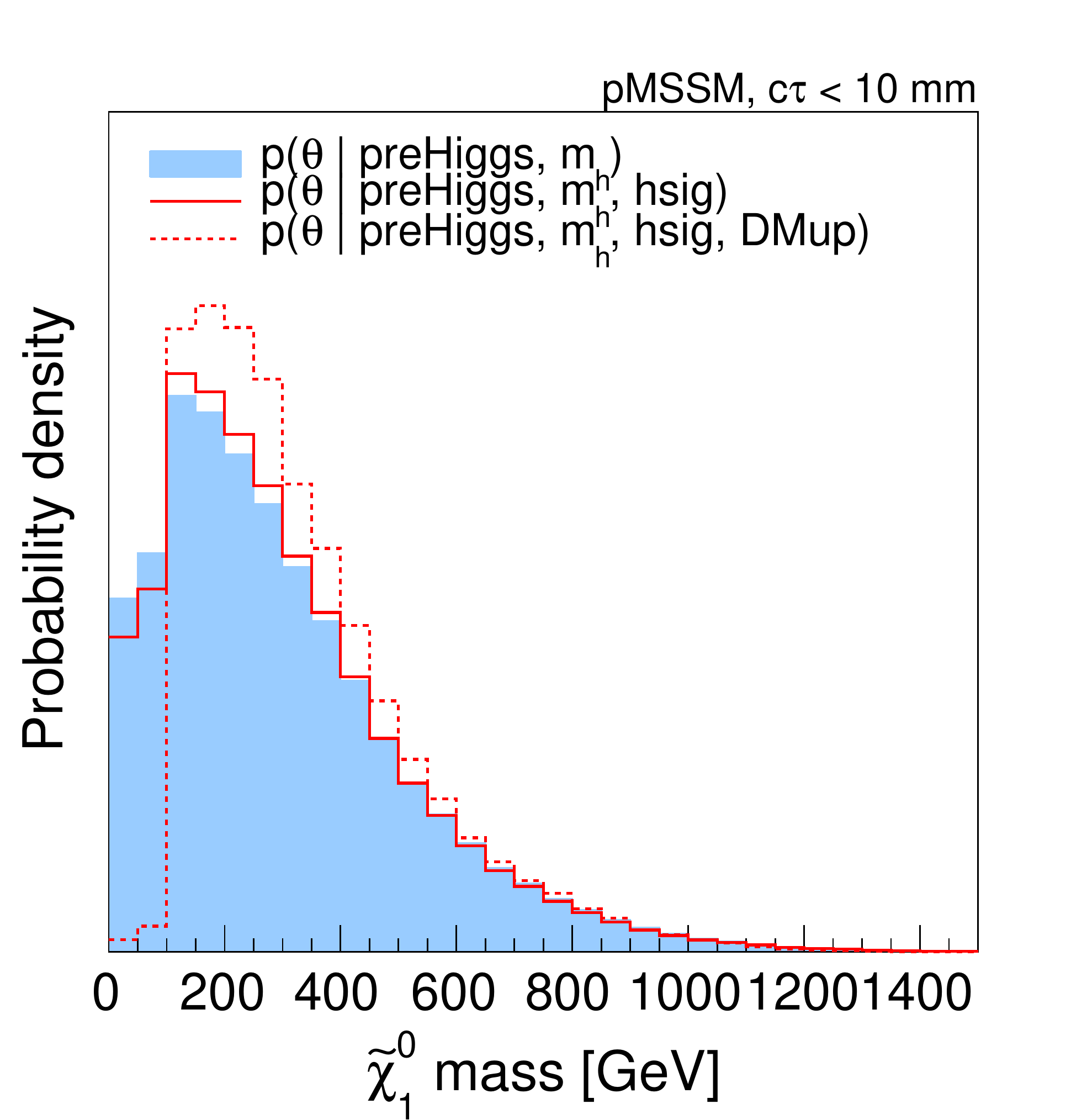}
\includegraphics[width=0.24\linewidth]{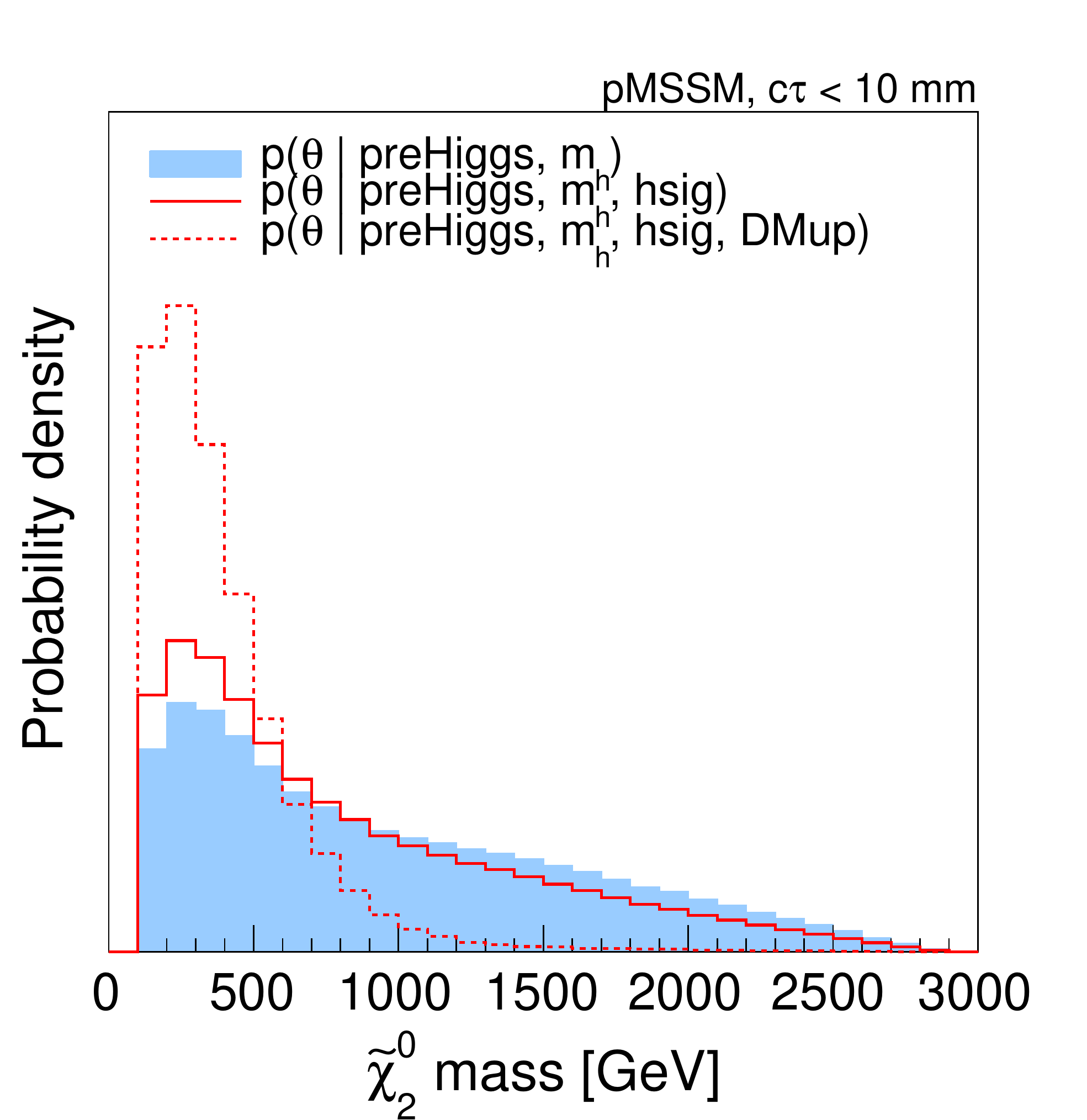}
\includegraphics[width=0.24\linewidth]{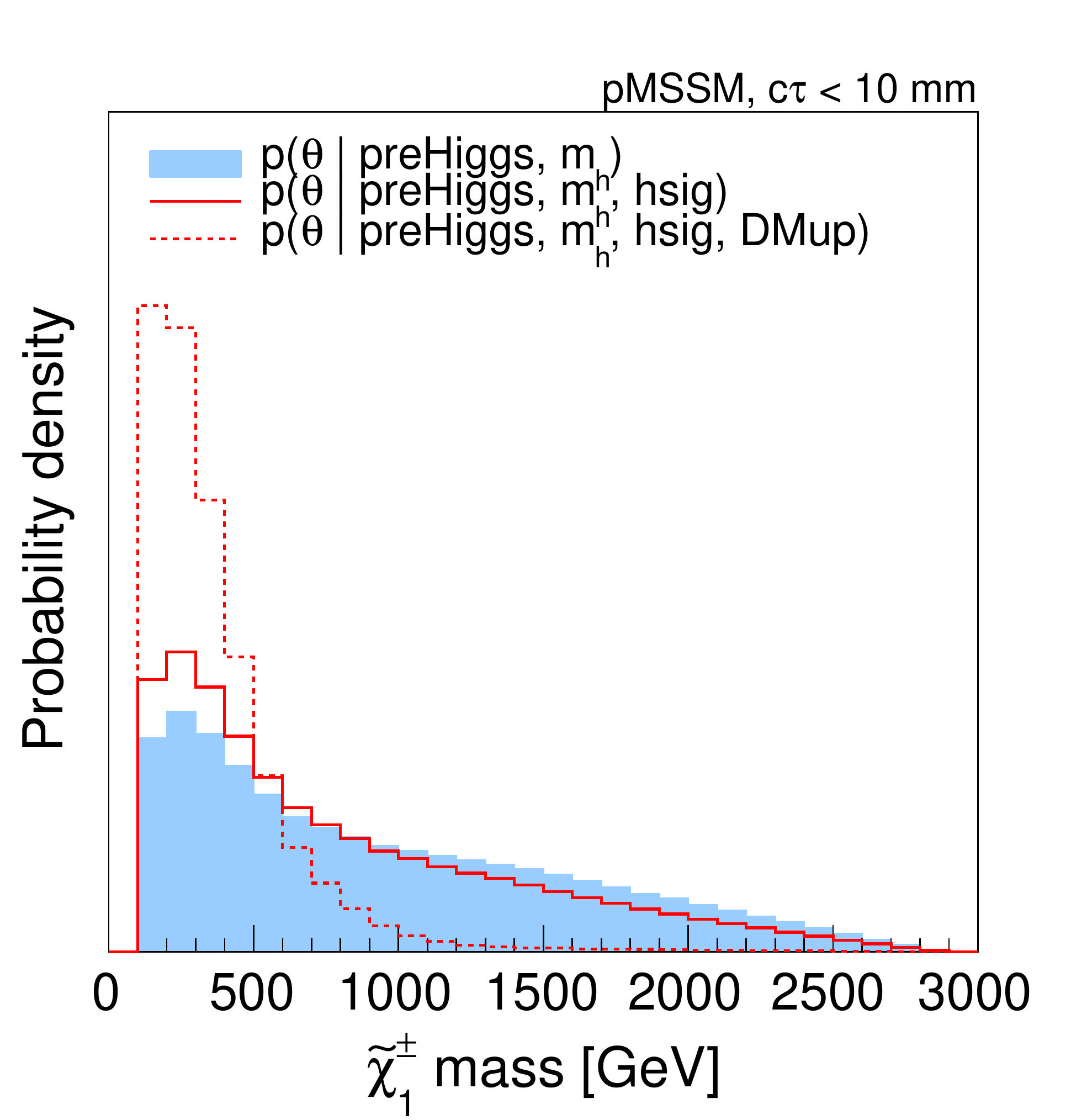}
\includegraphics[width=0.24\linewidth]{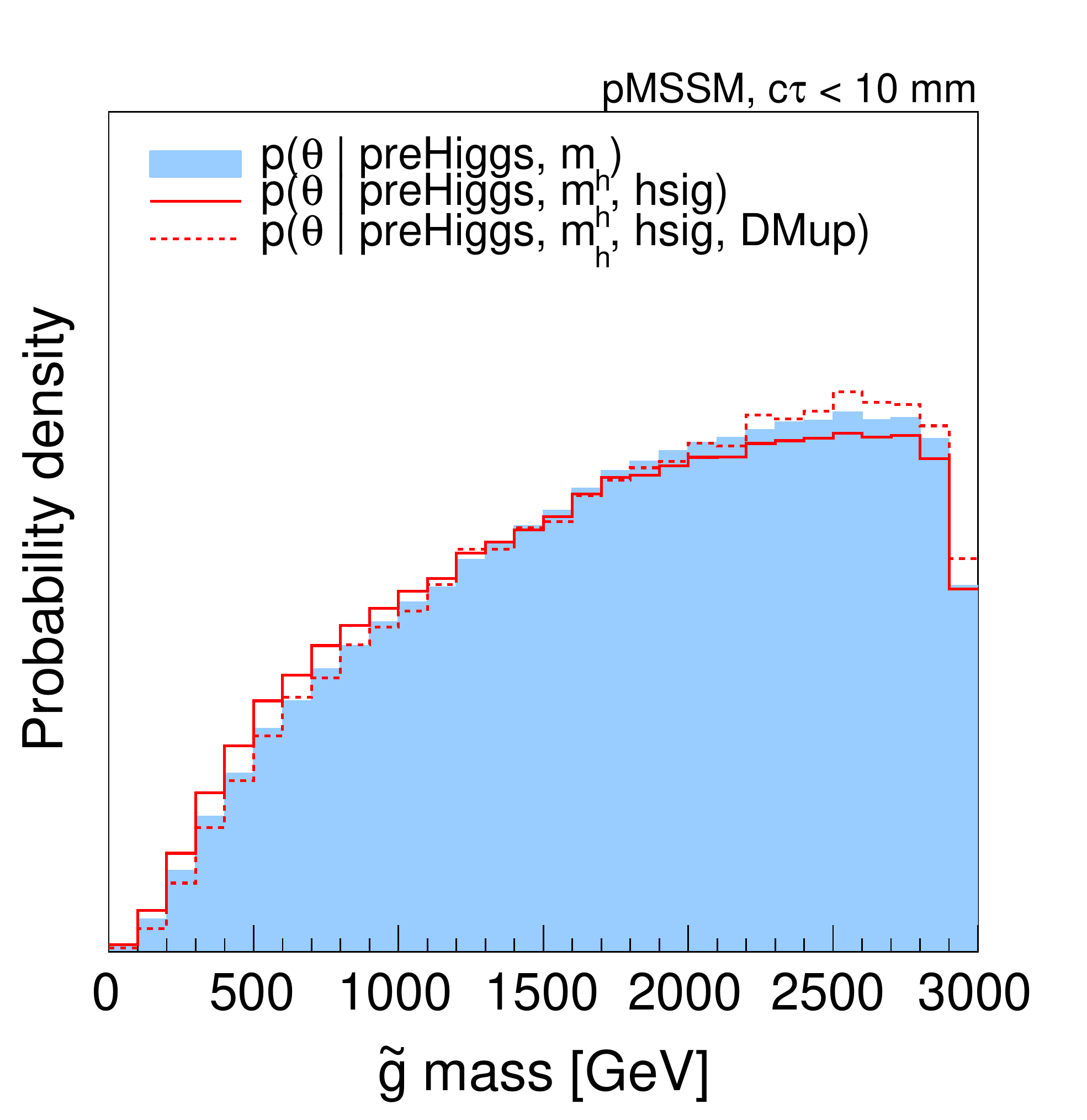}
\includegraphics[width=0.24\linewidth]{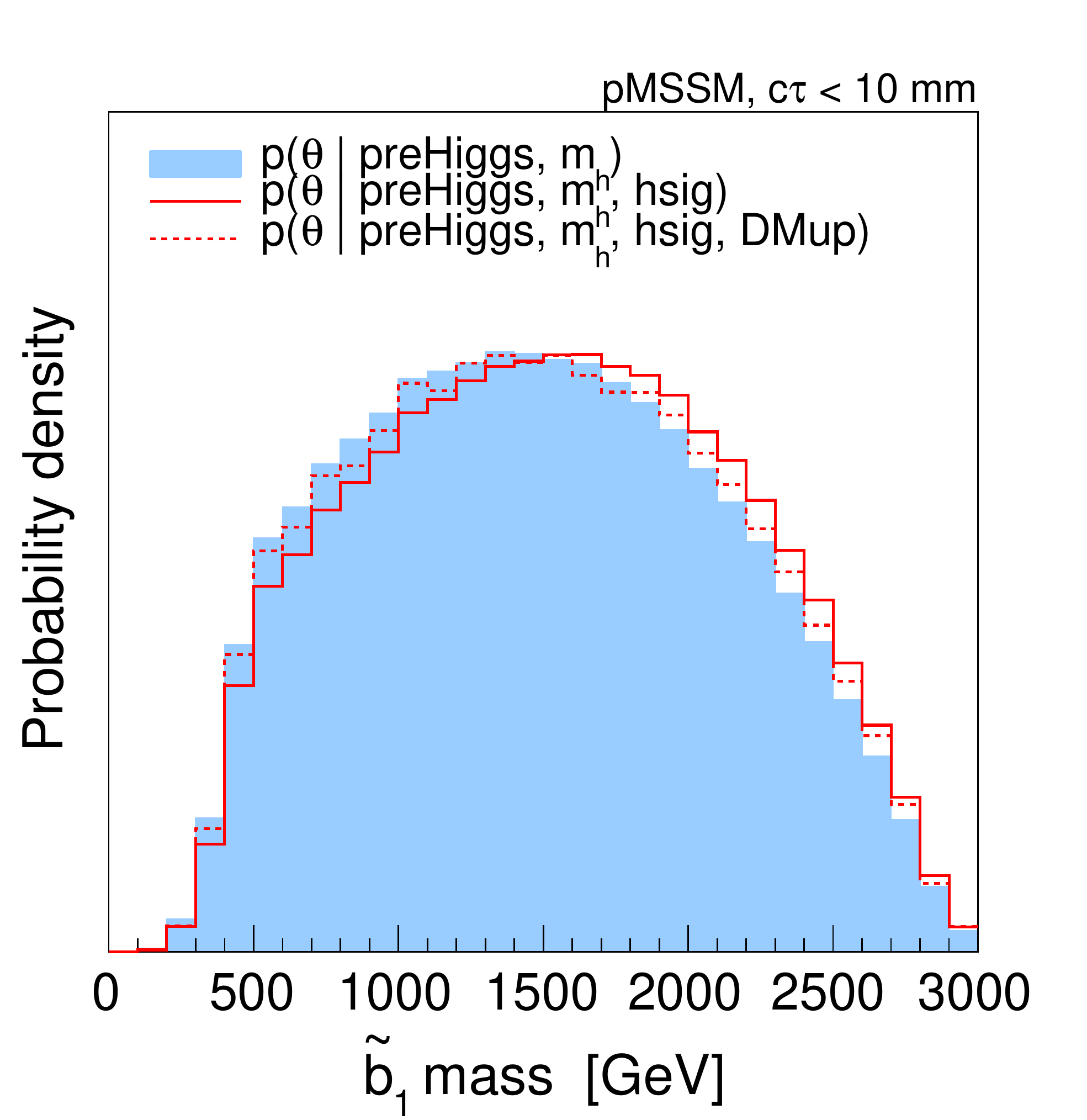}
\includegraphics[width=0.24\linewidth]{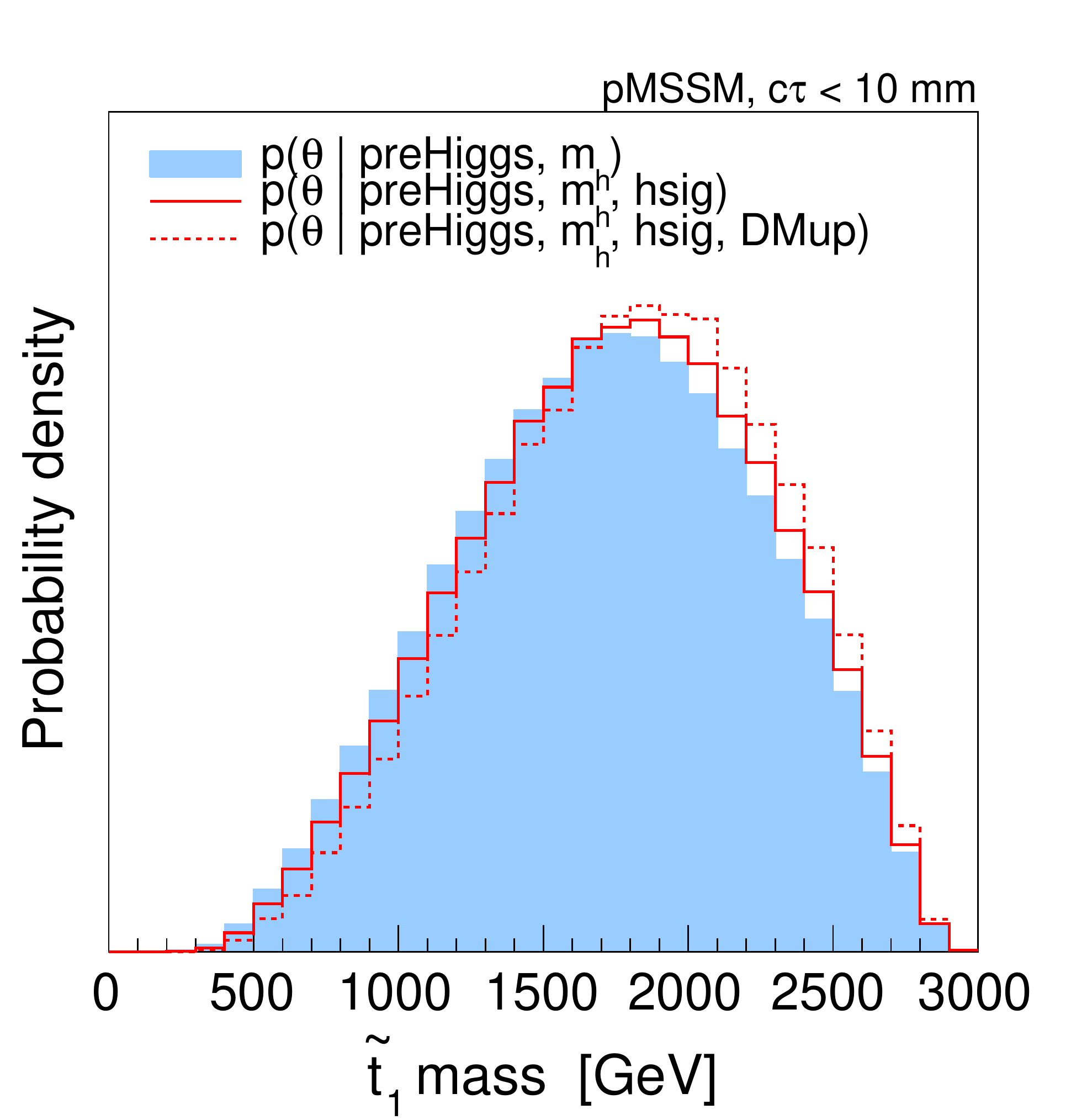}
\includegraphics[width=0.24\linewidth]{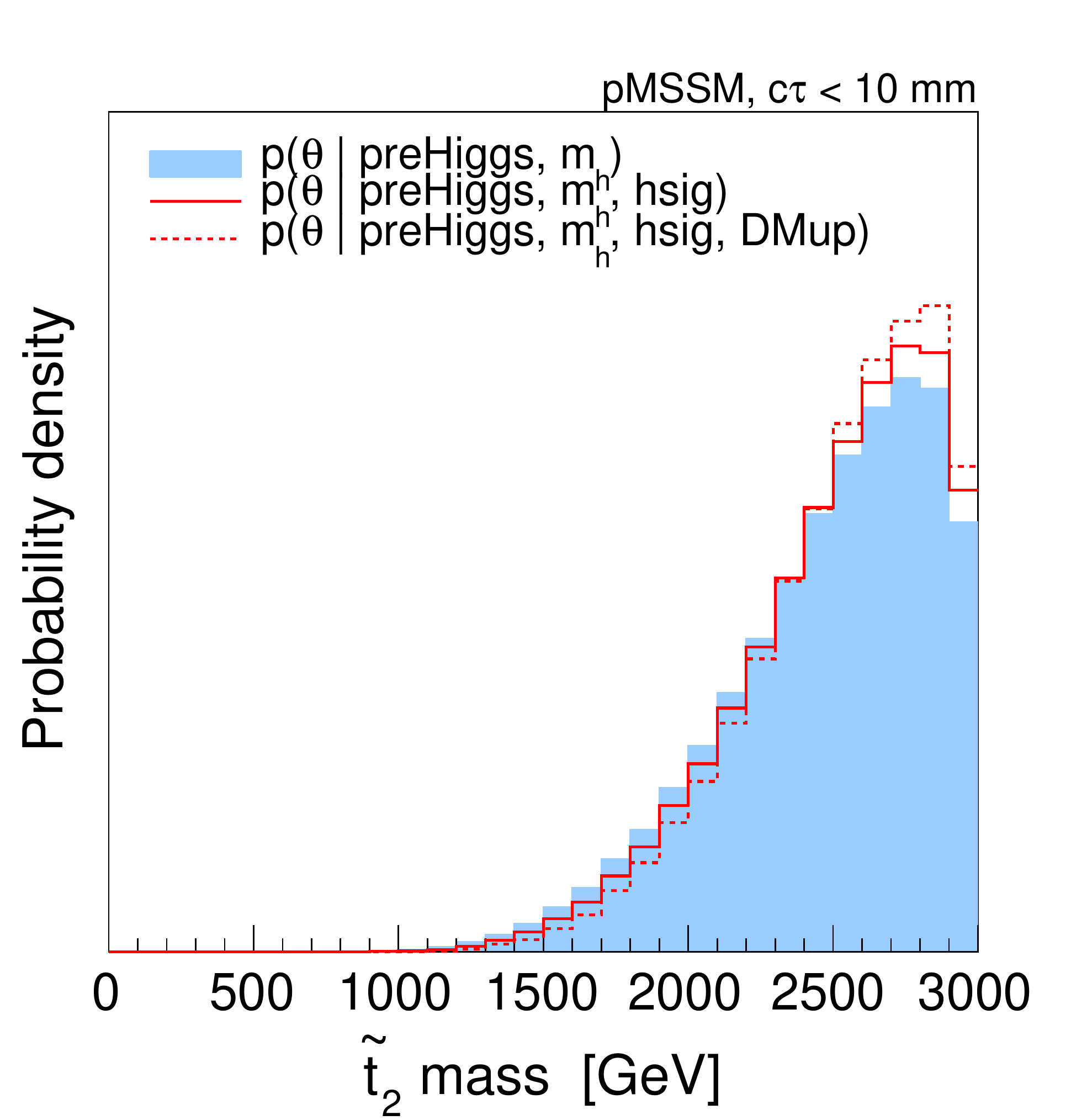}
\includegraphics[width=0.24\linewidth]{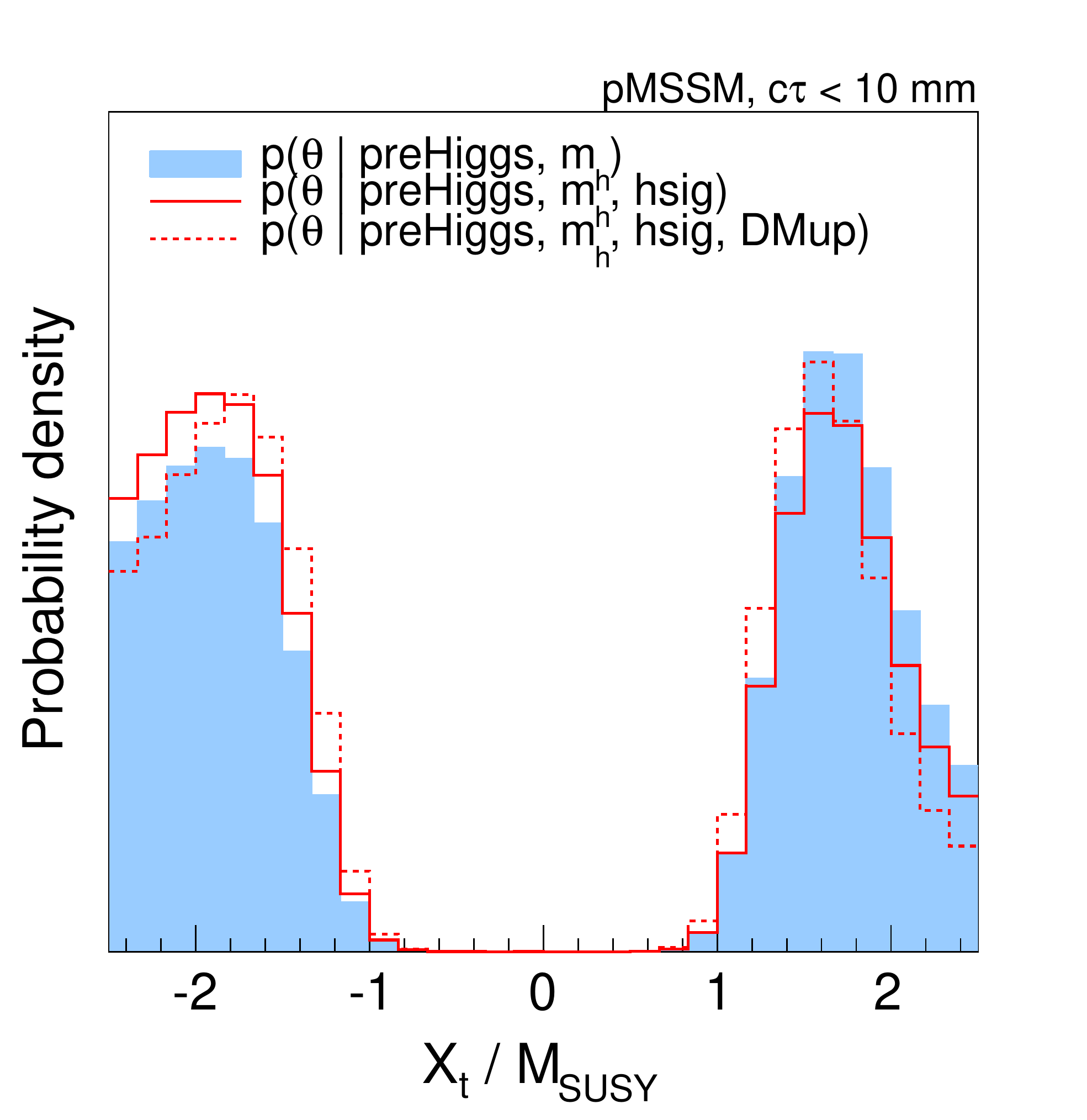}  
\caption{Marginalized 1D posterior densities for selected parameters and masses, showing the effect of the 
Higgs signal strength measurements. 
The light blue histograms show the distributions  based on the ``preHiggs'' measurements 
of Table~\ref{pmssm-tab:preHiggs} plus requiring in addition $m_h\in [123,\,128]$~GeV.
The solid red lines, labelled ``hsig'', are the distributions when moreover taking into account 
the measured Higgs signal strengths in the various channels. The limits from searches for the heavy Higgses ($H$ and $A$) 
are also included in the red line-histograms, but have a totally negligible effect. 
The dashed red lines, labelled ``DMup'', include in addition an upper limit on the 
neutralino relic density and the recent direct DM detection limit from LUX as explained in the text.}
\label{pmssm-fig:likehiggs1}
\end{center}
\end{figure}

As the next step, we include in addition the detailed properties of the $h$ signal in the computation of the likelihood as outlined in Section~\ref{pmssm-sec:higgslikeli}. 
The effects of the Higgs observations on the pMSSM parameters and on the particle masses are shown in Fig.~\ref{pmssm-fig:likehiggs1}.  
In these plots, the light blue histograms show the distributions  based on the ``preHiggs'' measurements 
of Table~\ref{pmssm-tab:preHiggs} plus requiring in addition $m_h\in [123,\,128]$~GeV, \ie\ they correspond 
to the blue line-histograms of Fig.~\ref{pmssm-fig:sampling1}.
The solid red lines are the distributions when moreover taking into account the measured Higgs signal strengths 
in the various channels as outlined in Section~2. Note that the limits from the MSSM $H,A\to \tau\tau$ searches,  
which are also included in the red line-histograms, have a negligible effect. 
(For completeness, a plot of the $\tan\beta$ versus $m_A$ plane is given in Fig.~\ref{pmssm-fig:heavierHiggses2D}.)
Finally, the dashed red lines also take into account upper limits from the DM relic density 
and direct DM searches, as explained in Section~\ref{pmssm-sec:dmconst}.

Let us first discuss the effect of the Higgs measurements, \ie\ consider the solid red lines only. 
We observe a significant preference for small or negative 
$\mu$ and smaller $\tan\beta$ values when including the Higgs signal strength likelihood. 
The main reason is the $\mu\tan\beta$ correction to the bottom Yukawa coupling~\cite{Carena:1999py,Eberl:1999he}, which for large $\tan\beta$ and large positive (negative) $\mu$ enhances (reduces) $\Gamma(h\to b\bar b)$ and the total $h$ width, hence reducing (increasing) all signal strengths except $\mu(Vh \to b\bar{b})$.
The preference for negative $\mu$ comes from the slight excess in the VBF and VH channels of $\gamma\gamma$ (mainly seen by ATLAS). In Section~\ref{sec:higgs2013}, $\mu({\rm VBF+VH}, \gamma\gamma) = 1.72 \pm 0.59$ is found, while other combined signal strengths are fully compatible with 1 at 68\%~CL. An overall excess (negative $\mu$) is therefore preferred over a general deficit (positive $\mu$). To a good approximation, the correction to the bottom Yukawa coupling is given by
\begin{eqnarray}
   \Delta_b \equiv \frac{\Delta m_b}{m_b} \simeq 
   \left[ \frac{2\alpha_s}{3\pi} \mu m_{\tilde{g}}\, I(m_{\tilde{g}}^2, m_{\tilde{b}_1}^2, m_{\tilde{b}_2}^2) +
            \frac{\lambda_t^2}{16\pi^2} A_t \mu \, I(\mu^2,m_{\tilde{t}_1}^2, m_{\tilde{t}_2}^2) \right] \tan\beta \,, \label{pmssm-Deltamb} 
\end{eqnarray}
where $I(x,y,z)$ is of order $1/{\rm max}(x,y,z)$~\cite{Djouadi:2005gj}. The shifts to higher values of all four stops and sbottoms masses and to lower values for the gluino mass also come from $\Delta_b$. In addition, negative values of $A_t$ are more likely after taking into account the Higgs likelihood. This comes from the second term of Eq.~(\ref{pmssm-Deltamb}): in order to compensate the first, dominant term, ${\rm sgn}(A_t \mu) = -{\rm sgn}(\mu)$ is required, hence a negative $A_t$.
The tree-level coupling $hbb$ also has an effect. It is given by
\beq
   g_{hbb} \simeq 1 - \frac{M_Z^2} {2 m_A^2} \sin 4\beta \tan \beta \,,
\eeq
for $m_A \gg M_Z$~\cite{Djouadi:2005gj}, and disfavors relatively light $A$ and $H$, with masses below about $700$~GeV (the effect from imposing the CMS $H,A \to \tau\tau$ limit is subdominant). Finally, $M_2$ shows a slight preference towards negative values. This is a direct consequence of the asymmetry in the distribution of $\mu$, since ${\rm sgn}(\mu M_2) > 0$ is required for $\Delta a^{\rm SUSY}_{\mu} > 0$ as suggested by the data. 

The DM constraints, on the other hand, have a dramatic effect on the bino and higgsino mass parameters and in turn on the chargino and neutralino masses. Since a mostly bino $\tilde \chi_1^0$  generically leads to a large $\Omega_{\tilde{\chi}^0_1}h^2$, low values of $M_1$ are strongly disfavored. 
The preferred solutions have a relevant higgsino or wino fraction of the LSP; 
therefore $\tilde\chi^\pm_1$ and $\tilde\chi^0_2$ masses  below about 1~TeV are strongly favored.  
At the same time, very light LSP masses below about 100~GeV are severely limited because of the LEP bound on the chargino mass. 
The preferred value of $\tan\beta$ is also affected; in fact, the preference for lower $\tanb$ coming from the Higgs signal strengths is removed by the DM constraints. The reason for this is an enhancement of $A$-funnel annihilation to comply with the upper limit on $\omhsq$.


\begin{figure}[t!]
\includegraphics[width=0.24\linewidth]{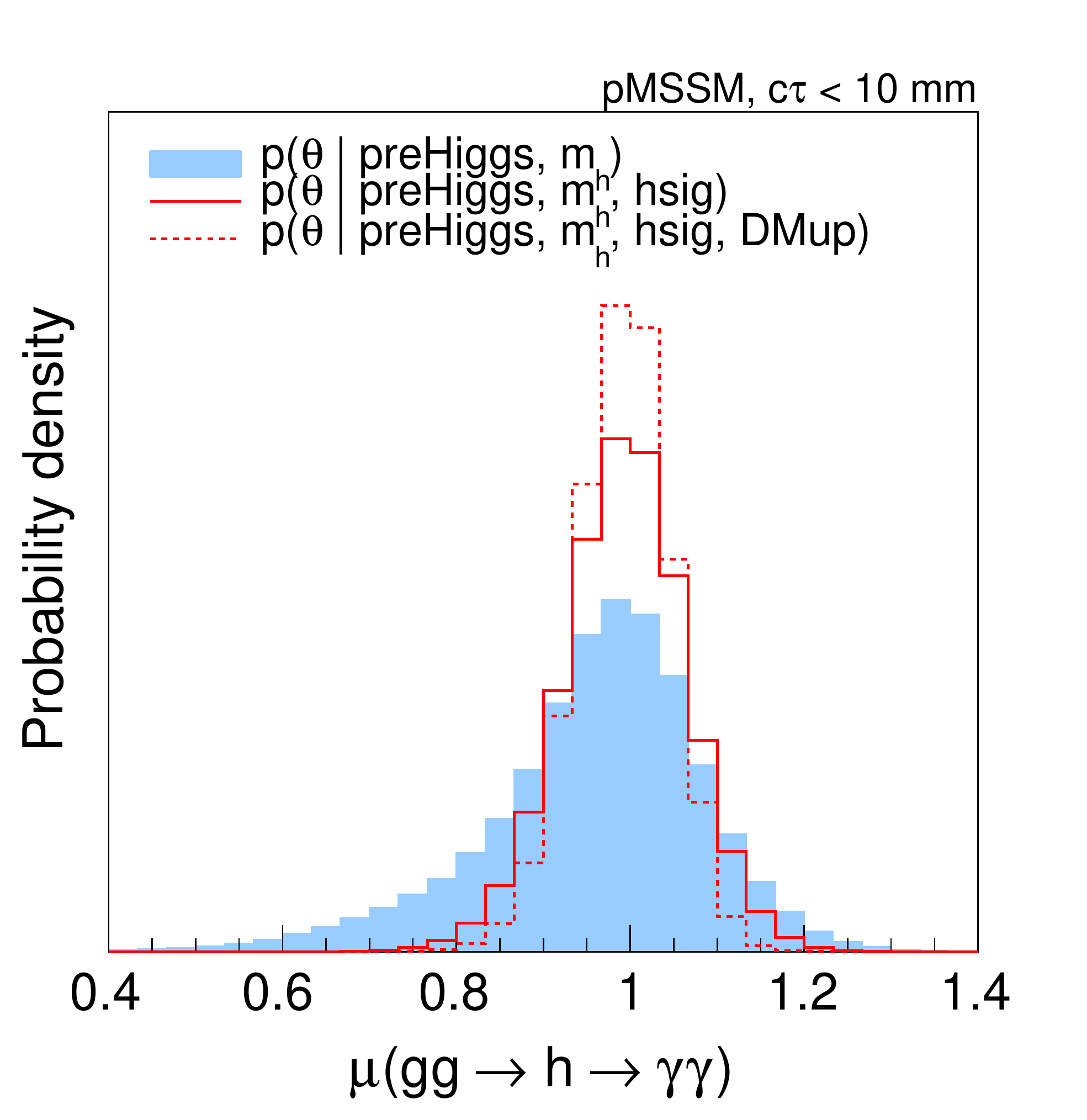}
\includegraphics[width=0.24\linewidth]{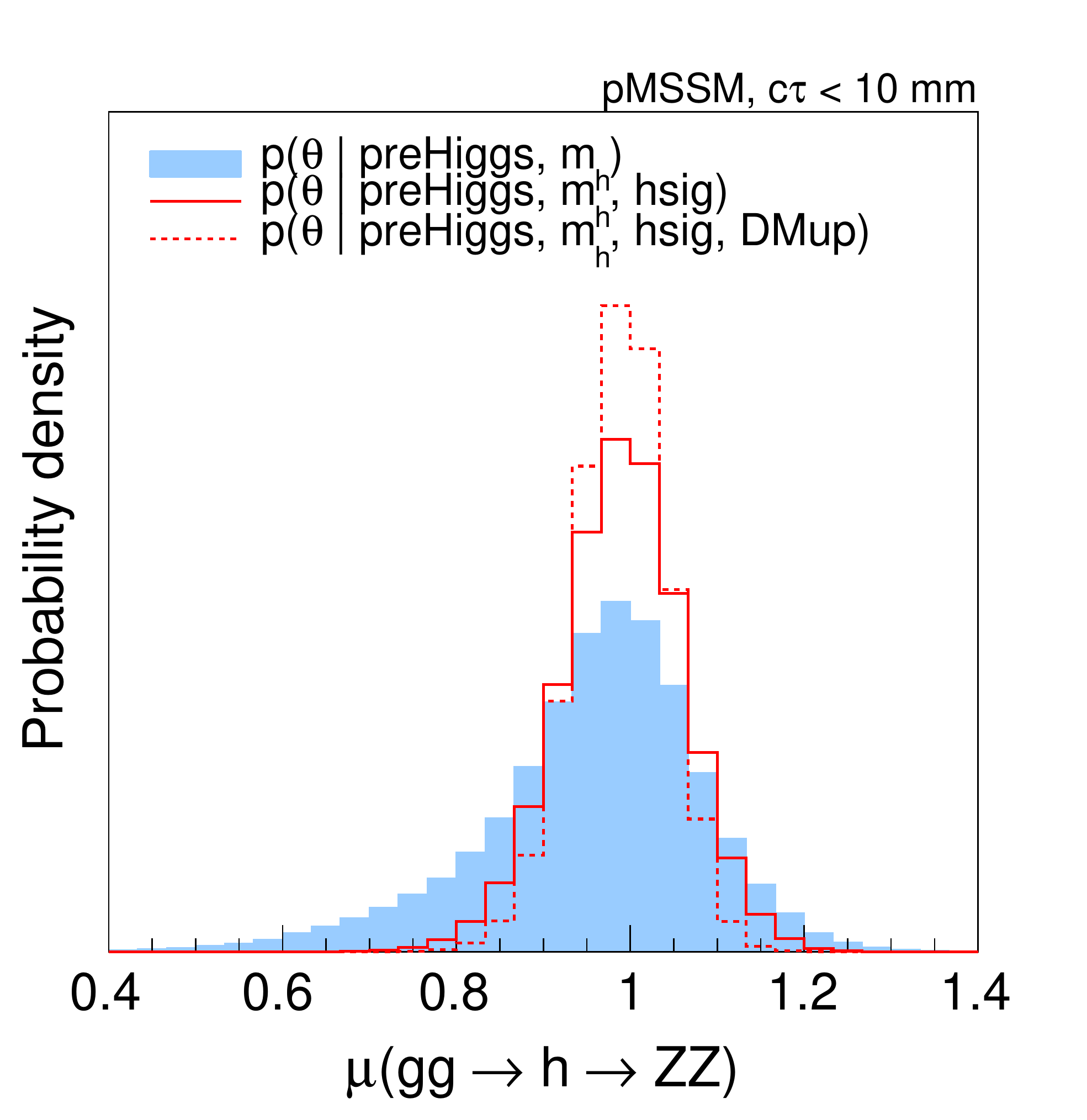}
\includegraphics[width=0.24\linewidth]{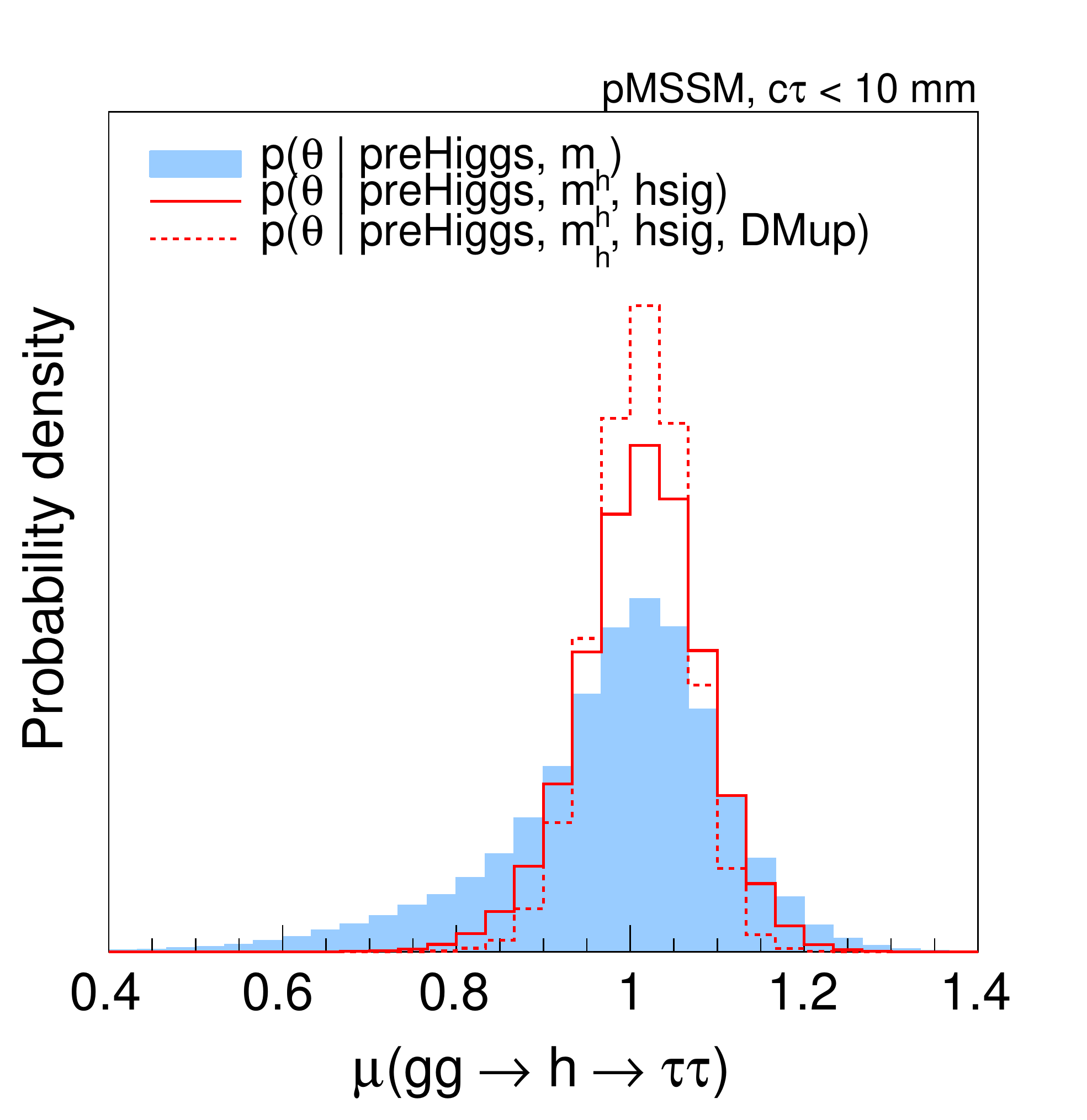}\\
\includegraphics[width=0.24\linewidth]{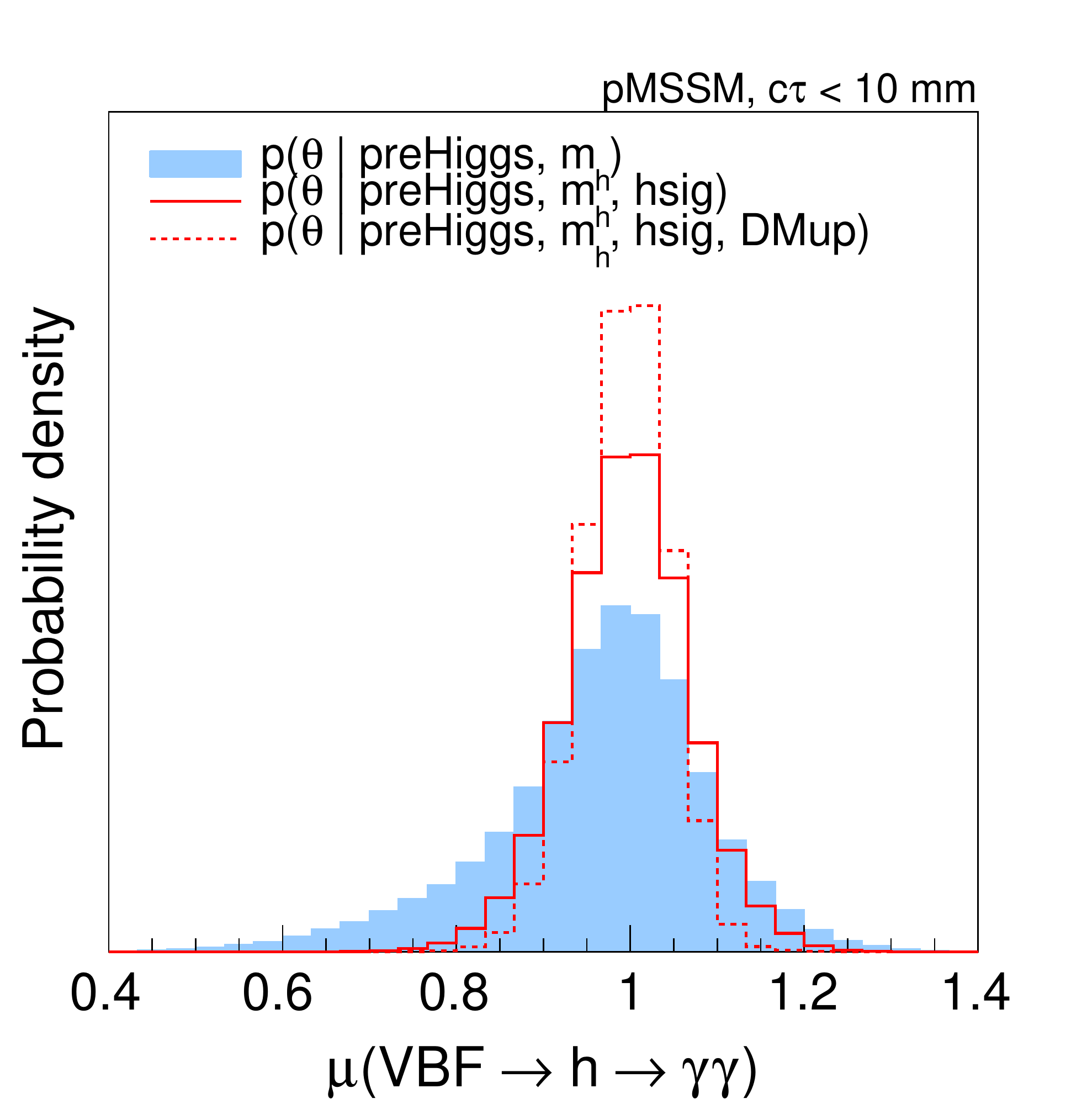}
\includegraphics[width=0.24\linewidth]{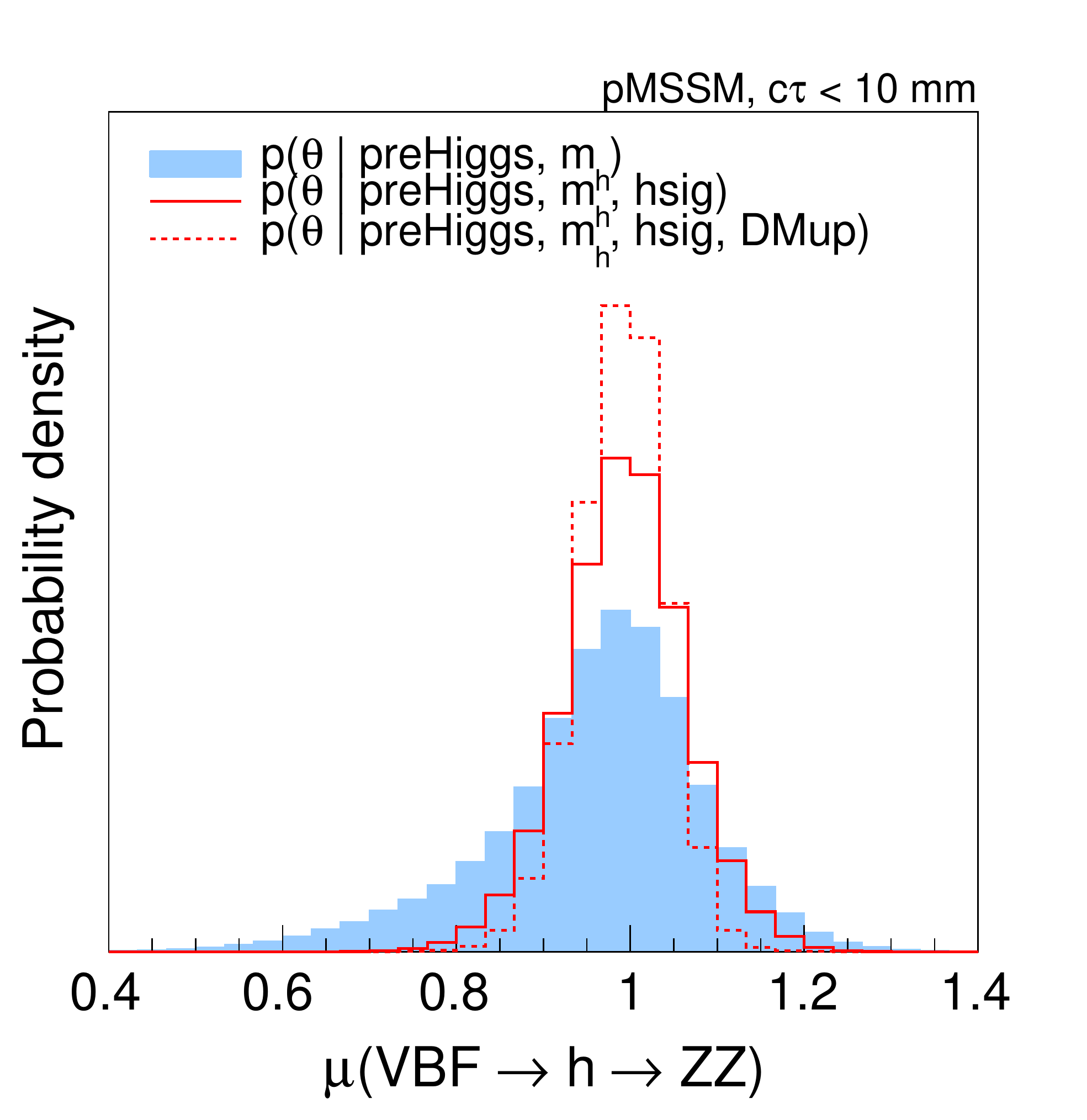}
\includegraphics[width=0.24\linewidth]{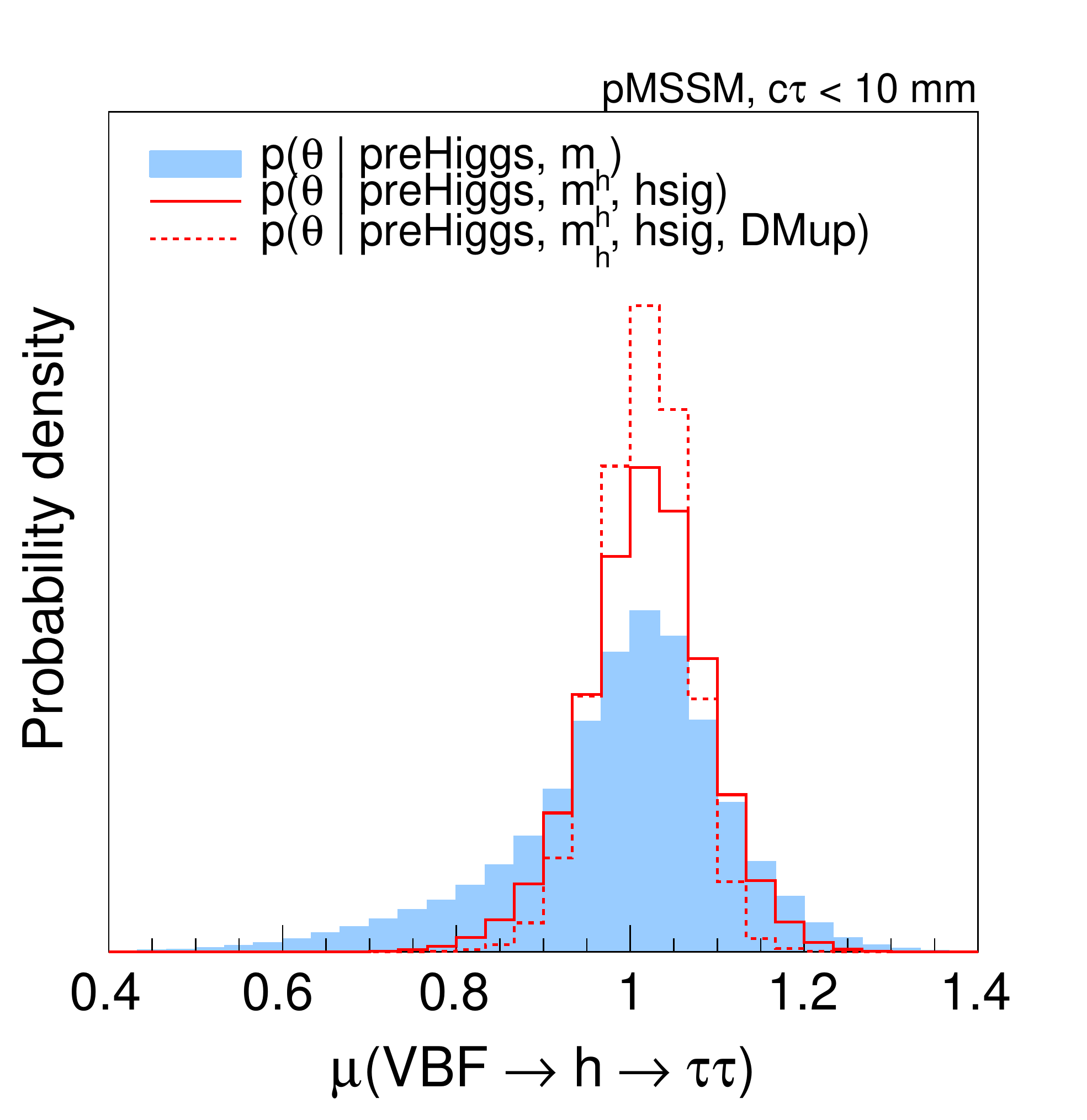}
\includegraphics[width=0.24\linewidth]{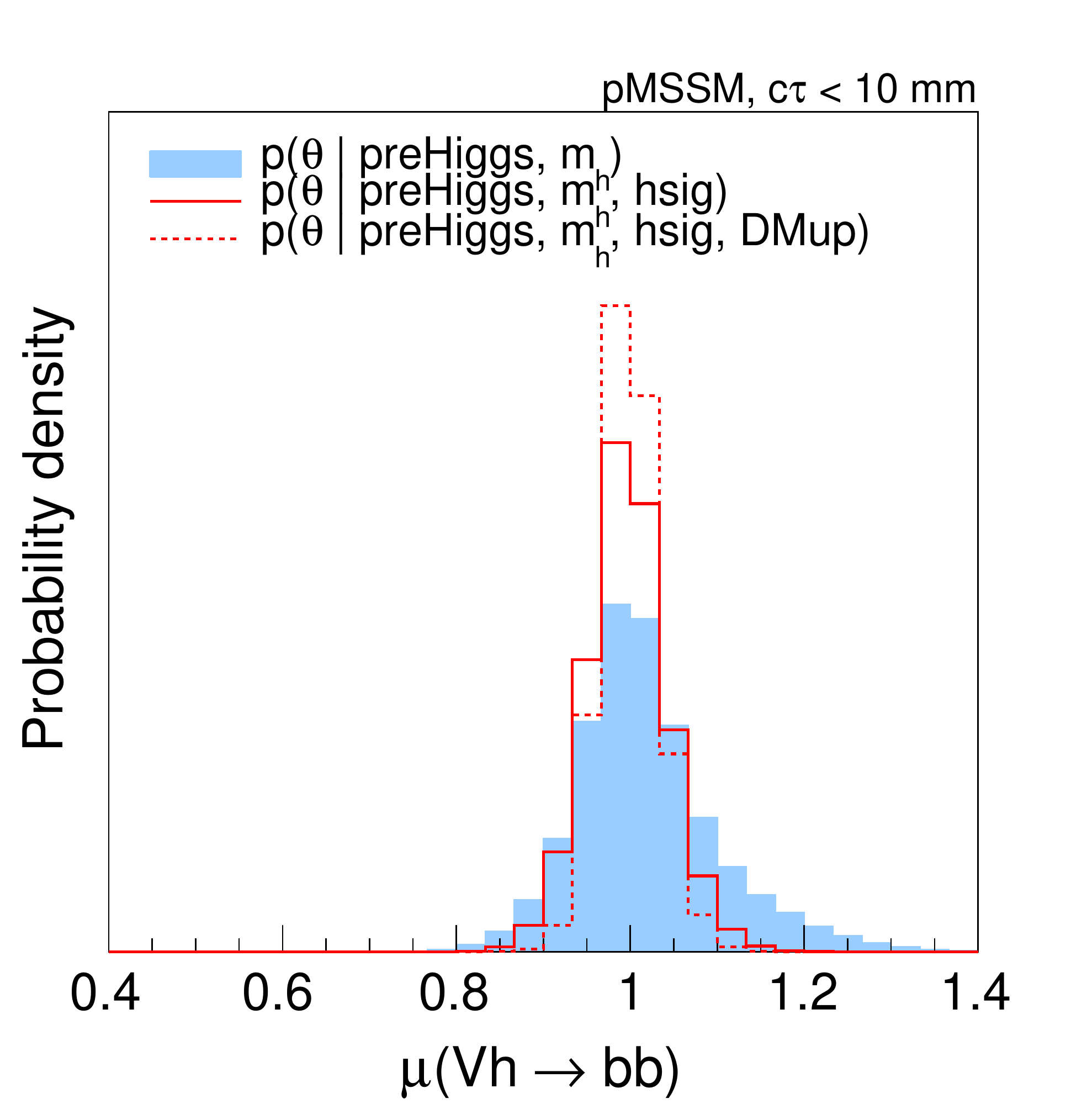}
\caption{Same as Fig.~\ref{pmssm-fig:likehiggs1} but for the relevant $h$ signal strengths.}
\label{pmssm-fig:likehiggs3}
\end{figure}

\begin{figure}[t!]
\begin{center}
\includegraphics[width=0.30\linewidth]{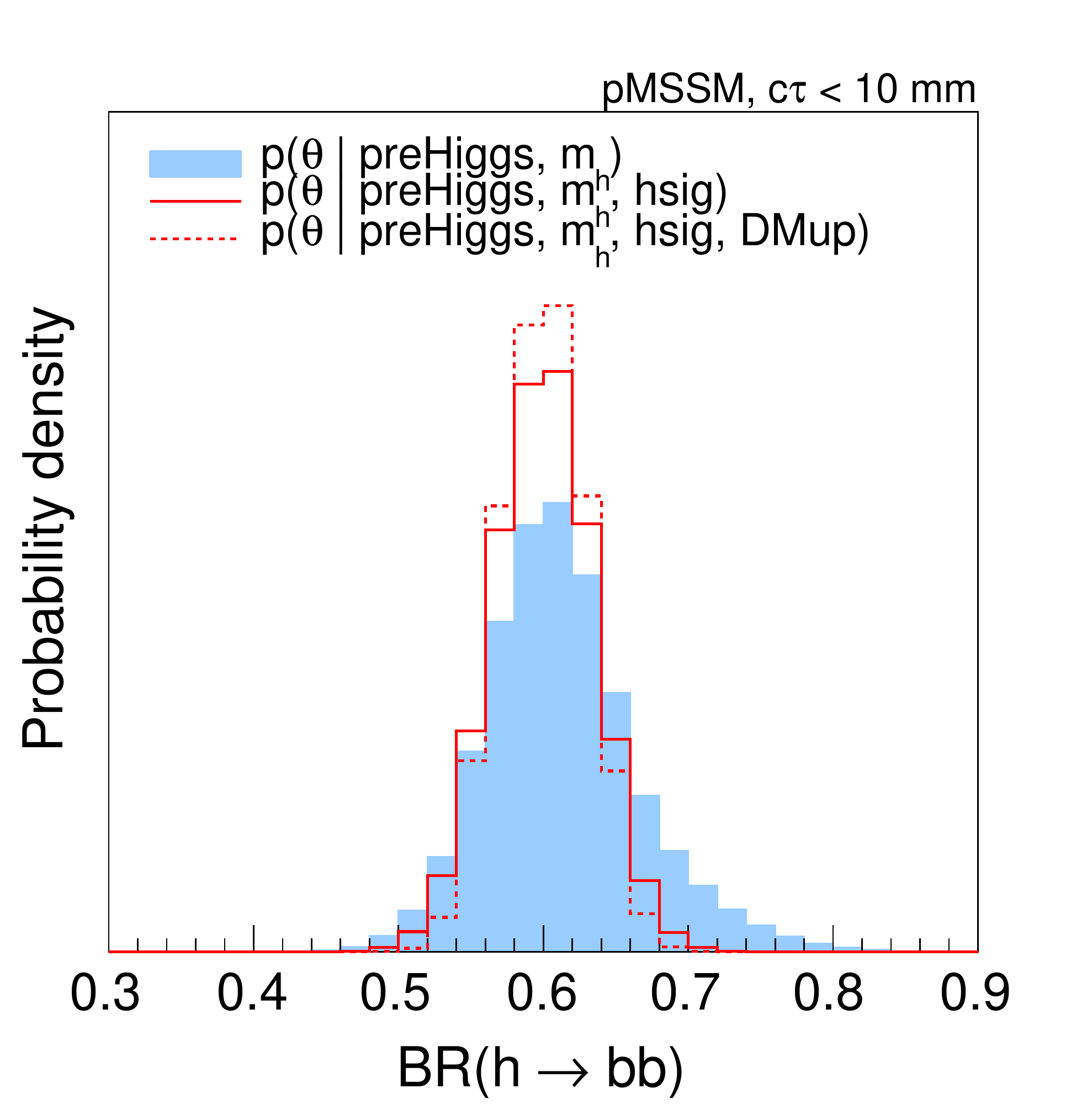}
\includegraphics[width=0.30\linewidth]{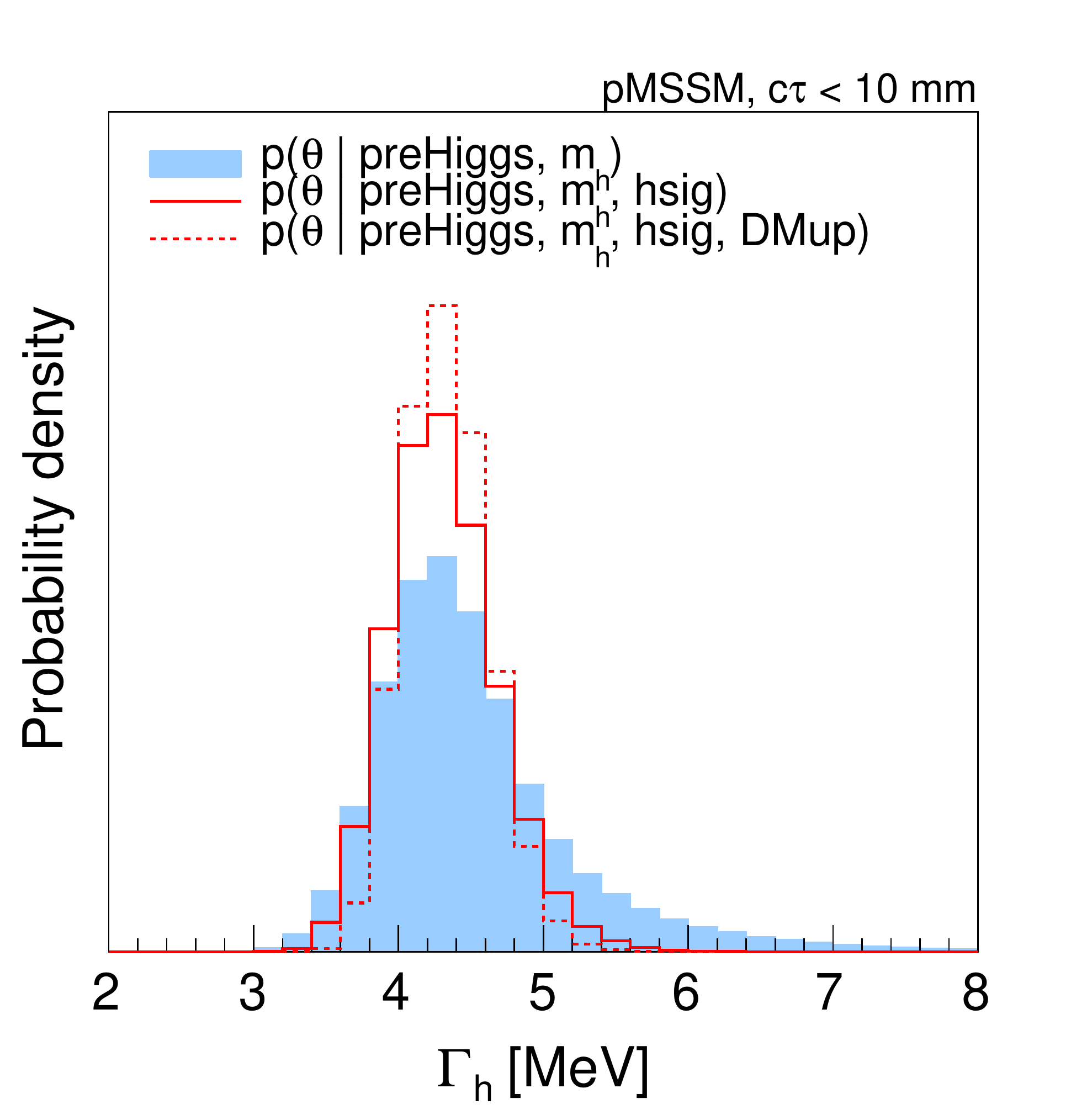}\\
\includegraphics[width=0.30\linewidth]{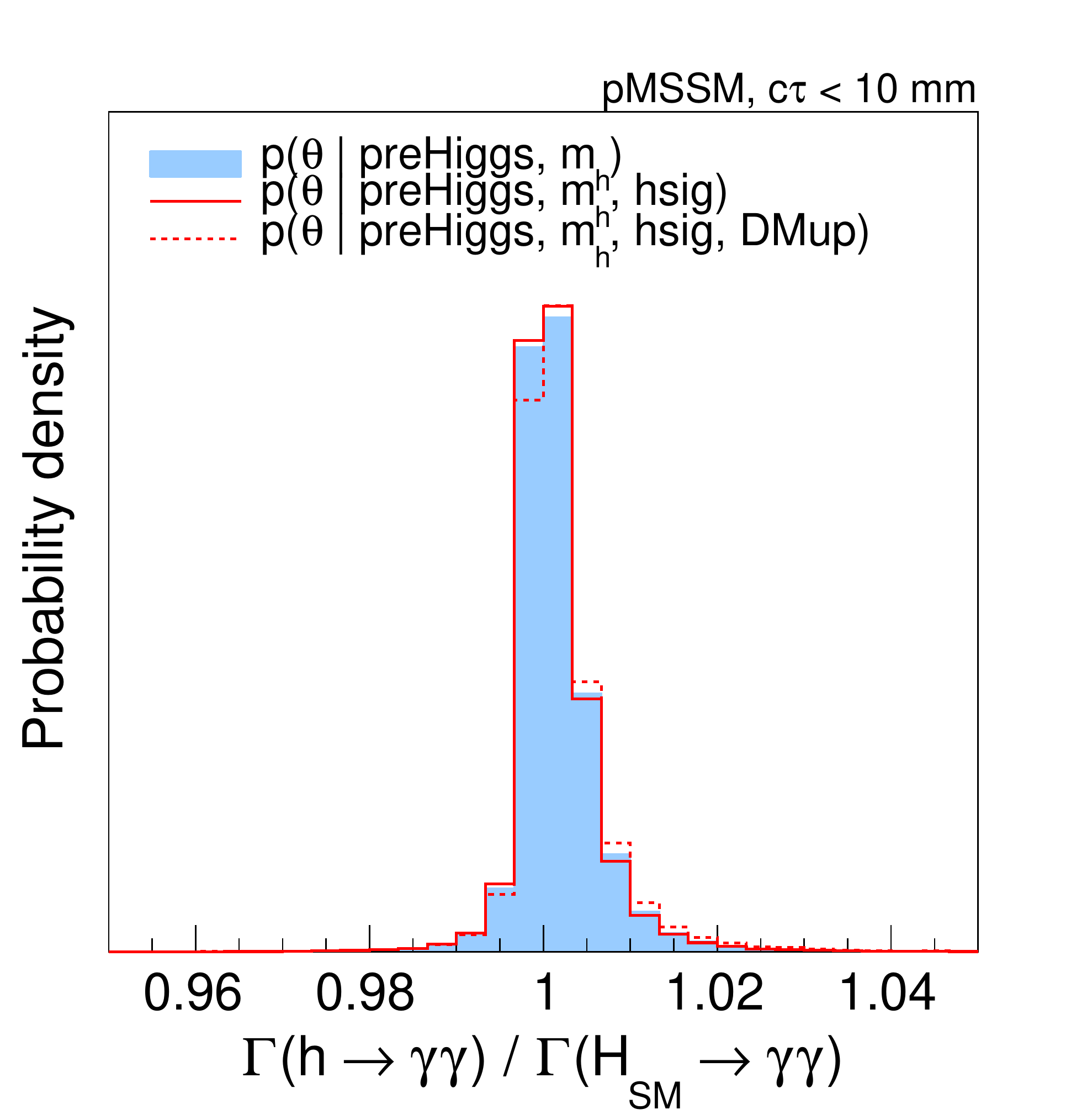}
\includegraphics[width=0.30\linewidth]{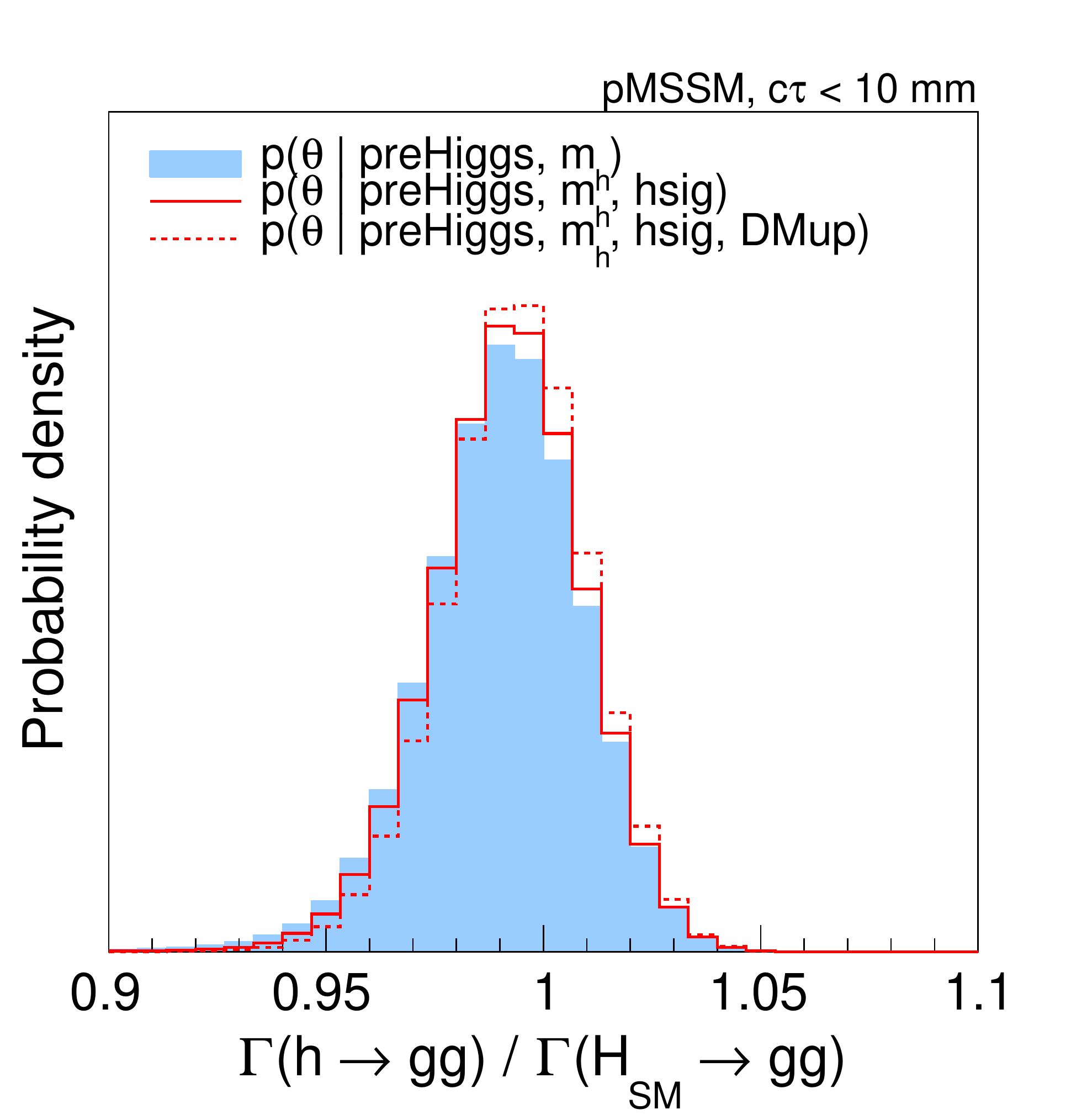}
\includegraphics[width=0.30\linewidth]{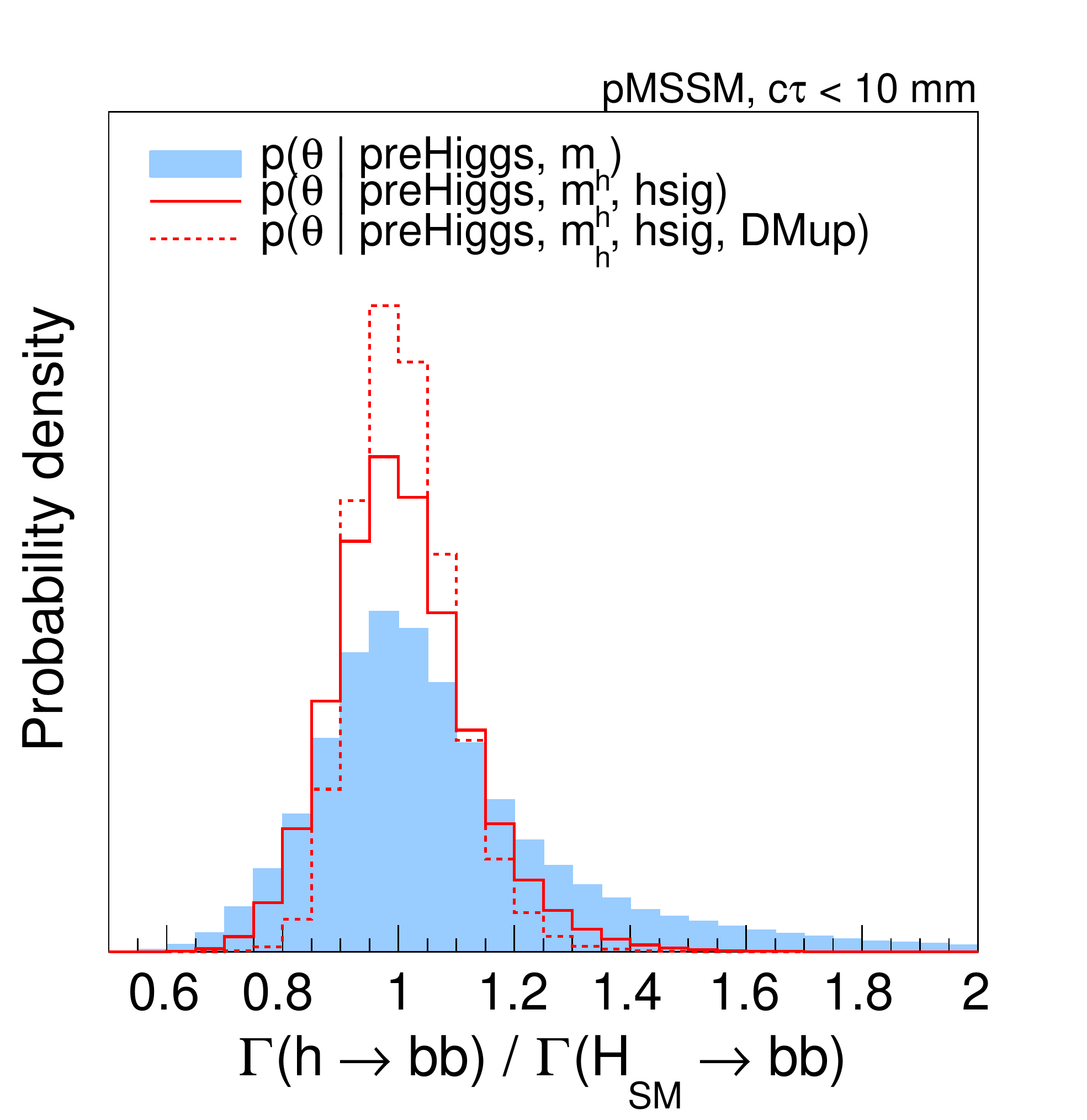}
\caption{Marginalized 1D posterior densities as in Fig.~\ref{pmssm-fig:likehiggs1}, in the top row for BR$(h\to b\bar b)$ and $\Gamma_h$, in the bottom row for $\Gamma(h\to Y)/\Gamma(H_{\rm SM}\to Y)$ with, from left to right, $Y=\gamma\gamma$, $gg$ and $b\bar b$.}
\label{pmssm-fig:hwidth}
\end{center}
\end{figure}

\begin{figure}[t!]
\includegraphics[width=0.33\linewidth]{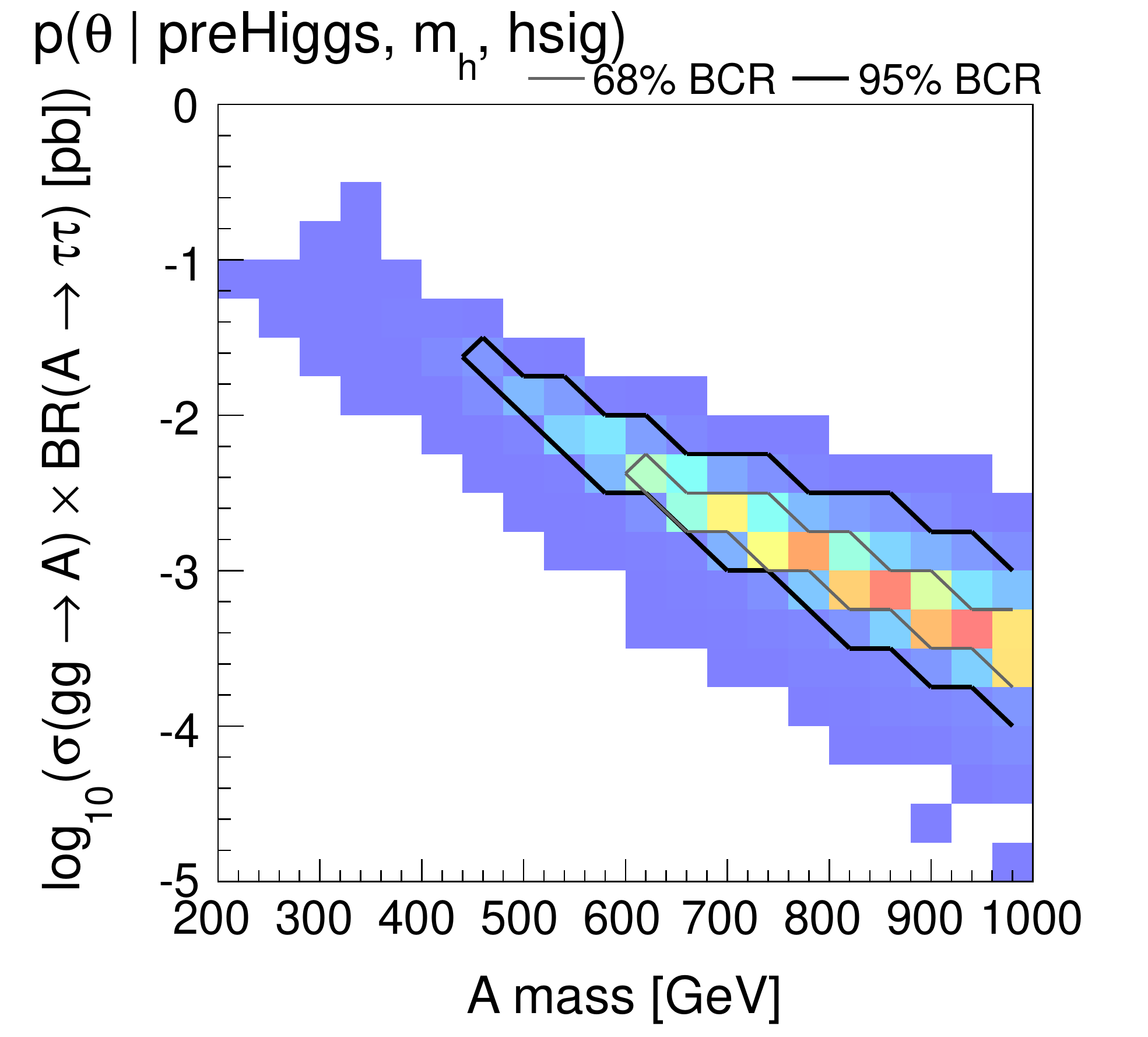}\includegraphics[width=0.33\linewidth]{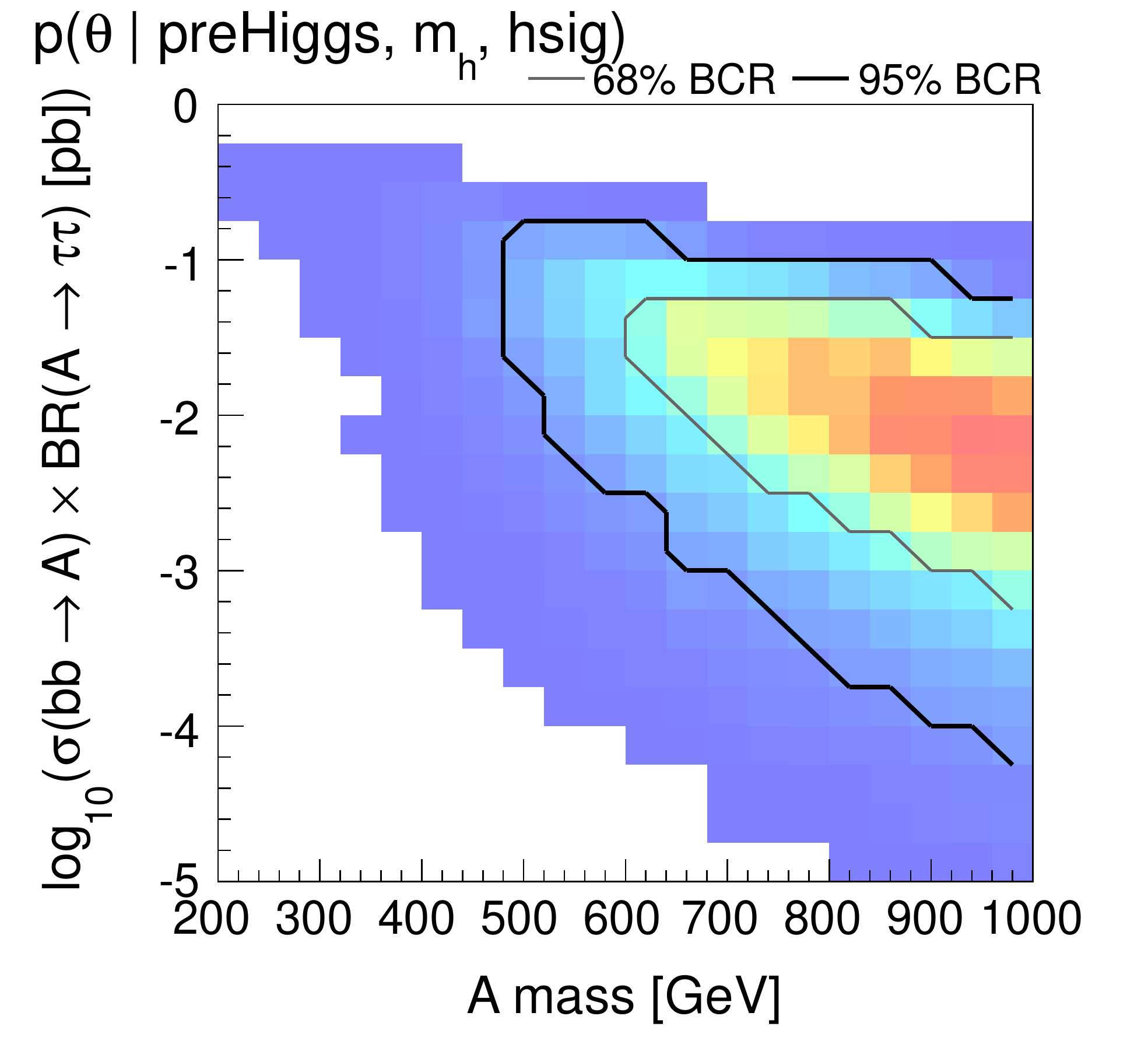}\includegraphics[width=0.33\linewidth]{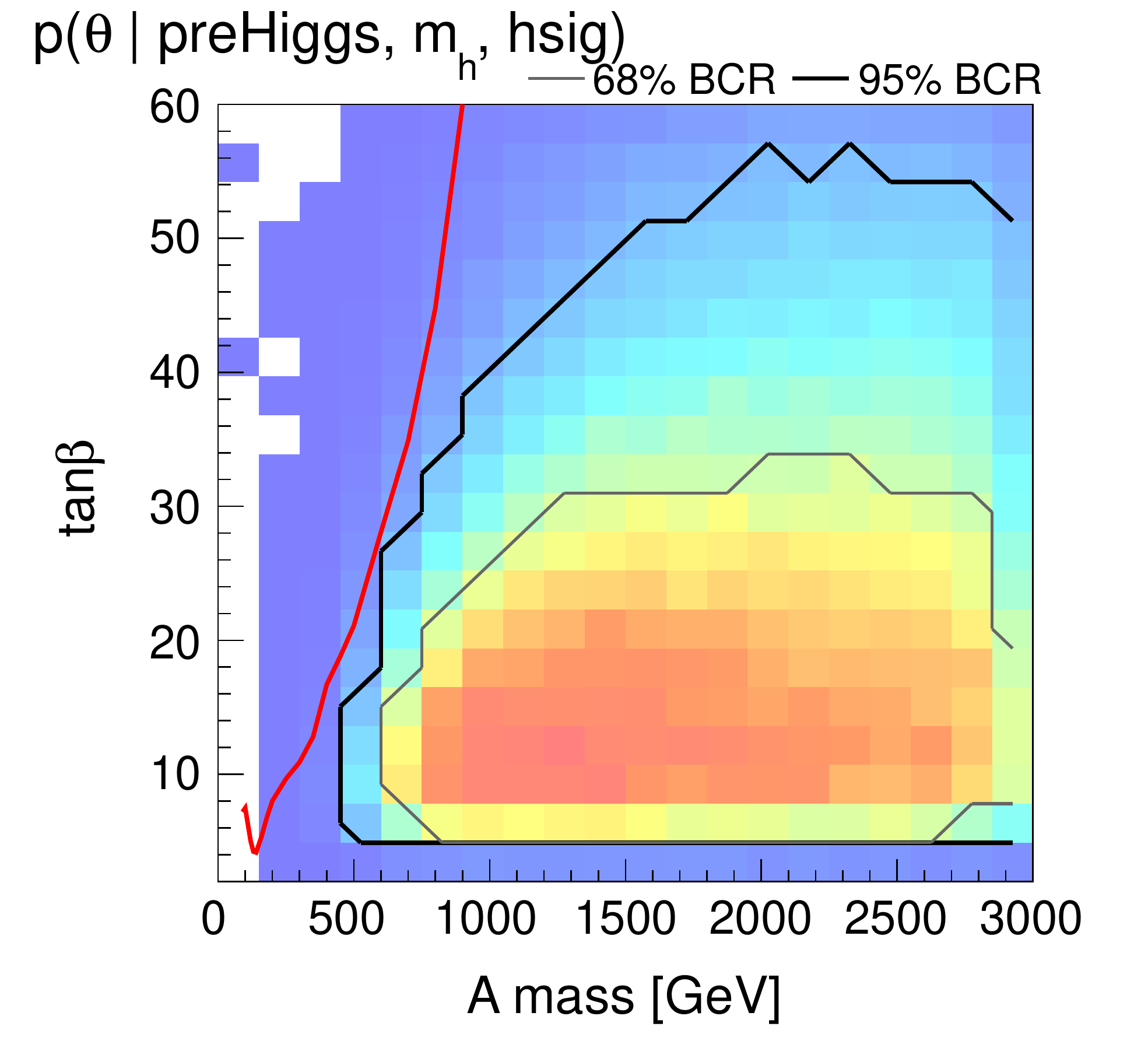}\\\includegraphics[width=0.33\linewidth]{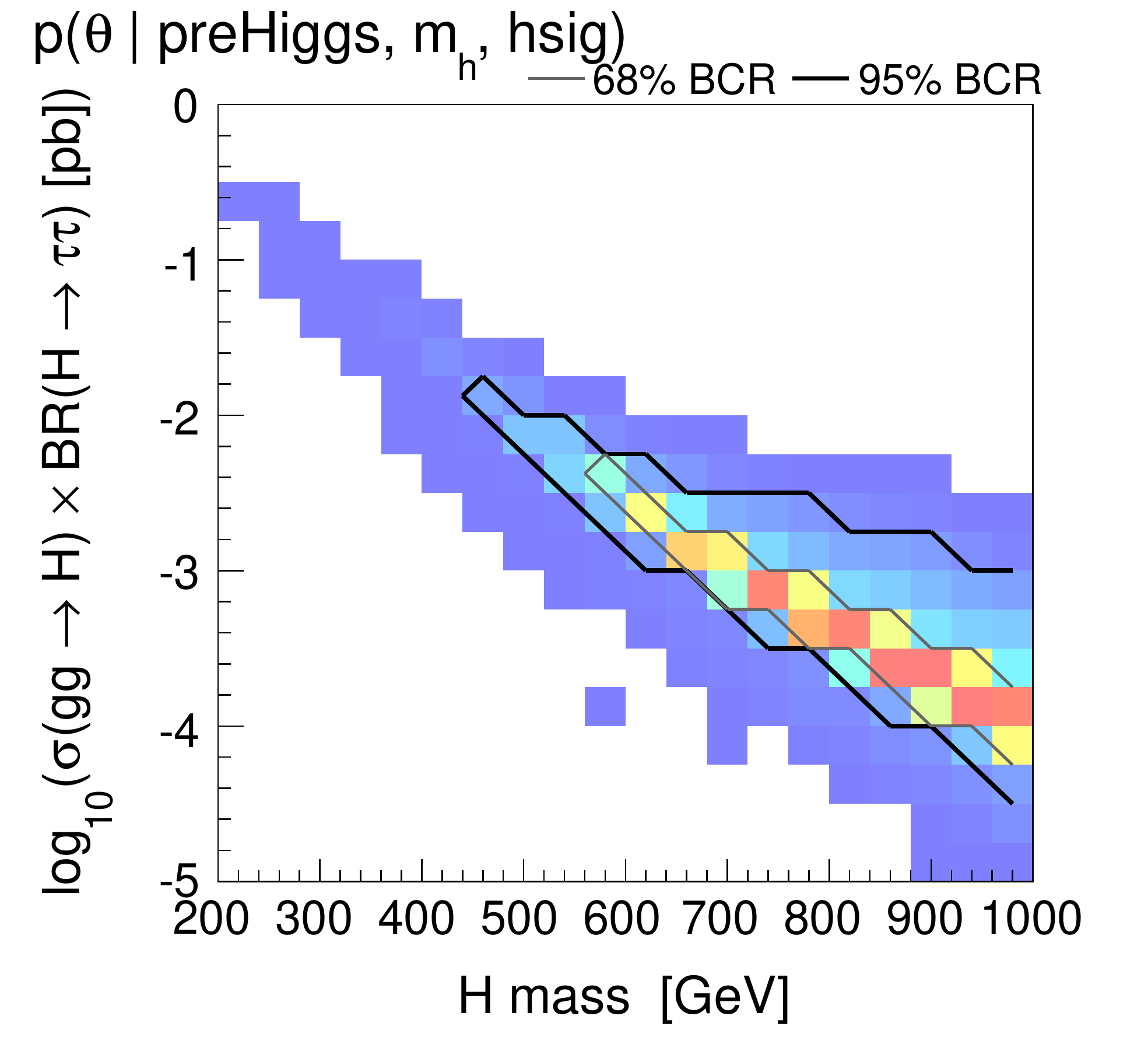}\includegraphics[width=0.33\linewidth]{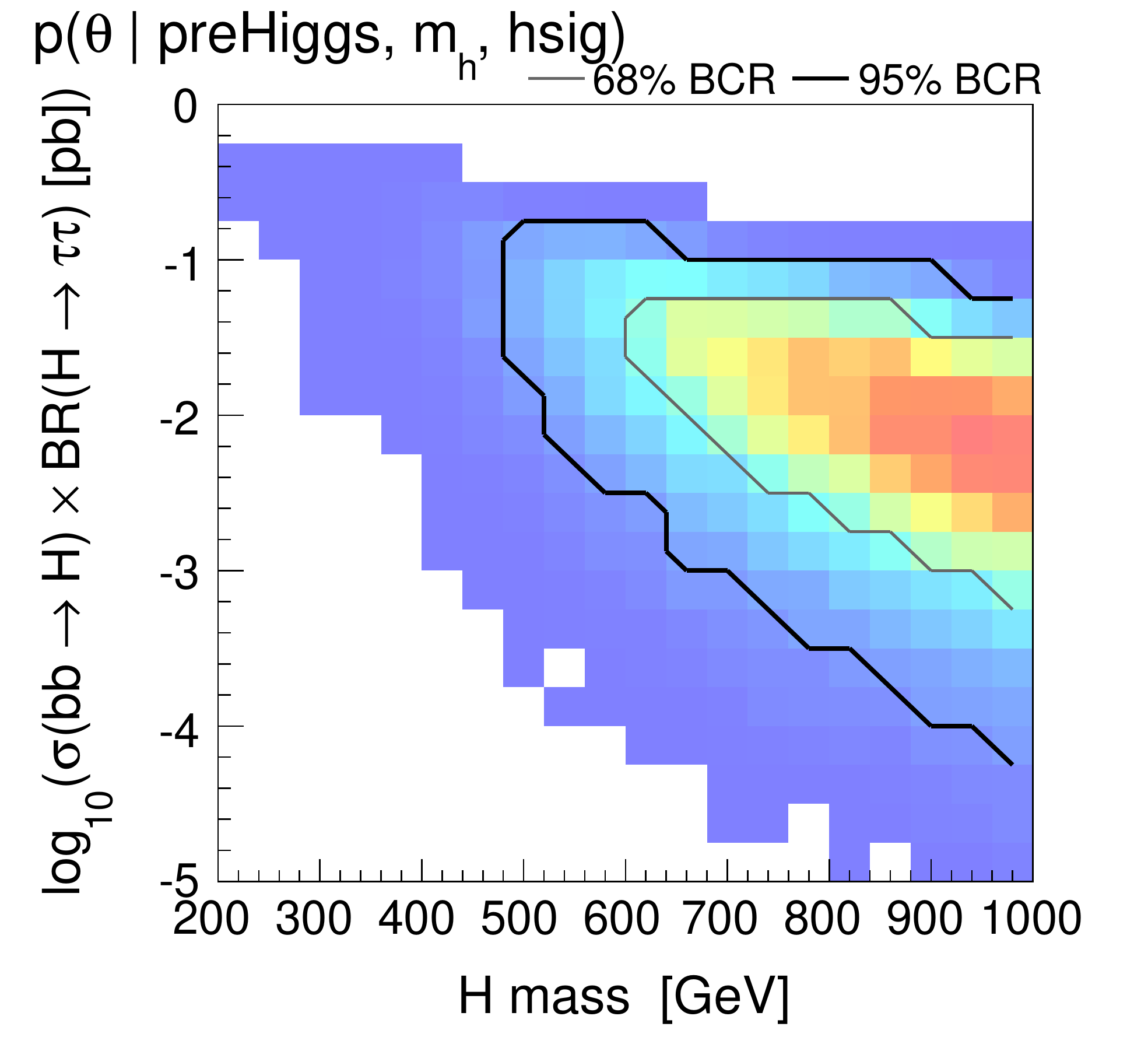}\includegraphics[width=0.33\linewidth]{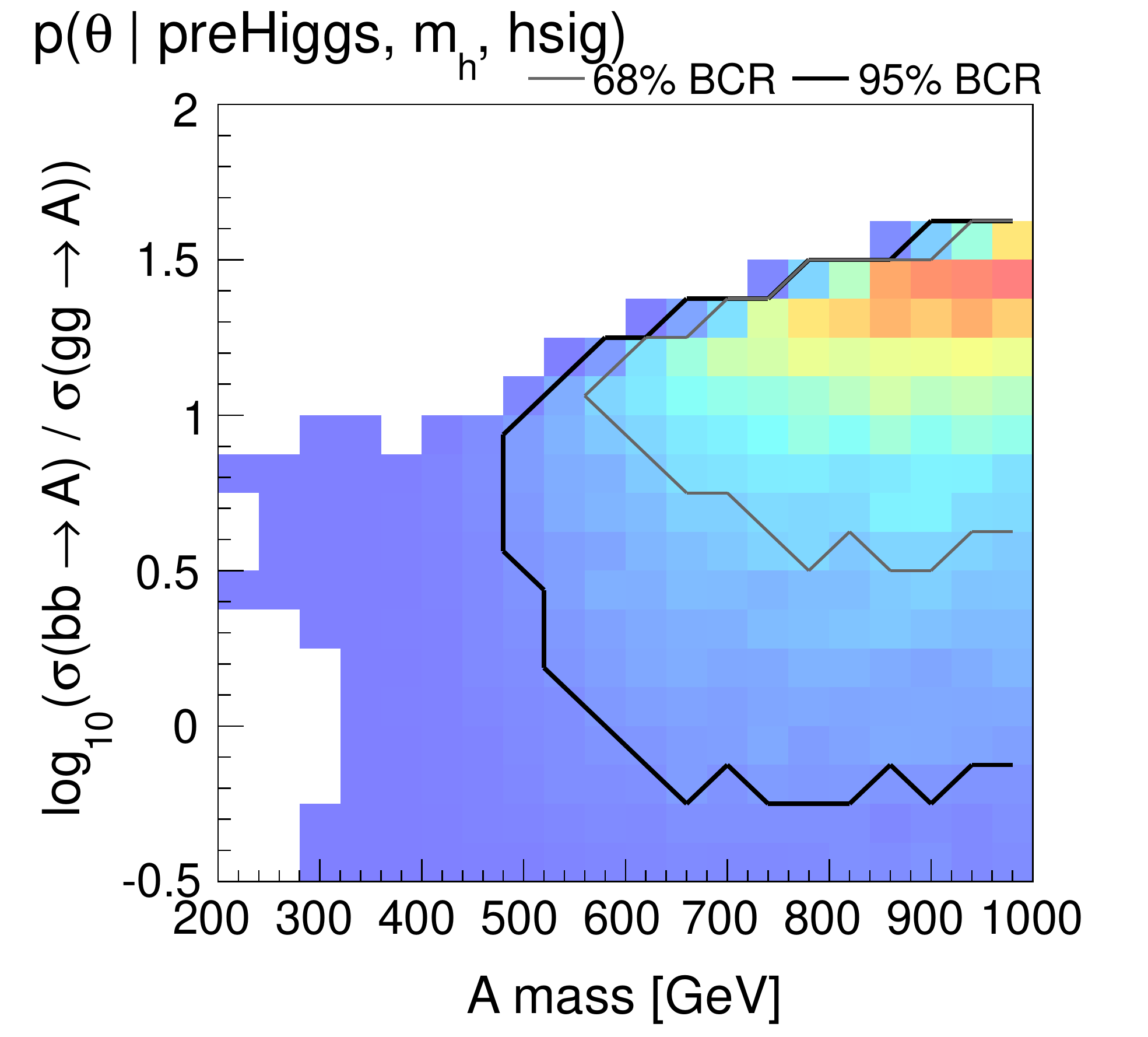}
\caption{Marginalized posterior densities in 2D for the heavy MSSM Higgses $A$ and $H$. The plots on the left and in the middle  show $\sigma \times \br$ in the $\tau\tau$ final state, from $bb$ and $gg$ production at $\sqrt{s} = 14$~TeV, versus the $A$ or $H$ mass. 
The top-right plot shows the posterior density in the $\tan\beta$ versus $m_A$ plane with the latest 95\%~CL from the CMS search for MSSM $H,A\to\tau\tau$~\cite{CMS-PAS-HIG-13-021} superimposed. 
The bottom-right plot compares $bb$ to $gg$ production as function of $m_A$.
In all plots, 
the probability density is represented by color shading, ranging from low values in blue to high values in red. The gray and black lines are contours of 68\% and 95\% Bayesian Credibility, respectively.}
\label{pmssm-fig:heavierHiggses2D}
\end{figure}

The posterior distributions of the $h$ signal strengths in the various channels are shown in Fig.~\ref{pmssm-fig:likehiggs3}. 
The red line-histograms correspond of course to the constraints which we used as experimental input. 
For the $\gamma\gamma$, $ZZ$ and $\tau\tau$ final states, we find signal strengths of about 
$1 \pm 0.15$ after the Higgs signal requirements, 
and about $1 \pm 0.10$ after the DM requirements, at 95\% Bayesian Credibility (BC).
For the $b\bar b$ final state, the distribution is much narrower than required by observations---we find that $\mu(Vh \to b\bar{b})$ is restricted to the 95\% BC interval $\mu(Vh \to b\bar{b}) \in [0.91,1.09]$ after Higgs signal requirements, and $[0.94,1.06]$ after DM requirements. This is an indirect effect of the constraint on $\br(h\to b\bar b)$ and the total $h$ width, $\Gamma_h$, in order to have large enough signal in the other channels, see Fig.~\ref{pmssm-fig:hwidth}. 
Interestingly, the constraints from the DM side narrow the signal strength distributions even more around the SM value of 1 because the higgsino mass $\mu$ tends to take on small values to fulfill the relic density requirement, leading to smaller $\Delta_b$.

Fig.~\ref{pmssm-fig:hwidth} also shows posterior distributions of  
$r_Y\equiv \Gamma(h\to Y)/\Gamma(H_{\rm SM}\to Y)$ for $Y=\gamma\gamma$, $gg$ and $b\bar b$. 
These ratios are equivalent to the ratios of the coupling strengths squared;    
$r_{\gamma\gamma}=C_\gamma^2$, $r_{gg}=C_g^2$, $r_{bb}=C_D^2$ in the notation of Section~\ref{sec:higgs2013}.
Our results for $r_Y$ can be compared to those for the neutralino LSP case in Ref.~\cite{Cahill-Rowley:2013vfa}. 
We observe that in our case
$r_{\gamma\gamma}$ peaks sharply at 1, the 95\% BC interval being [0.99,\,1.01], 
while $r_{gg}$ shows a wider distribution with a 95\% BC interval of [0.96,\,1.02].  
(The picture does not change if we remove the $c\tau$ cut.). 
These features are different from those in \cite{Cahill-Rowley:2013vfa}, where 
the $r_{\gamma\gamma}$ distribution peaks within $r_{\gamma\gamma}\approx 1$--$1.05$, 
and $r_{gg}$ exhibits an upper limit of $r_{gg}\lesssim 0.97$. 
Also, the $r_{bb}$ distribution is quite different. 
Some differences are of course expected as the distributions in \cite{Cahill-Rowley:2013vfa} come from 
a flat  random sampling and thus do not have the statistical meaning that underlies our approach. 
More importantly, however, the SM calculation of {\tt HDECAY} employed in \cite{Cahill-Rowley:2013vfa}  
includes additional radiative corrections which are not present  in the MSSM calculation.\footnote{We thank 
Ahmed Ismail and Matthew Cahill-Rowley for communication on this matter.} 
In our case, we avoid this problem by taking the MSSM decoupling 
case as the SM limit for computing  $\Gamma(H_{\rm SM}\to Y)$, {\it cf.}\ Section~\ref{pmssm-sec:higgslikeli}.   Of course, the $r_Y$ are not directly measurable at the LHC.  They become measurable only if it can be determined that the $h$ has no invisible (\eg\ $h\to\cnone\cnone$) or unseen (\eg\ $h\to 4\tau$) decay modes.


\begin{figure}[t!]
\begin{center}
\includegraphics[width=0.3\linewidth]{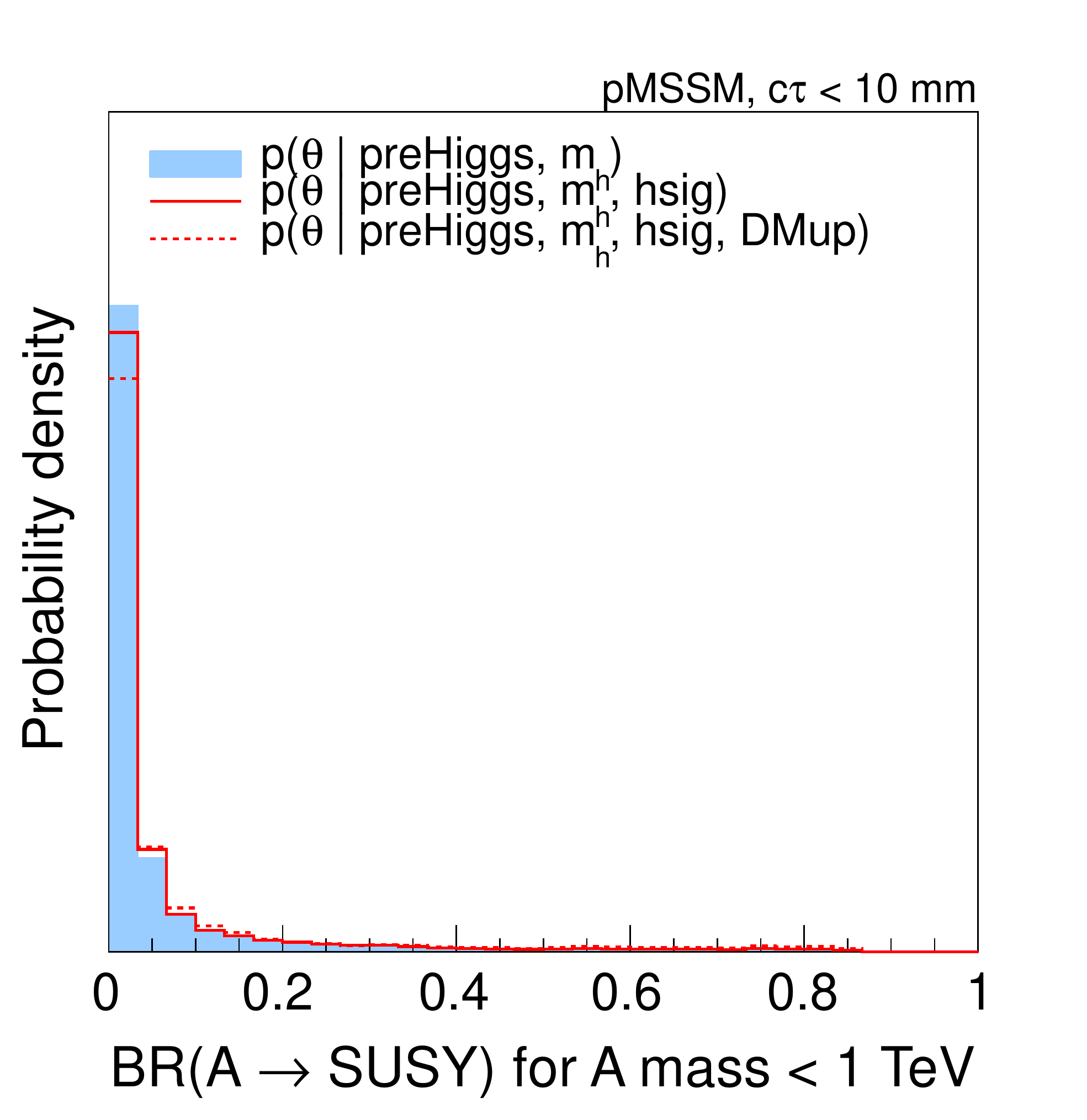}
\includegraphics[width=0.3\linewidth]{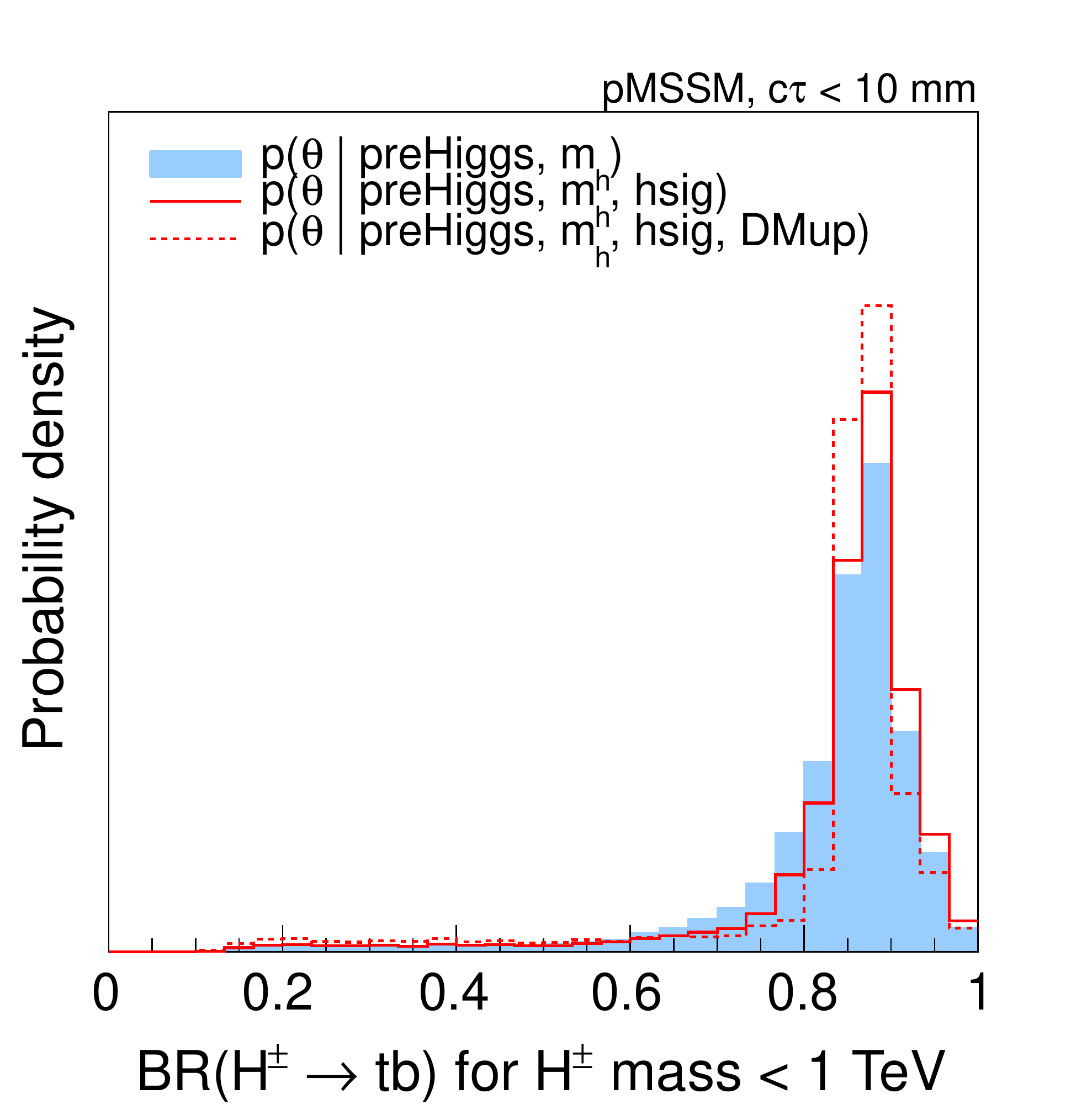}
\includegraphics[width=0.3\linewidth]{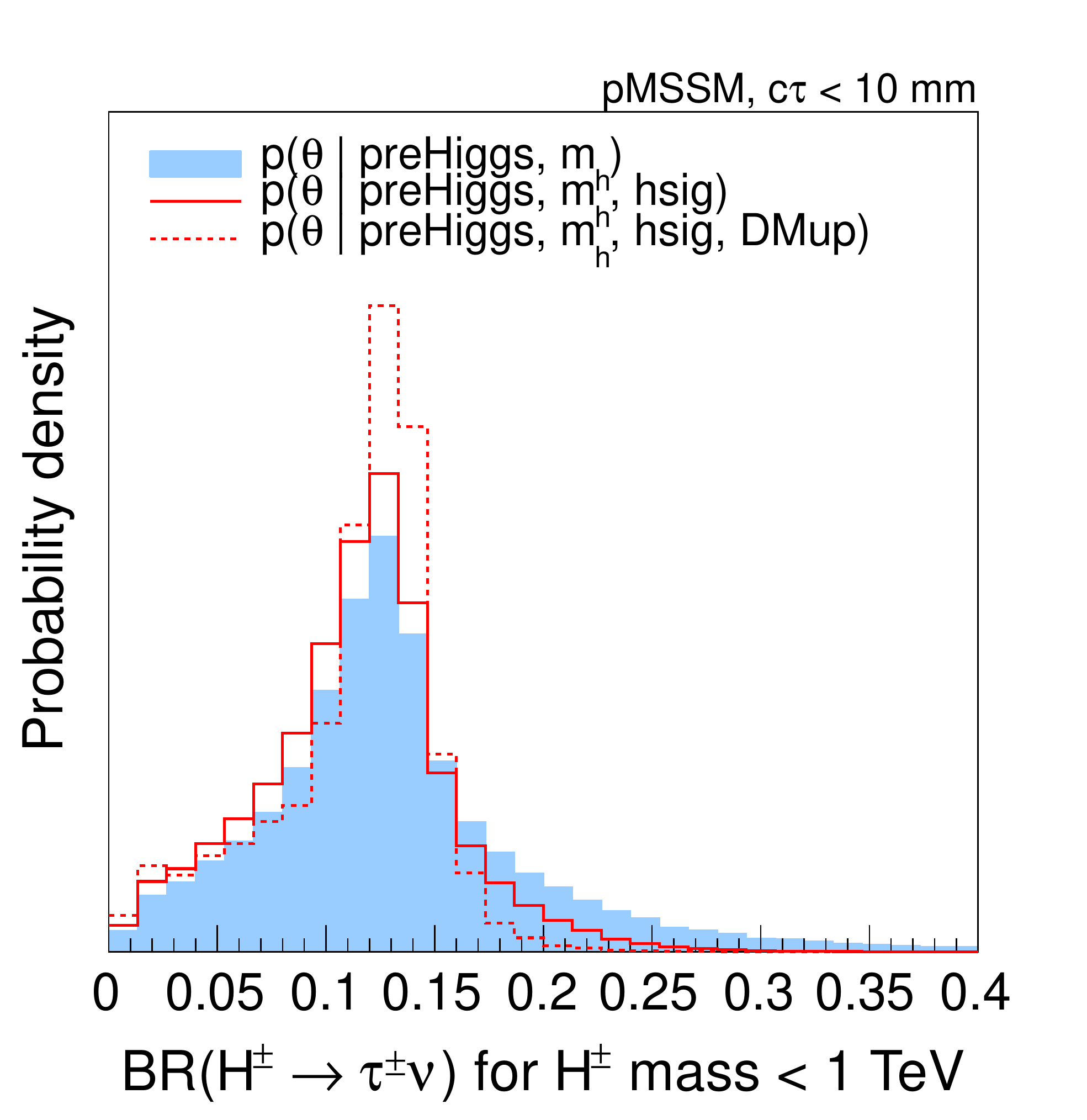}
\caption{Marginalized 1D posterior densities as in Fig.~\ref{pmssm-fig:likehiggs1}, here for the branching fractions of the heavy MSSM Higgses $A$ and $H^\pm$ with masses below 1~TeV.}
\label{pmssm-fig:heavierHiggses}
\end{center}
\end{figure}

Our procedure also allows us to derive predictions for the heavier MSSM Higgs states $H$, $A$ and $H^\pm$, as illustrated in Figs.~\ref{pmssm-fig:heavierHiggses2D} and \ref{pmssm-fig:heavierHiggses}. First, in the $\tan \beta$ versus $m_A$ plane, we show that the current CMS limit~\cite{CMS-PAS-HIG-13-021} interpreted in the $m_h^{\rm max}$ scenario has a negligible effect on our distributions, since after imposing constraints from low-energy observables and from Higgs measurements the likely region corresponds to $A$ masses above 500~GeV and moderate $\tan \beta$. (This observation remains valid when dark matter requirements are taken into account; in all cases we have checked that the current limits on $H \to ZZ$ are always satisfied.) We also show $\sigma(gg,b\bar b \to H,A) \times {\rm BR}(H,A \to \tau\tau)$ at $\sqrt s = 14~$TeV as a function of $m_{H,A}$, using {\tt SusHi\_1.1.1}~\cite{Harlander:2012pb} for the computation of the cross sections in the approximation of decoupled stops and sbottoms.\footnote{Neglecting contributions from stops and sbottoms in the computation of $gg, b\bar b \to H,A$ is a good approximation in most cases since the posterior densities of $m_{\tilde t_1}$ and $m_{\tilde b_1}$ peak around 2~TeV.}
These plots show that the signals from the CP-odd and CP-even Higgs bosons are very similar and that for high masses the dominant process is almost always $b\bar b \to H,A$ (see, in particular, the bottom right plot), where for a given mass $\sigma(b\bar b \to H,A)$ spans over about an order of magnitude due to its strong dependence on $\tan \beta$. Typical $\sigma \times \br$ values are of the order of 0.1 to 100~fb for $m_{H,A} < 1$~TeV and therefore most of this region should be probed during the next run of the LHC at 13--14~TeV.

Some more properties of the heavy Higgses (for masses $< 1$~TeV) are shown in Fig.~\ref{pmssm-fig:heavierHiggses}. We see that the decay branching fraction of $A$ into SUSY particles is often very small because most of the supersymmetric partners generally lie at the (multi-)TeV scale. Concretely, the probability for $\br(A \to {\rm SUSY}) > 10\%$ is only 1.6\% after the Higgs signal likelihood (2.1\% after DM requirement). Compared to the preHiggs distributions, decays into SUSY particles are however slightly enhanced by the Higgs likelihood and dark matter requirements because $\mu$, and hence neutralino and chargino masses, are pushed to lower values. Also shown are the dominant decay modes of the charged Higgs: $H^{\pm} \to tb$ and $H^{\pm} \to \tau^{\pm} \nu$. The dominance of hadronic decays over leptonic ones is strengthened when Higgs measurements are taken into account since small values of $m_A$ and large values of $\tan \beta$ are then disfavored.

\subsubsection{Impact of the $\bm c\tau$ cut}\label{pmssm-sec:GreenPlots}

\begin{figure}[t!]
\begin{center}
\includegraphics[width=0.25\linewidth]{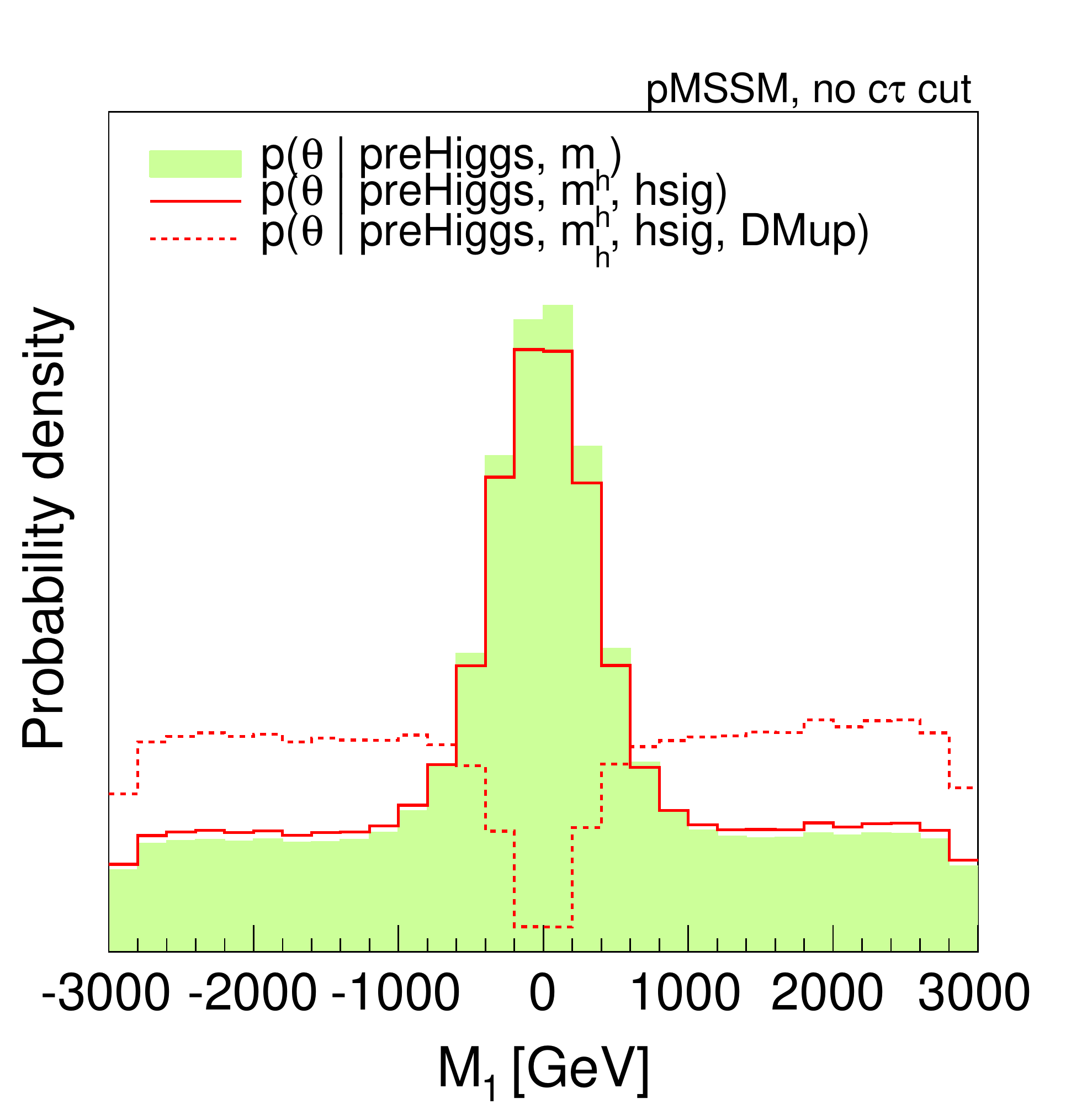}\includegraphics[width=0.25\linewidth]{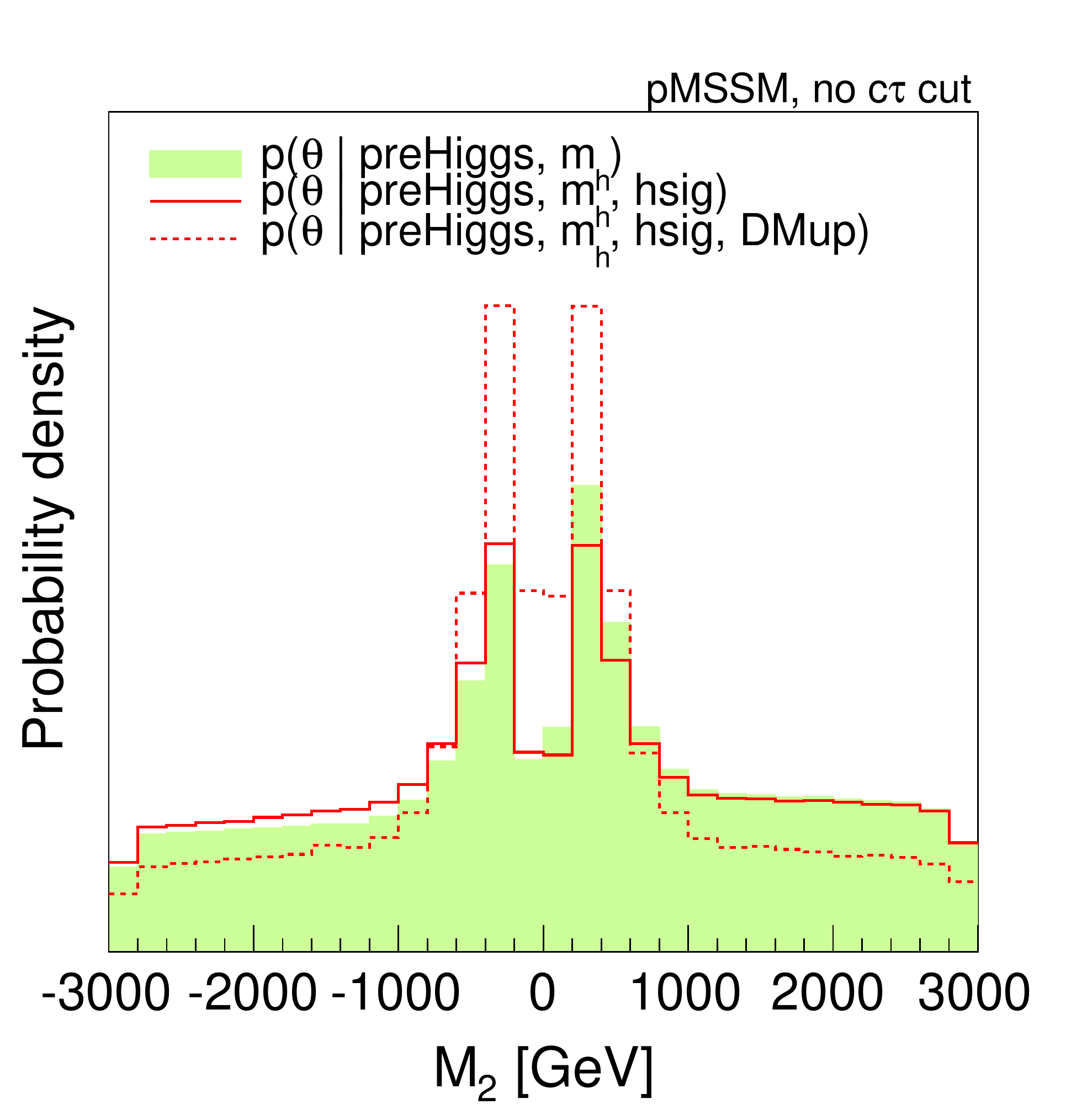}\includegraphics[width=0.25\linewidth]{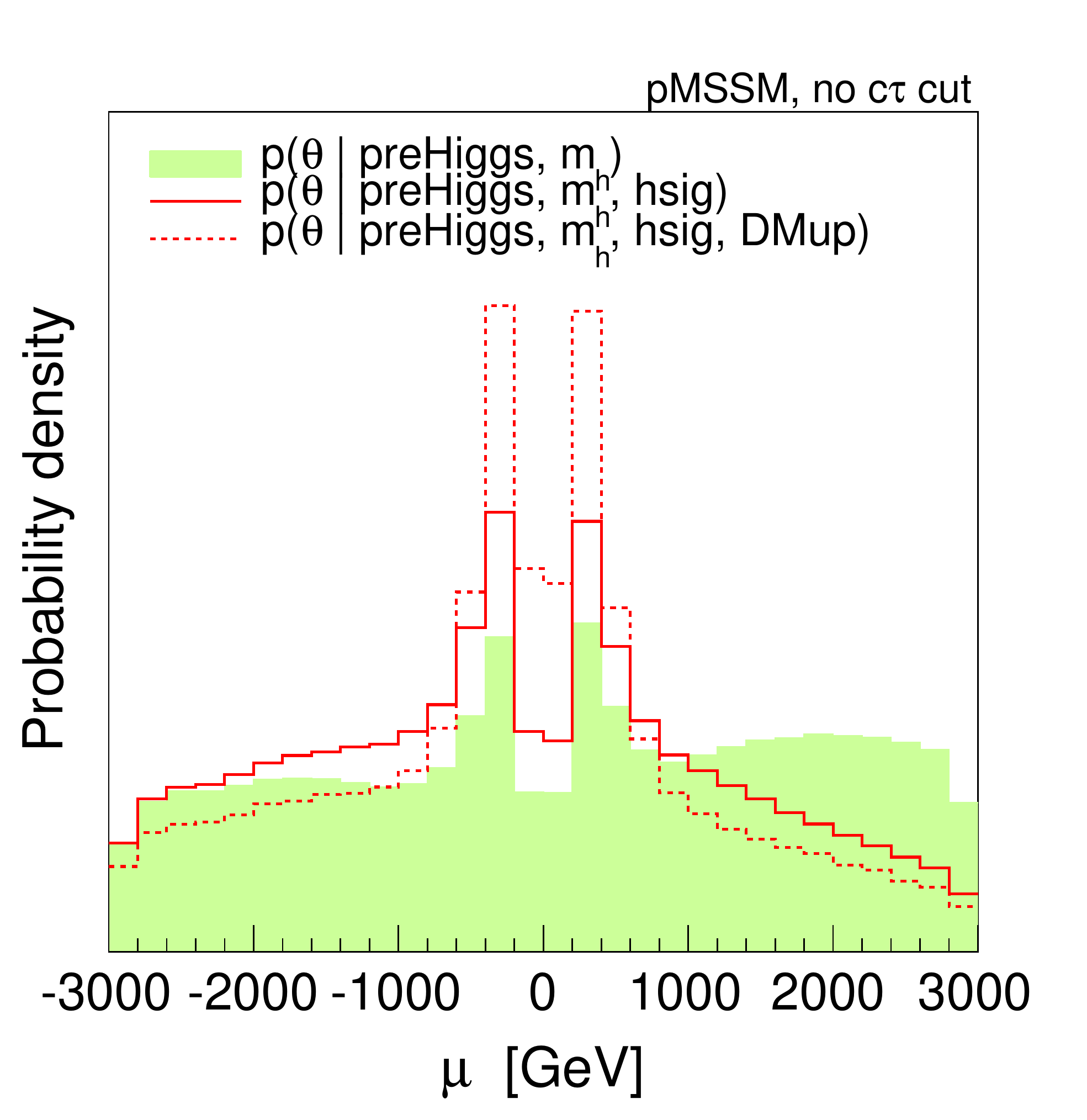}\includegraphics[width=0.25\linewidth]{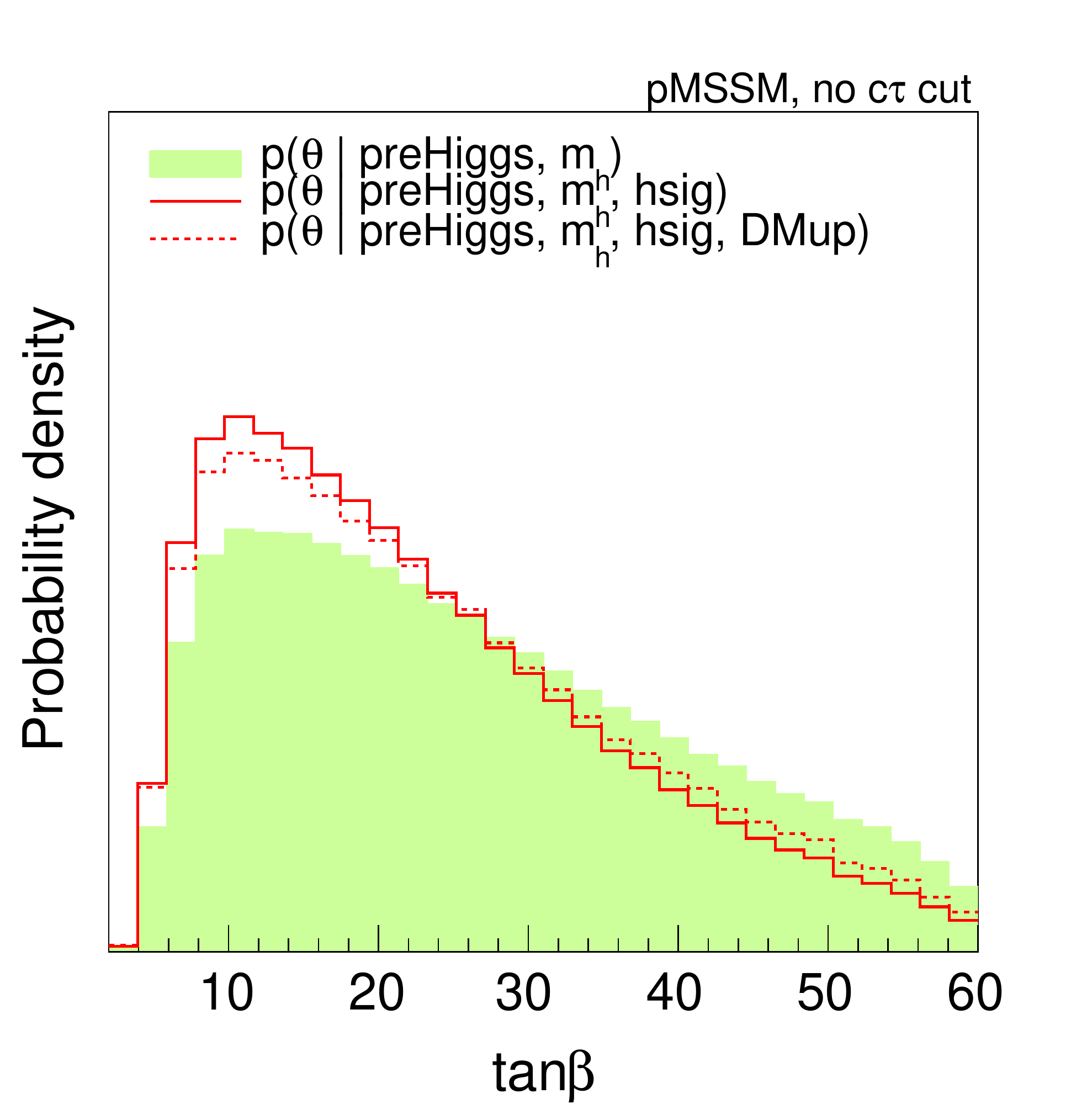}\\
\includegraphics[width=0.25\linewidth]{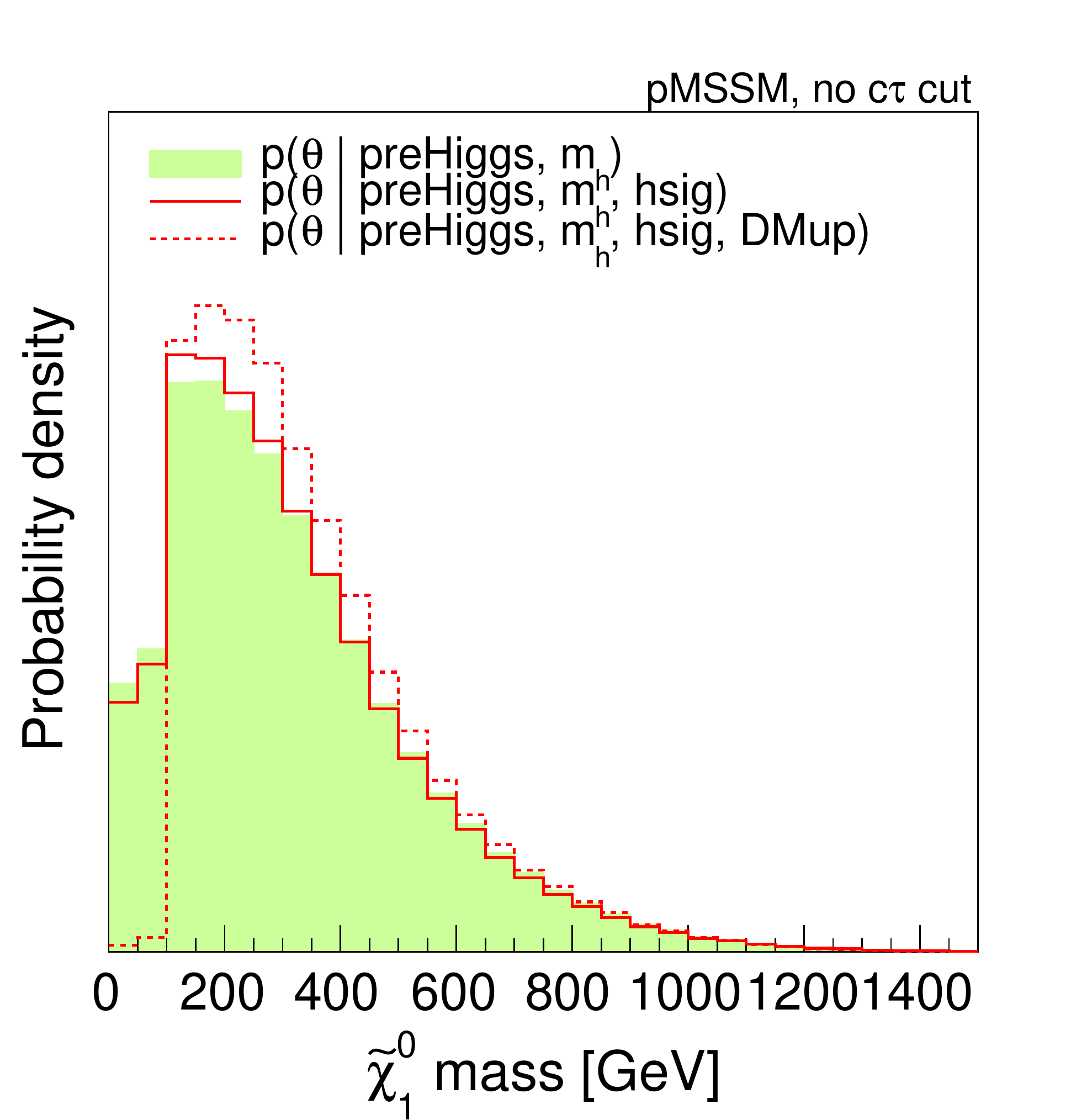}\includegraphics[width=0.25\linewidth]{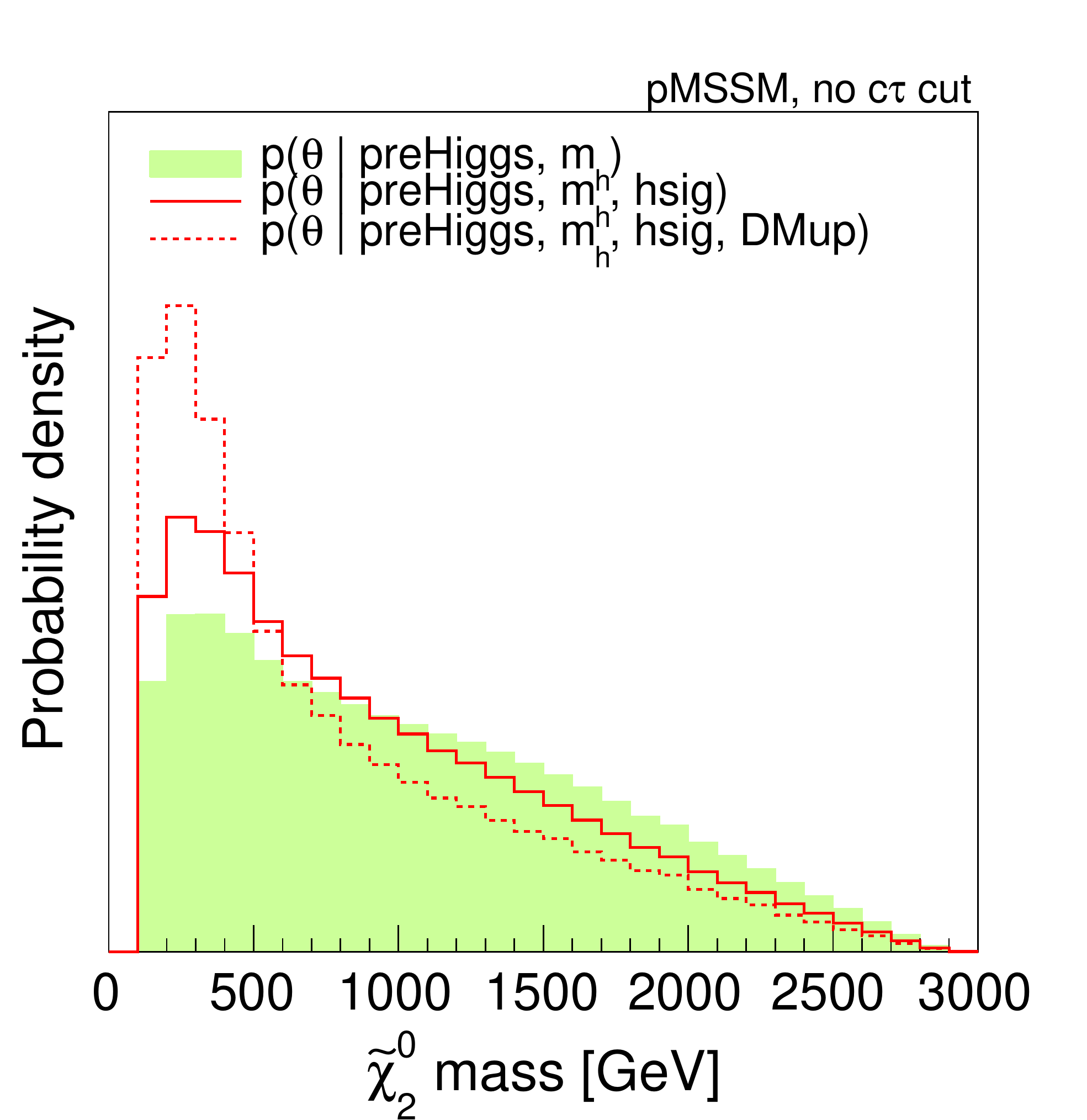}\includegraphics[width=0.25\linewidth]{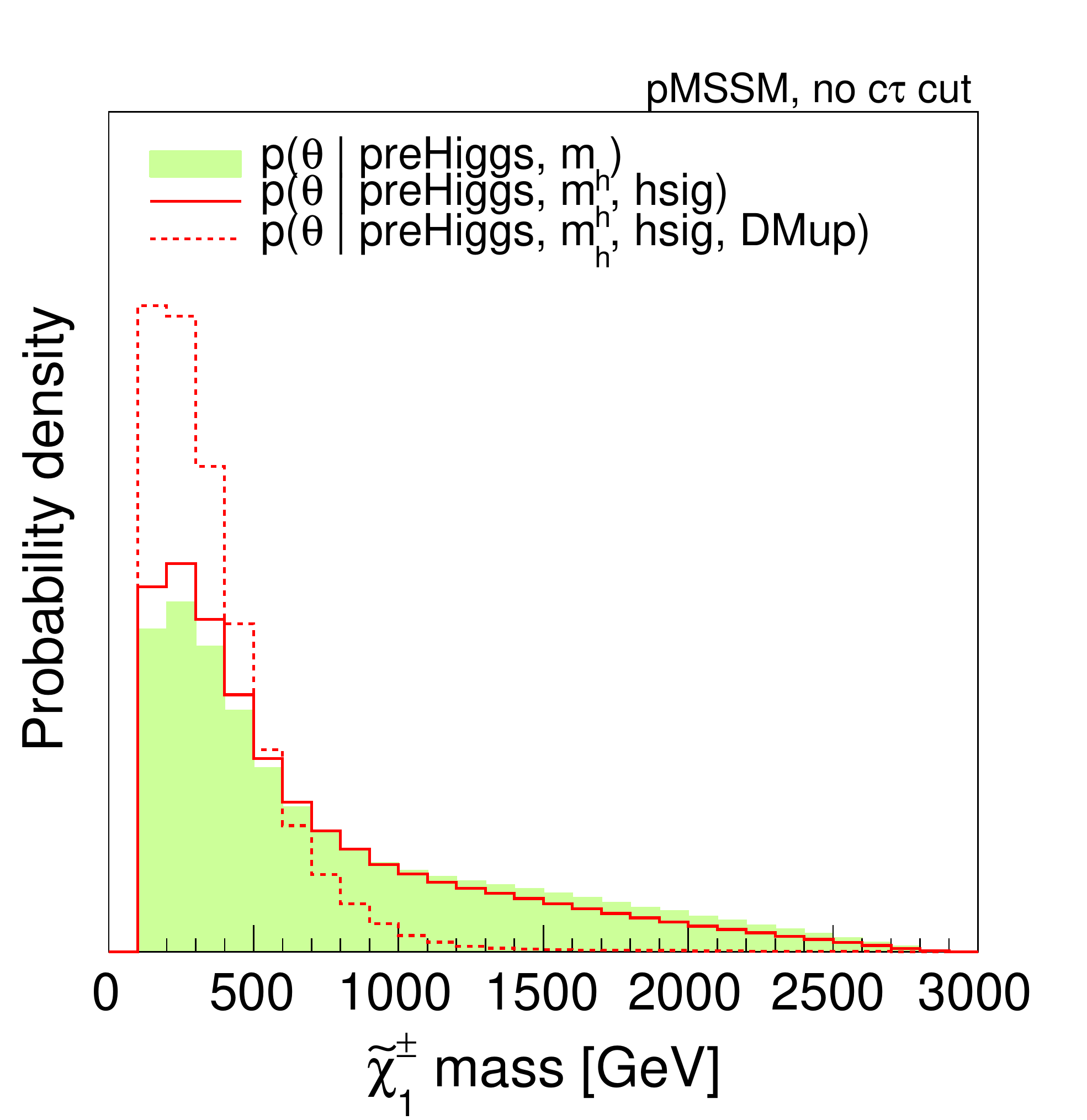}\includegraphics[width=0.25\linewidth]{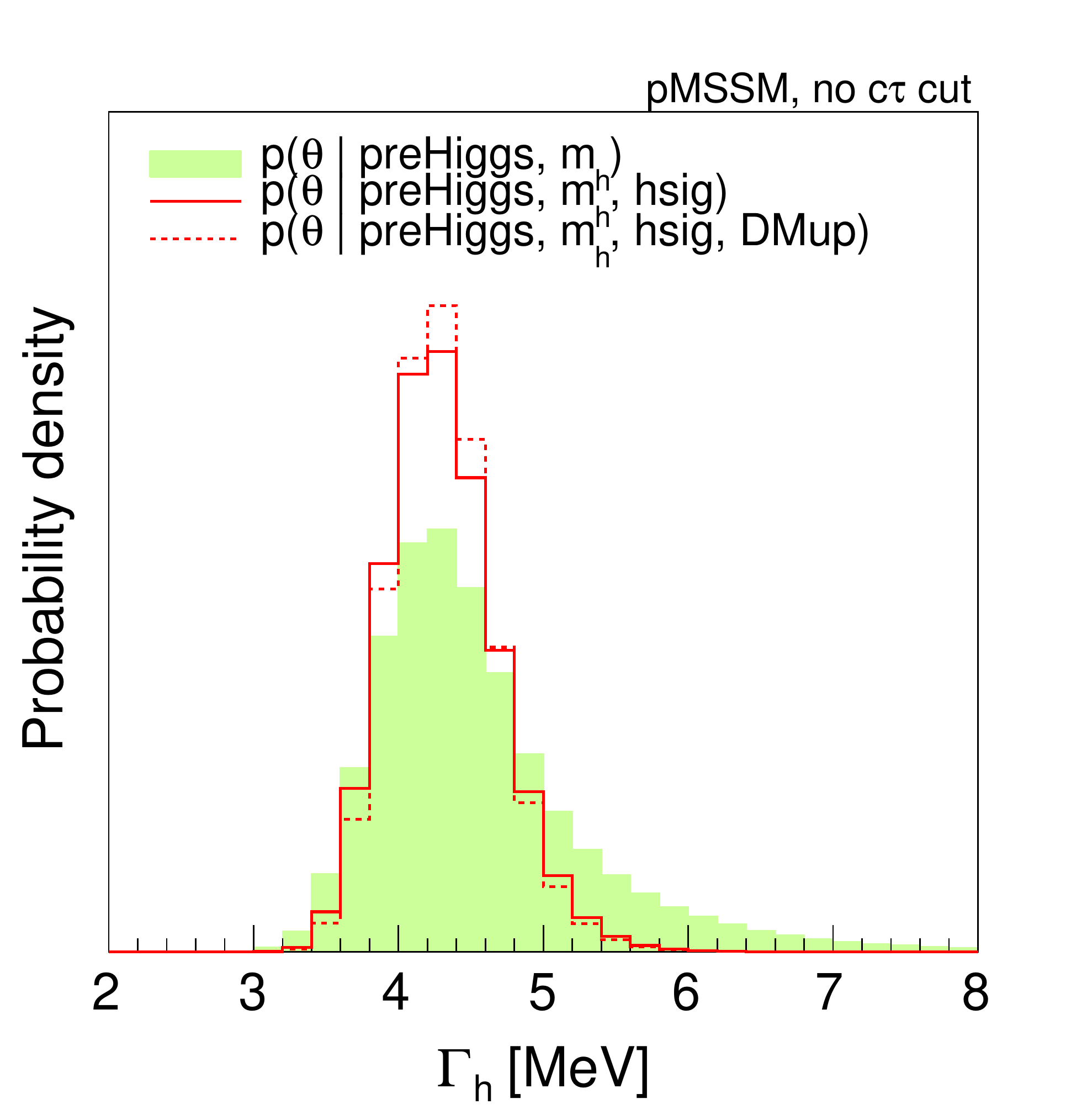}
\caption{Marginalized 1D posterior densities for selected parameters and derived quantities without the prompt chargino requirement.
The green histograms show the distributions  based on the ``preHiggs'' measurements 
of Table~\ref{pmssm-tab:preHiggs} plus requiring in addition $m_h\in [123,\,128]$~GeV, but without the $c\tau$ cut. 
The solid red lines are the distributions when taking into account in addition the measured Higgs signal strengths 
in the various channels, as well as the limits from the heavy MSSM Higgs searches. The dashed red lines include 
in addition an upper limit on the neutralino relic density and the recent direct DM detection limit from LUX.}
\label{pmssm-fig:likehiggsNoCtau}
\end{center}
\end{figure}

We saw from the plots in Section~\ref{pmssm-sec:YellowPlots} that the ``prompt chargino'' requirement has 
a strong effect on some of the distributions, above all on that of the wino mass parameter $M_2$. 
The influence on $\mu$ and $M_1$ is less dramatic but still quite strong. As a consequence, it is mostly the 
chargino and neutralino masses (and their gaugino--higgsino composition) which are affected by the 
$c\tau<10$~mm requirement. To assess the impact of this cut,
the relevant posterior densities {\em without} the $c\tau$ cut are shown in Fig.~\ref{pmssm-fig:likehiggsNoCtau}. 
Comparing these plots with their equivalents in Fig.~\ref{pmssm-fig:sampling1} of Section~\ref{pmssm-sec:BluePlots}, 
we see that, as expected, 
in both  the ``preHiggs+$m_h$'' and the ``preHiggs+$m_h$+hsig'' distributions, light charginos and neutralinos are more preferred. The effect is more pronounced for the $\tilde\chi^\pm_1$ and  $\tilde\chi^0_2$ than for the $\tilde\chi^0_1$. Note also that the preference for smaller $\mu$ through the Higgs signal strength measurements remains. Finally, note that the DM upper limits largely overrule the effect of the $c\tau$ cut: the red dashed line histograms are almost the same with or without the $c\tau$ cut. The exception is the  $\tan\beta$ distribution. (The $\tilde\chi^0_1$, $\tilde\chi^0_2$, $\tilde\chi^\pm_1$ mass differences can however be smaller  without the $c\tau$ cut.) 
The posterior densities of other quantities, which do not directly depend on $M_1$, $M_2$ or $\mu$ show hardly any sensitivity to the $c\tau$ cut. In particular our conclusions about the Higgs signals remain unchanged. 

\begin{figure}[h!]
\begin{center}
\includegraphics[width=0.35\linewidth]{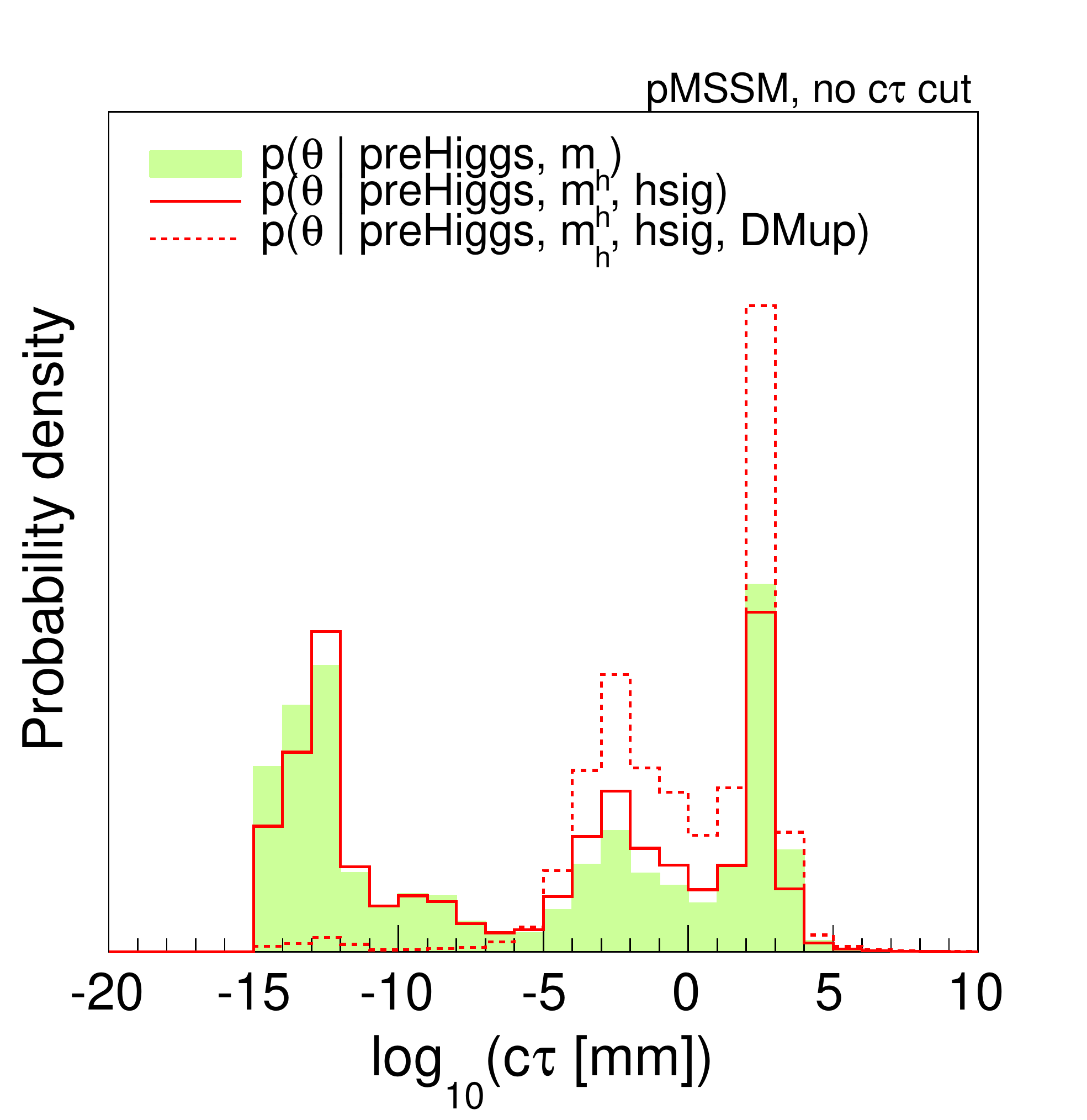}
\caption{Marginalized 1D posterior density of the average $\tilde\chi^\pm_1$ lifetime, $c\tau$ in mm. Color codes as in Fig.~\ref{pmssm-fig:likehiggsNoCtau}.}
\label{pmssm-fig:ctau}
\end{center}
\end{figure}

It is of course also interesting to ask how likely it is at all to have a long-lived chargino. 
To this end we show in Fig.~\ref{pmssm-fig:ctau} the marginalized posterior density of the average $\tilde\chi^\pm_1$ lifetime. 
We find that the probability of $c\tau>10$~mm is 28\%, 25\% and 47\% at the 
``preHiggs+$m_h$'', ``preHiggs+$m_h$+hsig'', and  ``preHiggs+$m_h$+hsig+DMup'' levels, respectively. 

\subsubsection{Interplay with dark matter searches}

\begin{figure}[ht]
\begin{center}
\includegraphics[width=0.3\linewidth]{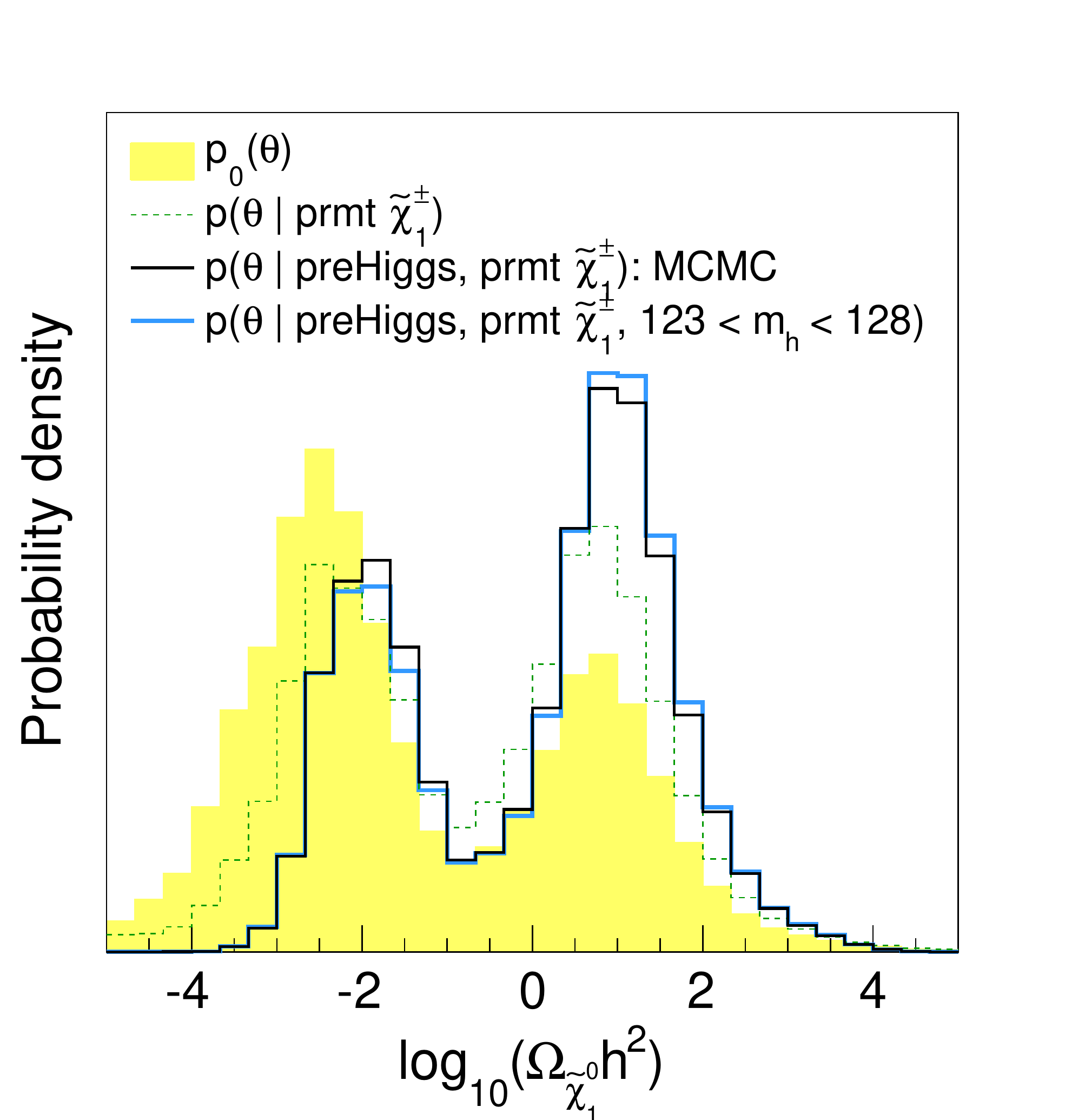}
\includegraphics[width=0.3\linewidth]{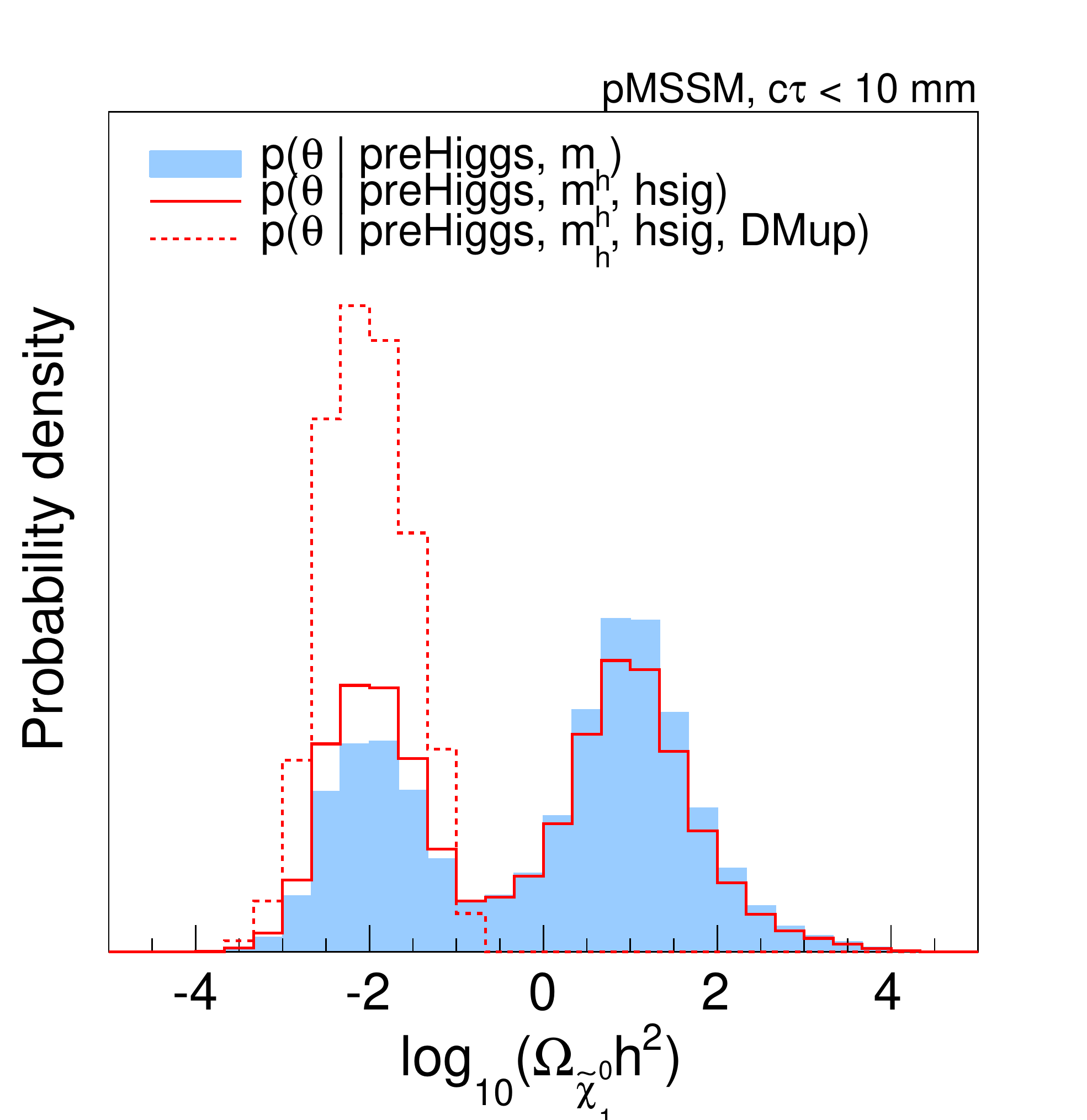}\includegraphics[width=0.3\linewidth]{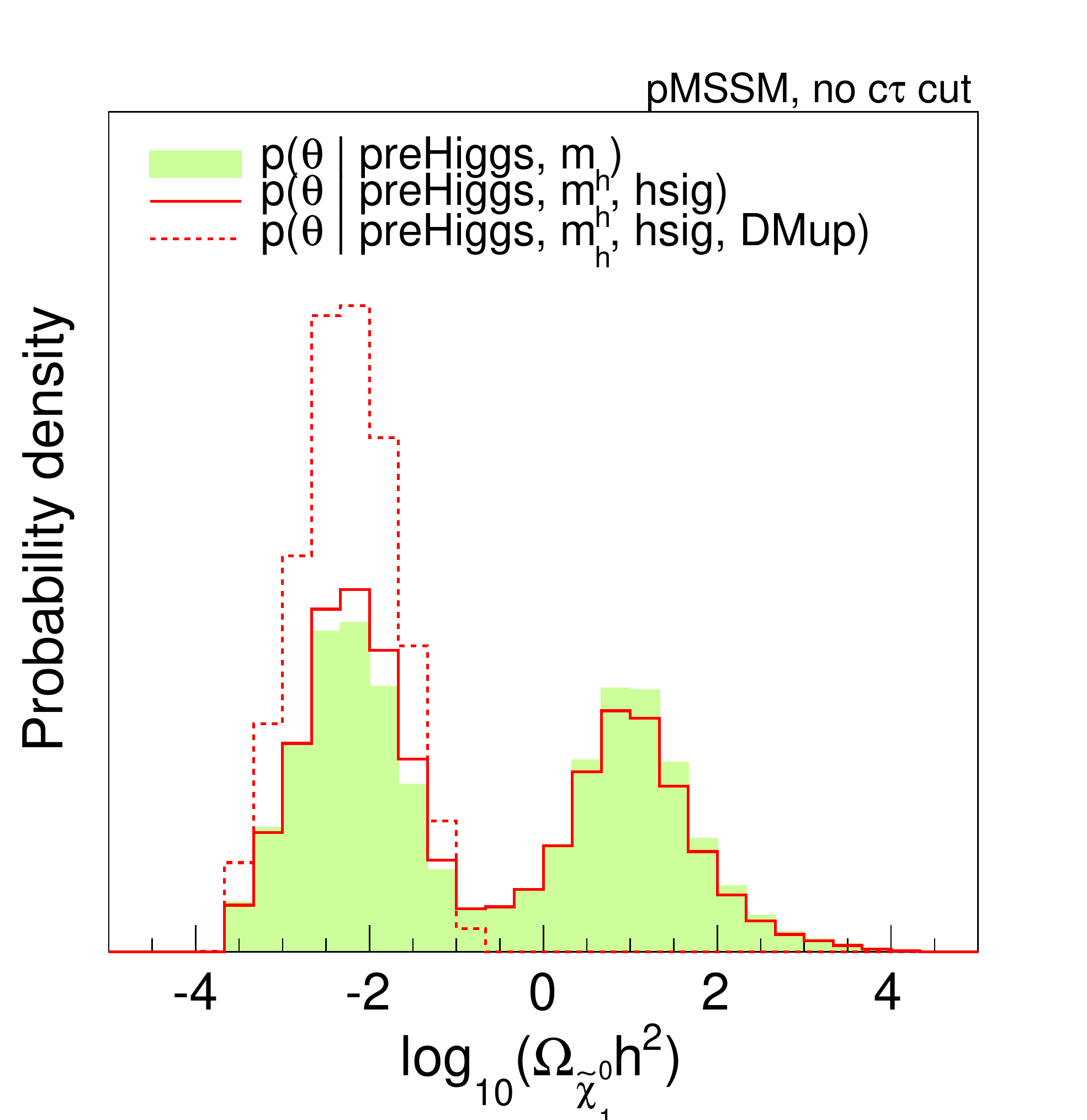}\\
\includegraphics[width=0.3\linewidth]{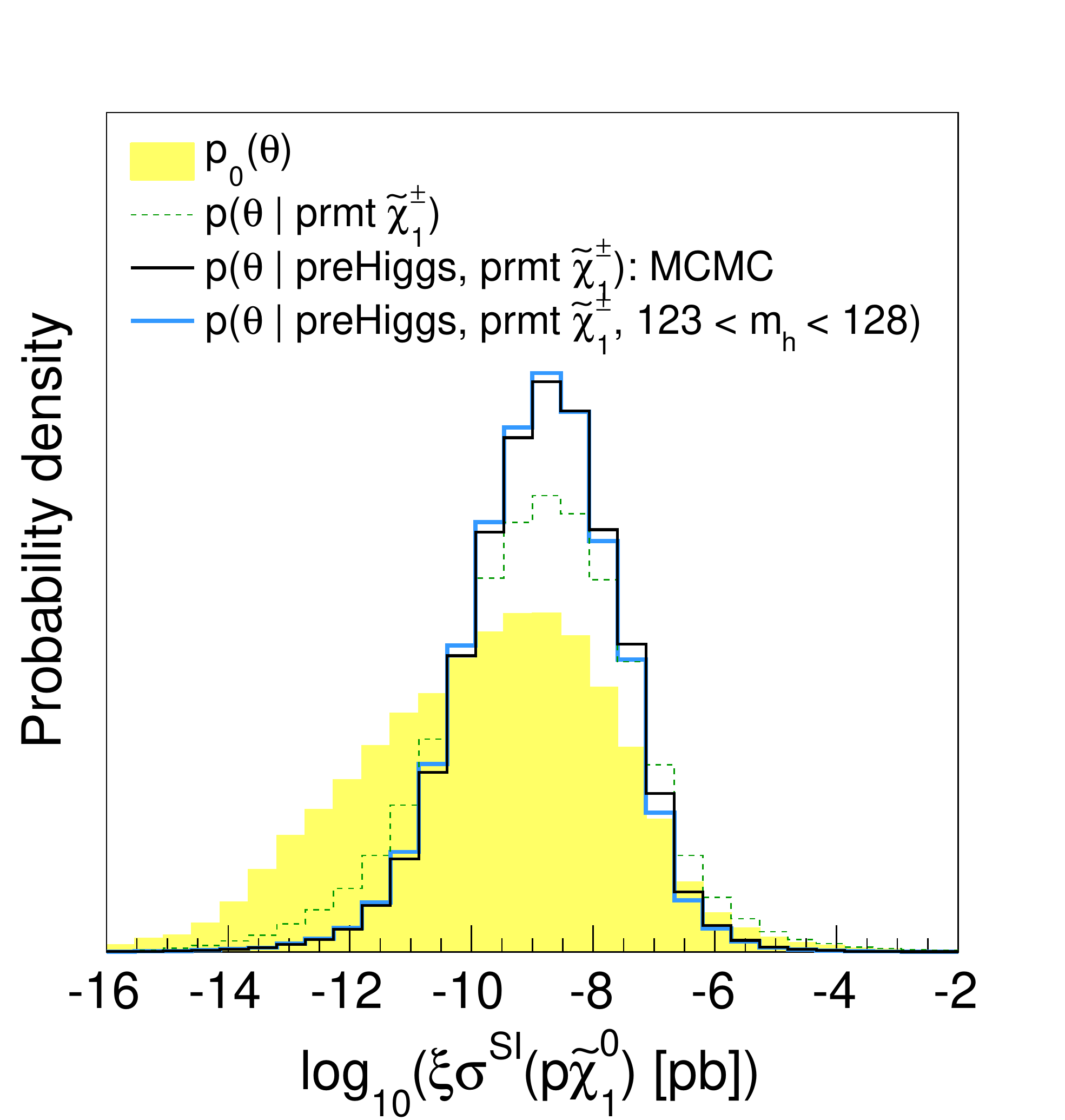}
\includegraphics[width=0.3\linewidth]{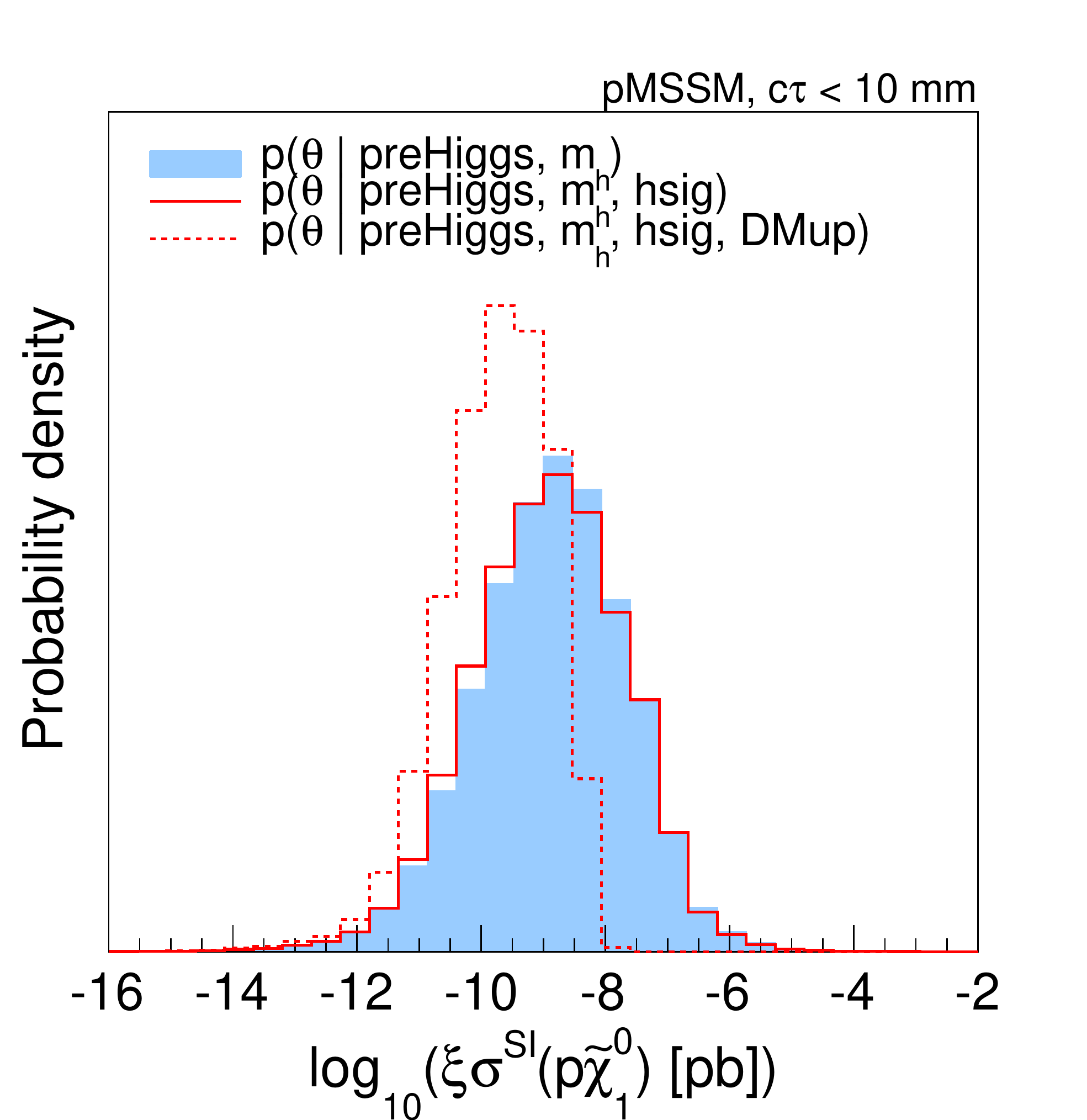}\includegraphics[width=0.3\linewidth]{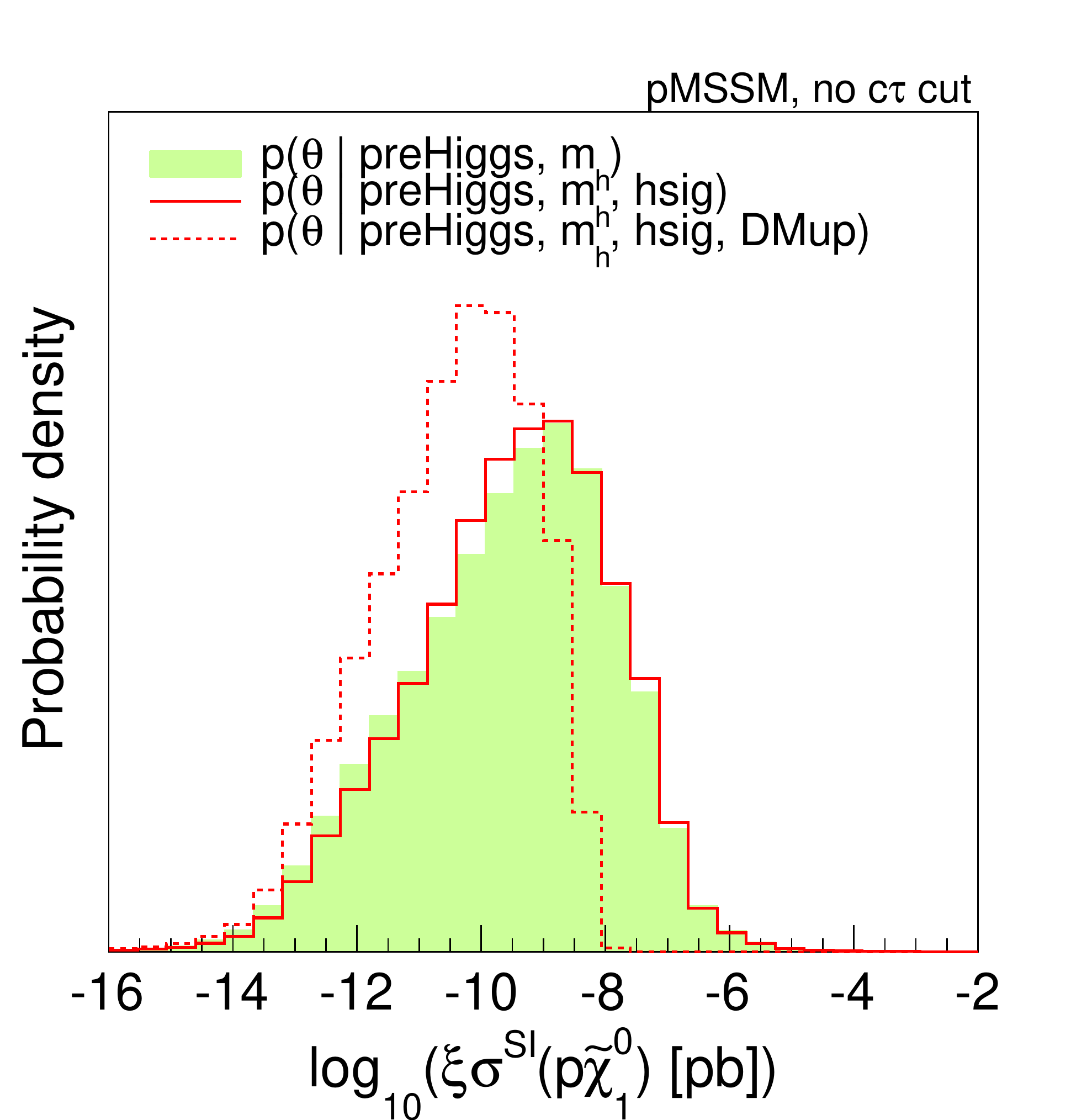}
\caption{Marginalized 1D posterior densities for dark matter quantities. 
Color codes as in Fig.~\ref{pmssm-fig:sampling1} (left), Fig.~\ref{pmssm-fig:likehiggs1} (middle) and Fig.~\ref{pmssm-fig:likehiggsNoCtau}
(right).  }
\label{pmssm-fig:darkmatter1D}
\end{center}
\end{figure}

\begin{figure}[ht]
\begin{center}
\includegraphics[width=0.38\linewidth]{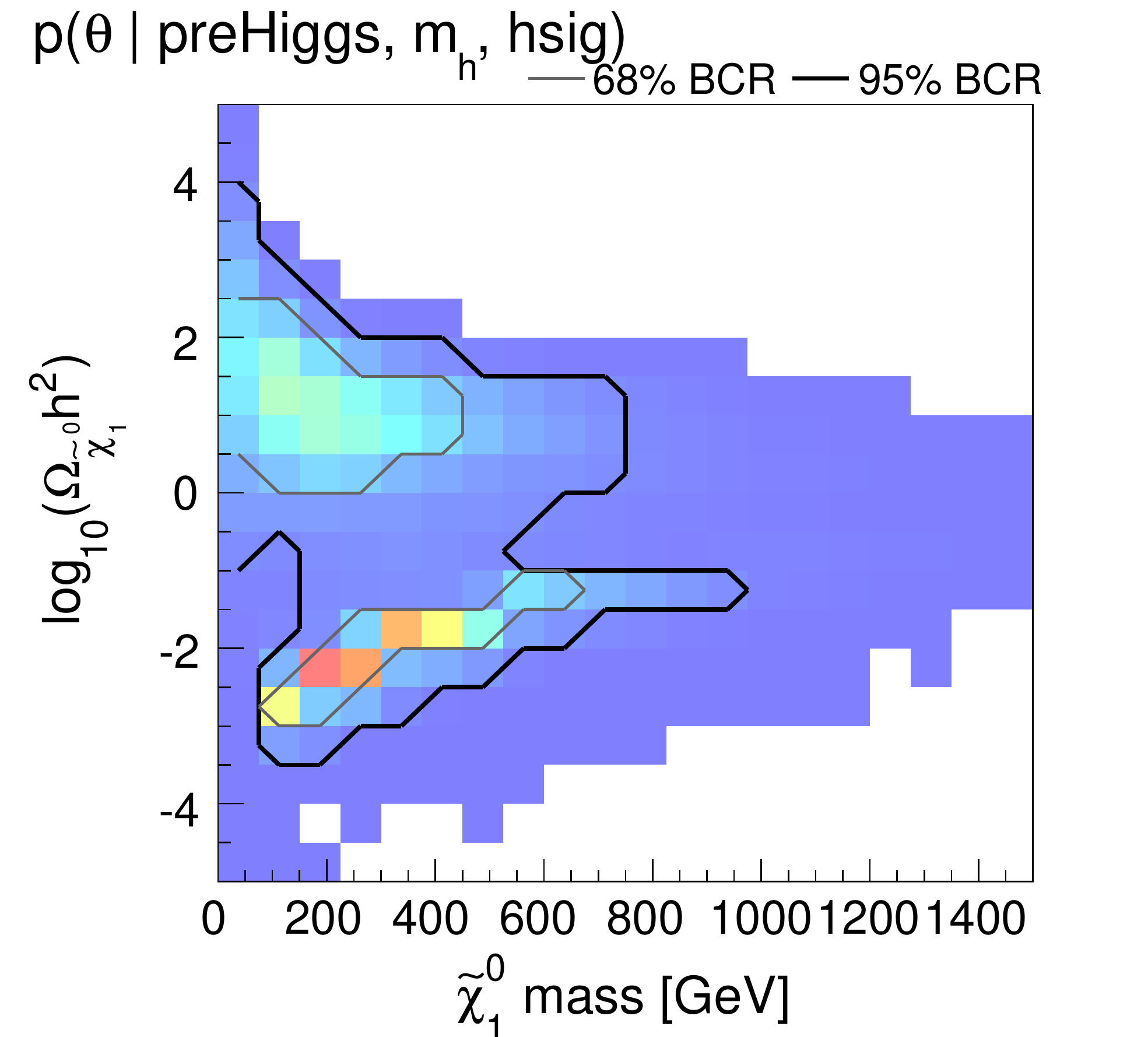}
\includegraphics[width=0.38\linewidth]{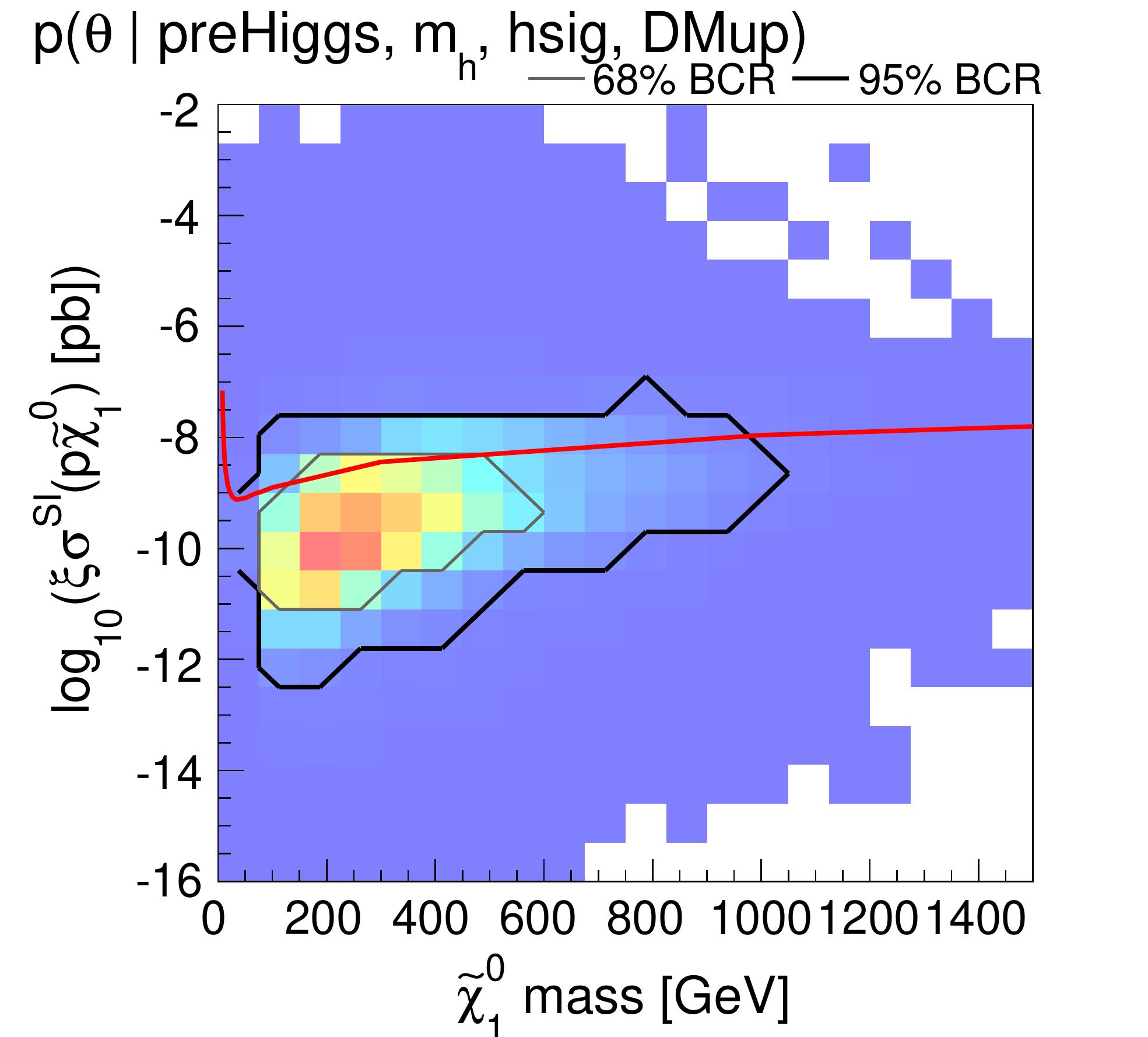}
\caption{Marginalized 2D posterior densities for dark matter quantities. The probability density is represented by color shading, ranging from low values in blue to high values in red. The gray and black lines are contours of 68\% and 95\% Bayesian Credibility, respectively. The red line in the right plot is the 90\%~CL limit from LUX.  }
\label{pmssm-fig:darkmatter2D}
\end{center}
\end{figure}

As discussed above, the dark matter requirements ({\it i.e.}, imposing upper limits on the relic density and on 
the spin-independent scattering cross section) have a significant impact on the MSSM parameters and masses, 
and even on the $h$ signal strengths. 
In this subsection, we now focus on dark matter observables themselves. 
Results for the neutralino relic density $\omhsq$ and the re-scaled spin-independent scattering cross section 
$\xi\sigsi$ are shown in Figs.~\ref{pmssm-fig:darkmatter1D} and\ \ref{pmssm-fig:darkmatter2D}. 

Let us start the discussion with the 1D distributions of $\log_{10}(\omhsq)$, shown in the upper row of plots in Fig.~\ref{pmssm-fig:darkmatter1D}. Already the $p_0(\theta)$ distribution shows a two-peak structure with the minimum actually lying near the cosmologically preferred value $\omhsq\approx 0.1$. This distribution is shifted to significantly higher values by the preHiggs constraints. Concretely, at preHiggs level, the probability for $\omhsq<0.14$ is 36\% (53\%) 
with (without) the prompt chargino requirement. This hardly changes when including also the requirement of $m_h=123-128$~GeV: $p(\omhsq<0.14)\simeq 34\%$ (53\%) in this case.  The Higgs signal likelihood has a larger effect, shifting the distribution towards lower $\omhsq$. This is mainly due to the preference for smaller $\mu$ induced by the 
Higgs signal likelihood. The effect is thus less pronounced without the $c\tau$ cut (RH-side plot) than with the $c\tau$ cut (middle plot). Concretely, we find $p(\omhsq<0.14)\simeq 43\%$ (57\%) with (without) the $c\tau$ cut. 
The peak at high $\omhsq$ values is of course completely removed by the DMup constraints. The probability of lying within the Planck window defined by $\omhsq=0.119\pm 0.024$ ($0.024$ being the $2\sigma$ error, dominated by theory uncertainties) is, for all three of the above cases, $\sim 1.1\%$ with the $c\tau$ cut and $\sim 0.9\%$ without the $c\tau$ cut.

Turning to the predictions for direct dark matter detection, we observe that the preHiggs constraints limit the probability of having very small values of $\xi\sigsi$. This is true with and without the $c\tau$ cut, though the effect is larger with the $c\tau$ cut. The latter is due to the fact that the prompt chargino requirement removes the pure wino-LSP scenarios which have extremely small $\omhsq$ and $\xi\sigsi$ (recall that $\xi=\omhsq/0.119$). Requiring consistency with the Higgs signal strengths has only a small effect, somewhat preferring smaller values of $\xi\sigsi$ 
because of the larger LSP higgsino component. 

The 2D distributions of $\omhsq$ and $\xi\sigsi$ versus the $\tilde\chi^0_1$ mass are shown in Fig.~\ref{pmssm-fig:darkmatter2D}. We observe that on the one hand the neutralino LSP can have mass up to 1~TeV at 95\% BC without conflicting with the DM constraints. Very low neutralino masses, on the other hand, are severely constrained by DM requirements. Note, moreover, that  
the most likely values lie around $m_{\tilde\chi^0_1}\approx200$--$300$~GeV, $\omhsq\approx 10^{-2}$ and $\xi\sigsi\approx 10^{-10}$~pb.

\FloatBarrier

\subsubsection{Consequences of future $h$ signal strength measurements}

\begin{figure}[h!]
\begin{center}
\includegraphics[width=0.3\linewidth]{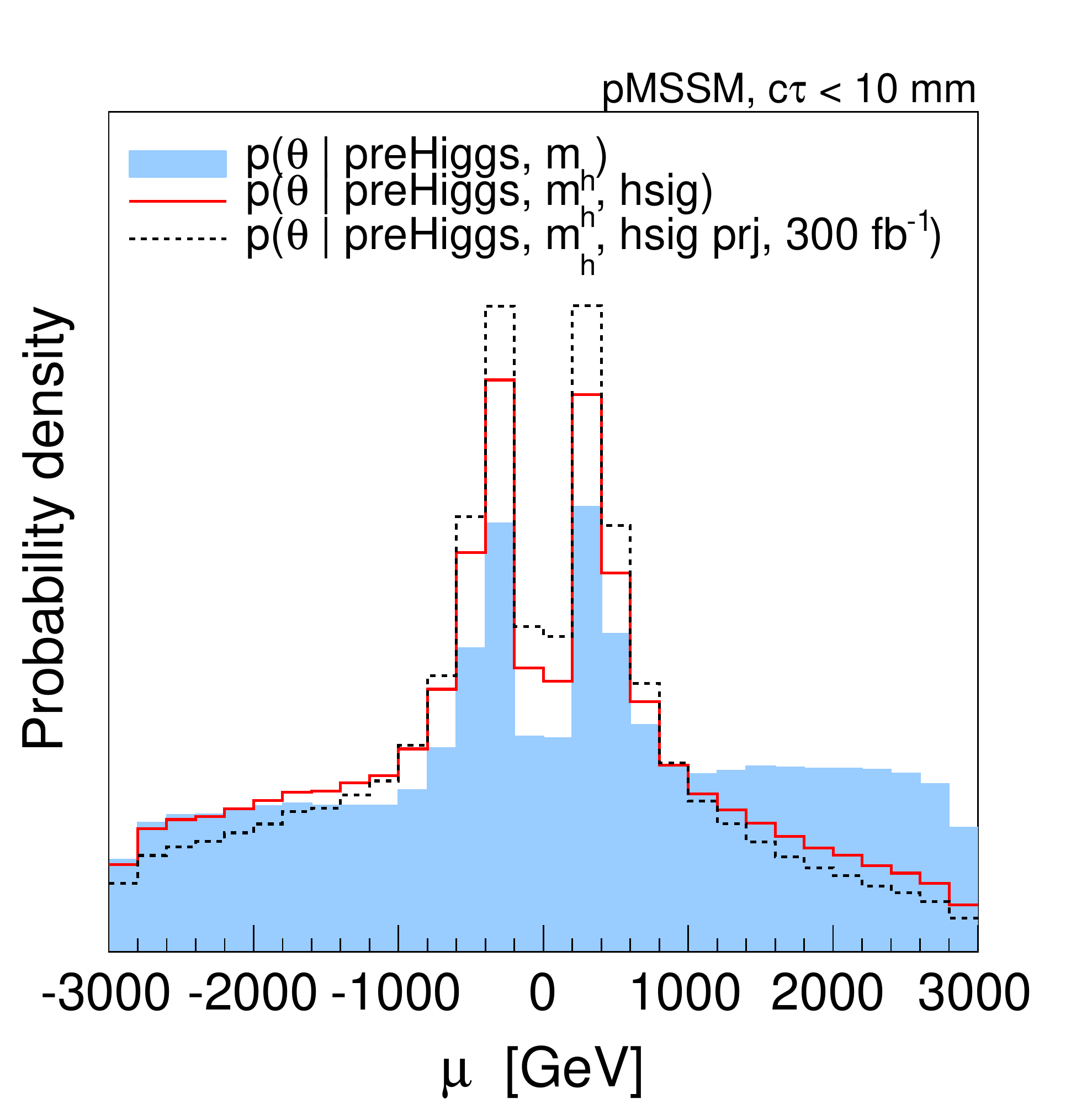}
\includegraphics[width=0.3\linewidth]{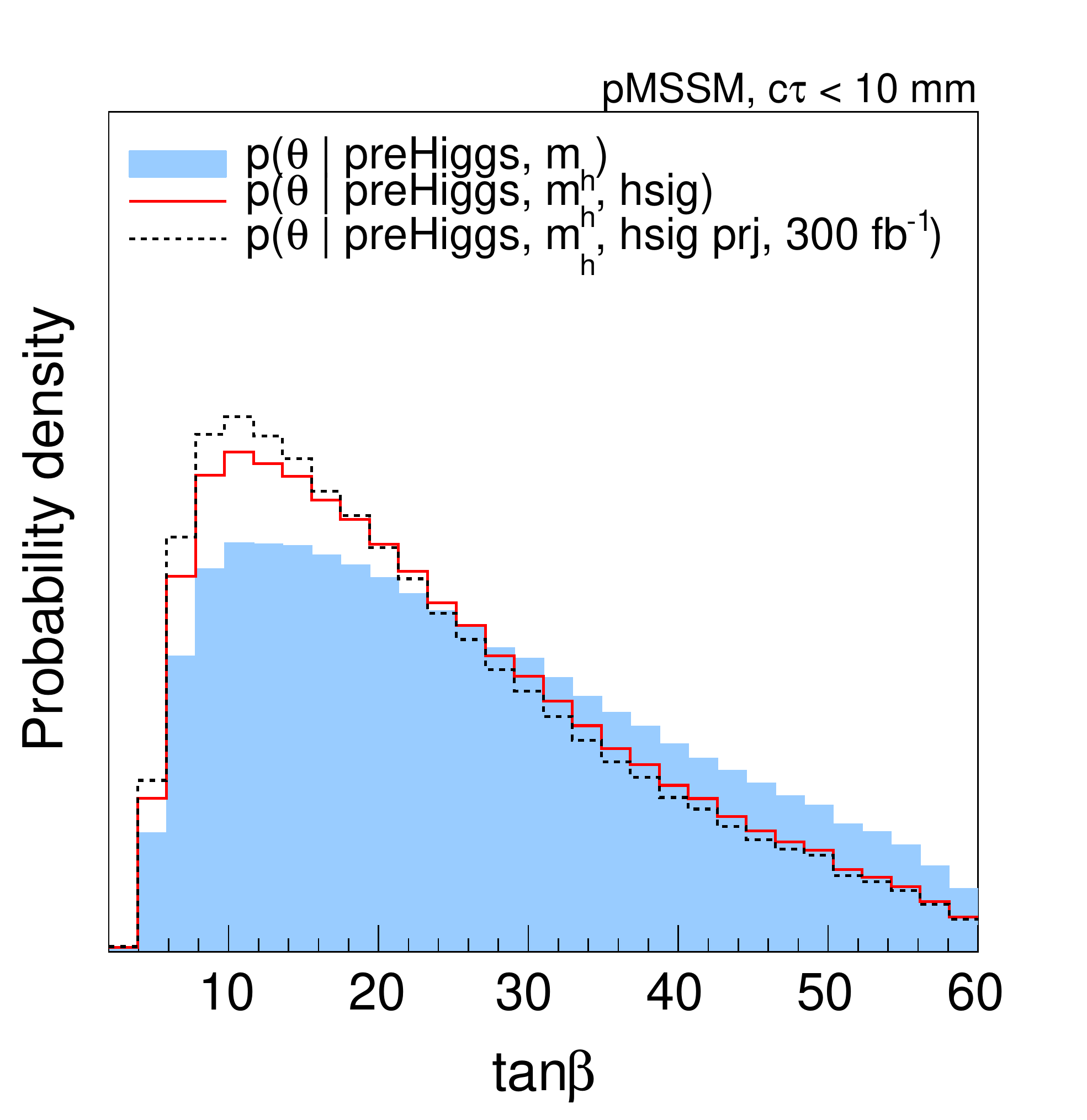}
\includegraphics[width=0.3\linewidth]{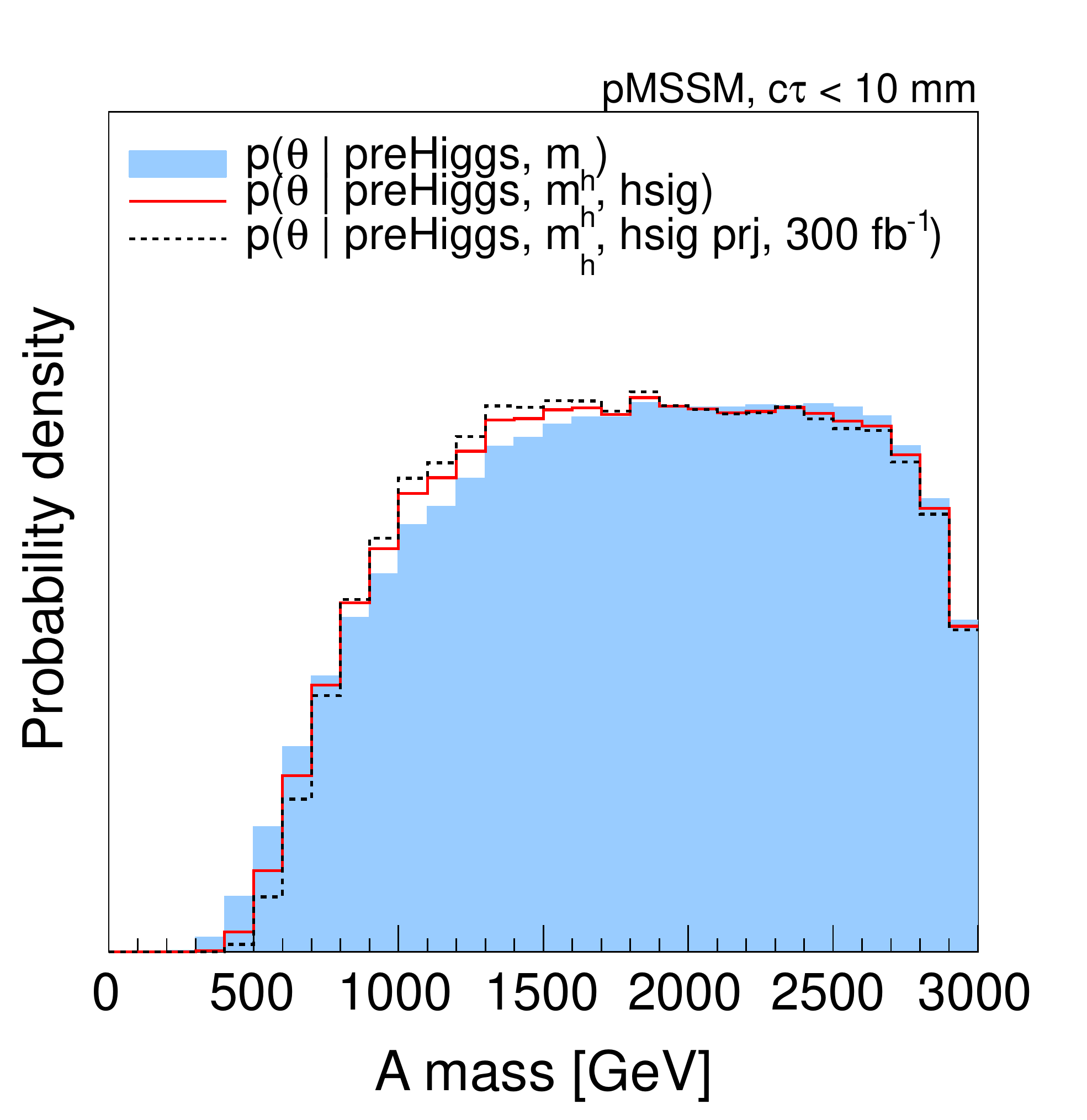}
\caption{Marginalized 1D posterior densities for some MSSM parameters, showing the effect of all $h$ signal strengths being $\approx 1$ with uncertainties as expected for 300$\fbi$ of data at 14~TeV, {\it cf.}\ Eq.~(\ref{pmssm-eq:projection}).}
\label{pmssm-fig:proj1d}
\end{center}
\end{figure}

It is also interesting to consider what happens if, with precision data at the next run of the LHC, 
the Higgs signal strengths have an even narrower probability distribution around unity. 
We estimate the precision attainable with 300$\fbi$ at 14~TeV based on \cite{ATL-PHYS-PUB-2012-004,CMS-NOTE-2012-006}
\begin{eqnarray}
 &\mu({\rm ggF+ttH},\gamma\gamma) = 1 \pm 0.1\,,  \qquad 
    \mu({\rm VBF+VH},\gamma\gamma) = 1 \pm 0.3\,,  \nonumber \\
 &\mu({\rm ggF+ttH},VV) = 1 \pm 0.1\,, \qquad 
    \mu({\rm VBF+VH},VV) = 1 \pm 0.6\,,  \nonumber \\
 &\mu({\rm ggF+ttH},b\bar b) = 1 \pm 0.6\,, \qquad 
    \mu({\rm VBF+VH},b\bar b) = 1 \pm 0.2\,,  \nonumber  \\
 &\mu({\rm ggF+ttH},\tau\tau) = 1 \pm 0.2\,, \qquad 
    \mu({\rm VBF+VH},\tau\tau) = 1 \pm 0.2 \,.
\label{pmssm-eq:projection}
\end{eqnarray}
The effect of these hypothetical results is illustrated in Fig.~\ref{pmssm-fig:proj1d}. We conclude that if the Higgs signal remains SM-like (but with smaller uncertainties), the effects already observed on some SUSY parameters are only slightly strengthened by more precise measurements.

\begin{figure}[t!]
\begin{center}
\includegraphics[width=0.3\linewidth]{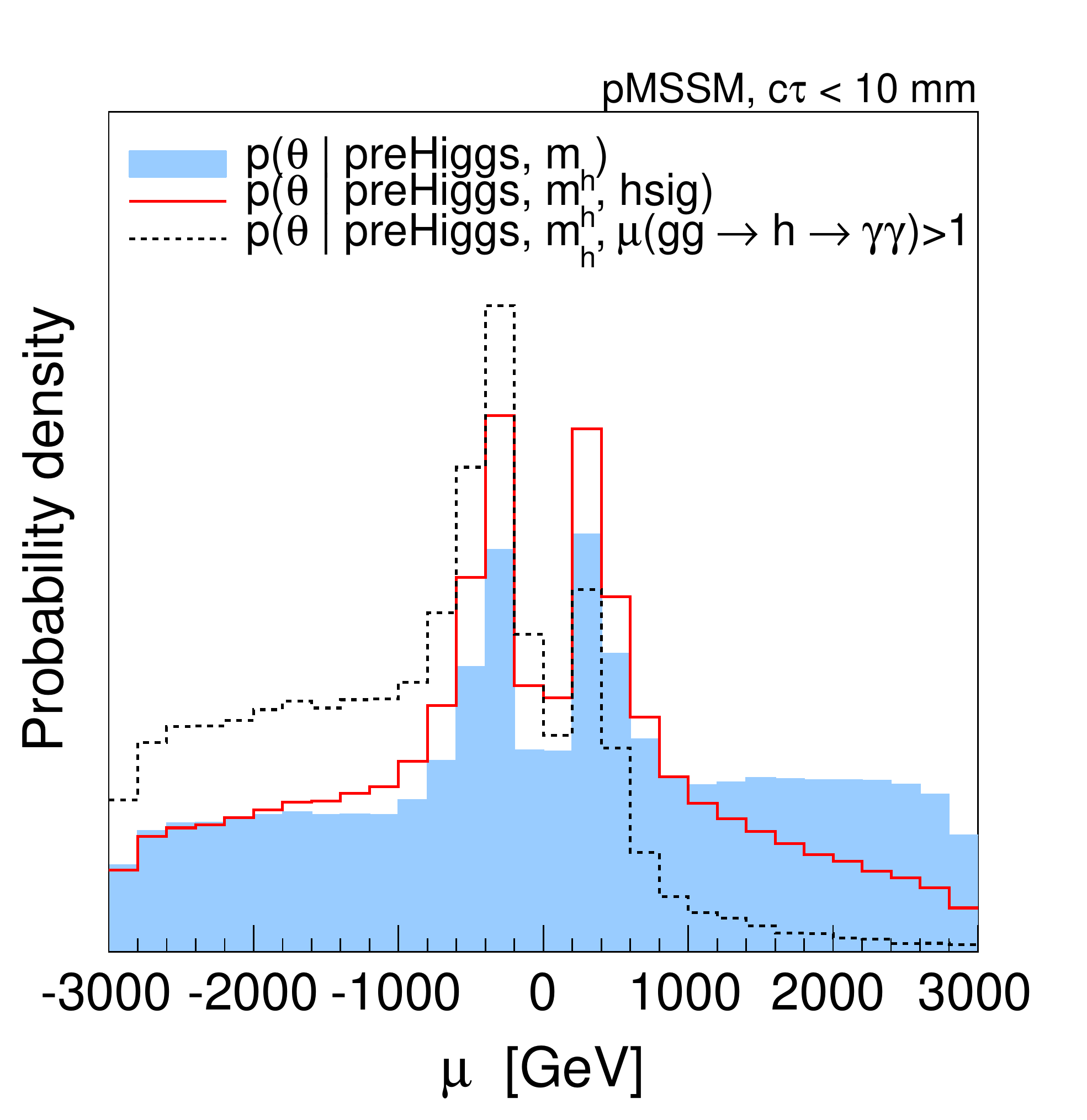}
\includegraphics[width=0.3\linewidth]{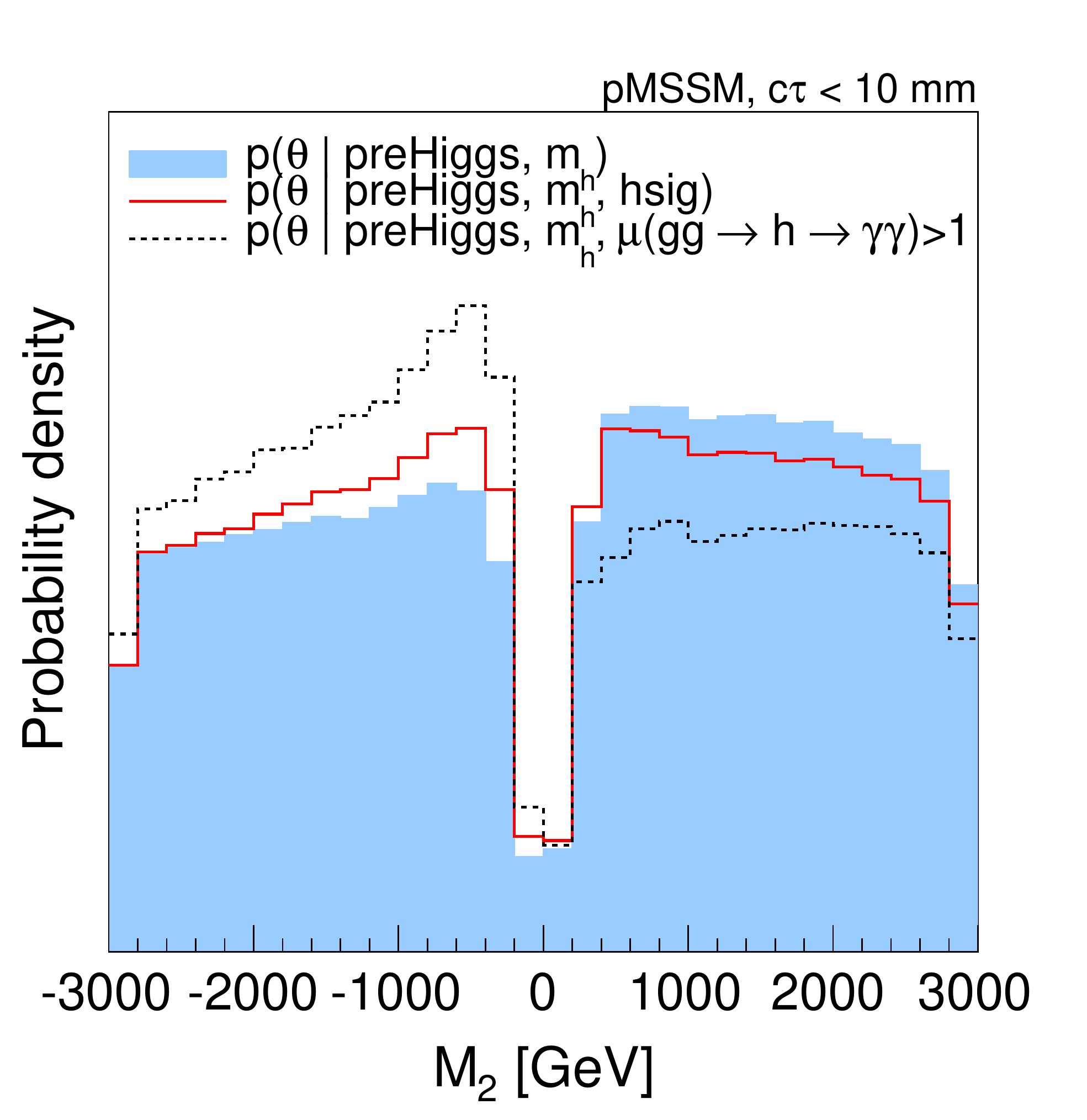}
\includegraphics[width=0.3\linewidth]{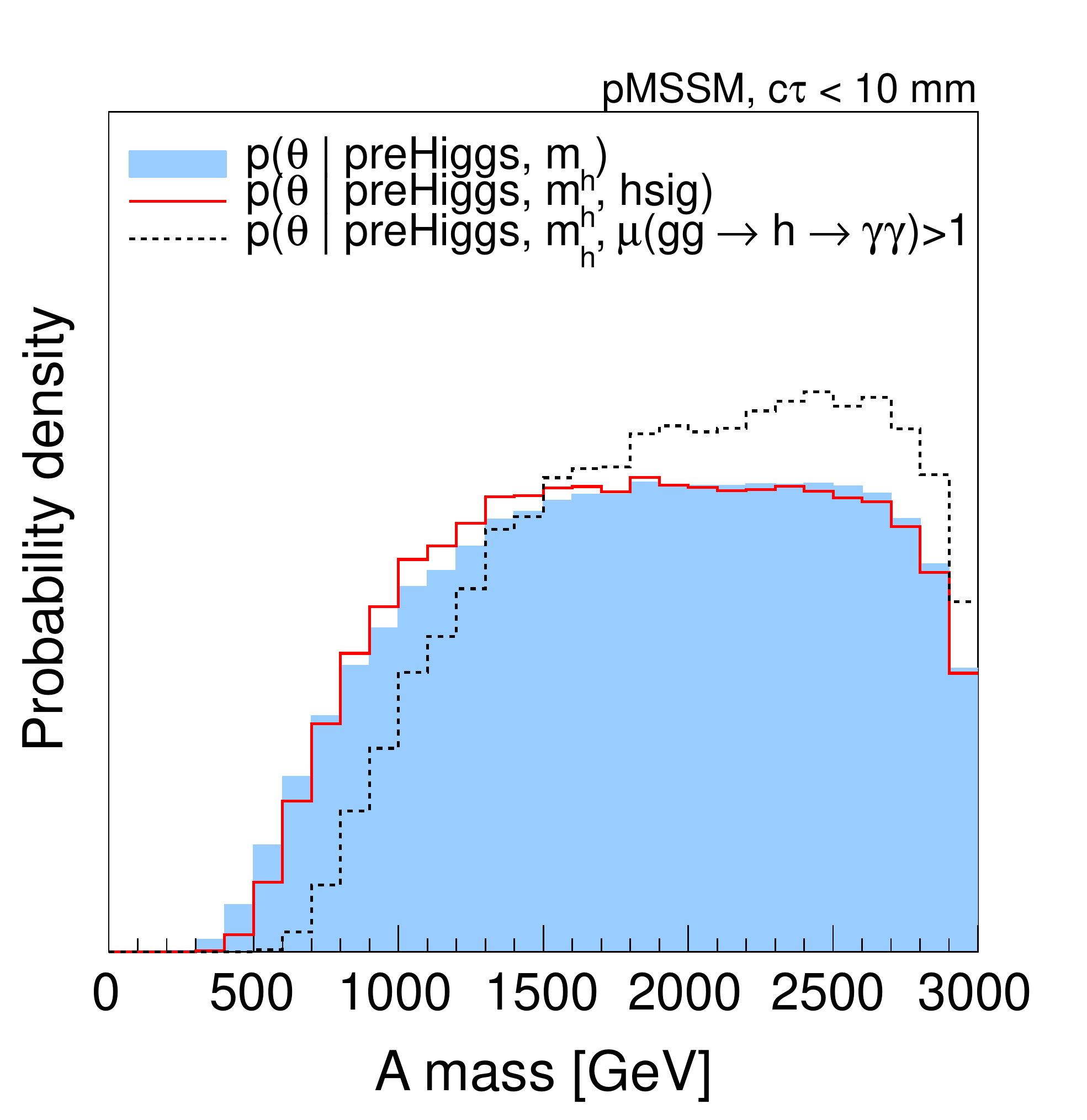}\\
\includegraphics[width=0.3\linewidth]{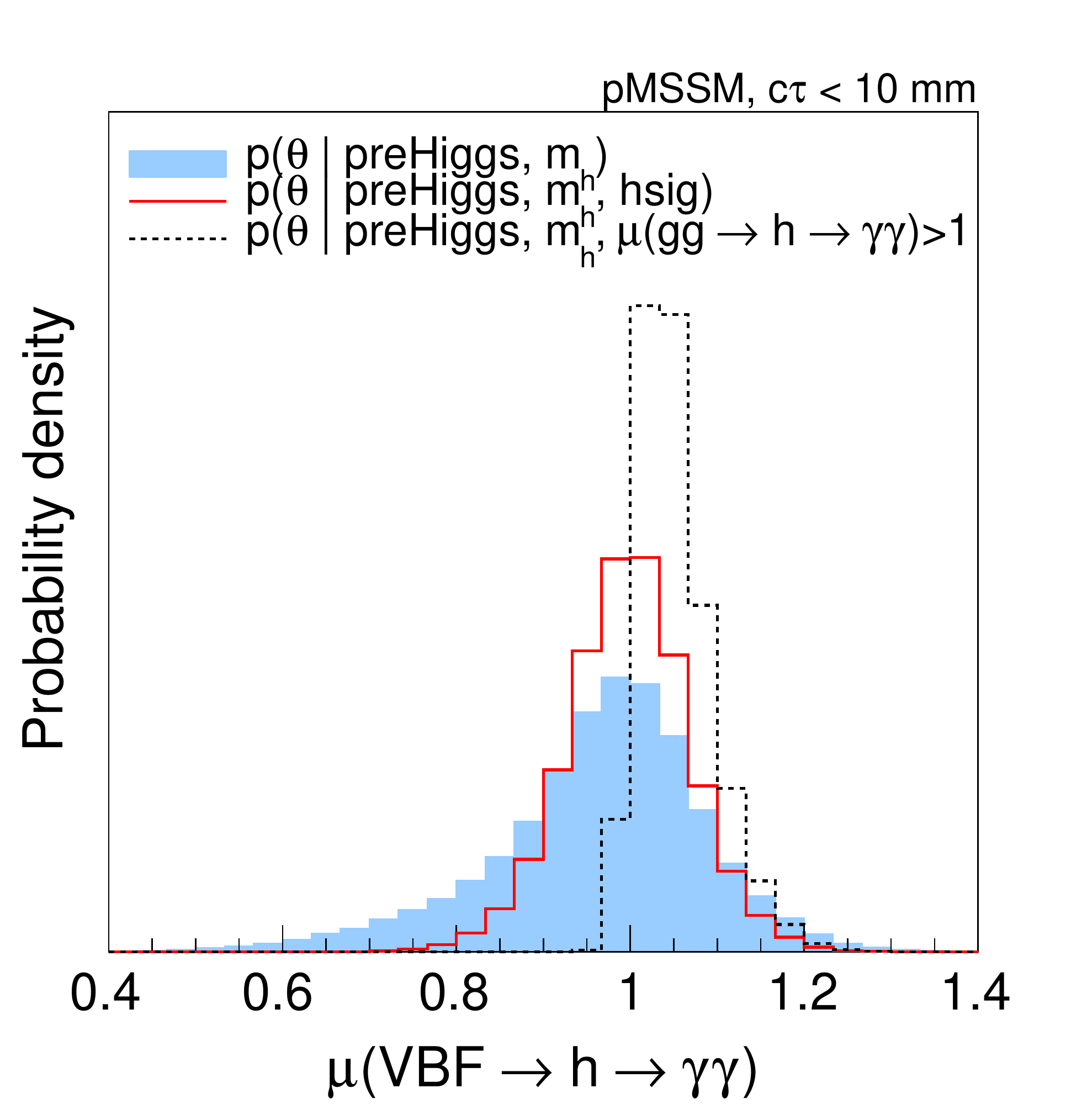}
\includegraphics[width=0.3\linewidth]{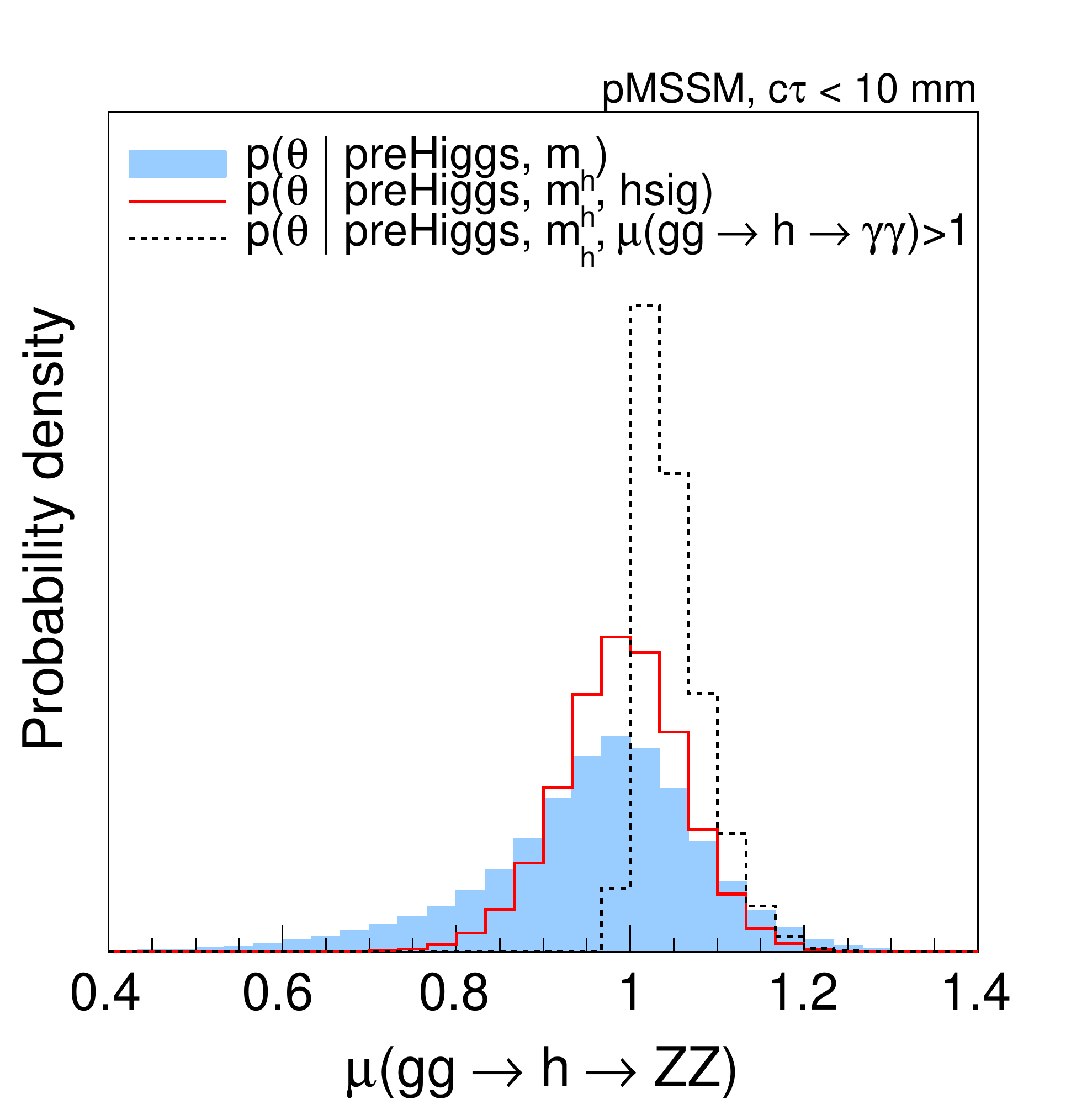}
\includegraphics[width=0.3\linewidth]{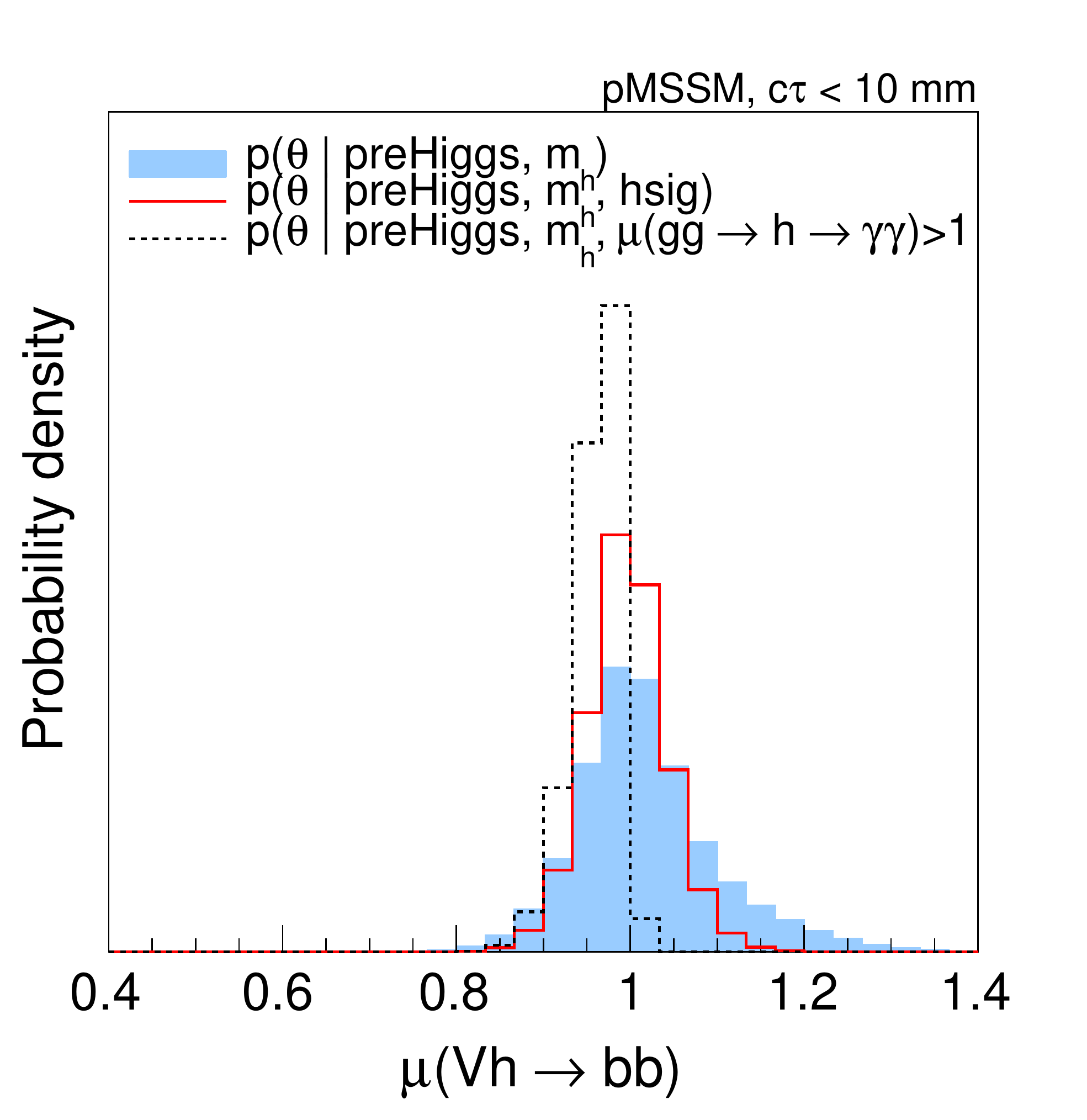}
\caption{Marginalized 1D posterior densities for  selected MSSM parameters and $h$ signal strengths, showing the effect of a hypothetical future determination of $\mu(gg\to h\to\gamma\gamma)>1$. }
\label{pmssm-fig:mugt1}
\end{center}
\end{figure}

The picture is quite different should the signal strength finally turn out to be larger than one. For illustration, we assume  
$\mu({\rm ggF},\gamma\gamma)>1$ and show in Fig.~\ref{pmssm-fig:mugt1} the impact on some other quantities.
As we have seen, $\Delta_b < 0$ corresponds to a suppression of $h \to b\bar b$ and, hence, to the enhancement of all other signal strengths. This is how one obtains $\mu({\rm ggF},\gamma\gamma)>1$ in our case. This leads to a strong preference for $\mu < 0$ and to an associated asymmetry for the  $M_2$ distribution. Moreover, strong evidence for $\mu({\rm ggF},\gamma\gamma)>1$ would strongly disfavor a CP-odd Higgs lying close to the current CMS bound because of the impact of $m_A$ on the tree-level coupling $hbb$.
Finally, $\mu({\rm ggF},\gamma\gamma)>1$ would also imply a preference for an enhancement of the diphoton signal in VBF production, as well as an enhancement of the $ZZ$ mode in both ggF and VBF. This is accompanied at the same time by the expected suppression of $Vh\to b\bar b$. Nonetheless, signal strength values close to 1 are still the most likely ones.

\FloatBarrier

\subsection{Conclusions}

We have performed a Bayesian analysis of the pMSSM taking into account the latest LHC results 
on the Higgs signal at $125.5$ GeV in addition to relevant low-energy observables and LEP constraints. 
We find that the requirement of obtaining the right $m_h$ strongly favors $|X_t/M_{\rm SUSY}|\approx 2$, 
\ie\ near-maximal (but not maximal) stop mixing.  Coincidently, such near-maximal mixing is also favored 
by naturalness arguments~\cite{Wymant:2012zp}. 

The constraints from the Higgs signal strengths in the various production$\times$decay modes, on the other hand, 
have an important influence on the posterior distributions of $\mu$ and $\tan\beta$, and hence on the electroweak-ino 
spectrum. 
Concretely, low values of $\mu$ and $\tan\beta\approx 10$ are favored. 
This is mainly due to radiative corrections to the bottom Yukawa coupling, which are proportional 
to $\mu\tan\beta$ and can significantly modify the total Higgs width. As a consequence, 
$\chitz$ and $\chipm$ masses below about 500~GeV are favored, as are LSPs with 
a significant higgsino fraction. 
While there is of course still a substantial tail at large masses, these results suggest that the Higgs data yield 
a certain preference for natural-SUSY-like scenarios. 

Regarding the heavy Higgs states, $H$ and $A$, we find that $m_{H,A}\gtrsim 500$~GeV mostly due to B-physics 
constraints. The 125~GeV Higgs data give only a small additional constraint; they mostly affect the heavy Higgses 
through their effect on $\tan\beta$. 
The limits from direct searches for $H,A\to \tau\tau$ at 7--8~TeV are less sensitive. 
If $m_A\lsim 1$~TeV, prospects for discovery of $H$ and $A$ at the next LHC run are substantial.  
Because $\tanb\gtrsim 10$ is preferred, we find that $b\anti b\to H,A$ typically dominates (by about a factor of 30) over  gluon fusion, with $\sigma(b\anti b \to H,A)\br(H,A\to \tau\tau)$ of the order of a few fb. 

 We have also explored the impact of DM limits associated with  $\omhsq$ and $\xi\sigsi$ on the Higgs bosons in the pMSSM context as well as the impact of the Higgs precision data on these same DM observables.  The most probable values for $\omhsq$ lie in the vicinity of $10^{-2}$, implying that DM would not consist entirely of the $\cnone$ 
(or that the missing abundance of $\cnone$ is substituted by non-thermal production). 
 The probability for obtaining $\omhsq$ within the Planck window is only of order 1\%: 
to get the correct annihilation rate, the $\cnone$ has to have a carefully balanced composition, 
or a  mass that is fine-tuned with respect to the $A$ or co-annihilating sparticles. Imposing the upper limit on $\omhsq$, we find $m_{\cnone}\in
[100,\,760]$~GeV and $\xi\sigsi\gtrsim 3.5\times 10^{-12}$~pb at 95\%
BC.

While we have not taken into account the recent LHC limits from direct SUSY searches, we have checked that 
our conclusions do not change when requiring gluino and squark masses above 1~TeV. The conclusions drawn 
from the Higgs sector are thus orthogonal to those from the SUSY searches. 
In particular, 
this makes our results directly comparable to the pMSSM interpretation of the CMS SUSY searches at 7--8~TeV~\cite{CMS-PAS-SUS-12-030,CMS-PAS-SUS-13-020}.

The 13--14~TeV run of the LHC will provide increased precision for Higgs measurements as well as 
a higher reach for SUSY particles.  
Particularly relevant in point of view of an interplay between Higgs and SUSY results is  
an improved sensitivity for higgsinos, gluinos and 3rd generation squarks.  
It will be interesting to see if a tension between Higgs results and SUSY limits arises or if there is a convergence as a result of the discovery of,  \eg,  light charginos and neutralinos.  Last but not least, if the Higgs boson is found in the end to have an enhanced $h\to \gam\gam$ rate compared to the SM, implications for $\mu$ and $M_2$ are substantial, $m_A$ is shifted to higher values and $\mu(Vh\to Vb\anti b)$ is suppressed --- allowing for some possibility of verifying consistency with or creating tension within the pMSSM.


 \section[Lilith: a new public tool for constraining BSM scenarios from Higgs measurements]{Lilith: a new public tool for constraining BSM scenarios from Higgs measurements%
\sectionmark{Lilith: constraining BSM physics from Higgs data}}
\sectionmark{Lilith: constraining BSM physics from Higgs data} \label{sec:lilith}
 
In the work presented in Sections~\ref{sec:higgs2012}--\ref{sec:pmssm}, the definition of the Higgs likelihood from the experimental results and its evaluation from a set of reduced couplings was done by a single {\tt FORTRAN} code. As more and more Higgs results were released, we found it necessary to develop a new, modular program for evaluating the Higgs likelihood, where the user input---specifying the modifications to the properties of the 125~GeV Higgs boson---and the experimental input entering the likelihood are decoupled from the main code.
Moreover, only measurements in the Gaussian approximation were taken into account in the {\tt FORTRAN} code, while the full likelihood in the 2D plane $(\mu({\rm ggF+ttH}, Y), \mu({\rm VBF+VH}, Y))$ recently became available for some final states $Y$ ($\gamma\gamma$, $WW^*$ and $ZZ^*$ for ATLAS~\cite{atlasgamgamgrid,atlasZZgrid,atlasWWgrid}, $\gamma\gamma$ for CMS~\cite{Khachatryan:2014ira}).
Finally, Higgs measurements are relevant for constraining a large variety of new physics models, while it is not trivial to define a good approximation to the Higgs likelihood.

This motivated us working on a public modular tool for applying the Higgs constraints on models of new physics. In collaboration with J\'er\'emy Bernon, another PhD student at the LPSC Grenoble, the development of a {\tt Python} program with this aim, called {\tt Lilith}, started at the end of 2013. The public release of the program and of its associated manual is in preparation~\cite{lilith}, but a beta version can already be found at~\cite{lilith-webpage}.
In addition to being a seductive female demon, {\tt Lilith} stands for ``\uline{li}ght \uline{li}kelihood fi\uline{t} for the \uline{H}iggs''. It is designed as a light and user-friendly program, where user and experimental input are stored in {\tt XML} files which are easy to modify.
The experimental results shipped with the program consist mostly in signal strengths in the theory space, when available, as was used in Sections~\ref{sec:higgs2012}--\ref{sec:pmssm}. This can easily be extended with any experimental result given in terms of signal strengths.
Below we briefly summarize the main features of {\tt Lilith}.

{\tt Lilith} works with two types of user inputs, provided in the form of {\tt XML} files. A first possibility is to provide directly signal strengths in the theory space ({\it i.e.}, in terms of $\mu(X,Y)$ with $X$ and $Y$ being production and decay modes of the SM Higgs). This is relevant if these are pre-calculated quantities, or if the user prefers to compute signal strengths himself/herself.
Alternatively, the user can provide as input to {\tt Lilith} a list of reduced couplings, again in {\tt XML} format. This list may or may not contain reduced couplings to gluons and photons; if not provided, they will be internally calculated from the reduced couplings to SM particles. The set of reduced couplings is used by the program to compute signal strengths in the theory space, either using analytic formulas at leading order or using grids for the production cross sections and decay widths, as function of the reduced couplings, taking into account (N)NLO QCD corrections as obtained from {\tt HIGLU}~\cite{Spira:1995rr,Spira:1995mt,Spira:1996if}, {\tt VBFNLO}~\cite{Arnold:2008rz} and {\tt HDECAY}~\cite{Spira:1996if,Djouadi:1997yw}.
In all cases (input in terms of signal strengths or reduced couplings), multi-particle labels are defined in order to simplify and improve readability of the input. For example, {\tt "VVH"} is the shortcut for a common scaling of the WH, ZH and VBF production modes, and {\tt "ff"} can be used to rescale all fermionic couplings or decays in the same way.

The experimental results are stored in a database consisting of {\tt XML} files (one per measurement). When {\tt Lilith} is called, a list of experimental results to be considered ({\it i.e.}, a list of {\tt XML} files) should be given. The program was designed to handle a variety of experimental input given in terms of signal strengths in a transparent way. First, measurements can be provided in the Gaussian approximation, either in 1D or in 2D. For a 1D measurement, the central value and the uncertainty at 68\%~CL should be provided, while for a 2D measurement the five parameters that appear in Eq.~\eqref{eq:mu2d} should be provided. Second, measurements beyond the Gaussian approximation can be used as input to the program, again in 1D or in 2D. In the 1D case, a grid making the correspondence between signal strength and likelihood values should be given; the same thing can be done in the 2D case by provided a 2D grid of signal strengths and their associated likelihood values.
This information is then interpolated internally by the program for evaluating the likelihood.
Note that the experimental results given as input may or may not correspond to signal strengths in the theory plane. It is always possible to associate efficiencies for different production and/or decay modes to a measurement, making it possible to give as input any kind of measurement expressed in terms of signal strengths.

The primary output of the program is a likelihood value for the tested scenario. An evaluation of the $p$-value is also given from a naive estimate of the number of degrees of freedom.
The functionalities of {\tt Lilith} can easily be integrated into any {\tt Python} code by importing {\tt Lilith} as a library; examples are provided in the beta version~\cite{lilith-webpage}. The integration of {\tt Lilith} into a {\tt C} and  {\tt C++} code has also been developed and is used for integrating the program into {\tt micrOMEGAs}~\cite{Belanger:2013oya}. Compared to system calls, calling {\tt Lilith} internally as a library has the advantage of being much lighter since the initialization of the program is only done once. This is particularly relevant in the case of large scans.
For couplings fits, the use of minimization algorithms such as those present in MINUIT~\cite{James:1975dr} is relevant for deriving constraints. This can easily be done in {\tt Python} by using, {\it e.g.}, {\tt iminuit}~\cite{iminuit} (a {\tt Python} module that passes low-level MINUIT functionality to {\tt Python} functions).

More details on the structure of {\tt Lilith} and instructions on how to use the program are given on the webpage~\cite{lilith-webpage}. It includes explicit examples for the two input modes and explanations for running {\tt Lilith} as a library in {\tt Python}, for which examples are shipped with the beta version.


\section[Some thoughts on the future of Higgs measurements and likelihoods]{Some thoughts on the future of Higgs measurements and likelihoods%
\sectionmark{Future Higgs measurements and likelihoods}}
\sectionmark{Future Higgs measurements and likelihoods} \label{sec:higgsfuture}

In this chapter, we derived an approximation to the Higgs likelihood based on signal strengths in the theory plane, most notably from the information given by the ATLAS and CMS collaborations in the $(\mu({\rm ggF+ttH}, Y), \mu({\rm VBF+VH}, Y))$ plane for $Y = \gamma\gamma$, $ZZ^*$, $WW^*$, $b\bar b$, and $\tau\tau$. This approach, presented in Section~\ref{sec:higgs-npconstlhc}, was shown to provide a good approximation to the full likelihood in fits to reduced Higgs couplings in Section~\ref{sec:higgs2013}, and was used to constrain new physics in Sections~\ref{sec:higgs2012}--\ref{sec:pmssm} in the context of effective parameterizations or concrete extensions of the SM.
The definition of the approximate Higgs likelihood was extended and refined in {\tt Lilith}, a new public tool that evaluates the likelihood for modified properties of the 125~GeV Higgs boson using the latest information given by the experimental collaborations, as was presented in Section~\ref{sec:lilith}. 

As we saw, the full likelihood in the 2D plane $(\mu({\rm ggF+ttH}, Y), \mu({\rm VBF+VH}, Y))$ has already been provided by the experimental collaborations for some final states. 
Hopefully, this information will be released systematically during Run~II of the LHC and provided in a numerical form.
There are however limitations when defining our approximation to the full Higgs likelihood from these 2D planes. It is therefore interesting to think of new ways of presenting the LHC Higgs results in the future. This was the motivation for the note ``On the presentation of the LHC Higgs Results'', Ref.~\cite{Boudjema:2013qla}, to which I actively contributed. This note was submitted to arXiv on July 22, 2013, and originated from the workshops ``Likelihoods for the LHC Searches'', 21-23 January 2013 at CERN, ``Implications of the 125~GeV Higgs Boson'', 18-22 March 2013 at LPSC Grenoble, and from the 2013 Les Houches ``Physics at TeV Colliders'' workshop. This was built upon the recommendations given in the ``Les Houches Recommendations for the Presentation of LHC Results''~\cite{Kraml:2012sg}, which stressed the importance of providing all relevant information, including the best-fit signal strengths, on a channel-by-channel basis for the independent production and decay processes. In this section, some of the discussion present in Ref.~\cite{Boudjema:2013qla} will be reproduced.

\subsection{Signal strengths}

First, let us go back to the likelihoods presented in the $(\mu({\rm ggF+ttH}, Y), \mu({\rm VBF+VH}, Y))$ plane. In the future, production of the Higgs boson in association with a top quark pair will be probed with a much better precision, making it necessary to disentangle $\mu({\rm ggF}, Y)$ from $\mu({\rm ttH}, Y)$. Moreover, even if rescaling the VBF, WH and ZH production mechanisms by a common factor is theoretically justified in models with custodial symmetry, this might not exactly hold and one might want to check precisely the extent to which custodial symmetry can be tested in a global Higgs coupling fit. 
Eventually, we want to test ggF, ttH, VBF, ZH and WH separately, which means that we need 
a more detailed break down of the channels beyond the 2D plane.

The optimum would of course be to have the full statistical model  
available,  and methods and tools are indeed being developed~\cite{publishL} to make this feasible,  
{\it e.g.}, in the form of {\tt RooFit} workspaces. 
However, it may still take a while until likelihoods will indeed be published in this way.  
We would therefore like to advocate as a compromise that the experiments give the likelihood for each final state $Y$ as a function of a full set of production modes, that is to say, in the 
\beq
 (m_H, \mu_{\rm ggF},\mu_{\rm ttH},\mu_{\rm VBF},\mu_{\rm ZH},\mu_{\rm WH}) 
\eeq
parameter space. By getting the likelihood function in this form for each decay mode, a significant step could be taken towards a more precise fit in the context of a given BSM theory.
Note that the signal strengths' dependence on $m_H$ is especially important  
for the high-resolution channels ($\gamma\gamma$ and $ZZ$, also $Z\gamma$ in the future). While the signal strengths seem to form a plateau in the case of $H \rightarrow \gamma\gamma$ (at least in ATLAS), there is a very sizable change in the $H \rightarrow ZZ$ channel if $m_H$ is changed by 1 or 2~GeV.
The likelihood could be communicated either as a standalone computer library or as a large grid data file. This choice is mostly meant to be an intermediate step between a full effective Lagrangian parameterization (which would be difficult to communicate) and simple 2D parameterizations which unfortunately do not cover all the theoretical possibilities.

Of course, approximations still have to be made in order to reconstruct a global Higgs likelihood from this six-dimensional information, even in the case of a simple scaling of the SM production cross sections and branching fractions. While all correlations are included for a given decay mode $Y$ in a given experiment, the global likelihood is simply defined as the product of the individual 6D likelihoods (analogously to Eq.~\eqref{eq:ourbestlike}), meaning that all experimental and theoretical correlations between different final states $Y$ (or the same final state $Y$ between ATLAS and CMS) are neglected. This is expected to become more and more problematic as measurements become limited by systematic uncertainties (see, {\it e.g.}, Ref.~\cite{Dawson:2013bba}). This includes experimental correlations (for instance, from the jet energy scale and resolution) and theoretical uncertainties on the SM predictions of the cross sections and branching fractions (the dominant one being gluon fusion, for which the uncertainties are currently estimated by the LHC Higgs Cross Section Working Group to be $^{+7.2}_{-7.8}\%$ and $^{+7.5}_{-6.9}\%$ at $m_H = 125.5$~GeV, from the variation of the QCD scale and uncertainties on PDF+$\alpha_s$, respectively~\cite{Heinemeyer:2013tqa}). Moreover, searches are not completely independent. For instance, $H \to WW^*$ events contribute to the signal in the search for $H \to \tau\tau$~\cite{Chatrchyan:2014nva}; also, in the search for $H \to b\bar b$ produced in association with a top-quark pair, there can be significant contributions from other decay modes (see Appendix~3 in Ref.~\cite{ATLAS-CONF-2014-011}), leading to correlations between the individual 6D likelihoods.

In the Gaussian approximation, the problem of missing correlations can be solved if the $n \times n$ covariance matrix $V$ (see Eq.~\eqref{eq:gausscovmatrix}) is provided by the experiments, either for the $n$ individual measurements or, even better and simpler, in theory space for the $n = n_X n_Y$ different $\mu(X,Y)$. In order to have all correlations, this covariance matrix $V$ should be published jointly by the ATLAS and CMS collaborations after combination of their results. In this simple and powerful approach, there are however limitations when constraining theories with the same coupling structure as the SM. First, the covariance matrix $V$ would be given at a fixed Higgs mass. It is possible to overcome this problem simply by publishing several covariance matrices corresponding to different Higgs masses (with, {\it e.g.},  a step of $\Delta m_H = 250$~MeV). Second, the Gaussian approximation might be well justified with a very large statistical sample but is not an extremely good approximation with the current data, as already discussed in Section~\ref{sec:higgs-npconstlhc}. Third, in this approach all correlations would be included in the covariance matrix and it is not straightforward, in particular, to change the theoretical uncertainties in a consistent way.

As was mentioned in Section~\ref{sec:higgs-npconstlhc}, there is no universal agreement on the treatment of theoretical uncertainties at the LHC, and it is time-dependent (it depends in particular on the status of the calculation of higher-order corrections and on the data included in the PDF sets). It would therefore be very valuable to be able to change it. Recently, an interesting proposal was made in this direction in Ref.~\cite{Cranmer:2013hia}. Provided experimental collaborations publish likelihoods that are not profiled over a set of theoretical nuisance parameters of interest, but instead given for a fixed scenario, it is possible to build a ``recoupled'' likelihood incorporating these uncertainties at the later stage. This has the advantage of not being restricted to the Gaussian approximation. It would certainly be of great interest if the information in the 2D plane $(\mu({\rm ggF+ttH}, Y), \mu({\rm VBF+VH}, Y))$, or even better in the 6D plane discussed above, could be given without profiling over the theoretical uncertainties on the Higgs signal. From the method presented in Ref.~\cite{Cranmer:2013hia}, one could then fully correlate the theoretical uncertainties between the different channels and experiments, and modify these uncertainties compared to what is done in ATLAS and CMS if wanted.

\subsection{Fiducial cross sections}

As we saw in Section~\ref{sec:higgsdim6}, a simple scaling of production cross sections and decay branching fractions (relative to the SM) is not sufficient in situations in which the kinematic distributions of the signal depend on model parameters. Specifically, one must account for the change in the signal selection efficiency.
In order to address this broader class of theories, we advocate the measurement of fiducial cross sections for specific final states, {\it i.e.}\ cross sections, whether total or differential, for specific final states within the phase space defined by the experimental selection and acceptance cuts. 
This is meant in addition to, not instead of,  fits for signal strength modifiers $\mu$. 
Indeed, the (largely model-independent) fiducial cross sections and signal strengths w.r.t.\ SM 
are complementary to each other and both provide very valuable information in their own right.

With the full dataset of the LHC Run~I, measurements of fiducial cross sections
with a precision of 20\% or so already became feasible in a number of channels.
In fact, ATLAS has already made the first attempt and released fiducial
cross sections for $H \to \gamma\gamma$~\cite{Aad:2014lwa} and $H \to ZZ^*$~\cite{ATLAS-CONF-2014-044} (preliminary).
Fiducial cross section measurements require no model-dependent extrapolations to the full phase space, 
nor do they acquire additional theoretical uncertainty associated with such extrapolations. 
With carefully defined ``fiducial volumes'', the model-dependence of signal efficiencies
within such ``fiducial volumes'' can also be minimized so as to make it smaller than
the overall experimental uncertainties. For example, cuts on lepton transverse momenta
can be raised well above the knee of the efficiency plateau---this would minimize the impact
of possible variations in leptons' $p_T$-spectra on the overall signal efficiency.
Including isolation of leptons into the ``fiducial volume'' definition would help minimize
the sensitivity of a measured fiducial cross section on assumptions
about the jet activity in signal events. In some cases this is more difficult, for instance when the  the fiducial volume is defined by a cut on missing transverse energy,  which often introduces sensitivity to the topology of the event.  In situations where there is residual model-dependence in the fiducial efficiency, 
a service such as RECAST~\cite{Cranmer:2010hk} provided by the collaborations for explicitly calculating the 
fiducial efficiency would be of great value.

Fiducial cross sections, both total and differential, 
are standard measurements in high energy physics 
and for some processes are the only experimental cross sections available. 
For example, $J/\psi$ and $\Upsilon$ production cross section measurements at hadron colliders 
are always performed in some specified ``fiducial volumes''. This has allowed for a variety of models, 
many of which appeared or were substantially updated {\it after} the measurements had been made, 
to be confronted with the fixed experimental results.
In the context of Higgs boson physics, 
the fiducial cross sections can be categorized according to:
\begin{itemize} 

\item ``target'' decay mode, {\it e.g.}, 
$H \to ZZ \to 4\ell$, 
$H \to \gamma\gamma$, 
$H \to WW \to \ell\nu\ell\nu$, {\it etc.}; 

\item ``target'' production mechanism signatures, {\it e.g.}, 
(VBF-like $jj$)+$H$, 
$(\ell\ell)+H$, 
$(\ell+E_{T}^{\mathrm{mis}})+H$, 
$(E_{T}^{\mathrm{mis}})+H$, 
($V$-like $jj$)+$H$, {\it etc.}; 

\item and signal purity, {\it e.g.}, 
0-jet, 
1-jet,
high-mass VBF-like $jj$,
low-mass VBF-like $jj$, {\it etc.}

\end{itemize}

Fiducial cross sections can be interpreted in the context of whatever theoretical model, 
provided it is possible to compute its predictions for the fiducial cross section at hand 
({\it i.e.}, if it is possible to include experimental selection/cuts into the model). 
Typically, if the cuts defining ``fiducial volume'' can be implemented in a MC generator, 
this is rather straightforward. 
Therefore, complicated ``fiducial volume'' criteria ({\it e.g.} MVA-based) are not well suited, unless 
the MVA function is provided and depends only on kinematic information available at the generator level. 
Some reduction in signal sensitivity due to simplifications
in the event selection and due to possibly tighter cuts 
(to minimize the dependence of a signal efficiency on model assumptions as discussed above)
is an acceptable price. 

If these requirements for ``fiducial volume'' definitions are satisfied, 
then theoretical parameters of interest can be extracted from a fit 
to the measured cross sections.  As more than one fiducial cross section become available, 
to make a proper fit for parameters of interest, 
it is important that experiments provide a complete covariance matrix of uncertainties between
the measured fiducial cross sections. 

The ultimate measurements of an ``over-defined'' set of fiducial cross sections 
$\sigma^{\mathrm{fid}}_i$
can be unravelled into total cross sections associated with specific production mechanisms 
$\sigma^{\mathrm{tot}}_j$ 
via a fit of the following set of linear equations:
\begin{equation}
\label{eq:CrossSections}
\sigma^{\mathrm{fid}}_i = \sum_{j} A^{\mathrm{th}}_{ij} \times \sigma^{\mathrm{tot}}_j \,,
\end{equation}
where $A^{\mathrm{th}}_{ij}$ are theoretical acceptances
of ``fiducial volumes'', in which 
fiducial cross sections $\sigma^{\mathrm{fid}}_i$ are measured.

The beauty of the concept of fiducial cross sections is that 
{\it experimental} uncertainties associated with 
measurements of fiducial cross sections $\sigma^{\mathrm{fid}}_i$ and
{\it theoretical} uncertainties associated with 
``fiducial volume'' acceptances $A^{\mathrm{th}}_{ij}$
are nicely factorized. Therefore, updates of theoretical acceptances/uncertainties
or a confrontation of emerging new models with experimental results
do not require a re-analysis of experimental data. 
One can also treat the total cross sections $\sigma^{\mathrm{tot}}_j$ 
as nuisance parameters and fit data for 
theoretical acceptances $A^{\mathrm{th}}_{ij}$ 
({\it e.g.}, a 0-jet veto acceptance), 
if it is these quantities that one is primarily interested in.

Finally, we note that measurements of differential fiducial cross sections, when they become possible, 
will be even more powerful (in comparison to just total exclusive fiducial cross sections)
for scrutinizing the SM Lagrangian structure of the Higgs boson interactions, including 
tests for new tensorial couplings, 
non-standard production modes, 
determination of effective form factors, {\it etc.}

\clearpage

\chapter{Interpreting LHC searches for new physics}

So the Higgs has been found---but where is new physics?
Beyond the discovery of the Higgs boson and the measurements of its properties, the LHC was designed as a discovery machine for TeV-scale physics. Guided by naturalness arguments, there were high hopes in finding new physics ``just around the (LEP) corner'', {\it i.e.}\ new particles in the $100-1000$~GeV mass range. Unfortunately, after Run~I of the LHC no significant excess was observed in the search for new physics in spite of the large variety of analyses performed by the ATLAS, CMS and LHCb collaborations. If new physics connected to electroweak symmetry breaking is indeed present, it is either well hidden, somewhat ``unnatural'' (in the case where the BSM particles are rather heavy), or it has experimental signatures not yet looked for or not yet thought of. It is however important to keep in mind that the 95\%~CL lower bounds set on the masses of BSM particles produced at the LHC depend a lot on the production cross section (hence on the quantum numbers of the initially produced BSM particle(s)), on the nature of the final state particles and on the kinematic configuration. For example, gluino production followed by $\tilde g \to qq\tilde\chi^0_1$ is excluded up to $m_{\tilde g} = 1.2-1.4$~TeV for a massless neutralino (while there is no limit above $m_{\tilde \chi^0_1} = 600$~GeV) in both ATLAS and CMS~\cite{Aad:2014wea,Chatrchyan:2014lfa}, while the limits on the pair production of staus followed by $\tilde \tau_1 \to \tau\tilde\chi^0_1$ have not been improved since LEP.

Of course, searches for new physics at the LHC cannot be completely general. Signals of new physics, in a given model, have specific kinematic features that can be used to discriminate against the SM background.\footnote{A very general search for events possibly including isolated electrons, photons and muons, as well as ($b$-)jets and missing transverse momentum has been performed by ATLAS~\cite{ATLAS-CONF-2014-006}. While valuable, such a search approach is much less sensitive than optimized searches for specific models.} The sensitivity to the expected signal can then be increased and, in the case of a significant deviation from the SM expectation, it can help distinguishing between the possible BSM explanations. Apart from the searches for Higgs-like states, the current BSM searches performed by the ATLAS and CMS collaborations are divided into two main categories: SUSY searches~\cite{atlassusytwiki,cmssusytwiki} and ``exotic'' searches~\cite{atlasexotictwiki,cmsexotictwiki}. The former category includes almost all searches where a significant amount of missing transverse momentum is required, while searches for new resonances (motivated, {\it e.g.}, by heavy $Z^0$-like states or gravitons in models with extra dimensions) are given in the latter category. In the rest of the chapter we will focus on the searches for supersymmetric particles, even though most of the general discussions remain valid to all BSM searches.

Within a given search for new physics performed at the LHC, a number of signal regions is defined. Each signal region corresponds to a unique set of selection criteria (also called cuts) designed to optimize the sensitivity to different new physics scenarios and to different kinematic configurations. For each signal region, the primary results are given as the number of observed events ($n_{\rm obs}$) and the SM expectation with its associated uncertainty at 68\%~CL ($n_b = \hat n_{b} \pm \Delta n_b$), from which the significance of possible excesses is assessed. If no significant excess is found, a 95\%~CL limit on the visible cross section ({\it i.e.}\ cross section after cuts, $\sigma_{\rm vis} = \sigma \times A \times \varepsilon$) is evaluated.
The impact of these results is then assessed for the models of new physics for which the analysis has been designed. If no significant excess is found, 95\%~CL limits are set on the parameter space of these models, possibly after combining the results of several signal regions. A short discussion on the statistical procedure for setting limits can be found at the end of Section~\ref{sec:analysisreimplementation}.

The first round of SUSY searches at the LHC, at $\sqrt{s} = 7$~TeV, mainly focused on the constrained MSSM (CMSSM), a popular restricted version of the general MSSM with only four GUT-scale parameters in addition to the sign of the $\mu$ parameter: the mass parameters $m_0$ and $m_{1/2}$ for the scalar particles and gauginos, respectively, the common trilinear coupling $A_0$, and $\tan \beta$. As LHC searches were pushing the bounds on $m_0$ and $m_{1/2}$ higher and higher with no evidence of a signal, it became evident that such a restricted framework---only covering a small subset of the possibilities of the general MSSM or even of the pMSSM---should no longer serve as a guidance for both the design and interpretation of SUSY searches. It is however difficult to represent and communicate the impact of a given search on less constrained models. More than two or three parameters can typically impact the expected number of events in the signal regions, making it difficult to represent the exclusion bounds on paper.\footnote{Attempts in this direction have however been made, see in particular Refs.~\cite{Sekmen:2011cz,CMS-PAS-SUS-12-030,CMS-PAS-SUS-13-020} for the impact of the CMS SUSY searches on the pMSSM.} In addition, given the large number of systematic uncertainties needed to realistically model the ATLAS or CMS detector, a lot of computing power is needed to test possible signals of new physics, hence new interpretations certainly do not come for free.

It is however possible to take a different approach and consider every new physics signal not as the mere result of the choice of model parameters, but instead as the superposition of different topologies contributing to the signal. In the case of topologies involving only two or three new particles, it is possible to define a simplified model with only these particle masses as free parameters, assuming that all other BSM particles are absent or too heavy to contribute to the signal. This is the simplified model approach, which is now systematically used by the ATLAS and CMS collaborations in the presentation of the results of SUSY searches (for a concise overview, see Refs.~\cite{Chatrchyan:2013sza,Okawa:2011xg}).

\begin{figure}[ht]\centering
	\includegraphics[width=0.248\textwidth]{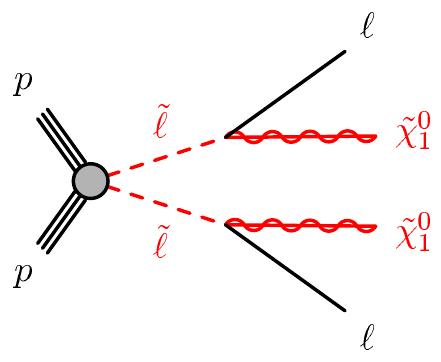}\hfil
	\includegraphics[width=0.248\textwidth]{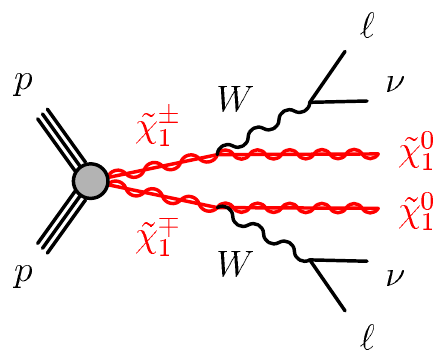}
	\includegraphics[width=0.248\textwidth]{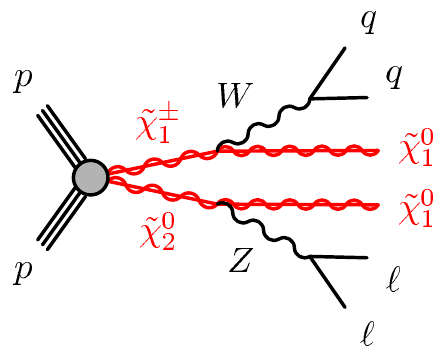}\hfil
	\includegraphics[width=0.248\textwidth]{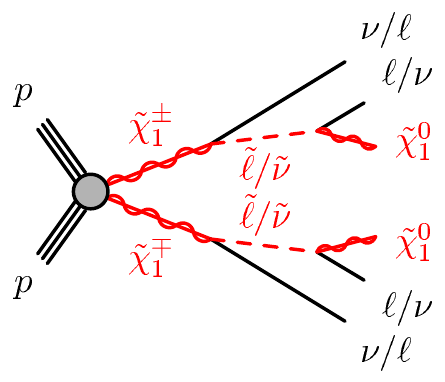}
	\caption{\label{simpmod2leptondiagrams}
		The four simplified model topologies considered in the ATLAS SUSY search for $\ell^+\ell^- + E_T^{\rm miss}$ at $\sqrt{s}=8$~TeV~\cite{Aad:2014vma}.}
\end{figure}

We illustrate the simplified model approach with the example of the ATLAS search for charginos, neutralinos and sleptons in final states with two leptons of opposite sign and missing transverse momentum ($\ell^+\ell^- + E_T^{\rm miss}$) with full luminosity at $\sqrt{s} = 8$~TeV~\cite{Aad:2014vma}. The SUSY simplified model topologies considered in this analysis are shown in Fig.~\ref{simpmod2leptondiagrams}. As a consequence of the assumption of conserved $R$-parity and neutralino LSP, the lightest neutralino, $\tilde\chi^0_1$, is a stable particle present at the end of all SUSY decay chains. The first two topologies involve only one new particle mass in addition to the one of the LSP, corresponding to the sleptons $\tilde\ell$ (SUSY partners of the electron and muon) and to the lightest chargino $\tilde\chi^{\pm}_1$. In the former case it is assumed that the first and second generation of sleptons are degenerate, {\it i.e.}\ that $m_{\tilde{L}_1} = m_{\tilde{L}_2}$ and $m_{\tilde{E}_1} = m_{\tilde{E}_2}$ or equivalently that $m_{\tilde e_R} = m_{\tilde\mu_R} \equiv m_{\tilde\ell_R}$ and $m_{\tilde e_L} = m_{\tilde\mu_L} \equiv m_{\tilde\ell_L}$.
Three extreme cases are then considered when setting limits on the simplified model: $m_{\tilde\ell_R} \ll m_{\tilde\ell_L}$, $m_{\tilde\ell_R} \gg m_{\tilde\ell_L}$, and $m_{\tilde\ell_R} = m_{\tilde\ell_L}$. As regards the last two simplified model topologies of Fig.~\ref{simpmod2leptondiagrams}, in order to represent the limits on a 2D plane it is subsequently assumed that $m_{\tilde\chi^\pm_1} = m_{\tilde \chi^0_2}$ and $m_{\tilde \nu} = m_{\tilde \ell_L} = (m_{\tilde \chi^\pm_1} + m_{\tilde \chi^0_1})/2$.\footnote{In the case of the production of a pair of charginos followed by slepton-mediated decays, it is assumed that the three flavors of (s)leptons contribute to the signal in this ATLAS analysis~\cite{Aad:2014vma}. This is in contrast with the usual definition of the symbol $\ell$ in this thesis.} The first assumption is well-motivated when $\tilde \chi^0_2$ and $\tilde \chi^\pm_1$ are both wino-like or higgsino-like, while the second one is an arbitrary choice. 

\begin{figure}[ht]\centering
	\includegraphics[width=0.495\textwidth]{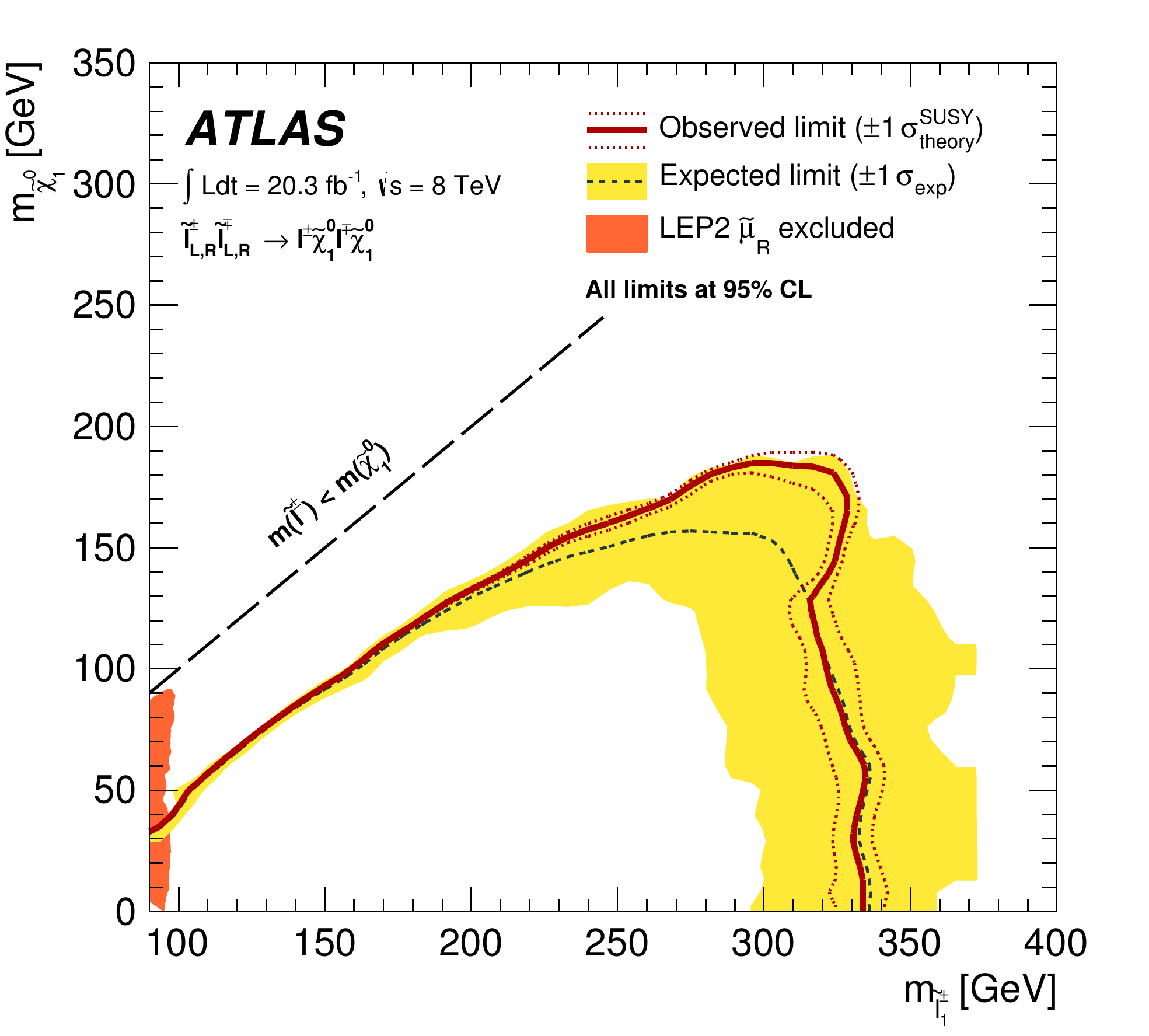}
	\includegraphics[width=0.495\textwidth]{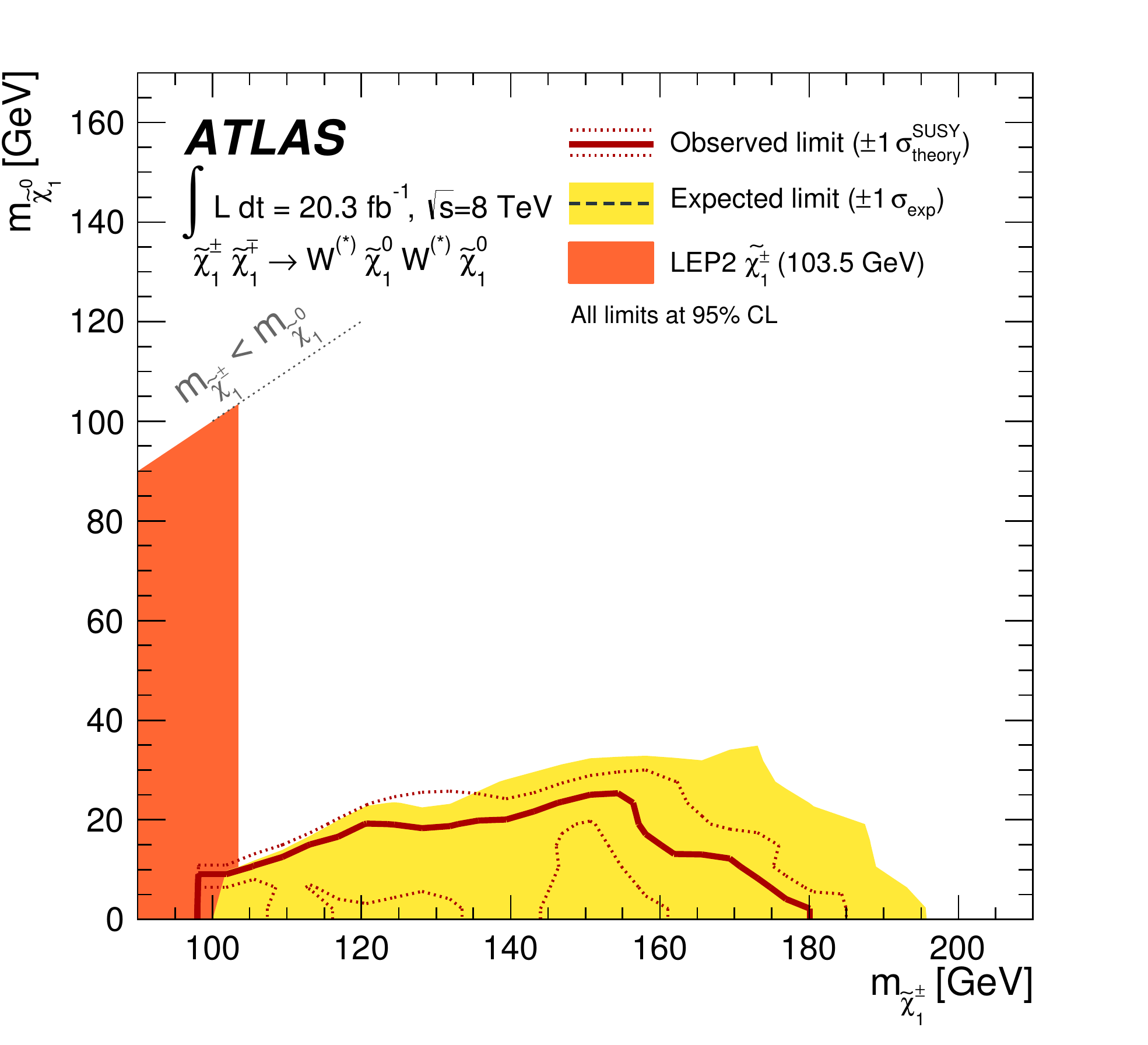}
	\caption{\label{interpretations-atlasdilepton}
		95\%~CL exclusion regions in the planes $(m_{\tilde\ell_{L,R}}, m_{\tilde\chi^0_1})$ (left)
		and $(m_{\tilde\chi^\pm_1}, m_{\tilde\chi^0_1})$ (right) for the ATLAS SUSY search for $\ell^+\ell^- + E_T^{\rm miss}$ at $\sqrt{s}=8$~TeV, taken from~\cite{Aad:2014vma}.
		Also illustrated are the LEP limits~\cite{lepsusy}.
		}
\end{figure}

The 95\%~CL exclusions in the parameter spaces of these simplified models are then given. Two examples are shown in Fig.~\ref{interpretations-atlasdilepton}, corresponding to the first two simplified model topologies of Fig.~\ref{simpmod2leptondiagrams}. For the direct production of sleptons, it is assumed that $m_{\tilde\ell_R} = m_{\tilde\ell_L}$. In both cases, the LEP limits on chargino and slepton production~\cite{lepsusy} are clearly improved by ATLAS in the case of a light neutralino. However, when the mass difference between the sleptons and the LSP is below $\sim 50$~GeV (in the case of slepton pair production) and when the mass of the LSP is above $\sim 20$~GeV (in the case of chargino-pair production followed by $\tilde\chi^\pm_1 \to W^{\pm}\tilde\chi^0_1$), the ATLAS limits vanish.

Presenting results in terms of simplified model scenarios is very interesting by itself as it gives the status of the constraints for a number of relevant topologies. However, this approach suffers from limitations by definition. First, in the exclusion regions delineated by the red lines in Fig.~\ref{interpretations-atlasdilepton} the branching fractions are always assumed to be 100\%. It is not difficult to find scenarios where this condition does not hold: for instance, in addition to the first topology of Fig.~\ref{simpmod2leptondiagrams}, slepton-pair production could be followed by $\tilde\ell^\pm \to \nu_{\ell} \tilde\chi^\pm_1 \to \nu_{\ell} W^{\pm} \tilde\chi^0_1$ if the chargino is sufficiently light. Also, the second topology of Fig.~\ref{simpmod2leptondiagrams} will always be present in the case of chargino-pair production, hence the fourth topology cannot yield 100\% branching fraction.
Second, the exclusion regions depend on the production cross sections, which are sensitive to the nature of the initially produced SUSY particles. This is especially relevant for the electroweak-ino pair production, where the cross section depends on the wino, higgsino and bino fractions of the produced particles. In the last three topologies of Fig.~\ref{simpmod2leptondiagrams} and on the limit shown in the right panel of Fig.~\ref{interpretations-atlasdilepton}, it is assumed that the $\tilde\chi^\pm_1$ and the $\tilde\chi^0_2$ are pure wino states. In the case of mostly higgsino states, the cross section would be smaller and the excluded regions might change drastically.

We have seen that the experimental collaborations at the LHC cannot cover all possibilities of new physics when interpreting the results of the BSM searches. Instead, limits on simplified models corresponding to a single topology can be derived, where only a handful of new parameters are needed to parameterize the model. This approach has been systematically adopted by the ATLAS and CMS collaborations for the SUSY searches, and gives a meaningful picture of the impact of these searches. However, the exclusion regions given in terms of simplified models as shown in Fig.~\ref{interpretations-atlasdilepton} cannot be used directly when testing most models of new physics because of the assumptions on the production cross section and on the branching fractions, assumed to be 100\% by definition of the simplified model. Therefore, interpretations beyond the ones of the experimental collaborations are needed to fully exploit the potential of these searches---even when testing ``standard'' MSSM scenarios that motivates the design of most SUSY analysis.

In Section~\ref{sec:simpmod-intro} we will give more details on the simplified models and explain how these experimental results can be used to constrain new physics signals beyond a simplified model. This is achieved through decomposition of the signal into simplified models, as is done by the programs {\tt SModelS}~\cite{Kraml:2013mwa} and {\tt FastLim}~\cite{Papucci:2014rja}. This approach will be used to constrain, using {\tt SModelS}, two specific scenarios of supersymmetric dark matter in Sections~\ref{sec:simpmod-lightneutralino} and~\ref{sec:simpmod-sneutrino}. As we will see, the simplified model approaches presented in Section~\ref{sec:simpmod-intro} are fast but have limitations---a more general way of constraining new physics from LHC searches consists in applying the analyses cuts on event samples generated by Monte Carlo simulation. The technical aspects of this approach and the related database of analyses we have initiated will be discussed in Section~\ref{sec:analysisreimplementation}. We will then present the latest modifications to the \madanalysis\ program relevant for the reimplementation of analyses in Section~\ref{sec:ma5delphes3}. Finally, two concrete examples of reimplemented LHC analyses will be given in Sections~\ref{sec:cmsvalid} and~\ref{sec:atlasvalid}.

\section{Constraining new physics with simplified models} \label{sec:simpmod-intro}

We argued above that LHC results presented in terms of simplified model topologies gives a meaningful picture of the impact of the BSM searches. However, the information given in Fig.~\ref{interpretations-atlasdilepton} makes it impossible to constrain models of new physics beyond the very constrained definition of the simplified model itself. This limitation can be overcome if, in addition to the exclusions indicated with a red line in Fig.~\ref{interpretations-atlasdilepton}, the experimental collaborations provide 95\%~CL upper bounds on the cross sections across the parameter space of the simplified model. From this information, one can constrain any model with the same topology (hence same final state particles with the same kinematic properties), but a different production cross section and/or branching fractions below 100\%. Taking the example of the $\tilde\chi^\pm_1 \tilde\chi^\mp_1 \to W^\pm \tilde\chi^0_1 W^\mp \tilde\chi^0_1$ topology, this information makes it possible to reinterpret the exclusion if, for instance, the chargino is higgsino-like instead of wino-like and/or the branching fraction is below 100\% because the chargino could also decay via, {\it e.g.}, a stau, $\tilde\chi^\pm_1 \to \nu_\tau \tilde\tau_1 \to \nu_\tau \tau \tilde\chi^0_1$.

\begin{figure}[ht]\centering
	\includegraphics[width=0.455\textwidth]{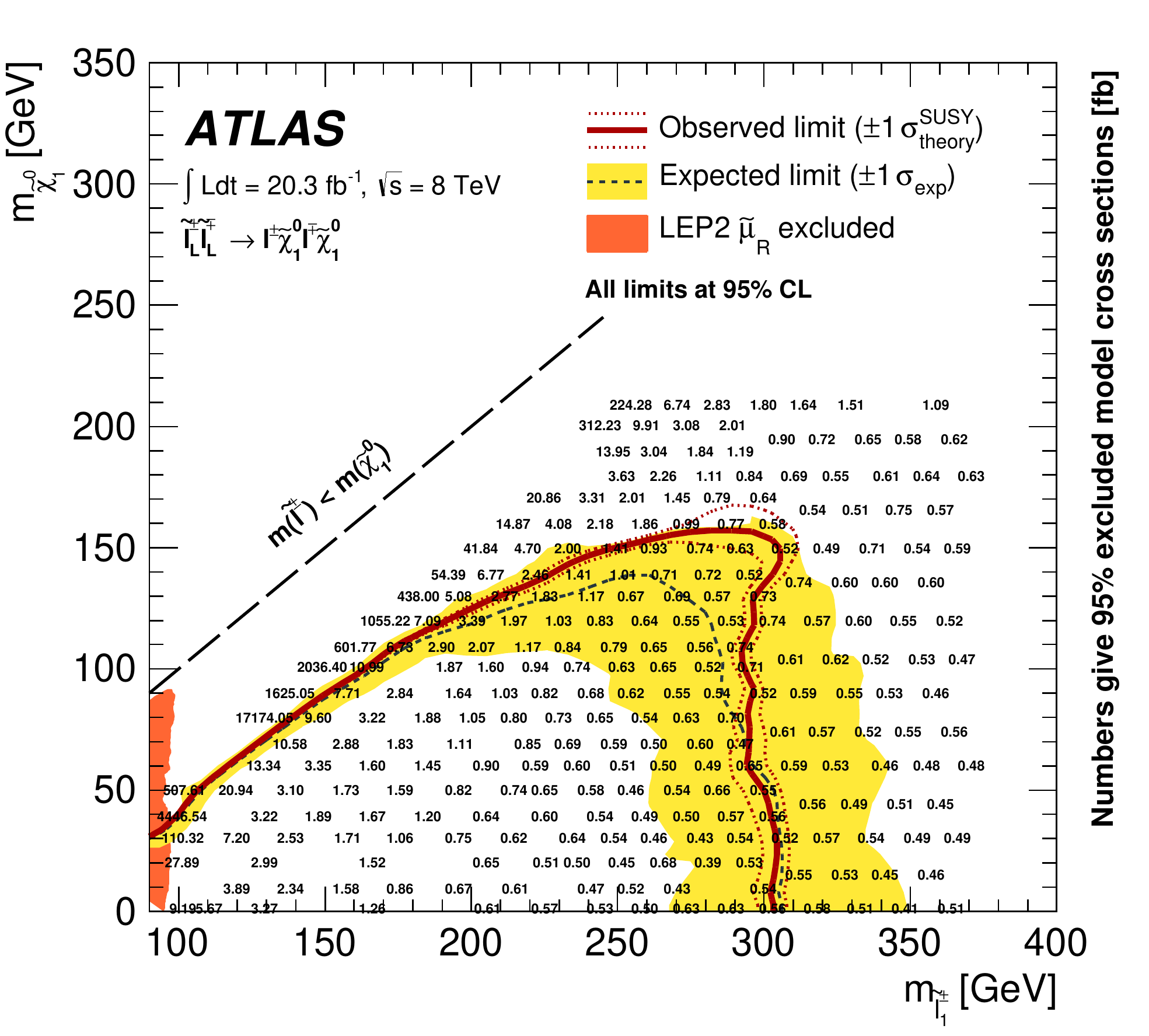}
	\includegraphics[width=0.535\textwidth]{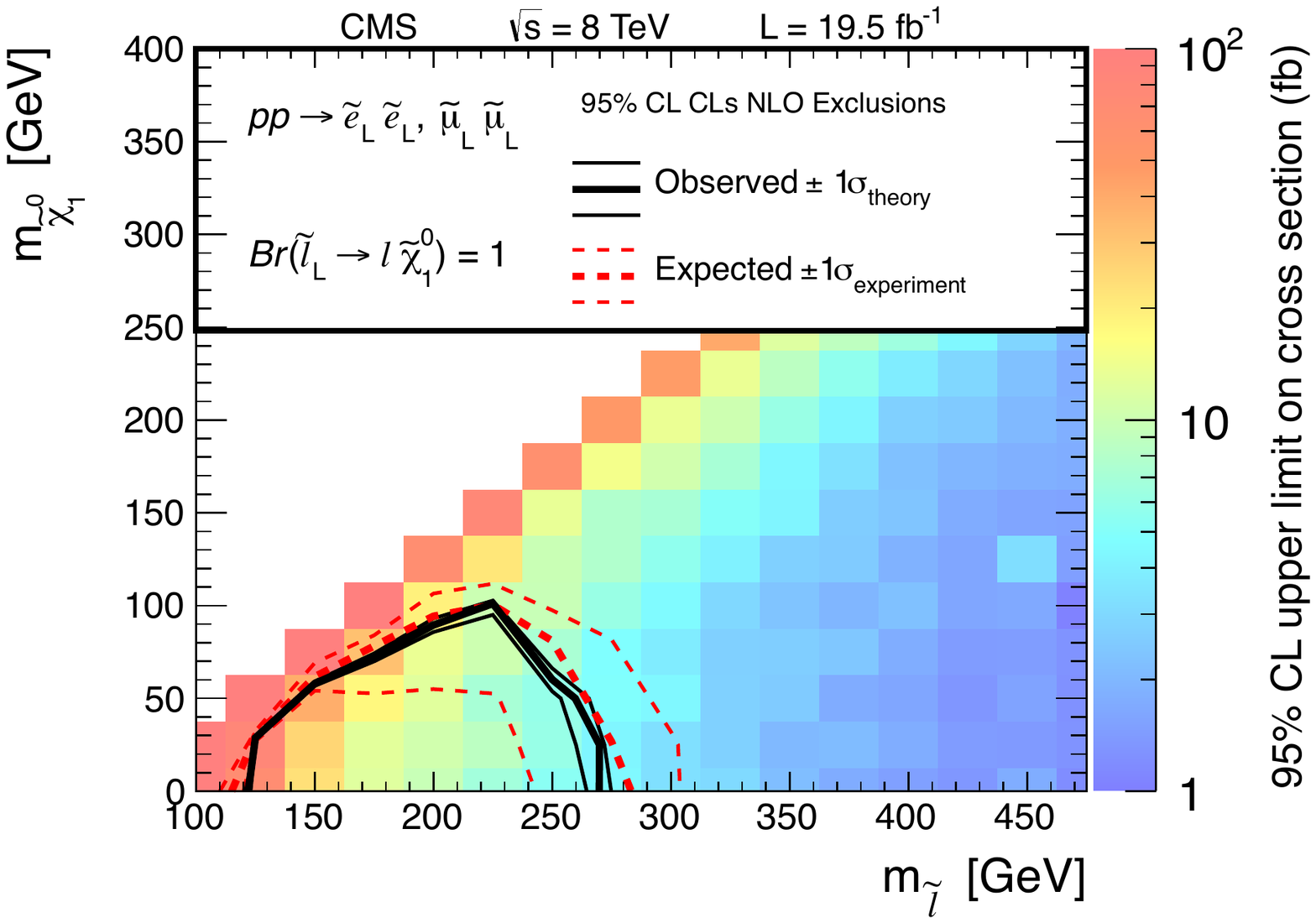}
	\caption{\label{interpretations-sm-withnumbers}
		95\%~CL exclusion regions in the plane $(m_{\tilde\ell_L}, m_{\tilde\chi^0_1})$ for the ATLAS SUSY search for $\ell^+\ell^- + E_T^{\rm miss}$ at $\sqrt{s}=8$~TeV from~\cite{Aad:2014vma} (left) and for the CMS SUSY search for charginos, neutralinos, and sleptons at $\sqrt{s}=8$~TeV from~\cite{Khachatryan:2014qwa} (right). The 95\%~CL upper bounds on the visible cross sections are indicated as numbers in the case of ATLAS, and is encoded in the color code for CMS.}
\end{figure}

Fortunately, this information is now systematically provided by the ATLAS and CMS collaborations and is present for all simplified model scenarios of the SUSY analyses with full luminosity at $\sqrt{s}=8$~TeV.
Two examples are given in Fig.~\ref{interpretations-sm-withnumbers} for the constraints on the pair production of left-handed sleptons with $\tilde\ell_L^{\pm} \to \ell^{\pm}\tilde\chi^0_1$ from the ATLAS $\ell^+\ell^- + E_T^{\rm miss}$ search aforementioned, and from the preliminary results of the CMS SUSY search for charginos, neutralinos, and sleptons at $\sqrt{s}=8$~TeV~\cite{Khachatryan:2014qwa}. In the case of ATLAS, the 95\%~CL upper bounds on the visible cross sections are given as numbers scattered along the 2D plane, while CMS encodes this information in the color code of a 2D histogram.
In the case of CMS, the information is systematically provided in a numerical form ({\tt ROOT} and/or plain text format) on the TWiki page of the analysis, while in the case of ATLAS this is less systematic but more and more often provided on {\tt HepData}. 

The signal of a model of new physics can be decomposed into topologies. For those topologies which are constrained by the experiments as in Fig.~\ref{interpretations-sm-withnumbers}, it is possible to test exclusion for any (known) cross section and branching fraction. This however requires to build a database of simplified model results, to decompose the signal into the relevant topologies, and to check that the assumptions made in the definition of simplified models are satisfied. For the latter, this includes checking that the assumptions on the masses of the intermediate $\tilde\ell$ and $\tilde\nu$ in the fourth topology of Fig.~\ref{simpmod2leptondiagrams} are (approximately) satisfied, {\it i.e.} a common slepton mass which is halfway between the chargino and the LSP mass. All this procedure has been automated in {\tt SModelS}~\cite{Kraml:2013mwa}, a new tool made in collaboration between Vienna, Grenoble and S\~ao Paulo. {\tt SModelS} can be used online at \url{http://smodels.hephy.at}, and a public version of the code is in preparation.

\begin{figure}[ht]\centering
	\includegraphics[width=.8\textwidth]{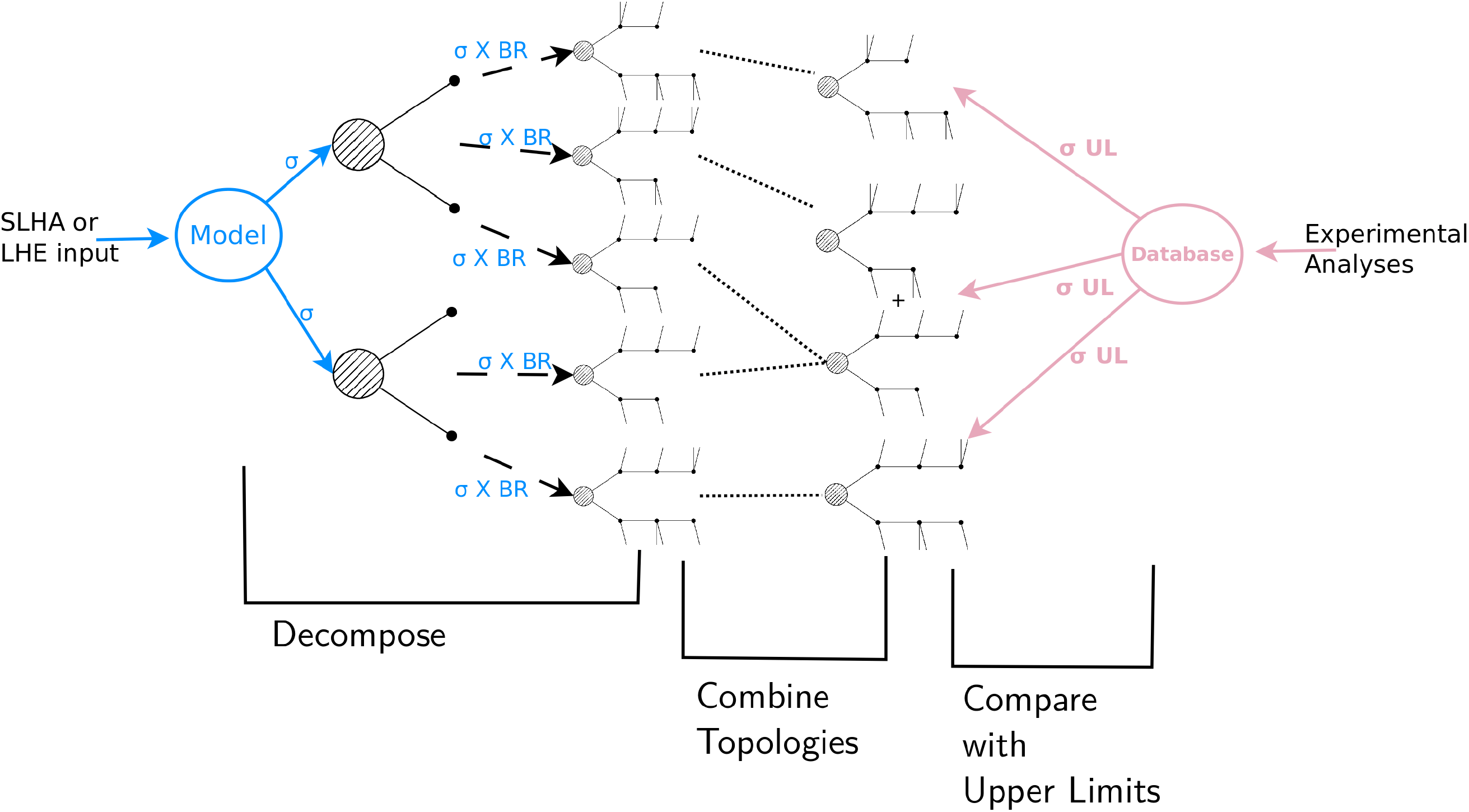}
	\caption{\label{SModelS-workingscheme}
		Schematic view of the working principle of {\tt SModelS}, from Ref.~\cite{Kraml:2013mwa}.}
\end{figure}

The working principle of {\tt SModelS} is illustrated in Fig.~\ref{SModelS-workingscheme}. The program can take as input an SLHA(-like) file~\cite{Skands:2003cj,Allanach:2008qq} containing the information on the cross sections and branching fractions (using the statements {\tt XSECTION}\footnote{The {\tt XSECTION} statement is not part of the SLHA standard yet but it has been proposed at the 2013 Les Houches workshop, see~\cite{slha-xsecinfo}.} and {\tt DECAY}, respectively), in addition to the masses of the BSM particles (in the {\tt MASS} block).
If not given in the SLHA file, the production cross sections of SUSY processes can be computed either at leading order through a Monte Carlo generator, at NLO using {\tt Prospino}~\cite{Beenakker:1996ed}, or at next-to-leading log (NLL) precision (for the production of squarks and gluinos) using {\tt NLL-fast}~\cite{nll-fast}.
The various combinations of production and decay are then matched to simplified model topologies, and compared to experimental limits if available.
Another possibility is to give as input an event file in Les Houches event ({\tt LHE}) format~\cite{Alwall:2006yp}; in this case, each individual event is mapped to a simplified model topology, and the sum of the weights given by the Monte Carlo generator are used to derive the various $\sigma \times {\rm BR}$.

A large number of simplified model interpretations has been performed by the ATLAS and CMS collaborations from the data collected during Run~I of the LHC. Hence, this approach is well-suited for constraining a variety of signals of new physics, beyond the very restricted definitions of the simplified models. In {\tt SModelS}, these signals could come from supersymmetric as well as non-supersymmetric models, as long as all topologies respect a $Z_2$ symmetry.
Compared to the full procedure of model testing, a clear advantage of this approach is that it is fast. Indeed, it requires at most to generate events at the parton level, and no detector simulation, implementation of cuts, or limit setting procedure is required. It is thus especially relevant in the context of large scans where the computing time and the disk space are limiting factors.

By definition, this procedure is conservative since the 95\%~CL upper limits are taken directly from the experimental publications and all simplified model topologies are tested independently.\footnote{However, the statistical interpretation of these simplified model results is not straightforward as only observed upper limits at 95\%~CL are available.} This is a desirable feature, but also a limitation of this approach. First, signal regions usually target one simplified model topology but do not veto completely the signal from other topologies. Second, combination of several signal regions is made in many analyses where the signal regions are non-overlapping ({\it i.e.}, mutually exclusive). Third, topologies involving more than two parameters are never completely covered by the experiments, hence many topologies simply cannot be confronted to the experimental results. For instance, the fourth topology of Fig.~\ref{simpmod2leptondiagrams} cannot be tested if $(m_{\tilde\chi^\pm_1}, m_{\tilde\ell_L,\tilde\nu}, m_{\tilde\chi^0_1}) = (150, 120, 20)$~GeV because the slepton mass is assumed to be $m_{\tilde\ell_L,\tilde\nu} = (m_{\tilde \chi^\pm_1} + m_{\tilde \chi^0_1})/2$ in the results given in Ref.~\cite{Aad:2014vma}.

\begin{figure}[ht]\centering
	\includegraphics[width=.49\textwidth]{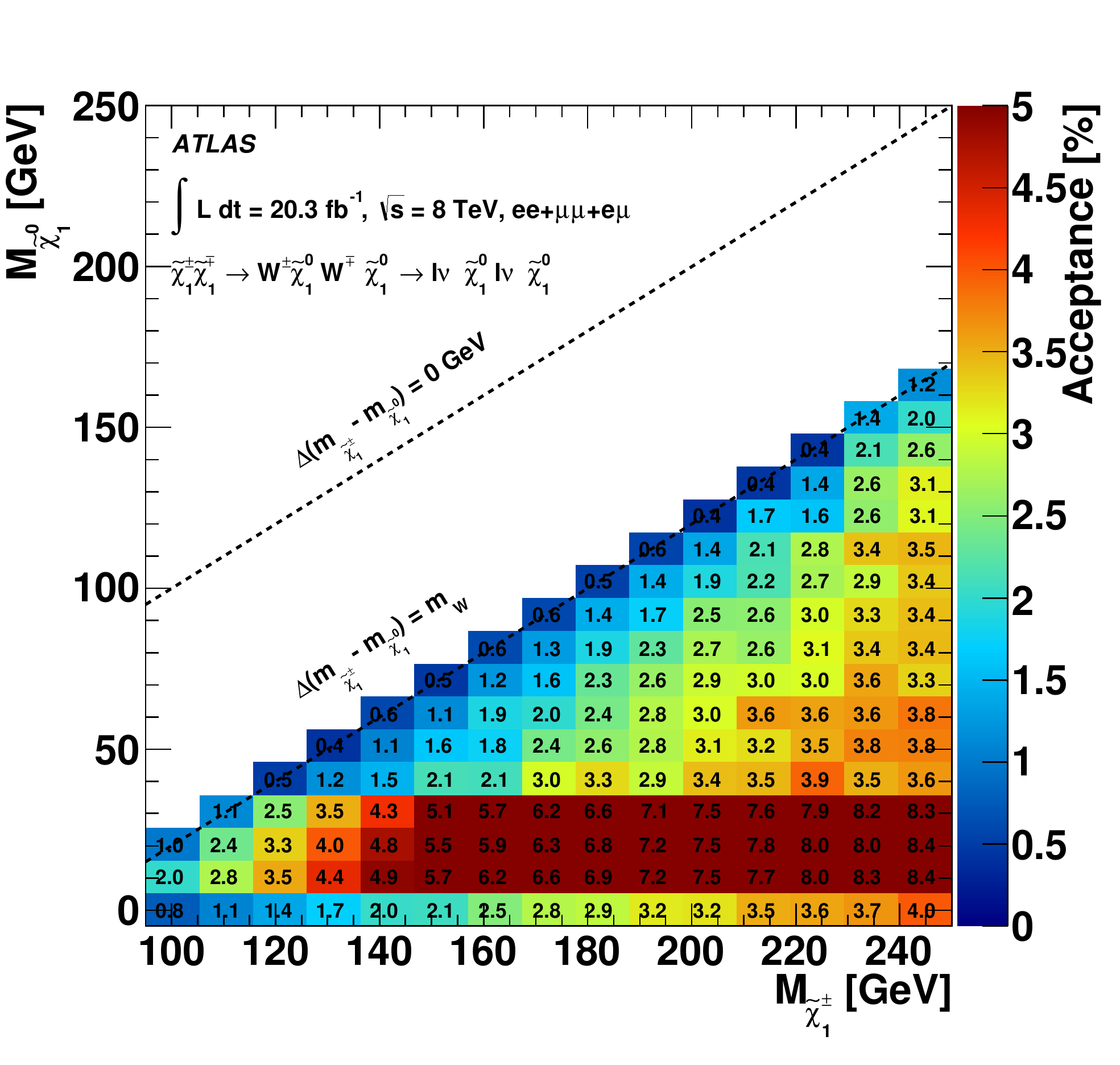}
	\includegraphics[width=.49\textwidth]{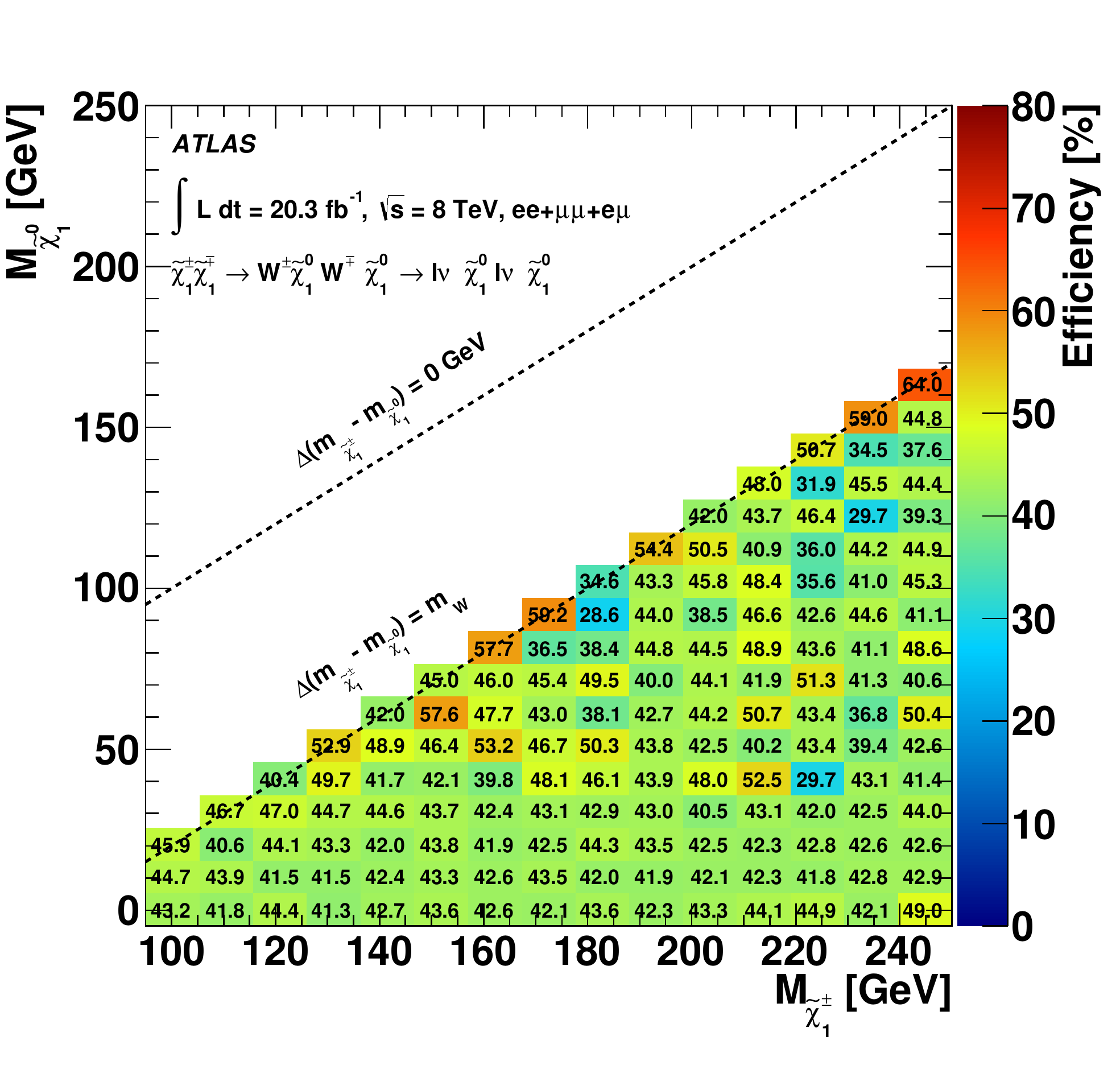}
	\caption{\label{acceptance-efficiency-dileptonmet}
		Acceptance (left) and efficiency (right) maps in the plane $(m_{\tilde\chi^\pm_1}, m_{\tilde\chi^0_1})$ for the signal region $WW$a of the ATLAS SUSY search for $\ell^+\ell^- + E_T^{\rm miss}$ at $\sqrt{s}=8$~TeV~\cite{Aad:2014vma}.}
\end{figure}

These limitations can be partly overcome by using simplified model topologies in a different way. We have seen that signals of new physics can be decomposed into $n$ topologies with a cross section $\sigma_i$ (including the branching fraction factors). Thus, the number of expected signal events in a given signal region can be written as
\beq
n_s = \sum_{i=1}^{n} \sigma_i \times (A \times \varepsilon)_i \times \mathscr{L} \,,
\eeq 
where $\mathscr{L}$ is the integrated luminosity. In order to confront a given model to the experimental results, the only missing piece of information is the acceptance and efficiency, $(A \times \varepsilon)_i$, for each of the $n$ simplified model topologies contributing to the signal in this signal region. This information is sometimes provided by the experimental collaborations, as in the case of the $\tilde\chi^\pm_1 \tilde\chi^\mp_1 \to W^\pm \tilde\chi^0_1 W^\mp \tilde\chi^0_1$ topology in the signal region $WW$a of the ATLAS SUSY search for $\ell^+\ell^- + E_T^{\rm miss}$, see Fig.~\ref{acceptance-efficiency-dileptonmet}.\footnote{The definitions of acceptance and efficiency as taken in Fig.~\ref{acceptance-efficiency-dileptonmet} do not match with the ones given in Section~\ref{sec:higgs-measlhc}, where $A$ is the geometrical acceptance of the detector and $\varepsilon$ is the efficiency of the cuts. Instead, $A$ is defined as the fraction of signal events which pass the analysis selection performed on Monte Carlo ``truth'' objects and $\varepsilon$ is a correcting factor for the reconstruction level cuts applied to reconstructed objects. Their product, $A \times \varepsilon$, is the same irrespective of the individual definitions of $A$ and $\varepsilon$.}
If the same information were available for other topologies in this signal region, it would be possible to overcome the first limitation mentioned above. Then, assuming that all the information on the acceptance$\times$efficiencies is available for the relevant topologies, it would in principle be possible to combine the results from different signal regions and/or analyses and go beyond the individual 95\%~CL upper bounds. The correlation between systematic uncertainties however makes it a difficult task, as will be discussed in Section~\ref{sec:analysisreimplementation}.  

\begin{figure}[ht]\centering
	\includegraphics[width=.55\textwidth]{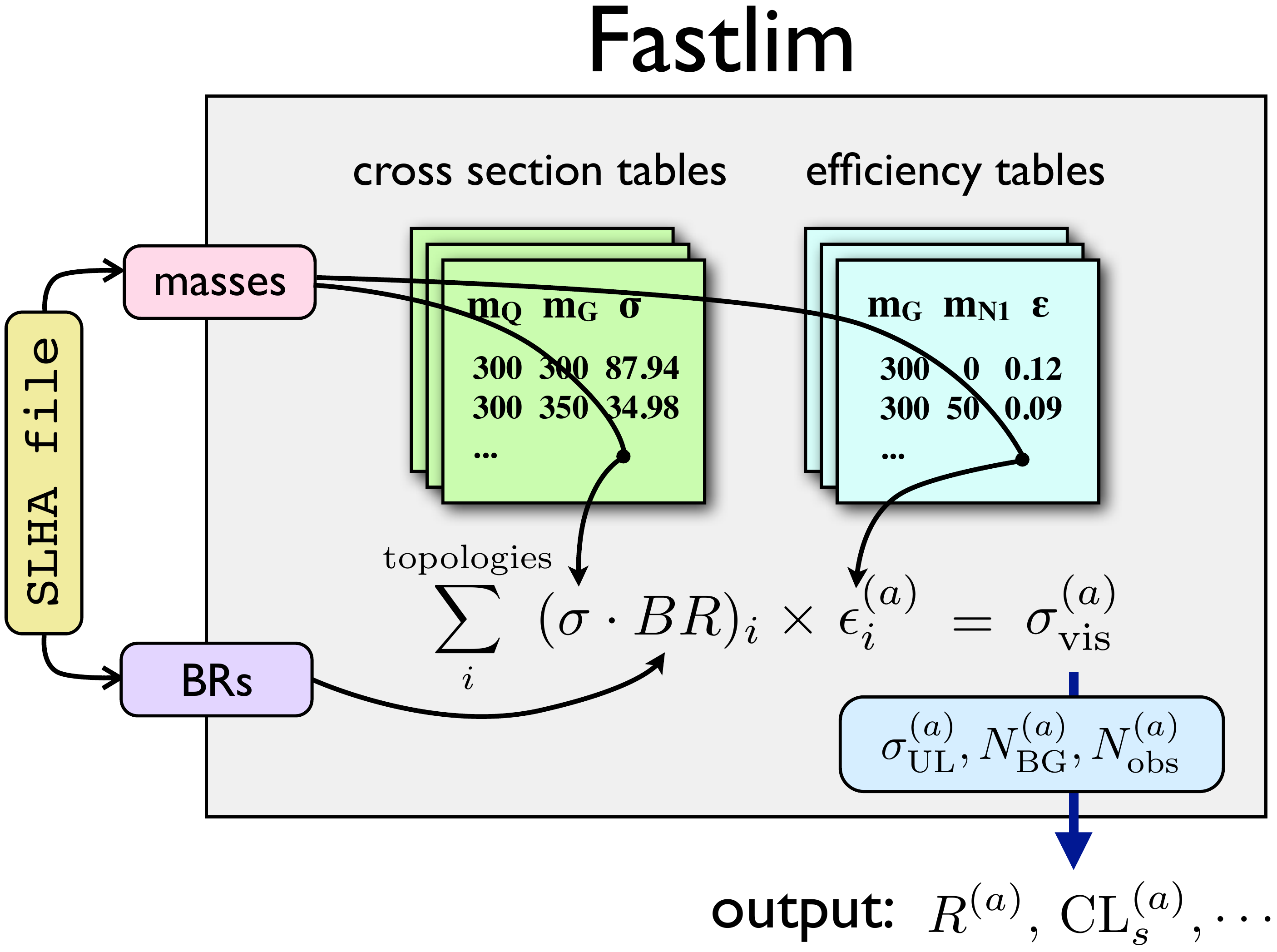}
	\caption{\label{FastLim-workingscheme}
		Schematic view of the working principle of {\tt FastLim}, from Ref.~\cite{Papucci:2014rja}.}
\end{figure}

This approach has been proposed and implemented in a new public tool, {\tt FastLim}~\cite{Papucci:2014rja}. Its working principle is given in Fig.~\ref{FastLim-workingscheme}. The program takes as input an SLHA file, and returns the ratio between the visible cross section and its 95\%~CL upper limit (as given in the experimental publication) in each signal region, or the confidence level with which the background$+$signal hypothesis is excluded in each signal region, using a simplified likelihood and the ${\rm CL}_s$ prescription~\cite{Read:2002hq}. As we saw, the acceptance$\times$efficiency maps are needed but are only rarely provided by the experimental collaborations. The solution found by the {\tt FastLim} collaboration is to use maps generated with {\tt ATOM}, a (private) tool where the analysis cuts are reproduced. After validation of the ``reimplemented'' analyses against the results given in the experimental publications, it is used to generate acceptance$\times$efficiency maps in every signal region for the simplified model topologies of interest. This is very interesting as it makes it possible to constrain simplified model topologies beyond the ones considered by the ATLAS and CMS collaborations, possibly covering simplified models with more than two parameters.

Regarding the comparison between {\tt SModels} and {\tt FastLim}, some comments are in order. First, only ten analyses from ATLAS are implemented in {\tt FastLim~1.0}, mostly the searches for stops and sbottoms, and the only topologies which are considered originate from squark or gluino-pair production.\footnote{Three preliminary ATLAS searches for electroweak-inos and sleptons~\cite{ATLAS-CONF-2013-035,ATLAS-CONF-2013-049,ATLAS-CONF-2013-093} are implemented in {\tt FastLim~1.0}, but cannot be used to constrain new physics because no efficiency map is given for any of the relevant topologies. Thus, only seven analyses are actually used in the current version of the program.} This is modest in comparison with more than fifty analyses currently implemented in {\tt SModels}, which includes all the 8~TeV SUSY analyses from ATLAS and CMS. Second, the results are given separately for the different signal regions and no combination is made. In the case where ATLAS or CMS results are given after combination of the signal regions, as in Ref.~\cite{ATLAS-CONF-2013-035}, this approach could lead to a {\it weaker} limit than the one derived from 95\%~CL limits on the cross section as given in Fig.~\ref{interpretations-sm-withnumbers} and used by {\tt SModelS}. Third, the acceptance$\times$efficiency maps depend on the reimplementation of the analysis in {\tt ATOM}, where the modeling of the ATLAS or CMS detector response has been done with a fast simulation software and not with the full simulation software from the ATLAS or CMS collaboration, being private (more information on issues when reimplementing analyses can be found in Section~\ref{sec:analysisreimplementation}). While validation materials are given in Ref.~\cite{Papucci:2014rja} for all used analyses, this usually do not cover all cases and do not always ensure that the derived acceptance$\times$efficiency maps will be a good approximation to the ones that could have been made by the experiments, especially in the regions where the mass splittings between BSM particles are small. Such limits should therefore be handled with some care and, contrary to {\tt SModelS}, they may not always be conservative. Finally, {\tt FastLim} can only take as input an SLHA file, hence is limited to testing supersymmetric extensions of the SM. {\tt SModelS} is much more general as it can take as input any event file in {\tt LHE} format.


\section[LHC constraints on light neutralino dark matter in the MSSM]{LHC constraints on light neutralino dark matter in the MSSM%
\sectionmark{LHC constraints on light neutralino DM in the MSSM}}
\sectionmark{LHC constraints on light neutralino DM in the MSSM} \label{sec:simpmod-lightneutralino}

The lightest neutralino, $\tilde\chi^0_1$, is a prime dark matter candidate in the MSSM. Its viability has already been assessed in Section~\ref{sec:pmssm} of this thesis, in the context of a global Bayesian analysis of the pMSSM. There, an upper bound on the relic density and the latest limits on the spin-independent scattering on nuclei were imposed. We found that the favored region typically has higgsino-like neutralinos that constitute only a small fraction of the observed relic density of dark matter, hence requiring the presence of another dark matter particle. While it was possible to accommodate the observed relic density in addition to all other constraints, it implied a significant bino-higgsino mixing, hence $M_1$ and $\mu$ to be close, which was found to be rather unlikely given the priors on these parameters. The observed relic density could also be achieved through co-annihilation with other supersymmetric particles or a precise tuning with respect to the mass of the $A$ boson. Both cases also require some fine-tuning and are disfavored in the Bayesian context.

In this section, we will go a step further and examine the possibility of having neutralino dark matter as light as possible in the MSSM. Such dark matter candidates are motivated by hints of signals in direct detection experiments found by CoGeNT~\cite{Aalseth:2010vx,Aalseth:2011wp}, DAMA~\cite{Bernabei:2008yi}, CDMS~\cite{Agnese:2013rvf}  and CRESST~\cite{Angloher:2011uu}, 
although the interpretation of these results in terms of dark matter is challenged by negative results obtained by XENON~\cite{Aprile:2012nq,Angle:2011th} and LUX~\cite{Akerib:2013tjd}. 
Hints of order 10~GeV dark matter might also be present in indirect dark matter searches, as discussed in, \eg, \cite{Hooper:2011ti,Hooper:2012ft,Berlin:2013dva}.
More generally, light dark matter candidates are allowed in many of the popular extensions of the SM 
and it is therefore interesting to explore this possibility irrespective of the direct detection results.

In the MSSM, several studies have shown that light neutralino dark matter  with mass of order 10~GeV 
can be compatible with collider data, in particular those from LEP, provided one allows for non-universality in gaugino masses~\cite{Hooper:2002nq,Bottino:2002ry,Belanger:2003wb,Bottino:2003iu,Bottino:2004qi}. 
Furthermore such light neutralinos can satisfy the recent constraints from 
B-physics observables, the muon anomalous magnetic moment, 
direct and indirect dark matter detection limits, as well as LHC limits, see~\cite{Dreiner:2009ic,Calibbi:2011ug, Arbey:2012na,Cumberbatch:2011jp,Kuflik:2010ah,Hooper:2008au,Vasquez:2010ru,
Belikov:2010yi,Boehm:2013qva,Han:2013gba,Calibbi:2013poa,Arbey:2013aba}. 

The connection between the invisible decays of the Higgs into a pair of neutralinos and the dark matter 
was explored before the discovery of the new boson, in the MSSM with non-universal gaugino masses as 
well as in the general MSSM (see for example~\cite{Belanger:2001am,AlbornozVasquez:2011aa}). The current 
precision determination of the relic density~\cite{Ade:2013zuv} and the possible constraints on the branching 
fraction of the Higgs into invisibles make further investigations of this connection very interesting. 
The precise determination of the relic density puts particularly strong constraints 
on the light dark matter.
Indeed, the mostly bino-like LSP that is found in the MSSM  typically requires some mechanism to enhance its annihilation in order not to overclose the Universe. Possible mechanisms include $s$-channel $Z$ or Higgs exchange, or $t$-channel slepton exchange (co-annihilation with sleptons is very much limited by slepton mass bounds from LEP).
For the Higgs exchange to be efficient, one has to be close to the (very narrow) $h^0$ resonance, {\it i.e.}\ $\mneut\simeq m_{h^0}/2\simeq 63$~GeV.
The $Z$ exchange is efficient for lighter neutralinos, but requires a non-negligible higgsino component. 
Hence ${\tilde{\chi}^0_2}$ and ${\tilde{\chi}^\pm_1}$ cannot be too heavy.  
For $t$-channel slepton exchange,  the sleptons must be light, close to the LEP mass bound. 
The light neutralino scenario can therefore be further probed by searching directly for electroweak-inos and/or sleptons at the LHC~\cite{Belanger:2012jn, Boehm:2013qva}.

In this section, we explore the parameter space of the MSSM, searching for scenarios with light neutralinos that are consistent with all relevant collider and dark matter constraints. We extend on previous studies in two main directions: 
first, we take into account the current LHC limits on sleptons and electroweak-inos in a simplified model approach, see section~\ref{sec:simpmod-intro}.  
Second, following Section~\ref{sec:higgs2013}, we include the fit to the properties of the 
observed 125--126~GeV Higgs boson in all production/decay channels, and we consider implications of the light neutralino dark matter scenario for this Higgs signal. 
These constraints were not taken into account in two recent studies~\cite{Boehm:2013qva,Han:2013gba}. 
Another recent paper \cite{Calibbi:2013poa} takes into account the most preliminary ATLAS limits from the di-tau plus $E_T^{\rm miss}$ searches~\cite{ATLAS-CONF-2013-028}, but does not discuss implications for the Higgs signal.

The work presented in this section has been conducted in collaboration with Genevi\`eve B\'elanger, Guillaume Drieu la Rochelle, Rohini M.~Godbole, Sabine Kraml, and Suchita Kulkarni in Spring and Summer 2013. This lead to the paper ``LHC constraints on light neutralino dark matter in the MSSM'', Ref.~\cite{Belanger:2013pna}, that was submitted to arXiv on August 16, 2013 and published in PLB in November 2013. It was also summarized from the Higgs point of view in a contribution to proceedings of the DIS~2014 conference~\cite{Dumont:2014ura}.  
The setup of the numerical analysis is described  in Section~\ref{lightneut-setup}. 
In Section~\ref{lightneut-expconst}, we discuss the various experimental constraints that are included in the analysis.
Our results are presented in Section~\ref{lightneut-results} and conclusions are given in Section~\ref{lightneut-concl}. 

\subsection{Setup of the numerical analysis} \label{lightneut-setup}

The model that we use throughout this study is the so-called pMSSM with parameters defined at the weak scale. 
We recall that the 19 free parameters of the pMSSM are the gaugino masses $M_1,\ M_2,\ M_3$, the higgsino parameter $\mu$, 
the pseudoscalar mass $M_A$,  the ratio of Higgs vev's, $\tan\beta=v_2/v_1$, 
the sfermion soft masses $M_{Q_i},M_{U_i},M_{D_i},M_{L_i},M_{R_i}$  ($i=1,\,3$ assuming degeneracy for the first two generations), and the trilinear couplings $A_{t,b,\tau}$.
In order to reduce the number of parameters to scan over, we fix a subset that is not directly relevant to our analysis to the following values: $M_3=1\tev$, $M_{Q_3}=750\gev$, $M_{U_i}=M_{D_i}=M_{Q_1}=2\tev$, and $A_{b}=0$.
This means that we take heavy squarks (except for stops and sbottoms) and a moderately heavy gluino. 
All the strongly interacting SUSY particles are thus above the LHC limits as of mid-2013; the gluino mass could be set to 2~TeV without changing our conclusions.
The parameters of interest are $\tanb$ and $\ma$ in the Higgs sector, the gaugino and higgsino mass parameters $M_1$, $M_2$ and $\mu$, the stop trilinear coupling $A_t$, the stau parameters $(M_{L_3}, M_{R_3}, A_\tau)$, and the slepton mass parameters $(M_{L_1}, M_{R_1})$. We allow these parameters to vary within the ranges shown in Table~\ref{lightneut-tab:scanrange}.\footnote{While the resulting pattern of heavy squarks and light sleptons is not the only possible choice, it seems well motivated from GUT-inspired models in which squarks typically turn out heavier than sleptons due to RGE running. Moreover, current LHC results indicate that squarks cannot be light. For a counter-example with light sbottoms, see Ref.~\cite{Arbey:2013aba}.}
The only free parameter in the squark sector, $A_t$, is tuned in order to match the mass of the lightest Higgs boson, $h^0$, with the observed state at the LHC.
\begin{table}[!h]
\begin{center}
\begin{tabular}{cc|cc}
 $\tanb$ & $[5,50]$ & $M_{L_3}$ & $[70,500]$ \\
 $\ma$ & $[100,1000]$ & $M_{R_3}$ & $[70,500]$ \\
 $M_1$ & $[10,70]$ & $A_\tau$\ & $[-1000,1000]$ \\
 $M_2$ & $[100,1000]$ & $M_{L_1}$ & $[100,500]$ \\
 $\mu$ & $[100,1000]$ & $M_{R_1}$ & $[100,500]$ \\
\end{tabular}
\end{center}
\vspace*{-5mm}
\caption{Scan ranges of free parameters. All masses are in GeV.}
\label{lightneut-tab:scanrange}
\end{table}

We have explored this parameter space by means of various flat random scans, some of them optimized to probe efficiently regions of interest. 
More precisely, two of our ``focused'' scans probe scenarios with light left-handed or light right-handed staus by fixing one of the stau soft mass to 500~GeV and varying the other in the $[70,150]$~GeV range. These two scans are subdivided according to the masses of the selectrons and smuons, by taking either fixed $M_{L_1}=M_{R_1}=500$~GeV or varying $M_{L_1}$ or $M_{R_1}$ within $[100,200]$~GeV. Another scan has been performed in order to probe scenarios with large stau mixing and light selectrons and smuons. In this case, $M_{L_3}$ and $M_{R_3}$ are varied within $[200,300]$~GeV and $M_{R_1}$ is tuned so that $m_{\tilde{e}_R}\in[100,200]$~GeV.  

In the following, we present the results for the combination of all our scans. 
The density of points has no particular meaning, as it is impacted by the arbitrary choice of regions of interest.
The computation of all the observables has been performed within \texttt{micrOMEGAs~3.1}~\cite{Belanger:2013oya}. \texttt{SuSpect 2.41}~\cite{Djouadi:2002ze} has been used for the 
computation of the masses and mixing matrices for Higgs particles and superpartners, while branching 
ratios for the decays of SUSY particles have been computed with 
\texttt{CalcHEP}~\cite{Belyaev:2012qa}.

\subsection{Experimental constraints} \label{lightneut-expconst}

The various experimental constraints that we use in the analysis are listed in 
Table~\ref{lightneut-tab:expconst}. A number of ``basic constraints'' are  imposed for a first selection. They 
include the LEP results for the direct searches for charginos and staus\footnote{Note that 
selectrons and smuons are safely above the LEP bound~\cite{lepsusy} since $M_{L_1} > 100\gev$ and $M_{R_1} > 100\gev$.}~\cite{lepsusy} and for invisible 
decays of the $Z$ boson~\cite{ALEPH:2005ab}, in addition to the OPAL limit on $e^+e^- \to \tilde{\chi}^0_{2,3} \tilde{\chi}^0_1 \to Z^{(*)}(\to q\bar{q})\tilde{\chi}^0_1$~\cite{Abbiendi:2003sc}. The anomalous magnetic moment of the muon is also 
required not to exceed the bound set by the E821 experiment~\cite{Bennett:2006fi,Hagiwara:2011af,Stockinger:2006zn}, 
and the flavor constraints coming from $b \to s\gamma$~\cite{Misiak:2006zs,Amhis:2012bh} and from $B_s \to 
\mu^+\mu^-$~\cite{CMSandLHCbCollaborations:2013pla,Akeroyd:2011kd} are taken into account. Finally, the ``basic constraints'' also require 
the 
lightest Higgs boson, $h^0$, to be within 3~GeV of the 2013 best fit mass from 
ATLAS~\cite{Aad:2013wqa} and CMS~\cite{CMS-PAS-HIG-13-005}. 
This range is completely dominated by the estimated theoretical uncertainties on the Higgs mass in 
the MSSM.

{\renewcommand{\arraystretch}{1.3}
\begin{table}[t]
\begin{center}
\begin{tabular}{c|c}
 LEP limits & $\mcha >100 ~\gev$ \\
                  & $\mstaua > 84-88 ~\gev$ (depending on $\mneut$) \\
                  & $\sigma(e^+e^- \to \tilde{\chi}^0_{2,3} \tilde{\chi}^0_1 \to Z^{(*)}(\to q\bar{q})\tilde{\chi}^0_1) \lesssim 0.05$ pb \\
 \hline
 invisible $Z$ decay & $\Gamma_{Z \to \neut\neut} < 3\ \mev$ \\
 \hline
 $\mu$ magnetic moment & $\Delta a_\mu < 4.5 \times 10^{-9}$ \\
 \hline
 flavor constraints & ${\rm BR}({b\to s\gamma}) \in [3.03,4.07] \times 10^{-4}$ \\
                             & ${\rm BR}({B_s\to\mu^+\mu^-}) \in [1.5,4.3] \times 10^{-9}$ \\
 \hline
 Higgs mass & $m_{h^0} \in [122.5,128.5]~\gev$ \\
 \hline
 $A^0, H^0 \to \tau^+\tau^-$ & CMS results for $\mathscr{L} = 17\rm{\ fb}^{-1}$, $m_h^{\rm max}$ scenario \\
 \hline
 Higgs couplings & ATLAS, CMS and Tevatron global fit, see text \\
 \hline
 relic density & $\Omega h^2<0.131$ or $\Omega h^2 \in [0.107,0.131]$ \\
 \hline
 direct detection & XENON100 upper limit \\
 \hline
 indirect detection & {\it Fermi}-LAT bound on gamma rays from dSphs \\
 \hline
 $pp \to \tilde{\chi}^0_2 \tilde{\chi}^\pm_1$ & Simplified Models Spectra approach, see text \\
 $pp \to \tilde{\ell}^+\tilde{\ell}^-$ & \\
\end{tabular}
\end{center}
\vspace*{-5mm}
\caption{Experimental constraints implemented in the analysis. For details, see text.}
\label{lightneut-tab:expconst}
\end{table}

In addition to the set of basic constraints, limits from searches for Higgs bosons at the LHC are taken into account. 
The heavier neutral Higgses, $A^0$ and $H^0$, are constrained by dedicated searches in the $\tau^+\tau^-$ 
channel. For these, we use the 2012 limit from CMS~\cite{CMS-PAS-HIG-12-050}, given in the 
($M_{A^0}$, $\tan \beta$) plane in the $m_h^{\rm max}$ scenario, which provides a conservative lower bound in the MSSM~\cite{Carena:2013qia}.\footnote{This is particularly the case in our study because our preferred very light neutralino scenarios have a small value for $\mu$ of order 200~GeV.}
The couplings of the observed Higgs boson at around 125.5~GeV, identified with $h^0$, are constrained following the procedure of  Section~\ref{2013c-sec:combinedss}, {\it i.e.}\ making use of the information given in the 2D plane $(\mu_{\rm ggF+ttH}, \mu_{\rm VBF+VH})$ for each final state provided by the LHC experiments. These ``signal strengths ellipses'' combine ATLAS and CMS results (plus results from Tevatron) for the  four effective final states that are relevant to the MSSM: $\gamma\gamma$, $VV = WW+ZZ$, $b\bar b$, and $\tau\tau$. As in Section~\ref{sec:higgs2013}, all the experimental results up to the LHCP 2013 conference are included in the present analysis. The signal strengths are computed from a set of reduced couplings %
($C_V,\, C_t,\, C_b,\, C_\tau,\, C_g$ and $C_\gamma$)
that are computed with leading order analytic formulas, except for the couplings of the Higgs to $b$ quarks, where loop corrections are included through $\Delta m_b$~\cite{Carena:1994bv}. A given point in parameter space is considered as excluded if one of these four 2D signals strengths falls outside the 95\%~CL experimental region.

Regarding dark matter limits, the following constraints are applied: direct detection with the 
spin-independent limit from XENON100~\cite{Aprile:2012nq} and relic density from the combined 
measurement released by Planck in Ref.~\cite{Ade:2013zuv}. 
The calculation of the spin-independent scattering cross section depends on nuclear parameters.
The light quark contents can be determined via the ratio of the masses of the light quarks, $m_u/m_d$ and $m_s/m_d$, and the light-quark sigma term $\sigma_{\pi N} = (m_u + m_d) \langle N | \bar{u}u + \bar{d}d | N \rangle/2$.  Moreover, we need the strange quark content of the nucleon,  $\sigma_s = m_s \langle N | \bar{s}s | N \rangle$. We use $m_u / m_d=0.553$, $m_s / m_d=18.9$, $\sigma_{\pi N}=44$~MeV and $\sigma_{s}=21$~MeV~\cite{Leutwyler:1996qg,Thomas:2012tg}. 
For the relic density, multiple ranges are given in~\cite{Ade:2013zuv}; 
we use the ``Planck+WP+BAO+highL'' best fit value of $\Omega h^2=0.1189$
assuming a theory dominated uncertainty of 10\% in order to account for unknown higher-order effects to the annihilation cross section. We will thus use $\Omega h^2<0.131$ as an upper bound or 
$0.107<\Omega h^2<0.131$ as an exact range.  
We also  consider indirect detection limits from dwarf spheroidal satellite galaxies (dSphs) released by {\it Fermi}-LAT based on measurements of the photon flux
\cite{Ackermann:2011wa,Ackermann:2013yva}; however, given that astrophysical uncertainties are still large and that current results do not strongly constrain scenarios of interest, we do not apply them to exclude parameter points but show the values of $\sigma v$ separately.

\subsubsection{LHC limits on sleptons, charginos and neutralinos}

Based on the data at $\sqrt{s} = 8\tev$, the ATLAS and CMS experiments have performed a number of searches for sleptons and electroweak-inos in final states with leptons and missing transverse energy, $E_T^{\rm miss}$. These have resulted in a significant improvement over the LEP limits and therefore need to be taken into account. Direct slepton production has been considered by ATLAS~\cite{ATLAS-CONF-2013-049} and CMS~\cite{CMS-PAS-SUS-12-022} in the $\ell^+\ell^- +\, E_T^{\rm miss}$ channel;\footnote{Shortly before completion of this study, Ref.~\cite{CMS-PAS-SUS-12-022} has been updated with full luminosity at 8~TeV~\cite{CMS-PAS-SUS-13-006}. This update has not been included in the present work.}
here only limits on  selectrons and smuons are currently available.
Electroweak-ino production is usually dominated by the $pp \to \tilde{\chi}^0_2 \tilde{\chi}^\pm_1$ process, which is searched for by ATLAS~\cite{ATLAS-CONF-2013-035} and CMS~\cite{CMS-PAS-SUS-12-022} in the trilepton + $E_T^{\rm miss}$ channel. The $\tilde{\chi}^0_2$ can decay either through an on-shell or off-shell $Z$ or a slepton, while the $\tilde{\chi}^\pm_1$ can decay through an on-shell or off-shell $W^\pm$ or a slepton.

Each scan point in the MSSM parameter space, which survives the basic constraints as well as the Higgs and dark
matter constraints discussed above, is decomposed into its relevant simplified model topologies (including the correct branching ratios) and compared against the limits given by the experiments using the {\tt SModelS} technology, as described in Section~\ref{sec:simpmod-intro} and in more detail in Ref.~\cite{Kraml:2013mwa}. A point is considered as excluded if one of the predicted $\sigma\times{\rm BR}$ exceeds the experimental upper limit, and allowed otherwise. 

In the present analysis, the simplified model results used are: {\it i)} $\tilde{\ell}^\pm_L \tilde{\ell}^\mp_L \to \ell^\pm \neut \ell^\mp \neut$ and $\tilde{\ell}^\pm_R \tilde{\ell}^\mp_R \to \ell^\pm \neut \ell^\mp \neut$  from both ATLAS~\cite{ATLAS-CONF-2013-049} and CMS~\cite{CMS-PAS-SUS-12-022}, {\it ii)} $\tilde{\chi}^0_2 \tilde{\chi}^\pm_1 \to Z^{(*)}\tilde{\chi}^0_1 W^{(*)}\tilde{\chi}^0_1$  again from ATLAS~\cite{ATLAS-CONF-2013-035} and CMS~\cite{CMS-PAS-SUS-12-022}, and {\it iii)} $\tilde{\chi}^0_2 \tilde{\chi}^\pm_1 \to \tilde{\ell}_R^\pm \nu \tilde{\ell}_R^\pm \ell^\mp \to \ell^\pm \neut \nu \ell^\pm \neut \ell^\mp$ from CMS~\cite{CMS-PAS-SUS-12-022}, where $\tilde{\ell}_R$ can be a selectron, a smuon or a stau. Note that the simplified model limits given by the experimental collaborations in terms of $\tilde{\chi}^0_2 \tilde{\chi}^\pm_1$ production apply for any $\tilde{\chi}^0_i \tilde{\chi}^\pm_j$ ($i=2,3,4$; $j=1,2$) combination.
Some more remarks are in order. First, for simplified model results involving more than two different SUSY particles, assumptions are made on their masses ({\it e.g.}\ degeneracy of $\tilde{\chi}^\pm_1$ and $\tilde{\chi}^0_2$ or specific relations between the masses in cascade decays) that are not always realized in the parameter space we consider. We allow up to 20\% deviation from this assumption in the analysis.
Second, the results for electroweak-ino production with decay through intermediate sleptons depend on the fractions  
of selectrons, smuons and staus in the cascade decay. 
When chargino/neutralino decays into staus as well as into selectrons/smuons are relevant, we use the results 
for the ``democratic'' case from \cite{CMS-PAS-SUS-12-022} if the branching ratios into the three flavors are nearly equal (within 20\%), and those for the ``$\tau$-enriched'' case otherwise.\footnote{We also apply the democratic case if decays into selectrons/smuons are more important then those into staus, but this hardly ever occurs for the scenarios of interest.}
Moreover, the results are provided by CMS for three specific values of $x=m_{\tilde{l}}/ (\mneut + m_{\tilde{\chi}^0_2})=0.05,0.5,0.95$; we use a quadratic interpolation to obtain a limit for other $x$ values. 
However, many of the scenarios we consider have light staus and heavy selectrons and smuons, 
for which the  ``$\tau$-dominated'' case applies.
Unfortunately, this has been provided by CMS only for $x=0.5$, corresponding to $m_{\tilde{\tau}_R}= (\mneut + m_{\tilde{\chi}^0_2})/2$. To get a limit for different mass ratios, we assume that the $x$ dependence is the same as in the $\tau$-enriched case; we estimate that the associated uncertainty is about a factor of 2, and we will flag the points affected by this uncertainty in the presentation of the results. 
Note that, unfortunately, we cannot use the ATLAS $2\tau$'s + $E_T^{\rm miss}$ analysis~\cite{ATLAS-CONF-2013-028}  in this approach, as it interprets the results only as left-handed staus and only for 
$m_{\tilde{\tau}_L} = (\mneut + m_{\tilde{\chi}^0_2})/2$. 
Parameter points with left-handed staus which are likely to be constrained by this analysis ({\it i.e.}, points satisfying $m_{\tilde{\chi}^{\pm}_1} < 350$~GeV and ${\rm BR}(\tilde{\chi}^{\pm}_1 \rightarrow \tilde \nu_\tau + \tau) > 0.3$) will also be flagged.

Finally, direct production of neutralino LSP's  can only be probed through mono-photon and/or mono-jet events. 
Limits from ATLAS and CMS have been given in \cite{CMS-PAS-EXO-12-048,ATLAS-CONF-2012-147} and interpreted as limits  on spin-independent interactions of dark matter with nucleons. We do not take into account these limits since they  can only be reliably interpreted in models where heavy mediators are responsible for the neutralino interactions with quarks.  This is not the case in the MSSM where the Higgs gives the dominant contribution to the neutralino interactions with nucleons.

\subsection{Results} \label{lightneut-results}

Let us now present the results of our analysis. 
Fig.~\ref{lightneut-fig:relic} shows the effect of the dark matter constraints. Here, 
the cyan points are all those which fulfill the ``basic constraints'' and also pass 
the limit on $A^0, H^0 \to \tau^+ \tau^-$ from CMS \cite{CMS-PAS-HIG-12-050}; 
blue points are in addition compatible at 95\%~CL with all Higgs signal strength measurements, based on the global fit of Section~\ref{sec:higgs2013}; 
red (orange) points obey moreover the relic density constraint $\Omega h^2<0.131$ ($0.107<\Omega h^2<0.131$) 
and abide the direct detection limits from XENON100 on $\sigma_{\rm SI}$~\cite{Aprile:2012nq}.\footnote{To account for the lower local density when the neutralino relic density is below the measured range, the predicted $\sigma_{\rm SI}$ is rescaled by a factor $\xi = \Omega h^2/0.1189$.} 
These red/orange points also pass the LHC limits  on charginos, neutralinos and sleptons; the set of points which fulfill all constraints including those from dark matter but are excluded by LHC searches are shown in gray (underlying the red/orange points).
We notice that typically the LHC limits reduce the density of points but do not restrict any further the 
range of masses that were allowed by the other constraints. 
The 90\%~CL limit from LUX is also shown in the right panel of Fig.~\ref{lightneut-fig:relic}. Taken at face value, this would translate into a lower limit on the LSP mass, $\mneut > 25$~GeV. This bound can however be weakened depending on the assumptions on the dark matter halo of our galaxy~\cite{Bozorgnia:2013pua}; therefore, we will only consider the XENON100 limit in the remainder of the section, as was done originally in Ref.~\cite{Belanger:2013pna}.

The upper bound on the relic density imposes a lower limit on the neutralino mass of approximately 15~GeV while the direct detection constraint does not modify the lower limit as will be discussed below. Moreover, the relic density  constrains the parameter space and the sparticles properties especially for neutralinos with mass below $\approx  30~{\rm GeV}$. 
These are  associated with light staus and light charginos as illustrated in Fig.~\ref{lightneut-fig:stau1chargino1}.
The light staus are  mostly right-handed to ensure efficient annihilation since the coupling of the bino LSP is proportional to the hypercharge which is largest for $\tilde\tau_R$. Furthermore, annihilation through stau exchange is not as efficient if staus are mixed since there is a destructive interference between the L--R contributions. The light charginos are mostly higgsino since a  small value for $\mu$ is required  to have an additional  contribution from $Z$ and/or Higgs exchange, both dependent on the LSP higgsino fraction. 

\begin{figure}[t!]\centering
\includegraphics[width=0.48\textwidth]{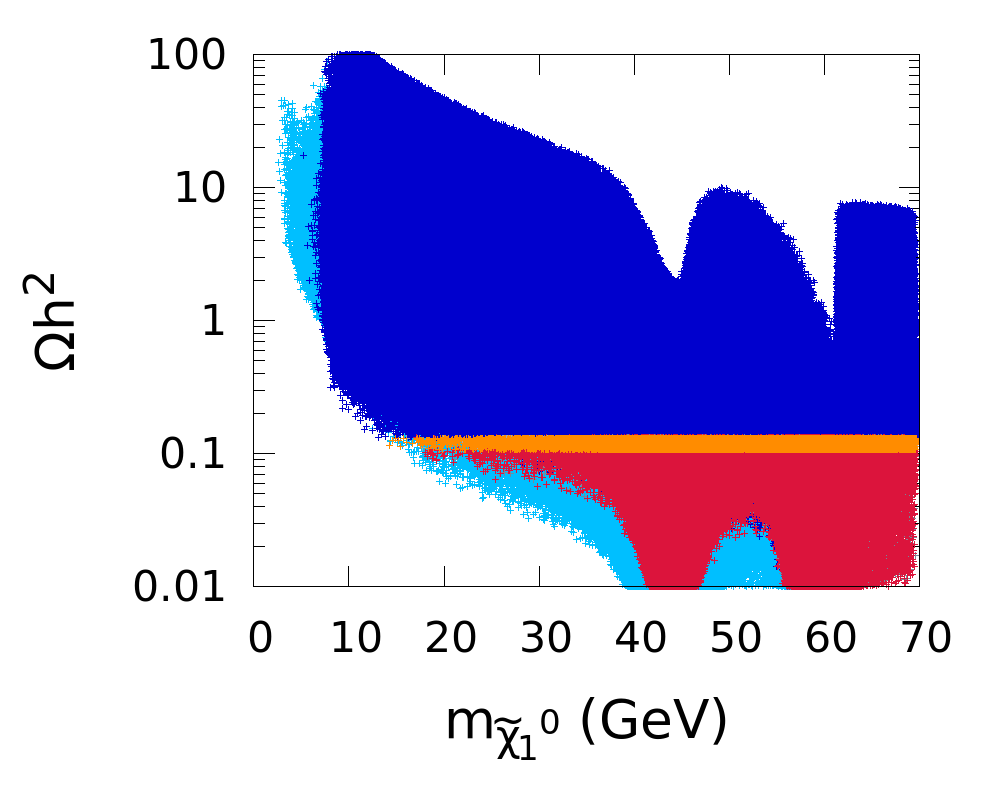}
\includegraphics[width=0.48\textwidth]{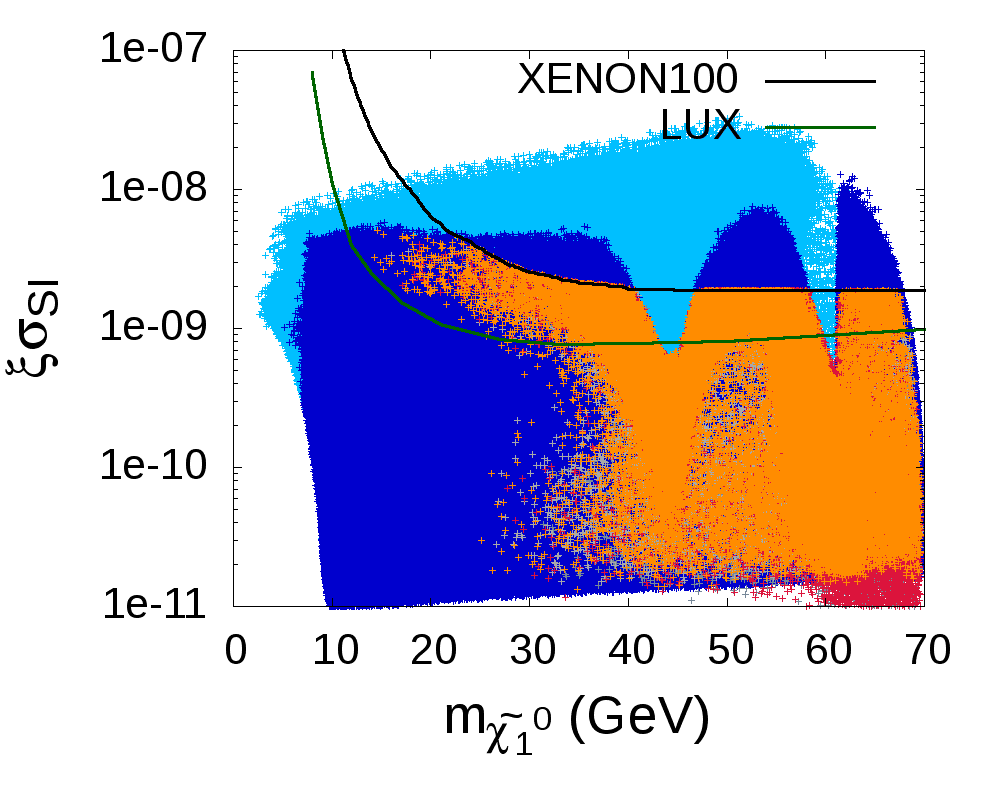}
\caption{Relic density $\Omega h^2$ (left) and rescaled spin independent scattering cross section $\xi 
\sigma_{\rm SI}$ (right) as function of the LSP mass, with $\xi = \Omega h^2/0.1189$.
Cyan points  fulfill the ``basic constraints'' and also pass 
the limit on $A^0, H^0 \to \tau^+ \tau^-$ from CMS; 
blue points are in addition compatible at 95\%~CL with all Higgs signal strengths based on the global fit of Section~\ref{sec:higgs2013}. Finally,
red (orange) points obey also the relic density constraint $\Omega h^2<0.131$ ($0.107<\Omega h^2<0.131$) 
and abide the direct detection limits from XENON100 on $\sigma_{\rm SI}$. The 2013 limit from the LUX experiment~\cite{Akerib:2013tjd}, which came out after the publication of~\cite{Belanger:2013pna}, is shown as a green line.
\label{lightneut-fig:relic} }
\end{figure}

\begin{figure}[t!]\centering
\includegraphics[width=0.48\textwidth]{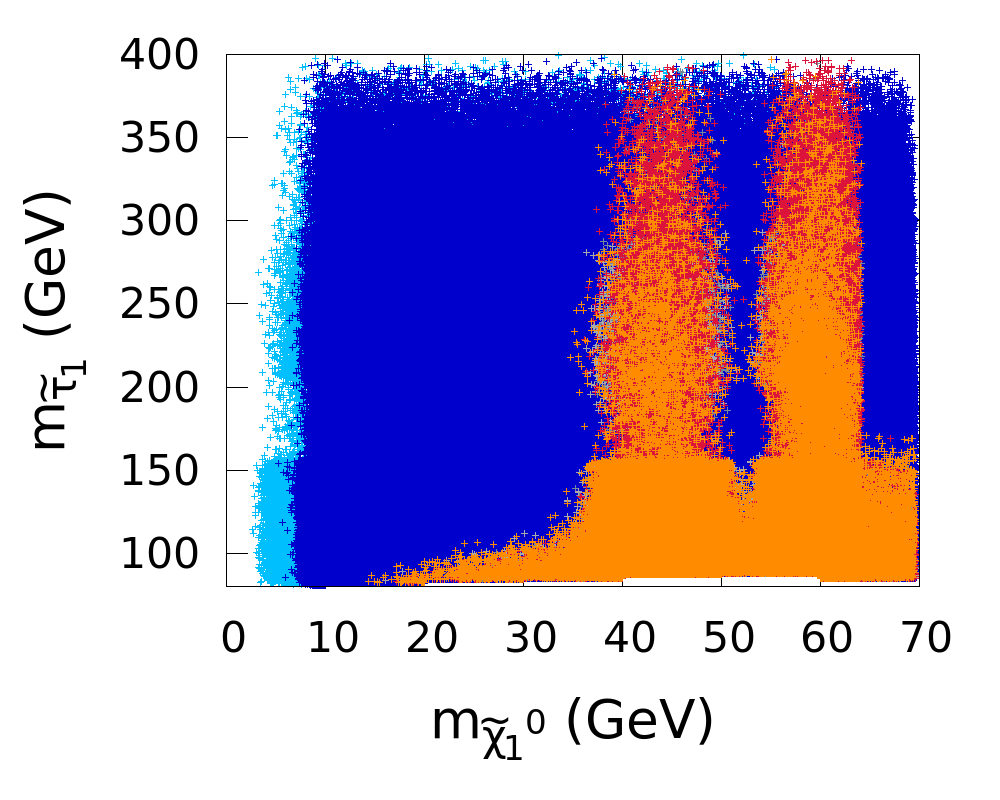}
\includegraphics[width=0.48\textwidth]{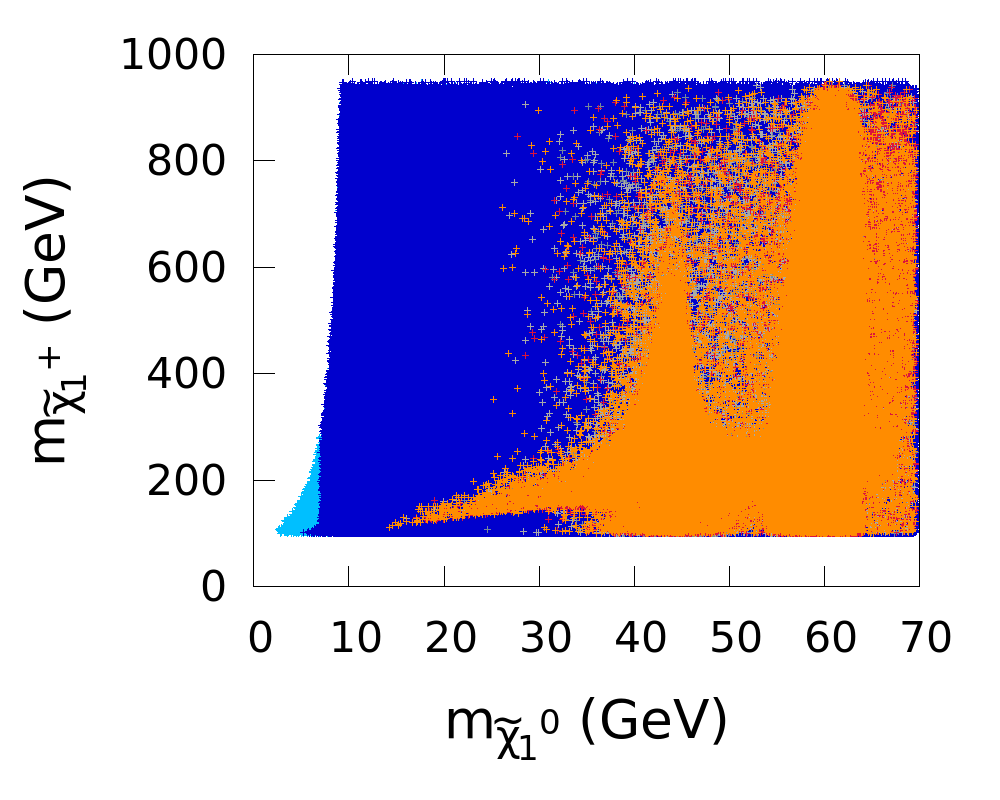}
\caption{Lighter stau mass (left) and chargino mass (right) versus $\mneut$; same color code as in Fig.~\ref{lightneut-fig:relic}. 
\label{lightneut-fig:stau1chargino1} }
\end{figure}

For neutralinos with masses above $\approx  30~{\rm GeV}$, the contribution of light selectrons/smuons in addition to that of the stau can bring  the relic density in the Planck range, in this case  it is not necessary to have a light chargino. These points  correspond to the scatter points with heavy charginos in Fig.~\ref{lightneut-fig:stau1chargino1} (right panel).
Finally, as  the LSP mass approaches $m_Z/2$ or  $m_h/2$ the higgsino fraction can be small because of the resonance enhancement in LSP annihilation---hence the chargino can be heavy. Moreover, for $m_{\neut}\gtrsim 35$~GeV the stau contribution to the LSP annihilation is not needed, so $m_{\tilde\tau_1}$ can be large. 
Fig.~\ref{lightneut-fig:muM2} summarizes the allowed parameter space in the $\mchar$ versus $\mstaua$ plane (left) as well as in the $M_2$ versus $\mu$ plane (right) for different ranges of LSP masses. The $M_2$ versus $\mu$ plot illustrates the fact that when the LSP is light, $\mu$ is small, hence $\charg$ and $\tilde\chi_2^0$ are dominantly higgsino as discussed above. 
In this plot also the points for which our implementation of LHC constraints in the simplified model approach has some 
significant uncertainty (from our extrapolation for the $\tau$-dominated case from \cite{CMS-PAS-SUS-12-022} 
or because the ATLAS di-tau + $E_T^{\rm miss}$ analysis~\cite{ATLAS-CONF-2013-028} is sensitive to this region in parameter space) become clearly visible. 
These points are flagged as triangles in a lighter color shade.  For $m_{\neut}<35$~GeV they concentrate 
in the region  $M_2,\mu\lesssim 320$~GeV (although a few such points have larger $\mu$). 
Most of these triangle points actually have a light $\tilde\tau_L$ and are thus likely to be excluded by the ATLAS 
result~\cite{ATLAS-CONF-2013-028}, see~\cite{Calibbi:2013poa}.
Note also that the production cross section for higgsinos is  low, so most of the points with low $\mu$ and larger 
$M_2$ are allowed. 

Another class of points that is strongly constrained by the LHC is characterized with light selectrons.   
The best limit comes from the ATLAS analysis~\cite{ATLAS-CONF-2013-049}; for LSP masses above 20~GeV, the ATLAS  searches are however insensitive to $\tilde{e}_R$ masses just above the LEP limit, more precisely in the range $m_{\tilde{e}_R} \approx 100$--120~GeV,
thus many points with light selectrons are still allowed. Furthermore, in many cases we have selectrons decaying into $\nu\tilde\chi_1^\pm$ and/or $e\tilde\chi_2^0$, thus avoiding the LHC constraint. All in all we find that 
for $\mneut>35$ GeV the whole selectron mass range considered in our scans is allowed (\textit{i.e.} either $[100,200]$ GeV or around $500$ GeV), while for $\mneut<35$ GeV, the ATLAS search 
imposes $m_{\tilde{e}_R} \approx 100$--120~GeV or ${\tilde{e}_R}$ being heavy, with the range $m_{\tilde{e}_R} \approx 120$--200~GeV being excluded. (Since we are mostly interested in how low the $\tilde\chi_1^0$ can go, we did not attempt to derive the upper end of the exclusion range for selectrons; note however that the bounds given by ATLAS vary between 230 and 450~GeV depending on the scenario.) 

\begin{figure}[t!]\centering
\includegraphics[width=0.48\textwidth]{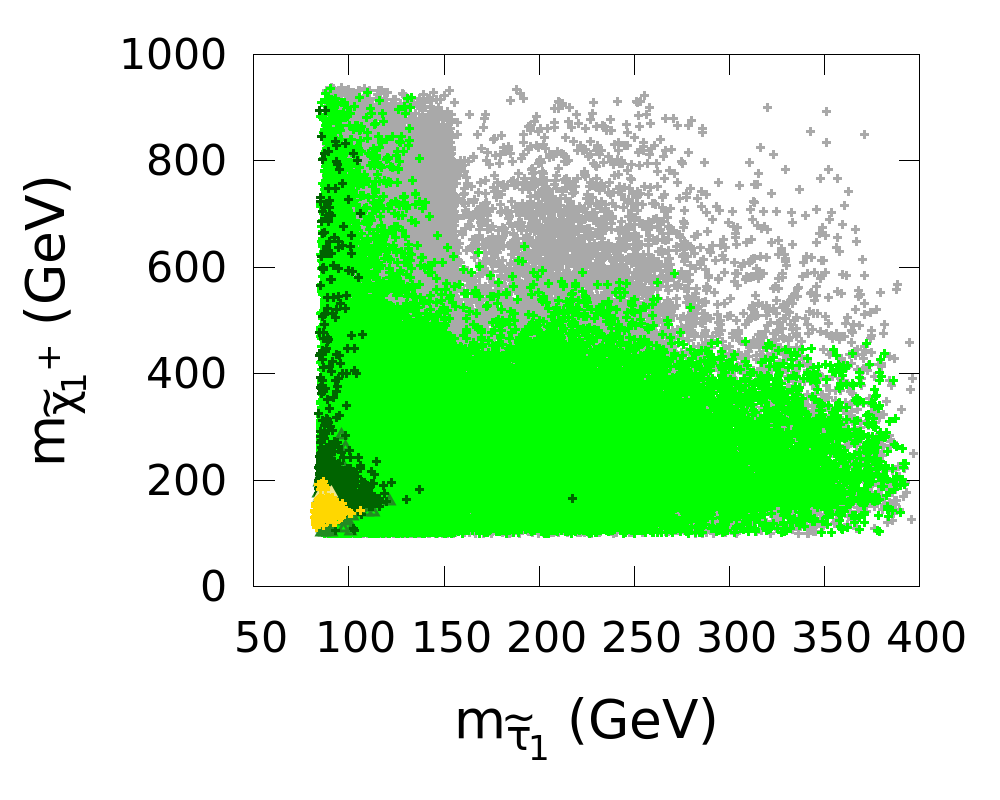}
\includegraphics[width=0.48\textwidth]{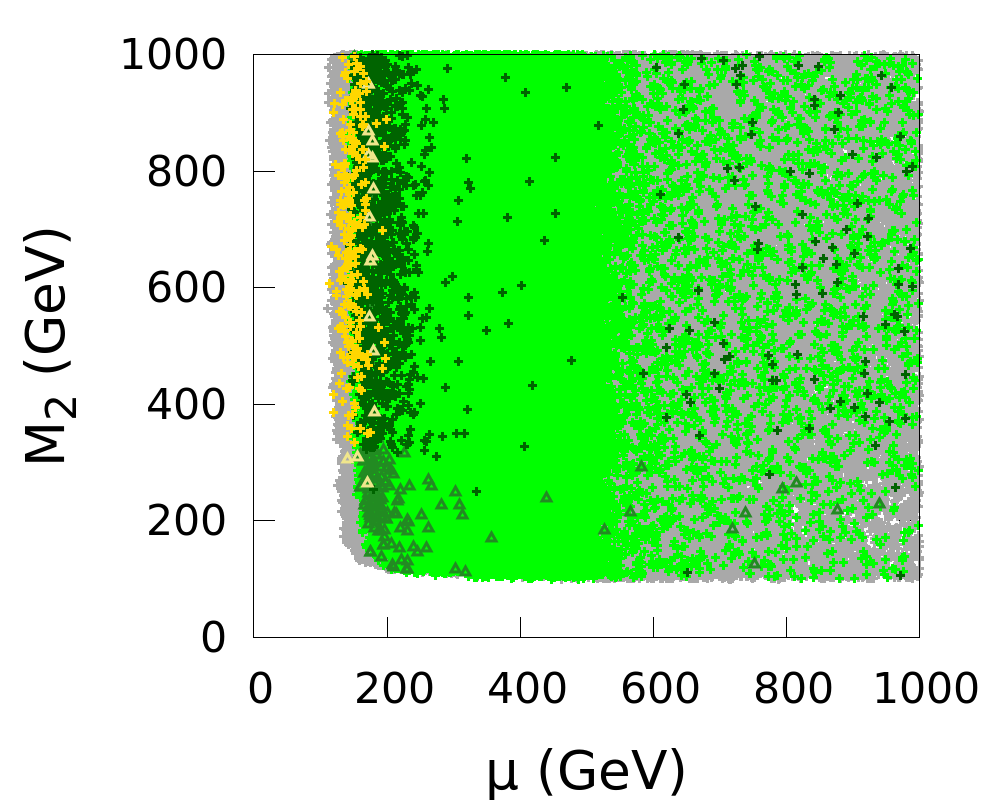}
\caption{Points passing all constraints, including $\Omega h^2<0.131$, XENON100 limits and simplified model 
limits from the LHC SUSY searches: on the left in the chargino versus stau mass plane, 
on the right in the $M_2$ versus $\mu$ plane. 
The yellow, dark green, light green and gray points have $\lsp$ masses of 15--25~GeV, 25--35~GeV, 
35--50~GeV and 50--60~GeV, respectively. Points which might be excluded either due to the factor 2 uncertainty 
in the implementation of the simplified model limit for the $\tau$-dominated case from  the CMS analysis \cite{CMS-PAS-SUS-12-022} or by the ATLAS $2\tau$'s + $E_T^{\rm miss}$ analysis~\cite{ATLAS-CONF-2013-028} are 
flagged as triangles in a lighter color shade.
\label{lightneut-fig:muM2} }
\end{figure}

The cross section for neutralino scattering on nucleons is dominated by the Higgs exchange diagram hence is driven by the higgsino fraction. For neutralinos below 30~GeV the cross section is mostly within one order of magnitude of the current XENON100 limit. It can however be much suppressed when the LSP has a small higgsino fraction. This occurs when the neutralino mass is near $m_Z/2$ or $m_h/2$ or when the light neutralino is purely bino and accompanied by light staus and light selectrons/smuons.
 
The interplay with indirect DM detection is also interesting. Fig.~\ref{lightneut-fig:indirect} shows $\sigma v$ corresponding to DM annihilation in the galaxy in either the $b\bar{b}$ or $\tau\tau$ channel. 
The latest upper limits  from {\it Fermi}-LAT indirect searches for photons produced from DM annihilation in dwarf spheroidal galaxies constrain a very small subset of the points with light  DM annihilating into $\tau\tau$.   Some of these points are also in the region probed by {\it Fermi}-LAT searches
for DM annihilation in subhalos~\cite{Berlin:2013dva} or from the Galactic Center~\cite{Hooper:2011ti}, the latter
bounds however depend  on the assumed DM profile. However, a large fraction of allowed points corresponding to  $\mneut>30 ~{\rm GeV}$ are several orders of magnitude below the current limits whether their main annihilation channel be into $\tau\tau$ or $b\bar{b}$. 
(For completeness we note that $\sigma v$ goes down to $\approx 10^{-34}~{\rm cm^2s^{-1}}$ in the $\tau\tau$ channel and down to $\approx10^{-31}~{\rm cm^2s^{-1}}$ in the $b\bar{b}$ channel.)

 \begin{figure}[t!]\centering
\includegraphics[width=0.48\textwidth]{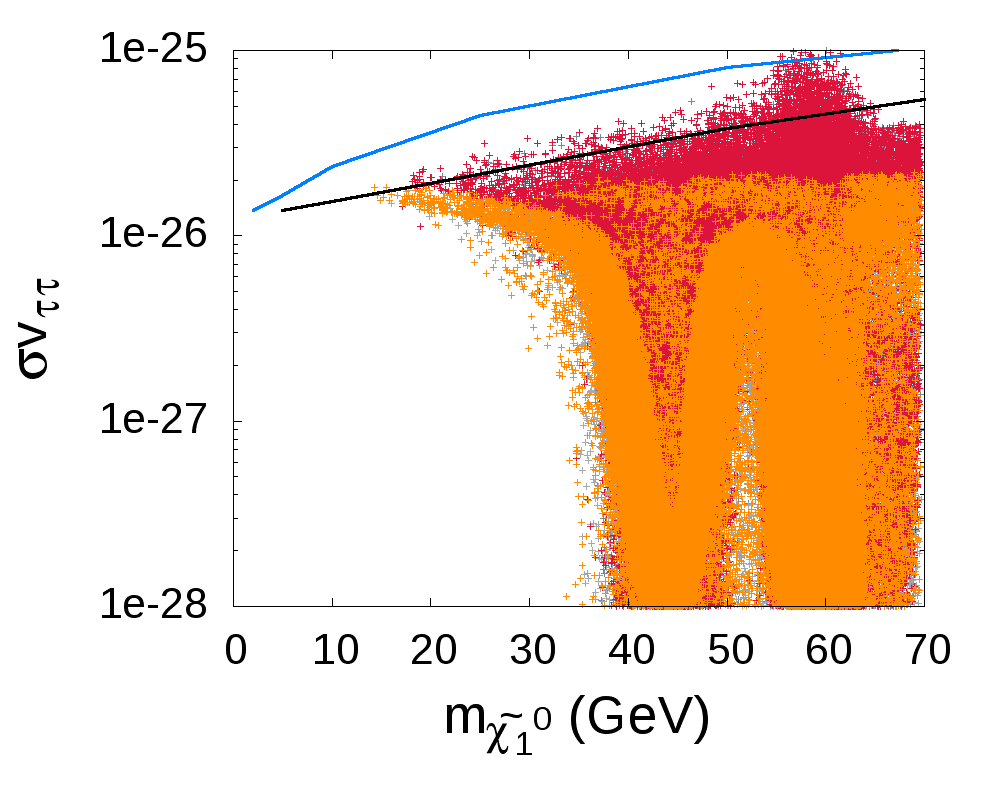}
\includegraphics[width=0.48\textwidth]{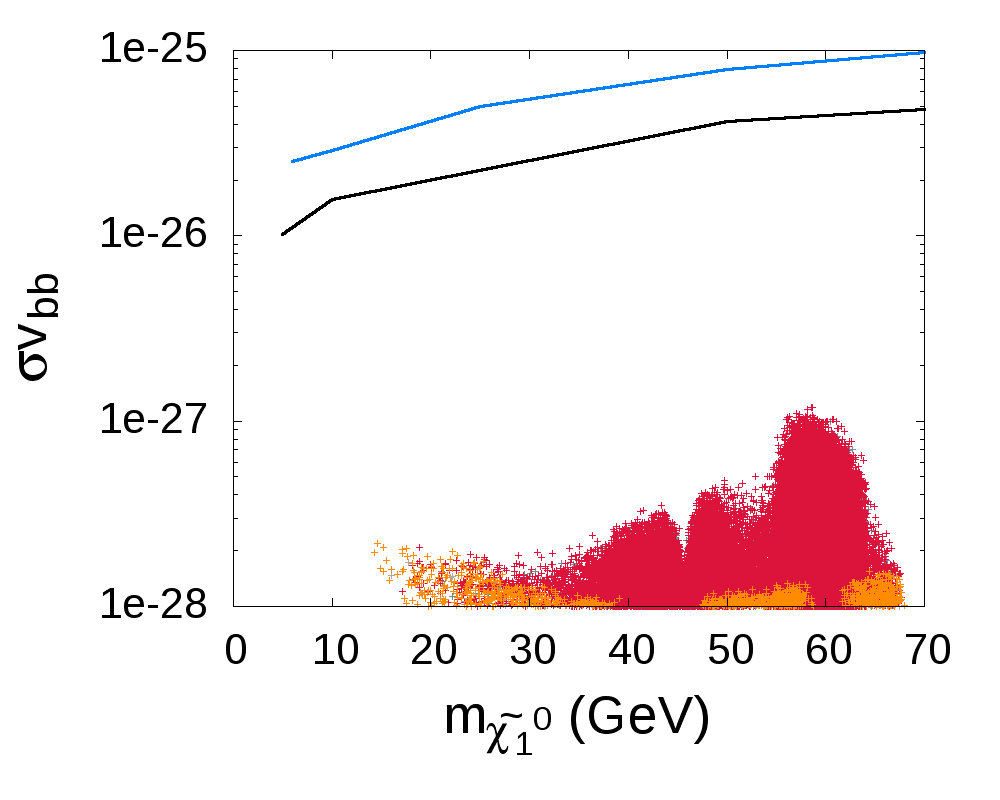}
\vspace*{-2mm}
\caption{Cross sections $\sigma v$ (in ${\rm cm^2s^{-1}}$) for indirect DM detection in the $\tau\tau$ (left) and $b\bar b$ (right) channels.  The black (blue) line shows the 2011 (2013) 95\%~CL {\it Fermi}-LAT bound~\cite{Ackermann:2011wa,Ackermann:2013yva}. 
Only points which satisfy the relic density and direct detection constraints are shown; 
following the color code of Fig.~\ref{lightneut-fig:relic}, red (orange) points have $\Omega h^2<0.131$ ($0.107<\Omega h^2<0.131$).
\label{lightneut-fig:indirect} }
\end{figure}

We next consider the implications for Higgs signal strengths $\mu$ relative to SM expectations in various channels. 
There are two features that can lead to modifications of the signal strengths in our scenario: a light neutralino 
and a light stau. The presence of a light neutralino can lead to a sizable invisible decay width, thus leading 
to reduced signal strengths in all channels.
A  light stau  can contribute to the loop-induced $h\gamma\gamma$ coupling~\cite{Carena:2012gp}. In particular, heavily mixed staus can lead to enhanced signal strengths in the diphoton channels, while not affecting the signal in other decay channels. This is illustrated in Fig.~\ref{lightneut-fig:mugamgam} (top panels). Here, the points with an enhanced $\mu(gg,\gamma\gamma)\equiv \mu(gg\to h\to \gamma\gamma)$ are the ones with light, maximally mixed staus; these points occur only for $m_{\lsp}>25$~GeV and their signal strengths in the $VV$ ($WW$ or $ZZ$) and $b\bar{b}/\tau\tau$ channels do not differ significantly from 1, as can be seen in the bottom panels of Fig.~\ref{lightneut-fig:mugamgam}. 
To achieve large stau mixing, we need $\mu\gtrsim 400$~GeV, so $\charg$ and $\tilde\chi^0_2$ are heavy in this case. 
Moreover, the scenarios with mixed staus require light selectrons/smuons in order to achieve low enough $\Omega h^2$.  Therefore these points are mostly constrained by the ATLAS results from direct slepton searches. 

\begin{figure}[t!]\centering
\includegraphics[width=0.48\textwidth]{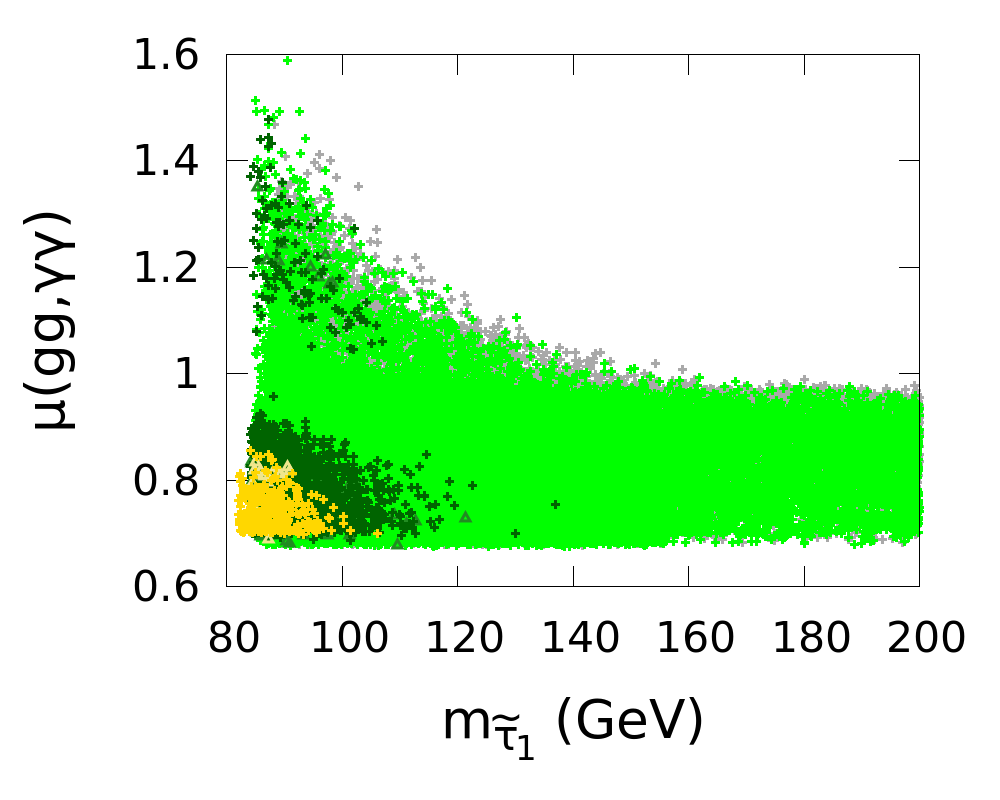}
\includegraphics[width=0.48\textwidth]{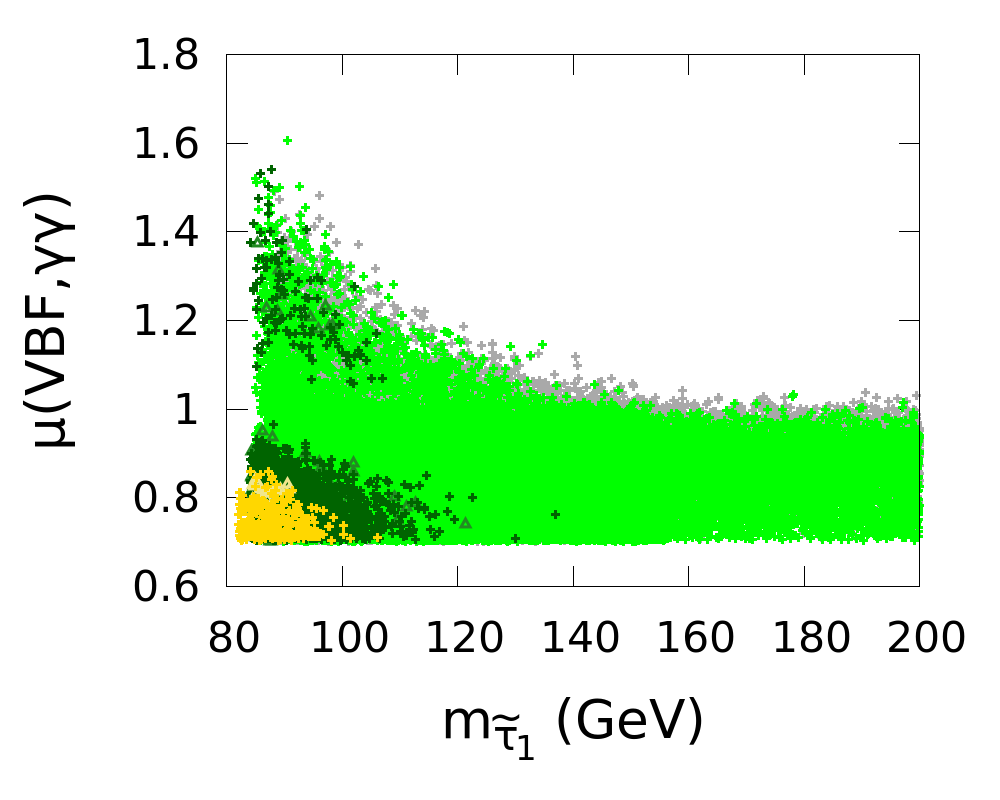} \\
\includegraphics[width=0.48\textwidth]{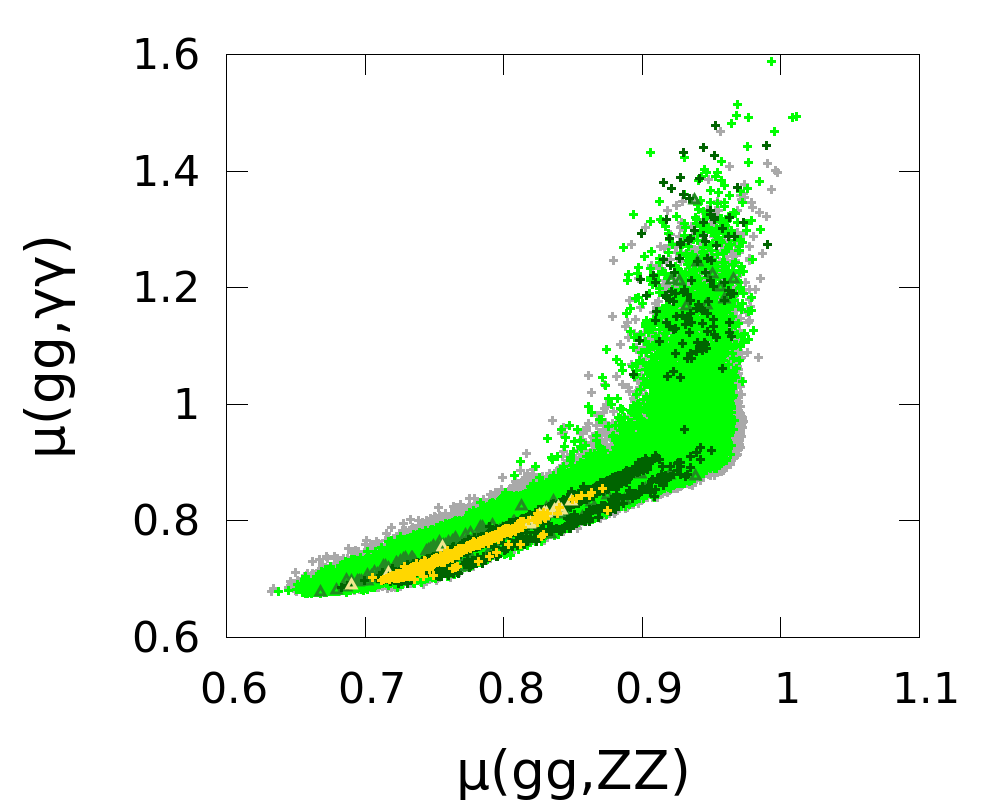}
\includegraphics[width=0.48\textwidth]{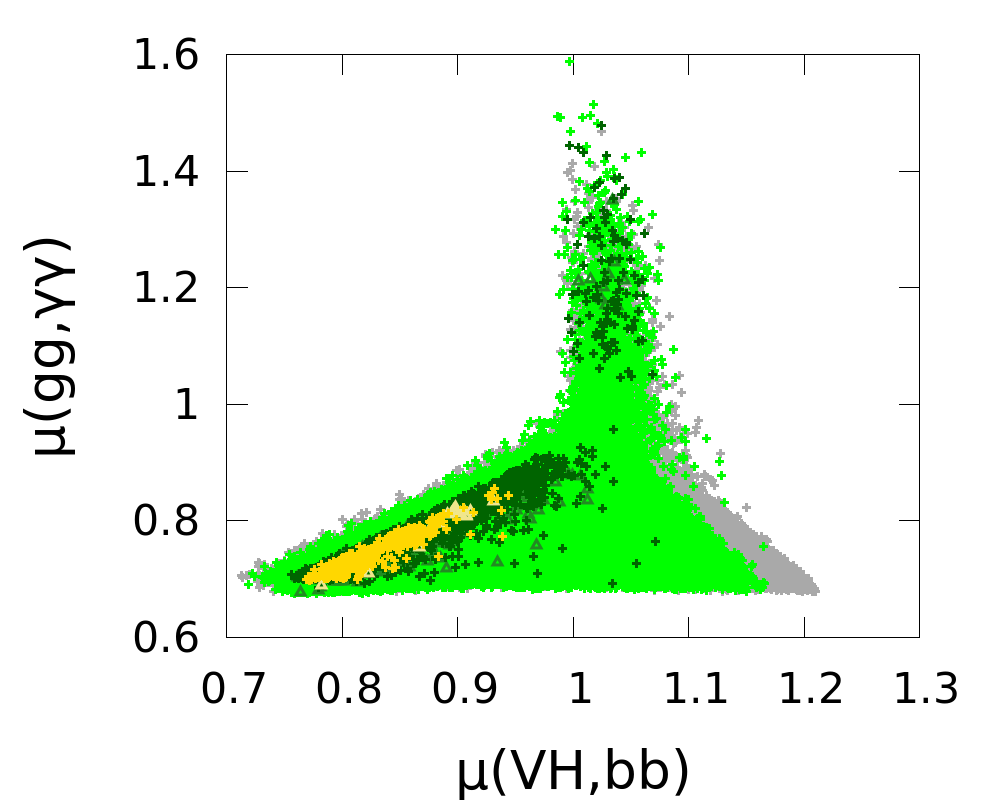} \\
\caption{Implications of the light neutralino dark matter scenario for Higgs signal strengths. 
Same color code as in Fig.~\ref{lightneut-fig:muM2}. 
\label{lightneut-fig:mugamgam} }
\end{figure}

\begin{figure}[t!]\centering
\includegraphics[width=0.48\textwidth]{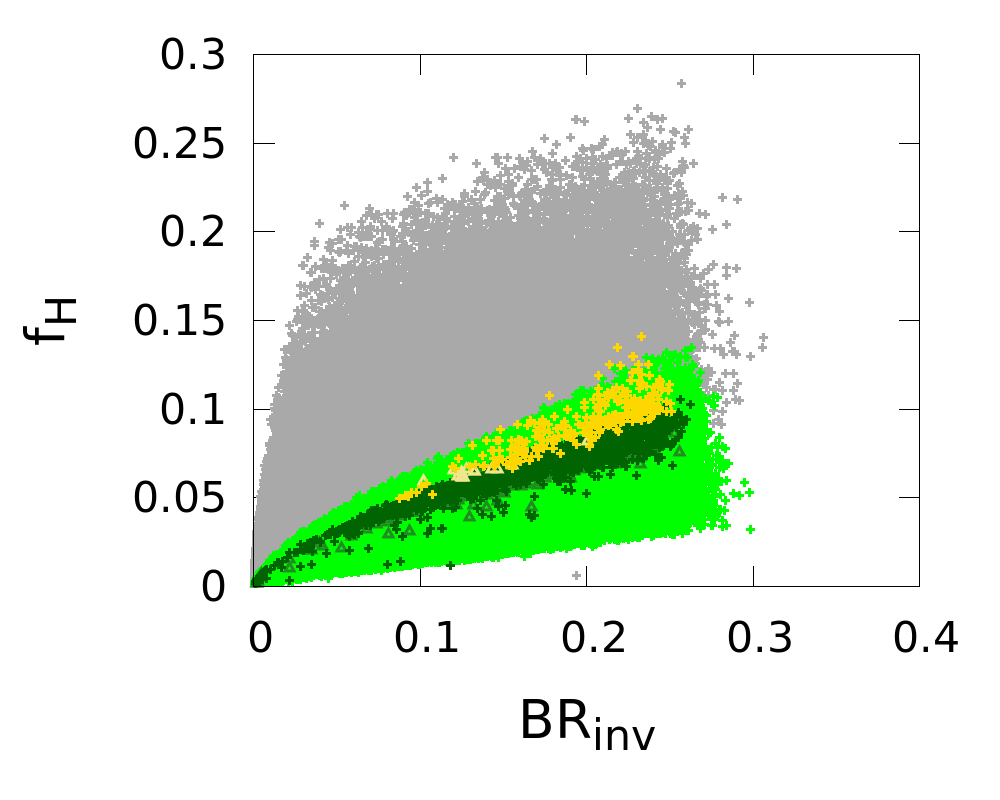} 
\includegraphics[width=0.48\textwidth]{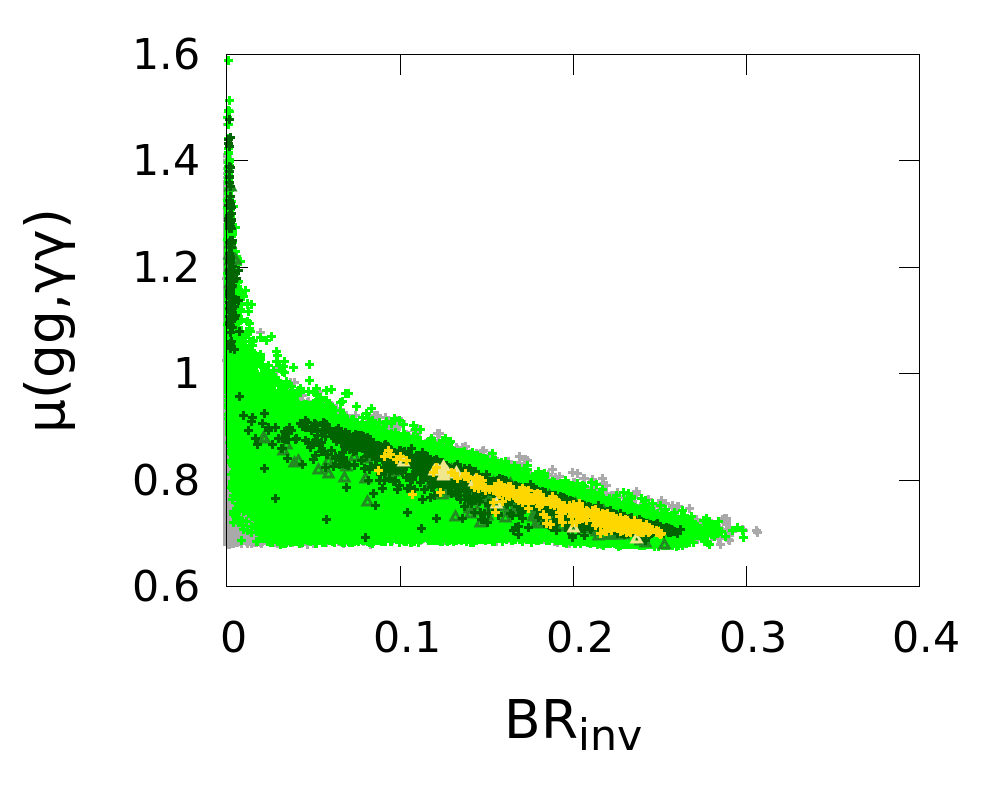}
\caption{Implications of the light neutralino dark matter scenario for invisible $h$ decays.
On the left panel, $f_H$ corresponds to the higgsino fraction.
Same color code as in Fig.~\ref{lightneut-fig:muM2}. 
\label{lightneut-fig:inv} }
\end{figure}

The bulk of the light $\lsp$ points
however features a  reduced $\mu({gg,\gamma\gamma})\approx 0.7-0.9$.  This occurs when the stau  has only a mild effect on  $h\gamma\gamma$ and the invisible decay is sizable, see Fig.~\ref{lightneut-fig:inv}. In particular, for the very light neutralino sample with $m_{\lsp}=15-25$~GeV the light  $\tilde\tau_R$ needed for DM constraints does not help in increasing the 
$h\gamma\gamma$  coupling, hence all these points (in yellow  in Fig.~\ref{lightneut-fig:mugamgam})
have a reduced signal strength.  Note also that for the points with $\mu({gg,\gamma\gamma})<1$, also  $\mu({gg,VV})$ is suppressed, see bottom-left panel in Fig.~\ref{lightneut-fig:mugamgam}. Here, suppression of the gluon-fusion process by the stop-loop contribution also plays a role on top of the effect from invisible decays of the Higgs boson. 
Associated VH production on the other hand is unaffected by this, and since the Higgs branching ratio into $b\bar{b}$ can be enhanced or suppressed $\mu({\rm VH},b\bar{b})$ can be above or below 1, as can be seen in the bottom-right panel of Fig.~\ref{lightneut-fig:mugamgam}.

The invisible branching ratio of the Higgs can vary up to $\approx 30\%$ (the maximum allowed by the global Higgs fit) and is large for a large higgsino fraction of the LSP modulo kinematic factors, as illustrated in Fig.~\ref{lightneut-fig:inv}. For this reason, the invisible width can be large for neutralinos below 35~GeV, leading to suppressed Higgs signals in all channels. Moreover, the points  with $\mu(gg,\gamma\gamma)>1$ have a small invisible width since they correspond to mixed staus and a small higgsino fraction (because large stau mixing requires large $\mu$) as mentioned above. The invisible width is also suppressed for $\mneut\approx m_Z/2$ because of the small higgsino fraction as well as near $m_h/2$ for kinematical reasons.  

Future experimental results on the various Higgs signals will help constraining MSSM scenarios with a light neutralino, as can be expected from the 14~TeV projections of ATLAS~\cite{ATL-PHYS-PUB-2012-004} and CMS~\cite{CMS-NOTE-2012-006} for $\mathscr{L} = 300\fbi$. The estimated precision on the signal strengths is of the order of $10$\% in several channels of interest, including $h \to \gamma\gamma$ and $h \to ZZ$. As can be seen in Fig.~\ref{lightneut-fig:mugamgam}, this will help discriminating between the various scenarios---in particular, the points with $m_{\lsp}=15$--$25$~GeV have $\mu^{\max}(gg,\gamma\gamma) \approx \mu^{\max}(gg,ZZ) \approx 0.86$. A better determination of the invisible decays of the Higgs boson should also probe further the remaining parameter space, both from a global fit to the properties of the Higgs and from direct searches for Higgs decaying invisibly. In the latter case, the projected upper bound is found to be $\brinv \lesssim 0.17$ at 95\%~CL for 100$\fbi$\ at 14~TeV~\cite{Ghosh:2012ep}.

\subsection{Conclusions} \label{lightneut-concl}

We found that although the most recent LHC limits on Higgs properties and on direct production of SUSY particles impose strong constraints on the model, neutralino LSPs as light as 15~GeV can be compatible with all data. These scenarios require light staus below about 100~GeV and light charginos below about 200~GeV. They will be further probed at the LHC at 13--14~TeV through searches for sleptons and electroweak-inos, as well as  through  a more precise determination of the Higgs couplings. Moreover, the 90\%~CL limit from LUX is putting pressure on the low-mass region, and the improvement of the direct detection limits by an order of magnitude will cover the whole range of $\lsp$ masses below $\approx 35$~GeV.


\section{Status of the mixed sneutrino dark matter model} \label{sec:simpmod-sneutrino}

In a certain class of models, small neutrino masses may naturally arise from F-term SUSY breaking~\cite{ArkaniHamed:2000bq,Borzumati:2000mc}. 
In addition to providing an explanation for neutrino masses, this class of SUSY models offers a particular DM candidate: a mainly right-handed (RH) mixed sneutrino. 
Mixed sneutrinos as thermal DM are thus a very interesting alternative to the conventional neutralino LSP of the MSSM. They have received much attention recently, in part because of their intriguing phenomenology and in part because they provide a possibility for light SUSY DM below 10~GeV. 
The crucial point of this model is that one can have a weak-scale trilinear sneutrino coupling 
$A_{\tilde\nu}$ that is not suppressed by a small Dirac-neutrino Yukawa coupling. It can hence induce a large mixing between left-handed and right-handed sneutrinos even though the Yukawa couplings may be extremely small.  The lightest sneutrino can thus become the LSP and a viable thermal DM candidate.  
Note that the mainly RH sneutrino LSP is not sterile but couples to SM gauge and Higgs 
bosons through the mixing with its LH partner. Sufficient mixing provides efficient 
annihilation so that the sneutrino relic density matches the one 
extracted from cosmological observations~\cite{Ade:2013zuv}. 

Direct detection experiments however pose severe constraints on Dirac or complex 
scalar, {\it i.e.}\ not self-conjugated, DM particles because the spin-independent 
elastic scattering cross section receives an important contribution from $Z$ 
exchange, which typically exceeds experimental bounds. 
In the mixed sneutrino model (MSSM+RH), this cross section is suppressed by the sneutrino 
mixing angle. Therefore, on the one hand a viable sneutrino DM candidate requires 
enough mixing to provide sufficient pair-annihilation, on the other hand the 
mixing should not be too large in order not to exceed the direct detection limits. 

In the present section, a wide range of sneutrino masses is explored, considering both light DM below 10~GeV motivated by hints of DM signals in direct detection experiments~\cite{Bernabei:2008yi,Aalseth:2010vx}
as well as heavier DM of the order of 100~GeV. 
The parameter space is explored by means of Markov Chain Monte Carlo sampling, using Bayesian statistics 
to confront the model predictions with the data. In taking into account the limits from direct detection experiments,  
special attention is paid to uncertainties stemming from astrophysical parameters (local DM density and velocity distribution) and to uncertainties in the quark contents of the nucleons (relevant in particular when there is a large Higgs-exchange contribution).
The results are presented as posterior probability densities of parameters and derived quantities, in particular of the DM mass and direct and indirect detection cross sections.

The study of the mixed sneutrino dark matter model was the first project of my PhD thesis, in collaboration with Genevi\`eve B\'elanger, Sylvain Fichet, Sabine Kraml and Thomas Schwetz. It was submitted to arXiv under the name ``Mixed sneutrino dark matter in light of the 2011 XENON and LHC results'' on June 7, 2012 and published in JCAP in the following September~\cite{Dumont:2012ee}. However, three important pieces of news from the experimental side came out after the publication of the paper. First, an SM-like Higgs boson with mass of about 125~GeV was discovered by the ATLAS and CMS collaborations at the LHC~\cite{Aad:2012tfa,Chatrchyan:2012ufa}. Second, significant improvements were made in direct detection experiments, resulting in improved limits on the spin-independent scattering cross section of dark matter, both in the low- and in the high-mass region. Third, new, more stringent constraints on the production of SUSY particles were set at the LHC with the full data set at $\sqrt{s} = 8$~TeV.
The work done in Ref.~\cite{Dumont:2012ee} will be presented first, in Sections~\ref{sn2012-sec:framework}--\ref{sn2012-sec:results} (where the LHC SUSY results are simply taken into account as a lower bound on the gluino mass). As we will see, the discovery of an SM-like Higgs boson ruled out sneutrino dark matter with mass below 45~GeV, which is a significant piece of news from the LHC. Nonetheless, the light sneutrino case was an important part of the study made in 2011--2012, and illustrates the discriminating power of the LHC. In Section~\ref{sn2012-sec:update}, an update of the analysis for the heavy sneutrino case will then be presented. The three major experimental updates mentioned above are taken into account. The latest direct detection limits from the LUX experiment~\cite{Akerib:2013tjd} are included using a private code based on Refs.~\cite{Bozorgnia:2013hsa,Bozorgnia:2013pua,Bozorgnia:2014dqa},\footnote{We thank Thomas Schwetz and Nassim Bozorgnia for providing this code.} and the LHC SUSY searches are taken into account in the simplified model approach, described in Section~\ref{sec:simpmod-intro}, using {\tt SModelS} as in the previous section.
Finally, conclusions are given in Section~\ref{sn2012-sec:conclusions}.

The phenomenology of the MSSM+RH neutrino model was previously investigated in detail in~\cite{ArkaniHamed:2000bq,Belanger:2010cd}. 
Indirect detection signatures were discussed in~\cite{Arina:2007tm,Choi:2012ap}, implications for $\Omega_b/\Omega_{\rm DM}$ in \cite{Hooper:2004dc}, and LHC signatures in~\cite{Thomas:2007bu,Belanger:2011ny,Arina:2013zca}.

\subsection{Framework}\label{sn2012-sec:framework}

The framework for our study is the model of \cite{ArkaniHamed:2000bq,Borzumati:2000mc} with only 
Dirac masses for neutrinos. In this case, the usual MSSM soft-breaking terms are extended by
\begin{equation}
  \Delta {\cal L}_{\rm soft} = m^2_{\tilde N_i}  |\tilde N_i |^2 +  
                                            A_{\tilde\nu_i} \tilde L_i \tilde N_i H_u + {\rm h.c.} \,,
\end{equation}
where ${m}^2_{\widetilde{N}}$ and $A_{\tilde\nu}$ are weak-scale soft terms, which we assume to 
be flavor-diagonal. Note that the lepton-number violating bilinear term, which appears 
in case of Majorana neutrino masses, is absent. 
Neglecting the tiny Dirac masses, the $2\times2$ sneutrino mass matrix for one generation is 
given by 
\begin{equation}
  m^2_{\tilde\nu} =
  \left( \begin{array}{cc}
   {m}^2_{\widetilde{L}} +\frac{1}{2} m^2_Z \cos 2\beta \quad &  \frac{1}{\sqrt{2}} A_{\tilde\nu}\, v \sin\beta\\
   \frac{1}{\sqrt{2}}    A_{\tilde\nu}\, v \sin\beta&  {m}^2_{\widetilde{N}}
  \end{array}\right) \,.
\label{sn2012-eq:sneutrino_tree}
\end{equation}
Here ${m}^2_{\widetilde{L}}$ is the SU(2) slepton soft term, $v^2=v_1^2+v_2^2=(246\;{\rm GeV})^2$ 
with $v_{1,2}$ the Higgs vacuum expectation values, and $\tan\beta=v_2/v_1$.  
The main feature of this model is that ${m}^2_{\widetilde{L}}$, ${m}^2_{\widetilde{N}}$
and $A_{\tilde\nu}$ are all of the order of the weak scale, and 
$A_{\tilde\nu}$ does not suffer any suppression from Yukawa couplings. 
In the following, we will always assume $m_{\widetilde N}<m_{\widetilde L}$ so that the lighter mass 
eigenstate, $\tilde\nu_1$, is mostly a $\tilde\nu_R$. 
This is in fact well motivated from renormalization group evolution, 
since for the gauge-singlet $\widetilde{N}$ the running at 1-loop is driven exclusively by the 
$A_{\tilde\nu}$ term:
\begin{equation}
  \frac{{\rm d}m_{{\widetilde N}_i}^2}{{\rm d}t} = \frac{4}{16\pi^2} A_{\tilde{\nu}_i}^2 \,,
\end{equation}
while
\begin{equation}
  \frac{{\rm d}m_{{\widetilde L}_i}^2}{{\rm d}t} = \rm{(MSSM\ terms)} + \frac{2}{16\pi^2}A_{\tilde{\nu}_i}^2 \,.
\end{equation}

\noindent
The renormalization group equation (RGE) for the $A$-term is
\begin{equation}
\label{sn2012-eq:runA}
  \frac{{\rm d}A_{\tilde{\nu}_i}}{{\rm d}t} = \frac{2}{16\pi^2}\left(-\frac{3}{2}g_2^2 - \frac{3}{10}g_1^2 + \frac{3}{2}y_t^2 + \frac{1}{2}y^2_{l_i}\right) A_{\tilde{\nu}_i} \,.
\end{equation} 
Here, $g_1$ and $g_2$ are the U(1) and SU(2) gauge couplings, and $y_t$ and $y_{l_i}$ are the top and charged lepton Yukawa couplings.

\noindent
A large $A_{\tilde\nu}$ term in the sneutrino mass matrix will induce a significant 
mixing between the RH and LH states, 
\begin{equation}
\label{sn2012-eq:mixingA}
  \left(\begin{array}{c}
    \tilde\nu_{1}\\
    \tilde\nu_{2}
  \end{array}\right) = 
  \left(\begin{array}{lr}
    \cos\theta_{\tilde\nu}\, & -\sin\theta_{\tilde\nu}\\
    \sin\theta_{\tilde\nu} & \cos\theta_{\tilde\nu}
  \end{array}\right) 
  \left(\begin{array}{c}
    \tilde\nu_{R}\\
    \tilde\nu_{L}
  \end{array}\right) ,
  \quad
  \sin2\theta_{\tilde\nu} = 
     \frac{\sqrt{2} A_{\tilde\nu} v \sin\beta}{m_{\tilde\nu_2}^2 - m_{\tilde\nu_1}^2}\,,
\end{equation}
and a sizable splitting between the two mass eigenstates $\tilde{\nu}_1$ and $\tilde{\nu}_2$ 
(with $m_{\tilde{\nu}_1}<m_{\tilde{\nu}_2}$). 

One immediate consequence of this mixing is that the mainly RH state, $\tilde{\nu}_1$, is no longer sterile. However, its left-handed couplings are suppressed by $\sin \theta_{\tilde{\nu}}$. This allows the $\tilde{\nu}_1$ to have a large enough pair-annihilation rate to be a viable candidate for thermal dark matter, while at the same time evading the limits from direct dark matter searches~\cite{ArkaniHamed:2000bq,Belanger:2010cd,Thomas:2007bu}. A mainly RH $\tilde{\nu}_1$ as the LSP will also have a significant impact on collider phenomenology, as it alters the particle decay chains as compared to the ``conventional'' MSSM. 
Moreover, it can have a significant impact on Higgs phenomenology: first, a light mixed sneutrino can give a large {\it negative} loop correction to $m_{h^0}$  which is $\propto |A_{\tilde{\nu}}|^4$ \cite{Belanger:2010cd}; second, a large $A_{\tilde{\nu}}$ can lead to dominantly invisible Higgs decays if $m_{\tilde{\nu}_1}<m_{h^0}/2$.

In the following, we will assume that electron and muon sneutrinos are mass-degenerate, 
$m_{\tilde{\nu}_{ie}}=m_{\tilde{\nu}_{i\mu}}$ with $i=1,2$. Moreover, by default we will assume that the tau-sneutrino, $\tilde{\nu}_{1\tau}$ is lighter than the $\tilde{\nu}_{1e}$ and is the LSP. This is motivated by the contribution in the running of the $A$-term coming from the Yukawa coupling,
see Eq.~\eqref{sn2012-eq:runA}. In this case, 
we take  $m_{\tilde\nu_1}$, $m_{\tilde\nu_2}$, $\sin\theta_{\tilde{\nu}}$ and $\tan \beta$ as input parameters 
in the sneutrino sector, from which we compute $m_{\widetilde{L}}$, $m_{\widetilde{N}}$, $A_{\tilde{\nu}}$ 
(all parameters are taken at the electroweak scale). 

\subsection{Analysis}\label{sn2012-sec:analysis}

\subsubsection{Method}

We choose to confront the sneutrino DM model to experimental constraints by means of Bayesian inference, as in Sections~\ref{sec:higgsdim6} and~\ref{sec:pmssm}. We recall that in this kind of analysis, one starts with an a priori probability density function (prior PDF) $p(\theta|\mathcal{M})$ for the parameters $\theta=\{\theta_{1\ldots n}\}$ of the model $\mathcal{M}$, and some experimental information enclosed in a likelihood function $p(d|\theta,\mathcal{M})\equiv L(\theta)$. The purpose is to combine these two pieces of knowledge, to obtain the so-called posterior PDF, possibly marginalized to some subset of parameters. Splitting the parameter set as $\theta=(\psi,\lambda)$, Bayesian statistics tells us that the  posterior PDF of the parameter subset $\psi$  is
\begin{equation}
p(\psi|\mathcal{M})\propto \int d\lambda\,\, p(\psi,\lambda|\mathcal{M}) L(\psi,\lambda)\,.
\end{equation}
That means one simply integrates over unwanted parameters to obtain the marginalized posterior PDFs. These unwanted parameters can be model parameters, but can also be nuisance parameters. 

In this work, we evaluate posterior PDFs by means of a MCMC method. The basic idea of a MCMC is setting a random walk in the parameter space such that the density of points tends to reproduce the posterior PDF. Any marginalisation is then reduced to a summation over the points of the Markov chain. We refer to~\cite{Allanach:2005kz,Trotta:2008qt} for details on MCMCs and Bayesian inference. Our MCMC method uses the Metropolis-Hastings algorithm with a symmetric, Gaussian proposal function, basically following the procedure explained in Section~\ref{sec:higgsdim6}. 
We use uniform (linear) priors for all parameters. The impact of logarithmic priors in the sneutrino sector is presented in Appendix~C of~\cite{Dumont:2012ee}. For each of the scenarios which we study,  
we run eight chains with $10^6$ iterations each, and we check their convergence using the Gelman and Rubin test with multiple chains~\cite{Gelman:1992zz}, requiring $\sqrt{\hat{R}} < 1.05$ for each parameter. First iterations are discarded (burn-in), until a point with $\log L >-5$ is found.

The likelihood function $L$ can be constructed as the product of the likelihoods $L_{i}$ associated to the $N$ observables $O_{i}$,
\begin{equation}
  L = \prod_{i=1}^N L_i \, .
  \label{sn2012-Ltot}
\end{equation}
Available experimental data fall into two categories: measurements of a central value, and upper/lower limits. In the former case, the central value $O_{\rm exp}$ comes with an uncertainty given at some confidence level CL. It is reasonable to assume that the likelihood function for this kind of measurement is a Gaussian distribution,
\begin{equation}
  L_i = \mathcal{N}(O- O_{\rm exp}, \Delta O) 
                 = \exp\left( \frac{-(O-O_{\rm exp})^2}{2(\Delta O)^2} \right) \, .
  \label{sn2012-Linterval}
\end{equation}
Here $\Delta O$ is the uncertainty at $1\sigma$.  
For combining experimental and theoretical uncertainties, we add them in quadrature. 
When $O_{\rm exp}$ is a (one-sided) limit at a given CL, it is less straightforward 
to account for the experimental uncertainty. 
Taking a pragmatic approach, we approximate the likelihood by a smoothed 
step function centered at the 95\%~CL limit $O_{\rm exp,\,95\%}$, 
\begin{equation}
  L_i = \mathbf{F}(O, O_{\rm exp,\,95\%}) 
                 = \frac{1}{1+\exp[\pm(O-O_{\rm exp,\,95\%})/\Delta O]}\,,
  \label{sn2012-Llimit}
\end{equation}
with $\Delta O = 1\% \times O_{\rm exp,\,95\%}$. 
The $\pm$ sign in the exponent is chosen depending on whether we are dealing with an upper or lower bound: for an upper bound the plus sign applies, for a lower bound the minus sign. Using a smeared step function rather than a hard cut also helps the MCMC to converge.
Finally, when the $\chi^2$ of the limit is available (this will be the case for the direct detection limits), we compute the likelihood as $L_i = e^{-\chi_i^2/2}$. 

To carry out the computations, we make use of a number of public tools. 
In particular, we use \texttt{micrOMEGAs~2.6.c}~\cite{Belanger:2008sj,Belanger:2010gh} for the calculation of the relic density and for direct and indirect detection cross sections. This is linked to an appropriately modified~\cite{Belanger:2010cd} version of \texttt{SuSpect 2.4}~\cite{Djouadi:2002ze} for the calculation of the sparticle (and Higgs) spectrum. Decays of the Higgs bosons are computed using a modified version of \texttt{HDECAY 4.40}~\cite{Spira:1996if,Djouadi:1997yw}, and 
Higgs mass limits are evaluated with \texttt{HiggsBounds 3.6.1beta}~\cite{Bechtle:2008jh,Bechtle:2011sb}. Regarding the computation of the direct detection limits, we  make use of a private code described in section~\ref{sn2012-sec:directdetection}.

\subsubsection{Parameters of the model}

We parametrize the model with twelve parameters as follows. 
The sneutrino sector is fixed by three parameters per generation (the two mass eigenvalues $m_{\tilde{\nu}_{1}}$, $m_{\tilde{\nu}_{2}}$ and the mixing angle $\sin\theta_{\tilde{\nu}}$, or the soft breaking parameters $m_{\widetilde L}$,  $m_{\widetilde N}$, ${A_{\tilde\nu}}$) plus $\tan\beta$.  
Assuming degeneracy between electron and muon sneutrinos, this gives seven parameters to scan over.  
The soft term for the LH sneutrino, $m_{\widetilde L}$, also defines the mass of the LH charged slepton (of each generation); the remaining free parameter in the slepton sector is $m_{\widetilde R}$, the soft mass of the RH charged slepton, which we fix by  
$m_{\widetilde R}=m_{\widetilde L}$ for simplicity. 

The chargino--neutralino sector is described by the gaugino mass parameters $M_1$, $M_2$ and the higgsino mass parameter $\mu$. Moreover, we need the gluino soft mass $M_3$. Motivated by gauge coupling unification, we assume [approximate] GUT relations for the gaugino masses,  $M_3=3M_2=6M_1$,\footnote{This assumption  is central when applying the gluino mass limits from LHC searches.}
so we have $M_2$ and $\mu$ as two additional parameters in the scan. 
For stops/sbottoms we assume a common mass parameter $m_{03}\equiv m_{\widetilde Q_3}=m_{\widetilde U_3}=m_{\widetilde D_3}$, which we allow to vary together with $A_t$ (other trilinear couplings are neglected). The masses of the 1st and 2nd generation squarks, on the other hand, are fixed at 2~TeV without loss of generality.
Finally, we need the pseudoscalar Higgs mass $M_A$ to fix the Higgs sector. 
The model parameters and their allowed ranges are summarized in Table~\ref{sn2012-tab:modelpars}. 

\begin{table}[t]
\begin{center}
\begin{tabular}{|c|c|c|c|}
\hline
$i$     & Parameter     & \multicolumn{2}{c|}{Scan bounds}\\
        & $p_i$       & light sneutrinos & HND sneutrinos \\
\hline\hline
1  & $m_{\tilde{\nu}_{1\tau}}$ & $[1,\,M_Z/2]$ & $[M_Z/2,\,1000]$ \\
\hline
2  & $m_{\tilde{\nu}_{2\tau}}$ & $[m_{\tilde{\nu}_{1\tau}}\!+1,\,3000]$ & $[m_{\tilde{\nu}_{1\tau}}\!+1,\,3000]$ \\
\hline
3  & $\sin\theta_{\tilde{\nu}_{\tau}}$ & $[0,\,1]$ & $[0,\,1]$ \\
\hline
\hline
4  & $m_{\tilde{\nu}_{1e}} = m_{\tilde{\nu}_{1\mu}}$ & $[m_{\tilde{\nu}_{1\tau}}\!+1,\,M_Z/2]$ & $[m_{\tilde{\nu}_{1\tau}}\!+1,\,3000]$ \\
\hline
5  & $m_{\tilde{\nu}_{2e}} = m_{\tilde{\nu}_{2\mu}}$ & $[m_{\tilde{\nu}_{1e}}\!+1,\,3000]$ & $[m_{\tilde{\nu}_{1e}}\!+1,\,3000]$ \\
\hline
6  & $\sin\theta_{\tilde{\nu}_{e}} = \sin\theta_{\tilde{\nu}_{\mu}}$ & $[0,\,1]$ & $[0,\,1]$ \\
\hline
\hline
7  & $\tan \beta$ &  \multicolumn{2}{c|}{$[3,\,65]$} \\
\hline
8  & $\mu$ &  \multicolumn{2}{c|}{$[-3000,\,3000]$} \\
\hline
9  & $M_2 = 2M_1 = M_3/3$ &  \multicolumn{2}{c|}{$[30,\,1000]$} \\
\hline
10 & $m_{\widetilde{Q}_3} = m_{\widetilde{U}_3} = m_{\widetilde{D}_3}$ &  \multicolumn{2}{c|}{$[100,\,3000]$} \\
\hline
11 & $A_t$ &  \multicolumn{2}{c|}{$[-8000,\,8000]$} \\
\hline
12 & $M_A$ &  \multicolumn{2}{c|}{$[30,\,3000]$} \\
\hline
\end{tabular}
\caption{Parameters and scan ranges for the light and the heavy non-democratic (HND) sneutrino cases. All masses and the $A$-term are given in GeV units. In the heavy democratic (HD) case, the same bounds as in the HND case are applied for quantities $i=1\mbox{--}3$ and $7\mbox{--}12$, but entries  $4\mbox{--}6$ are computed from  $m_{\widetilde{N}_e} \in m_{\widetilde{N}_\tau}\!\pm 5\%$, 
$m_{\widetilde{L}_e} \in m_{\widetilde{L}_\tau}\!\pm 5\%$,  and 
$A_{\tilde{\nu}_e} \in A_{\tilde{\nu}_\tau} \pm 5\%$,  with a flat distribution, see text.
\label{sn2012-tab:modelpars}}
\end{center}
\end{table}

The requirement of having enough sneutrino annihilation to achieve $\Omega h^2 \simeq 0.11$ while having a low enough scattering cross section off protons and neutrons to pass the direct detection limits, together with the constraints from the $Z$ invisible width, splits the parameter space into two disconnected regions with sneutrinos lighter or heavier than $M_Z/2$ (or more precisely, as we will see, ${m_{\tilde{\nu}_1}}\lesssim 7$~GeV and ${m_{\tilde{\nu}_1}} \gtrsim 50$~GeV). We call this the ``light''  and ``heavy'' cases in the following.
 
In the  ``light'' case, we assume that the $\tau$-sneutrino is the LSP, but the $e/\mu$ sneutrinos are not too different in mass from the $\tau$-sneutrino. More specifically, we assume that $m_{\tilde{\nu}_{1e}}$ lies within $[m_{\tilde{\nu}_{1\tau}}\! + 1$~GeV,~$M_Z/2]$, {\it i.e.}\ the tau sneutrino is the LSP and all the three sneutrinos are potentially in the region sensitive to the constraint on the invisible decays of the $Z$ boson. The 1~GeV minimal mass splitting is a quite natural assumption considering the sensitivity of ${m_{\tilde{\nu}_1}}$ to small variations in ${A_{\tilde\nu}}$, and suppresses co-annihilation effects (note that the degenerate case was previously studied in~\cite{Belanger:2010cd}).\footnote{We also performed MCMC sampling allowing $m_{\tilde{\nu}_{1e}}>M_Z/2$ up to 3~TeV, keeping only the $\tilde{\nu}_{1\tau}$ light, but the conclusions remain unchanged. So we will present our results only for the case $m_{\tilde{\nu}_{1\tau}}<m_{\tilde{\nu}_{1e}}<M_Z/2$.}

In the ``heavy'' case, we distinguish two different scenarios. First, in analogy to the light case, we assume that the $\tau$-sneutrino is the LSP, with $m_{\tilde{\nu}_{1\tau}} \in [M_Z/2,\,1000~{\rm GeV}]$, and we allow $m_{\tilde{\nu}_{1e}}$ to vary within $[m_{\tilde{\nu}_{1\tau}}\!+1,\,3000]$~GeV. We call this the ``heavy non-democratic'' (HND) case in the following. 
Second, we also consider a ``heavy democratic'' (HD) case, in which $m_{\tilde{\nu}_{1}}$, $m_{\tilde{\nu}_{2}}$ and $\sin\theta_{\tilde{\nu}}$ of the 3rd and the 1st/2nd generation are taken to be close to each other. As before,   
we use $m_{\tilde{\nu}_{1\tau}}$, $m_{\tilde{\nu}_{2\tau}}$ and $\sin\theta_{\tilde{\nu}_\tau}$ as input parameters, from which we compute $m_{\widetilde{N}_\tau}$, $m_{\widetilde{L}_\tau}$ and $A_{\tilde{\nu}_\tau}$.  
For the 1st/2nd generation, we then take 
$m_{\widetilde{N}_e} \in [m_{\widetilde{N}_\tau}\!-5\%,\,m_{\widetilde{N}_\tau}\!+5\%]$, 
$m_{\widetilde{L}_e} \in [m_{\widetilde{L}_\tau}\!-5\%,\,m_{\widetilde{L}_\tau}\!+5\%]$, and 
$A_{\tilde{\nu}_e} \in [A_{\tilde{\nu}_\tau}\!-5\%,\,A_{\tilde{\nu}_\tau}\!+5\%]$ 
with a flat distribution. 
This way either ${\tilde{\nu}_{1\tau}}$ or ${\tilde{\nu}_{1e,\mu}}$ can be the LSP; moreover ${\tilde{\nu}_{1\tau}}$ and ${\tilde{\nu}_{1e,\mu}}$ can be almost degenerate. 
In the latter case, co-annihilations have a sizable effect.\footnote{Note that if the electron/muon/tau sneutrinos are co-LSPs, this has important consequences for 
the relic density~\cite{Belanger:2010cd}. The $e,\mu,\tau$ sneutrino mass hierarchy moreover has important consequences for the LHC phenomenology (more electrons and muons instead of tau leptons from cascade decays), and for the annihilation channels for indirect detection signals. Furthermore, for a very light $\tau$-sneutrino, $m_{\tilde{\nu}_{1\tau}} < m_{\tau} \simeq 1.78$~GeV, annihilation into a pair of tau leptons is kinematically forbidden, while for $\tilde{\nu}_{1e,\mu}$ of the same mass annihilations into electrons or muons would be allowed.}
Nevertheless it turns out that the results for the HND and HD setups are almost the same, so we will take the HND scenario as our standard setup for the heavy case, see Table~\ref{sn2012-tab:modelpars}, and discuss only what is different in the HD case.

\subsubsection{Nuisance parameters}
\label{sn2012-nuisanceparams}

Nuisance parameters are experimentally determined quantities which are not of immediate interest to the analysis but which induce a non-negligible uncertainty in the (model) parameters which we want to infer. 
The Bayesian approach allows us to deal easily with nuisance parameters. 
In order to account for experimental uncertainties impacting the results, we choose ten nuisance parameters, listed in Table~\ref{sn2012-tab:nuipar}. They fall into three categories: astrophysical parameters (related to dark matter searches), nuclear uncertainties (related to the computation of the DM-nucleon scattering cross section) and Standard Model uncertainties.

In order to compute limits from direct detection experiments, we need to know the properties of the dark matter halo of our galaxy. We assume a Standard Halo Model, taking into account variations of the velocity distribution ($v_0$, $v_{\rm esc}$) and of the local dark matter density ($\rho_{\rm DM}$). To this end, we follow~\cite{McCabe:2010zh} and take the naive weighted average of the quoted values for each parameter (an alternative determination of $\rho_{\rm DM}$ can be found in Ref.~\cite{Catena:2009mf,Catena:2011kv,Bovy:2012tw}). Note that considering $v_0$ and $v_{\rm esc}$ as nuisance parameters is particularly important in the light DM case, because of its sensitivity to the tail of the velocity distribution; indeed a departure from the canonical value $v_0 = 220$~km/s may have a sizable impact on the direct detection limits at low masses. 

Turning to nuclear uncertainties, the Higgs exchange contribution to the elastic scattering cross section depends on the quark contents of the nucleons. We take the latest  results for $\sigma_{\pi N}$ and $\sigma_s$ from lattice QCD~\cite{Thomas:2012tg}. 
We stress that the new direct determinations of $\sigma_s$ lead to a much lower value as
compared to previous estimates based on octet baryon masses and SU(3) symmetry breaking effect.

The Standard Model uncertainties that we include as nuisance parameters in the MCMC sampling are $m_t$, the top pole mass, $m_b(m_b)$, the bottom mass at the scale $m_b$ in the $\overline{\rm MS}$ scheme, and $\alpha_s(M_Z)$, the strong coupling constant at the scale $M_Z$. They impact the derivation of the SUSY and Higgs spectrum. Moreover, the mass of the bottom quark is relevant in the light sneutrino case because if $m_{\tilde{\nu}_{1\tau}} < m_b$, annihilation into $b\bar{b}$ is kinematically forbidden. 

\begin{table}[t]
\begin{center}
\begin{tabular}{|c|c|c|c|}
\hline
$i$     & Nuisance parameter     & Experimental result     & Likelihood function \\
        & $\lambda_i$       & $\Lambda_i$                   &  $L_i$ \\
\hline\hline
1  & $m_u/m_d$ & $0.553 \pm 0.043$~\cite{Leutwyler:1996qg} & Gaussian \\
\hline
2  & $m_s/m_d$ & $18.9 \pm 0.8$~\cite{Leutwyler:1996qg} & Gaussian \\
\hline
3  & $\sigma_{\pi N}$ & $44 \pm 5$~MeV~\cite{Thomas:2012tg} & Gaussian \\
\hline
4  & $\sigma_s$ & $21 \pm 7$~MeV~\cite{Thomas:2012tg} & Gaussian \\
\hline
5  & $\rho_{\rm DM}$ & $0.3 \pm 0.1$~GeV/cm$^3$~\cite{Weber:2009pt} & Weighted Gaussian average \\
   &                 & $0.43 \pm 0.15$~GeV/cm$^3$~\cite{Salucci:2010qr} & \\
   &                 & $\Rightarrow 0.34 \pm 0.09$~GeV/cm$^3$ & \\
\hline
6  & $v_0$ & $242 \pm 12$~km/s~\cite{Ghez:2008ms} & Weighted Gaussian average \\
   &       & $239 \pm 11$~km/s~\cite{Gillessen:2008qv} & \\
   &       & $221 \pm 18$~km/s~\cite{Koposov:2009hn} & \\
   &       & $225 \pm 29$~km/s~\cite{McMillan:2009yr} & \\
   &       & $\Rightarrow 236 \pm 8$~km/s & \\
\hline
7  & $v_{\rm esc}$ & $550 \pm 35$~km/s~\cite{Smith:2006ym} & Gaussian \\
\hline
8  & $m_t$ & $173.3 \pm 1.1$~GeV~\cite{Group:2010ab} & Gaussian \\
\hline
9 & $m_b(m_b)$ & $4.19^{+0.18}_{-0.06}$~GeV~\cite{Nakamura:2010zzi} & Two-sided Gaussian \\
\hline
10 & $\alpha_s(M_Z)$ & $0.1184 \pm 0.0007$~\cite{Nakamura:2010zzi} & Gaussian \\
\hline
\end{tabular}
\caption{\label{sn2012-tab:nuipar} Nuisance parameters in the scan. The values of the astrophysical parameters are taken from Ref.~\cite{McCabe:2010zh}.}
\end{center}
\end{table}

\subsubsection{Experimental constraints entering the likelihood}
\label{sn2012-expconst}

We confront our model with the observables listed in Table~\ref{sn2012-tab:const}. 
Below we comment on the various constraints. 
 
\begin{table}[!ht]
\begin{center}
\begin{tabular}{|c|c|c|c|}
\hline
$i$     & Observable     & Experimental result     & Likelihood function \\
        & $\mu_i$        & $D_i$                   &  $L_i$ \\
\hline\hline
1  & $\Omega h^2$ & $0.1123 \pm 0.0035$~\cite{Komatsu:2010fb} & Gaussian \\
    & & (augmented by 10\% theory uncertainty) & \\
\hline
2  & $\sigma_N$ & $\left(m_{\rm DM},\sigma_N\right)$ constraints from & $L_{2} = e^{-\chi^2_{\rm DD}/2}$ \\
   &                   & XENON10~\cite{Angle:2011th}, XENON100~\cite{Aprile:2011hi}, & \\
   &                   & CDMS~\cite{Ahmed:2010wy} and CoGeNT~\cite{Aalseth:2011wp} & \\
\hline
3  & $\Delta\Gamma_Z$ & $< 2$~MeV (95\%~CL)~\cite{ALEPH:2005ab} & $L_{3} =\mathbf{F}(\mu_3, 2\rm{\ MeV})$ \\
\hline
4  & Higgs mass & from  & $L_4 = 1$ if allowed \\
   & limits   & {\tt HiggsBounds 3.6.1beta}~\cite{Bechtle:2008jh,Bechtle:2011sb} & $L_4 = 10^{-9}$ if not\; \\
\hline
5  & $m_{\tilde{\chi}^{\pm}_1}$ & $> 100$~GeV~\cite{lepsusy} & $L_5 = 1$ if allowed \\
   &                          &                                & $L_5 = 10^{-9}$ if not\; \\
\hline
6  & $m_{\tilde{e}_R} = m_{\tilde{\mu}_R}$ & $> 100$~GeV~\cite{lepsusy} & $L_6 = 1$ if allowed \\
   &                                       &                                & $L_6 = 10^{-9}$ if not\; \\
\hline
7  & $m_{\tilde{\tau}_1}$ & $> 85$~GeV~\cite{lepsusy} & $L_7 = 1$ if allowed \\
   &                      &                               & $L_7 = 10^{-9}$ if not\; \\
\hline
8  & $m_{\tilde{g}}$ & $> 750, \, 1000$~GeV~\cite{Aad:2011ib,CMS-PAS-SUS-11-008} & not included \\
   &                 & or none                                             & (a posteriori cut) \\
\hline
9 & ${\cal B}(b \rightarrow s\gamma)$ & $(3.55 \pm 0.34) \times 10^{-4}$~\cite{Asner:2010qj,Misiak:2006zs} & Gaussian \\
\hline
10 & ${\cal B}(B_s \rightarrow \mu^+\mu^-)$ & $< 1.26 \times 10^{-8}$~(95\%~CL)~\cite{cmsblhcbbsmumu,Akeroyd:2011kd} & $\mathbf{F}(\mu_{10}, 1.26 \times 10^{-8})$ \\
\hline
11  & $\Delta a_{\mu}$ & $(26.1 \pm 12.8) \times 10^{-10}$~\cite{Hagiwara:2011af,Bennett:2006fi,Stockinger:2006zn} & Gaussian \\
\hline
\end{tabular}
\caption{\label{sn2012-tab:const} Experimental constraints used to construct the likelihood. 
Where relevant, experimental and theoretical uncertainties are added in quadrature; 
in particular for $\Omega h^2$ we assume an overall uncertainty of 
$(0.0035^2 + 0.01123^2)^{1/2}=0.0118$.}
\end{center}
\end{table}

\subsubsection{Relic density of sneutrinos}\label{sn2012-sec:relic}

We assume the standard freeze-out picture for computing the sneutrino relic abundance. 
The main annihilation channels for mixed sneutrino dark matter are  
{\it i)}~${\tilde{\nu}_1}{\tilde{\nu}_1}\to \nu\nu$ (${\tilde{\nu}_1}^*{\tilde{\nu}_1}^*\to \bar\nu\bar\nu$)
through neutralino $t$-channel exchange,  
{\it ii)}~${\tilde{\nu}_1}{\tilde{\nu}_1}^* \to f\bar{f}$ through $s$-channel $Z$ exchange, and
{\it iii)}~${\tilde{\nu}_1}{\tilde{\nu}_1}^* \to b\bar{b}$ through $s$-channel exchange of a light Higgs.   
Moreover, if the ${\tilde{\nu}_1}$ is heavy enough, it can also annihilate into $W^+W^-$ (dominant),
$ZZ$ or $t\bar t$.  Note that for the heavy LSP the annihilation into neutrino pairs is always much suppressed while the annihilation into other channels can be enhanced by the heavy scalar Higgs resonance.

The annihilation into neutrino pairs proceeds mainly through the wino component of the 
$t$-channel neutralino and is proportional to $\sin^4\theta_{\tilde\nu}$; it is largest for light winos. 
The $Z$ exchange is also proportional to $\sin^4\theta_{\tilde\nu}$. 
The light Higgs exchange, on the other hand, is proportional to $({A_{\tilde\nu}}{\sin\theta_{\tilde\nu}})^2$.  
The dependence of $\Omega h^2$ on the sneutrino mass and mixing angle has been analyzed in \cite{Belanger:2010cd,Thomas:2007bu}. 

We assume a 10\% theory uncertainty on $\Omega h^2$, mostly to account for unknown higher-order effects.  
In the light DM cases, one also has to worry about the change in the number of effective degrees of freedom in the early Universe,  $g_{\rm eff}$, especially when $m_{\rm DM} \approx 20 \, T_{\rm QCD}$. While we do take into account the change of $g_{\rm eff}$ in the calculation of the relic density, the uncertainty related to it is not accounted for separately. Rather, we assume that it falls within the overall 10\% theory uncertainty.  (The issue of $g_{\rm eff}$ is discussed in more detail in Appendix~A of~\cite{Dumont:2012ee}.)

The same annihilation channels will be relevant for indirect DM detection experiments,  looking for gamma-rays ($Fermi$-LAT, H.E.S.S.),   charged particles (positrons, antiprotons; PAMELA, $Fermi$-LAT, AMS) 
or neutrinos (Super-Kamiokande, IceCube, ANTARES),   
that could be produced by annihilation of dark matter, especially in high density regions, see Section~\ref{sn2012-sec:indirect}.

\subsubsection{Direct detection limits}\label{sn2012-sec:directdetection}

The spin-independent (SI) scattering of ${\tilde{\nu}_1}$ on nucleons occurs
through $Z$ or Higgs exchange.  The $Z$ exchange is again suppressed
by the sneutrino mixing angle, while the Higgs exchange is enhanced by
the ${A_{\tilde\nu}}$ term.  A peculiarity of the $Z$-exchange contribution is
that the proton cross section is much smaller than the neutron one,
with the ratio of amplitudes $f_p/f_n=(1-4\sin^2\theta_W)$.  The Higgs
contribution on the other hand, which becomes dominant for large
values of ${A_{\tilde\nu}}$, is roughly the same for protons and neutrons.
The total SI cross section on a nucleus $N$ is obtained after averaging over the ${\tilde{\nu}_1} N$
and ${{\tilde{\nu}_1}}^*N$ cross sections, where we assume equal numbers of
sneutrinos and anti-sneutrinos. We note that the interference between
the $Z$ and $h^0$ exchange diagrams has opposite sign for ${\tilde{\nu}_1} N$ and
${{\tilde{\nu}_1}}^*N$, leading to an asymmetry in sneutrino and anti-sneutrino
scattering if both $Z$ and Higgs exchange are important.
All these effects are taken into account when we compute the normalized scattering 
cross section $\sigma_N$:
\begin{equation}
  \sigma_N = \frac{4 \mu_\chi^2}{\pi}\frac{\left( Z f_p+ (A-Z)f_n\right)^2}{A^2} \,,
\end{equation}
where $\mu_\chi$ is the sneutrino--nucleon reduced mass, $Z$ is the atomic number and $A$ the mass number.
This cross section can be directly compared to the experimental limits on $\sigma_p^{\rm SI}$, which are extracted from the observed 
limits on the LSP--nucleus scattering cross section assuming $f_p=f_n$.

We consider the limits coming from various direct detection experiments.  In
particular, we take into account the light dark matter results from
XENON10~\cite{Angle:2011th} and CDMS~\cite{Ahmed:2010wy}, as well as the
2011 XENON100~\cite{Aprile:2011hi} and CoGeNT~\cite{Aalseth:2011wp}
results.
We include the data from these experiments using a private code based on
Refs.~\cite{Kopp:2011yr, Schwetz:2011xm, HerreroGarcia:2012fu}, where
further details on the analysis can be found. For XENON100 we adopt the
best-fit light-yield efficiency $L_{\rm eff}$ curve from
\cite{Aprile:2011hi}. Especially for the low DM mass region, the energy
resolution close to the threshold is important. We take into account the
energy resolution due to Poisson fluctuations of the number of single
electrons. The XENON10 analysis is based on the so-called S2 ionization
signal which allows to go to a rather low threshold. In this case we follow
the conservative approach of \cite{Angle:2011th} and impose a sharp cut-off
of the efficiency below the threshold, which excludes the possibility of
upward fluctuations of a signal from below the threshold. Our analysis tries
to approximate as closely as possible the one performed in
\cite{Angle:2011th}.  From CDMS we use results from an analysis of Ge data
with a threshold as low as 2~keV~\cite{Ahmed:2010wy}. We use the binned data
from Fig.~1 of \cite{Ahmed:2010wy} and build a $\chi^2$, where we only take
into account bins where the predicted rate is larger than the observed data.
This ensures that only an upper bound is set on the cross section. We
proceed for CoGeNT in a similar way. We ignore the possibility that hints
for an annual modulation in CoGeNT are due to DM (see also
\cite{Ahmed:2012vq}), and use a similar $\chi^2$ method as for CDMS to set
an upper bound on the scattering cross section. The code makes it possible to
vary consistently the astrophysical parameters $v_0$, $v_{\rm esc}$
and $\rho_{\rm DM}$ for all considered experiments. 

The information from direct detection is included in the Bayesian analysis in the
following way. For XENON10 and XENON100 data, we apply the so-called
maximum-gap method~\cite{Yellin:2002xd} to calculate an upper bound on the
scattering cross section for a given mass. The probability returned by the
maximum-gap method as a function of the model parameters as well as
astrophysical parameters (appropriately normalized) is considered as the
likelihood function which then is converted into the posterior PDF within
the Bayesian analysis. This is an approximation to a pure Bayesian treatment
with the advantage that it allows us to use the maximum-gap method, which
offers a conservative way to set a limit in the presence of an unknown
background. Since the shape of the expected background distribution is
neither provided for XENON10 nor XENON100, it is not possible to construct a
``true'' likelihood from the data and we stick to the above mentioned
approximation based on the maximum-gap method.\footnote{In~\cite{Bertone:2011nj} XENON100 data has been implemented in a Bayesian study
by constructing a likelihood function from the Poisson distribution based on
the total number of expected signal and background events. We have checked
that such a procedure leads to similar results as our approach based on the
maximum-gap method.} For CDMS and CoGeNT, the likelihood is obtained from
the individual $\chi^2$ functions as $L \propto \exp(-\chi^2/2)$.
The method to construct the $\chi^2$ described in the previous paragraph
amounts to introducing the unknown background in each bin $i$ as a nuisance
parameter $b_i$ which is allowed to vary by maximizing the likelihood
function under the condition $b_i \ge 0$. Again this is an approximation to
a pure Bayesian approach (in which the posterior PDF would be integrated
over the nuisance parameters), which suffices for our purpose.

\subsubsection{$Z$ invisible width}

A light sneutrino with $m_{\tilde\nu}<M_Z/2$ will contribute to the invisible width of the $Z$ boson, well measured at LEP~\cite{ALEPH:2005ab}, thus putting a constraint on the sneutrino mixing:
\begin{equation}
    \Delta\Gamma_Z=\sum_{i=1}^{N_f} \Gamma_\nu\, \frac{\sin^4\theta_{\tilde{\nu}_i}}{2} 
      \left(1-\left(\frac{2 m_{\tilde{\nu}_i}}{M_Z} \right)^2 \right)^{3/2}< 2~{\rm MeV} \,,
\end{equation}
where $\Gamma_\nu= 166$~MeV is the partial width into one neutrino flavor.
For one light sneutrino with ${m_{\tilde{\nu}_1}}=5$ (20) GeV, this leads only to a mild constraint on the  
mixing angle of ${\sin\theta_{\tilde\nu}}<0.39$ ($0.43$). 
For $m_{\tilde{\nu}_{1\tau}} = 4$~GeV, $m_{\tilde{\nu}_{1e}} = m_{\tilde{\nu}_{1\mu}} = 5$~GeV and assuming a common mixing angle, this constraint becomes stricter: $\sin \theta_{\tilde{\nu}} < 0.3$.

On the other hand, a minimum amount of mixing is needed for light ${\tilde{\nu}_1}$'s to achieve large enough annihilation cross section. In \cite{Belanger:2010cd}, ${\sin\theta_{\tilde\nu}}\gtrsim 0.12$ was found for LSP masses above the $b$-threshold, where annihilation into $b\bar{b}$ through $Z$ or $h^0$ can contribute significantly, 
and ${\sin\theta_{\tilde\nu}}\gtrsim 0.25$ for $m_{{\tilde{\nu}_1}}<m_b$. Therefore, for light sneutrinos, the mixing angle should be not far from the limit imposed by the $Z$ invisible width. Such a large mixing is however in conflict with direct detection limits unless $m_{{\tilde{\nu}_1}}\lesssim 7$~GeV. For sneutrino LSPs with masses of, roughly, 7--40~GeV, the direct detection limits constrain ${\sin\theta_{\tilde\nu}}$ to be smaller than about 0.05--0.07, which makes it impossible to achieve low enough $\Omega h^2$. For heavier masses, one needs $m_{{\tilde{\nu}_1}}$ near the Higgs pole or above the $W^+W^-$ threshold to satisfy both direct detection and relic density constraints. 
This was also discussed in~\cite{Thomas:2007bu}. As mentioned, this splits our parameter space into two distinct regions where the Markov Chains converge, one with $m_{{\tilde{\nu}_1}}\lesssim 7$~GeV and one with $m_{{\tilde{\nu}_1}} > M_Z/2$ (more precisely, $m_{{\tilde{\nu}_1}} \gtrsim 50$~GeV).

\subsubsection{Higgs and SUSY mass limits}

In the MCMC sampling, we impose chargino and charged slepton mass limits~\cite{lepsusy} from LEP as listed in Table~\ref{sn2012-tab:const}.  We here choose conservative values because the LEP analyses in principle assumed a neutralino LSP, and hence the parametrization of the LEP limits in terms of {\it e.g.}\ the chargino--neutralino mass difference as implemented in {\tt micrOMEGAs} does not apply. 
To evaluate Higgs mass constraints based on LEP, Tevatron and LHC data, we use {\tt HiggsBounds\,3.6.1beta}. (The 2011 CMS limit on $A/H \rightarrow \tau\tau$~\cite{CMS-PAS-HIG-11-009} is also included via \texttt{HiggsBounds}.)
Here note that for large sneutrino mixing, which as detailed above is necessary for light $m_{{\tilde{\nu}_1}}$, the light Higgs mass receives an important negative correction from the sneutrino loop, which is proportional to $|{A_{\tilde\nu}}|^4/(m_{\widetilde{\nu}_2}^2 - m_{\widetilde{\nu}_1}^2)^2$. Thus the lower limit on $m_{h^0}$ also somewhat constrains the sneutrino sector. In order to take into account the theoretical uncertainty in $m_{h^0}$, we smear the Higgs mass computed with {\tt SuSpect}  by 
a Gaussian with a width of 1.5~GeV before feeding it to {\tt HiggsBounds}.
In the light sneutrino case, the Higgs decays into sneutrinos are always kinematically allowed, and they are enhanced as $A_{\tilde{\nu}}$; as a result the $h^0$ decays almost completely invisibly in this case. (In the heavy sneutrino case, only a small fraction of the points have $m_{{\tilde{\nu}_1}}<m_{h^0}/2$.) The Higgs decays into sneutrinos are properly taken into account in our {\tt HiggsBounds} interface.

An important point of our 2012 study was how SUSY mass limits from the 2011 LHC searches affect the sneutrino DM scenarios. Here note that squarks and gluinos undergo the same cascade decays into charginos and neutralinos as in the conventional MSSM. Since we assume gaugino mass unification, the gluino and squark mass limits derived in the CMSSM limits from jets$+E_T^{\rm miss}$ searches apply to good approximation. We have checked several ${\tilde{\nu}_1}$ LSP benchmark points and found $m_{\tilde g}\gtrsim 750$~GeV for $m_{\tilde q}\sim 2$~TeV based on a fast simulation of the ATLAS 0-lepton analysis for 1~fb$^{-1}$~\cite{Aad:2011ib}. This is in very good agreement with the corresponding gluino mass limit in the CMSSM for large $m_0$. For 5~fb$^{-1}$ of data, this limit should improve to $m_{\tilde g}\gtrsim 1$~TeV. 

However, a word of caution is in order. For $m_{\tilde q}\gg m_{\tilde g}$ we expect $\tilde g\to q\bar{q}\tilde\chi^0_{1,2}$ and $\tilde g\to q\bar{q}'\tilde\chi^\pm_{1}$ as in the MSSM with a neutralino LSP. In our model, the $\tilde{\chi}^0_{1,2}$ decay  further into the ${\tilde{\nu}_1}$ LSP; if this decay is direct, $\tilde{\chi}^0_{1,2}\to \nu{\tilde{\nu}_1}$, it is completely invisible. Indeed, the invisible $\tilde{\chi}^0_{1,2}$ decays often have close to 100\% branching ratio. We do not expect however that this has a large effect on the exclusion limits. The situation is different for chargino decays. In most cases, the $\tilde\chi^\pm_{1}$ decays dominantly into a sneutrino and a charged lepton ($e$, $\mu$ or $\tau$, depending on the sneutrino flavor). This can lead to a much larger rate of single lepton or di-lepton events. As a consequence, we expect the limits from 0-lepton  jets$+E_T^{\rm miss}$ searches to weaken, while  single lepton or di-lepton $+E_T^{\rm miss}$ searches should become more effective than in the CMSSM. Overall, assuming gaugino mass unification, the gluino mass limit should remain comparable to the limit derived in the CMSSM.

A detailed analysis of the SUSY mass limits in the sneutrino DM model is left for Section~\ref{sn2012-sec:update}. In the Section~\ref{sn2012-sec:results}, we simply consider the effect of the LHC pushing the gluino mass limit to $m_{\tilde g}\gtrsim 750$~GeV or $m_{\tilde g}\gtrsim 1000$~GeV, see above. In order to illustrate this effect without having to run the MCMC several times (which would have been too CPU intensive), we apply the gluino mass constraint a posteriori. As we will see, it is only relevant for the light sneutrino case.

\subsubsection{Low-energy observables}

Further important constraints on the model come from flavor physics and from the muon anomalous magnetic moment.
Regarding flavor physics constraints, we use the 2010 HFAG average value of ${\cal B}(b \rightarrow s\gamma)=(3.55\pm0.24\pm0.09)\times 10^{-4}$~\cite{Asner:2010qj} with a theoretical uncertainty of $0.23\times 10^{-4}$~\cite{Misiak:2006zs}. Moreover, we use the 2011 combined LHCb and CMS limit on ${\cal B}(B_s \rightarrow \mu^+\mu^-)$~\cite{cmsblhcbbsmumu}, augmented by a 20\% theory uncertainty (mainly due to $f_{B_s}$) as suggested in~\cite{Akeroyd:2011kd}. After completion of the MCMC runs, a new limit of ${\cal B}(B_s \rightarrow \mu^+\mu^-) < 4.5 \times 10^{-9}$ (95\%~CL)~\cite{Aaij:2012ac} became available. We impose this new limit a posteriori, again assuming 20\% theory uncertainty, but the effect of this on the posterior distributions is marginal.\footnote{Effectively, we impose ${\cal B}(B_s \rightarrow \mu^+\mu^-) < 5.4 \times 10^{-9}$ as a hard cut, but we have checked that this makes no difference as compared to reweighing the likelihood according to Eq.~(\ref{sn2012-Llimit}).} 

Regarding the supersymmetric contribution to the anomalous magnetic moment of the muon, $\Delta a^{\rm SUSY}_\mu$, we implement the 1-loop calculation taking into account the mixing between RH and LH $\tilde{\nu}_\mu$.  Then we compare this value to $\Delta a_{\mu} = a^{\rm exp}_\mu - a^{\rm SM}_\mu$, where for $a^{\rm exp}_\mu$ we take the experimental value reported by the E821 experiment~\cite{Bennett:2006fi}, and for $a^{\rm SM}_\mu$ we take the result of Ref.~\cite{Hagiwara:2011af} (note however the slightly lower $a^{\rm SM}_{\mu}$ reported in \cite{Davier:2010nc}). 
Guided by \cite{Stockinger:2006zn} and because of our ignorance of the 2-loop effects involving mixed sneutrinos, we assume a conservative theoretical uncertainty of $10 \times 10^{-10}$.  
This brings us to $\Delta a_{\mu}^{\rm SUSY}=(26.1 \pm 12.8) \times 10^{-10}$ in Table~\ref{sn2012-tab:const}.

\subsubsection{Indirect detection of photons and antiprotons}\label{sn2012-sec:indirect}

Dwarf Spheroidal galaxies (dSphs) in the Milky Way provide a good probe of DM through the observation of gamma-rays. Although the photon signal is weaker than from the Galactic center, the signal-to-noise ratio is more favorable since dSphs are DM dominated and the background from astrophysical sources is small.  From measurements of the gamma-rays from ten different dSphs~\cite{Ackermann:2011wa}, the {\it Fermi}-LAT collaboration has extracted
an upper limit on the  DM  annihilation cross section in three different channels: $W^+W^-$, $b\bar{b}$, and $\tau^+\tau^-$. 
For this one assumes a NFW dark matter profile~\cite{Navarro:1996gj}. For DM lighter than 40~GeV, both the $b\bar{b}$ and $\tau^+\tau^-$ channels have the sensitivity to probe the canonical DM annihilation cross section, $\sigma v> 3\times 10^{-26}~\rm{cm}^3 /{\rm s}$. We will not use these constraints in the fit but rather  compare our predictions for the annihilation cross section in different channels with the limit provided by {\it Fermi}-LAT. We will see in the next section that this measurement constrains  sneutrino DM in only a few scenarios for three reasons. 
First, for light sneutrinos we have a sizable ${\tilde{\nu}_1}$ (${\tilde{\nu}_1}^*$) pair annihilation into $\nu\nu$ ($\bar{\nu}\bar{\nu}$), which clearly cannot lead to a photon signal. Second, the {\it Fermi}-LAT collaboration has not published results for DM particles lighter than 5~GeV, where the bulk of our  light DM sample that survives direct detection constraints lies. Third, {\it Fermi}-LAT's sensitivity is still one order of magnitude above the canonical cross section for DM masses at the electroweak scale or above. 

Annihilation of DM  in the Milky Way will also, after hadronisation of the decay products of SM particles, lead to antiprotons. 
This antiproton flux has been measured by PAMELA~\cite{Adriani:2010rc} and fits rather well the astrophysics background~\cite{Maurin:2006hy}.  There is however a large uncertainty in  the background at low energies (below a few GeV) due to solar modulation effects that are not well known. Furthermore the antiprotons---as well as any other charged particle---propagate 
 through the Galactic halo  and  their energy spectrum at the Earth differs from the one produced at the source.
The propagation model  introduces additional model dependence in the prediction of the antiproton flux from DM annihilation.
As for photons above, we will not use the antiproton flux  as a  constraint in the fit, but compare our predictions  for different propagation model parameters with the measurements of PAMELA.  We will see that the largest flux, and the largest deviation from the background, are observed at low energies when the sneutrino DM has a mass of a few GeV, thus leading to an excess of events for some values of the propagation parameters.

Finally, a comment is in order regarding annihilation into neutrinos. Indeed, neutrino telescopes (Super-Kamiokande, IceCube, ANTARES) may probe sneutrino DM annihilation into neutrinos, {\it e.g.}\ from the Galactic Center or from accretion in the Sun. The neutrino flux from annihilation of DM captured by the Sun is determined by the cross section for sneutrino scattering on nucleons discussed in \cite{Belanger:2010cd} and Section~\ref{sn2012-sec:directdetection}. 
We do not include a possible neutrino signal in this analysis but leave it for a future study. 

\subsection{Status of June 2012} \label{sn2012-sec:results}

Let us now present the results of this analysis. This section focuses on the properties of the DM candidate, while an update
of the LHC constraints in a simplified model approach, also including the latest direct detection limits, will come in the next section.
As mentioned, for each of the three scenarios which we study,  we run eight Markov chains with $10^6$ iterations each. 
The distributions of the points in these chains map the likelihood of the parameter space. We hence present our results 
in terms of posterior probability distributions shown in the form of histograms (1-dimensional distributions) with 100 bins and of contour graphs (2-dimensional distributions) with $100 \times 100$ bins. Results based on alternative (logarithmic) priors in the sneutrino sector can be found in Appendix~C of~\cite{Dumont:2012ee}.

\subsubsection{Light sneutrino DM with mass below 10~GeV}

We begin with the case of light sneutrinos. 
Fig.~\ref{sn2012-fig:light-1d} shows the 1-dimensional (1D) marginalized posterior PDFs of various interesting quantities, in particular sneutrino masses and mixing angles, $A$ terms, squarks, gluino and Higgs masses, {\it etc.} 
The blue histograms are the posterior PDFs taking into account constraints 1--7 and 9--11 of Table~\ref{sn2012-tab:const}, 
while the black (red) lines show the posterior distributions after requiring in addition that the gluino be heavier than 750 (1000)~GeV. 
Note that a lower bound on the gluino mass not only cuts the peak of the gluino distribution but  also leads to a lower bound  on the chargino and neutralino masses, since $ 6 m_{\tilde\chi_1^0}\approx 3 m_{\tilde\chi^+}\approx m_{\tilde g}$. 
(We do not show the $m_{\tilde\chi^0_1}$, $m_{\tilde\chi^0_2}$, $m_{\tilde\chi^\pm_1}$ posterior probabilities in Fig.~\ref{sn2012-fig:light-1d}, because they follow completely the $m_{\tilde g}$ distribution.)

\begin{figure}[!ht] 
   \centering
   \includegraphics[width=4.96cm]{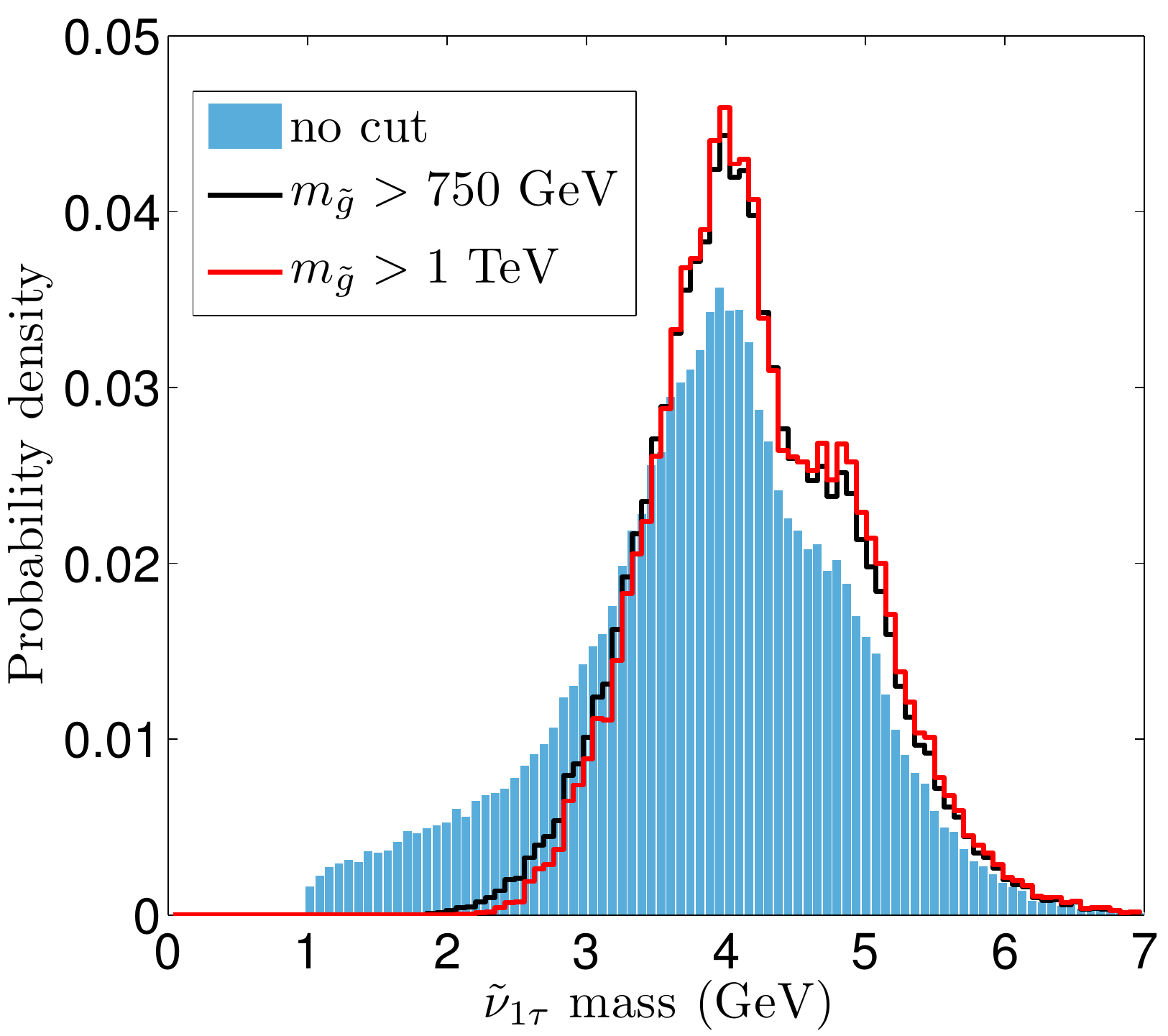}
   \includegraphics[width=4.96cm]{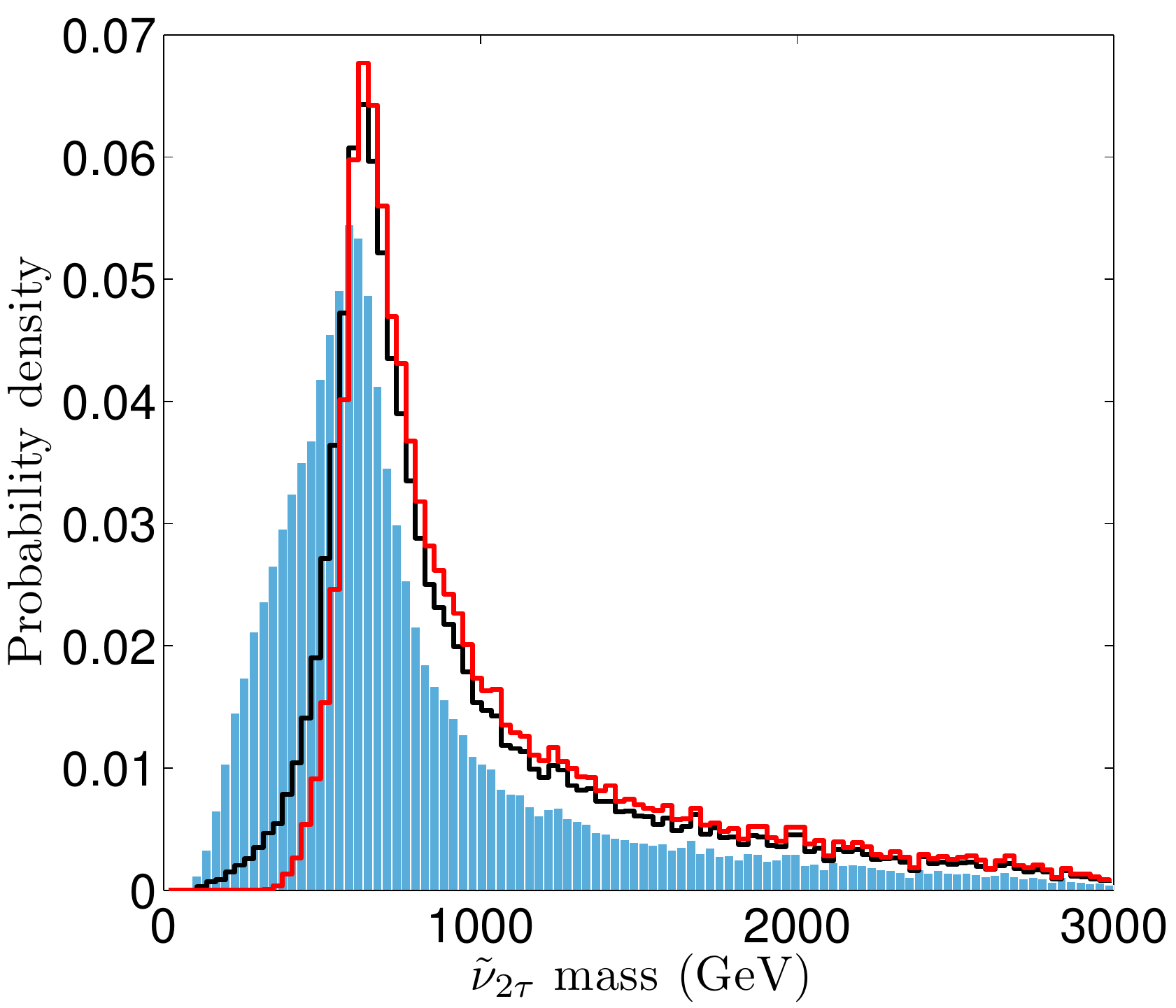}
   \includegraphics[width=4.96cm]{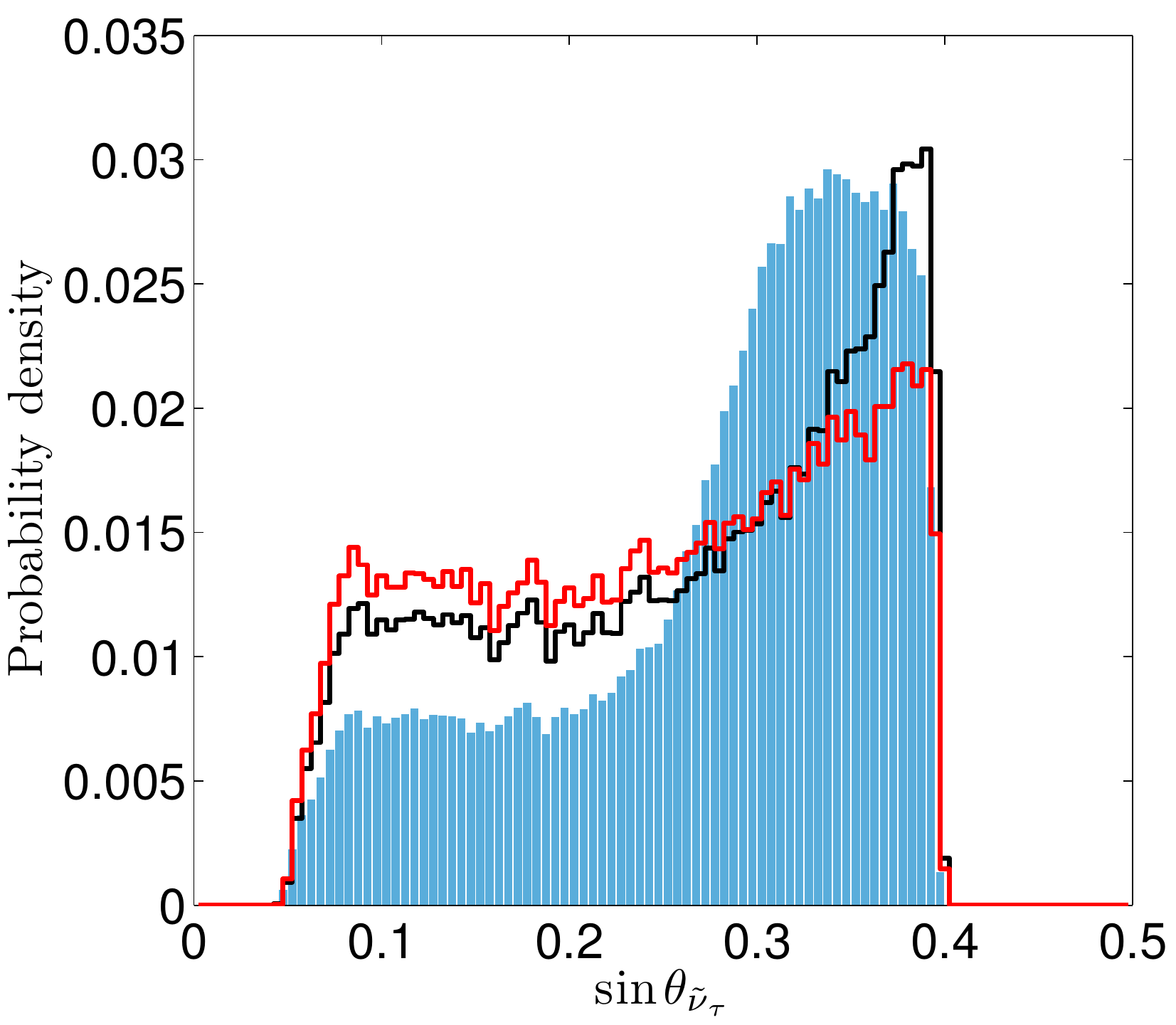}
   \includegraphics[width=4.96cm]{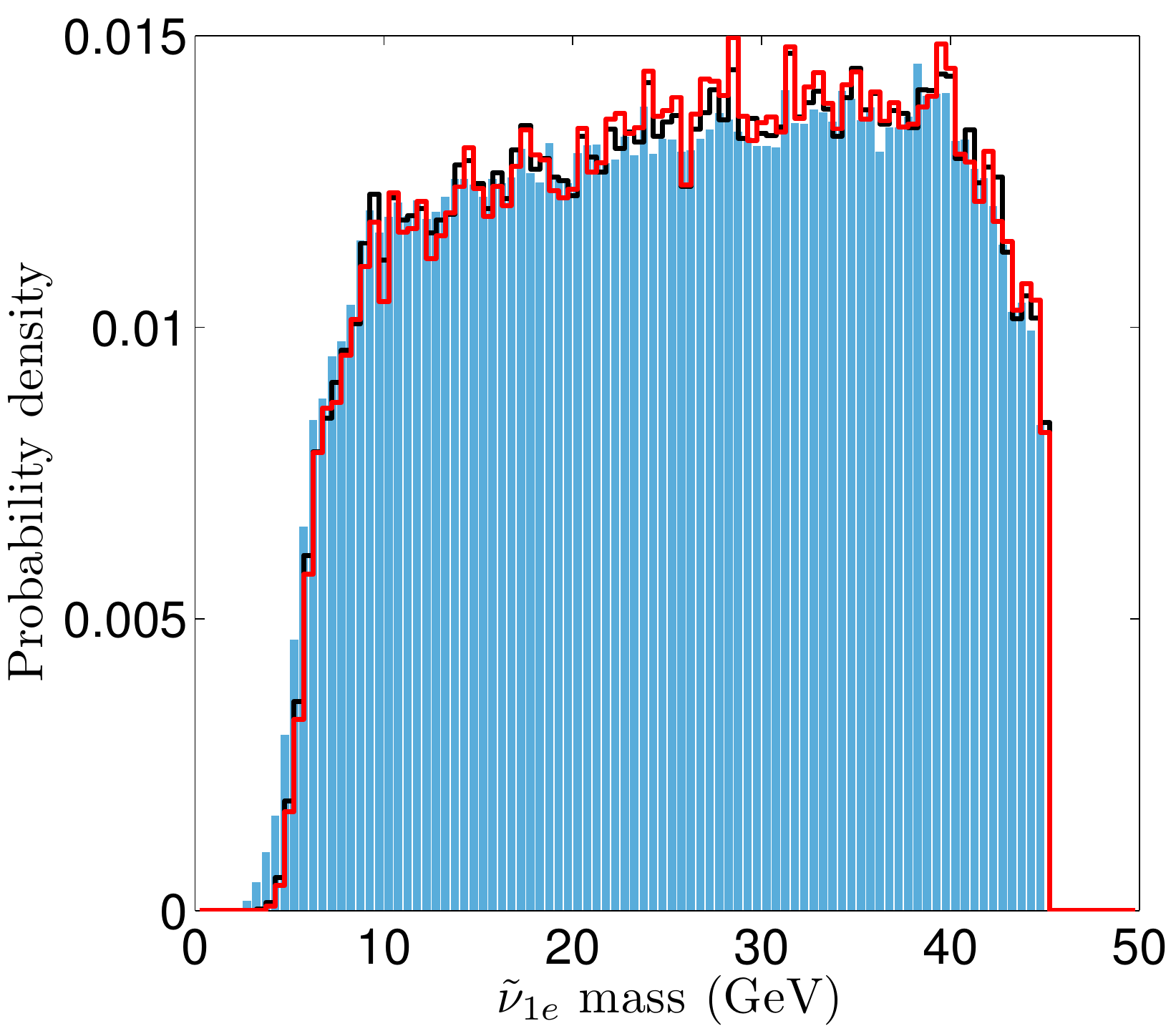}
   \includegraphics[width=4.96cm]{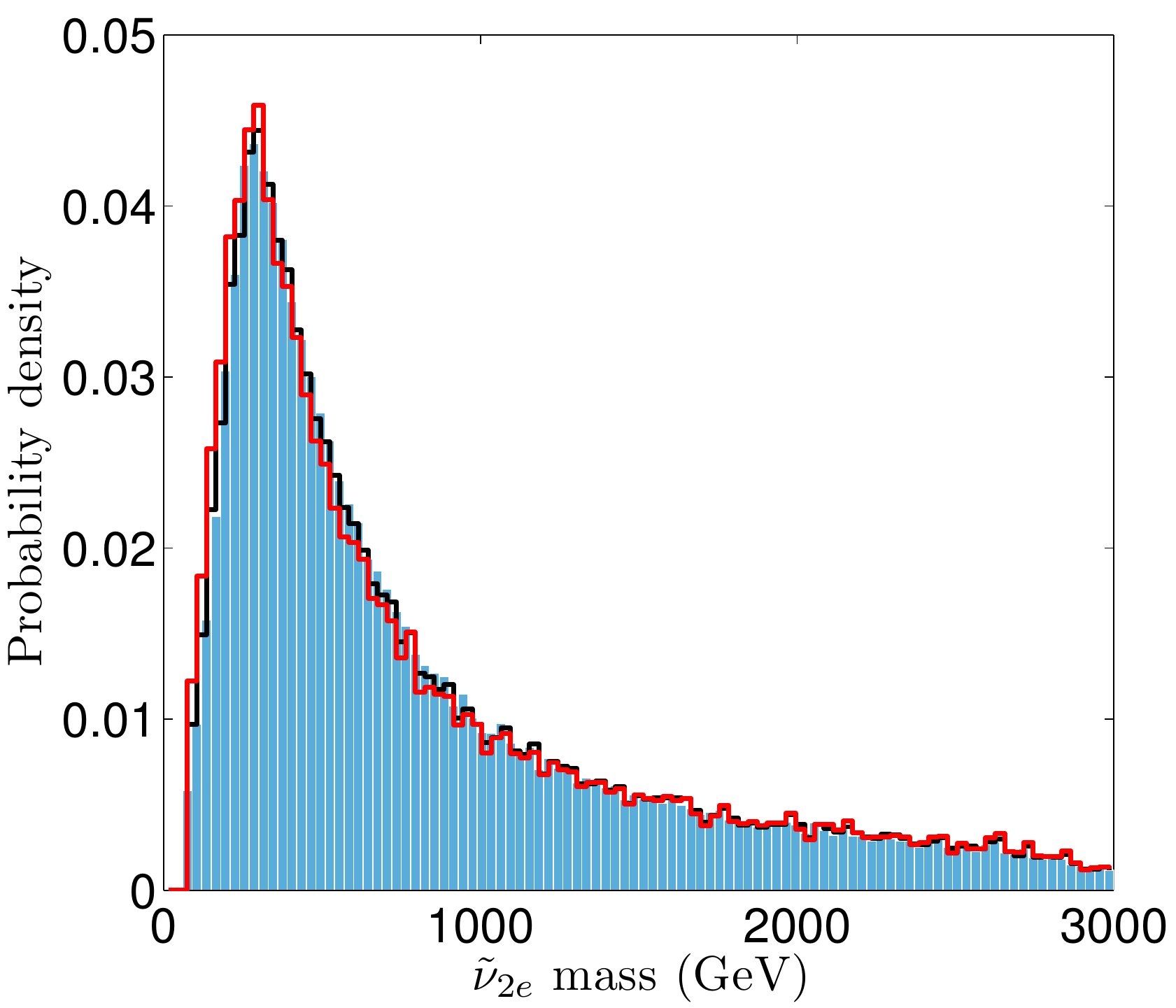}
   \includegraphics[width=4.96cm]{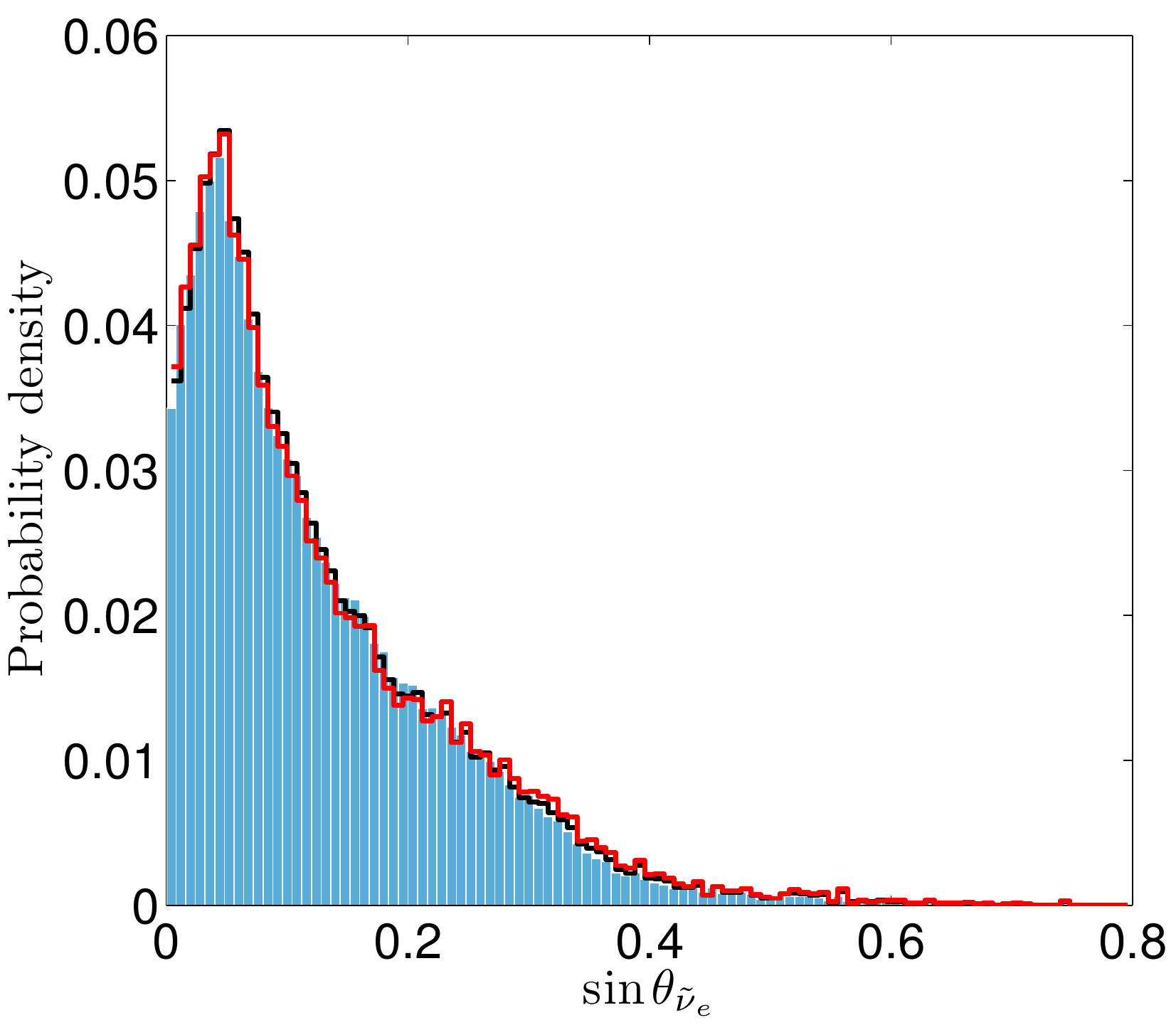}
   \includegraphics[width=4.96cm]{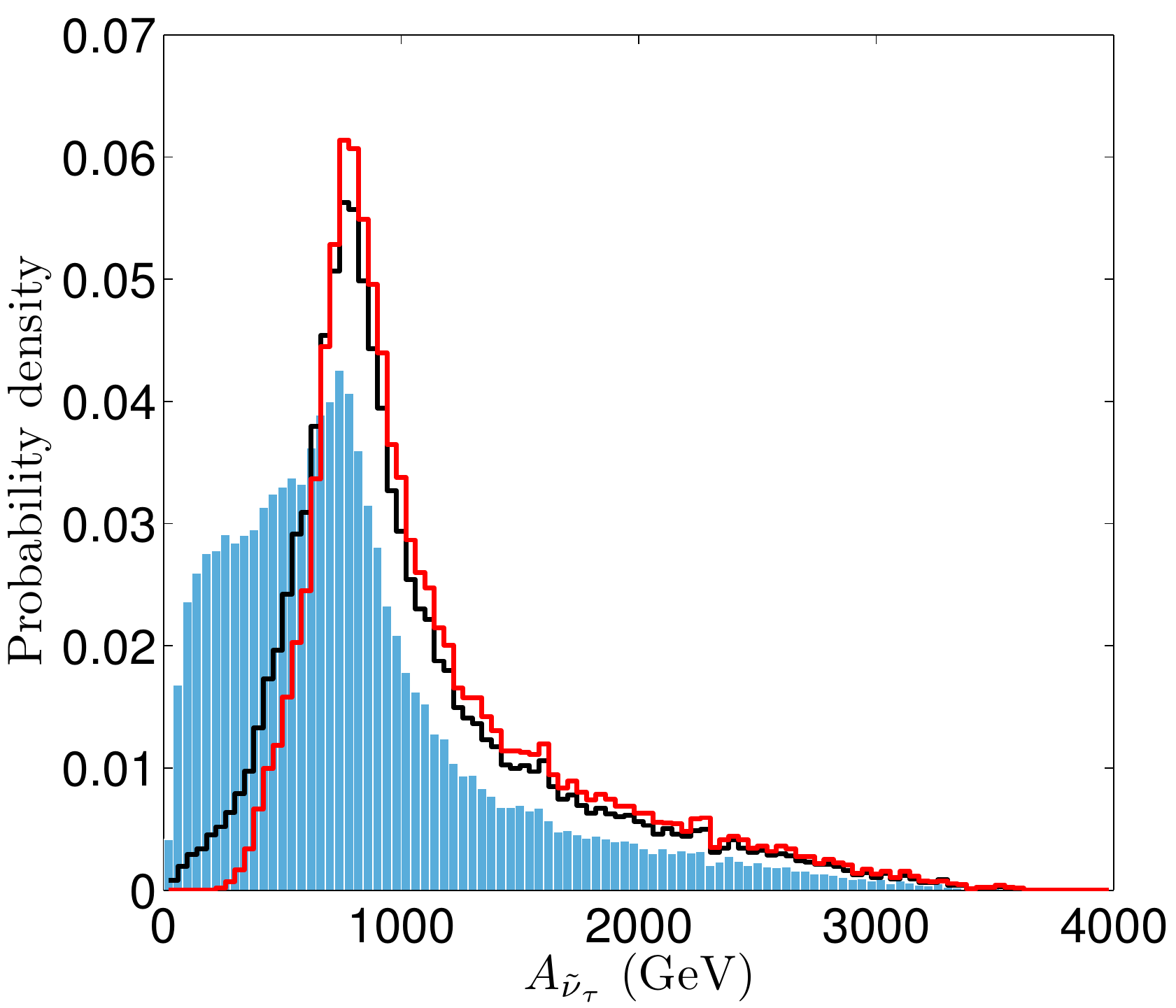}
   \includegraphics[width=4.96cm]{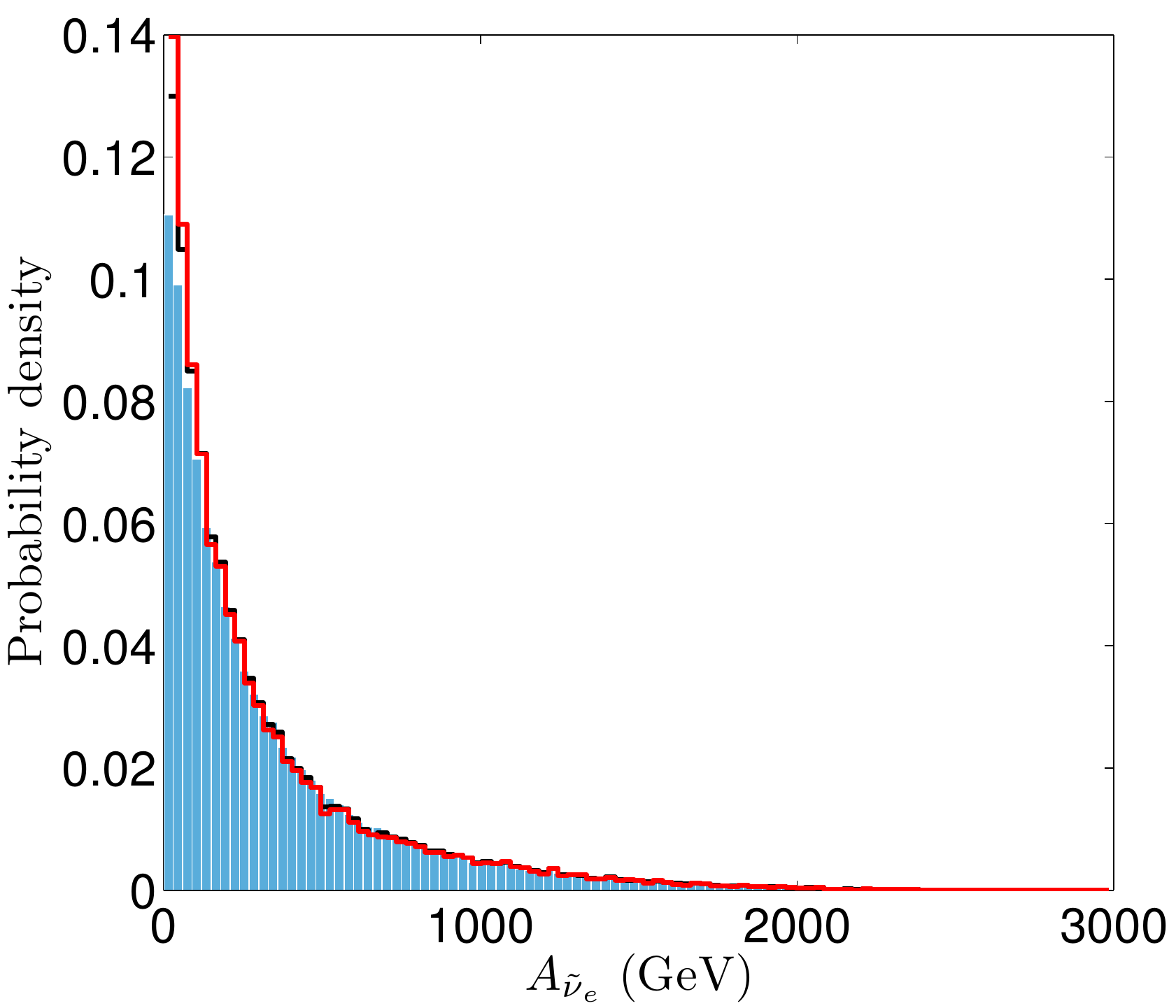}
   \includegraphics[width=4.96cm]{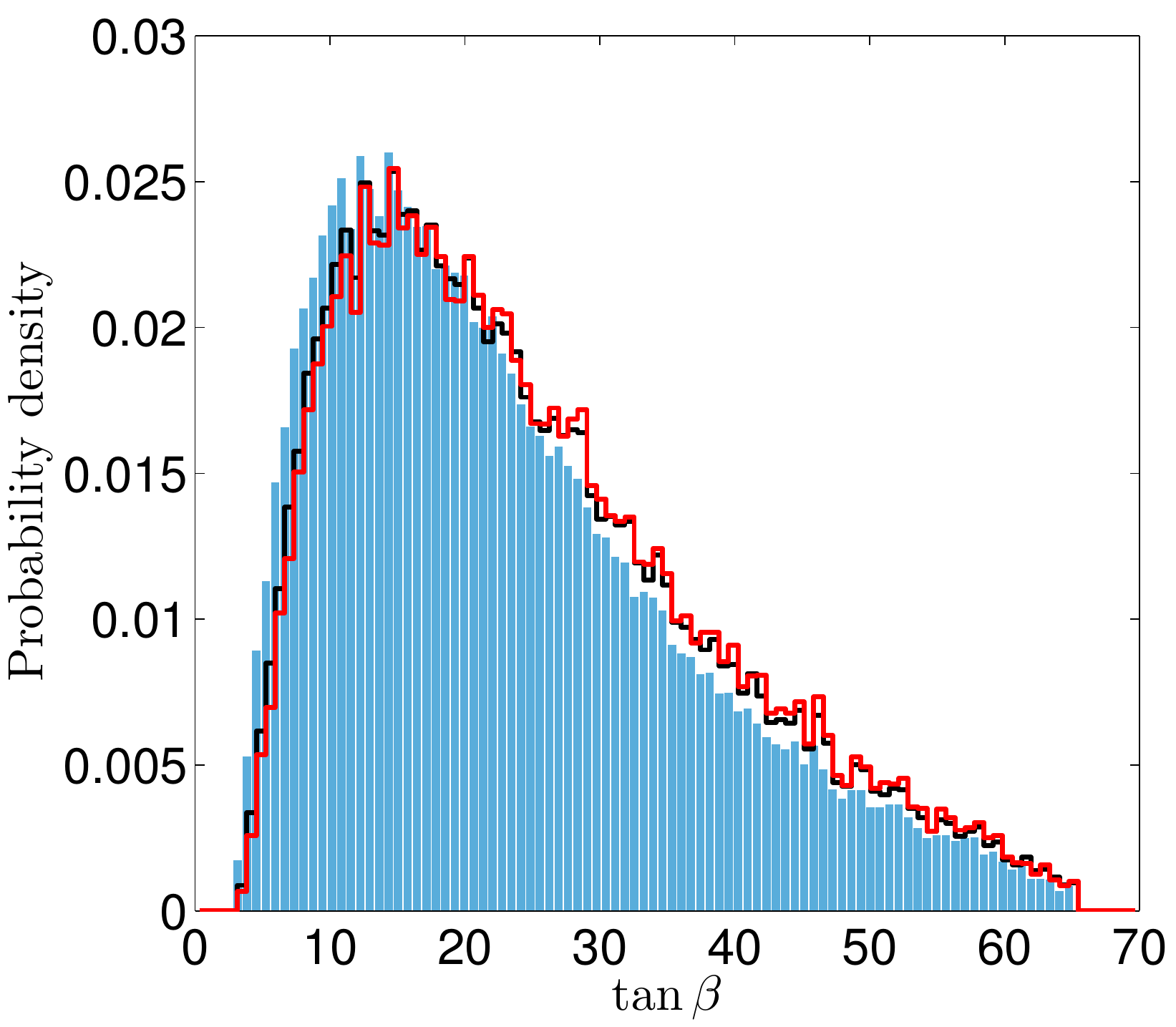}
   \includegraphics[width=4.96cm]{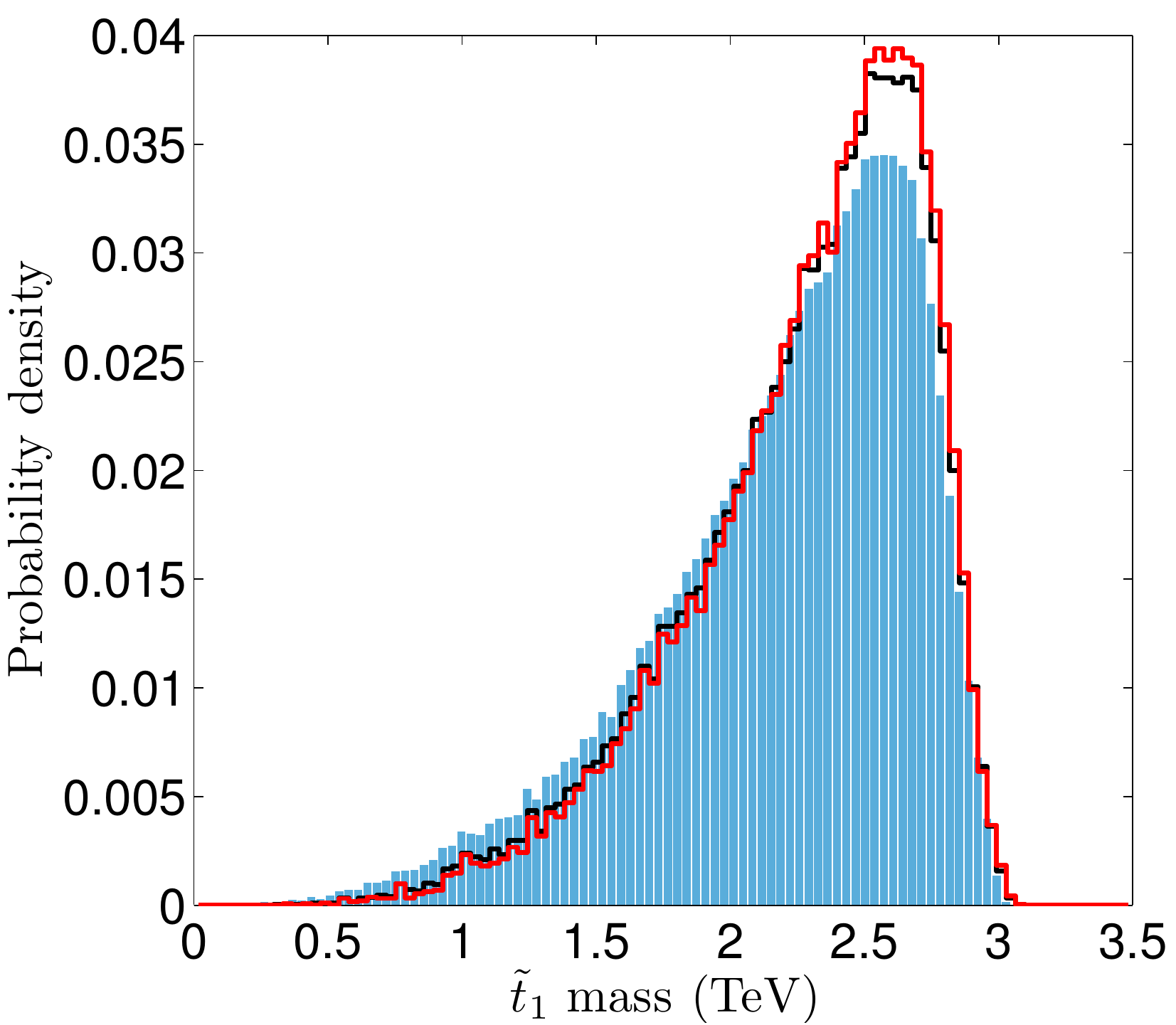}
   \includegraphics[width=4.96cm]{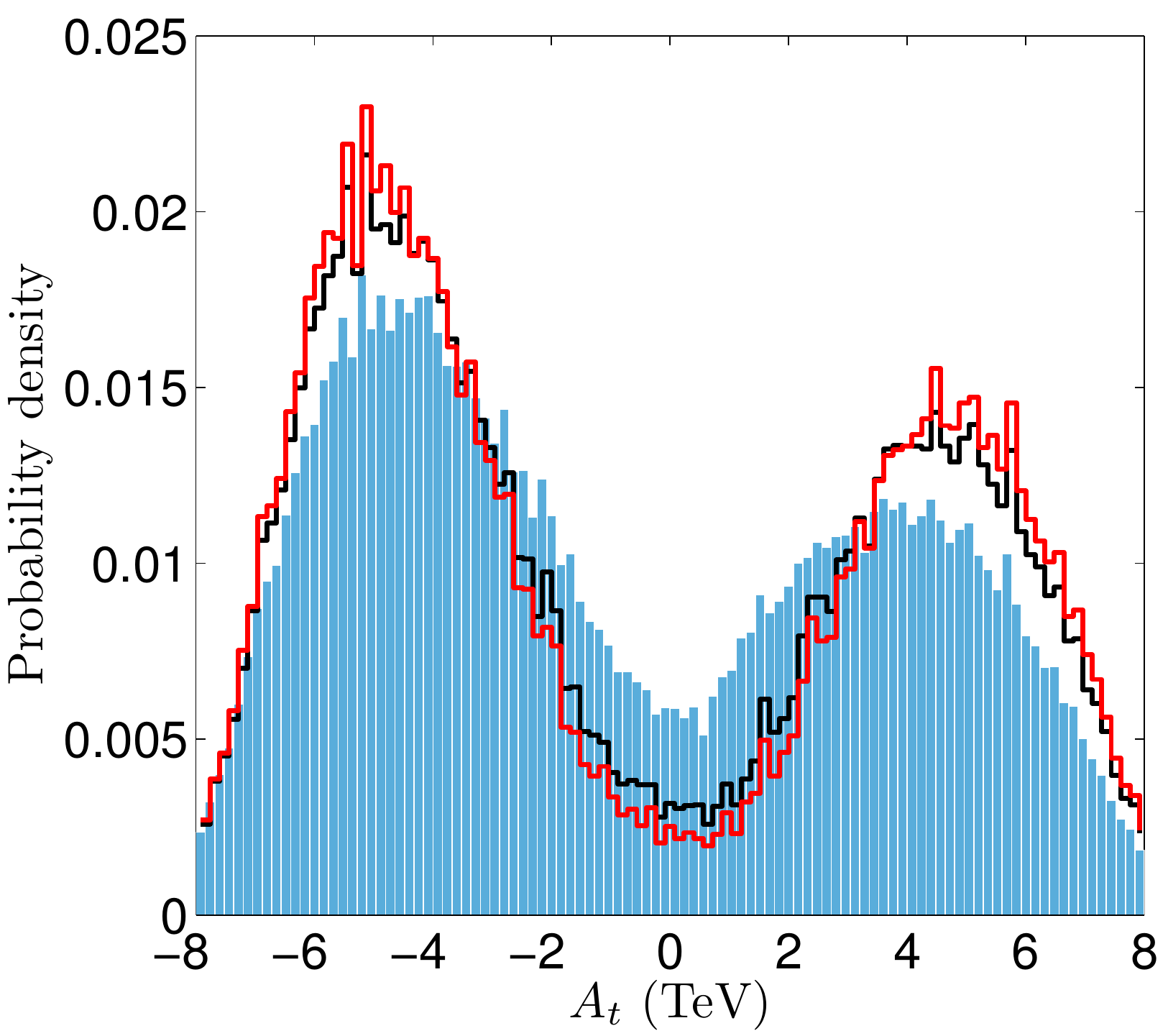}
   \includegraphics[width=4.96cm]{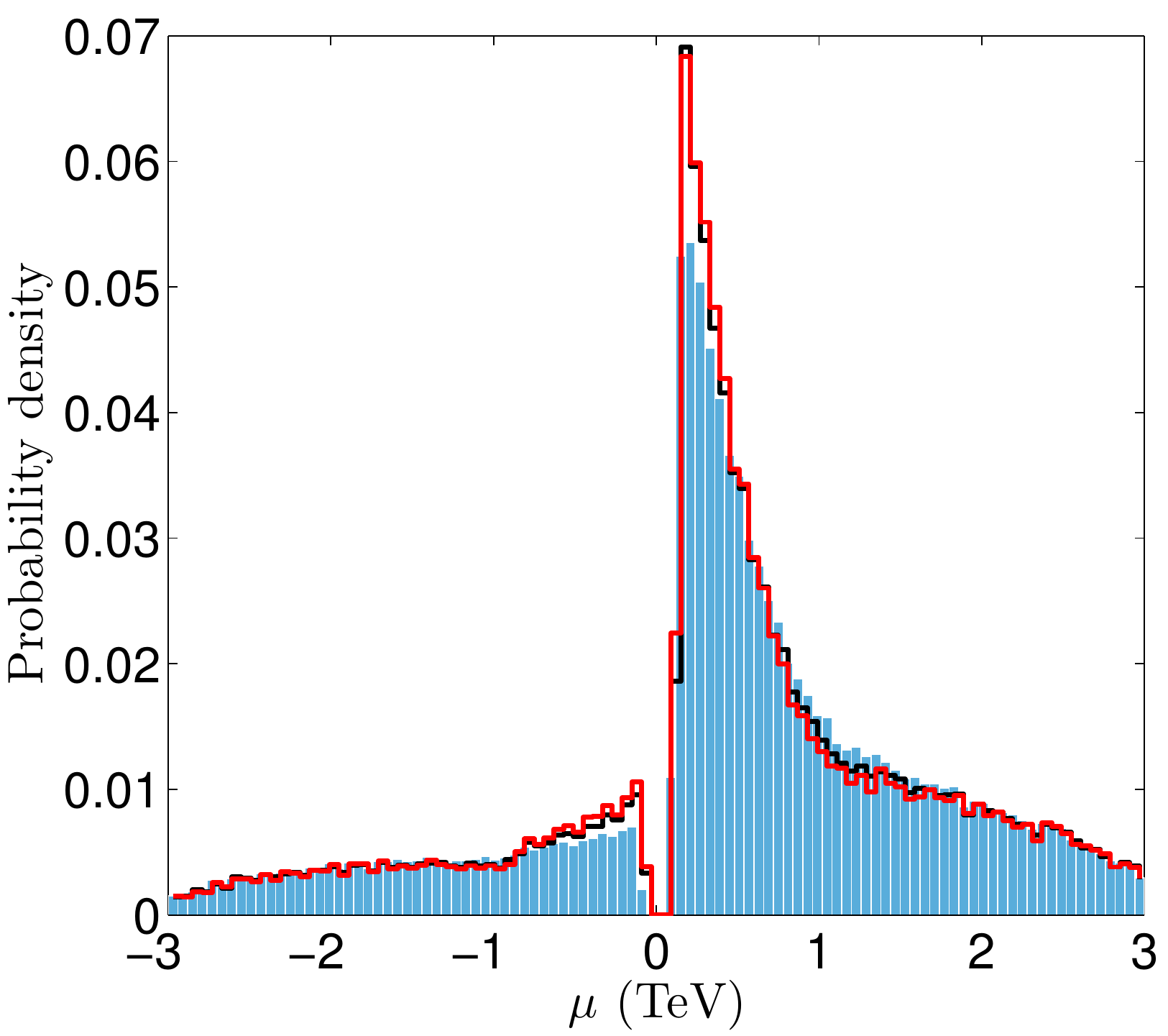}
   \includegraphics[width=4.96cm]{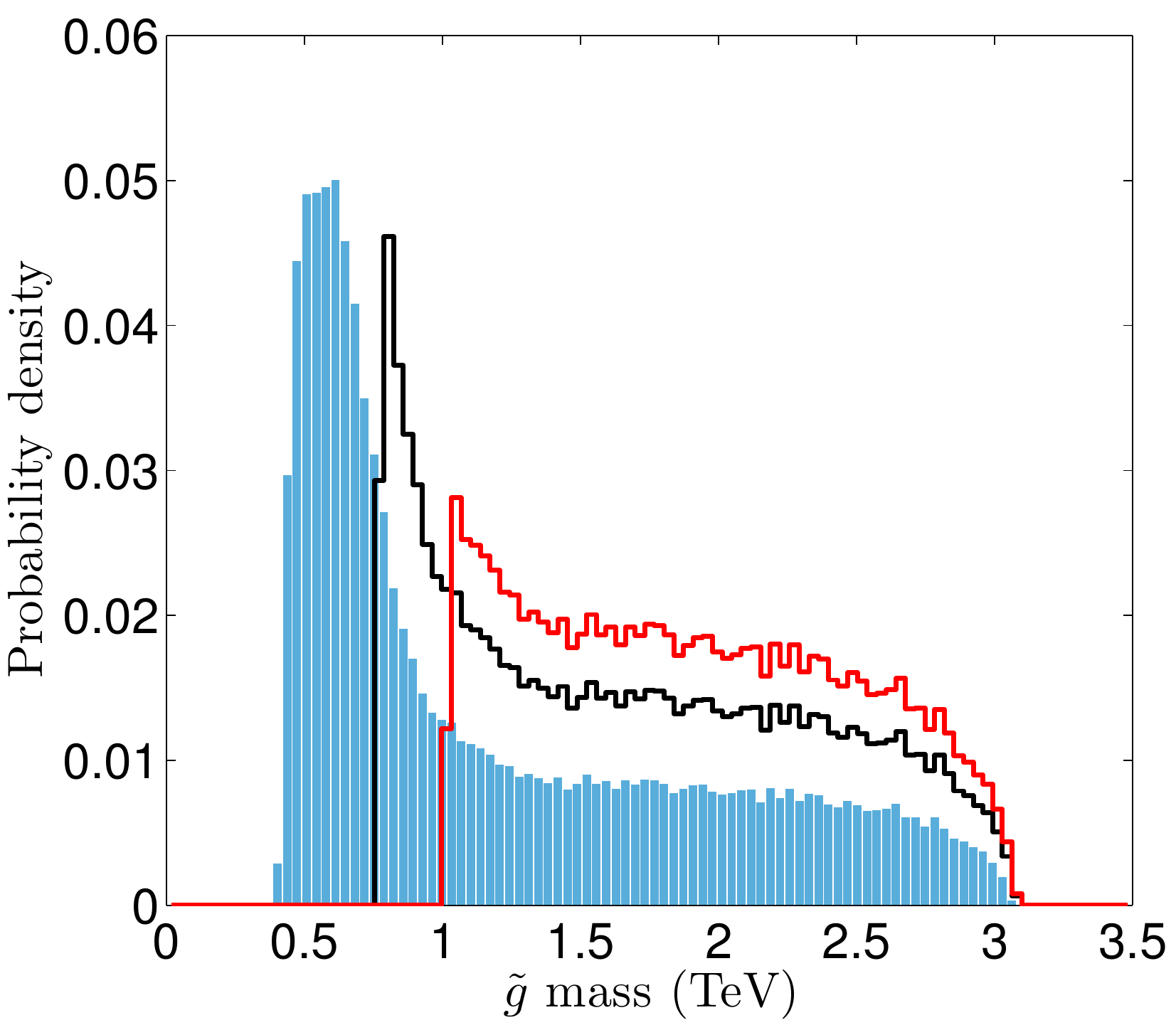}
   \includegraphics[width=4.96cm]{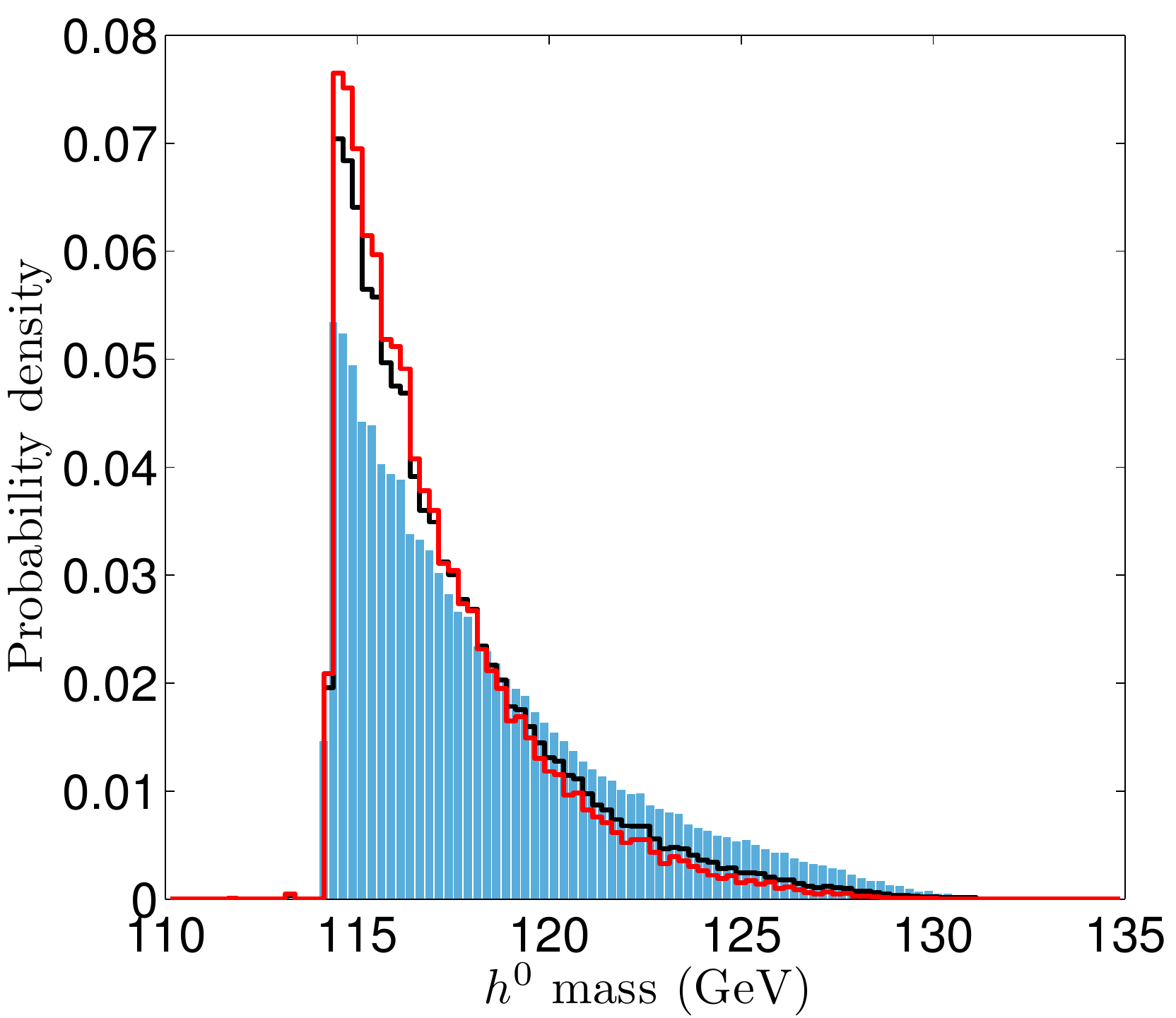}
   \includegraphics[width=4.96cm]{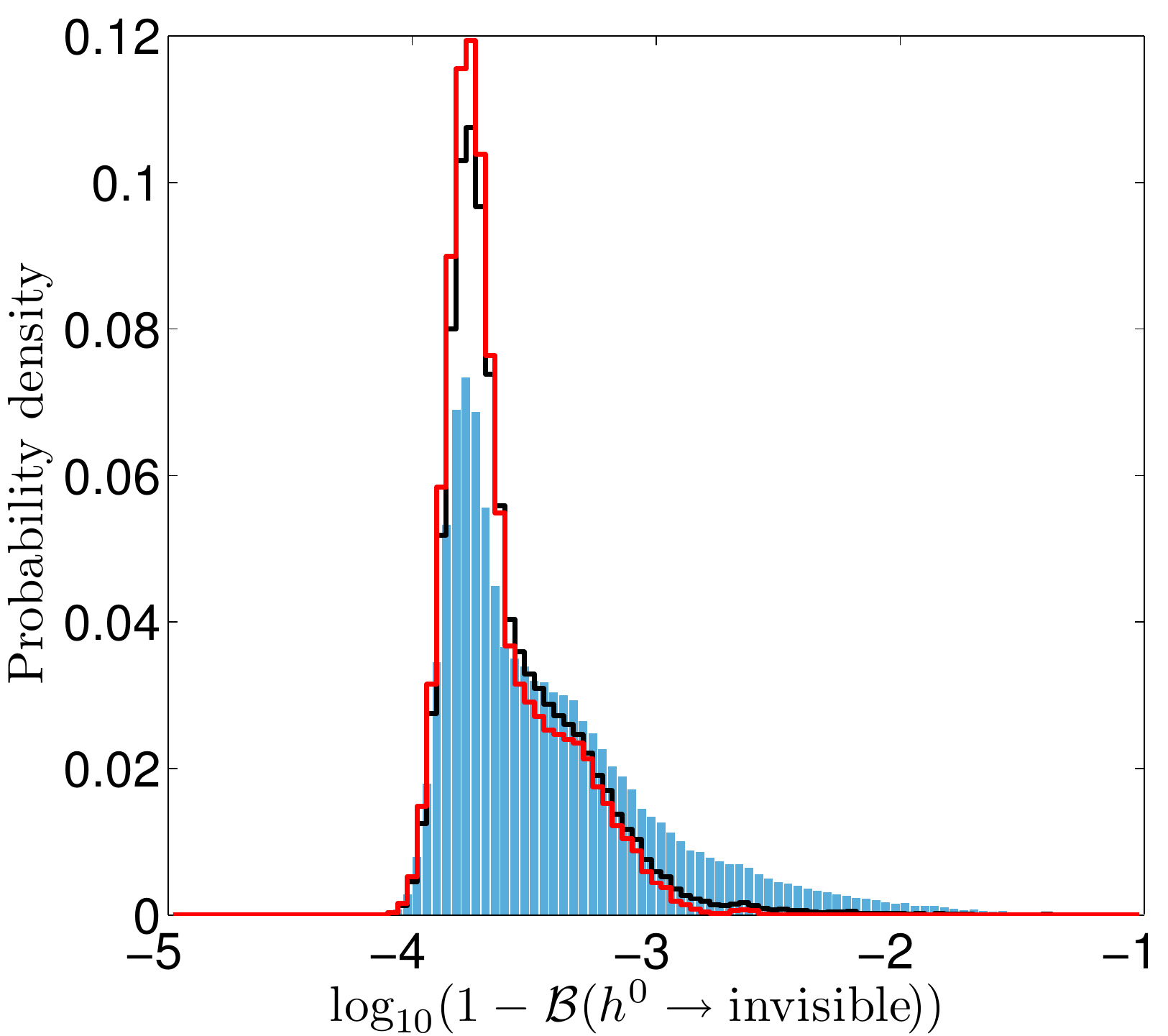}

   \caption{Posterior PDFs in 1D for the light sneutrino case. Specific values for best fit and quasi-mean points as well as the 68\% and 95\% BC intervals are given in Appendix~B of Ref.~\cite{Dumont:2012ee}.}
   \label{sn2012-fig:light-1d}
\end{figure}

As can be seen, the direct detection limits, in particular from XENON10, require the sneutrino LSP to be lighter than about 7~GeV, with the distribution peaking around 4~GeV. (The shoulder at $4.5\mbox{--}5$~GeV is due to the onset of the $b\bar b$ annihilation channel.) 
For LSP masses below 4~GeV, the direct detection limits are not important.  Indeed the  largest cross section, obtained with  the maximum value of $\sin\theta_{\tilde{\nu}_{\tau}}$ allowed by the $Z$ invisible width, is below the current experimental limits~\cite{Belanger:2010cd}.
The gluino mass bound from the LHC disfavors very light sneutrinos of about $1\mbox{--}3$~GeV, because the ${\tilde{\nu}_1}{\tilde{\nu}_1}\to \nu\nu$ and ${\tilde{\nu}_1}^*{\tilde{\nu}_1}^*\to \bar\nu\bar\nu$ annihilation channels get suppressed (recall that we assume GUT relations between gaugino masses).  This means one needs to rely on 
annihilation through $Z$ or Higgs exchange, as is reflected in the change of the $\sin\theta_{\tilde{\nu}_{\tau}}$
and $A_{\tilde\nu_\tau}$ probability densities in Fig.~\ref{sn2012-fig:light-1d}.

The other distributions are basically unaffected by the gluino mass cut, the exceptions being $A_t$ and $m_{h^0}$.  Larger values 
of $A_t$ are  preferred for $m_{\tilde g}>1$~TeV, because it is needed to compensate the negative loop correction to $m_{h^0}$ from the larger $A_{\tilde\nu_\tau}$ in order to still have $m_{h^0}>114$~GeV. Regarding $m_{h^0}$, the distribution is shifted towards the lower limit of 114~GeV because of this negative loop correction. Finally, we note that the light Higgs decays practically 100\% invisibly into sneutrinos. 
Therefore, the existence of an SM-like Higgs with mass of about 125.5~GeV would rule out the light sneutrino DM scenario.

Regarding the supersymmetric contribution to $\Delta a_\mu$,  
this is peaked towards small values. Nevertheless, the probability of falling within the experimental $1\sigma$  band is sizable, 
$p(\Delta a_\mu=(26.1 \pm 12.8) \times 10^{-10})= 31\%$. The larger values  of  $\Delta a_\mu$ are obtained when there is a large  contribution from the sneutrino exchange diagram.

Our expectations regarding the relation between mass and mixing angle are confirmed in
\clearpage
\noindent Fig.~\ref{sn2012-fig:light-2d-sinth}, which shows the 2-dimensional (2D) posterior PDF of $\sin\theta_{\tilde{\nu}_{\tau}}$ versus $m_{\tilde\nu_{1\tau}}$. To be more precise, what is shown are the 68\% and 95\% Bayesian credible regions (BCRs) before and after a gluino mass cut of $m_{\tilde g}>1$~TeV. As can be seen, the region of $m_{\tilde\nu_{1\tau}}\approx 1\mbox{--}3$~GeV, which requires $\sin\theta_{\tilde{\nu}_{\tau}}\approx 0.3\mbox{--}0.4$ to be consistent with WMAP, gets completely disfavored by a heavy gluino.\footnote{To be more precise, it gets disfavored by a heavy wino, since $m_{\tilde g}>1$~TeV implies $m_{\tilde\chi^0_2}\gtrsim 300$~GeV in our model.} 

\begin{figure}[t] 
   \centering
   \includegraphics[width=7cm]{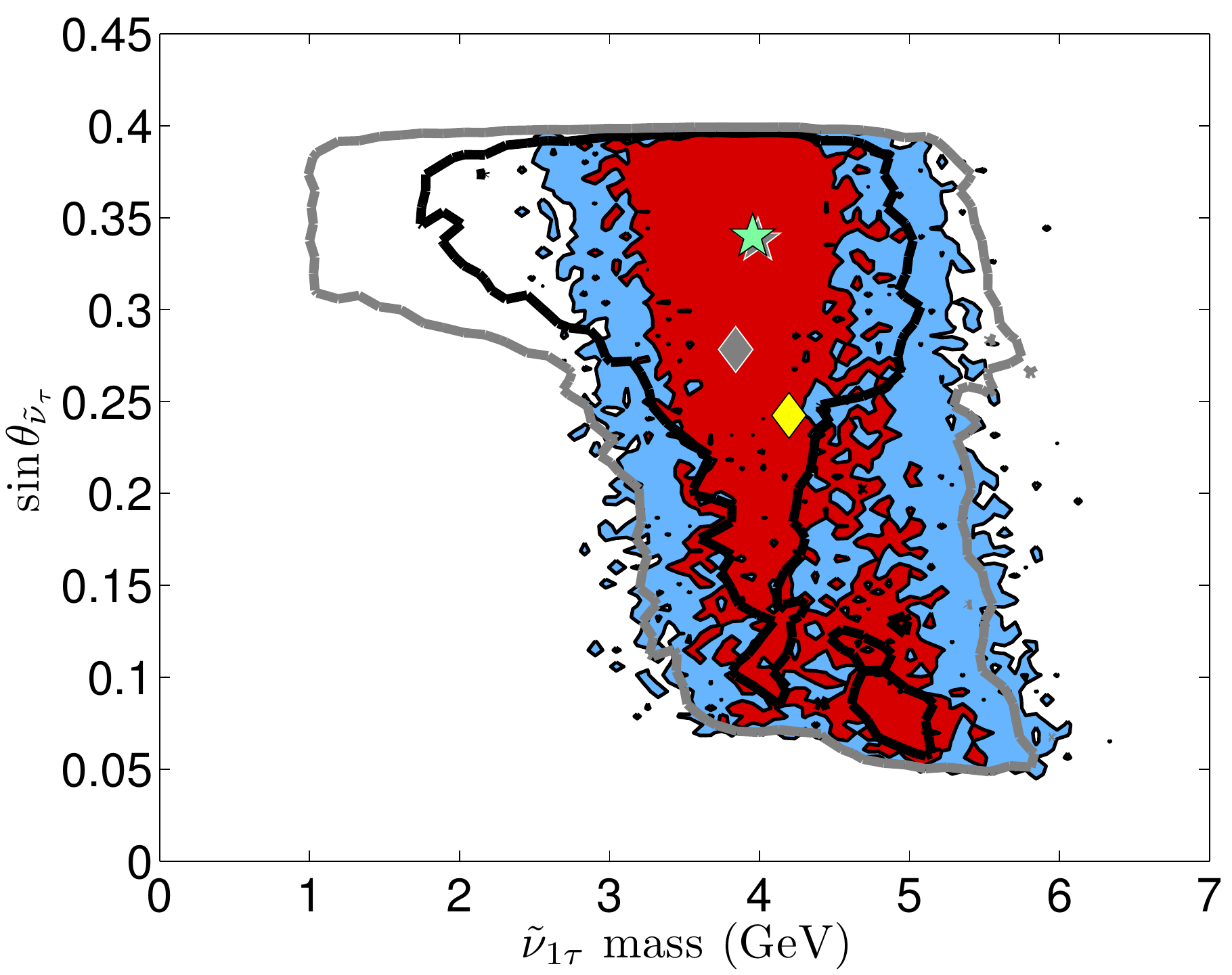}
   \caption{Posterior PDF of $\sin\theta_{\tilde{\nu}_{\tau}}$ versus $m_{\tilde\nu_{1\tau}}$ for the light sneutrino case. The black and gray lines show the 68\% and 95\% BCRs before gluino mass limits from the LHC. The red and blue regions are the 68\% and 95\% BCRs requiring $m_{\tilde g}>1$~TeV. The green star marks the bin with the highest posterior probability after the gluino mass limit, while the
 yellow diamond marks the  mean of the 2D PDF. The gray star/diamond are the highest posterior and  mean points before imposing the gluino mass limit.}
   \label{sn2012-fig:light-2d-sinth}
\end{figure}

In Fig.~\ref{sn2012-fig:light-2d-sigXe}, we show the influence of the gluino mass limit on the predicted direct detection cross section for Xenon (we display the Xenon cross section to directly compare with the best limit which comes from XENON10).
Imposing $m_{\tilde g}>1$~TeV has quite a striking effect, limiting $\sigma_{\rm Xe}$ to a small region just below the current limit. 
We recall that XENON10 only constrains the mass range above $\approx 4$~GeV; for lower ${\tilde{\nu}_1}$ masses, the direct detection cross section is constrained from above by the $Z$ invisible width. We also note that there is a lower limit on the direct detection cross section \cite{Belanger:2010cd}, so that if a lower threshold can be achieved to probe masses below 4~GeV, in principle the light sneutrino DM case can be tested completely. (For ${m_{\tilde{\nu}_1}}\approx 4\mbox{--}6$~GeV, an improvement of the current sensitivity by about a factor 3 is sufficient to cover the 95\% region, while an improvement by an order of magnitude will completely cover this mass range.) 

\begin{figure}[!t] 
   \centering
   \includegraphics[width=7cm]{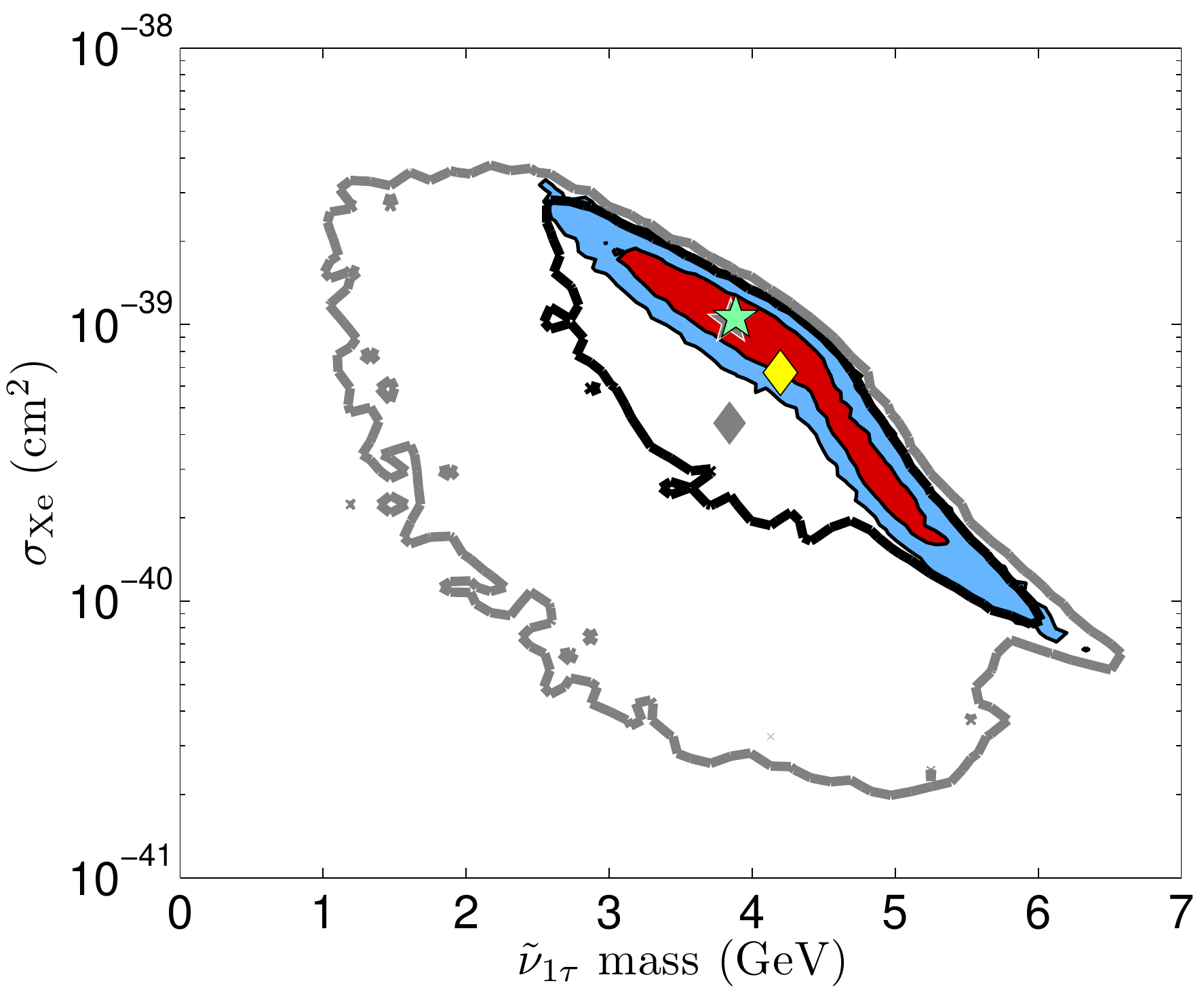}\quad
   \includegraphics[width=7cm]{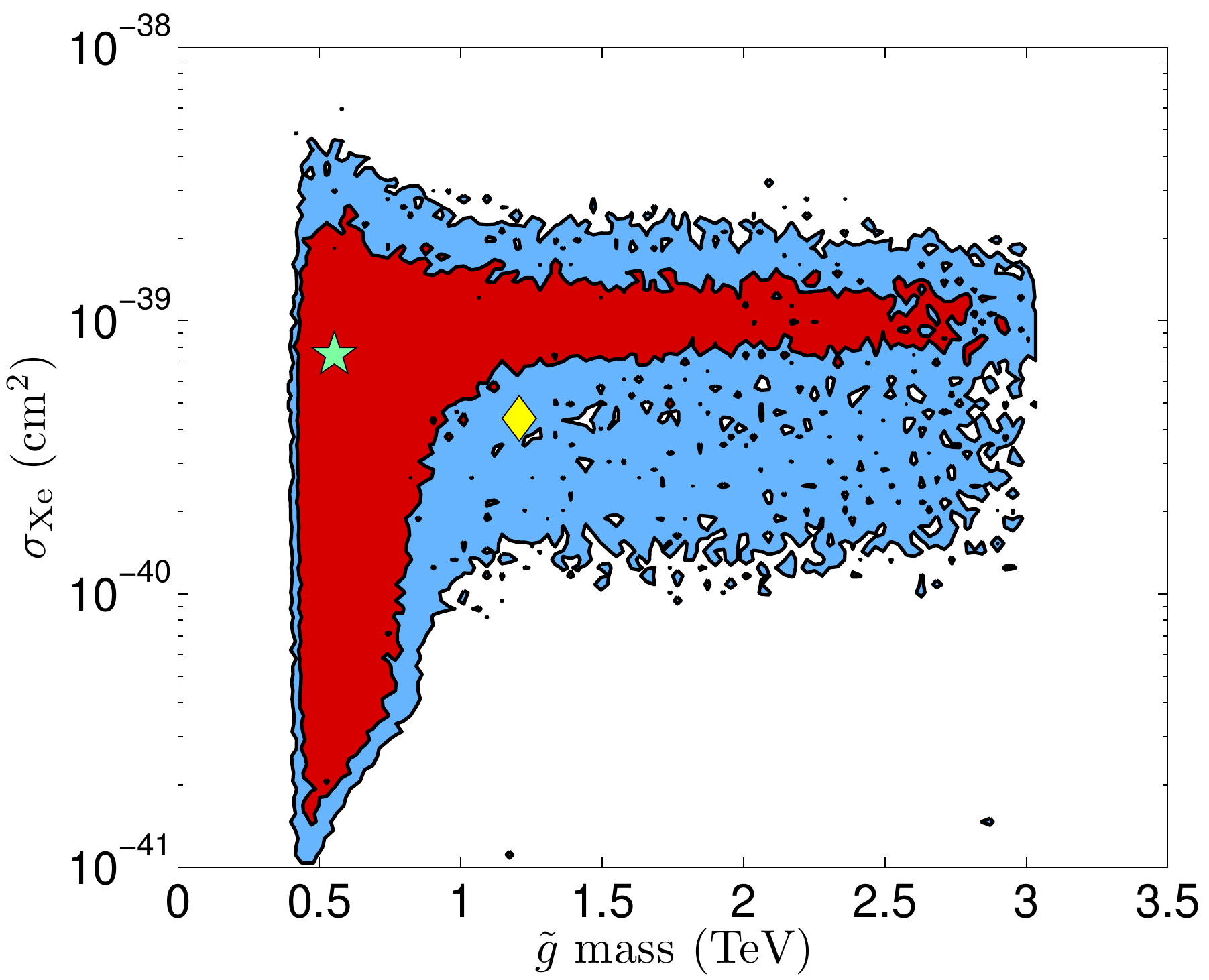}
   \caption{On the left, 2D posterior PDF of  $\sigma_{\rm Xe}$ versus $m_{\tilde\nu_{1\tau}}$ before and after imposing $m_{\tilde g}>1$~TeV; see the caption of Fig.~\ref{sn2012-fig:light-2d-sinth} for the meaning of colors and symbols. On the right, correlation between $\sigma_{\rm Xe}$ and gluino mass; the red and blue areas are the  68\% and 95\% BCRs.}
   \label{sn2012-fig:light-2d-sigXe}
\end{figure}

\begin{figure}[!t] 
   \centering
   \includegraphics[width=7cm]{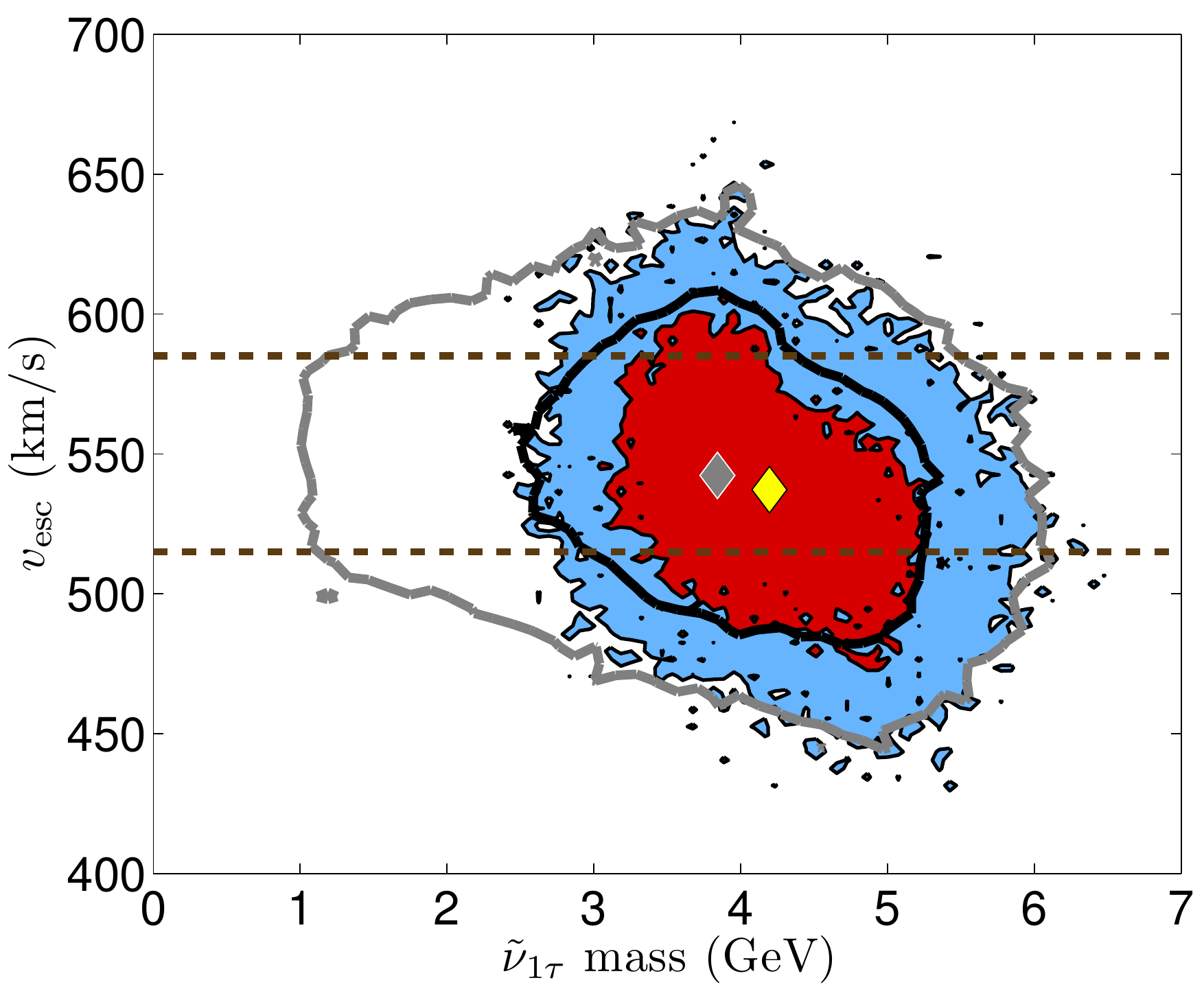}\quad
   \includegraphics[width=7cm]{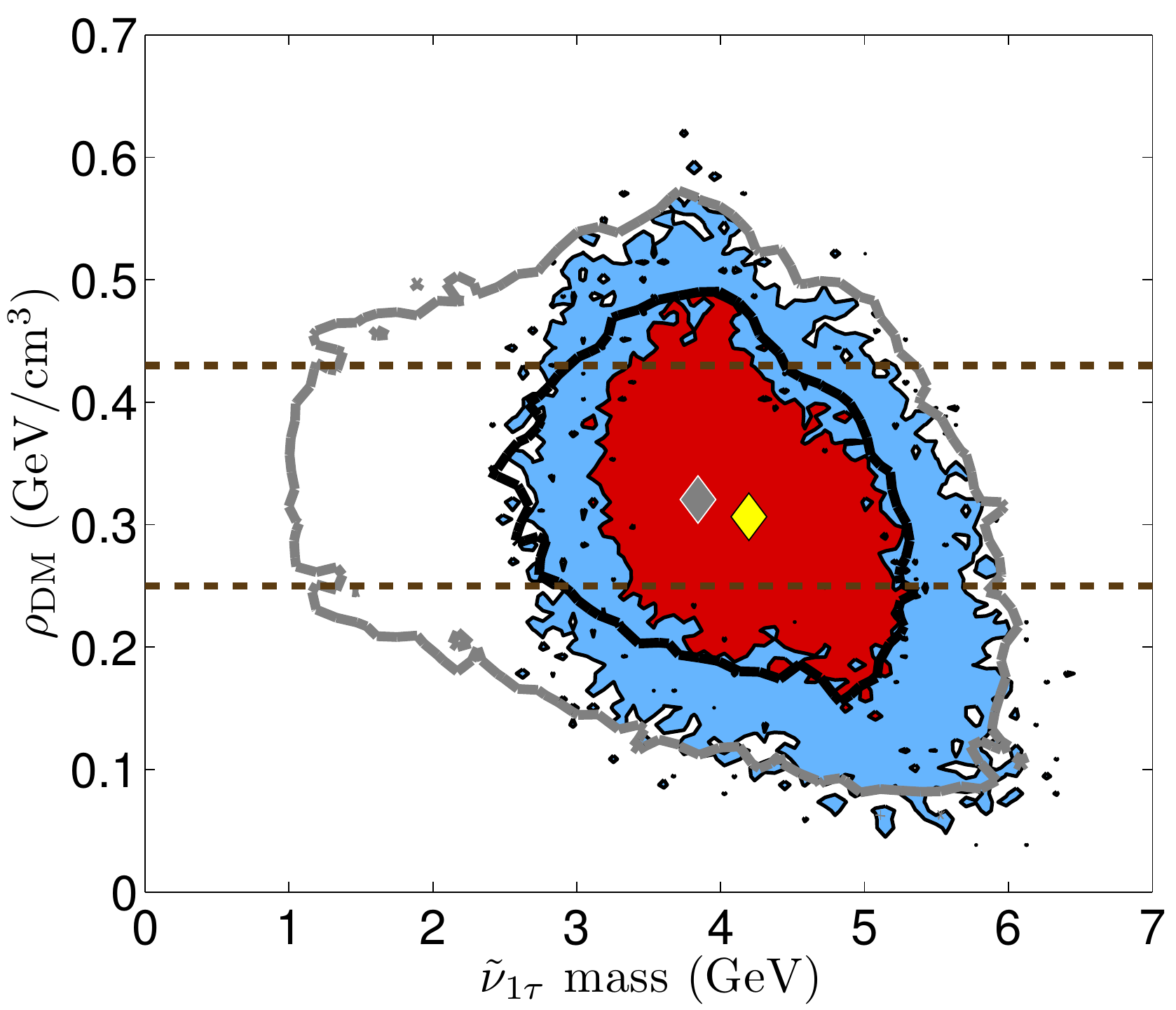}
   \caption{68\% and 95\% BCRs of $v_{\rm esc}$ versus $m_{\tilde\nu_{1\tau}}$ (left) and of $\rho_{\rm DM}$ versus $m_{\tilde\nu_{1\tau}}$ (right). The black (gray) contours are the 68\% (95\%) BCRs without gluino mass cut, while the red (blue) areas are the 68\% (95\%) BCRs for $m_{\tilde g}>1$~TeV. The dashed lines mark the $1\sigma$ experimental bounds for $v_{\rm esc}$ and $\rho_{\rm DM}$.}
   \label{sn2012-fig:light-2d-nuisance}
\end{figure}

The influence of the nuisance parameters is also interesting. For example, a low local DM density can bring points with high direct detection cross section in agreement with the XENON10 limits. Likewise, a small mixing angle at sneutrino masses around 4~GeV allows for higher $\rho_{\rm DM}$, because the direct detection cross section is low. 
Analogous arguments hold for $v_0$ and $v_{\rm esc}$, since for light DM one is very sensitive to the tail of the velocity distribution. 
The effect is illustrated in Fig.~\ref{sn2012-fig:light-2d-nuisance}.

\begin{figure}[!t] 
   \centering
   \includegraphics[width=7cm]{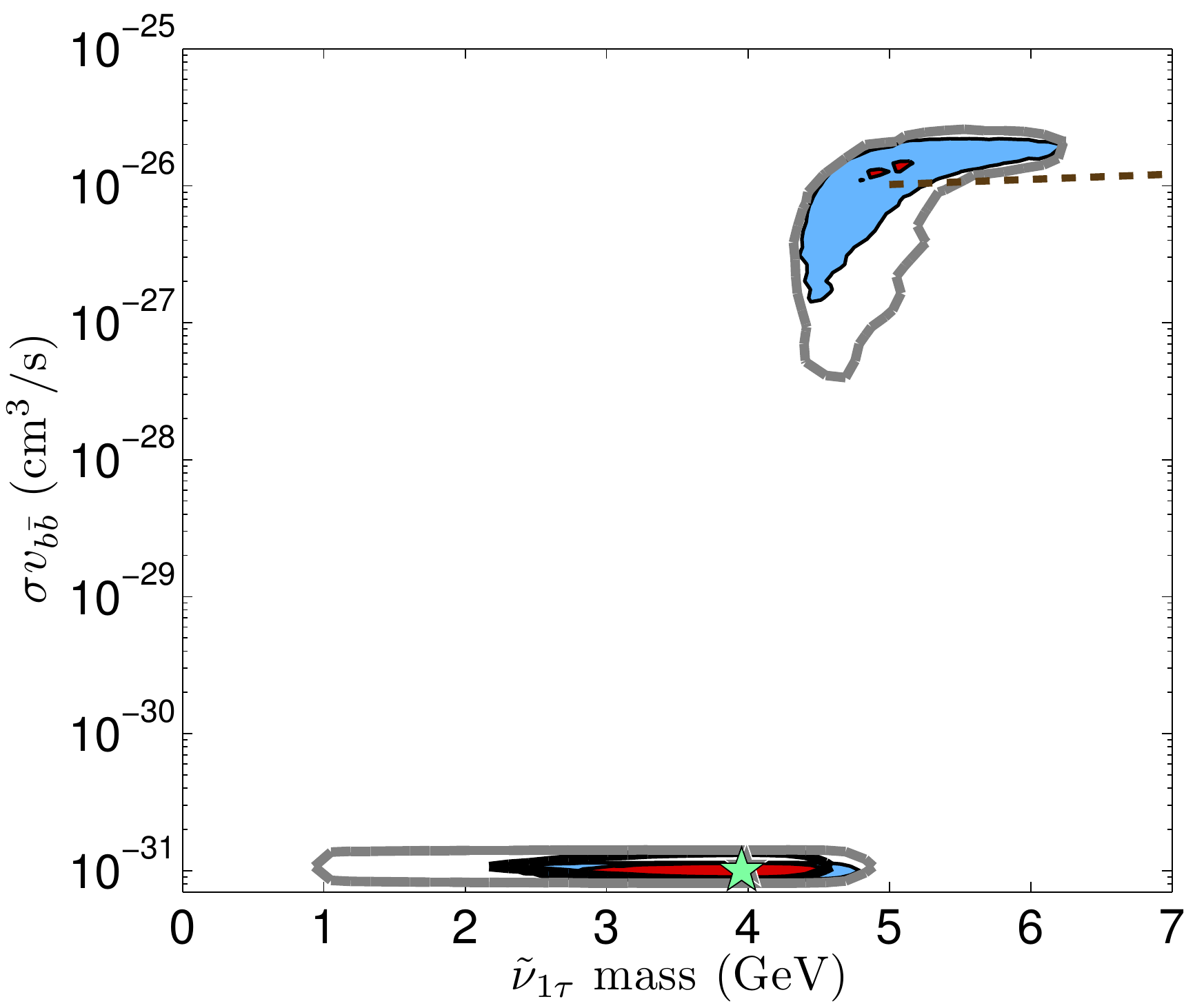}\quad 
   \includegraphics[width=7cm]{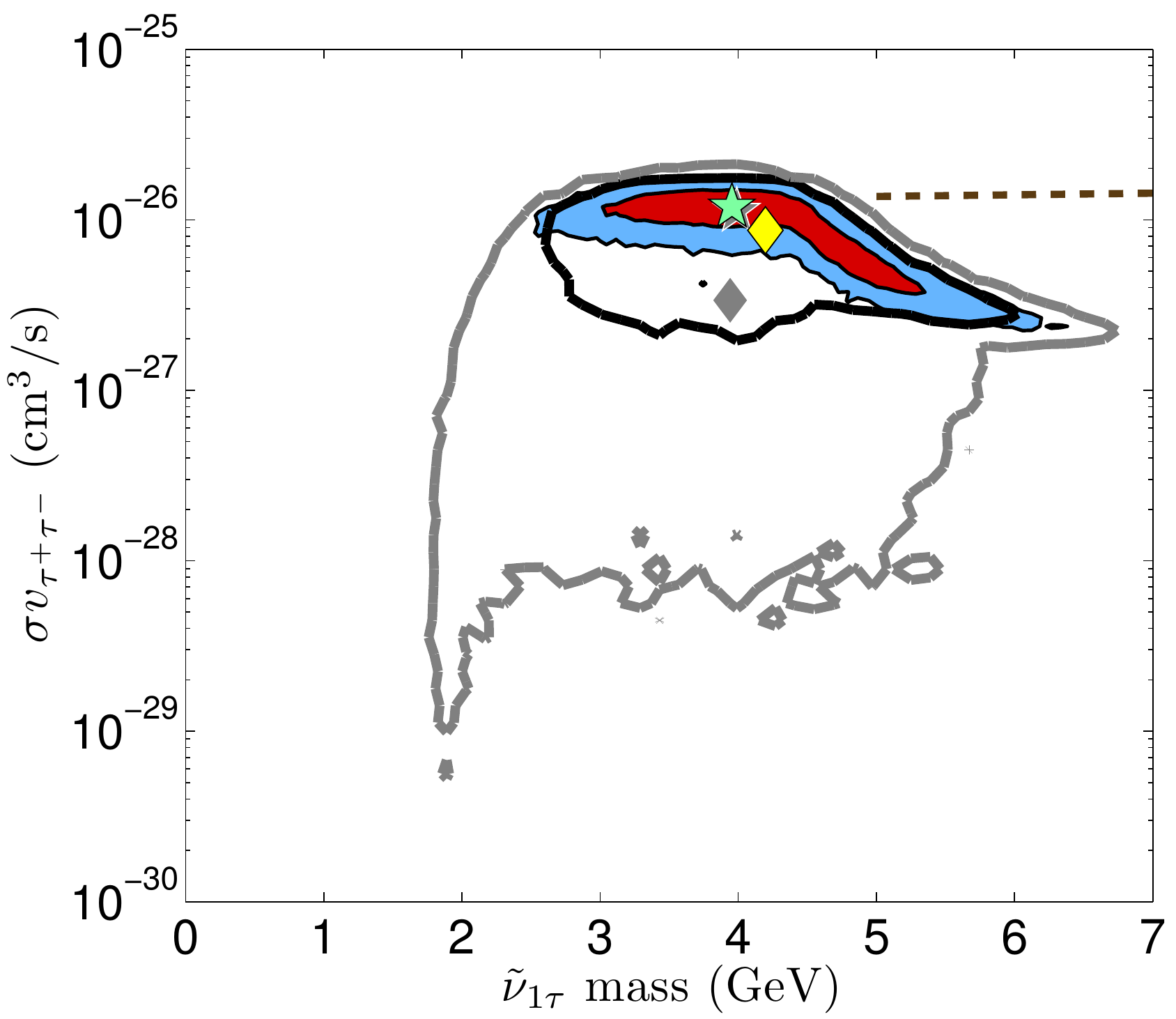}\\
   \includegraphics[width=7cm]{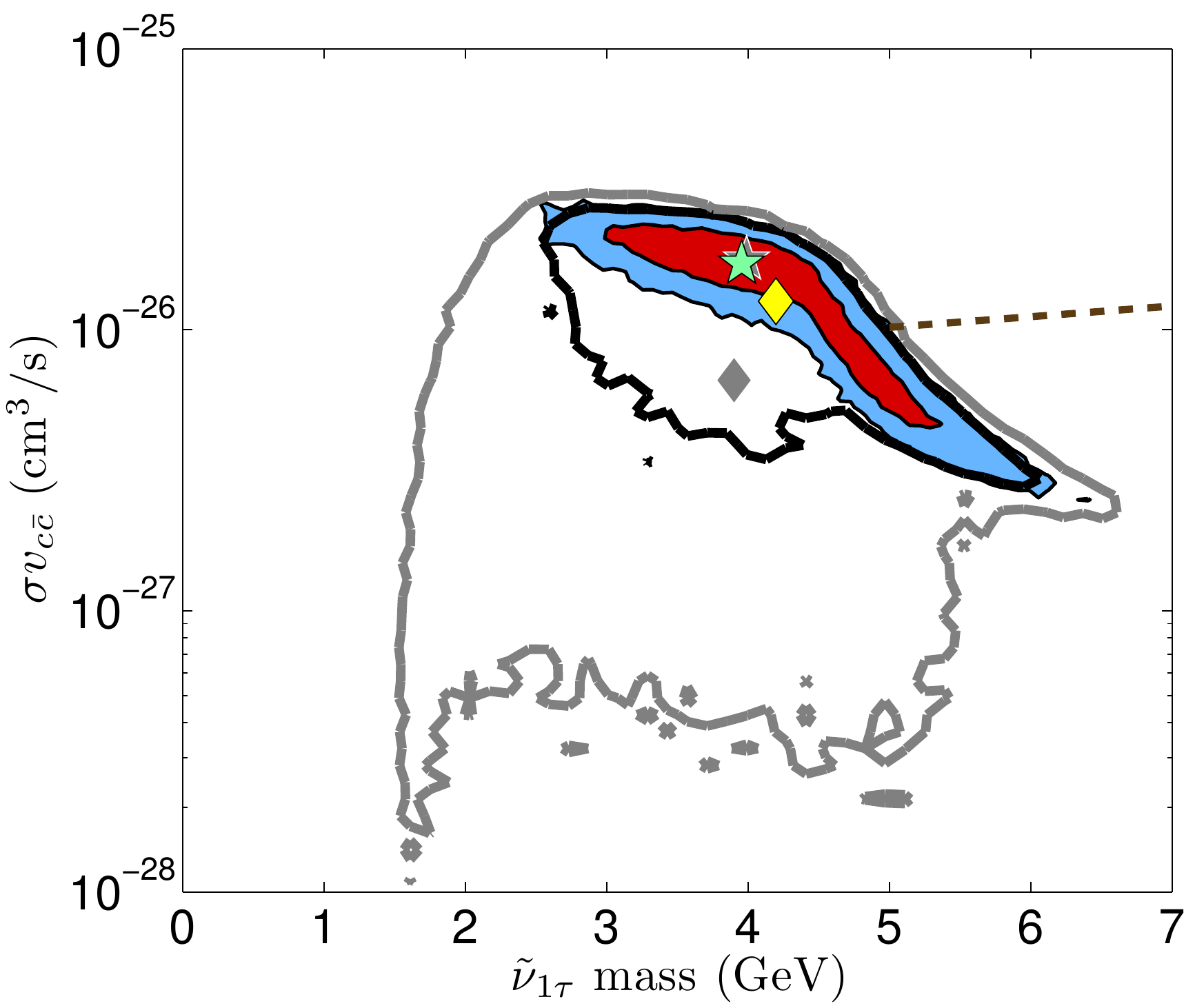}\quad
   \includegraphics[width=7cm]{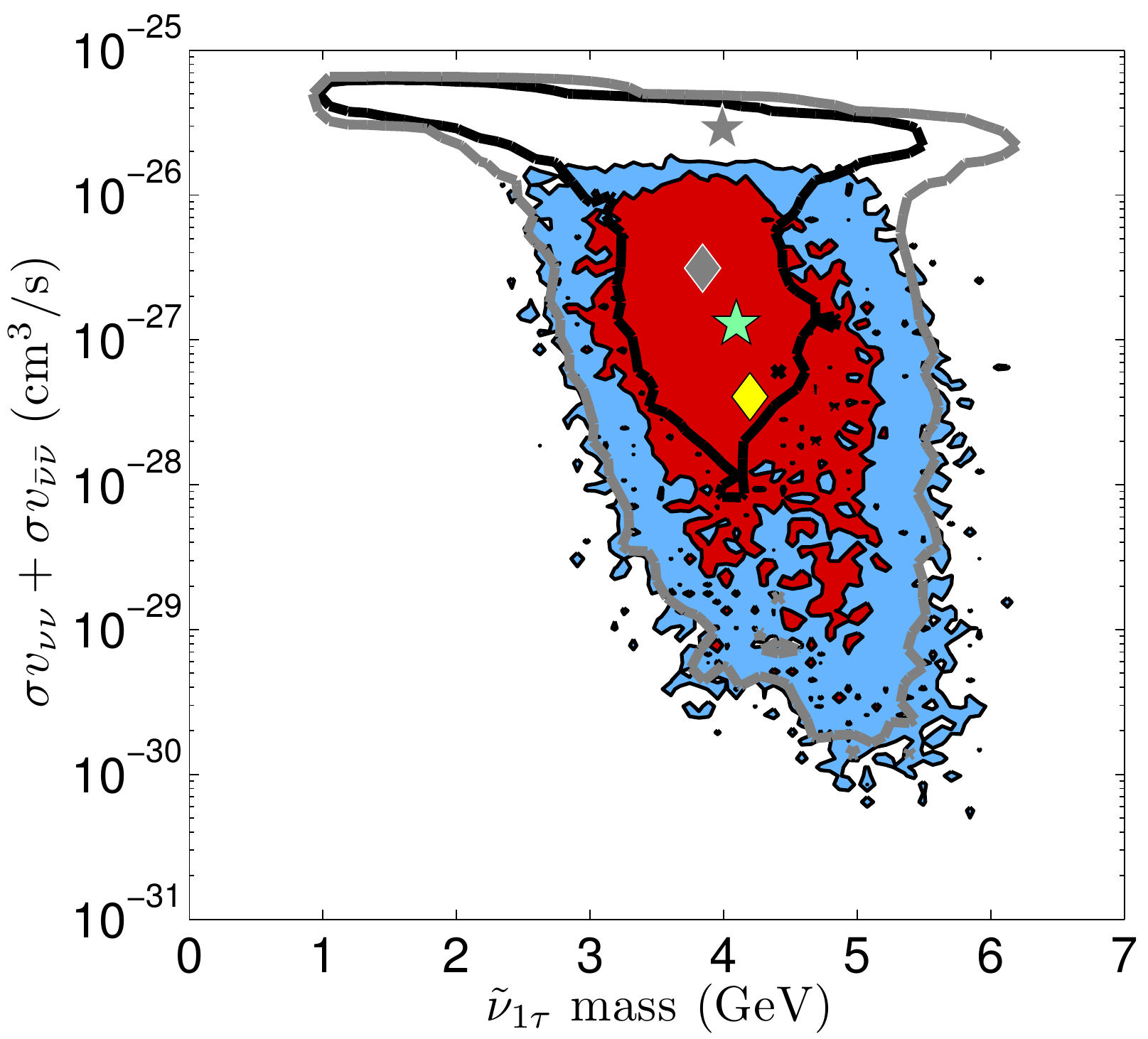}
   \caption{68\% and 95\% BCRs for $\sigma v$ versus sneutrino mass in various channels. Color code as in the previous figures. The dashed lines correspond to the {\it Fermi}-LAT limit~\cite{Ackermann:2011wa}, where  for $c\bar{c}$ we have used the same value as for $b\bar{b}$. Note that for $m_{{\tilde{\nu}_1}}<m_b$,  the cross section is zero, however to display this region we have arbitrarily set it to $\sigma v_{b\bar b}=10^{-31}~\rm{cm}^3 /{\rm s}$.}
   \label{sn2012-fig:light-2d-indirect}
\end{figure}

The MCMC approach also permits us to make predictions for the annihilation cross section  of light sneutrino dark matter into different final states, relevant for indirect DM searches, see Fig.~\ref{sn2012-fig:light-2d-indirect}.
When $m_{{\tilde{\nu}_1}}>m_b$,
the dominant DM annihilation channels are into $\nu\nu$ or   $b\bar{b}$ pairs. The latter will lead to a large photon flux---in fact the partial cross section into $b\bar{b}$  is always in the region constrained by {\it Fermi}-LAT when $m_{{\tilde{\nu}_1}}>5.2$~GeV.

For lighter DM, the charged fermions final states  giving photons are  $c\bar{c}$ and  $\tau^+\tau^-$. Here note that for a given  LSP mass, imposing the lower limit on the gluino mass  selects the upper range for both $\sigma v_{c\bar{c}}$  and $\sigma v_{\tau^+\tau^-}$ while having only a mild effect on $\sigma v_{b\bar{b}}$. 
In particular the $c\bar{c}$ channel typically has a large cross section of $\sigma v_{c\bar c}\gtrsim 10^{-26}~\rm{cm}^3 /{\rm s}$ throughout the 95\% BCR when $m_{\tilde g}>1$~TeV. This could hence be probed if  the {\it Fermi}-LAT search was extended to a lower mass range. 

Regarding annihilation into neutrinos, as mentioned earlier,  the gluino mass limit  strongly constrains scenario where annihilation into neutrino pairs is dominant, leading to an upper limit of $\sigma v_{\nu\nu} + \sigma v_{\bar{\nu}\bar{\nu}} \lesssim 1\times 10^{-26}~\rm{cm}^3 /{\rm s}$, see the bottom-right panel in Fig.~\ref{sn2012-fig:light-2d-indirect}. A discussion of the neutrino signal for light sneutrino DM can be found in~\cite{Belanger:2010cd}. 
As mentioned, we leave a more detailed analysis of neutrinos from the Sun for a future work.  

Dark matter annihilation in our galaxy can also lead to antiprotons. 
To illustrate the impact of the antiproton measurements on the parameter space of the model, we have computed the antiproton flux for some sample points and compared those to the flux measured by PAMELA~\cite{Adriani:2010rc}. To compute this flux we have used the  semi-analytical two-zone propagation model of~\cite{Maurin:2001sj,Donato:2001ms} with two sets of propagation parameters called MIN and MED, see~\cite{Belanger:2010gh}. For the background we have used the semi-analytical formulas of~\cite{Maurin:2006hy} with a solar modulation of $\phi=560~{\rm MeV}$,  which fit well the measured spectrum of PAMELA.

The first sample point has  a DM  mass of 4.8~GeV and  is dominated by annihilation into $b\bar{b}$  with $\sigma v_{bb}=1.1 
\times 10^{-26}~{\rm cm}^3/{\rm s}$.
The resulting antiproton flux is displayed as the blue band in Fig.~\ref{sn2012-fig:light-antiproton}.  A large excess is expected at 
energies below 1~GeV for MED propagation parameters, corresponding to the upper edge of the blue band. With MIN propagation parameters however,  the flux exceeds the $1\sigma$ range only in the lowest energy bin ($E_{\bar{p}}=0.28$~GeV). We therefore conclude that such sneutrino DM would  be compatible with the PAMELA measurements only for a restricted choice of propagation model parameters. Here note that the lowest energy bins are  the ones where the background is most affected by  solar modulation effects.

The second sample point has  lighter DM, $m_{{\tilde{\nu}_1}}=2.3$~GeV,  and annihilation into c-quarks dominates the hadronic channels ($\sigma v_{c\bar{c}}=1.7 \times 10^{-26}~{\rm cm}^3/{\rm s}$) although the dominant annihilation channel is into neutrinos. The antiproton flux is therefore expected to be both lower and shifted towards lower energies as compared to the previous case. 
We find that the antiproton flux again exceeds  the measured spectrum by more than $1\sigma$ only in the first energy bin.  Such a sneutrino DM is therefore not constrained by the antiproton measurements unless one chooses propagation parameters that lead to large fluxes. In this respect note that we can of course get even larger fluxes than those displayed in Fig.~\ref{sn2012-fig:light-antiproton} using the MAX set of propagation parameters.

\begin{figure}[ht] 
   \centering
   \includegraphics[width=7cm]{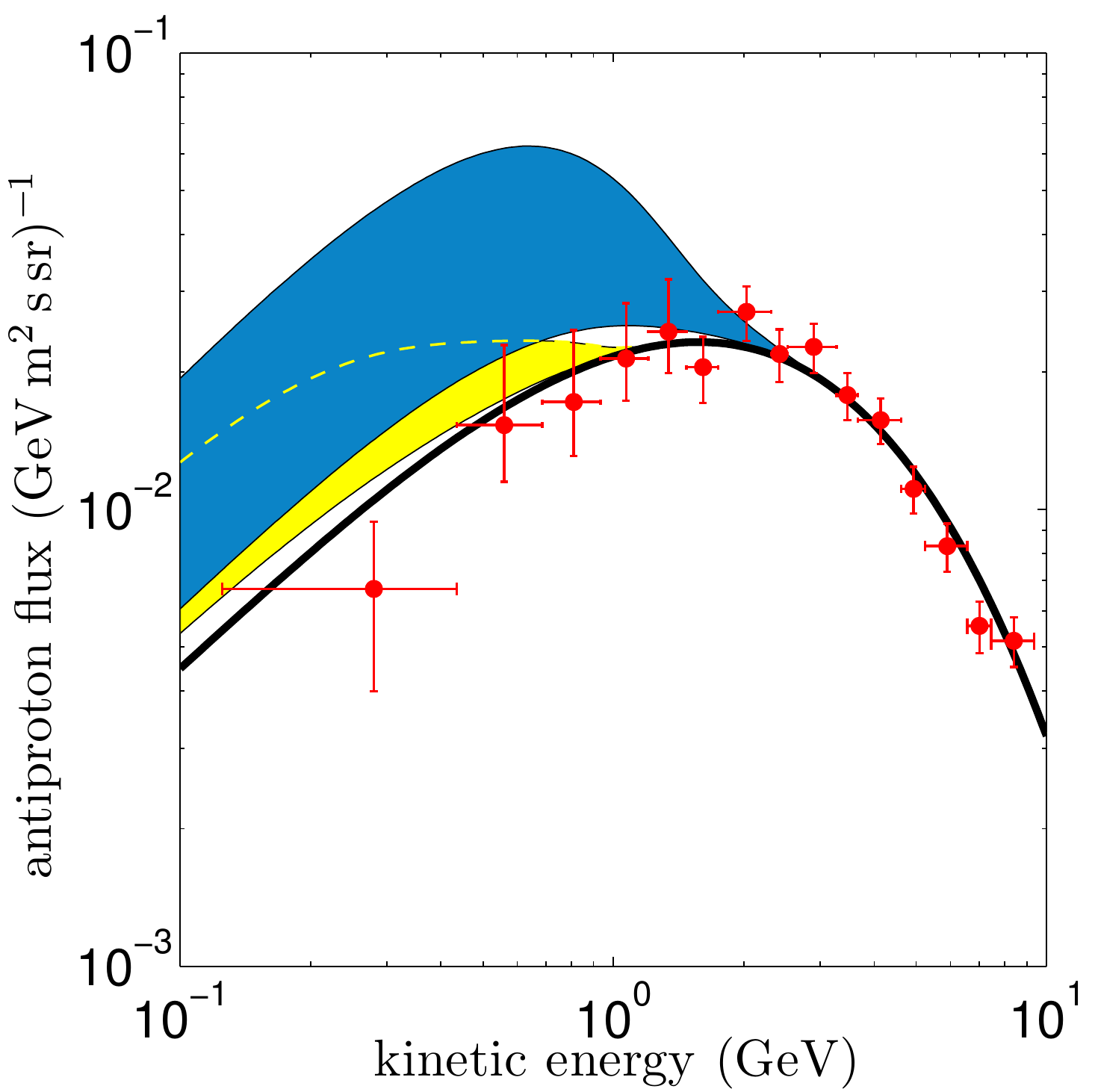}
   \caption{Antiproton flux as a function of the kinetic energy of the antiproton for two representative points as described in the text. The blue (yellow) band corresponds to $m_{{\tilde{\nu}_1}}= 4.8$ (2.3)~GeV, with the upper curve corresponding to MED and the lower curve corresponding to  MIN propagation parameters. We also display the background only (black line) and the PAMELA data for energies below 10~GeV  (red crosses).}
   \label{sn2012-fig:light-antiproton}
\end{figure}

\subsubsection{Heavy sneutrino DM}

Let us now turn to the case of heavy sneutrinos. We will first discuss the heavy non-democratic (HND) case, where the LSP is the $\tilde\nu_{1\tau}$, and then the heavy democratic (HD) case, where all three neutrinos are close in mass and any of them can be the LSP or co-LSP.

\begin{figure}[!ht] 
   \centering
   \includegraphics[width=4.96cm]{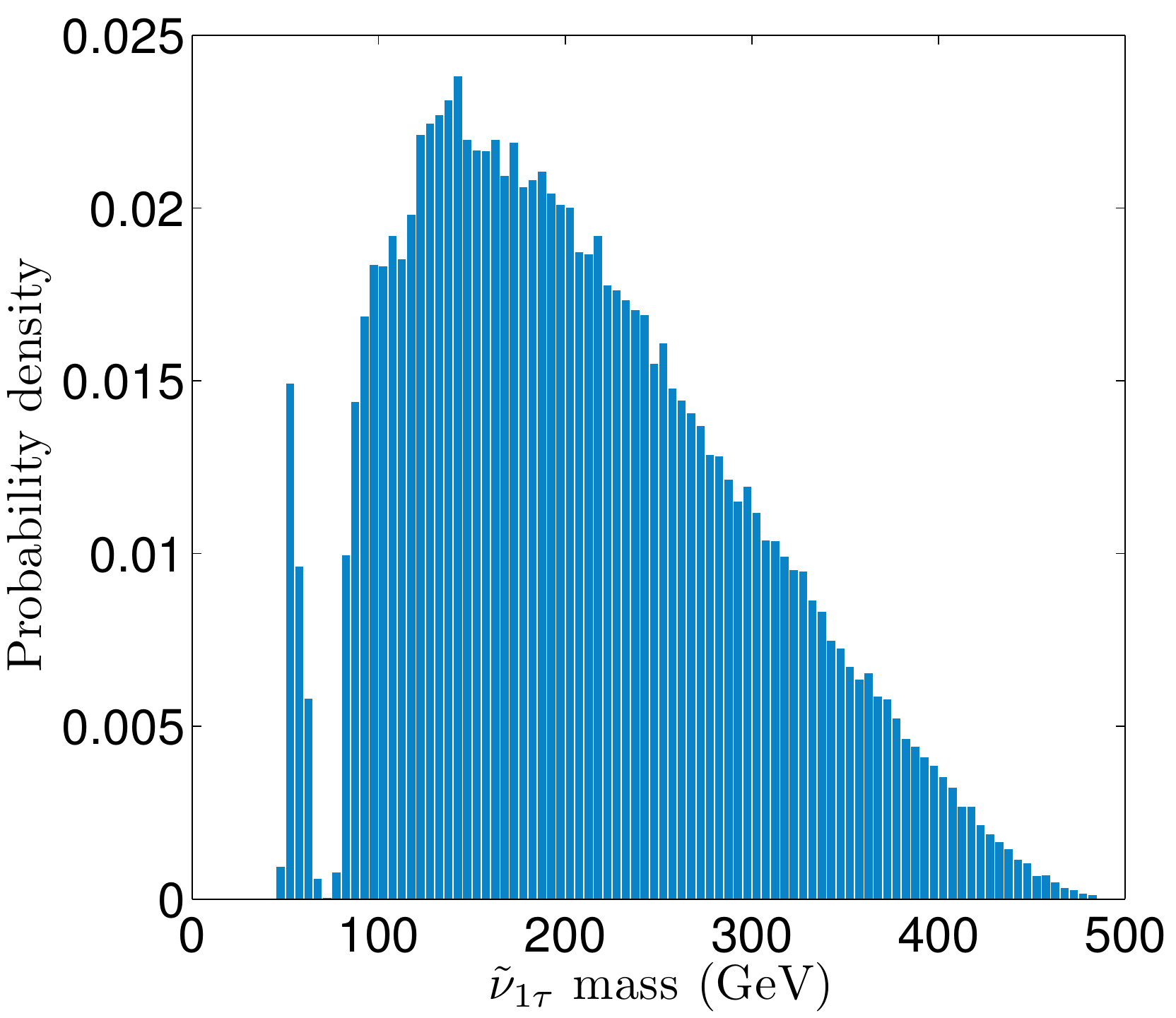}
   \includegraphics[width=4.96cm]{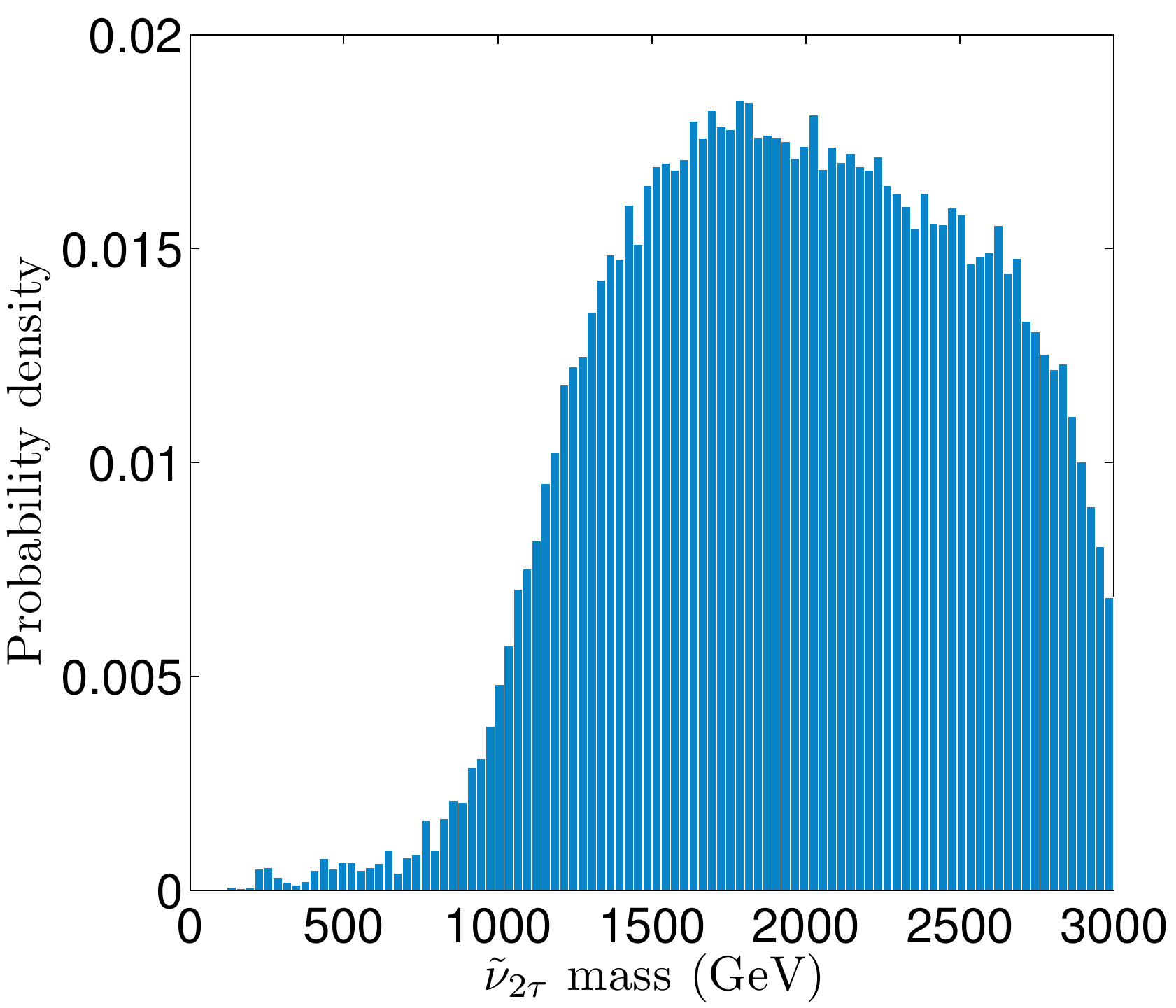}
   \includegraphics[width=4.96cm]{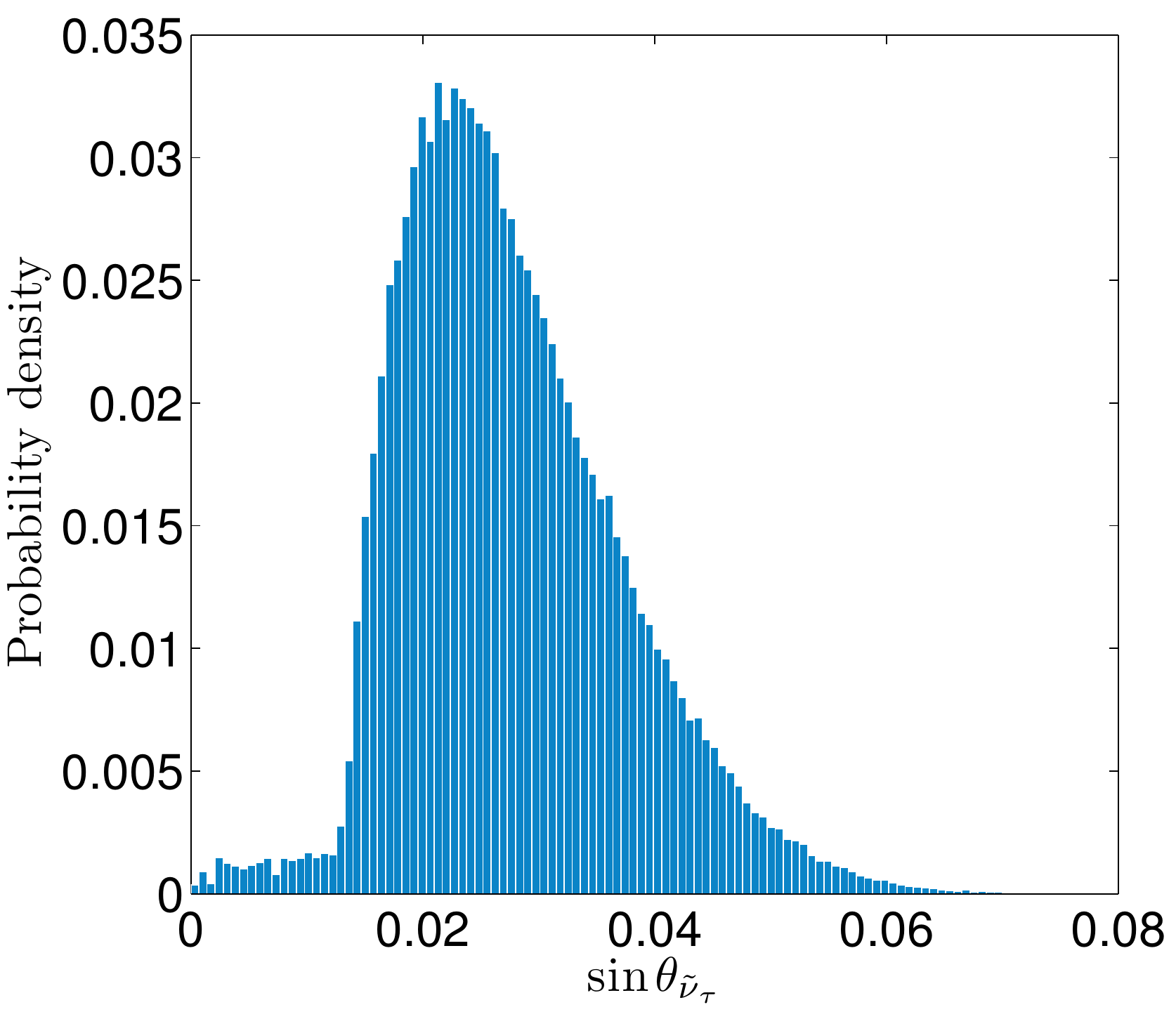}
   \includegraphics[width=4.96cm]{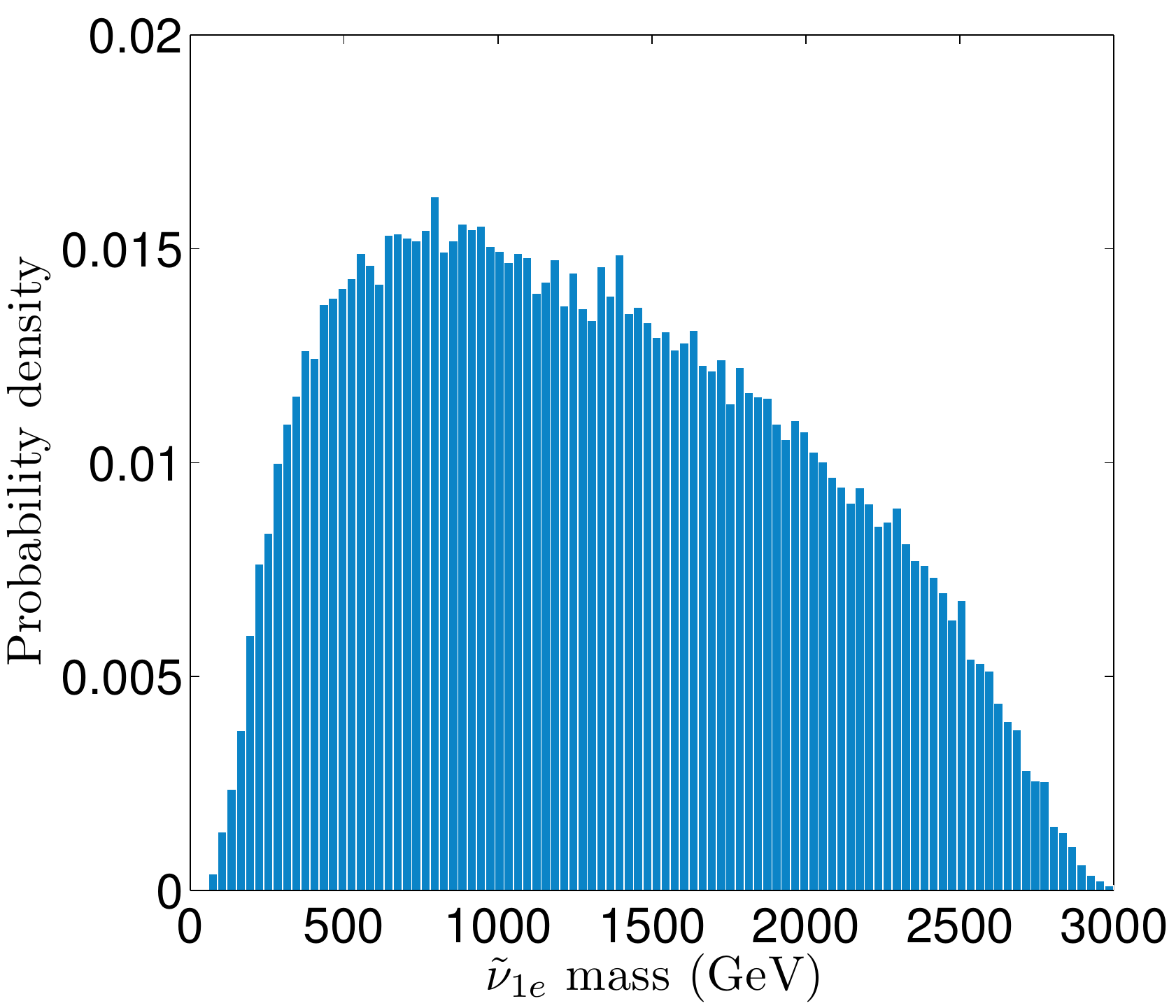}
   \includegraphics[width=4.96cm]{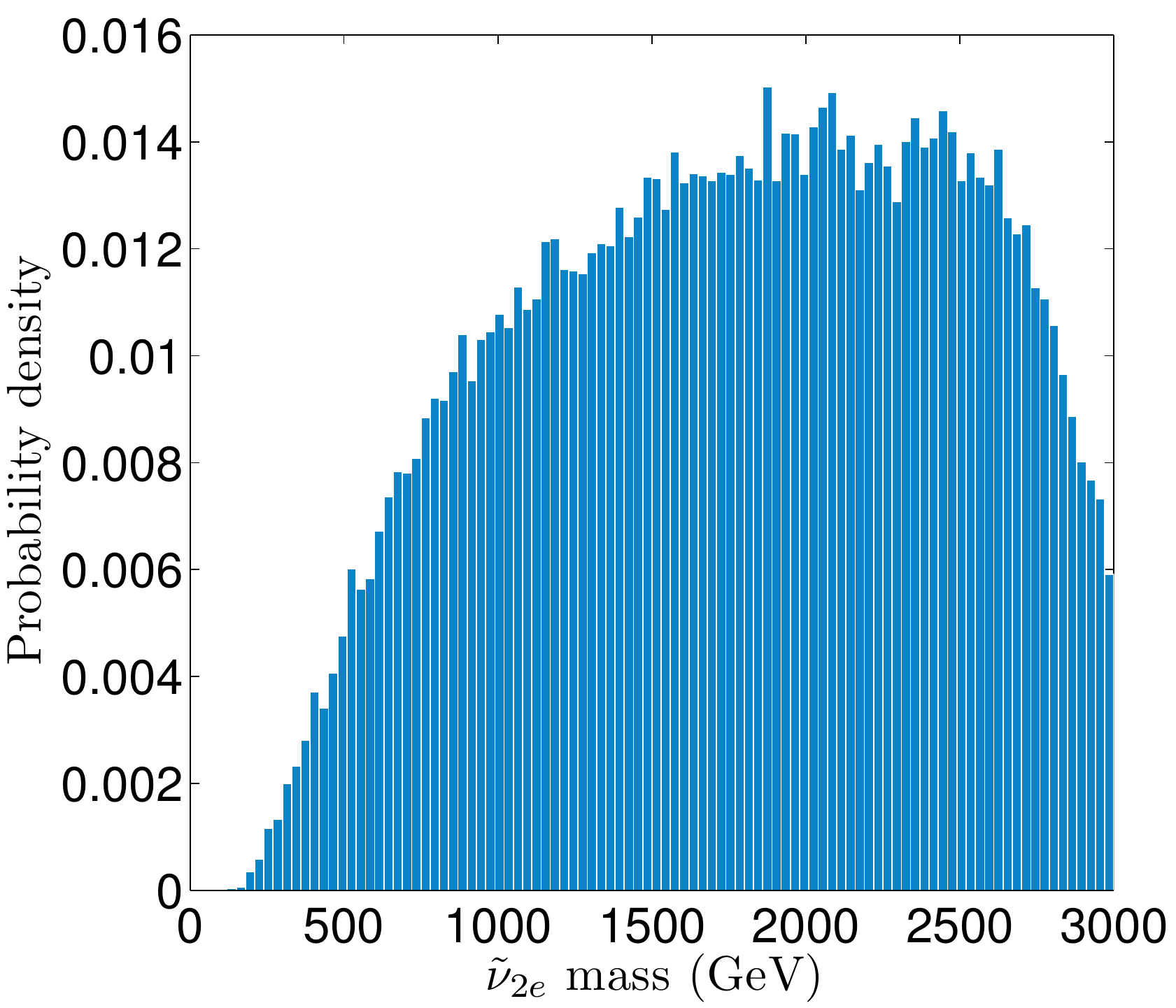}
   \includegraphics[width=4.96cm]{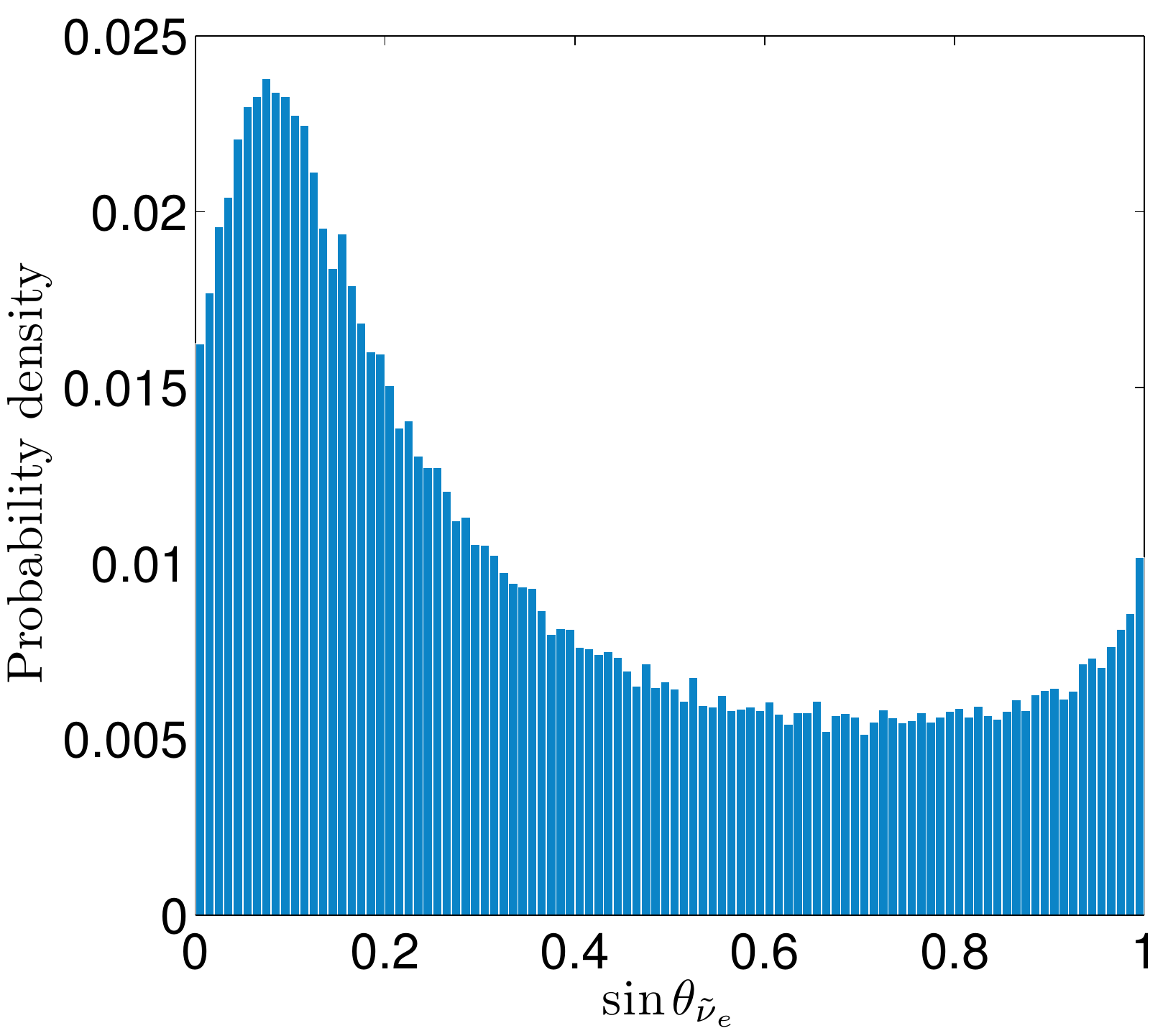}
   \includegraphics[width=4.96cm]{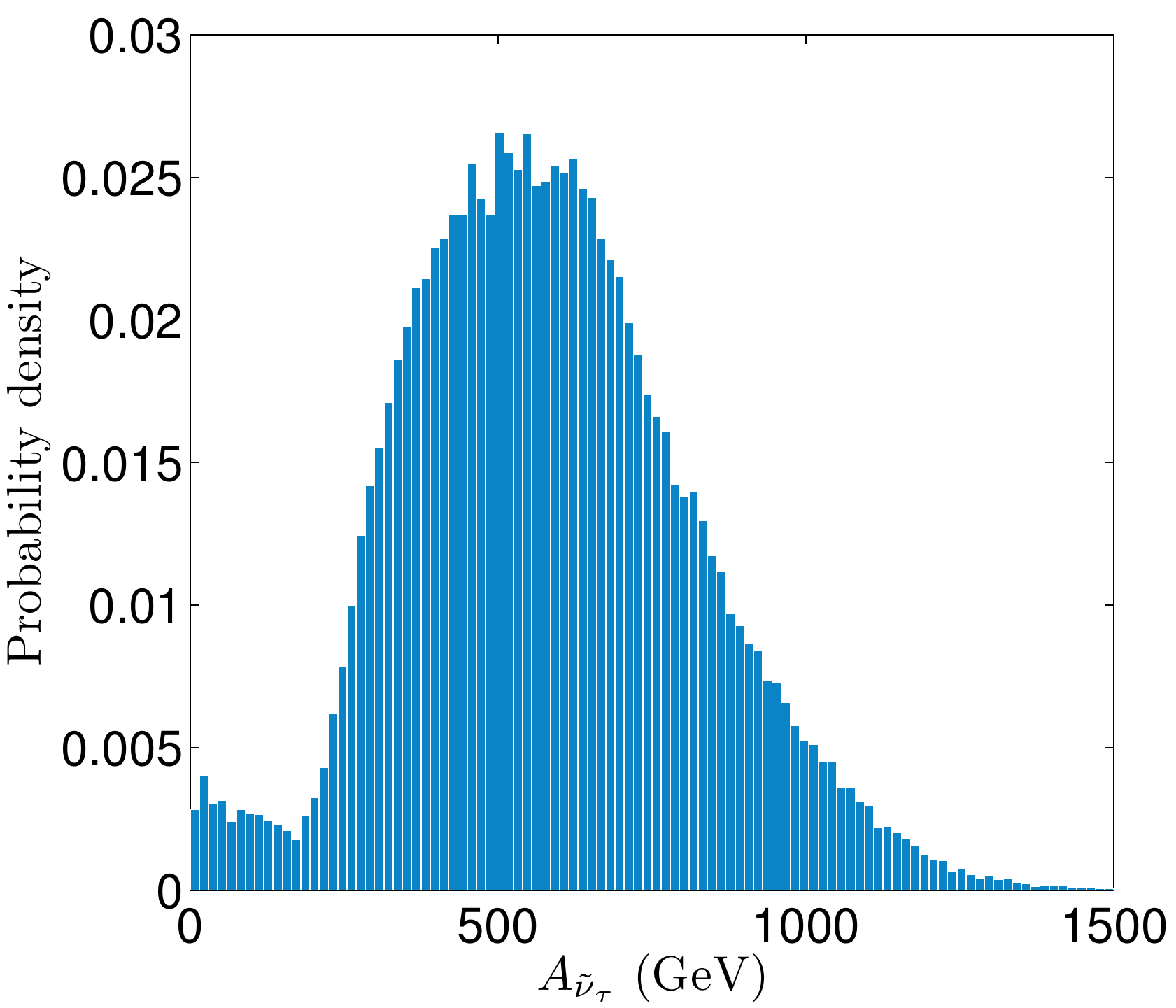}
   \includegraphics[width=4.96cm]{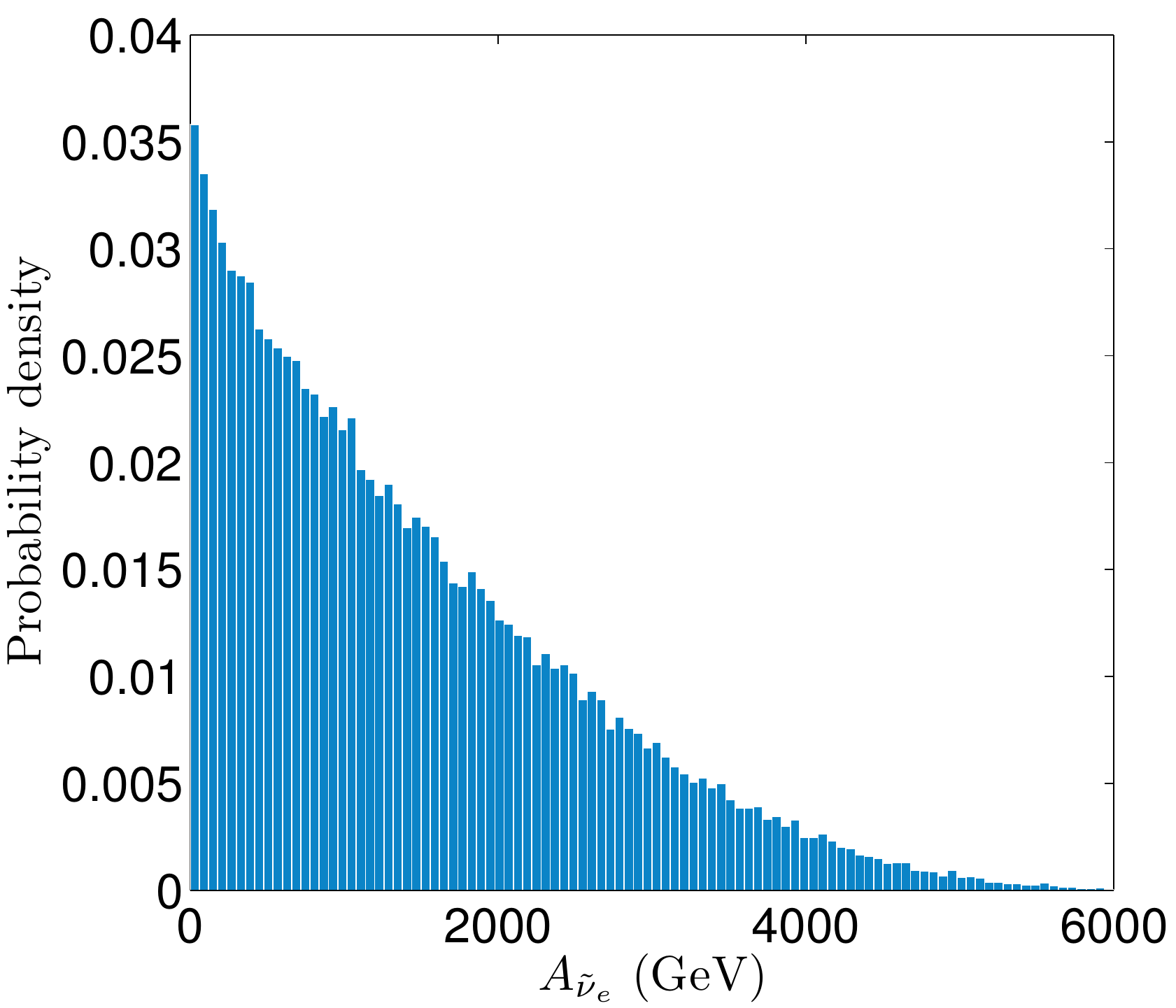}
   \includegraphics[width=4.96cm]{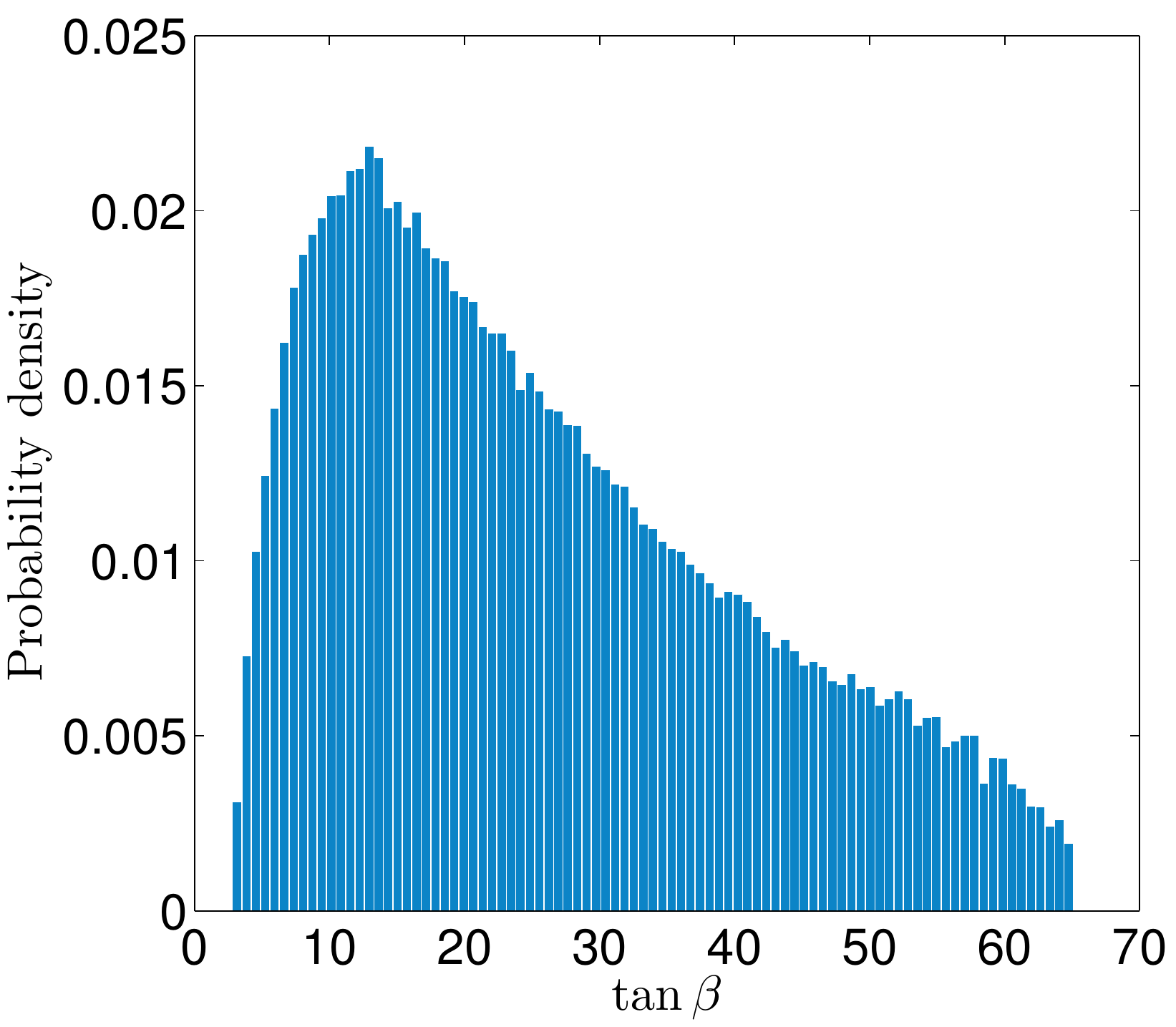}
   \includegraphics[width=4.96cm]{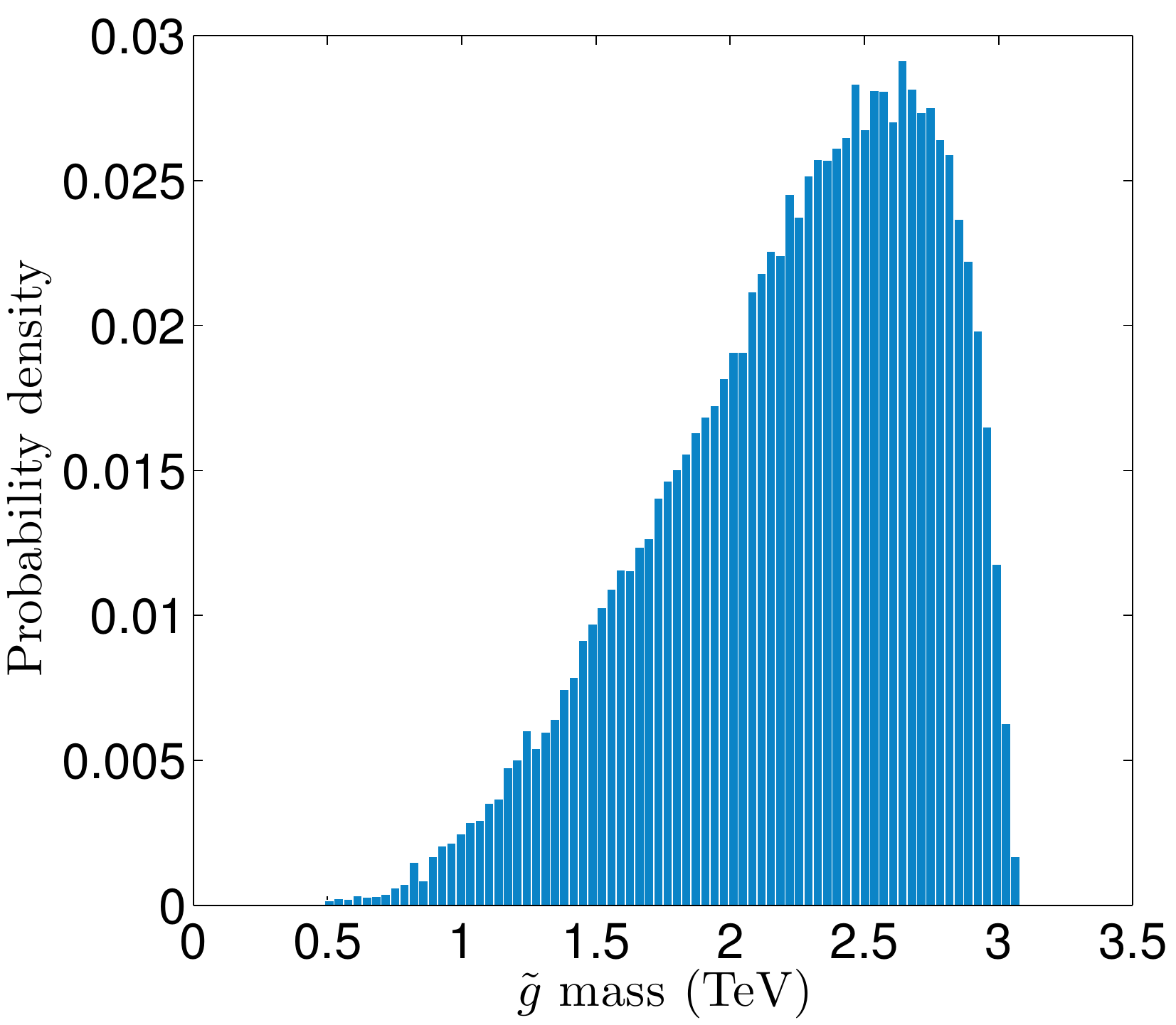}
   \includegraphics[width=4.96cm]{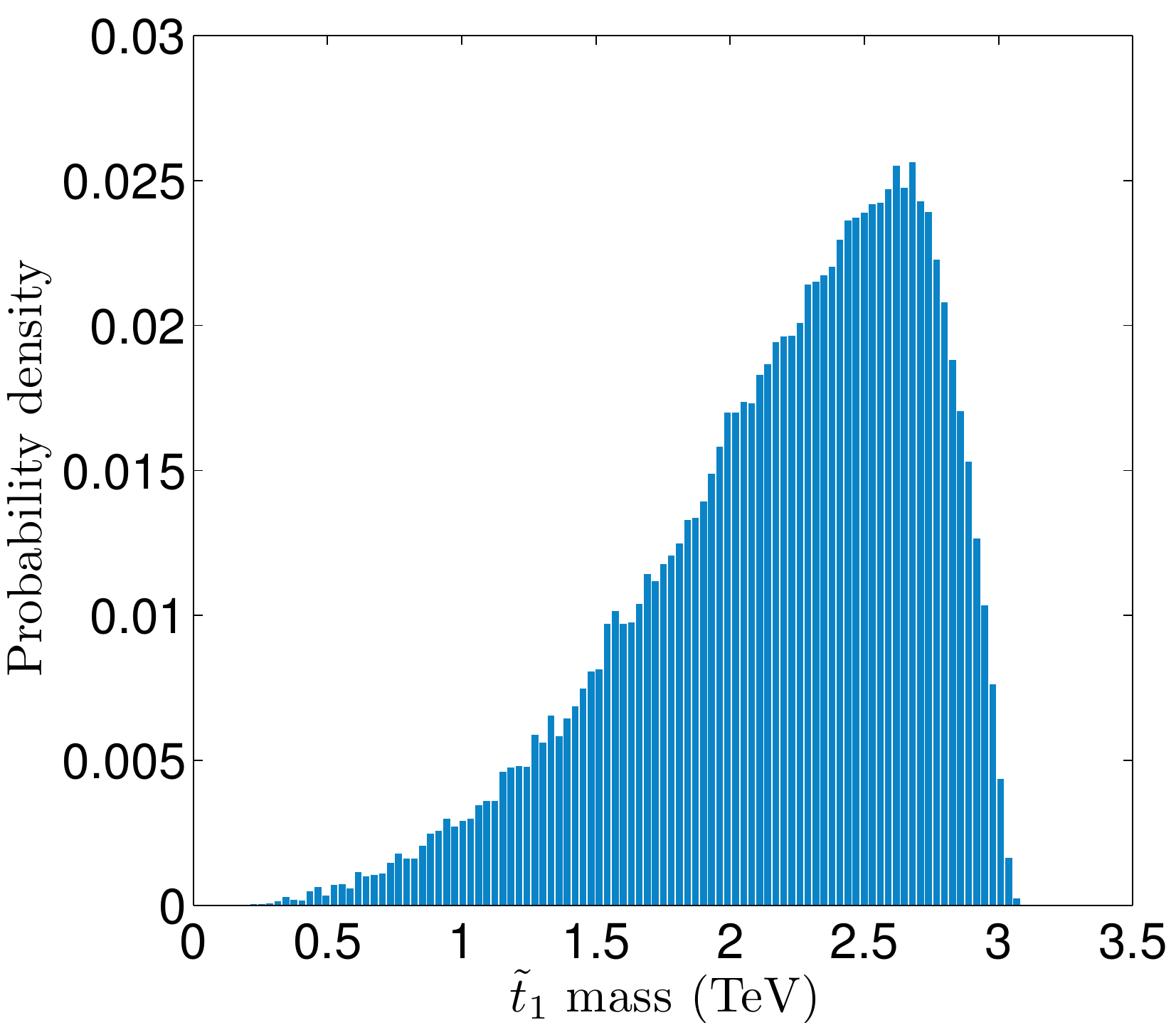}
   \includegraphics[width=4.96cm]{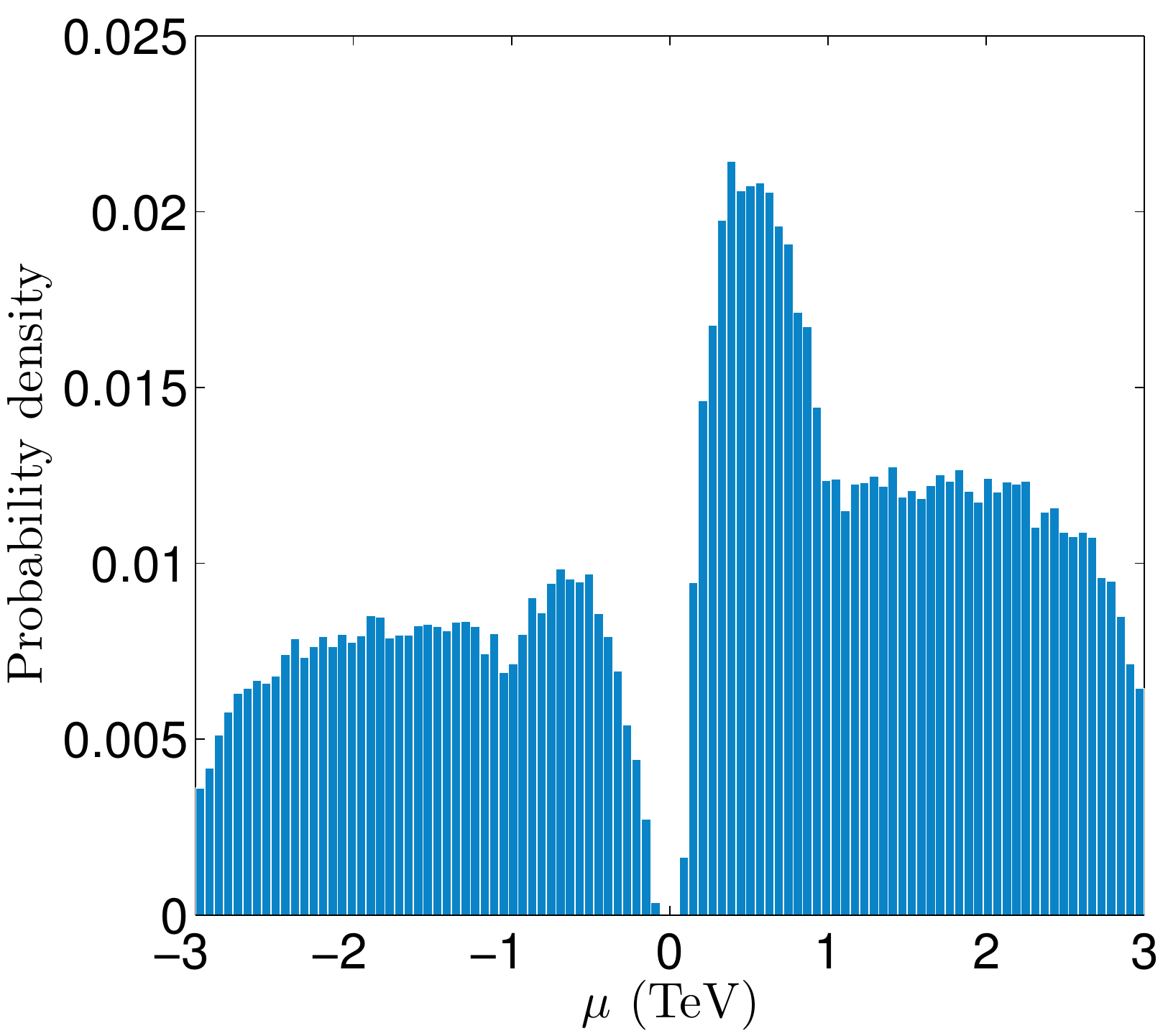}
   \includegraphics[width=4.96cm]{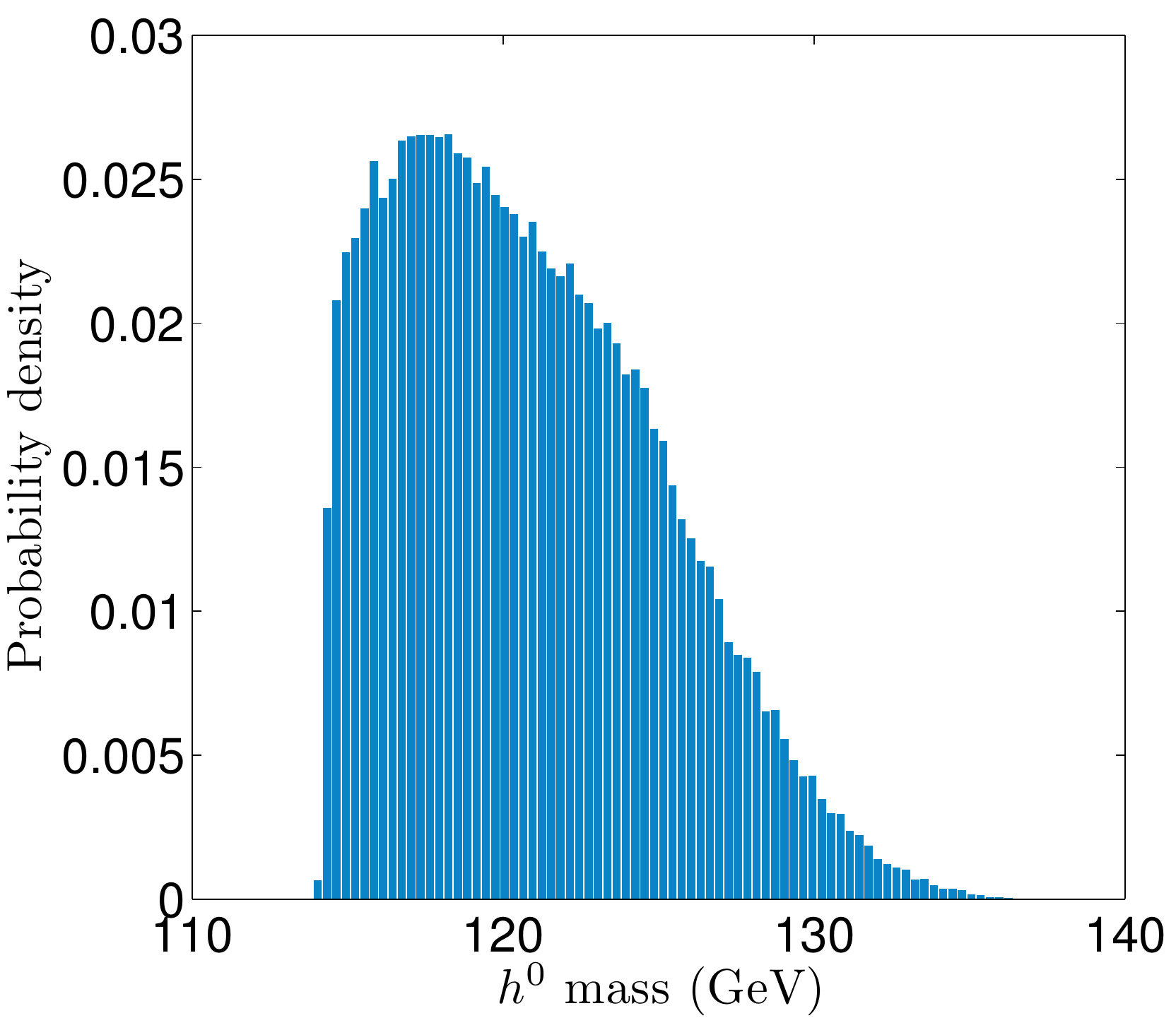}
   \includegraphics[width=4.96cm]{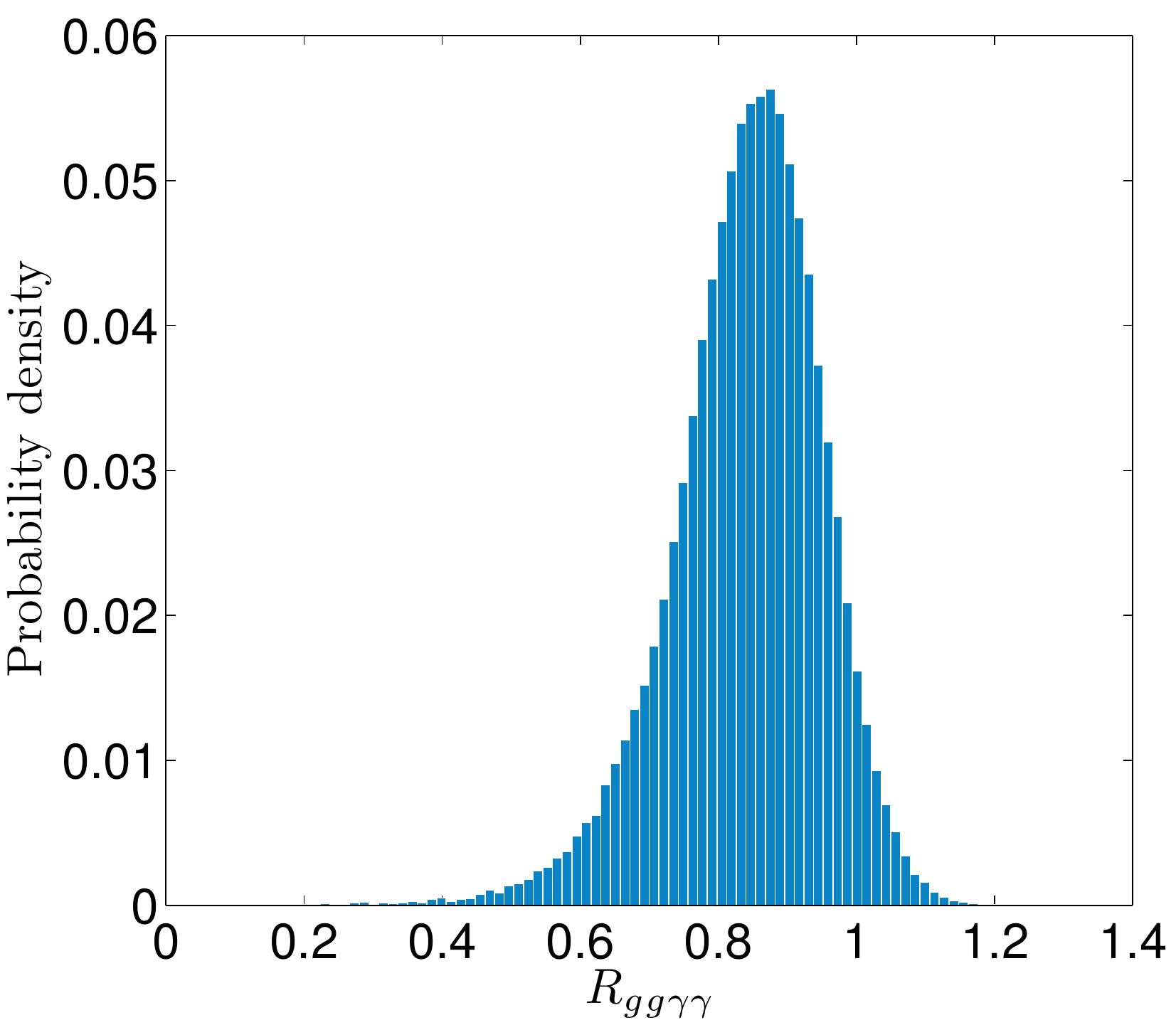}
   \includegraphics[width=4.96cm]{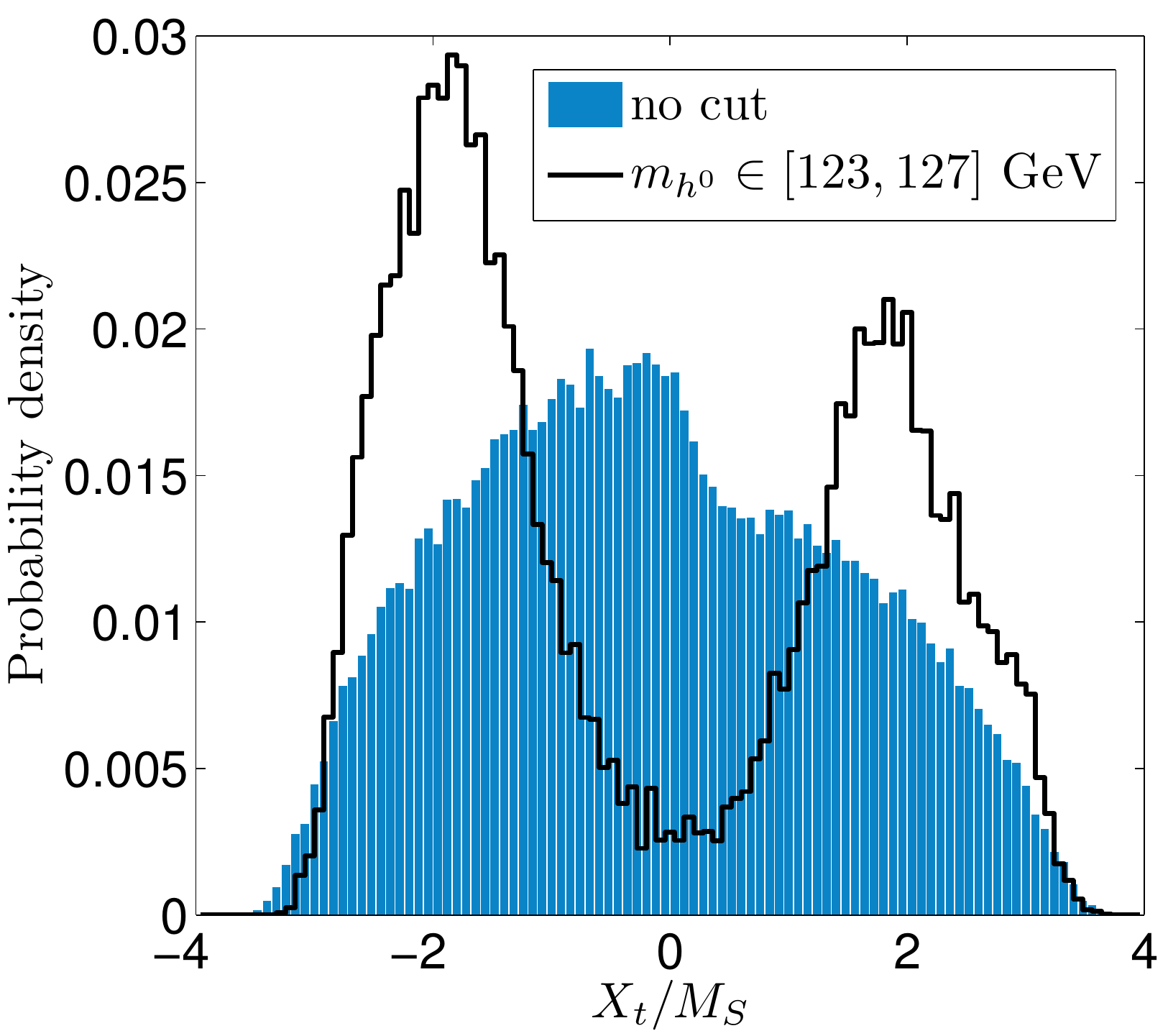}
   \caption{Posterior PDFs in 1D for the heavy non-democratic (HND) sneutrino case.}
   \label{sn2012-fig:heavy-1d}
\end{figure}

The posterior PDFs in 1D for the HND case are shown in Fig.~\ref{sn2012-fig:heavy-1d}. Here, we do not superimpose the distributions with $m_{\tilde g}>750$ or $1000$~GeV, because the gluino automatically \clearpage
\noindent turns out heavy, with 99\% probability above 1~TeV.
The $\tilde\nu_{1\tau}$ masses now range from 90 to 255 (80 to 375)~GeV at 68\% (95\%) Bayesian credibility.
There is also a small region near $m_{\tilde\nu_{1\tau}}\approx 60$~GeV, where the sneutrino annihilates through the light Higgs resonance; this region has 3\% probability.\footnote{As mentioned in Section~\ref{sn2012-sec:relic}, the sneutrino can also annihilate through the heavy scalar (not the pseudoscalar!) Higgs resonance. We have checked that this process does occur in our chains. However, it  turns out that it is statistically insignificant and does not single out any special region of parameter space.} 
See Table~6 in Appendix~B of~\cite{Dumont:2012ee} for more details.  
The $\tilde\nu_{2\tau}$ mass is typically very heavy, above 1~TeV, and the mixing angle is required to be very small to evade the direct detection limits, cf.\ the discussion in Section~\ref{sn2012-sec:directdetection}. Interestingly, the mixing can be almost vanishing; 
this happens either when $m_{\tilde\nu_{1\tau}}\simeq m_{h^0}/2$ so that the annihilation is on resonance, or when co-annihilation channels are important. In the first case, the ${A_{\tilde\nu}}$ term must be very small, otherwise the annihilation cross section would be too large and $\Omega h^2$ too small.  
Note that the upper limit on the sneutrino LSP mass is determined by the range for the gluino mass used in the scan which in turn sets an upper bound of 500~GeV on the lightest neutralino and hence on the sneutrino LSP. 

\begin{figure}[!ht] 
   \centering
   \includegraphics[width=7cm]{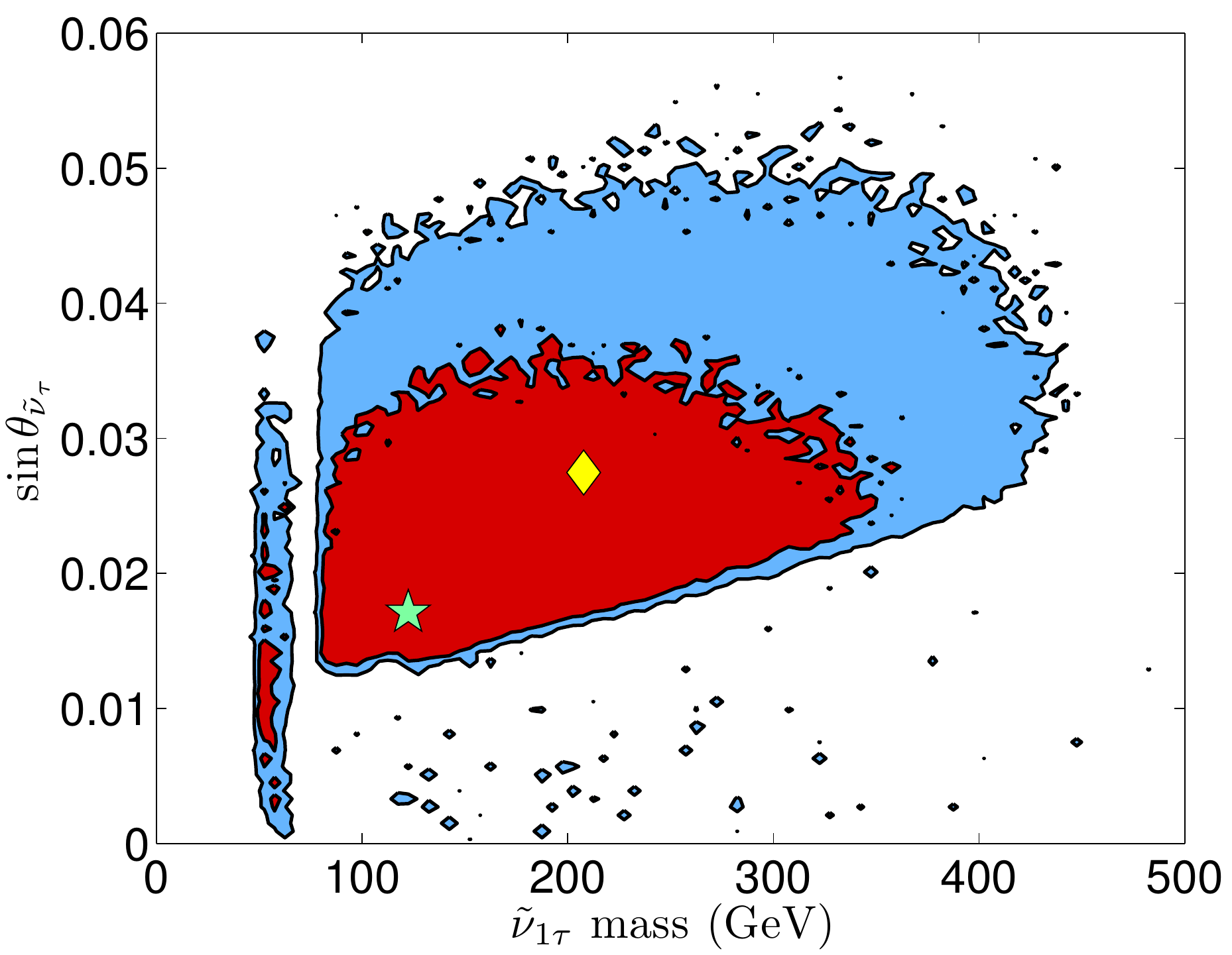} \quad
   \includegraphics[width=7cm]{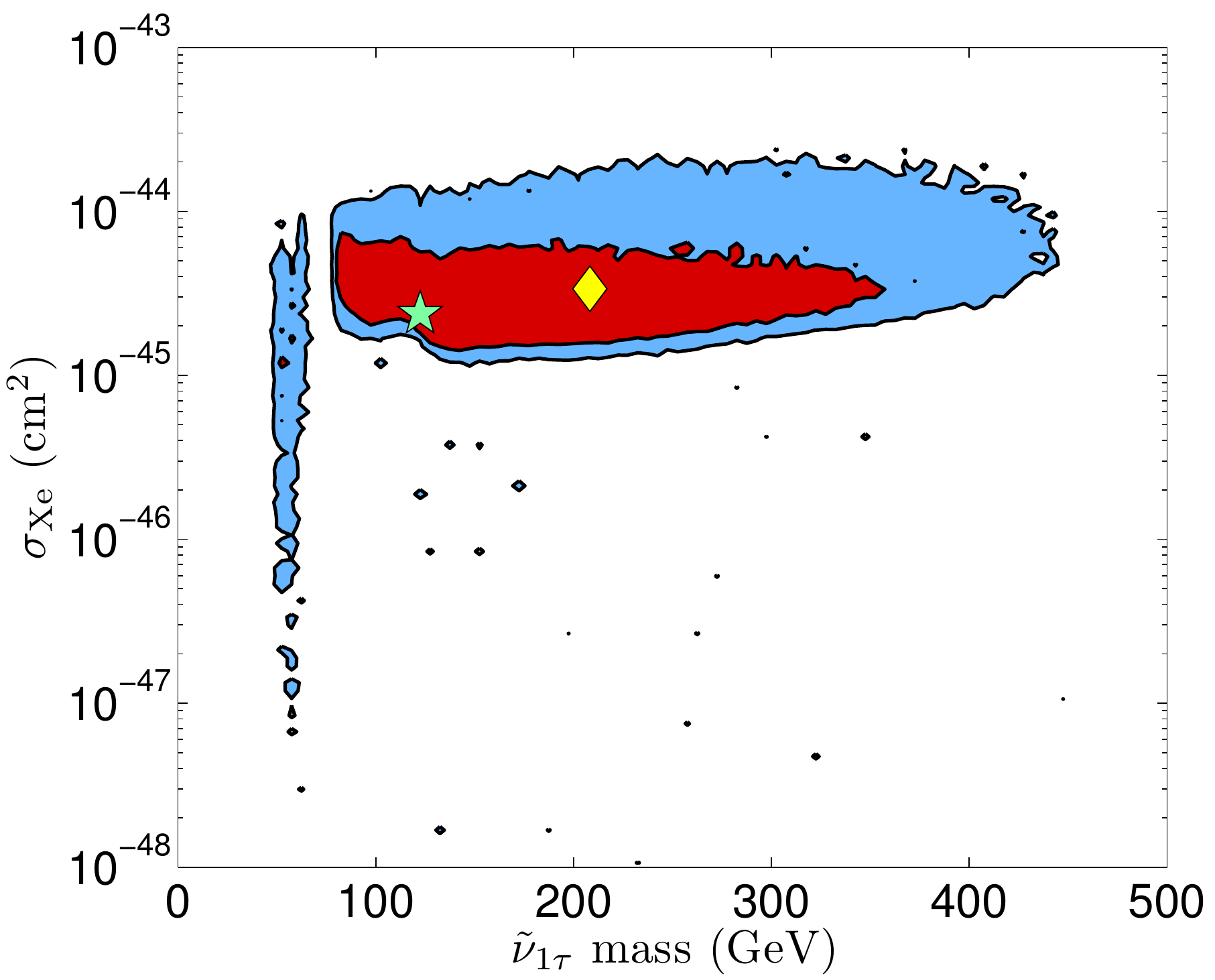}
   \caption{Posterior PDFs in 2D of $\sin\theta_{\tilde{\nu}_{\tau}}$ (left) and $\sigma_{\rm Xe}$ (right) versus $m_{\tilde\nu_{1\tau}}$ for the HND case. The red and blue areas are the  68\% and 95\% BCRs, respectively. The green stars mark the highest posterior, while the yellow diamonds mark the  mean of the  PDF.}
   \label{sn2012-fig:hnd-2d}
\end{figure}

The light Higgs mass is not much affected by radiative corrections from a heavy sneutrino, so the posterior PDF of $m_{h^0}$ is like in the conventional MSSM. (See the bottom row of Fig.~\ref{sn2012-fig:heavy-1d} for Higgs-related quantities.) A light Higgs in the $123\mbox{--}127$~GeV mass range has 21\% probability in this case. As in the MSSM, this mass range requires large mixing, see the distribution for $X_t/M_S$.\footnote{$X_t=A_t-\mu/\tan\beta$ and $M_S^2=m_{\tilde t_1}m_{\tilde t_2}$. In fact the distribution of $A_t$ is the only one that is significantly changed by requiring $m_{h^0} \in [123,127]$~GeV, see also Contribution~8 of~\cite{Brooijmans:2012yi} and Section~\ref{sec:pmssm}.}  
The signal strength in the $gg\to h\to\gamma\gamma$ channel relative to SM expectations ($\mu(gg \to h \to \gamma\gamma)$, denoted as $R_{gg\gamma\gamma}$ in Fig.~\ref{sn2012-fig:heavy-1d}) is also just like in the MSSM~\cite{Brooijmans:2012yi}, with the highest probability being around $\mu(gg \to h \to \gamma\gamma)\approx 0.9$. 
In this scenario, it is much more difficult to reach larger values of $\Delta a_\mu$ as the sneutrino contribution is never large.  
We find $\Delta a_\mu \le 8.6 \times 10^{-10}$ at 95\% BC.

\begin{figure}[!ht] 
   \centering
   \includegraphics[width=7cm,height=6cm]{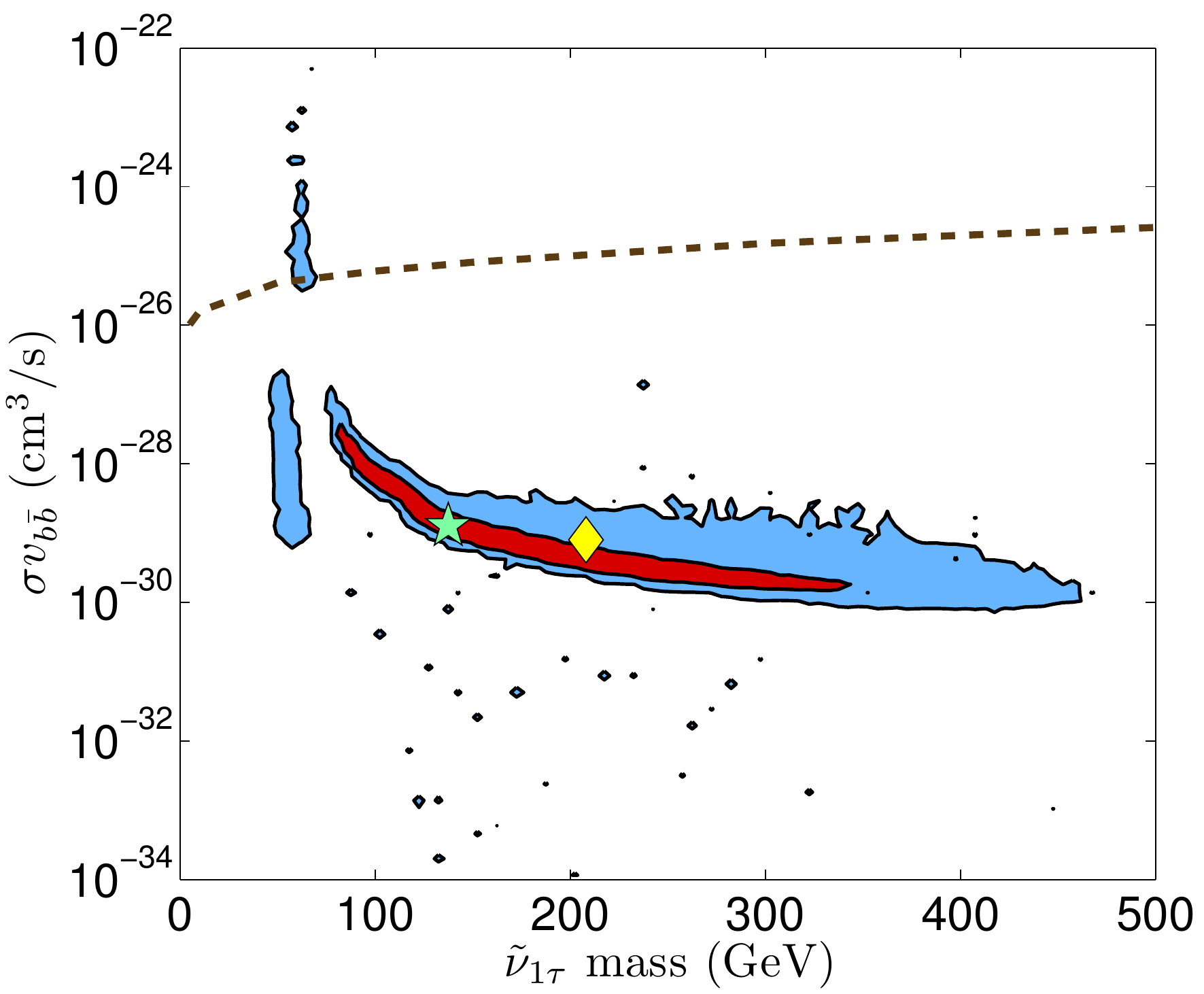} \quad
   \includegraphics[width=7cm,height=5.7cm]{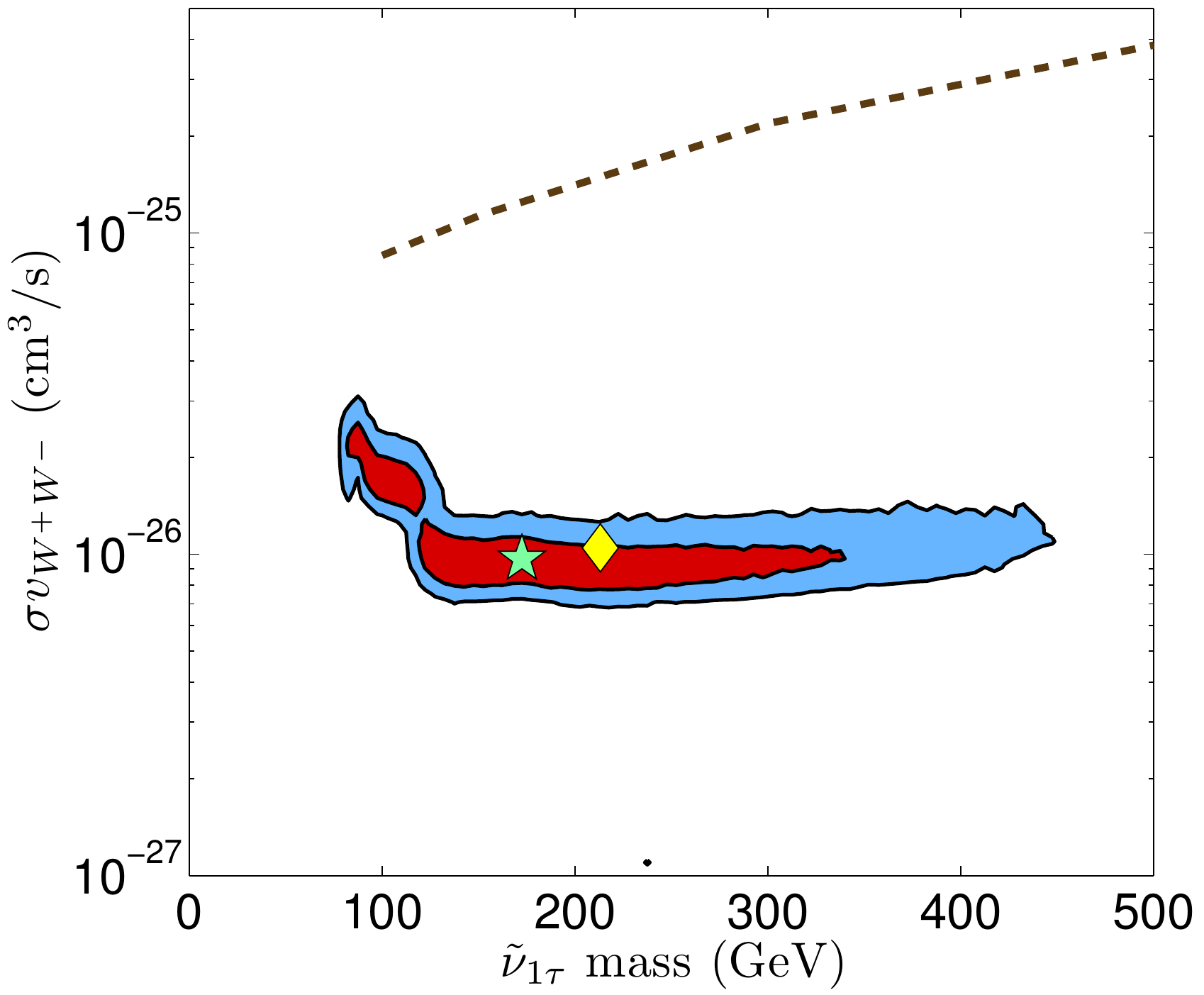}\\
   \includegraphics[width=7cm,height=6cm]{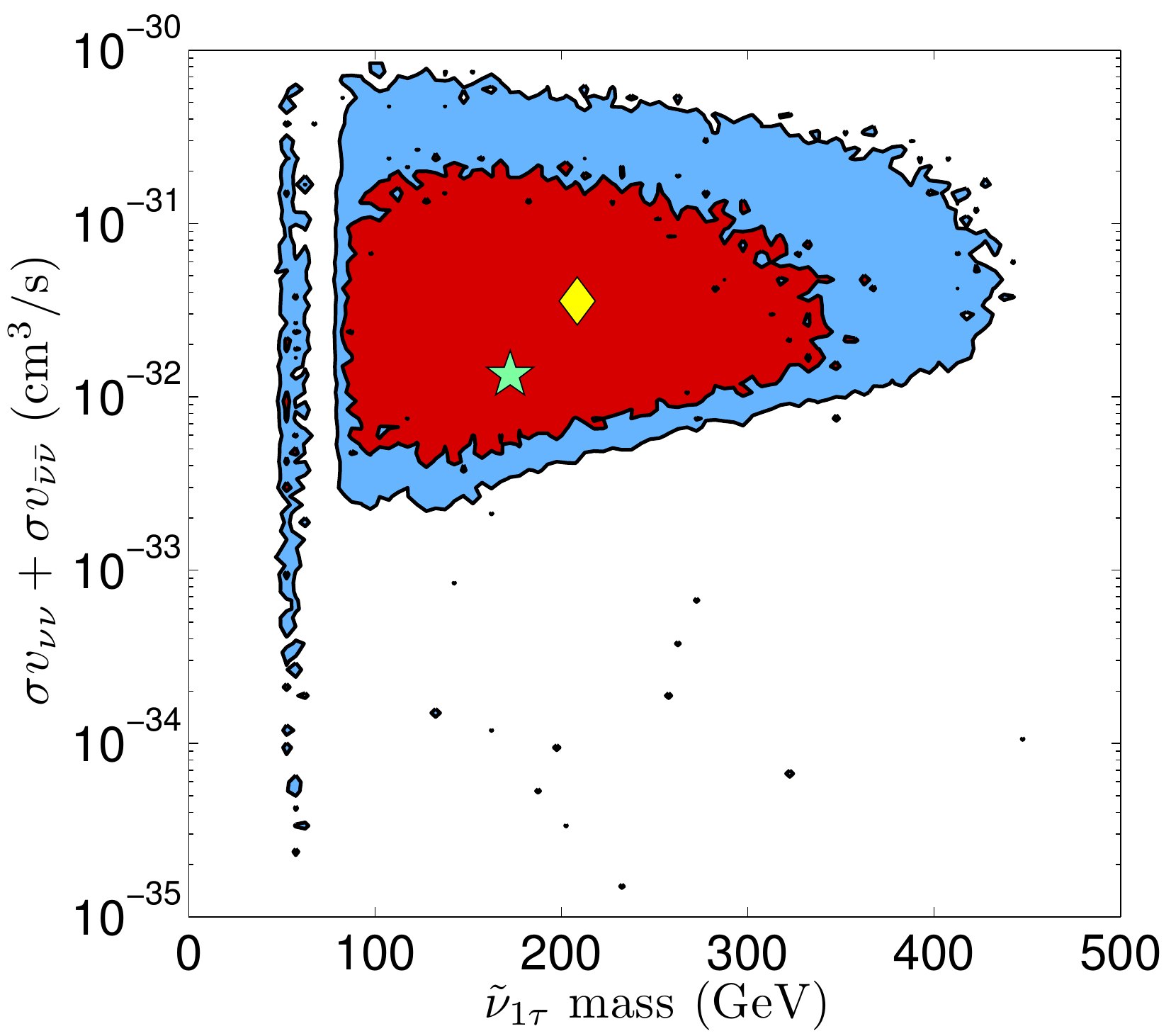} \quad 
   \includegraphics[width=7cm,height=6cm]{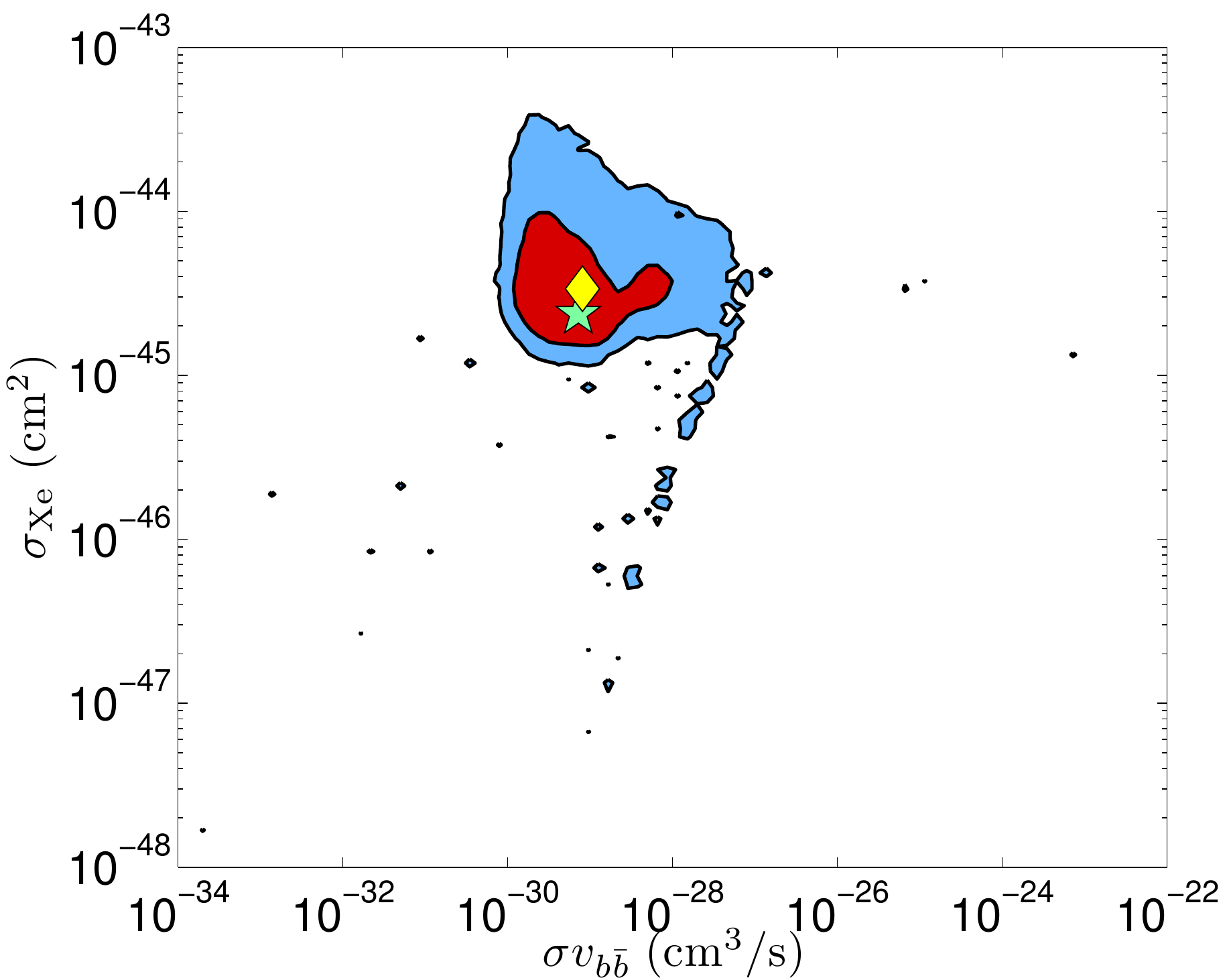}
   \caption{2D posterior PDFs for the HND case relevant for indirect DM detection; color codes {\it etc.}~as in Fig~\ref{sn2012-fig:hnd-2d}.}
   \label{sn2012-fig:hnd-2d:ann}
\end{figure}

In Fig.~\ref{sn2012-fig:hnd-2d}, we show the 2-dimensional posterior PDF of $\sin\theta_{\tilde{\nu}_{\tau}}$ versus $m_{\tilde\nu_{1\tau}}$. 
As can be seen, the mixing angle is always in the $\sin\theta_{\tilde{\nu}_{\tau}}\approx 0.01 -0.05$ region  except when $m_{{\tilde{\nu}_1}}\approx m_{h^0}/2$ or for a few scattered points with heavier LSP masses. The latter correspond to cases where the co-annihilation of pairs of  NLSPs nearly degenerate with the sneutrino LSP helps to increase the effective annihilation cross section, so that the  relic density of the sneutrino is in agreement with WMAP. 
The NLSP can be either a neutralino or a slepton. For the bulk of the points, however, the minimal value of the mixing increases with the sneutrino mass. 
 
The predictions for the SI cross section  are within one order of magnitude of the XENON and CDMS bounds except when $m_{\tilde\nu_{1\tau}}\simeq m_{h^0}/2$ and for the scattered point where coannihilation dominates,  see the right panel in Fig.~\ref{sn2012-fig:hnd-2d}. Indeed,  when the annihilation in the early Universe is enhanced by a resonance effect, the coupling of the LSP to the Higgs has to be small, hence  one needs a small mixing angle. This also means that the sneutrino coupling to the $Z$ is small, leading to  a small SI cross section. 

The precise relation between the LSP mass and the Higgs mass has important consequences when we consider annihilation channels in the galaxy.
In some cases, such annihilations can be strongly enhanced with respect to their values in the early Universe. This Breit-Wigner enhancement can occur when the annihilation proceeds through a $s$-channel exchange of a  Higgs particle near resonance, the cross section is then sensitive to the thermal kinetic energy: at small velocities, one gets the full resonance enhancement while at $v\approx c$, one only catches the tail of the resonance~\cite{Feldman:2008xs,Ibe:2008ye,Bi:2009uj,AlbornozVasquez:2011js}. 
This occurs when $1-m_{h^0}^2/4m^2_{\tilde\nu_{1\tau}}\ll 1$, thus when the annihilation is primarily into $b\bar{b}$. In the left panel in Fig.~\ref{sn2012-fig:hnd-2d:ann}, a small region at 95\% BC has a photon flux above the limit imposed by {\it Fermi}-LAT. Away from this special kinematical configuration, the annihilation cross section into $b\bar{b}$ is usually two orders of magnitude below the present limit. The dominant annihilation channel is rather into $W$-boson pairs.  Even for this channel, the predictions are  at least one order of magnitude below the {\it Fermi}-LAT limit except when $m_{\tilde\nu_{1\tau}}\approx 100$~GeV, where the predictions are only a factor 2--3 below the limit. The annihilation into neutrino pairs is always subdominant for heavy sneutrinos, with $\sigma v_{\nu\nu} + \sigma v_{\bar{\nu}\bar{\nu}} < 10^{-30}~\rm{cm}^3 /{\rm s}$.

Note that even after removing the points that are excluded by {\it Fermi}-LAT in the $b\bar{b}$ channel, the predictions for $\sigma_{\rm Xe}$ extend to small values. Indeed for these points  there is no large enhancement of the annihilation rate in the early Universe, hence no need to have small  couplings of the LSP to the Higgs. Therefore the predictions for the SI cross section covers a wide range and is not correlated with $\sigma v_{b\bar{b}}$, see the bottom right plot in Fig.~\ref{sn2012-fig:hnd-2d:ann}.

We have also computed the predictions for the antiproton flux for the heavy sneutrino case. The largest fluxes are expected for DM masses around 100~GeV where the annihilation cross section can reach $3\times 10^{-26}{\rm cm^3}/{\rm s}$.  We found that with the MED propagation parameters the flux is barely above the background and always within the 1$\sigma$ experimental error bars. Note that a large flux is also expected  for the few points that have a large annihilation into $b\bar{b}$, these points are however already excluded by {\it Fermi}-LAT as discussed above.
 
The results discussed above for the HND case also hold for the HD sneutrino case. 
In fact, most of the distributions in the HD case are practically the same as in the HND case. 
The only differences are observed for the LSP mass, and for the associated $A_{\tilde\nu}$, see Fig.~\ref{sn2012-fig:heavy-1d-demo}. 
We note a slightly higher probability of 6\% to be on the $h^0$ pole. Correspondingly, also small ${A_{\tilde\nu}}$ and small mixing angles have somewhat higher probability than in the HND case. Regarding the flavor of the LSP, we find that a $\tau$-sneutrino LSP has 55\% probability and is thus, as expected, somewhat preferred over $e/\mu$ sneutrino co-LSPs (45\% probability), see the right-most panel in Fig.~\ref{sn2012-fig:heavy-1d-demo}. 
The fact that the $\tilde\nu_{1e}$--$\tilde\nu_{1\tau}$ mass difference peaks within $\pm 10$~GeV is however just a consequence of our prior assumption for the HD case. 

\begin{figure}[ht] 
   \centering
   \includegraphics[width=4.96cm]{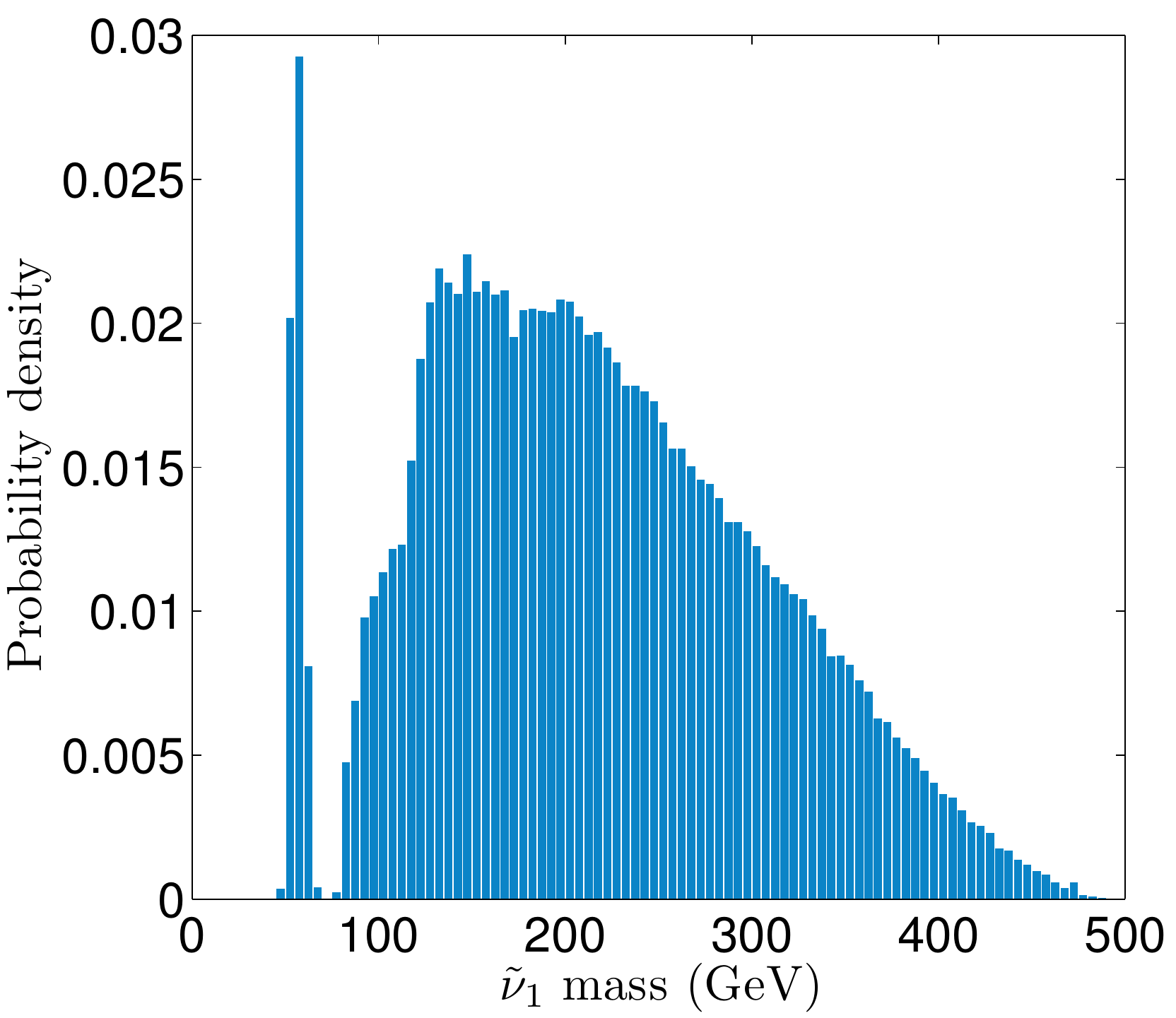}
   \includegraphics[width=4.96cm]{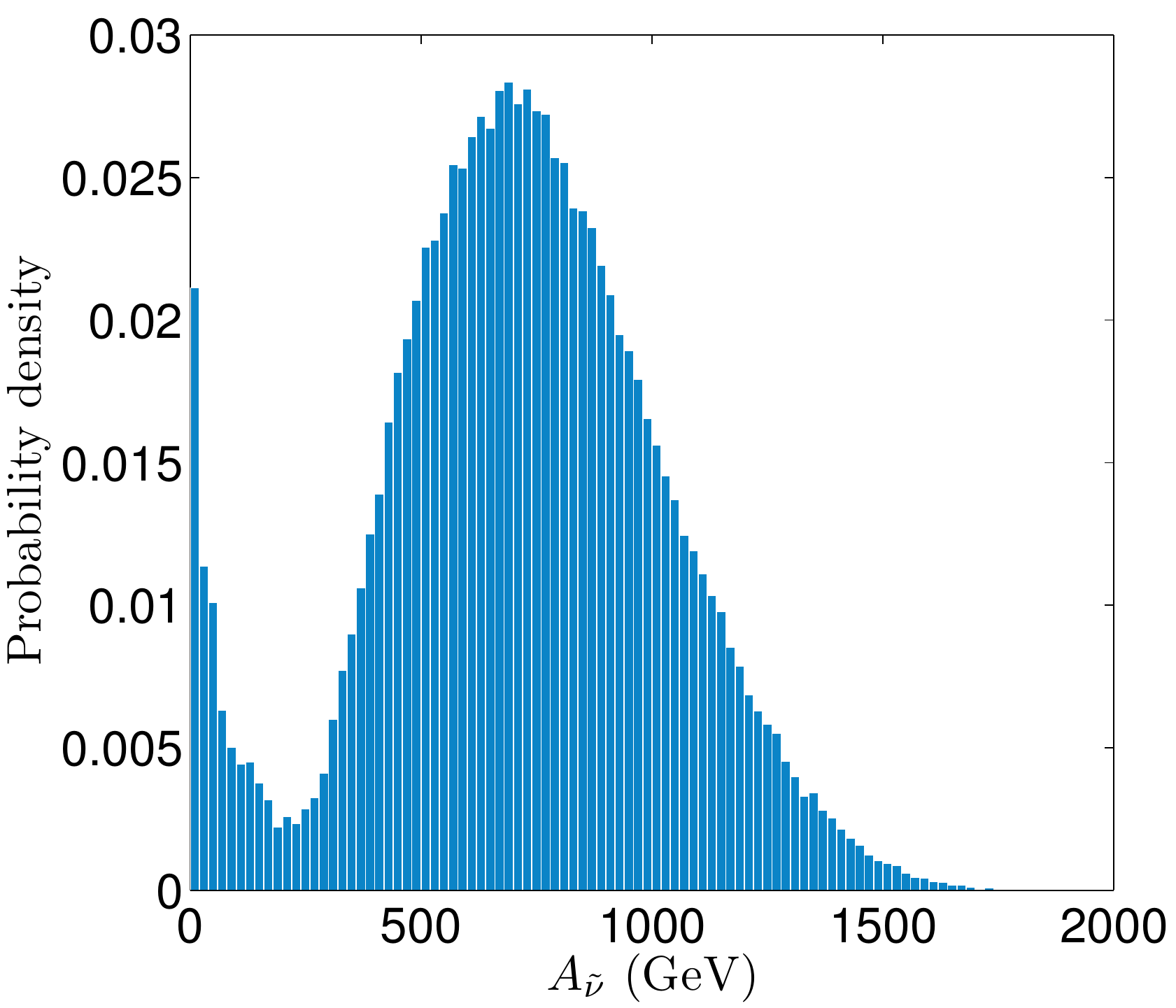}
   \includegraphics[width=4.96cm]{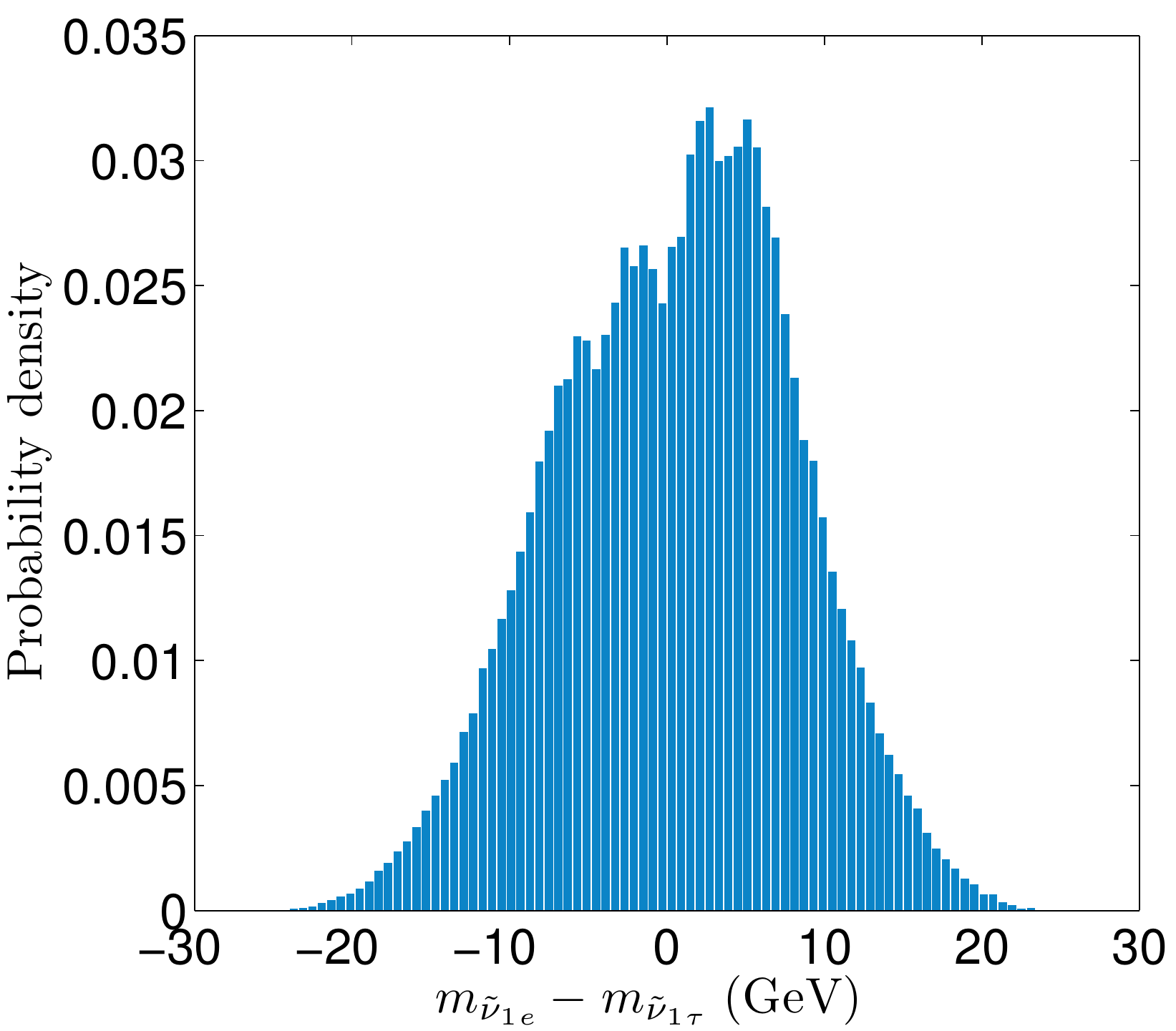}
   \caption{Posterior PDFs in 1D for the heavy democratic (HD) sneutrino case. All other distributions are practically the same as in the HND case.}
   \label{sn2012-fig:heavy-1d-demo}
\end{figure}

\subsection{Status of June 2014} \label{sn2012-sec:update}

After the completion of Ref.~\cite{Dumont:2012ee}, the ATLAS and CMS collaborations announced the discovery of an SM-like Higgs boson with mass around 125~GeV
(see Section~\ref{sec:higgs-measlhc}). This is clearly in contradiction with the predicted properties of the Higgs boson in the light sneutrino scenario considered in this work. Indeed, as can be seen on the bottom-right histogram of Fig.~\ref{sn2012-fig:light-1d}, invisible decays of the $h^0$ (mostly driven by $h^0 \to {\tilde{\nu}_1}{\tilde{\nu}_1}^*$) always dominate the width, which results in other decays being too much suppressed to account for the observed signal at the LHC.
This rules out the entire parameter space of the light case, while heavier sneutrinos (in the HND and HD cases) remain largely unaffected by the discovery of the Higgs boson.

In addition to the Higgs discovery, the final XENON100 results~\cite{Aprile:2012nq} from 225 live-days of data and the first results from the LUX experiment with 85.3 live-days of data~\cite{Akerib:2013tjd} became available. This represents an order of magnitude improvement in the limit on the spin-independent cross section compared to the 2011 XENON100 limit~\cite{Aprile:2011hi}, and is expected to have a sizable impact on sneutrino DM.
Moreover, more stringent bounds on the production of SUSY particles were derived by the ATLAS and CMS collaborations using the data collected at $\sqrt{s} = 8$~TeV. The variety of the SUSY searches performed at the LHC calls for a careful implementation of the obtained limits, beyond a simple lower bound on the gluino mass. Finally, improved determinations of $b \to s\gamma$~\cite{HFAG2013}  and $B_s \to \mu\mu$~\cite{CMSandLHCbCollaborations:2013pla} also appeared.

This motivated us updating the results for the HND and HD cases using the experimental results available in June~2014. The latest direct detection results from XENON100 and LUX were implemented using an updated version of the private code already mentioned in Section~\ref{sn2012-sec:analysis}, based on Refs.~\cite{Bozorgnia:2013hsa,Bozorgnia:2013pua,Bozorgnia:2014dqa}. It allowed us to define a likelihood that goes beyond the upper limit at 90\%~CL provided by the experimental collaboration, and to take into account variations of the DM halo parameters. Constraints from other DM experiments, as described in Section~\ref{sn2012-sec:analysis}, are also considered but found to be irrelevant in comparison with the latest results from LUX.

The negative results in the search for supersymmetric particles at the LHC were taken into account using {\tt SmodelS}, presented in Section~\ref{sec:simpmod-intro} (see also Ref.~\cite{Kraml:2013mwa}). After completion of the scans by means of MCMC, we generated {\tt SLHA} files containing cross section and decay tables for each point and passed them through {\tt SmodelS}.\footnote{We thank Ursula Laa for running {\tt SmodelS} on the {\tt SLHA} files.} The various combinations of production and decay were tested against the simplified model results from the more than fifty ATLAS and CMS SUSY analyses present in the {\tt SmodelS} database.
The information on the simplified model topology and the analysis yielding the strongest constraint was kept, along with the corresponding $r$-value, where $r = (\sigma \times \br) / \sigma^{95}_{\rm UL}$ and $\sigma^{95}_{\rm UL}$ is the upper limit at 95\%~CL on the cross section.

In addition to the constraints present in Table~\ref{sn2012-tab:const} and to the new experimental results mentioned above, we also consider an upper bound on invisible decays of the Higgs boson of  $\br(h^0 \to {\rm invisible}) < 20\%$, based on the upper limit at 95\%~CL obtained in Section~\ref{2013c-sec:coupfit} for an SM-like Higgs. Indeed, by definition the heavy case corresponds to $m_{\tilde\nu_1} > m_Z/2$, and significant decay of the $h^0$ into a pair of sneutrinos may be found for $m_{\tilde\nu_1} = 50 - 60$~GeV. Nonetheless, this requirement did not result in any significant modification of the probability distributions.
Finally, the new results on $b \to s\gamma$~\cite{HFAG2013} and $B_s \to \mu\mu$ are integrated to the likelihood as in Section~\ref{sec:pmssm} (see measurement 1b and 2b in Table~\ref{pmssm-tab:preHiggs}).

The new results on the properties of the DM candidate are shown in Fig.~\ref{sn2014-fig:update1} for the non-democratic case. This is in strong contrast with the 2012 results shown in Fig.~\ref{sn2012-fig:hnd-2d}. The formerly main region, with $\sin \theta_{\tilde\nu_\tau}$ in the $0.01 - 0.05$ range and $\sigma_{\rm Xe}$ in the $10^{-45} - 10^{-44}$~cm$^2$ range, shrunk because it is in tension with the new bounds from direct detection experiments. Consequently, other scenarios where the correct relic density of dark matter is achieved through co-annihilation and/or resonances become more likely, which explains why the 95\%~BCR extends down to vanishing values for the tau sneutrino mixing angle.
The corresponding results in the democratic case are shown in Fig.~\ref{sn2014-fig:update2}. The conclusion is the same, and in this case the necessity of new processes for achieving the observed relic density is more severe because of the decrease of the net annihilation of DM from the co-annihilation between different sneutrino flavors, see Ref.~\cite{Belanger:2010cd}.

Let us turn to the impact of the LHC SUSY searches on these sneutrino scenarios. First, in the HND case the probability of being excluded as given by {\tt SModelS} ({\it i.e.}, having at least one $r$-value $> 1$): is very small: 0.87\% in total, among which 78\% are excluded from the di-lepton $+ E_T^{\rm miss}$ topology corresponding to direct slepton-pair production ($\tilde e$ and $\tilde\mu$) followed by the decay into a lepton and the LSP, and 22\% from gluino-pair production followed by the decay into two ($b$-)jets and the LSP. Regarding the first simplified model results (where $pp \to \tilde\ell\tilde\ell$ followed by $\tilde\ell \to \ell\tilde\chi^0_1$ is targeted), the SUSY topology that is actually constrained in our sneutrino LSP scenarios is $pp \to \tilde\chi^+\tilde\chi^-$ followed by $\tilde\chi^\pm \to \ell^\pm\tilde\nu_\ell$. (The contribution from direct slepton production followed by $\tilde\ell^\pm \to \ell^\pm\tilde\chi^0_1 \to \ell^\pm\nu\tilde\nu$ is small as the production cross section is smaller.) Both processes yield the same final state and are not distinguished internally by {\tt SModelS}. The results for the second simplified topology, where a gluino is produced, put marginal constraints on light gluinos with masses below 700~GeV. This is a weak constraint because of the different decay possibilities of the gluino, where moreover the neutralino appearing in the cascade decay does not necessarily decay into invisible particles ({\it i.e.}, into $\nu\tilde\nu$).

The very small probability of being excluded by any LHC SUSY search in the HND case is easily understood as $pp \to \tilde\chi^+\tilde\chi^-$ followed by $\tilde\chi^\pm \to \ell^\pm\nu_\ell$ is only possible if an electron or muon sneutrino is rather light. This is not required for non-democratic sneutrinos, and has a small probability of being realized. On the other hand, in the HD scenario all flavors of sneutrinos are light. Therefore, the probability of being excluded as given by {\tt SmodelS} is much larger: 5.7\%. The gluino-induced topologies play a much smaller role as the probability of being excluded by such topologies is of 0.06\%.

The exclusion can be represented in the $(m_{\tilde\nu_1}, m_{\tilde\chi^+_1})$ plane. This is shown in Fig.~\ref{sn2014-fig:update3}, where excluded points are shown in gray on top of the 68\% and 95\%~BCR obtained without any requirement on the {\tt SmodelS} results. In both cases, the excluded points do not cover any region completely. In particular, while wino-like charginos are excluded, higgsino-like charginos typically evade the exclusion given their lower cross section.

Finally, a word of caution is in order. In {\tt SmodelS}, it is internally assumed that $\tilde\ell \to \ell \tilde\chi^0_1$ is equivalent to $\tilde\chi^{\pm}_1 \to \ell\tilde\nu_{\ell}$ from the kinematics point of view. This is an approximation since differences in the spin and helicity of particles between the two topologies should modify kinematic distributions. Then, the impact of the change in kinematics on the acceptance$\times$efficiency (hence on the exclusion) clearly depends on the cuts considered in the analyses of the ATLAS and CMS collaboration. For a given analysis, the validity of this approximation can be tested from the reimplementation of the analysis cuts (see Section~\ref{sec:analysisreimplementation}). This is work in progress based on the reimplementation of the ATLAS search for electroweak-inos and sleptons in the $2\ell + E^{\rm miss}_T$ final state at $\sqrt{s} = 8$~TeV~\cite{Aad:2014vma}, presented in Section~\ref{sec:atlasvalid}. A certain impact on the acceptance$\times$efficiency is expected as for this analysis the ATLAS limits (in terms of upper bounds on the cross section) for $pp \to \tilde\ell\tilde\ell$ are much stronger in the case of left-handed sleptons.

\begin{figure}[!ht] 
   \centering
   \includegraphics[width=7cm]{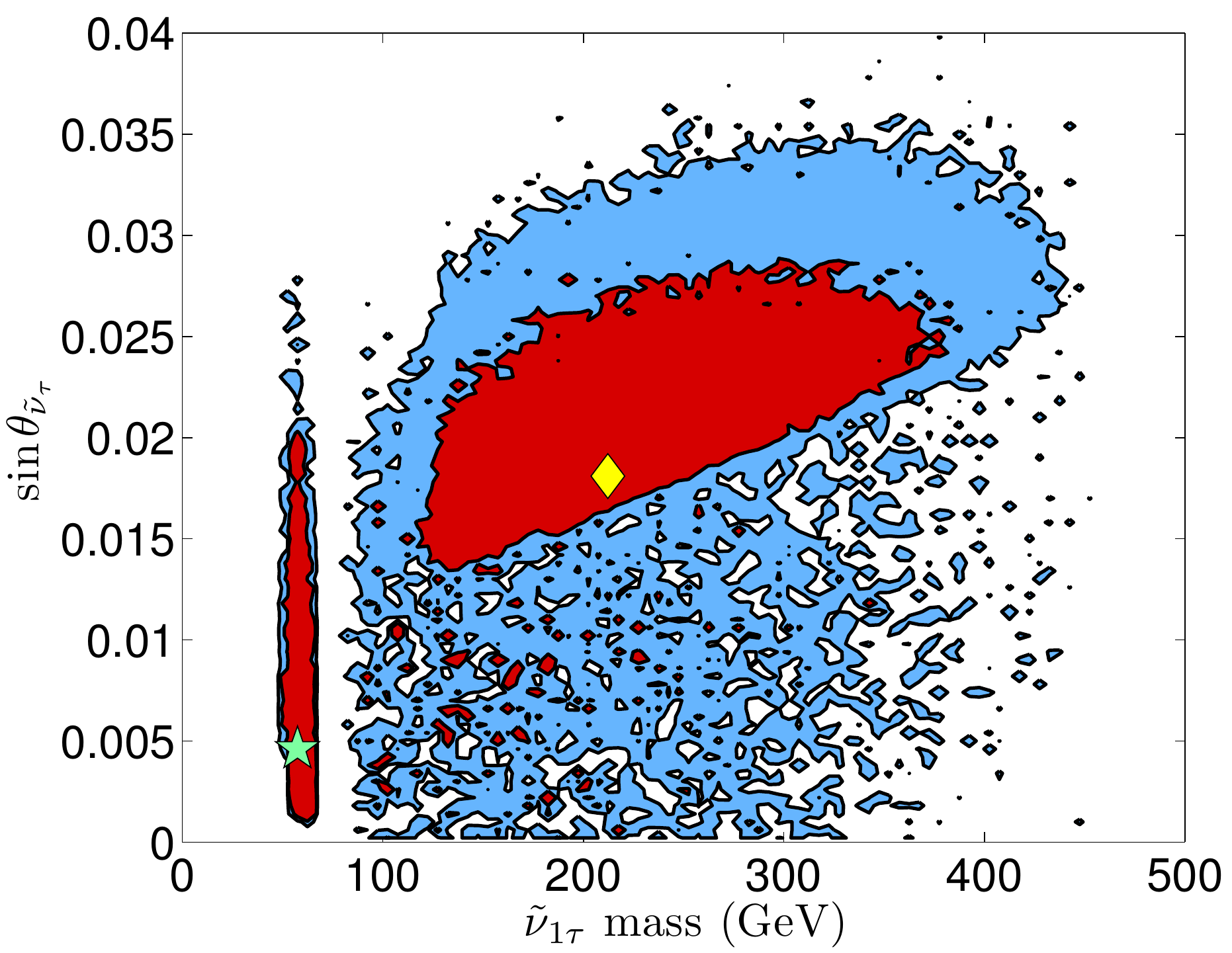} \quad
   \includegraphics[width=7cm]{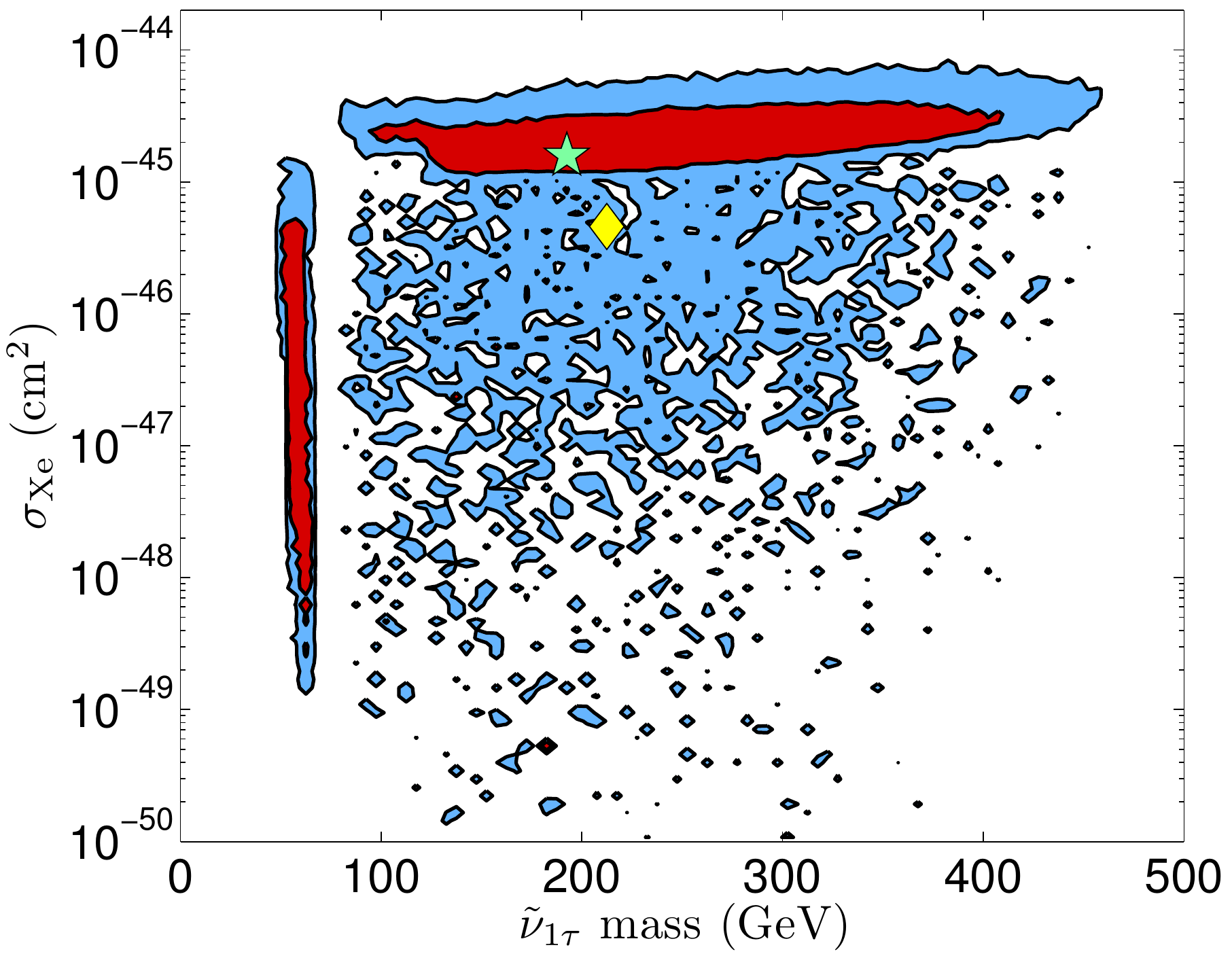}
   \caption{Posterior PDFs in 2D of $\sin\theta_{\tilde{\nu}_{\tau}}$ (left) and $\sigma_{\rm Xe}$ (right) versus $m_{\tilde\nu_{1\tau}}$ for the HND case after the 2014 update. The red and blue areas are the  68\% and 95\% BCRs, respectively. The green stars mark the highest posterior, while the yellow diamonds mark the  mean of the  PDF.}
   \label{sn2014-fig:update1}
\end{figure}

\begin{figure}[!ht] 
   \centering
   \includegraphics[width=7cm]{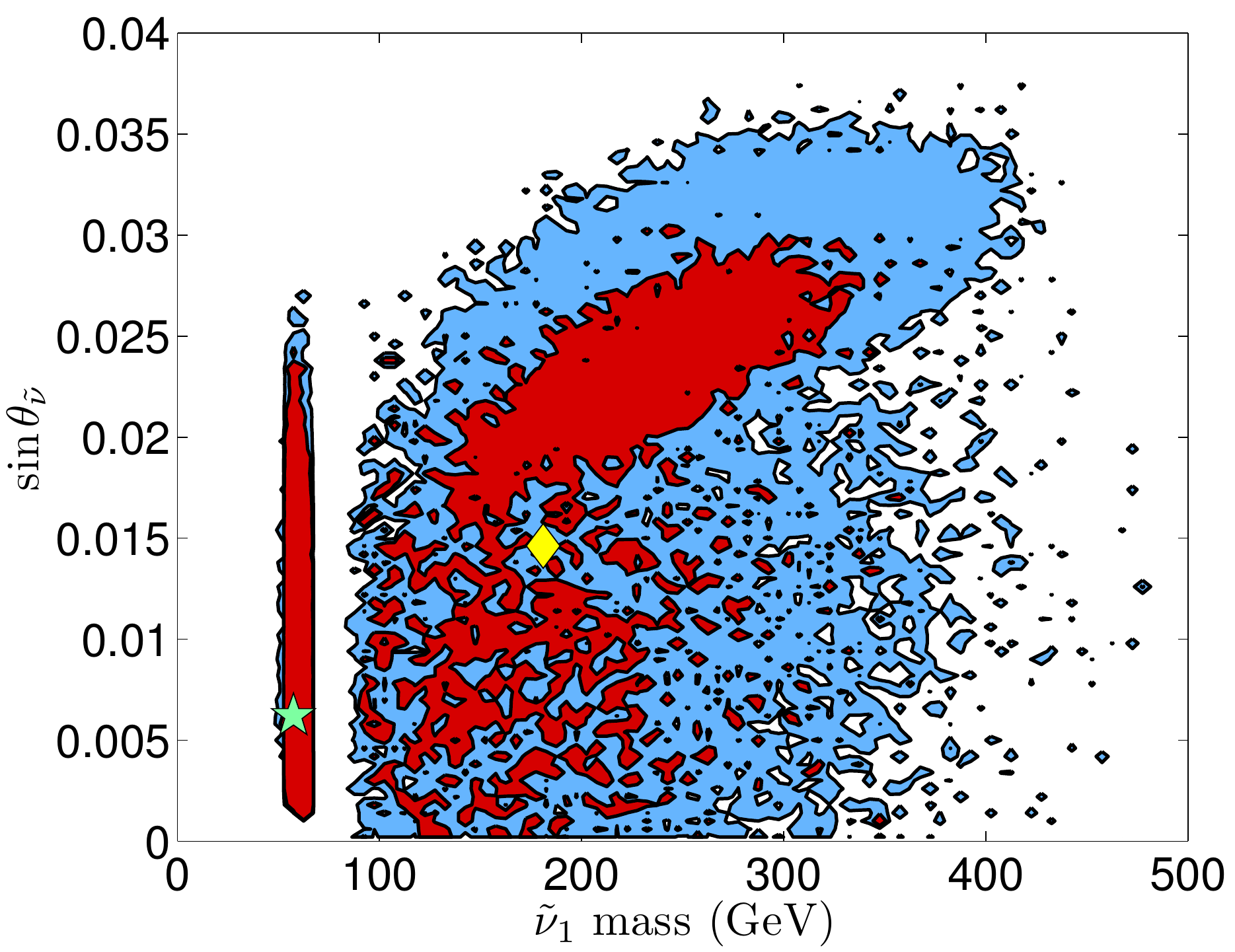} \quad
   \includegraphics[width=7cm]{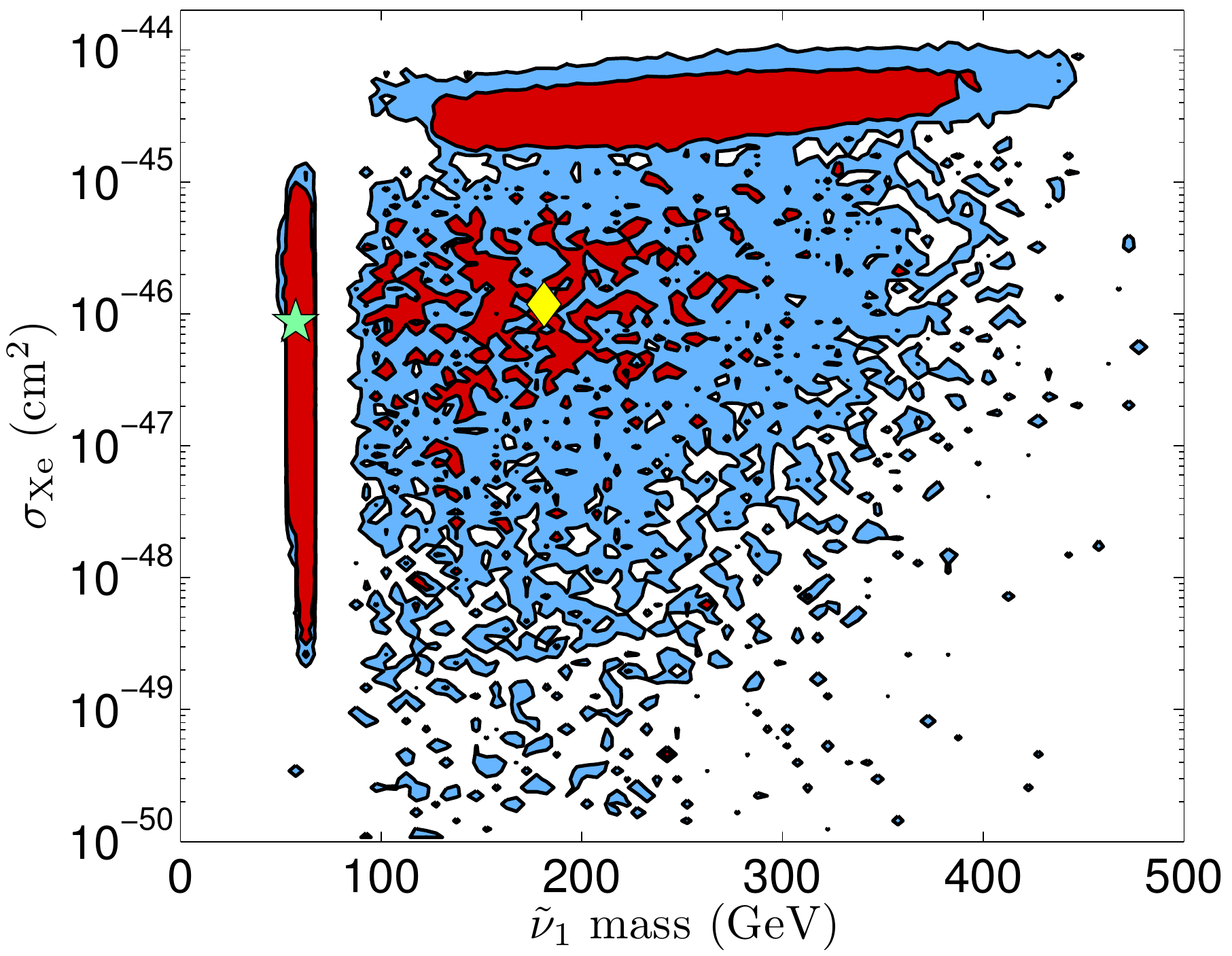}
   \caption{As in Fig.~\ref{sn2014-fig:update1}, for the HD case after the 2014 update.}
   \label{sn2014-fig:update2}
\end{figure}

\begin{figure}[!ht] 
   \centering
   \includegraphics[width=7cm]{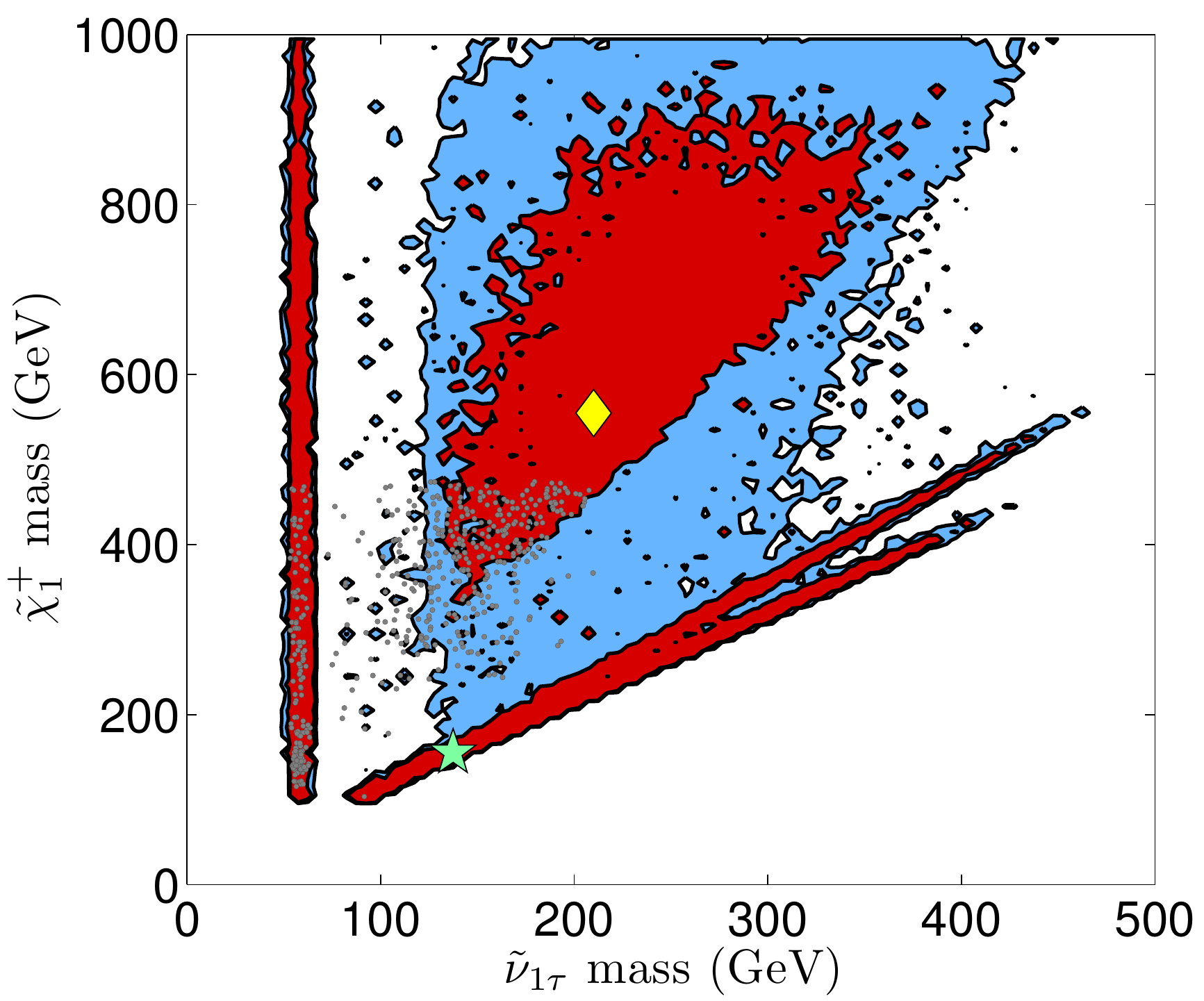} \quad
   \includegraphics[width=7cm]{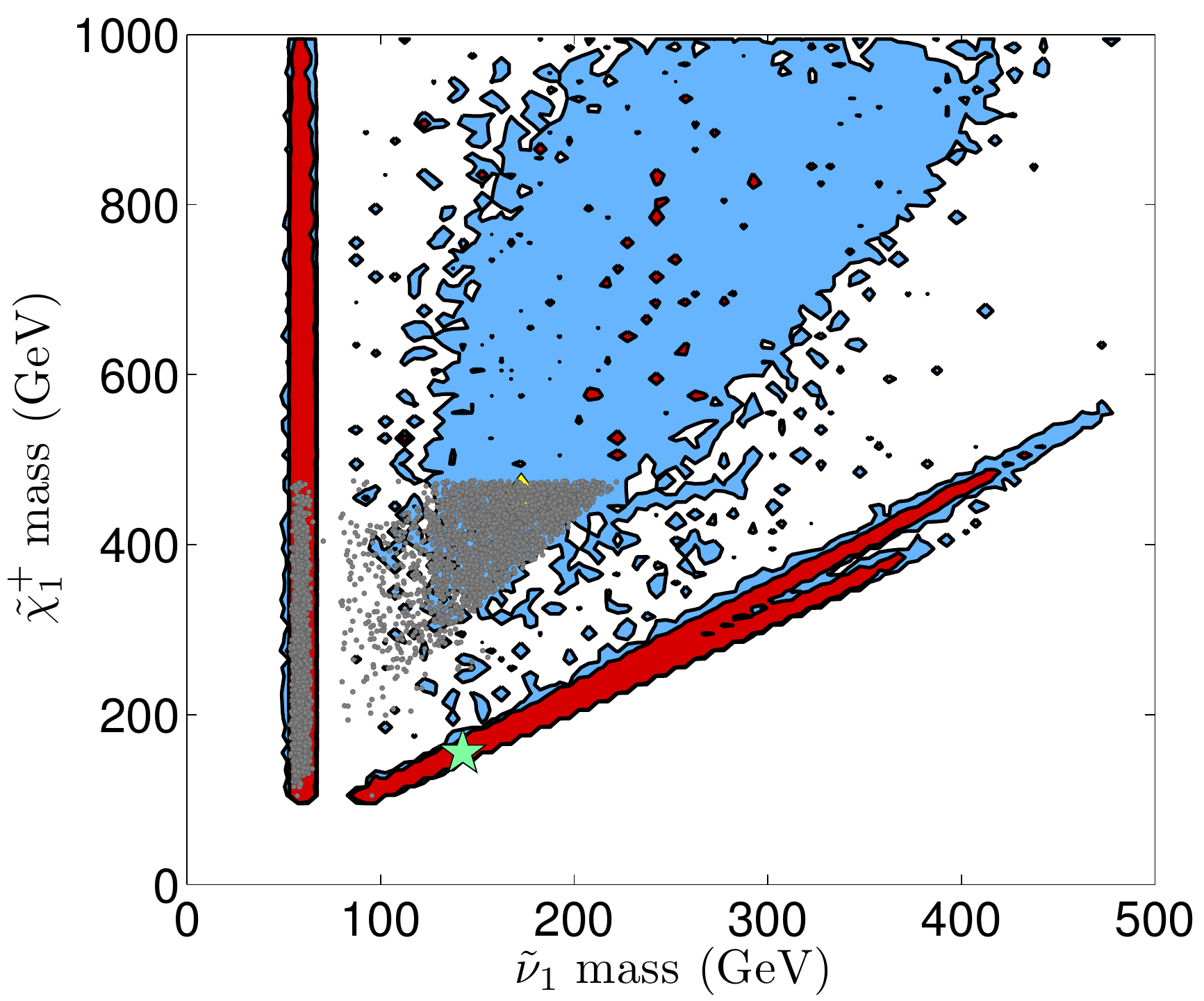}
   \caption{Posterior PDFs in 2D of $m_{\tilde\chi^+_1}$ versus $m_{\tilde\nu_{1}}$ for the HND (left) and the HD (right) case after the 2014 update, but without including results on the LHC SUSY searches. The red and blue areas are the  68\% and 95\% BCRs, respectively. The green stars mark the highest posterior, while the yellow diamonds mark the  mean of the  PDF.
   Points which are excluded by {\tt SmodelS} are shown in gray on top of the 68\% and 95\%~BCR.}
   \label{sn2014-fig:update3}
\end{figure}

\subsection{Conclusions}\label{sn2012-sec:conclusions}

We performed a global MCMC analysis of a sneutrino DM model with Dirac neutrino masses originating from supersymmetry breaking. The main feature of this model is a mainly RH mixed sneutrino as the LSP, which has a large coupling to the Higgs fields through a weak-scale trilinear $A$ term which is not suppressed by small Dirac-neutrino Yukawa couplings.  
We demonstrated that such a RH sneutrino can be consistent with all existing constraints for masses of about 50--500~GeV (the upper limit coming from the fact that we consider gluino masses only up to 3~TeV), while lighter sneutrinos are excluded by the discovery of an SM-like Higgs boson. 

Direct detection limits, in particular from LUX, significantly narrow down the possibility of viable sneutrino dark matter.
In imposing these constraints, we took special care to account for uncertainties arising from astrophysical parameters, like $v_0$,  $v_{\rm esc}$ and the local DM density $\rho_{\rm DM}$. Moreover, we accounted for  uncertainties from the quark contents of the nucleon, relevant for the Higgs exchange contribution to the direct detection cross section. 

The sneutrino scenarios offer distinctive LHC SUSY phenomenology. In particular, the dominant decay of charginos (with a branching fraction larger than 0.5) is into a charged lepton and the LSP 
with roughly 50\% probability.  The charged lepton is typically a $\tau$  for heavy non-democratic scenarios or a $e$/$\mu$ for the heavy democratic scenario. Also, neutralinos (typically $\tilde{\chi}^0_1$ and $\tilde{\chi}^0_2$) appearing in squark and gluino cascades can decay invisibly into the LSP. Indeed the probability for a 90\% invisible decay of the lightest (next-to-lightest) neutralino is  
close to 100\% (30--40\%) in the heavy sneutrino scenarios. This implies that there can be up to three different invisible 
sparticles in an event. 
The cascade decays of squarks, $\tilde q_R\rightarrow q \tilde{\chi}^0_1\rightarrow q \tilde\nu_1 \nu$, 
$\tilde q_L\rightarrow q \tilde{\chi}^0_2\rightarrow q \tilde\nu_1 \nu$, 
$\tilde q_L\rightarrow q' \tilde{\chi}^\pm_1\rightarrow q' l \tilde\nu_1$ therefore give different amount of missing energy as compared to the MSSM.
Furthermore the cascade decays of gluinos, $\tilde g\rightarrow \tilde{\chi}^0_i jj$ will also give a large contribution to the jets plus $E_T^{\rm miss}$ channel while the decay of gluino pairs via a chargino will give about the same amount of same-sign and opposite-sign lepton pairs.
Note that the alternative  dominant decay mode of the chargino  is $\tilde{\chi}^\pm_1\rightarrow W^\pm \tilde{\chi}^0_1$; in this case the mass of the invisible particle could be much larger than the DM mass.

In June~2014, an update of this analysis was performed for this thesis, taking into account the 8~TeV LHC SUSY searches in a simplified model approach via {\tt SModelS}. We found that current searches already constrain, in particular, a light chargino with mass below 500~GeV from the searches for sleptons in final states with two leptons and $E_T^{\rm miss}$, although it is highly dependent on the presence of light electron and muon sneutrinos, and on nature of the chargino. This represents a promising way for probing such scenarios. 

 
\section{Introduction to analysis reimplementation} \label{sec:analysisreimplementation}

The simplified model approach of {\tt SModelS} presented in Section~\ref{sec:simpmod-intro} and used in Sections~\ref{sec:simpmod-lightneutralino} and~\ref{sec:simpmod-sneutrino} is fast and conservative, hence well-suited for testing a model against the LHC BSM constraints in the context of large scans of a parameter space. However, in situations where the signal is shared among different topologies we expect significantly stronger limits than the ones obtained in this approach, where the various simplified model topologies are tested independently. Moreover, for many relevant topologies the results are not given in a complete form in the experimental publications (especially when the number of free parameters of the simplified model is greater or equal to 3).
One way to go beyond these limitations is to use acceptance$\times$efficiency maps, as done is {\tt FastLim} and also discussed in Section~\ref{sec:simpmod-intro}.
It makes it possible to combine the contributions from the various topologies before comparing to the experimental results. Of course, it relies on the availability of pre-calculated acceptance$\times$efficiency maps for all relevant topologies in each of the signal regions of interest. Since this information is only rarely provided in the experimental publications (see Fig.~\ref{acceptance-efficiency-dileptonmet} for a counterexample), {\tt FastLim} is using maps generated with an internal program called {\tt ATOM}, where the cuts defining the signal regions of the various analyses are implemented and applied to samples of signal events corresponding to a given simplified model, after fast simulation of the detector.

Interestingly, this approach is not limited to producing acceptance$\times$efficiency maps for simplified models. The analyses cuts can be applied directly on a MC sample of the model of interest (not corresponding to a single simplified model topology) and obtain a value of $A \times \varepsilon$. With in addition the information on the luminosity and on the cross section, it is possible to determine if the scenario of interest is excluded or not in light of the LHC results. This approach is more straightforward than the one of {\tt FastLim} since the identification of the relevant topologies is not needed. It is also much more powerful because all topologies, including the more complicated ones for which acceptance$\times$efficiency maps cannot reasonably be produced, are taken into account as part of the signal. In turn, it is dramatically slower and disk space-intensive than the simplified model approach. Indeed, the determination of $A \times \varepsilon$ requires to generate large event files (including showering and hadronization of the final-state partons), and to apply detector simulation and the analysis cuts on these files for every tested scenario.

So independent implementations of the analyses cuts are needed, both for a simplified model approach \`a la {\tt FastLim} and for testing directly scenarios of interest in a general way, {\it i.e.}\ without the decomposition into simplified model topologies.
This activity became part of my thesis shortly after the 2013 ``Physics at TeV Colliders'' Les Houches workshop to which I participated, in the context of the constraints on natural SUSY from the stop and sbottom search results at the LHC (see Contribution~14 of~\cite{Brooijmans:2014eja}). This project, involving theorists as well as experimentalists, consists in reimplementing all ATLAS and CMS SUSY searches relevant for natural SUSY. The derived constraints are then applied on a scan of a subset of the pMSSM with light stops and sbottoms. The complete status of natural SUSY in the context of the MSSM can then be drawn, in particular taking into account the various mixing possibilities for stops and sbottoms.
The implementation of the analyses cuts is done in the \madanalysis\ framework~\cite{Conte:2012fm,Conte:2013mea,Conte:2014zja}.
This project motivated the development of new features in \madanalysis, to which I significantly contributed and that will be presented in Section~\ref{sec:ma5delphes3}.
To date the natural SUSY project is still underway as it requires a substantial number of analyses to be implemented, which is a long and tedious task. 

Before proceeding with more details on \madanalysis, some comments on the general procedure for reimplementing LHC analyses are in order.
First of all, only searches where the signal regions are defined using a set of selection criteria (or cuts) described in an experimental publication can be reproduced. This represents the vast majority of the BSM searches at the LHC, but not the searches based on multivariate analyses (MVA) techniques  which generically cannot be externally reproduced and should be ignored at this stage. This is usually unproblematic as MVA-based searches are often presented along with a cut-based version of the search that is reproducible.
More importantly, simulation of the detector response is necessary to accurately reproduce the results from ATLAS and CMS. However, the official full and fast simulation softwares of the ATLAS or CMS detector are not public. It is therefore necessary to use other, public tools, such as {\tt PGS}~\cite{pgs} or {\tt Delphes}~\cite{deFavereau:2013fsa}. {\tt PGS} is the ``historical'' fast simulation software and has stopped developing several years ago. The development of {\tt Delphes}, on the other hand, is still active. The current version, {\tt Delphes~3}, provides a detailed description of the ATLAS and CMS detector and advanced features such as particle-flow reconstruction and the simulation of pile-up. Nonetheless, significant discrepancies with respect to the full simulation of the detector may be observed depending on the analyses requirements; hence agreement with the official results needs to be carefully checked case by case.

Besides the detector simulation itself, some other technical difficulties may arise, making it challenging to reach a good agreement with the ATLAS or CMS original analysis. Two important examples are as follows. First, the definition of the triggers is crucial. Trigger efficiencies can go much below 100\% even for events passing the various offline selections, depending on the analysis. The information on the trigger efficiencies as function of the properties of the relevant objects (usually the transverse momentum $p_T$ and the pseudorapidity $\eta$) is therefore crucial, but not always given in the experimental publications. Second, the definition of the ``candidate'' or ``signal'' reconstructed objects used in a given analysis may also be a source of uncertainty. From the information on the vertices and tracks in the inner detector, and from the energy deposits in the calorimeters and muon chambers, objects are reconstructed with efficiencies depending on the quality and isolation requirements. This include, for instance, the ``medium'' and ``tight'' quality of electrons defined in ATLAS~\cite{ATLAS-CONF-2014-032}, and the tagging of jets as originating from the fragmentation of a $b$-quark ($b$-jets). The experimental procedure can be difficult to reproduce, depending on the amount of information provided by the experimental collaboration. 
Sometimes, the information on the triggers and object definitions missing in the paper describing the analysis can be found in ``performance notes'' or on the ATLAS or CMS TWiki pages. If this is not the case, one has to contact the relevant conveners from the ATLAS or CMS collaboration.

After the implementation of an analysis based on the various cuts described in the experimental publication,
validation is a necessary step before deriving new interpretations of the results. The agreement of the reimplementation with the official analysis can usually be checked in a number of ways: while the first LHC SUSY publications based on the data at $\sqrt{s} = 7$~TeV typically provided very few information, making validation very challenging, the situation improved a lot recently thanks to the active communication between theorists and experimentalists (see also~\cite{Kraml:2012sg}). Of particular importance is the presence of cut flows, that give the number of events passing the cuts after each step of the analysis for some benchmark scenarios. An example is given in the left panel of Fig.~\ref{fig:cutflowhisto}, for the the signal region $Z$jets of the ATLAS SUSY search for $\ell^+\ell^- + E_T^{\rm miss}$ at $\sqrt{s}=8$~TeV already mentioned at the beginning of this chapter.
The benchmark points, denoted as S1 and S2, correspond to the $\tilde\chi^{\pm}_1 \tilde\chi^0_2 \to W^\pm \tilde\chi^0_1 Z^0 \tilde\chi^0_1$ simplified model (third topology of Fig.~\ref{simpmod2leptondiagrams}), with $(m_{\tilde\chi^{\pm}_1, \tilde\chi^0_2} ,m_{\tilde\chi^0_1}) = (250,0)$~GeV and $(350,50)$~GeV, respectively, wino-like $\tilde\chi^{\pm}_1$ and $\tilde\chi^0_2$, and bino-like $\tilde\chi^0_1$. The comparison of such a cut flow with the numbers obtained with the reimplemented version of the analysis (for the exact same benchmark points) is a powerful test as it shows explicitly the origin of possible discrepancies. The distribution of kinematic quantities for given benchmark points is another validation material of interest. An example is shown in the right panel of Fig.~\ref{fig:cutflowhisto}, for the CMS SUSY search for stops in the single lepton channel at $\sqrt{s}=8$~TeV~\cite{Chatrchyan:2013xna}. The variable shown, $p_T(b_1)$, corresponds to the transverse momentum of the leading-$p_T$ $b$-tagged jet after the preselection requirements. The magenta and gray lines correspond to the expectation for the signal of the $\tilde t_1 \tilde t_1 \to b \tilde\chi^\pm_1 b \tilde\chi^\pm_1 \to b W^\pm \tilde\chi^0_1 b W^\pm \tilde\chi^0_1$ and $\tilde t_1 \tilde t_1 \to t \tilde\chi^0_1 t \tilde\chi^0_1$ simplified models, respectively, with masses indicated on the caption of the figure. The shape of such distributions makes it possible to check the correct implementation of that variable, which is particularly relevant in the case of more complex discriminants. Possible discrepancies coming from the fast simulation of the detector (for instance, the treatment of the jet energy scale) can also be checked.

\begin{figure}[ht]
\begin{center}
\raisebox{0.050\textwidth}{\includegraphics[width=0.58\textwidth]{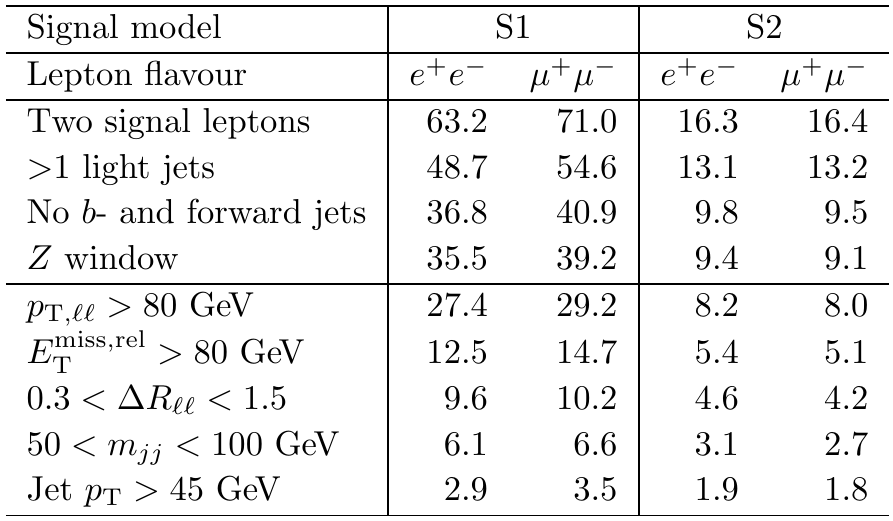}} \hspace{0.5pt}
\includegraphics[width=0.40\textwidth]{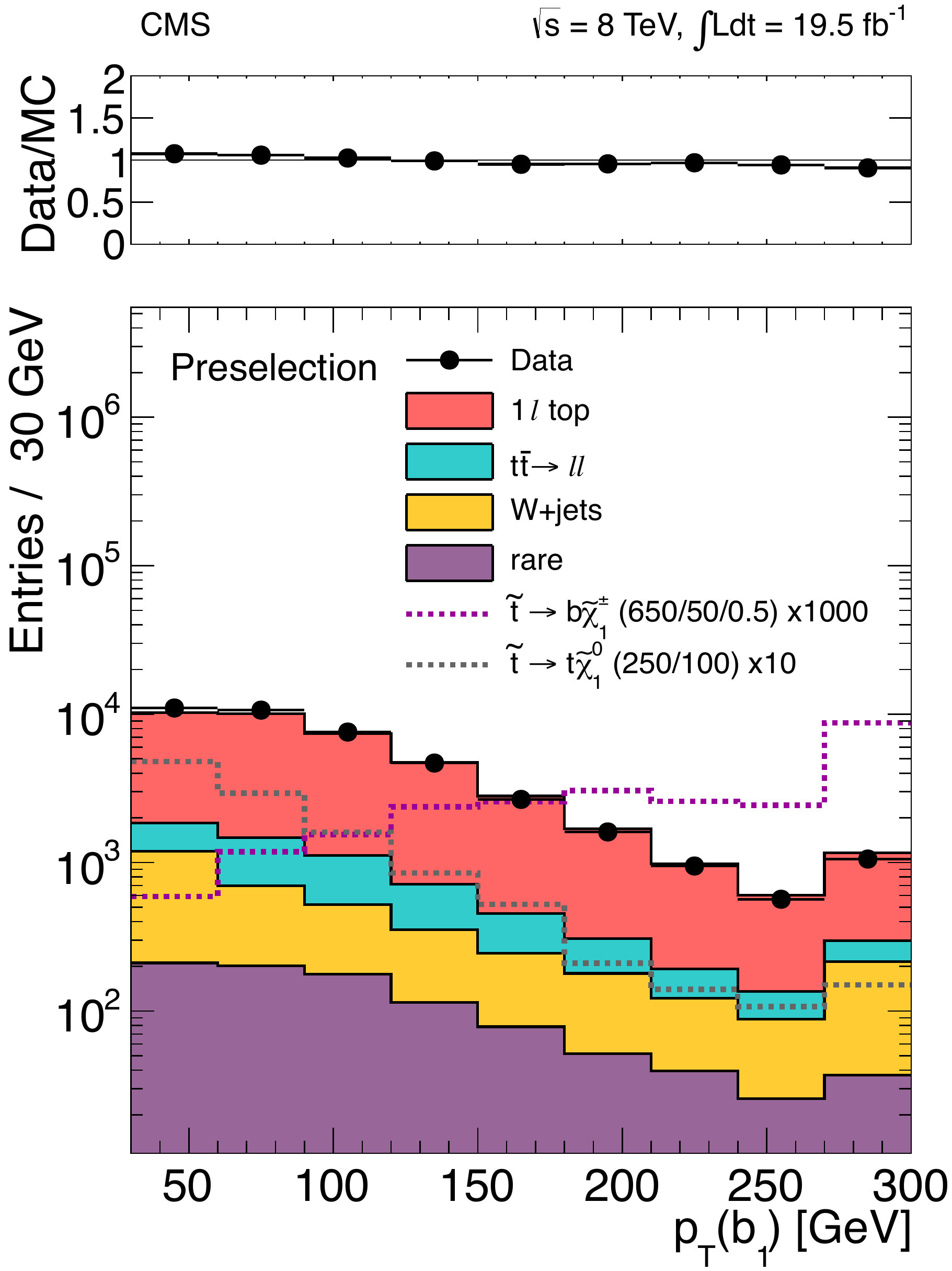}
\end{center}
\caption{Left: cut flows for benchmark points S1 and S2 in the signal region $Z$jets of the ATLAS SUSY search for $\ell^+\ell^- + E_T^{\rm miss}$ at $\sqrt{s}=8$~TeV~\cite{Aad:2014vma}. Right: distribution of the transverse momentum of the leading-$p_T$ $b$-tagged jet after preselection of the CMS SUSY search for stops in the single lepton channel at $\sqrt{s}=8$~TeV~\cite{Chatrchyan:2013xna}. For the $\tilde t_1 \to b\tilde\chi^\pm_1$ simplified model (magenta line), the masses are assumed to be $(m_{\tilde t_1}, m_{\tilde \chi^\pm_1}, m_{\tilde \chi^0_1}) = (650, 350, 50)$~GeV. For the $\tilde t_1 \to t\tilde\chi^0_1$ simplified model (gray line), the relevant masses are $(m_{\tilde t_1}, m_{\tilde \chi^0_1}) = (250, 100)$~GeV (since $m_{\tilde t_1} < m_{\tilde \chi^0_1} + m_t$ in this case, the simplified model corresponds to $\tilde t_1 \to bW \tilde \chi^0_1$).}
\label{fig:cutflowhisto}
\end{figure}

Other relevant validation materials include the number of expected signal events after all cuts for some benchmark points, sometimes given along with the results on the number of observed events and the expected number of background events in the various signal regions in the experimental publications. At a higher level, the acceptance$\times$efficiency maps in given signal regions, if provided by the experimental collaborations for a simplified model scenario as in Fig.~\ref{acceptance-efficiency-dileptonmet}, can be used to check that the reimplementation of a given analysis is a good approximation in the different kinematic regions, and in particular for small mass differences between particles that lead to softer final state objects on average (also, the treatment of the initial state radiation (ISR) is usually more critical in such regions).

All these validation materials are given for specific new physics scenarios. In order to validate an analysis, it thus crucial to consider the exact same definition of the benchmark scenarios and to use the same cross sections as the experimental collaboration. In the context of supersymmetry, this is non-trivial because the production cross sections and the various kinematic distributions depend not only on the masses of the BSM particles, but also on mixing matrices. Moreover, the NLO cross sections, obtained using {\tt Prospino}~\cite{Beenakker:1996ed} for instance, are potentially sensitive to all SUSY parameters---and depend on the choice of factorization and renormalization scales, on the set of parton distribution functions used, {\it etc.} Hence, for every analysis it is good practice to provide for each benchmark scenario the relevant SLHA files and the cross sections used to derive the results. SLHA files have been provided in some (but not all) LHC SUSY analyses via {\tt HepData}, while the production cross sections are tabulated in Ref.~\cite{susyxs} for some cases, and are given directly in the experimental publication or as auxiliary material on the TWiki page of the analysis in other cases. In addition, parton-level event files in {\tt LHE} format could be provided by the experimental collaboration for benchmark points of interest and used for the validation, as was done for a CMS analysis presented in Section~\ref{sec:cmsvalid}. This is useful as it guarantees that events are generated in the exact same way.

Assuming the successful validation of a given analysis, one is left with an analysis code and tunings of the fast simulation that guarantee that $[A \times \varepsilon]_{\rm reimplement} \approx [A \times \varepsilon]_{\rm official}$ in ``reasonable'' cases, {\it i.e.}~when the tested new physics signal is not dramatically different from the one in the SUSY scenarios given as interpretation to the search results. In particular, as this approach is based on the simulation of the signal events only (and not of the SM background events), possible contaminations of the control regions used to estimate the background with signal events will not be accounted for (but can be checked by implementing the cuts for the control regions).
Barring these limitations, after validation one can estimate the number of expected signal events $n_s = \sigma \times (A \times \varepsilon) \times \mathscr{L}$ for the BSM scenario of interest in each signal region, where $\sigma$ is the signal cross section given by the MC generator or other tools, and $\mathscr{L}$ is the integrated luminosity given in the experimental publication.
Then, the simplest way to test the model against the LHC results is to compare in relevant signal regions $\sigma \times (A \times \varepsilon)$ with $\sigma_{\rm vis}^{95}$, the upper limit on the visible cross section (after cuts) at 95\%~CL. Unfortunately, the information on $\sigma_{\rm vis}^{95}$ is not always present in the experimental publications. In such case, it is really necessary to build a procedure for setting limits. For a single signal region, a rather general likelihood could be defined as follows:
\begin{align} \label{eq:likelihood-countingexp}
L_{s+b}(\sigma, n_b, A \times \varepsilon, \mathscr{L}) &= {\rm Poisson}(n_{\rm obs} | n_b + \sigma \times (A \times \varepsilon) \times \mathscr{L}) \\
&\times {\rm Gauss}(n_b | \hat n_b, \Delta n_b) \times {\rm Gauss}(\sigma | \hat \sigma, \Delta \sigma) \nonumber \\
&\times {\rm Gauss}(A \times \varepsilon | \widehat{A \times \varepsilon}, \Delta (A \times \varepsilon)) \times {\rm Gauss}(\mathscr{L} | \widehat{\mathscr{L}}, \Delta \mathscr{L}) \nonumber \,.
\end{align}
In addition to the statistical Poisson term, the imperfect knowledge of the number of background events, of the signal cross section, of the acceptance$\times$efficiency and of the luminosity is modeled with Gaussian distributions.\footnote{The choice of Gaussian distributions is highly subjective. For instance, the Poisson distribution may model better the knowledge of the background if it is directly taken from an auxiliary measurement. Also, the uncertainty on the signal cross section usually includes the variation of the factorization and renormalization scales in a given range to account for the unknown higher-order effects. This does not have any well-defined statistical meaning. Finally, one might prefer to have a probability distribution function defined in $\mathbb{R}^+$ only.}
The nominal number of background events and its 68\%~CL uncertainty, $\hat n_b$ and $\Delta n_b$, can be taken from the experimental publication, as well as the nominal integrated luminosity and its uncertainty. The uncertainty on the cross section depends on the BSM scenario that is tested, and can be estimated using appropriate tools, such as {\tt Prospino} for SUSY models. Finally, the uncertainty on the acceptance$\times$efficiency, corresponding to uncertainties related to the Monte Carlo generator and to the modeling of the detector response, is difficult to estimate from outside the collaboration.

From the likelihood defined in Eq.~\eqref{eq:likelihood-countingexp}, the test static $t({\rm data})$, a function of the experimental data, can be computed. Following the Neyman--Pearson lemma, this is usually defined as the likelihood ratio of the two hypotheses,
\beq
t({\rm data}) = \frac{L_{s+b}}{L_b} \,, \label{eq:teststatistic}
\eeq
where $L_b$ is obtained from Eq.~\eqref{eq:likelihood-countingexp} by fixing $\sigma = 0$. This test statistic can then be used to compute $p_{s+b}$, the $p$-value for the signal+background hypothesis (corresponding to the probability, under the signal+background hypothesis, of observing a statistical under-fluctuation). Similarly, the $p$-value for the background-only hypothesis, $p_b$, can be computed and is required for deriving exclusions according to the ${\rm CL}_s$ prescription~\cite{Read:2002hq} that is used in LHC searches. The confidence level, denoted as $1-{\rm CL}_s$, with which a given signal hypothesis is excluded under the ${\rm CL}_s$ prescription is
\beq
1 - {\rm CL}_s = 1 - \frac{p_{s+b}}{p_b} \,.
\eeq
This conservative modification to the most simple exclusion procedure, based on ${\rm CL}_{s+b} = p_{s+b}$, was designed to prevent from excluding the signal+background hypothesis in the case of a statistical under-fluctuation when the two hypothesis are not clearly separated. Or, in other words, it prevents from rejecting the signal+background hypothesis when there is little sensitivity to the signal.

Some comments are in order. First, in practice the $p$-values are usually computed using toy Monte Carlo experiments. That is, under a given assumption (signal+background or background-only) giving the expectation for the observation, the distribution of the test statistic $t({\rm data})$ is built numerically from the generation of Monte Carlo pseudo-experiments. The computation of the $p$-value then corresponds to the fraction of generated toy experiments with $t({\rm data}) \le t_{\rm obs}$, where $t_{\rm obs}$ is the test statistic evaluated with the data actually observed at the LHC.
Second, this procedure is directly applicable only in the absence of nuisance parameters, and nuisance parameters are present in the likelihood defined in Eq.~\eqref{eq:likelihood-countingexp}. (This would amount to neglecting the systematic uncertainties, which is rarely found to be a good approximation at the LHC.) The treatment of nuisance parameters is a complex problem, and requires modifications to the test statistic and/or the way pseudo-experiments are generated. Different methods have been used over the time in high energy physics (see, {\it e.g.}, Ref.~\cite{Beringer:1900zz} and Appendix~A of Ref.~\cite{ATL-PHYS-PUB-2011-011}). The simplest such methods, used at LEP, is to evaluate the test statistic using the nominal values of the nuisance parameters, and to introduce the systematic uncertainties when generating toy Monte Carlo experiments by drawing random numbers for the nuisance parameters from their probability distribution functions. For instance, for every pseudo-experiment the number of background events $n_b$ would have a different value randomly generated from the distribution ${\rm Gauss}(n_b | \hat n_b, \Delta n_b)$. This is known as the hybrid Bayesian--frequentist approach.

All this procedure, based on an approximate likelihood, makes it possible to check if the LHC results in a given signal region exclude the new physics signal of interest at any confidence level. However, in almost all LHC SUSY searches there are not just one but multiple signal regions (up to a few dozens), requiring a dedicated treatment. Depending on the analysis, these signal regions could be overlapping or non-overlapping ({\it i.e.}, mutually exclusive or not). In the former case, a standard procedure for the exclusion at the LHC is to consider only the signal region that yields the best expected limit (or highest expected sensitivity). More technically, this corresponds to selecting the signal region with the highest $1-{\rm CL}_s$ for a given signal hypothesis under the assumption of $n_{\rm obs} = \hat n_b$. This avoids having the exclusion driven by an under-fluctuation in a signal region that is not expected to be very sensitive to the signal. In the latter case of non-overlapping signal regions, the results from different signal regions are combined into a single likelihood when deriving the exclusion bounds on benchmark scenarios. In principle, this combination could also be done externally. The most simple likelihood combining the information from two signal regions is the simple product of individual likelihoods of the form given in Eq.~\eqref{eq:likelihood-countingexp} (but not double-counting the constraints on the luminosity and on the signal cross section). However, this naive combination does not account for any correlation between the determination of the acceptance$\times$efficiency and of the number of background events in the two signal regions. The latter is probably the most important, as we expect the determination of the background to be strongly correlated from one signal region to the other in a given analysis. This is clearly a limitation of this approach, where only the signal events are generated and the prediction for the SM background is taken from the experimental publication.
Fortunately, in many practical cases the combined result is very close to the one obtained from the signal region with the best sensitivity (which is conservative) because the exclusion is mostly driven by only one signal region.

In this section, we have seen that testing models of new physics against the LHC results beyond the simplified model procedure of {\tt SModelS} is neither easy nor straightforward. Indeed, it requires to validate the implementation of the cuts and the treatment of the detector effects with the publicly available information, and also to define a reasonable statistical procedure for setting limits.
While implementing and validating analyses in order to study the constraints on natural SUSY, this observation motivated the idea of building a public database of validated reimplementations of LHC BSM analyses in the \madanalysis\ framework. This project was officially launched with the paper ``Towards a public analysis database for LHC new physics searches using \madanalysis''~\cite{Dumont:2014tja}, submitted to arXiv on July 11, 2014. It was lead by Sabine Kraml, Benjamin Fuks and myself and presented the implementation and validation of the first five analyses entering the database, to which Samuel Bein, Guillaume Chalons, Suchita Kulkarni, Dipan Sengupta and Chris Wymant significantly contributed. The paper also contains information on the modified and tuned version of the fast simulation software {\tt Delphes~3}, called {\tt Delphes-MA5tune}, that is present within \madanalysis\ and was used in the reimplementation of analyses. These improvements were made by Eric Conte, and will be presented in Section~\ref{ma5t-sec:delphestune}.

The wiki page listing the analyses present in the database is available at~\cite{padwiki}. Each analysis code, in the \cpp\ language used in \madanalysis, is submitted to INSPIRE (a web interface for that purpose is in preparation), and have an associated DOI~\cite{doi}, hence is searchable and citeable. For any given analysis, the information on the number of background and observed events is furthermore required for setting limits. This is provided in the form of an {\tt XML} file that is submitted to INSPIRE together with the analysis code. Finally, detector tunings (contained in the detector card for {\tt Delphes-MA5tune}) as well as detailed validation results for each analysis can be found on the wiki page~\cite{padwiki}.
To date, there are five SUSY analyses in the database, two from ATLAS and three from CMS. I was directly involved in the implementation and validation of the ATLAS search for electroweak-inos and sleptons in the di-lepton final state~\cite{Aad:2014vma} and in the CMS search for stops in the single-lepton final state~\cite{Chatrchyan:2013xna}, both being published analyses based on the full data set at $\sqrt{s} = 8$~TeV.
 
This database of analyses in the \madanalysis\ framework makes it easy to confront any model of new physics, taking as input an event file (in \texttt{StdHep}~\cite{stdhep} or \texttt{HepMc}~\cite{Dobbs:2001ck} format) and returning the information on the exclusion. While the execution of the analysis code itself produces, among other things, the information on the acceptance$\times$efficiency (see Section~\ref{ma5t-sec:saf}), the limit setting can subsequently be done with the {\tt Python} code {\tt exclusion\_CLs.py}. It reads the cross section and the acceptance$\times$efficiency from the output of \madanalysis, while the luminosity and the required information on the signal regions is taken from the {\tt XML} file mentioned above. This code can be found on the wiki~\cite{padwiki} and also installed as a module of \madanalysis; all details are given in Section~\ref{ma5t-sec:limitsetting}. The limit-setting procedure implemented in version 1 of this code is a simplification of what is discussed above in this section. First, the test statistic is defined as the number of events in a given signal region instead of a likelihood ratio. Second, only the uncertainty on the number of background events is taken into account. Nuisance parameters are implemented \`a-la-LEP as explained previously.
Limits obtained with this procedure were checked against the official exclusion bounds from ATLAS and CMS in several analyses; good agreement was observed~\cite{Dumont:2014tja}.

Note that a public tool with a similar scope, {\tt CheckMATE}, has recently been released~\cite{Drees:2013wra}, and is currently provided with nine validated SUSY analyses (eight from ATLAS and one from CMS). Being decoupled from the development of the tool and from possible applications, a major advantage of the public database we have initiated is that anyone from the community can contribute and have their work on analysis reimplementation and validation cited directly in future papers using their contribution to the database.


\section{Reimplementing analyses within MadAnalysis 5} \label{sec:ma5delphes3}

In this section, we focus on the expert mode of the
\madanalysis\ program~\cite{Conte:2012fm,Conte:2013mea,Conte:2014zja}
dedicated to the implementation of any
analysis based on a cut-and-count flow (in contrast to analyses relying on
multivariate techniques) and the investigation of the associated effects on
any Monte Carlo event sample.
This program follows the ``philosophy'' of the \texttt{MadGraph~5} event generator~\cite{Alwall:2011uj} (hence its name) and is lead by the two main developers, Eric Conte and Benjamin Fuks. In addition to the expert mode which we are interested in for the reimplementation of LHC analyses, simple cuts can be made using intuitive commands in a {\tt Python} interface; more information can be found in the first manual of the program from June 2012~\cite{Conte:2012fm}. Most of the recent developments in \madanalysis, to which I contributed, were aiming to facilitate the reimplementations of LHC searches. The remainder of the section is based on the paper ``Designing and recasting LHC analyses with \madanalysis'', Ref.~\cite{Conte:2014zja}, made in collaboration with Eric Conte, Benjamin Fuks, and Chris Wymant.
Moreover, details on the modified and tuned version of {\tt Delphes} and on limit-setting code {\tt exclusion\_CLs.py}, initially presented in Ref.~\cite{Dumont:2014tja}, will be given in Section~\ref{ma5t-sec:delphestune} and~\ref{ma5t-sec:limitsetting}, respectively.

In \madanalysis, the implementation of an analysis is facilitated by the
large number of predefined functions and methods included in the
\sampleanalyzer\ library shipped with the package, but
is however often complicated in cases where one has several
sub-analyses which we refer to as \textit{regions} (such as the
signal and control regions commonly used in BSM searches). The complication arose from
the internal format handled by \sampleanalyzer, which assumed the
existence of a single region. While this assumption is convenient for
prospective studies, \textit{i.e.}, the design of new analyses,
it is rarely fulfilled by
existing analyses that one may want to recast. In order to allow
the user to both design and recast analyses, we have consequently extended the
\sampleanalyzer\ internal format to support analyses with multiple regions
defined by different sets of cuts. We have also expanded the
code with extra methods and routines to facilitate
the implementation of more complex analyses by the user.

In the context of analyses which effectively contain sub-analyses, a further
useful classification of cuts can be made: namely into those which are
common/shared by different regions, and those which are not,
the latter serving to define the different sub-analyses themselves.
Fig.~\ref{ma5t-fig:DivergingRegions} schematically illustrates an
analysis containing four regions, which are defined by two
region-specific cuts imposed after two common cuts.
Some thought is required concerning
the best way to capture in an algorithm the set
of selection requirements shown in Fig.~\ref{ma5t-fig:DivergingRegions}.
For the common cuts (cuts 1 and 2 on the figure) this is clear: if the
selection condition is failed, the event is vetoed (\ie, we ignore it and
move on to analyzing the next event). Thereafter we have two conditions
to check (cuts 3 and 4), but they apply to
different regions. In terms of pseudo-code the most obvious, although
not the most efficient, method for implementing these third and
fourth cuts is
\begin{verbatim}
 count the event in region D
 if (condition 3)
 {
   count the event in region C
   if (condition 4)
   {
     count the event in region A
   }
 }
 if (condition 4)
 {
   count the event in region B
 }
\end{verbatim}
One important drawback of this naive approach is the duplication of
the check of the fourth condition. In the simple style of implementation
of the cuts above, this is unavoidable: condition 4 must be checked both
inside and outside the scope of condition 3. With the two region-specific
cuts that we have here, there is only one such clumsy duplication present in
the code. However, as the number of such cuts grows, the situation rapidly
gets worse. For instance, considering growing the decision tree shown in
Fig.~\ref{ma5t-fig:DivergingRegions} to include $N$ region-specific cuts,
combined in all possible permutations to define $2^N$ regions would
deepen the nesting of the above pseudo-code and lead to $2^N -(N+1)$
unnecessary duplications of checks. Moreover, each of those needs to be
carefully implemented by the user in the correct scope, a task becoming
less and less straightforward for large values of $N$.

\begin{figure}
\begin{center}
 \includegraphics[width=0.5\textwidth]{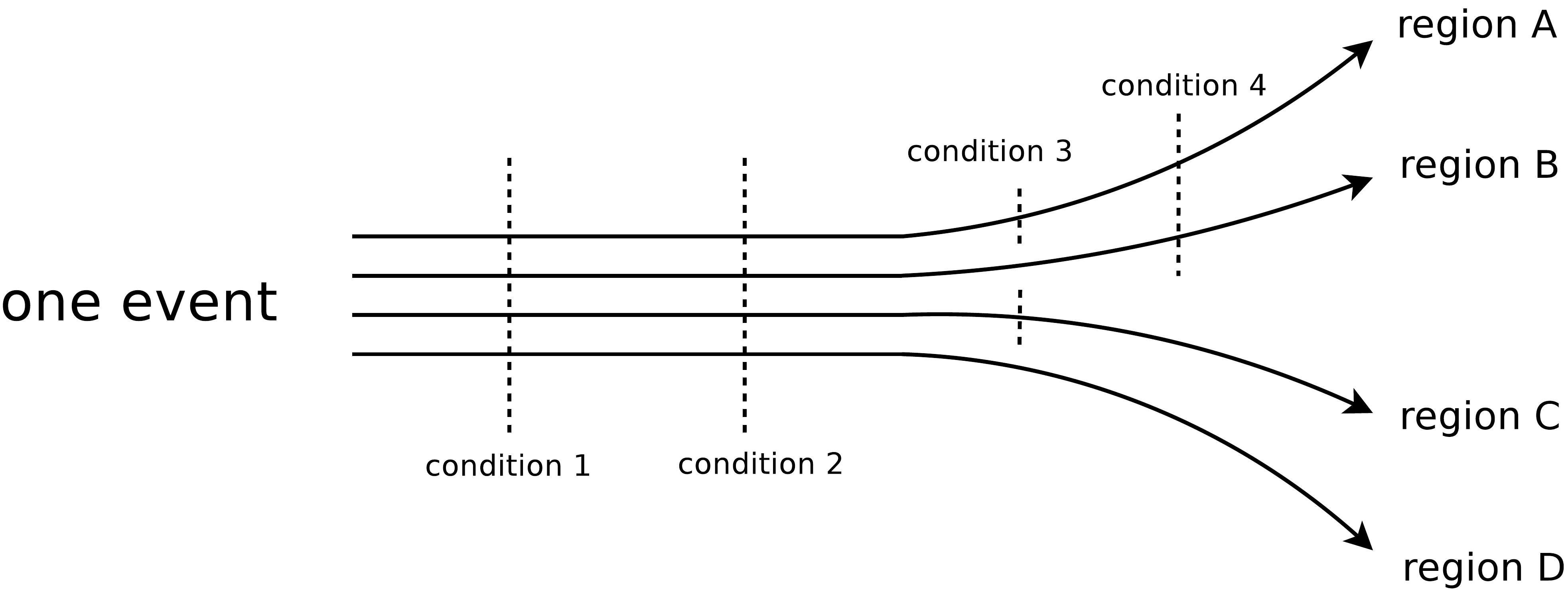}
\end{center}
\caption{Schematic illustration of the definition of different regions in
  which a given event can be counted (or not), based on different
  combinations of selection cuts.}
\label{ma5t-fig:DivergingRegions}
\end{figure}

Ideally the algorithm should be structured so that there is {\it no}
unnecessary duplication, which is
one of the new features of the latest version
of \sampleanalyzer, the \cpp\ core of the \madanalysis\ program. Both
can be obtained from the \madanalysis\ website,\\
  \verb+  https://launchpad.net/madanalysis5+\\
and all the features described in this section are available
from version 1.1.10 of the code onwards. This document supersedes
the previous version of the manual for the expert mode of the program~\cite{Conte:2012fm}.

The remainder of this section is organized as follows. In Sections~\ref{ma5t-sec:template}--\ref{ma5t-sec:saf},
we recall the basic functionalities
of \madanalysis\ for the implementation of physics analyses
in the expert mode of the program, which has been
extended according to the needs of the users. Moreover, we
introduce the new features of the \sampleanalyzer\ kernel.
Our conclusions are then given in Section~\ref{ma5t-sec:conclusions}.

\subsection{Creation of an analysis template}
\label{ma5t-sec:template}
In the \textit{expert mode} of the program,
the user is asked to write his/her analysis in \cpp, using all the classes
and methods of the \sampleanalyzer\ library. To begin implementing
a new analysis, the user is recommended to use the \python\ interpreter
of \madanalysis\ to create a working directory. This is achieved
by starting \madanalysis\ with the command
\begin{verbatim}
 ./bin/ma5 <mode> -E
\end{verbatim}
where the value of \texttt{<mode>} refers to
an analysis of events generated at
the parton level (\texttt{-P}), hadron level (\texttt{-H})
or reconstructed level (\texttt{-R}). It is then enough to
follow the instructions displayed on the screen---the user is
asked for the names of the working directory and of his/her analysis, which
we denote by \texttt{name} in the rest of the section.
The directory that has been created contains
three subdirectories: the \texttt{Input}, \texttt{Output}
and \texttt{Build} directories.

Use of the \texttt{Input} directory
is optional. It has been included in the analysis template in order to
have a unique structure for both the normal and expert modes
of \madanalysis. In the normal mode, its purpose is to collect text files
with the lists of paths to the event samples to analyze.
The \texttt{Output} directory has been conceived to store the results of each
execution of the analysis. The \texttt{Build} directory includes a series
of analysis-independent files organized into several sub-directories,
together with files to be modified by the user.

At the root of the \texttt{Build} directory, one finds one {\tt bash}
script together with its {\tt tcsh} counterpart. These
scripts set appropriately the environment variables necessary
for the compilation and execution of
an analysis within the \madanalysis\ framework.
They are initiated by typing in a shell the respective commands
\begin{verbatim}
 source setup.sh     source setup.csh
\end{verbatim}

A \texttt{Makefile} is also available so that the standard
commands
\begin{verbatim}
 make     make clean     make mrproper
\end{verbatim}
can be used to (re)compile the analysis (see
Section~\ref{ma5t-sec:exec}).
The final executable is obtained from two
pieces---a library and the main program. The library originates from
the merging of
the \sampleanalyzer\ library and the analysis of the user, and is stored
in the subdirectory \texttt{Build/Lib}. The main program is located in the
\texttt{Build/Main} subdirectory and has a simple structure.
It first initializes the analysis, then runs the analysis over all events
(possibly collected into several files) and eventually
produces the results in the \texttt{Output}
directory previously mentioned.

The \texttt{Build} directory contains moreover the \texttt{SampleA\-na\-ly\-zer}
subdirectory that stores the source and header files associated
with the analysis being implemented (\texttt{A\-na\-ly\-zer/na\-me.cpp} and
\texttt{A\-na\-ly\-zer/na\-me.h}), together with a \python\ script,
\texttt{new\-A\-na\-ly\-zer.py}, dedicated to the implementation of
several analyses into a single working directory.
The \texttt{Analyzer} subdirectory additionally includes a list with all
analyses implemented in the current working directory (\texttt{analysisList.h}).
More information about those files is provided
in the next subsections.

\subsection{Merging several analyses in a single working directory}
\label{ma5t-sec:multianalyses}
In Section~\ref{ma5t-sec:template}, we have explained how to create a working
directory containing a single (empty) analysis that is called, in our example,
\texttt{name}. The analysis itself is
implemented by the user in a pair of
files \texttt{name.cpp} and \texttt{name.h},
which should be consistently referred to in the file \texttt{analysisList.h}.
In addition, the main program (the file \texttt{Build/Main/main.cpp})
takes care of initializing and executing the analysis.
The structure of this analysis provides guidelines for the implementation of
any other analysis---\texttt{newname} for
the sake of the example---in the same working directory. This new analysis
has to be written in
the two files \texttt{newname.cpp} and \texttt{newname.h} (stored in the
\texttt{Build/SampleAnalyzer/Analyzer} directory)  and referred to in the
\texttt{analysisList.h} file.
The main program also needs to be
modified in order to initialize and execute the new analysis, in addition to
the first analysis (\texttt{name}).

All these tasks have
been automated (with the exception of the implementation of the analysis
itself) so that the user is only required to run the {\tt Python} script
\texttt{newAnalyzer.py} by typing in a shell the command
\begin{verbatim}
 ./newAnalysis.py newname
\end{verbatim}
from the \texttt{Build/SampleAnalyzer} directory.

\subsection{Coding of an analysis}
\label{ma5t-sec:sa}
\subsubsection{General features}
As briefly sketched above, the implementation of a specific
analysis within the \madanalysis\ framework consists of providing the
analysis \cpp\ source and header files \texttt{name.h} and \texttt{name.cpp}.

The header file contains the declaration of a class dedicated to the
analysis under consideration. This class is defined as a child class inheriting from
the base class \texttt{AnalysisBase}, and includes, in addition to constructor
and destructor methods, three functions to be implemented by the user (in the
source file \texttt{name.cpp}) that define the analysis itself. The first
of these, dubbed \texttt{Initialize}, is executed just once prior to
the reading of the user's set of events. In particular, it enables one both to declare
selection regions and to associate them with a series of cuts and histograms. It returns
a boolean quantity indicating whether the initialization procedure has been achieved
properly. If not, the execution of the main program is stopped.
The second method, named \texttt{Execute}, is the core of
the analysis and is applied to each simulated event provided by the user.
Among others things, it takes care of the application of the selection cuts
and the filling of the various histograms. This function returns a boolean
quantity that can be used according to the needs of the user, although
it is by default not employed.
Finally, the last function, a function of void type called
\texttt{Finalize}, is called once all events have been read
and analyzed.
Moreover, the user is allowed to define his/her own set of functions
and variables according to his/her purposes.

\renewcommand{\arraystretch}{1.5}%
\begin{table*}
  \centering
  \begin{tabular*}{\textwidth}{@{\extracolsep{\fill}}p{.32\textwidth}p{.63\textwidth}@{}} \hline\hline
    \texttt{AddCut(...)} & Declares a cut and links it to a set of regions.
     A single string must be passed as an argument, corresponding to the user-defined
     name of one of the selection cuts of the analysis. If no other argument is
     provided, the cut is associated with
     all declared signal regions. Otherwise, an additional
     single string or an array of strings, corresponding to
     the name(s) of the region(s) associated with the cut, can optionally be specified.\\
    \texttt{AddHisto(...)} & Declares a histogram. The first
     argument is the name of the histogram, the second one is
     the numbers of bins (an integer number), the third and fourth arguments
     define the lower
     and upper bounds of the $x$-axis (given as floating-point numbers), respectively.
     The last argument is optional and links all or
     some of the declared regions to the histogram (see the \texttt{AddCut} method
     for more information on this feature).\\
    \texttt{AddRegionSelection(...)} & Declares a new region. This method takes a string,
      corresponding to a user-defined name for the region, as its argument.\\
    \texttt{ApplyCut(...)} & Applies a given cut. This method
      takes two mandatory arguments. The first is a boolean variable and
      indicates whether
      the selection requirement associated with a given cut is satisfied.
      The second argument is the name of
      the considered cut, provided as a string. The method returns \texttt{true}
      if at least one region defined anywhere in the analysis is
      still passing all cuts so far, or \texttt{false} otherwise.\\
    \texttt{FillHisto(...)} & Fills a histogram. The first argument is a string specifying the
      name of the considered histogram, and the second is a floating-point number providing
     the value of the observable being histogrammed.\\
    \texttt{InitializeForNewEvent(...)} & To be called prior to the analysis of each event
     at the beginning of the \texttt{Execute} function. This
      method tags all regions as surviving the cuts, and initializes the weight associated with
      the current event to
      the value defined by the user passed as an argument (given as a floating-point number).\\
    \texttt{IsSurviving(...)} & Takes as an argument the name of a region (a string).
      The method returns {\tt true} if the region survives all cut applied so far, {\tt false} otherwise.\\
    \texttt{SetCurrentEventWeight(...)} & Modifies the weight of the current event to a user-defined value
      passed as an argument (given as a floating-point number).\\
\hline\hline
  \end{tabular*}
  \caption{\small \label{ma5t-tab:RSMfcts}Methods of the \texttt{RegionSelectionManager} class.}
\end{table*}
\renewcommand{\arraystretch}{1}%

The splitting of the analysis into regions, the application of the selection
criteria, and the filling of histograms are all controlled through the automatically
initialized object \texttt{Manager()}, a pointer to an instance
of the class \texttt{RegionSelectionManager}. The member methods
of this class are listed in Table~\ref{ma5t-tab:RSMfcts} and will be detailed
in the next subsections, in which we also
provide guidelines for the implementation of the functions
\texttt{Initialize}, \texttt{Execute} and  \texttt{Finalize}
in the \cpp\ source file \texttt{name.cpp}.

\subsubsection{Initialization}
\label{ma5t-sec:init}
When the analysis is executed from a shell, the program first calls the
\texttt{Initialize} method
before starting to analyze one or several event samples.
Prior to the declaration of regions, histograms and cuts, we first
encourage the user to include an electronic signature to
the analysis being implemented and to ask the program
to display it to the screen. Although this is neither
mandatory nor standardized, it improves the traceability of a
given analysis and provides information to the community about
who has implemented the analysis and which reference works
have been used. In particular for analyses that are being made public,
we strongly recommend including at least
the names and e-mail addresses of the authors,
a succinct description of the analysis and related experimental notes or
publications. Taking the example of the CMS stop search
in monoleptonic events~\cite{Chatrchyan:2013xna} presented in
Section~\ref{sec:cmsvalid}, an electronic signature could be
\begin{verbatim}
 INFO << "Analysis: CMS-SUS-13-011, arXiv:1308.1586"
      << " (stop search, single lepton)" << endmsg;
 INFO << "Recast by: Conte, Dumont, Fuks, Wymant"
      << endmsg;
 INFO << "E-mails: " << "conte@iphc.cnrs.fr, "
      <<                "dumont@lpsc.in2p3.fr, "
      <<                "fuks@cern.ch, "
      <<                "wymant@lapth.cnrs.fr"
      << endmsg;
 INFO << "Based on MadAnalysis 5 v1.1.10" << endmsg;
 INFO << "DOI: xx.yyyy/zzz" << endmsg;
 INFO << "Please cite arXiv:YYMM.NNNN [hep-ph]"
      << endmsg;
\end{verbatim}
where the last three lines refer to the Digital Object Identifier~\cite{doi}
of the analysis code (if available) and the physics publication
for which this analysis reimplementation has been
developed.
The sample of code above
also introduces the \texttt{INFO} message service
of the \sampleanalyzer\ framework, which is presented in
Section~\ref{ma5t-sec:msg_services}.

As already mentioned, each analysis region must be properly declared
within the function \texttt{Initialize}. This is achieved by making use
of the \texttt{AddRegionSelection} method of the
\texttt{Re\-gi\-on\-Se\-lec\-tion\-Ma\-na\-ger} class (see Table~\ref{ma5t-tab:RSMfcts}).
This declaration requires
provision of a name (as a string) which serves as a unique identifier
for this region within both the code itself
(to link the region to cuts and histograms) and the output files
that will be generated by the program. For instance, the
declaration of two regions,
dedicated to the analysis of events with a missing transverse energy
$E_T^{\rm miss} > 200$~GeV and 300~GeV could be implemented as
\begin{verbatim}
 Manager()->AddRegionSelection("MET>200");
 Manager()->AddRegionSelection("MET>300");
\end{verbatim}
As shown in these lines of code,
the declaration of the two regions is handled by
the \texttt{Manager()} object, an instance
of the \texttt{RegionSelectionManager} class that is
automatically included with any given analysis. As a result,
two new regions are created and the program internally assigns the intuitive
identifiers
\mbox{\texttt{"MET>200"}} and \texttt{"MET>300"} to the respective regions.

Once all regions have been declared, the user can continue with the
declaration of cuts and histograms.
As for regions, each declaration requires
a string name which acts as an
identifier in the code and the output. Histogram declaration also asks for
the number of bins (an integer number) and the lower and upper
bounds defining the range of the $x$-axis
(two floating-point numbers) to be specified.
Both histograms and cuts are associated with one or more regions.
In the case of cuts, this finds its source at the conceptual level:
each individual region is
{\it defined} by its unique set of cuts.
In the case of histograms, this enables one to establish the
distribution of a particular observable {\it after} some region-specific
cuts have been applied.
The association of both types of objects to their regions follows a similar syntax,
using an optional argument in their declaration. This argument is either a string
or an array of strings, each being the name of one of the previously declared regions.
If this argument is absent, the cut/histogram is automatically associated with all regions.
This feature can be used, for example, for preselection cuts that are requirements
common to all regions.

\renewcommand{\arraystretch}{1.5}%
\begin{table*}
  \centering
  \begin{tabular*}{\textwidth}{@{\extracolsep{\fill}}p{.35\textwidth}p{.60\textwidth}@{}} \hline\hline
    \texttt{mc()->beamE().first} & Returns, as a floating-point number, the energy of the first
     of the colliding beams.\\
    \texttt{mc()->beamE().second} & Same as  \texttt{mc()->beamE().first} but for the second
     of the colliding beams.\\
    \texttt{mc()->beamPDFauthor().first} & Returns, as an integer number, the identifier of the
     group of parton densities that have been used for the first of the colliding beams. The numbering
     scheme is based on the \texttt{PdfLib}~\cite{PlothowBesch:1992qj} and \texttt{LhaPdf}~\cite{Giele:2002hx}
     packages.\\
    \texttt{mc()->beamPDFauthor().second} & Same as \texttt{mc()->beamPDFauthor().first} but for the second
     of the colliding beams.\\
    \texttt{mc()->beamPDFID().first} & Returns, as an integer number, the code associated with
     the parton density set (within a specific group of parton densities) that has been used for the first
     of the colliding beams. The numbering
     scheme is based on the \texttt{PdfLib}~\cite{PlothowBesch:1992qj} and \texttt{LhaPdf}~\cite{Giele:2002hx}
     packages.\\
    \texttt{mc()->beamPDFID().second} & Same as \texttt{mc()->beamPDFID().first} but for the second
     of the colliding beams.\\
    \texttt{mc()->beamPDGID().first} & Returns, as an integer number, the Particle Data Group
     identifier defining the nature of the first of the colliding beams. The numbering
     scheme is based on the Particle Data Group review~\cite{Beringer:1900zz}.\\
    \texttt{mc()->beamPDGID().second} & Same as \texttt{mc()->beamPDGID().first} but for the second
     of the colliding beams.\\
    \texttt{mc()->processes()} & Returns a vector of instances of the \texttt{ProcessFormat} class
     associated with the set of subprocesses described by the sample. A \texttt{ProcessFormat} object
     contains information about the process identifier fixed by the generator (an integer number accessed
     via the \texttt{processId()} method), the associated cross section in pb (a floating-point number
     accessed via the \texttt{xsection()} method) and the related uncertainty (a floating-point
     number accessed via the \texttt{xsection\textunderscore error()} method), and the maximum
     weight carried by any event of the sample (a floating-point number accessed
     via the \texttt{maxweight()} method).\\
    \texttt{mc()->xsection()} & Returns, as a floating-point number, the cross section in pb 
     linked to the event sample.\\
    \texttt{mc()->xsection\textunderscore error()} & Returns, as a floating-point number,
      the (numerical) uncertainty on the cross section associated with the event sample.\\
\hline\hline
  \end{tabular*}
  \caption{\small \label{ma5t-tab:sampleformat}Methods of the \texttt{SampleFormat} class.}
\end{table*}
\renewcommand{\arraystretch}{1}%

As an illustrative example, the code
\begin{verbatim}
 Manager()->AddCut("1lepton");
 std::string SRlist[] = {"MET>200","MET>300"};
 Manager()->AddCut("MET>200 GeV",SRlist);
\end{verbatim}
would create two preselection cuts,
\texttt{"1lepton"} and \texttt{"MET>200 GeV"}, and
assign them to the two previously declared regions
\mbox{\texttt{"MET>200"}} and \texttt{"MET>300"}.
Although both cuts are associated with both regions, for
illustrative purposes we have shown two methods of doing this --
using the syntax for automatically linking to all regions (here
there are two) and explicitly stating both regions.
As a second example, we consider
the declaration of a histogram of 20 bins
representing the transverse momentum distribution of the leading
lepton, $p_T(\ell_1)$, in the range $[50,500]$~GeV. In the case
where the user chooses to associate it with
the second region only, the line
\begin{verbatim}
  Manager()->AddHisto("ptl1",20,50,500,"MET>300");
\end{verbatim}
should be added to the analysis code.

Finally, the \texttt{Initialize} method
can also be used for the initialization of one or several user-defined
variables that have been previously declared in the header file \texttt{name.h}.

\subsubsection{Using general information on Monte Carlo samples}

Simulated events can be classified into two categories: Monte Carlo events
either at the parton or at the hadron level, and reconstructed
events after object reconstruction.\footnote{Strictly speaking, there exists
a third class of events once detector simulation has been included. In this case,
the event final state consists of tracks and calorimeter deposits.
\madanalysis\ has not been designed to analyze those events and
physics objects such as (candidate) jets and electrons must be reconstructed prior to be
able to use the program.} Contrary to reconstructed
event samples, Monte Carlo samples in general contain
global information on the generation process,
such as cross section, the nature of the parton density set
that has been used, \textit{etc}. In the \madanalysis\
framework, these pieces of information are collected under the form of instances
of the \texttt{SampleFormat} class and can be retrieved by means
of the methods provided in Table~\ref{ma5t-tab:sampleformat}.

The function \texttt{Execute}
takes, as a first argument, a \texttt{Sam\-ple\-For\-mat} object associated
with the current analyzed sample. In this way, if
the sample is encoded in the \texttt{LHE}~\cite{Boos:2001cv,Alwall:2006yp},
\texttt{StdHep}~\cite{stdhep} or \texttt{HepMc}~\cite{Dobbs:2001ck} format,
the user may access most of the available information passed by the
event generator. In contrast, the other event formats supported by
\madanalysis, namely the
\texttt{LHCO}~\cite{lhco} and (\texttt{ROOT}-based~\cite{Brun:1997pa})
\texttt{Delphes~3}~\cite{deFavereau:2013fsa} format\footnote{In order
to activate the support of \madanalysis\ for the output
format of {\tt Delphes~3}, the user is requested to start
the \madanalysis\ interpreter
(in the normal execution mode of the program) and to type
\texttt{install delphes}.}, do not include
any information of this kind so that the first argument of the
\texttt{Execute} function is a null pointer. In the case where
the user may need such information, it will have to be
included by hand.

\renewcommand{\arraystretch}{1.5}%
\begin{table*}
  \centering
  \begin{tabular*}{\textwidth}{@{\extracolsep{\fill}}p{.18\textwidth}p{.80\textwidth}@{}} \hline\hline
    \texttt{alphaQCD()}  & Returns, as a floating-point number, the employed value for the
                           strong coupling constant.\\
    \texttt{alphaQED()}  & Returns, as a floating-point number, the employed value for
                           the electromagnetic coupling constant.\\
    \texttt{particles()} & Returns, as a vector of
                           \texttt{MCParticleFormat} objects, all the
                           final-, intermediate- and initial-state particles of the event.\\
    \texttt{processId()} & Returns, as an integer number, the identifier of the physical process related to
                           the considered event.\\
    \texttt{scale()}     & Returns, as a floating-point number, the employed value for the factorization scale.\\
    \texttt{weight()}    & Returns, as a floating-point number, the weight of the event.\\ \hline
    \texttt{MET()}       & Returns, as an \texttt{MCParticleFormat} object, the missing
                           transverse momentum $\vec{p}_T^{\rm miss}$ of the event. The particles relevant for the
                           calculation must be properly
                           tagged as invisible (see Section~\ref{ma5t-sec:phys_services}).\\
    \texttt{MHT()}       & Returns, as an \texttt{MCParticleFormat} object, the missing
                           transverse hadronic momentum $\vec{H}_T^{\rm miss}$ of the event.
                           The particles relevant for the
                           calculation must be properly
                           tagged as hadronic, and not tagged as invisible (see Section~\ref{ma5t-sec:phys_services}).\\
    \texttt{TET()}       & Returns, as a floating-point number, the total visible transverse energy of
                           the event $E_T$. The particles relevant for the
                           calculation must not be
                           tagged as invisible (see Section~\ref{ma5t-sec:phys_services}).\\
    \texttt{THT()}       & Returns, as a floating-point number, the total visible transverse hadronic energy of
                           the event $H_T$. The particles relevant for the
                           calculation must be properly
                           tagged as hadronic, and not tagged as invisible
                           (see Section~\ref{ma5t-sec:phys_services}).\\
  \hline\hline
  \end{tabular*}
  \caption{\small \label{ma5t-tab:MCEventFormat}Methods of the \texttt{MCEventFormat} class.}
\end{table*}
\renewcommand{\arraystretch}{1}%

\renewcommand{\arraystretch}{1.5}%
\begin{table*}
  \centering
  \begin{tabular*}{\textwidth}{@{\extracolsep{\fill}}p{.20\textwidth}p{.78\textwidth}@{}} \hline\hline
  \texttt{ctau()}       & Returns, as a floating-point number, the lifetime of the particle
    in millimeters.\\
  \texttt{daughters()}  & Returns, as a vector of pointers to \texttt{MCParticleFormat} objects, a list
    with the daughter particles that are either produced from the decay of the considered particle
    or from its scattering with another particle.\\
  \texttt{momentum()}   & Returns, as a (\texttt{ROOT}) \texttt{TLorentzVector} object~\cite{Brun:1997pa},
    the four-momentum of the particle. All the properties of the four-momentum can be accessed
    either from the methods associated with the \texttt{TLorentzVector} class, or
    as direct methods of the \texttt{MCParticleFormat} class, after changing the method
    name to be entirely lower case.
    For instance, \texttt{pt()} is equivalent
    to \texttt{momentum().Pt()}. In addition, the methods \texttt{dphi\textunderscore 0\textunderscore 2pi(...)}
    and  \texttt{dphi\textunderscore 0\textunderscore pi(...)} return the difference in azimuthal
    angle normalized in the $[0,2\pi]$ and $[0,\pi]$ ranges, respectively, between the particle
    and any other particle passed as an argument, whereas \texttt{dr(...)} returns their angular distance, the second
    particle being provided as an argument as well.\\
  \texttt{mothers()}    & Returns, as a vector of pointers to \texttt{MCParticleFormat} objects, a list with all the mother
    particles of the considered particle. In the case of an initial particle, this list is empty, while for a decay and a
    scattering process, it contains one and two elements, respectively.\\
   \texttt{mt\textunderscore met()} & Returns, as a floating-point number, the transverse
     mass obtained from a system comprised of the considered particle and the invisible
     transverse momentum of the event. The particles relevant for the
                           calculation must be properly
                           tagged as invisible (see Section~\ref{ma5t-sec:phys_services}).\\
  \texttt{pdgid()}      & Returns, as an integer number, the Particle Data Group identifier defining the
    nature of the particle. The numbering scheme is based on the Particle Data Group
    review~\cite{Beringer:1900zz}. \\
  \texttt{spin()}       & Returns, as a floating-point number, the cosine of the angle
    between the three-momentum of the particle and its spin vector. This quantity is computed
    in the laboratory reference frame.\\
  \texttt{statuscode()} & Returns, as an integer number, an identifier fixing the initial-,
    intermediate- or final-state nature of the particle. The numbering scheme is based on
    Ref.~\cite{Boos:2001cv}.\\
  \texttt{toRestFrame(...)} & Boosts the four-momentum of the particle
      to the rest frame of a second particle (an \texttt{MCParticleFormat} object given as argument).
      The method modifies the momentum of the particle.\\
  \hline\hline
  \end{tabular*}
  \caption{\small \label{ma5t-tab:MCParticleFormat}Methods of the \texttt{MCParticleFormat} class.}
\end{table*}
\renewcommand{\arraystretch}{1}%

For instance, assuming that an event sample containing
$N = 10000$ events ($N$ being stored as a double-precision number
in the \texttt{nev} variable) is analyzed, the weight of each event
could be calculated (and stored in the \texttt{wgt} variable
for further use within the analysis) by means of the code sample
\begin{verbatim}
 double lumi = 20000.;
 double nev  = 10000.;
 double wgt = MySample.mc()->xsection()*lumi/nev;
\end{verbatim}
The \texttt{MySample} object is an instance of the
\texttt{SampleFormat} class associated with the sample
being analyzed and we impose the results to be
normalized to 20~fb$^{-1}$ of simulated collisions
(stored in pb$^{-1}$ in the \texttt{lumi} variable).
For efficiency purposes, such a computation should be performed once
and for all at the time of the initialization of the
analysis, and not each time an event is analyzed.
The variable \texttt{wgt} is then promoted
to a member of the analysis class being implemented.

\subsubsection{Internal data format for event handling}\label{ma5t-sec:dataformat}
\renewcommand{\arraystretch}{1.45}%
\begin{table*}
  \centering
  \begin{tabular*}{\textwidth}{@{\extracolsep{\fill}}p{.23\textwidth}p{.75\textwidth}@{}} \hline\hline
    \texttt{electrons()} & Returns, as a vector of \texttt{RecLeptonFormat} objects,
                           all the reconstructed electrons of the event.\\
    \texttt{jets()}      & Returns, as a vector of \texttt{RecJetFormat} objects,
                           all the reconstructed jets of the event.\\
    \texttt{muons()}     & Returns, as a vector of \texttt{RecLeptonFormat} objects,
                           all the reconstructed muons of the event.\\
    \texttt{photons()}   & Returns, as a vector of \texttt{RecPhotonFormat} objects,
                           all the reconstructed photons of the event.\\
    \texttt{taus()}      & Returns, as a vector of \texttt{RecTauFormat} objects,
                           all the reconstructed hadronic taus of the event.\\
    \texttt{tracks()}    & Returns, as a vector of \texttt{RecTrackFormat} objects,
                           all the reconstructed tracks of the event.\\ \hline
    \texttt{genjets()}   & Returns, as a vector of \texttt{RecJetFormat} objects,
                           all the parton-level jets of the event.\\
    \texttt{MCBquarks()}   & Returns, as a vector of pointers to \texttt{MCParticleFormat} objects,
                           all the parton-level $b$-quarks of the event.\\
    \texttt{MCCquarks()}   & Returns, as a vector of pointers to \texttt{MCParticleFormat} objects,
                           all the parton-level $c$-quarks of the event.\\
    \texttt{MCElectronicTaus()} & Returns, as a vector of pointers to \texttt{MCParticleFormat} objects,
                           all the parton-level
                           tau leptons that have decayed into an electron and a pair of neutrinos.\\
    \texttt{MCHadronicTaus()} & Returns, as a vector of pointers to \texttt{MCParticleFormat} objects,
                           all the parton-level tau leptons that have decayed hadronically.\\
    \texttt{MCMuonicTaus()} & Returns, as a vector of pointers to \texttt{MCParticleFormat} objects,
                            all the parton-level tau leptons that have decayed into a muon and a pair of neutrinos.\\
\hline
    \texttt{MET()}       & Returns, as a \texttt{RecParticleFormat} object,
                           the missing transverse momentum $\vec{p}_T^{\rm miss}$ of the event as stored in the event file.\\
    \texttt{MHT()}       & Returns, as a \texttt{RecParticleFormat} object, the missing
                           transverse hadronic momentum $\vec{H}_T^{\rm miss}$ of the event, computed from reconstructed jets.\\
    \texttt{TET()}       & Returns, as a floating-point number, the total visible transverse energy of
                           the event $E_T$.\\
    \texttt{THT()}       & Returns, as a floating-point number, the total visible transverse hadronic energy of
                           the event $H_T$, computed from reconstructed jets.\\
  \hline\hline
  \end{tabular*}
  \caption{\small \label{ma5t-tab:RecEventFormat}Methods of the \texttt{RecEventFormat} class.}
\end{table*}
\renewcommand{\arraystretch}{1}%

In the \sampleanalyzer\ framework, both Monte Carlo and reconstructed
events are internally handled as instances of a class named
\texttt{EventFormat}. At the time of execution of the analysis
on a specific event, the \texttt{Execute} function receives such an
\texttt{EventFormat} object as its second argument. The properties
of this object reflect those of the current event and can
be retrieved via the two methods
\begin{verbatim}
       event.mc()       event.rec()
\end{verbatim}
which return a pointer to an
\texttt{MCEventFormat} object encompassing
information at the Monte Carlo event level, and a pointer to
a \texttt{RecEventFormat} object specific for managing
information at the reconstructed event level, respectively.

Focusing first on Monte Carlo events, the properties of all
initial-state, in\-ter\-me\-di\-a\-te-state and
final-state particles can be retrieved by means of the \texttt{MCEventFormat} class
(see Table~\ref{ma5t-tab:MCEventFormat}).
Particles are encoded as instances of the
\texttt{MCParticleFormat} class whose associated methods are
shown in Table~\ref{ma5t-tab:MCParticleFormat}.
Additionally, general event information,
such as the values for the gauge couplings or the factorization
scale used, is also available if properly stored in the event file.
Finally, the \texttt{MCEventFormat} class also contains
specific methods for the computation of four global event observables:
the amount of (missing) transverse energy $E_T$ ($E_T^{\rm miss}$)
and of (missing) transverse hadronic energy $H_T$ ($H_T^{\rm miss}$). 
These quantities are calculated from the transverse momentum
of the final-state particles according to
\be\bsp
  E_T = \sum_{\text{visible particles}} \big| \vec p_T \big|
    \ ,\qquad 
  H_T = \sum_{\text{hadronic particles}} \big| \vec p_T \big|
    \ , \\
  E_T^{\rm miss} = \big|\vec{p}_T^{\rm miss}\big| = \bigg| -\sum_{\text{visible particles}} \vec p_T \bigg|
     \ ,\hspace{1.75cm} \\
  H_T^{\rm miss} = \big|\vec{H}_T^{\rm miss}\big| = \bigg| -\sum_{\text{hadronic particles}} \vec p_T
    \bigg| \ ,\hspace{1.50cm}
\esp\label{ma5t-eq:mcmettet}\ee
once the user has defined, in the initialization part
of the analysis, which particles are invisible and which ones are
hadronizing (by means of the configuration functions
described in Section~\ref{ma5t-sec:phys_services}).
However, the definitions of
Eq.~\eqref{ma5t-eq:mcmettet} may be not appropriate
if the user wants to
include only specific visible/ha\-dro\-nic particles in the sums.
In this case, he/she should perform their implementation within the
\texttt{Execute} function of the analysis according to his/her needs.
The entire set of properties that can be employed
to analyze a Monte Carlo event is shown in Table~\ref{ma5t-tab:MCEventFormat}.

For example, the selection of
all the final-state electrons and positrons that are
present in an event and whose transverse momentum
is larger than 50 GeV could be implemented as
\begin{verbatim}
 std::vector<const MCParticleFormat*> electrons;

 for(unsigned int i=0;
     i<event.mc()->particles().size(); i++)
 {
   const MCParticleFormat* prt =
     &event.mc()->particles()[i];

   if(prt->statuscode() != 1) continue;

   if(std::abs(prt->pdgid()) == 11)
   {
     if(prt->momentum().Pt()>50)
       electrons.push_back(prt);
   }
 }
\end{verbatim}
The first line of the code above indicates the declaration of
a vector, dubbed \texttt{electrons}, of pointers to (constant)
\texttt{MC\-Par\-ti\-cle\-For\-mat} objects that contain the selected electrons.
With the next block of \cpp\ commands, we
loop over all the event particles (the \texttt{for} loop)
and store the current particle into a temporary variable \texttt{prt}.
We then discard non-final-state particles,
which have a status code different from one (the first \texttt{if} statement).
Finally, we fill the \texttt{electrons} vector with all electrons and positrons
(with a Particle Data Group
code equal to $\pm 11$, as shown in the second \texttt{if} statement)
whose transverse momentum is greater than 50~GeV (the third \texttt{if}
statement).

We next present the methods that have been designed
for the analysis of reconstructed events and which are
part of the \texttt{RecEventFormat} class. This class
contains functions (see Table~\ref{ma5t-tab:RecEventFormat})
allowing access to two series of containers,
the first ones gathering final state objects of a given nature and
the second ones collecting specific generator-level (or equivalently parton-level)
objects. All these containers can be further employed within an analysis
so that the properties of the different objects can be retrieved and
subsequently used, \textit{e.g.}, for cuts and histograms.
All the available methods associated with reconstructed objects
have been collected in Table~\ref{ma5t-tab:RecParticleFormat} and~\ref{ma5t-tab:RecParticleFormat2}, while we recall
that the \texttt{MCParticleFormat} class has been described in
Table~\ref{ma5t-tab:MCParticleFormat} (necessary
for the handling of generator-level objects).
In the case where some pieces of information
(either specific properties of a given particle species or a given container
itself) are absent from the event file, the related methods return
null results.

\renewcommand{\arraystretch}{1.5}%
\begin{table*}
  \centering
  \begin{tabular*}{\textwidth}{@{\extracolsep{\fill}}p{.11\textwidth}p{.84\textwidth}@{}} \hline\hline
    \texttt{btag()}      & This method is specific to \texttt{RecJetFormat} objects and returns a boolean quantity
                           describing whether the jet has been tagged as a $b$-jet.\\
    \texttt{ctag()}      & This method is specific to \texttt{RecJetFormat} objects and returns a boolean quantity
                           describing whether the jet has been tagged as a $c$-jet.\\
    \texttt{charge()}    & Returns, as an integer number, the electric charge of the object
                           (relative to the fundamental unit of electric charge $e$).
                           This method is available
                           for the \texttt{RecLeptonFormat}, \texttt{RecTauFormat} and \texttt{RecTrackFormat} classes.\\
    \texttt{etaCalo()}   & This method is specific to the \texttt{RecTrackFormat} class and returns,
                           as a floating-point number,
                           the pseudorapidity corresponding to the entry point of the track in the calorimeter.\\
    \texttt{isolCones()} & Returns a vector of pointers to instances of the \texttt{IsolationConeType} class.
       This class allows one to retrieve information about the isolation of the object
       after defining a cone of a given size
       (a floating-point number accessed via the \texttt{deltaR()} method of the class) centered on it. The (integer)
       number of tracks in the cone is obtained by means of the \texttt{ntracks()} method, the sum of the
       transverse momenta of these tracks by means of the \texttt{sumPT()} method and the amount of
       calorimetric (transverse) energy in the cone
       by means of the \texttt{sumET()} method. The \texttt{isolCones()}
       method has only been implemented for the \texttt{RecTrackFormat}, \texttt{RecLeptonFormat},
       \texttt{RecPhotonFormat} and \texttt{RecJetFormat} classes. A modified version of \texttt{Delphes~3}
       that supports this structure has been introduced in
       Ref.~\cite{Brooijmans:2014eja}.\\
    \texttt{momentum()}  & Returns, as a (\texttt{ROOT}) \texttt{TLorentzVector} object~\cite{Brun:1997pa},
        the four-momentum of the particle. This method is available for all types of
        reconstructed objects. All the properties of the four-momentum can
        be accessed either from the methods associated with the \texttt{TLorentzVector}
        class, or as direct methods of the different classes of objects, after changing the
        method name to be entirely lower case.
        For instance, the method \texttt{pt()} is equivalent
                           to \texttt{momentum().Pt()}. In addition, the methods
                           \texttt{dphi\textunderscore 0\textunderscore 2pi(...)}
                           and  \texttt{dphi\textunderscore 0\textunderscore pi(...)} return the difference in azimuthal
                           angle normalized in the $[0,2\pi]$ and $[0,\pi]$ ranges, respectively, between the object
                           and any other object passed as an argument, whereas \texttt{dr(...)} returns their
                           angular distance, the second
                           object being provided as an argument as well.\\
    \texttt{mt\textunderscore met()} & Returns, as a floating-point number, the transverse
     mass obtained from a system comprised of the considered particle and the missing
     transverse momentum of the event.\\
  \hline\hline
  \end{tabular*}
  \caption{\small \label{ma5t-tab:RecParticleFormat}Methods giving access the properties of
    the reconstructed objects represented as instances of the \texttt{RecLeptonFormat},
    \texttt{RecJetFormat}, \texttt{RecPhotonFormat}, \texttt{RecTauFormat}, \texttt{RecTrackFormat}
    and \texttt{RecParticleFormat} classes.}
\end{table*}
\renewcommand{\arraystretch}{1}%

\renewcommand{\arraystretch}{1.5}%
\begin{table*}
  \centering
  \begin{tabular*}{\textwidth}{@{\extracolsep{\fill}}p{.15\textwidth}p{.80\textwidth}@{}} \hline\hline
    \texttt{ntracks()}   & Returns, as an integer number, the number of charged tracks associated with
                           the reconstructed object. This method has been implemented for the  \texttt{RecTauFormat}
                           and \texttt{RecJetFormat} classes.\\
    \texttt{pdgid()}     & This method is specific to the \texttt{RecTrackFormat} class and returns,
                           as an integer number, the Particle Data Group
                           identifier defining the nature of the particle giving rise to the track. The numbering
                           scheme is based on the Particle Data Group review~\cite{Beringer:1900zz}.\\
    \texttt{phiCalo()}   & This method is specific to the \texttt{RecTrackFormat} class and returns,
                           as a floating-point number,
                           the azimuthal angle with respect to the beam direction corresponding to the entry point
                           of the track in the calorimeter.\\
    \texttt{sumET\textunderscore isol()} & Returns, as a floating-point number, the amount of
       calorimetric (transverse) energy lying in a specific
       cone centered on the object. The cone size is fixed at the level of
       the detector simulation and this method is available for the \texttt{RecLeptonFormat} class
       (this information is available in the \texttt{LHCO} format).\\
    \texttt{sumPT\textunderscore isol()} & Returns, as a floating-point number, the sum of the
       transverse momenta of all tracks lying in a given cone centered on
       the object. The cone size is fixed at the level of
       the detector simulation and this method is available for the \texttt{RecLeptonFormat} class
       (this information is available in the \texttt{LHCO} format).\\
    \texttt{EEoverHE()}  & Returns, as a floating-point number, the ratio of the electromagnetic to hadronic
                           calorimetric energy associated with the object. This method is available
                           for the \texttt{RecLeptonFormat}, \texttt{RecTauFormat} and \texttt{RecJetFormat} classes.\\
    \texttt{ET\textunderscore PT\textunderscore isol()} & Returns, as a floating-point number, the amount
       of calorimetric (transverse) energy lying in a given cone centered on the object
       calculated relatively to the sum
       of the transverse momentum of all tracks in this cone. The cone size is fixed at the level of
       the detector simulation and this method is available for the \texttt{RecLeptonFormat} class
       (this information is available in the \texttt{LHCO} format).\\
    \texttt{HEoverEE()}  & Returns, as a floating-point number, the ratio of the hadronic to electromagnetic
                           calorimetric energy associated with the object. This method is available
                           for the \texttt{RecLeptonFormat}, \texttt{RecTauFormat} and \texttt{RecJetFormat}
                           classes.\\
  \hline\hline
  \end{tabular*}
  \caption{\small \label{ma5t-tab:RecParticleFormat2} (continuation of Table~\ref{ma5t-tab:RecParticleFormat}) Methods giving access the properties of
    the reconstructed objects represented as instances of the \texttt{RecLeptonFormat},
    \texttt{RecJetFormat}, \texttt{RecPhotonFormat}, \texttt{RecTauFormat}, \texttt{RecTrackFormat}
    and \texttt{RecParticleFormat} classes.}
\end{table*}
\renewcommand{\arraystretch}{1}%

Finally, as for the \texttt{MCEventFormat} class, specific functions (see
Table~\ref{ma5t-tab:RecEventFormat}) have been implemented
to access the (missing) transverse energy and (missing) hadronic transverse
energy of the event. While the value of the $E_T^{\rm miss}$ variable is taken
from the event file and not calculated on the fly, the other variables
are computed from the information on the reconstructed objects,
\be\bsp
  E_T =&\ \sum_{{\rm jets}, \, \ell^\pm, \, \gamma} \big| \vec p_T \big|
    \ , \\
  H_T =&\ \sum_{\text{jets}} \big| \vec p_T \big|
    \ , \\
  H_T^{\rm miss} = &\ \big|\vec{H}_T^{\rm miss}\big| = \bigg| -\sum_{\text{jets}} \vec p_T
    \bigg| \ .
\esp\label{ma5t-eq:recglobal}\ee

As an example, we show how
an isolation requirement on final-state muons can be implemented.
To do this we define an isolation variable $I_{\rm rel}$ as
the amount of transverse energy,
relative to the transverse momentum of the muon,
present in a cone of radius $R = 0.4$ centered on the muon.
We constrain this quantity to satisfy $I_{\rm rel} < 20\%$. A possible
corresponding sample of \cpp\ code is
\begin{verbatim}
 std::vector<const RecLeptonFormat*> MyMuons;
 for(unsigned int i=0;
     i<event.rec()->muons().size(); i++)
 {
   const RecLeptonFormat *Muon =
     &event.rec()->muons()[i];

   for(unsigned int j=0;
       j<Muon->isolCones().size(); j++)
   {
     const IsolationConeType *cone = 
       &Muon->isolCones()[j];

     if(std::fabs(cone->deltaR()-0.4)<1e-3)
     {
       if(cone->sumET()/Muon->momentum().Pt()<.20)
         MyMuons.push_back(Muon);
     }
   }
 }
\end{verbatim}
With those lines of code, we start by declaring the \texttt{MyMuons}
variable, a vector of pointers to
\texttt{RecLeptonFormat} objects, that will refer
to the reconstructed muons tagged as isolated. Then, we proceed with
a \texttt{for}-loop dedicated to the computation of
the $I_{\rm rel}$ variable for each of the final state muons. In the case where
$I_{\rm rel}$ is smaller than 20\%, the muon is added to the \texttt{MyMuons}
container. In more detail, this \texttt{for}-loop works as follows. The current muon is
stored in a temporary variable called \texttt{Muon}. The calculation
of $I_{\rm rel}$ relies, first, on the amount of calorimetric energy in
a cone of radius \mbox{$R=0.4$} centered on the muon and second,
on the transverse momentum of the current muon.
The first of these two quantities is evaluated via the \texttt{isolCones()} method
of the \texttt{RecLeptonFormat} class (see Table~\ref{ma5t-tab:RecParticleFormat} and~\ref{ma5t-tab:RecParticleFormat2})
whereas the second one is derived from the
muon four-momentum (obtained from the \texttt{momentum()} me\-thod of the
\texttt{RecLeptonFormat} class). In the example above,
we assume that information on muon isolation associated with several cone sizes
is available, including the choice \mbox{$R=0.4$}. The second
\texttt{for}-loop that has been implemented selects the desired
value of $R$. The subsequent computation of the $I_{\rm rel}$ quantity
is immediate.
We refer to Ref.~\cite{Brooijmans:2014eja} for more detailed
examples on this topic, in cases where event simulation is
based on a modified version of \texttt{Delphes~3} properly handling
such a structure for the isolation information.

\renewcommand{\arraystretch}{1.5}%
\begin{table*}
  \centering
  \begin{tabular*}{\textwidth}{@{\extracolsep{\fill}}p{.18\textwidth}p{.78\textwidth}@{}} \hline\hline
    \texttt{DisableColor()} & Switches off the display of messages in color.
                              Colors are switched on by default, and the color scheme is hard-coded.\\
    \texttt{EnableColor()}  & Switches on the display of messages in color.
                              Colors are switched on by default, and the color scheme is hard-coded.\\
    \texttt{Mute()}         & Switches off a specific message service. Services are switched on by default.\\
    \texttt{SetStream(...)}  & Takes a pointer of type \texttt{ofstream} as an
      argument and redirect the output of a given service to a file.\\
    \texttt{UnMute()}       & Switches on a specific message service. Services are switched on by default.\\
\hline\hline
  \end{tabular*}
  \caption{\small \label{ma5t-tab:msg}Methods associated with a given message service. The available services are
   \texttt{INFO}, \texttt{WARNING}, \texttt{ERROR} and \texttt{DEBUG}.}
\end{table*}
\renewcommand{\arraystretch}{1}%

\subsubsection{Applying cuts and filling histograms}\label{ma5t-sec:cutshistos}
The cuts for the analysis, having been declared in the \texttt{I\-ni\-ti\-a\-li\-ze}
function (see Section~\ref{ma5t-sec:init}),
are applied in the \texttt{Execute} function by means
of the \texttt{RegionSelectionManager} method \texttt{ApplyCut}
(see Table~\ref{ma5t-tab:RSMfcts}). Its two arguments consist
of a boolean quantity governing the cut condition (\textit{i.e.}, it indicates
whe\-ther the current event satisfies this cut) and a string which should be the name of one
of the declared cuts.

This method starts by cycling through all regions
associated with this cut. For each region, it checks whether the region is still
surviving all cuts applied so far by evaluating an internal boolean variable.
If a given region is found to be already failing one of the preceding
cuts (indicated by the internal surviving variable having the value
\texttt{false}), the \texttt{ApplyCut}
method continues with the next region associated with the considered cut.
On the other hand if the region is surviving, the cut-flow
for this region is updated according to
the cut condition (the boolean argument of the
\texttt{ApplyCut} method) and the internal
surviving variable will be kept as \texttt{true} or changed to \texttt{false}
as appropriate.
The aforementioned internal boolean variables indicating the survival of each region
should all be initialized to \texttt{true} 
when starting to analyze a given event.
This is achieved by adding,
at the beginning of the \texttt{Execute} function,
\begin{verbatim}
 Manager()->InitializeForNewEvent(myWeight);
\end{verbatim}
where \texttt{MyWeight} is a floating-point number representing the weight of the event.
The weight is used when histograms are filled and cut-flow charts
calculated, and can be modified within the analysis by making use
of the \texttt{Set\-Cur\-rent\-E\-vent\-Weight}
method of the \texttt{RegionSelectionManager} class.

The analysis manager also stores internally the total number of surviving regions,
which is updated when a specific region fails a cut. This enables the \texttt{ApplyCut}
method to determine and return,
after cycling through the associated \texttt{Re\-gion\-Se\-lec\-ti\-on} instances,
a boolean quantity which is set to \texttt{false} in the case where not a single surviving region
remains. The output of the \texttt{ApplyCut} method
is thus equal to the boolean value of the statement {\it there is at least one region in the analysis,
not necessarily one of those associated with this specific cut, which is still passing all cuts so far}.
When it switches from \texttt{true} to \texttt{false},
the present event should no longer be analyzed, and one should move on with the next event.
It is therefore recommended, for efficiency purposes,
to always call the \texttt{ApplyCut} method in the following schematic manner,
\begin{verbatim}
 if ( !ApplyCut(...) )
   return;
\end{verbatim}
with the \texttt{return} command terminating the analysis of the current event
if all regions are failing the set of cuts applied so far.

Since, trivially, cuts keep some events and reject others, the distribution of an
observable is affected by the placement of its histogram-filling command
within the full sequence of cuts.
Then since each region has its own unique set of cuts (by definition), the
distribution of any observable is in general different for any two regions.
However, it is meaningful to consider a single histogram as associated with multiple
regions, {\it if} it is filled before any cuts are made that distinguish the regions.
As an example, a typical format for processing an event would be a set of common
preselection cuts, then the filling of various histograms (which are thus
associated with all regions), then the application of the region-specific cuts
(possibly followed by some further histogramming).

In \madanalysis, we deal with this within \texttt{Fill\-His\-to},
the histogram-filling method of the \texttt{RegionSelectionManager}
class, which takes as arguments a string and
a floating-point number. The string
should be the name of one of the declared histograms, and the floating-point
number represents the value of the histogrammed observable for the event under
consideration. This method can be called as in
\begin{verbatim}
 Manager()->FillHisto("ptl1", val);
\end{verbatim}
where \texttt{"ptl1"} is the name of the considered histogram
(continuing with the example from Section~\ref{ma5t-sec:init})
and \texttt{val} is the value of the observable of interest,
namely the transverse momentum of the leading lepton in our case.
The \texttt{FillHisto} method begins by
verifying whether each of the regions associated with this histogram
is surviving all cuts applied so far (via the internal surviving variable
above-mentioned). In the case where all the associated regions are found
surviving (failing) the cuts, the histogram is (not) filled.
If a mixture of surviving and non-surviving regions is found,
the program stops and displays an error message to the screen, as
this situation implies that the histogram filling command has been called
{\it after} at least one cut yields a distinction among the associated
regions. This indicates an error in the design of the analysis.

\subsubsection{Finalizing an analysis}
Once all the events have been processed, the program calls
the function \texttt{Finalize}. The user can make use
of it for drawing histograms or
deriving cut-flow charts as indicated in the manual for older versions of the
program~\cite{Conte:2012fm}; however, from the version
of \madanalysis\ introduced in this section onwards,
the \texttt{Finalize} function does not need to be implemented
anymore. Output files written according to
the \texttt{SAF} format (see Section~\ref{ma5t-sec:saf}) are automatically generated.

\subsubsection{Message services}
\label{ma5t-sec:msg_services}

The \cpp\ core of \madanalysis\ includes a class of functions
dedicated to the display of text on the screen at the time of the
execution of the analysis.
Whereas only two distinct levels of message are accessible
by using the standard \cpp\ streamers
(\texttt{std::cout} and \texttt{std:cerr} for normal
and error messages), the \sampleanalyzer\ library enables the user
to print messages that can be classified into four categories.
In this way,
information (the \texttt{INFO} function),
warning (the \texttt{WARNING} function),
error (the \texttt{ERROR} function) and
debugging (the \texttt{DEBUG} function)
messages can be displayed as in the following sample of code,
\begin{verbatim}
 INFO    << "..." << endmsg;
 WARNING << "..." << endmsg;
 ERROR   << "..." << endmsg;
 DEBUG   << "..." << endmsg;
\end{verbatim}
Additionally, warning and error messages provide information on the
line number of the analysis code that is at the source of the message.
The effect of a given message service can finally be modified by
means of the methods presented in Table~\ref{ma5t-tab:msg}.

\subsubsection{Physics services}
\label{ma5t-sec:phys_services}

\renewcommand{\arraystretch}{1.5}%
\begin{table*}
  \centering
  \begin{tabular*}{\textwidth}{@{\extracolsep{\fill}}p{.46\textwidth}p{.52\textwidth}@{}} \hline\hline
    \texttt{mcConfig().AddHadronicId(...)} & Adds a particle species, identified
       via its Particle Data Group code (an integer number given as argument), to the list of
       hadronizing particles. Mandatory for the computation of $H_T$
       and $H_T^{\rm miss}$ in the case of Monte Carlo events (see Section~\ref{ma5t-sec:dataformat}).\\
    \texttt{mcConfig().AddInvisibleId(...)} & Adds a particle species, identified
       via its Particle Data Group code (an integer number given as argument),
       to the list of invisible particles. Mandatory for the computation of $E_T$
       and $E_T^{\rm miss}$ in the case of Monte Carlo events
       (see Section~\ref{ma5t-sec:dataformat}).\\
    \texttt{mcConfig().Reset()} & Reinitializes the lists of invisible and hadronizing particles to empty lists.\\
    \texttt{recConfig().Reset()} & Defines (reconstructed) leptons as isolated when no jet is present in a cone
       of radius $R = 0.5$ centered on the lepton.\\
    \texttt{recConfig().UseDeltaRIsolation(...)} & Defines (reconstructed) leptons as isolated when no jet is present
       in a cone, with a radius given as a floating-point number in argument, centered on the lepton.\\
    \texttt{recConfig().UseSumPTIsolation(...)} & Defines (reconstructed) leptons as isolated when both
       the sum $\Sigma_1$ of the transverse momenta of all tracks in a cone
       (of radius fixed at the level of the detector simulation)
       centered on the lepton is smaller than a specific threshold (the first argument) and
       the amount of calorimetric energy in this cone, relative to $\Sigma_1$,
       is smaller than another threshold (the second argument).
       This uses the information provided
       by the \texttt{sumPT\textunderscore isol()}
       and \texttt{ET\textunderscore PT\textunderscore isol()} methods
       of the \texttt{RecLeptonFormat} class (see Table~\ref{ma5t-tab:RecParticleFormat} and~\ref{ma5t-tab:RecParticleFormat2}).\\
    \texttt{Id->IsBHadron(...)} & Returns a boolean quantity indicating whether an \texttt{MCParticleFormat}
      object passed as argument is a hadron originating from the fragmentation of a $b$-quark.\\
    \texttt{Id->IsCHadron(...)} & Returns a boolean quantity indicating whether an \texttt{MCParticleFormat}
      object passed as argument is a hadron originating from the fragmentation of a $c$-quark.\\
\hline\hline
  \end{tabular*}
  \caption{\small \label{ma5t-tab:physics}Physics service methods.}
\end{table*}
\renewcommand{\arraystretch}{1}%

\renewcommand{\arraystretch}{1.5}%
\begin{table*}
  \centering
  \begin{tabular*}{\textwidth}{@{\extracolsep{\fill}}p{.33\textwidth}p{.65\textwidth}@{}} \hline\hline
    \texttt{Id->IsFinalState(...)} & Returns a boolean quantity indicating whether an \texttt{MCParticleFormat} object
      passed as argument is one of the final-state particles of the considered event.\\
    \texttt{Id->IsHadronic(...)} & Returns a boolean quantity indicating whether an \texttt{MCParticleFormat}
      or a reconstructed object passed as argument yields any hadronic activity in the event.\\
    \texttt{Id->IsInitialState(...)} & Returns a boolean quantity indicating whether an \texttt{MCParticleFormat} object
      passed as argument is one of the initial-state particles of the considered event.\\
    \texttt{Id->IsInterState(...)} & Returns a boolean quantity indicating whether an \texttt{MCParticleFormat} object
      passed as argument is one of the intermediate-state particles of the considered event.\\
    \texttt{Id->IsInvisible(...)} & Returns a boolean quantity indicating whether an \texttt{MCParticleFormat}
      or a reconstructed object passed as argument gives rise to missing energy.\\
    \texttt{Id->IsIsolatedMuon(...)} & Returns a boolean quantity indicating whether a \texttt{RecLeptonFormat}
      object passed as a first argument is isolated within a given reconstructed event, passed as a
      second argument (under the format of a \texttt{RecEventFormat} object).\\
    \texttt{Id->SetFinalState(...)} & Takes an \texttt{MCEventFormat} object as argument and defines
      the status code number associated with final-state particles.\\
    \texttt{Id->SetInitialState(...)} & Takes an \texttt{MCEventFormat} object as argument and defines
      the status code number associated with initial-state particles.\\
    \texttt{Transverse->AlphaT(...)} & Returns the value of the $\alpha_T$ variable~\cite{Randall:2008rw},
      as a floating-point number,
      for a given (Monte Carlo or reconstructed) event passed as argument.\\
    \texttt{Transverse->MT2(...)} & Returns, as a floating-point number, the value of the $M_{T2}$
      variable~\cite{Lester:1999tx,Cheng:2008hk} computed from a system of two visible objects (the first two arguments, any particle
      class being accepted),
      the missing momentum (the third argument) and a test mass (a floating-point number given as the last argument).\\
    \texttt{Transverse->MT2W(...)} & Returns, as a floating-point number, the value of the $M_{T2}^W$
      variable~\cite{Bai:2012gs} computed from a system of jets (a vector of \texttt{RecJetFormat}
      objects in the first argument), a visible particle (given as the second argument, any particle class being accepted) and
      the missing momentum (the third argument). Only available for reconstructed events.\\
\hline\hline
  \end{tabular*}
  \caption{\small \label{ma5t-tab:physics2}(continuation of Table~\ref{ma5t-tab:physics}) Physics service methods.}
\end{table*}
\renewcommand{\arraystretch}{1}%

The \sampleanalyzer\ core includes a series of built-in functions aiming to facilitate
the writing of an analysis from the user viewpoint. More precisely, these
functions are specific for
particle identification or observable calculation and have
been grouped into several subcategories of the \cpp\ pointer
\texttt{PHY\-SICS}. All the available methods are listed in Table~\ref{ma5t-tab:physics} and~\ref{ma5t-tab:physics2},
and we provide, in the rest of this section, a few more details,
together with some illustrative examples.

As mentioned in Section~\ref{ma5t-sec:dataformat}, \madanalysis\ can
compute the (missing) transverse energy and (missing) hadronic transverse
energy associated with a given Monte Carlo event. This calculation however
relies on a correct identification of the invisible and hadronizing particles.
This information must be provided
by means of the \texttt{mcConfig()} category of physics services, as
for instance, in
\begin{verbatim}
 PHYSICS->mcConfig().AddInvisibleId(1000039);
 PHYSICS->mcConfig().AddHadronicId(5);
\end{verbatim}
These intuitive lines of code indicate to the program that the gravitino
(whose Particle Data Group identifier is 1000039)
yields missing energy and that the bottom quark (whose Particle Data
Group identifier is 5) will eventually hadronize.

An important category of methods shipped with the physics services
consists of functions dedicated to the identification of particles
and to the probing of their nature (invisible, hadronizing,
\textit{etc.}). They are collected within the \texttt{Id} structure
attached to the \texttt{PHYSICS} object. For instance (see Table~\ref{ma5t-tab:physics} and~\ref{ma5t-tab:physics2}
for the other methods),
\begin{verbatim}
 PHYSICS->Id->IsInvisible(prt)
\end{verbatim}
allows one to test the (in)visible nature
of the particle referred to by the pointer \texttt{prt}. Also, basic
isolation tests on \texttt{RecLeptonFormat} objects can be performed when analyzing
reconstructed events. Including in the analysis
\begin{verbatim}
 PHYSICS->Id->IsIsolatedMuon(muon, event)
\end{verbatim}
yields a boolean value related to the (non-)isolated
nature of the reconstructed lepton \texttt{muon}, \texttt{event} being here a
\texttt{Rec\-E\-vent\-For\-mat} object.
Two isolation algorithms can be employed.
By default, the program verifies that no reconstructed jet lies in a
cone of radius $R=0.5$ centered on the lepton. The value of
$R$ can be modified via the \texttt{recConfig()} category of physics
services,
\begin{verbatim}
 PHYSICS->recConfig().UseDeltaRIsolation(dR);
\end{verbatim}
where \texttt{dR} is a floating-point variable with the chosen cone size.
The user can instead require the program to tag leptons as isolated when
both the sum of the transverse momenta of all tracks in a
cone (of radius fixed at the level of the detector simulation)
centered on the lepton is smaller than a specific threshold
and when the amount of calorimetric energy in this cone, calculated
relative to the sum of the transverse momenta of all tracks in the cone,
is smaller than another threshold. This uses the information provided
by the \texttt{sumPT\textunderscore isol()}
and \texttt{ET\textunderscore PT\textunderscore isol()} methods
of the \texttt{RecLeptonFormat} class (see Table~\ref{ma5t-tab:RecParticleFormat} and~\ref{ma5t-tab:RecParticleFormat2}) and
can be activated by implementing
\begin{verbatim}
 PHYSICS->recConfig().UseSumPTIsolation(sumpt,et_pt);
\end{verbatim}
where \texttt{sumpt} and \texttt{et\textunderscore pt} are the two mentioned
thresholds.
For more sophisticated isolation tests,
such as those based on the information encompassed
in \texttt{IsolationConeType} objects possibly provided for reconstructed
jets, leptons and photons (see Section~\ref{ma5t-sec:dataformat}), it is left to the user
to manually implement the corresponding routines in his/her analysis.

In addition to identification routines, physics services include
built-in functions allowing one to compute global event
observables, such as several
transverse variables that are accessible through the
\texttt{Transverse} structure attached to the \texttt{PHY\-SICS} object.
More information on the usage of these methods are provided in Table~\ref{ma5t-tab:physics} and \ref{ma5t-tab:physics2}.

\subsubsection{Sorting particles and objects}
In most analyses, particles of a given species are identified
according to an ordering in their transverse momentum or energy.
In contrast, vector of particles as returned after the reading
of an event are in general unordered and therefore need to be sorted.
This can be achieved by means of sorting routines that can be called
following the schematic form:
\begin{verbatim}
  SORTER->sort(parts, crit)
\end{verbatim}
In this line of code, \texttt{parts} is a vector of
(Monte Carlo or reconstructed) objects and \texttt{crit}
consists of the ordering criterion. The allowed choices
for the latter
are \texttt{ETAordering} (ordering in pseudorapidity),
\texttt{ETordering} (ordering in transverse energy),
\texttt{Eordering} (ordering in energy), \texttt{Pordering}
(ordering in the norm of the three-momentum),
\texttt{PTordering} (ordering in the transverse momentum), \texttt{PXordering}
(ordering in the $x$-component of the three-momentum), \texttt{PYordering}
(ordering in the $y$-component of the three-momentum) and
\texttt{PZordering} (ordering in the $z$-component of the three-momentum).
The objects are always sorted in terms of decreasing values of the
considered observable.

\subsection{Compiling and executing the analysis}
\label{ma5t-sec:exec}
In Section~\ref{ma5t-sec:template}, we have pointed out that the \texttt{Build}
subdirectory of the analysis template contains
a \texttt{Makefile} script readily to be used.
In this way, the only task left to the user after having implemented his/her analysis
is to launch this script in a shell, directly from the \texttt{Build} directory.
This leads first to the creation of a library that is stored in the \texttt{Build/Lib}
subdirectory, which includes all the analyses implemented
by the user and the set of classes and methods of the \sampleanalyzer\
kernel. Next, this library is linked to the main program and
an executable named \texttt{MadAnalysis5Job} is generated (and stored in the
\texttt{Build} directory).

The program can be run by issuing in a shell the command
\begin{verbatim}
 ./MadAnalysis5Job <inputfile>
\end{verbatim}
where \texttt{<inputfile>} is a text file with a list of paths to
all event files to analyze. All implemented analyses are sequentially
executed and the results, generated according to the \texttt{SAF}
format (see Section~\ref{ma5t-sec:saf}), are stored in the \texttt{Output}
directory.

\subsection{The structure of the output of an analysis}
\label{ma5t-sec:saf}
As indicated in the previous section, the program stores, after its execution,
the results of the analysis or analyses that have
been implemented by the user in the \texttt{Output} subdirectory of
the working directory. First, a subdirectory with the same name as the input file
(\texttt{<inputfile>} in the schematic example of Section~\ref{ma5t-sec:exec}) is created.
If a directory with this name exists already, the code uses
it without deleting its content. It contains a \texttt{SAF} file (updated if
already existing) with global information
on the analyzed event samples organized following an \texttt{XML}-like syntax that supports positive as well as negative event weights:
\begin{verbatim}
 <SampleGlobalInfo>
  # xsection  xsec_error  nevents  sum_wgt+  sum_wgt-
  0.00e+00    0.00e+00    0        0.00e+00  0.00e+00
 </SampleGlobalInfo>
\end{verbatim}
where we have set the numerical values to zero for the sake of the illustration.
In reality these values are extracted from the event file that is read;
they are kept equal to zero
if not available. In addition, the format includes header and footer tags
(\texttt{SAFheader} and \texttt{SAFfooter}) omitted for brevity.

Secondly, a subdirectory specific to each of the executed analyses is created within
the \texttt{<inputfile>} directory. The name of the subdirectory
is the name of the associated analysis followed by an integer number chosen in such
a way that the directory name is unique. This directory contains a \texttt{SAF}
file with general information
on the analysis (\texttt{name.saf}, \texttt{name} denoting
a generic analysis name), a directory with histograms
(\texttt{Histograms}) and a directory with cut-flow charts (\texttt{Cutflows}).

In addition to a header and a footer, the \texttt{name.saf} file, still encoded
according to an \texttt{XML}-like structure, contains a list
with the names of the regions that have been declared in the analysis implementation.
They are embedded in a \texttt{Re\-gi\-on\-Se\-lec\-ti\-on} \texttt{XML} structure, as in
\begin{verbatim}
 <RegionSelection>
  "MET>200"
  "MET>300"
 </RegionSelection>
\end{verbatim}
when taking the example of Section~\ref{ma5t-sec:init}.

The \texttt{Histograms} subdirectory contains a unique \texttt{SAF} file
with, again in addition to a possible header and footer, all the histograms
implemented by the user. The single histogram declared in Section~\ref{ma5t-sec:init} would
be encoded in the \texttt{SAF} format as in the following self-explanatory lines of code:
\begin{verbatim}
 <Histo>
  <Description>
   "ptl1"
   # nbins        xmin           xmax
   20             50             500
   # associated RegionSelections
   MET>300   # Region nr. 1
  </Description>
  <Statistics>
   0 0 # nevents
   0 0 # sum of event-weights over events
   0 0 # nentries
   0 0 # sum of event-weights over entries
   0 0 # sum weights^2
   0 0 # sum value*weight
   0 0 # sum value^2*weight
   0 0 # sum value*weight^2
 </Statistics>
 <Data>
  0 0 # number of nan
  0 0 # number of inf
  0 0 # underflow
  0 0 # bin 1 / 20
  ...
  0 0 # bin 20 / 20
  0 0 # overflow
  </Data>
</Histo>
\end{verbatim}
where the first and second columns in the \texttt{Statistics} and \texttt{Data} blocks correspond to events with a positive and negative weight, respectively. The dots stand for the other bins that we have omitted for brevity.
Again, for the
sake of the example we have set all values to zero.

Finally, the \texttt{Cutflows} directory contains one
\texttt{SAF} file for each of the declared regions, the filename
being the name of the region followed by the \texttt{saf}
extension. Each of these files contains
the cut-flow chart associated with the considered region encoded
by means of two types of \texttt{XML} tags. The first one is only
used for the initial number of events (\texttt{I\-ni\-ti\-al\-Coun\-ter})
whereas the second one is dedicated to
each of the applied cuts. Taking
the example of the first of the two cuts declared in Section~\ref{ma5t-sec:init},
the \texttt{MET\textunderscore gr\textunderscore200.saf} file (the \texttt{>}
symbol in the region name has been replaced by \texttt{\textunderscore gr\textunderscore})
would read
\begin{verbatim}
 <InitialCounter>
  "Initial number of events"    #
  0          0                  # nentries
  0.00e+00   0.00e+00           # sum of weights
  0.00e+00   0.00e+00           # sum of weights^2
 </InitialCounter>
 <Counter>
  "1lepton"                     # 1st cut
  0          0                  # nentries
  0.00e+00   0.00e+00           # sum of weights
  0.00e+00   0.00e+00           # sum of weights^2
 </Counter>
\end{verbatim}
where again the first and second columns in the \texttt{InitialCounter} and \texttt{Counter} blocks correspond to events with a positive and negative weight, respectively.

\subsection{Modifications to Delphes: Delphes-MA5tune} \label{ma5t-sec:delphestune}

{\tt Delphes}~\cite{deFavereau:2013fsa} is a {\tt C++} framework dedicated to
the simulation of a generic detector such as those used in collider experiments.
Contrary to full detector simulation software, {\tt Delphes} does not simulate
the particle-matter interactions, but uses instead
a parameterization of the detector response and reconstructs the main
physics objects considered in the analyses.
This simplified picture results in much faster simulations, while the accuracy level is 
maintained suitable for realistic phenomenological investigations. 
From the computing side, {\tt Delphes} is a modular framework where developers can
both add their own contributions and tune the default parameterization according
to their needs. This modularity is based on a division of the simulation process
into modules inspired by the \texttt{TTask} \texttt{ROOT} class, and the addition and
removal of new elements are easily achievable through a \texttt{TCL}
configuration file. Similarly, the content of the output
\texttt{ROOT} files can be configured at will.

In order to properly recast ATLAS and CMS analyses, a tuning of the version 3 of
{\tt Delphes} has been performed.
In the original version of {\tt Delphes}, an isolation criterion is applied
to both leptons and photons, and only particles satisfying this requirement
are stored in the output files. We have designed a new {\tt Delphes} module
named \texttt{CalculationIsolation} that allows one to move the isolation
requirements in the analysis selection. 
This module computes several variables useful for the implementation of
isolation cuts. Defining cone sizes of $\Delta R=0.2, 0.3, 0.4$ and $0.5$,
the number of tracks with a transverse momentum larger than a given
threshold, the scalar sum of the transverse momenta of these
tracks and the scalar sum of the calorimetric transverse energy deposits
lying in the cones are evaluated and saved. In addition, the 
default module of {\tt Delphes} dedicated to the
filtering of non-i\-so\-la\-ted lepton and photon candidates
is switched off so that all candidates are kept in the output \texttt{ROOT}
files. For consistency reasons, the {\tt Delphes} module \texttt{U\-ni\-que\-Ob\-ject\-Find\-er}
giving a unique identification to all reconstructed objects is bypassed.
Isolation selection cuts can then be performed at the analysis
level by means of the \texttt{isolCones} method of the
\texttt{RecLeptonFormat} class of \madanalysis, described in Section~\ref{ma5t-sec:dataformat}.

Adding the isolation information to the output format yields an
increase of the size of the output files. A cleaning of all collections is
therefore in order to reduce the file sizes. First, collections such as
calorimeter towers and particle-flow objects are not stored. Next,
the (heavy) collection of all particles that have been generated at the
different level of the simulation chain (hard scattering process, parton
showering and hadronization) is pruned, while all reconstructed objects are kept. Only particles produced
at the hard-scattering process level, as well as final-state leptons
and $b$ quarks present after parton showering, are stored. In addition,
the relations between generated and reconstructed leptons have been
retained, together with information on the origin (the mother
particle) of each lepton. All these changes result in a reduction
of the size of the produced \texttt{ROOT} files by about a factor of ten when
compared to the files produced with the original configuration of {\tt Delphes}.

This tailored version of {\tt Delphes~3}, which we internally call
{\tt Delphes-MA5tune} to avoid confusion with the original version,
can conveniently be installed locally from the interpreter of \madanalysis\  by
typing in the command
\begin{verbatim}
  install delphesMA5tune
\end{verbatim}
Even if {\tt Delphes~3} is already installed on a given system, one will
need this modified `MA5tune' version of the program in order to run the
\madanalysis\ analyses present in the public database~\cite{padwiki}. 
Note however that for the moment  \madanalysis\ is  not able to run with 
both {\tt Delphes}\ and {\tt Delphes-MA5tune}  installed in parallel.  
This means that  
the user must take care 
that only the directory \texttt{tools/delphesMA5tune} (but not the directory \texttt{tools/delphes}) 
be available in his/her local installation of \madanalysis.

In order to process an (hadronized) event sample with
the `MA5tune' of {\tt Delphes}, it is sufficient to start \madanalysis\
in the reconstructed mode, import the considered sample and type
\begin{verbatim}
  set main.fastsim.package = delphesMA5tune
  set main.fastsim.detector = cms
  submit
\end{verbatim}
where \texttt{cms} can be replaced by \texttt{atlas} according to the needs of
the user. Default detector parameters are employed and can be modified by the user, following
the guidelines displayed on the screen. The output \texttt{ROOT} file can then be
retrieved from the automatically generated working directory.

\subsection{Limit setting} \label{ma5t-sec:limitsetting}

For the statistical interpretation of the results obtained when recasting an analysis, we provide on~\cite{padwiki} {\tt exclusion\_CLs.py}, a {\tt Python} code 
for computing exclusions using the $\mathrm{CL}_s$ prescription~\cite{Read:2002hq}.\footnote{The
{\tt Python} code requires {\tt SciPy} libraries to be installed.}
This code can also be installed on a user system by typing in, from the \madanalysis\ interpreter, the command
\begin{verbatim}
 install RecastingTools
\end{verbatim}
which results in the file \texttt{exclusion\_CLs.py} being present at the
root of any working directory created in the expert mode of \madanalysis. 

The \texttt{exclusion\_CLs.py} code takes as input the acceptance$\times$efficiency information 
from the cut flow {\tt SAF} files generated when executing
an analysis implemented in \madanalysis\ (see Section~\ref{ma5t-sec:saf}).
Moreover, an {\tt XML} file named {\tt analysis\_name.info} (\texttt{analysis\_name} stands for a generic
analysis name), needs to be provided by the user in the {\tt Build/Sam\-ple\-Analyzer/U\-ser/A\-na\-ly\-zer} directory, 
specifying the luminosity {\tt <lumi>}, 
the number of observed events {\tt <nobs>}, 
the nominal number of expected SM background events {\tt <nb>}, 
and its uncertainty at 68\%~CL {\tt <deltanb>} in each of the regions, as given in the experimental publication.
The syntax of this file is as follows:

\begin{verbatim}
 <analysis id="cms_sus_13_011">
   <lumi>19.5</lumi> <!-- in fb^-1 -->

   <region type="signal" id="SRname">
     <nobs>227</nobs>
     <nb>251</nb>
     <deltanb>50</deltanb>
   </region>
   ...
   ...
 </analysis>  
\end{verbatim}

\noindent
The attribute {\tt type} of the root tag {\tt <analysis>} can be {\tt signal} or {\tt control} and is optional (the default value is {\tt signal}). 
The {\tt id} of each {\tt <region>} tag has to match the exact name of the SR used in the analysis code. 
When results are given after combining several SRs (for example, for same-flavor leptons instead of $ee$ and $\mu\mu$ separately), the relevant SRs should all be listed  in the attribute {\tt id} separated by semicolons (without extra space). 
Taking the example of the ATLAS analysis that will be presented in Section~\ref{sec:atlasvalid}, this would read

\begin{verbatim}
 <region id="MT2-90 ee;MT2-90 mumu">
\end{verbatim}

The last piece of information essential for calculating exclusions is the signal cross section.
It can be provided by the user in the {\tt SAF} file {\tt mypoint.txt.saf} (automatically generated when
executing an analysis, see Section~\ref{ma5t-sec:saf}), 
where {\tt mypoint.txt}, stored in the \texttt{Input} folder of the working directory, 
is the input file for running the analysis under consideration. 
Alternatively, the cross section can be given as argument when calling {\tt exclusion\_CLs.py}.
Concretely, the limit-setting code is called as
\begin{verbatim}
 ./exclusion_CLs.py analysis_name mypoint.txt \
         [run_number] [cross section in pb]
\end{verbatim}
where the run number and cross section value are optional arguments. 
The run number $x$ (default zero) identifies the output directory to use, 
as each execution of the analysis code yields the creation of a new output directory, 
\texttt{analysis\_name\_}$x$, for the $x^{\rm th}$ execution of the analysis code (starting from 0). 

The procedure of {\tt exclusion\_CLs.py} starts by selecting
the most sensitive SR ({\it i.e.}, the one that yields the best expected exclusion, assuming that
the number of observed events is equal to the nominal number of background events).
This is a standard procedure at the LHC whenever the SRs defined in the analysis are overlapping; 
here we use it as the default for all analyses. 
Then the actual exclusion is calculated, and the confidence level with which the tested
scenario is excluded using the CL$_s$ prescription~\cite{Read:2002hq} is printed on the screen 
together with the name of the most sensitive SR. The same information is also
stored in the file {\tt a\-na\-ly\-sis\_na\-me\_}$x${\tt .out} located in the working directory 
of the  \texttt{Output} folder.
Last but not least, if a negative number is given for the cross section, the code returns instead the nominal cross section
that is excluded at 95\%~CL, computed using a root-finding algorithm.

The core of the calculation works as follows. First, the number of signal events ($n_s$) is obtained as the
product of the luminosity, signal cross section and acceptance$\times$efficiency for the SR of interest.
This is used, together with the number of observed events ($n_{\rm obs}$) and the nominal number of background
events ($\hat n_b$) and its uncertainty ($\Delta n_b$) to compute the exclusion.
A large number of toy MC experiments ($10^5$ by default) are then generated from the Poisson distribution
${\rm poiss}(n_{\rm obs} | n_{\rm expected})$, 
corresponding to the distribution of the total number of events in the SR under the
background-only hypothesis on the one hand ($n_{\rm expected} = n_b$), and under the
signal $+$ background hypothesis ($n_{\rm expected} = n_s + n_b$) on the other hand.
We assume that the uncertainty on the number of background events is modeled as ${\rm gauss}(\hat n_b, \Delta n_b)$,
and for each toy MC the number of background events $n_b$ is randomly generated from this normal distribution.
Under the two different hypotheses, $p$-values are then calculated using the number of events actually observed at the LHC, and finally used to compute the CL$_s$ value.

\subsection{Conclusions}
\label{ma5t-sec:conclusions}
We have presented a major extension of
the expert mode of the \madanalysis\ package. Both designing a prospective
new physics analysis and recasting an experimental search featuring multiple
signal regions can now be achieved in a user-friendly fashion that relies
on a powerful handling of regions, histogramming and selection cuts.


\section{CMS search for stops in the single-lepton final state} \label{sec:cmsvalid}

The CMS search for stops in the single lepton and missing energy, $\ell + E^{\rm miss}_T$, final state with full luminosity at
$\sqrt{s} = 8$~TeV~\cite{Chatrchyan:2013xna} has been taken as a ``template analysis'' to develop a common language and framework for the analysis implementation. It also allowed us to test the new developments in \madanalysis\ which were necessary for carrying out this project.

\begin{figure*}[ht]\centering
\includegraphics[width=6cm]{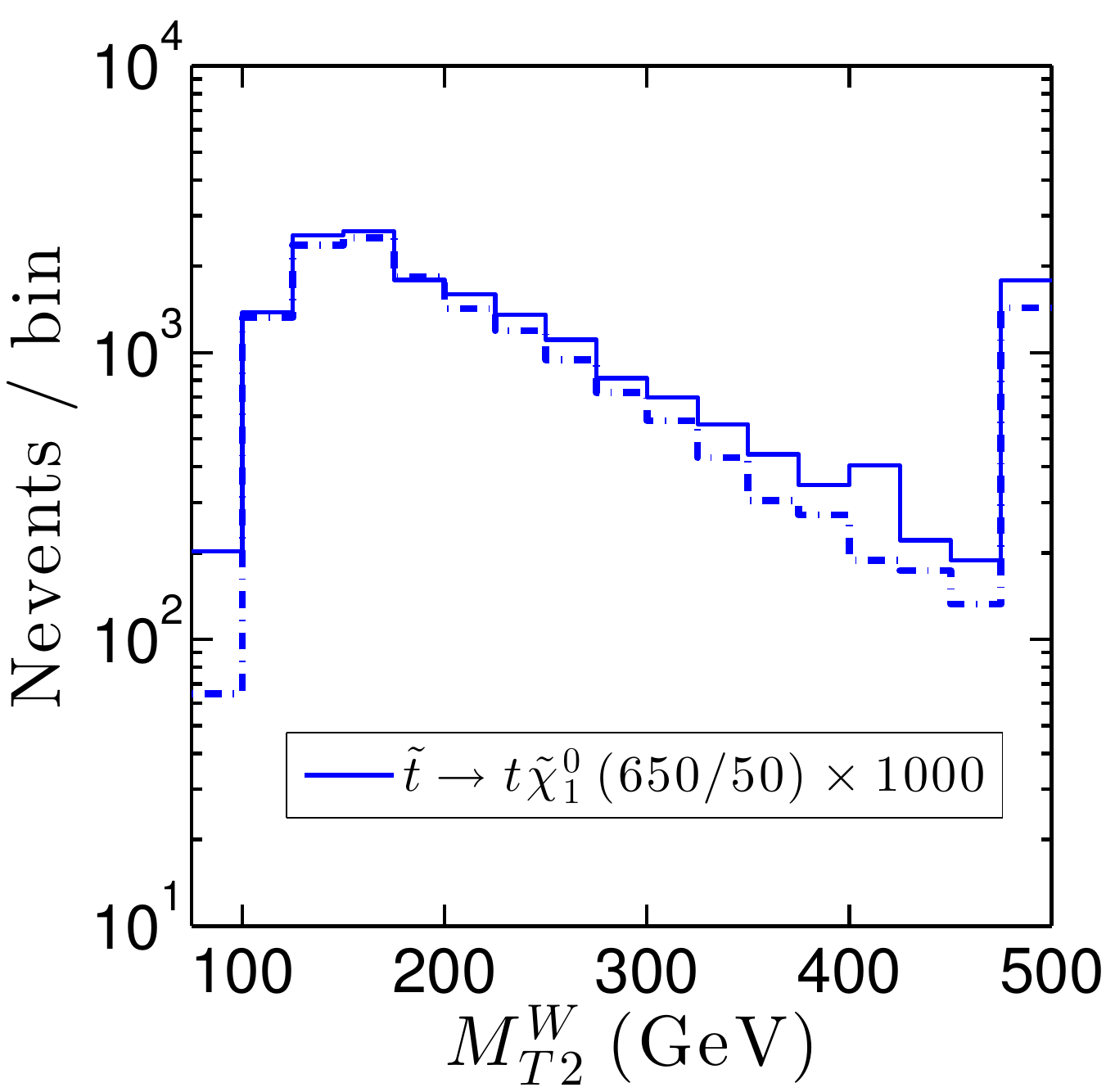}\quad
\includegraphics[width=6cm]{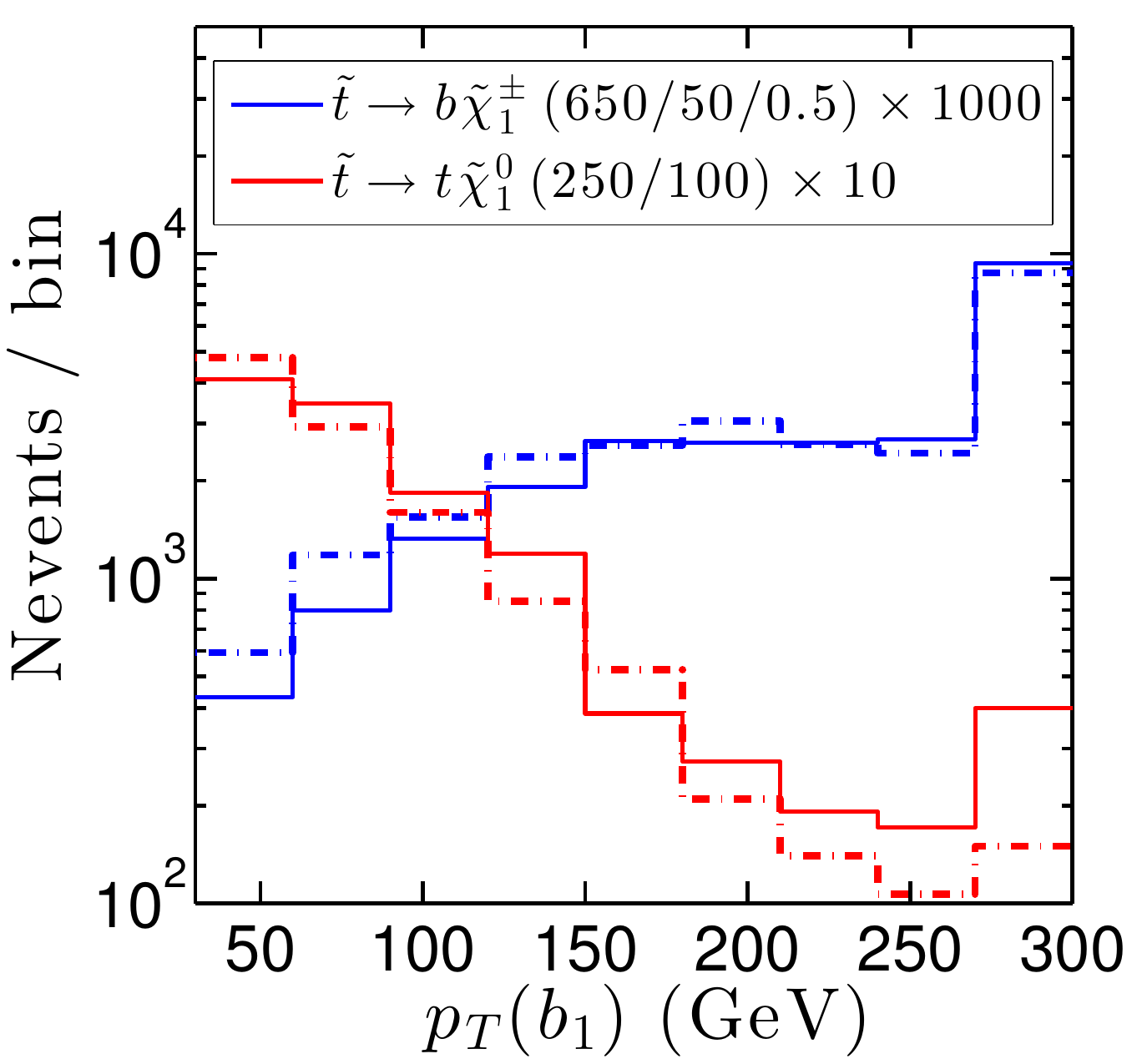}
\caption{Distributions of the kinematic variable $M^W_{T2}$ (left) and of the $p_T$ of the leading $b$-tagged jet (right) after the preselection cuts of the analysis CMS-SUS-13-011. The solid lines are obtained from our re-interpretation within \madanalysis, while the dash-dotted lines correspond to the CMS results, given in Fig.~2 of~\cite{Chatrchyan:2013xna}. See captions of Tables~\ref{tab:NevStopbcharginoLowDeltaMMET2501} and \ref{tab:NevStopTneutralinoLowDeltaMMET2502} for the notation of the benchmark points.} \label{fig:kinvarsus13011}
\end{figure*}

The analysis targets two possible decay modes of the stop: $\tilde{t} \to t \tilde{\chi}^{0}_1$ (or $\tilde{t} \to bW \tilde{\chi}^{0}_1$ when $m_{\tilde t_1} < m_{\tilde \chi^0_1} + m_t$) and 
$\tilde{t} \to b \tilde{\chi}^{\pm}_1$.  
Since the stops are pair-produced, their decays give rise to two $W$ bosons in each event, one of which is assumed to  decay leptonically, whilst the other one is assumed to decay hadronically. 
In the cut-based version of the analysis, two sets of signal regions with different cuts, each  dedicated to one of the two decay modes, are defined.\footnote{The search also contains an analysis based on multivariate analysis techniques (MVA); such analyses generically cannot be externally reproduced unless the final MVA is given. As this is not the case so far, we here only use the cut-based version of the analysis.} These two sets are further divided into ``low $\Delta M$'' and ``high $\Delta M$'' categories, targeting small and large mass differences with the lightest neutralino $\tilde\chi_1^0$, respectively. Finally, each of these four categories are further sub-divided using four different $E^{\rm miss}_T$ requirements. In total, 16 different, potentially overlapping SRs are defined. 
Two cuts are based on rather complex and specific kinematic variables designed to reduce the dilepton $t\bar{t}$ background: a $\chi^2$ resulting from the full reconstruction of the hadronic top and $M^W_{T2}$---a variant of the $m_{T2}$ observable. The implementation of the $\chi^2$ quantity in our code was straightforward thanks to the {\tt C++} {\tt ROOT} code provided on the CMS Twiki page.  
The $M_{T2}^W$ variable is calculated with the standard \madanalysis\ method, see~\cite{Conte:2014zja}, 
according to the algorithm presented in~\cite{Bai:2012gs}. 

Overall, this analysis is very well documented. Some important pieces of information were however missing, in particular the detailed trigger efficiencies and the identifi\-cation-only efficiencies for electron and muons. These were provided by the CMS collaboration upon request and are now available on the analysis Twiki page~\cite{cms-sus-13-011-twiki} in the section ``Additional Material to aid the Phenomenology Community with Reinterpretations of these Results''. In addition,  the $b$-tagging efficiency as a function of $p_T$ is not given in the paper, but was taken from~\cite{Chatrchyan:2013fea}.
Another technical difficulty came from the isolation criteria. Indeed, the CMS analysis considers the sum of transverse momenta of so-called `Particle Flow' particles in a cone of given $\Delta R$. This is difficult to reproduce in our case. Instead, we only use tracks in the inner detector for the isolation. From the two benchmark points for which cut flows are available (see Table~\ref{tab:cms-13-011-cutflow}) we found that a weighting factor of $0.885$, applied on the events at the same time as the isolation, is sufficient to correct our track-only isolation. Therefore we incorporate this correction to our analysis code.

\begin{table}[ht]
\begin{center}
  \begin{tabular}{l|c|c}
  benchmark point & CMS result & {\tt MA}5 result \\ 
  \hline\noalign{\smallskip}
        \multicolumn{3}{c}{$\tilde{t} \rightarrow b \tilde{\chi}^{\pm}_1, {\rm low\,} \Delta M, E^{\rm miss}_T > 150~{\rm GeV}$} \\ 
  $(250/50/0.5)$ & $157 \pm 9.9$ & $141.2$ \\ 
  $(250/50/0.75)$ & $399 \pm 18$ & $366.8$ \\ 
  \noalign{\smallskip}
  \hline\noalign{\smallskip}
        \multicolumn{3}{c}{$\tilde{t} \rightarrow b \tilde{\chi}^{\pm}_1, {\rm high\,} \Delta M, E^{\rm miss}_T > 150~{\rm GeV}$} \\ 
  $(450/50/0.25)$ & $23 \pm 2.3$ & $23.4$ \\ 
  \noalign{\smallskip} 
  \hline\noalign{\smallskip}
        \multicolumn{3}{c}{$\tilde{t} \rightarrow b \tilde{\chi}^{\pm}_1, {\rm high\,} \Delta M, E^{\rm miss}_T > 250~{\rm GeV}$} \\ 
  $(600/100/0.5)$ & $6.1 \pm 0.5$ & $5.4$ \\ 
  $(650/50/0.5)$ & $6.7 \pm 0.4$ & $5.8$ \\ 
  $(650/50/0.75)$ & $6.3 \pm 0.4$ & $5.7$ \\ 
  \noalign{\smallskip}\hline
\end{tabular}
\caption{Final number of events for $\tilde{t} \rightarrow b \tilde{\chi}^{\pm}_1$  in three SRs of the analysis
CMS-SUS-13-011. The benchmark points are given in the format 
$(m_{\tilde t},m_{\tilde\chi^0_1},x)$ in GeV, with $x$ setting the chargino mass according to 
$m_{\tilde{\chi}^\pm_1} = x \cdot m_{\tilde{t}} + (1 - x) m_{\tilde{\chi}^0_1}$.
\label{tab:NevStopbcharginoLowDeltaMMET2501}}
\end{center}
\end{table}

\begin{table}[ht]
\begin{center}
\begin{tabular}{c|c|c}
  benchmark point & CMS result & {\tt MA}5 result \\
  \hline\noalign{\smallskip}
        \multicolumn{3}{c}{$\tilde{t} \rightarrow t \tilde{\chi}^0_1, {\rm low\,} \Delta M, E^{\rm miss}_T > 150~{\rm GeV}$} \\ 
  $(250/50)$ & $108 \pm 3.7$ & $100.1$ \\ 
\noalign{\smallskip}
\hline\noalign{\smallskip}
        \multicolumn{3}{c}{$\tilde{t} \rightarrow t \tilde{\chi}^0_1, {\rm high\,} \Delta M, E^{\rm miss}_T > 300~{\rm GeV}$} \\ 
  $(650/50)$ & $3.7 \pm 0.1$ & $3.6$ \\ 
  \noalign{\smallskip} 
  \hline\noalign{\smallskip}
\end{tabular}
\caption{Final number of events for $\tilde{t} \rightarrow t \tilde{\chi}^0_1$ in two SRs of the analysis CMS-SUS-13-011. 
For each benchmark point, the first number indicates the stop mass, the second the LSP mass (in GeV).
\label{tab:NevStopTneutralinoLowDeltaMMET2502}}
\end{center}
\end{table}

The validation of the reimplementation of the analysis can be done using the eleven benchmark points 
presented in the experimental paper: four for the ``T2tt'' simplified model (in which the stop always decays as $\tilde{t} \to t \tilde{\chi}^{0}_1$), and seven for the ``T2bW'' simplified model (in which the stop always decays as $\tilde{t} \to b \tilde{\chi}^{\pm}_1$), with different assumptions on the various masses. The distributions of the kinematic variables used in the analysis are given in Fig.~2 of~\cite{Chatrchyan:2013xna} after the preselection cuts, with at least one benchmark point for illustration. Also provided are the corresponding histograms after the \mbox{$M_T > 120$~GeV} cut, as supplementary material on the CMS Twiki page \cite{cms-sus-13-011-twiki}. We use this information, together with the final number of events in the individual SRs ({\it i.e.}, after all selection cuts) for given benchmark points provided in Tables~4 and 6 of~\cite{Chatrchyan:2013xna}. 

The validation material both before and after cuts defining the SRs is truly valuable information since one can separately check on the one hand the implementation of the kinematic variables and the preselection/cleaning cuts, 
and on  the other hand the series of cuts defining the SRs. Furthermore, the large number of benchmark points allows us to check in detail the quality of the reimplementation in complementary regions of phase space. 

The validation process was based on (partonic) event samples, in LHE format~\cite{Boos:2001cv,Alwall:2006yp}, provided by the CMS collaboration. 
The provision of such event files greatly reduced the uncertainties in the first
stage of validation since it avoided possible differences in the configuration of the used
Monte Carlo tools. In the case of this CMS analysis, the setup of {\tt MadGraph~5}~\cite{Alwall:2011uj}---the event generator employed for generating the necessary hard scattering matrix elements---is crucial, in particular
with respect to the merging of samples with different (parton-level) jet multiplicities. 
The LHE files were passed through {\tt Pythia~6.4}~\cite{Sjostrand:2006za} for parton showering and hadronization, then processed by our modified version of {\tt Delphes~3} (see Ref.~\cite{Dumont:2014tja}) for the simulation of the detector effects. The number of events after cuts and histograms produced by \madanalysis\ were then normalized to the correct luminosity after including cross sections at the next-to-leading
order and next-to-leading logarithmic (NLO+NLL) accuracy~\cite{Kramer:2012bx}, as tabulated by the LHC SUSY Cross Section Working Group~\cite{8tevxs_susy}.

\begin{table*}[ht]
\begin{center}
\begin{tabular}{ l ||c|c||c|c}
\hline\noalign{\smallskip}
& \multicolumn{2}{|c||}{$m_{\tilde t}=650$~GeV} & \multicolumn{2}{c}{$m_{\tilde t}=250$~GeV}  \\
cut & CMS result & {\tt MA}5 result & CMS result & {\tt MA}5 result \\ 
\hline\noalign{\smallskip}
$1\ell\, + \ge 4{\rm jets} + E_T^{\rm miss}>50$~GeV & $31.6\pm0.3$ & $29.0$ & $8033.0\pm38.7$ &  $7365.0$  \\ 
+ $E_T^{\rm miss}>100$~GeV   & $29.7\pm0.3$ & $27.3$ & $4059.2\pm 27.5$  & $3787.2$ \\
+ $n_b\ge1$        & $25.2\pm0.2$ & $23.8$ & $3380.1\pm25.1$ & $3166.0$ \\
+ iso-track veto   & $21.0\pm0.2$ & $19.8$ & $2770.0\pm22.7$ & $2601.4$ \\
+ tau veto             & $20.6\pm0.2$ & $19.4$ & $2683.1\pm22.4$ & $2557.2$ \\
+ $\Delta\phi_{\rm min}>0.8$  & $17.8\pm0.2$ & $16.7$ & $2019.1\pm19.4$ & $2021.3$ \\
+ hadronic $\chi^2<5$  & $11.9\pm0.2$ & $9.8$ & $1375.9\pm16.0$ & $1092.0$ \\
+  $M_T>120$~GeV & $9.6\pm0.1$ & $7.9$ & $355.1\pm8.1$ & $261.3$ \\
${\rm high\,} \Delta M, E^{\rm miss}_T > 300~{\rm GeV}$ & $4.2\pm0.1$ & $3.9$ & --- & ---\\
${\rm low\,} \Delta M, E^{\rm miss}_T > 150~{\rm GeV}$ & --- & --- & $124.0\pm4.8$ & $107.9$\\
\noalign{\smallskip}\hline
\end{tabular}
\caption{Summary of yields for the $\tilde{t} \rightarrow t \tilde{\chi}^0_1$ model for two benchmark points with 
$m_{\tilde\chi^0_1}=50$~GeV, as compared to official CMS-SUS-13-011 results given on \cite{cms-sus-13-011-twiki}. 
The next-to-last (last) line corresponds to the most sensitive
signal region for the benchmark point with $m_{\tilde t}=650$ (250) GeV as in the official CMS cut flow, while all other cuts are common to all signal regions targeting the $\tilde{t} \to t \tilde\chi^0_1$ decay mode. The uncertainties given for the CMS event numbers are statistical only.  In contrast to Tables~\ref{tab:NevStopbcharginoLowDeltaMMET2501} and \ref{tab:NevStopTneutralinoLowDeltaMMET2502}, no trigger efficiency or ISR reweighting is applied here. 
See \cite{cms-sus-13-011-twiki} for more details on the definition of the cuts.  
\label{tab:cms-13-011-cutflow}}
\end{center}
\end{table*}

Some examples of histograms reproduced for the validation are shown in Fig.~\ref{fig:kinvarsus13011}. The shapes of the distributions shown---as well as all other distributions that we obtained but do not show here---follow closely the ones from CMS, which indicates the correct implementation of the analysis and all the kinematic variables. 
(Note that discrepancies in bins where the number of events is relatively small, as seen on a logarithmic scale, suffers from larger statistical uncertainties and hence should not be over-interpreted.)
The expected yields for several benchmark points in their relevant SRs are given in Tables~\ref{tab:NevStopbcharginoLowDeltaMMET2501} and~\ref{tab:NevStopTneutralinoLowDeltaMMET2502}. 
The agreement is good for all tested benchmark points.

Upon our request, the CMS SUSY group furthermore provided detailed cut-flow tables, which are now also available at \cite{cms-sus-13-011-twiki}. 
These proved extremely useful because they allowed us to verify our implementation step-by-step in the analysis. 
A comparison of our results with the official CMS ones is given in Table~\ref{tab:cms-13-011-cutflow}. 
(Note that here no trigger efficiency or initial state radiation, ISR, reweighting is applied.)  
For both cases shown, CMS results are reproduced within about 20\%.  
On the whole, we conclude that our implementation gives reasonably accurate results 
(to the level that can be expected from fast simulation) and declare it as validated. 
As mentioned, the \madanalysis\ code for this analysis, including extensive comments, is published as \cite{ma5code:cms-sus-13-011}.
More detailed validation material, including extra histograms and validation of the limit-setting procedure (see Section~\ref{ma5t-sec:limitsetting}), is available at~\cite{padwiki}.


\section[ATLAS search for electroweak-inos and sleptons in the di-lepton final state]{ATLAS search for electroweak-inos and sleptons in the di-lepton final state%
\sectionmark{ATLAS SUSY search in the di-lepton final state}}
\sectionmark{ATLAS SUSY search in the di-lepton final state} \label{sec:atlasvalid}

We consider the ATLAS search for the electroweak production of 
charginos, neutralinos and sleptons in final states with two leptons (electrons and muons) 
and missing transverse momentum based on $20.3$~fb$^{-1}$  of data at 8 TeV~\cite{Aad:2014vma}.
The event selection requires two signal leptons of opposite charge, with $p_T > 35$~GeV and $p_T > 20$~GeV. 
Two kind of final states are considered:  same flavor (SF = $e^+e^-$ or $\mu^+\mu^-$) and 
different flavors (DF = $e^\pm\mu^\mp$).

Three types of signal regions are defined in this analysis. First, the $m_{T2}$ and $WW$ signal regions
require the invariant mass of the lepton pair to be outside the $Z$ window, and jets are vetoed.
The $m_{T2}$ signal regions (SR-$m_{T2}$) target direct slepton-pair production and chargino-pair production
followed by slepton-mediated decays. 
Each $m_{T2}$ signal region is defined by its threshold on the $m_{T2}$ (``stransverse mass'') variable~\cite{Lester:1999tx,Cheng:2008hk} that is used for reducing the $t\bar t$ and $Wt$ backgrounds: $m_{T2} > 90$, $> 120$ and $> 150$~GeV, for SR-$m^{90}_{T2}$, SR-$m^{120}_{T2}$, and SR-$m^{150}_{T2}$, respectively. The implementation of this requirement is straightforward as the $m_{T2}$ variable is available as a standard method in \madanalysis.

Next, the $WW$a, $WW$b and $WW$c signal regions (referred to as SR-$WW$) are designed to provide sensitivity
to $\tilde\chi_1^+\tilde\chi_1^-$ production followed by leptonic $W$ decays. Each of these three regions is optimized for a given kinematic configuration, using cuts on the invariant mass and/or transverse momentum of the lepton pair ($m_{\ell\ell}$ and $p_{T,\ell\ell}$, respectively), possibly combined with cuts on $m_{T2}$ and on the ``relative missing transverse momentum'' $E_T^{\rm miss,rel}$. Here, $E_T^{\rm miss,rel}$ is defined as the missing transverse momentum $E_T^{\rm miss}$, multiplied by $\sin \Delta \phi_{\ell,j}$ (where $\Delta \phi_{\ell,j}$ is the azimuthal angle between the direction of ${\bf p}_T^{\rm miss}$ and that of the closest lepton or jet) if $\Delta \phi_{\ell,j}$ is below $\pi/2$. This modified $E_T^{\rm miss}$ aims at suppressing events where missing transverse momentum is likely to come from mis-measured jets and leptons.  

\begin{figure*}[!th]\centering
\includegraphics[width=5.9cm]{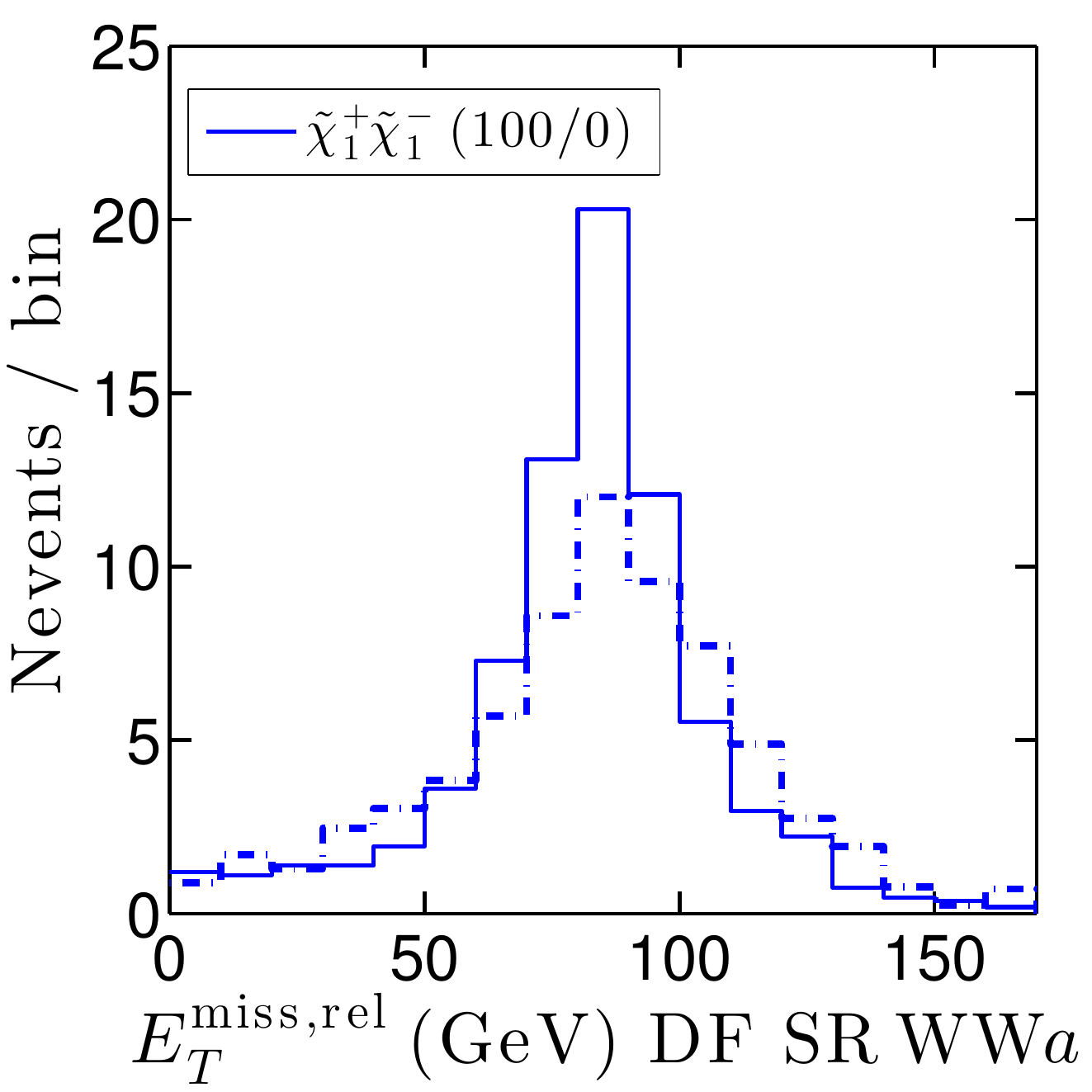}\qquad
\includegraphics[width=6cm]{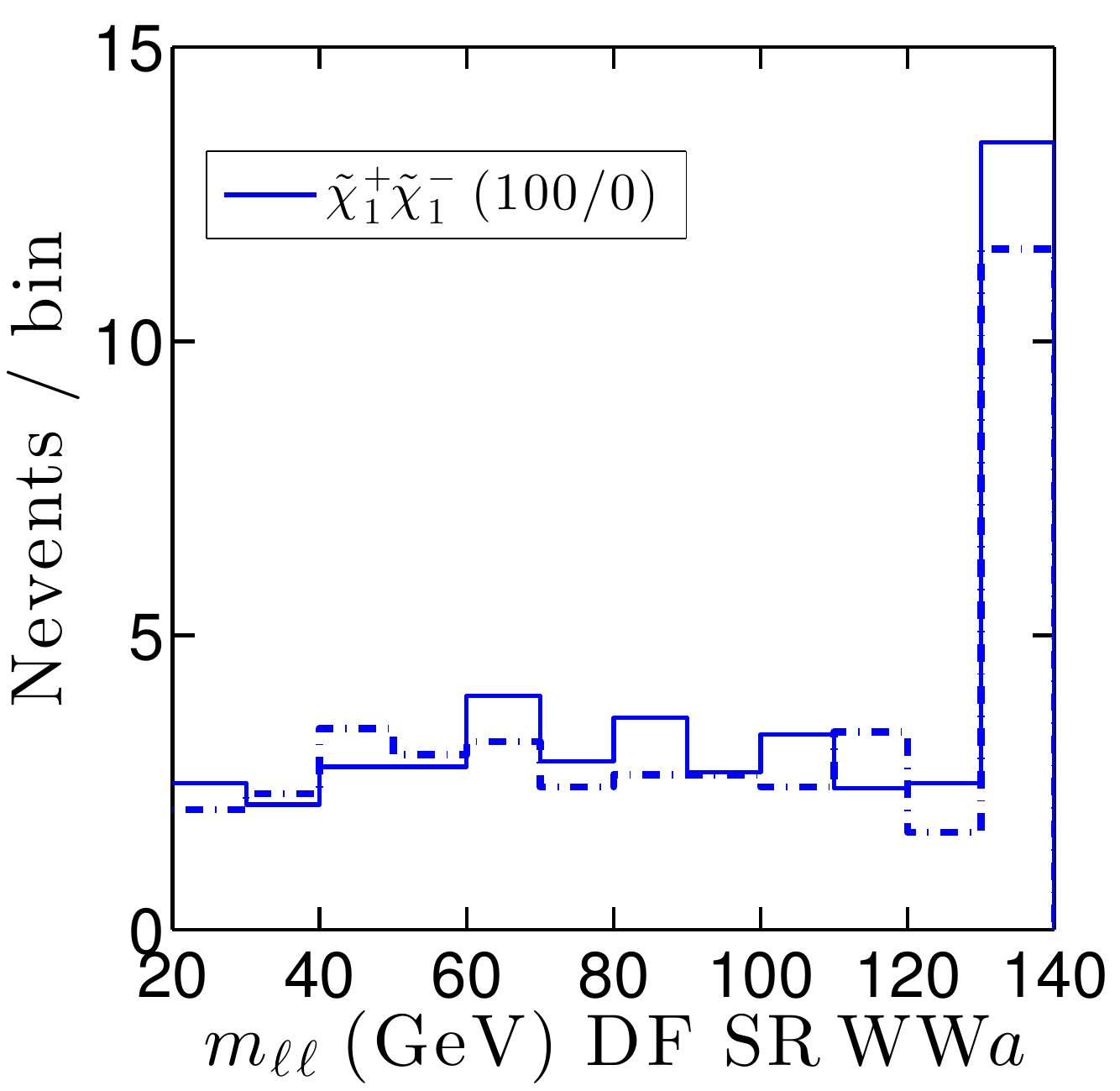}
\vspace*{1mm}
\caption{Distributions of $E_T^{\rm miss,rel}$ (left) and $m_{\ell\ell}$ (right) in the DF SR~$WW{\rm a}\,ee$ 
of ATLAS-SUSY-2013-11, for the benchmark point with $(m_{\tilde{\chi}^{\pm}_1}, m_{\tilde{\chi}^0_1})=(100,0)$~GeV,  
after all cuts except the ones on $m_{\ell\ell}$ and on $E_T^{\rm miss,rel}$ (left), or all cuts except the one on $m_{\ell\ell}$ (right). The solid lines are obtained from our re-interpretation within \madanalysis, while the dash-dotted lines correspond to the official ATLAS results in~\cite{Aad:2014vma}.} 
\label{fig:atlas-11-figA}
\end{figure*}

\begin{figure*}[!th]\centering
\includegraphics[width=6cm]{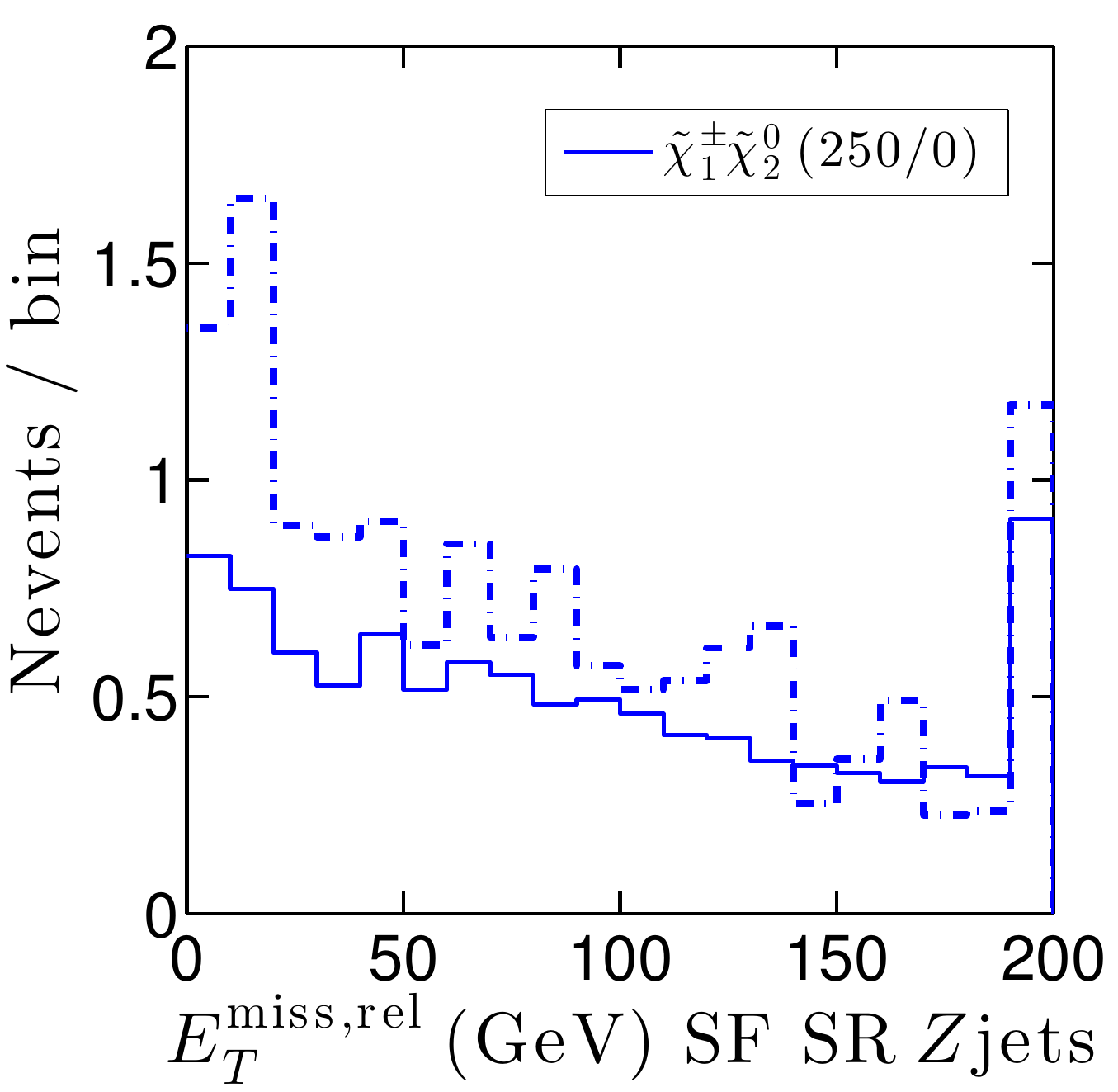}\qquad
\includegraphics[width=6cm]{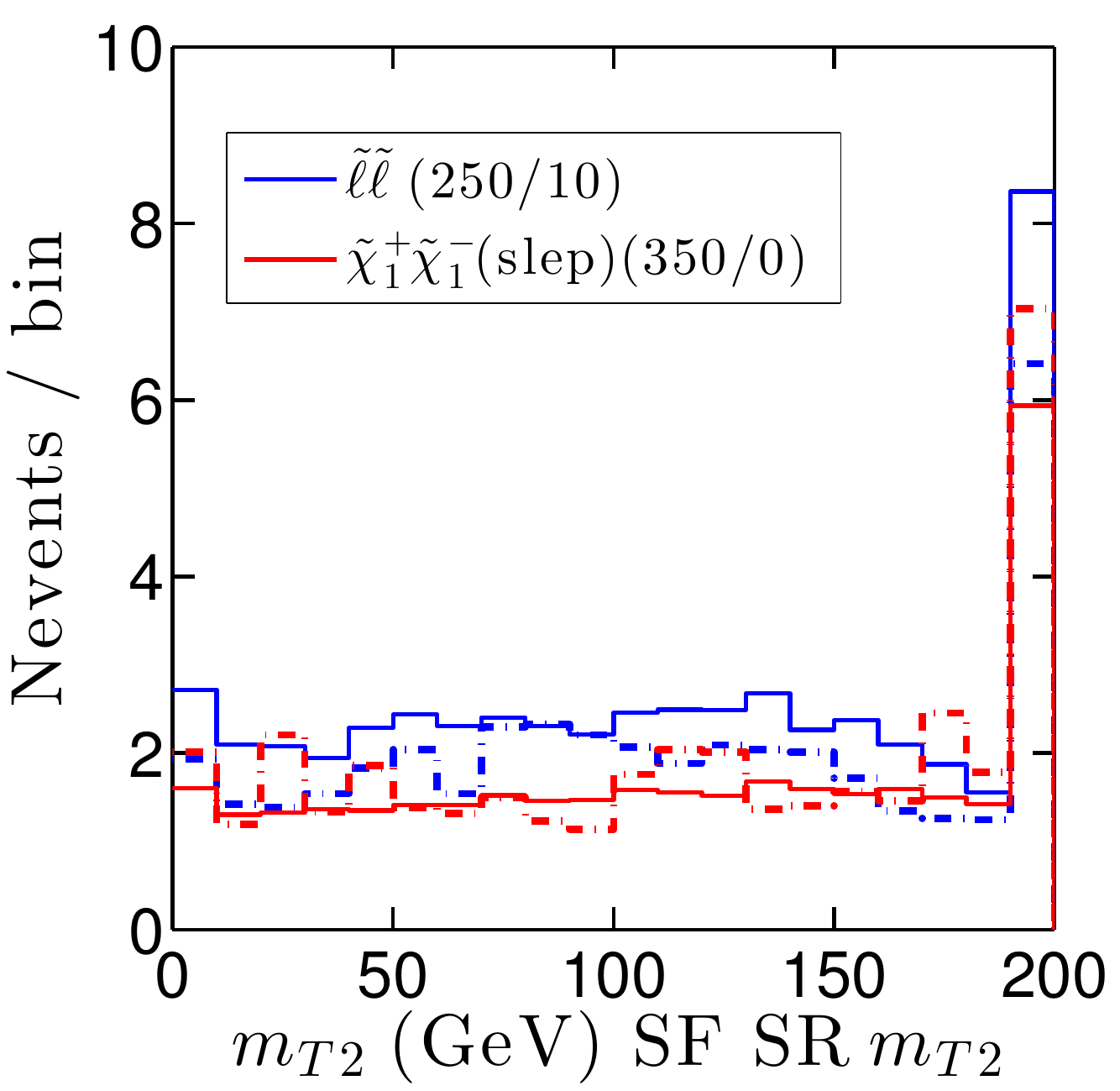}
\vspace*{1mm}
\caption{Distributions of $E_T^{\rm miss,rel}$ in the SF SR~$Z{\rm jets}$ (left) and $m_{T2}$ in the SF SR~$m_{T2}$ (right)  of ATLAS-SUSY-2013-11, after all cuts except the one on the variable plotted. The solid lines are obtained from our re-interpretation within \madanalysis, while the dash-dotted lines correspond to the official ATLAS results in~\cite{Aad:2014vma}.} 
\label{fig:atlas-11-figB}
\end{figure*}

Finally, the $Z$jets signal region (SR-$Z$jets) targets $\tilde\chi^\pm_1 \tilde\chi^0_2$ production, followed by $\tilde\chi^\pm_1 \to W^\pm \tilde\chi^0_1$ and $\tilde\chi^0_2 \to Z \tilde\chi^0_1$, with 
hadronic $W$ and leptonic $Z$ decays. Unlike in the other regions, jets are not vetoed; instead at least two central ``light'' jets (non-$b$-tagged with $|\eta| < 2.4$) are required.
In addition to $m_{\ell\ell}$ being consistent with leptonic $Z$ decays, requirements are made on $E_T^{\rm miss,rel}$, $p_{T,\ell\ell}$, on the invariant mass of the two leading jets ($m_{jj}$) and on the separation between the two leptons ($\Delta R_{\ell\ell}$) in order to suppress, in particular, the $Z$ + jets background.

All signal regions separately consider SF and DF leptons, except SR-$Z$jets where only SF leptons are considered. In total, 20 potentially overlapping signal regions are defined (considering $ee$ and $\mu\mu$ signal regions separately, as required for comparison with the official ATLAS cut flows). Detailed electron efficiencies as a function of $p_T$ and $\eta$ are available in~\cite{ATLAS-CONF-2014-032}; we used the electron efficiencies as a function of $p_T$ for $|\eta| < 2.47$, while muon efficiencies were taken to be 100\% as a good approximation.
The analysis is very well-documented and gives clearly the various preselection criteria and signal region cuts. Moreover, an effort was made in the definition of the tested new physics scenarios: a whole section of the experimental publication is dedicated to the description of the different SUSY scenarios. Furthermore, SLHA files were uploaded to {\tt HepData} in May 2014 after discussion with the ATLAS SUSY conveners.

For validation, at least one cut-flow table is given for every signal region and type of scenario tested, which is very good practice. In addition, several histograms are given and can be used to validate the distribution of, in particular, $E_T^{\rm miss,rel}$ and $m_{T2}$. Finally, regarding the interpretations in terms of simplified models, not only the information on the 95\%~CL upper bound on the visible cross section is given, but also the CL$_s$ value, which is useful for validation of the limit-setting procedure.
The only difficulty came from the benchmark points for direct slepton production. Given the SLHA files provided on {\tt HepData}, it was not clear whether the slepton masses given as $m_{\tilde\ell}$ in the cut-flow charts and histograms really correspond to the physical masses or to the slepton soft terms. The difference can be of several GeV, inducing some uncertainty in the kinematic distributions and in the production cross sections for these scenarios.

Event samples used for the validation were generated with {\tt Herwig++~2.5.2}~\cite{Bahr:2008pv}, using as input the SLHA files provided on {\tt HepData}. For each of the nine benchmark points we considered, $10^5$ events were generated. In the case of chargino-pair production, non-leptonic decays of the intermediate $W$-boson were filtered to increase statistics. Similarly, for chargino--neutralino production, non-leptonic decays of the intermediate $Z$-boson were filtered. The cross sections for the benchmark points, evaluated at the NLO+NLL accuracy~\cite{Fuks:2012qx,Fuks:2013vua,Fuks:2013lya},  were taken from the {\tt HepData} entry.

\begin{table}[!t]
\label{tab:cutflowWWaee_C1C1noslep1000}
\begin{center}
\begin{tabular}{l|c|c}
cut & ATLAS result & {\tt MA}5 result \\ 
\hline
Initial number of events & & $12301.5$  \\ 
2 OS leptons & & $1666.5$  \\ 
$m_{\ell\ell} > 20$~GeV & & $1637.5$  \\ 
$\tau$ veto & & $1637.5$  \\ 
$ee$ leptons & $402.1$ & $392.9$ \\ 
jet veto & $198.6$ & $257.0$  \\ 
$Z$ veto & $165.0$ & $215.9$ \\ 
$p_{T,\ell\ell} > 80$~GeV & $28.0$ & $35.3$  \\ 
$E_T^{\rm miss,rel} > 80$~GeV & $14.7$ & $18.9$  \\ 
$m_{\ell\ell} < 120$~GeV & $9.2$ & $10.1$  \\ 
\hline 
\end{tabular}
\caption{Cut flow for chargino-pair production in SR-$WW{\rm a}\,ee$ of ATLAS-SUSY-2013-11, 
for the benchmark point with $(m_{\tilde{\chi}^{\pm}_1}, m_{\tilde{\chi}^0_1})=(100,0)$~GeV.}
\end{center}
\vspace*{-5mm}
\end{table}

\begin{table}[!ht]
\label{tab:cutflowZjetsmumu_C1N235050}
\begin{center}
\begin{tabular}{l|c|c}
cut & ATLAS result & {\tt MA}5 result \\ 
\hline
Initial number of events & & $152.2$  \\ 
2 OS leptons & & $47.0$ \\ 
$m_{\ell\ell} > 20$~GeV & & $46.9$  \\ 
$\tau$ veto & & $46.9$  \\ 
$\mu\mu$ leptons & $16.4$ & $24.2$  \\
$\ge 2$ central light jets & $13.2$ & $15.5$  \\ 
$b$ and forward jet veto & $9.5$ & $12.5$  \\ 
$Z$ window & $9.1$ & $11.7$  \\ 
$p_{T,\ell\ell} > 80$~GeV & $8.0$ & $10.2$  \\ 
$E_T^{\rm miss,rel} > 80$~GeV & $5.1$ & $7.0$  \\
$0.3 < \Delta R_{\ell\ell} < 1.5$ & $4.2$ & $5.9$  \\ 
$50 < m_{jj} < 100$~GeV & $2.7$ & $3.6$  \\ 
$p_T(j_1,j_2) > 45$~GeV & $1.8$ & $1.7$  \\ 
\hline 
\end{tabular}
\caption{Cut flow for $\tilde{\chi}^{\pm}_1\tilde{\chi}^0_2$ associated production 
in SR-$Z{\rm jets}\,\mu\mu$ of ATLAS-SUSY-2013-11, 
for the benchmark point with $(m_{\tilde{\chi}^{\pm}_1}, m_{\tilde{\chi}^0_1})=(350,50)$~GeV.}
\end{center}
\end{table}

\begin{table}[!ht]
\label{tab:cutflowMT2120ee_slep25010}
\begin{center}
\begin{tabular}{l|c|c}
cut & ATLAS result & {\tt MA}5 result \\ 
\hline
Initial number of events & & $96.8$  \\ 
2 OS leptons & & $65.3$  \\ 
$m_{\ell\ell} > 20$~GeV & & $65.1$ \\ 
$\tau$ veto & & $65.1$  \\
$ee$ leptons & $51.2$ & $32.1$  \\ 
jet veto & $19.4$ & $17.5$  \\ 
$Z$ veto & $18.7$ & $16.9$  \\ 
$m_{T2} > 120$~GeV & $9.1$ & $8.2$  \\ 
\hline
\end{tabular}
\caption{Cut flow for slepton-pair production in SR-$m^{120}_{\rm T2}\,ee$ of ATLAS-SUSY-2013-11, 
for the benchmark point with $(m_{\tilde\ell}, m_{\tilde{\chi}^0_1})=(250,10)$~GeV.}
\end{center}
\end{table}

Tables~\ref{tab:cutflowWWaee_C1C1noslep1000}--\ref{tab:cutflowMT2120ee_slep25010} 
give some examples of cut flows for different benchmark points and signal regions, comparing 
the results obtained with our \madanalysis\ implementation to the official ATLAS numbers. 
(The complete list of cut flows for all nine benchmark points is available at~\cite{padwiki}.)
We systematically find the jet veto to be less efficient than it should be, but did not find
any explanation for this effect. This was also noted in Ref.~\cite{Drees:2013wra}.
Still, reasonably good agreement is observed for the available benchmark points.
Distributions of $E_T^{\rm miss,rel}$, $m_{\ell\ell}$ and $m_{T2}$ in some signal regions are shown in Figs.~\ref{fig:atlas-11-figA} and \ref{fig:atlas-11-figB}. Good agreement is observed. Note that the fluctuations in the ATLAS results in the left panel of Fig.~\ref{fig:atlas-11-figB} may correspond to statistical fluctuations and/or uncertainties when digitizing the ATLAS histogram (the results are extracted from a logarithmic scale that spans over six orders of magnitude).

We conclude that our \madanalysis\ implementation of ATLAS-SUSY-2013-11 reproduces well the experimental results. 
Our {\tt C++} code for this analysis is published as~\cite{ma5code:atlas-susy-2013-11}; complete validation materials including validation of the limit-setting procedure (see Section~\ref{ma5t-sec:limitsetting}) can be found at~\cite{padwiki}.
The reimplementation of this analysis can be used to constrain alternative scenarios with final states with two leptons and missing transverse energy. If the sneutrino is the LSP, see Section~\ref{sec:simpmod-sneutrino}, this final state can be obtained from chargino pair-production followed by $\tilde\chi^{\pm}_1 \to \ell\tilde\nu_{\ell}$. An estimation of the change in acceptance$\times$efficiency compared to slepton pair-production followed by $\tilde \ell \to \ell \tilde\chi^0_1$ is work in progress.

\clearpage

\chapter{Conclusions}

Current times are crucial for the future of high-energy physics. Run~II of the LHC may lead to the discovery of new particles at the TeV scale or deviations from SM expectations in the properties of known particles, opening great opportunities for a complete understanding of electroweak symmetry breaking. Indeed, the presence of new physics at or close to the electroweak scale is most naturally understood in relation with the hierarchy problem in the SM Higgs sector.
If, on the other hand, no BSM physics is found during the next running years of the LHC, it will progressively become clear that nature is somehow ``unnatural'' or that the hierarchy problem itself is a mirage after all (possibly accepting the fine-tuning of the Higgs mass or explaining it by the anthropic principle).
This would, unfortunately, be the main---and not even definitive---conclusion: among the alleged problems and known limitations of the SM, discussed in Section~\ref{sec:intro-bsm},
only the hierarchy problem  gives a strong physics case for new physics at a scale that is accessible at the LHC.
This latter ``nightmare'' scenario has became more and more present in high-energy physicists' minds after all negative results at Run~I of the LHC, and represents the least we can learn from the LHC.

It is, however, too early to give up on naturalness. Since supersymmetry is a prime candidate for solving the hierarchy problem, the status of supersymmetric models should be investigated in light of the LHC results. This is not a trivial task since the minimal supersymmetric extension of the Standard Model, the MSSM, has more than 100 parameters, while hundreds of potentially relevant searches have been performed by the ATLAS, CMS and LHCb collaborations at the LHC. The interpretation of the positive and negative results obtained at the LHC during Run~I and the interplay with other measurements (in particular related to dark matter) was the main focus of this PhD~thesis. This was done in the context of supersymmetric models, but not only: effective parametrizations of new physics, applicable to a wide class of new physics models, were also considered. Throughout this thesis, a particular effort was made in order that the phenomenological work---made for experiments and theory to meet---can be reused by the whole community and applied to other new physics scenarios. This has been leading, in particular, to the development of public tools.

Starting with the positive results at the LHC, clearly the most important piece of news was the discovery on July 4th, 2012 of a Higgs boson with mass of about 125~GeV, nearly fifty years after its theoretical prediction. This was the subject of Chapter~2. In addition to representing the ultimate triumph of the SM, it sheds light on the hierarchy problem and opens up new ways of probing new physics. Indeed, the various measurements, reviewed in Section~\ref{sec:higgs-measlhc}, performed at the LHC on the observed Higgs boson constrain its couplings to SM particles as well as invisible and undetected decays. This can be used to constrain BSM physics with a modified Higgs sector and/or new particles coupling to the Higgs, the latter possibility inducing modifications to the loop-induced processes (mainly gluon fusion and the decay into two photons) or the opening of new decay modes.

The results, given in terms of signal strengths $\mu = \sigma / \sigma_{\rm SM}$, have correlated systematic uncertainties which make it difficult to constrain models of new physics in a precise way from outside the experimental collaboration.
Fortunately, the ATLAS and CMS collaborations systematically present results in the $(\mu({\rm ggF+ttH}, Y), \mu({\rm VBF+VH}, Y))$ plane, where the five production modes of the SM are grouped into just two effective modes (gluon fusion $+$ associated production with top quarks, ${\rm ggF+ttH}$, and vector boson fusion $+$ associated production with a vector boson, ${\rm VBF+VH}$) and where $Y$ are the decay modes of the SM Higgs boson (currently $\gamma\gamma$, $ZZ^*$, $WW^*$, $b\bar b$, and $\tau\tau$). This information can be used directly to constrain a very wide class of new physics models, as was discussed in detail in Section~\ref{sec:higgs-npconstlhc}.

These signal strengths in the theory space were the primary experimental input used in all studies presented in Chapter~2. In Sections~\ref{sec:higgs2012} and~\ref{sec:higgs2013}, different scenarios with modified Higgs coupling strengths were tested. While an excess was observed in the diphoton channels in 2012, updated measurements using all data collected during Run~I pointed to an SM-like Higgs boson. This had severe consequences on the allowed ranges of the reduced couplings in the different scenarios. The impact was illustrated in models with an extended Higgs sector, and the possibility of invisible and undetected decays was evaluated.
Generic modifications to the couplings of the Higgs boson were also considered in the context of an effective field theory with dimension-6 operators, and confronted to LHC Higgs data in Section~\ref{sec:higgsdim6}. This unraveled interesting interplays between Higgs and electroweak measurements (via the Peskin--Takeuchi $S$ and $T$ parameters and the triple gauge couplings). Probabilities were derived for the coefficients of the higher-dimensional operators, and the possibilities for having large deviations from the SM after Run~I measurements were discussed.

In parallel to these studies focusing on generic new physics modifications to the properties of the observed Higgs boson, the impact of the LHC Higgs measurements on specific models of new physics was assessed, taking into account all other relevant experimental constraints (low-energy observables, searches for BSM particles and for dark matter). In Section~\ref{sec:pmssm}, the phenomenological MSSM, a 19-dimensional parametrization of the weak-scale Lagrangian of the MSSM, was considered. In this work, a Bayesian approach was taken and we investigated how the latest LHC results on the properties of the 125 GeV Higgs state impact the probability distributions of the pMSSM parameters, masses and other observables.
We found that the current LHC results on the Higgs boson have a significant impact on the posterior distributions of $\mu$ and $\tan\beta$ because of the SUSY radiative corrections to the bottom Yukawa coupling, which can be large for large $\tan\beta$.
During this thesis, the parameter spaces of the two-Higgs-Doublet Models of Type~I and~II were also investigated in light of the LHC Higgs results and other constraints in Ref.~\cite{Dumont:2014wha}.

In all these studies, a Higgs likelihood was defined to take into account the Higgs constraints. It was shown in Section~\ref{sec:higgs2013} that the constraints from ATLAS, CMS and the Tevatron can be combined and the resulting likelihood expressed with a simple $\chi^2$ formula, for which numerical values were given based on all experimental results up to the LHCP 2013 conference. This can directly be used as a good approximation to the full Higgs likelihood to constrain a large variety of new physics models. It is however possible to build a better approximation to the Higgs likelihood, in particular based on the full likelihood information (instead of one or two likelihood contours) in the 2D plane $(\mu({\rm ggF+ttH}, Y), \mu({\rm VBF+VH}, Y))$, which is already provided by the experimental collaborations for some final states. This is the goal of {\tt Lilith}, a tool for testing new physics models against the LHC Higgs data in a user-friendly way. A short presentation of {\tt Lilith} is given in Section~\ref{sec:lilith}; the program is intended for public release in September~2014.
Still, as Higgs measurements will become more and more precise new ways will have to be found for presenting the LHC Higgs results and using them to constrain new physics from outside the collaboration. Some thoughts on this matter are given in Section~\ref{sec:higgsfuture}.

Let us now turn to the negative results in the search for new physics at the LHC. While the Higgs boson has been found, no sign of BSM physics was observed at Run~I in spite of the large number of BSM searches performed by the ATLAS and CMS collaborations. Many of these searches are motivated by ($R$-parity conserving) supersymmetry, and target final states with transverse missing momentum. The implications of the negative results obtained in these searches, on the MSSM and possibly beyond, were the main focus of Chapter 3.
For any given search, implications on new physics cannot be covered by the ATLAS or CMS collaboration, thus results are primarily given in terms of simplified model scenarios in the experimental publication. These simplified models are defined with only a few particle masses as free parameters, assuming that all other BSM particles are absent or too heavy to contribute to the signal. It is therefore challenging to evaluate the impact of these searches on all other, more complicated, scenarios.

A first possibility for constraining more generically new physics from these negative results is to consider every BSM signal as the superposition of different topologies contributing to the signal. The most simple topologies define simplified models on which constraints are known, either directly as upper limits on the cross section or as acceptance$\times$efficiency maps.
These constitute two simplified model approaches, which were explained in Section~\ref{sec:simpmod-intro}. A comparison of these two ways of deriving limits on new physics from simplified models, and of the available tools built on these ideas, was given in Section~\ref{sec:simpmod-intro}. In Section~\ref{sec:simpmod-lightneutralino} and~\ref{sec:simpmod-sneutrino} we used one of these tools, {\tt SModelS}, to test scenarios with a supersymmetric dark matter candidate against the LHC SUSY searches. These BSM scenarios with a WIMP dark matter candidate are, in addition, challenged by the negative results in direct detection experiments.

First, the possibilities for having the neutralino as light as possible and a viable dark matter candidate within the general phenomenological MSSM were examined in Section~\ref{sec:simpmod-lightneutralino}. The Higgs constraints were also taken into account using the results from Section~\ref{sec:higgs2013}. We found that the upper bound on the relic density sets a lower bound on the neutralino mass of about 15 GeV. These very light neutralinos are a mixture of bino and higgsino, and come with rather light staus, but are put under strong pressure by the latest limits from direct direction experiments. LHC BSM searches were found to exclude already some parts of the parameter space, in particular from the searches for electroweak-ino pair production followed by stau-mediated decays.
Finally, we found that sizable deviations are possible in the properties of the Higgs boson, from the decay into two neutralinos or from the stau contribution to diphoton decays. These represent good prospects for probing these light neutralino scenarios during Run~II of the LHC.

Second, an alternative supersymmetric dark matter candidate, the mixed sneutrino, was studied in Section~\ref{sec:simpmod-sneutrino}. This was the first project of my PhD thesis; the results obtained in 2012 were updated with the most recent experimental results for this document. In this Bayesian study, we took special care to account for uncertainties arising from astrophysical parameters and from the quark contents of the nucleon, which have an impact on the results from direct detection experiments. We found that the discovery of a Higgs boson excludes a light sneutrino with mass below about 50~GeV, while heavier sneutrinos remain viable although a large portion of the parameter space is challenged by the negative results in the direct detection of dark matter. The limits from BSM searches at the LHC affect only marginally the considered scenarios, but light wino-like charginos may already be excluded by the results of analyses targeting slepton-pair production.

In the second part of Chapter~3, another approach for constraining new physics from the BSM LHC results was considered. Instead of relying on the decomposition into simplified models, the selection criteria used to discriminate signal from background in a given BSM search can be implemented and applied on event samples corresponding to the new physics scenario of interest. This is a more direct and powerful way of constraining new physics as it does not rely on the decomposition into simple topologies.
In turn, it is dramatically slower and disk space-intensive than the simplified model approach as event samples need to be generated and handled for each tested signal.
This approach was presented in detail in Section~\ref{sec:analysisreimplementation}. Re-implementing analyses is a tedious and difficult task because non-collaboration members do not have access to the experimental data, nor the Monte Carlo event set simulated with an official collaboration detector simulation. Instead, we simulate detector effects with the public tool {\tt Delphes}. This makes it necessary to validate the implementation of the analyses cuts with the information provided in the experimental publications.

While implementing searches for stops and sbottoms at the LHC in the \madanalysis\ framework, it became evident that new developments were needed to handle multiple signal regions in a single analysis code without unnecessary duplications. This motivated us working on a new version of the \madanalysis\ program, presented in Section~\ref{sec:ma5delphes3}. Moreover, we found it important that the work done for implementing the selection criteria and validating such a reimplementation be accessible to the whole community. This was the starting point of the public database of reimplemented analysis in the \madanalysis\ framework, presented at the end of Section~\ref{sec:analysisreimplementation}. In Sections~\ref{sec:cmsvalid} and~\ref{sec:atlasvalid}, the implementation and validation of two SUSY analyses done during this thesis, and integrated to the public database, was presented.
Hopefully, the growing number of validated reimplementations of LHC BSM analyses in the public database will be used to fully exploit the potential of these searches, and will also give useful feedback to the experiments on the impact of their searches.

The work done during this PhD thesis on the Higgs boson measurements on the one hand, and on the limits from BSM searches on the other hand, aimed at understanding where we stand with new physics at the TeV scale. It is, hopefully, useful in (developing tools for) assessing the impact of Run~I of the LHC in a number of scenarios of new physics, in particular but not only on supersymmetric models. But the main, fundamental question remains open: does nature care about naturalness? If it is the case, a new golden age of particle physics might come at Run~II of the LHC. If not, the LHC will only probe one order of magnitude in the scale of new physics for typical scenarios, which is tiny: there are still sixteen orders of magnitude to go between the TeV and the Planck scale. Let us hope nature will have a surprise for us in between, of the kind that can be discovered in our lifetimes.

\clearpage

\backmatter
\bibliography{thesis}

\clearpage

\invisiblesection{Abstract / R\'esum\'e}

Two major problems call for an extension of the Standard Model (SM): the hierarchy problem in the Higgs sector and the dark matter in the Universe.
The discovery of a Higgs boson with mass of about 125~GeV was clearly the most significant piece of news from CERN's Large Hadron Collider (LHC). In addition to representing the ultimate triumph of the SM, it shed new light on the hierarchy problem and opened up new ways of probing new physics. The various measurements performed at Run~I of the LHC constrain the Higgs couplings to SM particles as well as invisible and undetected decays.
In this thesis, the impact of the LHC Higgs results on various new physics scenarios is assessed, carefully taking into account uncertainties and correlations between them. Generic modifications of the Higgs coupling strengths, possibly arising from extended Higgs sectors or higher-dimensional operators, are considered. Furthermore, specific new physics models are tested. This includes, in particular, the phenomenological Minimal Supersymmetric Standard Model.
While a Higgs boson has been found, no sign of beyond the SM physics was observed at Run~I of the LHC in spite of the large number of searches performed by the ATLAS and CMS collaborations. The implications of the negative results obtained in these searches constitute another important part of this thesis. First, supersymmetric models with a dark matter candidate are investigated in light of the negative searches for supersymmetry at the LHC using a so-called ``simplified model'' approach. Second, tools using simulated events to constrain any new physics scenario from the LHC results are presented. Moreover, during this thesis the selection criteria of several beyond the SM analyses have been reimplemented in the {\tt MadAnalysis~5} framework and made available in a public database.

Deux probl\`emes majeurs requi\`erent une extension du Mod\`ele Standard (MS) : le probl\`eme de hi\'erarchie dans le secteur de Higgs, et la mati\`ere noire de notre Univers.
La d\'ecouverte d'un boson de Higgs avec une masse d'environ 125~GeV est clairement l'\'ev\'enement majeur en provenance du Large Hadron Collider (LHC) du CERN. Cela repr\'esente le triomphe d\'efinitif du MS, mais cela met \'egalement en lumi\`ere le probl\`eme de hi\'erarchie et ouvre de nouvelles voies pour sonder la nouvelle physique. Les diff\'erentes mesures effectu\'ees pendant le run~I du LHC contraignent les couplages du Higgs aux particules du MS ainsi que les d\'esint\'egrations invisibles et non-d\'etect\'ees.
Dans cette th\`ese, l'impact des r\'esultats sur le boson de Higgs au LHC est \'etudi\'e dans le cadre de diff\'erents mod\`eles de nouvelle physique, en prenant soigneusement en compte les incertitudes et leurs corr\'elations. Des modifications g\'en\'eriques \`a la force des couplages du Higgs (pouvant provenir de secteurs de Higgs \'etendus ou d'op\'erateurs de dimension sup\'erieure) sont \'etudi\'ees. De plus, des mod\`eles de nouvelle physique sp\'ecifiques sont test\'es, notamment, mais pas seulement, le Mod\`ele Standard Supersym\'etrique Minimal ph\'enom\'enologique.
Alors qu'un boson de Higgs a \'et\'e trouv\'e, il n'y a toutefois nulle trace de physique au-del\`a du MS au run~I du LHC en d\'epit du grand nombre de recherches effectu\'ees par les collaborations ATLAS et CMS. Les cons\'equences des r\'esultats n\'egatifs obtenus lors de ces recherches constituent un autre volet important de cette th\`ese. Tout d'abord, des mod\`eles supersym\'etriques avec un candidat \`a la mati\`ere noire sont \'etudi\'es \`a la lumi\`ere des r\'esultats n\'egatifs dans les recherches de supersym\'etrie au LHC, en utilisant une approche bas\'ee sur les \guillemotleft~mod\`eles simplifi\'es~\guillemotright. Ensuite, des outils pour contraindre un mod\`ele de nouvelle physique quelconque \`a partir des r\'esultats du LHC et d'\'ev\'enements simul\'es sont pr\'esent\'es. De plus, au cours de cette th\`ese, les crit\`eres de s\'election de plusieurs analyses au-del\`a du MS ont \'et\'e r\'eimpl\'ement\'es dans le cadre de {\tt MadAnalysis~5} et ont \'et\'e int\'egr\'es \`a une base de donn\'ees publique.

\end{document}